\documentclass[a4paper,fleqn]{mnras}

\usepackage[T1]{fontenc}
\usepackage{ae,aecompl}

\usepackage{graphicx}	
\usepackage{amsmath}	
\usepackage{amssymb}	
\graphicspath{ {./} {R2/} }

\title[Catalog of magnetic phase curves of stars]{Catalog of averaged magnetic
phase curves of stars: the second edition.}

\author[V.D. Bychkov et al.]{
V.D. Bychkov$^{1}$,\thanks{E-mail: vbych@sao.ru (VB)}
L.V. Bychkova$^{1}$, J. Madej$^{2}$ \\
$^{1}$Special Astrophysical Observatory, Russian Academy of Sciences, Nizhnii Arkhyz, 369167 Russia\\
$^{2}$Astronomical Observatory, University of Warsaw, Al. Ujazdowskie 4, 00-478 Warszawa, Poland \\
}

\date{Accepted XXX. Received YYY; in original form ZZZ}
\pubyear{2020}

\begin{document}
\label{firstpage}
\pagerange{\pageref{firstpage}--\pageref{lastpage}}
\maketitle

\begin{abstract}
Magnetized stars exhibit periodic variations of their longitudinal 
global magnetic fields, $B_e$, owing to rotation. Here, we present  
the second catalog of averaged stellar magnetic rotational phase curves
and their parameters derived from a compilation of the published
observational data and personal communications for 350 stars of various 
spectral types, which were published up to the end of December 2019. 
Magnetic Ap and Bp stars constitute the most numerous 
subset in the catalog (215 objects). Phase curves were obtained by 
fitting either a sinusoid or a double sine wave to series of the observed 
$B_e$ measurements using the least squares method. For some stars, we present
magnetic phase curves derived from time series of the surface magnetic 
field, $B_s$, or obtained improved values of the rotational period, $P_{\rm rot}$.
We have also identified eight stars in our catalog that host planets or planetary
systems. 
\end{abstract}


\section{Introduction}

The global magnetic fields of stars are typically studied by measuring the
Zeeman splitting of spectral lines in circularly polarized light. The resulting
observable quantity is the longitudinal (effective) magnetic field, $B_e$, 
which usually exhibits periodic variations caused by the star's rotation.
Apparent magnetic variations of this type were analyzed by Stibbs (1950),
who studied the effect of the rotation of the dipole magnetic field on the
displacement of the Zeeman components in a spectral line.

Longitudinal magnetic field strength, $B_e$, is a projection of the
local vector of the surface stellar magnetic field on the line of sight
integrated over the visible stellar disc, see Appendix in Stibbs (1950).
Periodic variations of the $B_e$ are caused by the star's rotation since,
commonly the field configuration is not symmetrical about the 
axis of rotation (as in the inclined dipole model). Period of the apparent
magnetic $B_e$ variations is then equal to the rotational period.

Periodic variability of the longitudinal magnetic field correlated with the
rotation is a purely geometrical effect and it commonly occurs when the true
surface magnetic field is intrinsically constant during the rotational period.
Intrinsic variations of the surface magnetic field (owing to
solar-like cycles) are not considered in this work. For example, we
neglect physical changes to the global magnetic fields of
M dwarfs, which proceed over a timescale of 1 year, while presenting $B_e$ periodic
variations, which proceed within the time scale of the rotation
(several days).

Another observable scalar magnetic quantity is the surface magnetic field,
$B_s$, which can be measured from the separation of the Zeeman-split
line components seen in
natural light in high-dispersion spectra of stars with magnetically
resolved line components (Preston 1971).

Both observables, $B_e$ and $B_s$, were defined and applied to observations
of spectral lines in early-type stars over 60 years ago, when the available
instruments and observational techniques were much simpler or less advanced
than now. From those years on, the growing interest toward stellar magnetism
caused the inflow of new magnetic observational data for stars of various
spectral types, which also was possible by use of new high resolution 
spectropolarimeters which were very productive in this area.

\subsection{ New magnetic measurements }

Table 1 lists the principal types of stars included in the catalog 
of MPCs.

\begin{table}
\label{table:classes}
\caption{ Number of stars in various classes with known magnetic phase
   curves }
\vspace{0mm}
\begin{tabular}{lrlr}
\hline
\hspace{-2mm}Ap/Bp                      & 215 &  Stars hosting planets       &  8  \\
\hspace{-2mm}Var. $\beta$ Cep type      &  17 &  Normal chem. comp.stars     &  5  \\
\hspace{-2mm}Slowly Pulsating B stars   &   9 &  Be stars                    &  9  \\
\hspace{-2mm}High Proper Motion stars   &  11 &  Var. $\delta$ Sct type      &  2  \\
\hspace{-2mm}Var. $\delta$ Cep type     &   1 &  Semi-regular var.pulsating  &  2  \\
\hspace{-2mm}Multiple stars             &  14 &  Flare stars                 & 14  \\
\hspace{-2mm}Pulsating stars            &   7 &  Ae/Be Herbig stars          & 10  \\
\hspace{-2mm}Var. BY Dra                &   8 &  T Tau stars                 &  6  \\
\hspace{-2mm}Var. Ori type              &   3 &  Pre-main sequence           &  3  \\
\hspace{-2mm}Rotationally var.stars     &   9 &  EB Algol type               &  1  \\
\hline                    
\end{tabular}
\end{table}

Table 2 lists the maximum values of the half-amplitudes, $B_1$ and $B_2$, of
the longitudinal field variability for all subclasses of stars discussed
in this paper. 

\begin{table*}
\label{table:classes}
\caption{ Maximum values of the half-amplitudes $B_1$ and $B_2$ for various
   types of stars }
\vspace{2mm}
\begin{tabular}{|l|r|r|r|r|r|}
\hline
Type of star & n stars with & $B_1 \, {\rm max}$ & n stars with & $B_1
     \, {\rm max}$ & $B_2 \, {\rm max}$ \\
  &simple wave& G\,\,\, & double sinusoid & G\,\,\, & G\,\,\, \\
\hline
Ap/Bp                       & 160 &  4300 &  58 & 5450 &  1550 \\   %
Var. $\beta$ Cep type       &  15 &   500 &   2 &  450 &   220 \\   %
Slowly Pulsating B stars    &   6 &   350 &   3 &  260 &    80 \\   %
High Proper Motion stars    &   6 &    85 &   5 &   70 &    45 \\   %
Var. $\delta$ Cep type      &   1 &    80 &     &      &       \\   %
Multiple stars              &  11 &  1050 &   3 & 5000 &   420 \\   %
Pulsating stars             &   4 &   550 &   2 &  260 &    60 \\   %
Var. BY Dra                 &   4 &    20 &   5 &   25 &    15 \\   %
Var. Ori type               &   4 &  1100 &     &      &       \\   %
Rotationally var.star       &   6 &  3050 &   5 & 3700 &   700 \\   %
Stars hosting planets       &   7 &    10 &   2 &    3 &     2 \\   %
Normal chemical comp.stars  &   5 &  1350 &   2 &  940 &   470 \\   %
Be stars                    &   5 &   400 &   1 &  620 &   300 \\   %
Var. $\delta$ Sct type      &   2 &  3000 &     &      &       \\   %
Semi-regular var.pulsating  &   2 &     7 &   1 &    2 &     1 \\   %
Flare stars                 &  10 &   400 &   5 &  700 &   150 \\   %
Ae/Be Herbig stars          &   9 &  1200 &   1 &  620 &   300 \\   %
T Tau stars                 &   3 &   310 &   4 &  350 &    80 \\   %
Pre-main sequence           &   3 &   600 &   1 &   20 &    11 \\   %
EB Algol type               &     &       &   1 &   35 &    10 \\   %
\hline                                                              
\end{tabular}
\end{table*}

The situation with magnetic measurements for the last 15 years has changed 
significantly primarily due to the increased accuracy and number of 
measurements. So in the period from 2005 to 2019 more than 30,000 magnetic 
field (MF) estimates were received. The greatest number of new high-precision 
measurements were obtained in the last few years on the following
telescopes:

\begin{itemize}
\item[1.] ESPaDOnS CFHT (Canada-France-Hawaii Telescope 3.6m) ~$\sim$ 38.5\%
\item[2.] FORS1 / 2 (ESO, 8m, telescopes UT1, UT2, UT3) ~$\sim$ 23.1\%
\item[3.] NARVAL TBL (Telescope Bernard Lyot, 2m) ~$\sim$ 17.7\%
\item[4.] HARPS (ESO, 3.6m) ~$\sim$ 7.4\%,
\item[5.] MMS or NES (SAO RAS, 6m) ~$\sim$ 6.3\%,
\item[6.] MuSiCoS (TBL 2m, Isaac Newton Telescope 2.5m, South African
    Astronomical Observatory 1.9m) ~$\sim$ 3.1\%.
\end{itemize}

The remaining measurements were obtained on a number of other instruments, 
but the contribution of each did not exceed one percent.
Note, that the FORS1/2 telescope is mainly intended for viewing. Viewing tasks involve 
obtaining estimates for weak objects or observing an object with high temporal 
resolution. In some cases, the telescope can obtain MF estimates with high accuracy. 

      Naturally, this displays only the current state.

Thanks to these measurements it was possible to specify already known magnetic 
phase curves (MPC) and make new observations too. Therefore, out of 136 stars of the 
first catalog, only 23 MPCs remained unchanged. The remaining 113 were supplemented 
by new measurements, and therefore new parameters were obtained for there MF 
variability. MF estimates obtained with low accuracy (mostly ``photographic'') 
were often discarded and only highly accurate estimates obtained by modern methods used.

  For example, for building an MPC in the first edition of the catalog relatively
low-resolution ``photographic'' MF estimates amounted to 54\%, and for the MPC of 
the second catalog, this share is already only 6.6\%.

Table 1 provides data on the number of stars each type in the first and second 
versions of the catalog. As can be seen from Table 1, in the second directory shows 
information about MPC for Flare stars, Multiple stars, Hosting planets stars, 
Be stars and a number of other star types that were not in the first directory. 
For some types of stars, the number of MPC was noticeable and increased, allowing 
you to arrive at more confident conclusions about the variability of magnetic 
fields in these stars.

The number of MPCs for CP stars increased from 127 up to 215 in the new catalog. 
However, most importantly, the higher accuracy allowed us to identify more 
deviations from purely harmonic phase dependencies. The percentage of 
multifunctional compounds with double curves increased from the first catalog 
with 18 out of 127 (14\%), and in the second catalog with 62 out of 215 (29\%). 
It is very important for the study of the origin and evolution of magnetic field 
stars. 

In the first directory, we paid attention to a noticeable deviation of 
the $B_0$ coefficient for stars with a double wave (Fig. 5, the average $B_0 = -473 \pm
296$ G)  which indicated a possible explanation for this - small statistics - on
18 Ap objects. This assumption confirmed. The second directory contains 49 Ap 
stars $B_0$ coefficient distribution practically not shifted (Fig. 5, 
the average $B_0 = -250 \pm 206$ G). Note that in the second directory unchanged MPC 
(and their parameters) remain only for 23 stars (7\%) from the first catalog. 

Both the high number of new, precise MF measurements and recent
discovery of the global magnetic field in stars of various types
motivated us to compile and submit the entire second catalog and notin 
addition to the first. The second edition of larger size also represents
an expansion into previously unexplored classes of stellar magnetic activity.

\subsection{ The second catalog }

The first edition of this catalog (Bychkov et al. 2005) presented magnetic
phase curves, $B_e(\phi)$, for 136 stars, most of which were Ap/Bp or related
stars in the Main Sequence with strong surface fields. Recently, 
application of the newest instrumentation and data reduction procedures
allowed one for the detection of the global magnetic fields and the partial investigation of
their behaviors among stars of many other spectral classes.
 
In this work, we present and analyze results of the longitudinal magnetic field
measurements, $B_e$, and data on the surface field, $B_s$, for 350 stars.
Data were obtained from various bibliographic sources and originally were
measured using both old photographic plates and the newest high resolution spectrographs.
We appended data compiled
from personal communications, as well as our own determinations obtained 
at the Special Astrophysical Observatory (Russian Academy of Sciences)
and magnetic measurements of stars in the Orion OB1 Association
(Borra 1981) by permission of Dr. E.F. Borra.

In this catalog Ap and Bp stars still
are the most well-investigated group of stars; here, the magnetic
phase curves (MPCs) for 215 objects of this type are collected.
We did not include here neither white dwarf stars nor other degenerate
objects and selected only stars built up of normal matter.

The catalog of MPCs in the present version
can be applied to review and compare the magnetic behaviors of a wide 
range of types. The MPCs and their parameters are presented here 
in a homogeneous graphical and numerical forms, respectively.

\section{Averaged magnetic rotational phase curves}

Periodic variations of the longitudinal (effective) or surface magnetic 
field, $B_e$ or $B_s$, with the rotational phase, $\phi$, were aproximated
here by the two lowest terms in the Fourier series expansion.

{\bf 1.} For all stars with an adequate number of $B_e$ determinations and
for which the period of magnetic variability, $P_{mag}$, is known, we
determined the best fit for $B_e$ vs. phase:
\begin{equation}
B_{ei} (\phi) = B_0 + B_1 \cos (\phi - \pi ) \, ,
\label{equ:harm1}
\end{equation}
where
\begin{equation}
\phi = 2\pi \, \left( {{T_i - T_0} \over P} \right) \, ,
\end{equation}
using the least squares method.
Here, $B_0$ is the average field and $B_1$ equals half the amplitude,
$T_i$ is the time at which the measurement was taken, $P$ denotes the period, and $T_0$ is the 
time corresponding to the zero phase, $\phi$. Half-amplitudes $B_1$ in
Eq. (1) are positive numbers by the assumption. We selected the zero epoch, 
$T_0$, in such a way that the phase $\phi=0$ corresponds
to the minimum of the best fit MPC.

{\bf 2.}
In cases where the shape of the magnetic phases curve is more complex than
a simple cosine, we included the second harmonic wave:
\begin{equation}
B_{ei} (\phi) = B_0 + B_1 \cos (\phi+z_1) + B_2 \cos ({2\phi}+z_2) \, ,
\label{equ:harm2}
\end{equation}
where phases $z_1$ and $z_2$ are also parameters of fitting.
Half-amplitudes $B_2$ can be either positive or negative.

{\bf 3.}
Some stars have magnetic variations of unknown periods. In some cases,
if it was possible to compile a sufficient number of momentary $B_e$ 
measurements from the available papers, then we also attempted to 
search for the magnetic period, $P$, using standard methods. We are aware that
the periods determined in those cases require additional $B_e$ measurements, or
new spectral, photometric, or polarimetric observations in order to confirm
our magnetic periods and to reduce their errors.

{\bf 4.}
There are also exceptional stars, which exhibit still more complex magnetic
phase curves, and in such cases Eq. (6) does not provide an adequate fit.
A very good example is the star HD 37776, which exhibits the MPC
of exceptional complexity. For this star we present various forms of
fitting curves and regard it as a unique case.

{\bf 5.}
Average MPCs in the catalog were obtained by the 
least squares fitting of a sine wave or a double wave to the observed 
$B_e$ points for each star. This is the same method as that used in 
the first version of the catalog (Bychkov et al. 2005). In general,
this catalog was built following criteria very similar to those adopted
for its first version.

Tables A1--A5 present the parameters of the MPCs, 
and also the parameter $r$, which was defined by Stibbs (1950).
Parameter $r$ relates both the angle $\beta$ between the magnetic dipole axis
and the rotational axis, and the angle $i$ between the rotational axis and
the line of sight:
\begin{equation}
r = {{\cos\beta \cos i - \sin\beta \sin i} \over
     {\cos\beta \cos i + \sin\beta \sin i}} 
  = {B_e (\min) \over {B_e (\max)}}  \, .
\end{equation}

\section{Discussion: stars in our catalog}

The following subsections detail the statistical properties of various 
parameters in the catalog: spectral types, rotational
periods and fitting coefficients $B_0$, $B_1$, and $B_2$.

\subsection{ Distribution of spectral and variability types  }

\begin{figure}
\resizebox{0.98\hsize}{!}{\includegraphics{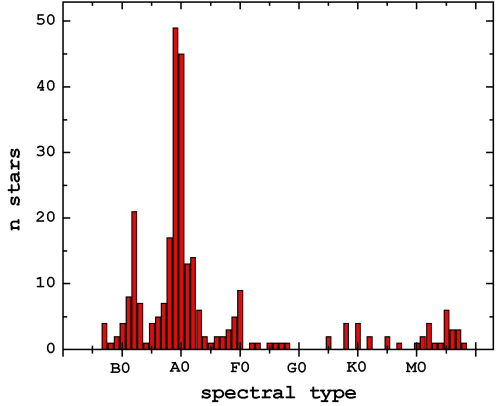}}
\caption{ Distribution of stars with known $B_e(\phi)$ phase curves
   by spectral types. }
\label{fig:spc}
\end{figure}

Fig.~\ref{fig:spc} shows the distribution of stars in our catalog by
spectral types. Magnetic rotational phase curves, $B_e(\phi)$, are
determined mostly for stars of early spectral types. Fig.~\ref{fig:spc}
shows that the highest concentration of stars is located around the 
spectral type A. 

Therefore, the present knowledge of the global magnetic fields across
the Hertzsprung-Russell diagram is still poor and selective, and the
completeness of this catalog is weak.

\subsection{ Ap/Bp stars }

Ap stars are the most well investigated and numerous subclass in
this catalog: 215 stars. This subclass spans a very wide range
from late O to early F spectral types.

There are 153 Ap stars with sinusoid phase curves (71 per cent)
and 62 stars with double wave MPC (29 per cent). The proportion of stars with
complex MPCs increased with respect to the first edition (Bychkov
et al. 2005), which reflects the increased accuracy of the new measurements. 

\subsubsection{ Distribution of periods }

\begin{figure}
\resizebox{\hsize}{\hsize}{\rotatebox{0}{\includegraphics{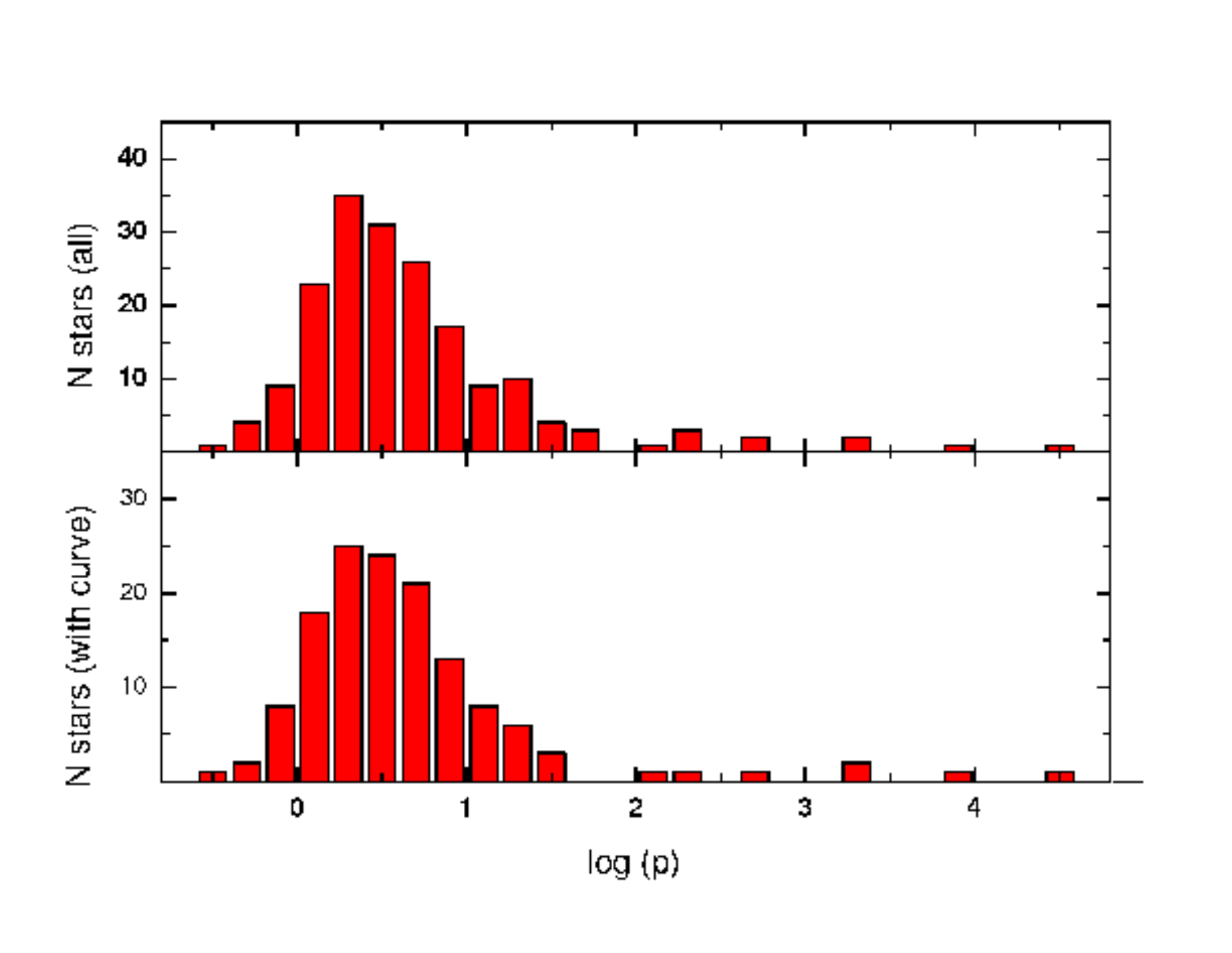}}}
\caption[]{ Distribution of Ap/Bp stars vs. decimal logarithm of period 
(in days), $\log P_{\rm mag}$, for all stars with known
periods (upper panel) and for stars with known phase curves (lower panel). }
\label{fig:plog}
\end{figure}

Fig.~\ref{fig:plog} presents the distribution of stars in our catalog vs.
the decimal logarithm of the period, $P_{\rm mag}$, separately for all CP stars with known
periods (upper panel) and for stars whose period enables creation of a
rotational phase curves (lower panel). The width of a single bin equals 0.3 dex.

The majority of Ap/Bp stars exhibit periods, $P_{\rm mag}$, between 1 and 10 days.
The same is true for stars with known phase curves.

\subsubsection{ Coefficients $B_0$, $B_1$, and $B_2$ }

\begin{figure}
\resizebox{\hsize}{0.7\hsize}{\rotatebox{0}{\includegraphics{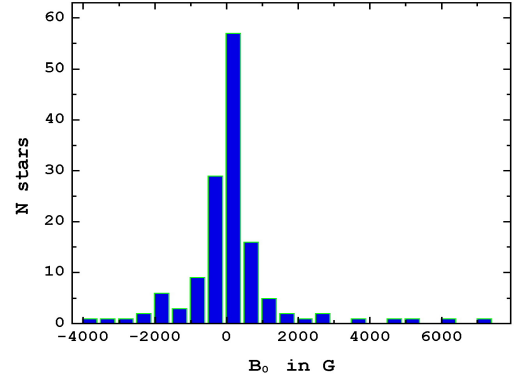}}}
\caption[]{Number distribution of stars vs. the coefficient $B_0$
for the subset of stars in which the phase curve $B_e(\phi)$ was approximated by
a sine wave. }
\label{fig:b01}
\end{figure}

Fig.~\ref{fig:b01} shows the number distribution vs. coefficient $B_0$ for
the catalogued stars whose rotational phase curves, $B_e(\phi)$,
were approximated by sine waves. Fig.~\ref{fig:b01} does not include
HD 215441, whose exceptionally strong $B_e$ field exceeds the
scale of the figure. The average value of $B_0$ is $+125\pm 116$ G.
For stars with double wave MPCs the average value of $B_0$ is $-250 \pm 206$ G,
We conclude, that the average longitudinal magnetic field of the MPCs
in the catalog, $B_0$, equals zero and does not show meaningful bias
towards neither positive nor negative values.

\begin{figure}
\resizebox{\hsize}{0.7\hsize}{\rotatebox{0}{\includegraphics{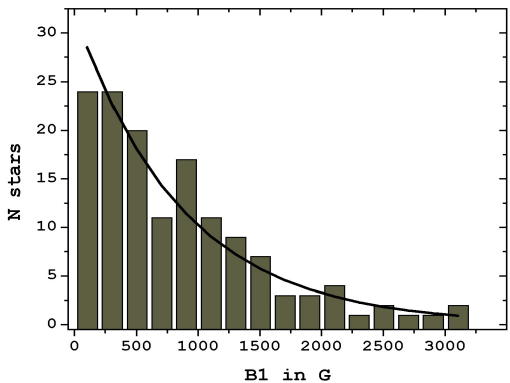}}}
\caption{\small{ Number distribution of stars vs. the coefficient
$B_1$ (half-amplitude of $B _e$ variability). Only
stars with the sine wave phase curves were included here.  }}
\label{fig:b111}
\end{figure}

Fig.~\ref{fig:b111} displays an analogous number distribution vs.
coefficient $B_1$ (half-amplitude of the $B_e$ variation) for  
stars with the sine wave rotational phase curves, $B_e(\phi)$.
Half-amplitudes, $B_1$, usually do not exceed 3000 G. Number of 
stars in a single bin can be approximated by the exponential function
\begin{equation}
n \approx 126.449 \times \exp(0.001448 B_1) \, .
\end{equation}

Note, that the number of stars with the lowest amplitude of $B_e$ variations
is underestimated. The accuracy of magnetic measurements is still limited
and many stars with low magnetic variability were not detected as variable
stars and therefore were not included in our catalog. 

\begin{figure}
\resizebox{\hsize}{0.7\hsize}{\rotatebox{0}{\includegraphics{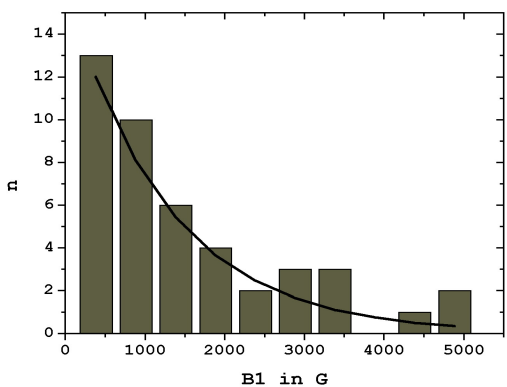}}}
\caption{\small{ Number distribution of stars vs. coeffcient $B_1$ for
those stars that exhibit the magnetic rotational phase curve with the double
wave.  }}
\label{fig:b120}
\end{figure}

Fig.~\ref{fig:b120} displays an analogous number distribution vs.
coefficient $B_1$ (half amplitude of the $B_e$ variation) for the 
catalogued stars with double-wave phase curves, $B_e(\phi)$.
Half-amplitudes $B_1$ do not usually exceed 5000 G. The
occurence of stars in a bin can be approximated by the exponential function
\begin{equation}
n \approx 54.652 \times \exp(0.000742 B_1) \, .
\end{equation}

For stars with the double wave phase curve
half-amplitudes $B_2$ usually are lower than $B_1$ and do not exceed 1200 G.
Expected number of stars \bf in a bin also \rm can be approximated by the exponential function
\begin{equation}
n \approx 109.957 \times \exp(0.003044 \,\,\, \rm{abs}(B_2)) \, .
\end{equation}


\section{ Measurements for stars in Orion OB1 association }

In 1994 Dr. E.F. Borra kindly provided us original 
estimates of the longitudinal magnetic fields of 13
stars from the Orion OB1 association which were analyzed in his 
1981 publication. Measurements
were obtained using the Laval University
Pockels cell polarimeter, which is similar to the instrument
described by Borra and Landstreet (1980).
The author used is the Las Campanas 2.5-m du Pont telescope.
The observational techniques and reduction procedures are
the same as those reported by Borra and Landstreet (1980).
The results of these measurements are presented in Table 6 
by permission of Dr. E.F. Borra.
We used these estimates in both the first catalog (Bychkov et al.
2005) and this second catalog  
(Borra 1994, private communication).

\section{Summary}

We have compiled a catalog of magnetic rotational phase curves,
$B_e(\phi)$, for 350 stars that exhibit periodic variations of the
effective magnetic field, $B_e$. Most of the catalogued objects (215 stars)
are chemically peculiar A and B type stars. The catalog consists of figures
that display individual $B_e$ measurements, error bars, and phase curves,
$B_e(\phi)$, approximated either by a sine wave or a double wave.

The catalog also presents a listing of the following parameters of 
MPCs: coefficients $B_0$, $B_1$, and $B_2$ of the harmonic
expansion of $B_e(\phi)$, the magnetic period, $P_{mag}$ (in days),
which is usually equal to the rotational period, $P_{rot}$, the Julian Day
of the zero phase, $T_0$, and the coefficient $r$, defined by Stibbs (1950).

There are a number of magnetic stars with
strong effective magnetic field, $B_e$, and very low or zero
variation. Such stars were not included in our catalog, partly because
it was impossible to determine the magnetic period, $P_{\rm mag}$, and phases
of individual $B_e$ measurements.

This paper is based on observational data published or posted in
astronomical databases up to the end of December 2019. 

\section*{Acknowledgements}
We are grateful to Stefano Bagnulo for his criticism
and numerous suggestions improving our paper.
Our research is based on data compiled and posted in the SIMBAD, ADS,
and CDS databases.



\appendix

\section{Our tables}

This section presents set of tables of our catalog in the
following order:

\begin{description}
\item {\bf A1.} List of stars in the catalog.
\item {\bf A2.} Stars with single wave $B_e$ phase curves: parameters
   of phase curves. 
\item {\bf A3.} Stars with single wave $B_e$ phase curves: parameters
   of the magnetic dipole. 
\item {\bf A4.} Stars with double wave $B_e$ phase curves: part 1. 
\item {\bf A5.} Stars with double wave $B_e$ phase curves: part 2. 
\item {\bf A6.} Measurements for 13 stars in Orion OB1 by EF Borra (unpublished).
\end{description}

Note, that some entries in Table A3 and A5 present either huge or
unrealistically low values of the reduced chi-square for fits to
the actual time series. Such a chi-square can imply low quality of the 
corresponding magnetic phase curve due to scattered data points
or overestimated errors. On the other hand,
one can not exclude that in some stars the global magnetic
field exhibited intrinsic changes during the observing time period.
In such cases parameters given in the Tables are not of any use.

\begin{table*}
	\caption{List of stars in the catalog. }
	\label{tab:list}
\begin{tabular}{rllll}
        \hline
     HD  &      Name           &Sp. type & comments                   & mv    \\
        \hline
     108  &HIP505               &O7pe      &   multiple star              &  7.382   \\
     358  &HR15        Alph.And &B9pHgMn   &     SB                       &  2.1     \\
     886  &HR39         Gam Peg &B2IV      &   var.beta Cep type          &  2.83    \\
     965  &HIP 1127             &A8pSr     &                              &  8.624   \\
    1237  &HIP1292              &G8V       &   HPMS host Planet           &  6.578   \\
    2453  &BD+31 59      GR And &A1SrEuCr  &                              &  6.9     \\
    3229  &HR143   14 Ceti      &F5V       &                              &  5.949   \\
\hline
     \end{tabular}
\end{table*}

\begin{table*}
	\contcaption{List of stars in the catalog. }
	\label{tab:list}
	\begin{tabular}{rllll}
        \hline
     HD  &      Name           &Sp. type & comments                   & mv    \\
        \hline
   34736  &BD-07 1036           &B9 Si  &      double system              &  7.82    \\
   34859  &HIP24922 Renson8900  &Ap Si  &                                 &  9.14    \\
   35177  &HIP25163 Renson 8980 &B9Vp Si &                                &  8.13    \\
   35298  &PPM148877  V1156 Ori &B6V He-w &  Rotationally var.star        &  7.91    \\
   35456  &PPM188029            &B6Vp He-w &    multiple star             &  6.94    \\
   35502  &PPM188035            &B6V Si He-w &  double system             &  7.331   \\
   35881  &BD+00 1081           &B8V   &                                  &  7.77    \\
\hline
     \end{tabular}
\end{table*}

\begin{table*}
	\contcaption{List of stars in the catalog. }
	\label{tab:list}
	\begin{tabular}{rllll}
        \hline
     HD  &      Name           &Sp. type & comments                   & mv    \\
        \hline
   66665  &HIP39540             &B0.5V& normal chem.comp.                 &  7.81    \\
   66765  &HIP39371             &B1/B2V&                                  &  6.622   \\
   67621  &HIP39691             &B2IV   &                                 &  6.316   \\
   68351  &HR3215        BM Cnc &B9pSiCr&                                 &  5.61    \\
   70331  &CD-47 3803           &B8pSi  &                                 &  8.898   \\
   71866  &PPM50637      TZ Lyn &A0pSiSrEu&                               &  6.72    \\
   72106  &PPM285296            &A0IV & B comp.-Ae/Be Herbig              &  8.58    \\
\hline
     \end{tabular}
\end{table*}

\begin{table*}
	\contcaption{List of stars in the catalog. }
	\label{tab:list}
	\begin{tabular}{rllll}
        \hline
     HD  &      Name           &Sp. type & comments                   & mv    \\
        \hline
  125823  &HR5378      V761 Cen &B7IIIpv Si He-w &                        &  4.40    \\
  126515  &BD+1 2927     FF Vir &A2pCrSr  &                               &  7.094   \\
  127381  &HR5425       sig Lup &B1/2V &Early B type Eclips.var.star      &  4.416   \\
  128898  &HR5463     Alpha Cir &A9pSrCrEu  &                             &  3.2     \\
  129333  &BD+64 1017  EK Dra   &G1.5V    &  var.BY Dra type              &  7.61    \\
  130144  &HR5512        EK Boo &M6III  & semi-reguliar var.pulsating     &  5.88    \\
  130322  &BD+00 3243  HIP72339 &K0V  & HPMS host Planet                  &  8.04    \\
\hline
     \end{tabular}
\end{table*}

\begin{table*}
	\contcaption{List of stars in the catalog. }
	\label{tab:list}
	\begin{tabular}{rllll}
        \hline
     HD  &      Name           &Sp. type & comments                   & mv    \\
        \hline
  181615  &HR7342       ups Sgr &B8pl &   Be Star                         &  4.578   \\
  182180  &HR7355               &B2Vn &He-w                               &  6.023   \\
  183056  &HR7395         4 Cyg &B9pSi&                                   &  5.12    \\
  183339  &HR7401               &B8IV &Si He-weak                         &  6.58    \\
  184471  &BD+32 3471           &A9 SrCrEu &                              &  8.99    \\
  184927  &PPM83182   V1671 Cyg &B2V  He-r&var.                           &  7.44    \\
  186205  &PPM168470            &B5  He-r &                               &  8.5     \\
\hline
     \end{tabular}
\end{table*}

\begin{table*}
	\contcaption{List of stars in the catalog. }
	\label{tab:list}
	\begin{tabular}{rllll}
        \hline
     HD  &      Name           &Sp. type & comments                   & mv    \\
        \hline
          &BD+61 195   GJ 49               &M1.5&   HPMS                  &  9.56    \\  
          &YZ CMi  HIP37766                &M4.5&   Flare star            & 11.12    \\  
          &EQ Peg A HIP116132 BD+19 5116A  &M3.5&  double system          & 10.35    \\  
          &EQ Peg B  GJ 896 B  BD+19 5116B &M4.5V&   double system        & 12.4     \\  
          &WD1953-011  GJ772               &DA  &                         & 13.69    \\  
          &V2129 Oph  NSV7733              &K5e & CTTS                    & 11.60    \\  
          &GJ 51    V388 Cas               &M5ep&   Flare star            & 13.66    \\
\hline
     \end{tabular}
\end{table*}

\def\ha{\hskip 5.0 mm} \def\hb{\hskip 2.5 mm}

\begin{table*}
\caption{Single wave $B_e$ phase curves and their parameters.
   Julian Day of zero phase equals JD = 2400000. + $T_0$.
   In the first column, HD/Name(*), an asterisk denotes the number of optional phase curve. }
\label{tab:col}
\renewcommand{\arraystretch}{1.1}
\begin{tabular}{|l|rrrrrllc|}
\hline
HD/Name(*) &$B_0$&$\sigma_{B_0}$&$B_1$&$\sigma_{B_1}$&$\ha P$ (days)&$\ha\sigma_P$&$T_0$&$\hb \sigma_{T_0}$ \\
\hline
HD   108   &     -146.93&  10.2&   -73.97&  12.7&    46.296300&0.0710   & 54276.160&   1.633 \\[-1.0pt]
HD   358   &      -32.52&   8.3&    39.16&  11.6&     2.383270&0.0039   & 51890.142&    .131 \\[-1.0pt]
HD   886   &        9.35&   2.1&   -17.23&   3.0&     6.653800&0.0016   & 50675.911&    .222 \\[-1.0pt]
HD  1237   &        -.28&    .3&    -9.47&    .4&     6.694336&0.70     & 56115.870&    .044 \\[-1.0pt]
HD  2453   &     -706.72&  12.7&  -288.52&  18.7&   520.450000&2.0      & 33364.980&   4.921 \\[-1.0pt]
HD  3229   &      -27.73&    .3&    -9.43&    .5&   242.777000&2.50     & 54200.066&   1.496 \\[-1.0pt]
HD  3360(1)&       -6.23&    .3&   -14.33&    .4&     5.370447&0.120    & 52263.676&    .020 \\
\hline
\end{tabular}
\end{table*}

\begin{table*}
\contcaption{Single wave $B_e$ phase curves. }
\label{tab:col}
\renewcommand{\arraystretch}{1.1}
\begin{tabular}{|l|rrrrrllc|}
\hline
HD/Name(*) &$B_0$&$\sigma_{B_0}$&$B_1$&$\sigma_{B_1}$&$\ha P$ (days)&$\ha\sigma_P$&$T_0$&$\hb \sigma_{T_0}$ \\
\hline
HD 34859   &      160.22&  62.6&   416.40&  79.7&     1.046200&0.04     & 56639.738&    .038 \\[-1.0pt]
HD 35177   &    -1723.55& 702.8&  3165.07& 735.2&      .549600&0.02     & 56590.370&    .030 \\[-1.0pt]
HD 35298   &      -73.74&  47.4&  3030.68&  61.4&     1.854580&0.00003  & 44591.797&    .007 \\[-1.0pt]
HD 35456   &      668.29&  57.3&  -492.97&  94.6&     2.518143&0.0002   & 44580.908&    .058 \\[-1.0pt]
HD 35881   &       45.87&  66.4&  -205.42& 118.9&      .699800&0.0010   & 55553.163&    .059 \\[-1.0pt]
HD 35912   &     -323.10& 101.4& -1334.85& 139.9&      .897860&0.00006  & 39571.019&    .014 \\[-1.0pt]
HD 36313   &       24.94&  74.6& -1556.37& 106.6&     1.416282&0.0002   & 44589.341&    .018 \\
\hline
\end{tabular}
\end{table*}

\begin{table*}
\contcaption{Single wave $B_e$ phase curves. }
\label{tab:col}
\renewcommand{\arraystretch}{1.1}
\begin{tabular}{|l|rrrrrllc|}
\hline
HD/Name(*) &$B_0$&$\sigma_{B_0}$&$B_1$&$\sigma_{B_1}$&$\ha P$ (days)&$\ha\sigma_P$&$T_0$&$\hb \sigma_{T_0}$ \\
\hline
HD 62658   &       -1.25&  17.7&   311.97&  42.5&     4.724900&0.0009   & 58556.529&    .051 \\[-1.0pt]
HD 64740   &     -228.81&  24.3&  -678.15&  32.9&     1.330260&0.00006  & 43498.438&    .011 \\[-1.0pt]
HD 66665(1)&      -46.85&   6.2&    93.01&   9.4&    24.500000&2.60     & 55249.780&    .380 \\[-1.0pt]
HD 66765   &     -258.63&  38.0&  -862.11&  50.0&     1.620000&0.15     & 55907.881&    .018 \\[-1.0pt]
HD 67621   &      212.77&   9.8&  -295.27&  13.7&     3.600000&0.26     & 55906.463&    .028 \\[-1.0pt]
HD 68351   &      100.96&  14.1&   -84.44&  20.1&     4.259000&0.016    & 52250.743&    .152 \\[-1.0pt]
HD 70331(1)&    -2515.26&  53.5&  -198.86&  74.8&     1.998120&0.0056   & 49020.840&    .132 \\
\hline
\end{tabular}
\end{table*}

\begin{table*}
\contcaption{Single wave $B_e$ phase curves. }
\label{tab:col}
\renewcommand{\arraystretch}{1.1}
\begin{tabular}{|l|rrrrrllc|}
\hline
HD/Name(*) &$B_0$&$\sigma_{B_0}$&$B_1$&$\sigma_{B_1}$&$\ha P$ (days)&$\ha\sigma_P$&$T_0$&$\hb \sigma_{T_0}$ \\
\hline
HD130144   &       -4.97&    .2&    -4.91&    .4&    62.173600&1.92     & 54546.262&    .537 \\[-1.0pt]
HD130322   &        5.14&   1.5&    -4.03&   1.8&    26.100000&3.5      & 54463.370&   1.638 \\[-1.0pt]
HD131156   &        7.64&    .2&    -3.52&    .3&     6.248010&0.0072   & 52812.045&    .081 \\[-1.0pt]
HD133652   &     -382.97&  60.0& -1904.73& 113.9&     2.304000&0.00016  & 47312.624&    .014 \\[-1.0pt]
HD134793   &     -296.98&  60.0& -1108.83&  98.1&     2.798612&0.0001   & 35938.481&    .027 \\[-1.0pt]
HD137909(1)&      101.48&   5.4&  -671.23&   7.8&    18.486800&0.003    & 40018.535&    .033 \\[-1.0pt]
HD137909(2)&      129.52&  12.1&  -534.37&  15.3&    18.486800&0.003    & 42495.937&    .106 \\
\hline
\end{tabular}
\end{table*}

\begin{table*}
\contcaption{Single wave $B_e$ phase curves. }
\label{tab:col}
\renewcommand{\arraystretch}{1.1}
\begin{tabular}{|l|rrrrrllc|}
\hline
HD/Name(*) &$B_0$&$\sigma_{B_0}$&$B_1$&$\sigma_{B_1}$&$\ha P$ (days)&$\ha\sigma_P$&$T_0$&$\hb \sigma_{T_0}$ \\
\hline
HD184471(2)&      251.02&  71.3&  -295.15&  87.5&    50.800000&0.55     & 52132.711&   3.534 \\[-1.0pt]
HD184927(1)&      925.35&  11.5&   810.58&  13.7&     9.531328&0.00731  & 44416.702&    .032 \\[-1.0pt]
HD184927(2)&      588.42&   6.5&   491.27&   8.5&     9.531328&0.00731  & 55692.248&    .030 \\[-1.0pt]
HD184927(3)&     1036.73&  11.0&   497.48&  15.3&     9.531328&0.00731  & 55692.248&    .052 \\[-1.0pt]
HD184927(4)&      785.84&  16.5&   405.05&  22.9&     9.531328&0.00731  & 55692.248&    .096 \\[-1.0pt]
HD184927(5)&      501.50&  15.7&   419.27&  18.7&     9.531328&0.00731  & 55692.248&    .085 \\[-1.0pt]
HD184927(6)&      198.68&   6.7&   923.24&   8.0&     9.531328&0.00731  & 55692.336&    .017 \\
\hline
\end{tabular}
\end{table*}

\begin{table*}
\contcaption{Single wave $B_e$ phase curves. }
\label{tab:col}
\renewcommand{\arraystretch}{1.1}
\begin{tabular}{|l|rrrrrllc|}
\hline
HD/Name(*) &$B_0$&$\sigma_{B_0}$&$B_1$&$\sigma_{B_1}$&$\ha P$ (days)&$\ha\sigma_P$&$T_0$&$\hb \sigma_{T_0}$ \\
\hline
EQ Peg A   &      272.32&  12.9&  -163.69&  15.6&     1.060000&0.045    & 53950.108&    .012 \\[-1.0pt]
GJ772      &    86861.27& 557.6& -8081.50& 889.4&     1.447949&0.0001   & 50674.662&    .022 \\[-1.0pt]
V2129 Oph(3)&     923.04&  90.6&  -311.55& 109.7&     6.530000&0.75     & 53541.696&    .443 \\[-1.0pt]
GL Vir     &     -153.51&  39.8&  -263.43&  43.6&      .491000&0.002    & 54161.347&    .013 \\[-1.0pt]
DX Cnc     &       84.22&  11.2&   -64.92&  15.1&      .412788&0.01     & 54160.016&    .019 \\[-1.0pt]
LHS292     &      -31.51&   5.1&   -37.17&   7.7&     1.467955&0.2      & 54544.646&    .044 \\[-1.0pt]
GJ 1289    &       28.33&   3.2&    84.90&   5.7&    54.675000&5.5      & 57584.713&    .416 \\
\hline
\end{tabular}
\end{table*}


\def\eol{\hfil\break}
\def\va{\vskip 5.0 mm} \def\vb{\vskip 2.5 mm} \def\vc{\vskip 1.5 mm}
\def\ha{\hskip 5.0 mm} \def\hb{\hskip 2.5 mm} \def\hc{\hskip 1.5 mm}
\def\vd{\vskip 1.0 mm} \def\hd{\hskip 1.0 mm}

\def\deg{\ifmmode^\circ \else$^\circ$ \fi}      
\def\dper{\ifmmode \buildrel d\over . \else $\buildrel d\over
      .$\fi}                   

\begin{table*}
\caption{Single wave $B_e$ phase curve stars: parameter $r$ of the dipole magnetic field.
Hl - from hydrogen \eol lines, all -- both previous methods,
Bs - surface magnetic field.
In HD/Name (*) - asterisk denotes the number of optional phase curve. }
\label{tab:col}
\renewcommand{\arraystretch}{1.1}
\begin{tabular}{|l|rrrcrrll|}
\hline
HD/Name(*)&$N$& $\sigma$  &$r$&$\sigma_r$& $N_V$ &$\chi^2\hd$ &Note/Method & References   \\
\hline
HD   108   &     38&   9.8&    .334&  .067&    35&   1.27&                   &499,514     \\[-1.0pt]
HD   358   &     19&   8.9&   -.084&  .177&    16&   6.41& all lines         &387,826     \\[-1.0pt]
HD   886   &     20&   2.7&   -.297&  .116&    17&   3.21&                   &475,478,638 \\[-1.0pt]
HD  1237   &     15&    .3&   -.935&  .041&    12&   1.62&                   &709         \\[-1.0pt]
HD  2453   &     33&  16.6&    .421&  .024&    30&   2.86&                   &1,26,256,752   \\[-1.0pt]
HD  3229   &     31&    .8&    .493&  .021&    28&  12.16&                   &588         \\[-1.0pt]
HD  3360(1)&     82&    .3&   -.394&  .020&    79&   1.36& all lines         &730         \\
\hline
\end{tabular}
\end{table*}

\clearpage

\begin{table*}
\contcaption{Single wave $B_e$ stars: the dipole magnetic field. }
\label{tab:col}
\renewcommand{\arraystretch}{1.1}
\begin{tabular}{|l|rrrcrrll|}
\hline
HD/Name(*)&$N$& $\sigma$  &$r$&$\sigma_r$& $N_V$ &$\chi^2\hd$ &Note/Method & References   \\
\hline
HD 34859   &      4&  19.9&   -.455&  .161&     1&    .31&                   &776         \\[-1.0pt]
HD 35177   &      4& 286.6&   -.316&  .169&     1&   1.54& H line            &776         \\[-1.0pt]
HD 35298(1)&     28&  65.4&   -.951&  .027&    25&   3.77& met line          &713,769     \\[-1.0pt]
HD 35456   &     13&  23.2&    .157&  .095&    10&    .77&                   &201,427,734 \\[-1.0pt]
HD 35881   &      6&  11.7&   -.499&  .366&     3&    .06&                   &734         \\[-1.0pt]
HD 35912   &     12&  66.2&   -.615&  .099&     9&    .84&                   &55,427,327,769 \\[-1.0pt]
HD 36313   &     13& 115.3&   -.925&  .057&    10&  18.49& H bet             &201,734     \\
\hline
\end{tabular}
\end{table*}

\clearpage

\begin{table*}
\contcaption{Single wave $B_e$ stars: the dipole magnetic field. }
\label{tab:col}
\renewcommand{\arraystretch}{1.1}
\begin{tabular}{|l|rrrcrrll|}
\hline
HD/Name(*)&$N$& $\sigma$  &$r$&$\sigma_r$& $N_V$ &$\chi^2\hd$ &Note/Method & References   \\
\hline
HD 62658   &      5&   9.0&   -.914&  .063&     2&    .56&                   &824         \\[-1.0pt]
HD 64740   &     21&  30.9&   -.496&  .044&    18&   1.73&                   &24,135      \\[-1.0pt]
HD 66665(1)&     21&   4.8&   -.329&  .077&    18&    .76& only Si lines     &769         \\[-1.0pt]
HD 66765   &      7&  27.6&   -.539&  .058&     4&    .67&                   &682         \\[-1.0pt]
HD 67621   &      7&   8.4&   -.162&  .035&     4&    .94&                   &682         \\[-1.0pt]
HD 68351   &     16&  14.5&    .096&  .155&    13&   1.27&                   &423         \\[-1.0pt]
HD 70331(1)&     16&  48.5&    .854&  .050&    13&   1.31&                   &256,752     \\
\hline
\end{tabular}
\end{table*}

\clearpage

\begin{table*}
\contcaption{Single wave $B_e$ stars: the dipole magnetic field. }
\label{tab:col}
\renewcommand{\arraystretch}{1.1}
\begin{tabular}{|l|rrrcrrll|}
\hline
HD/Name(*)&$N$& $\sigma$  &$r$&$\sigma_r$& $N_V$ &$\chi^2\hd$ &Note/Method & References   \\
\hline
HD130144   &     10&    .3&    .007&  .039&     7&   3.40&                   &523         \\[-1.0pt]
HD130322   &      9&    .4&    .135&  .104&     6&   1.22&                   &653         \\[-1.0pt]
HD131156   &    104&    .5&    .370&  .038&   101&  16.88&                   &385,586     \\[-1.0pt]
HD133652   &     10&  73.8&   -.665&  .048&     7&   2.71&                   &230,712     \\[-1.0pt]
HD134793   &      9&  54.1&   -.581&  .056&     6&   1.98&                   &1,402       \\[-1.0pt]
HD137909(1)&     17&   3.2&   -.737&  .013&    14&    .35& met.l             &310         \\[-1.0pt]
HD137909(2)&     39&  17.8&   -.610&  .030&    36&  10.41& H line            &2,76,77,621 \\
\hline
\end{tabular}
\end{table*}

\clearpage

\begin{table*}
\contcaption{Single wave $B_e$ stars: the dipole magnetic field. }
\label{tab:col}
\renewcommand{\arraystretch}{1.1}
\begin{tabular}{|l|rrrcrrll|}
\hline
HD/Name(*)&$N$& $\sigma$  &$r$&$\sigma_r$& $N_V$ &$\chi^2\hd$ &Note/Method & References   \\
\hline
HD184471(2)&     12&  48.6&   -.076&  .189&     9&   1.06& H lines           &402,621     \\[-1.0pt]
HD184927(1)&     21&  11.3&    .066&  .010&    18&   1.23& H lines           &769         \\[-1.0pt]
HD184927(2)&     26&   5.3&    .090&  .011&    23&   1.00& all He from 676   &676         \\[-1.0pt]
HD184927(3)&     26&  15.3&    .352&  .014&    23&   1.55& He 4471 676       &676         \\[-1.0pt]
HD184927(4)&     26&  15.4&    .320&  .028&    23&   1.15& He 6678 676       &676         \\[-1.0pt]
HD184927(5)&     21&  10.4&    .089&  .027&    18&    .60& all He from 769   &676         \\[-1.0pt]
HD184927(6)&     21&  10.8&   -.646&  .010&    18&   4.41& met lines         &769         \\
\hline
\end{tabular}
\end{table*}

\clearpage

\begin{table*}
\contcaption{Single wave $B_e$ stars: the dipole magnetic field. }
\label{tab:col}
\renewcommand{\arraystretch}{1.1}
\begin{tabular}{|l|rrrcrrll|}
\hline
HD/Name(*)&$N$& $\sigma$  &$r$&$\sigma_r$& $N_V$ &$\chi^2\hd$ &Note/Method & References   \\
\hline
YZ CMi(1)  &      7&  14.5&    .059&  .030&     4&   2.43&                   &439         \\[-1.0pt]    
EQ Peg A   &     15&   6.3&    .250&  .065&    12&   1.35&                   &439         \\[-1.0pt]    
GJ772      &     29& 453.5&    .830&  .017&    26&    .98& Bs                &456         \\[-1.0pt]    
V2129 Oph(3)&     8&  43.3&    .504&  .131&     5&    .84&                   &502         \\[-1.0pt]    
GL Vir     &     20&  10.5&   -.274&  .076&    17&    .62&                   &507         \\[-1.0pt]    
DX Cnc     &     21&   9.6&    .136&  .142&    18&    .79&                   &507         \\[-1.0pt]    
LHS292     &     14&   4.2&   -.077&  .128&    11&    .84&                   &507         \\
\hline
\end{tabular}
\end{table*}

\clearpage


\begin{table*}
\caption{Double wave $B_e$ variations.
In HD/Name(*) - asterisk denotes the number of optional phase curve. }
\label{tab:col}
\renewcommand{\arraystretch}{1.1}
\begin{tabular}{|l|rrrrrrr|}
\hline
HD/Name(*) &$B_0$ &$\sigma_{B_0}$ &$B_1$& $\sigma_{B_1}$ &$B_2$ &$\sigma_{B_2}$&$T_0$\\
\hline
HD   965    &  -304.95&     6.48&  -840.25&    10.10&   -81.40&     7.80&  50767.100  \\[-1.0pt]
HD  4778    &    51.00&    76.67& -1317.50&    79.52&   132.85&   231.71&  46671.647  \\[-1.0pt]
HD  5737    &    97.96&    24.83&  -400.82&    43.16&   -63.21&    52.01&  43695.627  \\[-1.0pt]
HD  9996    &  -200.78&    14.29&  -886.46&    17.24&  -299.76&    20.90&  33240.989  \\[-1.0pt]
HD 18078    &   162.27&    13.10& -1077.86&    21.76&  -215.46&    21.22&  51068.729  \\[-1.0pt]
HD 24712    &   506.70&    15.15&  -480.45&    21.76&   145.49&    21.09&  50848.849  \\[-1.0pt]
HD 25267    &  -310.07&    61.93&  -308.80&   107.15&  -106.62&    97.36&  43497.093  \\
\hline
\end{tabular}
\end{table*}

\begin{table*}
\contcaption{Double wave $B_e$ variations. }
\label{tab:col}
\renewcommand{\arraystretch}{1.1}
\begin{tabular}{|l|rrrrrrr|}
\hline
HD/Name(*) &$B_0$ &$\sigma_{B_0}$ &$B_1$& $\sigma_{B_1}$ &$B_2$ &$\sigma_{B_2}$&$T_0$\\
\hline
HD 93030    &   -22.35&    16.05&  -133.43&    23.22&   -83.33&    23.67&  18099.905  \\[-1.0pt]
HD 93507(1) &  2133.48&    63.06&  -797.08&    91.62&  -262.48&    78.18&  48931.158  \\[-1.0pt]
HD 93507(2) &  7166.26&      .74&  -298.23&     1.04&   -30.93&     1.01&  48931.158  \\[-1.0pt]
HD 94660(1) &  6231.13&     1.33&  -162.15&     2.09&   -35.84&     1.97&  47028.     \\[-1.0pt]
HD 95650(1) &      .09&      .67&   -15.59&      .89&    -8.59&      .80&  54098.764  \\[-1.0pt]
HD 95650(2) &     5.60&      .59&   -14.18&      .84&    -5.10&      .80&  56666.974  \\[-1.0pt]
HD 99563    &     6.19&    69.48&  -855.68&   109.43&  -336.91&   108.43&  52031.115  \\
\hline
\end{tabular}
\end{table*}

\begin{table*}
\contcaption{Double wave $B_e$ variations. }
\label{tab:col}
\renewcommand{\arraystretch}{1.1}
\begin{tabular}{|l|rrrrrrr|}
\hline
HD/Name(*) &$B_0$ &$\sigma_{B_0}$ &$B_1$& $\sigma_{B_1}$ &$B_2$ &$\sigma_{B_2}$&$T_0$\\
\hline
HD187474(2) &  5427.78&    12.76&  -628.68&    17.73&   208.03&    18.25&  46676.5    \\[-1.0pt]
HD189775(1) &   364.24&    12.03&  -619.22&    16.55&  -212.97&    18.02&  54959.380  \\[-1.0pt]
HD189775(2) &   460.09&    13.71&  -765.54&    19.10&  -228.90&    21.11&  56249.979  \\[-1.0pt]
HD189775(3) &   -17.82&    34.07&  -819.45&    45.66&  -142.57&    52.42&  56252.469  \\[-1.0pt]
HD189775(4) &   816.27&    24.98&  -783.11&    36.48&   183.78&    38.09&  56250.524  \\[-1.0pt]
HD192678(1) &  4745.08&    17.50&  -120.27&    24.18&   -24.05&    47.80&  48158.941  \\[-1.0pt]
HD192678(2) &  1337.36&    34.15&  -284.08&    43.74&   -77.39&    79.89&  36118.89   \\
\hline
\end{tabular}
\end{table*}


\clearpage

\begin{table*}
\caption{Double wave $B_e$ variations.
In HD/Name(*) - asterisk denotes the number of optional phase curve. }
\label{tab:col}
\renewcommand{\arraystretch}{1.1}
\begin{tabular}{|l|llcrrrrrrll|}
\hline
HD/Name(*)&$\ha\hc P$&$\sigma_{P}$&$z_1$& $\sigma_{Z_1}$ & $z_2$& $\sigma_{Z_2}$ & $N_V$&$\chi^2\hd$& Method/Note & References \\
\hline
HD   965    &  6212.      &194.        & .9882   & .0015&    .0433&    .0197&   78&   7.14&             &678,704,732,752,768,774,814   \\[-1.0pt]
HD  4778    &     2.561641&0.0001      & .0528   & .0127&    .0863&    .1709&   25&  14.68&             &171,575   \\[-1.0pt]
HD  5737    &    21.653786&0.020       & .0001   & .0136&    .0254&    .1150&   19&   1.34&             &37,181,256   \\[-1.0pt]
HD  9996    &  7961.8     &22.         & .9465   & .0042&    .0750&    .0097&  126&   8.11&             &1,19,208,209,270,327,727  \\[-1.0pt]
HD 18078    &  1358.      &12.         & .0083   & .0026&    .9907&    .0129&   20&   1.62&             &705   \\[-1.0pt]
HD 24712    &    12.461   &0.00019     & .0131   & .0075&    .9960&    .0258&   28&   1.22&             &310,382,417   \\[-1.0pt]
HD 25267    &     9.389671&1.03        & .0467   & .0351&    .9701&    .1112&    2&    .01&             &2   \\
\hline
\end{tabular}
\end{table*}

\begin{table*}
\contcaption{Double wave $B_e$ variations. }
\label{tab:col}
\renewcommand{\arraystretch}{1.1}
\begin{tabular}{|l|llcrrrrrrll|}
\hline
HD/Name(*)&$\ha\hc P$&$\sigma_{P}$&$z_1$& $\sigma_{Z_1}$ & $z_2$& $\sigma_{Z_2}$ & $N_V$&$\chi^2\hd$& Method/Note & References \\
\hline
HD 93030    &     0.61111 &0.0000018   & .0180   & .0292&    .9897&    .0499&   21&   1.54&             &592   \\[-1.0pt]
HD 93507(1) &   568.6     &40.5        & .0007   & .0174&    .0004&    .0689&    7&   1.54&          &256,752  \\[-1.0pt]
HD 93507(2) &   568.6     &40.5        & .9805   & .0006&    .1098&    .0046&   25& 807.41&          &254,752  \\[-1.0pt]
HD 94660(1) &  2764.0     &454.        & .0348   & .0017&    .9227&    .0069&   20&  51.51&          &254,752  \\[-1.0pt]
HD 95650(1) &    13.368805&0.474       & .9803   & .0097&    .0270&    .0176&   31&   6.90&          &438       \\[-1.0pt]
HD 95650(2) &    13.831067&0.474       & .0840   & .0101&    .8810&    .0292&   37&   2.75&          &438,758 \\[-1.0pt]
HD 99563    &     2.91179 &0.0012      & .9759   & .0300&    .0325&    .0715&    4&  10.41&          &370,389,477 \\
\hline
\end{tabular}
\end{table*}

\begin{table*}
\contcaption{Double wave $B_e$ variations. }
\label{tab:col}
\renewcommand{\arraystretch}{1.1}
\begin{tabular}{|l|llcrrrrrrll|}
\hline
HD/Name(*)&$\ha\hc P$&$\sigma_{P}$&$z_1$& $\sigma_{Z_1}$ & $z_2$& $\sigma_{Z_2}$ & $N_V$&$\chi^2\hd$& Method/Note & References \\
\hline
HD187474(2) &  2345.      &59.         & .0078   & .0044&    .9733&    .0129&   41&    .93& BS       &218,254,752 \\[-1.0pt]
HD189775(1) &     2.6071  &0.0003      & .9488   & .0044&    .0865&    .0117&   21&   5.20& H lines  &769   \\[-1.0pt]HD189775(2) &     2.6071  &0.0003      & .9568   & .0040&    .0658&    .0120&   21&   2.58& met lines     &769  \\[-1.0pt]
HD189775(3) &     2.6071  &0.0003      & .0130   & .0097&    .9776&    .0591&   21&    .81& He lines      &769  \\[-1.0pt]
HD189775(4) &     2.6071  &0.0003      & .0004   & .0068&    .0016&    .0287&   21&    .62& Si lines      &769  \\[-1.0pt]
HD192678(1) &     6.4186  &0.0098      & .8959   & .0340&    .0541&    .1632&   29&   1.19& BS            &255  \\[-1.0pt]
HD192678(2) &     6.4186  &0.0098      & .9195   & .0275&    .0821&    .1220&   13&  10.28& all           &255,427 \\
\hline
\end{tabular}
\end{table*}

\begin{table*}
\caption{Measurements of 1994 for stars in Orion OB1 by E.F. Borra (personal communication). }
\label{tab:col}
\renewcommand{\arraystretch}{1.1}
\begin{tabular}{|rrr|rrr|rrr|}
\hline
    HD         &         &               &   HD          &         &               &   HD          &         &           \\
$JD2400000.+$ & $B_{e}$ & $\sigma B_{e}$& $JD2400000.+$ & $B_{e}$ & $\sigma B_{e}$& $JD2400000.+$ & $B_{e}$ & $\sigma B_{e}$\\
\hline
HD35298     &          &        &  HD36526     &          &        &   HD37210     &          &            \\[-1.0pt]
44597.558   &   -2808. &    447.&  43507.6510  &    3387. &    453.&   44591.5490  &    -761. &    511.    \\[-1.0pt]
44598.591   &    2919. &    447.&  43508.5970  &    -464. &    460.&   44592.5550  &     403. &    508.    \\[-1.0pt]
44599.719   &     469. &    450.&  43509.6320  &    3060. &    458.&   44594.6990  &     156. &    496.    \\[-1.0pt]
44600.544   &    2714. &    447.&  43509.7880  &    1701. &    561.&   44598.7330  &    -473. &    499.    \\[-1.0pt]
44601.695   &    1381. &    425.&  43510.6550  &    2216. &    458.&               &          &            \\
\hline
\end{tabular}
\end{table*}

\clearpage
\newpage

\section{Comments on individual stars }

\begin{description}

\item {\sl
HD 108} MPC is a simple sine wave. Open
circles - ref.499, filled circles - ref.514.

\item {\sl
HD   358}  MPC is a simple sine wave.  We did not use low-precision
estimates from 140, 327, 330, 387(FORS1), 389, 760.
We used only high-precision estimates received with 
MUSICOS and ESPaDoNS of 387, 826.   We used the period 2.38327 days close 
to the period obtained by Adelman et al., ApJ, 575, 449, 2002.

\item {\sl
HD 886} MPC is a simple sine wave. Open squares - ref.638, filled
circles - ref.478, open circles - ref.475.

\item {\sl
HD 965} MPC is a double sinusoid. We used $B_e$ estimates: open
circles - ref.704, filled circles - ref.752. We determined the
magnetic period of $ 6212. \pm 194. $ days (17 years).

\item {\sl
HD 1237} MPC is a simple sine wave. $B_e$ estimates were taken only
from ref.709. We determined the magnetic period 6.69 days.

\item {\sl
HD 2453} MPC is a simple sine wave. Open squares - ref.1, open
circles - ref.26, filled squares - ref.256, filled circles - ref.752.
We determined the magnetic period equal 520.45 days. There exists a
systematic shift between $B_e$ values measured by the old photographic
method, refs.1 and 26, and more modern method of ref.752.

\item {\sl
HD 3229} MPC is a simple sine wave, points $B_e$ were taken from ref.588.
Magnetic period was determined by us, $P_{\rm mag} = 242.777$ days. 
Additional high-precision observations are necessary for a reliable period determination.
\item {\sl
HD 3360 (1) } MPC is a simple sine wave, which was obtained with the period 5.370447
days. Only high-precision measurements from ref.730,  all lines.

\item {\sl
HD 3360 (2) }  MPC is a simple sine wave, which was obtained with the period 5.370447
days. All except He lines.

\item {\sl 
HD 3360 (3) }  MPC is a simple sine wave, which was obtained with the period 5.370447
days - N lines.

\item {\sl
HD 3379} MPC is a simple sine wave. We determined the period of 21.91 days
using only the best $B_e$ measrements. Open circles - ref.40, filled circles - ref.457, filled squares - ref.475.

\item {\sl
HD 3980} MPC is a simple sine wave. Open circles - ref.121, filled circles - ref.389.

\item {\sl
HD 4128} MPC is a simple sinusoid according open circles - to data from ref.578, 
filled circles - ref.801. We found
the period equal to 115.856598 days using only the best measurements.

\item {\sl
HD 4778} MPC is a double sinusoid. The period determined by us, $P_{\rm mag}
=2.561641$ days is very close to the period 2.56171 from ref.575. 
Open circles - ref.171, filled circles - ref.575.

\item {\sl
HD 5550} MPC is a simple sine wave following measurements from ref.706.
We found the most probable period of 5.764 days and rejected previously
determined value 6.84 days of ref.706. Additional measurements are necessary
to clarify the period, shape and parameters of MPC.

\item {\sl
HD 5601 (1)} MPC is a simple sinusoid based on estimates from ref.0735. We
determined the period of 1.114129 day, close to the period $1.110\pm 0.002$ days
found by Hensberge et al. (1981), Astron. and Astroph. Suppl. 46, 151.
Ref.735 reported period 1.756 days, which was determined from the photometric
data of {\sc HIPPARCOS}. Photometric variability of the star is low and the
amplitude is comparable to the accuracy of measurements. Therefore, determination 
of the period from the photometric data yields not very confident results.
Additional measurements of the longitudinal magnetic field are necessary to
improve the period. Note, that variability of the magnetic field is much
larger than the measurement accuracy ($B_1 / \sigma ~ 40$). Currently the
number of available $B_e$ measurements is low. Presumably 2-3 additional
estimates of the magnetic field will allow to obtain a unequivocal value of
the period.

\item {\sl
HD 5601 (2)} MPC is a simple sinusoid. Estimates of $B_e$ in ref.735 were 
obtained by the method of regression.

\item {\sl
HD 5737} MPC is a double sinusoid. We have refined magnetic period to
$P_{\rm mag}=21.653786$ days. This value is close to the period 21.654 from 
ref.181. Open squares - ref.37, open circles - ref.181, open circles - ref.256.

\item {\sl
HD 5797} The MPC is a simple sine wave. We used $B_e$ estimates from ref.534.
Period 68.046 days was taken from Adelman and Duker, (2014), AAS, 224, id.322.16.
Complementary measurements are needed to clarify shape and parameters of
the phase curve.

\item {\sl
HD 6757} MPC is a simple sinusoid with the period of 6.166 days. Low number
of available magnetic measurements. There are two probable periods: 1.1678 days
and a slightly less probable 6.166 days. Additional measurements are
necessary to clarify the period and parameters of the MPC. Open squares 
- ref.402, open circles - ref.621, filled squares - ref.677, filled circles
- ref.710,768,774.

\item {\sl
HD 8441} MPC is a simple sine wave. We used a period of 69.51 days
from work by Pyper and Adelman, PASP, v.129:104203 (20pp), 2017.
Refs. 1 and 327 were not used due to low accuracy.
Additional precision measurements needed for clarification of MFB parameters.
filled circles - ref.774, open circles - ref.423;

\item {\sl
HD 8890} MPC is a simple sine wave. Period 3.97586 days was adopted from ref.536.
Additional $B_e$ observations are needed to clarify parameters of
the phase curve. Open squares - ref.76, filled circles - ref.536.

\item {\sl
HD 9996} MPC is double sinusoid. Filled circles - ref.1,19,208,209; open
circles - ref.270; filled squares - ref.327,727.
The period from ref.727 equals 7961.8 days.

\item {\sl
HD 10783} MFC is a simple sinusoid. Open circles - ref.1, open squares -
ref.35, filled circles - ref.424.
We found the period equal 4.099335 days using best $B_e$ measurements and
rejected the photometric UBV period (4.13282 days).

\item {\sl
HD 11503} MPC is a simple sinusoid. Open circles - ref.2, filled 
circles - ref.327. We determined the period 1.60935 day replacing 1.6093 of ref.2.

\item {\sl
HD 11753} MPC is a simple sine wave.
Spectral binary, $P_{\rm orb} = 41.49$ d,
following Korhonen et al., 2013, A\&A, 553, A27.
$P_{\rm rot} = 9.53077 \pm 0.00011$ d.
We used the period 9.53077 days. Phase curve is uncertain.
Open circles - ref.550; filled circles - ref.611.

\item {\sl
HD 12098} The MPC is a simple sine wave.
Rotational period $P_{\rm rot} = 5.460$ days - from ref.382;
Very good phase curve.

\item {\sl
HD 12288} MPC is a simple sine wave.
$P_{\rm mag}=34.79$ days from ref.312.
Open circles - $B_e$ measured in H lines, filled circles - in metal lines.

\item {\sl
HD 12447} MPC is a simple sine wave. It was plotted using only 
data from ref.2. Estimates from ref.25 were not used due to low accuracy.
The best period for this star equals 1.4907 days. It is almost twice the
photometric period of Wraight et al. (2012), MNRAS, 420, 757.
The other photometric period of this paper, 6.650 days, is not
suitable for a reasonable phase curve.

\item {\sl
HD 12767} MPC is a simple sinusoid following measurements from
ref.137. Period $P = 1.892$ day.
Number of $B_e$ observations is low.
Additional $B_e$ measurements are needed to clarify the MPC.

\item {\sl
HD 14437} The MPC is a simple sine wave. We determined the period 
equal 26.867 days. Additional observations are needed to clarify
period and MPC parameters. Filled circles - ref.41,142; open
circles - ref.312.

\item {\sl
HD 15144} MPC is a simple sine wave. Measurements from ref.1.87 
were not used due to low accuracy. Measurements from ref.423 
are of high accuracy, but they appended only 6 points. MPC was
plotted using only points from ref.423 with the period
2.99787 days. Additional observations are needed to clarify the 
period and parameters of the phase curve.

\item {\sl
HD 16582} MPC is a simple sine wave. Estimates from ref. 96 are not
used due to low accuracy. Periods suitable for magnetic measurements:
87.389671 days - the most probable period, 9.434674 days - a possible value.
Open squares - ref.327; open circles - ref.475; filled squares -
ref.478, filled circles - ref.457. Additional observations are 
needed to clarify period and MPC parameters.

\item {\sl
HD 16605} MPC is a simple sine wave. Very little data is available. 
Additional measurements are necessary to clarify value of the period 
and parameters of the MPC.
Currently the most probable period is 1.327173 day.
Open squares - ref.402, filled circles - ref.459, filled squares - ref.621.

\item {\sl
HD 17051} MPC is a simple sine wave. We used the period 7.70 days and 
measurements from ref.805.

\item {\sl
HD 17330} MPC is a simple sine wave. Very little data are available.
Additional measurements of the magnetic field are necessary to
clarify the period and parameters of the MPC.
The most probable period is 3.228024 days according to ref.713.

\item {\sl
HD 18078} MPC is a double sine wave.
Period equals $1358 \pm 12 $ days. Measurements were taken from ref.705.

\item {\sl
HD 18296} MPC is a simple sine wave. Data from refs.1,14,373 were not
used in consequence of low accuracy. Open circles - ref.2,243,
filled circles - ref.423, filled squares - ref.826. Magnetic period 2.88416 from ref.826.
Additional measurements are needed 
to clarify the period and parameters of the MPC.

\item {\sl
HD 19712} The MPC is a simple sinusoid. Period from Catalano \& Renson
(1998) is not suitable here. For magnetic measurements, the best period 
equals P = 2.675728 days.
Open squares - ref.389, filled circles - ref.402, open circles - ref.621.

\item {\sl
HD 19832} MPC is a simple sine wave. The period P=0.727893 days. Open circles -
ref.2. To clarify MPC we need additional high-precision magnetic observations.

\item {\sl
HD 21190} MPC is a simple sine wave. The period 1.02471 days from ref.786.
Filled circles - ref.760,786 - all lines, filled circles - ref.786 - hydrogen lines.

\item {\sl
HD 21699} The MPC is a simple sine wave. The period and data from ref.252
were used to plot the phase curve. Data from ref.40 were not used due to 
low accuracy.

\item {\sl
HD 22316} MPC is a simple sine wave.
P = 2.9761 days from ref.309.

\item {\sl
HD 22470} The MPC is a simple sine wave. Data from ref.25 were not used 
due to low accuracy. MPC was drawn according to data from ref.37, period 
1.9387 from Adelman \& Boyce 1995. Additional observations are needed, 
especially in phases close to min-max of $B_e$.

\item {\sl
HD 23478 (1)} The MPC is a simple sinusoid, $B_e$ measured in $H_\beta$ line.
Filled circles - ref.769, open circles - ref.724.

\item {\sl
HD 23478 (2)} The MPC is a simple sinusoid. Filled circles - ref.724, 
in He + metal lines.

\item {\sl
HD 24155} MPC is a simple sinusoid with the period of 2.53465 days
and measurements from ref.230. More observations are needed, especially 
in phases close to min - max of $B_e$.

\item {\sl
HD 24587 (1)} MPC is a simple sinusoid, in all lines. $P_{\rm rot}$ 
from ref.513. Open circles - ref.513, filled circles - ref.571.

\item {\sl
HD 24587 (2)} The MPC is a simple sinusoid. $B_e$ measured in H lines were used.
$P_{\rm rot}$ from ref.513. Filled circles - ref.513.

\item {\sl
HD 24712} MPC is a double sine wave. Measurements not used: ref.21,
47, 111, 120, 195, 256, 413 due to low accuracy.
Measurements used here: open squares - ref.310, open circles - ref.382,
filled circles - ref.417.

\item {\sl
HD 25267} MPC is a double sinusoid. The most probable period 
equals 9.389671 days according to ref.2. Few $B_e$ estimates
are available. Additional observations are needed for 
refinement of the period and parameters of the phase curve.

\item {\sl
HD 25354} MPC is a simple sinusoid with the period 3.90072 days
by Schoneich et al. (1976). We used measurements from ref. 1.
Few estimates of the magnetic field strength are available.
Additional observations are needed for refinement of the period 
and parameters of the MPC.

\item {\sl
HD 25558}  MPC is a simple sine wave. The period 1.233(1) days from ref.769.
Filled circles - ref.760, open circles - ref.769.

\item {\sl
HD 25823} MPC is a simple sine wave. Open circles - ref.1,
filled circles - ref.71.

\item {\sl
HD 27309} MPC is a simple sine wave. We did not use estimates from
ref.2,41,142 due to low accuracy. Phase curve was plotted using 
high-precision measurements from ref.423. The best period for 
$B_e$ measurements in ref.423 equals P = 4.525092 days. Photometric
period 1.56889 day from the paper by Adelman and Dukes (2014), AAS,
224, id.322.16, was neglected because phase curve could not be
obtained with this period.

\item {\sl
HD27404 (1)} MPC is a double sinusoid. Magnetic field measurements
were obtained by fitting a gaussian to line profiles. Data and 
period 2.77929 days were taken from ref.767.

\item {\sl
HD27404 (2)} MPC is a double sinusoid obtained by the regression
method using measurements and period 2.77929 days from ref.767.

\item {\sl
HD 27536} MPC is a double sinusoid. Star is a yellow dwarf; the best 
period for magnetic measurements from ref.559 equals P = 300.3003 days.

\item {\sl
HD 27962} MPC is a simple sine wave. Open squares - ref.1,
open circles - ref.98, filled circles - ref.77.

\item {\sl
HD 28305}  MPC is a simple sine wave. We found the most likely magnetic period
4.76 days received from an estimated 485.

\item {\sl
HD 28843} MPC is a simple sine wave. Open circles - ref.37. Few estimates
are available. Additional observations are needed to clarify the period and
parameters of the MPC.

\item {\sl
HD 29009} MPC is a simple sine wave. The best magnetic period equals 3.043584
days. Filled circles - ref.230, open circles - ref.548. Additional observations
are necessary to clarify parameters of the phase curve.

\item {\sl
HD 29248} MPC is a simple sine wave. The best magnetic period found by us
equals 21.557 days. Open squares - ref.419, open circles - ref.406, filled
squares - ref.457, filled circles - ref.475. Additional observations are
needed to clarify the period and parameters of the phase curve.

\item {\sl
HD 30466} The MPC is a simple sine wave. The magnetic period was found 
equal 3.867814 days. Open squares - ref.1, open circles - ref.142,327, 
filled circles - ref.483.

\item {\sl
HD 32549 (1)} The MPC is a simple sinusoid for $B_e$ points obtained from 
metal lines. We have determined magnetic period of 4.654161 days, which 
is close to the period 4.64 days from ref.710.
Open circles - ref.423, filled circles - ref.710.

\item {\sl
HD 32549 (2)} The MPC also is a simple sinusoid for $B_e$ points obtained 
from hydrogen lines. Open circles - ref.2, filled circles - ref.710.
Few measurements are available. Additional measurements are needed
to clarify the period and parameters of the MPC.

\item {\sl
HD 32633} MPC is a double sine wave. Measurements from 
ref.1,2,8,175,285,301,310,427 were not used here due to low accuracy.
Phase curve with the period of 6.430 days was obtained from
high-precision measurements of ref.575.

\item {\sl
HD 32650} The MPC is a simple sine wave. We used $B_e$ measurements
from ref.423. Period 2.7347 days is not suitable for analysing of
magnetic data. The best period equals 2.117836 days.

\item {\sl
HD 33328 (1)} MPC is a simple sinusoid. Filled circles - ref.474. Magnetic
field was measured in all lines.

\item {\sl
HD 33328 (2)} MPC is a simple sinusoid. Filled circles - ref.474 (hydrogen
lines only). Period from ref.474 equals 21.12 min. (rapid variability), 
Phase curve was plotted using the latter period.

\item {\sl
HD 33798} The MPC is a simple sine wave. Filled circles - ref.635,
open circles - ref.587. $P_{\rm rot} = 9.8254$ days.

\item {\sl
HD 34452} MPC is a simple sine wave. We used only values of the magnetic
field strength determined from hydrogen lines. Open circles - ref.2,
filled circles - ref.230. Photographic estimates of low accuracy from 
ref.41 were not used in this figure.

\item {\sl
HD 34736} MPC is a simple sine wave, but a more complex structure is
likely. We found the most probable period 0.92898 day.
We used measurements from ref.713. Additional observations are needed
to clarify the period and parameters of the phase curve.

\item {\sl
HD 34859} MPC is a simple sine wave. Were used by us
period 1.0462002 days and measurements from 776. It is not enough
measurements. To specify parameters and shape of MPC
additional MF measurements are needed.

\item {\sl
HD 35177} MPC is a simple sine wave. Were used by us
period 0.5496 days and measurements on hydrogen lines from 776.
It is not enough measurements. To specify parameters and shape of MPC
additional MF measurements are needed.

\item {\sl
HD 35298(1) } MPC is a simple sinusoid. The period 1.85458 days
was taken from Shultz et al., 2018, MNRAS, 475, 5144.
The MPC was determined from metal lines. Open circles - ref.769, filled circles - ref.713.

\item {\sl
HD 35298(2) } MPC is a double sine wave determined from hydrogen lines.
Open circles - ref.201, filled circles - ref.769, filled squares - ref.774.

\item {\sl
HD 35298(3) } MPC is a double sine wave deterined from Fe lines. 
Filled squares - ref.769.

\item {\sl
HD 35456} MPC is a simple sinusoid. Open circles - ref.201 (H line), 
open squares - ref.427 (photographic observations), filled squares 
- ref.734. Period 4.9506 days which was obtained from {\sc HIPPARCOS}
photometry in ref.735 does not correspond to magnetic measurements from
ref.734. We determined period P = 2.518143 days corresponding to all
all magnetic data. Additional measurements of the magnetic field are 
required, because photometric data unfortunately do not yield a 
reliable result.

\item {\sl
HD 35502 (1)}  MPC is a double sine wave. Phase curve and its parameters 
were derived from magnetic measurements in $H_\beta$ line. 
Rotation period has been defined Sikora et al. ref.717 = 0.853807(3).
filled circles - ref.201, open circles - ref.769.

\item {\sl
HD 35502 (2)}  MPC is a double sine wave. Phase curve and parameters were derived 
from measurements in H+metal lines from ref.769.

\item {\sl
HD 35502 (3)} MPC is a double sine wave.  Phase curve and parameters were obtained 
from He lines from ref.769.

\item {\sl
HD 35502 (4)}  MPC is a double sine wave. Phase curve and parameters were derived 
from measurements on only Si lines from ref.769.

\item {\sl
HD 35502 (5)}  MPC is a double sine wave. Phase curve and parameters were derived 
from measurements on Fe lines from ref.769.

\item {\sl
HD 35881} MPC is a simple sine wave. Period of 0.6998 day and magnetic 
measurements obtained by the regression method were taken from ref.734.

\item {\sl
HD 35912} MPC is a double sine wave. We found probable period of 0.896059 days
close to the period 0.89786 days of ref.580. For determination of the MPC parameters
and the period more additional precise $B_e$ observations are necessary.
Open squares - ref.55, open circles - ref.427, filled squares - ref.327,
filled circles  - ref.769.

\item {\sl
HD 36313} MPC is a simple sine wave. We used only measurements of the
magnetic field obtained in the wings of $H_\beta$ line from ref.201 
and ref.734. According to those estimates, we found the period 
$P = 1.4162824 $ day. Filled circles - ref.201, open circles - ref.734.
All previously found periods are not consistent with those field strength
estimates. Additional measurements are necessary to clarify the period.

\item {\sl
HD 36395} MPC is a double sinusoid with the period 33.63 days
from ref.758. All data were taken from ref.758.

\item {\sl
HD 36485 (1) } MPC is a simple sine wave. We used the period
1.47775 day from ref.492. We used only H lines from ref.769.

\item {\sl
HD 36485 (2) } MPC is a simple sine wave. We used the period
1.47775 day from ref.492. We used only metal lines from ref.769.
The phase curve is very uncertain. For specification additional 
high-precision measurements are required for MPC parameters.

\item {\sl
HD 36526 (1) } MPC is a simple sine wave. Probable magnetic period we found
1.54191 (4) days close to the photometric period 1.5405 days presented by
North in A\&ASS, 55, 259 (1984). Only precise estimates from ref.769 were used.
Open squares - ref.201, open circles - ref.734, filled circles - ref.769.

\item {\sl
HD 36526 (2) } MPC is a simple sine wave by all metal lines.
Only precision estimates from ref.769 were used.

\item {\sl
HD 36526 (3) } MPC is a double sinusoid by He+met lines.
Only precision estimates from ref.769 were used.

\item {\sl
HD 36526 (4) } MPC is a double sinusoid by only Fe lines.
Only precision estimates from ref.769 were used.

\item {\sl
HD 36526 (5) } MPC is a double sinusoid by only Si lines.
Only precision estimates from ref.769 were used.

\item {\sl
HD 36540} MPC is a simple sine wave. Probable magnetic period found
by us equals 2.706946 days. This period does not coincide with the 
photometric period 2.172 days found by North (1984), A\&ASS, 55, 259,
as well as with the period 1.8437 day, the latter was determined from
ASAS3 photometric database. Open squares - ref.201, open circles - ref.388,
filled squares - ref.762. Additional measurements of the magnetic field
and polarimetric observations are necessar.

\item {\sl
HD 36629} MPC is a simple sine wave. We determined probable magnetic 
period P = 5.011753 days. Filled circles - ref.53, open circles - ref.201,
open squares - ref.388, filled squares - ref.549. Additional measurements
of the magnetic field and polarimetry are necessary to improve the period.

\item {\sl
HD 36668} MPC is a simple sine wave. Total time length of the magnetic
measurements of this star equals almost 35 years. Therefore, we improved
magnetic period to 2.125588 days, which is close to the photometric period
in Adelman (2000), A\&A, 357, 548. Filled circles - ref.201, open circles - ref.762.

\item {\sl
HD 36916} MPC is a simple sine wave. Phase curve was computed using
the photometric period 1.5652386 days. MPC is rather uncertain, new
$B_e$ determintions are necessary for this star. Open circles - ref.37, 
filled circles - ref.388, filled squares - ref.762.

\item {\sl
HD 36982 (1) } MPC is a simple sine wave by metal lines. For MPC it is used
photometric period 1.8551 (5) of ref.769. The MPC is very uncertain.
Additional measurements of the magnetic field and polarimetry are necessary 
to improve the period.    open squares - ref.760,  filled circles - ref.568,  
filled squares - ref.768,  open circles - ref.769.
 
\item {\sl
HD 36982(2) } MPC is a simple sine wave from metal lines with the
photometric period 1.8551(5) days of ref.769. The MPC is very uncertain.
Additional measurements of the magnetic field and polarimetry are necessary 
to improve the period.  filled circles - ref.388, open circles - ref.769.

\item {\sl
HD 37017 (1) } MPC is a simple sine wave on H lines. The period 0.901186
days from ref.769. Open  circles - ref.135, filled circles - ref.24.

\item {\sl
HD 37017 (2) } MPC is a simple sine wave on H lines. The period 0.901186 from ref.769.
MPC was obtained from hydrogen lines received LSD methods by spectra with ESPaDOnS. Filled circles - ref.769.
Amplitude of variability obtained from hydrogen lines two methods match, but the constant 
component shifted by -450 Gs.

\item {\sl
HD 37017 (3) } MPC is a simple sine wave on met+He lines. The period 0.901186 from ref.769.
Filled circles - ref.769.

\item {\sl
HD 37017 (4) } MPC is a simple sine wave on Al lines. The period 0.901186 from ref.769.
Filled circles - ref.769.

\item {\sl
HD 37022 (1)}
Measurements of the longitudinal magnetic field were obtained by
FORS2.  Open circles - ref.453.

\item {\sl
HD 37022 (2)}
Filled circles - ref.0338, open circles - ref.386, filled circles - ref.582 - LSD.

\item {\sl
HD 37022 (3)}
Filled circles - ref.435 - absorption line CIV5801 + 5812.

\item {\sl
HD 37041} MPC is a simple sine wave. Open circles - ref.69,
filled circles - ref.75, filled squares - ref.388. We determined
the period equal 3.296685 days.
Additional polarimetric and magnetic field measurements are
necessary to clarify value of the period.

\item {\sl
HD 37058 (1) } MPC is a simple sine wave on H lines.  We settled the period equal
14.673498 days. It is close to the previus period of 14.581(2) days
presented ref.769.
Open squares - ref.37, open circles - ref.338,762, filled squares - ref.769,
filled circles - ref.774.
Additional polarimetric and magnetic field measurements are
necessary to clarify value of the period.

\item {\sl
HD 37058 (2) }  MPC is a simple sine wave from metal lines. No two low-precision 
estimates out of ref.762 were used.   
Open squares - ref.388, open circles - ref.762, filled squares - ref.769, 
filled circles - ref.774.

\item {\sl
HD 37058 (3) }  MPC is a simple sine wave from Fe lines.  Filled circles - ref.769.

\item {\sl
HD 37061 (1) } MPC is a simple sine wave from H lines with the period 
1.09478 days from 797.  Filled circles - ref.797.

\item {\sl
HD 37061 (2) } MPC is a simple sine wave on met+He lines width period 
1.09478 day from ref.797.  Filled circles - ref.797.

\item {\sl
HD 37061 (3) } MPC is a simple sine wave on component C lines width period 
1.09478 day from ref.797.  Filled circles - ref.797

\item {\sl
HD 37140} MPC is a simple sine wave. Filled circles - ref.549, 
open circles - ref.201.

\item {\sl
HD 37151} MPC is a simple sine wave. Open circles - ref.201,204,457
(H lines), filled circles - ref.549. Additional magnetic and polarimetric
measurements are necessary to clarify the period.

\item {\sl
HD 37210} MPC is a simple sine wave. Open circles - ref.201, filled 
circles - ref.388. 1 measurement was appended from ref.388. Additional
measurements of the magnetic field and polarimetric measurements are
necessary to improve the period.

\item {\sl
HD 37479 (1) } MPC is a double sine wave on H lines. We determined
the period, $P= 1.19084$ days, which is very close to period of Shore 
\& Brown (1990). Estimates from ref.549 have inverse sign and were not
used here. Open squares - ref.28, open circles - ref.135, 
filled squares - ref.565, filled circles - ref.765.

\item {\sl
HD 37479 (2) } MPC is a double sine wave on He lines.
Open squares - ref.135, open circles - ref.565, filled circles - ref.769.

\item {\sl
HD 37479 (3) } MPC is a double sine wave on all metal lines.
Open squares - ref.565, filled squares - ref.768, open circles - ref.769.

\item {\sl
HD 37479 (4) } MPC is a double sine wave on only O lines.
           Filled circles - ref.769.

\item {\sl
HD 37490} MPC is a simple sine wave. We did not use $B_e$ estimates 
from ref.375,409 due to low accuracy. Only high-precision data from
ref.577 obtained by the LSD method were used to plot the figure.
Two points from ref.577 were neglected due to low accuracy (140 G).
The period was determined here, $P=1.4002$ days, points from ref.577.

\item {\sl
HD 37642} MPC is a simple sine wave. MPC via hydrogen lines. 
Open circles - ref.201, filled circles - ref.774.
We have defined the period for magnetic measurements - 1.582807 day.
Additional magnetic and polarimetric observations are necessary 
to improve parameters of variability.

\item {\sl
HD 37776 (1) }  Unique star, MPC consists of few waves.
Very unusual magnetic behaviour. MPC is a double sine wave on only H lines.
Period 1.53876 is taken from ref.769. 
Filled circles - ref.24, open circles - ref.174, filled squares - ref.769.

\item {\sl
HD 37776 (2) } Unique star, MPC consists of few waves.
MPC is a double sine wave on only C lines.  Filled circles - ref.769.

\item {\sl
HD 37776 (3) } Unique star, MPC consists of few waves.
MPC is a double sine wave on only Fe lines.  Filled circles - ref.769.

\item {\sl
HD 37776 (4) } Unique star, MPC consists of few waves.
MPC is a double sine wave on only He lines.  Filled circles - ref.769.
The MPC badly describes measurements.

\item {\sl
HD 37776 (5) } Unique star, MPC consists of few waves.
MPC is a double sine wave on only Mg lines.  Filled circles - ref.769.
The MPC badly describes measurements.

\item {\sl
HD 37776 (6) } Unique star, MPC consists of few waves.
MPC is a double sine wave on only N lines.  Filled circles - ref.769.
The MPC badly describes measurements.

\item {\sl
HD 37776 (7) }  Unique star, MPC consists of few waves.
MPC is a double sine wave on only O lines.  Filled circles - ref.769.
The MPC badly describes measurements. 

\item {\sl
HD 37776 (8) }  Unique star, MPC consists of few waves.
MPC is a double sine wave on only Si lines.  Filled circles - ref.769.
The MPC badly describes measurements.

\item {\sl
HD 37776 (9) } Unique star, MPC consists of few waves.
MPC is a double sine wave on only S lines.  Filled circles - ref.769.
The MPC badly describes measurements.

\item {\sl
HD 37776 (10) }  Unique star, MPC consists of few waves.
MPC is a double sine wave on only met lines.  Filled circles - ref.769.
The MPC badly describes measurements.

\item {\sl
HD 38823 }  MPC is a simple sine wave. We used a period of 8.6756 days
close to the period of ref.621. Filled circles - ref.402, open circles - ref.621.

\item {\sl
HD 39317} MPC is a simple sine wave. We have specified the period 2.59757 (13)
close to the period of 710 2.6571 days. It is not enough data. MPC is not fully covered.
Additional measurements are necessary to specify period and parameters of MPC.
Open squares - ref.2, open circles - ref.423, filled squares - ref.710, filled circles - ref.768.

\item {\sl
HD 39801 } MPC is a double sine wave.  We determi an estimate of the possible period
of MF variability - 1612. days. Additional magnetic and polarimetric observations are necessary 
to improve parameters of variability.  Open circles - ref.504, filled circles - ref.802.

\item {\sl
HD 40312 (1) }  MPC is a double sine wave. Measurements obtained in hydrogen lines.  
We determined the period equal 3.618636 days. Magnetic field estimates
from ref.25,226 were not used here due to low accuracy.
Open circles - ref.2, filled circles - ref.60.

\item {\sl
HD 40312 (2)}  MPC is a double sine wave. High-precision estimates were obtained 
by the LSD method in lines of metals. Open circles - ref.310, 
open squares - ref.575, filled circles - ref.415.

\item {\sl
HD 40535}  MPC is a simple sine wave. Phase curve is very uncertain, 
since it is unevenly covered by estimates of the magnetic field.
One of the probable magnetic periods equals 0.431351 days. This period 
does not coincide neither with the period of pulsations nor with the 
period of Blazhko effect. Additional measurements of the magnetic
field and polarimetric data are necessary for determination of period
and MPC parameters. Open circles - ref.362.

\item {\sl
HD 41403}  MPC is a simple sine wave but it is very uncertain, since
is not reasonably covered by $B_e$ estimates. One of the probable 
magnetic periods equals 4750 days; variability of the magnetic field
is very slow. Additional observations are necessary for the determination
of period and MPC parameters. Open circles - ref.402

\item {\sl
HD 43317 (1) } MPC is a double sinusoid. The period 0.897673 days from ref.792.
Measurements in all lines were used.
Filled circles - ref.793, open circles - ref.794.

\item {\sl
HD 43317 (2) } MPC is a double sinusoid. The period 0.897673 days from ref.792.
Measurements in He lines were used. Open circles - ref.794.

\item {\sl
HD 43819}   MPC is a double sinusoid. We did not use here low precise
measurements from ref.120,427. We have a refined period of 15.022459 
close to the period from Adelman and Yuong,  2005, AaA, 429, p.317.
Filled circles - ref.402, open circles - ref.423.

\item {\sl
HD 44743}    MPC is a simple sine wave, determined using the data and period 
of 13.6 days from ref.689. Number of magnetic field estimates is low, new
high-precision observations are necessary for the determination of MPC parameters.

\item {\sl
HD 45348}    MPC is a simple sine wave. Filled circles - ref.42, open circles - ref.273.
High-precision magnetic measurements are necessary for determination of 
period and parameters of MF variability.

\item {\sl
HD 45530}   MPC is a simple sine wave with the period of 1.585 days from
Manfroid \& Renson, 1983, IBVS, No. 2311, p.1. We used here nonnumerous data 
from ref.402. New measurements are necessary for determination of the form
and parameters of the MPC.

\item {\sl
HD 45583 (1)}    MPC is a double sinusoid, measurements were done in
hydrogen lines. We have clarified the period 1.17706 (13) which
close to a period of ref.428.
Open squares - ref.388, open circles - ref.621, filled squares - ref.710, 
filled circles - ref.732,768.

\item {\sl
HD 45583 (2)}    MPC is a double sinusoid, obtained from metal lines.
Open squares - ref.402, open circles - ref.428,
filled circles - ref.388,760, filled squares - ref.621.

\item {\sl 
HD 46328 (1) } MPC is a simple sine wave. MF estimates and period 10950. (30 years)
are taken from ref.807. It is not enough data. To refine the rotation period
additional parameters of MFB variability are required high-precision measurements.

\item {\sl
HD 46328 (2) } MPC is a double sinusoid. MF estimates and pulsation period 0.2095769 days
are taken from ref.807. Just as in 807 in each set the mean of
the MPC was relative to each average. According to us,
a double sinusoid describes a variable with a period of pulsation
better than a simple wave.

\item {\sl
HD 47103}   MPC is a simple sine wave. We determined the period
18.6 days, but this value is very uncertain. For determination of the period
new magnetic, photometric and polarimetric observations are necessary.
Filled circles - ref.259, open circles - ref.752.

\item {\sl
HD 47129 (1)}  MPC is a simple sine wave. We found the probable period
0.9918 day for the A1 component. MPC is constructed using $B_e$ estimates
from ref.733. We ignored here low-precise estimates obtained before
JD2456010.3588 and 2456264.1275. Not all phases MPC are covered by
magnetic estimates. For specification
of MPC parameters new high-precision observations are required.

\item {\sl
HD 47129 (2)}  MPC is a simple sine wave. We found the probable period
equal 0.8847 days for the component A2. Phase curve was plotted using
$B_e$ points from ref.733.

\item {\sl
HD 47777 }  MPC is a simple sine wave. We used the period 2.6415 days from ref.792.
Filled circles - ref.427, open circles - ref.492.

\item{\sl
HD 49333}   MPC is a simple sine wave. We used only estimates from ref.0230.
New magnetic field and polarimetric measurements are necessary for determination
of the period and parameters of phase curve.

\item {\sl
HD 49606}  MPC is a simple sine wave. We did not use estimates 
from ref.38,230,241,427 due to low accuracy. We found the new probable 
period 1.457025 days. The period and MPC parameters were determined 
using the following data: open circles - ref.267, filled circles - ref.330,
open squares - ref.389.
New magnetic field and polarimetric data are necessary for the determination
of period and parameters of phase curve.

\item {\sl
HD 49976}   MPC is a double sinusoid. We determined the period 2.976616 days.
Open circles - ref.92, filled circles - ref.1,91, open squares - ref.285, 427.
New magnetic data are necessary for this star.

\item {\sl
HD 50169 (1) }  MPC is a double sinusoid. We found the most probable period
10870.7 days (~ 29 years). Open squares - ref.256, open circles - ref.677, 
filled squares - ref.768, filled circles - ref.779.

\item {\sl
HD 50169 (2) }   MPC is a double sinusoid. MPC is built according to BS measurements.
Open squares - ref.218, open circles - ref.254, filled squares - ref.752, 
filled circles - ref.779.

\item {\sl
HD 50461}   MPC is a simple sine wave. The period 0.894 days and 
measurements were taken from ref.710. 

\item {\sl
HD 50707}   MPC is a simple sine wave.
Open circles - ref.513 field strength was determined from all lines,
filled circles - ref.457 from hydrogen lines. Measurements from ref.571 
do not agree with the MPC, but is not critical. New high-precision 
measurements are necessary to confirm variability of the longitudinal
magnetic field.

\item {\sl
HD 50896}  MPC is a simple sine wave.
The period 3.766 days and measurements were taken from ref.718.

\item {\sl
HD 51418}  MPC is a simple sine wave. Photometric period 5.4379 daes from
ref.768 does not satisfy magnetic measurements. For magnetic
measurements relatively suitable period 2.2908, approx.
twice well than photometric. To refine the period and parameters
additional high-precision measurements are needed.
Open circles - ref.29, filled circles - ref.768.

\item {\sl
HD 54118}  MPC is a simple sine wave. 
We determined the period 3.221691 days which is close to the period
in Bohlender et al. (1993, A\&A, 269, 355).
Open squares - ref.81, open circles - ref.230, filled circles - ref.425.

\item {\sl
HD 55719}  MPC is a simple sine wave. Only high-precision measurements 
from refs.256,752 were used. We determined probable period 357.3 days 
using $B_e$ points from these papers. The period is very uncertain.
New magnetic and polarimetric measurements are necessary for determination
of the period and parameters of the phase curve.
Filled circles - ref.256, open circles - ref.752.

\item {\sl
HD 57682} MPC is a simple sine wave. Filled circles - ref.576.

\item {\sl
HD 58260} MPC is a simple sine wave.
Open circles - ref.24, filled circles - ref.135.

\item {\sl
HD 59435}  MPC is a double sinusoid, which was plotted using only 
surface field measurements, $B_s$.
Open squares - ref.253, open circles - ref.274,
filled squares - ref.752. We found the period 1365.9 days using these 
estimates. Unfortunately, measurements of the longitudinal field
strength $B_e$ are not available.

\item {\sl
HD 61468 (1)}   MPC is a simple sine wave. Phase curve was plotted using
only surface field measurements, $B_s$. Using these estimates we found the period 319.68 days.
Filled circles - ref.253, open circles - ref.752.

\item {\sl
HD 61468 (2)}   MPC is a simple sine wave. We found the period 319.68 days 
using data from ref.752. Few measurements are available. Additional measurements
are necessary for improvement of appearance and parameters of phase curve.

\item {\sl
HD 62140}    MPC is a simple sine wave. Low-precision measurements 
from ref.62,91 were not used here. We determined the period equal 4.286788 days
using high-precision data: open circles - ref.310, filled circles -
ref.575.

\item {\sl
HD 62509}   MPC is a simple sine wave.
We did not use data from ref.56,299,321,322,623, due to low accuracy.
We determined the period equal 144.36 days using high-precision data
from ref.480.

\item {\sl
HD 62658}   MPC is a simple sine wave. We used the period 4.7249 days
and data from ref.824. Additional measurements
are necessary for improvement of appearance and parameters of the phase curve.

\item {\sl
HD 64740}   MPC is a simple sine wave. Period 1.33026 day was taken 
from Shore \& Brown (1990). Open circles - ref.24, filled circles 
- ref.135 ($H_\beta$ data), filled circles - ref.135 (HeI 5876).

\item {\sl
HD 65339 (1)} MPC is a double wave, but is very close to a single
sinusoid. We determined the period 8.026973 days.
MPC was obtained using high-precision LSD data from
ref.310 measured in metal lines.

\item {\sl
HD 65339 (2)}   MPC is a double sinusoid obtained from hydrogen
lines, ref.33, but is very close to a simple sine wave for 
the mean moment JD24542633.4. Time length for series of measurements
in ref.33 equals 275.3 days.
 
\item {\sl
HD 65339 (3)}   MPC is a double sine wave, obtained in hydrogen 
lines. Filled circles - ref.427, open circles - ref.623.
phase curve is very close to a simple sinusoid for the
mean moment JD24548794.6. Time length of the series of measurements
in ref.427,623 equals 5200. days.
Difference between HD65339(2) and HD65339(3) figures is caused 
most likely by intrinsic variability of the magnetic field.

\item {\sl
HD 65339 (4)} MPC is a double sinusoid which was obtained using
surface field estimates, $B_s$. Open squares - ref.254, open circles - ref.752.

\item {\sl
HD 66665 }  MPC is a simple sine wave. The period and measurements from 
ref.517 were used.  Too little data are available for determination 
of a credible MPC.
New magnetic observations are necessary to determine the period and
parametrs of phase curve.

\item {\sl
HD 66665 (1)}  MPC is a simple sine wave. The period used for the MPC construction
24.5(1) days from ref.769. 
MPC was determined from Si lines in ref.769.
 
\item {\sl
HD 66665 (2)}  MPC is a simple sine wave. MPC was determined from metal lines in ref.769.
 
\item {\sl
HD 66665 (3)}  MPC is a simple sine wave. MPC was determined from N lines in ref.769.

\item {\sl
HD 66665 (4)}  MPC is a simple sine wave. MPC was determined from O lines in
ref.769.

\item {\sl
HD 66765} MPC is a simple sine wave, plotted using the period
1.62 days and data from ref.682.

\item {\sl
HD 67621} MPC is a simple sine wave, following data and period 
3.60 days from ref.682.

\item {\sl
HD 68351} MPC is a simple sine wave. We determined period equal
4.259 days. We did not use low-precise measurements from ref.230,
only those from ref.423.

\item {\sl
HD 70331 (1) } MPC is a simple sine wave with the period 1.99812 days 
from Bagnulo et al, 2002, A\&A, 394, 1023. MPC is very uncertain. 
Additional $B_e$ measurements are necessary for the improvement
of MPC parameters. Filled circles - ref.256, open circles - ref.752.

\item {\sl
HD 70331 (2) }  MPC is a simple sine wave and was constructed using
surface field $B_s$ estimates. Filled circles - ref.254, open circles - ref.752.

\item {\sl
HD 71866}  MPC is a simple sine wave. Low-precise measurements
from ref.3,6,196 were not used. We applied hig-precision data:
filled circles - ref.310, open circles - ref.575.
Period equals 6.80024 days.

\item {\sl
HD 72106} MPC is a simple sine wave. We determined the period 
$P = 0.6389$ days. Open squares - ref.397, filled circles - ref.581, 
filled squares - ref.409, open circles - ref.462.

\item {\sl
HD 72968} MPC is a simple sine wave. We did not use measurements of
low preciseion from ref.1,91,327. Only high-precision estimates from 
ref.423 were used.
Photometric period P =11.305 days (two-wave photometric curve from 
Adelman \& Kaewkornmaung, 2005, A\&A, 435, 1099). This period is possibly
equal to doubled rotational period. However, high-precision measurements
of the magnetic field do not exhibit phase curve with the latter period.
The most probable magnetic period 5.693464 days equals almost accurately
half of the photometric period 11.305 days.
The above magnetic period is close to the value 5.6525 days reported in
Auriere et al., 2007, A\&A, 475, 1058A. Additional measurements are 
necessary for dtermination of the period and parameters of MPC.

\item {\sl
HD 73340}  MPC is a simple sine wave. Open circles - ref.230.

\item {\sl
HD 74521}  MPC is a simple sine wave.
We determined the period $P = 7.148554$ days.
This period is very close to the period from
from Adelman \& Dukes, 2014, AAS, 224, id.322.16 (7.0505 days).
Open squares - ref.230, 
open circles - ref.184, filled squares - ref.333, filled circles - ref.424.
Additional measurements are necessary for determination of MPC parameters.

\item {\sl
HD 74560 (1)} MPC is a double sine wave. MPC was measured in all lines using the period 1.643633 days.
It is one of the probable periods. 
Open circles - ref.513, filled circles - ref.660.

\item {\sl
HD 74560 (2)} MPC is a double sine wave. MPC was obtained from measurements in hydrogen lines. 
Open circles - ref.513, filled circles - ref.457.
Additional measurements are necessary for determination of the period,
form and parameters of the MPC.

\item {\sl
HD 74575}  MPC is a simple sine wave with the period 3.19779 days from ref.513.
Open squares - ref.513 all lines, open circles - ref.513 hydr.lines,
filled circles - ref.571 metall lines.
Measurements from ref.571 do not match MPC.
New high-precision measurements are necessary for the confirmation
of lack of MPC variability. 

\item {\sl
HD 75049 (1)} MPC is a double sinusoid. Period 4.048267 days was taken from ref.691.
Measured by FORS1 in all lines from ref.466.
Average phase curve is close to a simple sinusoid.

\item {\sl
HD 75049 (2)} MPC is a double sinusoid. MPC is close to a simple
sine wave. Measurements were obtained from H lines, ref.466.

\item {\sl
HD 75049 (3)} MPC is a double sinusoid. Measurements obtained by the LSD method in ref.691.
The average curve is close to a simple sine wave.

\item {\sl
HD 75049 (4)}    MPC is a double sinusoid.
We used $B_s$ measurements from ref.691.

\item {\sl
HD 77350}   MPC is a simple sine wave. Filled circles - ref.230, 
open circles - ref.327, filled squares - ref.333.

\item {\sl
HD 79158 } MPC is a double sinusoid. We determined period 3.834898 
days, which is close to the period from ref.407.

\item {\sl
HD 79158 (1)} MPC is a double sinusoid.  We determined period 3.834898 
days, which is close to the period from ref.407.
The MPC was plotted using high-precision LSD measurements 
(metal lines) from ref.407.

\item {\sl
HD 79158 (2)} MPC is a double sinusoid. $B_e$ points were determined from hydrogen lines. 
Open squares - ref.37, open circles - ref.181,
filled squares - ref.327, filled circles - ref.407.

\item {\sl
HD 81009 (1)} MPC is a simple sine wave. We used period 33.984 days 
taken from Bagnulo et al., 2002, A\&A, 394, 1023.
Low-precision photographic field estimates from ref.91 were 
not used, since they show an average offset of -1000 G.
Open circles - ref.256, filled squares - ref.310, filled circles - ref.752.

\item {\sl
HD 81009 (2)}   MPC is a simple sine wave obtained from measurements of the surface 
magnetic field $B_s$, determined using the period 
33.984 days from Bagnulo et al., 2002, A\&A, 394, 1023.
Open squares - ref.218, open circles - ref.254, filled squares - ref.752.

\item {\sl
HD 83368 (1) } MPC is a simple sine wave. MPC was measured in metal lines, ref.285. 
Additional observations are necessary for determination of phase curve parameters.

\item {\sl
HD 83368 (2) }   MPC is a simple sine wave.  Magnetic field data were measured in hydrogen lines, ref.285. 
New observations are necessary for determination the MPC parameters.

\item {\sl
HD 90044}    MPC is a simple sine wave. Open circles - ref.324, 
filled circles - ref.230.

\item {\sl
HD 90569}    MPC is a simple sine wave. We found the new period 3.237356 days.
Only high-precision estimates from ref.423 were used.

\item {\sl
HD 92664}     MPC is a double sinusoid.
We used new magnetic period 1.66558 day. Filled circles - ref.230.

\item {\sl
HD 93030} MPC is a double sinusoid. Period equals 8.8 minutes = 0.006111 days.
Period and $B_e$ estimates were taken from ref.592.
Additional measurements of high time resolution are necessary for 
determination of the period and shape of MPC.

\item {\sl
HD 93507 (1)}    MPC is a double sinusoid. We applied period 568.6 days
taken from Bagnulo et al., 2002, A\&A, 394, 1023.
Open circles - ref.752, filled circles - ref.256.

\item {\sl
HD 93507 (2)}    MPC is a double sinusoid plotted using 
surface field $B_s$ points.
and the period 568.6 days from Bagnulo et al., 2002, A\&A, 394, 1023.
Open circles - ref.254, filled circles - ref.752.

\item {\sl
HD 94660 (1)}    MPC is a double sinusoid drawn using 
surface field meaurements, $B_s$.
Filled squares - ref.254 (FeII 6147/6149),  open circles - ref.752.

\item {\sl
HD 94660 (2)}   MPC is a simple sine wave, phase curve
was obtained from measurements in metal lines. MPC is very uncertain.
Open circles - ref.752, filled circles - ref.754.
Additional measurements are necessary for determination of the period, 
shape and parameters of MPC.

\item {\sl
HD 94660 (3)}    MPC is a simple sine wave, phase curve is very uncertain.
It was obtaied from hydrogen lines. Open squares - ref.230, 
open circles - ref.341,  filled squares - ref.388, filled circles - ref.409, 629.
Additional measurements are necessary for determination of the period, 
shape and parameters of MPC.

\item {\sl
HD 95650 (1)}   MPC is a double sinusoid. We determined the period 
13.880205 days using measurements from ref.438, SET1 therein.

\item {\sl
HD 95650 (2)}   MPC is a double sinusoid with the period 13.831067 days obtained
from measurements of ref.758, SET2 therein.
It is clearly seen, that during about 6 years MPC considerably changed, 
also the period changed (this is the average time length between
SET1 from ref.438 and SET2 from ref.758).
This effect possibly occurred as the result of differential rotation.

\item {\sl
HD 96446 } MPC is a simple sine wave.

\item {\sl
HD 96446 (1)} MPC is a simple sine wave. MPC was plotted using the period 23.38 days from ref.740.  
Measurements were obtained in $H_\beta$ line. Filled circles - ref.24, 
open circles - ref.135. MPC is not satisfactorily filled over phases. 
Additional high-precision measurements are necessary for determination 
of the period and parameters of phase curve.

\item {\sl
HD 96446 (2)} MPC is a simple sine wave.  We found the period 23.381367 days drawn using measurements 
from all lines, ref.184, 740. 
Open circles - ref.184, filled circles - ref.740.
MPC is not well filled over phases. 
Additional high-precision measurements are necessary for 
determination of the period and parameters of MPC.

\item {\sl
HD 96446 (3)}  MPC is a simple sine wave.  We determined period 
23.381367 days using measurements from ref.184,740. 
Open circles - ref.184, filled circles - ref.740.
MPC is not adequately filled over phases. 
Additional high-precision measurements are necessary for determination 
of the period and parameters of MPC.

\item {\sl
HD 96707}  MPC is a simple sine wave.
We did not use low-precise field estimates from ref.91,260,324.
Only high-precision measurements from ref.423 were used here.
We determined magnetic period equal 4.301927 days.
Our period does not agree with photometric periods 7.0286 or 3.516 days 
from Leone \& Catanzaro (2001).
Additional high-precision measurements are necessary for determination 
of the period and parameters of MPC.

\item {\sl
HD 97048 (1)}   MPC is a simple sine wave. We determined period
$P= 0.69471$ days using field estimates obtained in all lines.
Filled circles - ref.409, open circles - ref.609.

\item {\sl
HD 97048 (2)}   MPC is a simple sine wave.  
Measurements of the longitudinal magnetic field were obtained in hydrogen lines
Filled circles - ref.409, open circles - ref.609. 

\item {\sl
HD 98088} MPC simple sinusoid. We used only estimates from ref.615. 
We adopted here the period of 5.90513 days from Catalano \& Leone (1994).
MPC is given here only for primary component. Secondary component
probably has no magnetic field.

\item {\sl
HD 99563} MPC is a double sinusoid.
Filled circles - ref.370,389, open circles - ref.477.
Additional measurements are necessary for determination of the  phase
curve parameters.

\item {\sl
HD101065 } MPC is a simple sine wave with period 188 years according to estimates
from 773. This is currently the longest period
known real time. Naturally for period clarification
and MFB parameters require magnetic and polarimetric
monitoring this unique object.

\item {\sl
HD101412 (1)} MPC is a double sinusoid. Magnetic field measurements 
were obtained in all lines.
Open circles - ref.409,471,509,760,  filled circles - ref.721,782.

\item {\sl
HD101412 (2)}  MPC is a double sinusoid. Magnetic field
measurements were obtained in hydrogen lines.
Open circles - ref.471, filled circles - ref.509, 
open squares - ref.409.

\item {\sl
HD101412 (1)} MPC is a double sinusoid. Magnetic field measurements 
were obtained in BS. filled circles - ref.509.

\item {\sl
HD102195} MPC is a double sinusoid.
Spectral class K0V, magnetic field data were taken from ref.653.
Filled circles - ref.2, filled circles - ref.230.

\item {\sl
HD103192} MPC is a simple sine wave.
Filled circles - ref.2, open circles - ref.230.
Magnetic field estimates are scarce. Additional measurements are necessary
for determination of the phase curve parameters.

\item {\sl
HD103498} MPC is a simple sine wave according to measurements from ref.423.
We determined the most probable period of 18.970293 days,
which differs from that given in ref.423 (15.830 days).
Additional high-precision measurements are necessary for filling of
the phase curve, determination of the period and parameters of the MPC.

\item {\sl
HD104237 } MPC is a simple sine wave according to measurements from ref.781
These measurements are derived for the second component, a star type
TTS. We tried to define the rotation period of the second component
by measurements of magnetic field obtained from ref.781. One of the most
probable periods equals 5.029 days. For determination of the period
additional precise measurements are needed.

\item {\sl
HD105382} MPC is a simple sine wave. 
Period of 1.295 days was adopted from ref.563.
Open circles - ref.760, filled circles - ref. 563.
filled square  - ref.769.

\item {\sl
HD107000 } MPC is a simple sine wave. The period of 5.638 days is taken from ref.710.
Open circles - ref.677,710, filled circles - ref.795. For clarification of MPC parameters
Additional high-precision MF measurements are needed.

\item {\sl
HD107612} MPC is a simple sine wave.
We used the period 1.988 days from ref.402.
Additional measurements are necessary for specification of the period, 
shape and parameters of MPC.

\item {\sl
HD108662} MPC is a double sinusoid. We used a period of 5.07735 day
from ref.826. The MPC is built only on high-precision estimates of ref.826.

\item {\sl
HD108945 (1) }  MPC is a simple sine wave. We used a period of 2.05186 of ref.826.
MPC is built along hydrogen lines. For adjustment of MPC parameters
additional precision measurements are trailed.
           Filled circles - ref.2, open circles - ref.425.

\item {\sl
HD108945 (2) }  MPC is a simple sine wave. We used the period 2.05186
days from ref.826. The MPC is determined only from metal lines. 
Open squares - ref.333, open circles - ref.423, filled square - ref.459, 
filled squares - ref.826.

\item {\sl
HD109026} MPC is a simple sine wave. Only high-precision magnetic field
estimates from ref.682 were used. We found the period 2.864 days, which
differs from 2.84 days in ref.682.

\item {\sl
HD110066} MPC is a simple sine wave. We determined the period 1.92349 days 
using remaining magnetic field measurements.
Open squares - ref.1, filled squares - ref.77, 752, open circles - ref.768.
Additional high-precision measurements are necessary for filling of
the phase curve, determination of the period and parameters of MPC.

\item {\sl
HD110379} MPC is a simple sine wave with the period 12.92696 days.
Open circles - ref.1,61; filled circles - ref.77,327; filled squares - ref.333.

\item {\sl
HD111133} MPC is a simple sine wave. Open squares - ref.1, 
open circles - ref.23, filled squares - ref.142, filled circles - ref.427.

\item {\sl
HD111812} MPC is a double sinusoid.
We determined the period 5.26 days.
Filled circles - ref.685, open circles - ref.731.

\item {\sl
HD112185 (1)}   MPC is a simple sine wave.
High-precision measurements from ref.310 were used here, obtained
from metal lines and LSD method, ref.310.

\item {\sl
HD112185 (2)}    MPC is a simple sine wave.
Magnetic field measurements used here were obtained in hydrogen lines.
Open squares - ref.2, open circles - ref.182, filled squares - ref.271.

\item {\sl
HD112381} MPC is a simple sine wave. Open circles - ref.230.
Additional measurements are necessary for determination of the MPC 
parameters.
We used high-precision measurements from ref.575, obtained in metal
lines by LSD method. Phase curves were plotted using high-precision data 
from ref.310 and ref.575 and significantly differ. Time interval between
measurements presented in ref.310 and ref.575 equals to about 10 years. 
It is necessary to check possible variability of the phase curve over time. 

\item {\sl
HD112413 (1)}   MPC is a double sinusoid.
High-precision measurements were used here, obtained by the
LSD method from metal lines, ref.310. 

\item {\sl
HD112413 (2)}   MPC is a double sinusoid. High-precision measurements 
were used here, obtained by the LSD method from metal lines, ref.575. 
Phase curves obtained from high-precision data from ref.310 and ref.575
significantly differ. Time interval between receipts
data ref.310 and ref.575 are about 10 years. It is necessary to check
possible variation of the phase curve over time.

\item {\sl
HD112413 (3)}   MPC is a double sinusoid. Magnetic field measurements 
were obtained in hydrogen lines. Open squares - ref.33, open circles 
- ref.77, filled squares - ref.215, filled circles - ref.621.

\item {\sl
HD112989} MPC is a double sinusoid.
We found the best magnetic period of 115.3 days 
which is close to the period 111. days from ref.743.
Phase curve was plotted using the period 115.3 days according to ref.743.

\item {\sl
HD115735 } MPC is a simple sine wave. We found a probable magnetic period of 1.297811 days
by measurements from ref.826. For specification of the period
and the MPC parameters require additional high-precision measurements.

\item {\sl
HD116114}     MPC is a simple sine wave.
We found the best magnetic period 2.9406 days which does not coincide 
with the period 4.1156 days from ref.677 and with the photometric period
from Wraight et.al., 2012, MNRAS, 420, 757.
Additional measurements are necessary for determination of the period, 
and parameters of MPC.

\item {\sl
HD116458}      MPC is a simple sine wave.
Only high-precision estimates of the magnetic field were used. 
Low-precision photographic measurements from ref.93, 187 were omitted.
MPC was plotted using the period 148.39 days from
Landstreet and Mathys, 2000, A\&A, 359, 213.
Open squares - ref.187, filled squares - ref.184, 256.

\item {\sl
HD117555 (1)}   MPC is a double sinusoid. We used measurements 
of magnetic field from ref.458 obtained in hydrogen lines
and period 2.4002455(56) from Jetsu et al,A\&A,v.278, p.449.

\item {\sl
HD117555 (2)}   MPC is a double sine wave, very uncertain. 
We used measurements from ref.675 obtained in metal lines. 
Additional measurements are necessary for determination of the
MPC parameters.

\item {\sl
HD118022 (1) } MPC is a simple sine wave. Low-precision 
measurements from ref.1,13,22,64,81,84,324,327 were not used.
MPC was derived from the high-precision 
estimates obtained in metal lines, see ref.310,575.

\item {\sl
HD118022 (2)} MPC is a simple sine wave plotted using measurements
obtained from hydrogen lines. Open squares - ref.2, filled circles - ref.76,
filled squares - ref.77. 
Low-precision measurements. New high-precision magnetic field observations
are necessary for determination of MPC.

\item {\sl
HD119213 (1) }  MPC is a double sinusoid. We used the period
2.4499141 taken by Mikulasek Z.and Ziznovsky J., IBVS 5188, 2001.
The MPC was obtained only from hydrogen lines. Open circles - ref.147.

\item {\sl
HD119213 (2) } MPC is a double sinusoid. 
MPC was plotted using only data from hydrogen lines. Open circles - ref.826.
New high-precision measurements are necessary for determination of 
the reliable MPC.

\item {\sl
HD119419 (1)}   MPC is a double sinusoid plotted using measurements obtained
in hydrogen lines. We used period 2.60059 day from ref.813.
Open circles - ref.230, filled circles - ref.168.

\item {\sl
HD119419 (2)} MPC is a double sinusoid. MPC is built according to BQ estimates
from ref.752.

\item {\sl
HD119419 (3)} MPC is a double sinusoid. MPC is built according to Fe-peak estimates
from ref.813.

\item {\sl
HD119419 (4)} MPC is a double sinusoid. MPC is built according to RRE-lines estimates
from ref.813.

\item {\sl
HD120136 } MPC is a simple sine wave. We determined the best period
3.0342 days according to data from ref.653.  Open circles - ref.653.

\item {\sl
HD120198} MPC is a double sinusoid, which we determined using
measurements and the period 1.38576 days from ref.763.

\item {\sl
HD121743 (1) }  MPC is a simple sine wave. MPC built with the period 1.130170
days from ref.769 and estimated from ref.769 by hydrogen lines.

\item {\sl
HD121743 (2) } MPC is a simple sine wave. MPC built using estimates from ref.769 by met+He lines.

\item {\sl
HD121743 (3) } MPC is a simple sine wave. MPC built using estimates from ref.769 by met lines.

\item {\sl
HD122451 } MPC is a simple sine wave. MPC using met lines of ref.769 
with the period 2.885 days from ref.769.  
Filled circles - ref.184.

\item {\sl
HD122532 (1)}   MPC is a double sinusoid obtained using $B_e$ values 
measured in hydrogen lines.
Filled circles - ref.168, open circles - ref.230.

\item {\sl
HD122532 (2)}   MPC is a double sinusoid obtained from metal lines.
Filled circles - ref.184.

\item {\sl
HD122970}  MPC is a simple sine wave. We applied the period 3.877 days.
Open circles - ref.382, filled circles - ref.370.
We did not use low-precise data from ref.389,398.

\item {\sl
HD124224 (1)} MPC is a double sinusoid measured in hydrogen lines.
Open circles - ref.2,  filled circles - ref.60.

\item {\sl
HD124224 (2)}    MPC is a double sinusoid obtained from metal 
lines, ref.753.

\item {\sl
HD125248 (1)}   MPC is a simple sine wave measured in hydrogen lines,
as in ref.2.

\item {\sl
HD125248 (2)}   MPC is a double sinusoid.
MPC was plotted using high-precision estimates from ref.333 and ref.749.
MPC is a double sinusoid but is very close to a single wave.
Open circles - ref.749, filled circles - ref.333.

\item {\sl
HD125823 (1)}   MPC is a simple sine wave. Points measured in
hydrogen lines from ref.37 - open squares, open circles - 757, filled squares -769.

\item {\sl
HD125823 (2)}   MPC is a simple sine wave, ref.769, metal lines.

\item {\sl
HD126515}      MPC is a double sinusoid.
Open squares - ref.1,18,91; open circles - ref.184; 
filled squares - ref.256,324; filled circles - ref.310,752.

\item {\sl
HD127381}     MPC is a simple sine wave.
We used the period 3.01972 days adopted from ref.570.

\item {\sl
HD128898}  MPC is a simple sine wave. We used the period 4.4790 days 
taken from Kurtz et al. (1994). We did not use estimates from ref.093 
due to apparently low accuracy.
Measurements in hydrogen lines: open circles - ref.2,81,369.
Measurements in metal lines: filled circles - ref.184,256.
Uncertain phase curve. 
Additional high-precision measurements are necessary for determination 
of the period and parameters of MPC.

\item {\sl
HD129333 (1)} MPC is a double sinusoid. MPC and the period were obtained 
from the first set of measurements (2007).
Uncertain MPC. We found the probable period 2.78 days. 
Configuration of the magnetic field is very complex according to ref.716. 
The period is very indefinite since the star rotates differentially 
according to ref.745. 

\item {\sl
HD129333 (2)} MPC is a double sinusoid. MPC and the period was obtained 
from the second set of measurements (2012). 
Uncertain MPC. We found the period 3.10 days. 
Configuration of the magnetic field is very complex according to ref.716.
The period is very indefinite since the star rotates differentially 
according to ref.745. 

\item {\sl
HD130144} MPC is a simple sine wave. We determined the best 
magnetic period of 62.1736 days according to data from ref.523.
Additional high-precision measurements are necessary for determination 
of the period and parameters of MPC.

\item {\sl
HD130322}  MPC is a simple sine wave.
Magnetic period 26.1 days was taken from ref.654.
Additional high-precision measurements are necessary for determination 
of phase curve parameters.
 
\item {\sl
HD131156}  MPC is a simple sine wave.
MPC was drawn using high-precision measurements from ref.385,586. 
We did not use low-precision data from ref.47,61,78,82,298,299,311.
We found the best period 6.24801 days.
Filled circles - ref.385, open circles - ref.586.

\item {\sl
HD133029 (1)} MPC is a double sinusoid. Data were obtained in hydrogen lines.
We applied period P=2.88756 days from Adelman, 2008, PASP, 120, 595.
Measurements of low precision from ref.25 were not used here.
Open circles - ref.2, filled circles - ref.427.

\item {\sl
HD133029 (2)} MPC is a double sinusoid. Data were obtained in metal lines.
We did not use low precision measurements from ref.1.
Open circles - ref.94, filled circles - ref.327.

\item {\sl
HD133652}  MPC is a simple sine wave, measurements in H lines.
We applied here period 2.3040 days.
Open circles - ref.230, filled circles - ref.712.

\item {\sl
HD133880}  MPC is a double sinusoid.
High-precision magnetic measurements from ref.712 allowed us
to determine period P = 0.877475 days.
Filled circles - ref.230, open circles - ref.712.

\item {\sl
HD134793} MPC is a simple sine wave.
We determined the best magnetic period 2.798612 days, which
is close to previously known period 2.78 days.
Filled circles - ref.1, open circles - ref.402.

\item {\sl
HD137509} MPC is a double sinusoid.
Open circles - ref.184,256, metal lines; filled circles - ref.230, H lines.
Relatively old period 4.4912 days from Lanz et al. (1991).

\item {\sl
HD137909 (1)} MPC is a simple sine wave.  MPC was plotted using 
measurements from ref.310, obtained in metal lines. We did not use data 
from ref.1,17,22,31,39,47,51,56,57,59,63,72,84,125,131,184,190,217,256,
324,327,350,362,373,427,621, due to low accuracy..

\item {\sl
HD137909 (2)}    MPC is a simple sine wave.
Measurements were obtained in hydrogen lines.
Open circles - ref.2,76,77, filled circles - ref.621.
We did not use data from ref.25,190 of low accuracy.

\item {\sl
HD137949}  MPC is a simple sine wave.
We found the period for existing $B_e$ measurements
P = 7.151 days. which is close to previously determined
period from ref.677. MPC is very uncertain. 
Additional high-precision measurements are necessary for determination 
of the period and phase curve parameters.
Magnetic field data from ref.1,26,91 were not used due to low accuracy.
Filled circles - ref.256, open circles - ref.677, open squares - ref.752.

\item {\sl
HD140160}  MPC is a simple sine wave.
Filled circles - ref.2, open circles - ref.25, filled squares - ref.333.

\item {\sl
HD142070 (1)} MPC is a double sinusoid. MPC was drawn using surface field 
$B_s$ data from ref.254,752 and period 3.3719 days from ref.677.
Open circles - ref.254; filled circles - ref.752.

\item {\sl
HD142070 (2)} MPC is a double sinusoid.
MPC was drawn using the period 3.3719 days from ref.677.
Open circles - ref.606, open squares - ref.677, filled squares - ref.752.

\item {\sl
HD142184}  MPC is a double sinusoid.
We used measurements from ref.566 obtained by ESPaDOnS and LSD method.
FORS data from ref.566 were not used due to low accuracy.

\item {\sl
HD142301}  MPC is a double sinusoid. 
Data from ref.89.

\item {\sl
HD142990}  MPC is a double sinusoid,
Data from ref.37,230.
Additional high-precision measurements are necessary to determine 
phase curve parameters.

\item {\sl
HD143473}      MPC is a simple sine wave.
Open circles - ref.230, filled circles - ref.184.
Additional high-precision measurements are necessary for determination 
of the MPC parameters.

\item {\sl
HD144334}   MPC is a double sinusoid,
Data from ref.37.
Additional high-precision measurements are necessary to determine 
the MPC parameters.

\item {\sl
HD144897 (1)}  MPC is a simple sine wave. We used period 48.62 days taken from 
Bagnulo et al., 2002, A\&A, 394, 1023.
Open circles - ref.752, filled circles - ref.256.

\item {\sl
HD144897 (2)}  MPC is a simple sine wave. We used period 48.62 days taken from 
Bagnulo et al., 2002, A\&A, 394, 1023.
MPC is a sinusoid, drawn using measurements of the surface field $B_s$.
Open circles - ref.752, filled circles - ref.254.

\item {\sl
HD147010 (1)}    MPC is a double sinusoid.
Magnetic data measured in H lines, from ref.168.

\item {\sl
HD147010 (2)}    MPC is a double sinusoid.
Data from ref.184,256.

\item {\sl
HD147394}   MPC is a simple sine wave.
Most probable magnetic period equals 1.692683 day.
Open circles - ref.478, filled circles - ref.77, filled squares - ref.333.

\item {\sl
HD147513}  MPC is a simple sine wave.
Accepted magnetic period equals 9.174 days determined using data from ref.707.
Low number of points. Additional high-precision measurements are necessary
to determine period and MPC parameters.

\item {\sl
HD148112}  MPC is a simple sine wave.
MPC was plotted using high-accuracy LSD measurements from ref.423.
Phase curve from H lines is very uncertain. 
Data from ref.2,25,389,389,333,427 were not used here.
Period of 3.04296 days was adopted from Hatzes,
MNRAS, 253, 89, 1991.

\item {\sl
HD148199}  MPC is a simple sine wave. 
Filled circles - ref.168, open circles - ref.230.

\item {\sl
HD148330}  MPC is a simple sine wave.
Very uncertain phase curve.
Additional high-precision measurements are necessary for determination 
of the period and MPC parameters.
Open squares - ref.179, open circles - ref.327,427; filled squares - ref.333.

\item {\sl
HD148937}  MPC is a simple sine wave.
We used measurements: 0pen circles - ref.567, filled circles - ref.605.
Single point 2455401.776 from ref.567 was not used.
We applied the period 7.0323 days from ref.624

\item {\sl
HD149438}  MPC is a double sinusoid.
Low-accuracy data from ref.77 were not used, we plotted points
from ref.418 of high accracy.
Star type B0.2V of normal chemical composition.

\item {\sl  
HD149757}   MPC is a simple sine wave.
We determined the best magnetic period 1.38271 days.
Open squares - ref.200, filled circles - ref.605, open circles - ref.681.
Additional high-precision measurements are necessary for determination 
of the period and MPC parameters.

\item {\sl
HD149822} MPC is a simple sine wave.
MPC was plotted with the best magnetic period 1.534013 days.
Note, that  Burke and Barr, 1981, PASP, 93, 344, found different
photometric period $P_{\rm phot}=1.45876 $ days.
Number of magnetic field measurements is low. 
Open squares - ref.230, filled circles - ref.398, open circles - ref.402.
Additional high-precision measurements are necessary for determination 
of the period and MPC parameters.

\item {\sl
HD150193 (1)}   MPC is a simple sine wave.
We only used measurements from ref.609, in all lines.

\item {\sl
HD150193 (2)}   MPC is a simple sine wave.
We only used measurements from ref.609, in H lines.

\item {\sl
HD151199  } MPC is a double sinusoid. We used the period 1.833166 days close 
to the period from ref.826. Filled circles - 402, open circles - 826.

\item {\sl
HD151965}  MPC is a simple sine wave.
Measurements from ref.230, in H lines.

\item {\sl
HD152107 (1)} MPC is a simple sine wave.
Period equals 3.86122 days from ref.630. 
MPC was plotted using data from ref.630, but is very unreliable.
Additional high-precision measurements are necessary for improvement 
of MPC parameters.

\item {\sl
HD152107 (2)} MPC is a simple sine wave.   
Data from ref.2,220,290,427 obtained in H lines.

\item {\sl
HD153882}   MPC is a simple sine wave with the period 6.00858 days.
Open squares - ref.184, open circles - ref.256, filled circles - ref.310,
filled circles - ref.350.
Probably the MPC has more complex shape.
Additional measurements are necessary to clarify shape and parameters
of the phase curve.

\item {\sl
HD154708 (1)}    MPC is a simple sine wave.
Filled circles - ref.473, open circles - ref.389.
Additional measurements are necessary to clarify shape and parameters
of the phase curve.

\item {\sl
HD154708 (2)}    MPC is a simple sine wave.
Measurements obtained in H lines, ref.473.

\item {\sl
HD156324 (1) }  MPC is a double sinusoid.  MPC is built along hydrogen lines 
from ref.769 with period 1.5805(3) days out of ref.769.

\item {\sl
HD156324 (2) }  MPC is a double sinusoid.  MPC is built along lines He of ref.769.
Component A.

\item {\sl
HD156324 (3) } MPC is a double sinusoid. MPC is built along met lines of ref.769.
for Component A.

\item {\sl
HD156424 (1) }  MPC is a simple sine wave. We used the period 2.8721 days from ref.769.
The MPC is built along hydrogen lines from ref.769.
The MPC is very uncertain. For clarification of the MPC parameters
additional precise measurements are required.

\item {\sl
HD156424 (2) }  MPC is a simple sine wave.  MPC is built along met lines from ref.769.
The MPC is very uncertain. For clarification of the MPC parameters
additional high-precision measurements are required.  

\item {\sl
HD156424 (3) } MPC is a simple sine wave. MPC is built along Si lines from ref.769.
The MPC is very uncertain. For clarification of the MPC parameters
additional high-precision measurements are required.

\item {\sl
HD163472 (1) } MPC is a simple sine wave.
We plotted only high-precision measurements in all lines from ref.584 using the period
3.638833 days from ref.333. Data from ref.333,379,406 were not used.

\item {\sl
HD163472 (2) } MPC is a simple sine wave. 
We plotted only high-precision measurements only in He lines from ref.584

\item {\sl
HD163472 (3)} MPC is a simple sine wave. 
We plotted only high-precision measurements by all lines without He from ref.584

\item {\sl
HD164258}         MPC is a simple sine wave. 
Filled squares - ref.98, open circles - ref.327, open squares - ref.41.

\item {\sl
HD164492c (1) }  MPC is a simple sine wave.  MPC is built along metal lines ref.738 with period
1.59689 days from ref.738. C2 component. 

\item {\sl           
HD164492c (2) }  MPC is a simple sine wave. MPC is built along metal lines ref.739 with period
1.36986 days from ref.739. C2 component.

\item {\sl
HD165474}  MPC is a simple sine wave.
One of the probable magnetic periods equals 1.381082 days.
Other periods also are possible. The MPC is uncertain.
New measurements of the magnetic field + polarimetry are necessary 
to determine the period and the MPC parameters.
Open squares - ref.184,256, open circles - ref.677, filled squares - ref.762.

\item {\sl
HD166473 (1)} MPC is a simple sine wave
obtained from measurements of the surface magnetic field $B_s$.
We determined the period 3892.565 days.
Open circles - ref.444, open squares - ref.752.

\item {\sl
HD166473 (2)}  MPC is a simple sine wave, obtained
with our period 3892.565 days, points from ref.584. LSD method.

\item {\sl
HD168733}  MPC is a simple sine wave.
Weakly probable period is 14.523906 days.
Filled squares - ref.256, open circles - ref.86, filled circles - ref.184.

\item {\sl
HD169842}  MPC is a simple sine wave.
Open squares - ref.402, open circles - ref.459, filled squares - ref.621.
We found the most probable period 3.426 days. 
Additional measurements are necessary to improve the period and MPC parameters.

\item {\sl
HD170000 (1) }   MPC is a simple sine wave.
Data of low accuracy from ref.25 were not used. 
We used old magnetic field estimates from ref.60 in hydrogen lines.

\item {\sl
HD170000 (2) }  MPC is a double sine wave. 
The MPC is built from all lines from ref.826 and with the period 1.71649 days of ref.826.

\item {\sl
HD170153 }  MPC is a double sine wave, which was drawn using only 
high-precision $B_e$ observations from ref.766.
We used the period 23.39 days from ref.766.

\item {\sl
HD170397}  MPC is a simple sine wave.
Data were obtained in hydrogen lines. 
Open squares - ref.2, metal lines; open circles - ref.184, filled squares - ref.230.

\item {\sl
HD170973}  MPC is a simple sine wave.
Open circles - ref.230, H lines; filled circles - ref.285,402, metal lines.
We determined the period 7.0666 days. 
Additional measurements are necessary to improve the period and MPC parameters.

\item {\sl
HD171586}  MPC is a simple sine wave
using data and period 2.1308 days from ref.423.
Number of available $B_e$ points is low.
Additional measurements are necessary to improve the period and MPC parameters.

\item {\sl
HD171782}  MPC is a simple sine wave.
Measurements and the period 4.4674 days were taken from ref.423.
Few data, additional measurements are necessary to improve the period and MPC parameters.

\item {\sl
HD172167}  MPC is a simple sine wave.
Data from Plachinda S. and Butkovskaya V. (personal communication). 
The period $P = 0.62255$  days.

\item {\sl
HD173650}  MPC is a simple sine wave.
We determined the best magnetic period 17.473658 days. 
Filled circles - ref.1.
New measurements are necessary to improve the period and MPC parameters.

\item {\sl
HD174638}  MPC is a simple sine wave.
The star $\beta$ Lyr, we applied the orbital period 12.9374 days.
Only $B_e$ measurements in Si6371 line show a weak phase dependence 
of $B_e$ for this period.
Filled circles - ref.344.

\item {\sl
HD175156}  MPC is a simple sine wave, was
plotted using data from ref.37. We determined the probable period
3.081 days. Low number of measurements. 
Additional measurements are necessary to refine the shape and MPC parameters.

\item {\sl
HD175362 (1) }  MPC is a double sinusoid with the period 3.67381 days from ref.769.
The MPC was built from hydrogen lines. Open squares - ref.37, open circles - ref.135, 
filled squares - ref.769.

\item {\sl
HD175362 (2)}  MPC is a double sinusoid. MPC is built along He lines from ref.769.

\item {\sl
HD175362 (3)}  MPC is a double sinusoid. MPC is built along met. lines from ref.769.

\item {\sl
HD175362 (4)}  MPC is a double sinusoid. MPC is built along Si lines from ref.769.

\item {\sl
HD175362 (5)}  MPC is a double sinusoid. MPC is built along Fe lines from ref.769.

\item {\sl
HD176386 (1)}  MPC is a simple sine wave.
Measurements in all lines.
Filled circles - ref.471, open circles - ref.609.

\item {\sl
HD176386 (2)}  MPC is a simple sine wave.
Measurements in H lines, from ref.609.

\item {\sl
HD176582 (1) }  MPC is a simple sine wave. We used the period 1.58984 days 
from ref.527. MPC is built on metal lines from ref.769.

\item {\sl
HD176582 (2) }  MPC is a simple sine wave.
MPC is built on H lines from ref.769.

\item {\sl
HD176582 (3) }  MPC is a simple sine wave.
MPC is built on He lines from ref.769.

\item {\sl
HD176582 (4) }  MPC is a simple sine wave.
MPC is built on Si lines from ref.769.

\item {\sl
HD178892 (1)}      MPC is a simple sine wave.
Data were obtained in metal lines.
Open squares - ref.361, open circles - ref.402, 
filled squares - ref.621, filled circles - ref.710.

\item {\sl
HD178892 (2)}      MPC is a simple sine wave.
Data were obtained in H lines.
Open squares - ref.621, filled circles - ref.677, filled squares - ref.710.

\item {\sl
HD179527}   MPC is a simple sine wave. 
We determined the most probable period 19.68 days.
Filled circles - ref.2, open circles - ref.423.

\item {\sl
HD180642}      MPC is a simple sine wave
with the period 13.893 days from ref.537.
Open circles - ref.537, open squares - ref.457, filled squares - ref.571.
The latter measurements contradict the MPC.
New observational data are necessary to eliminate this contradiction.

\item {\sl
HD181615}  MPC is a simple sine wave.
Data are too scarce to construct credible phase curve.
We determined one of the probable periods P = 1.44 days.
Additional measurements are necessary to clarify the period and MPC parameters.
Open circles - ref.591,474; filled circles - ref.475.

\item {\sl
HD182180}  MPC is a simple sine wave.
Filled circles - ref.491, open circles - ref.614.

\item {\sl
HD183056}  MPC is a simple sine wave.
Open circles - ref.423.
The period equals 2.9919 days following ref.423.

\item {\sl
HD183339}  MPC is a simple sine wave.
We determined the period P = 2.423361 days.
Open circles - ref.40, filled circles - ref.427.

\item {\sl
HD184471 (1)} MPC is a simple sine wave.
Measurements obtained in metal lines.
Open circles - ref.402, filled circles - ref.621.

\item {\sl
HD184471 (2)} MPC is a simple sine wave.
Measurements obtained in H lines.
Open circles - ref.402, filled circles - ref.621.
MPC is very uncertain.
Additional measurements are necessary to clarify the period and MPC parameters.

\item {\sl
HD184927 (1) } MPC is a simple sine wave. The period specified by us is
9.531328 days, which is close to the period from ref.281.
Only high-precision estimates from ref.769 were used to generate 
the MPC from hydrogen lines.

\item {\sl
HD184927(2) } MPC is a simple sine wave. MPC is built according to 
measurements in He lines from ref.676.

\item {\sl
HD184927(3) } MPC is a simple sine wave. MPC is built according to 
measurements in He 4471 line from ref.676.

\item {\sl
HD184927(4) } MPC is a simple sine wave. MPC is built according to 
measurements in He 6678 line from ref.676.
MF points were measured in lines He 4471 - He 6678.

\item {\sl
HD184927(5) } MPC is a simple sine wave. MPC is built according to 
measurements in all He lines from ref.676.

\item {\sl
HD184927(6) } MPC is a simple sine wave. MPC is built according to 
measurements in met lines from ref.676.

\item {\sl
HD184927(7) } MPC is a simple sine wave. MPC is built according to 
measurements in Si line from ref.676.

\item {\sl
HD186205 (1) }   MPC is a simple sine wave. We have specified the period 
37.11657 days close to the period of ref.769. MPC is built along metal lines.
filled circles - 427, open circles - 769.

\item {\sl
HD186205 (2) } MPC is a simple sine wave. MPC is built along hydrogen lines from ref.769.
The MPC is very uncertain.

\item {\sl
HD186205 (3) } MPC is a simple sine wave. MPC is built along Si lines from ref.769.

\item {\sl
HD186205 (4) } MPC is a simple sine wave. MPC is built along He lines from ref.769.
The MPC is very uncertain.

\item {\sl
HD187474 (1) }   MPC is a double sine wave. The MPC was plotted
using the period 2345.8 days.
Open squares - ref.1,184,  open circles - ref.256, 
filled squares - ref.510,760, filled circles - ref.752,826.

\item {\sl
HD187474 (2) }  MPC is a double sinusoid. The MPC was plotted 
using $B_s$ estimates.
Filled circles - ref.218, open circles - ref.254, filled squares - ref.752.

\item {\sl
HD188041}  MPC is a simple sine wave, plotted using
magnetic field measurements from ref.4. This is the
longest homogeneous time series for this star. All
other measurements from ref.1,15,16,232,256,756,760
were not used, because they were obtained using various
methods and consist of lower number of observations.
We applied the period 224.5 days from Musielok (1986), 
Acts Astronomica, 36, 131.
MPC shows a large scatter of individual points.
Additional high-precision measurements are necessary to 
determine the period, shape and the MPC parameters.

\item {\sl
HD188209 } MPC is a simple sine wave. Our estimated period is 3.17 day.
filled circles - ref.770, open circles - ref.733. To refine the period 
and parameters the MPC needs additional high-precision observations. 

\item {\sl
HD189733}   MPC is a simple sine wave.
Estimated period equals 7.667536 days. 
Filled circles - ref.485, open circles - ref.498.

\item {\sl
HD189775 (1) }  MPC is a double sinusoid. The period 2.6071 days from ref.769 was used.   
The MPC was built along hydrogen lines from ref.769. 
Filled squares - ESPaDOnS, open circles - Narval.

\item {\sl
HD189775 (2) }  MPC is a double sinusoid. 
The MPC was built along metal lines from ref.769. 
Filled squares - ESPaDOnS, open circles - Narval.

\item {\sl
HD189775(3) } MPC is a double sinusoid. 
The MPC was built along He lines from ref.769. 
Filled squares - ESPaDOnS, open circles - Narval.

\item {\sl
HD189775 (4) }  MPC is a double sinusoid. 
The MPC was built along Si lines from ref.769. 
Filled squares - ESPaDOnS, open circles - Narval.

\item {\sl
HD189849}   MPC is a simple sine wave. Uncertain MPC. For specification
additional parameters of MPC and period are required
high-precision measurements.
Open squares - ref.10, open circles - ref.97,98,  filled circles - ref.327.

\item {\sl
HD190073}  MPC is a simple sine wave,
was plotted using data from ref.583. 
Additional high-precision measurements are necessary to 
determine period, shape and MPC parameters.
Open squares - ref.400, open circles - ref.583, filled squares - ref.782, 
filled circles - ref.822.

\item {\sl
HD191612}  MPC is a simple sine wave,
plotted using data from ref.560. 
The period equals 527.2 days from ref.560.

\item {\sl
HD192678 (1)} MPC is a double sinusoid,
according to measurements of the surface field $B_s$ and the period 
6.4186 days from ref.255.

\item {\sl
HD192678 (2)} MPC is a double sinusoid.
The period P = 6.4186 days from ref.255.
Open circles - ref.255, filled circles - ref.427.

\item {\sl
HD194093}  MPC is a simple sine wave.
Filled squares - ref.76,140,189, filled circles - ref.299, filled squares - ref.516.

\item {\sl
HD196178}  MPC is a simple sine wave.
Open circles - ref.2, filled circles - ref.327.

\item {\sl
HD196502}  MPC is a simple sine wave.
Open squares - ref.1,7,22, open circles - ref.190, filled squares - ref.327.

\item {\sl
HD200311 (1)}  MPC is a simple sine wave for the surface field
$B_s$ measurements with the period 52.0084 days from Adelman (1997), A\&AS, 122, 249.
Open squares - ref.254, open circles - ref.752.
Some phases are not adequately covered by observations.
Additional high-precision measurements are necessary to 
improve the period, shape and MPC parameters.

\item {\sl
HD200311 (2)}  MPC is a simple sine wave
with the period 52.0084 days from Adelman (1997), A\&AS, 122, 249.
Filled circles - ref.291.
Some phases are not well covered by observations.
Additional high-precision measurements are necessary to 
improve the period, shape and MPC parameters.

\item {\sl
HD200775}    MPC is a simple sine wave
according to data from ref.454 with the period P=4.328 days.

\item {\sl
HD201091 (1)}   MPC is a simple sine wave.
The period equals 36.618 days, measurements were taken from ref.623.

\item {\sl
HD201091 (2)}   MPC is a simple sine wave.
New high-precision measurements were presented in ref.746.
Authors studied therein the magnetic cycle equal 7.1 years in this star.
Reliable phase curve can be obtained only using measurements of
the set 2015.54. We determined the period 36.889 days using these data.

\item {\sl
HD201174 (1) }  MPC is a double sinusoid. MFB is built along hydrogen lines.
We found the most probable period equal 2.430301 days.
Filled circles - ref.732, open circles - ref.774.

\item {\sl
HD201174 (2) }  MPC is a double sinusoid. MFB is built along metal lines.
Open squares - ref.548, open circles - ref.732, filled squares - ref.768, 
filled circles - ref.774.

\item {\sl
HD201601} MPC is a simple sine wave.
We used new data and determined the period of magnetic variability,
$P=38312. /pm 3506.$ days. This period differs from the previous
value 35462.5 days, see MNRAS, 455, 2567 (2016).
open squares - ref.1,46,47,48,105,146,286, open circles - ref.184,373,384,702,768,774,826,
filled squares - ref.327, filled circles - ref.775.

\item {\sl
HD205021 (1) }  MPC is a simple sine wave  with the period 12.00075 days from ref.655.
MPC is built along metal lines. Open circles - ref.655, filled circles - ref.769.

\item {\sl
HD205021 (2) }  MPC is a simple sine wave.
MPC is built along He lines - ref.769.

\item {\sl
HD205021 (3) }  MPC is a simple sine wave.
MPC is built along O lines - ref.769.

\item {\sl
HD205021 (4) }  MPC is a simple sine wave.
MPC is built along Si lines - ref.769.

\item {\sl
HD206860}  MPC is a double sinusoid
with the period 4.66 days. Measurements from ref.716.
Large scatter of points indicate probable variations of phase
curve. Additional high-precision observations are necessary
to solve this problem.

\item {\sl
HD207330}  MPC is a simple sine wave.
Estimated magnetic period equals 1.290502 days.
Additional high-precision measurements are necessary to 
improve the period and MPC parameters.
Filled circles - ref.475, open circles - ref.478.

\item {\sl
HD208057 (1) }  MPC is a double sinusoid. Period is 1.3678 days out of ref.769.
          MPC is built along metal lines from ref.769. 

\item {\sl
HD208057 (2) }  MPC is a double sinusoid.  
The MPC is built along He lines from ref.769.

\item {\sl
HD208217}    MPC is a simple sine wave
with the period 8.44475 days and measurements from ref.752, see
Manfroid \& Mathys, 1997, A\&A, 320, 497.
Unfortunately some phases of the MPC are not covered by observations.
Additional data are necessary to improve the shape and MPC parameters.

\item {\sl
HD209290}  MPC is a double sinusoid.
The period equals 10.73 days, data were taken from ref.758.
Points exhibit large scatter around the average phase curve, which 
can be caused by its intrinsic variations.
Additional magnetic measurements are necessary to clarify this issue. 

\item {\sl
HD210873}  MPC is a simple sine wave. 
The MPC was drawn assuming the period 1.686426 days for data from ref.427.
Additional high-precision observations are needed to  
improve the period and MPC parameters.

\item {\sl
HD214680}  MPC is a simple sine wave 
for the assumed period 4.670431 days. Only high-precision observations
from ref.681 were used here.
Additional high-precision measurements are necessary to 
improve the period and MPC parameters.

\item {\sl
HD215441 (1)}    MPC is a simple sine wave
for the surface magnetic field points $B_s$ from ref.12.
The period equals 9.4871 days.

\item {\sl
HD215441 (2)}     MPC is a simple sine wave
obtained using measurements of low accuracy in H lines.
Filled circles - ref.30, open circles - ref.190, filled squares - ref.775.
The MPC cannot be determined using data obtained in metal lines.
New high-precision magnetic observations are necessary for 
this star. It is possible that there exists a slower secular
evolution of the magnetic field overlapped by the common
periodic rotational variability of its longitudinal component. 
Note, that the MPC of this famous star still is not credibly
determined. Additional photometric studied are needed for
refinement of the period, possibly using database of ASAS3 
or other robotic survey.

\item {\sl
HD217522 } MPC is a simple sine wave. We used the period 0.183009 days
close to period from Mellon et al., ApJSS, 244:15, 2019.
Filled circles - ref.256, open circles - ref.760, filled squares - ref.826.
It is not enough measurements. For clarification of MPC parameters and period
Additional precision measurements are needed.

\item {\sl
HD217833 (1)}  MPC is a simple sine wave,
we used the period 5.393074 days from Monin (personal communication).
Magnetic field measurements and the MPC were obtained in metal lines.
Filled circles - ref.40, open circles - ref.233.

\item {\sl
HD217833 (2)}  MPC is a simple sine wave.
Data were measured in H lines. 
Filled circles - ref.230, open circles - ref.427.

\item {\sl
HD218376}  MPC is a simple sine wave.
The probable period equals 7.071686 days.
Open circles - ref.478, filled circles - ref.475.

\item {\sl
HD219134 }  MPC is a simple sine wave determined using high-precision measurements ref.798.
We used the period 22.247 days found from magnetic measurements.
The MPC was determined at the accuracy limit.

\item {\sl
HD220825}  MPC is a double sinusoid
for high-precision measurements from ref.423.

\item {\sl
HD221760 } MPC is a simple sine wave. We used the period 5.941 days close
to the period of ref.826.   Open circles - ref.2, filled circles - ref.760,
filled squares - ref.826.  Additional determinations of the MPC and the period 
from high-precision measurements are necessary.

\item {\sl
HD221936}  We used data from ref.548. MPC is a double sinusoid
with the period 1.716906 days, which corresponds best to those data.

\item {\sl
HD223460}  MPC is a double sinusoid,
it is very close to a simple sinusoid, with the period 24.2 days
from ref.731. We did not include two $B_e$ points from ref.731 of exceptionally
large deviation: JD2456553.52  -345.0 G and JD2456597.38  -145.0 G.
All measurements from ref.685 were shifted in phase by 0.505.
Filled circles - ref.683, open circles - ref.731.

\item {\sl
HD223640 (1) }  MPC is a double sinusoid. We used the improved period 3.758272 days.
The MPC is built along H lines of ref.224. Low-exact estimates. 
High-precision measurements are required for determination of the MPC parameters.

\item {\sl
HD223640 (2) }  MPC is a double sinusoid. MPC is built along metal lines.
Filled circles - ref.389, open circles - ref.826.
It is not enough MF determinations. To specify the MPC parameters,
additional high-precision measurements are necessary.

\item {\sl
HD224085}  MPC is a simple sine wave.
We determined the most probable magnetic period and the period
of rotation equal 6.759127 days. It is longer than the period
6.7242078 from ref.664.
Open squares - ref.616, open circles - ref.664, filled squares - ref.693.

\item {\sl
HD226868}  MPC is a sinusod of the period 2.799918 days (half period 5.599836 days from 
Brocksopp et al., 1999, A\&A, v.343, p.861)  and data from ref.595.
This is the unique binary star Cyg X-1. The MPC is very uncertain.
Additional high-precision measurements are necessary to 
improve the shape and the MPC parameters.

\item {\sl
HD258686}  MPC is a simple sine wave.
We determined one of the probable periods equal 1.597674 days.
Open circles - ref.402, filled circles - ref.621.
Additional measurements are necessary to improve the shape 
and the MPC parameters.

\item {\sl
HD265866 }   MPC is a simple sine wave. We determined one of the probable periods
21.012 days. Filled circles - ref.821. The MPC is not fully filled. Additional
observations are necessary for clarification of the MPC parameters.

\item {\sl
HD281934}  MPC is a simple sine wave
obtained in all lines, Data points and the period 7.65 days
were taken from ref.445.

\item {\sl
HD283518 } MPC is a double sinusoid. The period 1.87197 days and high-precision 
estimates from ref.823.

\item {\sl
HD293764}  MPC is a simple sine wave.
Open circles - ref.402, filled circles - ref.621.
We determined the most probable period equal 2.112164 days.

\item {\sl
HD318107 (1)} MPC is a simple sine wave
with the period 9.7085 days adopted from
Manfroid \& Mathys, (2000), A\&A, 364, 689.
Open squares - ref.256, open circles - ref.561, filled squares - ref.752,
filled circles - ref.760. Additional measurements of $B_e$ are 
necessary to clarify the shape and the MPC parameters.

\item {\sl
HD318107 (2)} MPC is a simple sine wave,
but possibly it has a more complex shape. The MPC was plotted 
using surface field $B_s$ measurements. Large scatter of points.
Open squares - ref.561, open circles - ref.752.

\item {\sl
HD343872 (1)} MPC is a double sinusoid.
Measurements were obtained in metal lines.
Open circles - ref.402, filled circles - ref.732,768,
open squares - ref.621, filled squares - ref.710.

\item {\sl
HD343872 (2)} The MPC is a double sinusoid.
Measurements were obtained in hydrogen lines.
Filled circles - ref.732, open squares - ref.621,
open circles - ref.710., filled circles - ref.768.

\item {\sl
HD345439 (1)} The MPC is a double sinusoid
with the period 0.77018 days from ref.736.
Data points were measured in hydrogen lines, from ref.736.

\item {\sl
HD345439 (2)} The MPC is a double sinusoid
with the period 0.77018 days from ref.736.
Data poins were measured in all lines, from ref.736.

\item {\sl
HD349321} The MPC is a double sinusoid.
We determined the most probable period 2.503768 days.
Open circles - ref.402, filled circles - ref.621.
Additional measurements are necessary to improve the period
and the MPC parameters.

\item {\sl
V1005 Ori} 
The MPC is a simple sine wave
with the period 4.35 days from ref.438 and data from ref.438.
Few estimates, very uncertain MPC.
Additional measurements are necessary to improve the shape and
the MPC parameters.

\item {\sl
V1079 Tau (1) } 
The MPC is a double sinusoid. MPC is built on all lines. The best period
$P=5.736 \pm 0.460$. Filled circles - ref.803.

\item {\sl
V1079 Tau (2) } 
The MPC is a double sinusoid. MPC is built on CaII 8542 line. 
The MPC has large amplitude and is shifted
relative to the MPC obtained from all lines.
Filled circles - ref.803.

\item {\sl
V380 Ori}  
The MPC is a simple sine wave
Open circles - ref.468, filled circles - ref.409, filled squares - ref.397.

\item {\sl
BD+40 175B } 
The MPC is a simple sine wave.
Open ircles - ref.272, filled squares - ref.774, filled circles - ref.710.
We found one of the probable periods P = 1.803 days.
Additional measurements are necessary to improve the shape 
and the MPC parameters.

\item {\sl
BD+40 175A }            
The MPC is a simple sine wave.
We found one of the probable periods of 5.367.
Additional measurements are required to clarify the period and parameters of MFB.
Open circles - ref.289, filled circles - ref.547, filled squares - ref.710, 
open squares - ref.774.

\item {\sl
CD-48 11051 (1) } 
The MPC is a simple sine wave. 
We used a period of 1.678 of ref.769.
MPC is built along hydrogen lines. 
filled circles - ref.388, open circles - ref.769.

\item {\sl
CD-48 11051 (2) } 
The MPC is a simple sine wave. 
MPC is built along metal lines. 
filled circles - ref.388, open circles - ref.769.

\item {\sl
BD-19 5044L} 
The MPC is a simple sine wave
obtained using magnetic data from ref.748 and from ref.388.
We refined magnetic period to $P_{\rm mag}= 5.0382$ days, which
is very close to the period 5.040 days from ref.748. Magnetic
measurements refer to the primary component of the SB2 binary system,
which is a Ap star. Orbital period equals 17.63 days.

\item {\sl
DT Vir (1)} 
The MPC is a simple sine wave 
with the period 2.85 days. Average JD2555163.36 - dt = 9.96 days, from ref.438.

\item {\sl
DT Vir (2)} 
The MPC is a simple sine wave 
with the most probable period 3.2015 days determined by us. 
Average JD2555488.18 - dt = 50.96 days, data from ref.438.
The average value of the field did not change, but the period and
amplitude of variations apparently changed.

\item {\sl
CE Boo } 
The MPC is a simple sine wave 
plotted using data from ref.438. Note, that the best magnetic
period equals 14.049 days and not 14.7 days.

\item {\sl
OT Ser (1)} 
The MPC is a double sinusoid, Set 1 from ref.438.

\item {\sl
OT Ser (2)} 
The MPC is a double sinusoid, Set 2 from ref.438.
During half year time interval between both sets of measurements,
parameters of magnetic phase curve considerably changed, including
the average longitudinal field strength and amplitude of variations.

\item {\sl
DS Leo (1)} 
The MPC is a double sinusoid.
Period equals 14.0 days from ref.438. Measurements were carried out
in two sets. The first set: mean JD2454131.64, del t = 10. days.  
In the figure, MPC based on Set 1 is drawn by the solid curve. 
Filled circles - ref.438.

\item {\sl
DS Leo (2)} 
The MPC is a double sinusoid.
The second set: mean JD2454488.14, del t = 51. days.
Time difference between both sets of observations is 356.5 days,
or 1 year. MPC significantly changed during this time period.
Minimum of the field strength has shifted and the shape of phase
curve changed, which is characteristic for red dwarfs.
Filled circles - ref.438. 

\item {\sl
DS Leo (3)} 
The MPC is a double sinusoid.
MPC includes both sets of measurements from ref.438.

\item {\sl
BD+61 195 } 
The MPC is a double sinusoid.
MPC was drawn using the period 18.6 days and measurements from ref.438.

\item {\sl
YZ CMi (1)} 
The MPC is a simple sine wave, Set 1 from ref.439.

\item {\sl
YZ CMi (2)} 
The MPC is a double sinusoid, Set 2 from ref.439. 
During half year time interval between both sets of measurements,
parameters of magnetic phase curve considerably changed, including
the average longitudinal field strength and amplitude of variations.

\item {\sl
EQ Peg A } 
 MPC is a simple sine wave.
Phase curve is very incomplete, ref.439.

\item {\sl
EQ Peg B } 
 MPC is a double sinusoid.
Data from ref.439.

\item {\sl
WD1953-011 } 
 MPC is a simple sine wave
for measurements from ref.456 and the period 1.4479487 days.

\item {\sl
V2129 Oph (1) } 
 MPC is a double sinusoid
obtained using averaged field estimates. 
Few measurements are available,  MPC is incomplete. 
We determined and applied period 6.696 days, close to 
the period 6.53 days from ref.502.
Open circles - ref.502, filled circles - ref.610.

\item {\sl
V2129 Oph (2) } 
 MPC is a double sinusoid,
measurements were obtained in emission cores of Ca lines, ref.502.

\item {\sl
V2129 Oph (3) } 
The MPC is a simple sine wave,
measurements were obtained in helium lines, ref.502.

\item {\sl
V388 Ori }  
 MPC is a double sinusoid.
Data from ref.507,538, period equals 1.0209 days.
Study of the star is difficult, since the period is close to 1 day.

\item {\sl
GL Vir }  
 MPC is a simple sine wave
with the period 0.491 days from ref.507.
MPC is scarcely filled.

\item {\sl
LHS 3495 (1) } 
 MPC is a double sinusoid
with the period 0.704911 days, which is close to the period
given in ref.507.  MPC changes from set to set, i.e. from
year to year. On the HD500255 figure we plotted all three phase
curves: 2006 - filled squares - ref., 2007 - filled circles, 2008 - open circles.

\item {\sl
LHS 3495 (2) }  
 MPC is a double sinusoid. 2007 - filled circles.

\item {\sl
LHS 3495 (3) }  
 MPC is a double sinusoid. 2008 - open circles.  

\item {\sl
WX UMa }  
 MPC is a double sinusoid
with the period 0.781598 days, which is close to the period from ref.507.

\item {\sl
DX Cnc }  
 MPC is a simple sine wave
with the period 0.412788 days, which is close to the period from ref.507.
Large scatter of points.

\item {\sl
LHS 292 }  
 MPC is a simple sine wave
with the period 1.467955 days, which is close to the period from ref.507.
Large scatter of points.

\item {\sl
GJ 1154 A }  
 MPC is a double sinusoid
with the probably period 0.406029 days determined by us. 
Filled circles - ref.507.

\item {\sl
GJ 1224 }  
 MPC is a simple sine wave
with the period 0.399347 day determined by us. Data from ref.507.
MPC simple wave with period 54. dayes out of ref.821 and estimated
received in ref.821.

\item {\sl
GJ 644 C } 
MPC is a simple sine wave
with the period 0.543972 day determined by us. Data from ref.507.

\item {\sl
V1298 Aql }  
 MPC is a simple sine wave
with the period 0.53775 day, which is close to the period from ref.507.
MPC is scarcely filled, additional measurements are necessary here.

\item {\sl
CPD-28 2561 (1) }  
 MPC is a simple sine wave.
with the period 73.41 days from ref.679 by all lines.
Filled circles - ref.679, open circles - ref.680. 

\item {\sl
CPD-28 2561 (2) } 
 MPC is a simple sine wave.
with the period 73.41 days from ref.679. No emission lines were used here.

\item {\sl
BD+53 1183 } 
 MPC is a simple sine wave.
Estimated period is 2.7 days. It is not enough measurements. 
To refine the period and parameters MPC requires additional measurements.
Filled circles - ref.548, open circles - ref.768. 

\item {\sl
Tr16-22 }  
MPC is a simple sine wave.
Period 54.4 days, which was found from X-ray variability in 0.5-10 keV 
range from ref.742, does not allow to construct MPC.
We determined the best period 4.97 days using the available 
$B_e$ estimates. Magnetic field measurements are scarce.
Additional magnetic and, preferably, polarimetric measurements 
are necessary to improve the period and MPC parameters.
Filled circles - ref.742.

\item {\sl
NGC1624-2 }  
 MPC is a simple sine wave
drawn with the period 157.99 days from ref.574, using data from ref.733.
Many phases in MPC are not covered by observations.  
Additional high-precision measurements are necessary to improve the shape 
and the MPC parameters.
Filled circles - ref.733.

\item {\sl
CPD-57 3509 (1) }  
 MPC is a simple sine wave
using measurements from ref.764 (in hydrogen lines) and the
period 6.36093 days from ref.764. Large scatter of points.
Additional measurements are necessary to improve the period 
and the MPC parameters.
Filled circles - ref.764.

\item {\sl
CPD-57 3509 (2) } 
 MPC is a simple sine wave
using meaurements from ref.764 (in all lines) and the
period 6.38255 days from ref.764. There is a large scatter 
of measurements around magnetic phase curve.
Additional measurements are necessary to improve the period 
and the MPC parameters.
Filled circles - ref.764.

\item {\sl
CPD-62 2124 (1) } 
The MPC is a double sinusoid
using the period 1.45031 days from ref.737. Magnetic field
measurements were obtained in hydrogen lines, ref.737.

\item {\sl
CPD-62 2124 (2) }  
The MPC is a double sinusoid
using the period 1.45031 days from ref.737. Magnetic field
measurements were obtained in all lines, ref.737.

\item {\sl
CD-51 6859 } 
The MPC is a double sinusoid.
We applied the period 24.04 days from ref.758 and used data from ref.758. 

\item {\sl
CD-40 5404 } 
The MPC is a simple sine wave.
We applied the period 25.37 days from ref.758 and used data from ref.758.

\item {\sl
UV Cet } 
The MPC is a simple sinusoid.
We applied the period 0.11345 day which is half of the 
period 0.2269 days from ref.759 and used data from ref.759.

\item {\sl
CPD-83 64B } 
The MPC is a simple sinusoid with the period 
1.84933 days of ref.786 using measurements from ref.786.

\item {\sl
V1000 Sco } 
The MPC is a simple sinusoid.
Period 2.58 days from Adams et al., 1998, (ApJ, 116, 237) does 
not satisfy magnetic measurements.
The most likely magnetic period is 4.26 days. For specification of the period
and the MPC parameters require additional measurements.
MPC for measurements ref.799 with the period 4.26 days.

\item {\sl
V1156 Sco } 
The MPC is a simple sinusoid.
The period 2.15 days is from Adams et al., 1998 (ApJ, 116, 237). 
The MPC used data from ref.799 with the period 2.15 days.

\item {\sl
V1239 Cen } 
The MPC is a double sinusoid with the period 5.1 days from ref.804 
following measurements from ref.804.

\item {\sl
GJ 793 } 
The MPC is a simple sinusoil with the period of 33.5 days close to the period in ref.821.
Available data are scarce. Additional measurements are necessary 
to specify the period and parameters of the MPC
The MPC is determined according to data ref.821.

\end{description}

\clearpage
\newpage

\section{Cross-reference list }

\begin{table}
\caption{Cross--reference list }
\label{tab:col4}
\renewcommand{\arraystretch}{1.1}
\begin{tabular}{|r|l|}
\hline
  1& Babcock H.W., 1958, ApJSS, 30, 141                  \\
  2& Borra E.F. \& Landstreet J.D., 1980, ApJSS, 42, 421 \\
  4& Babcock H.W., 1954, ApJ, 120, 66                    \\
  5& Babcock H.W., 1960, ApJ, 132, 521                   \\
  7& Preston G.W., 1967, ApJ, 150, 871                   \\
  8& Preston G.W. \& Stepien K., 1968, ApJ, 151, 577     \\
 10& Conti P.S., 1969, ApJ, 156, 661                     \\
 11& Preston G.W. et al., 1969, ApJ, 156, 653            \\
 14& Preston G.W., 1969, ApJ, 158, 251                   \\
 15& Wolff S.C., 1969, ApJ,157, 253                      \\
 16& Wolff S.C., 1969, ApJ, 158, 1231                    \\
 18& Preston G.W., 1970, ApJ, 160, 1059                  \\
 19& Preston G.W. \& Wolff S.C., 1970, ApJ, 160, 1071    \\
 20& Conti P.S., 1970, ApJ, 160, 1077                    \\
 21& Preston G.W., 1972, ApJ, 175, 465                   \\
 22& Wolff S.C. \& Bonsack W.K., 1972, ApJ, 176, 425     \\
 23& Wolff S.C. \& Wolff R.J., 1972, ApJ, 176, 433       \\
 24& Borra E.F. \& Landstreet J.D., 1979, ApJ, 228, 809  \\
 25& Landstreet J.D. et al., 1975, ApJ, 201, 624         \\
 26& Wolff S.C., 1975, ApJ, 202, 127                     \\
 27& Borra E.F. \& Vaughan A.H., 1978, ApJ, 220, 924     \\
 28& Landstreet J.D. \& Borra E.F., 1978, ApJ, 224, L5   \\
 29& Jones T.J. et al., 1974, ApJ, 190, 579              \\
 30& Borra E.F. \& Landstreet J.D., 1978, ApJ, 222, 226  \\
 31& Vogt S.S. et al., 1980, ApJ, 236, 308               \\
 32& Bonsack W.K., 1976, ApJ, 209, 160                   \\
 33& Borra E.F. \& Landstreet J.D., 1977, ApJ, 212, 141  \\
 35& Preston G.W. \& Stepien K., 1968, ApJ, 154, 971     \\
 37& Borra E.F et al., 1983, ApJSS, 53, 151              \\
 40& Glagolevskij Yu.V. \& Chunakova N.M., 1985, BSAO, 19, 36 \\
 41& Glagolevskij Yu.V. et al., 1985, BSAO, 19, 26       \\
 42& Rakos K.D. et al., 1977, A\&A, 56, 453              \\
 44& Huchra J., 1972, ApJ, 174, 435                      \\
 46& Bonsack W.K., \& Pilachowski C.A., 1974, ApJ, 190, 327 \\
 47& Brown D.N. \& Landstreet J.D., 1981, ApJ, 246, 899  \\
 48& Scholz G., 1979, AN, 300, 213                       \\
 49& Scholz G., 1971, AN, 292, 281                       \\
 53& Sargent W.L.W. et al., 1967, ApJ, 147, 1185         \\
 55& Conti P.S., 1970, ApJ, 159, 723                     \\
 60& Landstreet J.D. \& Borra E.F., 1977, ApJ, 212, L43  \\
 61& Boesgaard A.M., 1974, ApJ, 188, 567                 \\
 62& Bonsack W.K. et al., 1974, ApJ, 187, 265            \\
 63& Borra E.F. \& Vaughan A.H., 1977, ApJ, 216, 462     \\
 69& Kemp J.C. \& Wolstencroft R.D., 1973, ApJ, 182, L43 \\
 71& Wolff S.C., 1973, ApJ, 186, 951                     \\
 72& Slovak M.H., 1982, ApJ, 262, 282                    \\
 74& Angel J.R.P. et al., 1973, ApJ, 184, L79            \\
 75& Borra E.F., 1975, ApJ, 196, L109                    \\
 76& Borra E.F. et al., 1981, ApJ, 247, 569              \\
 77& Landstreet J.D., 1982, ApJ, 258, 639                \\
 81& Borra E.F. \& Landstreet J.D., 1975, PASP, 87, 961  \\
 83& Wolff S.C. \& Preston G.W., 1978, PASP, 90, 406     \\
 86& Jones T.J. \& Wolff S.C., 1974, PASP, 86, 67        \\
 87& Bonsack W.K., 1981, PASP, 93, 756                   \\
 89& Landstreet J.D. et al., 1979, MNRAS, 188, 609       \\
 91& Heuvel E.P.J. van den, 1971, A\&A, 11, 461          \\
 92& Pilachowski C.A. et al., 1974, A\&A, 37, 275        \\
 93& Wood H.J. \& Campusano L.B., 1975, A\&A, 45, 303    \\
 94& Bonsack W.K., 1977, A\&A, 59, 195                   \\
\hline
\end{tabular}
\end{table}

\begin{table}
\contcaption{Cross--reference list }
\label{tab:col4}
\renewcommand{\arraystretch}{1.1}
\begin{tabular}{|r|l|}
\hline
 96& Rudy R.J. \& Kemp C.J., 1978, MNRAS, 183, 595       \\
 97& Kuvshinov V.M., 1972, Astron.Tsirk., 682, 3         \\
 98& Kuvshinov V.M. et al., 1976, AN, 297, 181           \\
105& Scholz G., 1975, AN, 296, 31                        \\
108& Hildebrandt G. et al., 1973, AN, 294, 175           \\
111& Gollnow H., 1964, PASP, 74, 163                    \\
118& Rustamov Yu.S., Khotnyanskij A.N., 1980, Pis.Astr.Zh., 6, 364 \\
120& Gollnow H., 1971, Observatory, 91, 37              \\
121& Maitzen H.M. et al., 1980, A\&A, 81, 323           \\
126& Wolstencroft R.D. et al., 1981, MNRAS, 195, 39     \\
135& Bohlender D.A. et al., 1987, ApJ, 323, 325         \\
136& Scholz G. \&  Gerth E., 1980, AN, 301, 211         \\
137& Rudiger G. \& Scholz G., 1988, AN, 309, 181        \\
140& Glagolevskij Yu.V. et al., 1989, BSAO, 27, 34      \\
142& Glagolevskij Yu.V. et al., 1982, Pis'ma Astron.Zh., 8, 26 \\
146& Zverko J. et al., 1989, Contr.Astr.Obs.Skalnate Pleso, 18, 71 \\
147& Mikulasek Z. et al., 1984, mast.conf, 4, 13 \\
153& Romanov Yu.S. et al., 1988, mast.conf, 51 \\
159& Ryabchikova T.A. et al., 1988, mast.conf, 40 \\
168& Thompson I.B. et al., 1987, ApJSS, 64, 219          \\
171& Bohlender D.A., 1989, A\&A, 220, 215                \\
172& Landstreet J.D., 1990, ApJ, 352, L5                \\
174& Thompson I.B. \& Landstreet J.D., 1985, ApJ, 289, L9 \\
175& Renson P., 1984, A\&A, 139, 131                    \\
176& Wolff S.C. \& Morrison N.D., 1974, PASP, 86, 935   \\
178& Gerth E., 1990, AN, 311, 41                        \\
179& Ziznovsky J. \& Romanyuk I.I., 1990, BAICz, 41, 118 \\
181& Shore S.N. et al., 1990, ApJ, 348, 242             \\
182& Bohlender D.A. \& Landstreet J.D., 1990, ApJ, 358, L25 \\
184& Mathys G., 1991, A\&ASS, 89, 121                    \\
186& Thompson I.B., 1983, MNRAS, 205, 43                 \\
187& Albrecht R. et al., 1977, A\&A, 58, 93              \\
189& Plachinda S.I., 1990, Mitt.Astr.Obs.Krim, 81, 112   \\
190& Weiss W.W. et al., 1990, Mitt.Astr.Obs.Krim, 82, 69 \\
191& Skulskij M.Yu., 1990, MiTau., 125, 146  \\
194& Gerth E. et al., 1991, AN, 312, 107                \\
195& Bychkov V.D. et al., 1991, Pis'ma Astron.Zh., 17, 43 \\
197& El'kin V.G. et al., 1991, BSAO, 25, 22              \\
201& Borra E.F., 1994, private communication             \\
204& Landstreet J.D., 2000, private communication       \\
208& Scholz G., 1983, ApSS, 94, 159                   \\
209& Scholz G., 1978, AN, 299, 81                      \\
215& Shtol' V.G. et al., 1992, pres.conf, 190 \\
219& Gerth E. et al., 1992, pres.conf, 60 \\
220& Bychkov V.D. et al., 1992, pres.conf, 211 \\
224& El'kin V.G., 1992, pres.conf, 67 \\
227& Skulskij M.Yu. et al., 1992, pres.conf, 207\\
229& Skulskij M.Yu. et al., 1992, pres.conf, 217 \\
230& Bohlender D.A. et al., 1993, A\&A, 269, 355        \\
237& Plachinda S.I. et al., 1993, Mitt.Astr.Obs.Krim, 87, 91 \\
239& Johnstone R.M. \& Penston M.V., 1987, MNRAS, 227, 797 \\
243& Glagolevskij Yu.V. et al., 1995, Pis'ma Astron.Zh., 21, 190 \\
252& Brown D.N. et al., 1985, AJ, 90, 1354               \\
255& Wade G.A. et al., 1996, A\&A, 313, 209              \\
256& Mathys G. \& Hubrig S., 1997, A\&ASS, 124, 475      \\
260& Bychkov V.D. et al., 1997, smf.proc, 110 \\
267& Bychkov V.D. et al., 1997, smf.proc, 197 \\
268& Bychkov V.D. \& Shtol' V.G., 1997, smf.proc, 200 \\
270& Bychkov V.D. et al., 1997, smf.proc, 204 \\
\hline
\end{tabular}
\end{table}

\begin{table}
\contcaption{Cross--reference list }
\label{tab:col4}
\renewcommand{\arraystretch}{1.1}
\begin{tabular}{|r|l|}
\hline
271& El'kin V.G. et al., 1997, smf.proc, 207 \\
273& Weiss W.W., 1986, A\&A, 325, 195                   \\
286& Glagolevskij Yu.V., Chuntonov G.A., 1998, BSAO, 45, 105 \\
290& Gerth E., 1994, private communication           \\
299& Borra E.F. et al., 1984, ApJ, 284, 211            \\
310& Wade G.A. et al., 2000, MNRAS, 313, 851            \\
311& Plachinda S.I. \& Tarasova T.N., 2000, ApJ, 533, 1016 \\
312& Wade G.A. et al., 2000, A\&A, 355, 1080            \\
324& Leone F. \& Catanzaro G., 2001, A\&A, 365, 118     \\
325& Verdugo E. et al., 2002, Proc.IAU Symp. 212, 255   \\
326& Elkin V.G., 2000, private communication            \\
327& Bychkov V.D., Bychkova L.V., 2002, private comm. \\
329& Elkin V.G. et al., 2002, Pis'ma Astron.Zh., 28, 195 \\
330& Chuntonov G.A., 2001, BSAO, 51, 112                \\
333& Shorlin S.L.S. et al., 2002, A\&A, 392, 637        \\
334& Monin D.N. et al., 2002, A\&A, 396, 131            \\
342& Skulskij M.Yu., 1982, Pis'ma Astron.Zh., 8, 238    \\
343& Skulskij M.Yu., 1985, Pis'ma Astron.Zh., 11, 51    \\
344& Skulskij M.Yu., Plachinda S.I., 1993, Astron.Zh., 19, 517 \\
348& Kudryavtsev D.O. et al., 2000, mfcp.proc, 64 \\
350& Panchuk V.E. et al., 2000, mfcp.proc, 75        \\
354& Tarasova T.N. et al., 2001, Astron.Zh., 78, 550    \\
360& Neiner C. et al., 2003, A\&A, 406, 1019         \\
361& El'kin V.G. et al., 2003, Pis'ma Astron.Zh., 29, 455 \\
362& Romanov Y.S. et al., 1985, Pis'ma Astron.Zh., 11, 378 \\
367& Leone F. \& Kurtz D., 2003, A\&A, 407, L67 \\
369& Hubrig S. et al., 2004, A\&A, 415, 661 \\
370& Hubrig S. et al., 2004, A\&A, 415, 685 \\
373& Hildebrandt G. et al, 2000, AN, 321, 115 \\
382& Ryabchikova T. et al., 2005, A\&A, 429, L55 \\
384& Bychkov V., 2006, MNRAS, 365, 585-598 \\
385& Petit P. et al., 2005, MNRAS, 361, 837-849 \\
386& Wade G. et al., 2006, A\&A, 451, 196 \\
388& Bagnulo S. et al., 2006, A\&A, 450, 777 \\
389& Hubrig S. et al., 2006, AN, 327, 289 \\
397& Wade G. et al., 2005, A\&A, 442, L31 \\
398& Kochukhov O. \& Bagnulo S., 2006, A\&A, 450, 763 \\
401& Briquet M. et al., 2007, AN, 328, 41 \\
402& Kudryavtsev D.O. et al., 2006, MNRAS, 372, 1804 \\
406& Hubrig S. et al., 2006, MNRAS, 369, L61 \\
407& Wade G. et al., 2006, A\&A, 458, 569 \\
409& Wade G. et al., 2007, MNRAS, 376, 1145 \\
415& Kim, Kang-Min et al., 2007, PASP, 119, 1052 \\
417& Ryabchikova T. et al., 2007, A\&A, 462, 1103 \\
418& Donati J.-F. et al., 2006, MNRAS, 370, 629 \\
419& Schnerr R.S. et al., 2006, A\&A, 452, 969 \\
423& Auriere M. et al., 2007, A\&A, 475, 1053 \\
424& Leone F., 2007, MNRAS, 382, 1690 \\
425& Donati J.-F. et al., 1997, MNRAS, 291, 658 \\
427& Bychkov V.D.\& Bychkova L.V., 2005, private comm. \\
428& Semenko E.A. et al., 2008, BSAO, 63, 136 \\
435& Chuntonov G.A., 2007, pms.conf, 214 \\
438& Donati J.-F. et al., 2008, MNRAS, 390, 545 \\
439& Morin J. et al., 2008, MNRAS, 390, 567 \\
444& Mathys G. et al., 2007, MNRAS, 380, 181 \\
445& Donati J.-F. et al., 2008, MNRAS, 386, 1234 \\
453& Hubrig S. et al., 2008, A\&A, 490, 793 \\
454& Alecian E. et al., 2008, MNRAS, 385, 391 \\
456& Valyavin G. et al., 2008, ApJ, 683, 466 \\
\hline
\end{tabular}
\end{table}

\begin{table}
\contcaption{Cross--reference list }
\label{tab:col4}
\renewcommand{\arraystretch}{1.1}
\begin{tabular}{|r|l|}
\hline
457& Hubrig S. et al., 2009, AN, 330, 317 \\
458& Korhonen H. et al., 2009, MNRAS, 395, 282 \\
459& Landstreet J.D. et al., 2008, A\&A, 481, 465 \\
462& Folsom C. et al., 2008, MNRAS, 391, 901 \\
466& Elkin V.G. et al., 2010, MNRAS, 402, 1883 \\
468& Alecian E. et al., 2009, MNRAS, 400, 354  \\
471& Hubrig S. et al., 2009, A\&A, 502, 283 \\
473& Hubrig S. et al., 2009, MNRAS, 396, 1018 \\
474& Hubrig S. et al., 2009, AN, 330, 708 \\
475& Silvester J. et al., 2009, MNRAS, 398, 1505 \\
477& Elkin V.G. et al., 2008, IBVS, 5851 \\
478& Schnerr R.S. et al., 2008, A\&A, 483, 857 \\
480& Auriere M. et al., 2009, A\&A, 504, 231 \\
483& Glagolevskij Yu.V. et al., 2005, Astron.Letters, 31, 327 \\
485& Moutou C. et al., 2007, A\&A, 473, 651 \\
491& Oksala M.E. et al., 2010, MNRAS, 405, L51 \\
492& Leone F. et al., 2010, MNRAS, 401, 2739 \\
498& Fares R. et al., 2010, MNRAS, 406, 409 \\
499& Martins F. et al., 2010, MNRAS, 407, 1423 \\
502& Donati J.-F. et al., 2007, MNRAS, 380, 1297 \\
504& Auriere M. et al., 2010, A\&A., 516, id.L2 \\
507& Morin J. et al., 2010, MNRAS, 407, 2269 \\
509& Hubrig S. et al.,  2011, A\&A, v.525, L4 \\
510& Hubrig S. et al., 2011, A\&A, 528, id.A151 \\
513& Hubrig S. et al., 2011,  ApJ, 726, L5 \\
514& Hubrig S. et al., 2010, AN, 331, 781 \\
516& Grunhut J.H. et al., 2010, MNRAS, 408, 2290 \\
517& Petit V. et al., 2011, MNRAS, 412, L45 \\
523& Konstantinova-Antova R. et al., 2010, A\&A, 524, A57 \\
527& Bohlender D.A. \& Monin D., 2011, AJ, 141, id.169 \\
534& Semenko E.A. et al., 2011, Astr.Letters, 37, 20 \\
536& Usenko I.A. et al., 2010, OAP, 23, 140 \\
537& Hubrig S. et al., 2011, A\&A, 531, id.L20 \\
547& Semenko E.A. et al., 2011, AN, 332, 948 \\
548& Kudryavtsev D.O. et al., 2011, AN, 332, 961 \\
549& Yakunin I.A. et al., 2011, AN, 332, 974 \\
550& Hubrig S. et al.,  2012, A\&A, 547, id.A90, 24 pp. \\
559& Auriere M. et al., 2011, A\&A, 534, id.A139 pp.8 \\
560& Wade G. et al., 2011, MNRAS, 416, 3160 \\
561& Bailey J.D. et al., 2011, A\&A, 535, id.A25, 11 pp \\
563& Alecian E. et al., 2011, A\&A, 536, id.L6, 4 pp \\
565& Oksala M.E. et al., 2012, MNRAS, 419, 959 \\
566& Grunhut J.H. et al., 2012, MNRAS, 419, 1610 \\
567& Wade G. et al., 2012, MNRAS, 419, 2459 \\
570& Henrichs H.F. et al., 2012, A\&A, 545, id.119, 10 pp \\
571& Shultz M. et al., 2012, ApJ, 750, id.2, 10 pp \\
572& Bailey J.D. et al., 2012, MNRAS, 423, 328 \\
575& Silvester J. et al., 2012, MNRAS, 426, 1003 \\
576& Grunhut J.H. et al., 2012, MNRAS, 426, 2208 \\
577& Neiner C. et al., 2012, MNRAS, 426, 2738 \\
578& Tsvetkova S. et al., 2013, A\&A, 556,id.A43, 9 pp. \\
581& Alecian E. et al., 2013, MNRAS, 429, 1001 \\
582& Petit V. et al., 2008, MNRAS, 387, L23 \\
584& Neiner C. et al., 2012, A\&A, 537, A148 \\
586& Morgenthaler A. et al., 2012, A\&A, 540, id.A138, 15 pp. \\
587& Konstantinova-Antova R. et al., 2012, A\&A, 541, id.A44, 7 pp. \\
588& Auriere M. et al., 2012, A\&A, 543, id.A118, 6 pp. \\
591& Hubrig S. et al.,  2007, AN, 328, 1133 \\
592& Hubrig S. et al.,  2008, A\&A, 488, pp.287 \\
\hline
\end{tabular}
\end{table}

\begin{table}
\contcaption{Cross--reference list }
\label{tab:col4}
\renewcommand{\arraystretch}{1.1}
\begin{tabular}{|r|l|}
\hline
595& Karitskaya E.A. et al., 2010, IBVS 5950 \\
605& Hubrig S. et al.,  2013, A\&A, 551, id.A33, 13pp. \\
606& Mathys G. et al., 2012, AN, 333, 30 \\
609& Hubrig S. et al.,  2011, A\&A, 536, id.A45, 8 pp. \\
610& Johns-Krull C.M. et al., 2013, ApJ, 765, 11, 11 pp. \\
611& Makaganiuk V. et al., 2012, A\&A, 539, id.A142,15 pp. \\
614& Rivinius Th. et al, 2010, MNRAS, 405, L46 \\
615& Folsom C.P. et al., 2013, MNRAS, 431, 1513 \\
616& Kochukhov O. et al., 2013, A\&A, 550, id.A84, 19 pp. \\
621& Kudryavtsev D.O. \& Romanyuk I.I., 2012, AN, 333, 41 \\
623& Baklanova D. et al., 2011, AN, 332, 939 \\
629& Leone F. et al., 2011, ApJ, 731, id. L33 \\
630& Bychkov V.D. et al., 2012, Astrophysical Bulletin, 67, 207 \\
635& Konstantinova-Antova R. et al., 2008, A\&A, 480, 475 \\
641& Yakunin I.A., 2013, Astrophysical Bulletin, 68, 226 \\
643& Hill G.M. et al., 1998, MNRAS, 297, 236 \\
653& Fares R. et al., 2009, MNRAS, 398, 1383 \\
655& Henrichs H.F. et al., 2013, A\&A 555, A46 \\
660& Schoeller M. et al., 2014, ArXiv:1309.5497  \\
664& Rosen L. et al., 2013, MNRAS, 436, L10 \\
675& Puzin V.B. et al., 2014, Astrophysical Bulletin, 69, 321 \\
676& Yakunin I. et al., 2015, MNRAS, 447, 1418 \\
677& Romanyuk I.I. et al., 2014, Astrophysical Bulletin, 69, 451 \\
679& Hubrig S. et al, 2015, MNRAS, 447, 1885 \\
680& Wade G. et al., 2015, MNRAS, 447, 2551 \\
681& David-Uraz A. et al., 2014, MNRAS, 444, 429 \\
682& Alecian E. et al., 2014, A\&A, 567, id.A28, 19pp. \\
685& Auriere M. et al., 2015, A\&A, 574, id.A90, 30 pp. \\
689& Fossati L. et al., 2015, A\&A 574, A20 \\
691& Kochukhov O. et al., 2015, A\&A, 574, id.A79, 12pp. \\
693& Rosen L. et al., 2015, ApJ, 805, id.169, 17pp \\
702& Wade G., 2015, private communication \\
704& Romanyuk I.I. et al., 2015, Astrophysical Bulletin, 70, N4, 482 \\
705& Mathys G. et al., 2016, A\&A, 586, id.A85, 6pp. \\
706& Alecian E. et al., 2016, A\&A, 589, id.A47, 13pp. \\
707& Hussain G.A.J. e al., 2016, A\&A, 585, id.A77, 7 pp. \\
709& Alvarado-Gomez J.D. et al., 2015, A\&A, 582, id.A38, 12 pp. \\
710& Romanyuk I.I. et al., 2015, Astrophysical Bulletin, 70, 469 \\
712& Landstreet J.D. et al., 2015, A\&A, 580, id.A120, 8 pp. \\
713& Yakunin I. et al., 2015, ASP conf.ser., 494, 86 \\
716& Rosen L. et al., 2016, A\&A, 593, id.A35, 24pp. \\
717& Sikora J. et al., 2016, MNRAS, 460, 1811 \\
718& Hubrig S. et al., 2016, MNRAS, 458, 3381 \\
724& Sikora J. et al., 2015, MNRAS, 451, 1928 \\
727& Karitskaya E.A. et al., 2009, ArXiv:0908.2719 \\
730& Briquet M. et al., 2016, A\&A, 587, id.A126, 12 pp. \\
731& Borisova A. et al., 2016, A\&A, Vol.591, id.A57, 15 pp. \\
733& Grunhut J.H. et al., 2017, MNRAS, 465, 2432 \\
734& Romanyuk I.I. et al., 2016, Astrophysical Bulletin, 71, 468 \\
735& Romanyuk I.I. et al., 2016, Astrophysical Bulletin, 71, 480 \\
736& Hubrig S. et al., 2017, MNRAS, 467, L81 \\
738& Gonzalez J.F. et al., 2017, MNRAS, 467, 437 \\
739& Wade G. et al., 2017, MNRAS, 465, 2517 \\
740& Jarvinen S.P. et al., 2017, MNRAS, 464, L85 \\
742& Naze Y. et al., 2016, A\&A, 596, id.A44, 7 pp. \\
743& Tsvetkova S. et al., 2017, A\&A, 599, id.A72, 13pp. \\
746& Boro Saikia S. et al., 2016, A\&A, 594,id.A29, 19pp. \\
748& Landstreet J.D. et al., 2017, A\&A, 601,id.A129, 10pp. \\
749& Rusomarov N. et al., 2016, A\&A, 588, id.A138, 20 pp. \\
\hline
\end{tabular}
\end{table}

\begin{table}
\contcaption{Cross--reference list }
\label{tab:col4}
\renewcommand{\arraystretch}{1.1}
\begin{tabular}{|r|l|}
\hline
752& Mathys G., 2017, A\&A, 601, id.A14, 90 pp. \\
753& Kochukhov O. et al., 2014, A\&A, 565, id.A83 14 pp. \\
754& Bagnulo S., 2017, A\&A, 601, id.A136, 10 pp. \\
758& Hebrard E.M. et al., 2016, MNRAS, 461, 1465 \\
759& Kochukhov O. \& Lavail A., 2017, ApJ, 835, id.L4 5 pp. \\
760& Bagnulo S. et al., 2015, A\&A, 583, id.A115, 37 pp. \\
762& Romanyuk I.I. et al., 2017, Astrophysical Bulletin, 72, 183 \\
763& Wade G. et al., 1998, A\&A, 335, 973 \\
764& Hubrig S. et al., 2017, MNRAS, 471, 1543 \\
765& Fares R. et al., 2017, MNRAS, 471, 1246 \\
766& Lee Byeong-Cheol et al, 2018, MNRAS, 473, L41 \\
767& Semenko E. et al, 2017, A., Astrophysical Bulletin, 72, 422 \\
768& Romanyuk I. et al., 2017, Astrophysical Bulletin, 72, 429 \\
769& Shultz M.E. et al, 2018, MNRAS, 475, 5144 \\
773& Hubrig S. et al., 2018, MNRAS, 477, 3791 \\
774& Romanyuk I. et al., 2018, Astrophysical Bulletin, 73, 185 \\
775& Bagnulo S., Landstreet J.D., 2018, A\&A 618, A113 \\
776& Romanyuk I.I. et al, 2019, Astrophysical Bulletin, 74, 60 \\
779& Mathys G. et al., 2019, A\&A, v.624, id.A32, 10 pp. \\
781& Jarvinen S.P. et al., 2019, MNRAS, 486, 5499 \\
782& Jarvinen S.P. et al., 2015, A\&A, 584, id.A15, 9 pp. \\
786& Jarvinen S.P. et al., 2018, MNRAS, 481, 5163 \\
792& Shultz E.M. et al., 2019, MNRAS, 485, 1508 \\
793& Briquet M. et al., 2013, A\&A, 557, id.L16, 4 pp. \\
794& Buysschaert B. et al., 2017, A\&A, 605, id.A104, 16 pp. \\
795& Buysschaert B. et al., 2018, MNRAS, 478, 2777 \\
797& Shultz M. et al., 2019, MNRAS, 482, 3950 \\
798& Folsom C.P. et al., 2018, MNRAS, 481, 5286 \\
801& Tsvetkova S. et al., 2019, Bulgarian Astronomical Journal, 30, 67 \\
802& Mathias P. et al., 2019, A\&A, 615, id.A116, 14 pp. \\
803& S.H.P.Alencar S.H.P. et al., 2018, A\&A, 620, id.A195, 15 pp. \\
804& Nicholson B.A. et al., 2018, MNRAS, 480, 1754 \\
805& Alvarado-Gomez J.D. et al., 2018, MNRAS, 473, 4326 \\
807& Shultz M. et al., MNRAS, 471, 2286 \\
813& Rusomarov N. et al., 2018, A\&A, 609, id.A88, 16 pp. \\
814& Mathys G.et al., 2019, A\&A, 629, id.A39, 7 pp. \\
821& Moutou C. et al., 2017, MNRAS, 472, 4563 \\
822& Jarvinen S.P.et al., 2019, MNRAS, 489, 886 \\
826& Sikora J. et al., 2019, MNRAS, 483, 3127 \\
828& de la Chevrotiere A. et al., 2013, ApJ, 764, id.171, 25 pp. \\
834& Butkovskaya V., Plachinda S., privat comm., 2015.  \\
\hline
\end{tabular}
\end{table}

\section{Magnetic rotational phase curves}

The catalog of magnetic rotational phase curves, $B_e(\phi)$, for 
350 stars, is presented in graphical form in the appendix.  All phase curves
are expressed in a homogeneous manner with the effective magnetic field,
$B_e$ (in G), against the rotational phase, $\phi$. Note that in
some cases several phase curves correspond to the same star. This occured
in cases when different observational techniques produced significantly 
different series of $B_e$ values for that star.

Note that in a few cases the rotational phase curves of the surface field, 
$B_s$, are presented. MPCs of this type were computed for stars where
the longitudinal magnetic field data, $B_e$, were not available.

\clearpage
\newpage

\begin{figure}
\resizebox{0.98\hsize}{!}{\includegraphics{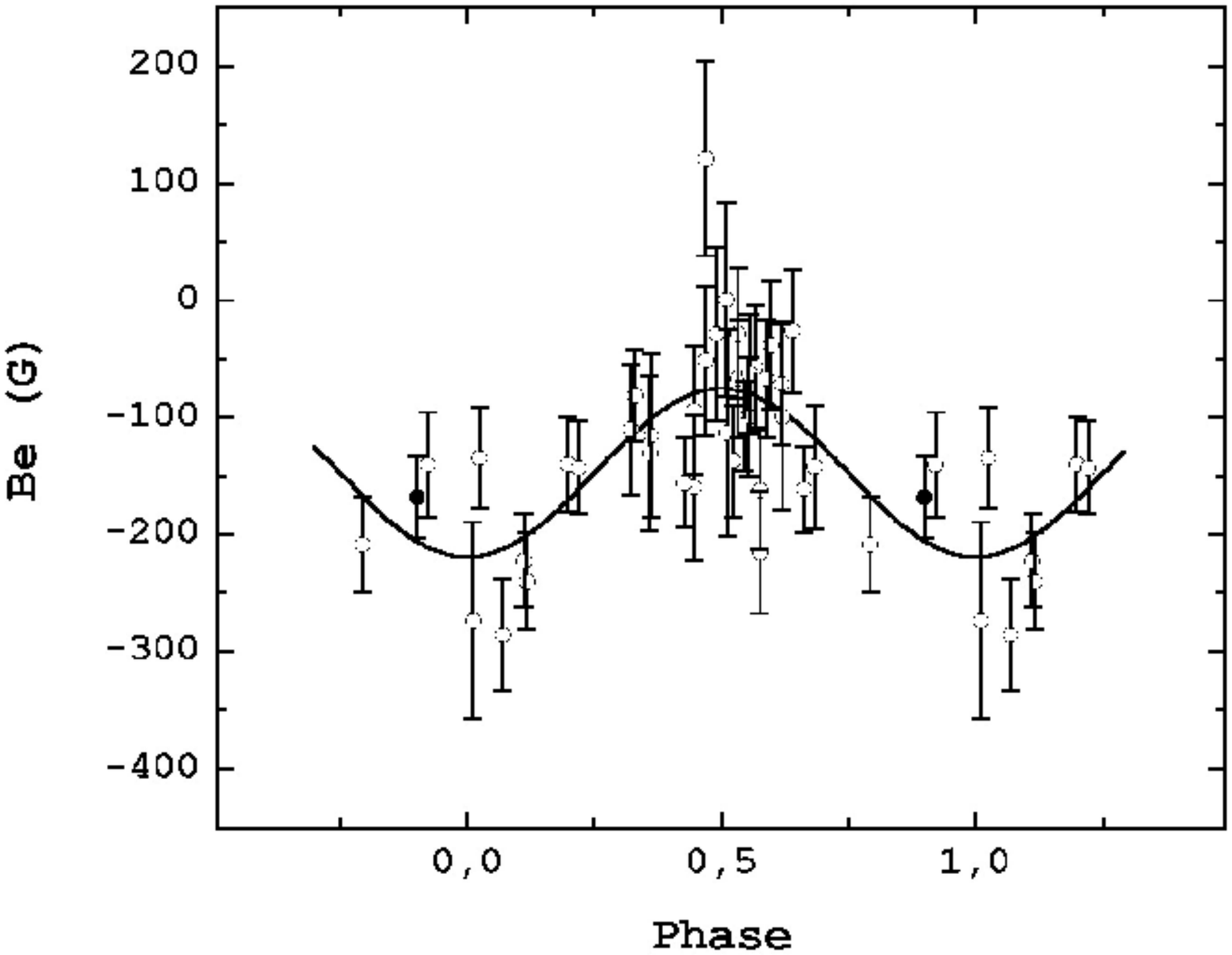}}
\vspace{-3.5mm}
\caption{ HD 108 }
\label{fig:fig1}
\end{figure}

\begin{figure}
\resizebox{0.98\hsize}{!}{\includegraphics{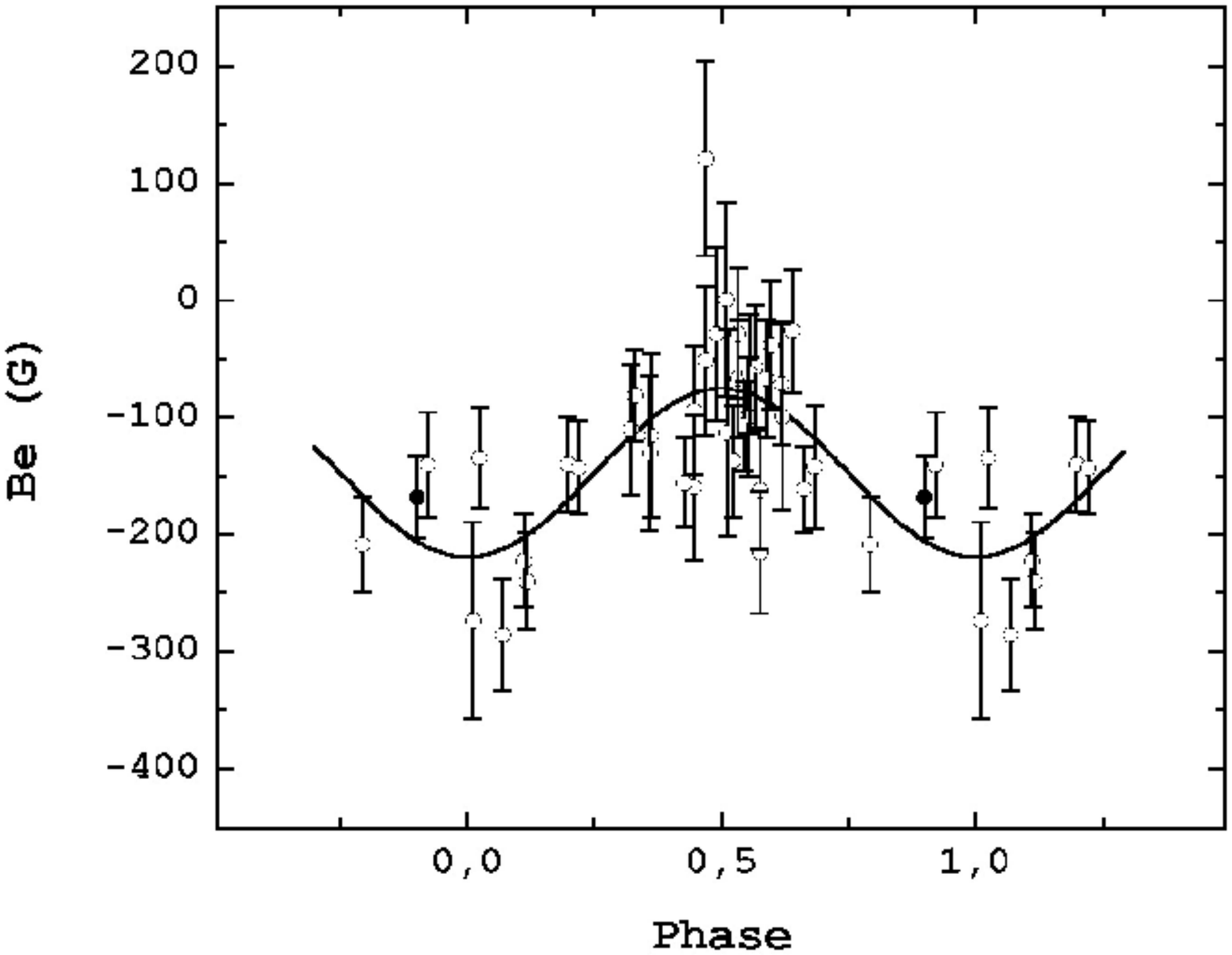}}
\vspace{-3.5mm}
\caption{ HD 358 }
\label{fig:fig2}
\end{figure}

\begin{figure}
\resizebox{0.98\hsize}{!}{\includegraphics{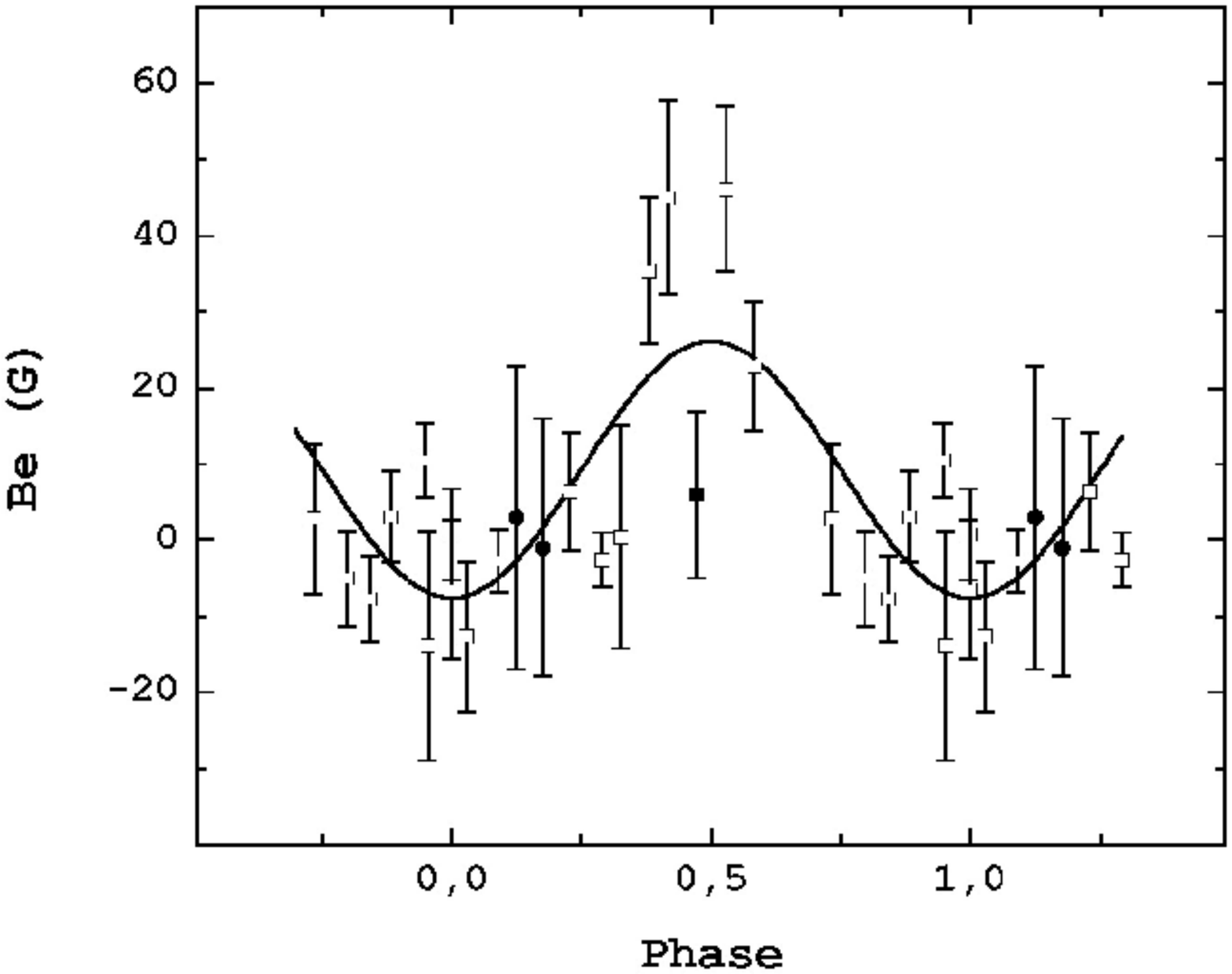}}
\vspace{-3.5mm}
\caption{ HD 886 }
\label{fig:fig2}
\end{figure}

\begin{figure}
\resizebox{0.98\hsize}{!}{\includegraphics{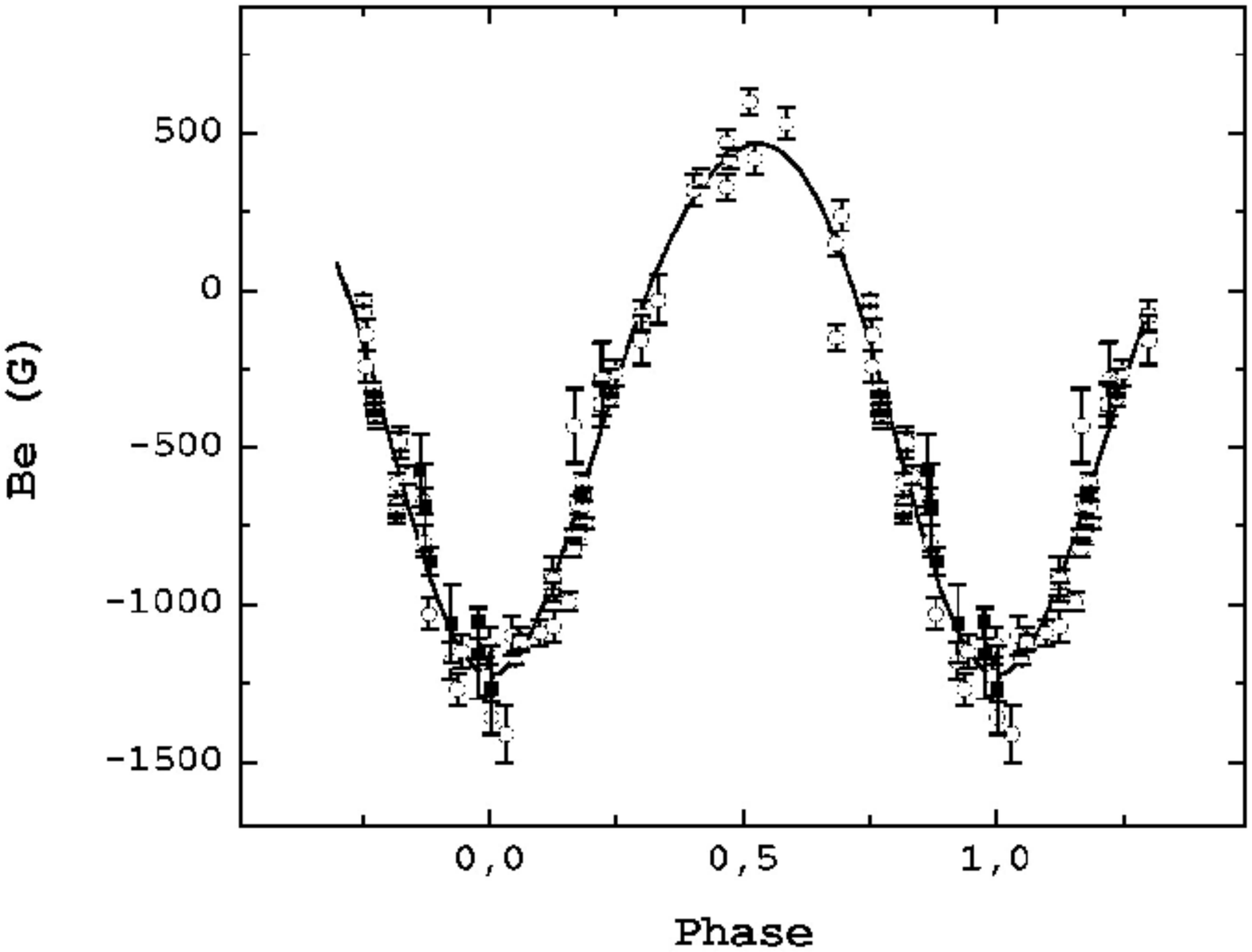}}
\vspace{-3.5mm}
\caption{ HD 965 }
\label{fig:fig3}
\end{figure}

\begin{figure}
\resizebox{0.98\hsize}{!}{\includegraphics{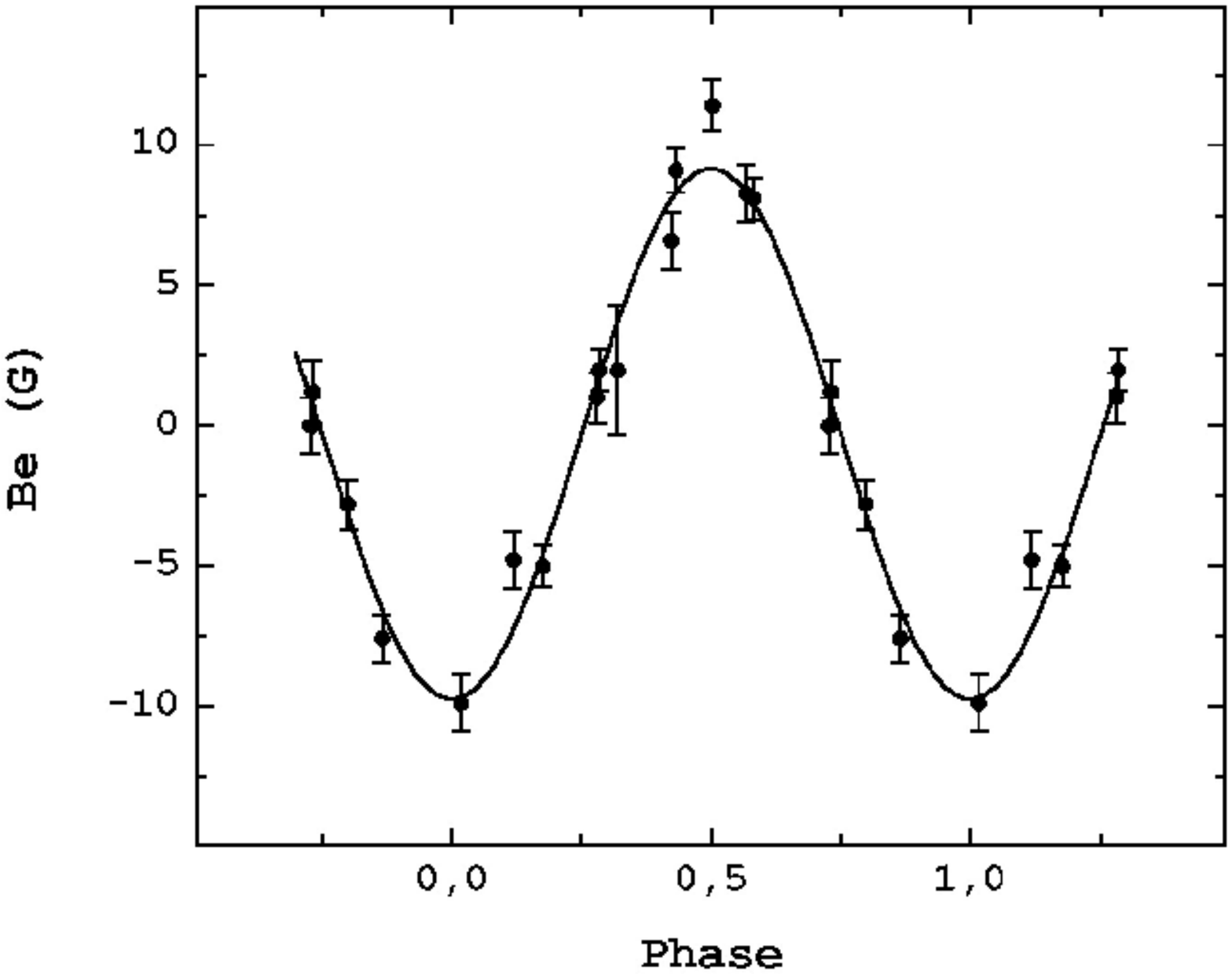}}
\vspace{-3.5mm}
\caption{ HD 1237 }
\label{fig:fig4}
\end{figure}

\begin{figure}
\resizebox{0.98\hsize}{!}{\includegraphics{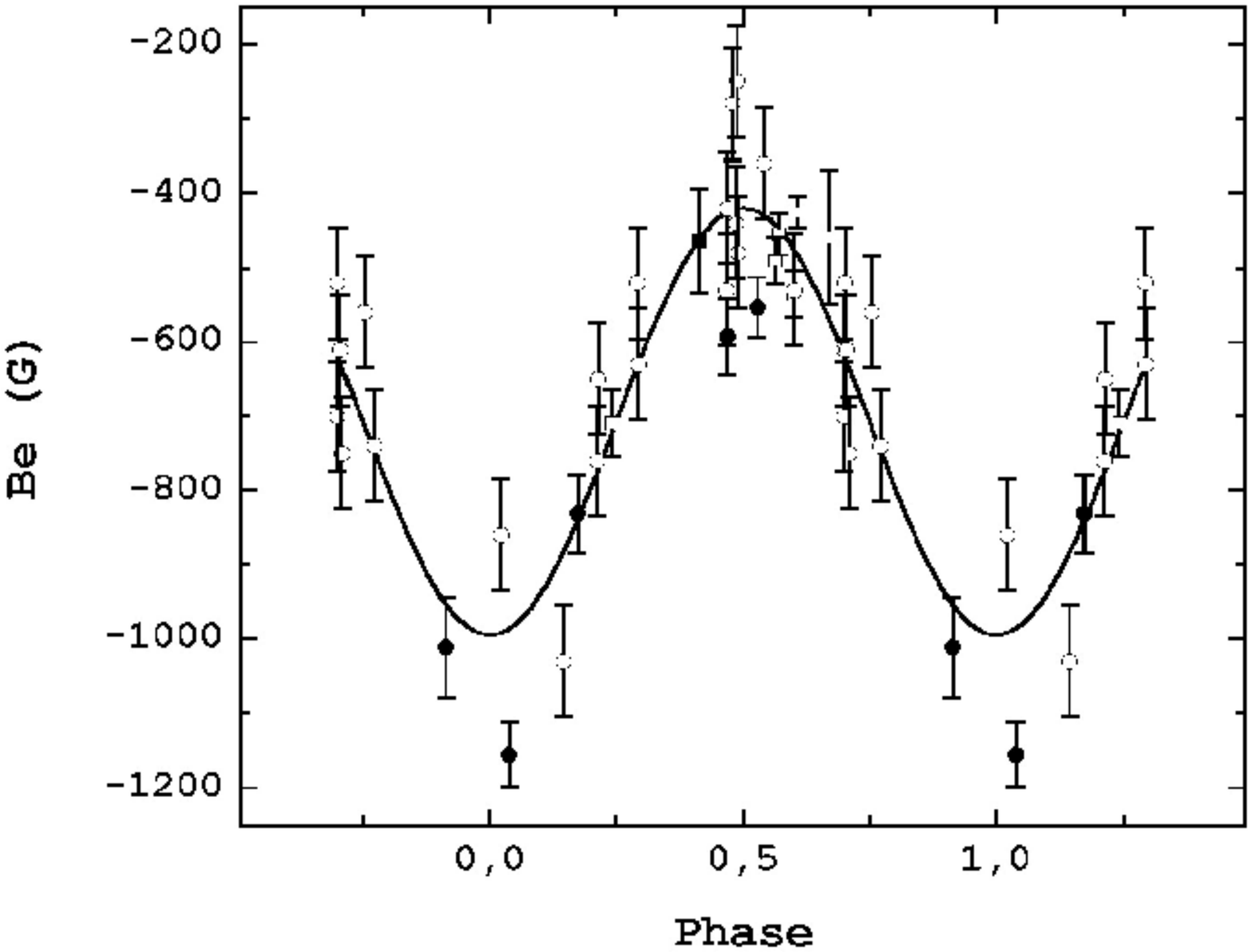}}
\vspace{-3.5mm}
\caption{ HD 2453 }
\label{fig:fig5}
\end{figure}

\clearpage
\newpage

\begin{figure}
\resizebox{0.98\hsize}{!}{\includegraphics{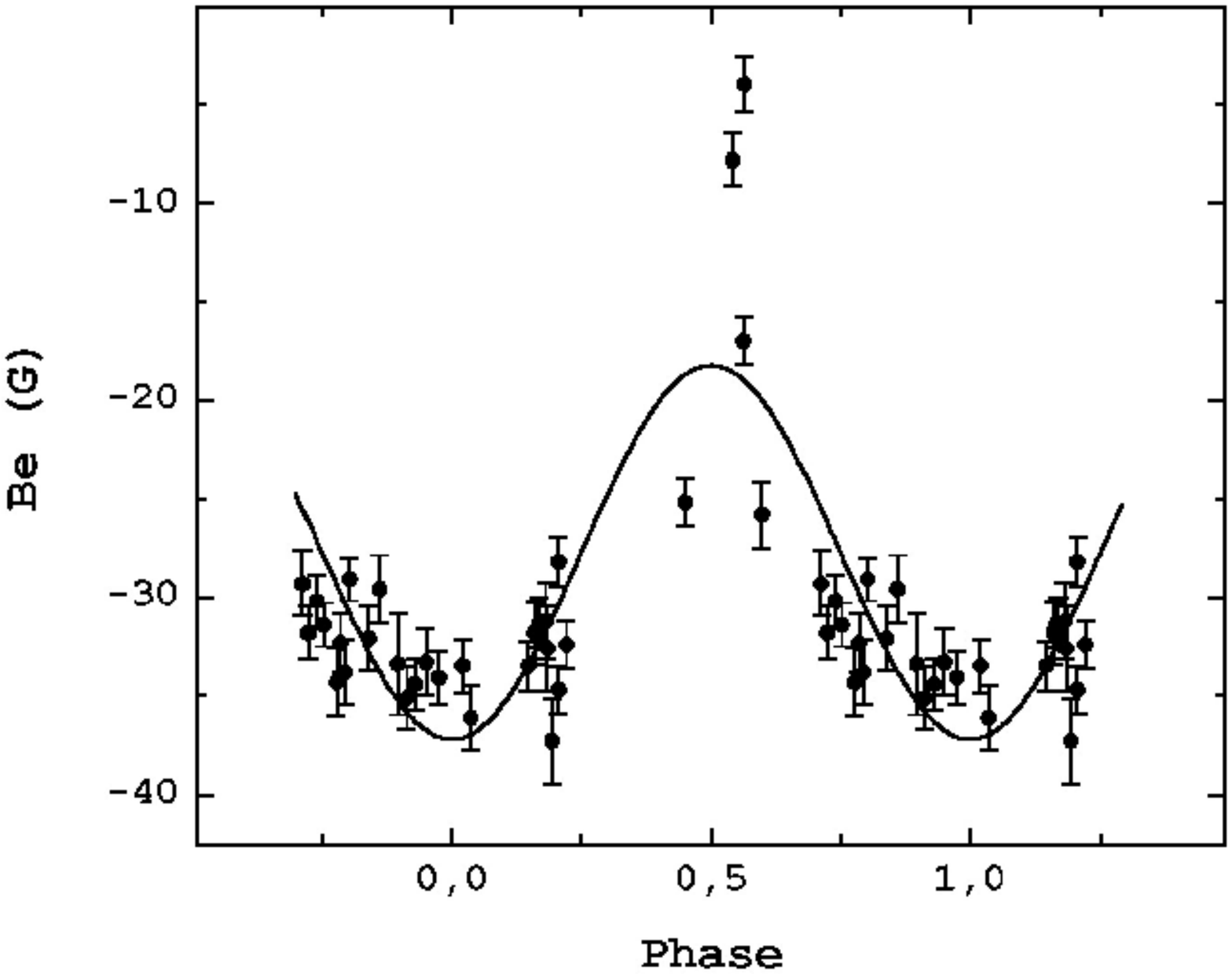}}
\vspace{-3.5mm}
\caption{ HD 3229 }
\label{fig:fig6}
\end{figure}

\begin{figure}
\resizebox{0.98\hsize}{!}{\includegraphics{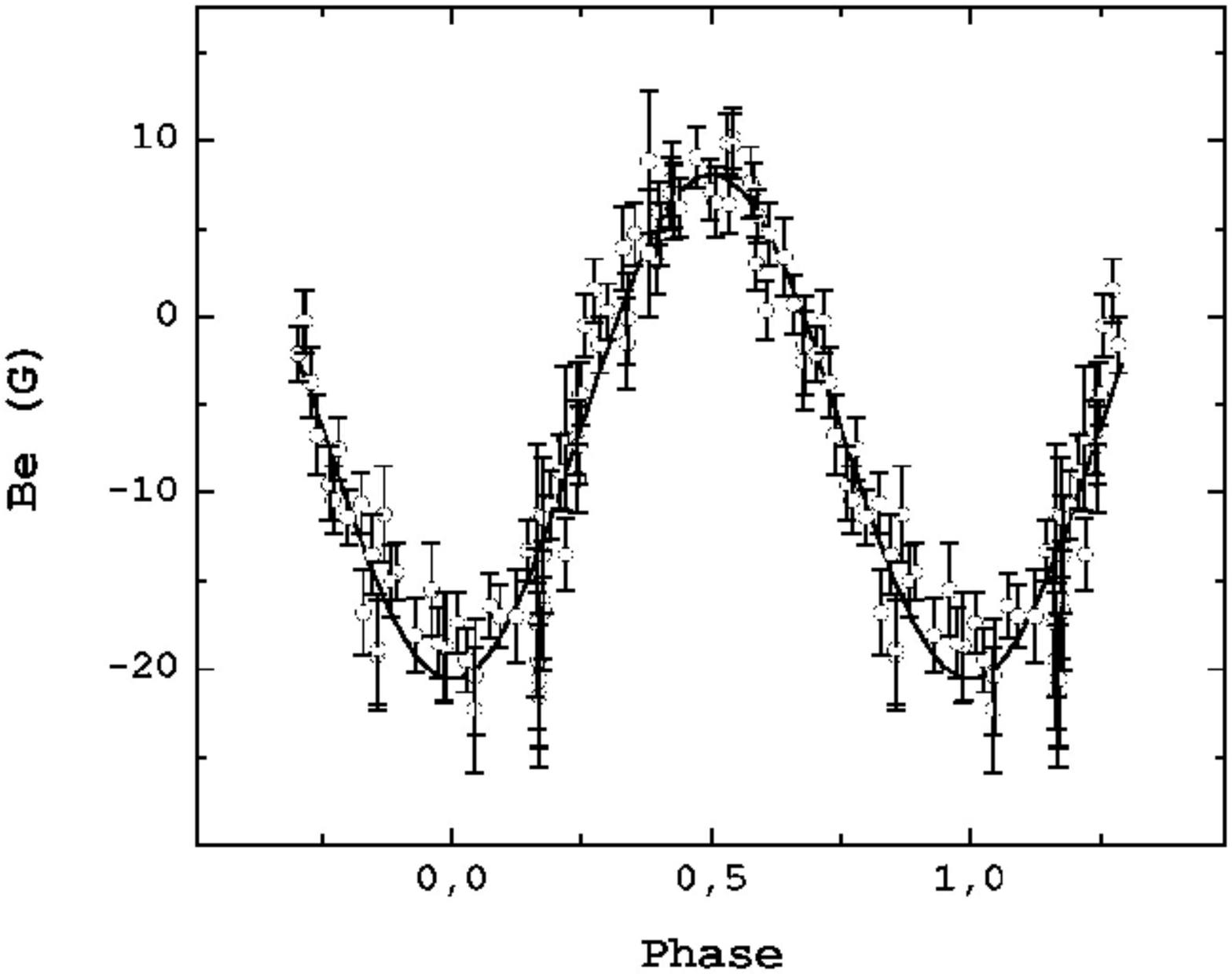}}
\vspace{-3.5mm}
\caption{ HD 3360 (1) }
\label{fig:fig7}
\end{figure}

\begin{figure}
\resizebox{0.98\hsize}{!}{\includegraphics{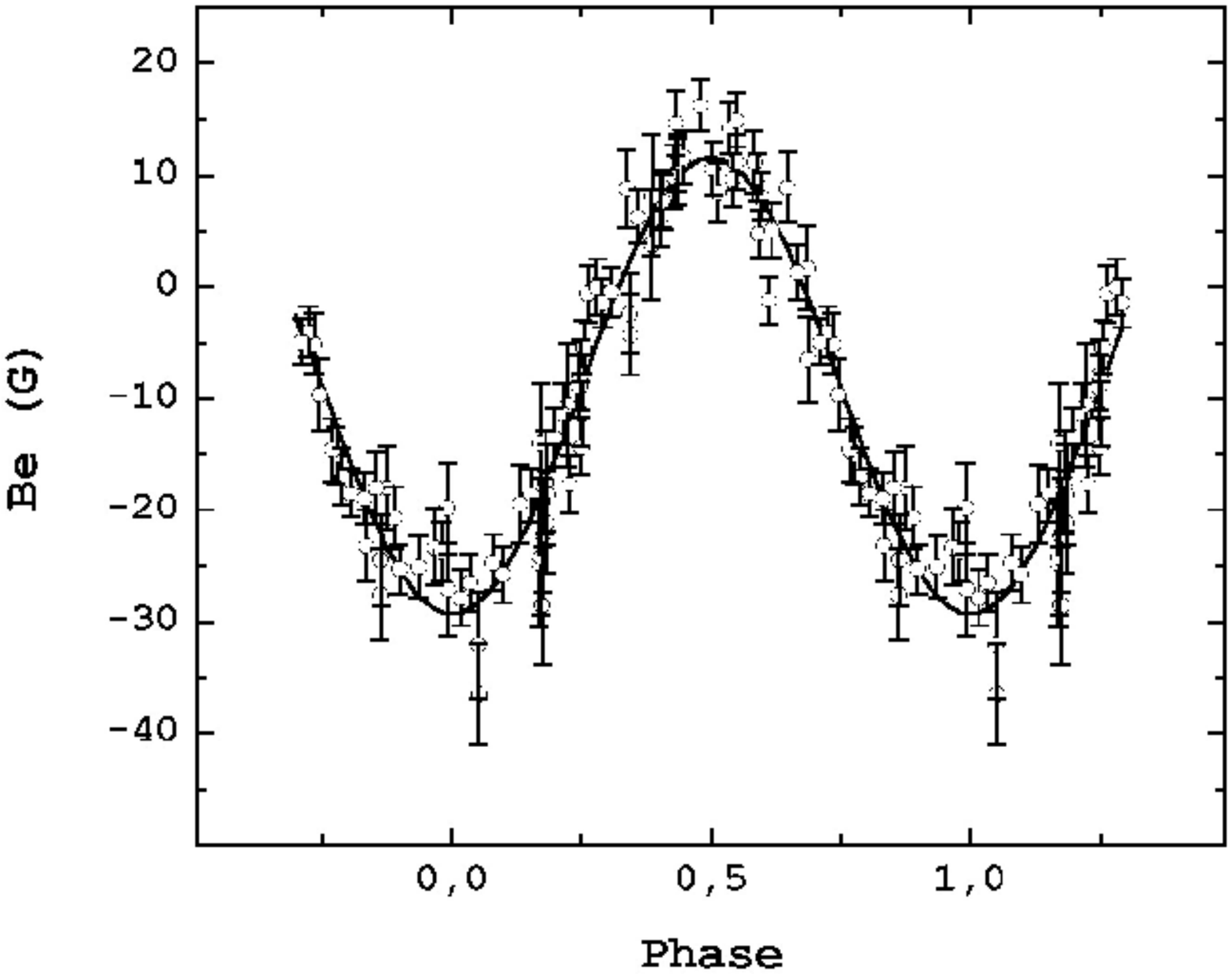}}
\vspace{-3.5mm}
\caption{ HD 3360 (2) }
\label{fig:fig8}
\end{figure}

\begin{figure}
\resizebox{0.98\hsize}{!}{\includegraphics{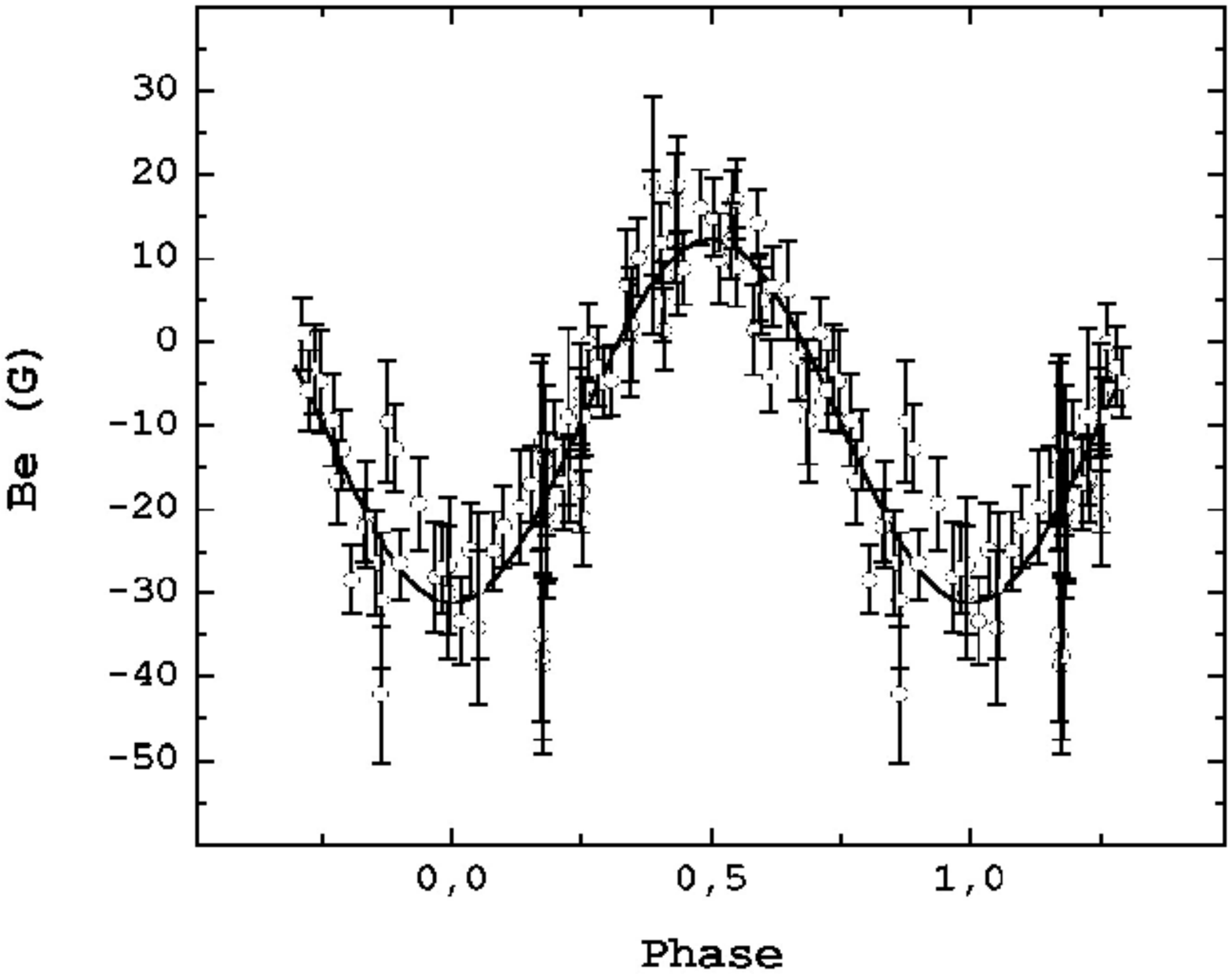}}
\vspace{-3.5mm}
\caption{ HD 3360 (3) }
\label{fig:fig9}
\end{figure}

\begin{figure}
\resizebox{0.98\hsize}{!}{\includegraphics{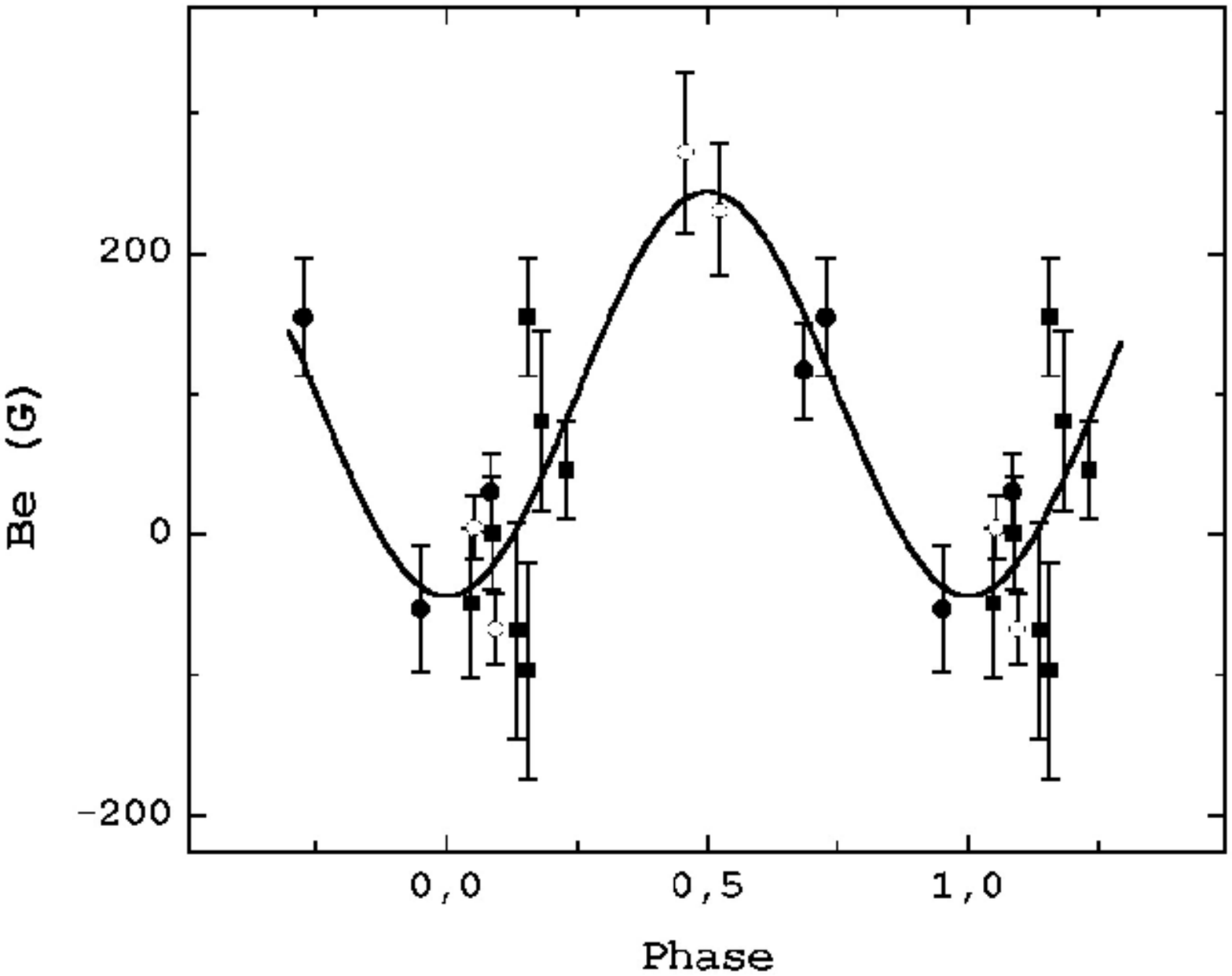}}
\vspace{-3.5mm}
\caption{ HD 3379 }
\label{fig:fig10}
\end{figure}

\begin{figure}
\resizebox{0.98\hsize}{!}{\includegraphics{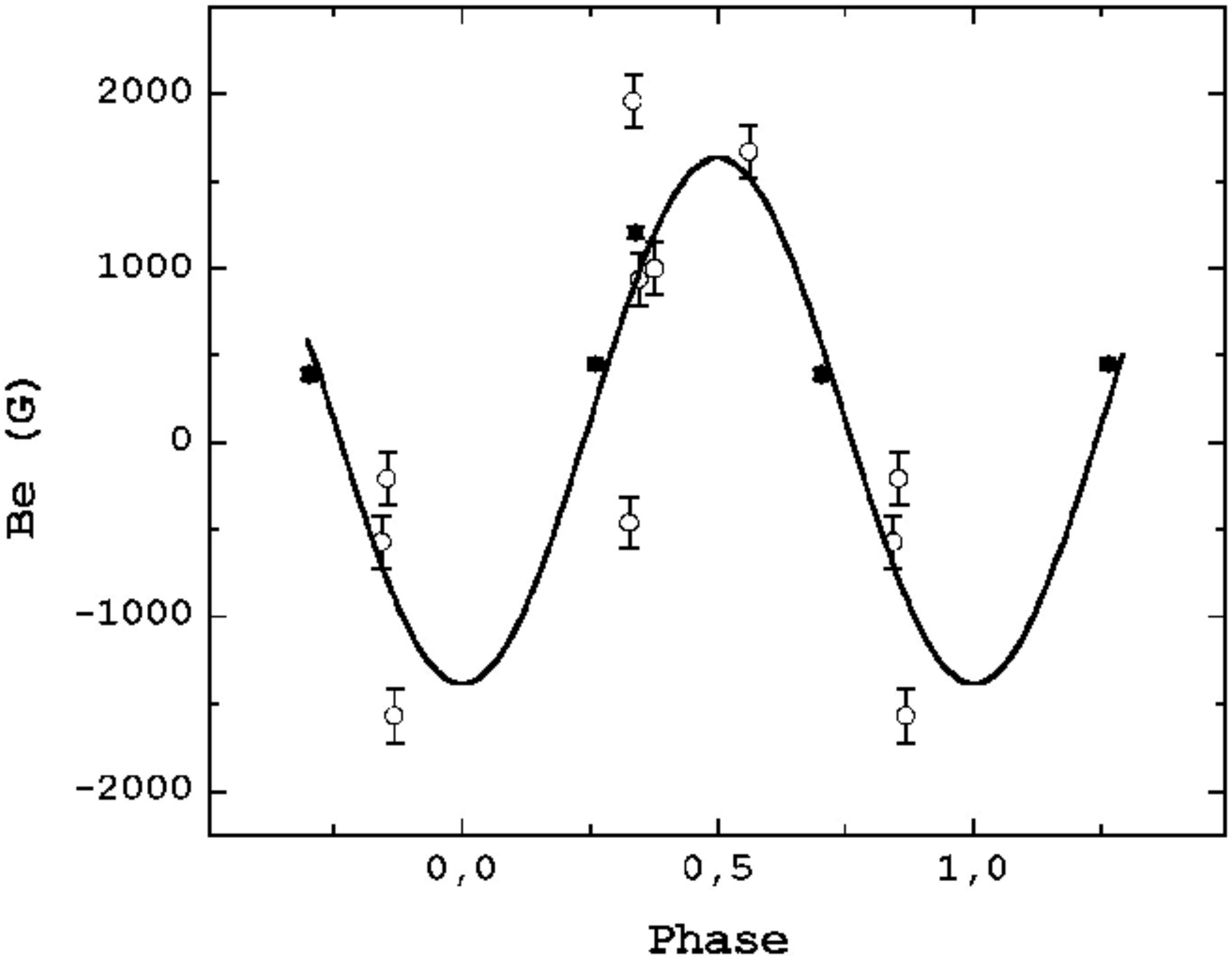}}
\vspace{-3.5mm}
\caption{ HD 3980 }
\label{fig:fig11}
\end{figure}

\clearpage
\newpage

\begin{figure}
\resizebox{0.98\hsize}{!}{\includegraphics{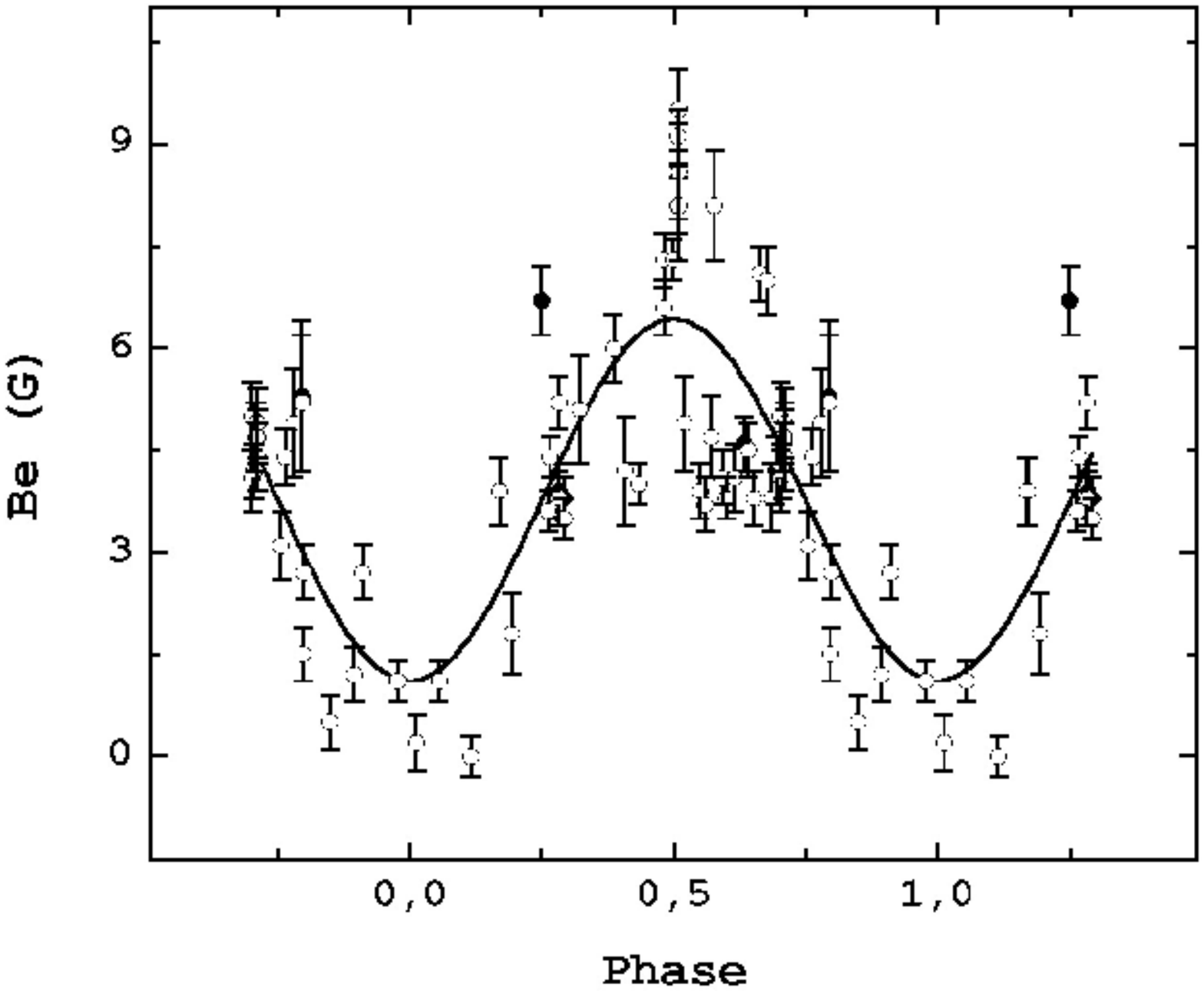}}
\vspace{-3.5mm}
\caption{ HD 4128 }
\label{fig:fig12}
\end{figure}

\begin{figure}
\resizebox{0.98\hsize}{!}{\includegraphics{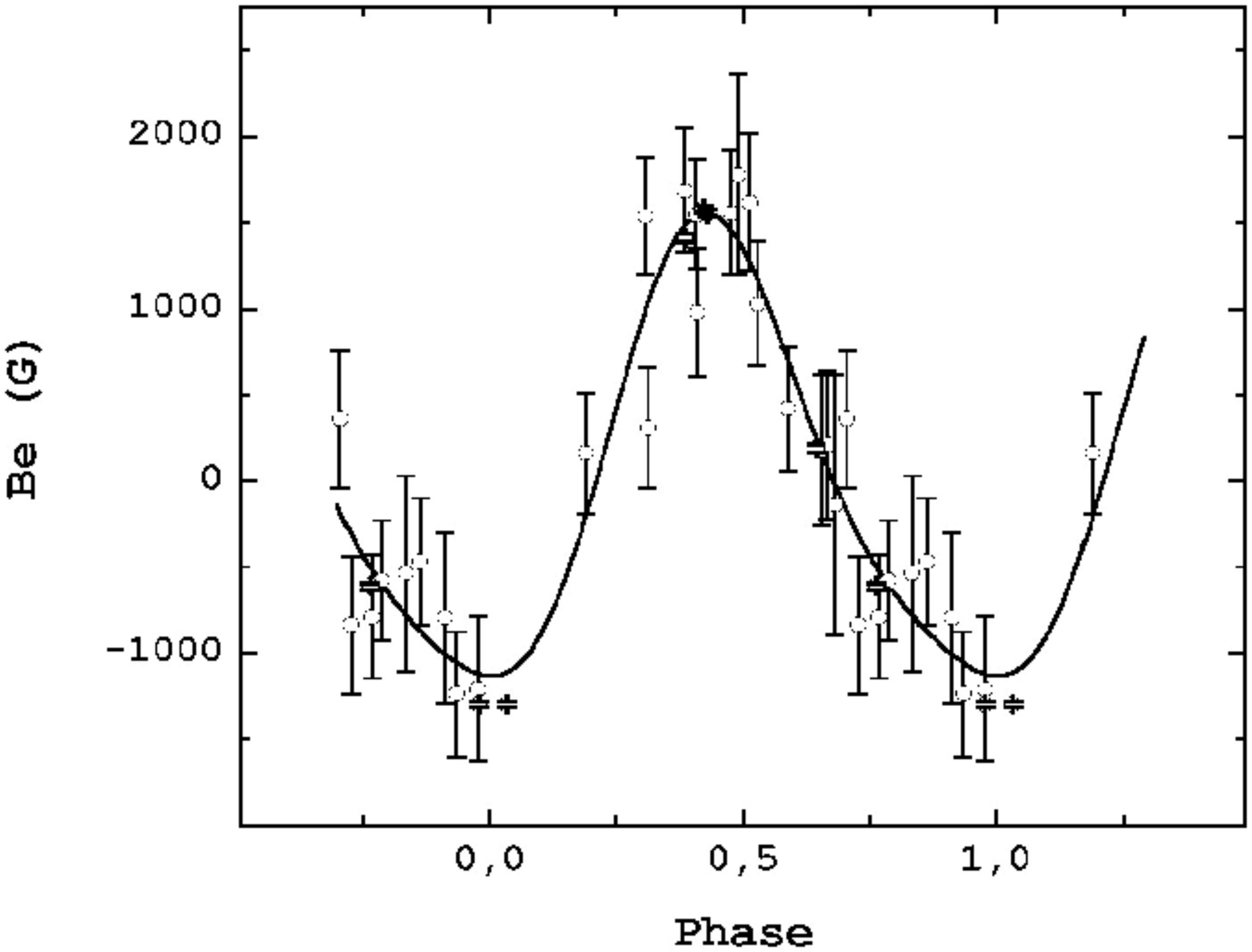}}
\vspace{-3.5mm}
\caption{ HD 4778 }
\label{fig:fig13}
\end{figure}

\begin{figure}
\resizebox{0.98\hsize}{!}{\includegraphics{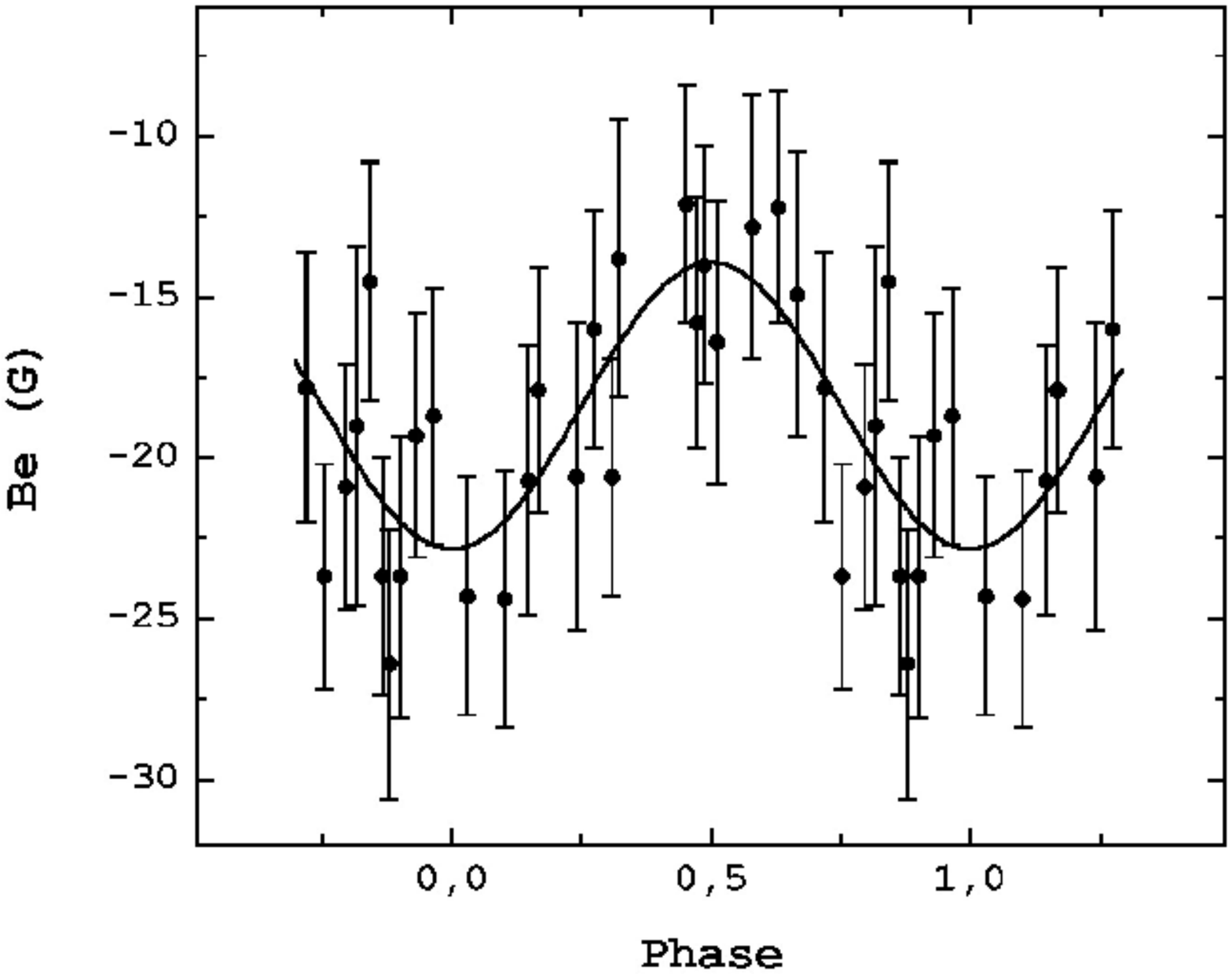}}
\vspace{-3.5mm}
\caption{ HD 5550 }
\label{fig:fig14}
\end{figure}

\begin{figure}
\resizebox{0.98\hsize}{!}{\includegraphics{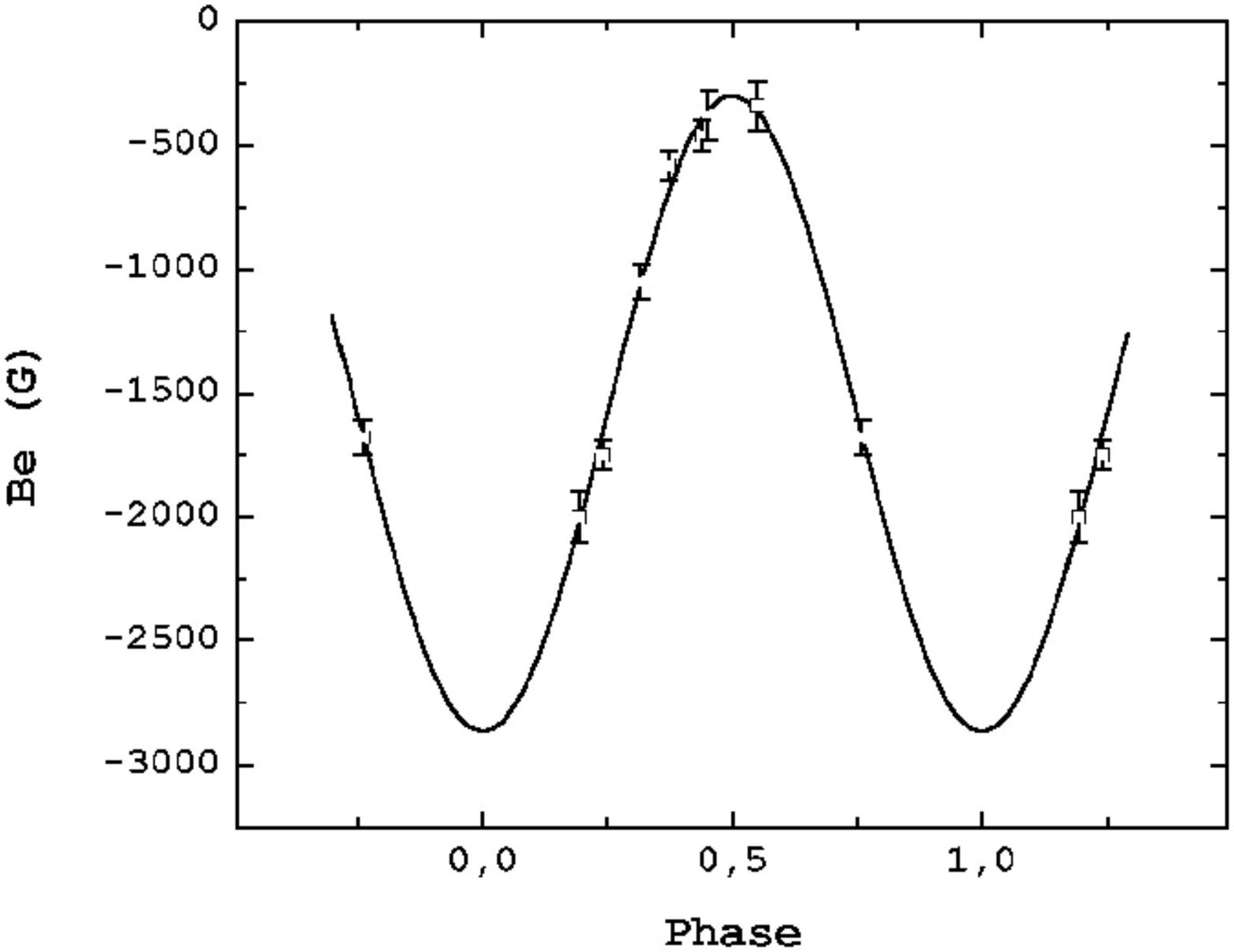}}
\vspace{-4.0mm}
\caption{ HD 5601 (1) }
\label{fig:fig15}
\end{figure}

\begin{figure}
\resizebox{0.98\hsize}{!}{\includegraphics{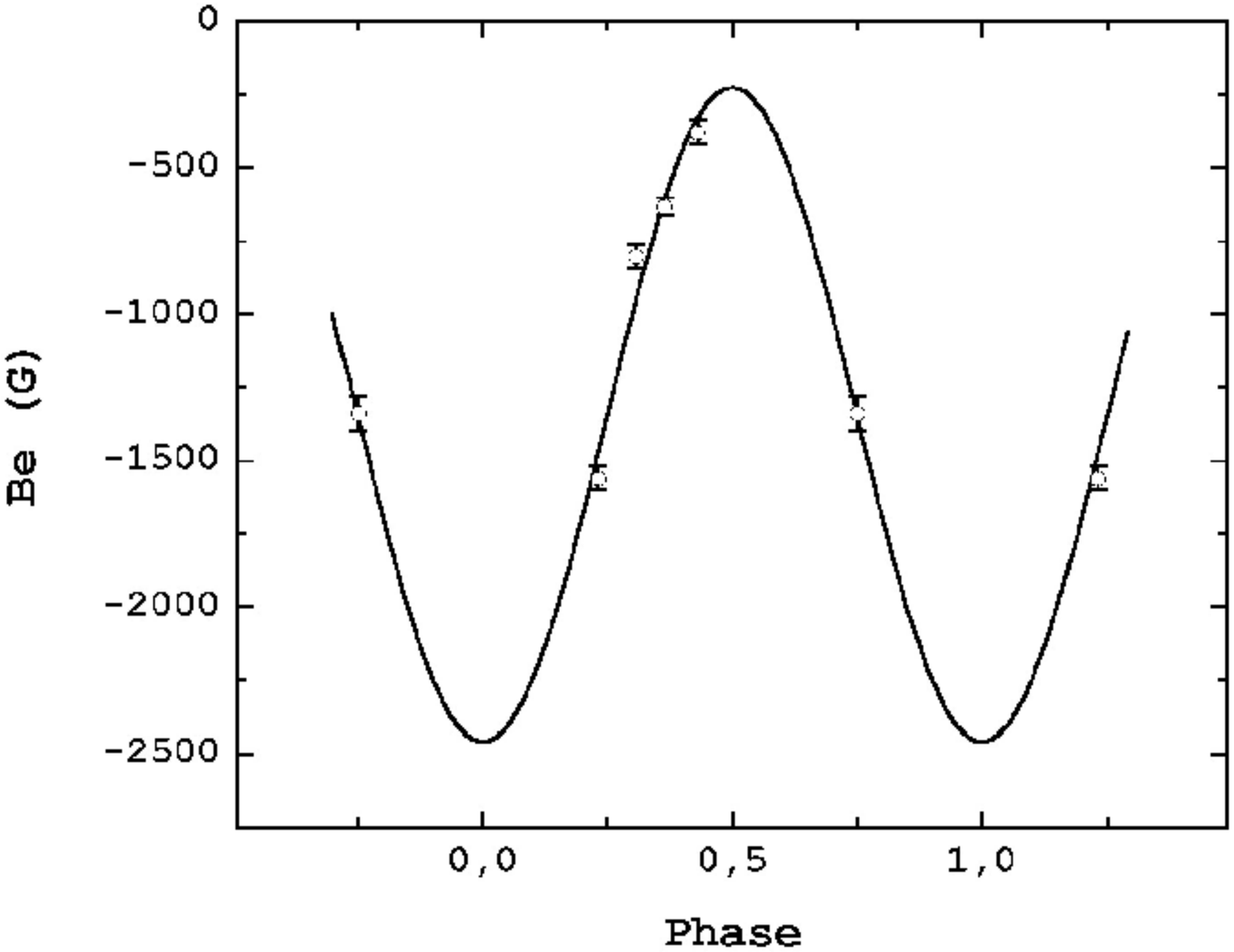}}
\vspace{-3.5mm}
\caption{ HD 5601 (2) }
\label{fig:fig16}
\end{figure}

\begin{figure}
\resizebox{0.98\hsize}{!}{\includegraphics{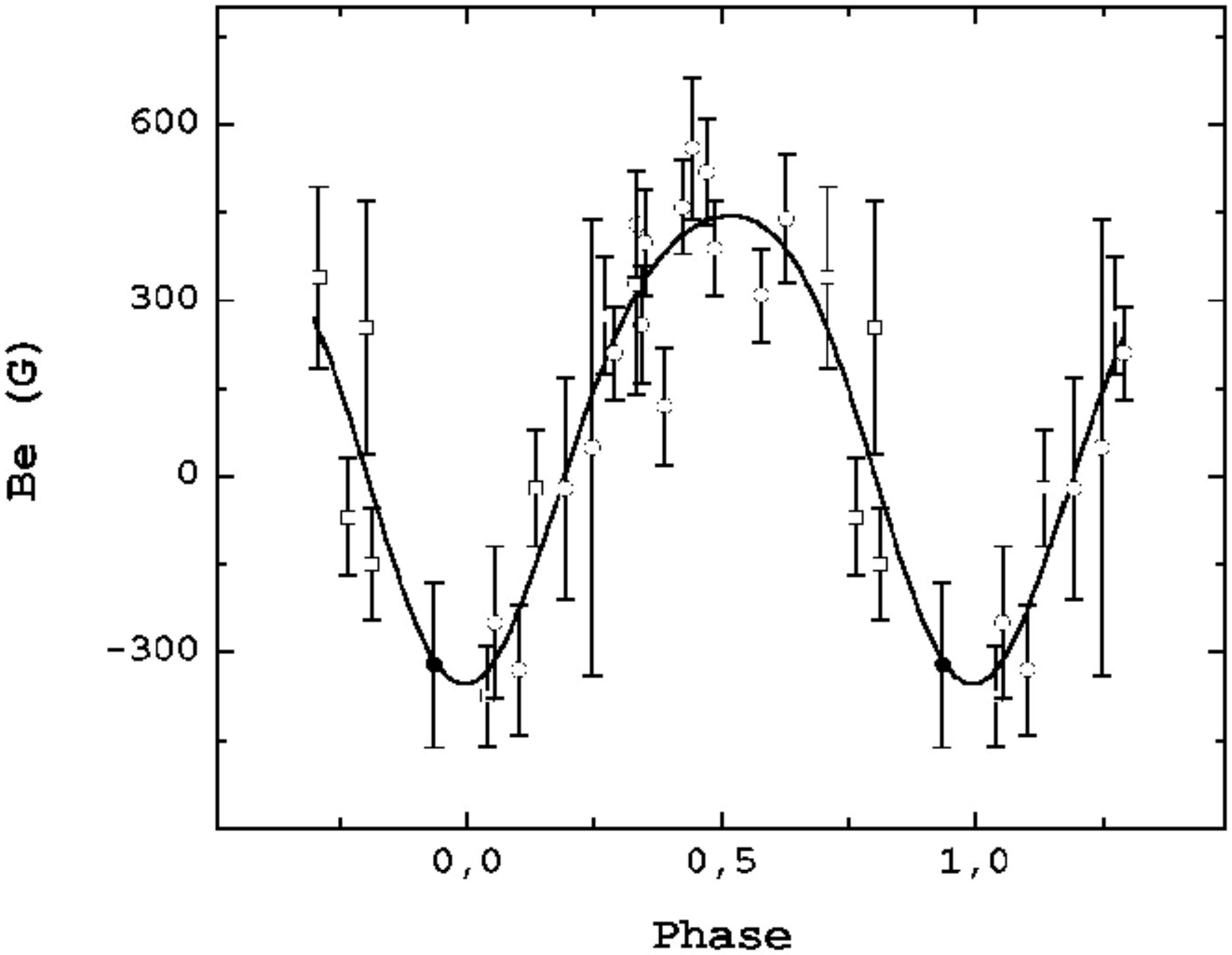}}
\vspace{-3.5mm}
\caption{ HD 5737 }
\label{fig:fig17}
\end{figure}

\clearpage
\newpage

\begin{figure}
\resizebox{0.98\hsize}{!}{\includegraphics{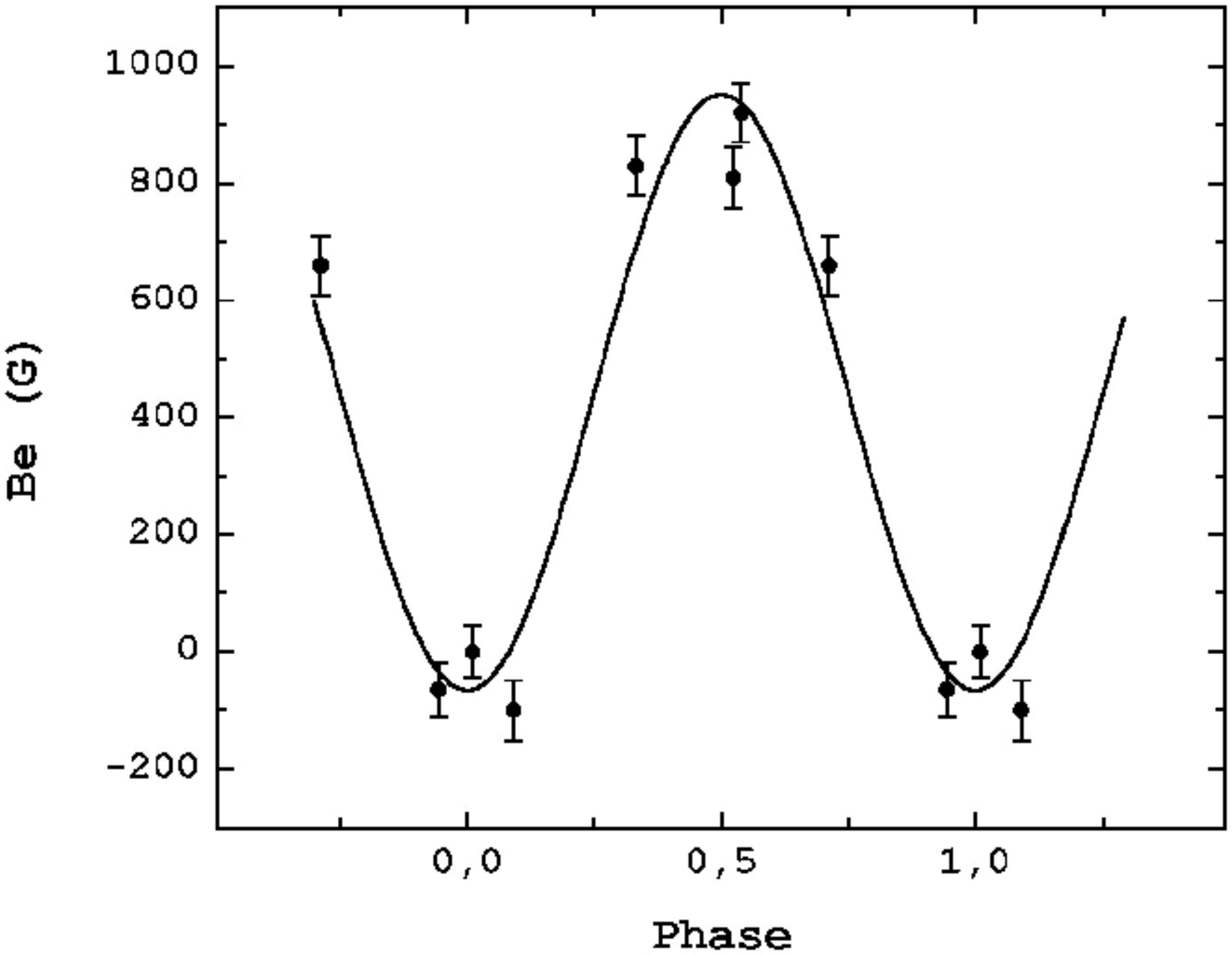}}
\vspace{-3.5mm}
\caption{ HD 5797 }
\label{fig:fig18}
\end{figure}

\begin{figure}
\resizebox{0.98\hsize}{!}{\includegraphics{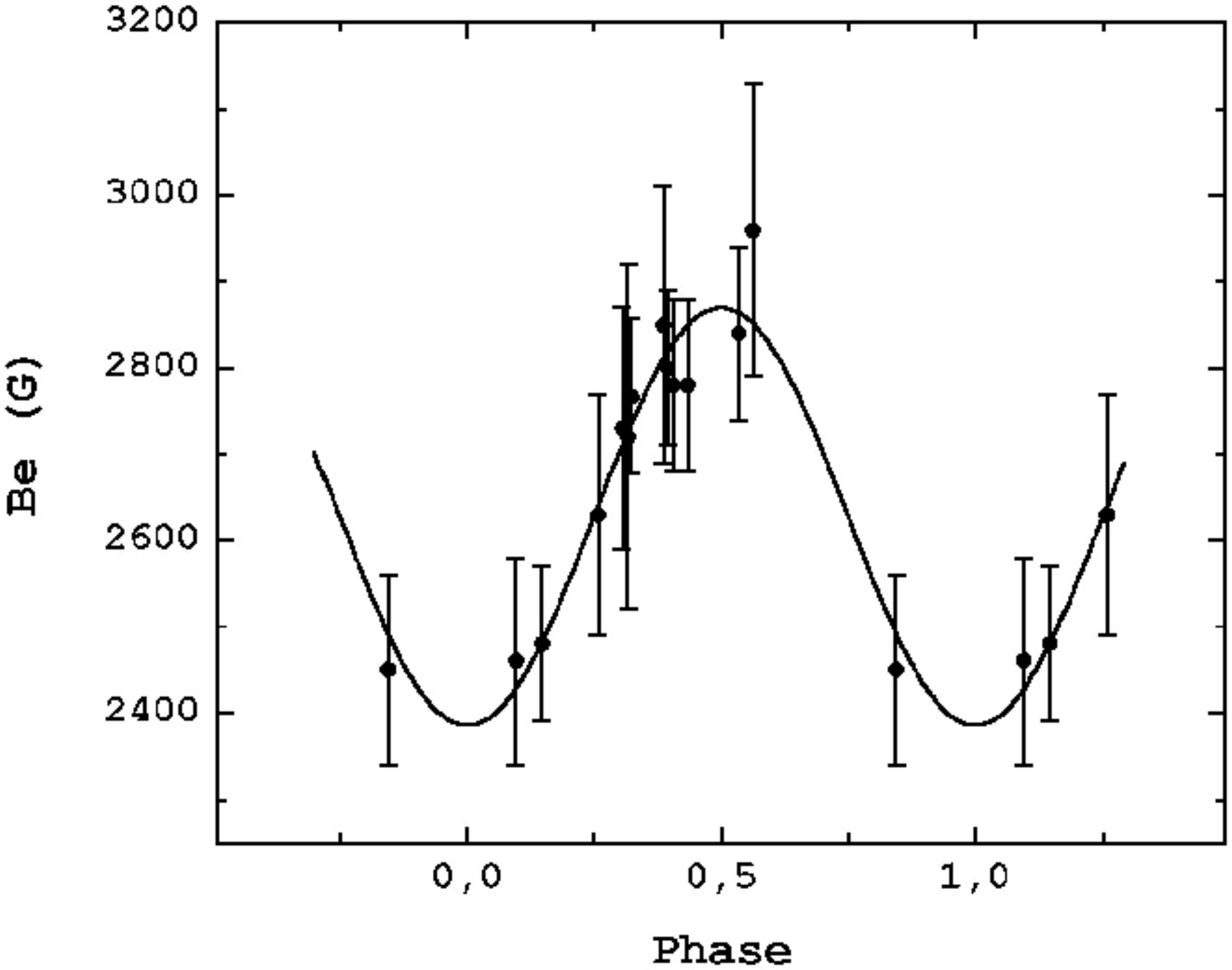}}
\vspace{-3.5mm}
\caption{ HD 6757 }
\label{fig:fig19}
\end{figure}

\begin{figure}
\resizebox{0.98\hsize}{!}{\includegraphics{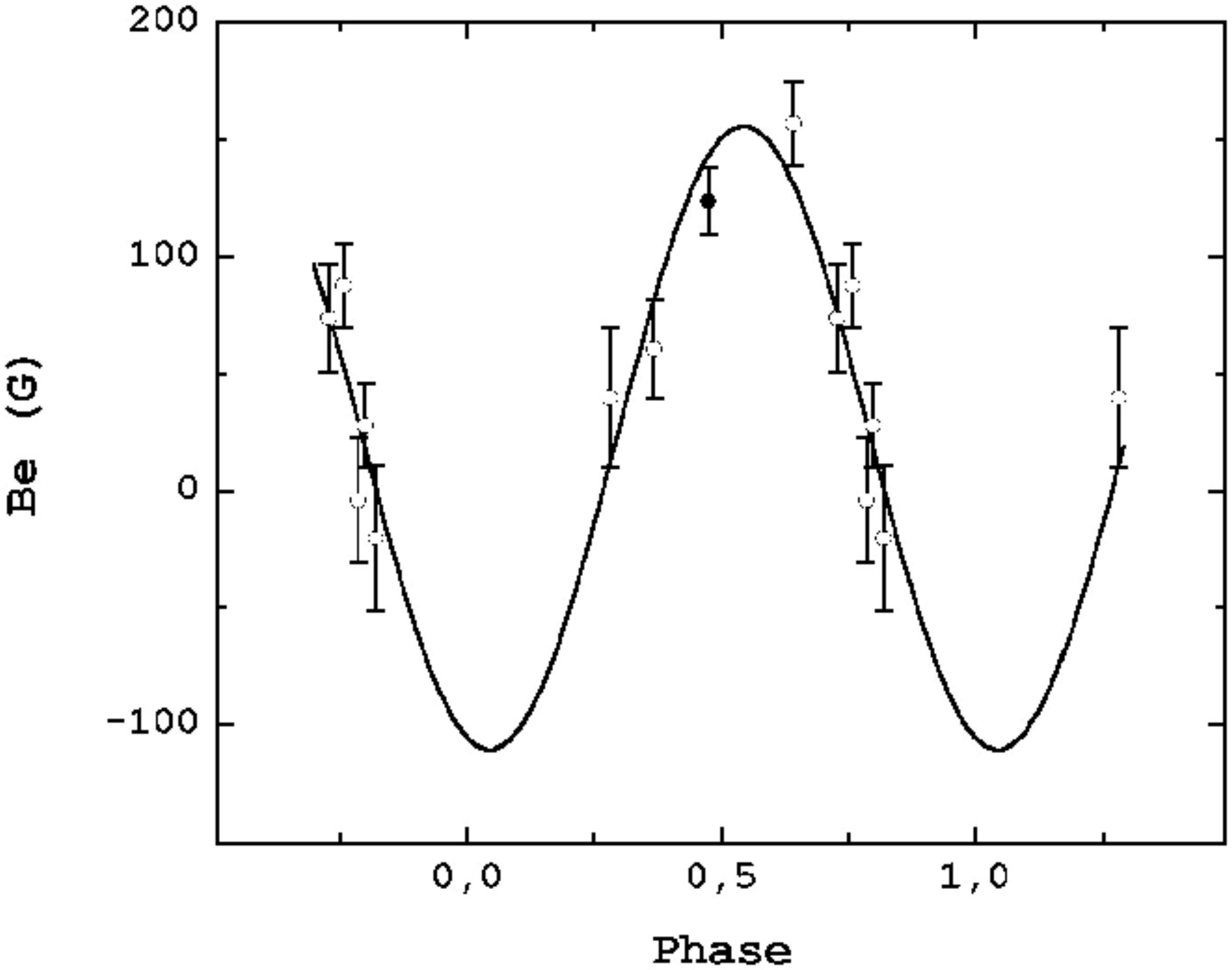}}
\vspace{-3.5mm}
\caption{ HD 8441 }
\label{fig:fig20}
\end{figure}

\begin{figure}
\resizebox{0.98\hsize}{!}{\includegraphics{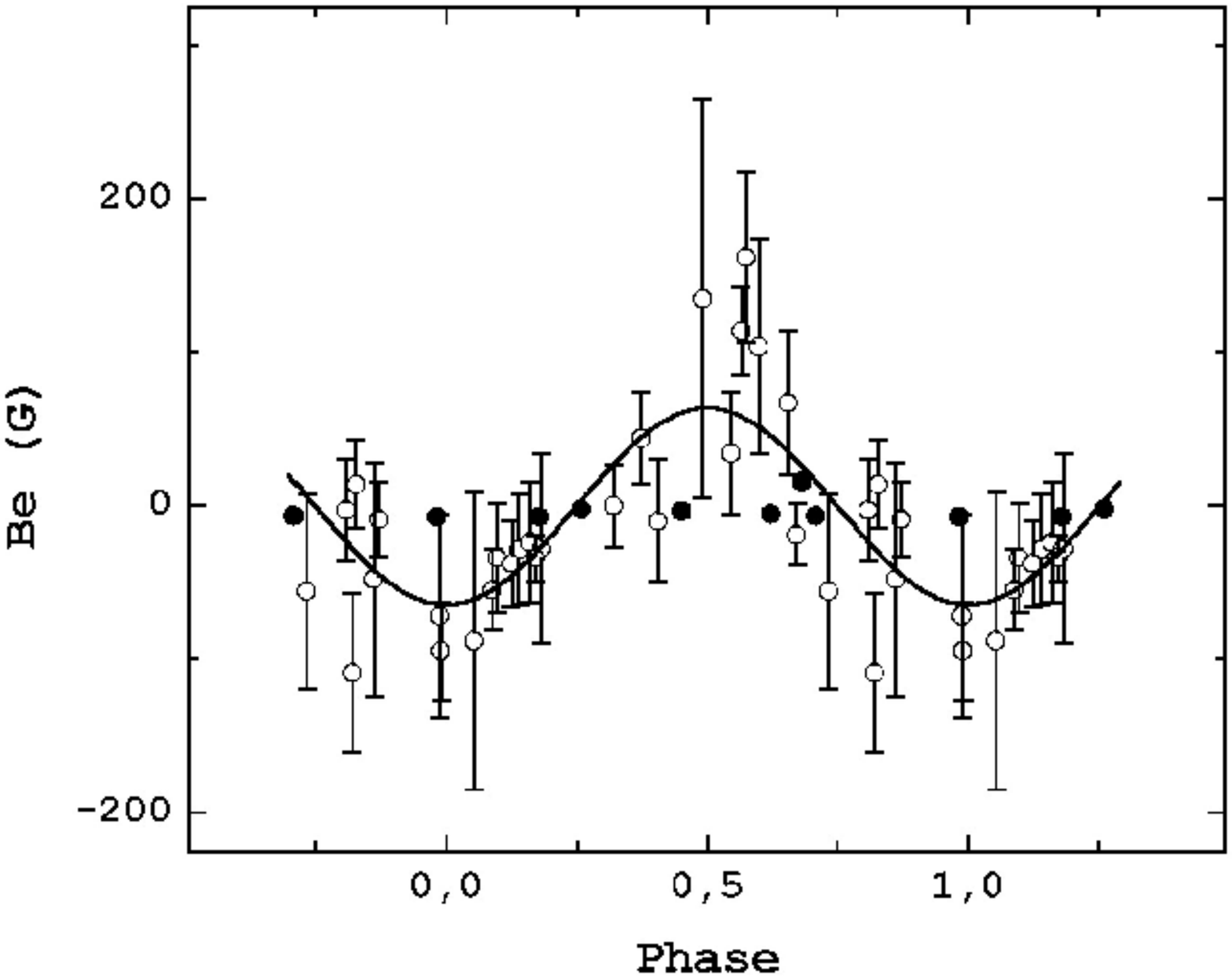}}
\vspace{-3.5mm}
\caption{ HD 8890 }
\label{fig:fig21}
\end{figure}

\begin{figure}
\resizebox{0.98\hsize}{!}{\includegraphics{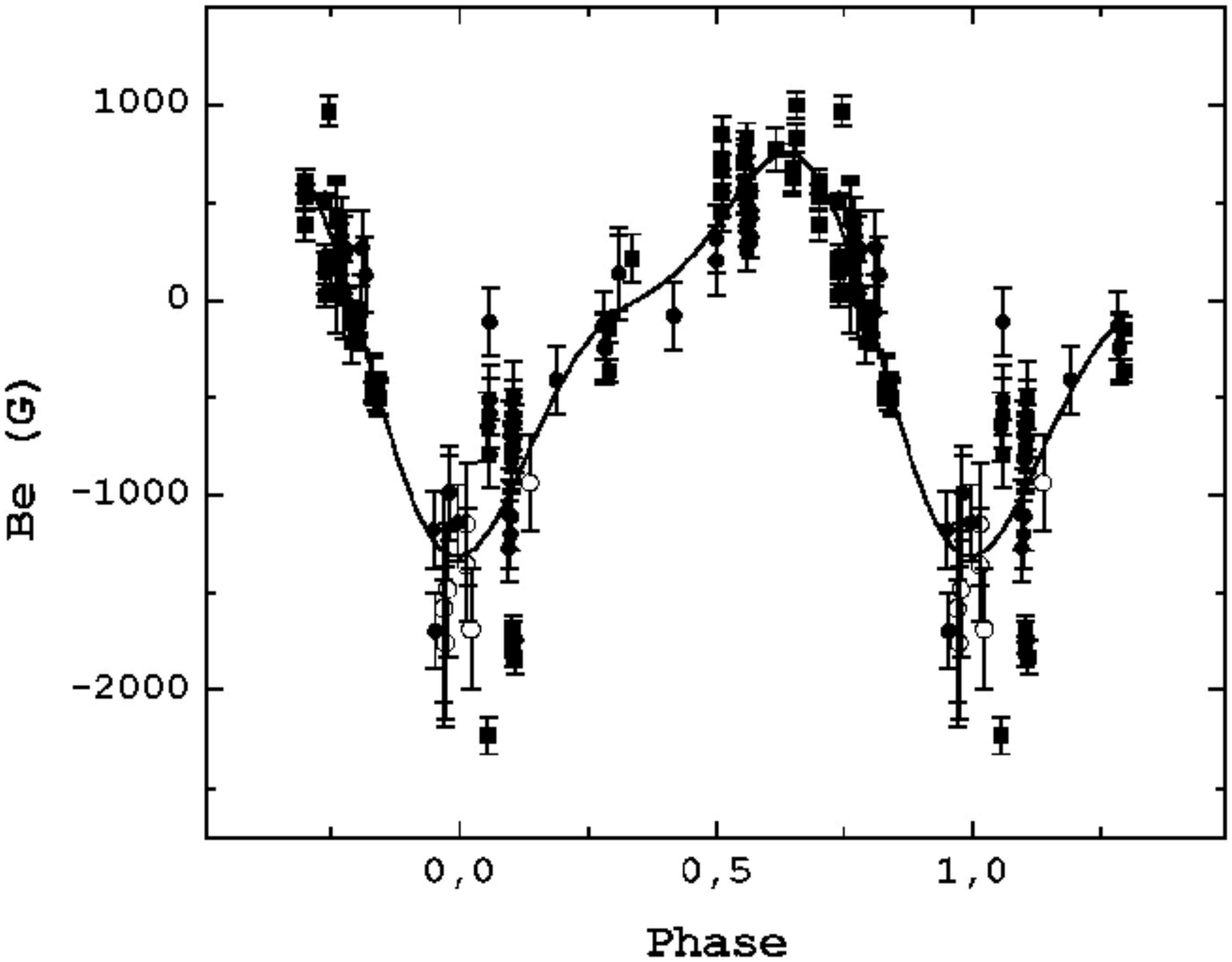}}
\vspace{-3.5mm}
\caption{ HD 9996 }
\label{fig:fig22}
\end{figure}

\begin{figure}
\resizebox{0.98\hsize}{!}{\includegraphics{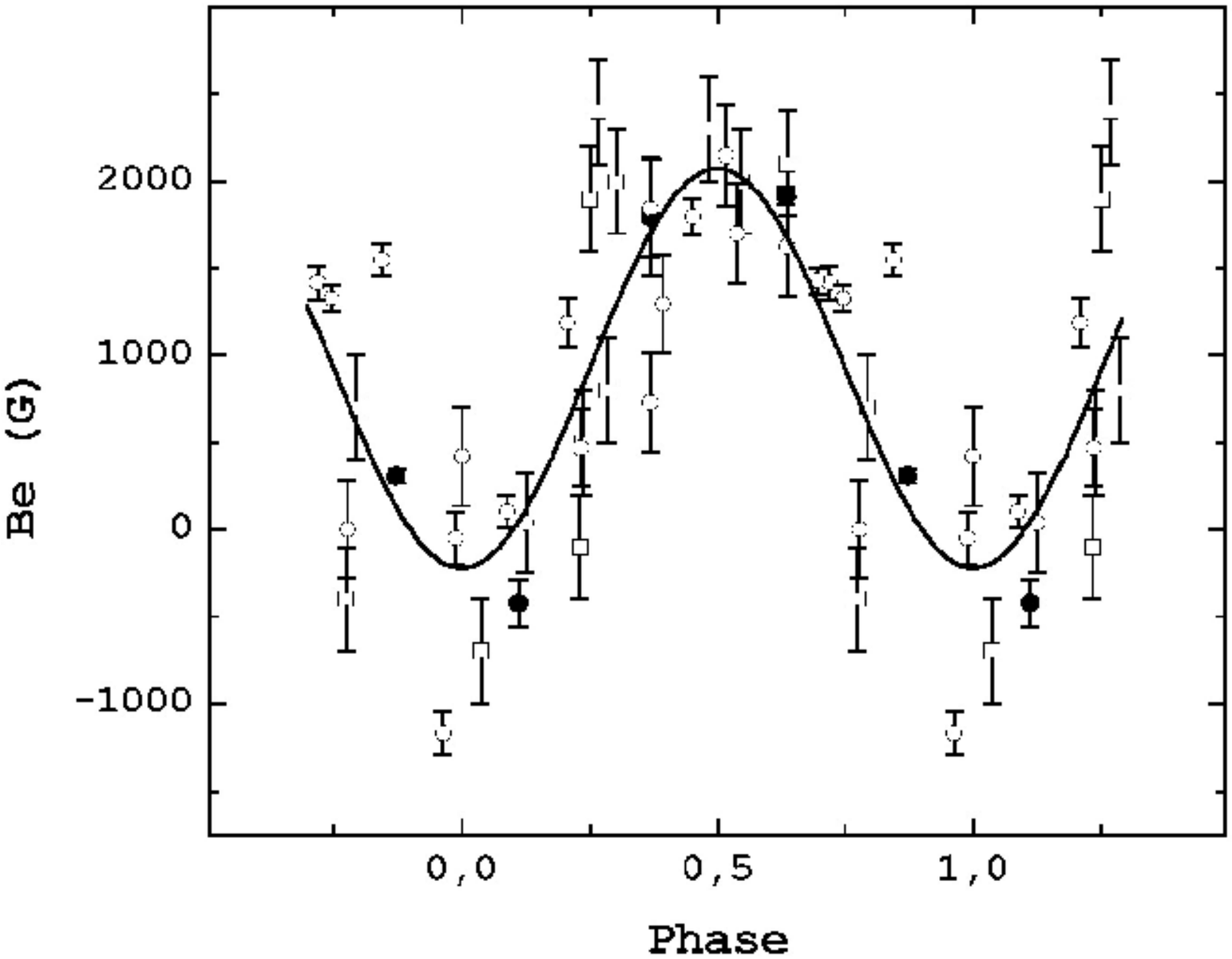}}
\vspace{-3.5mm}
\caption{ HD 10783 }
\label{fig:fig23}
\end{figure}

\clearpage
\newpage

\begin{figure}
\resizebox{0.98\hsize}{!}{\includegraphics{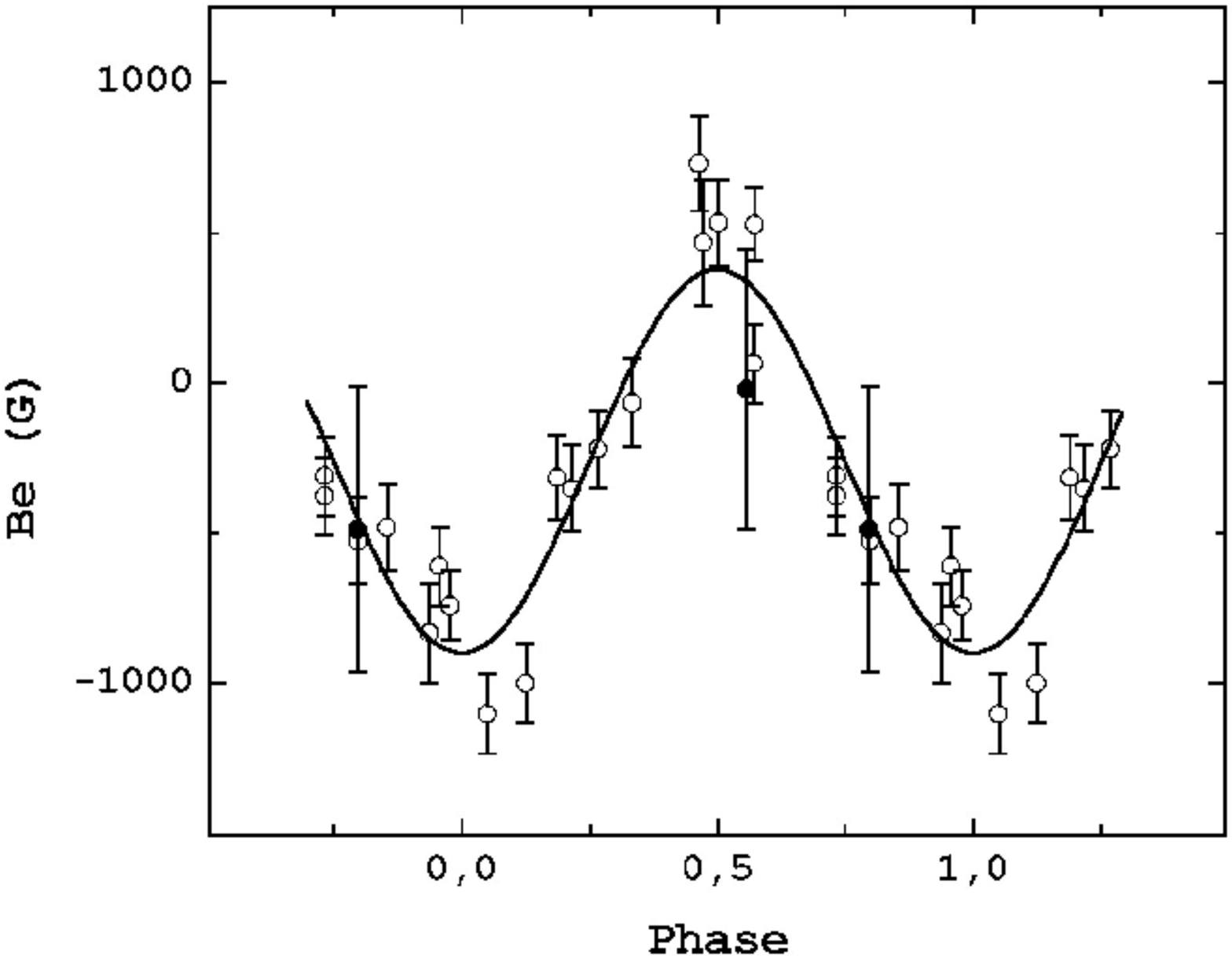}}
\vspace{-3.5mm}
\caption{ HD 11503 }
\label{fig:fig24}
\end{figure}

\begin{figure}
\resizebox{0.98\hsize}{!}{\includegraphics{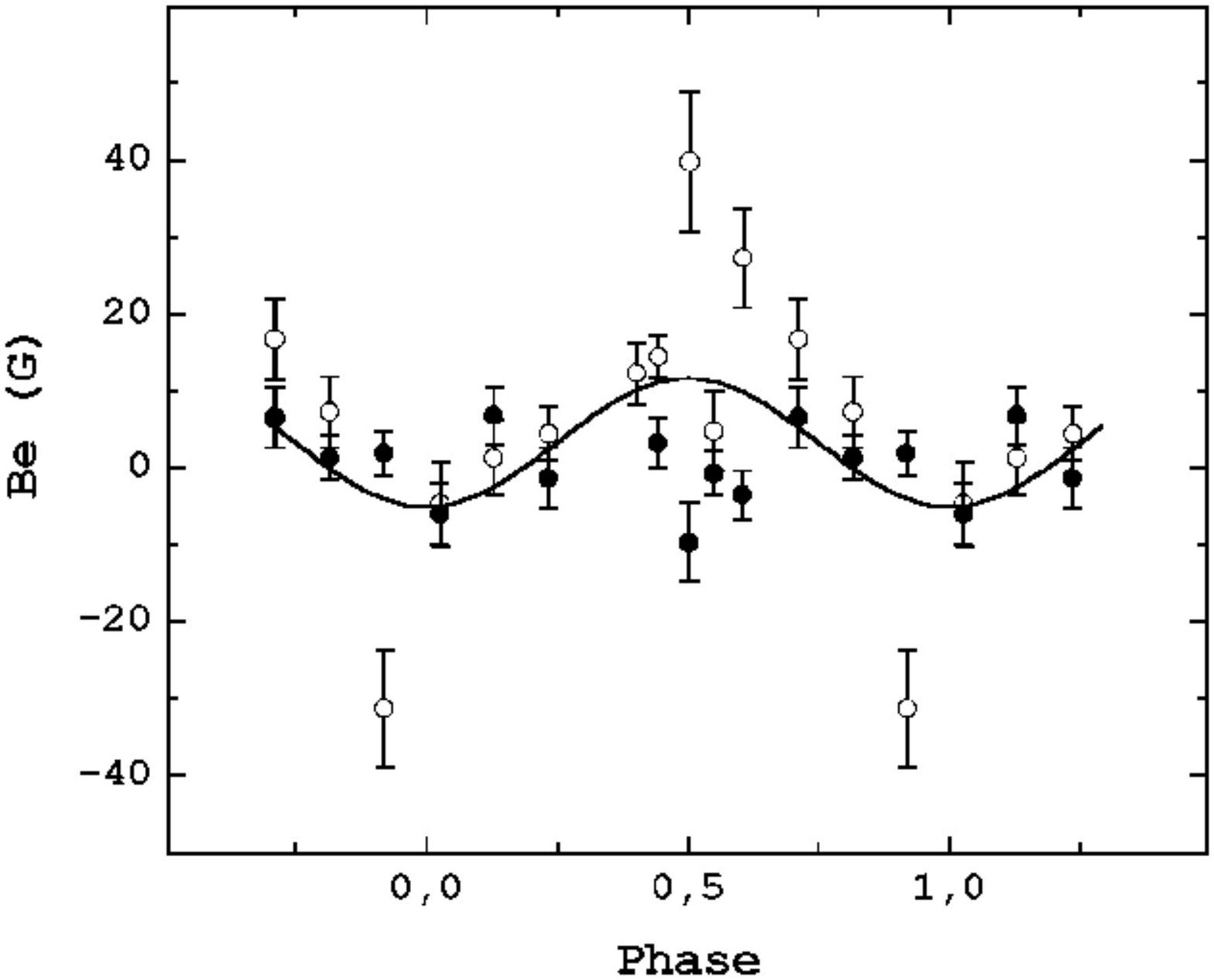}}
\vspace{-3.5mm}
\caption{ HD 11753 }
\label{fig:fig25}
\end{figure}

\begin{figure}
\resizebox{0.98\hsize}{!}{\includegraphics{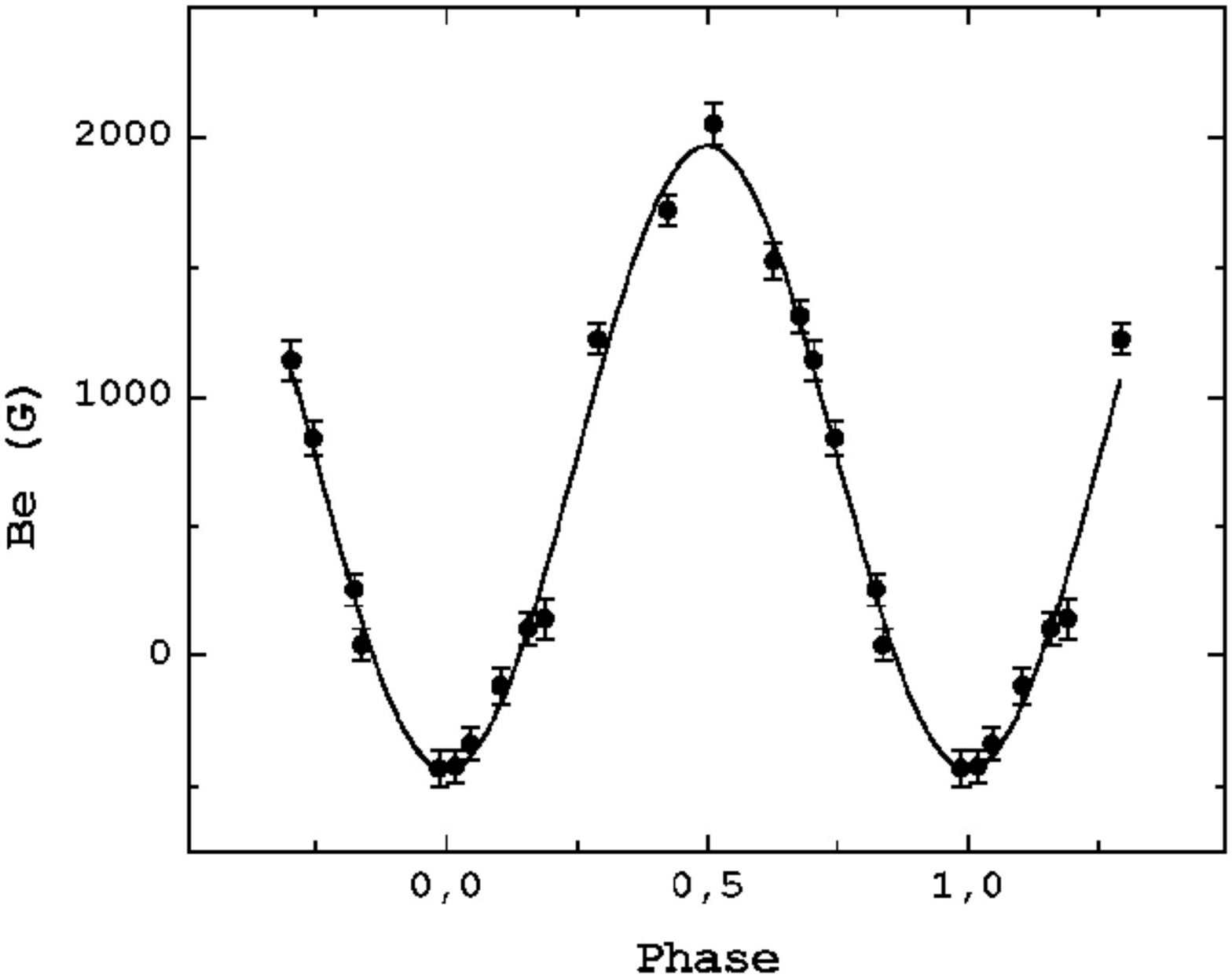}}
\vspace{-3.5mm}
\caption{ HD 12098 }
\label{fig:fig26}
\end{figure}

\begin{figure}
\resizebox{0.98\hsize}{!}{\includegraphics{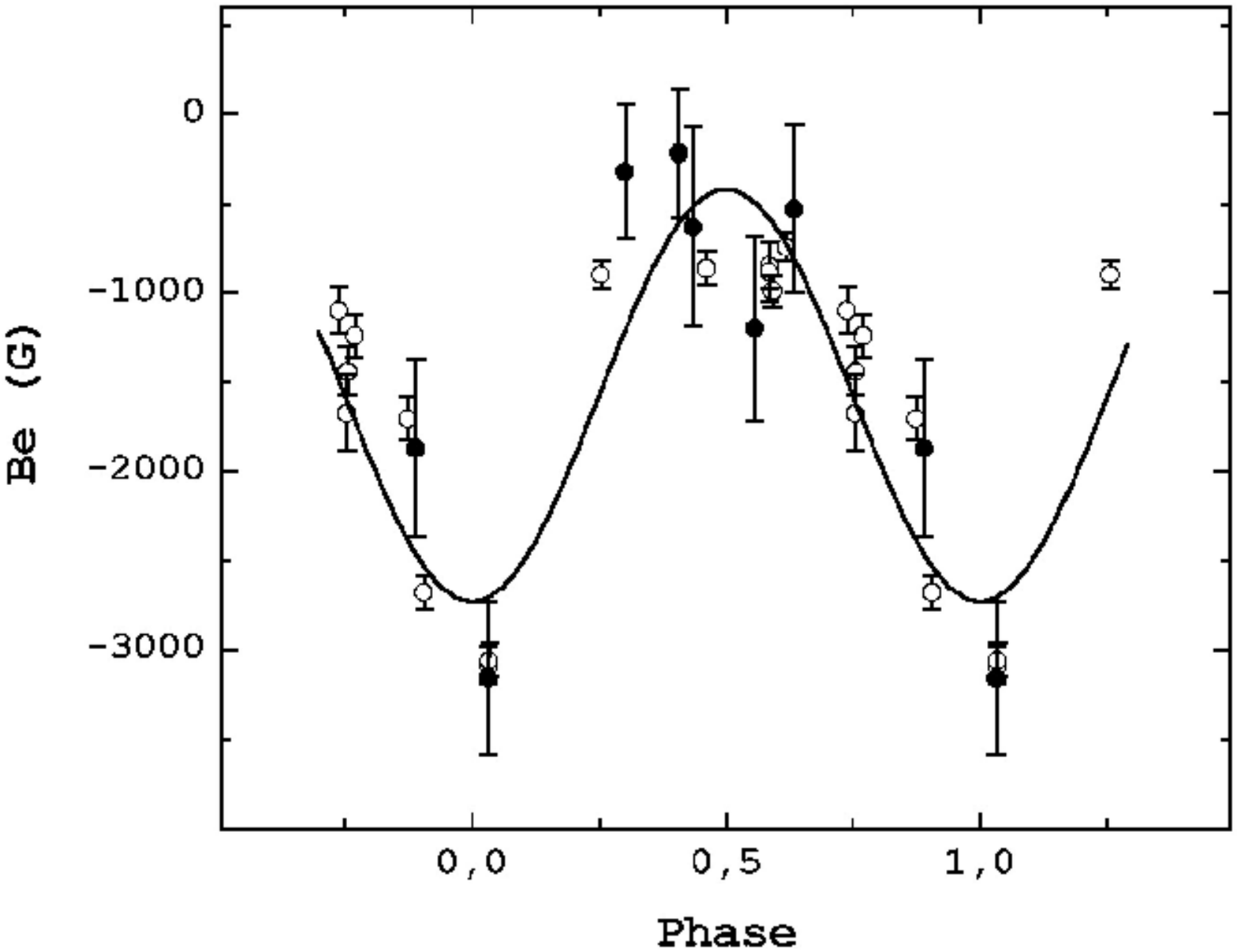}}
\vspace{-3.5mm}
\caption{ HD 12288 }
\label{fig:fig27}
\end{figure}

\begin{figure}
\resizebox{0.98\hsize}{!}{\includegraphics{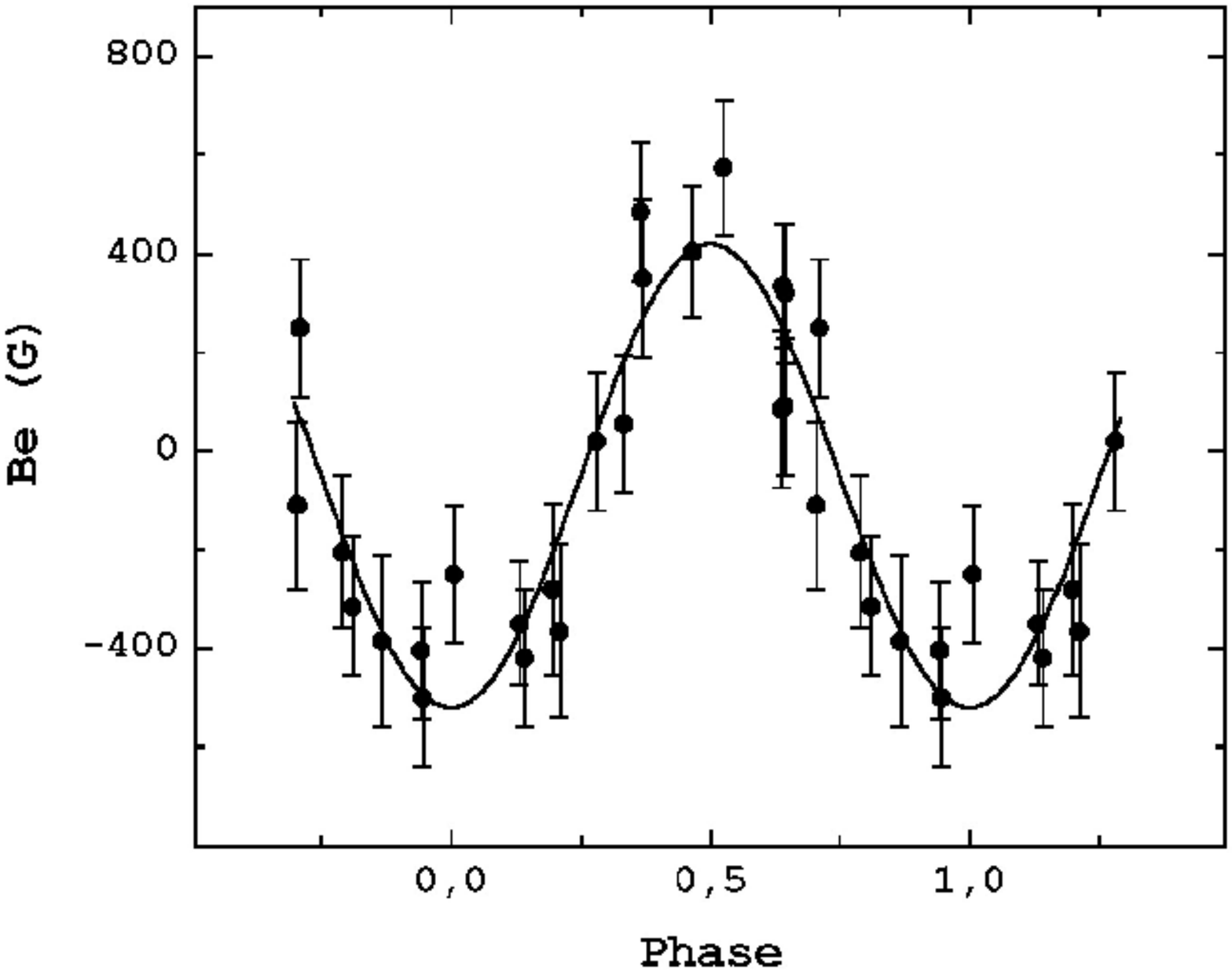}}
\vspace{-3.5mm}
\caption{ HD 12447 }
\label{fig:fig28}
\end{figure}

\begin{figure}
\resizebox{0.98\hsize}{!}{\includegraphics{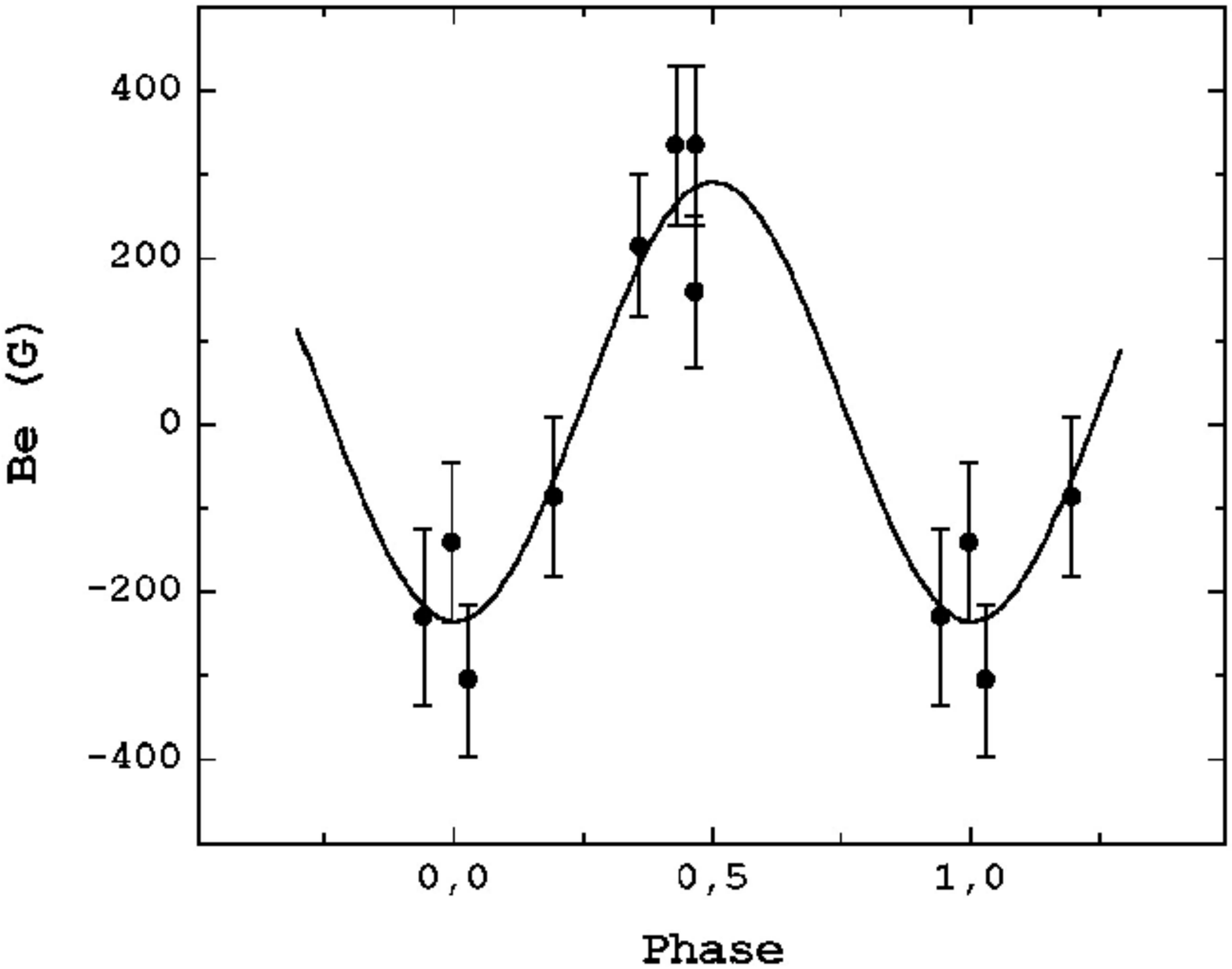}}
\vspace{-3.5mm}
\caption{ HD 12767 }
\label{fig:fig29}
\end{figure}

\clearpage
\newpage

\begin{figure}
\resizebox{0.98\hsize}{!}{\includegraphics{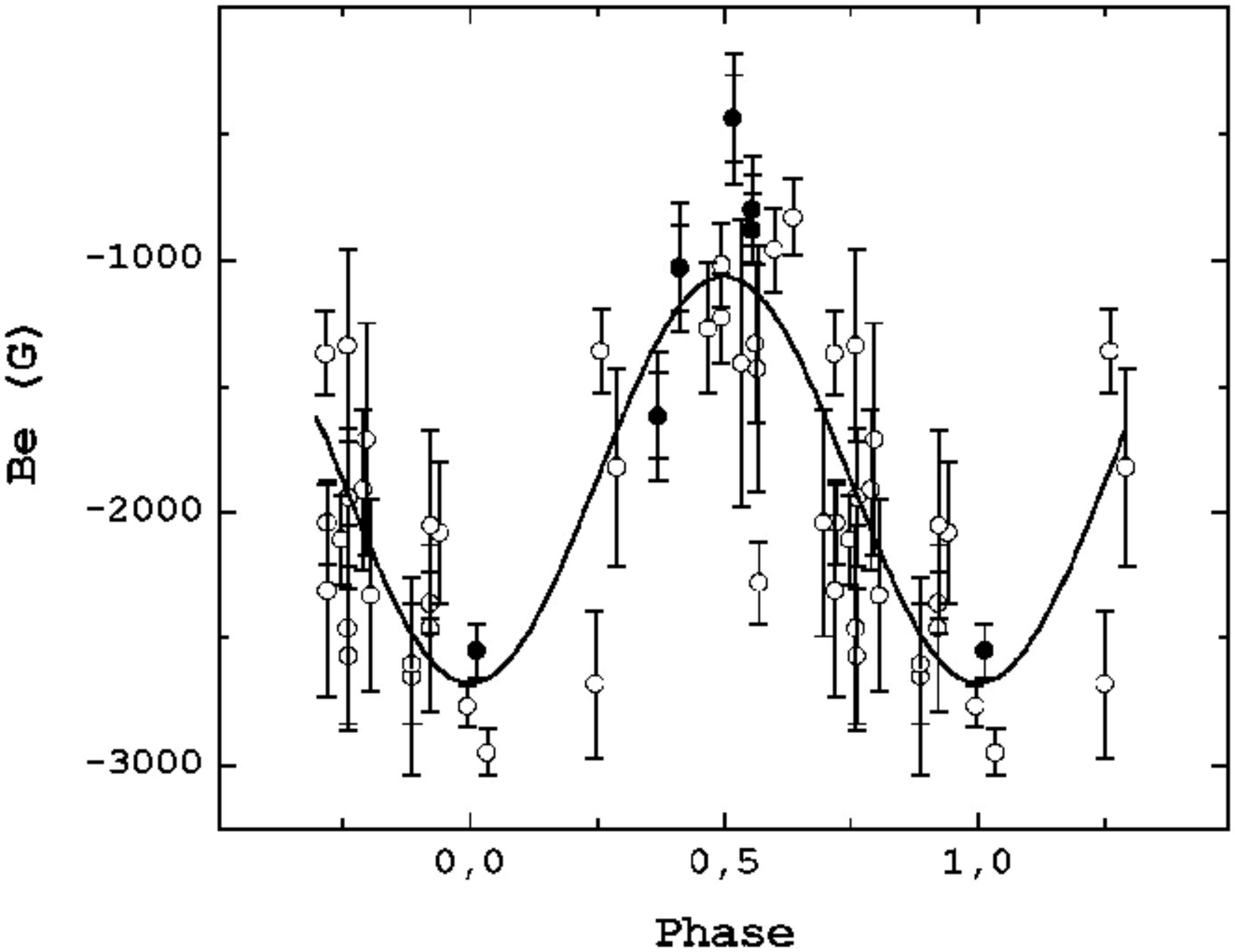}}
\vspace{-3.5mm}
\caption{ HD 14437 }
\label{fig:fig30}
\end{figure}

\begin{figure}
\resizebox{0.98\hsize}{!}{\includegraphics{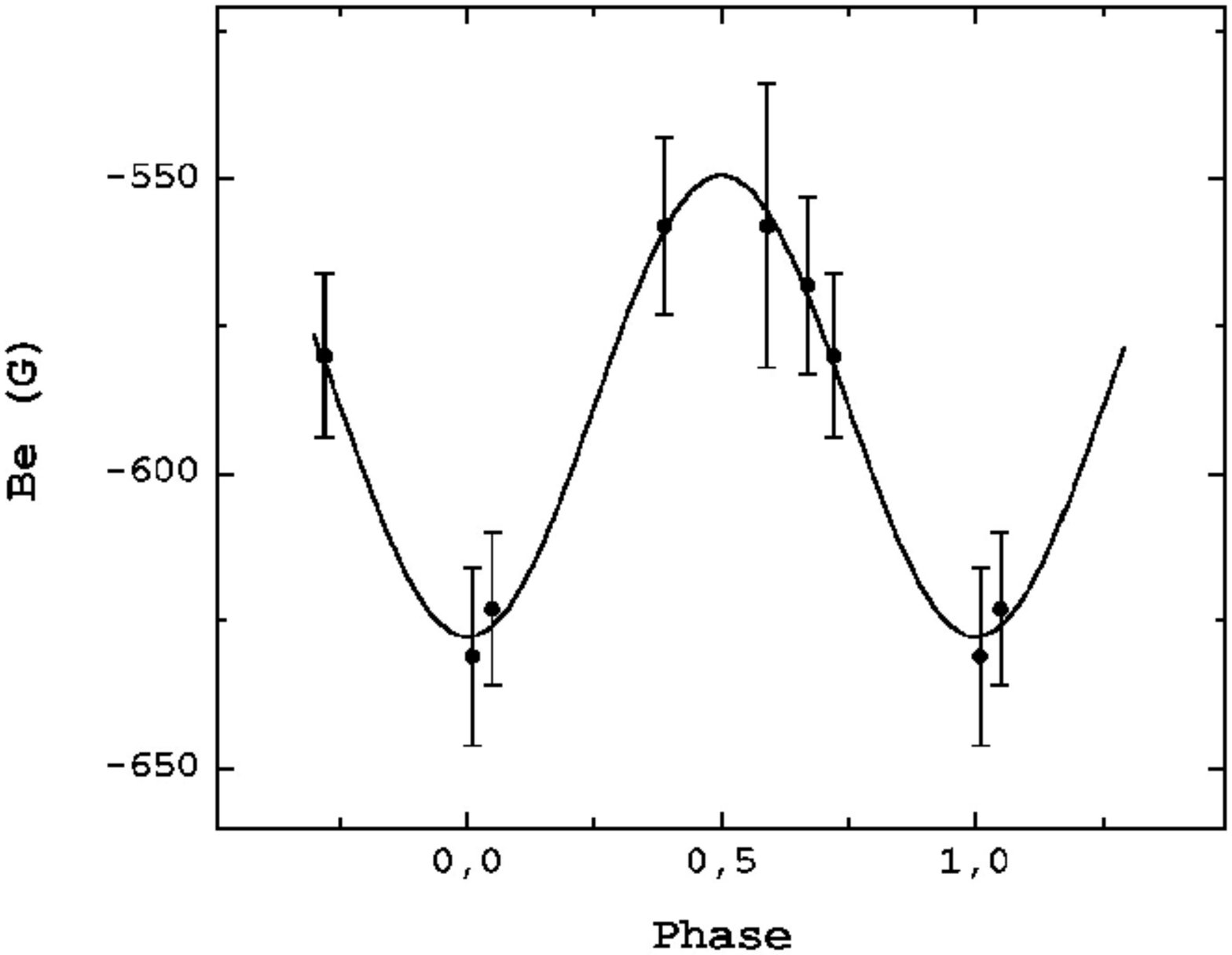}}
\vspace{-3.5mm}
\caption{ HD 15144 }
\label{fig:fig31}
\end{figure}

\begin{figure}
\resizebox{0.98\hsize}{!}{\includegraphics{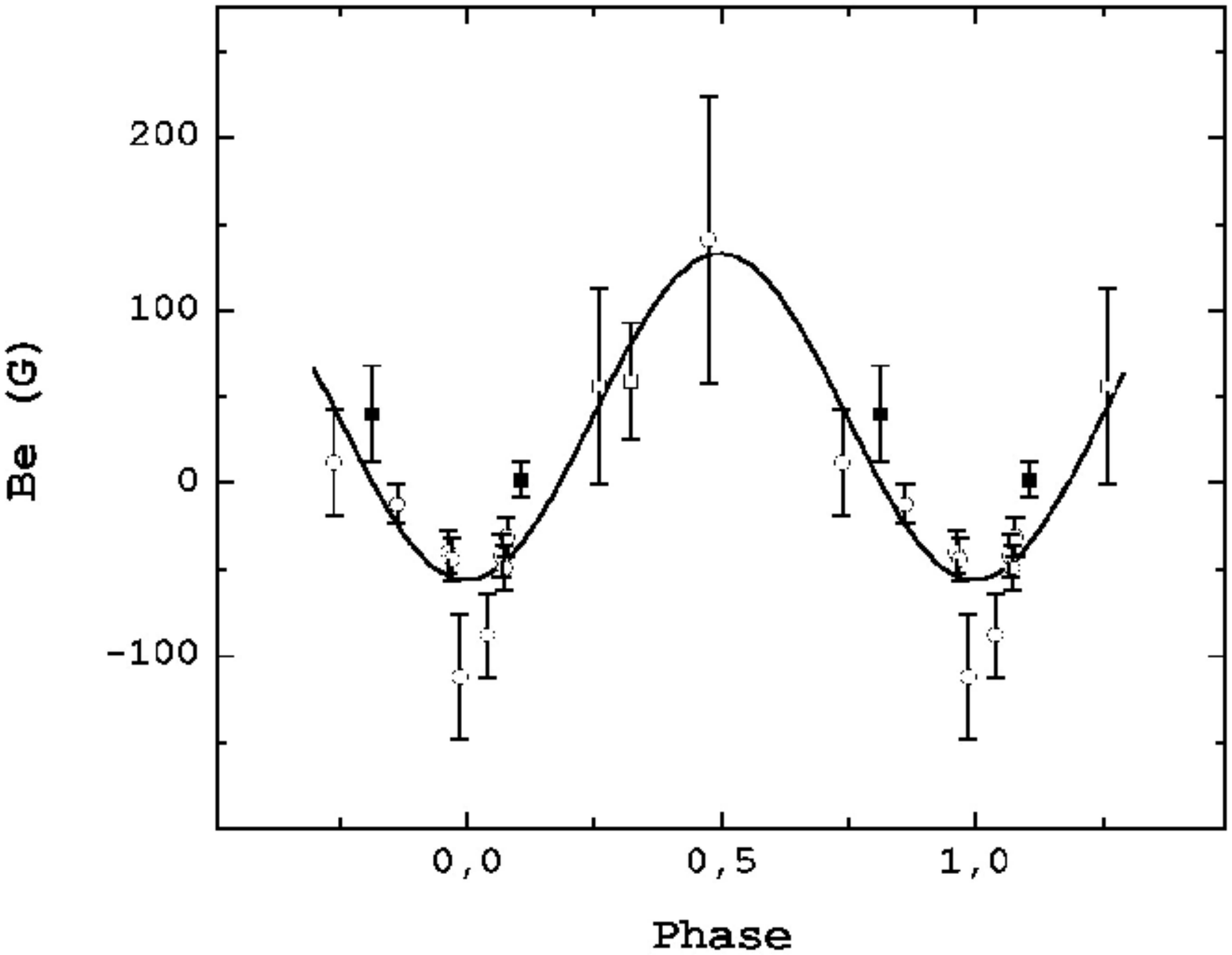}}
\vspace{-3.5mm}
\caption{ HD 16582 }
\label{fig:fig32}
\end{figure}

\begin{figure}
\resizebox{0.98\hsize}{!}{\includegraphics{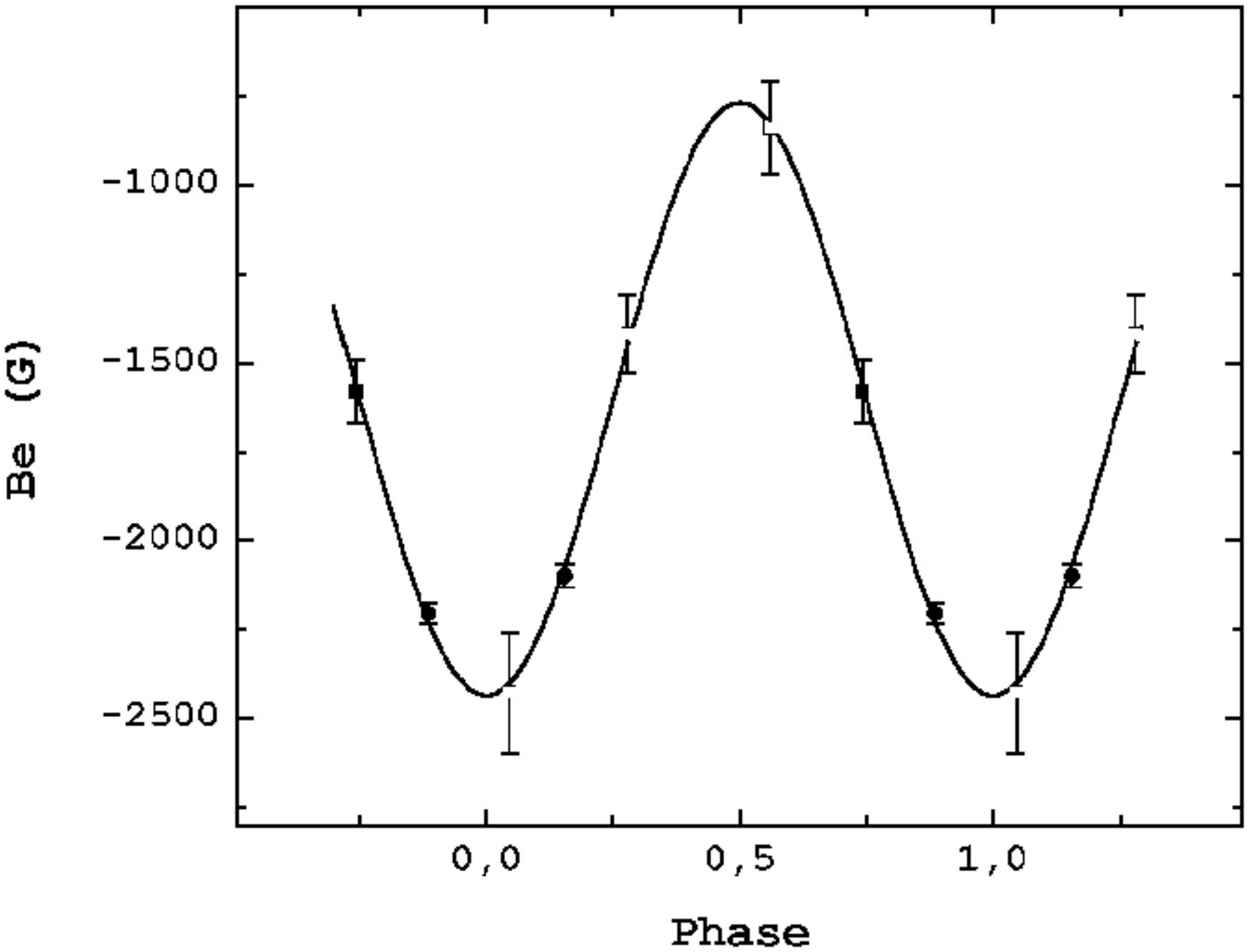}}
\vspace{-3.5mm}
\caption{ HD 16605 }
\label{fig:fig33}
\end{figure}

\begin{figure}
\resizebox{0.98\hsize}{!}{\includegraphics{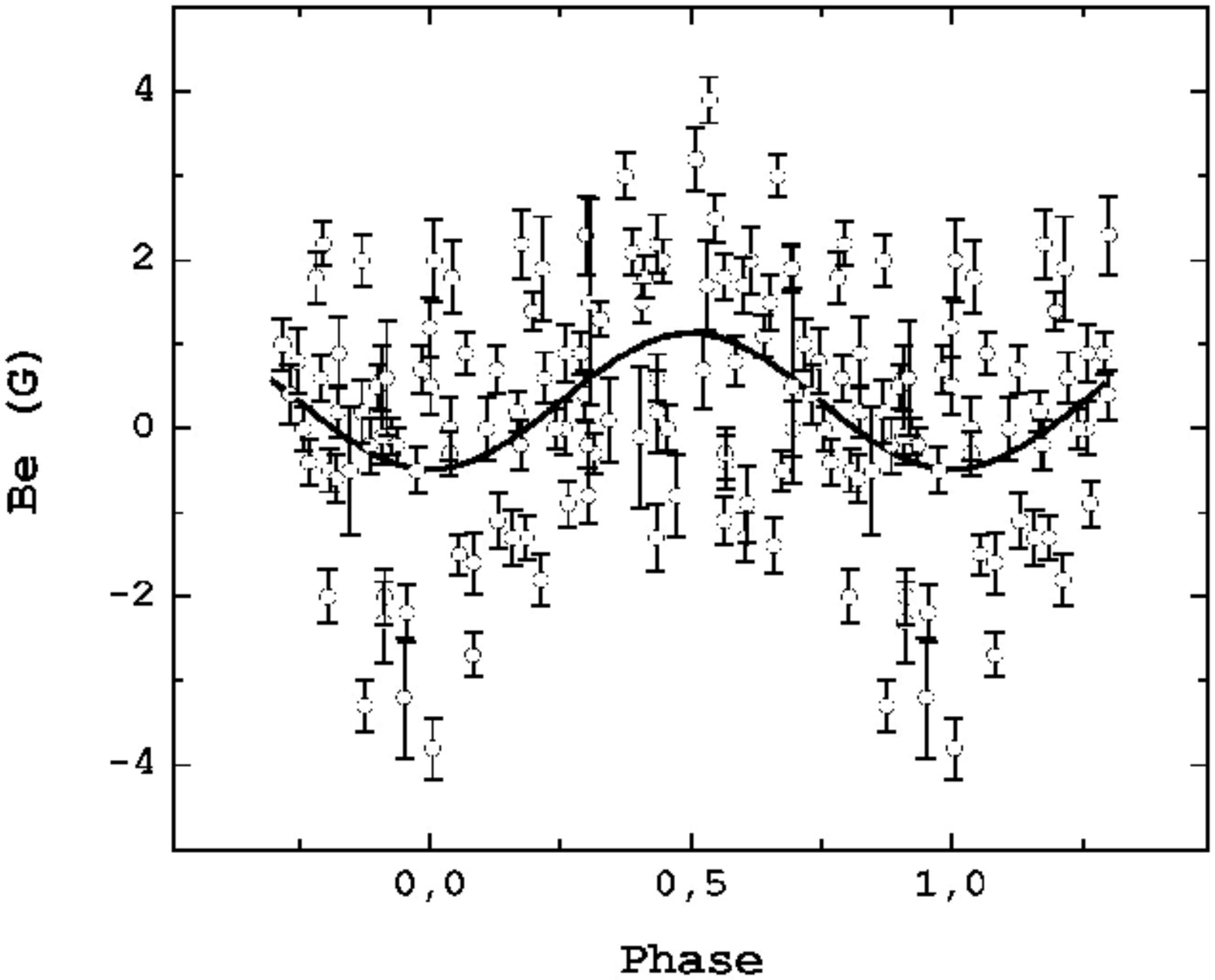}}
\vspace{-3.5mm}
\caption{ HD 17051 }
\label{fig:fig33}
\end{figure}

\begin{figure}
\resizebox{0.98\hsize}{!}{\includegraphics{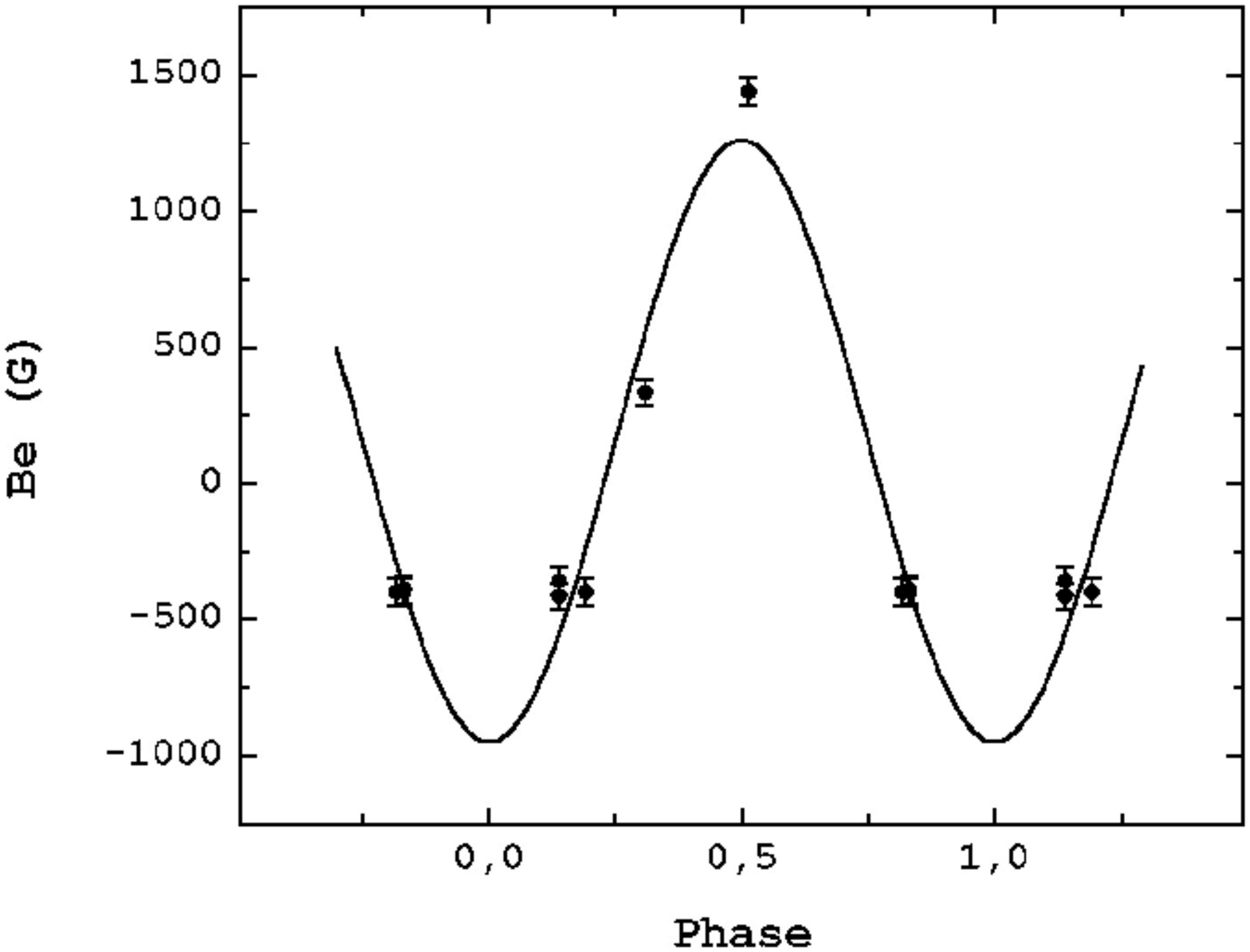}}
\vspace{-3.5mm}
\caption{ HD 17330 }
\label{fig:fig34}
\end{figure}

\clearpage
\newpage

\begin{figure}
\resizebox{0.98\hsize}{!}{\includegraphics{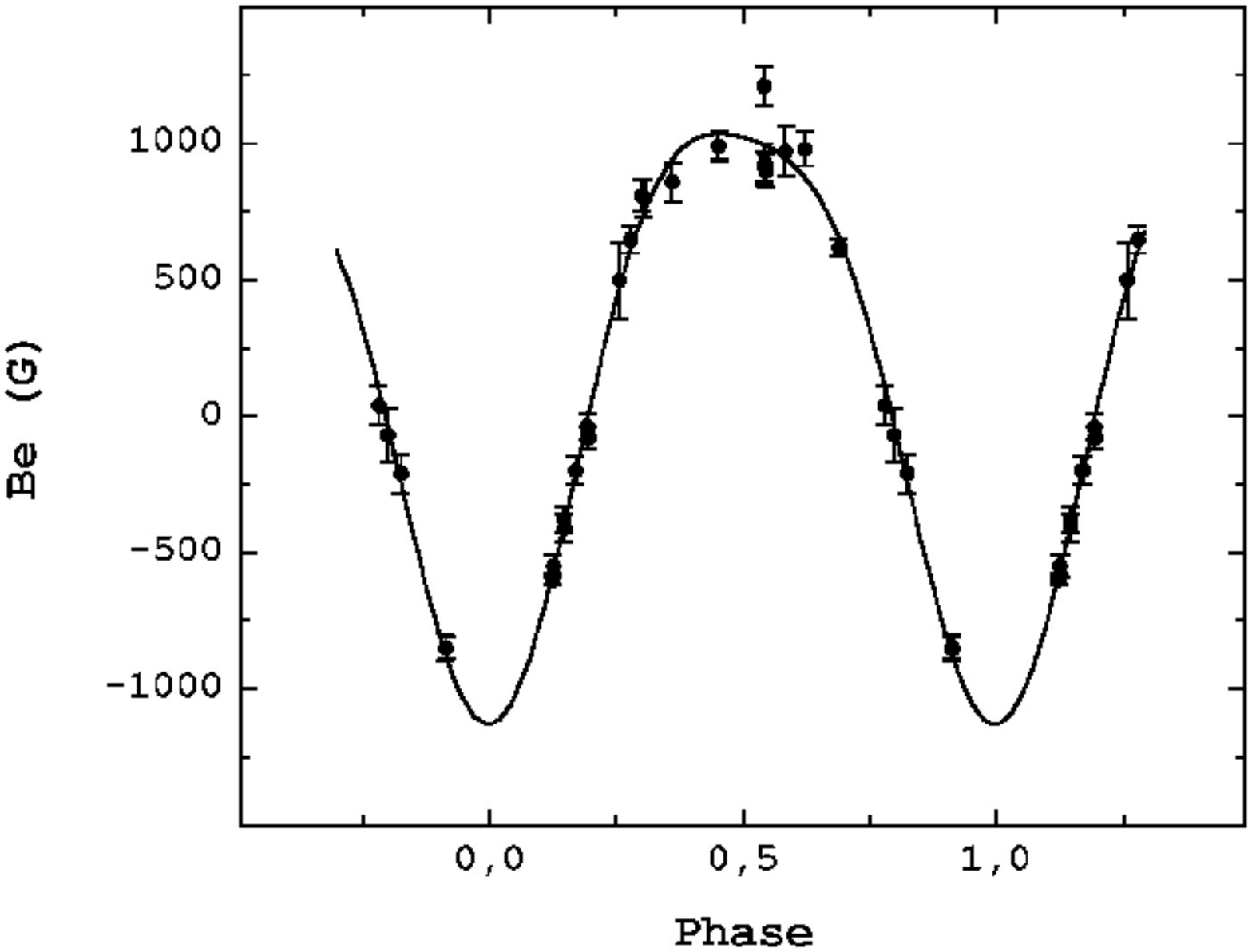}}
\vspace{-3.5mm}
\caption{ HD 18078 }
\label{fig:fig35}
\end{figure}

\begin{figure}
\resizebox{0.98\hsize}{!}{\includegraphics{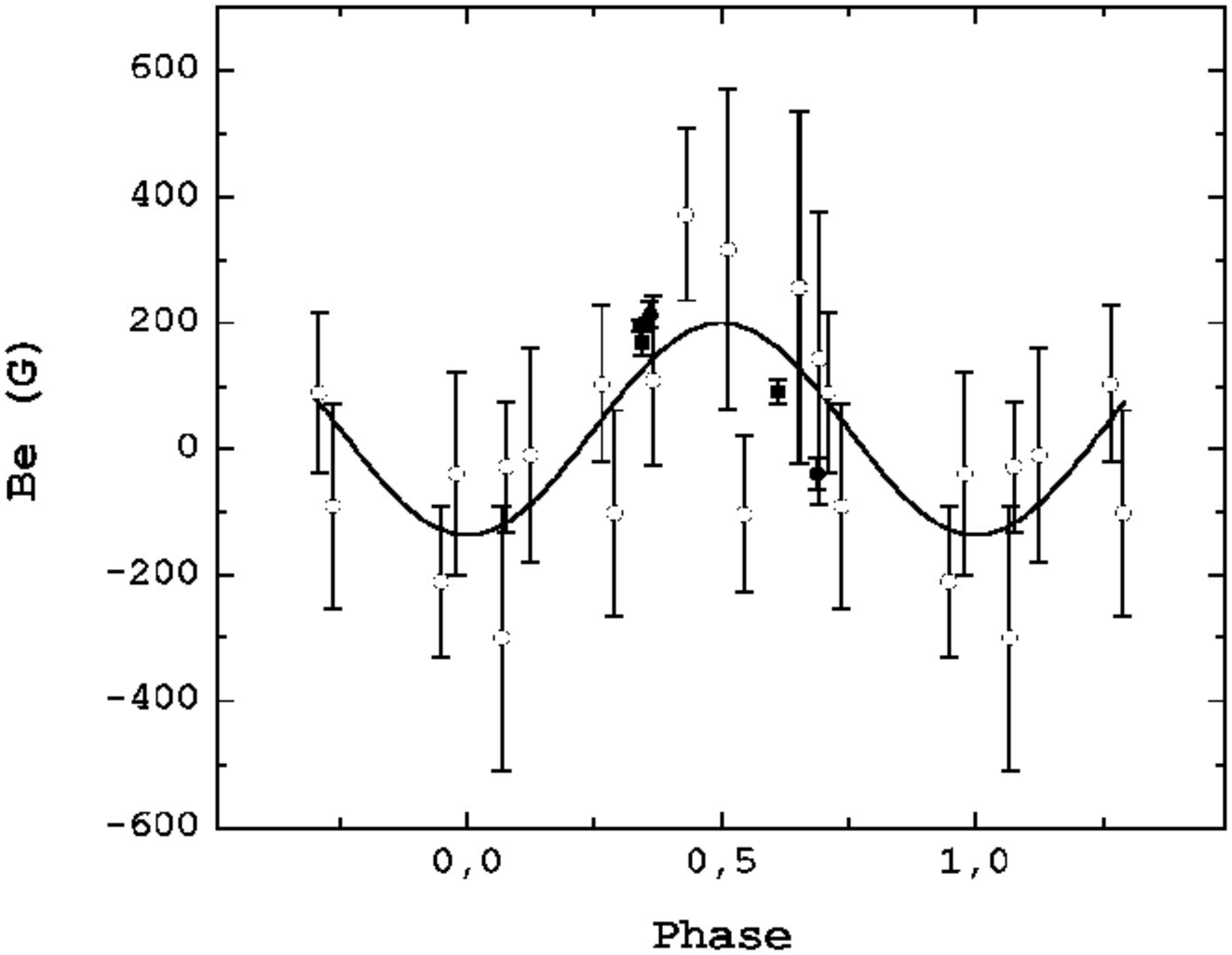}}
\vspace{-3.5mm}
\caption{ HD 18296 }
\label{fig:fig36}
\end{figure}

\begin{figure}
\resizebox{0.98\hsize}{!}{\includegraphics{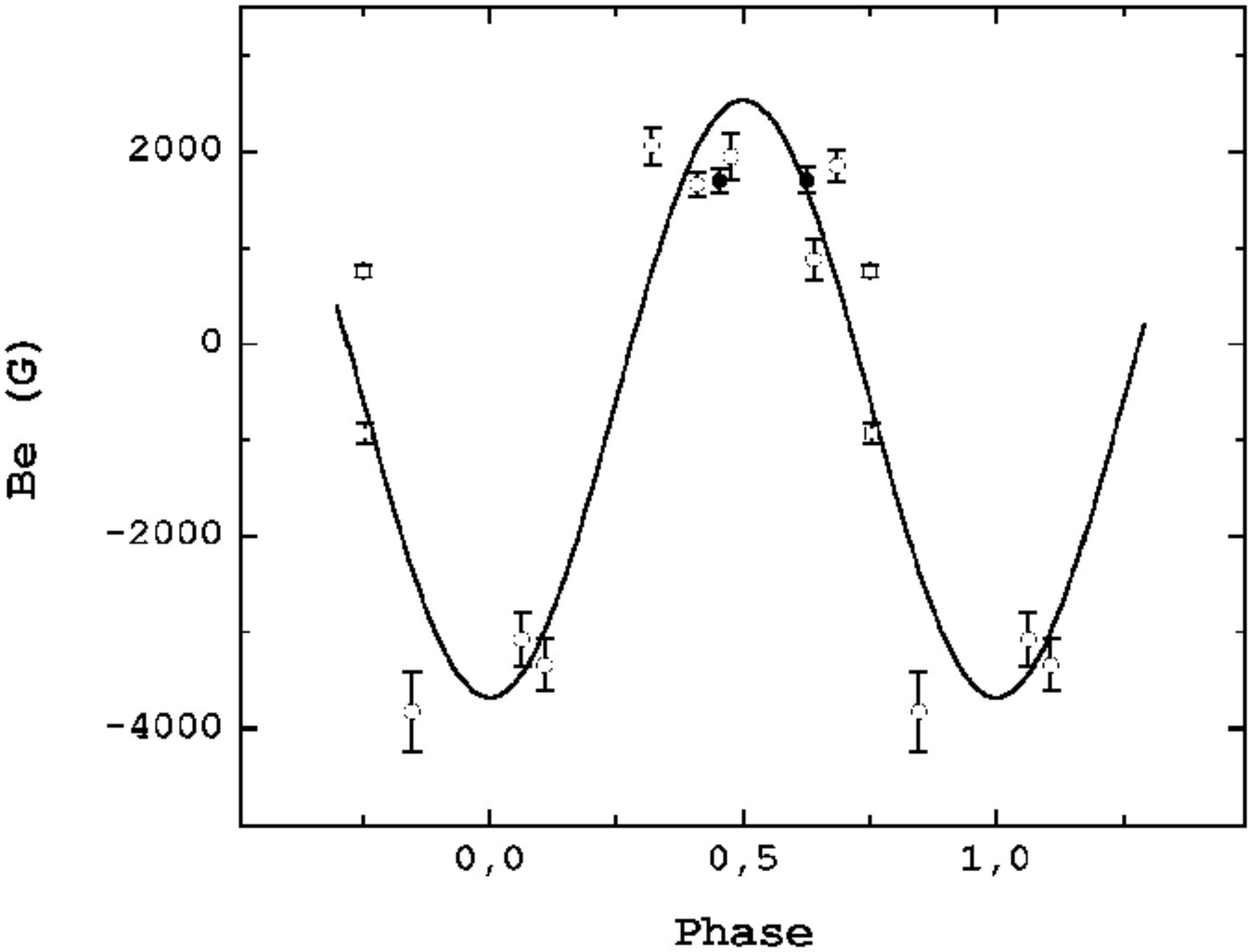}}
\vspace{-3.5mm}
\caption{ HD 19712 }
\label{fig:fig37}
\end{figure}

\begin{figure}
\resizebox{0.98\hsize}{!}{\includegraphics{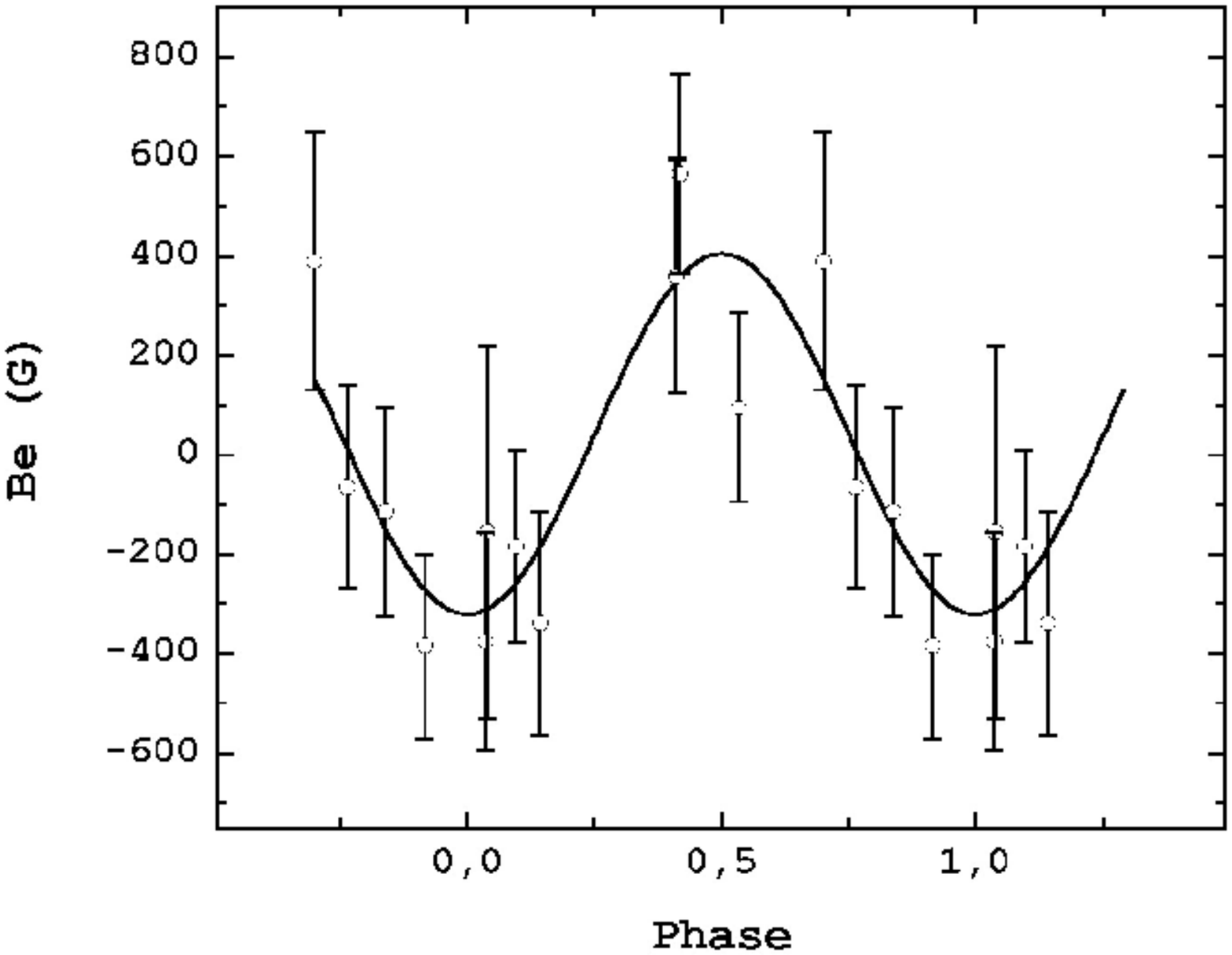}}
\vspace{-3.5mm}
\caption{ HD 19832 }
\label{fig:fig38}
\end{figure}

\begin{figure}
\resizebox{0.98\hsize}{!}{\includegraphics{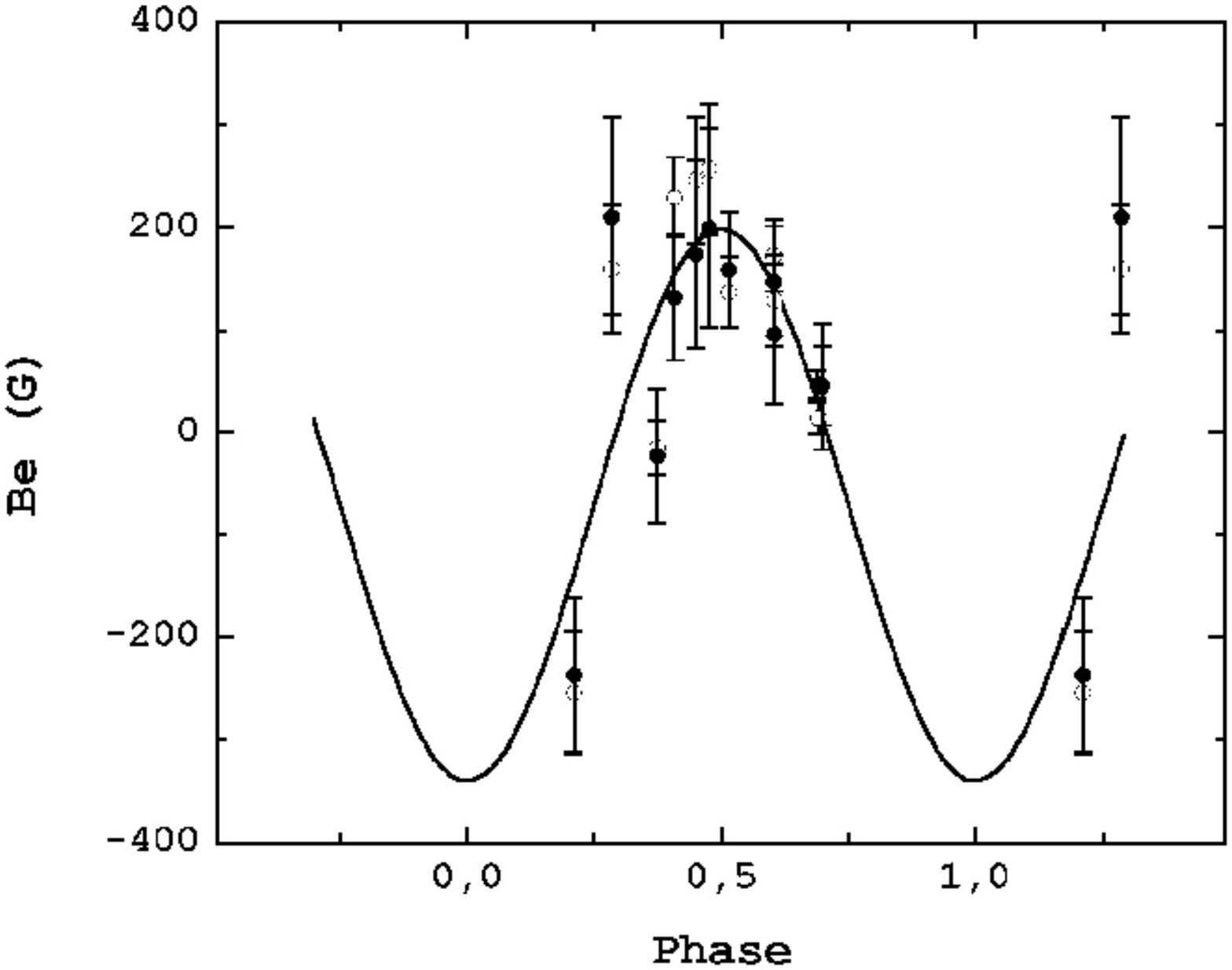}}
\vspace{-3.5mm}
\caption{ HD 21190 }
\label{fig:fig33}
\end{figure}

\begin{figure}
\resizebox{0.98\hsize}{!}{\includegraphics{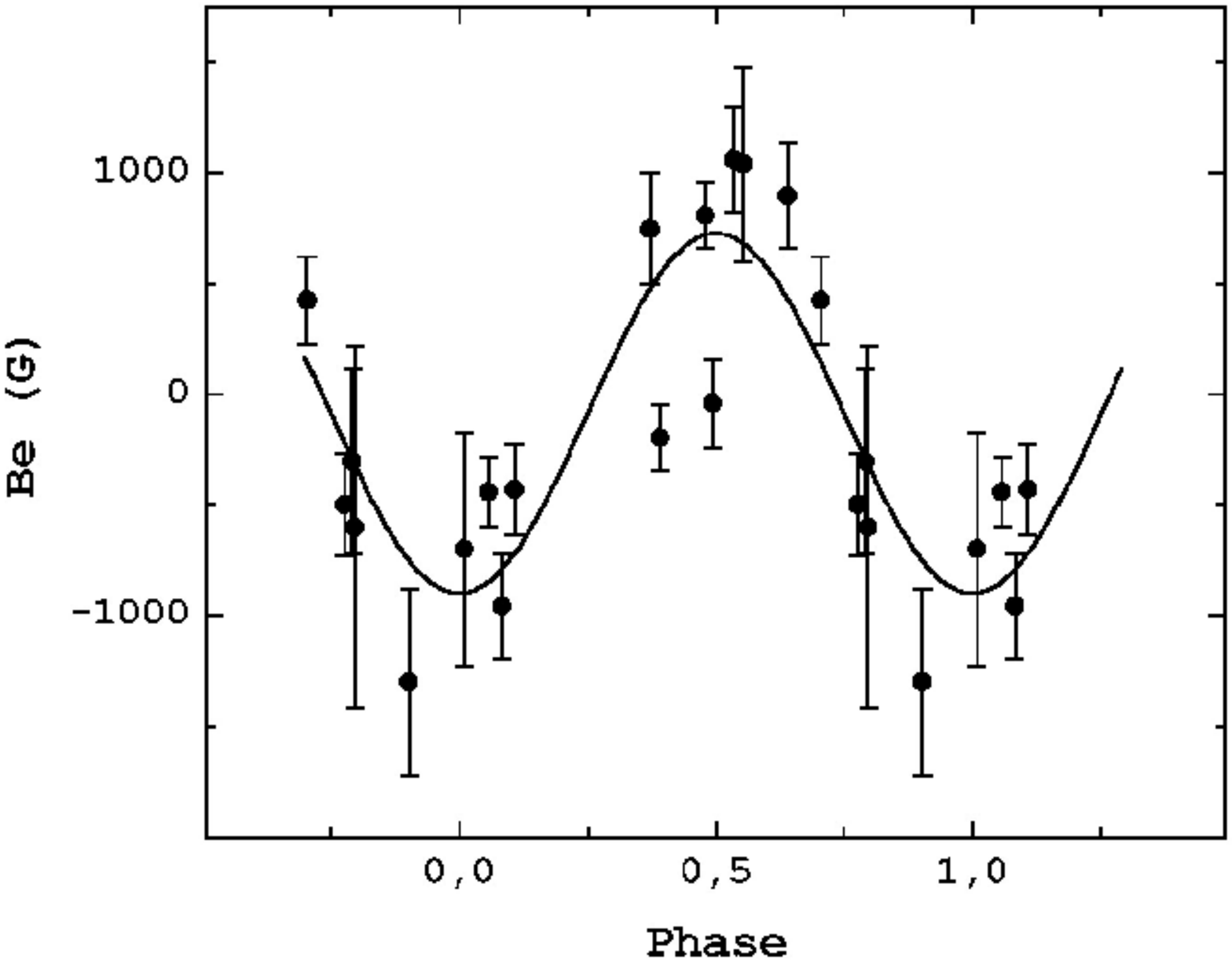}}
\vspace{-3.5mm}
\caption{ HD 21699 }
\label{fig:fig39}
\end{figure}

\clearpage
\newpage

\begin{figure}
\resizebox{0.98\hsize}{!}{\includegraphics{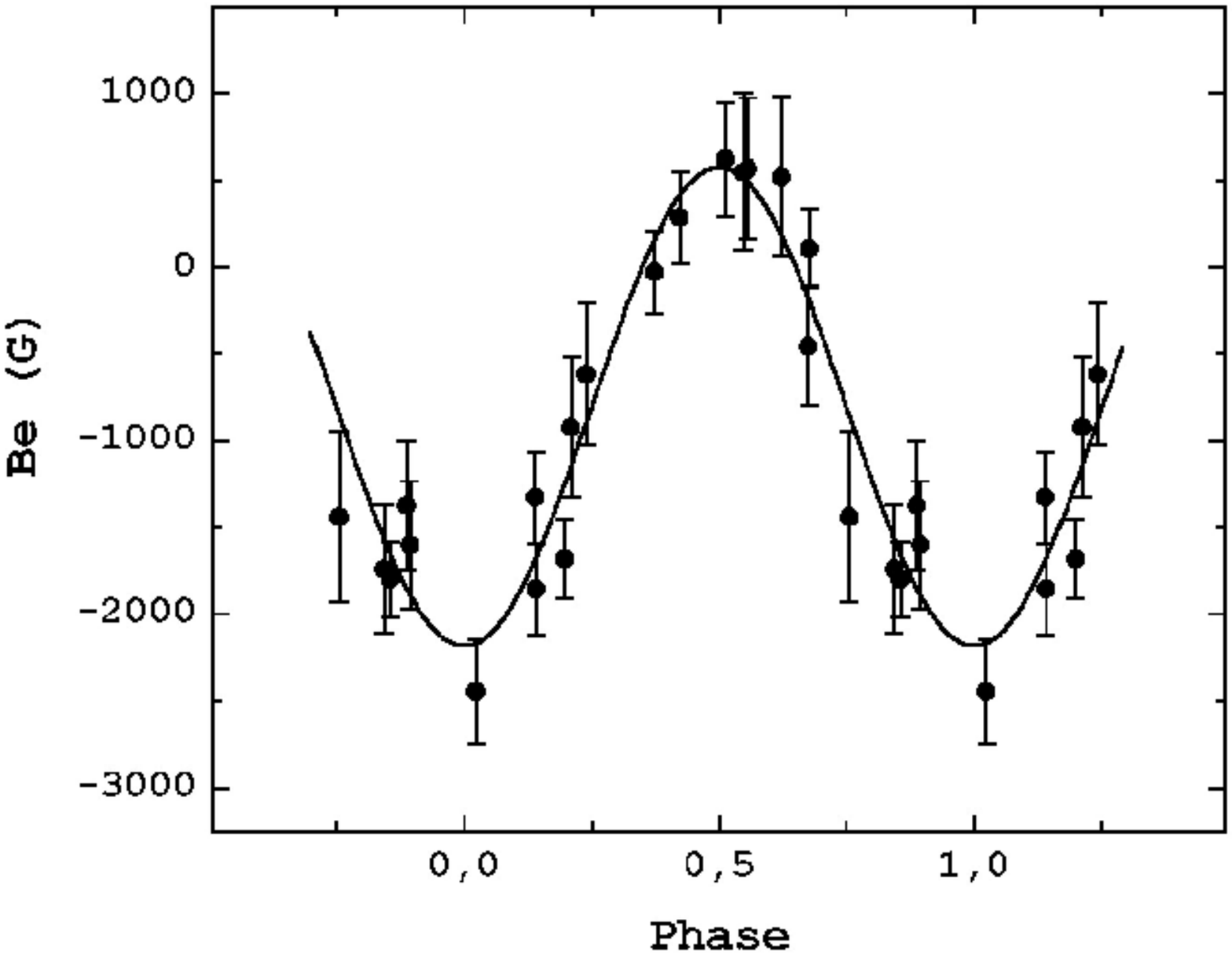}}
\vspace{-3.5mm}
\caption{ HD 22316 }
\label{fig:fig40}
\end{figure}

\begin{figure}
\resizebox{0.98\hsize}{!}{\includegraphics{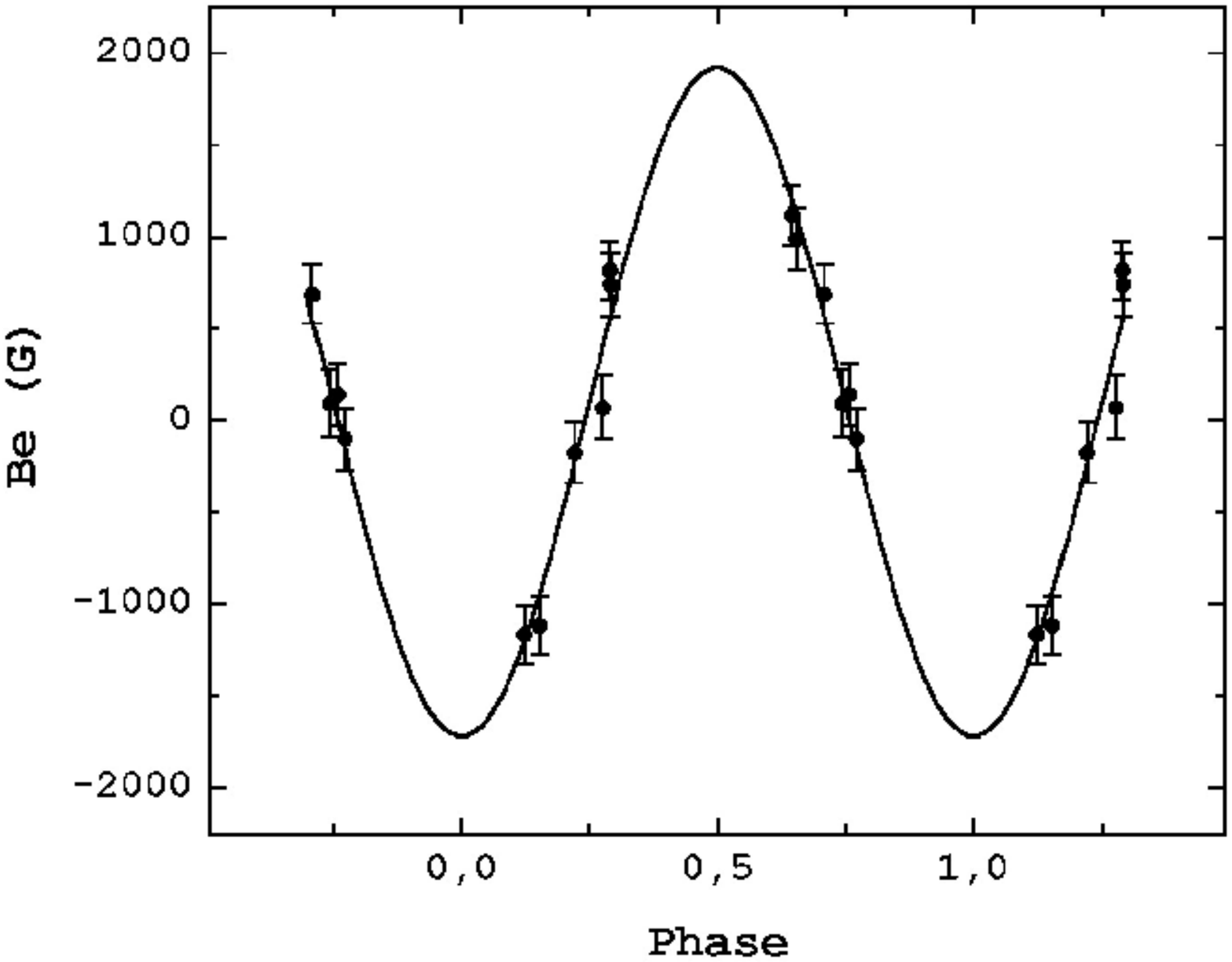}}
\vspace{-3.5mm}
\caption{ HD 22470 }
\label{fig:fig41}
\end{figure}

\begin{figure}
\resizebox{0.98\hsize}{!}{\includegraphics{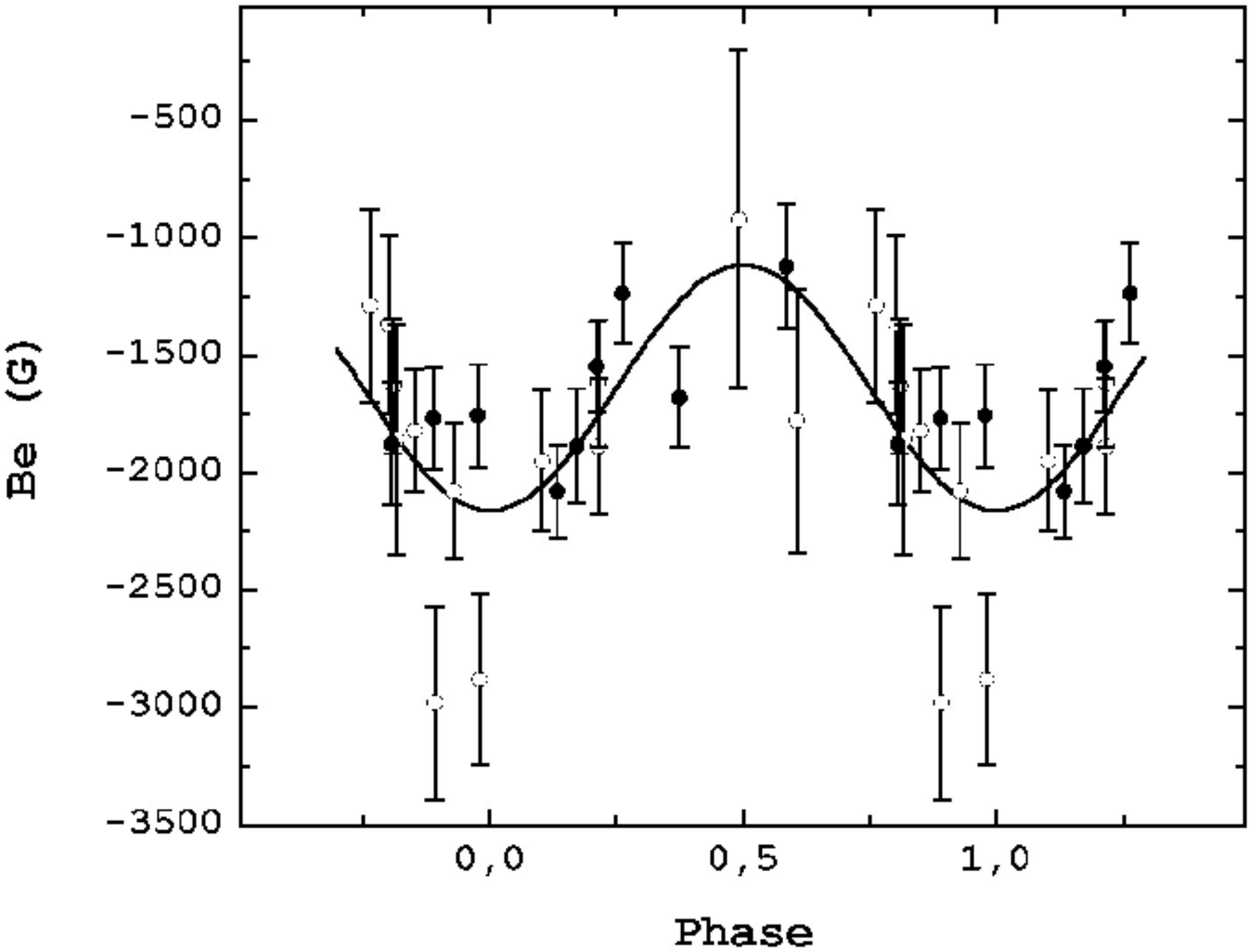}}
\vspace{-3.5mm}
\caption{ HD 23478 (1)}
\label{fig:fig42}
\end{figure}

\begin{figure}
\resizebox{0.98\hsize}{!}{\includegraphics{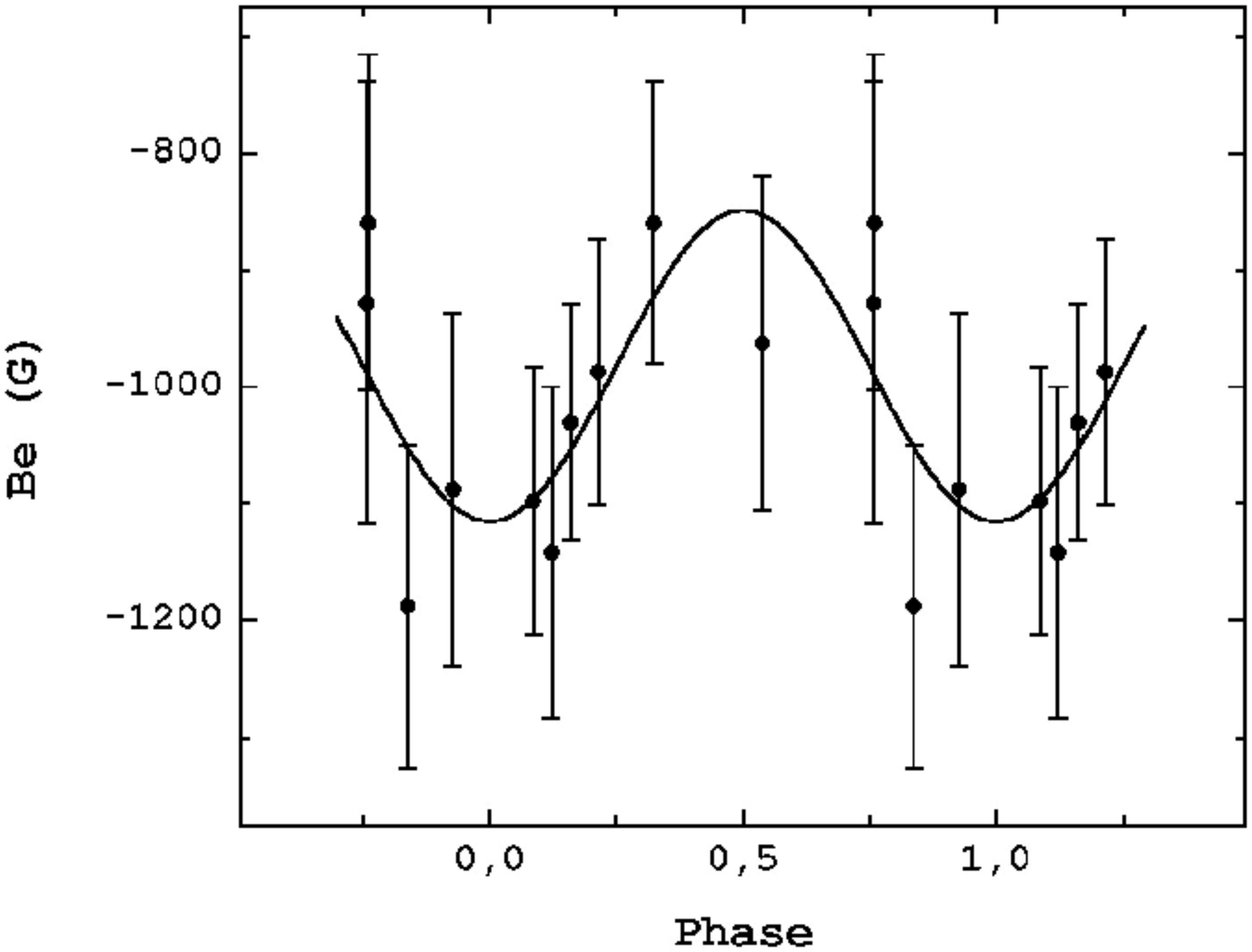}}
\vspace{-3.5mm}
\caption{ HD 23478 (2) }
\label{fig:fig43}
\end{figure}

\begin{figure}
\resizebox{0.98\hsize}{!}{\includegraphics{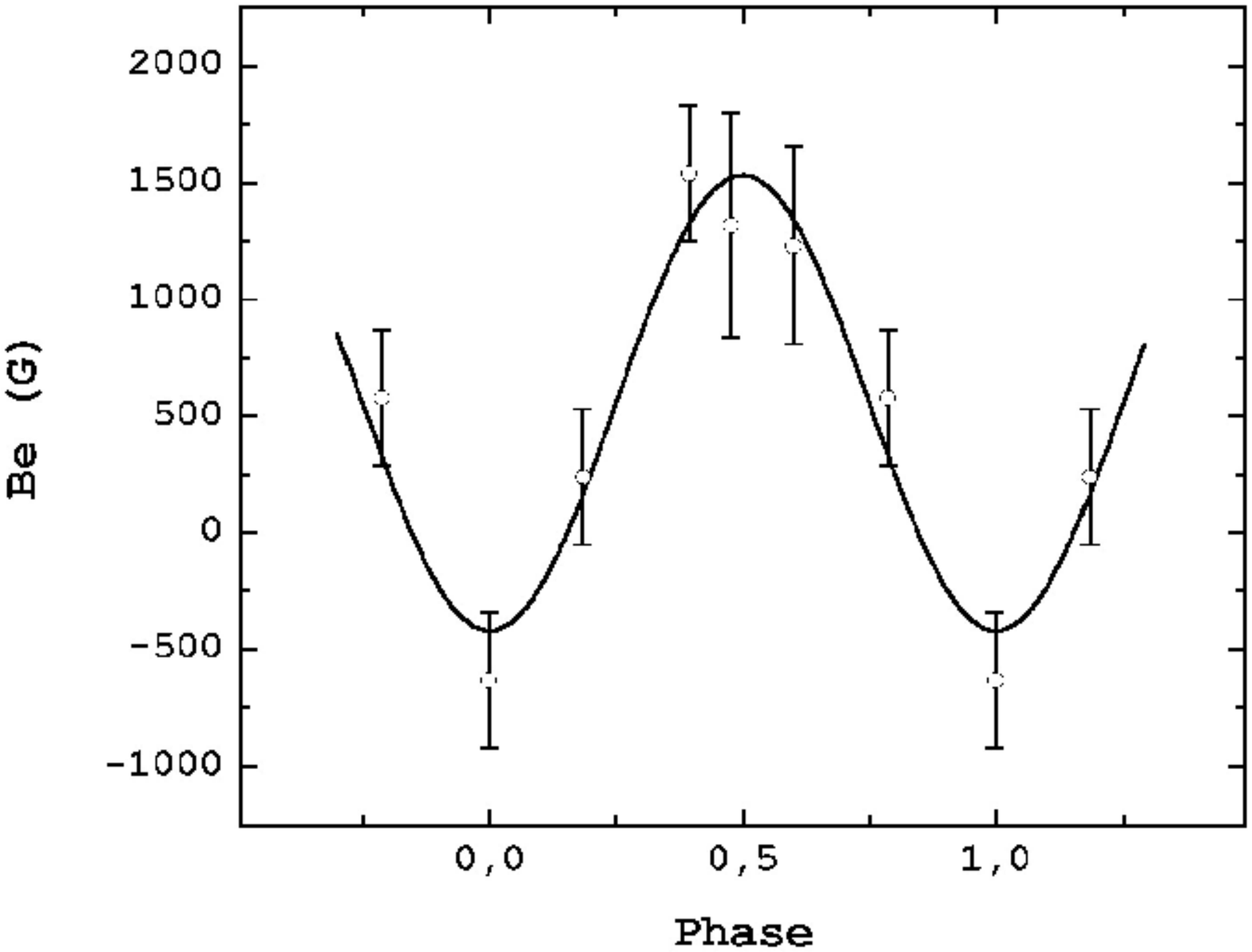}}
\vspace{-3.5mm}
\caption{ HD 24155 }
\label{fig:fig44}
\end{figure}

\begin{figure}
\resizebox{0.98\hsize}{!}{\includegraphics{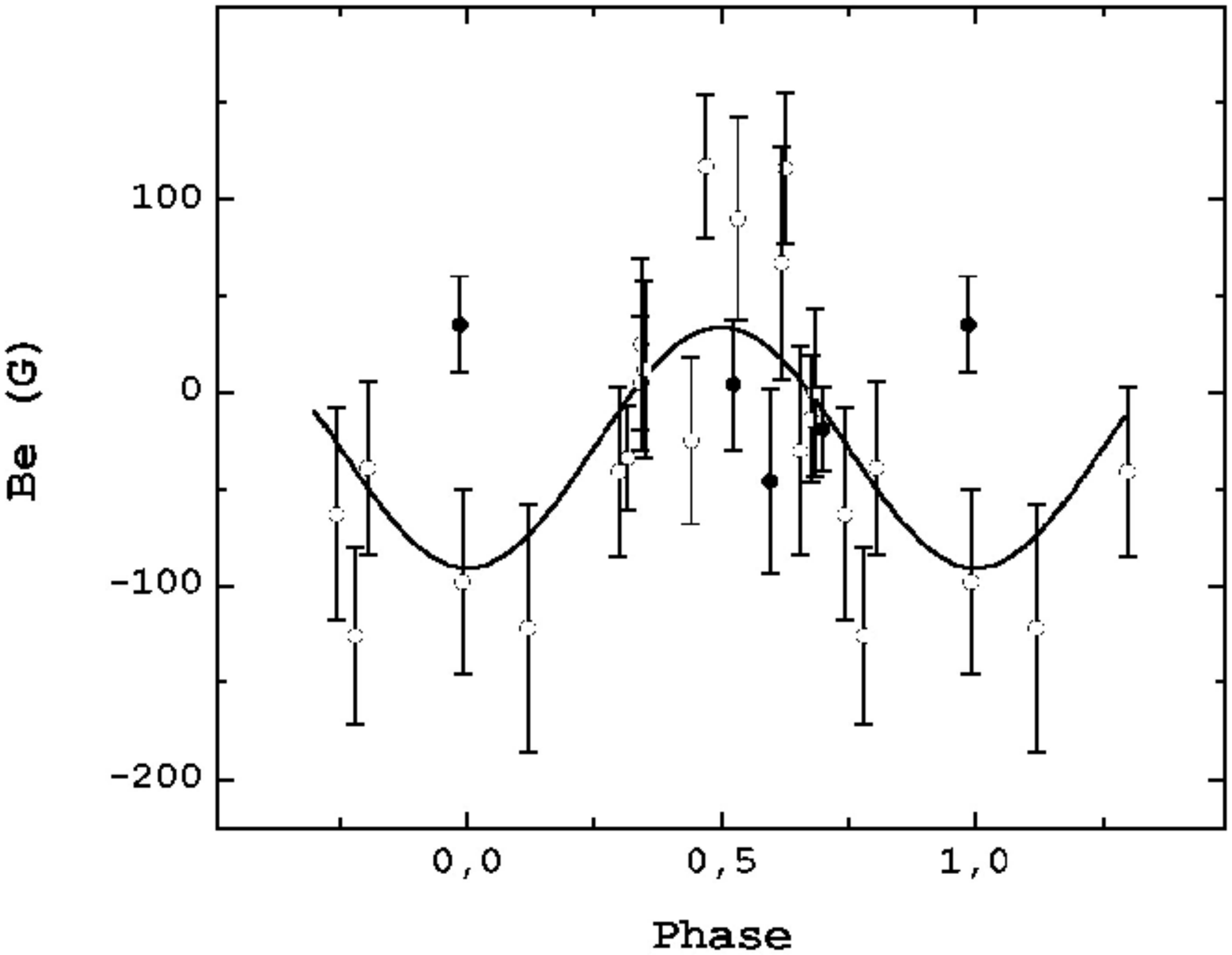}}
\vspace{-3.5mm}
\caption{ HD 24587 (1) }
\label{fig:fig45}
\end{figure}

\clearpage
\newpage

\begin{figure}
\resizebox{0.98\hsize}{!}{\includegraphics{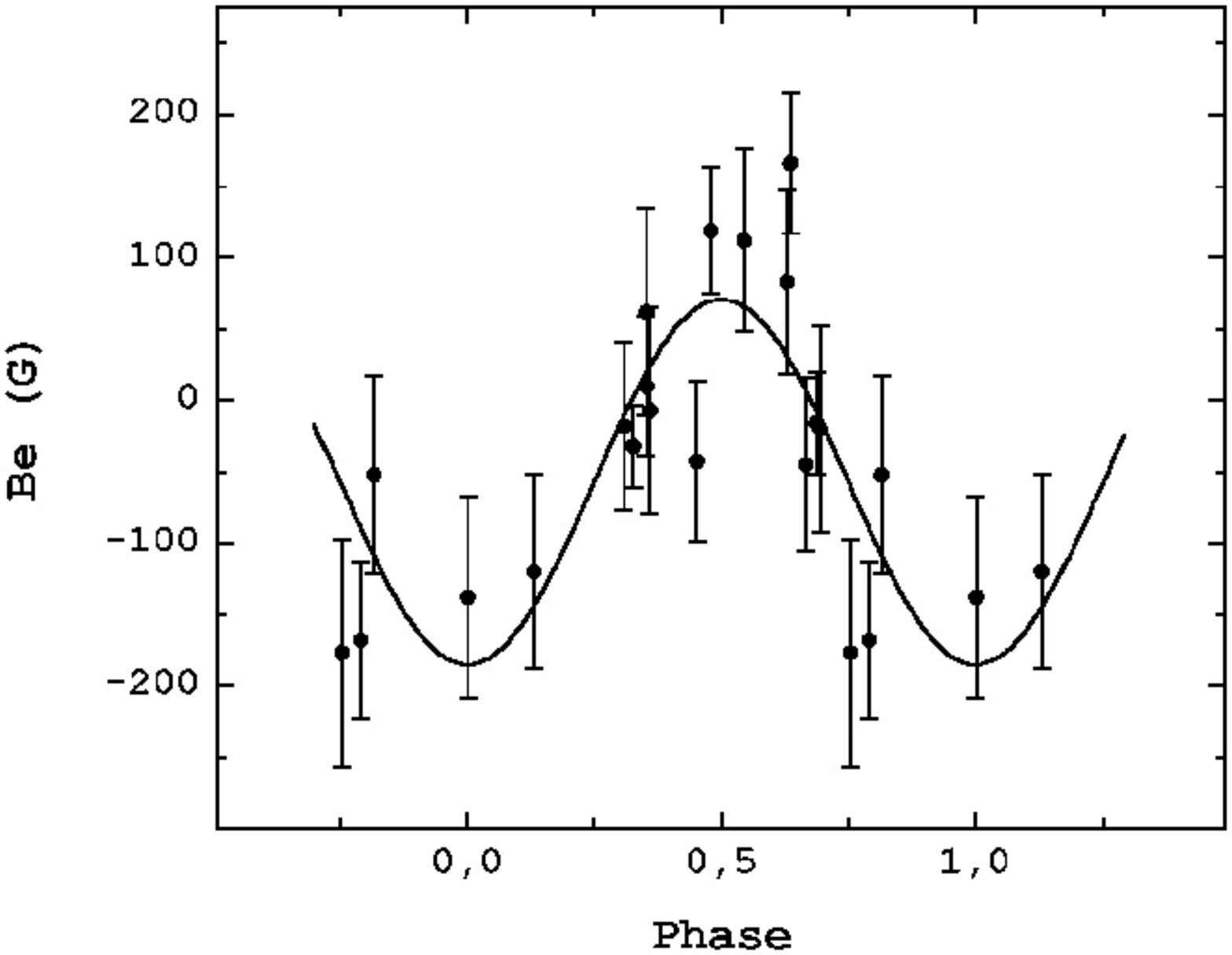}}
\vspace{-3.5mm}
\caption{ HD 24587 (2) }
\label{fig:fig46}
\end{figure}

\begin{figure}
\resizebox{0.98\hsize}{!}{\includegraphics{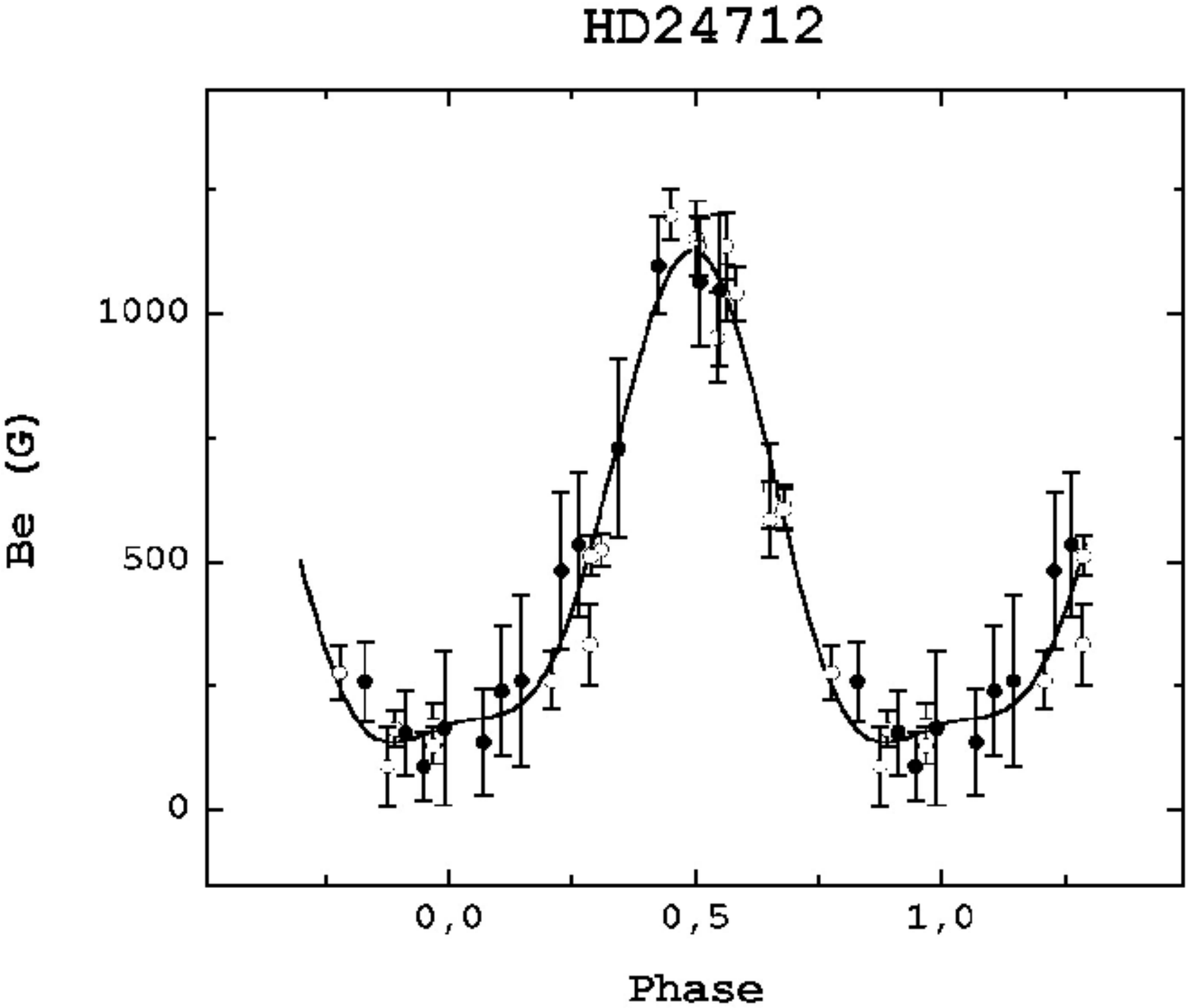}}
\vspace{-3.5mm}
\caption{ HD 24712 }
\label{fig:fig47}
\end{figure}

\begin{figure}
\resizebox{0.98\hsize}{!}{\includegraphics{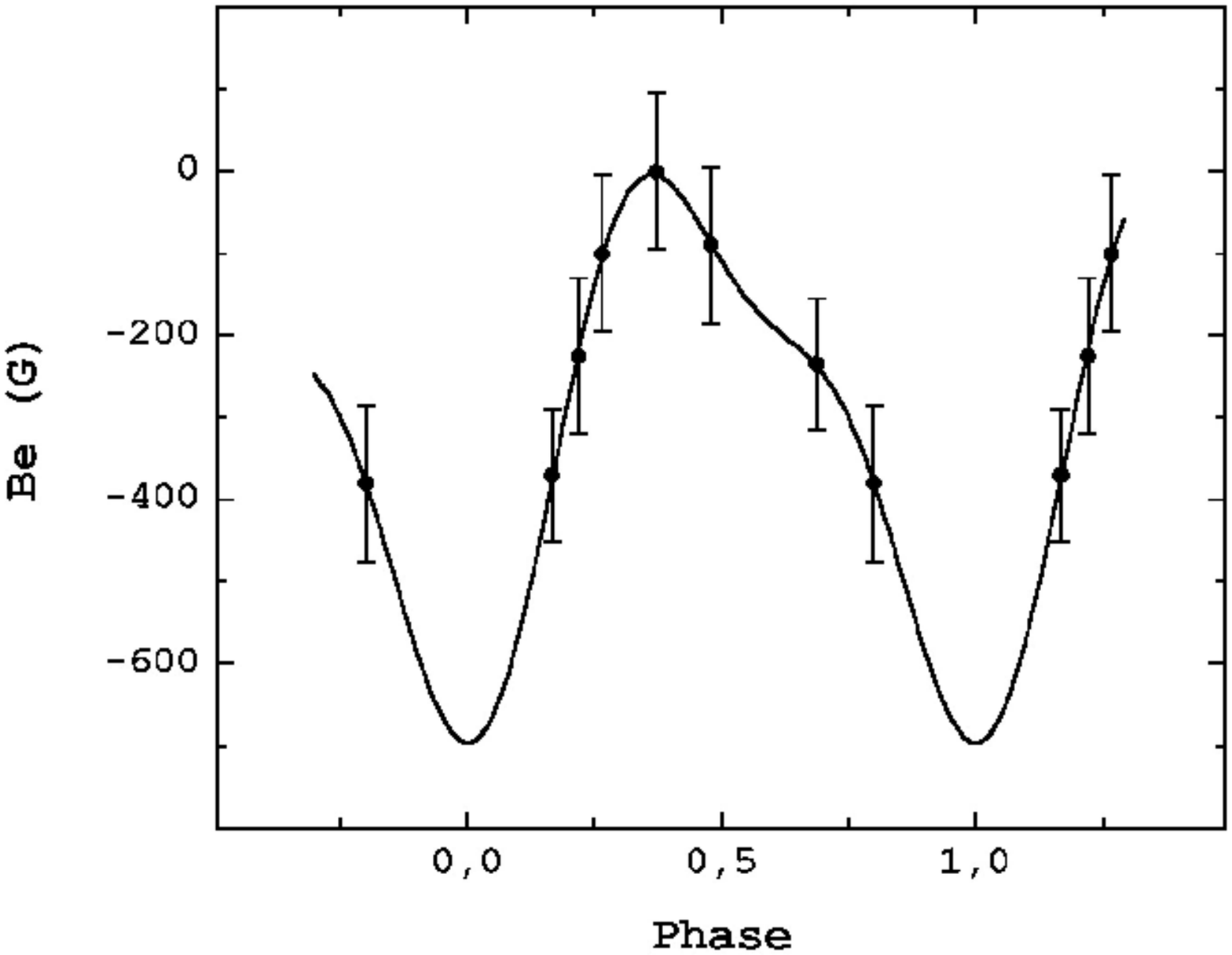}}
\vspace{-3.5mm}
\caption{ HD 25267 }
\label{fig:fig48}
\end{figure}

\begin{figure}
\resizebox{0.98\hsize}{!}{\includegraphics{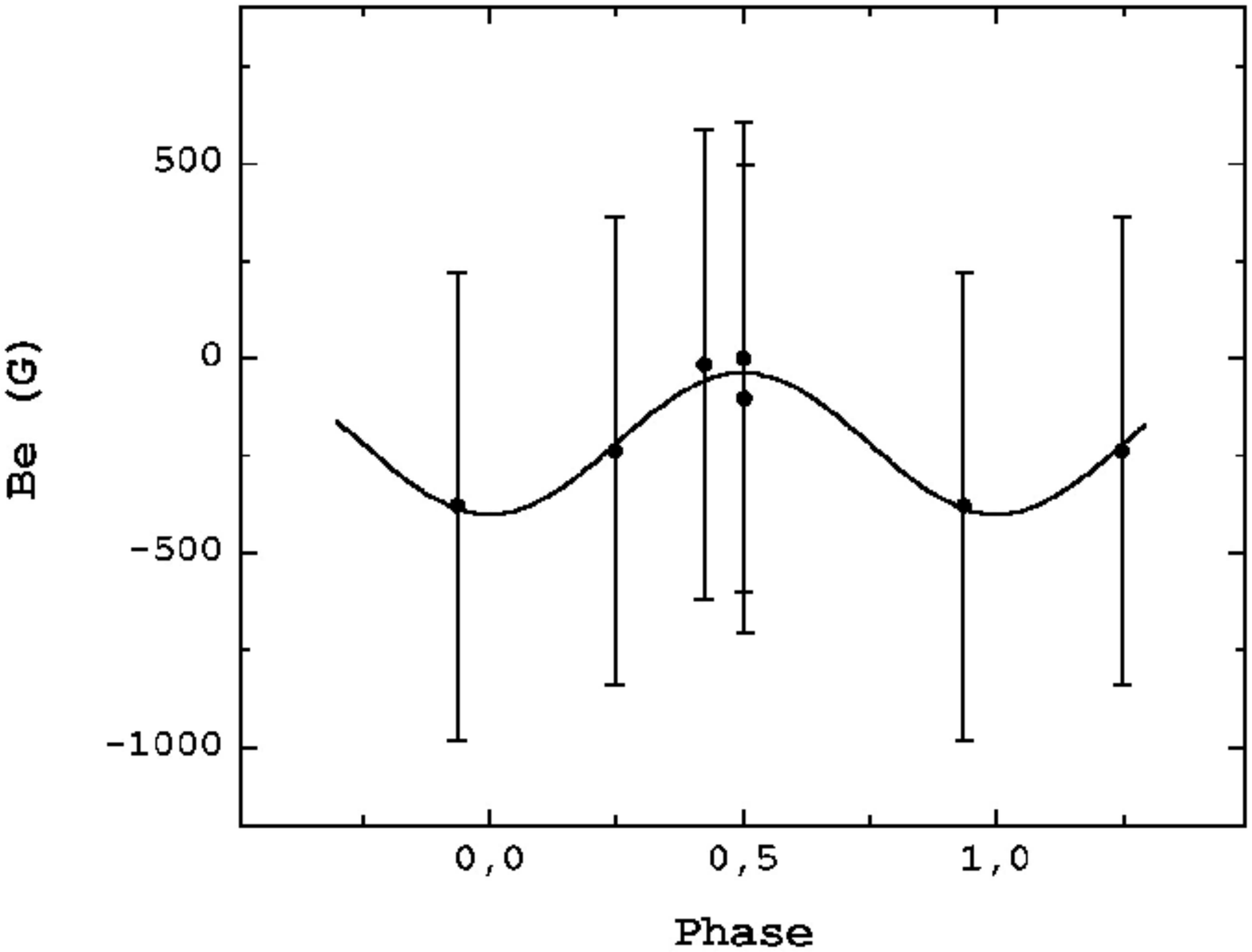}}
\vspace{-3.5mm}
\caption{ HD 25354 }
\label{fig:fig49}
\end{figure}

\begin{figure}
\resizebox{0.98\hsize}{!}{\includegraphics{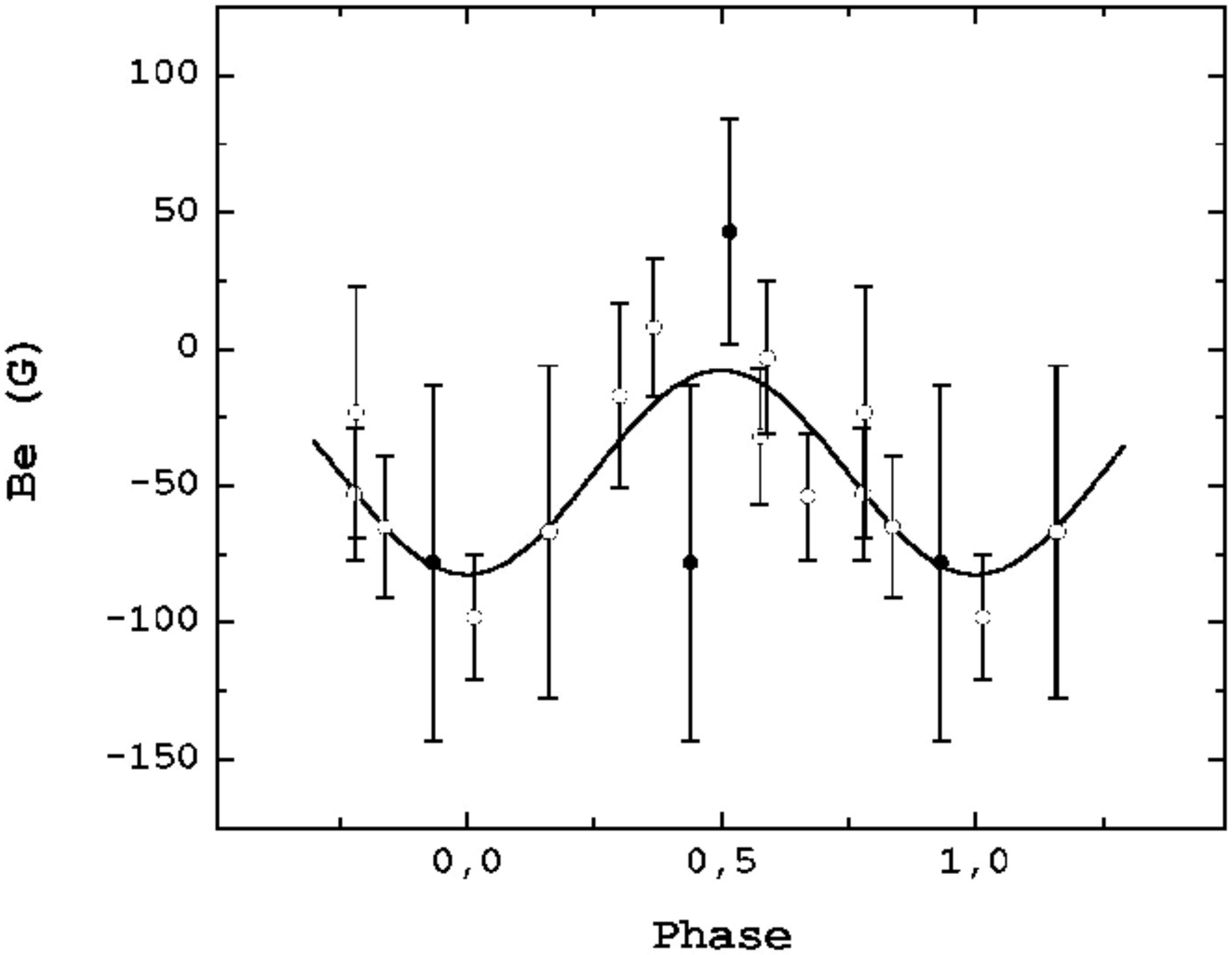}}
\vspace{-3.5mm}
\caption{ HD 25558 }
\label{fig:fig49}
\end{figure}

\begin{figure}
\resizebox{0.98\hsize}{!}{\includegraphics{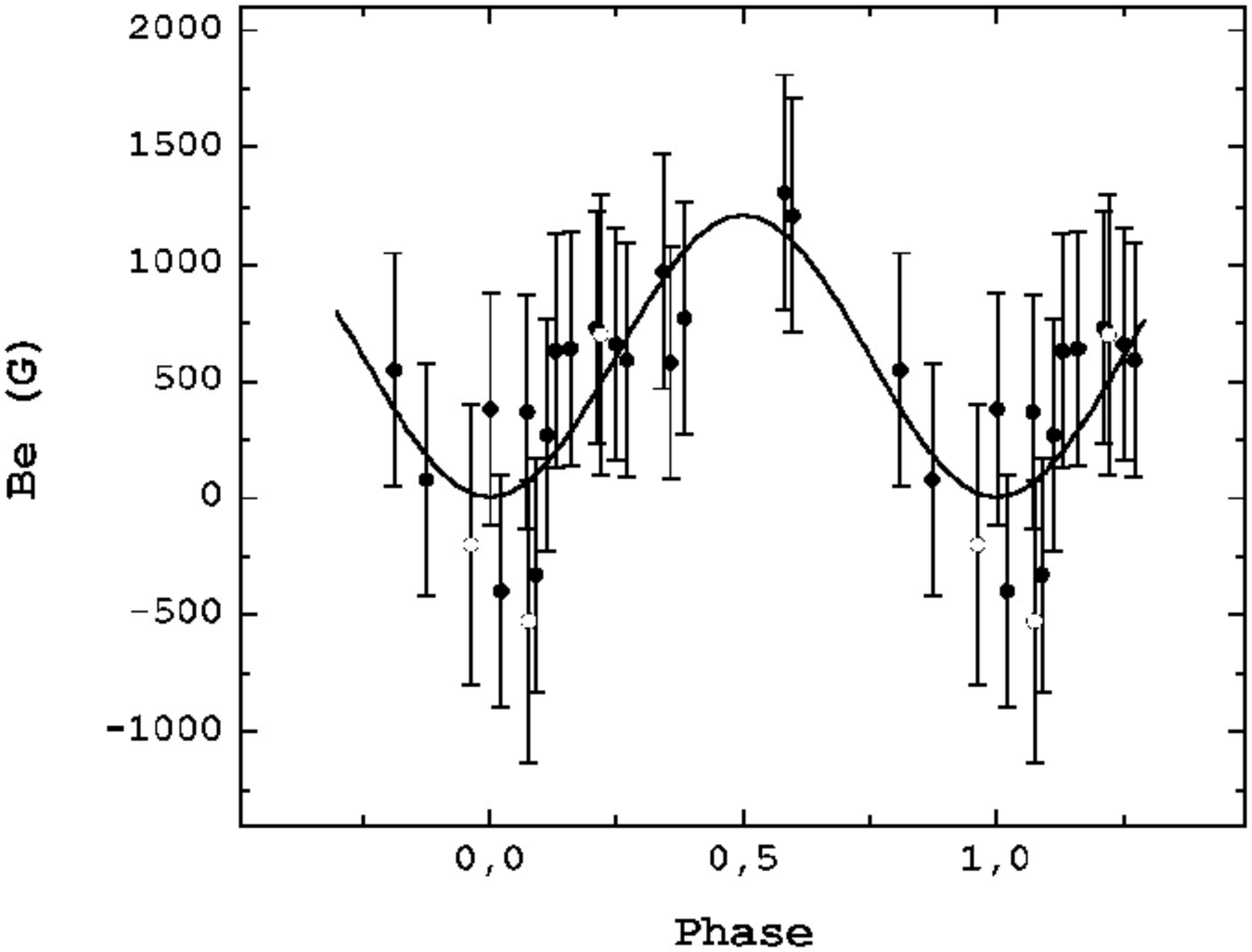}}
\vspace{-3.5mm}
\caption{ HD 25823 }
\label{fig:fig50}
\end{figure}

\clearpage
\newpage

\begin{figure}
\resizebox{0.98\hsize}{!}{\includegraphics{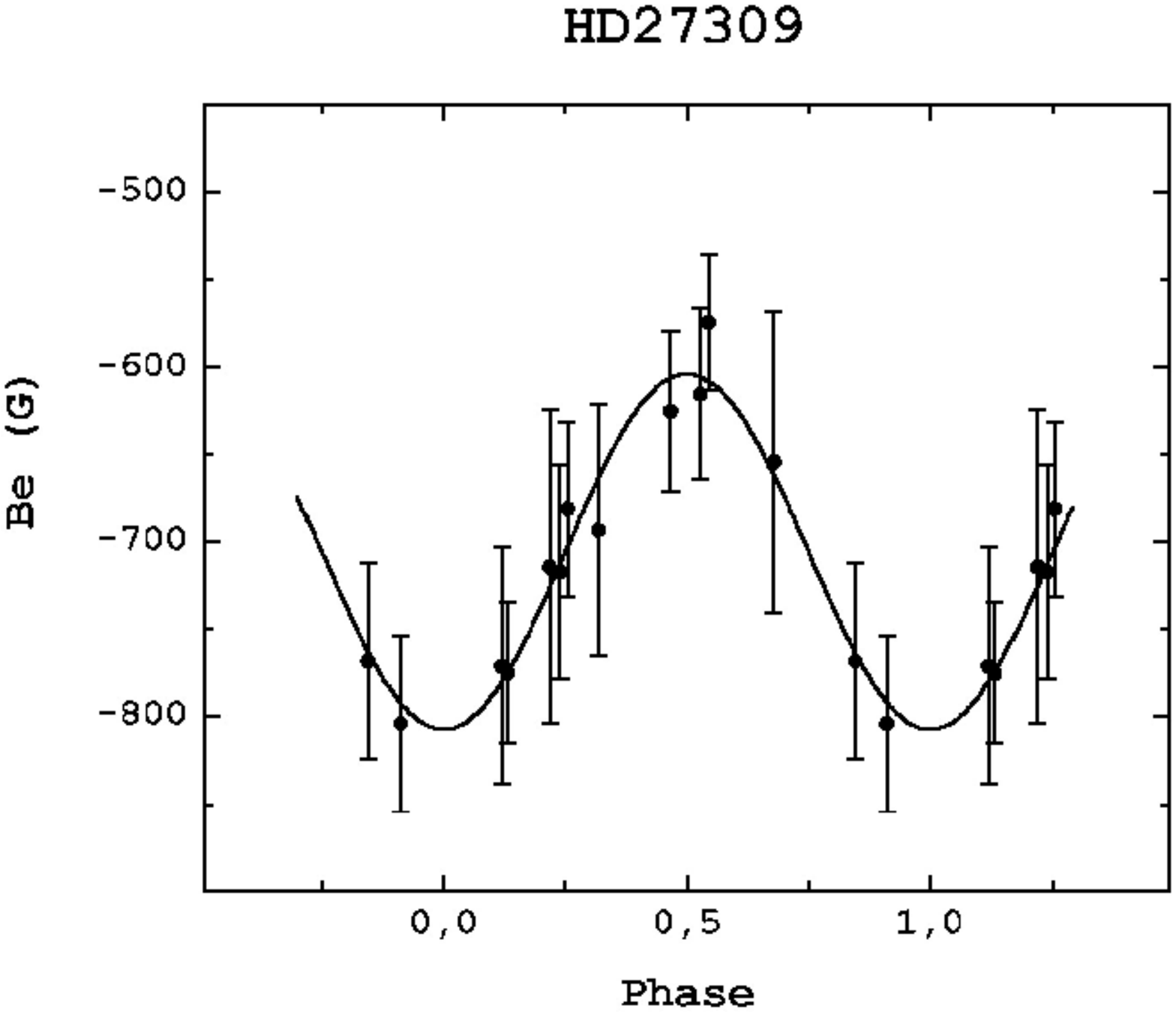}}
\vspace{-3.5mm}
\caption{ HD 27309 }
\label{fig:fig51}
\end{figure}

\begin{figure}
\resizebox{0.98\hsize}{!}{\includegraphics{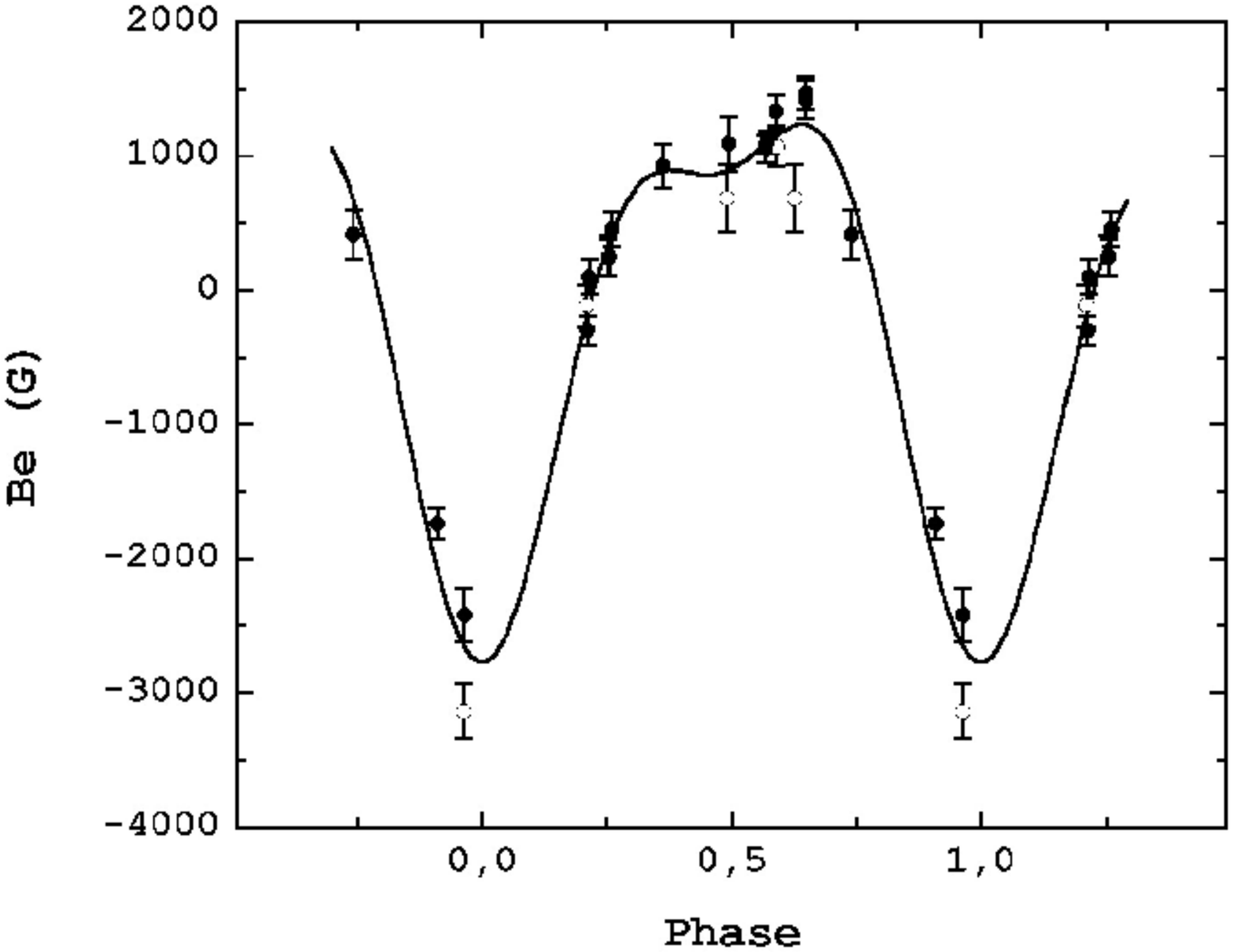}}
\vspace{-3.5mm}
\caption{ HD 27404 (1) }
\label{fig:fig52}
\end{figure}

\begin{figure}
\resizebox{0.98\hsize}{!}{\includegraphics{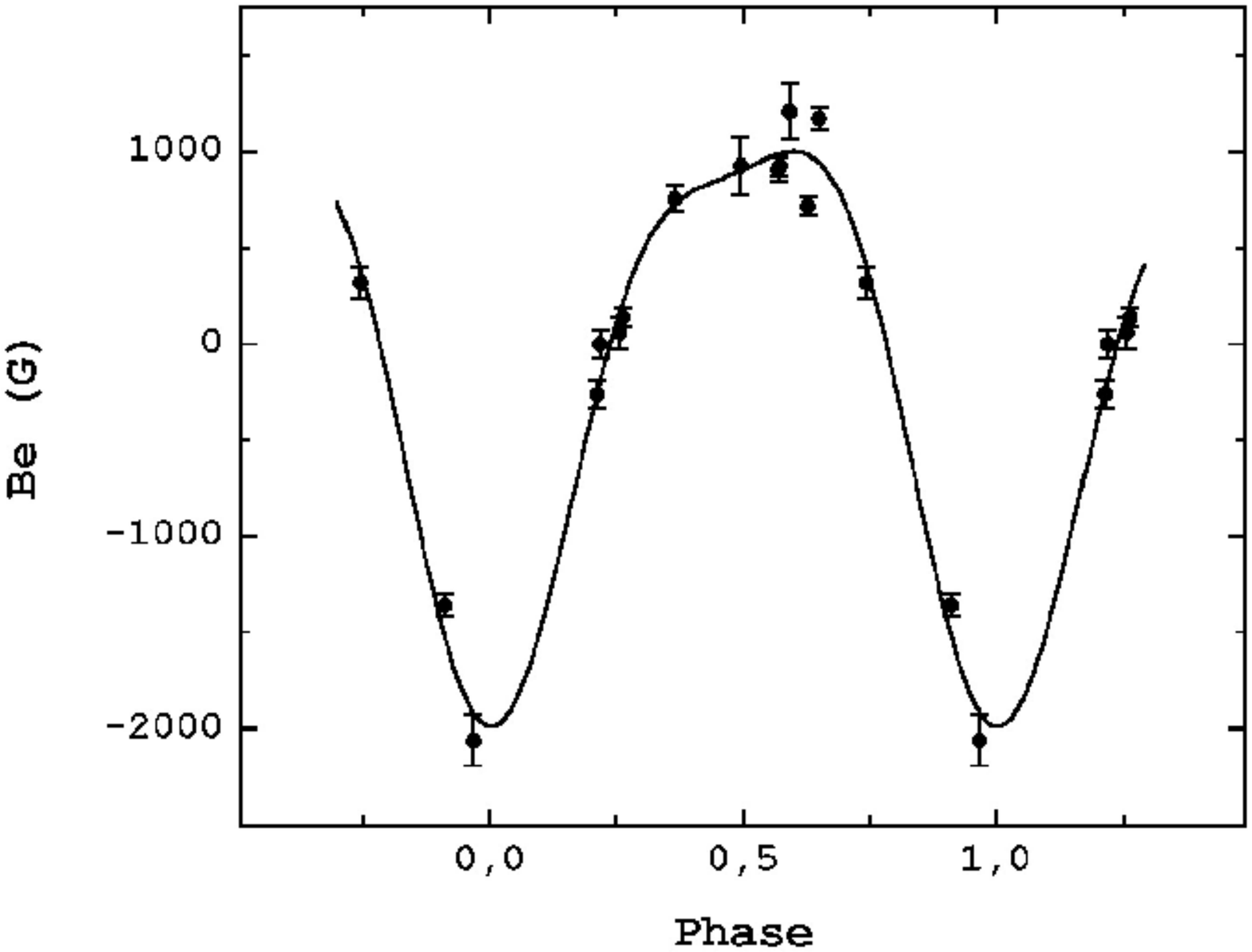}}
\vspace{-3.5mm}
\caption{ HD 27404 (2) }
\label{fig:fig53}
\end{figure}

\begin{figure}
\resizebox{0.98\hsize}{!}{\includegraphics{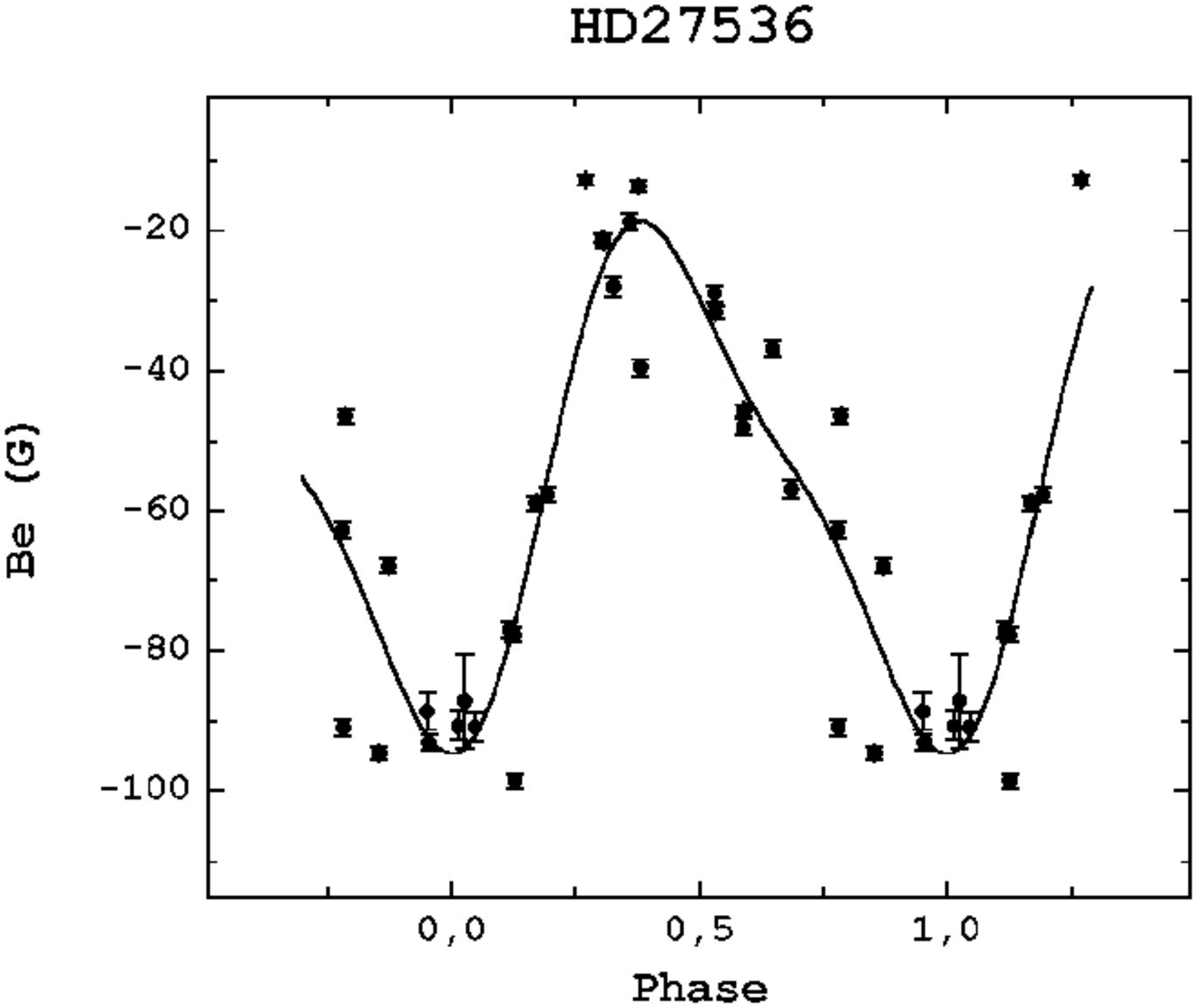}}
\vspace{-3.5mm}
\caption{ HD 27536 }
\label{fig:fig54}
\end{figure}

\begin{figure}
\resizebox{0.98\hsize}{!}{\includegraphics{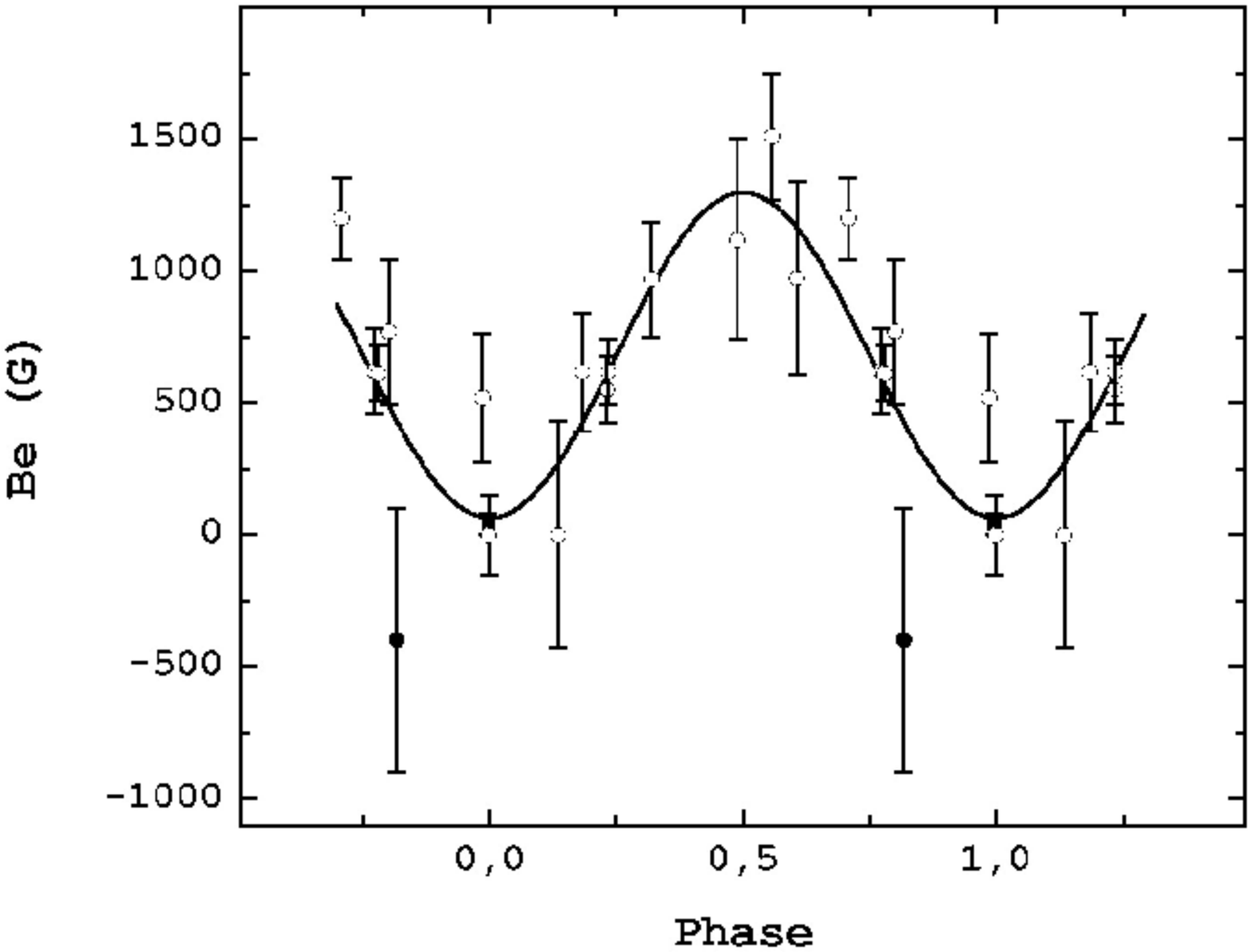}}
\vspace{-3.5mm}
\caption{ HD 27962 }
\label{fig:fig55}
\end{figure}

\begin{figure}
\resizebox{0.98\hsize}{!}{\includegraphics{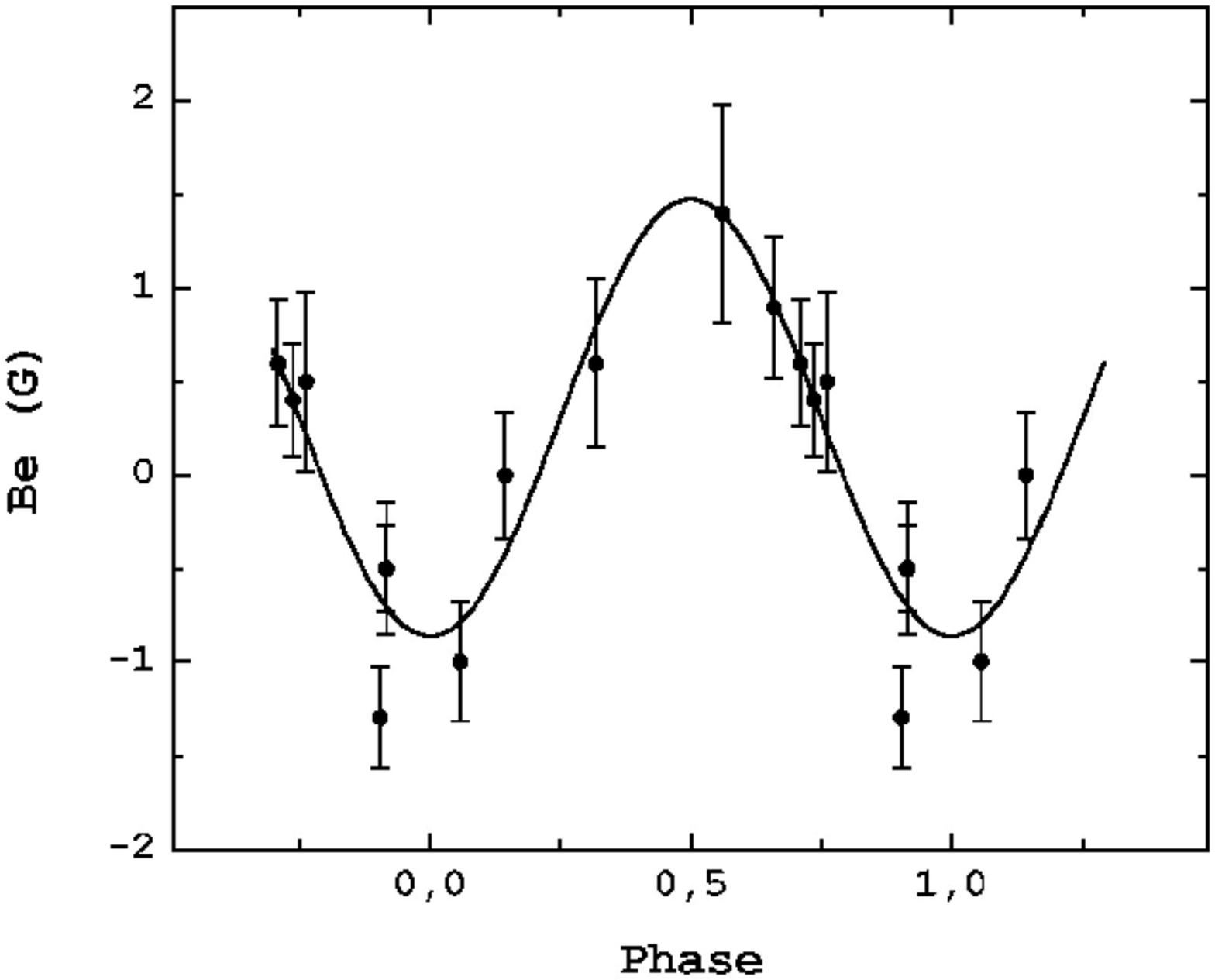}}
\vspace{-3.5mm}
\caption{ HD 28305 }
\label{fig:fig49}
\end{figure}

\clearpage
\newpage

\begin{figure}
\resizebox{0.98\hsize}{!}{\includegraphics{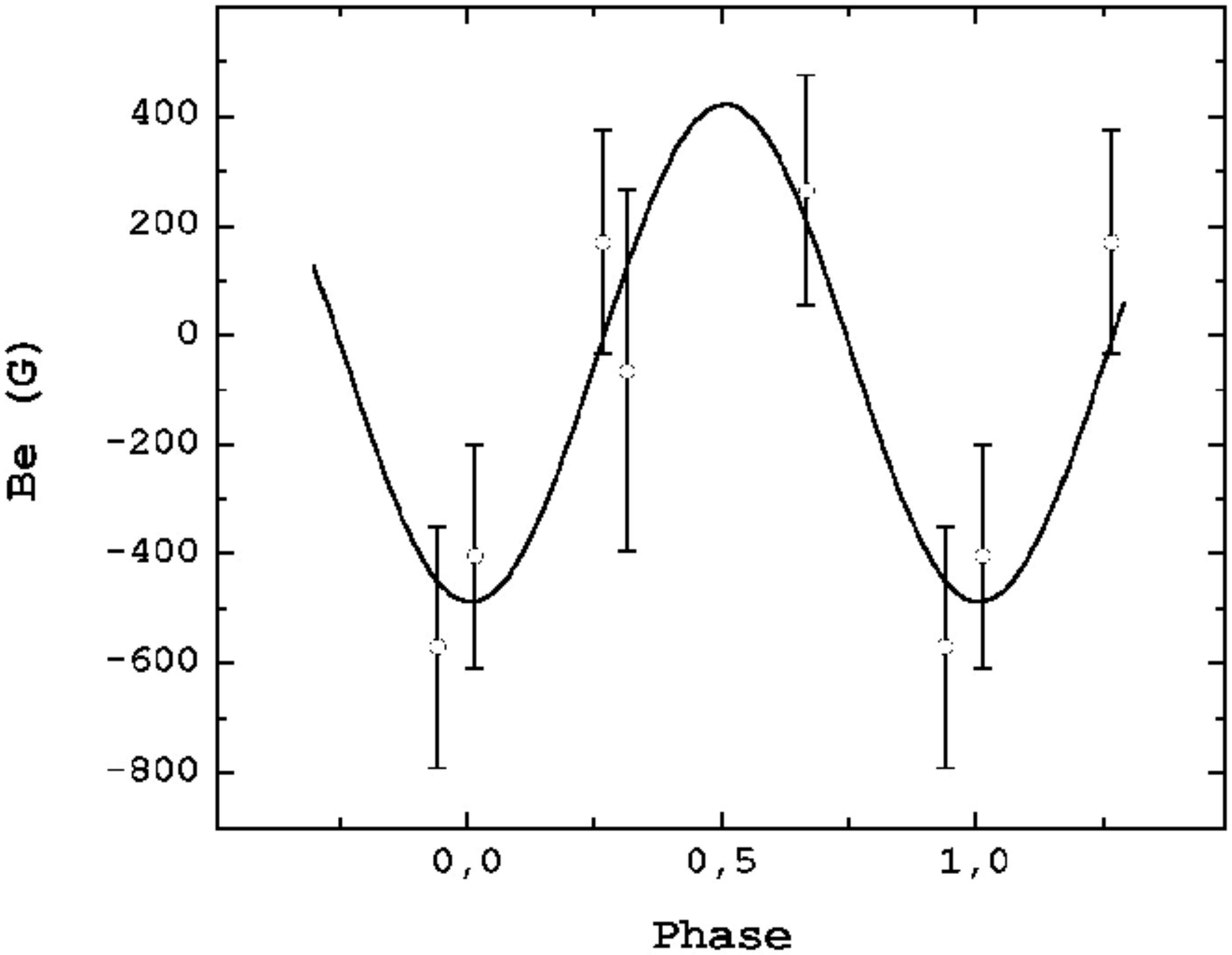}}
\vspace{-3.5mm}
\caption{ HD 28843 }
\label{fig:fig56}
\end{figure}

\begin{figure}
\resizebox{0.98\hsize}{!}{\includegraphics{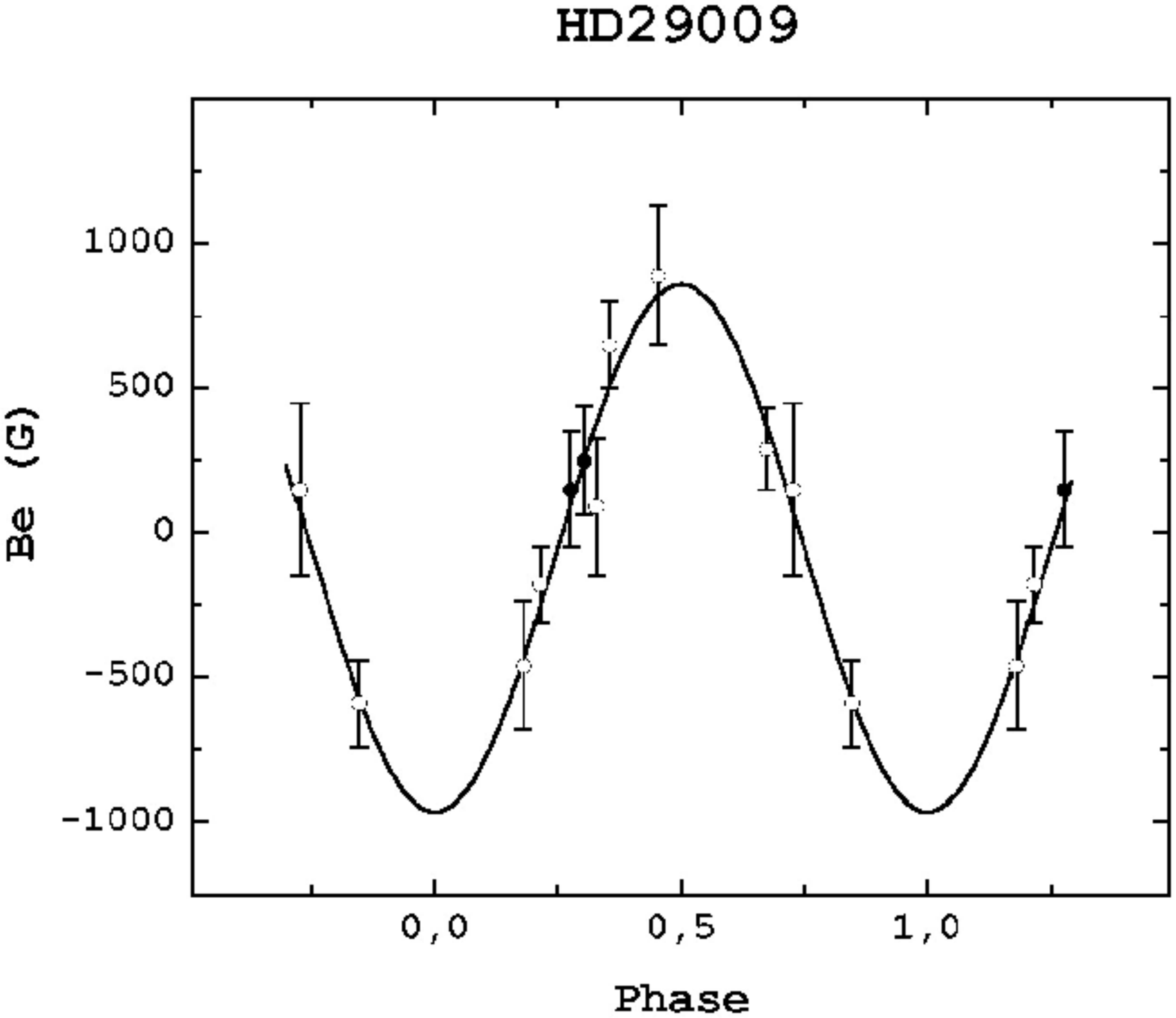}}
\vspace{-3.5mm}
\caption{ HD 29009 }
\label{fig:fig57}
\end{figure}

\begin{figure}
\resizebox{0.98\hsize}{!}{\includegraphics{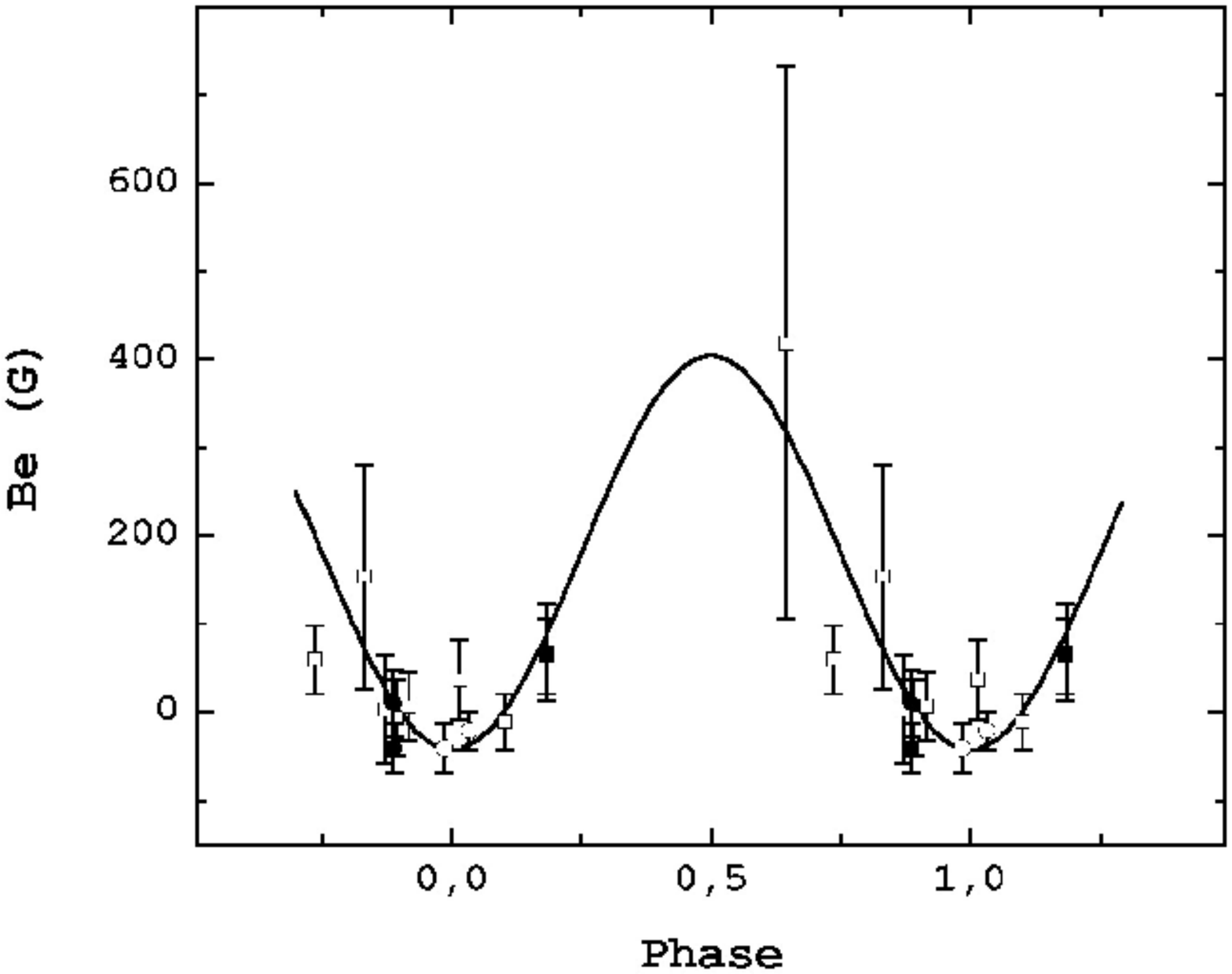}}
\vspace{-3.5mm}
\caption{ HD 29248 }
\label{fig:fig58}
\end{figure}

\begin{figure}
\resizebox{0.98\hsize}{!}{\includegraphics{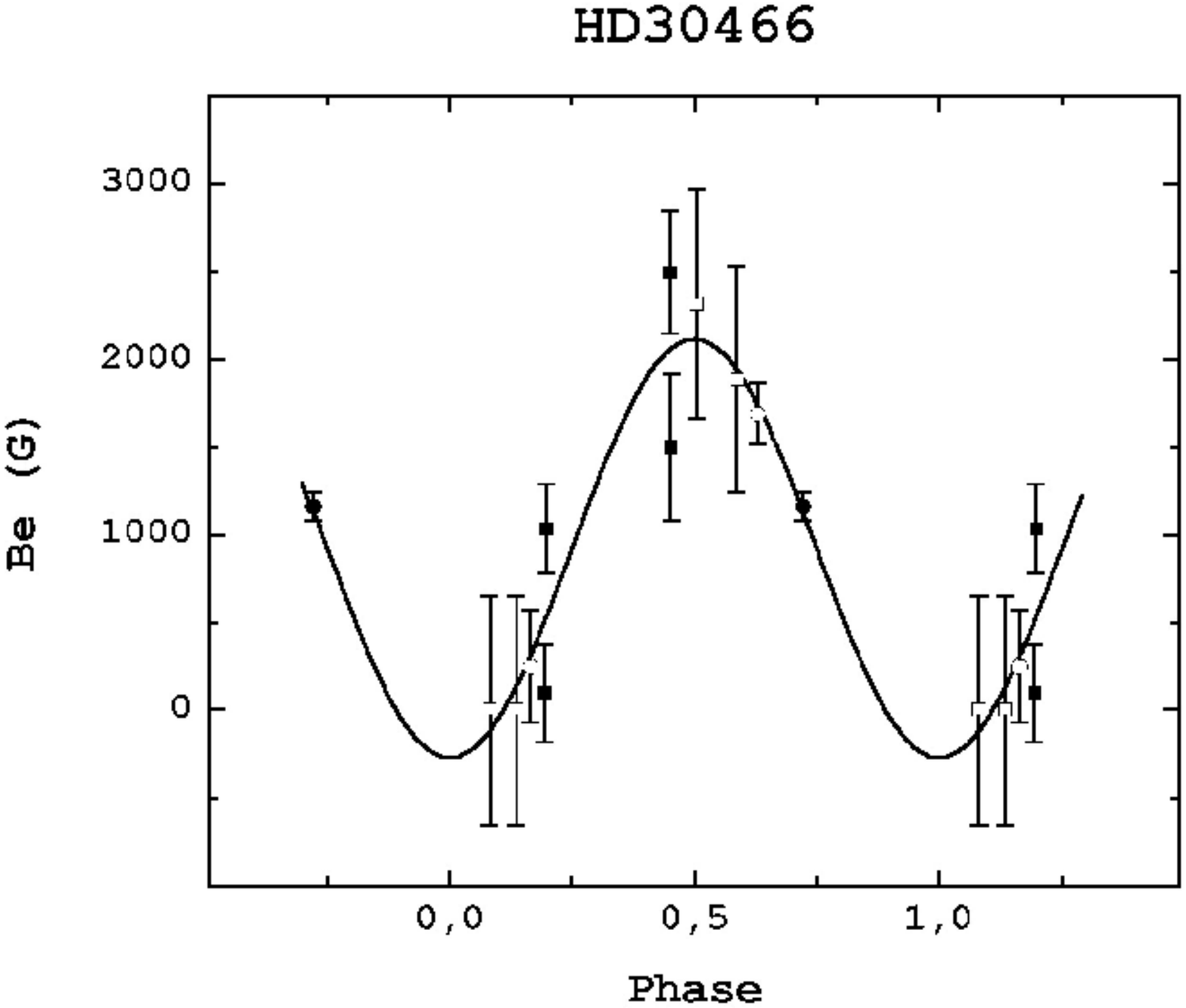}}
\vspace{-3.5mm}
\caption{ HD 30466 }
\label{fig:fig59}
\end{figure}

\begin{figure}
\resizebox{0.98\hsize}{!}{\includegraphics{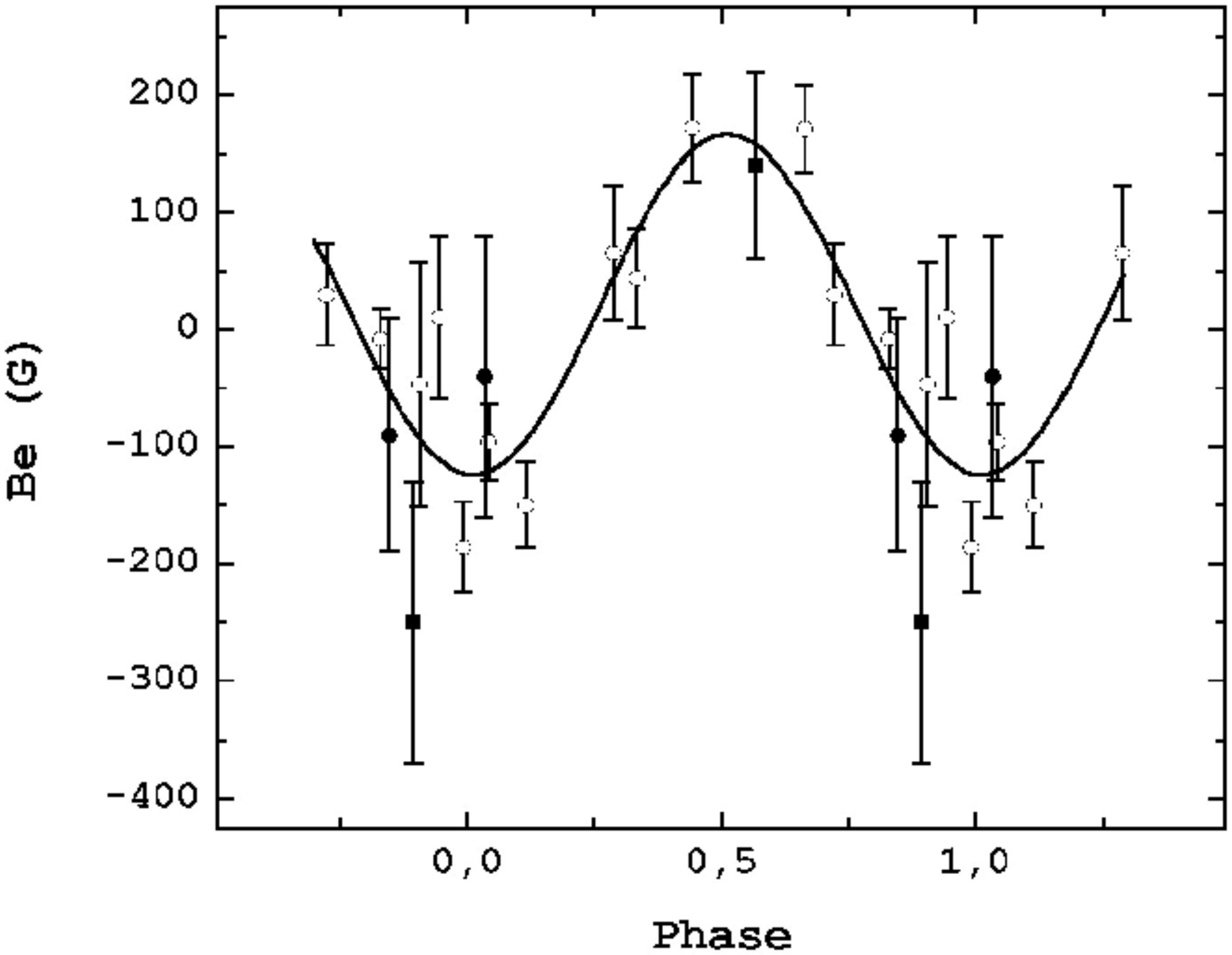}}
\vspace{-3.5mm}
\caption{ HD 32549 (1) }
\label{fig:fig60}
\end{figure}

\begin{figure}
\resizebox{0.98\hsize}{!}{\includegraphics{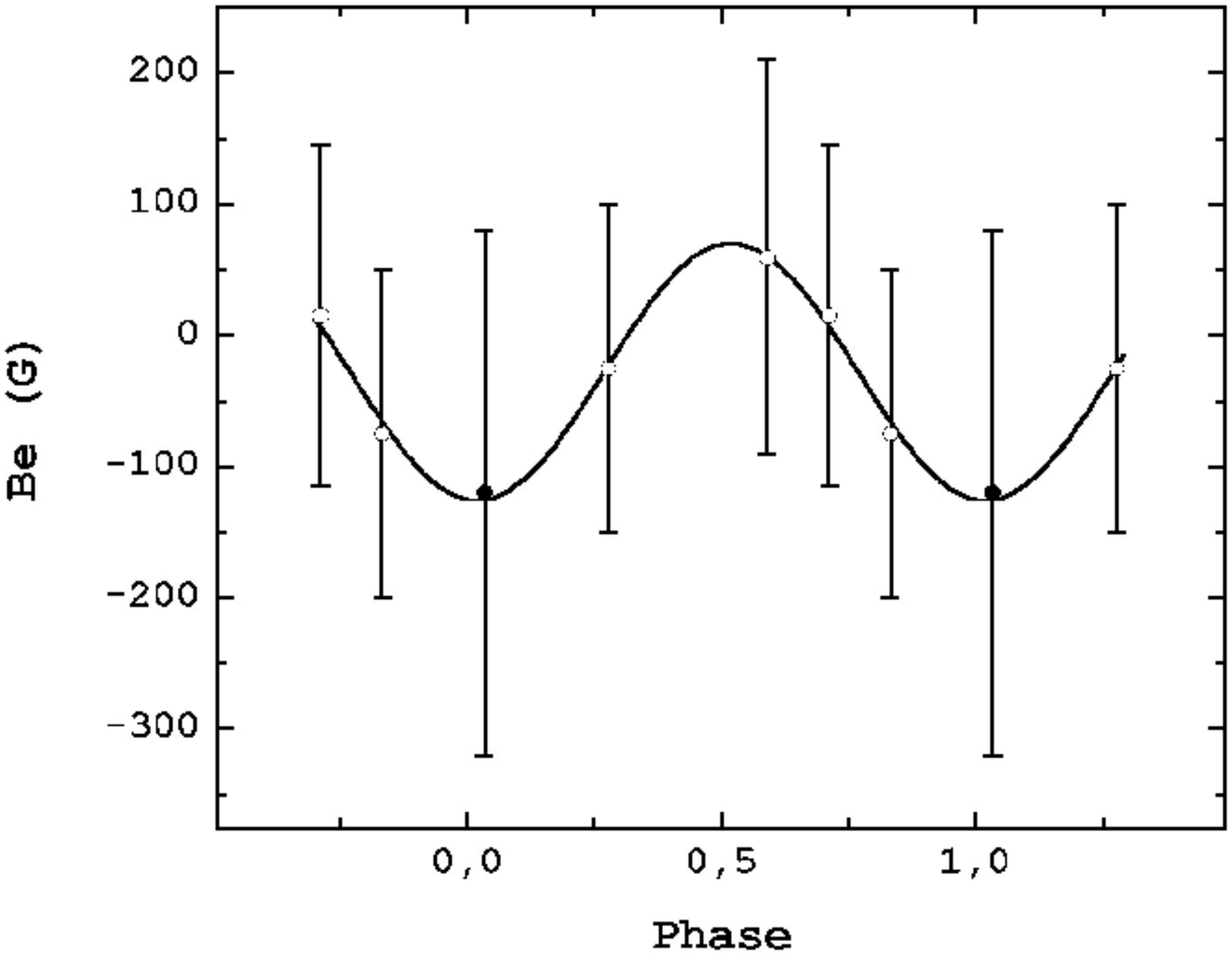}}
\vspace{-3.5mm}
\caption{ HD 32549 (2) }
\label{fig:fig61}
\end{figure}

\clearpage
\newpage

\begin{figure}
\resizebox{0.98\hsize}{!}{\includegraphics{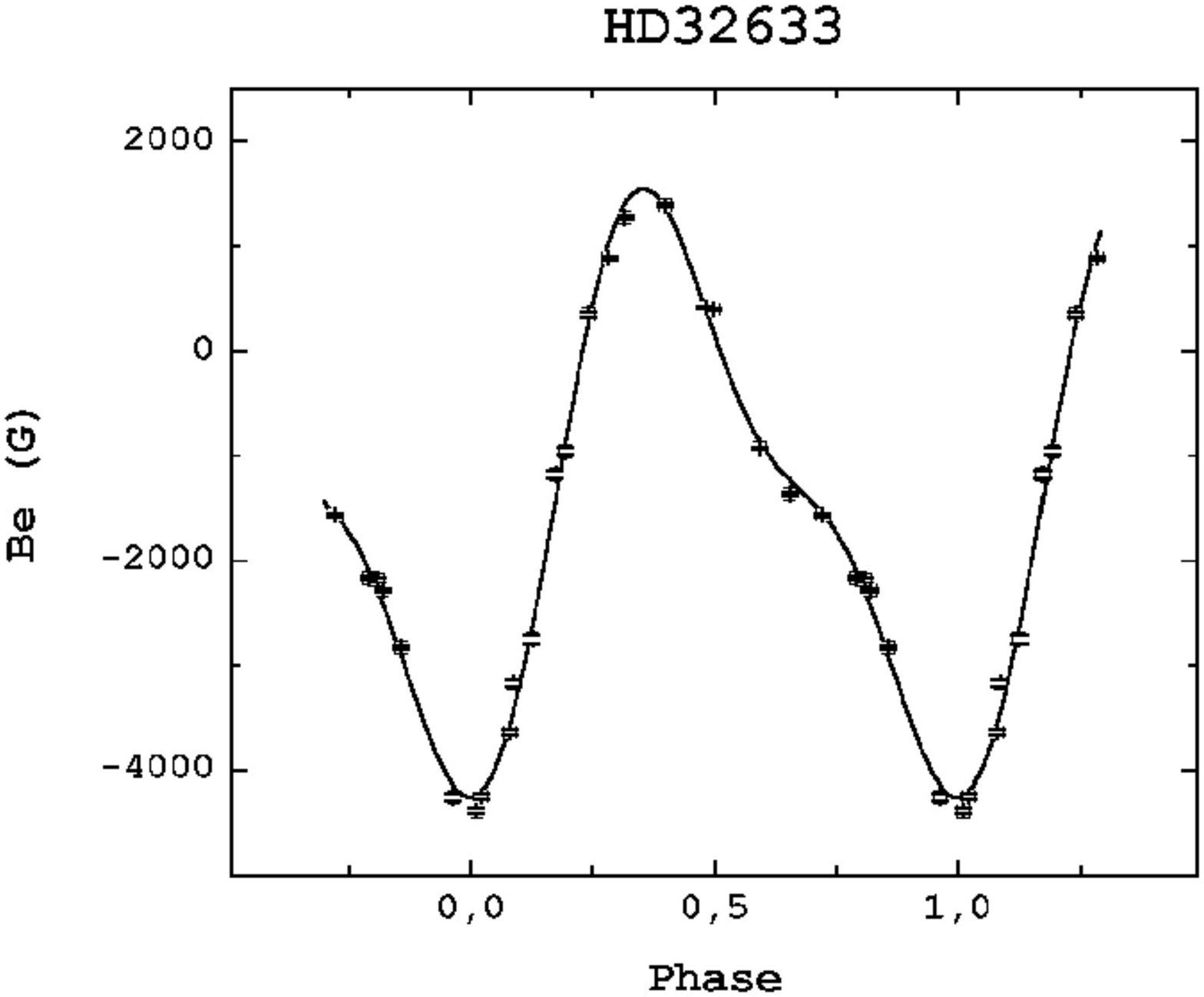}}
\vspace{-3.5mm}
\caption{ HD 32633 }
\label{fig:fig62}
\end{figure}

\begin{figure}
\resizebox{0.98\hsize}{!}{\includegraphics{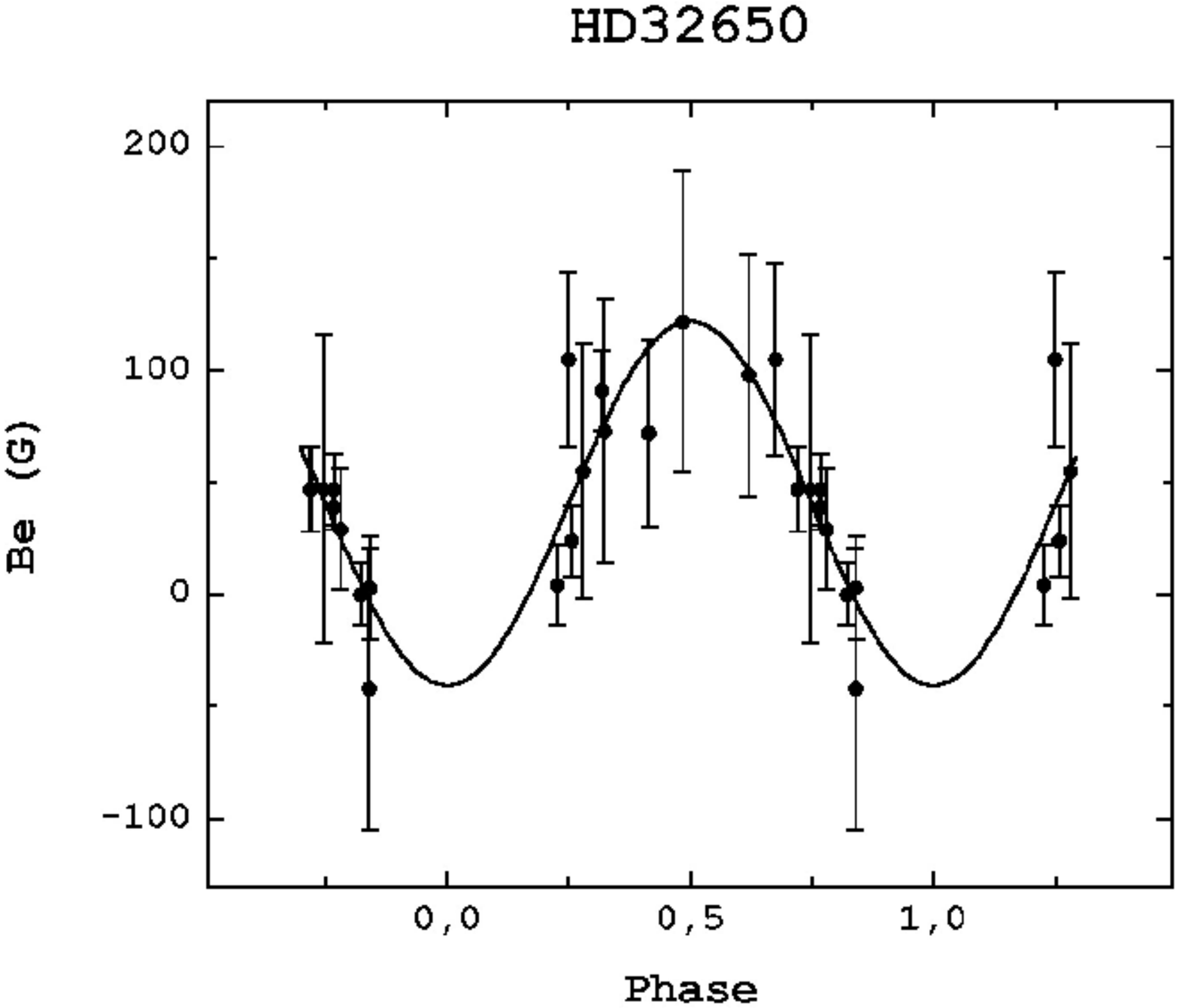}}
\vspace{-3.5mm}
\caption{ HD 32650 }
\label{fig:fig63}
\end{figure}

\begin{figure}
\resizebox{0.98\hsize}{!}{\includegraphics{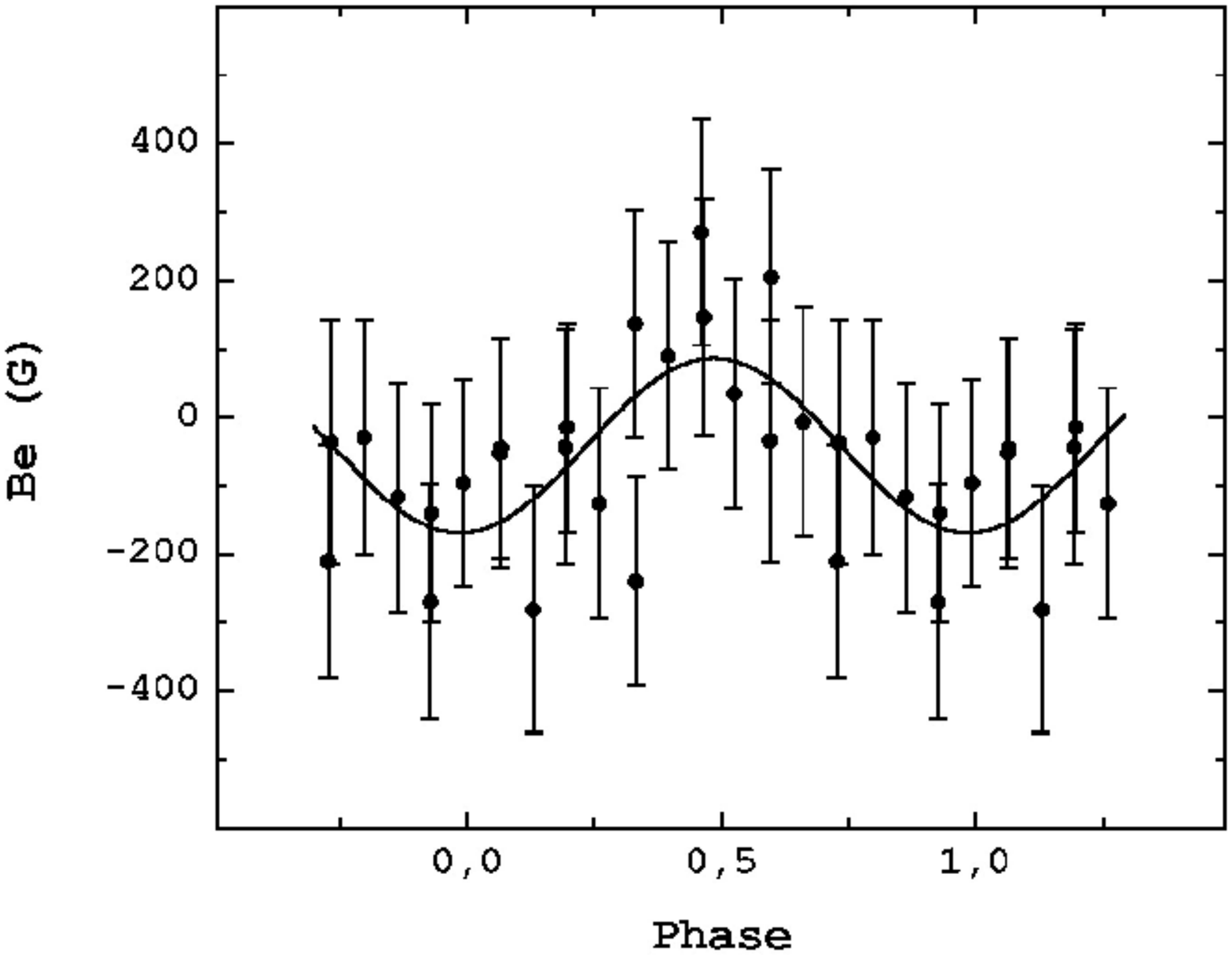}}
\vspace{-3.5mm}
\caption{ HD 33328 (1) }
\label{fig:fig64}
\end{figure}

\begin{figure}
\resizebox{0.98\hsize}{!}{\includegraphics{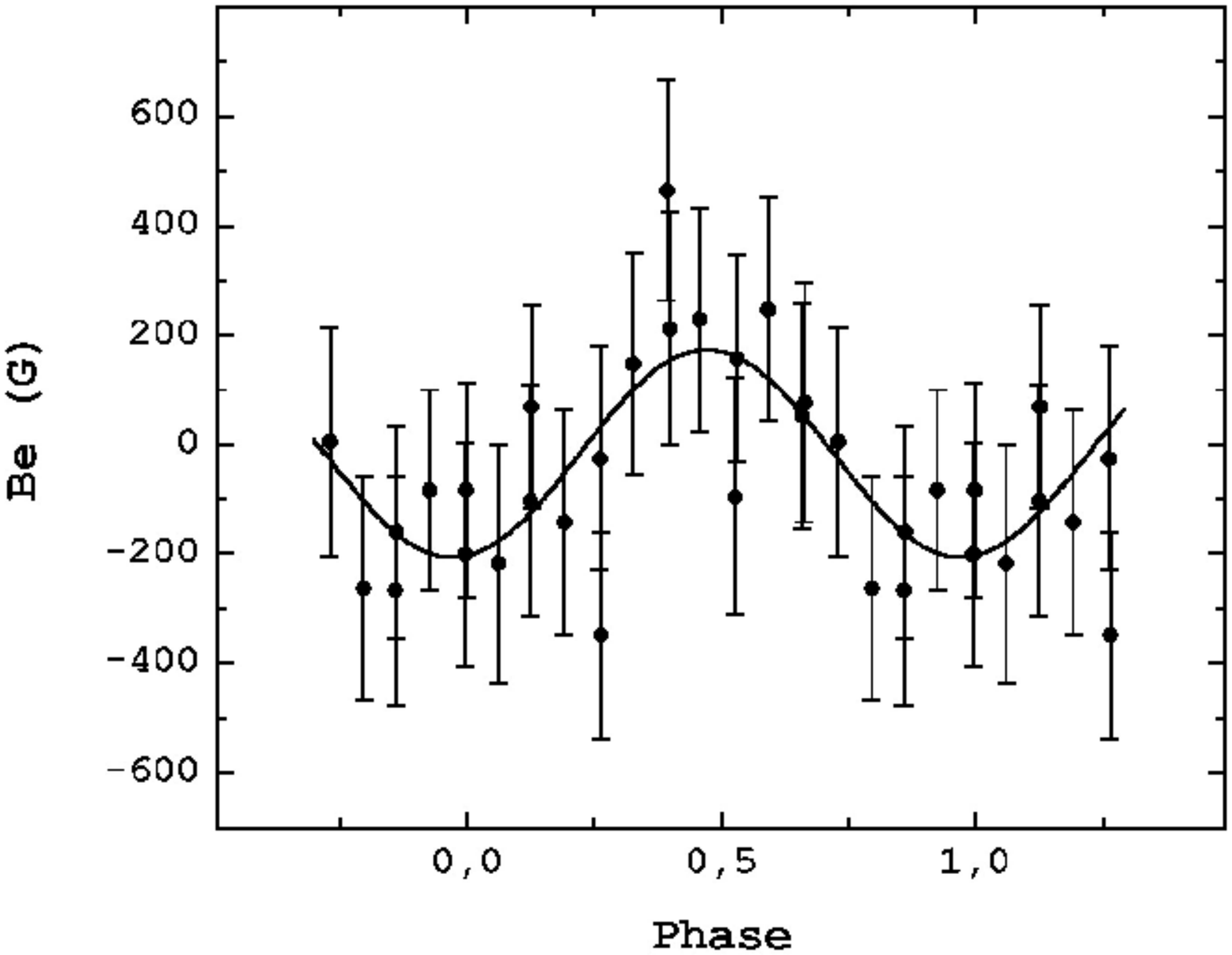}}
\vspace{-3.5mm}
\caption{ HD 33328 (2) }
\label{fig:fig65}
\end{figure}

\begin{figure}
\resizebox{0.98\hsize}{!}{\includegraphics{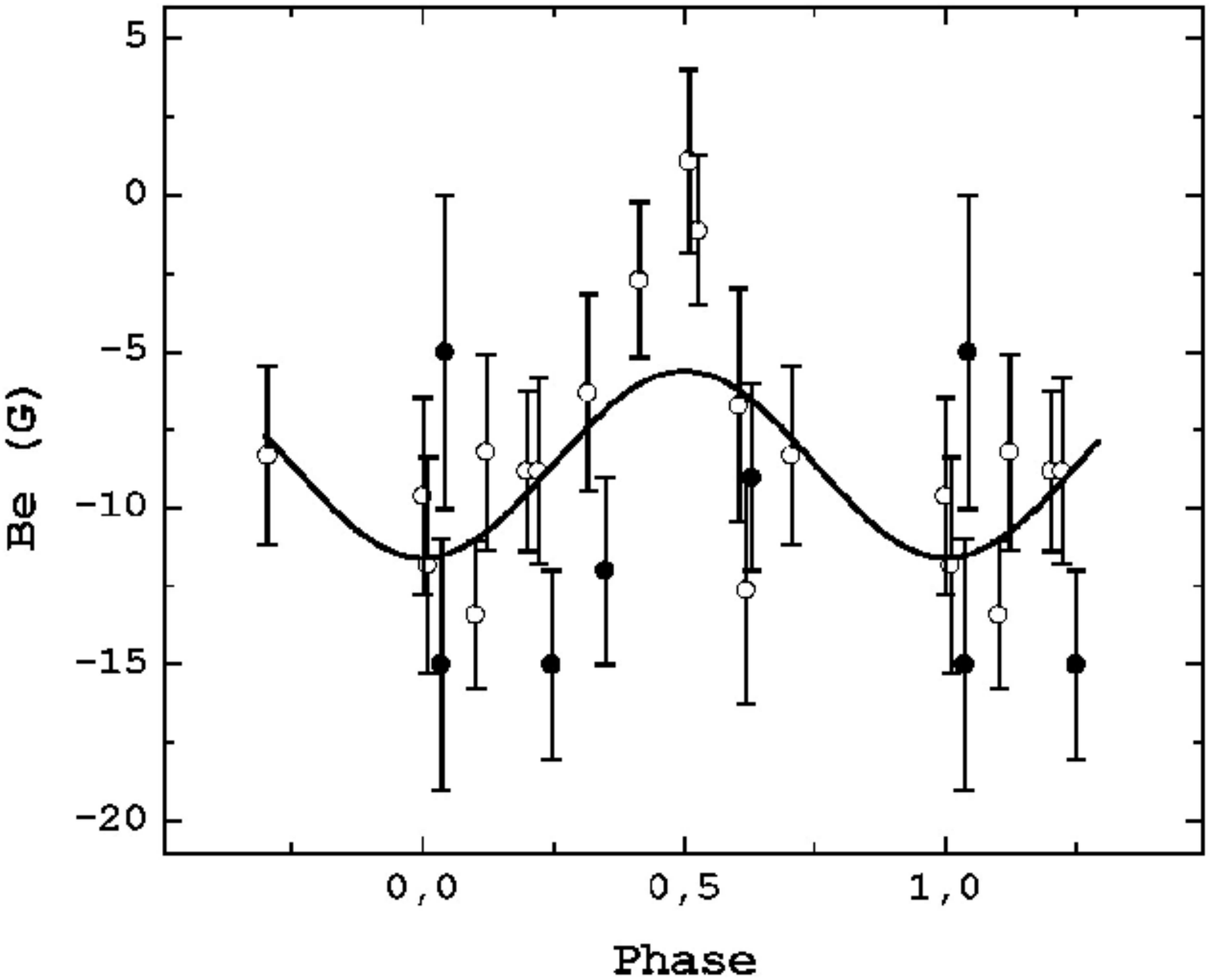}}
\vspace{-3.5mm}
\caption{ HD 33798 }
\label{fig:fig66}
\end{figure}

\begin{figure}
\resizebox{0.98\hsize}{!}{\includegraphics{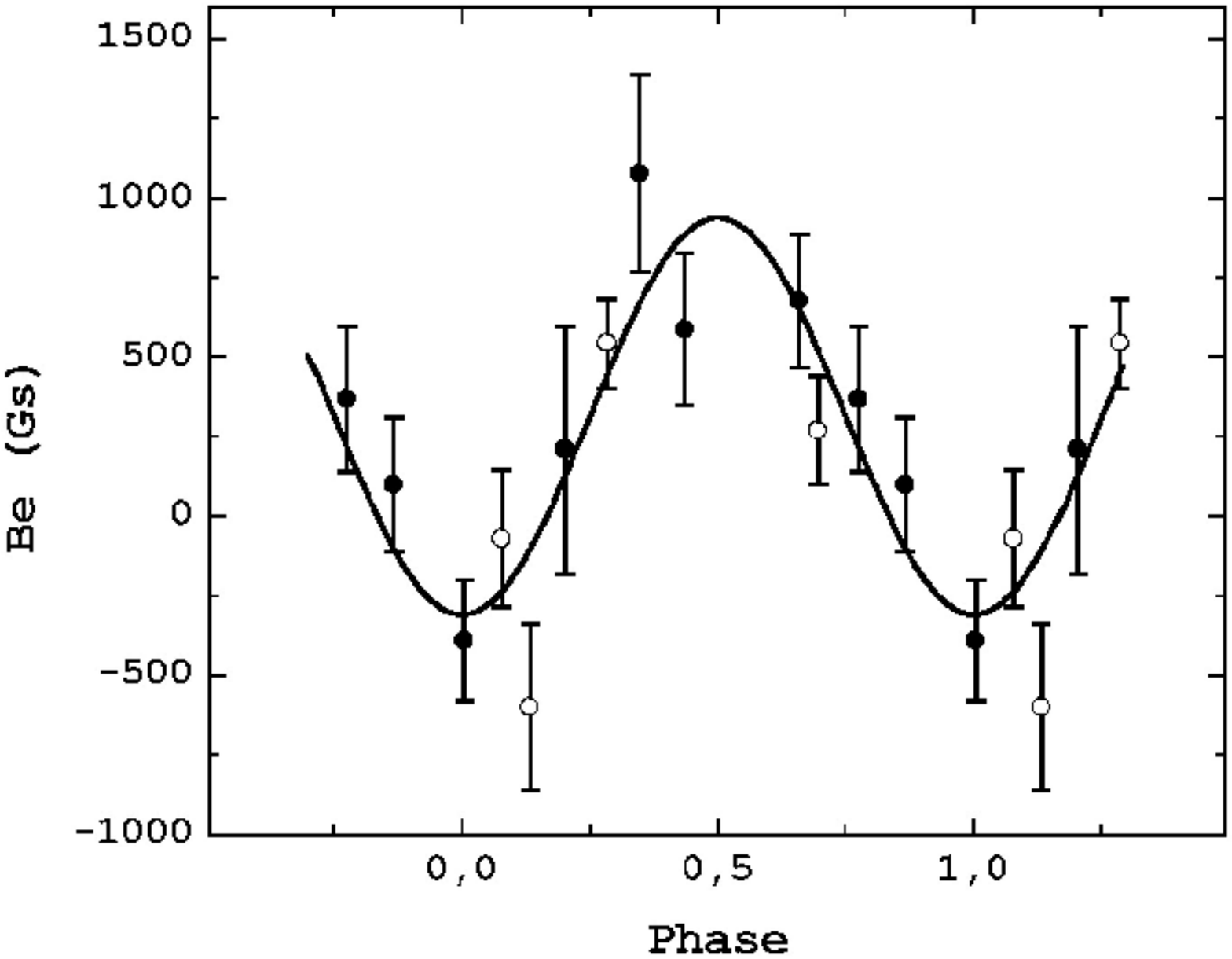}}
\vspace{-3.5mm}
\caption{ HD 34452 }
\label{fig:fig67}
\end{figure}

\clearpage
\newpage

\begin{figure}
\resizebox{0.98\hsize}{!}{\includegraphics{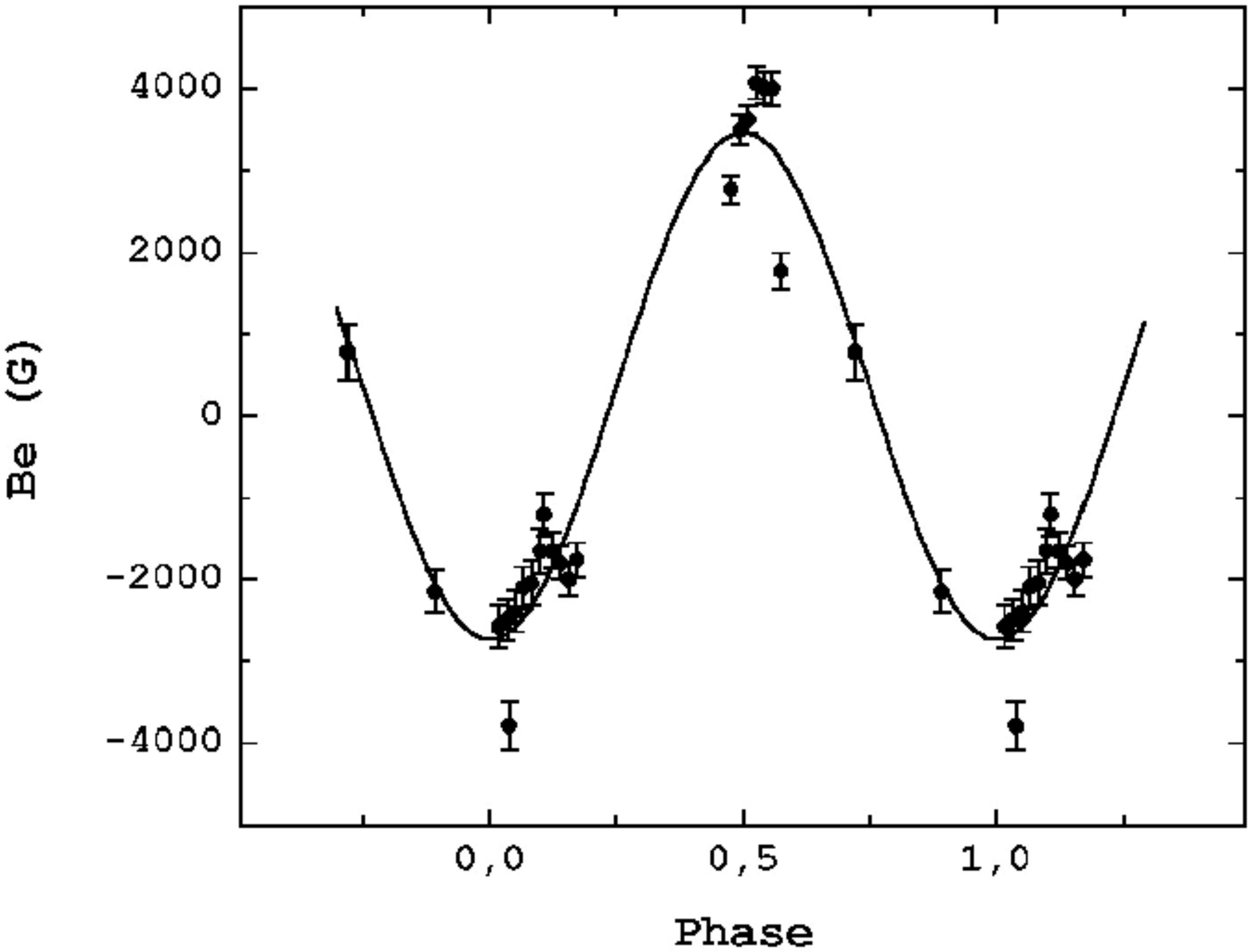}}
\vspace{-3.5mm}
\caption{ HD 34736 }
\label{fig:fig68}
\end{figure}

\begin{figure}
\resizebox{0.98\hsize}{!}{\includegraphics{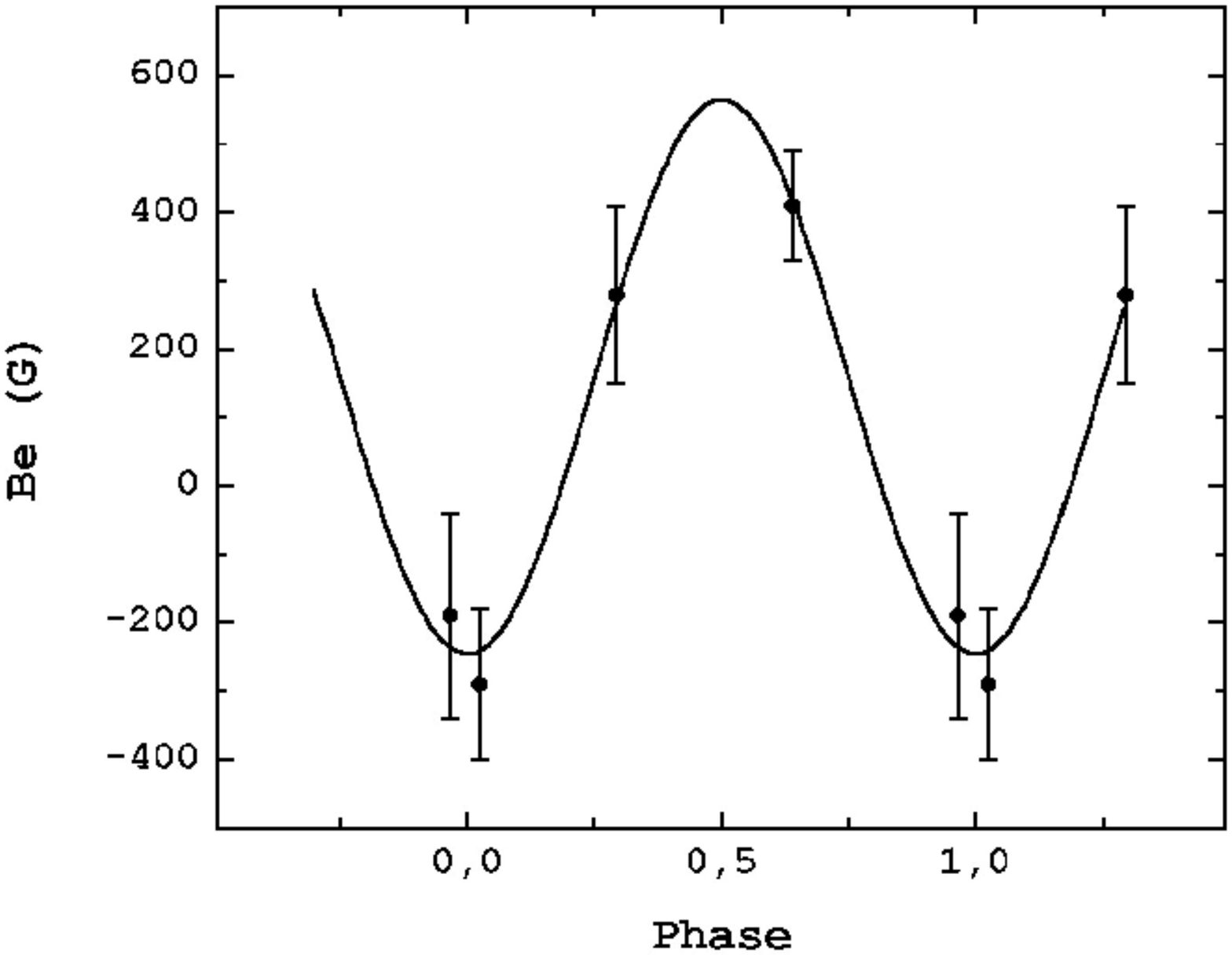}}
\vspace{-3.5mm}
\caption{ HD 34859 }
\label{fig:fig49}
\end{figure}

\begin{figure}
\resizebox{0.98\hsize}{!}{\includegraphics{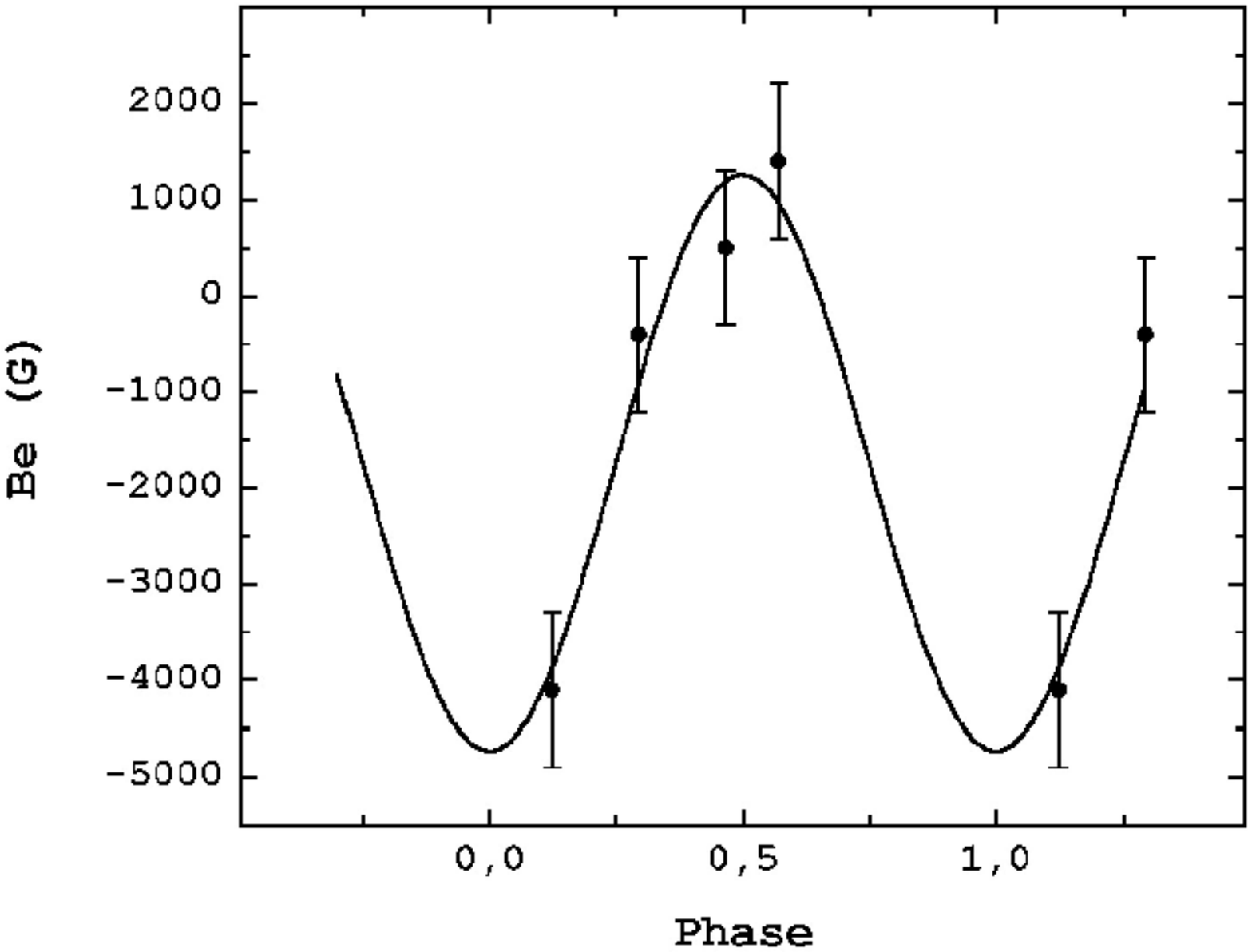}}
\vspace{-3.5mm}
\caption{ HD 35177 }
\label{fig:fig49}
\end{figure}

\begin{figure}
\resizebox{0.98\hsize}{!}{\includegraphics{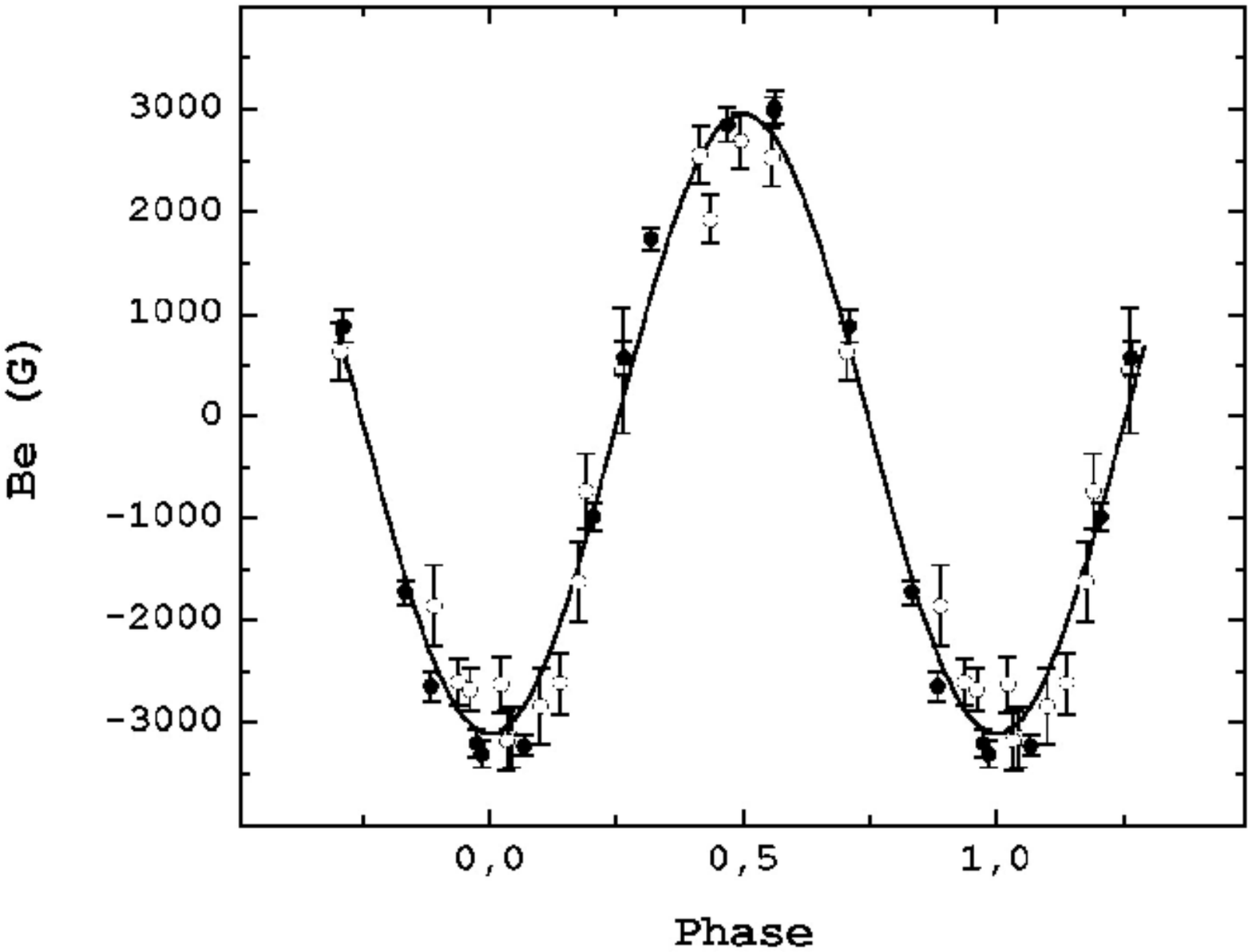}}
\vspace{-3.5mm}
\caption{ HD 35298(1) }
\label{fig:fig69}
\end{figure}

\begin{figure}
\resizebox{0.98\hsize}{!}{\includegraphics{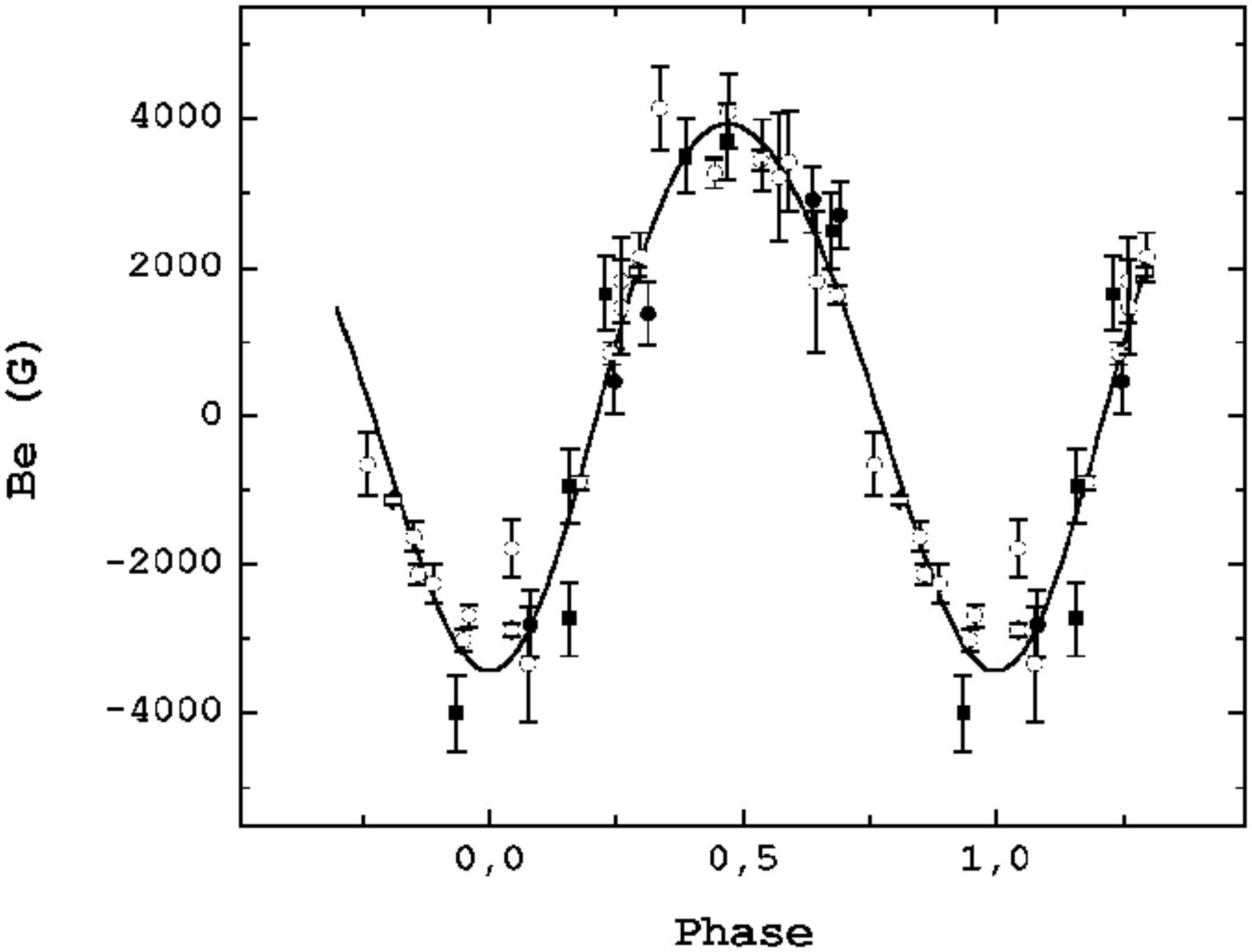}}
\vspace{-3.5mm}
\caption{ HD 35298(2) }
\label{fig:fig69}
\end{figure}

\begin{figure}
\resizebox{0.98\hsize}{!}{\includegraphics{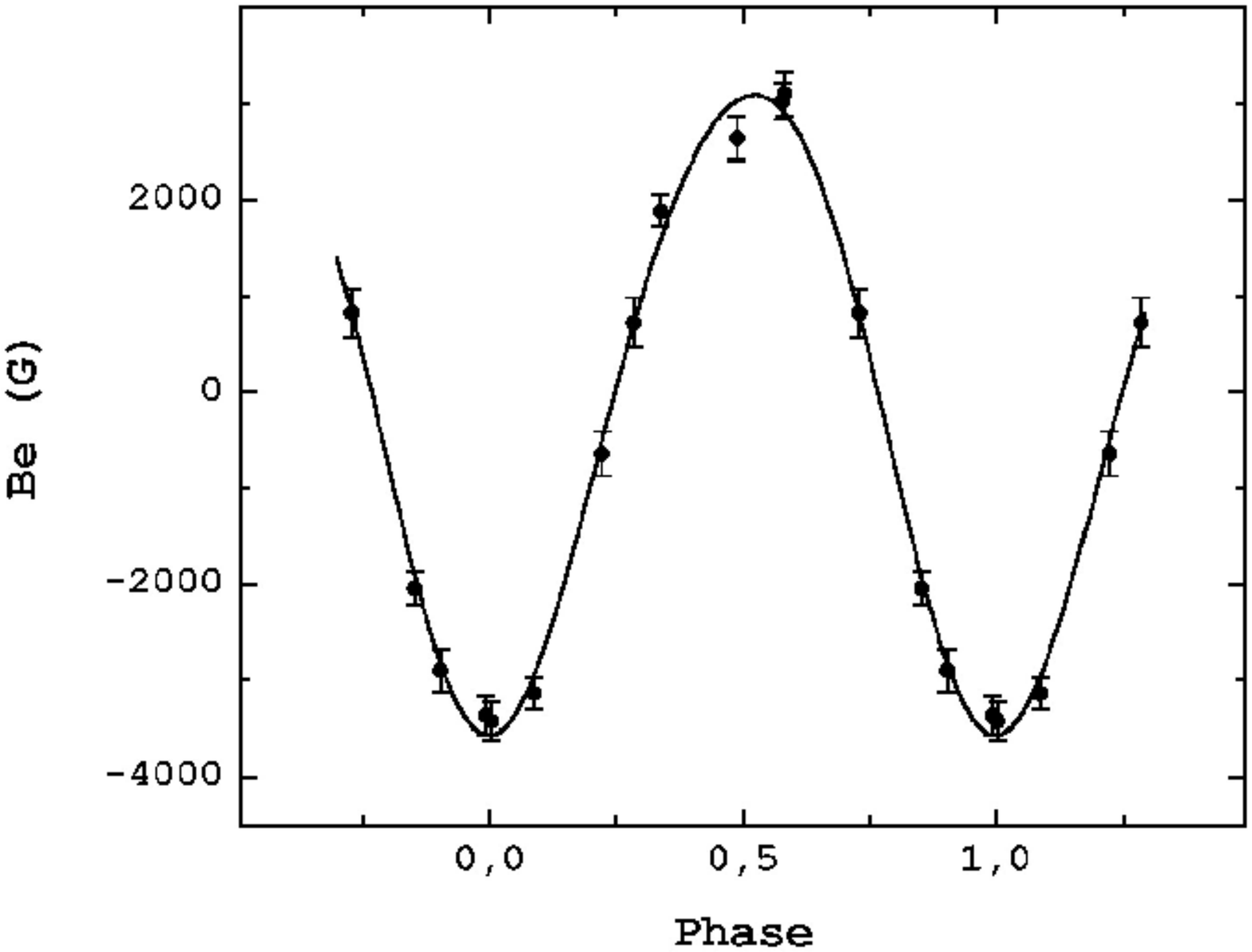}}
\vspace{-3.5mm}
\caption{ HD 35298(3) }
\label{fig:fig69}
\end{figure}

\clearpage
\newpage

\begin{figure}
\resizebox{0.98\hsize}{!}{\includegraphics{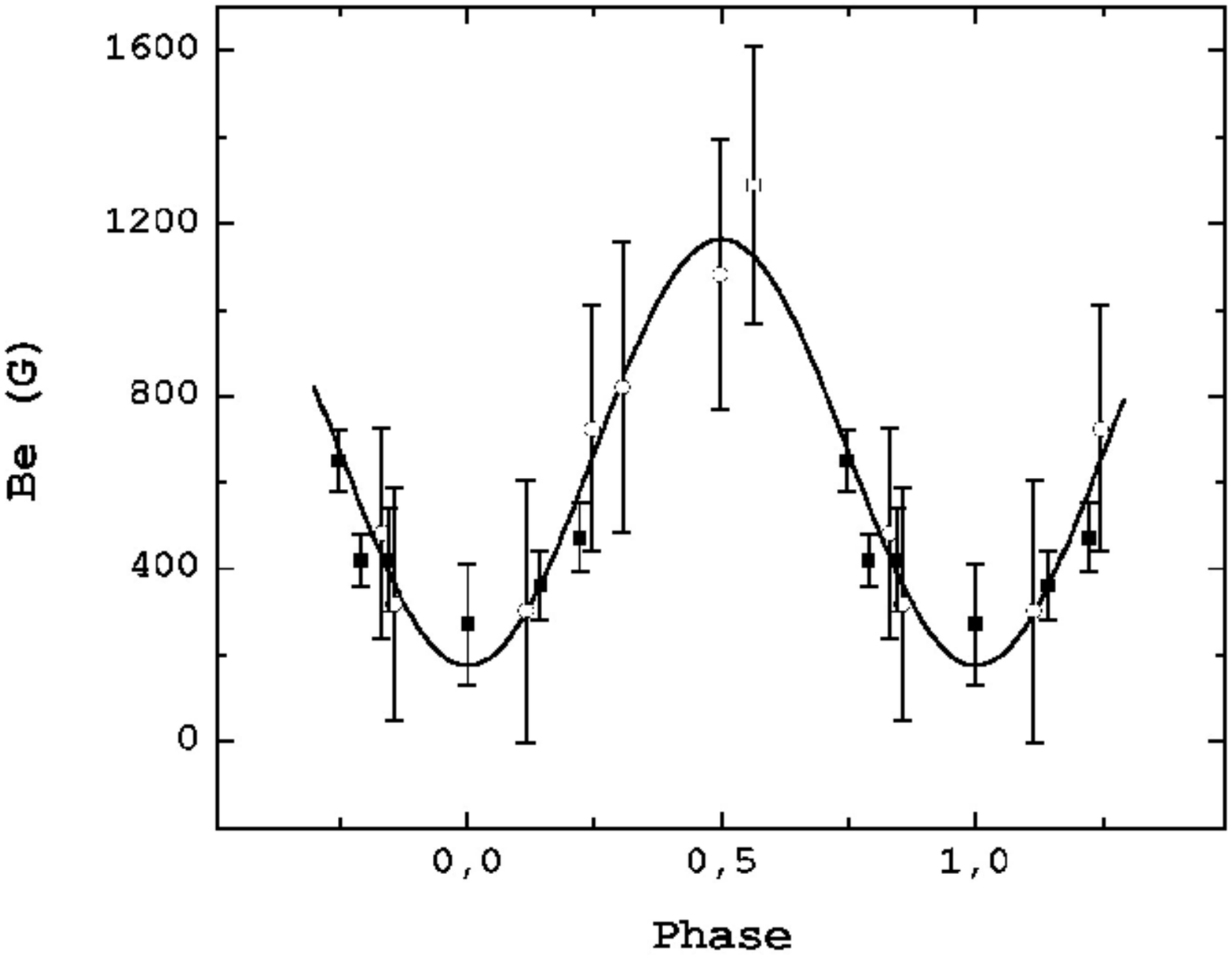}}
\vspace{-3.5mm}
\caption{ HD 35456 }
\label{fig:fig70}
\end{figure}

\begin{figure}
\resizebox{0.98\hsize}{!}{\includegraphics{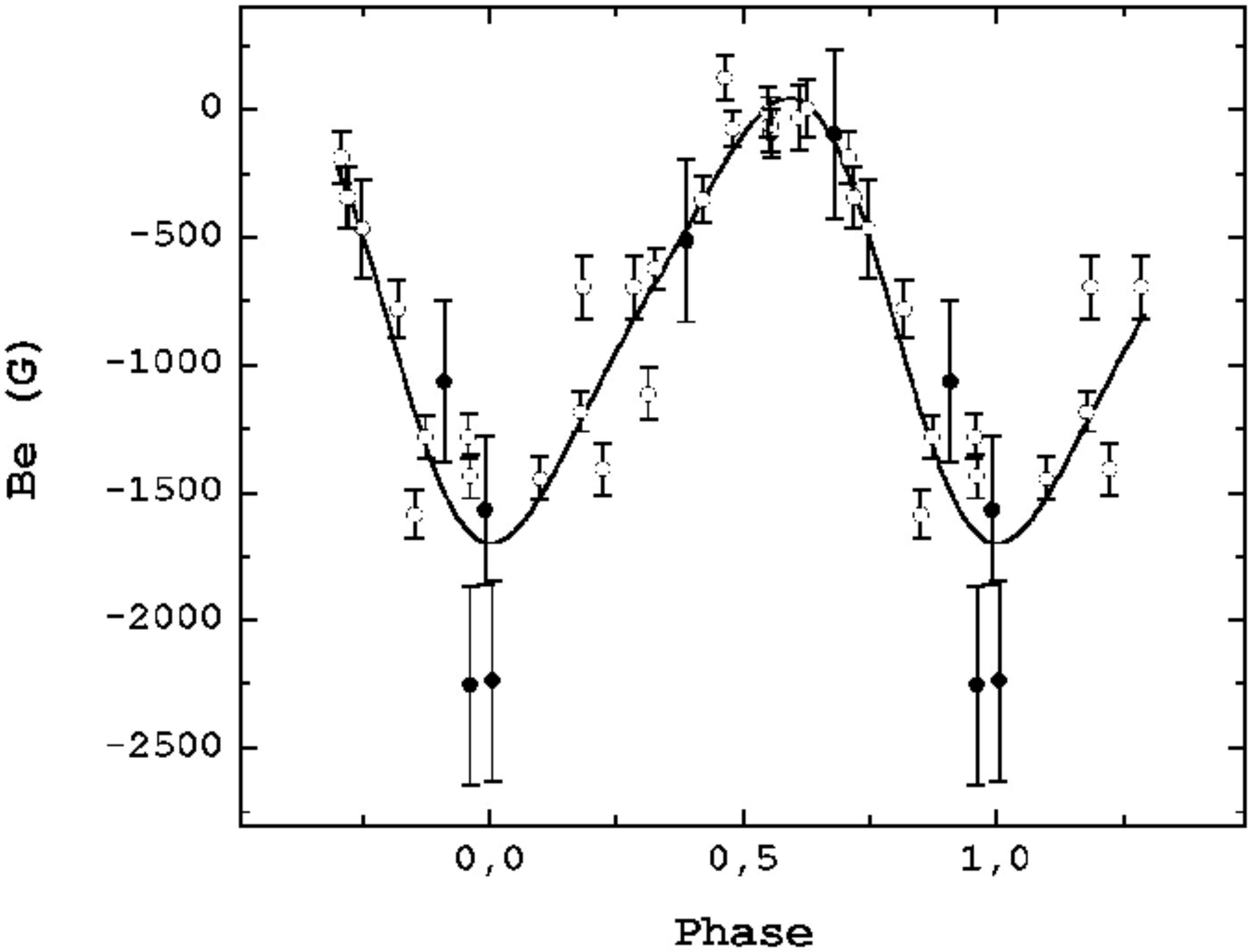}}
\vspace{-3.5mm}
\caption{ HD 35502 (1) }
\label{fig:fig71}
\end{figure}

\begin{figure}
\resizebox{0.98\hsize}{!}{\includegraphics{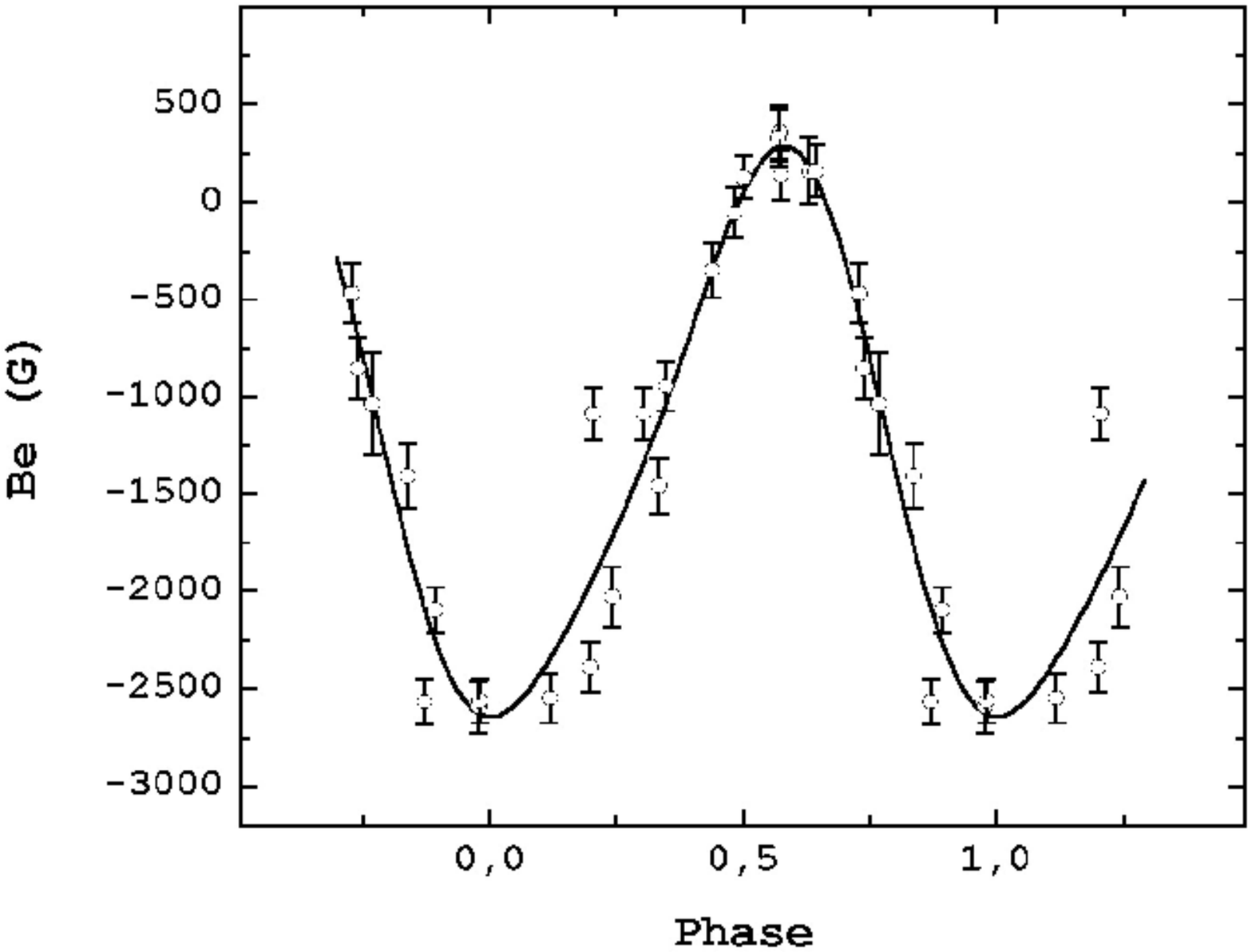}}
\vspace{-3.5mm}
\caption{ HD 35502 (2) }
\label{fig:fig72}
\end{figure}

\begin{figure}
\resizebox{0.98\hsize}{!}{\includegraphics{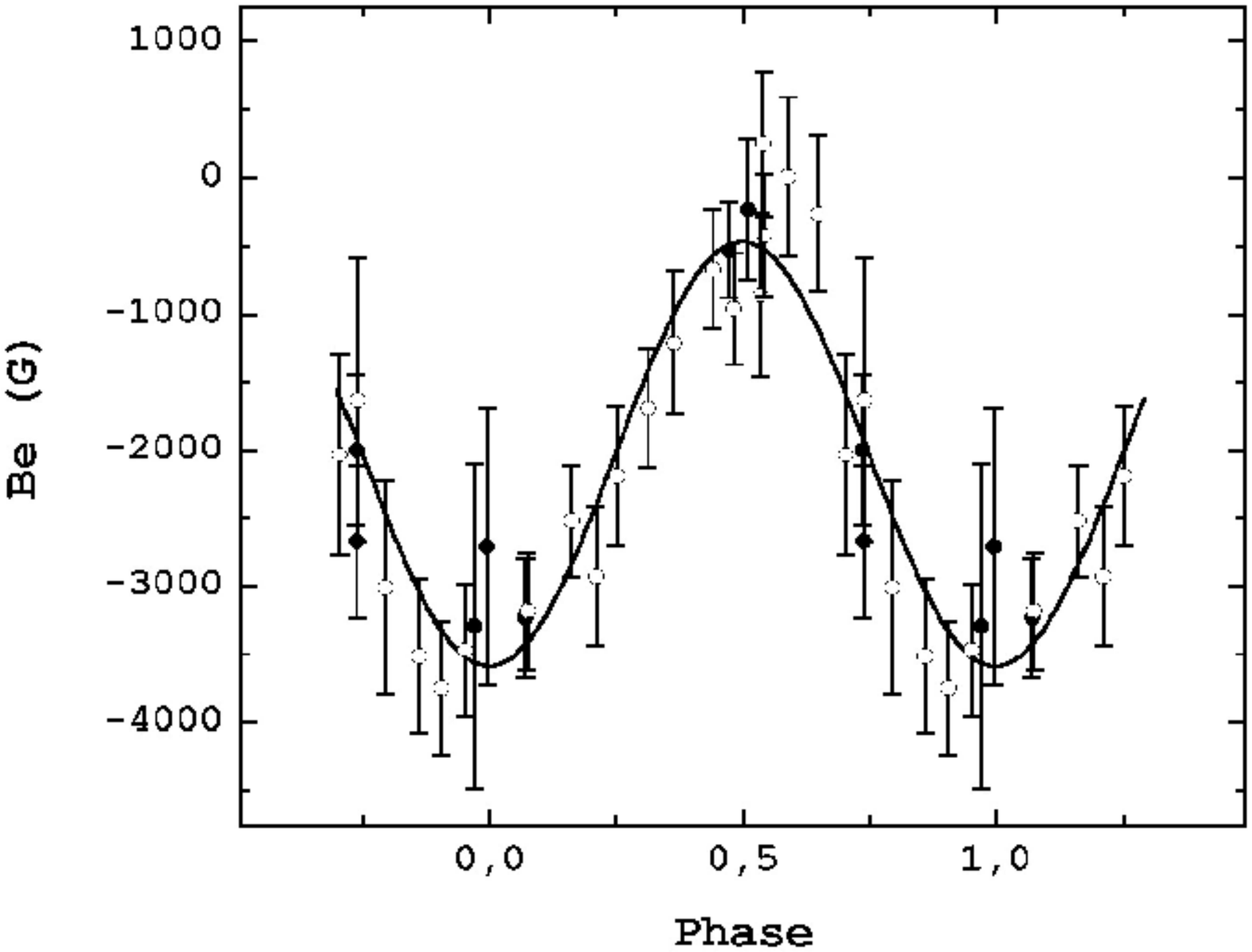}}
\vspace{-3.5mm}
\caption{ HD 35502 (3) }
\label{fig:fig73}
\end{figure}

\begin{figure}
\resizebox{0.98\hsize}{!}{\includegraphics{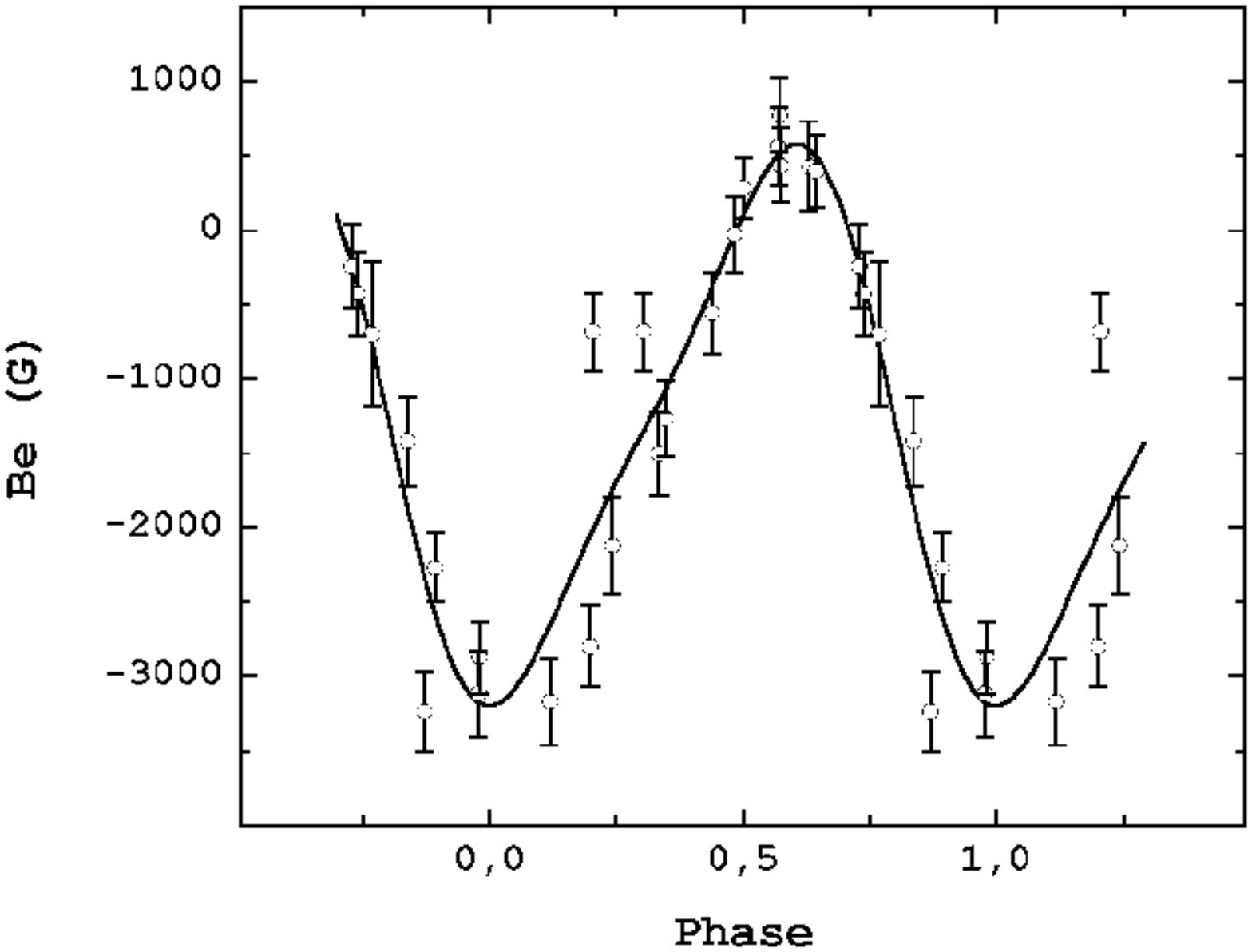}}
\vspace{-3.5mm}
\caption{ HD 35502 (4) }
\label{fig:fig74}
\end{figure}

\begin{figure}
\resizebox{0.98\hsize}{!}{\includegraphics{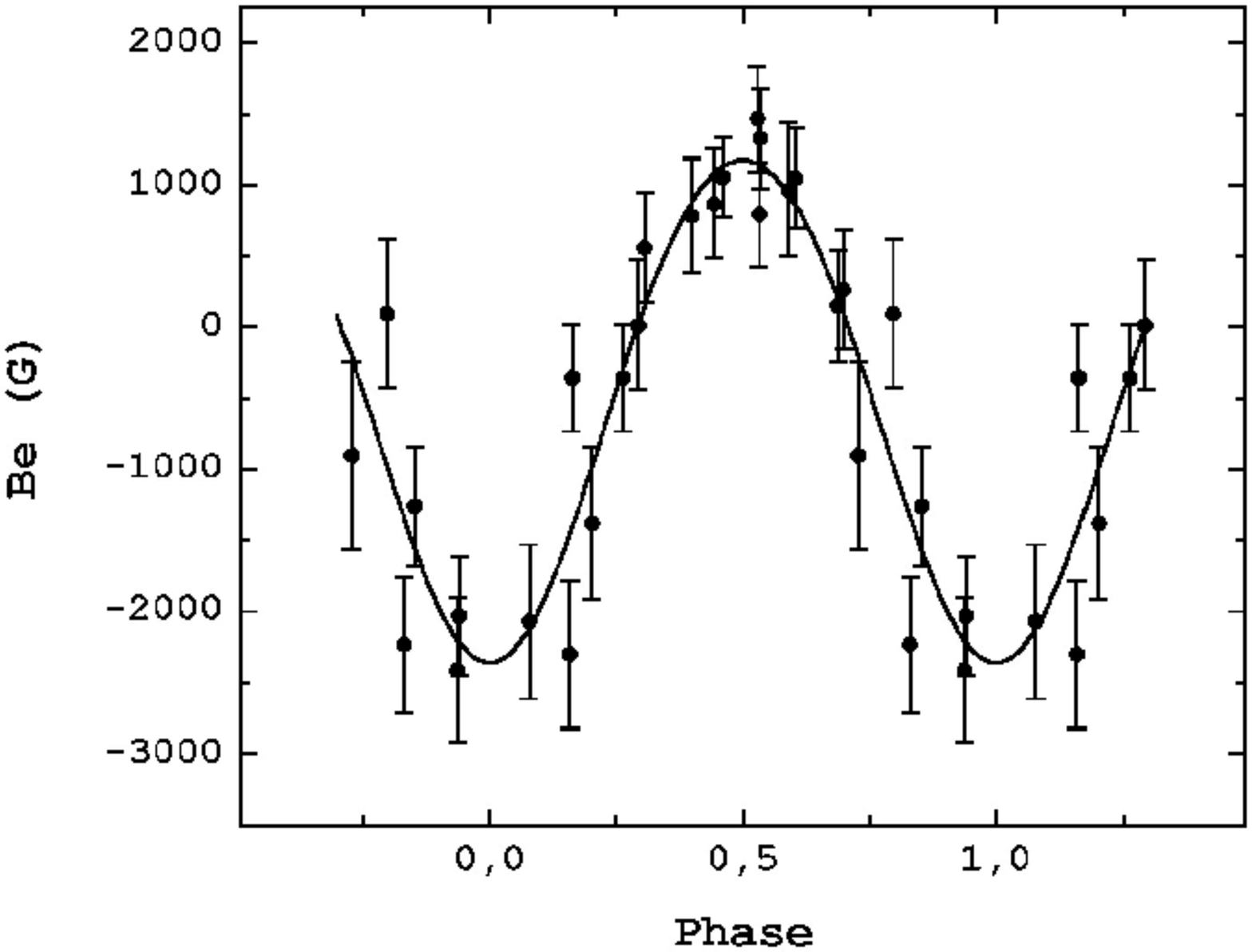}}
\vspace{-3.5mm}
\caption{ HD 35502 (5) }
\label{fig:fig75}
\end{figure}

\clearpage
\newpage

\begin{figure}
\resizebox{0.98\hsize}{!}{\includegraphics{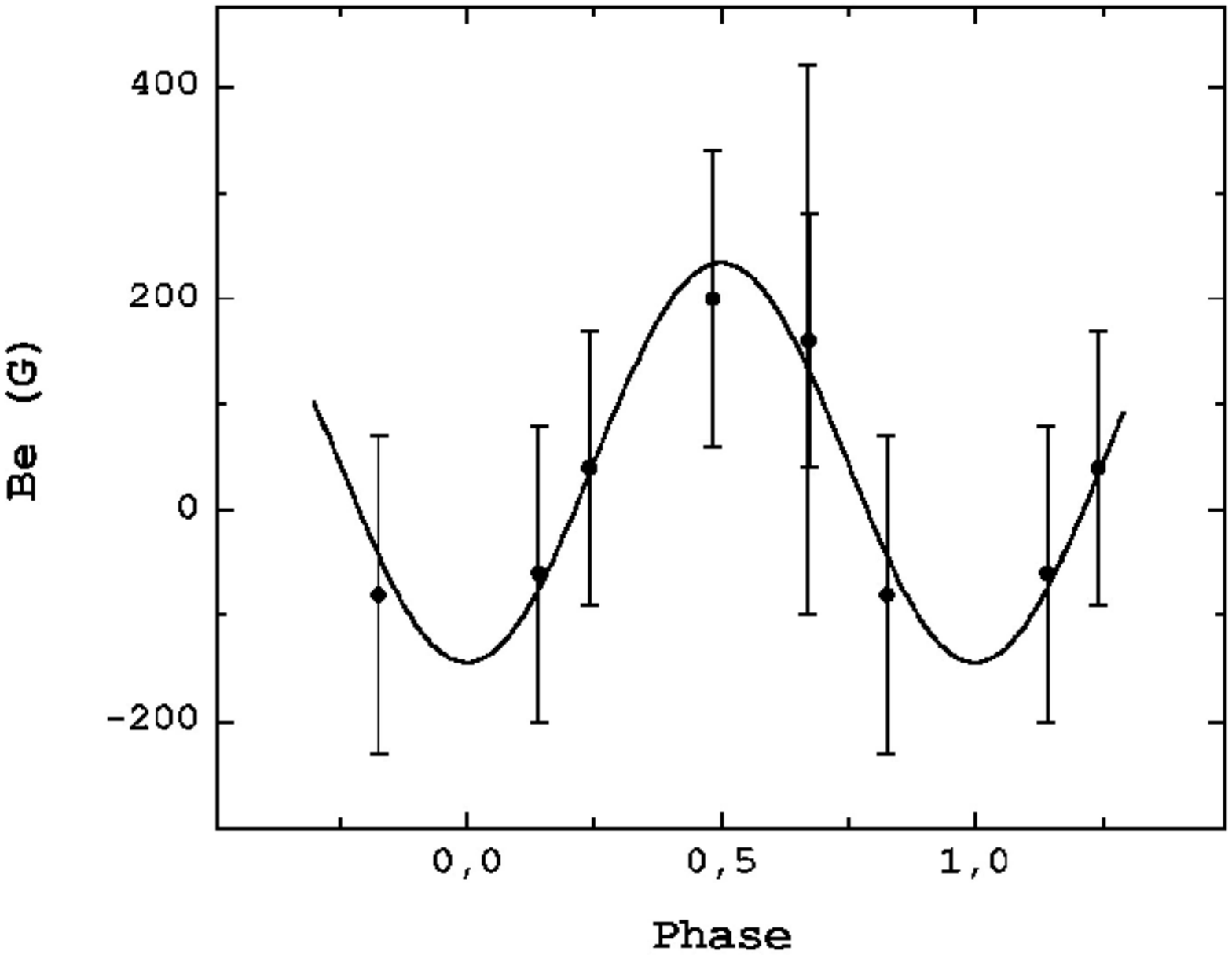}}
\vspace{-3.5mm}
\caption{ HD 35881 }
\label{fig:fig79}
\end{figure}

\begin{figure}
\resizebox{0.98\hsize}{!}{\includegraphics{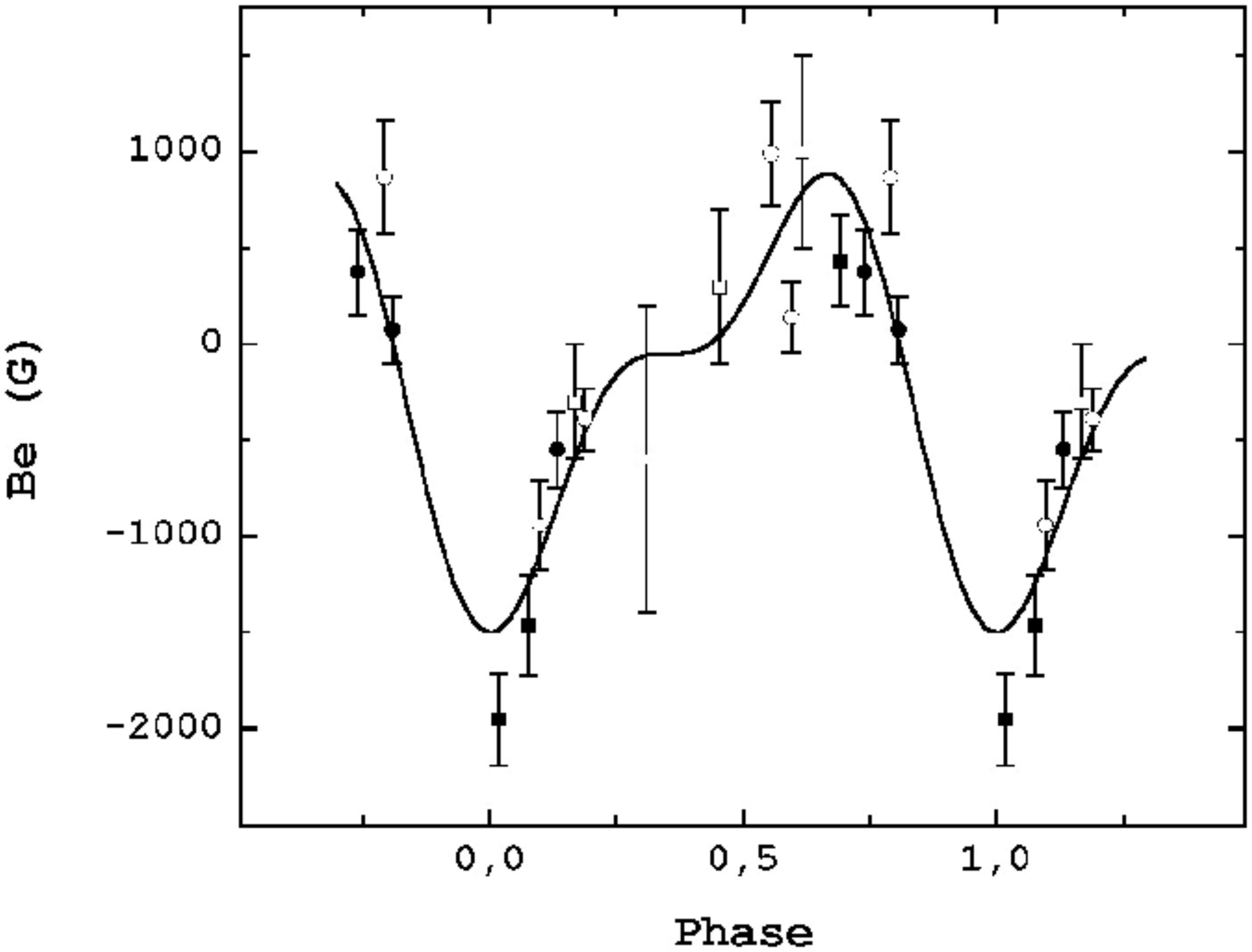}}
\vspace{-3.5mm}
\caption{ HD 35912 }
\label{fig:fig80}
\end{figure}

\begin{figure}
\resizebox{0.98\hsize}{!}{\includegraphics{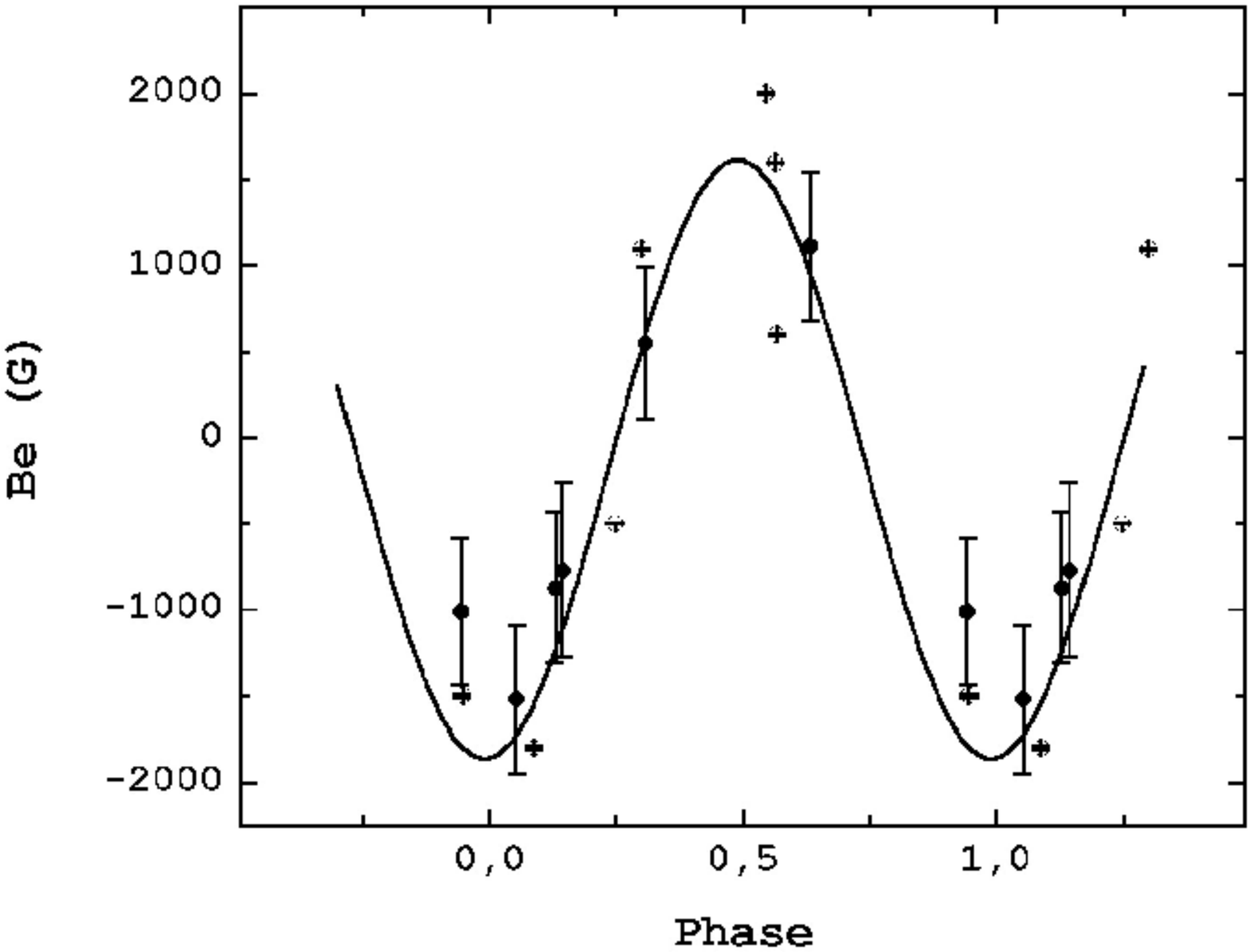}}
\vspace{-3.5mm}
\caption{ HD 36313 }
\label{fig:fig81}
\end{figure}

\begin{figure}
\resizebox{0.98\hsize}{!}{\includegraphics{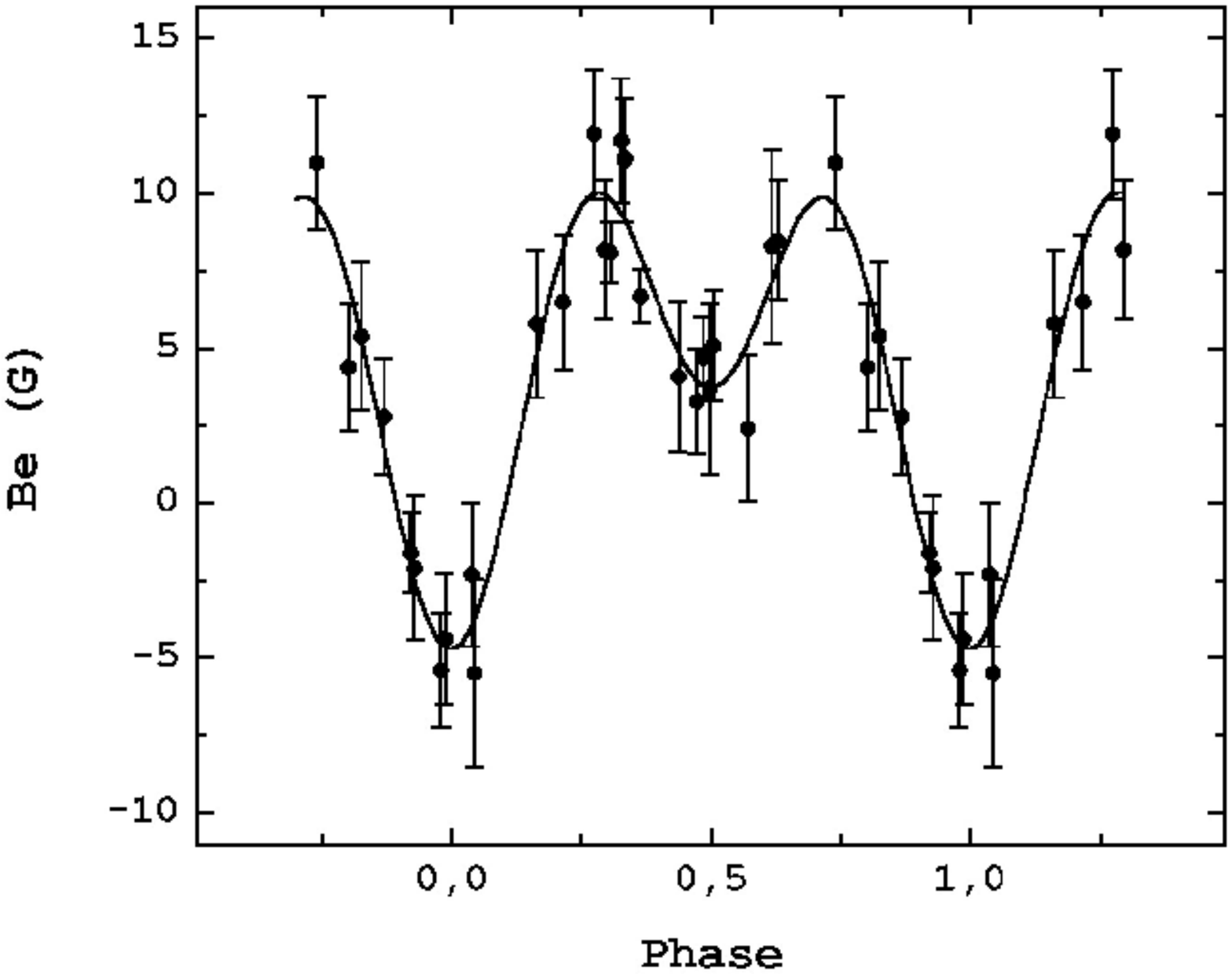}}
\vspace{-3.5mm}
\caption{ HD 36395 }
\label{fig:fig82}
\end{figure}

\begin{figure}
\resizebox{0.98\hsize}{!}{\includegraphics{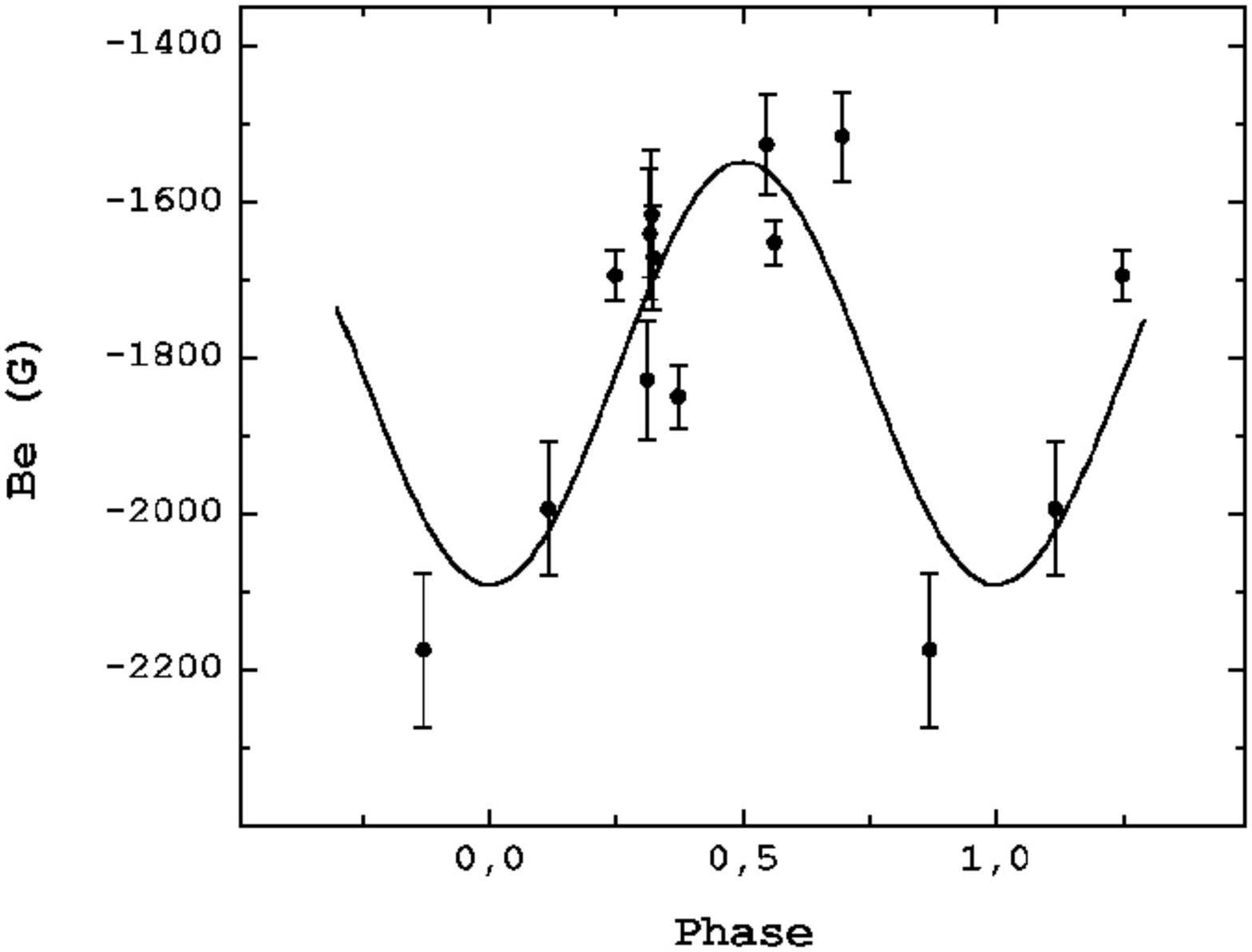}}
\vspace{-3.5mm}
\caption{ HD 36485 (1) }
\label{fig:fig83}
\end{figure}

\begin{figure}
\resizebox{0.98\hsize}{!}{\includegraphics{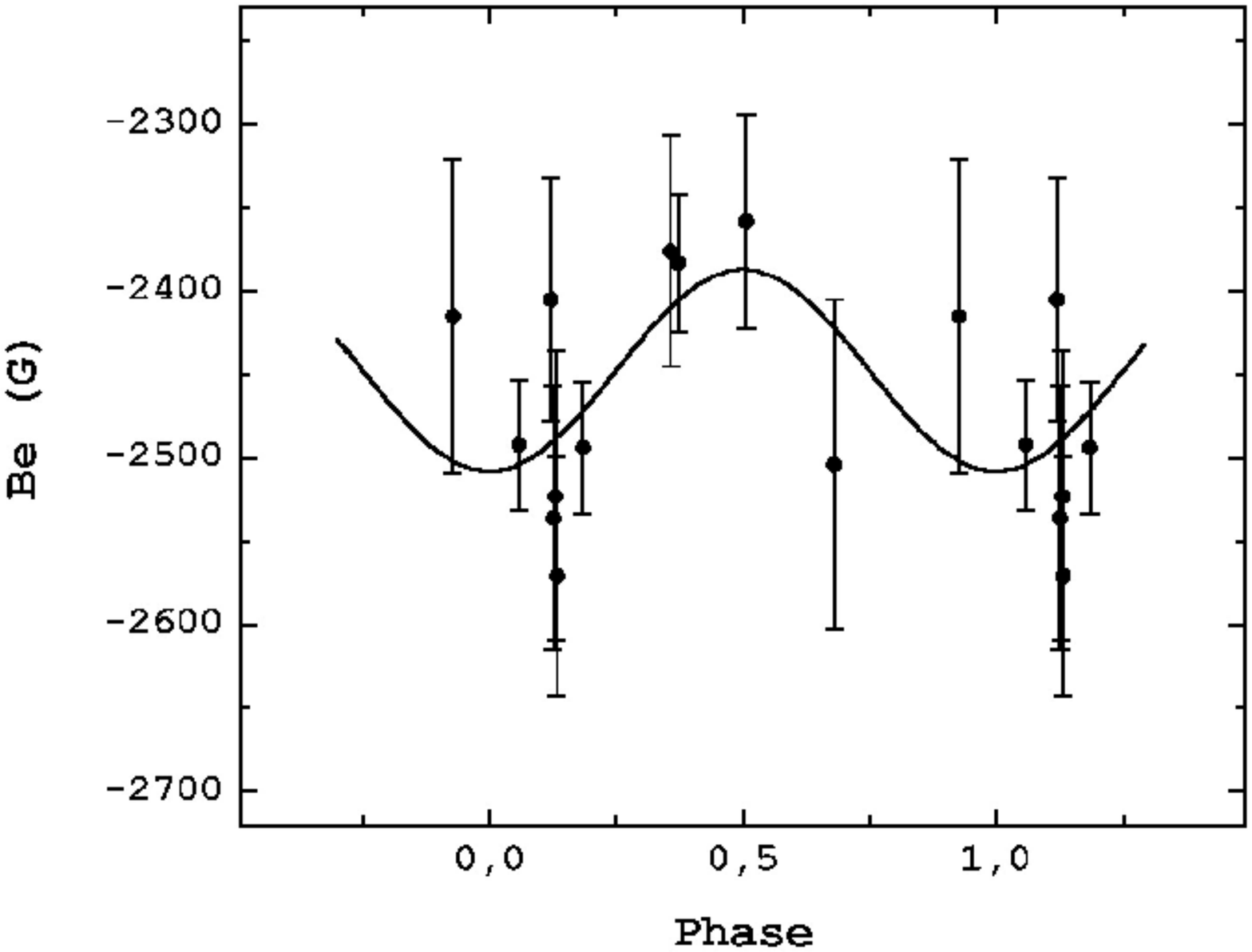}}
\vspace{-3.5mm}
\caption{ HD 36485 (2) }
\label{fig:fig83}
\end{figure}

\clearpage
\newpage

\begin{figure}
\resizebox{0.98\hsize}{!}{\includegraphics{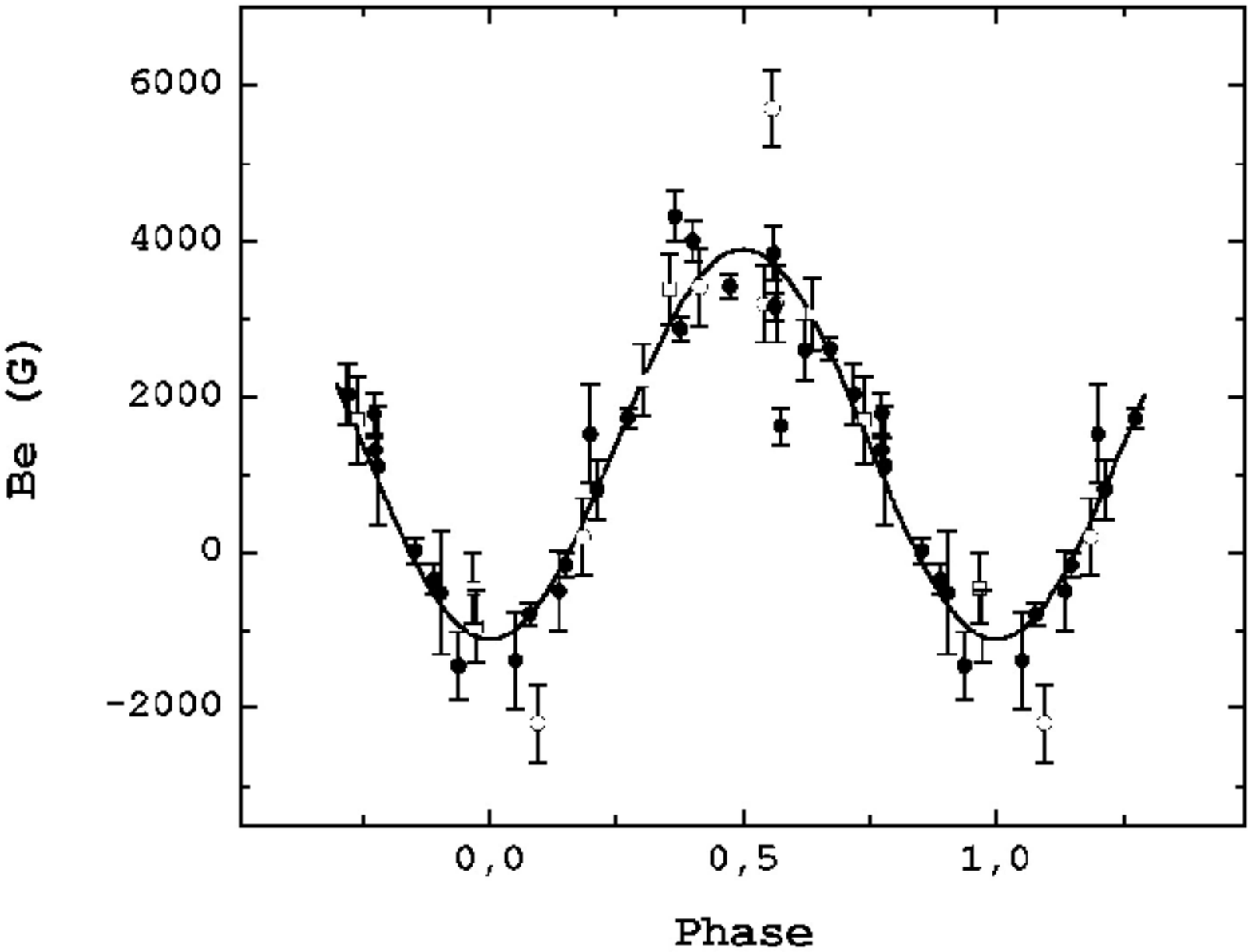}}
\vspace{-3.5mm}
\caption{ HD 36526 (1) }
\label{fig:fig83}
\end{figure}

\begin{figure}
\resizebox{0.98\hsize}{!}{\includegraphics{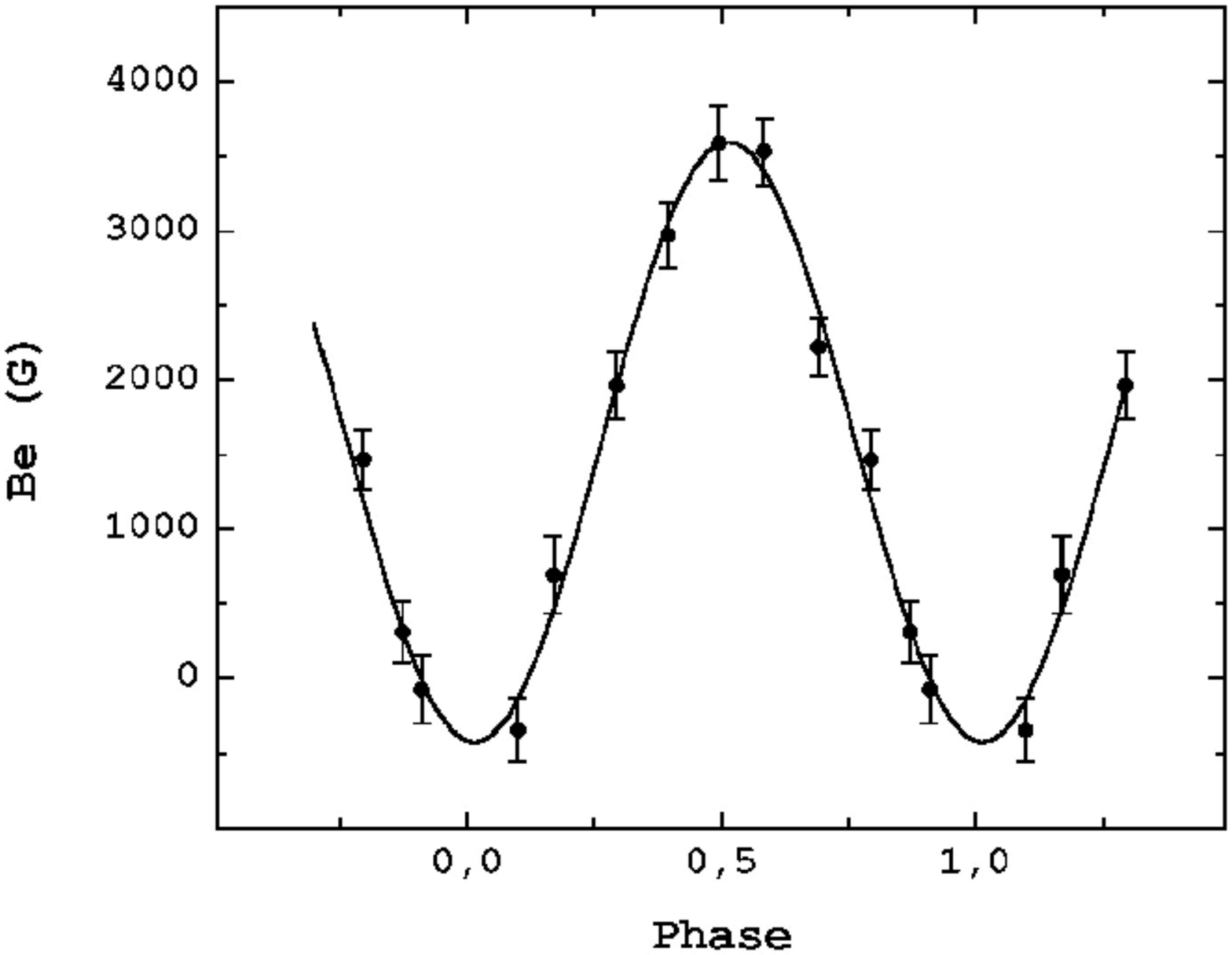}}
\vspace{-3.5mm}
\caption{ HD 36526 (2) }
\label{fig:fig84}
\end{figure}

\begin{figure}
\resizebox{0.98\hsize}{!}{\includegraphics{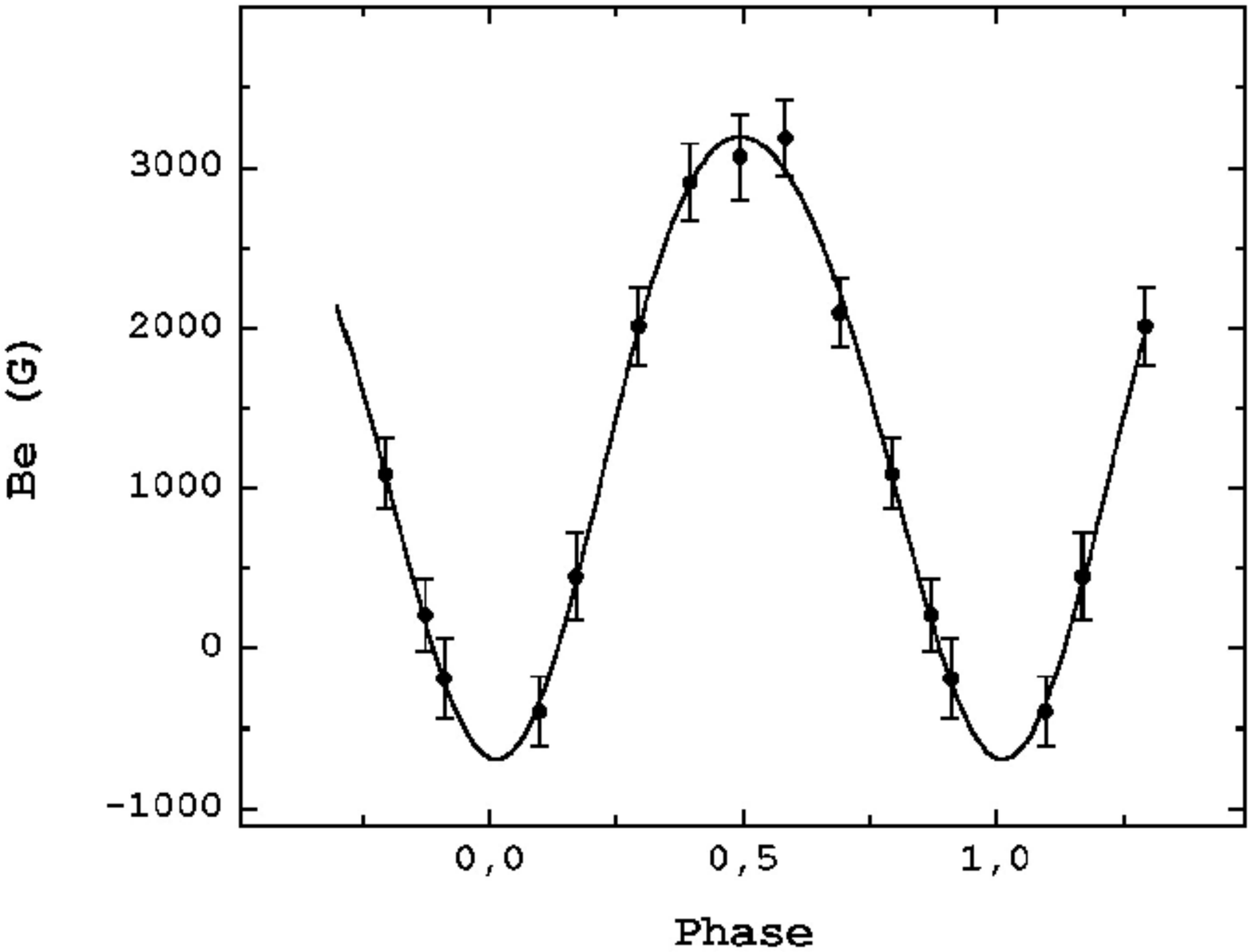}}
\vspace{-3.5mm}
\caption{ HD 36526 (3) }
\label{fig:fig83}
\end{figure}

\begin{figure}
\resizebox{0.98\hsize}{!}{\includegraphics{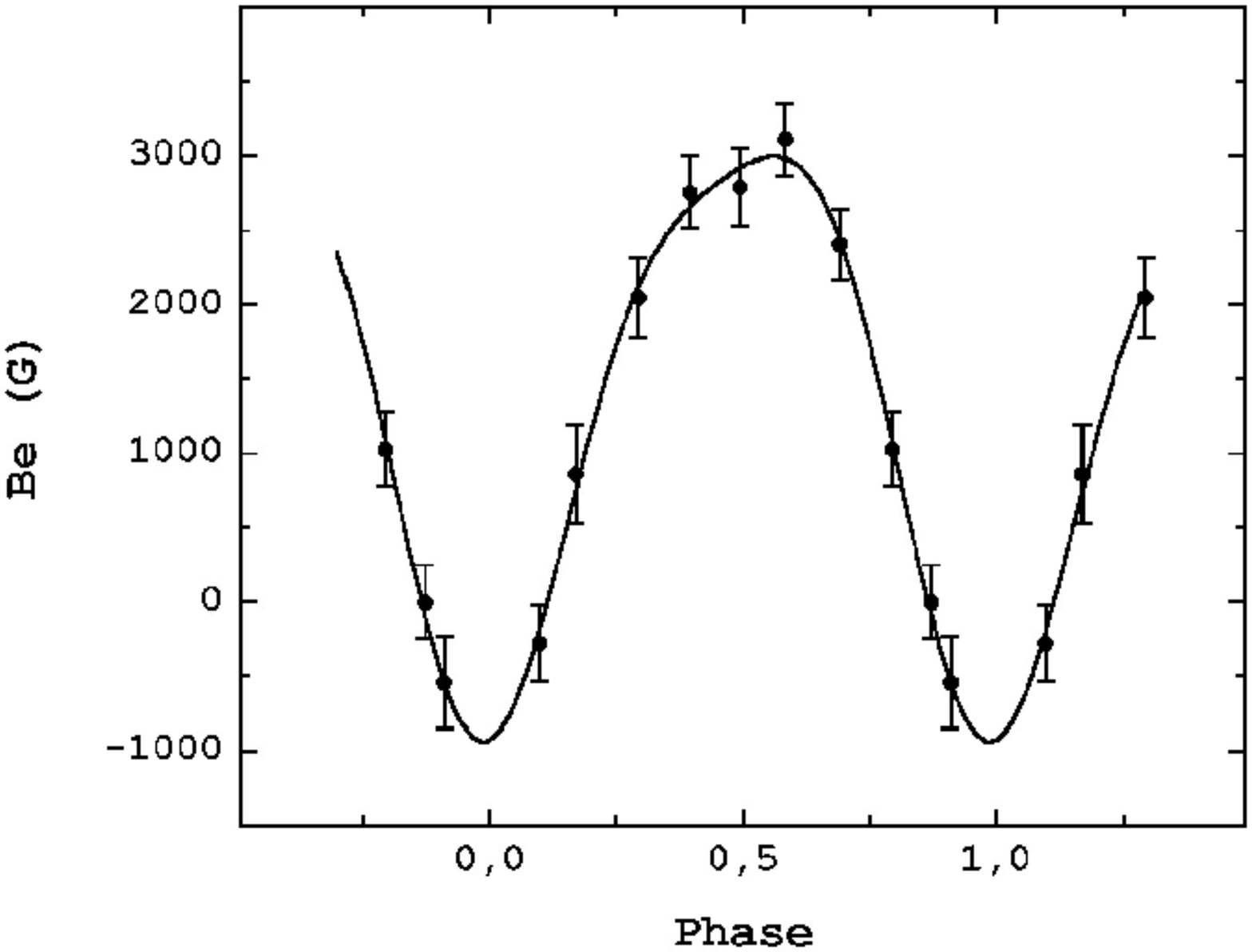}}
\vspace{-3.5mm}
\caption{ HD 36526 (4) }
\label{fig:fig83}
\end{figure}

\begin{figure}
\resizebox{0.98\hsize}{!}{\includegraphics{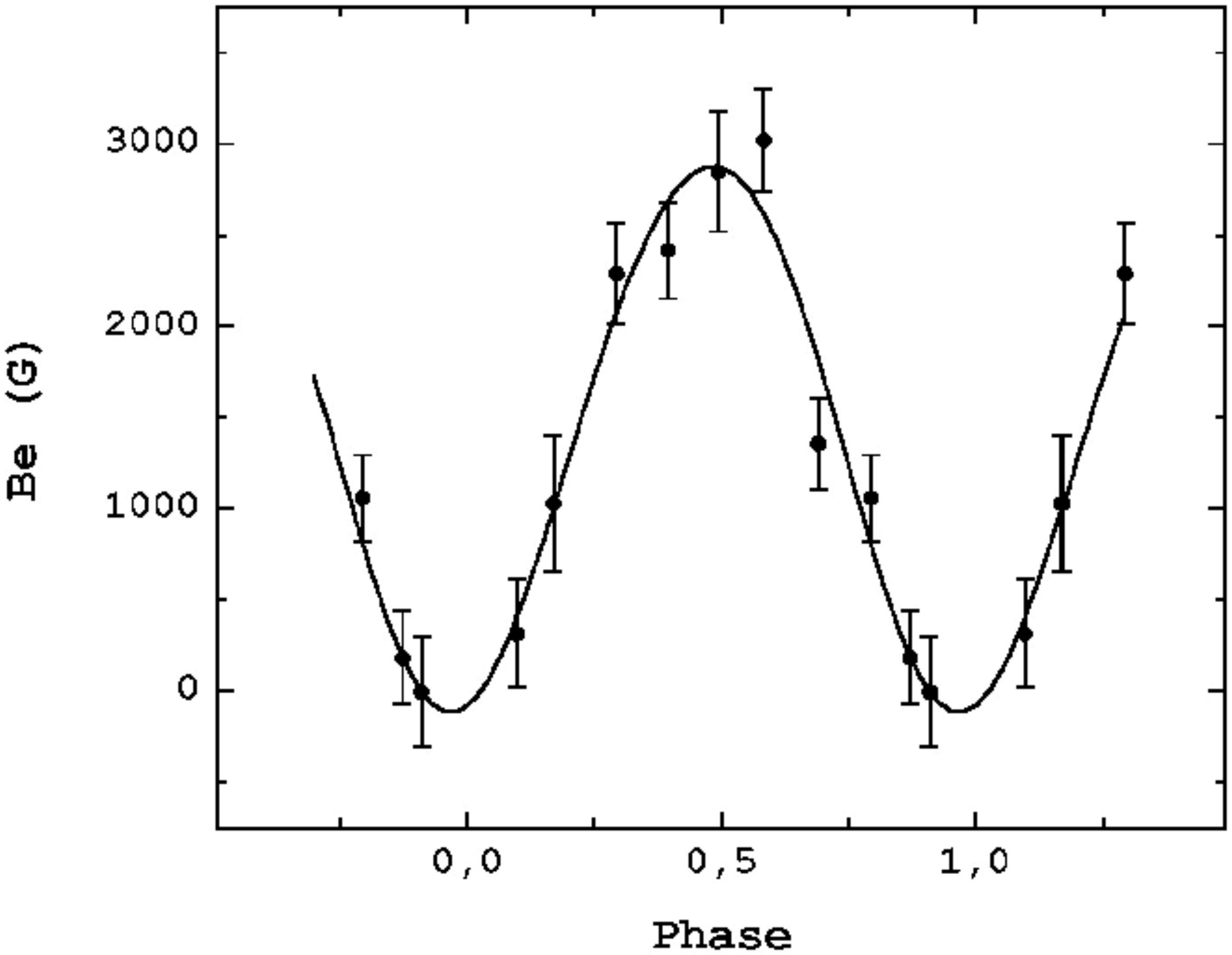}}
\vspace{-3.5mm}
\caption{ HD 36526 (5) }
\label{fig:fig83}
\end{figure}

\begin{figure}
\resizebox{0.98\hsize}{!}{\includegraphics{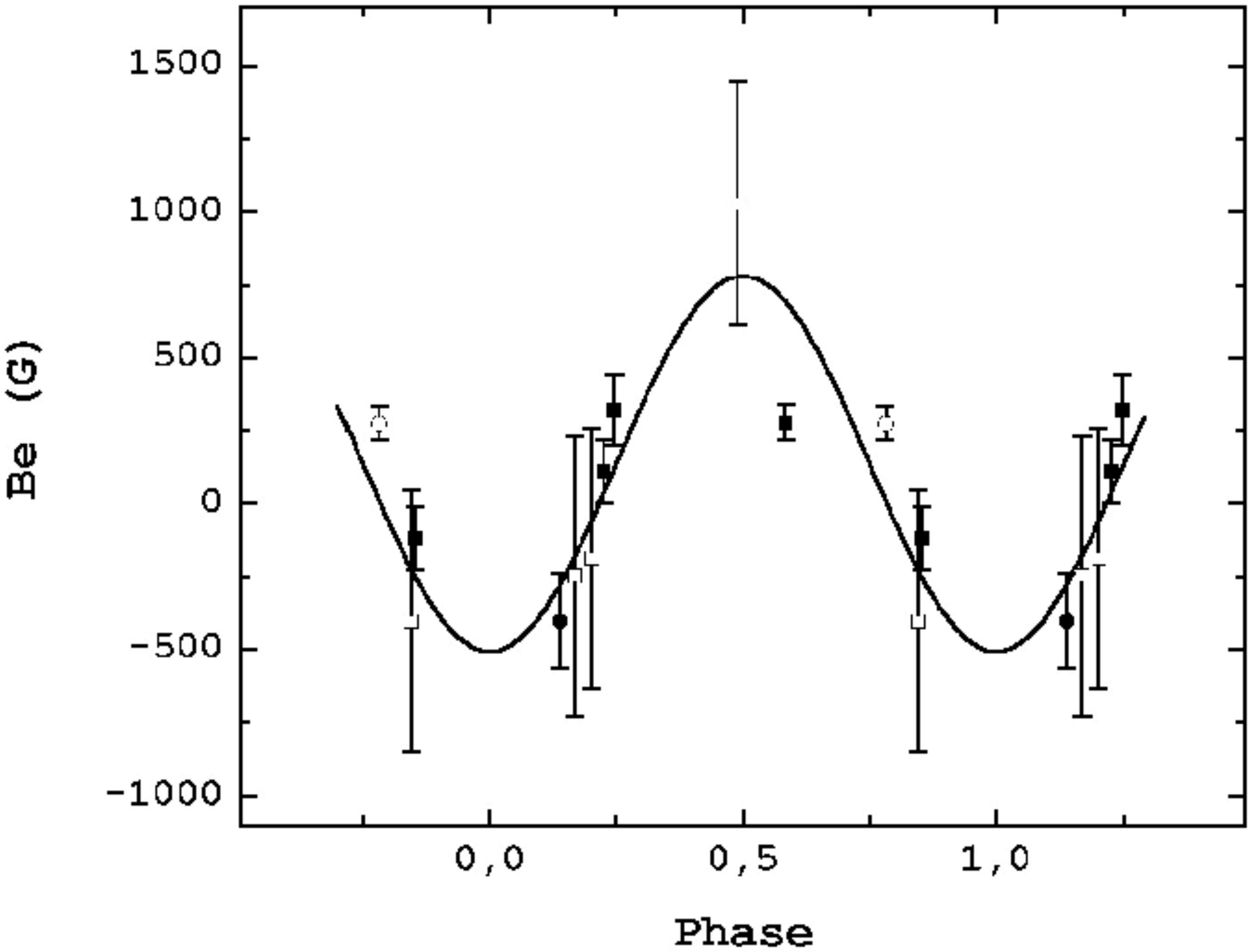}}
\vspace{-3.5mm}
\caption{ HD 36540 }
\label{fig:fig85}
\end{figure}

\clearpage
\newpage

\begin{figure}
\resizebox{0.98\hsize}{!}{\includegraphics{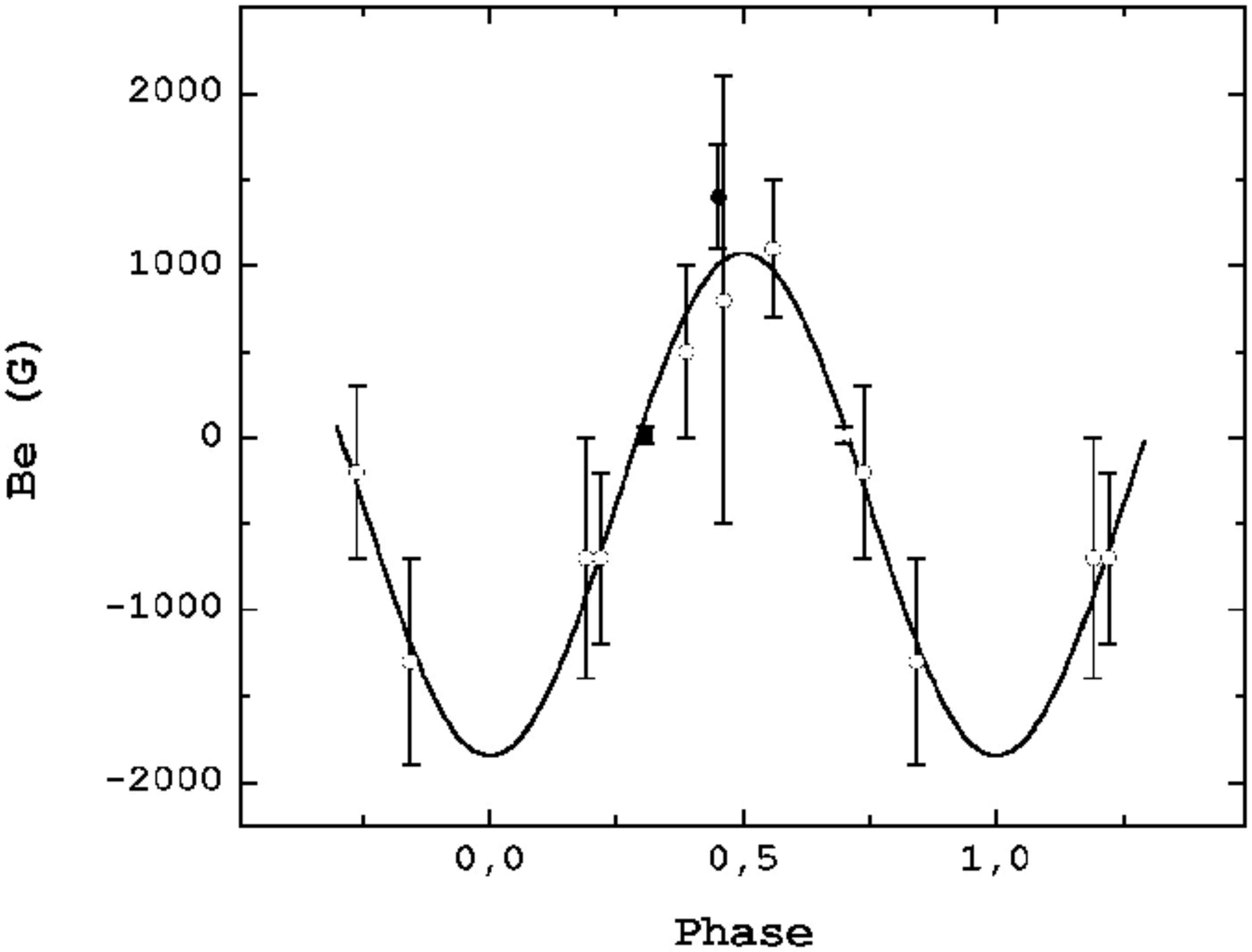}}
\vspace{-3.5mm}
\caption{ HD 36629 }
\label{fig:fig86}
\end{figure}

\begin{figure}
\resizebox{0.98\hsize}{!}{\includegraphics{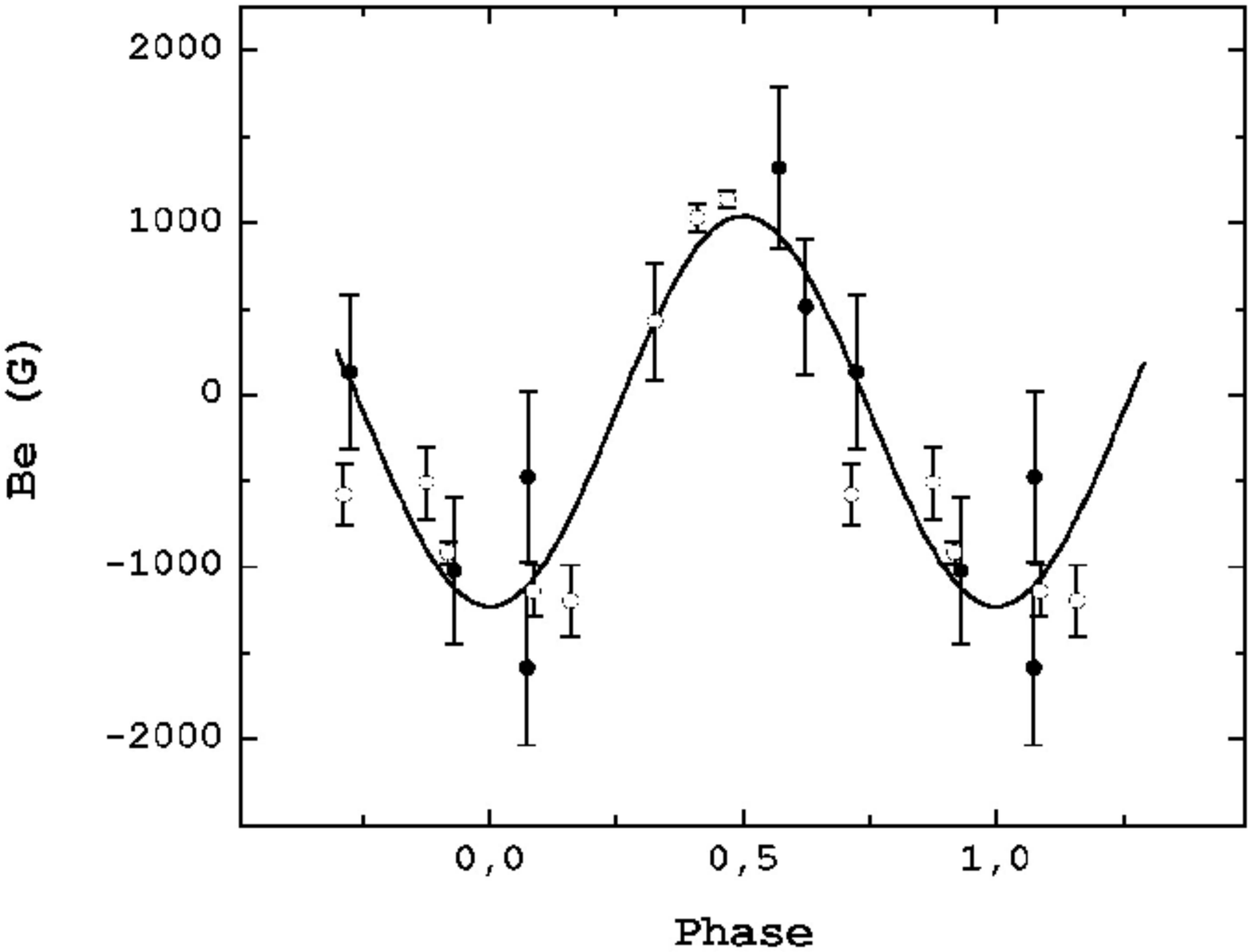}}
\vspace{-3.5mm}
\caption{ HD 36668 }
\label{fig:fig87}
\end{figure}

\begin{figure}
\resizebox{0.98\hsize}{!}{\includegraphics{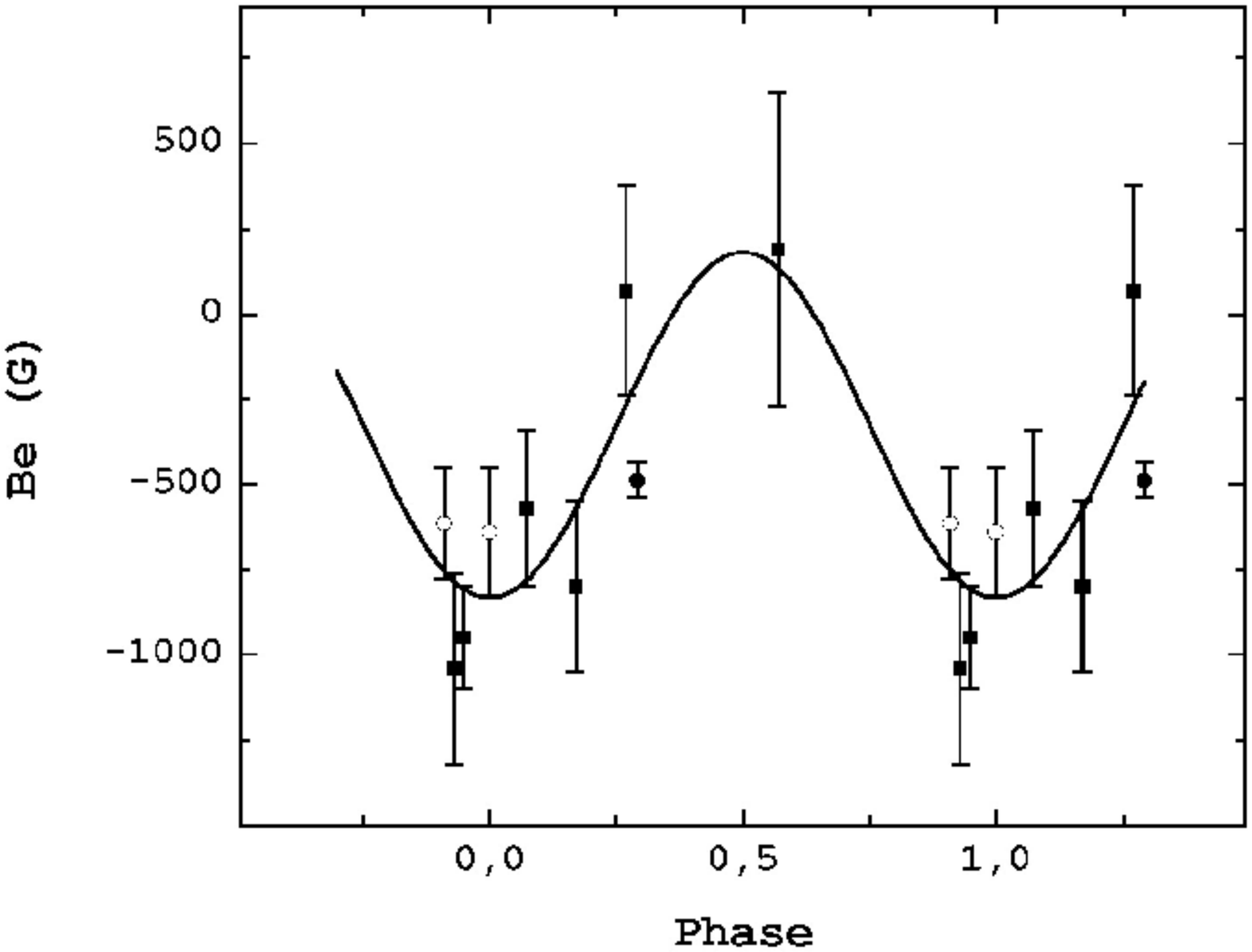}}
\vspace{-3.5mm}
\caption{ HD 36916 }
\label{fig:fig88}
\end{figure}

\begin{figure}
\resizebox{0.98\hsize}{!}{\includegraphics{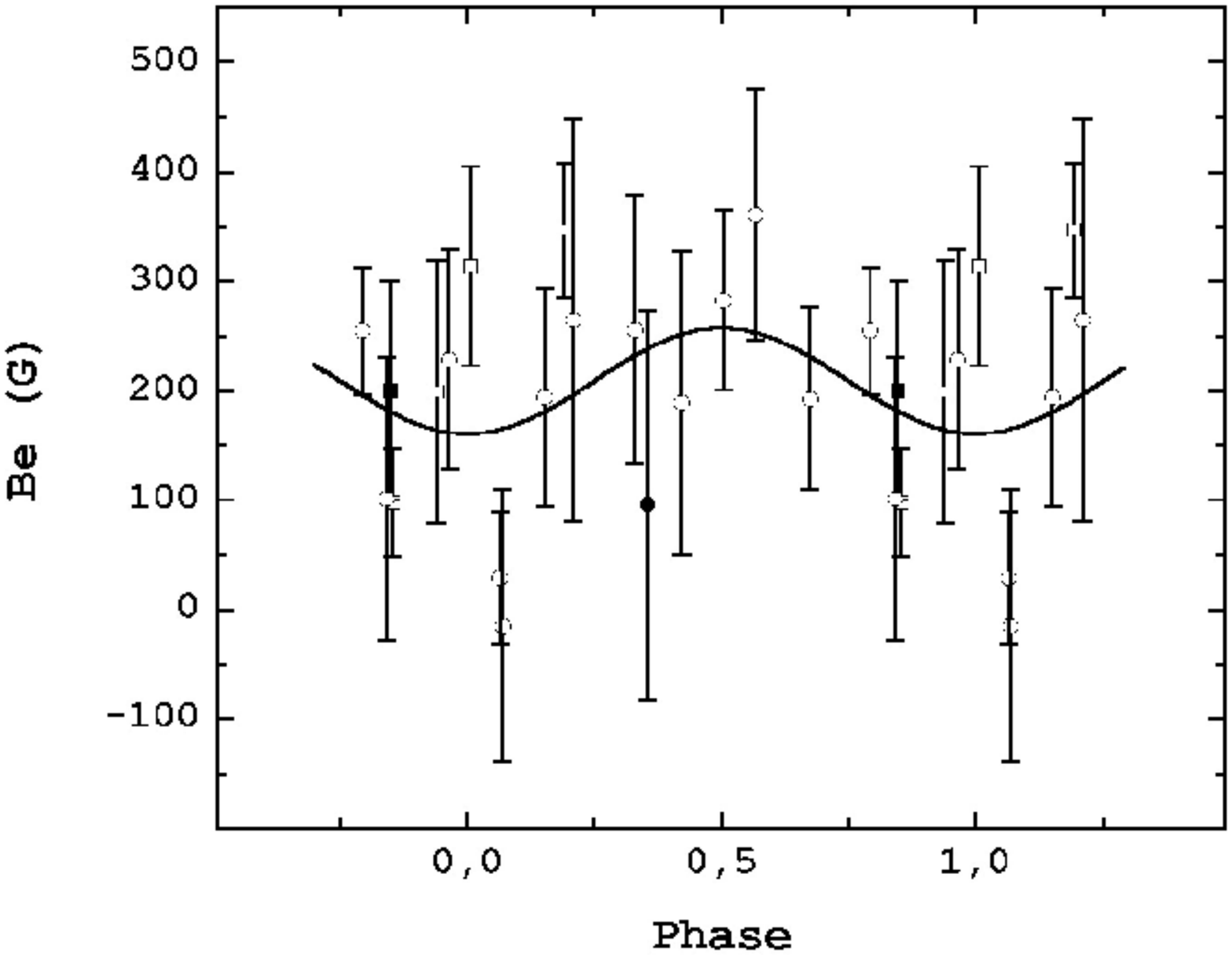}}
\vspace{-3.5mm}
\caption{ HD 36982 (1) }
\label{fig:fig83}
\end{figure}

\begin{figure}
\resizebox{0.98\hsize}{!}{\includegraphics{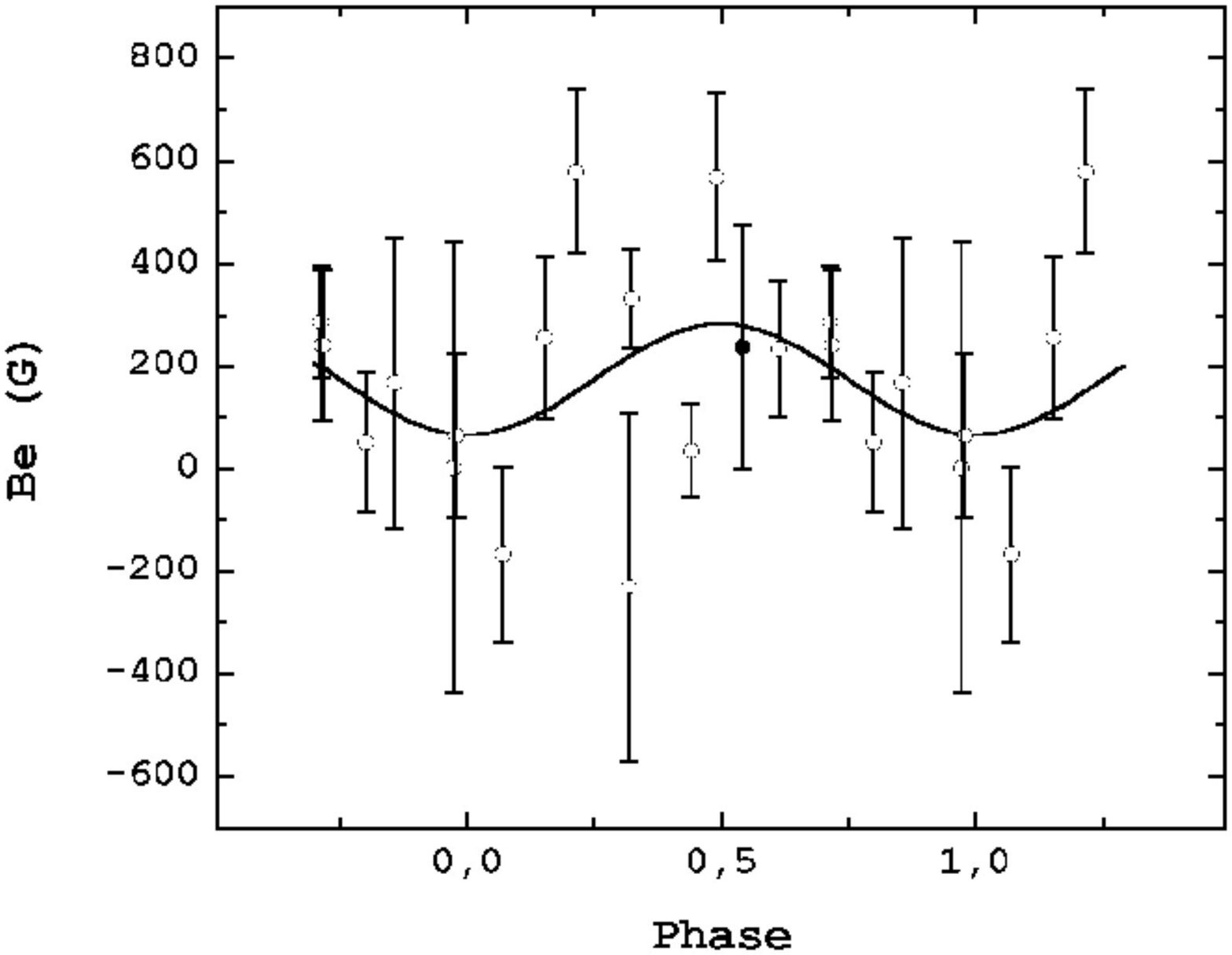}}
\vspace{-3.5mm}
\caption{ HD 36982 (2) }
\label{fig:fig83}
\end{figure}

\begin{figure}
\resizebox{0.98\hsize}{!}{\includegraphics{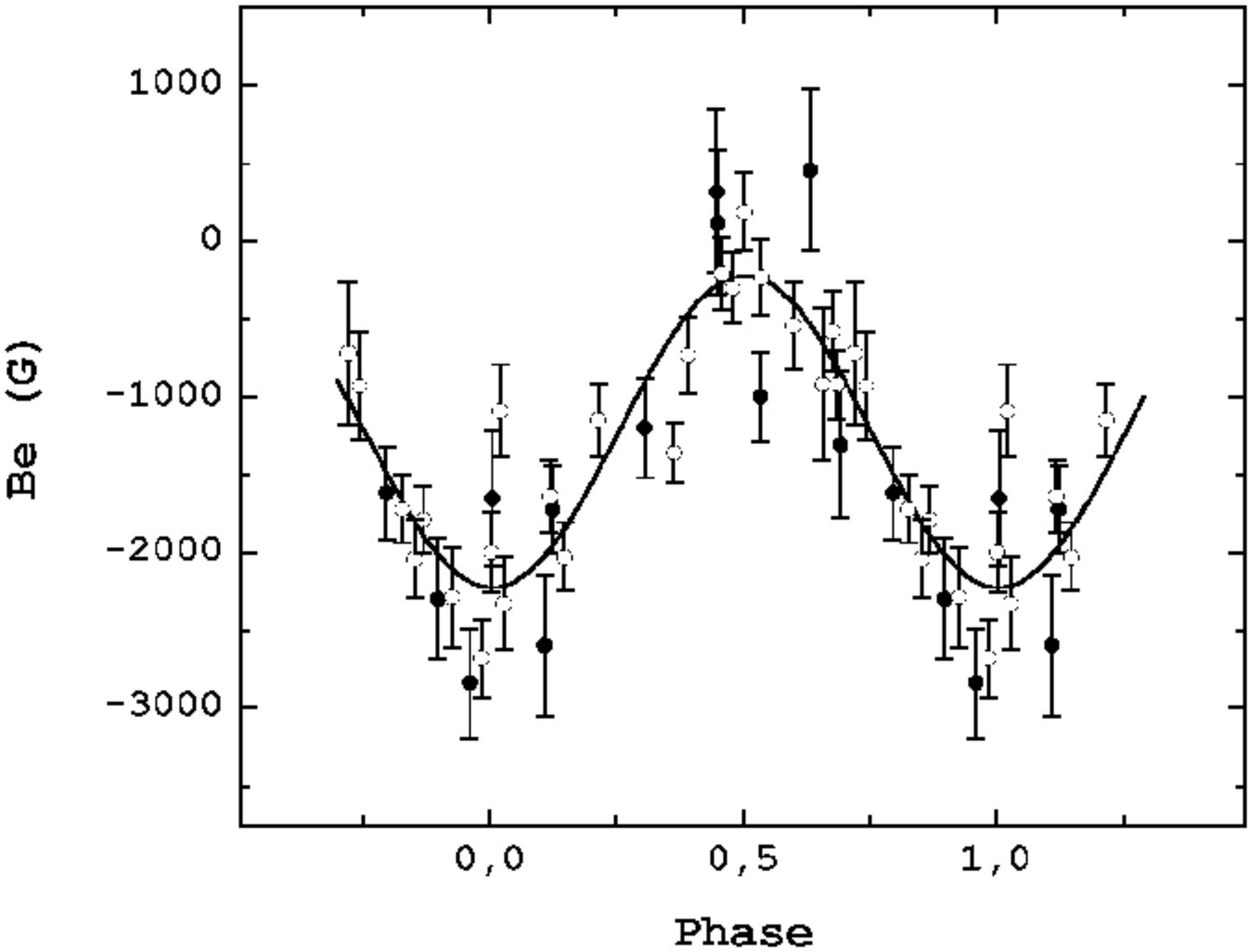}}
\vspace{-3.5mm}
\caption{ HD 37017 (1) }
\label{fig:fig89}
\end{figure}

\clearpage
\newpage

\begin{figure}
\resizebox{0.98\hsize}{!}{\includegraphics{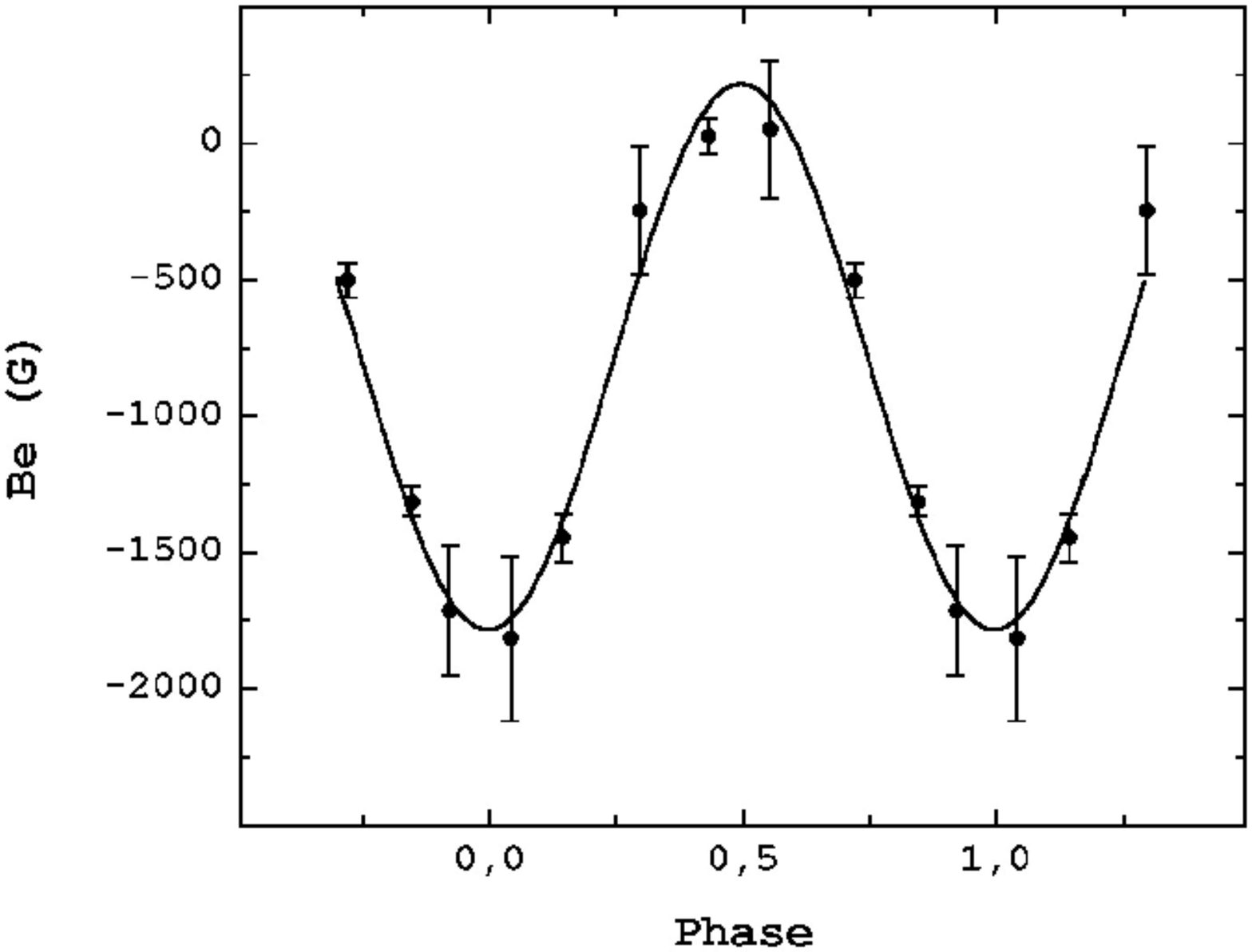}}
\vspace{-3.5mm}
\caption{ HD 37017 (2) }
\label{fig:fig89}
\end{figure}

\begin{figure}
\resizebox{0.98\hsize}{!}{\includegraphics{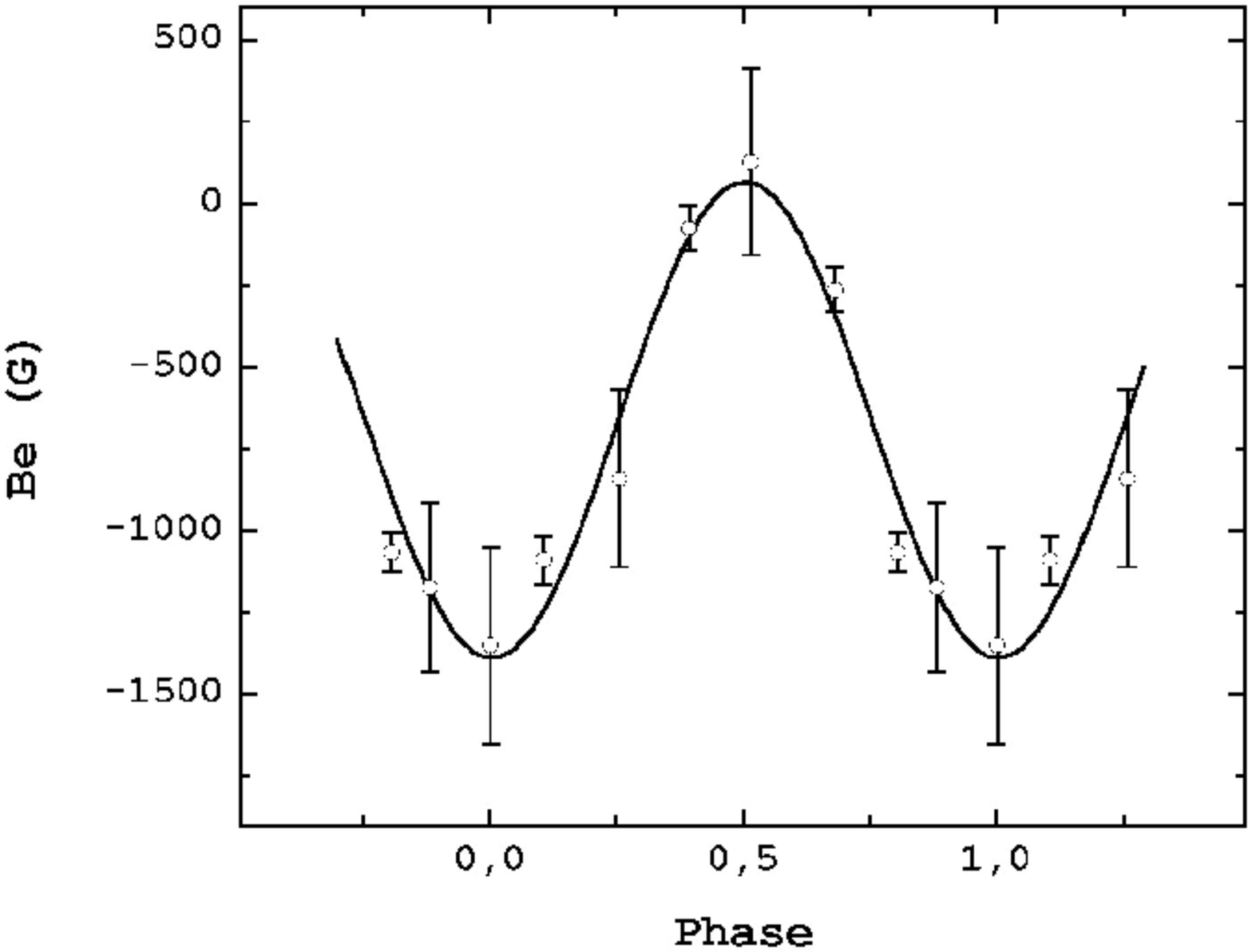}}
\vspace{-3.5mm}
\caption{ HD 37017 (3) }
\label{fig:fig89}
\end{figure}

\begin{figure}
\resizebox{0.98\hsize}{!}{\includegraphics{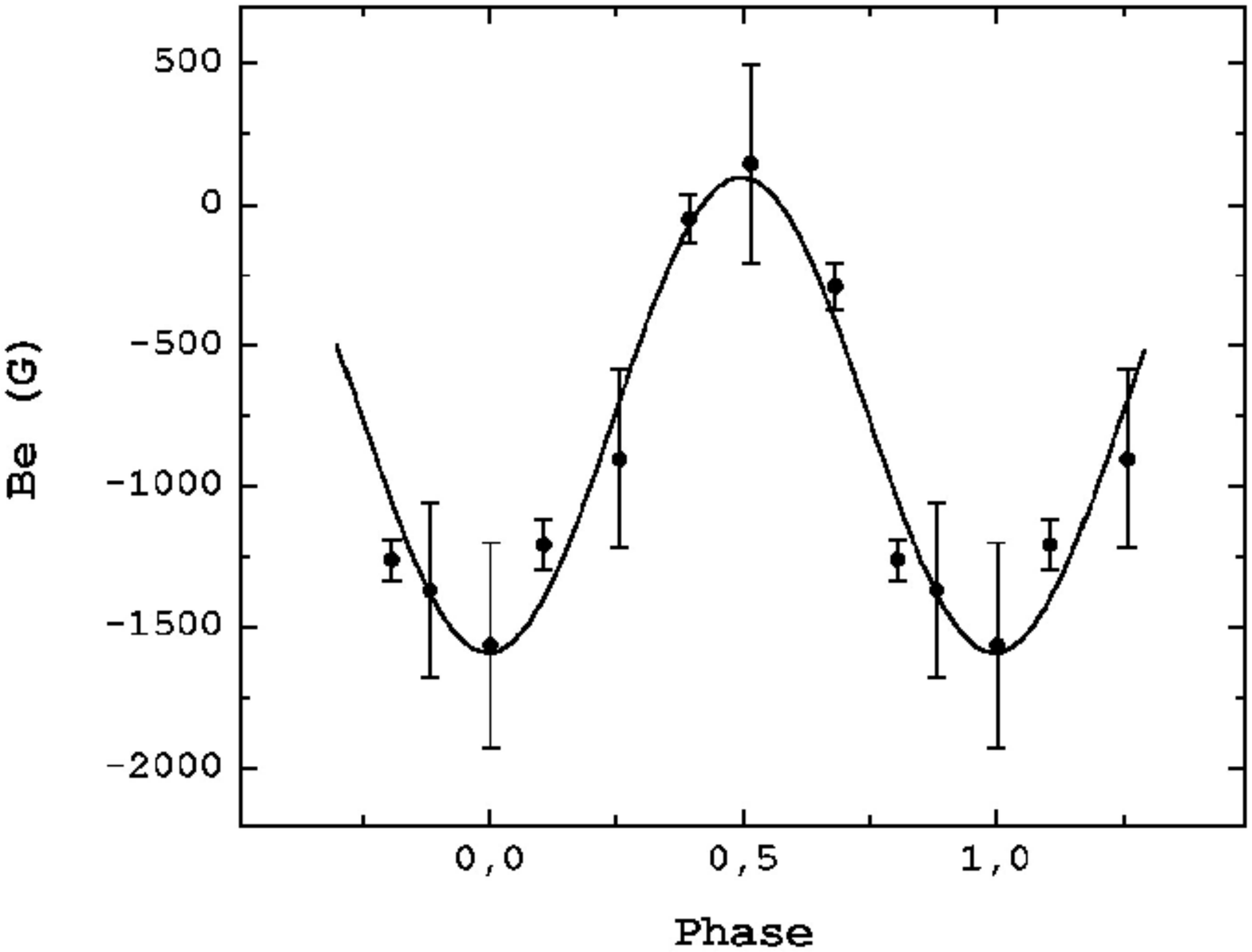}}
\vspace{-3.5mm}
\caption{ HD 37017 (4) }
\label{fig:fig89}
\end{figure}

\begin{figure}
\resizebox{0.98\hsize}{!}{\includegraphics{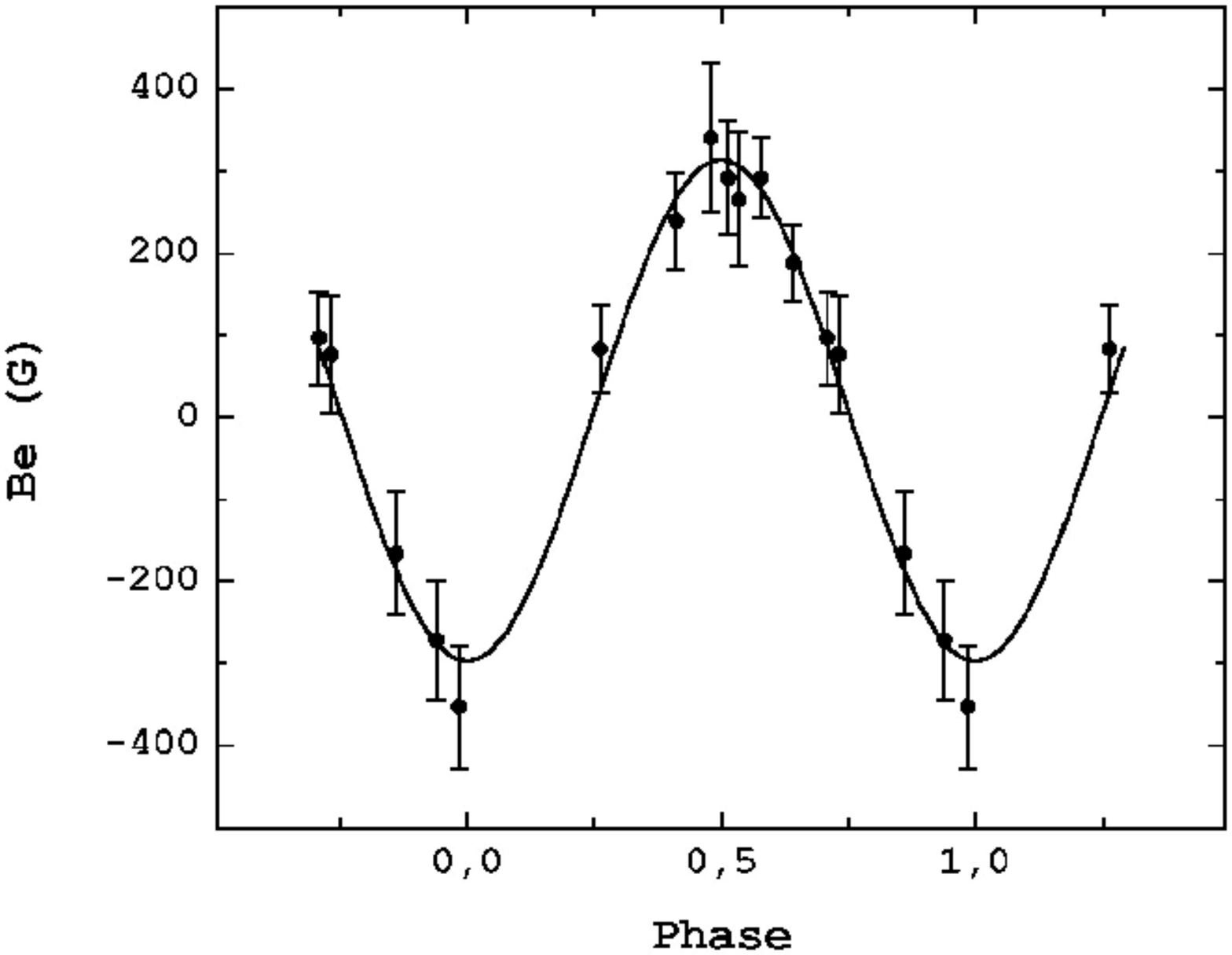}}
\vspace{-3.5mm}
\caption{ HD 37022 (1) }
\label{fig:fig90}
\end{figure}

\begin{figure}
\resizebox{0.98\hsize}{!}{\includegraphics{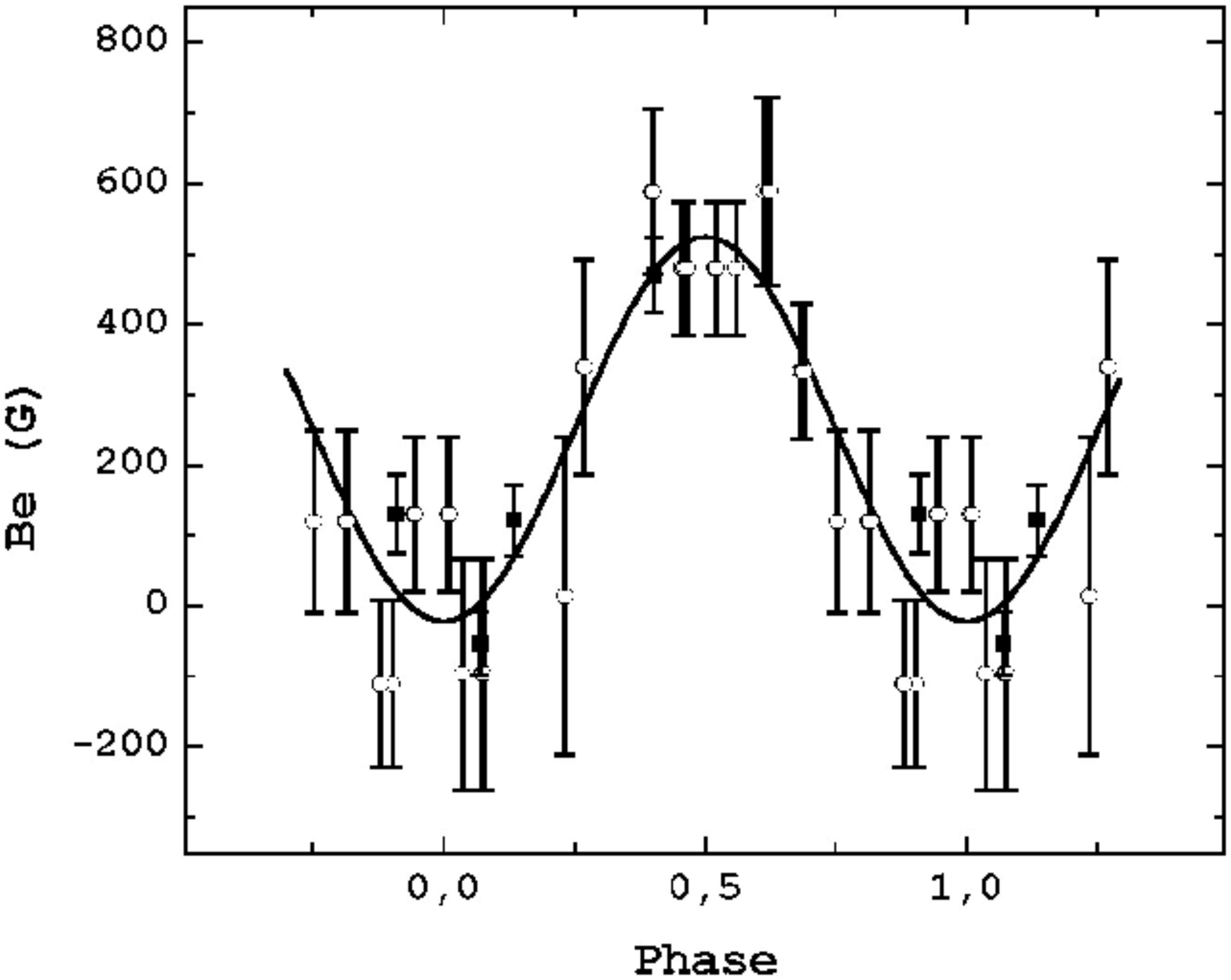}}
\vspace{-3.5mm}
\caption{ HD 37022 (2) }
\label{fig:fig91}
\end{figure}

\begin{figure}
\resizebox{0.98\hsize}{!}{\includegraphics{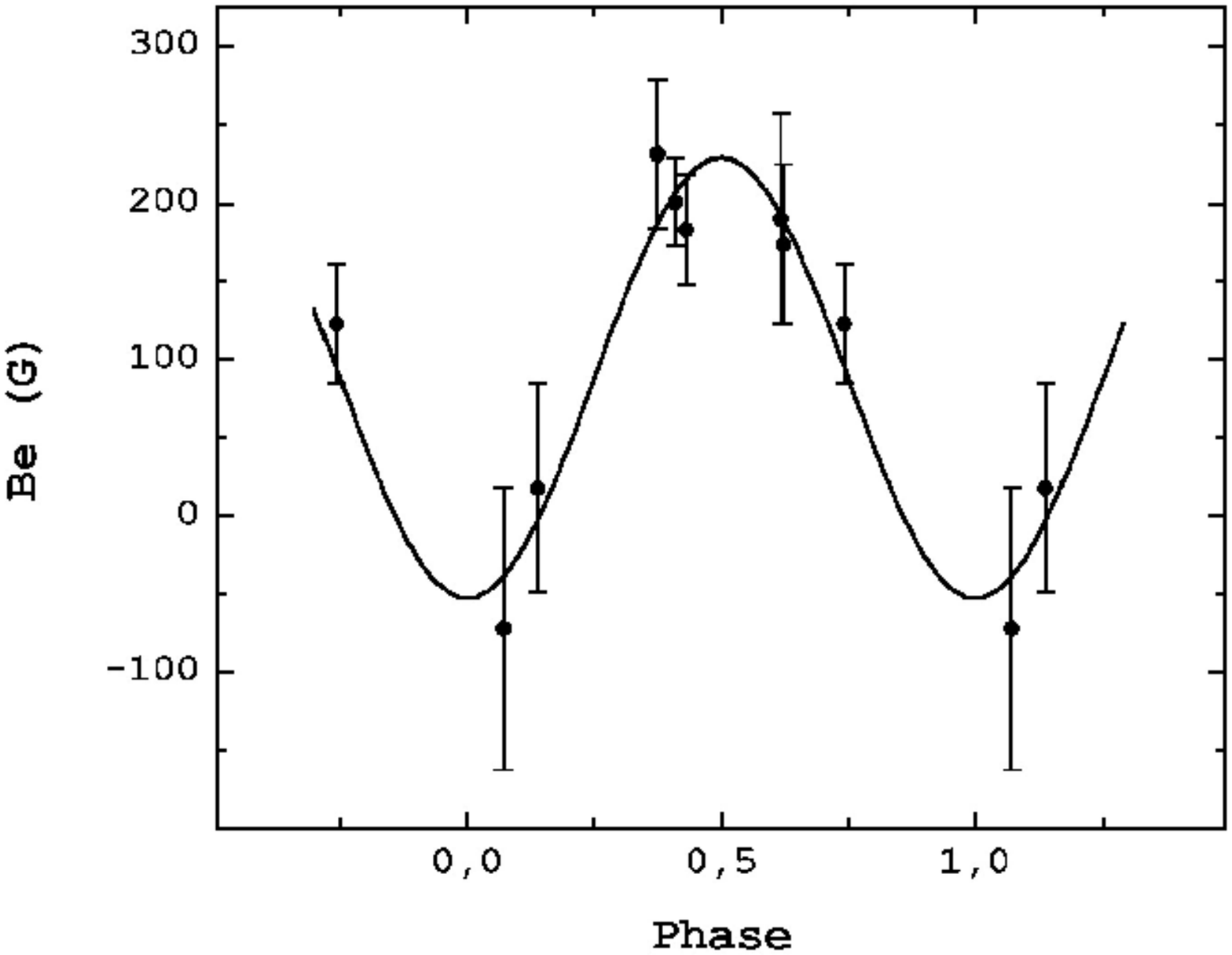}}
\vspace{-3.5mm}
\caption{ HD 37022 (3) }
\label{fig:fig92}
\end{figure}

\clearpage
\newpage

\begin{figure}
\resizebox{0.98\hsize}{!}{\includegraphics{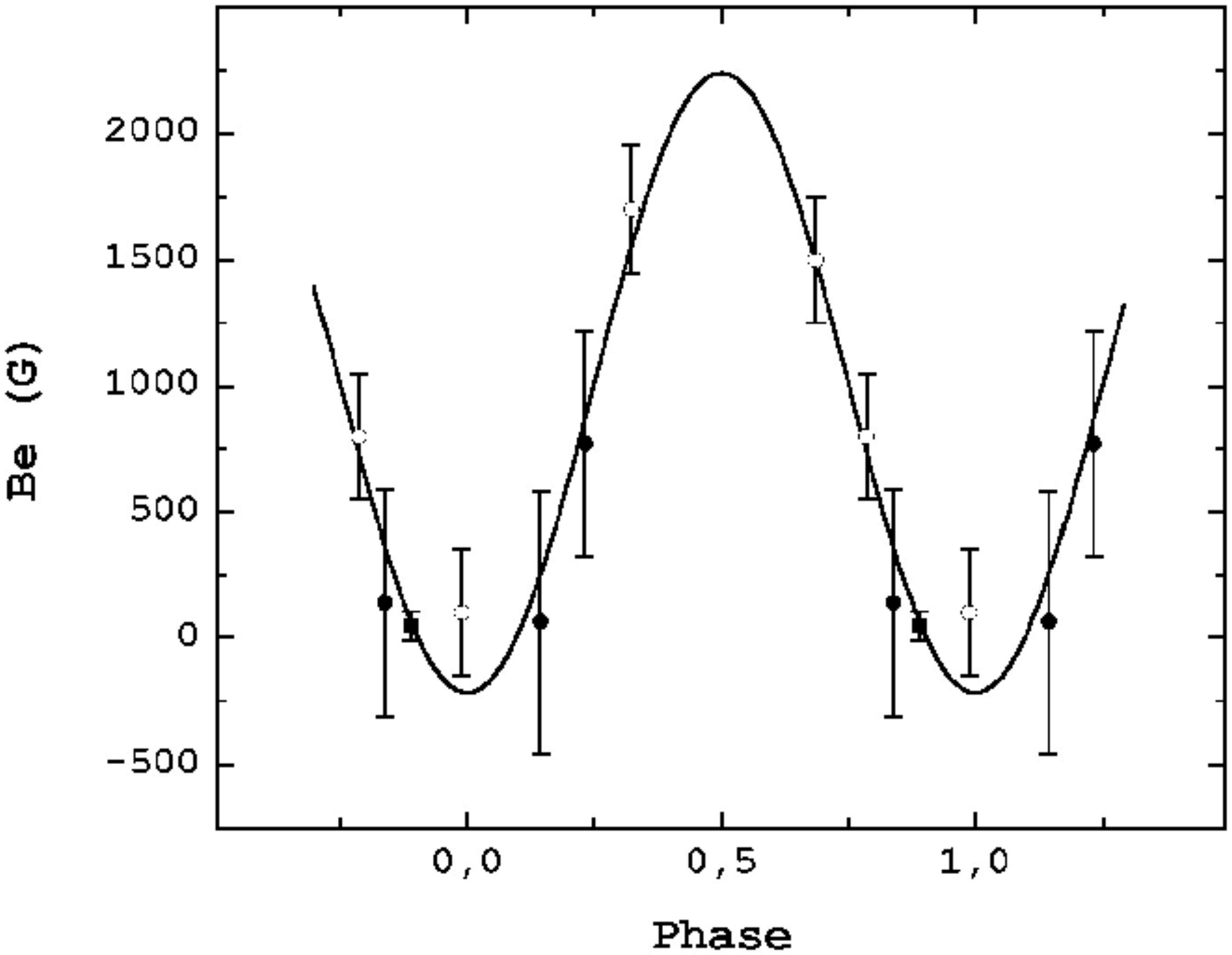}}
\vspace{-3.5mm}
\caption{ HD 37041 }
\label{fig:fig93}
\end{figure}

\begin{figure}
\resizebox{0.98\hsize}{!}{\includegraphics{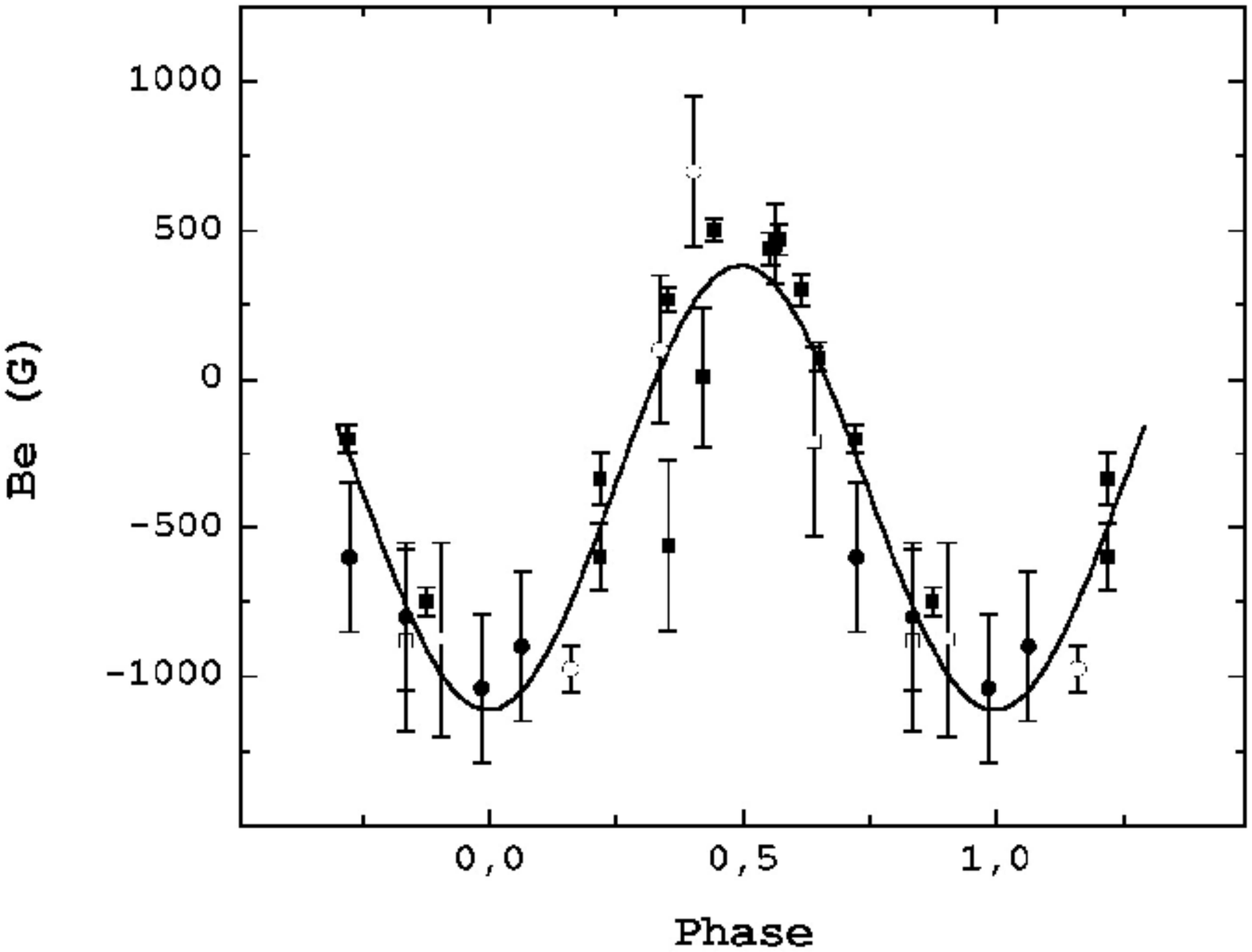}}
\vspace{-3.5mm}
\caption{ HD 37058 (1) }
\label{fig:fig94}
\end{figure}

\begin{figure}
\resizebox{0.98\hsize}{!}{\includegraphics{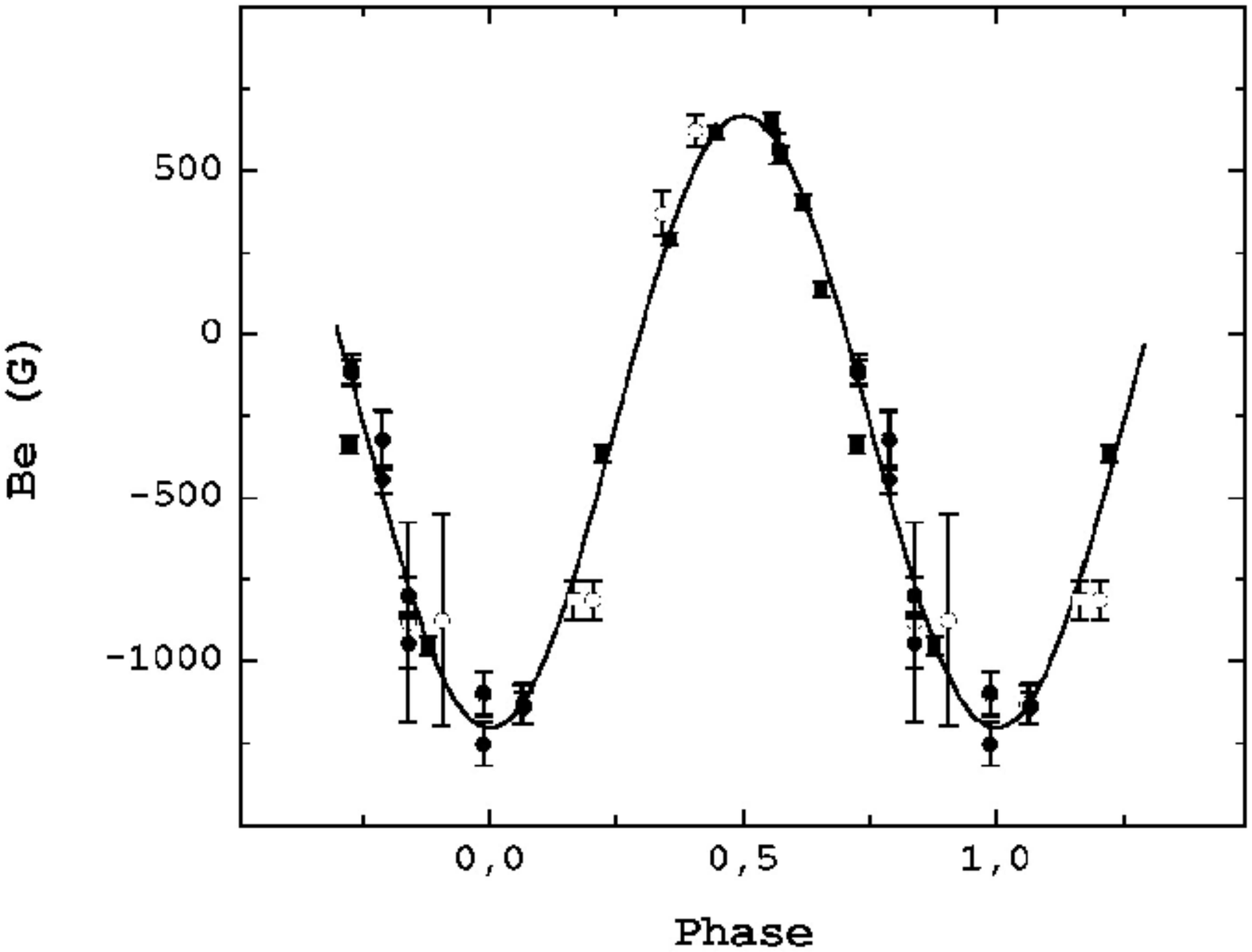}}
\vspace{-3.5mm}
\caption{ HD 37058 (2) }
\label{fig:fig94}
\end{figure}

\begin{figure}
\resizebox{0.98\hsize}{!}{\includegraphics{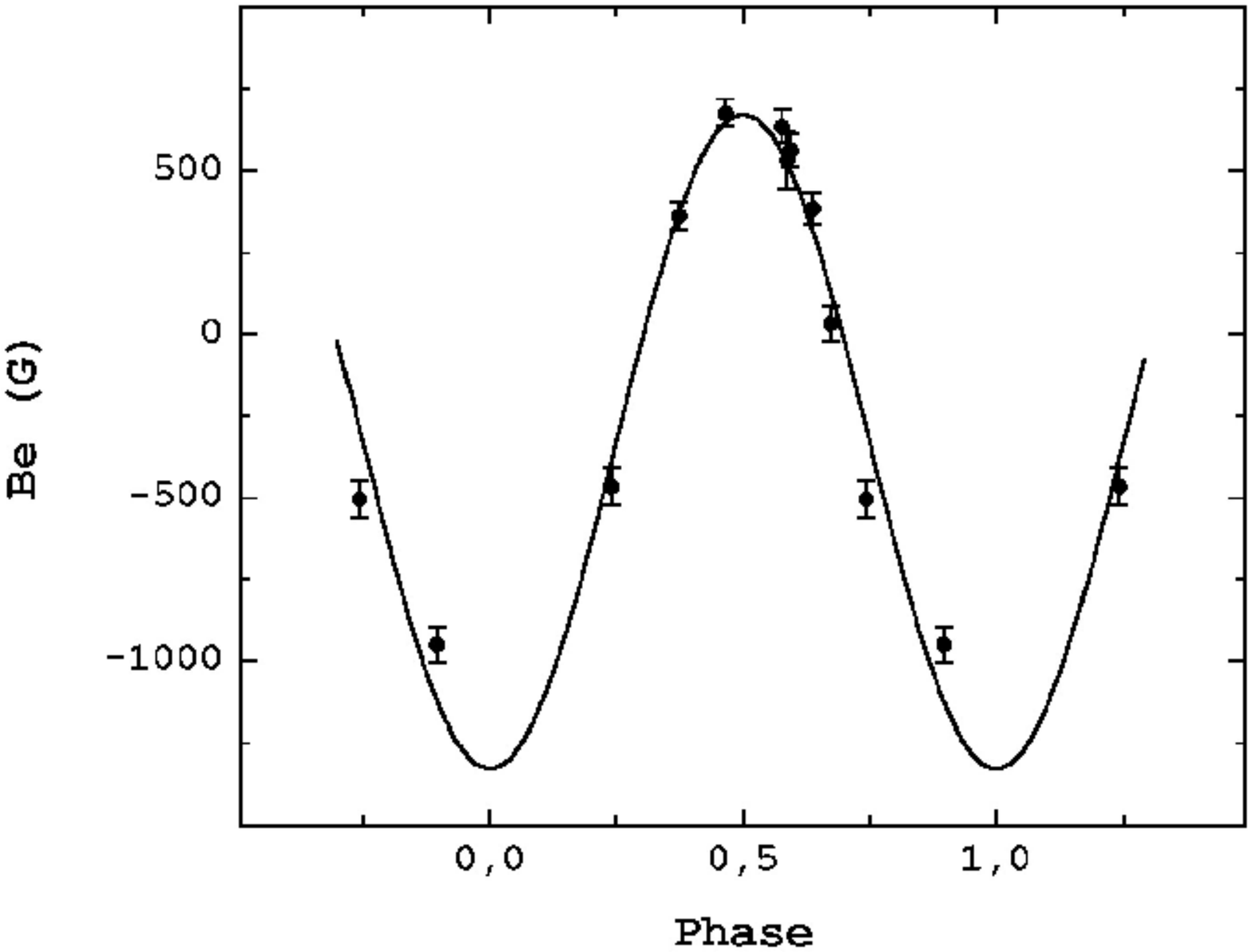}}
\vspace{-3.5mm}
\caption{ HD 37058 (3) }
\label{fig:fig94}
\end{figure}

\begin{figure}
\resizebox{0.98\hsize}{!}{\includegraphics{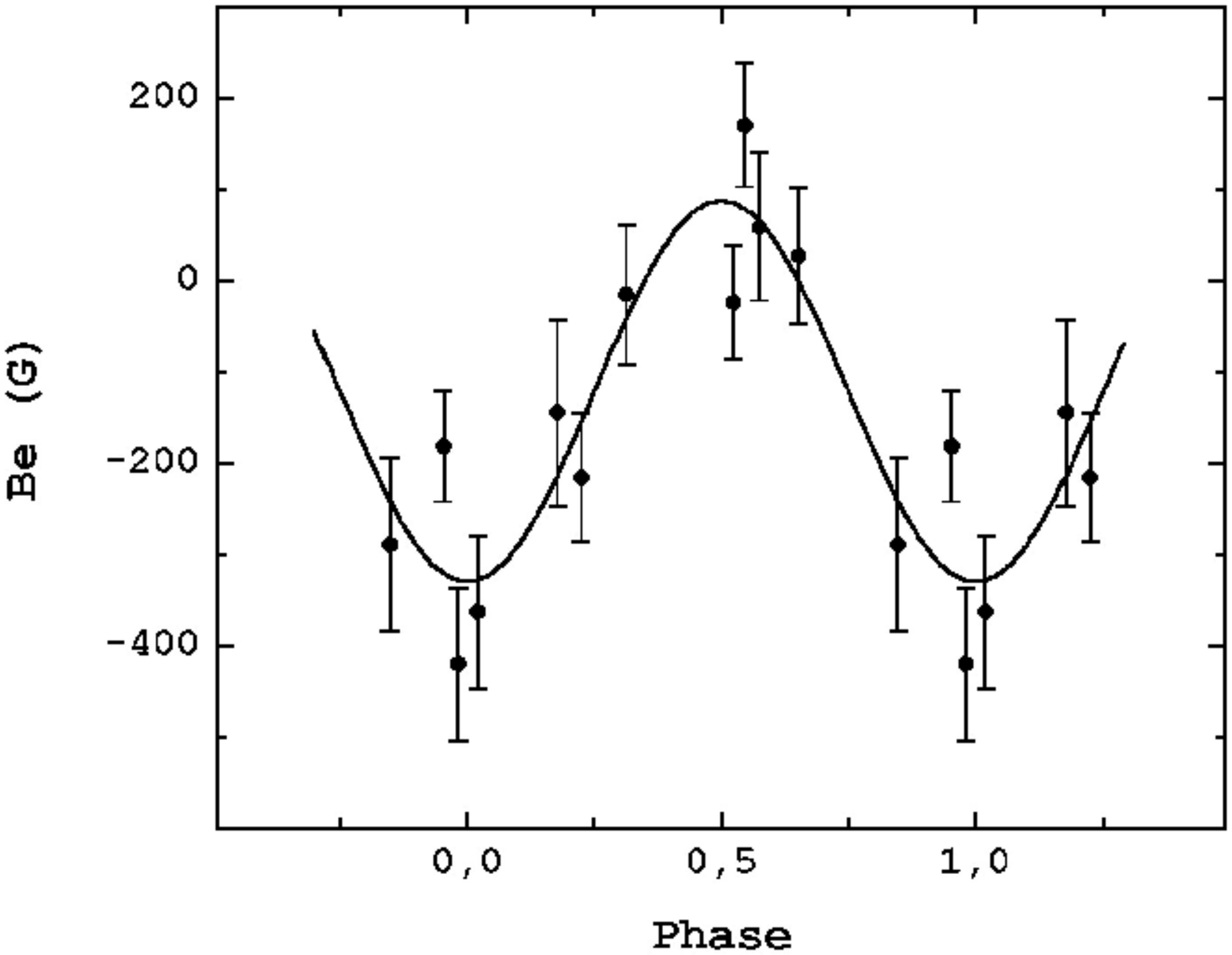}}
\vspace{-3.5mm}
\caption{ HD 37061 (1) }
\label{fig:fig94}
\end{figure}

\begin{figure}
\resizebox{0.98\hsize}{!}{\includegraphics{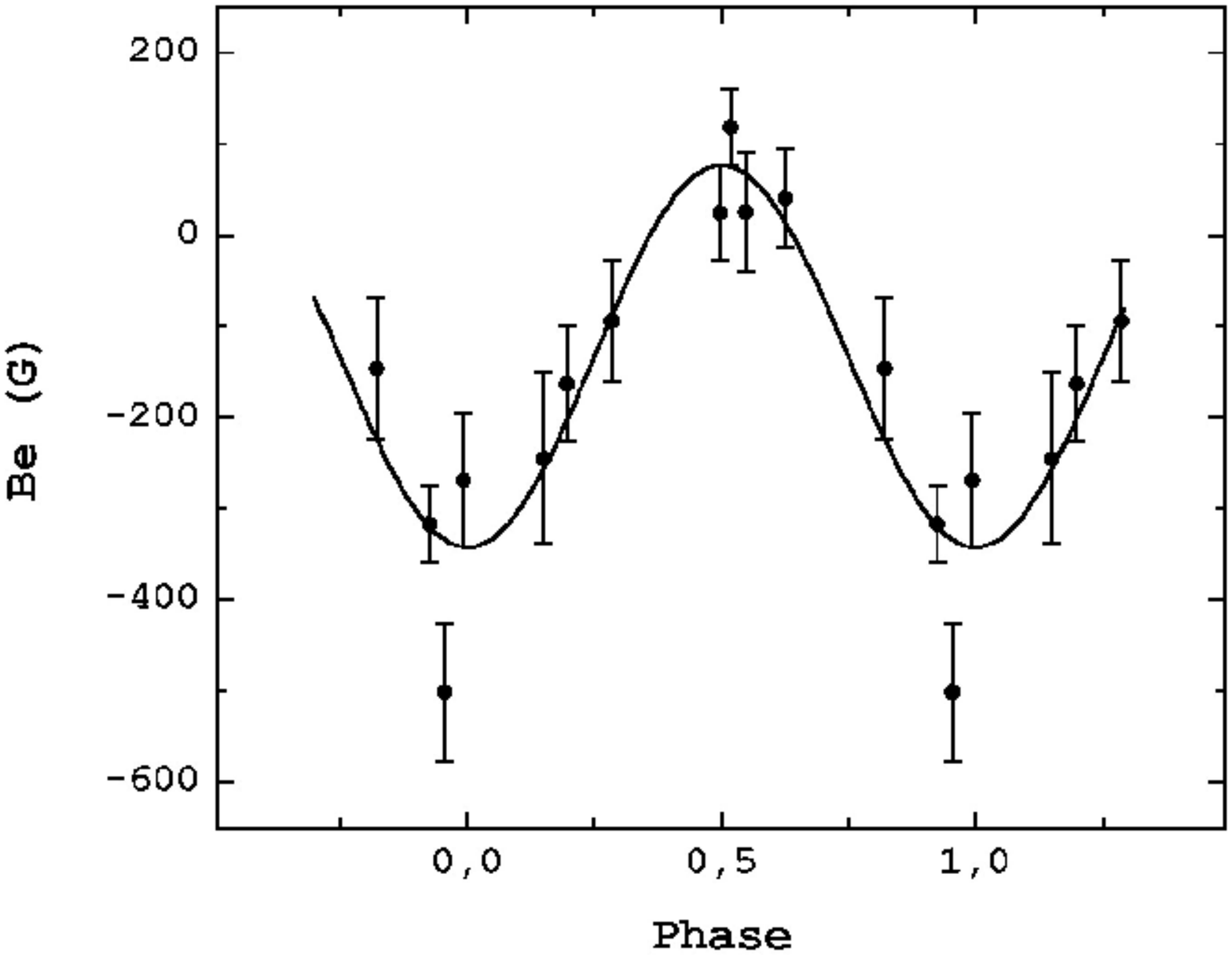}}
\vspace{-3.5mm}
\caption{ HD 37061 (2) }
\label{fig:fig94}
\end{figure}

\clearpage
\newpage

\begin{figure}
\resizebox{0.98\hsize}{!}{\includegraphics{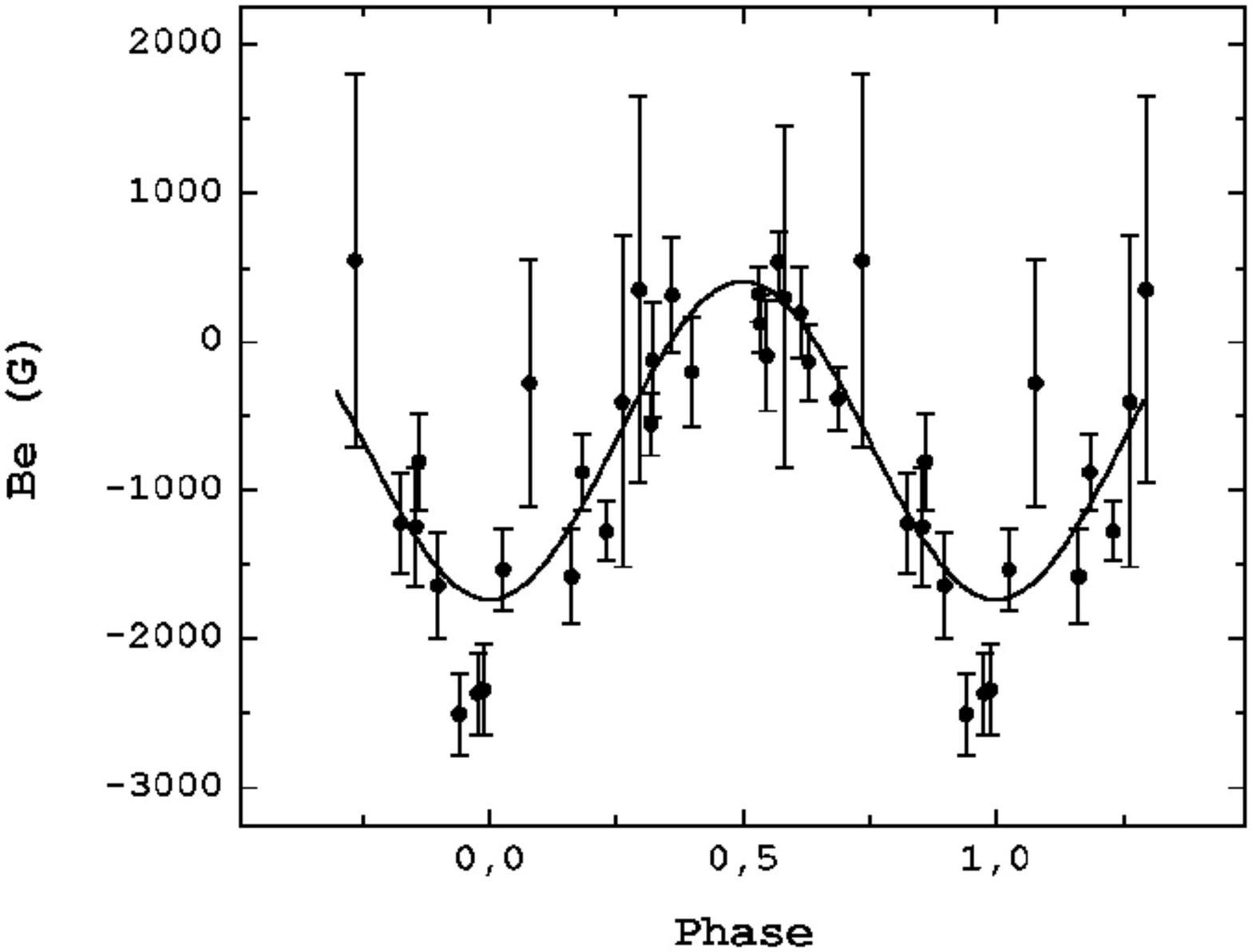}}
\vspace{-3.5mm}
\caption{ HD 37061 (3) }
\label{fig:fig94}
\end{figure}

\begin{figure}
\resizebox{0.98\hsize}{!}{\includegraphics{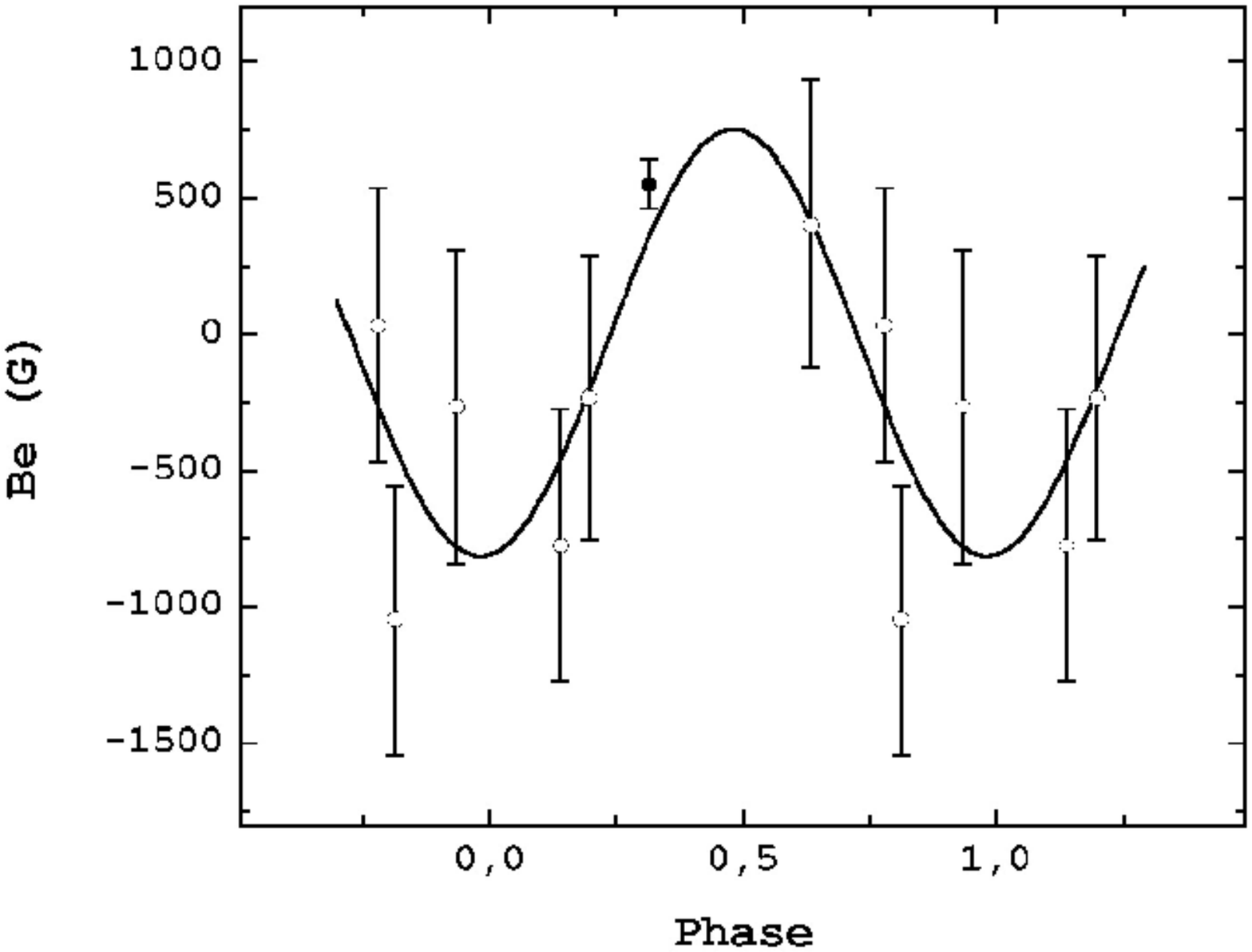}}
\vspace{-3.5mm}
\caption{ HD 37140 }
\label{fig:fig95}
\end{figure}

\begin{figure}
\resizebox{0.98\hsize}{!}{\includegraphics{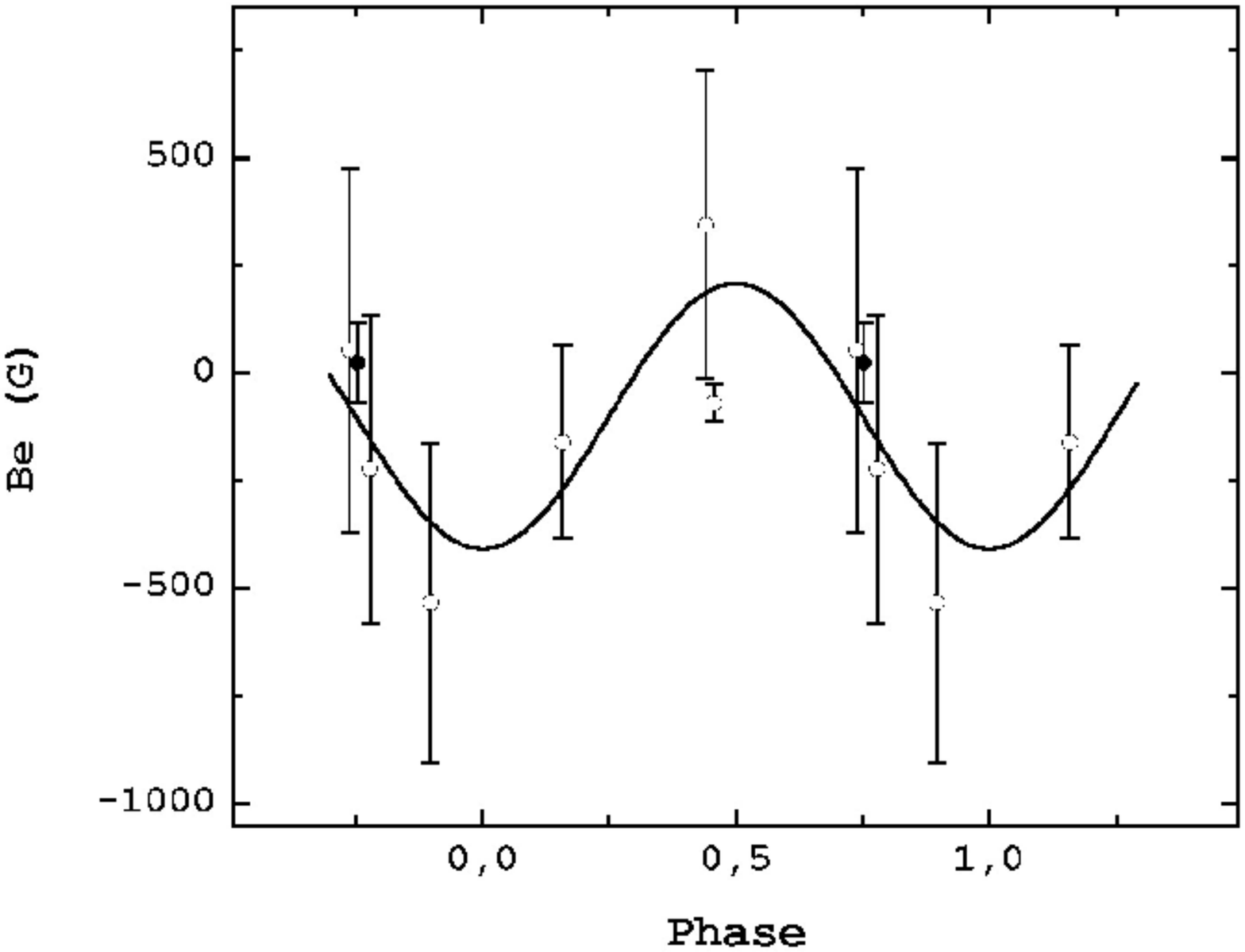}}
\vspace{-3.5mm}
\caption{ HD 37151 }
\label{fig:fig96}
\end{figure}

\begin{figure}
\resizebox{0.98\hsize}{!}{\includegraphics{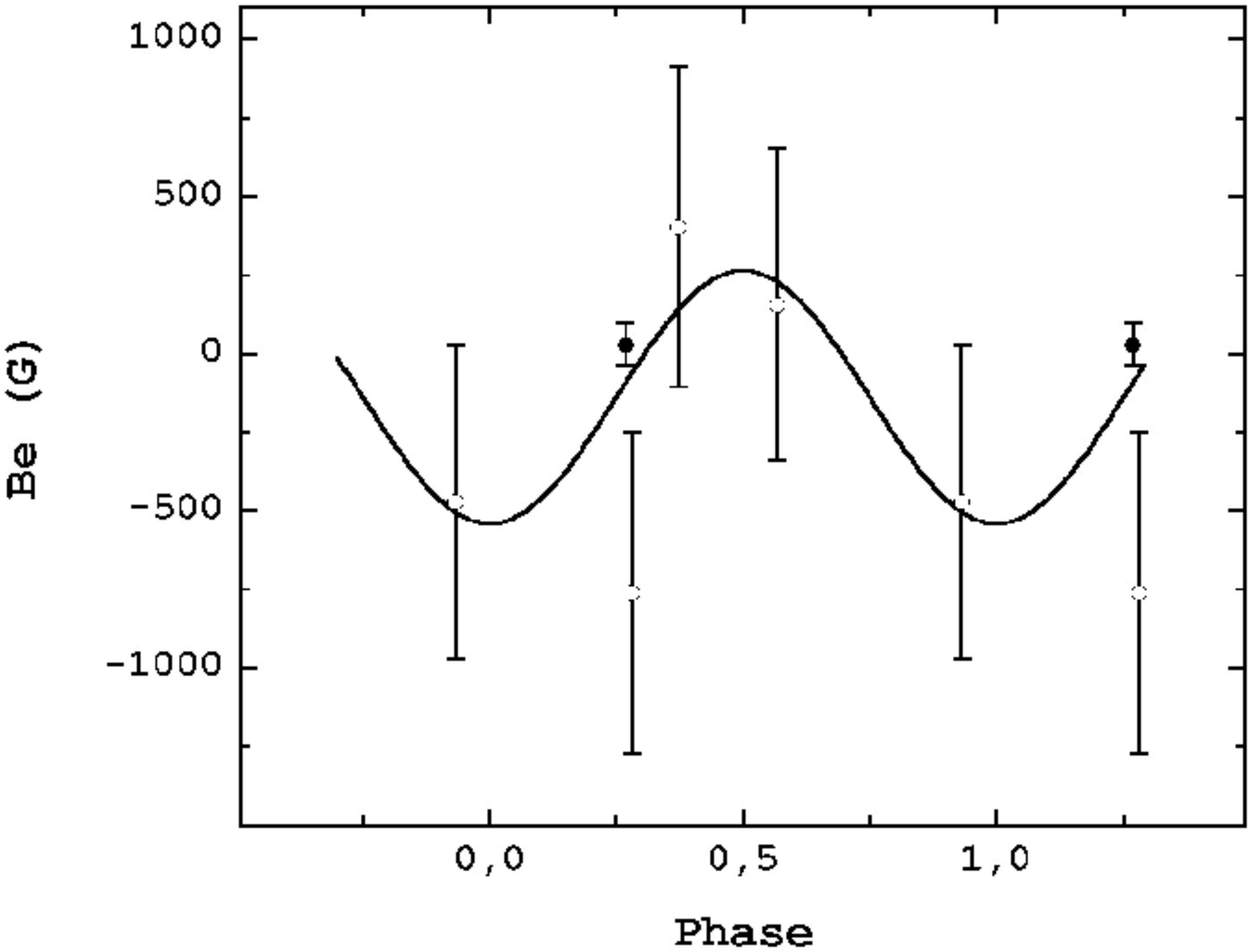}}
\vspace{-3.5mm}
\caption{ HD 37210 }
\label{fig:fig97}
\end{figure}

\begin{figure}
\resizebox{0.98\hsize}{!}{\includegraphics{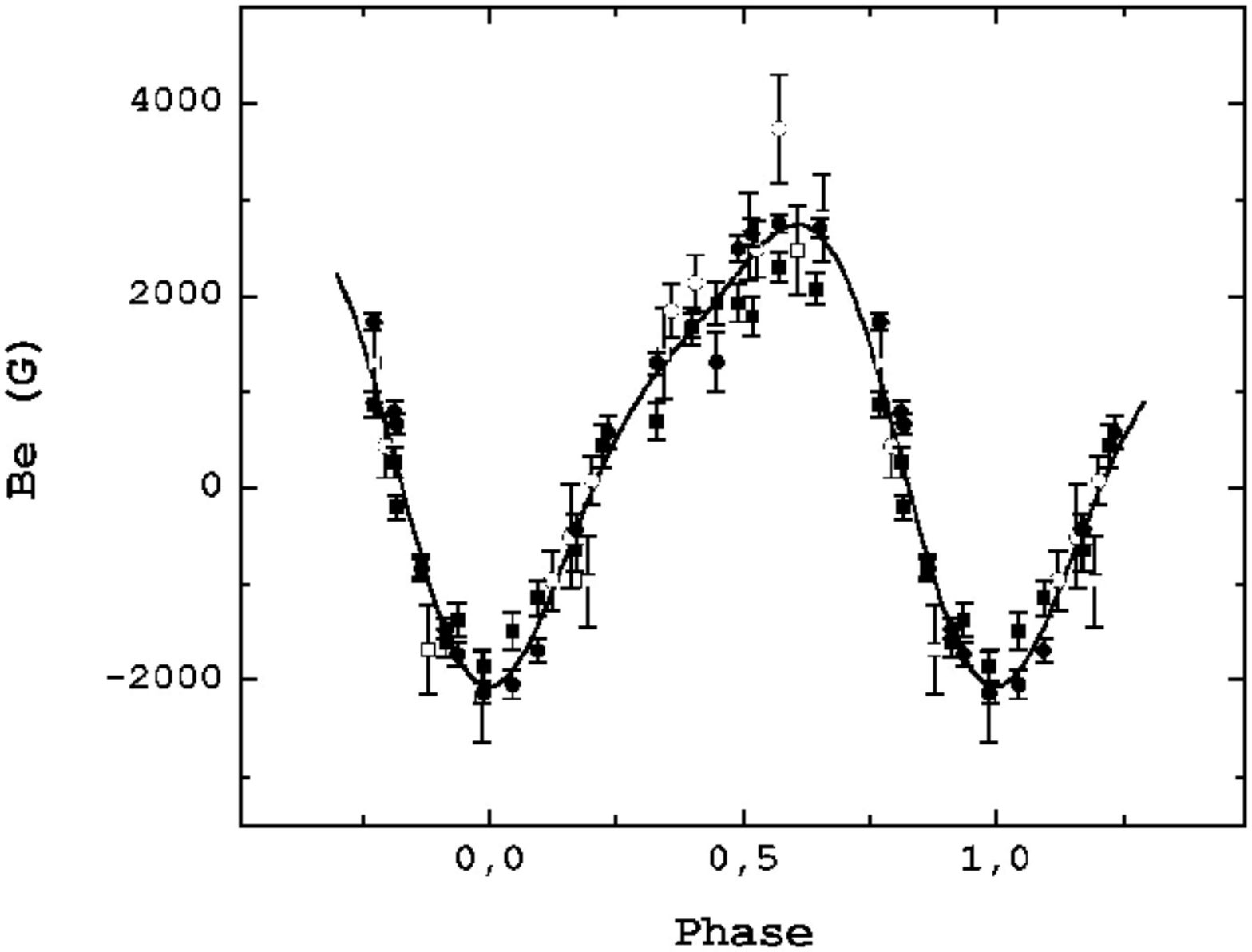}}
\vspace{-3.5mm}
\caption{ HD 37479 (1)}
\label{fig:fig98}
\end{figure}

\begin{figure}
\resizebox{0.98\hsize}{!}{\includegraphics{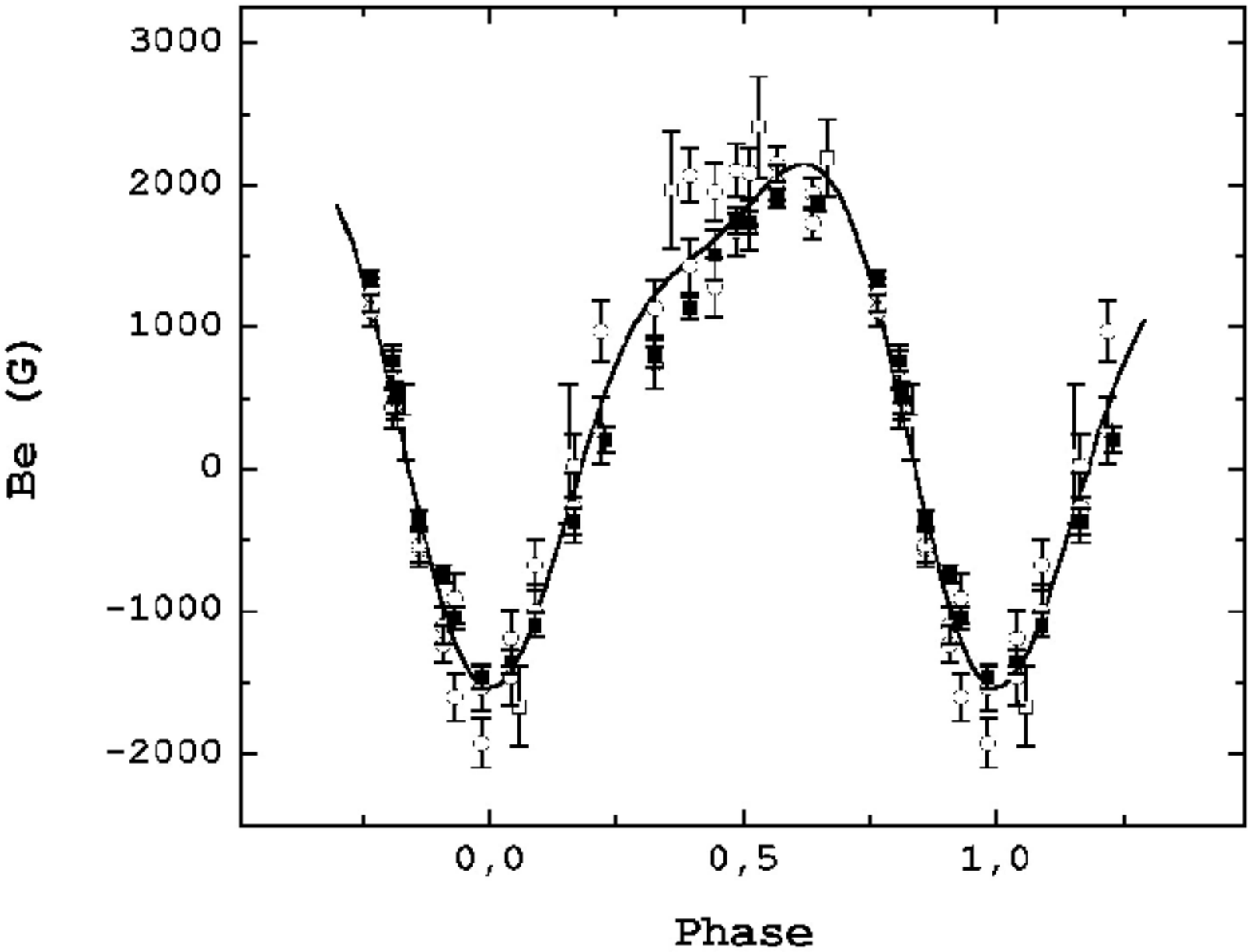}}
\vspace{-3.5mm}
\caption{ HD 37479 (2)}
\label{fig:fig99}
\end{figure}

\clearpage
\newpage

\begin{figure}
\resizebox{0.98\hsize}{!}{\includegraphics{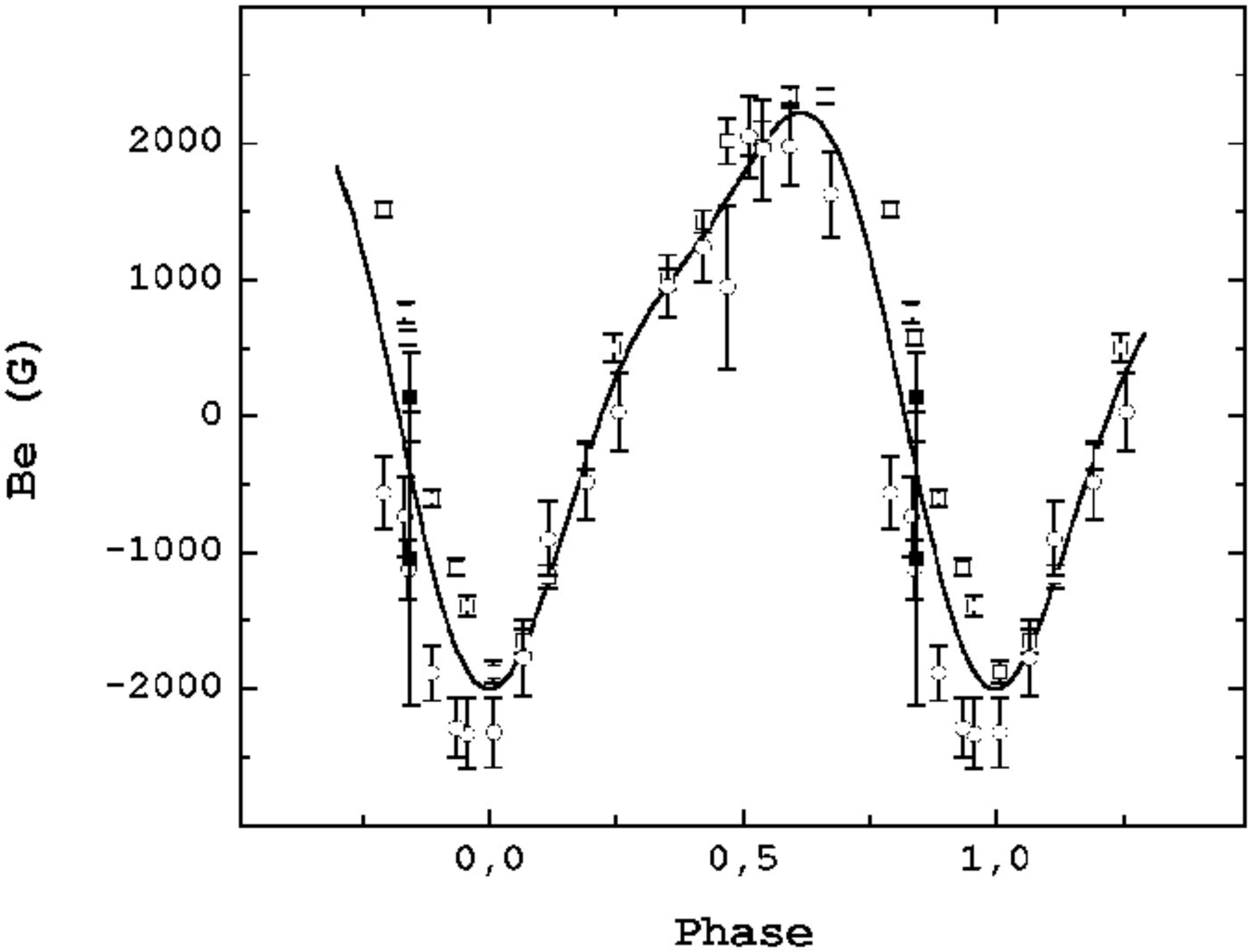}}
\vspace{-3.5mm}
\caption{ HD 37479 (3)}
\label{fig:fig100}
\end{figure}

\begin{figure}
\resizebox{0.98\hsize}{!}{\includegraphics{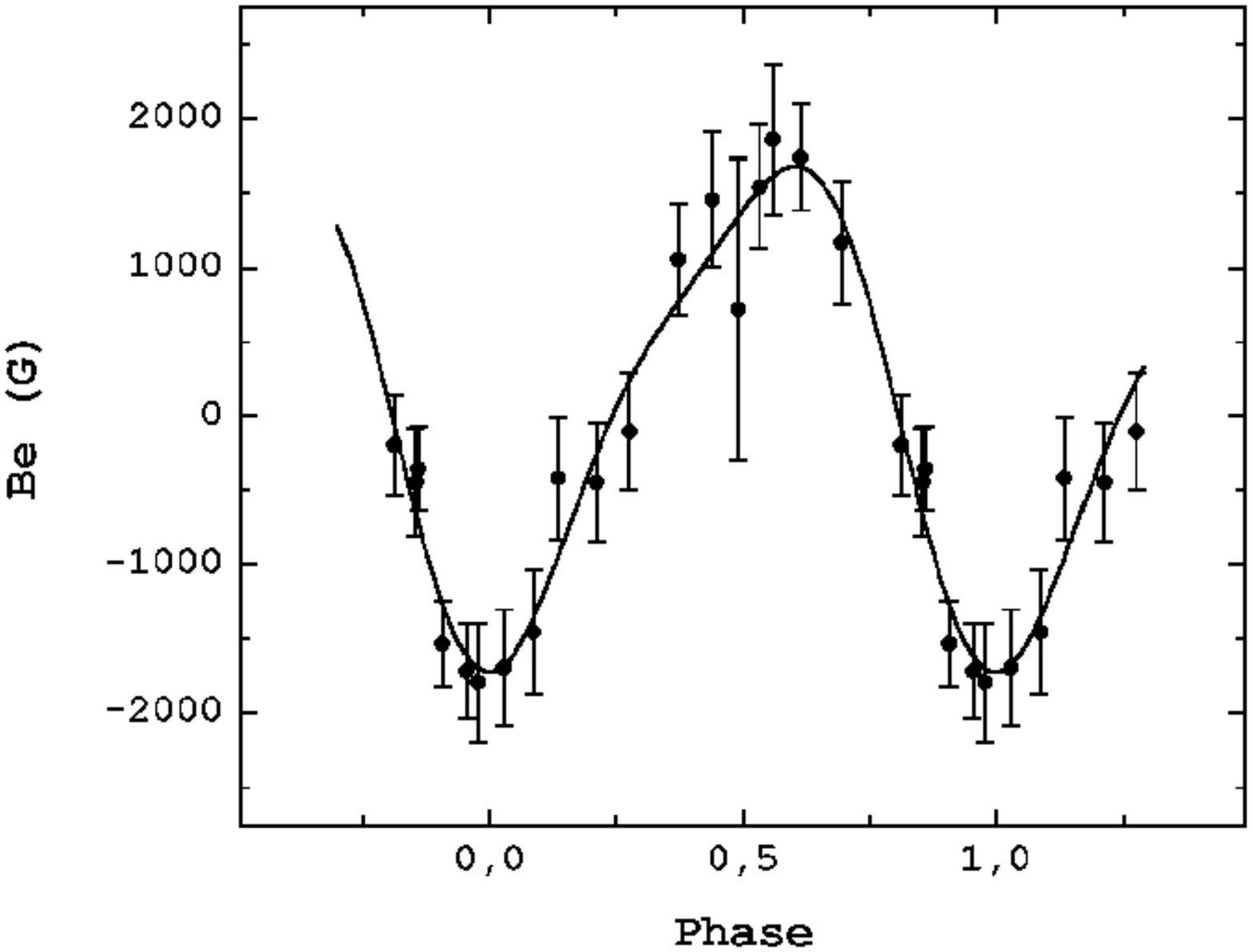}}
\vspace{-3.5mm}
\caption{ HD 37479 (4)}
\label{fig:fig101}
\end{figure}

\begin{figure}
\resizebox{0.98\hsize}{!}{\includegraphics{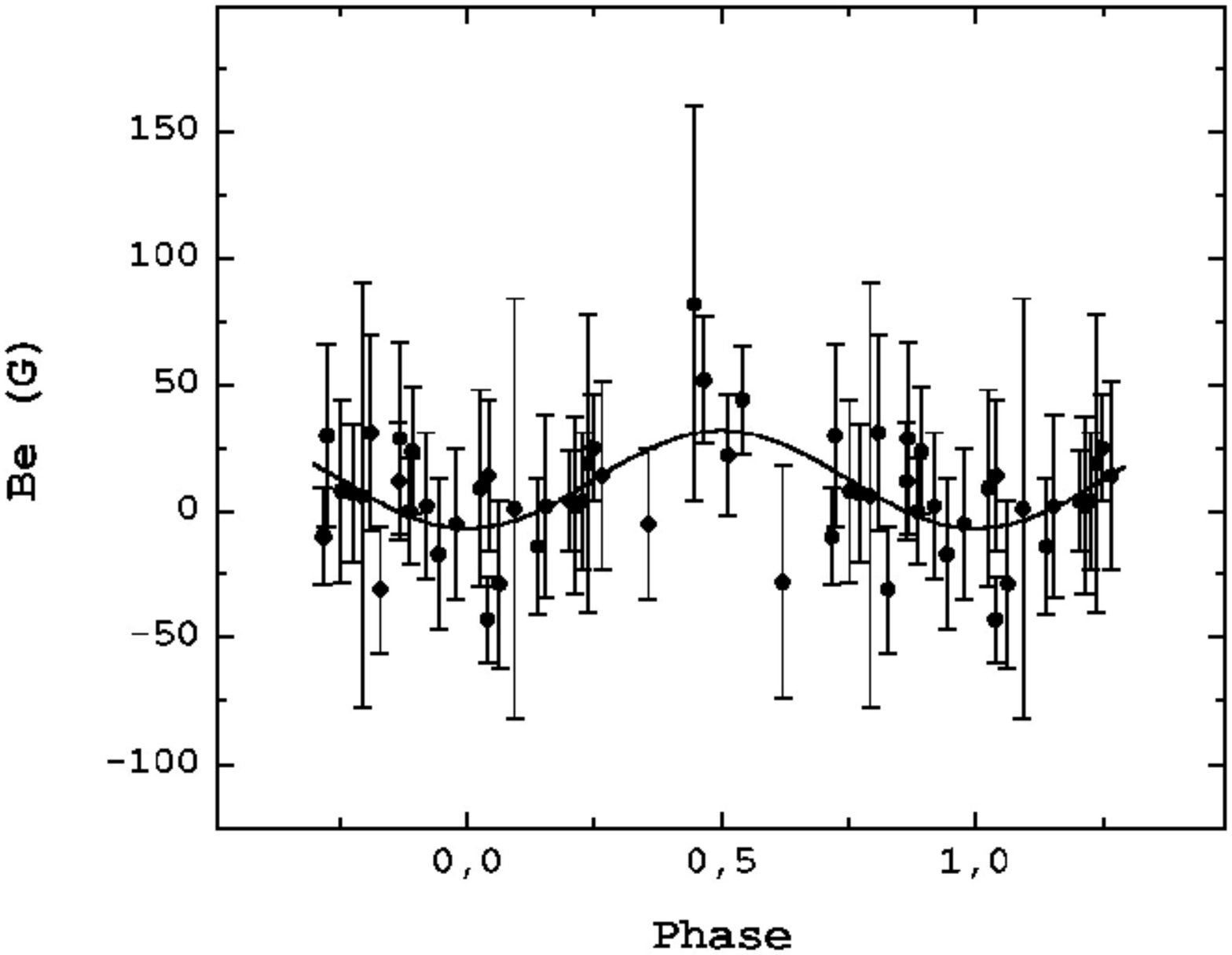}}
\vspace{-3.5mm}
\caption{ HD 37490 }
\label{fig:fig102}
\end{figure}

\begin{figure}
\resizebox{0.98\hsize}{!}{\includegraphics{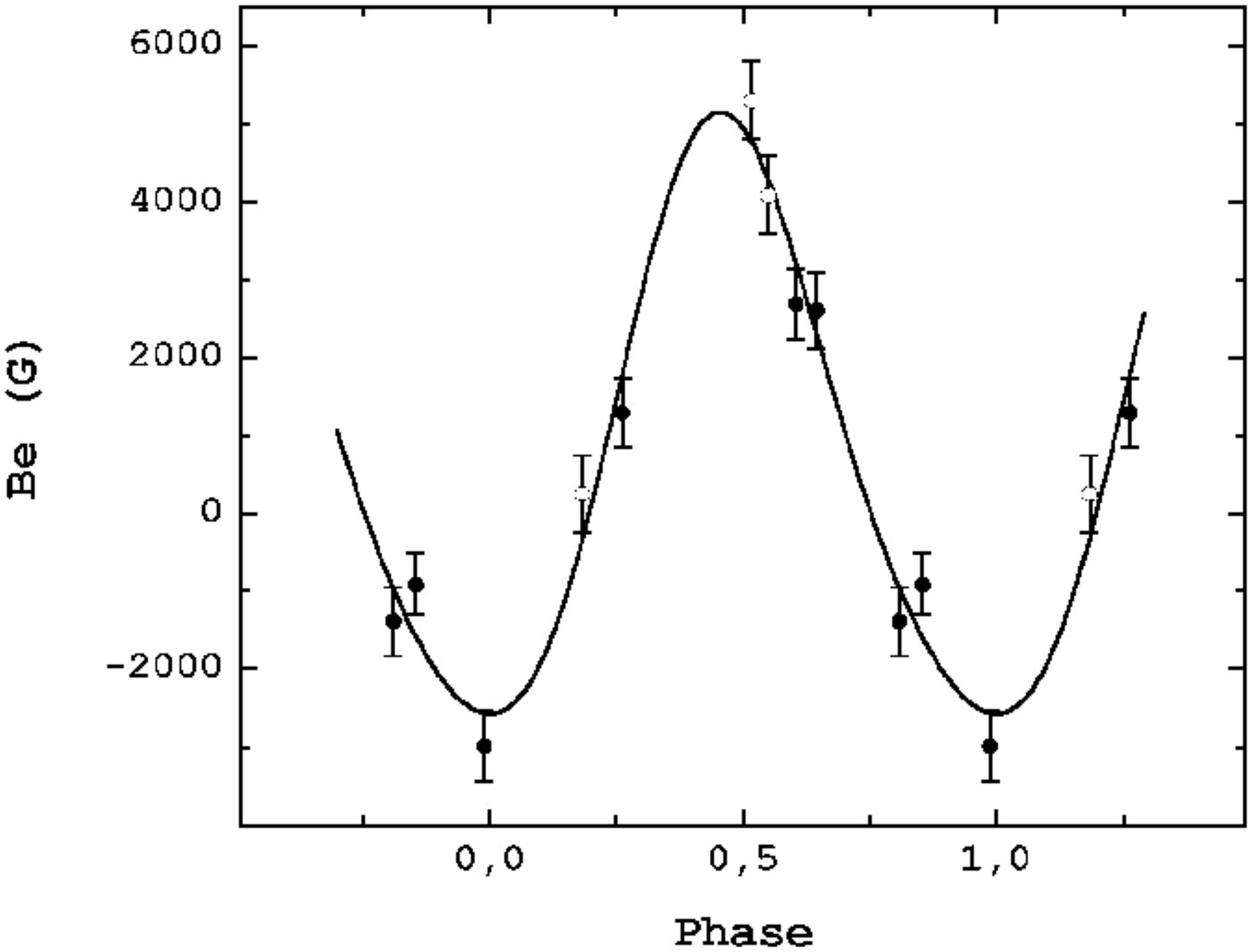}}
\vspace{-3.5mm}
\caption{ HD 37642 }
\label{fig:fig103}
\end{figure}

\begin{figure}
\resizebox{0.98\hsize}{!}{\includegraphics{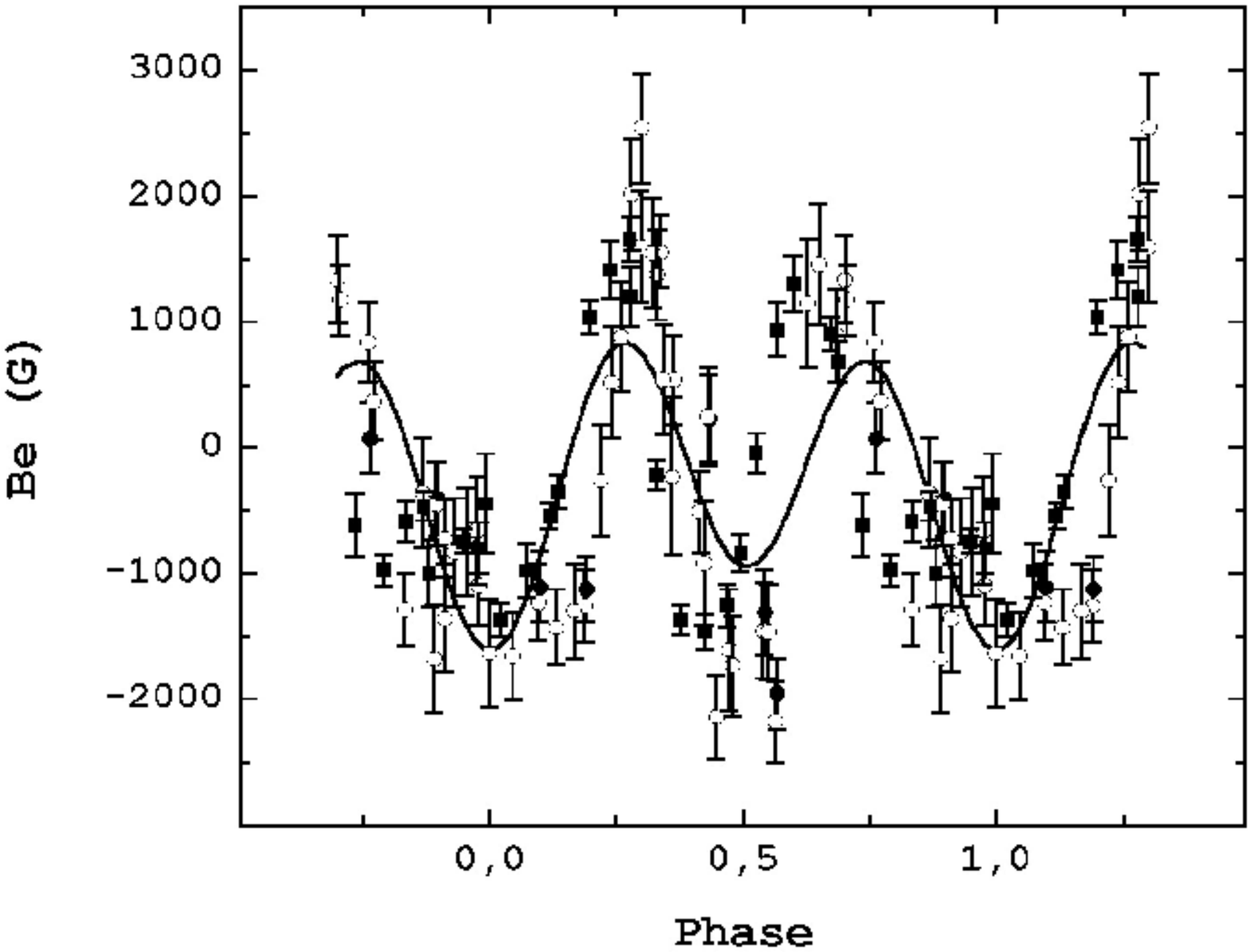}}
\vspace{-3.5mm}
\caption{ HD 37776 (1) }
\label{fig:fig104}
\end{figure}

\begin{figure}
\resizebox{0.98\hsize}{!}{\includegraphics{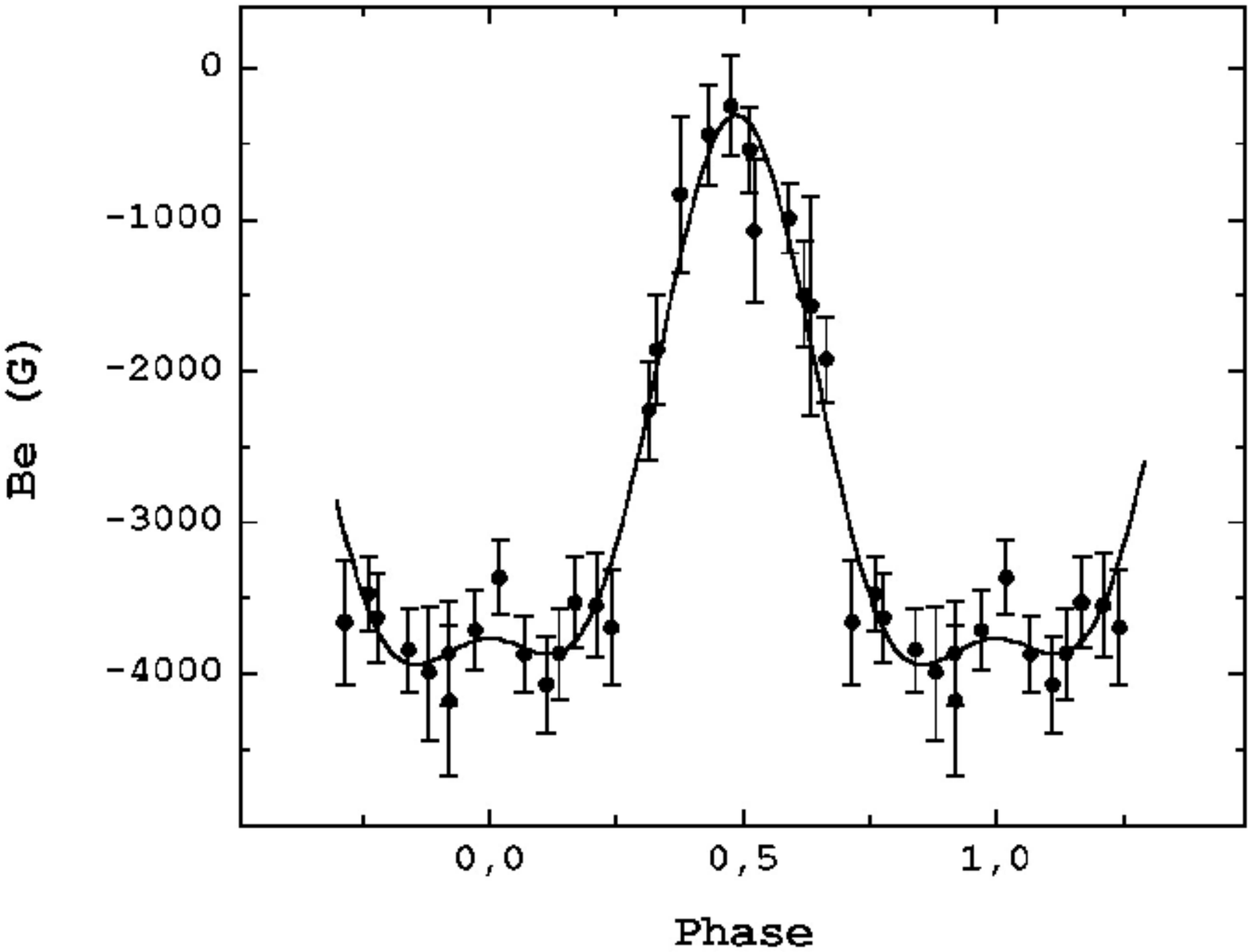}}
\vspace{-3.5mm}
\caption{ HD 37776 (2) }
\label{fig:fig104}
\end{figure}

\clearpage
\newpage

\begin{figure}
\resizebox{0.98\hsize}{!}{\includegraphics{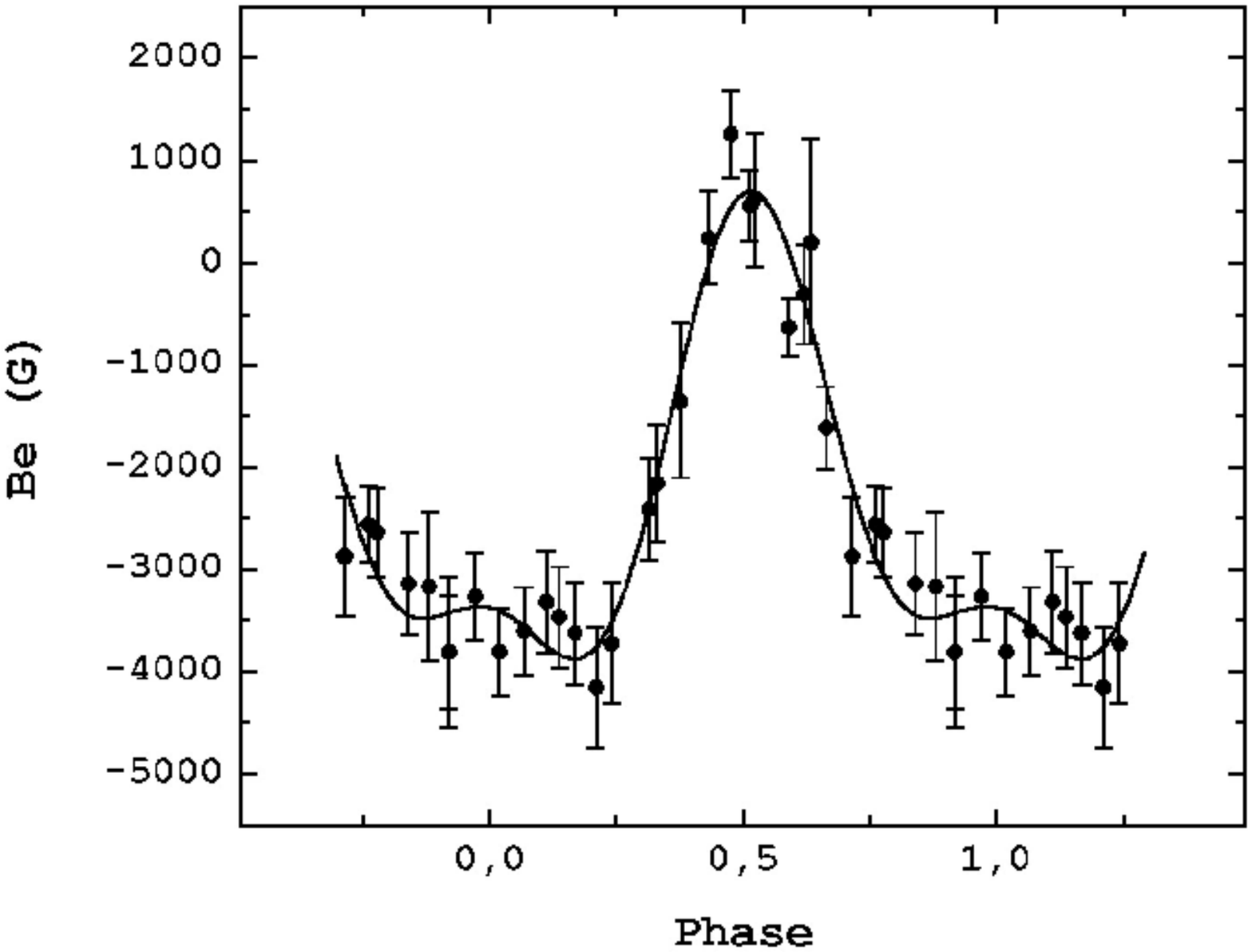}}
\vspace{-3.5mm}
\caption{ HD 37776 (3) }
\label{fig:fig104}
\end{figure}

\begin{figure}
\resizebox{0.98\hsize}{!}{\includegraphics{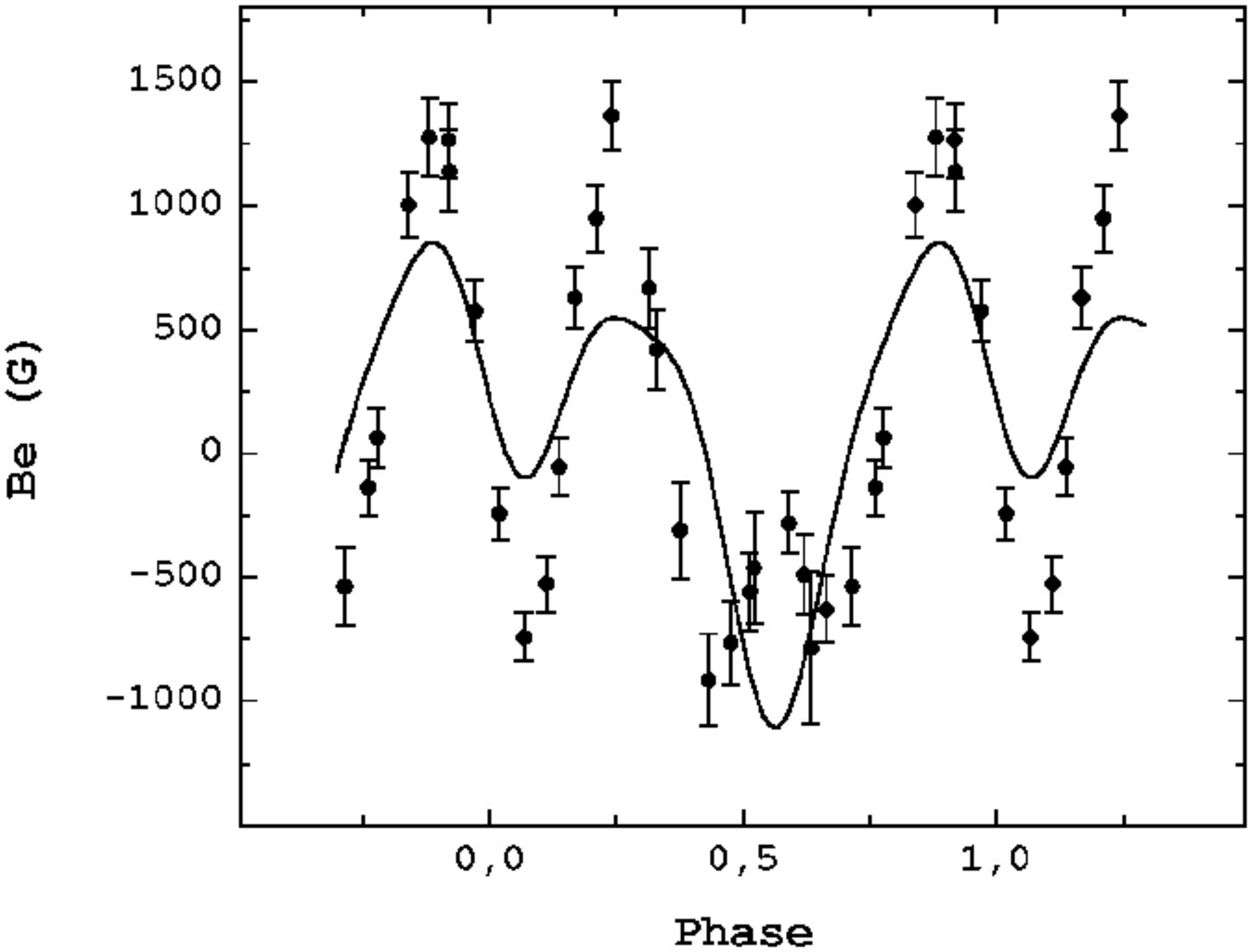}}
\vspace{-3.5mm}
\caption{ HD 37776 (4) }
\label{fig:fig104}
\end{figure}

\begin{figure}
\resizebox{0.98\hsize}{!}{\includegraphics{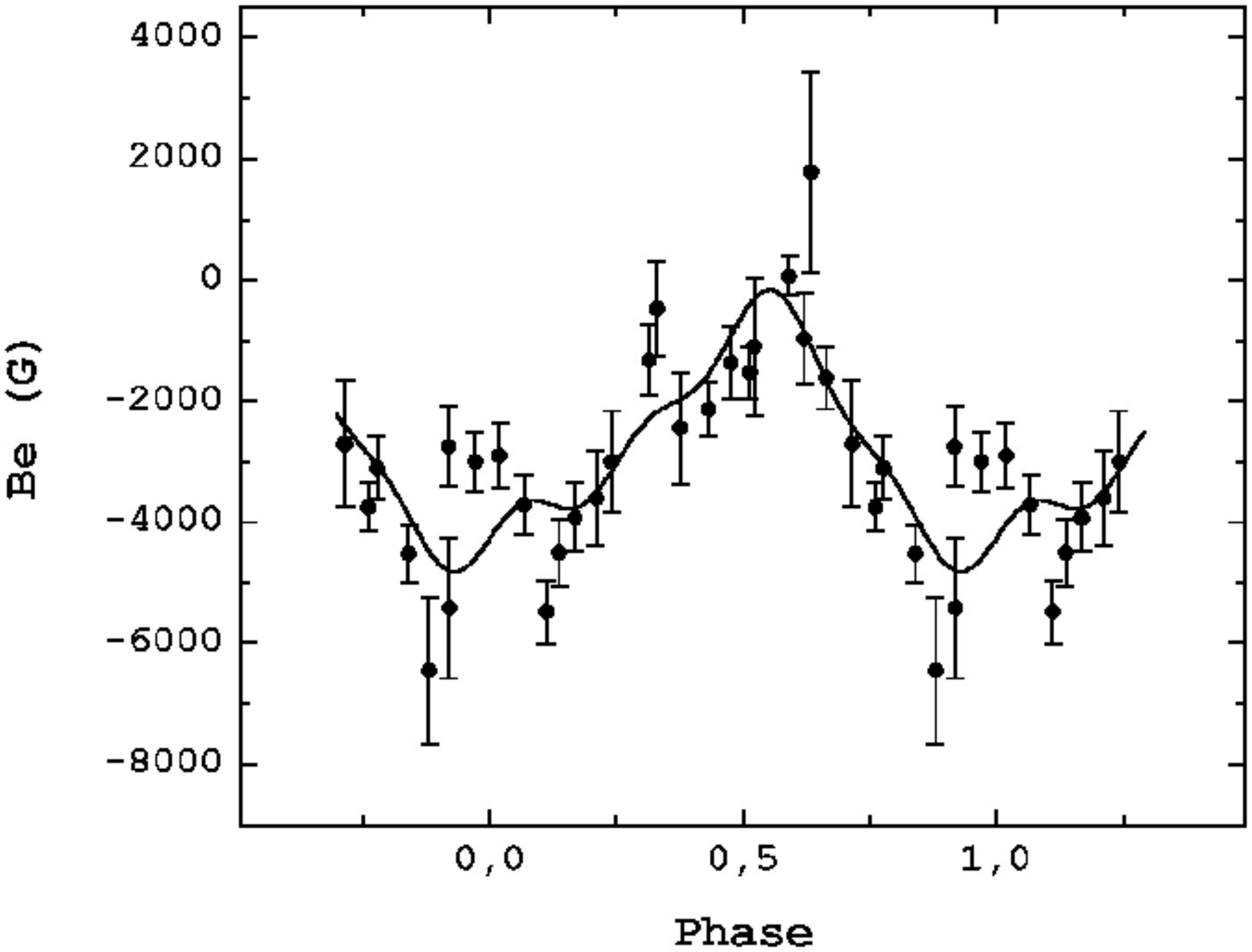}}
\vspace{-3.5mm}
\caption{ HD 37776 (5) }
\label{fig:fig104}
\end{figure}

\begin{figure}
\resizebox{0.98\hsize}{!}{\includegraphics{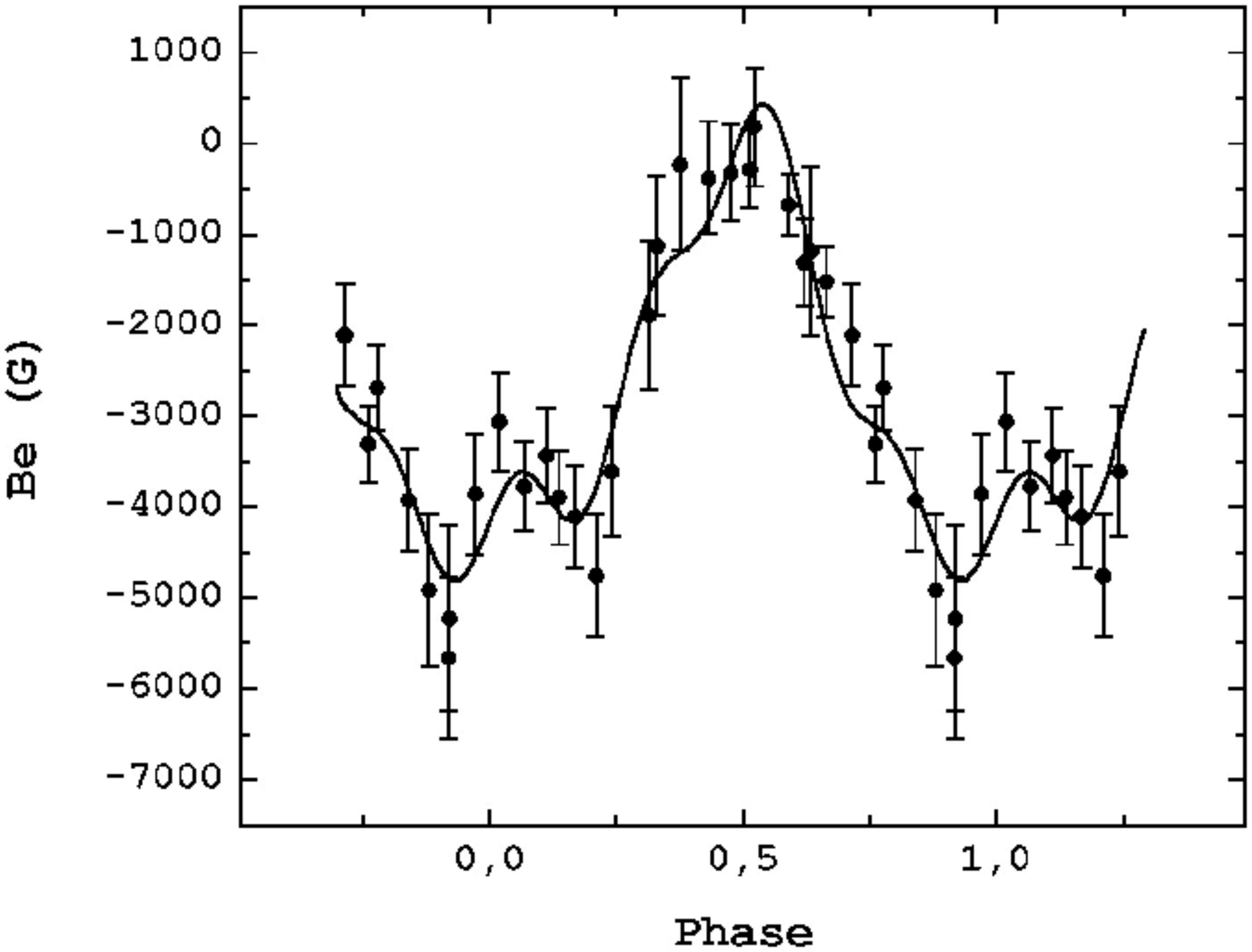}}
\vspace{-3.5mm}
\caption{ HD 37776 (6) }
\label{fig:fig104}
\end{figure}

\begin{figure}
\resizebox{0.98\hsize}{!}{\includegraphics{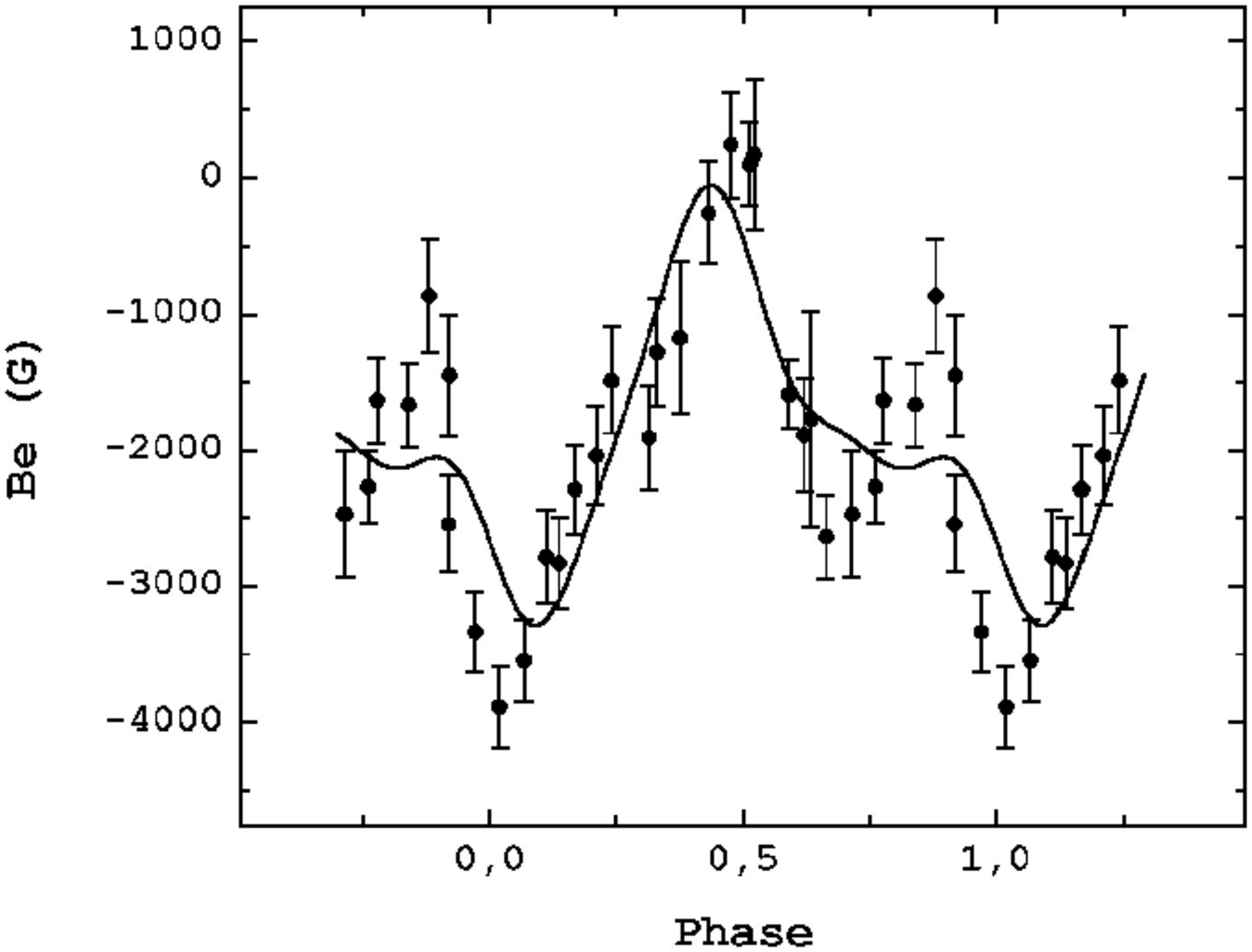}}
\vspace{-3.5mm}
\caption{ HD 37776 (7) }
\label{fig:fig104}
\end{figure}

\begin{figure}
\resizebox{0.98\hsize}{!}{\includegraphics{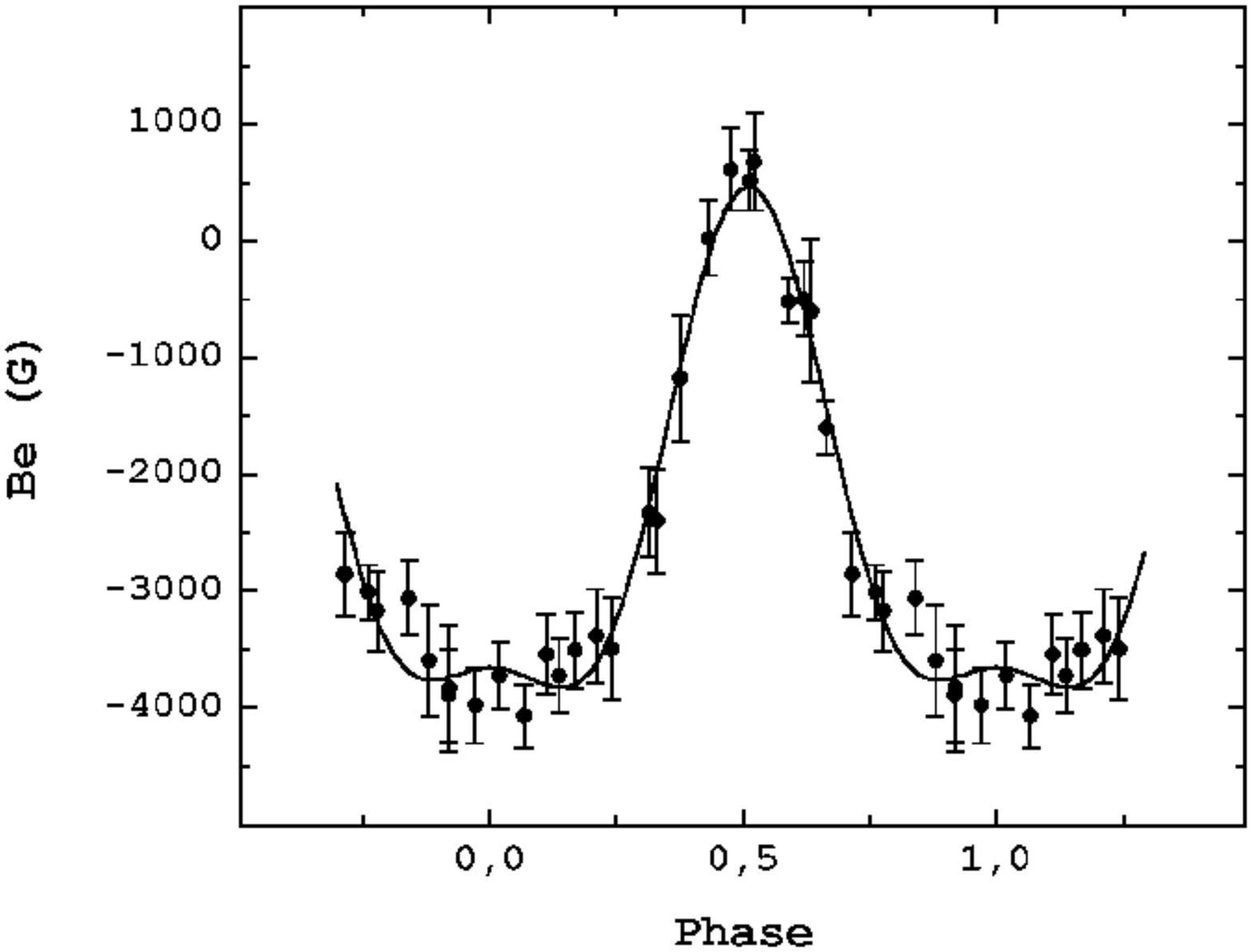}}
\vspace{-3.5mm}
\caption{ HD 37776 (8) }
\label{fig:fig104}
\end{figure}

\clearpage
\newpage

\begin{figure}
\resizebox{0.98\hsize}{!}{\includegraphics{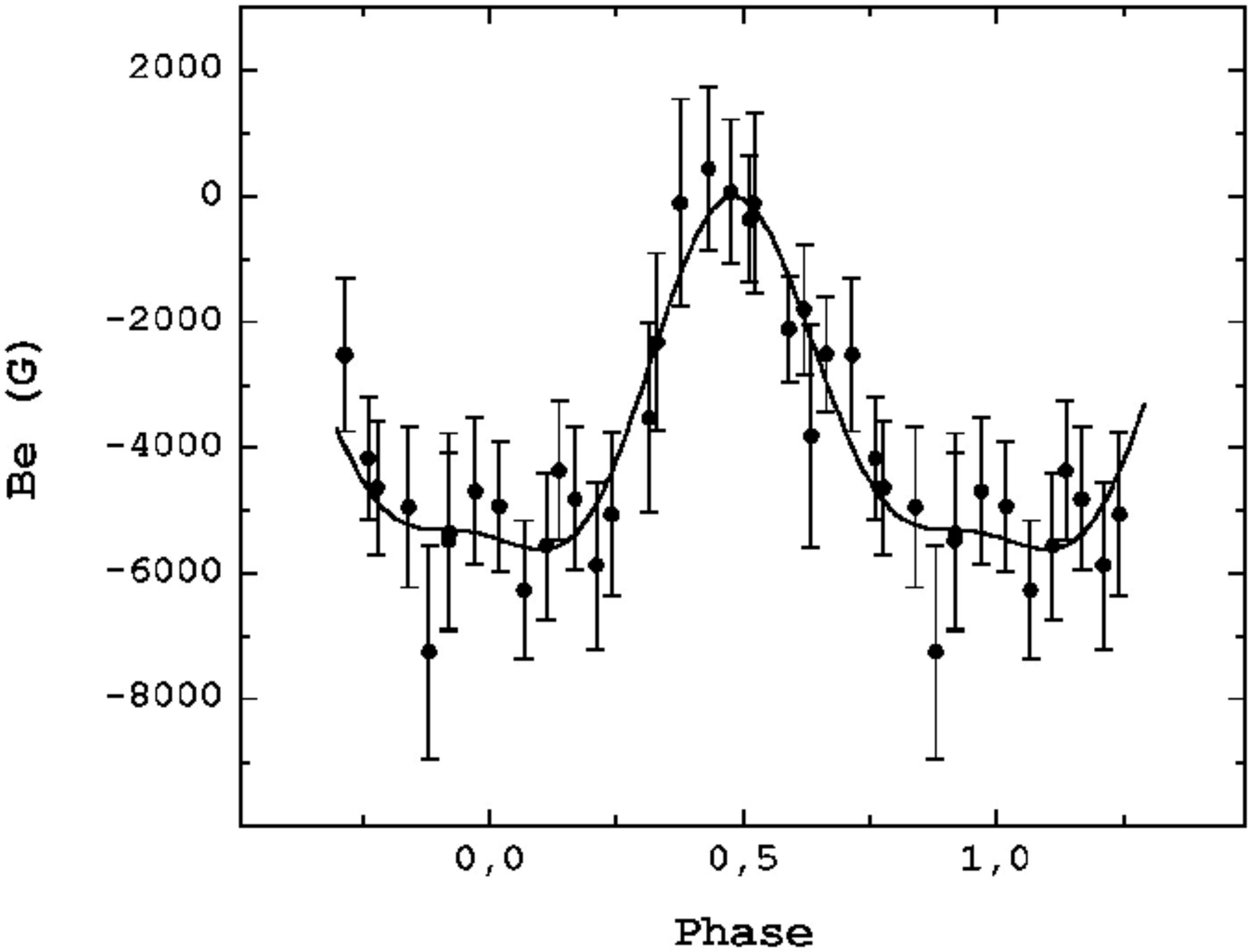}}
\vspace{-3.5mm}
\caption{ HD 37776 (9) }
\label{fig:fig104}
\end{figure}

\begin{figure}
\resizebox{0.98\hsize}{!}{\includegraphics{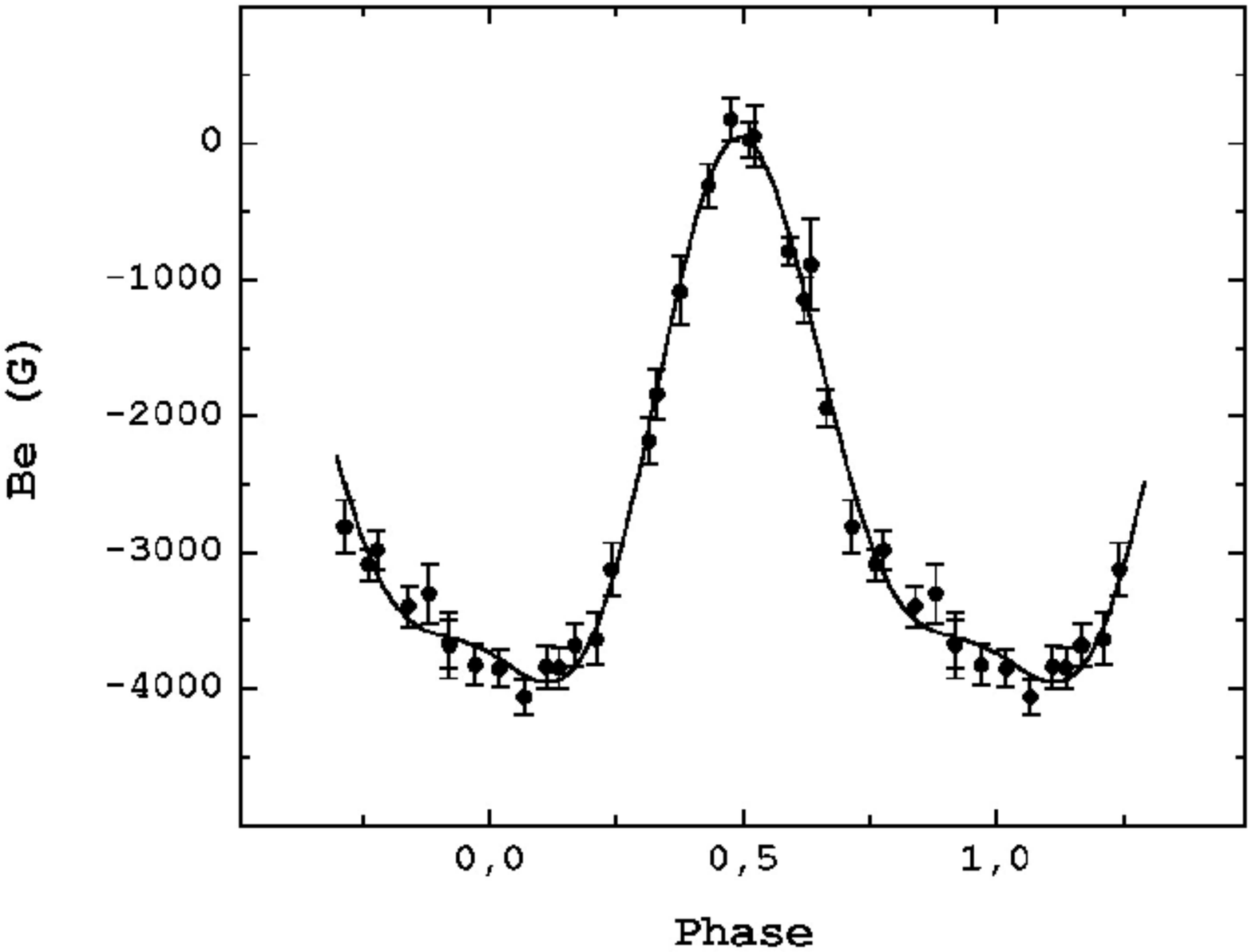}}
\vspace{-3.5mm}
\caption{ HD 37776 (10) }
\label{fig:fig104}
\end{figure}

\begin{figure}
\resizebox{0.98\hsize}{!}{\includegraphics{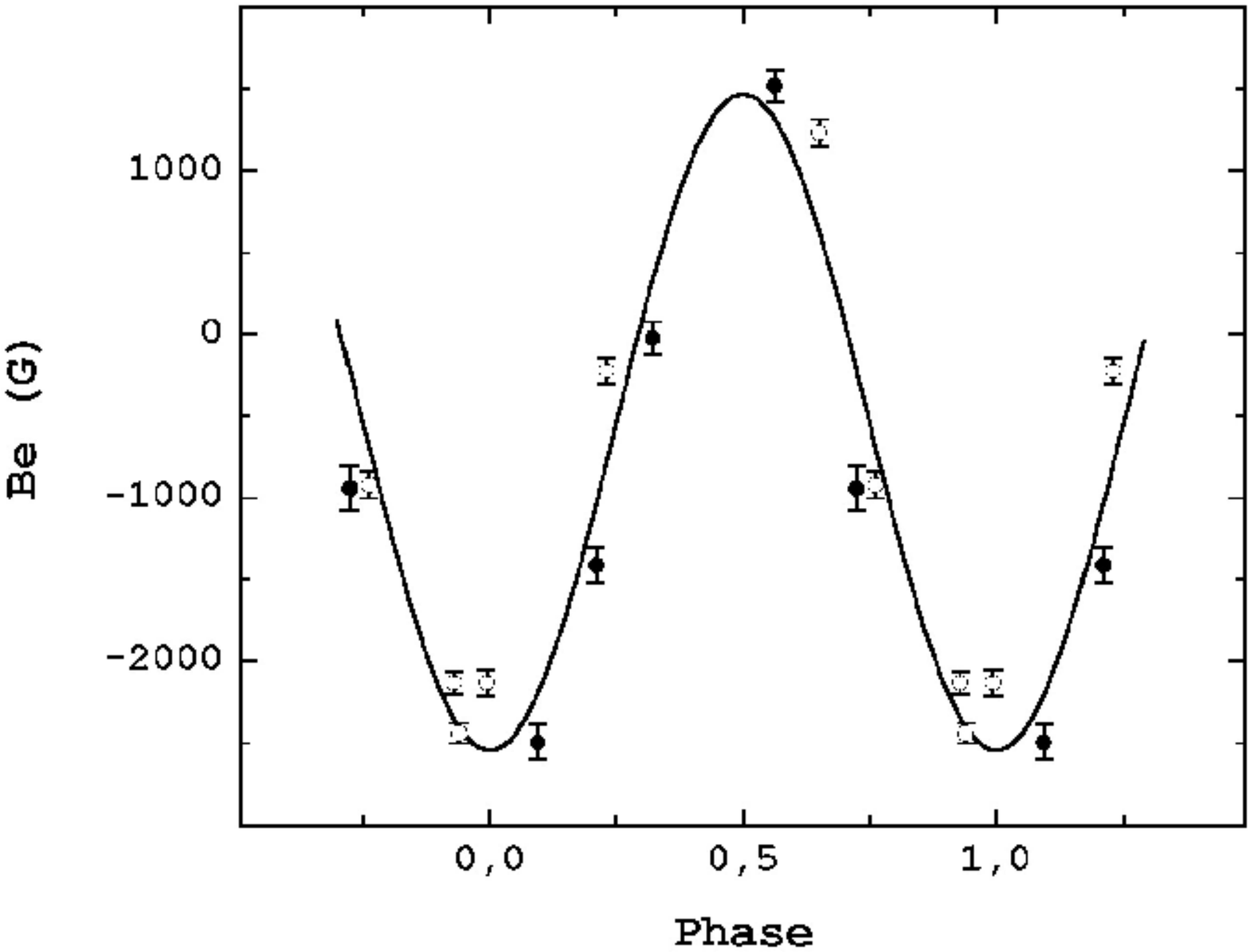}}
\vspace{-3.5mm}
\caption{ HD 38823 }
\label{fig:fig105}
\end{figure}

\begin{figure}
\resizebox{0.98\hsize}{!}{\includegraphics{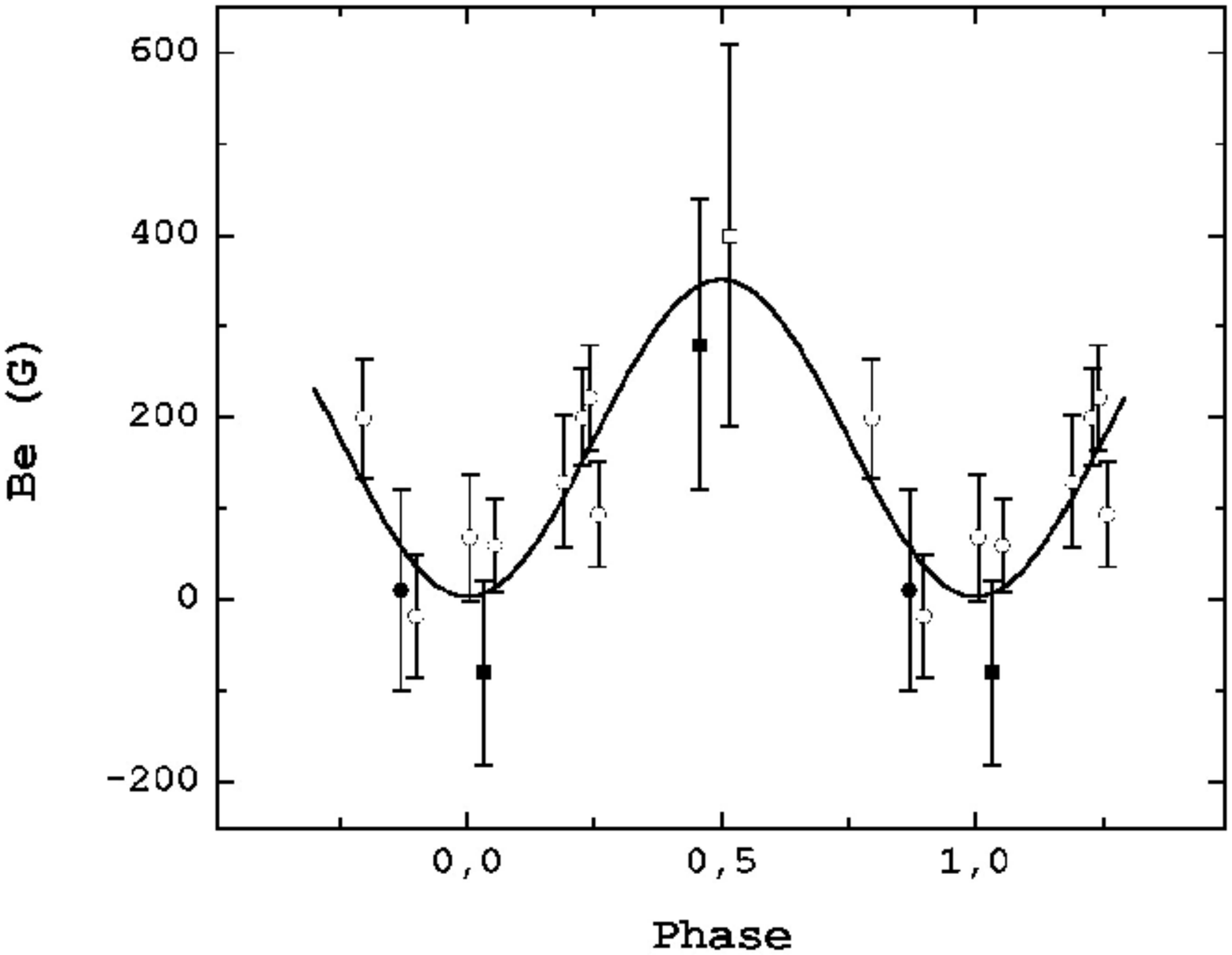}}
\vspace{-3.5mm}
\caption{ HD 39317 }
\label{fig:fig107}
\end{figure}

\begin{figure}
\resizebox{0.98\hsize}{!}{\includegraphics{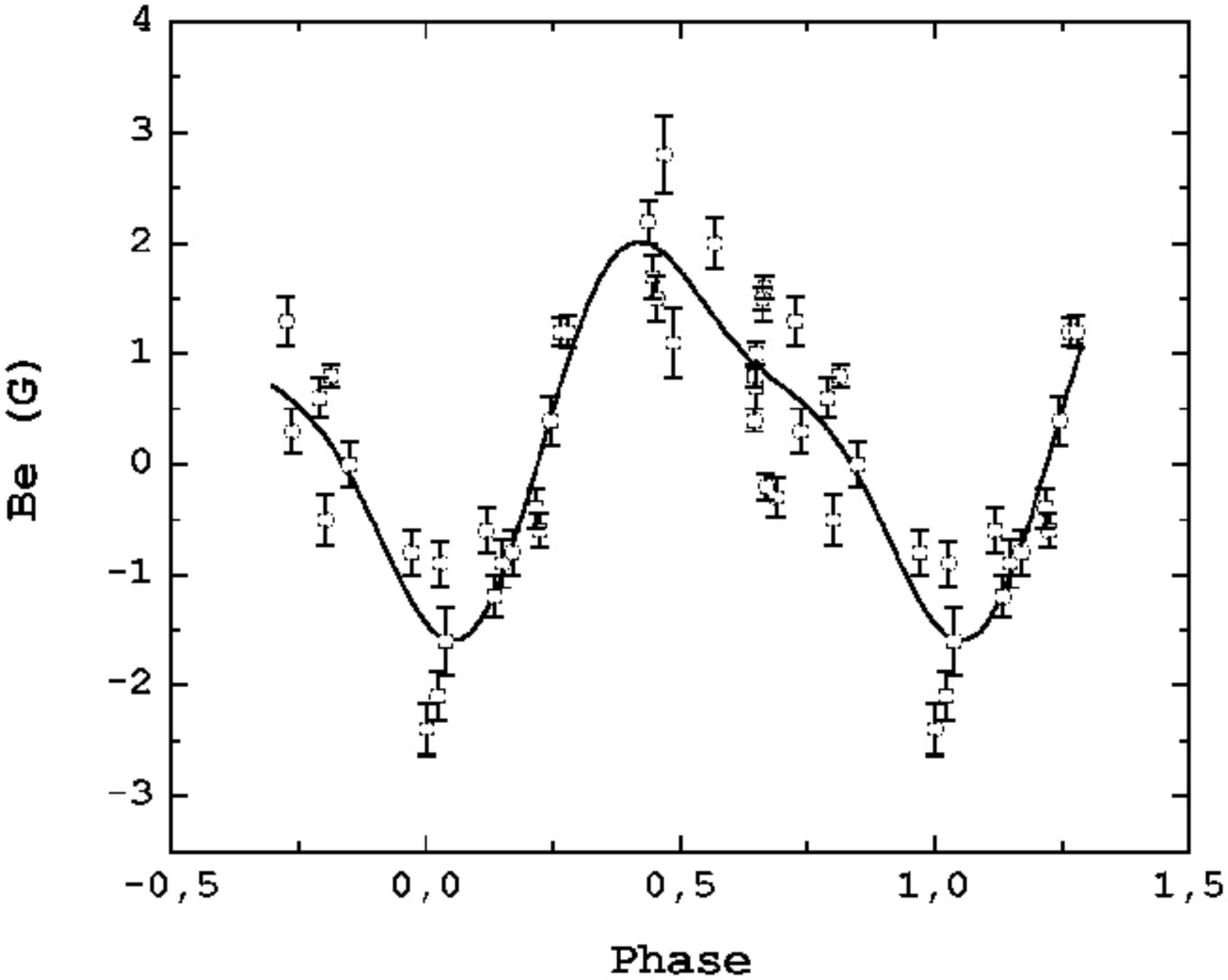}}
\vspace{-3.5mm}
\caption{ HD 39801 }
\label{fig:fig108}
\end{figure}

\begin{figure}
\resizebox{0.98\hsize}{!}{\includegraphics{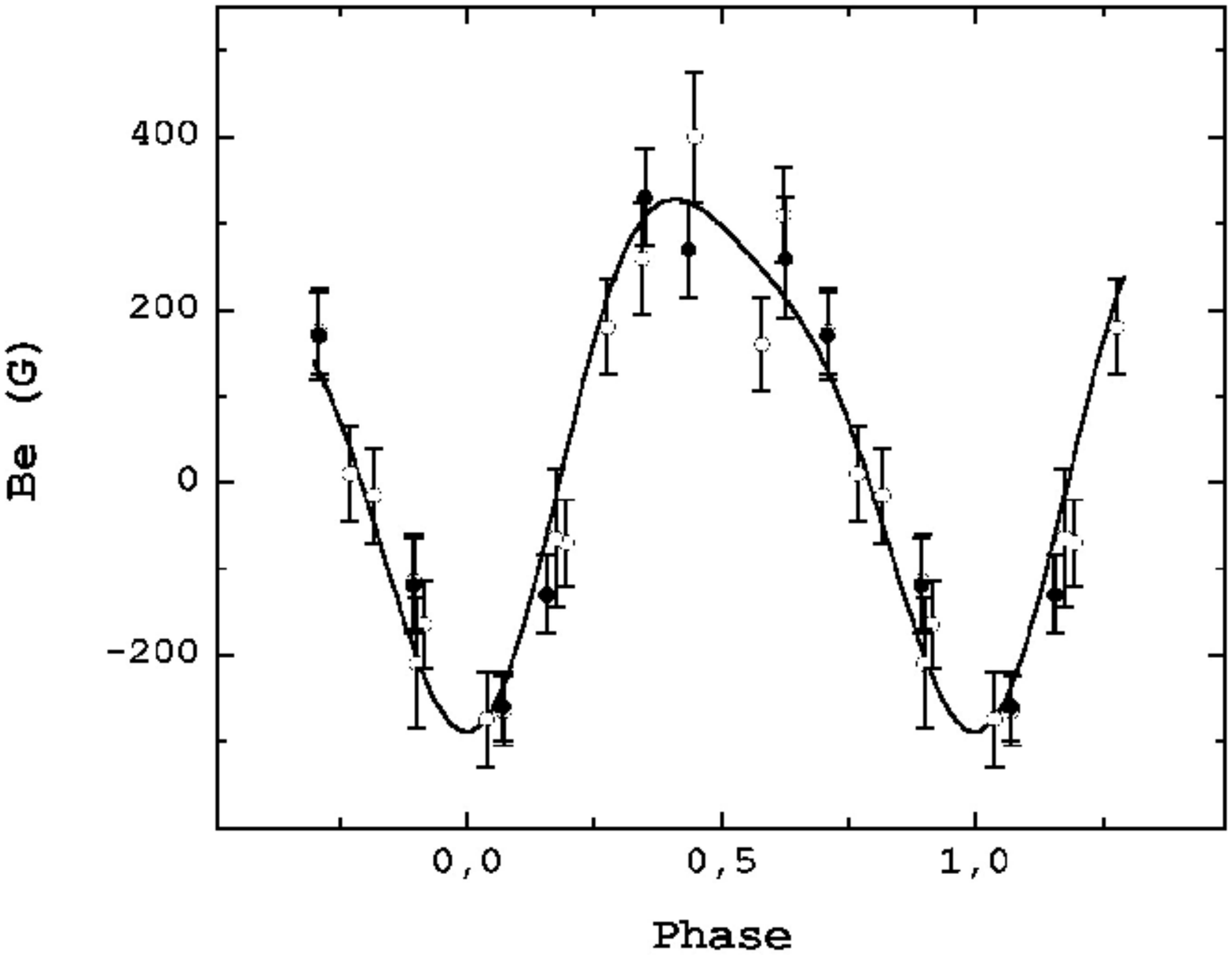}}
\vspace{-3.5mm}
\caption{ HD 40312 (1) }
\label{fig:fig109}
\end{figure}

\clearpage
\newpage

\begin{figure}
\resizebox{0.98\hsize}{!}{\includegraphics{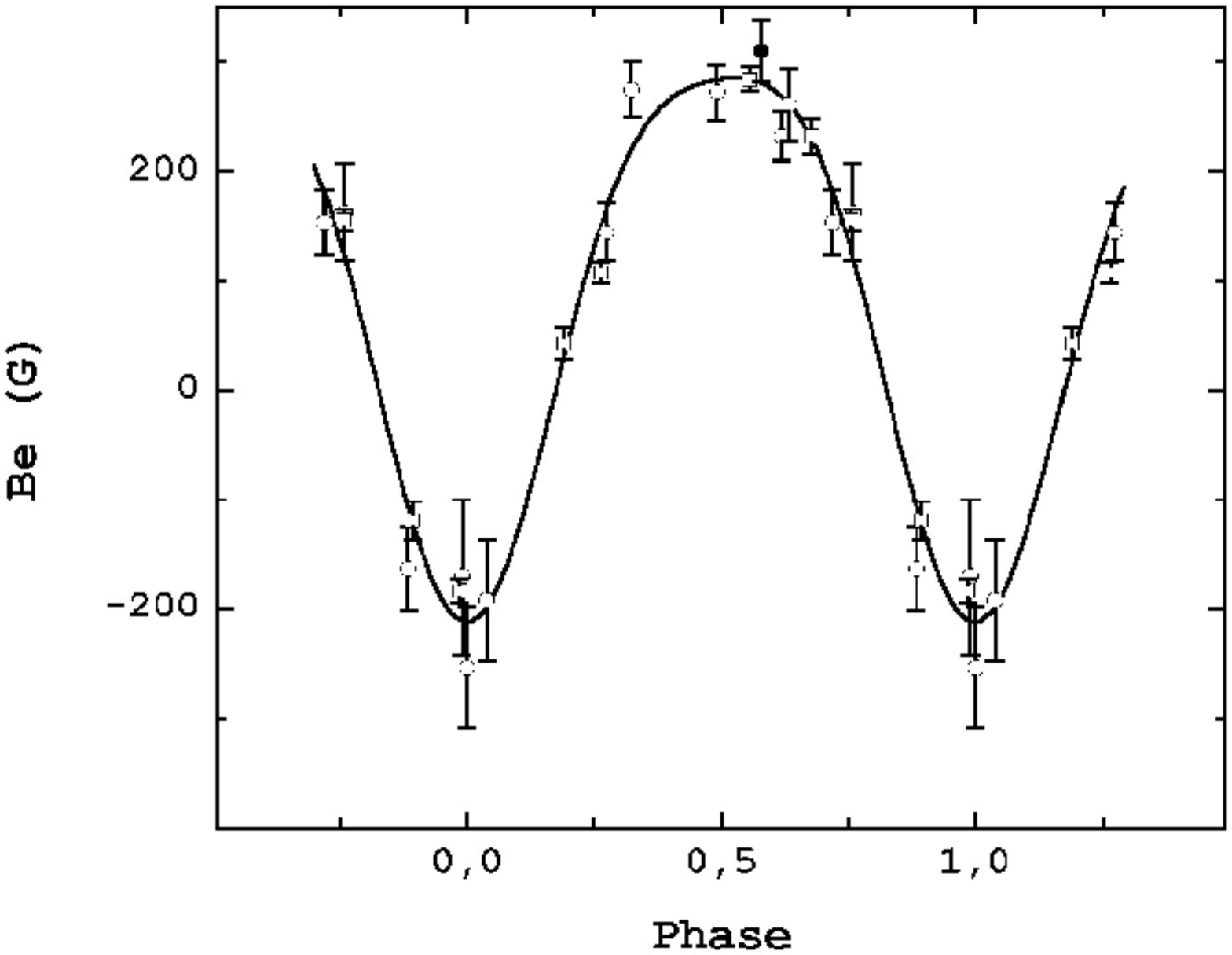}}
\vspace{-3.5mm}
\caption{ HD 40312 (2) }
\label{fig:fig110}
\end{figure}

\begin{figure}
\resizebox{0.98\hsize}{!}{\includegraphics{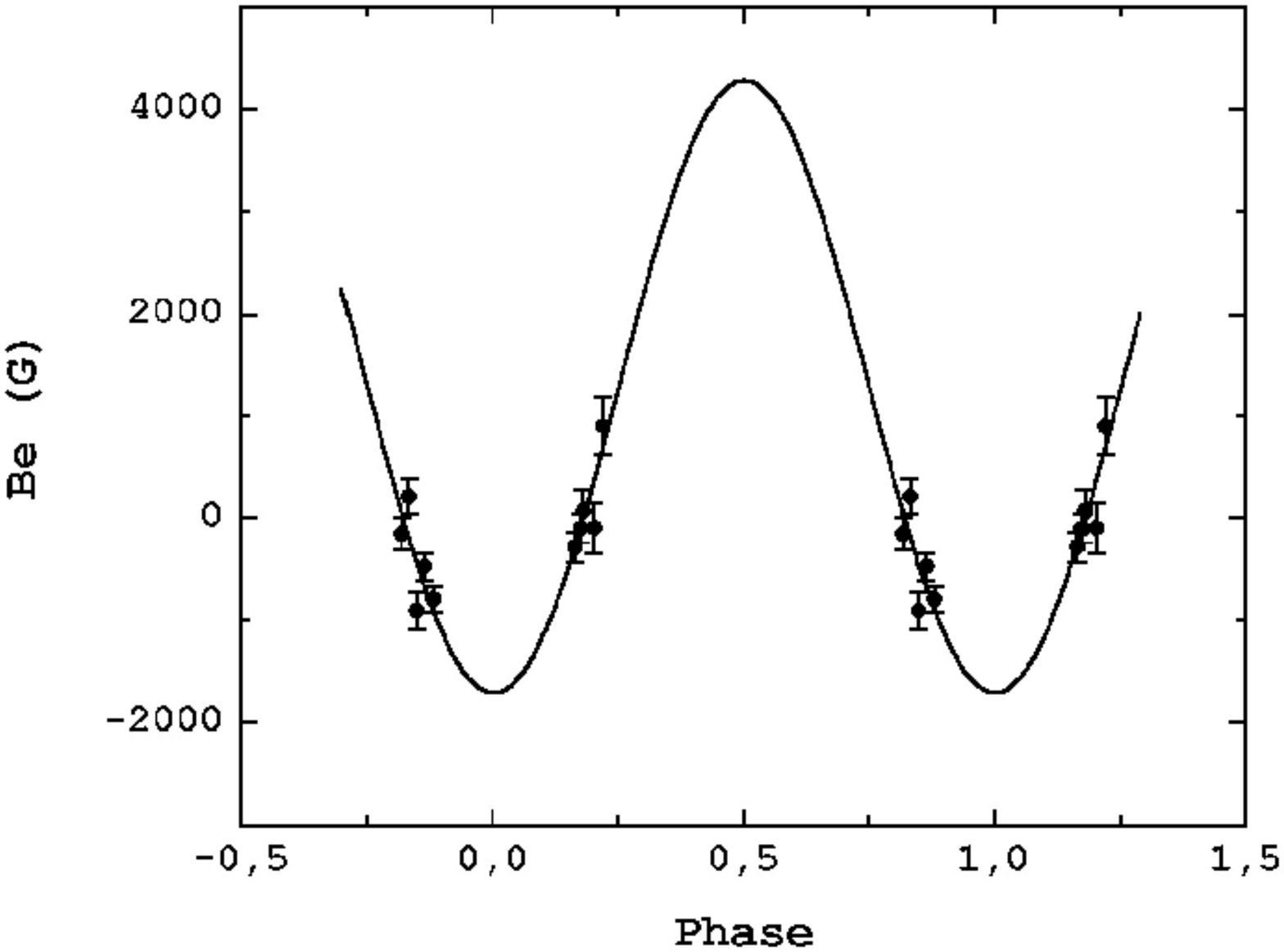}}
\vspace{-3.5mm}
\caption{ HD 40535 }
\label{fig:fig111}
\end{figure}

\begin{figure}
\resizebox{0.98\hsize}{!}{\includegraphics{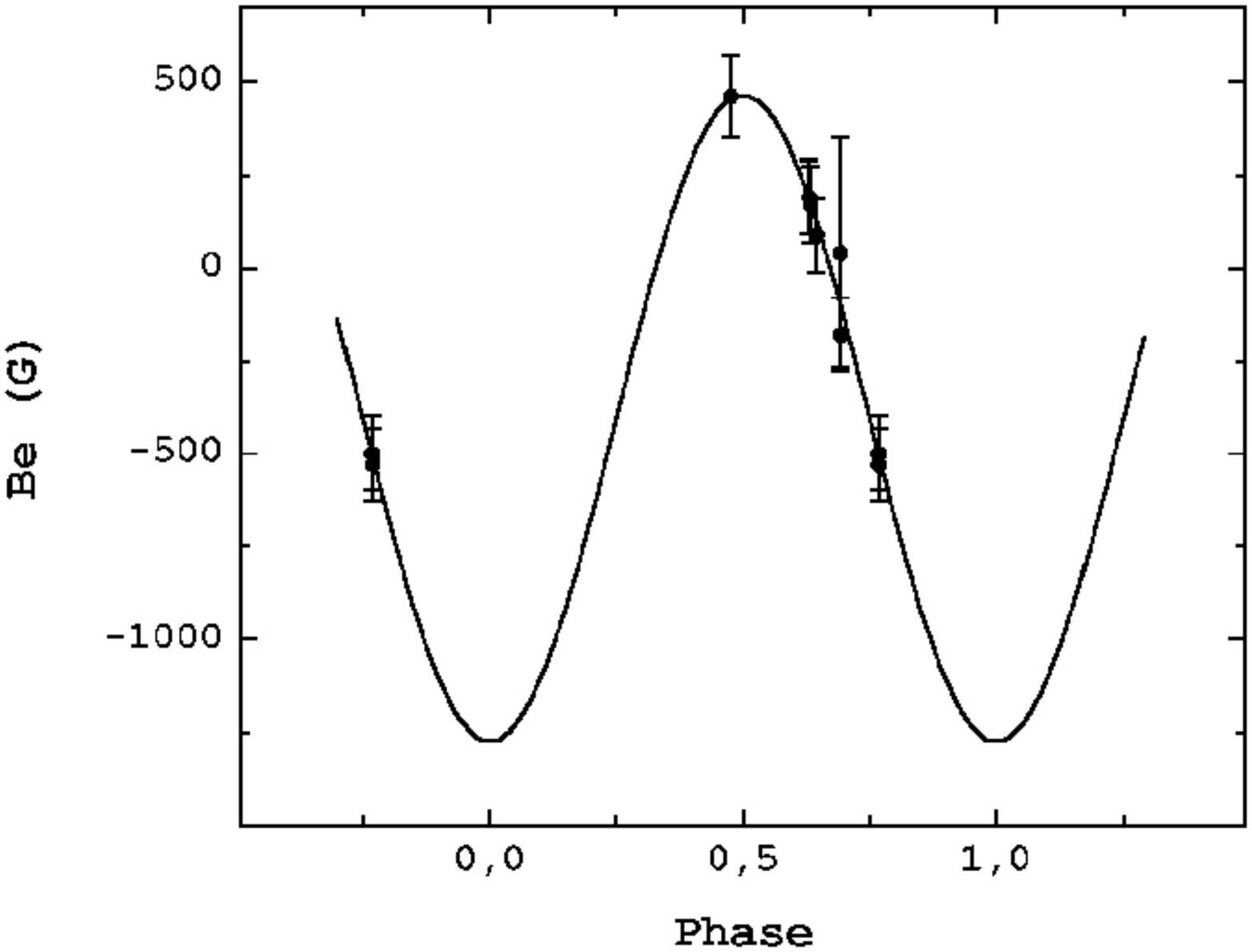}}
\vspace{-3.5mm}
\caption{ HD 41403 }
\label{fig:fig112}
\end{figure}

\begin{figure}
\resizebox{0.98\hsize}{!}{\includegraphics{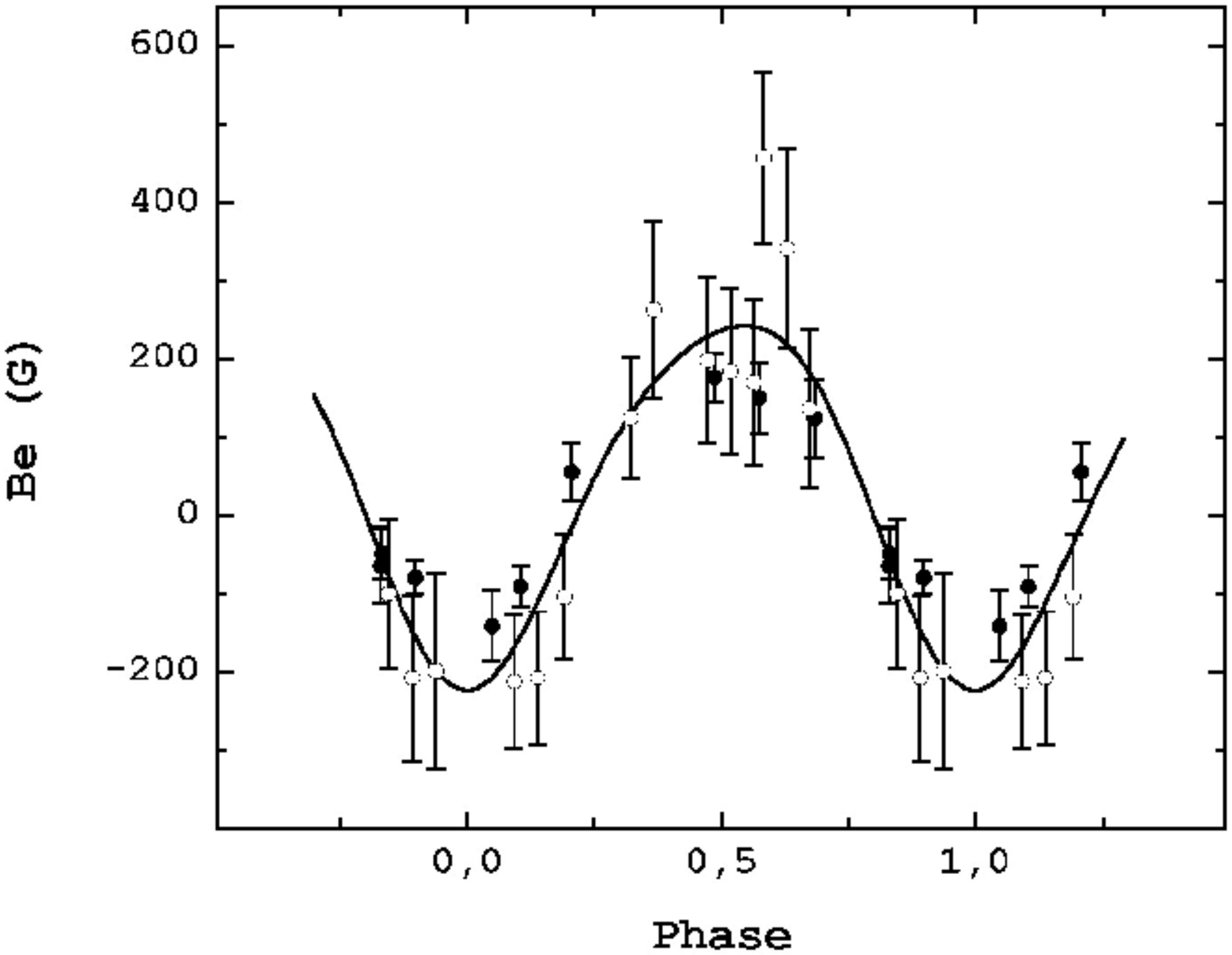}}
\vspace{-3.5mm}
\caption{ HD 43317 (1) }
\label{fig:fig104}
\end{figure}

\begin{figure}
\resizebox{0.98\hsize}{!}{\includegraphics{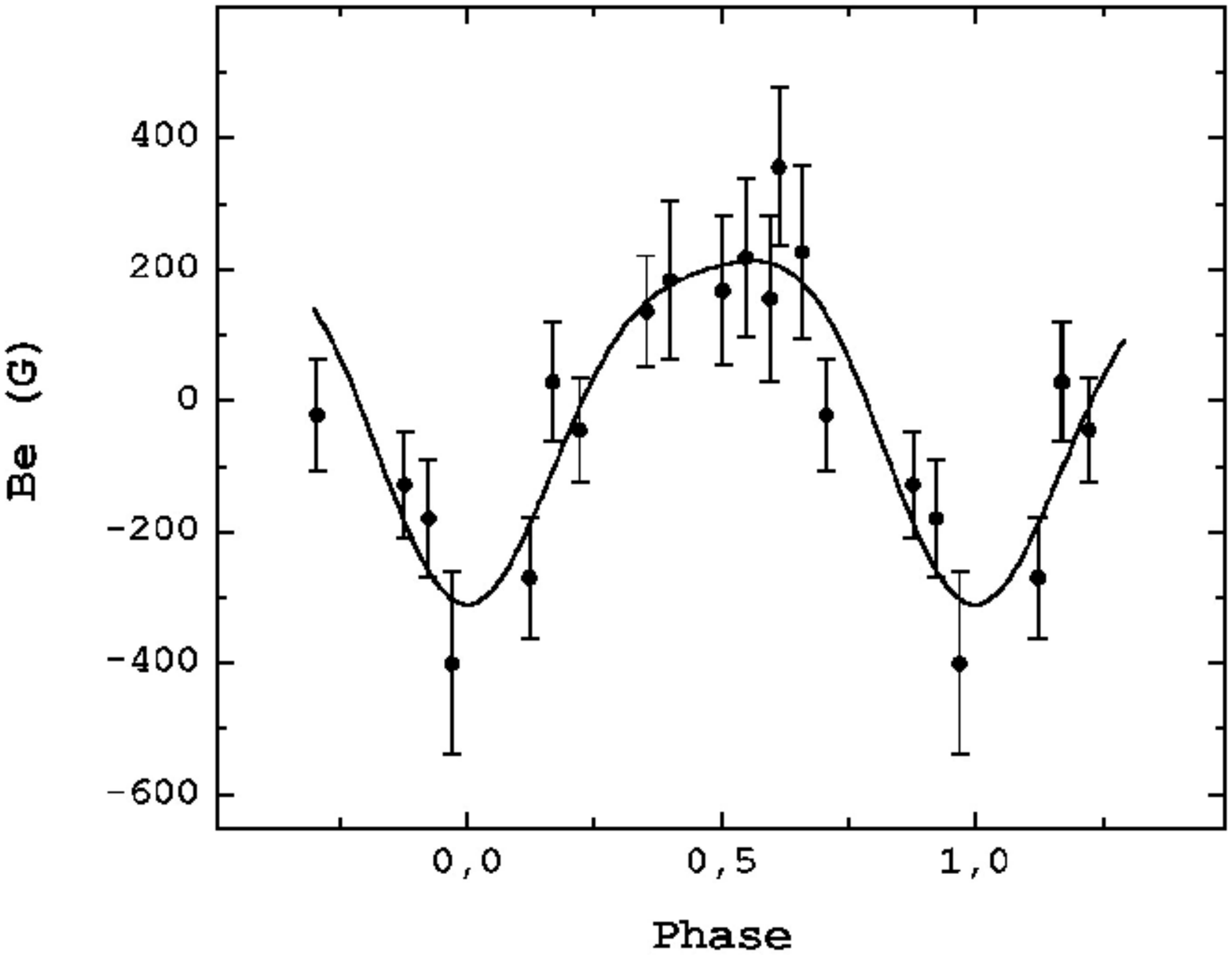}}
\vspace{-3.5mm}
\caption{ HD 43317 (2) }
\label{fig:fig104}
\end{figure}

\begin{figure}
\resizebox{0.98\hsize}{!}{\includegraphics{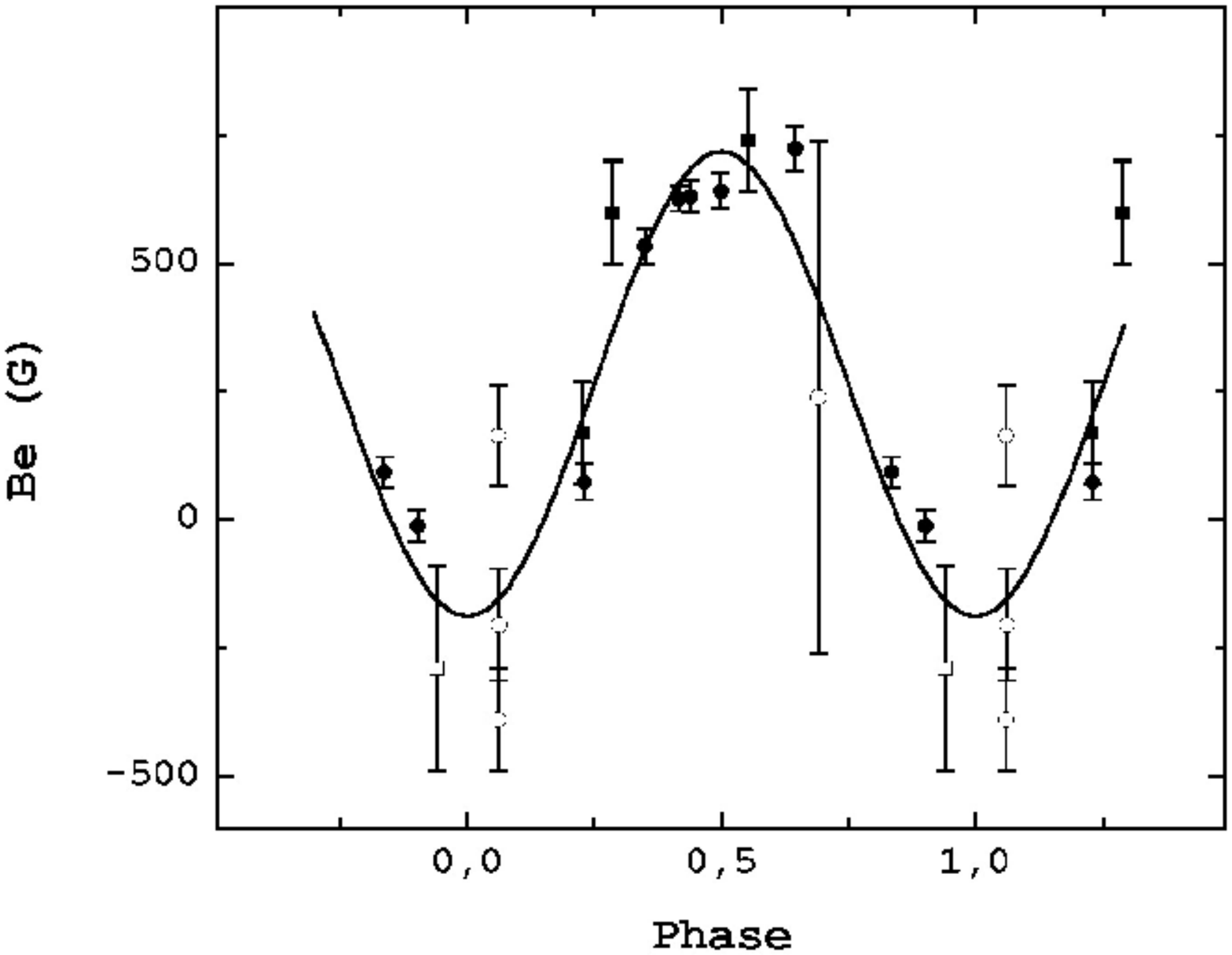}}
\vspace{-3.5mm}
\caption{ HD 43819 }
\label{fig:fig113}
\end{figure}

\clearpage
\newpage

\begin{figure}
\resizebox{0.98\hsize}{!}{\includegraphics{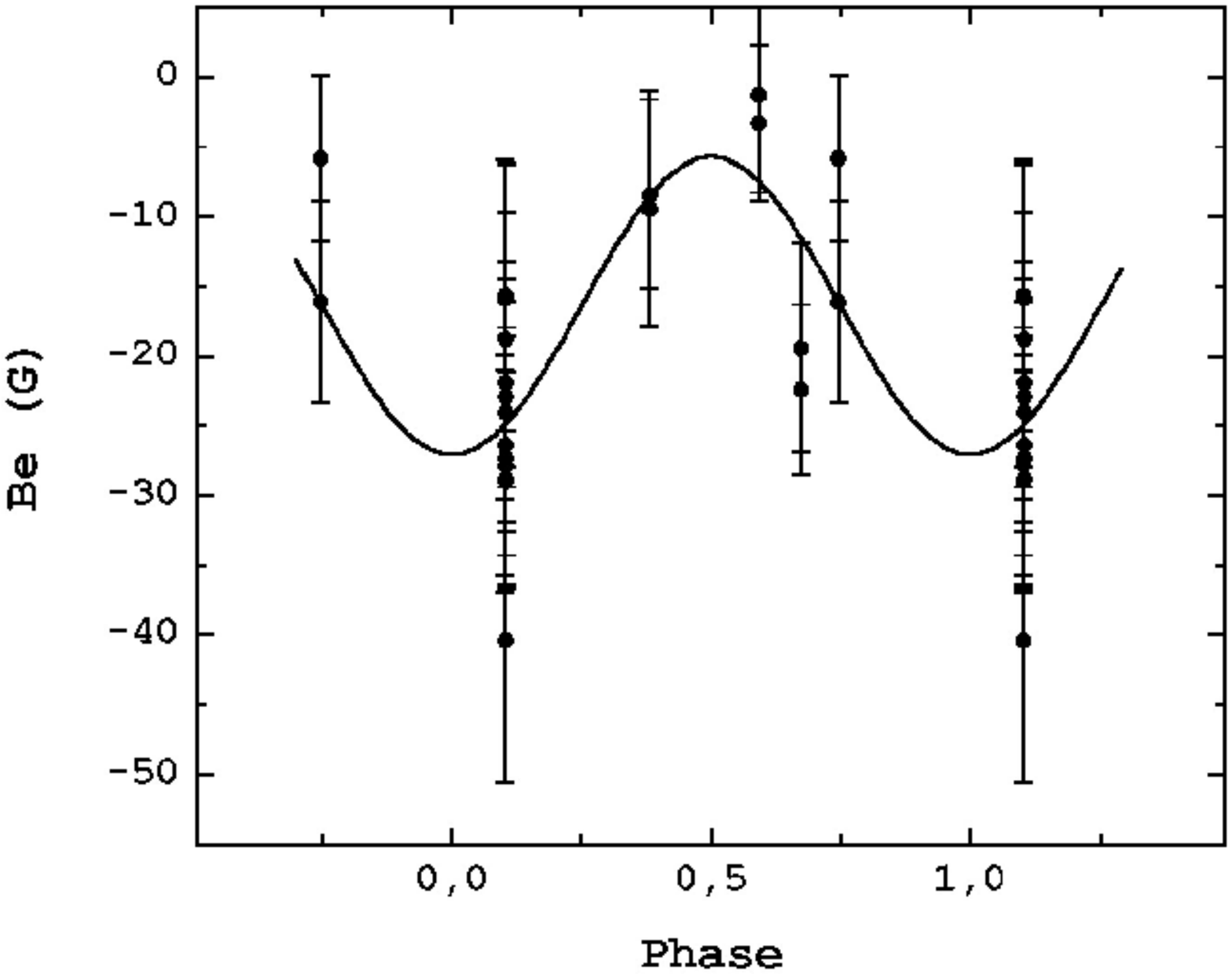}}
\vspace{-3.5mm}
\caption{ HD 44743 }
\label{fig:fig114}
\end{figure}

\begin{figure}
\resizebox{0.98\hsize}{!}{\includegraphics{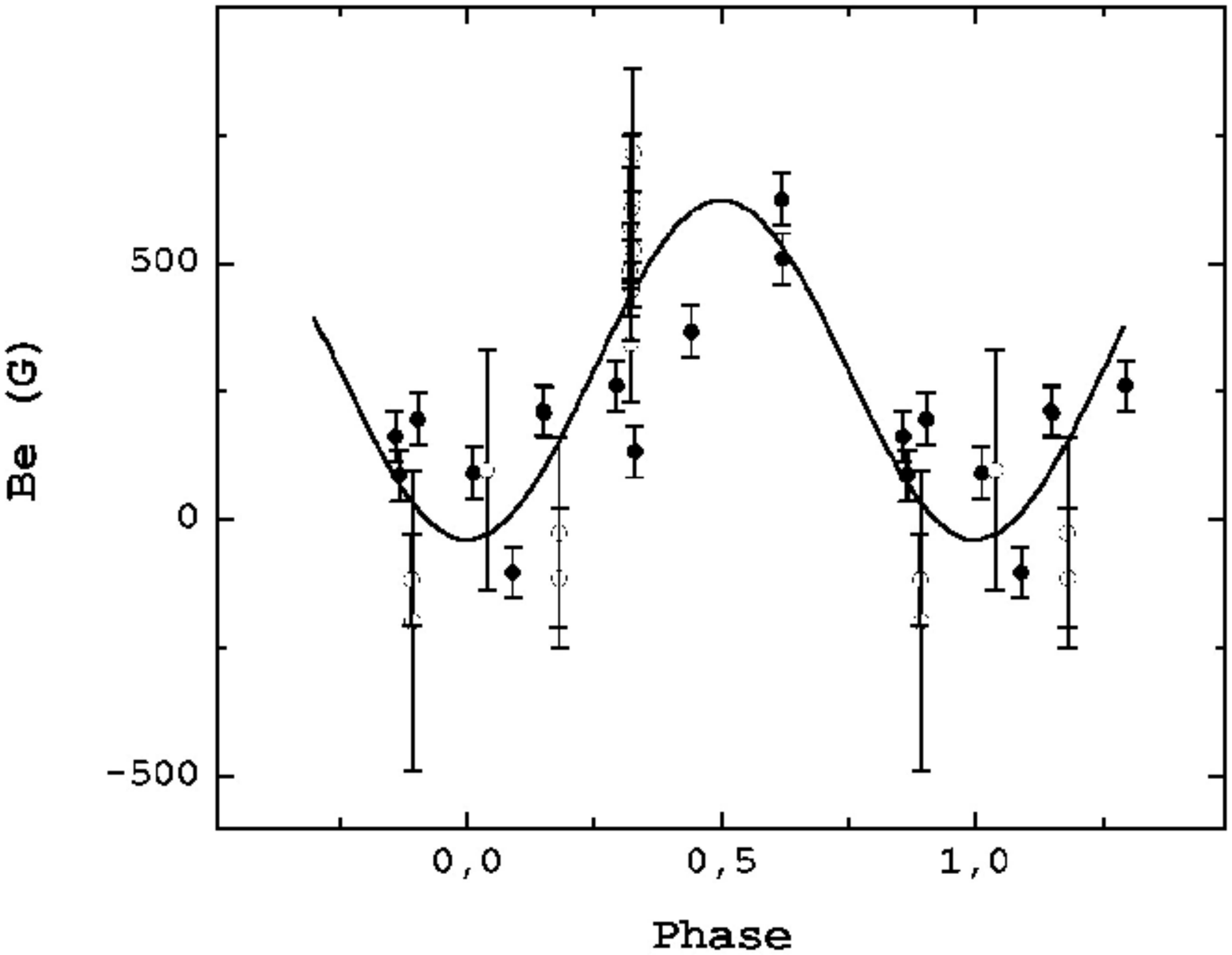}}
\vspace{-3.5mm}
\caption{ HD 45348 }
\label{fig:fig115}
\end{figure}

\begin{figure}
\resizebox{0.98\hsize}{!}{\includegraphics{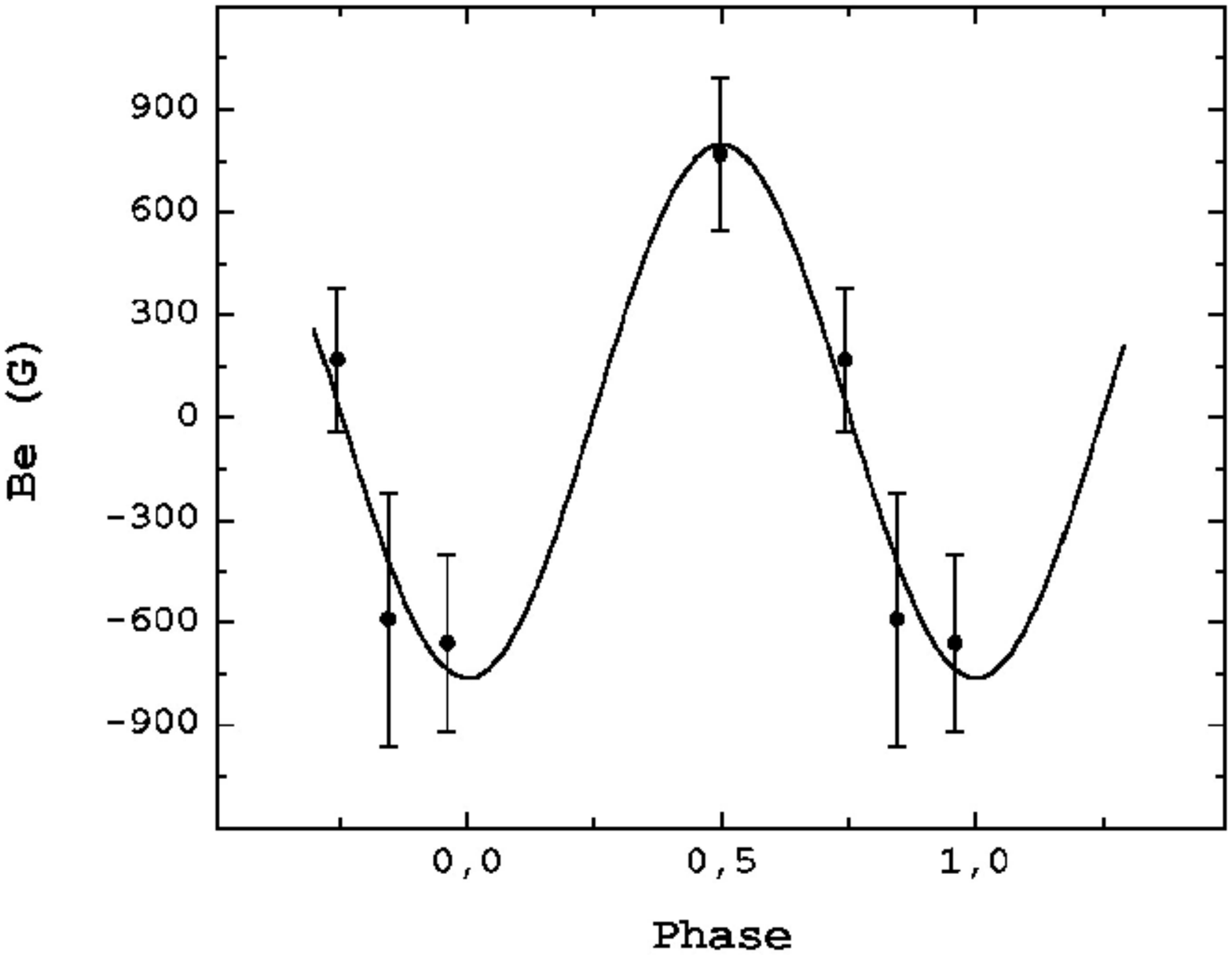}}
\vspace{-3.5mm}
\caption{ HD 45530 }
\label{fig:fig116}
\end{figure}

\begin{figure}
\resizebox{0.98\hsize}{!}{\includegraphics{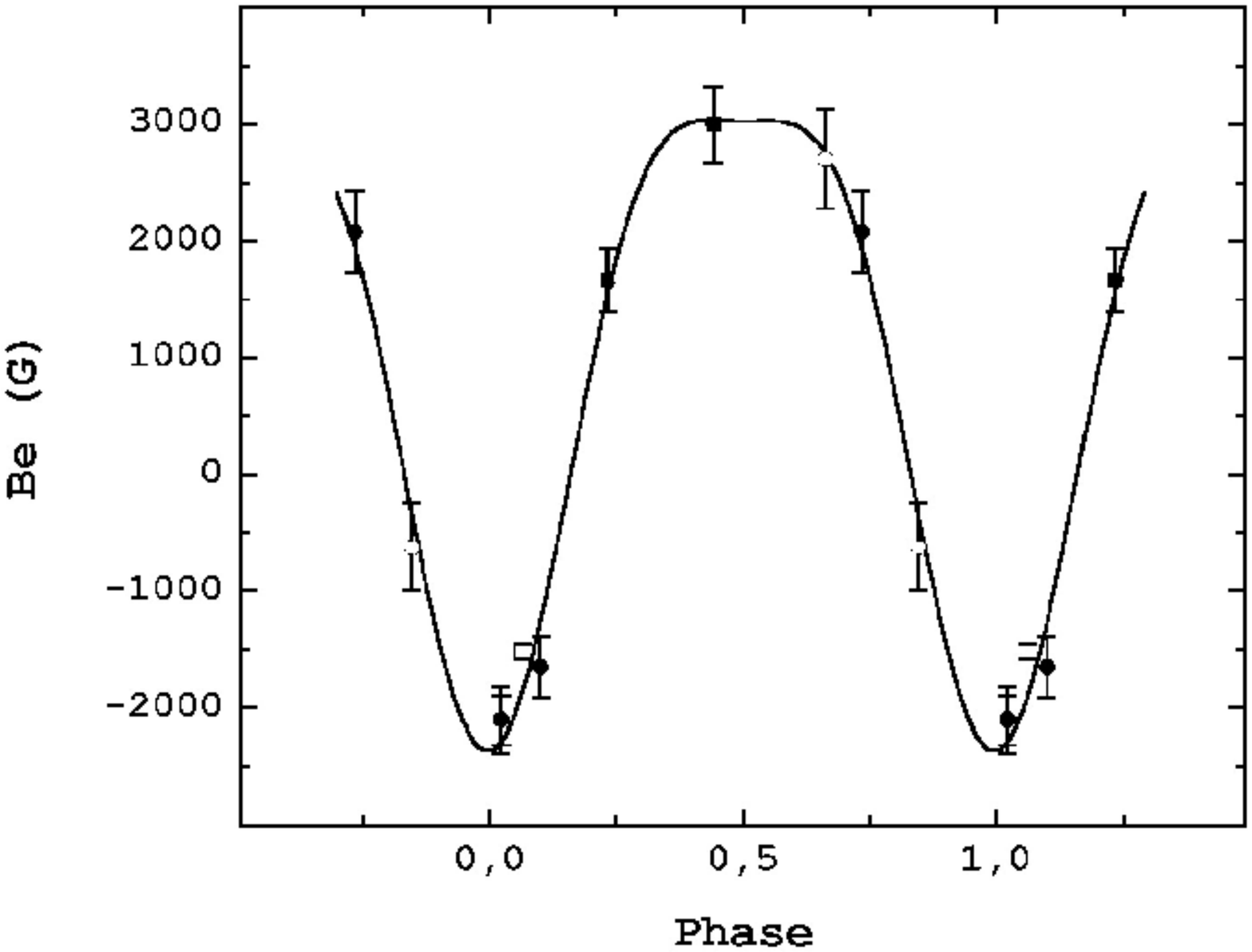}}
\vspace{-3.5mm}
\caption{ HD 45583 (1) }
\label{fig:fig117}
\end{figure}

\begin{figure}
\resizebox{0.98\hsize}{!}{\includegraphics{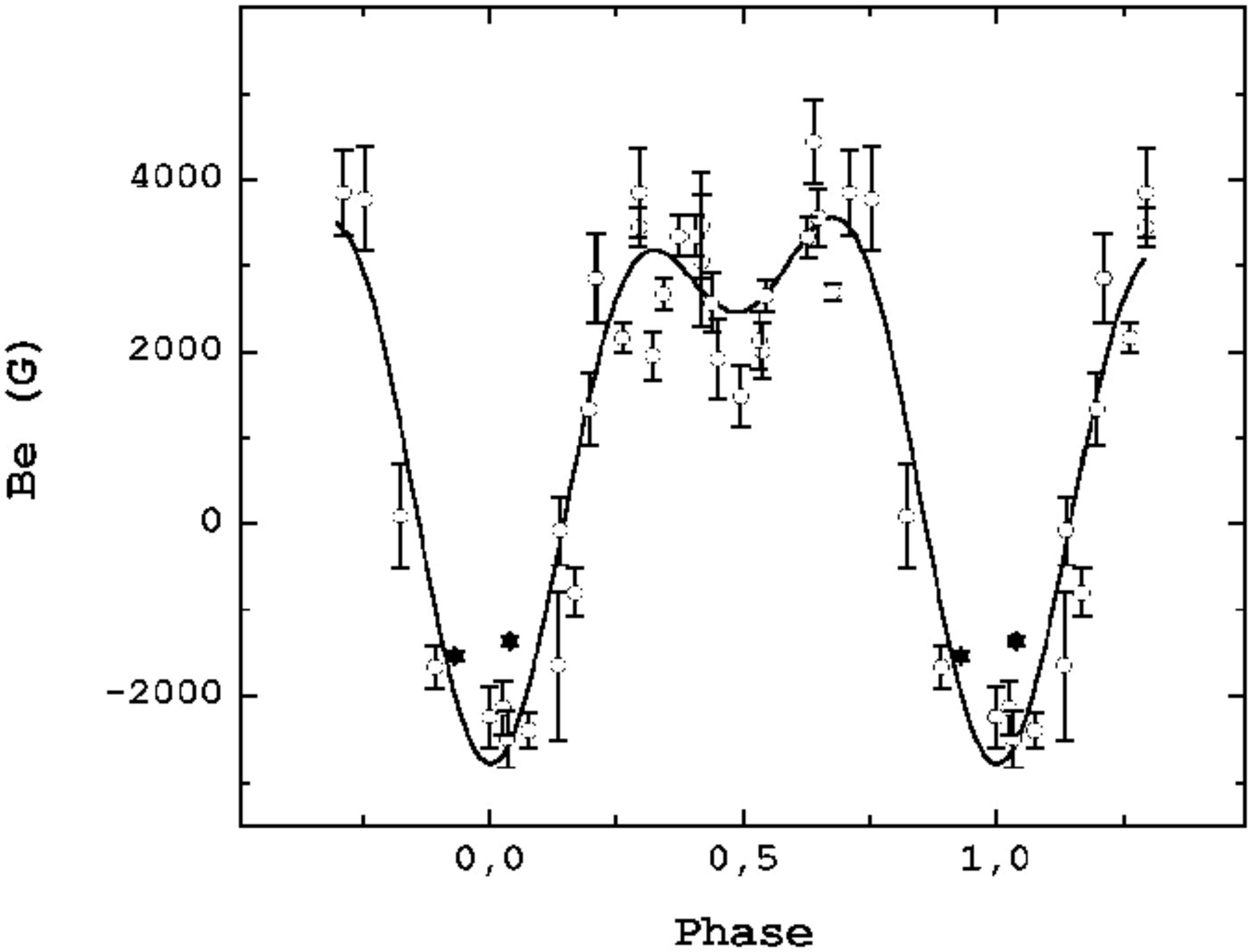}}
\vspace{-3.5mm}
\caption{ HD 45583 (2) }
\label{fig:fig118}
\end{figure}

\begin{figure}
\resizebox{0.98\hsize}{!}{\includegraphics{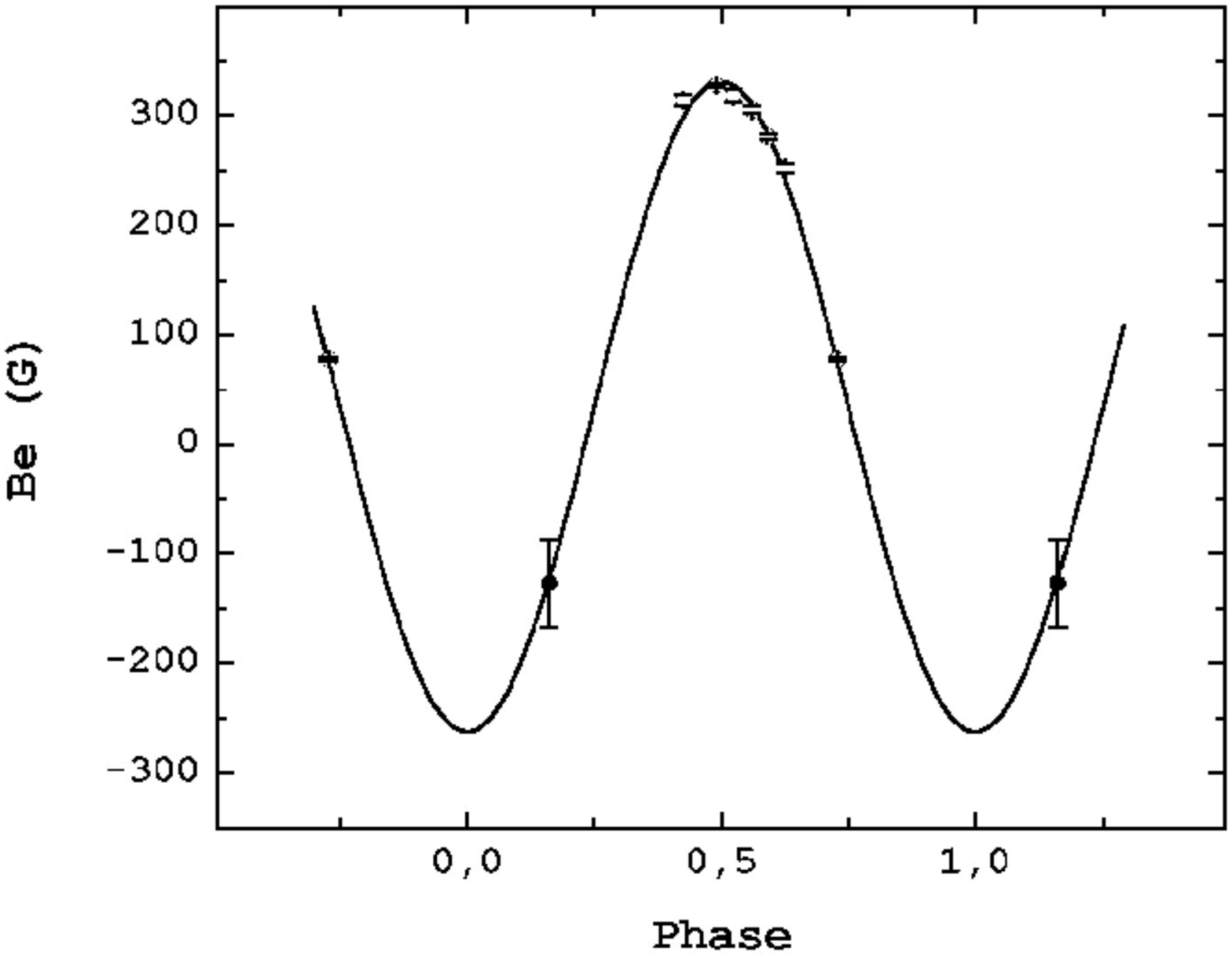}}
\vspace{-3.5mm}
\caption{ HD 46328 (1)}
\label{fig:fig119}
\end{figure}

\clearpage
\newpage

\begin{figure}
\resizebox{0.98\hsize}{!}{\includegraphics{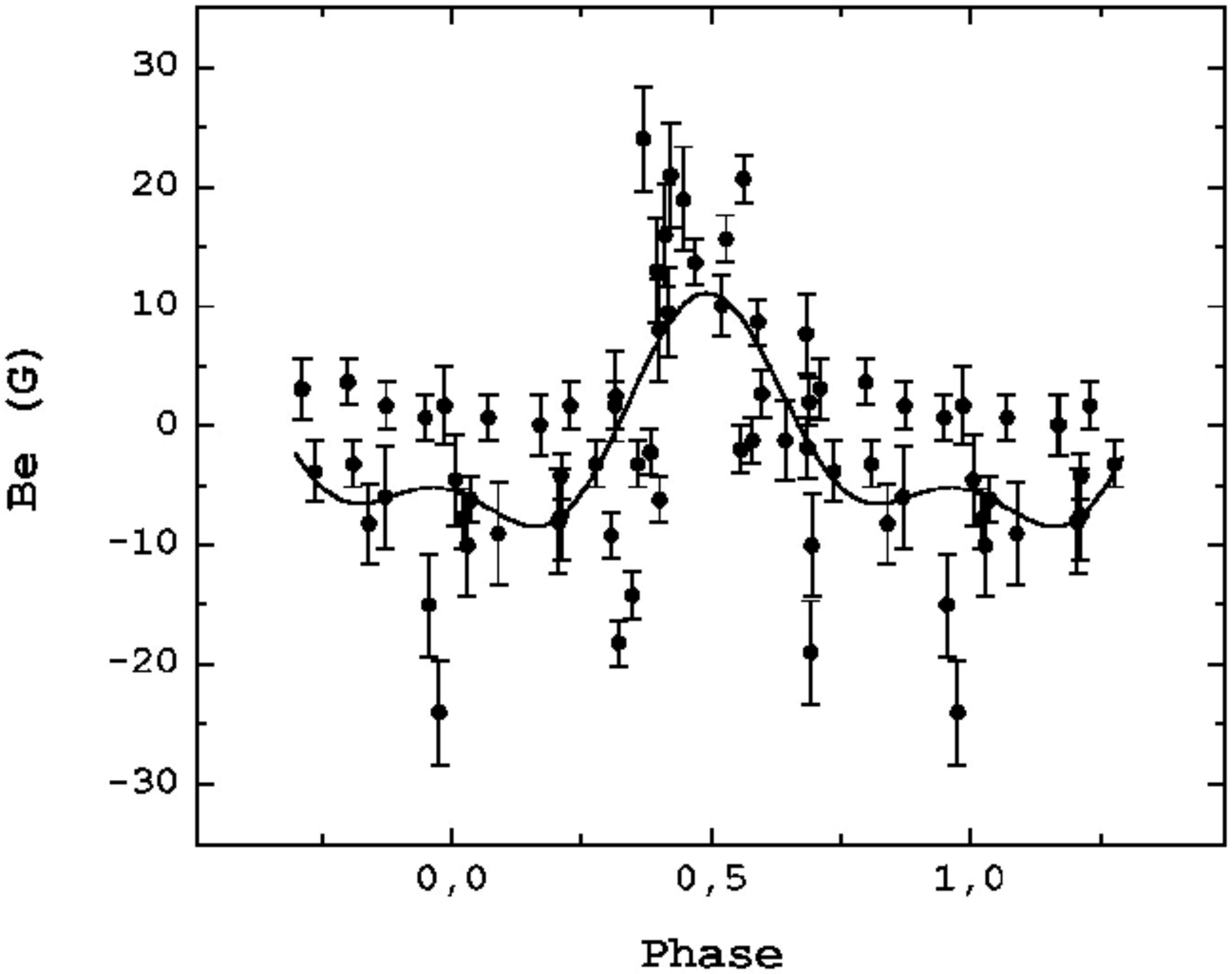}}
\vspace{-3.5mm}
\caption{ HD 46328 (2)}
\label{fig:fig120}
\end{figure}

\begin{figure}
\resizebox{0.98\hsize}{!}{\includegraphics{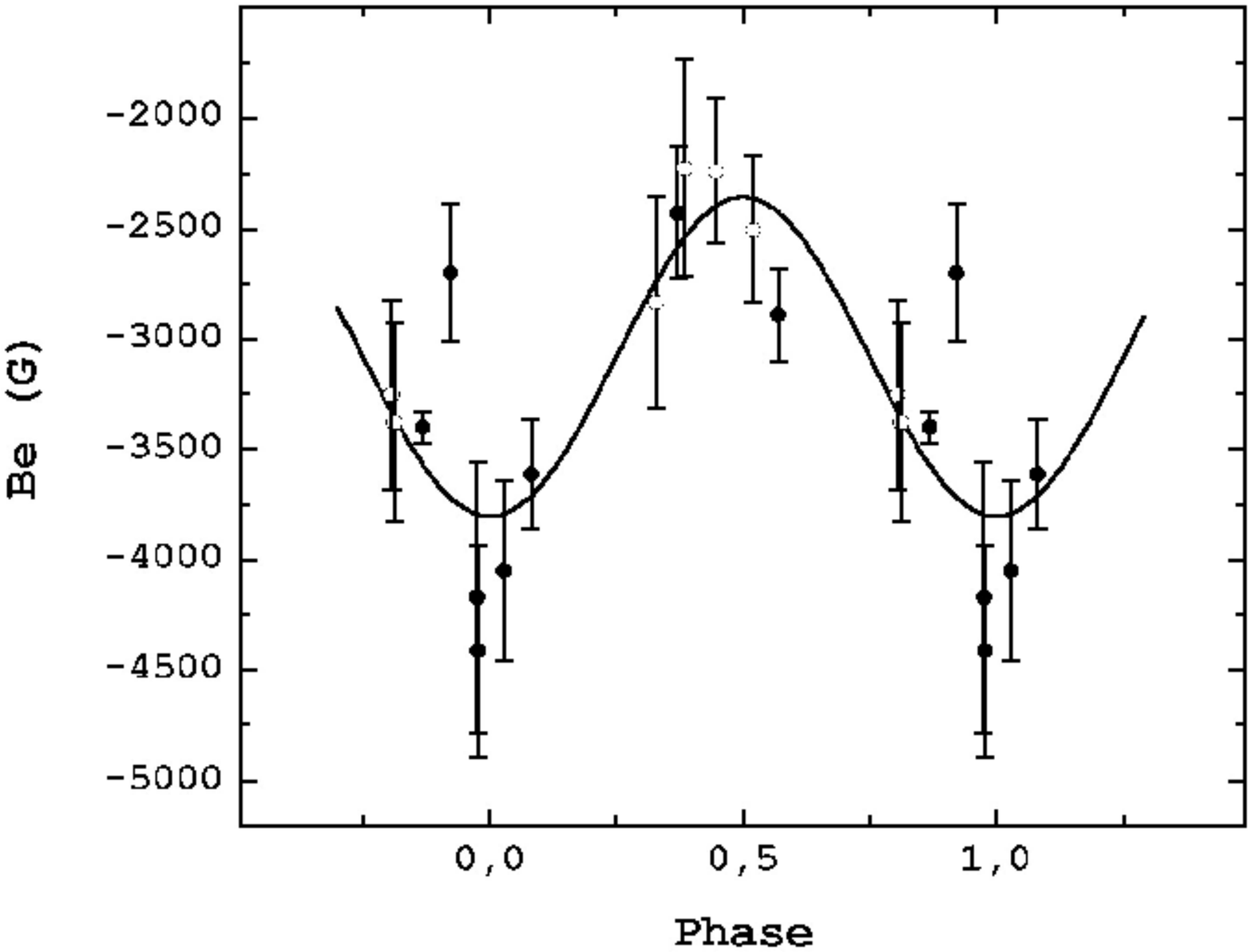}}
\vspace{-3.5mm}
\caption{ HD 47103 }
\label{fig:fig121}
\end{figure}

\begin{figure}
\resizebox{0.98\hsize}{!}{\includegraphics{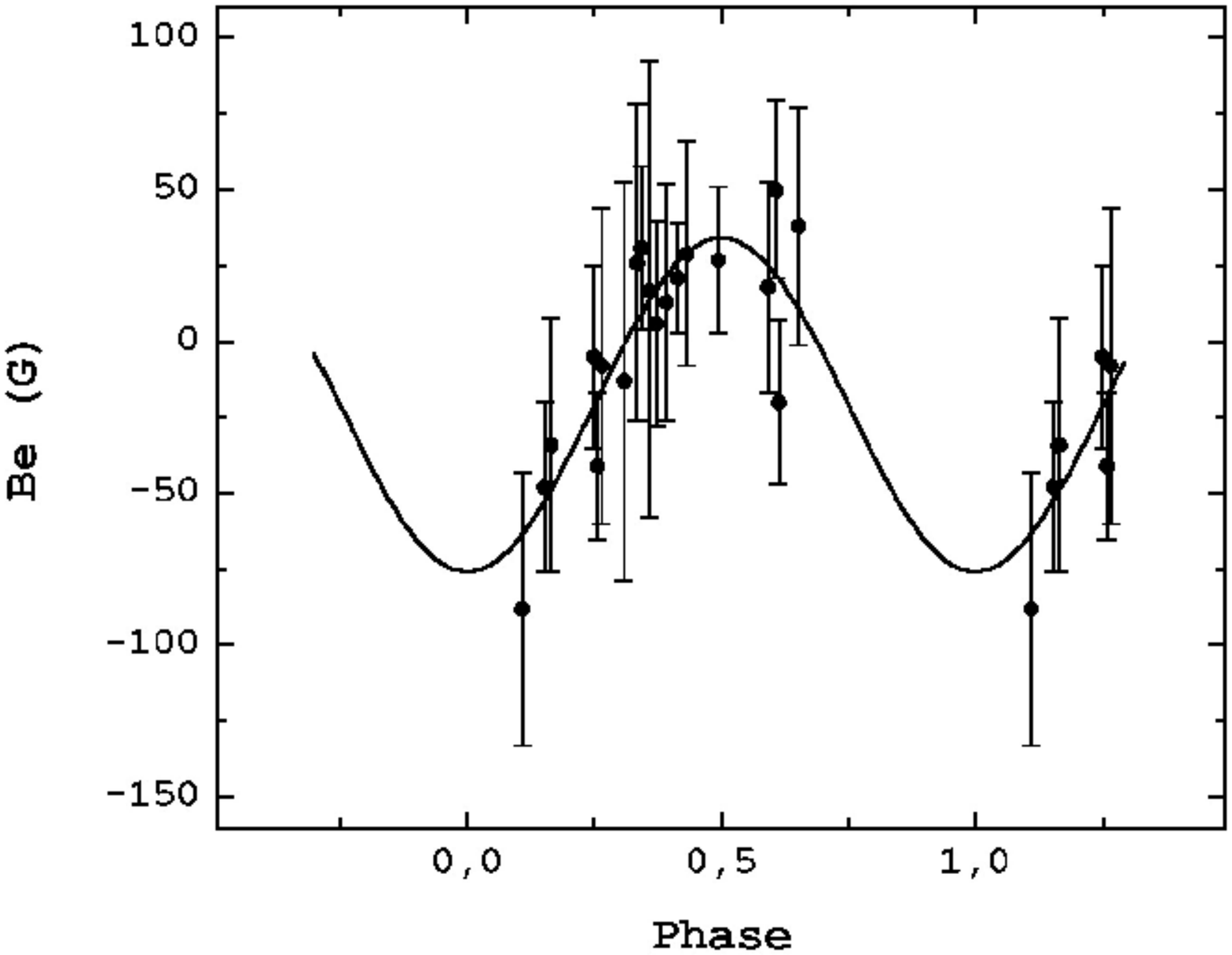}}
\vspace{-3.5mm}
\caption{ HD 47129 (1) }
\label{fig:fig122}
\end{figure}

\begin{figure}
\resizebox{0.98\hsize}{!}{\includegraphics{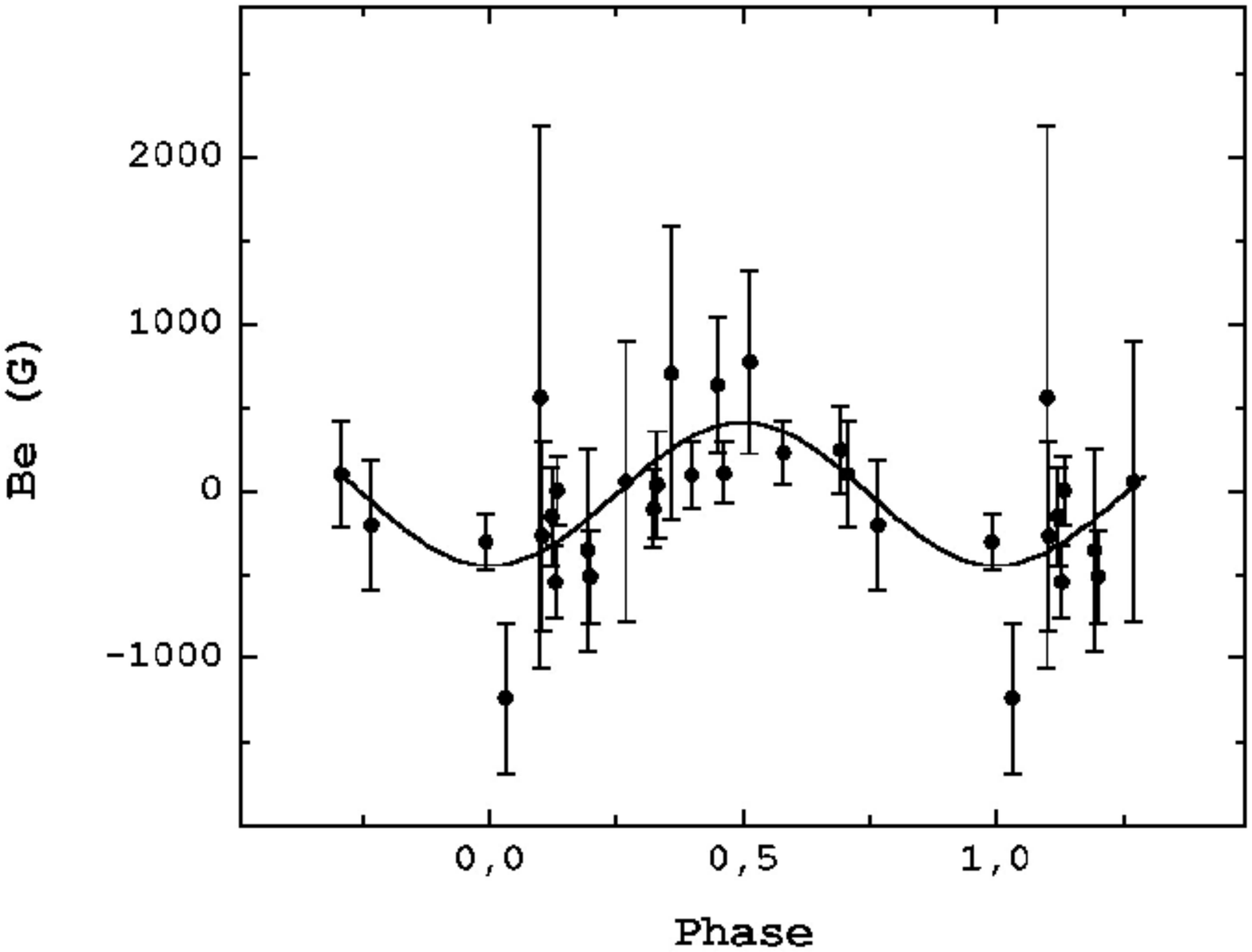}}
\vspace{-3.5mm}
\caption{ HD 47129 (2) }
\label{fig:fig123}
\end{figure}

\begin{figure}
\resizebox{0.98\hsize}{!}{\includegraphics{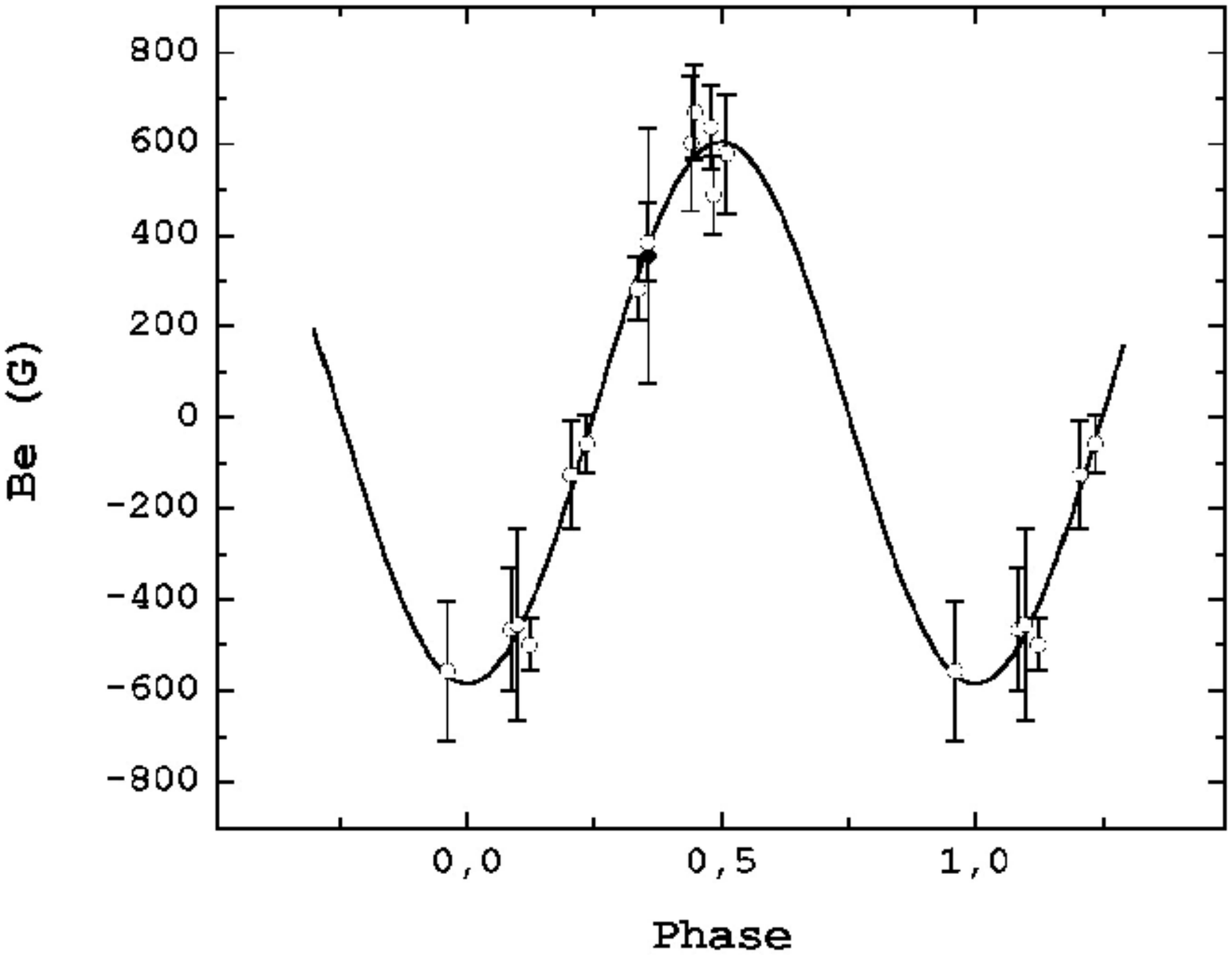}}
\vspace{-3.5mm}
\caption{ HD 47777 }
\label{fig:fig104}
\end{figure}

\begin{figure}
\resizebox{0.98\hsize}{!}{\includegraphics{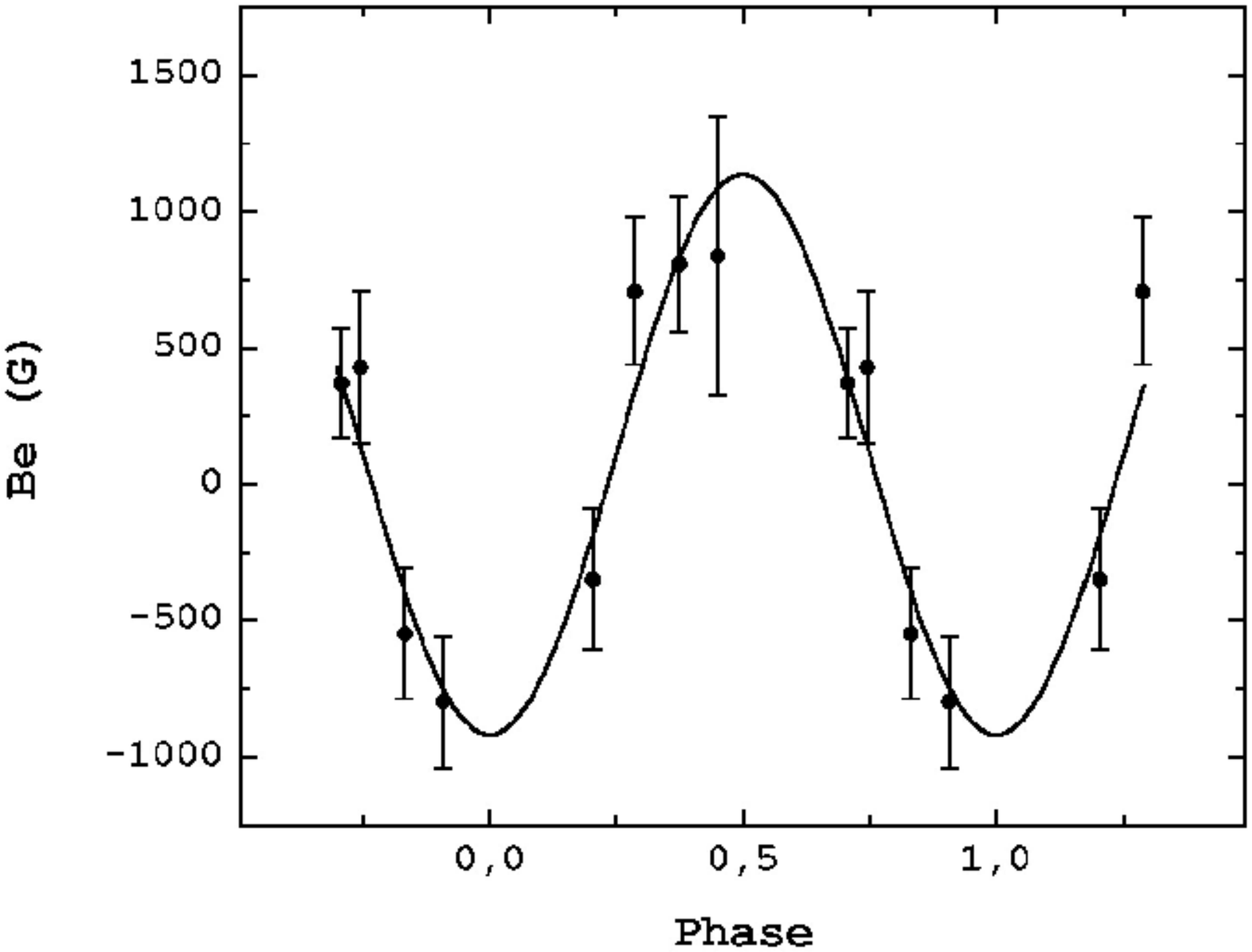}}
\vspace{-3.5mm}
\caption{ HD 49333 }
\label{fig:fig124}
\end{figure}

\clearpage
\newpage

\begin{figure}
\resizebox{0.98\hsize}{!}{\includegraphics{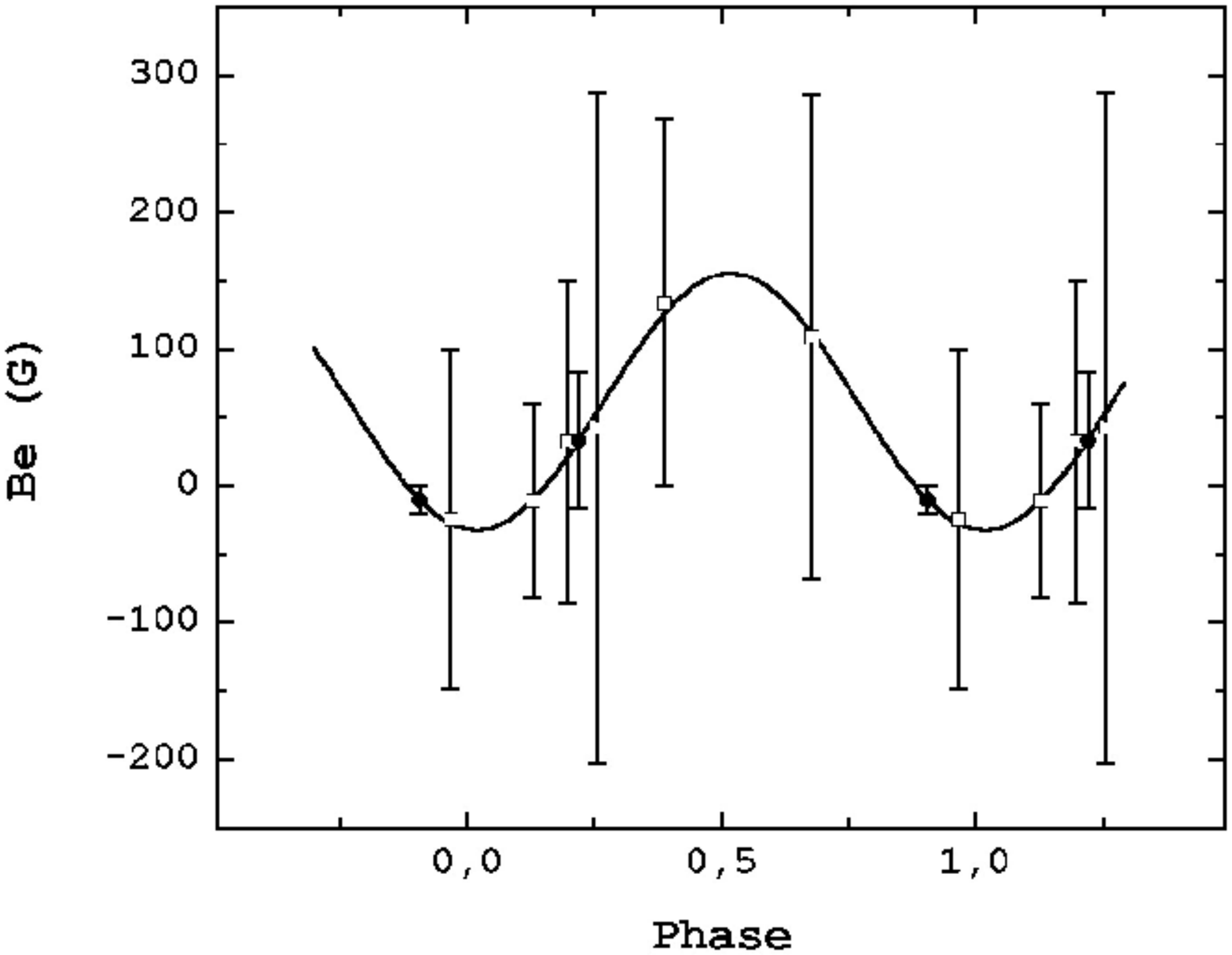}}
\vspace{-3.5mm}
\caption{ HD 49606 }
\label{fig:fig125}
\end{figure}

\begin{figure}
\resizebox{0.98\hsize}{!}{\includegraphics{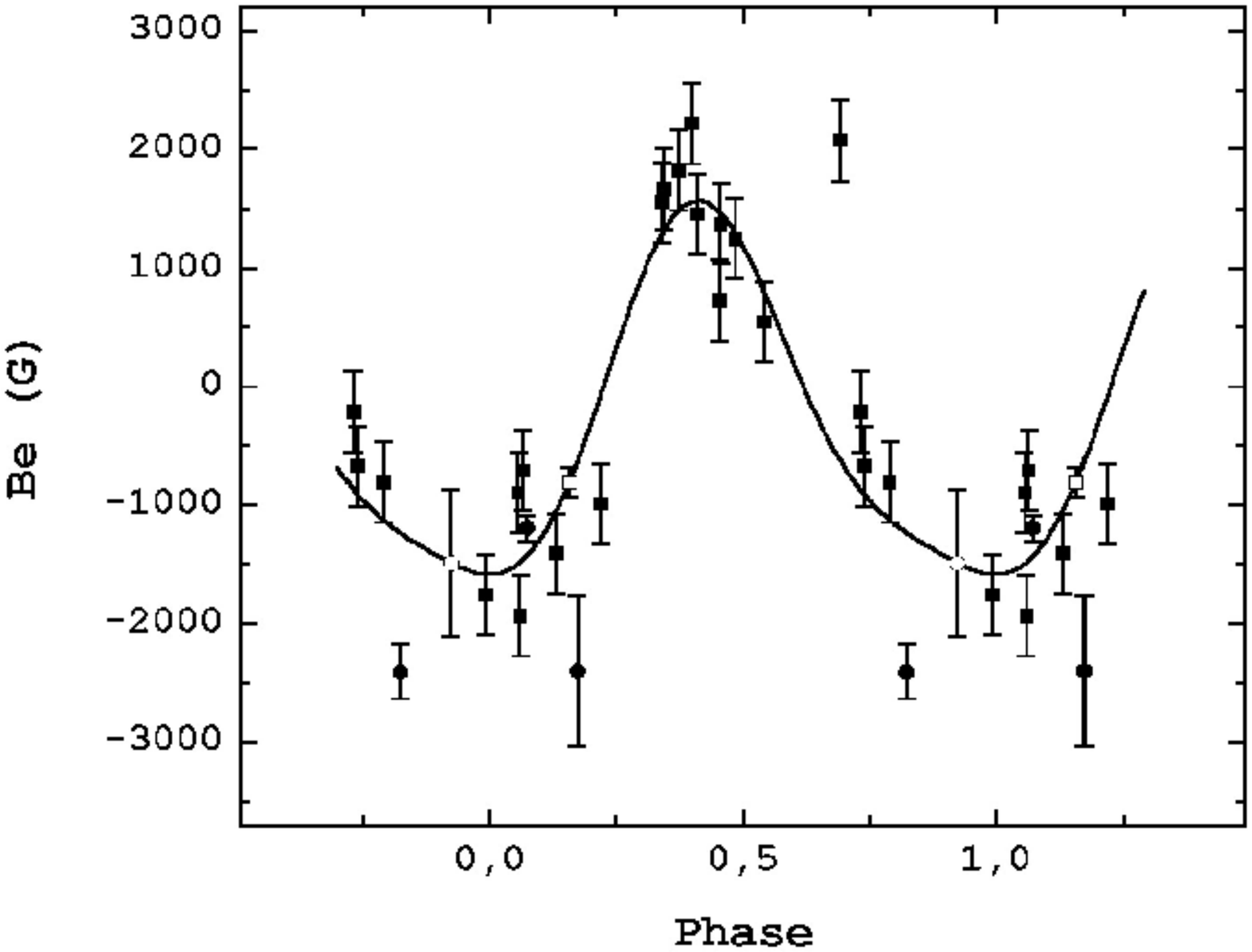}}
\vspace{-3.5mm}
\caption{ HD 49976 }
\label{fig:fig126}
\end{figure}

\begin{figure}
\resizebox{0.98\hsize}{!}{\includegraphics{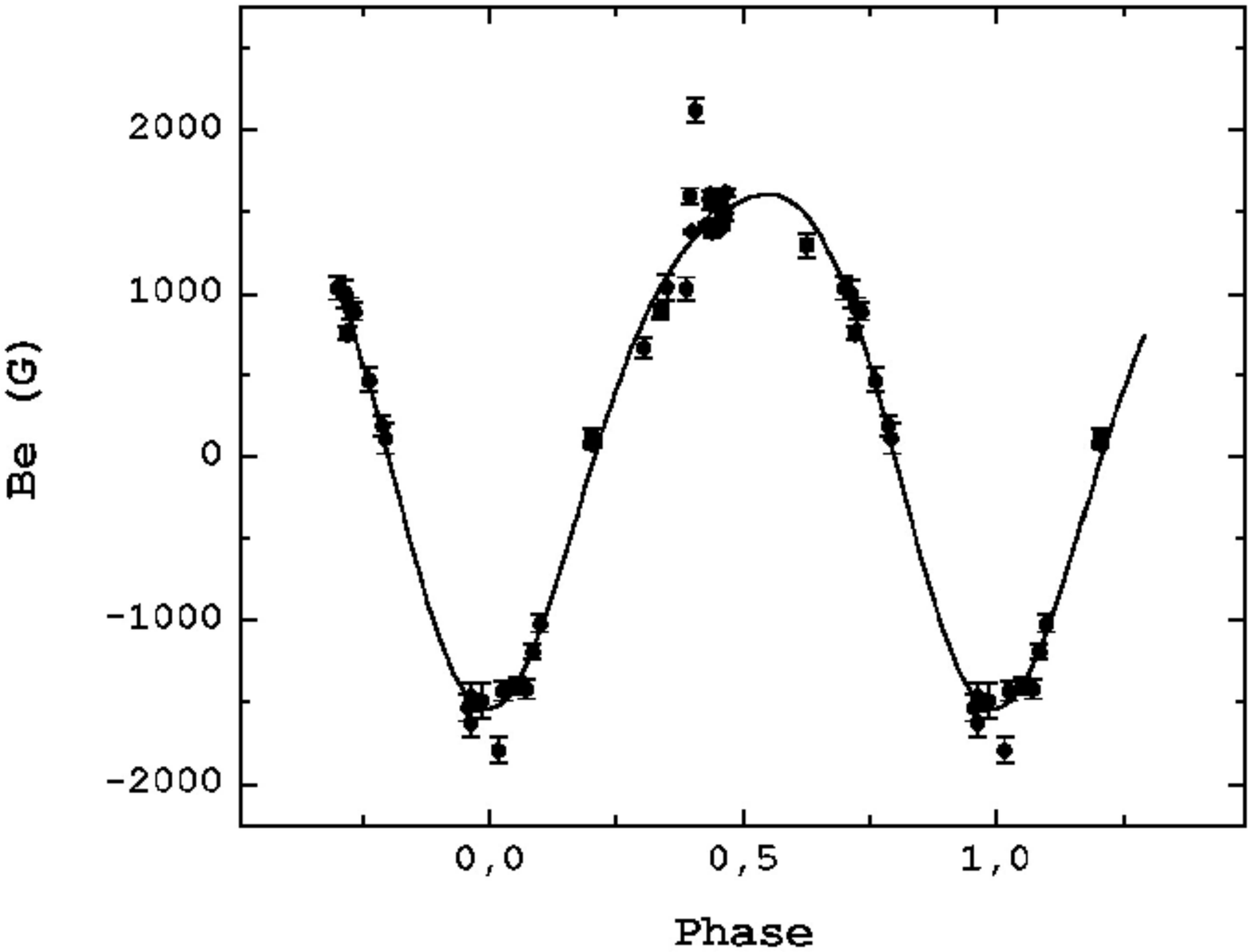}}
\vspace{-3.5mm}
\caption{ HD 50169 (1)}
\label{fig:fig127}
\end{figure}

\begin{figure}
\resizebox{0.98\hsize}{!}{\includegraphics{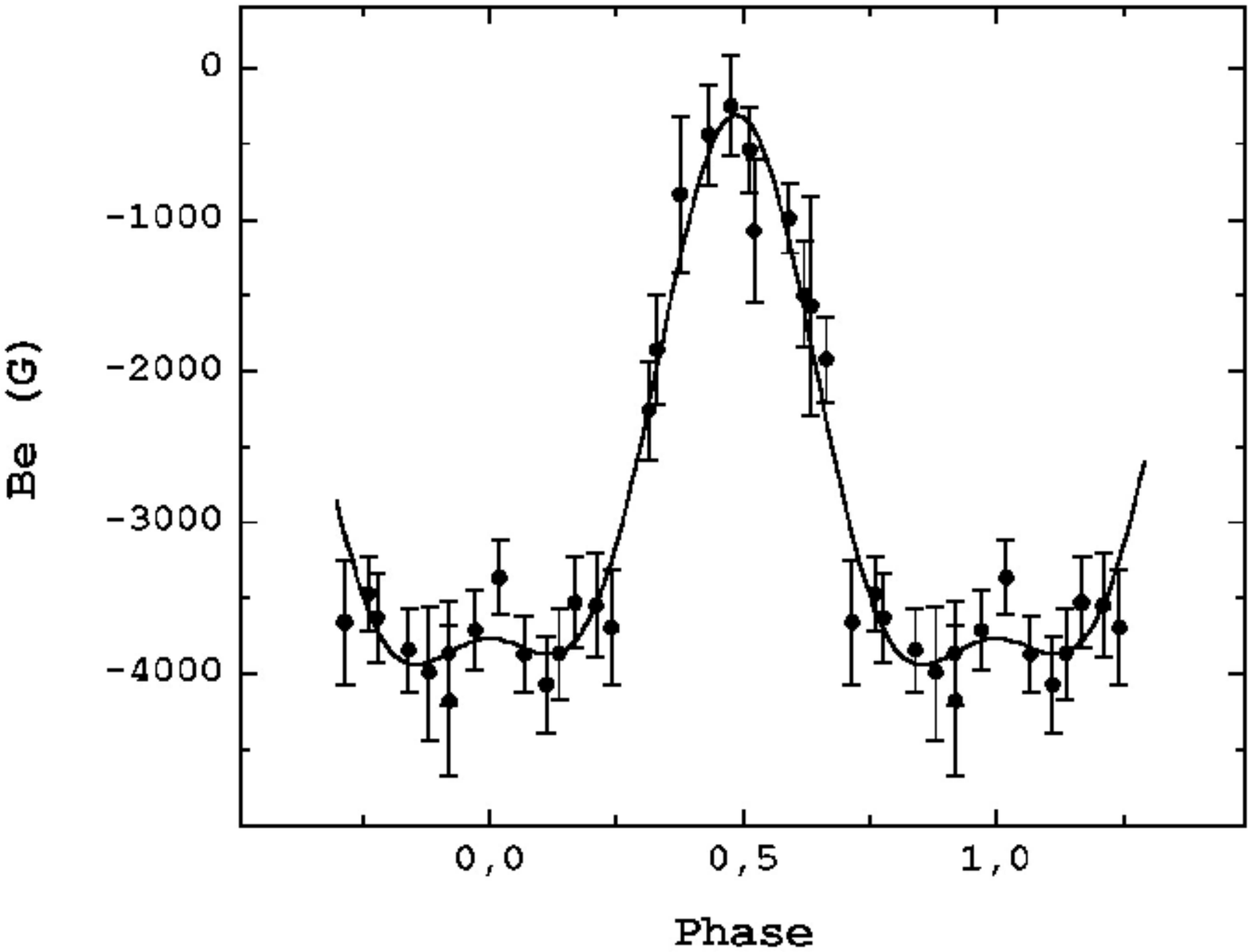}}
\vspace{-3.5mm}
\caption{ HD 50169 (2) }
\label{fig:fig104}
\end{figure}

\begin{figure}
\resizebox{0.98\hsize}{!}{\includegraphics{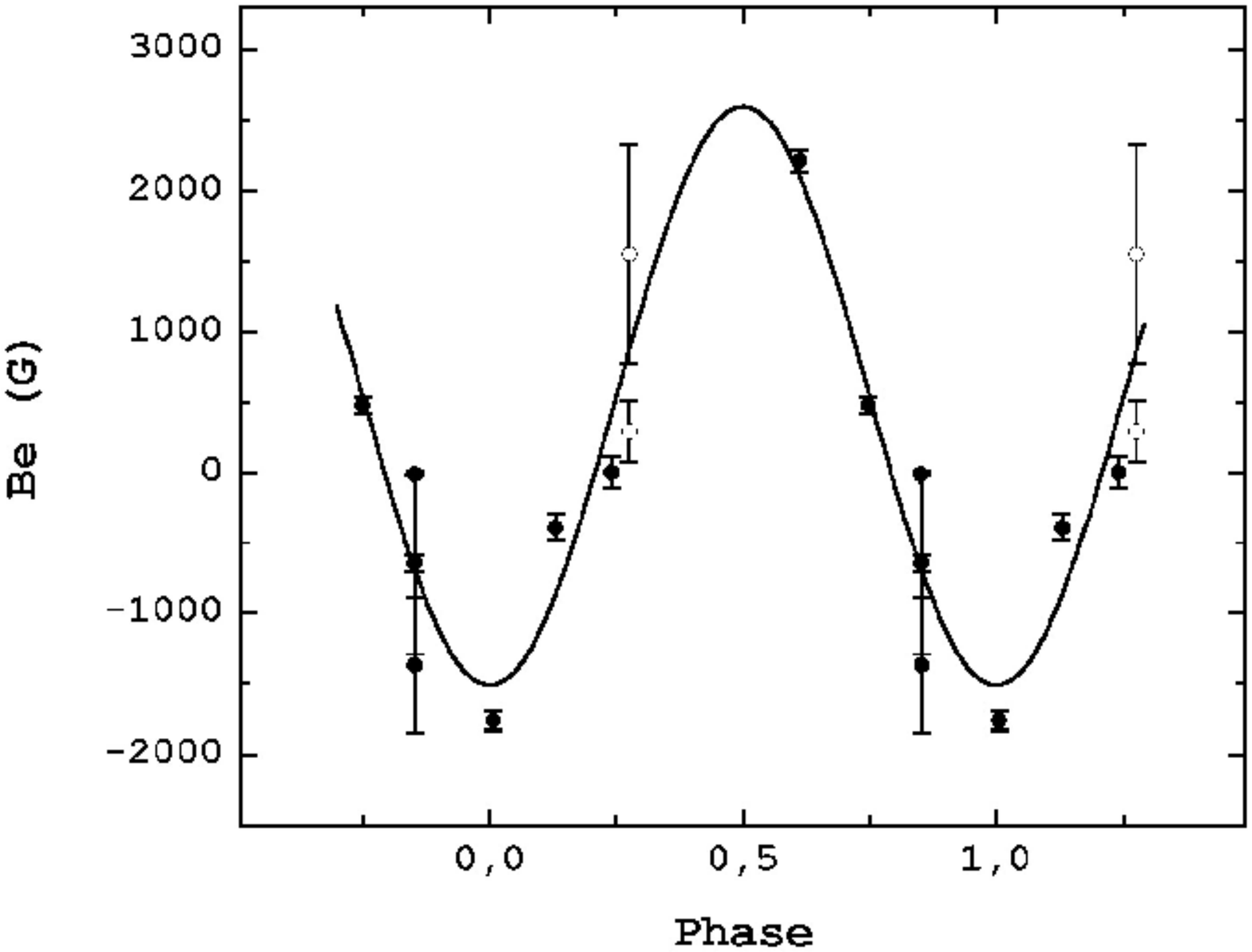}}
\vspace{-3.5mm}
\caption{ HD 50461 }
\label{fig:fig128}
\end{figure}

\begin{figure}
\resizebox{0.98\hsize}{!}{\includegraphics{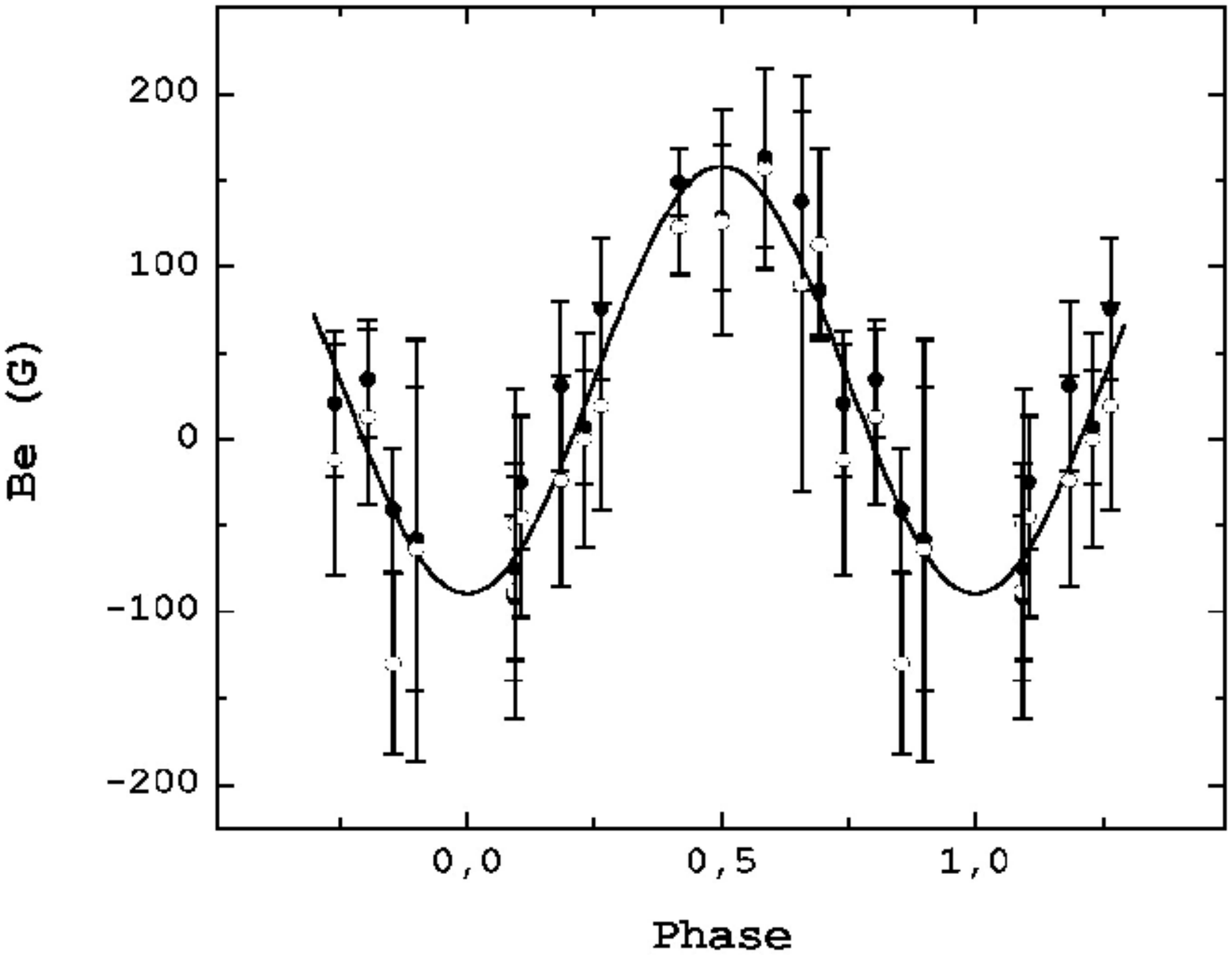}}
\vspace{-3.5mm}
\caption{ HD 50707 }
\label{fig:fig129}
\end{figure}

\clearpage
\newpage

\begin{figure}
\resizebox{0.98\hsize}{!}{\includegraphics{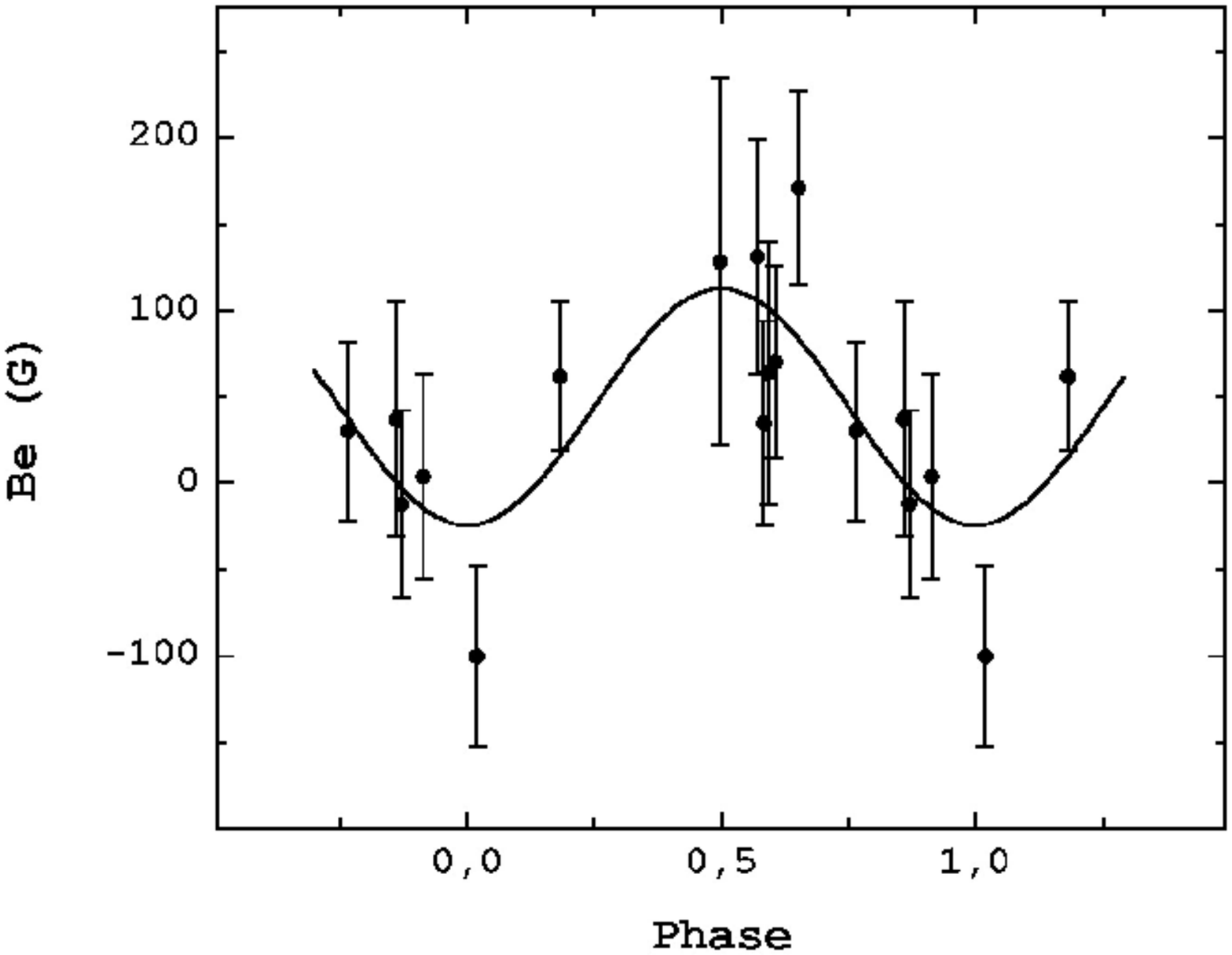}}
\vspace{-3.5mm}
\caption{ HD 50896 }
\label{fig:fig130}
\end{figure}

\begin{figure}
\resizebox{0.98\hsize}{!}{\includegraphics{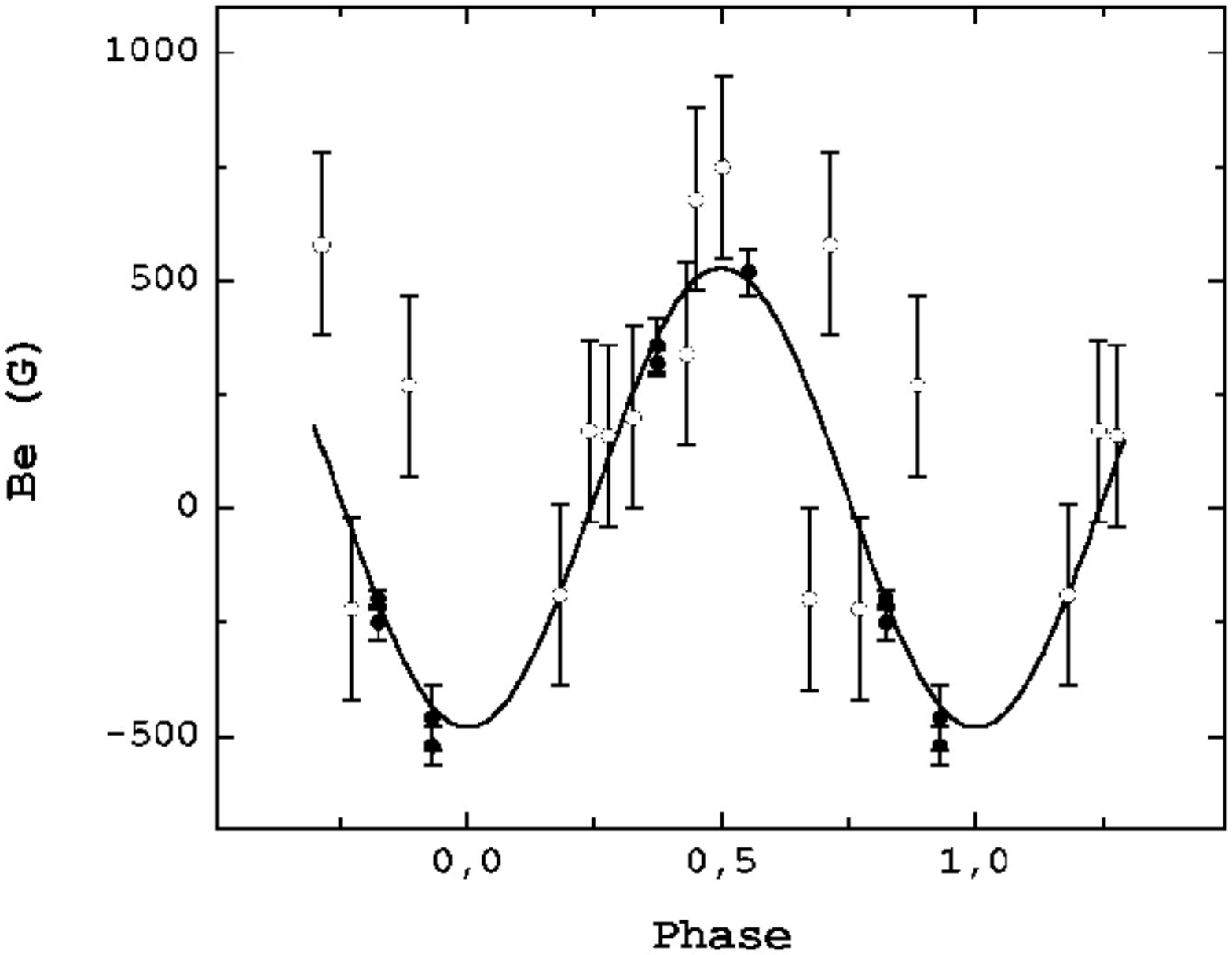}}
\vspace{-3.5mm}
\caption{ HD 51418 }
\label{fig:fig131}
\end{figure}

\begin{figure}
\resizebox{0.98\hsize}{!}{\includegraphics{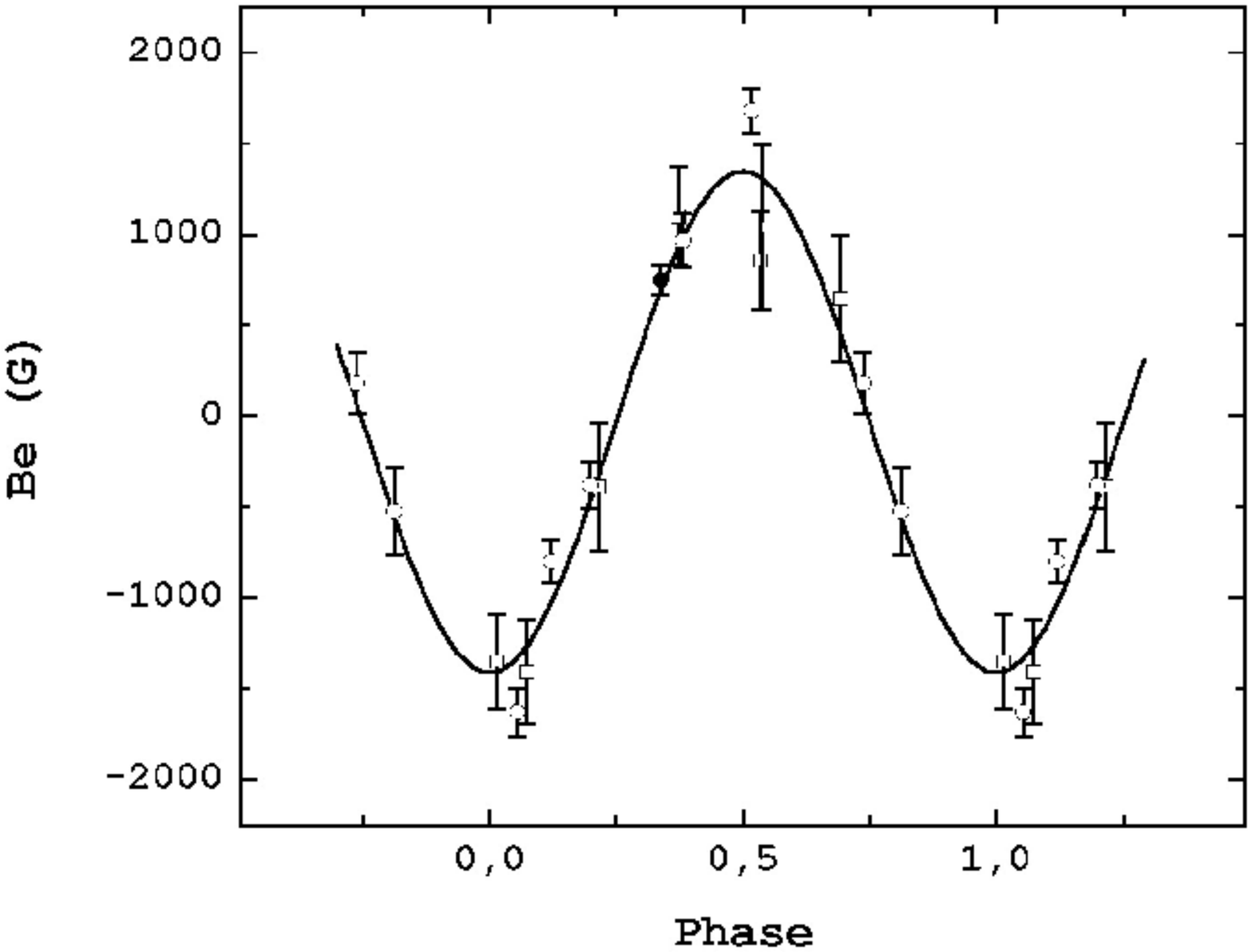}}
\vspace{-3.5mm}
\caption{ HD 54118 }
\label{fig:fig131}
\end{figure}

\begin{figure}
\resizebox{0.98\hsize}{!}{\includegraphics{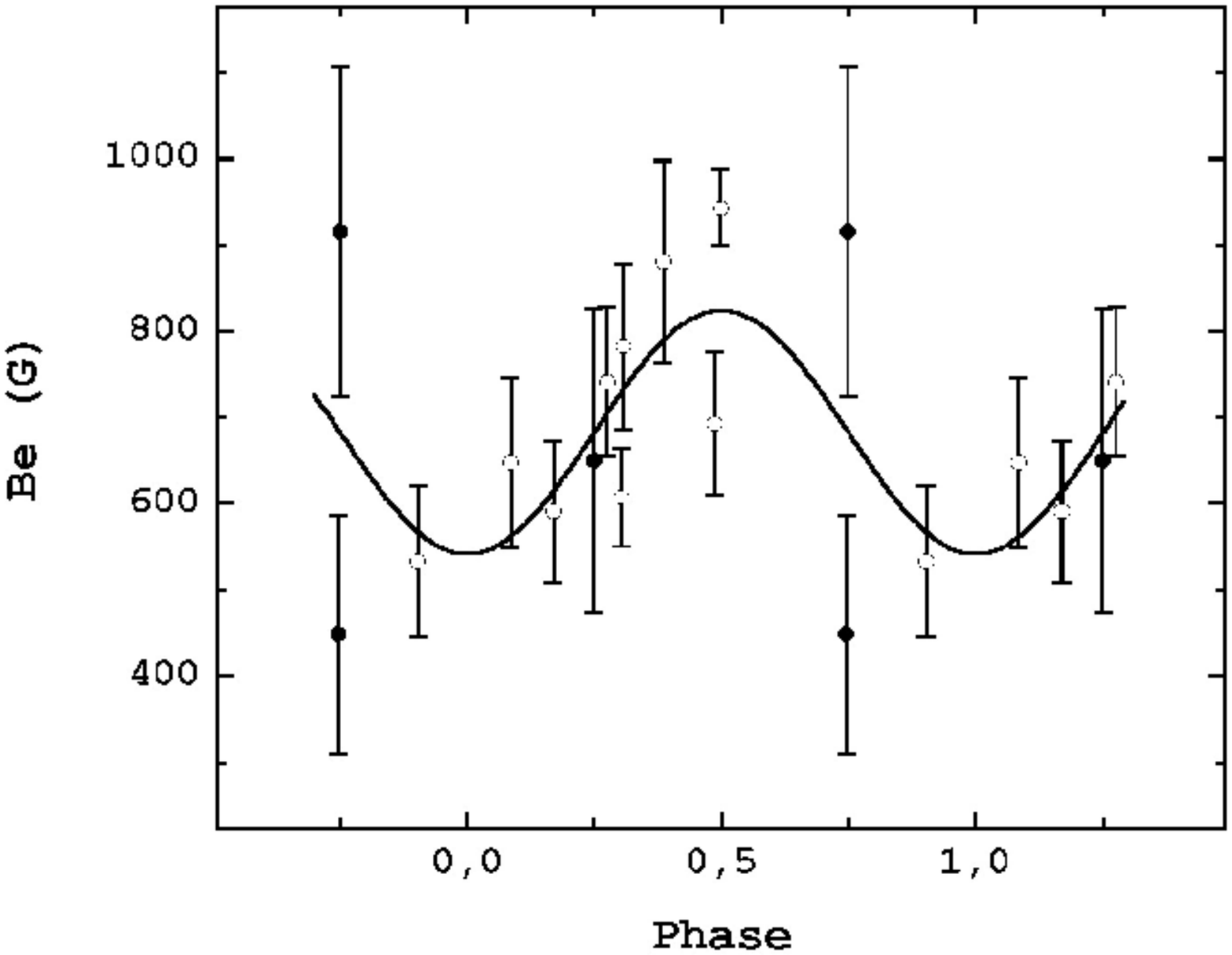}}
\vspace{-3.5mm}
\caption{ HD 55719 }
\label{fig:fig132}
\end{figure}

\begin{figure}
\resizebox{0.98\hsize}{!}{\includegraphics{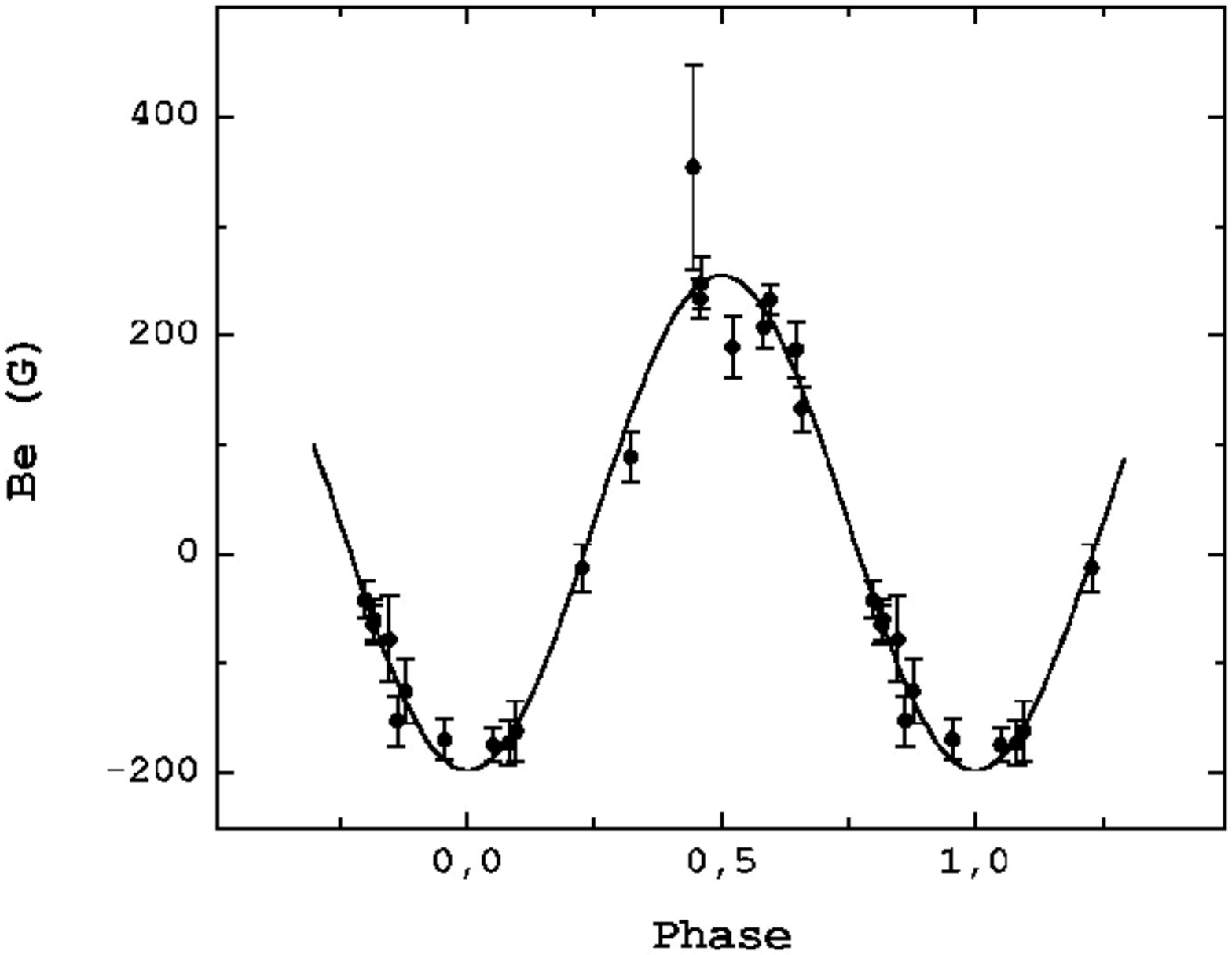}}
\vspace{-3.5mm}
\caption{ HD 57682 }
\label{fig:fig133}
\end{figure}

\begin{figure}
\resizebox{0.98\hsize}{!}{\includegraphics{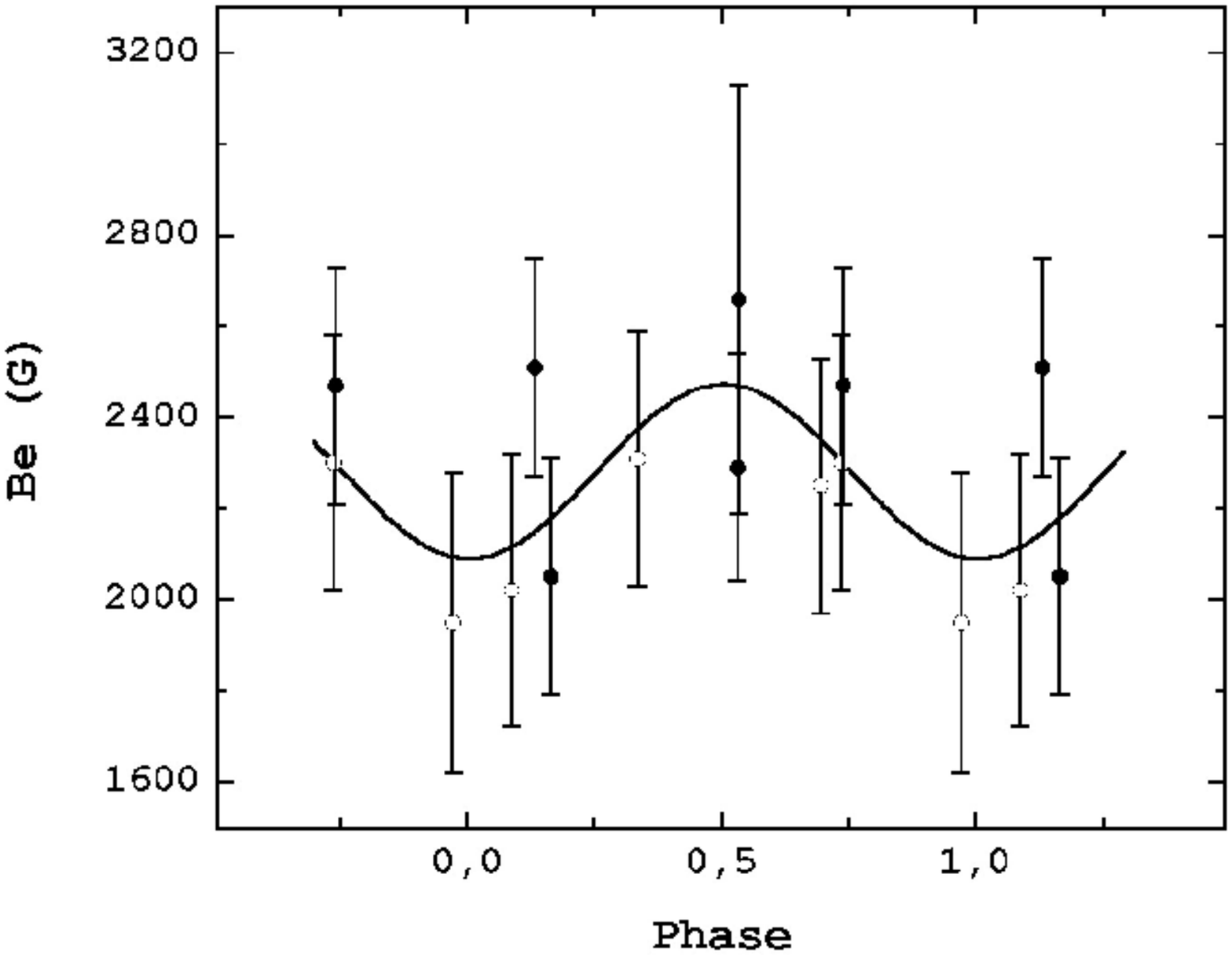}}
\vspace{-3.5mm}
\caption{ HD 58260 }
\label{fig:fig134}
\end{figure}

\clearpage
\newpage

\begin{figure}
\resizebox{0.98\hsize}{!}{\includegraphics{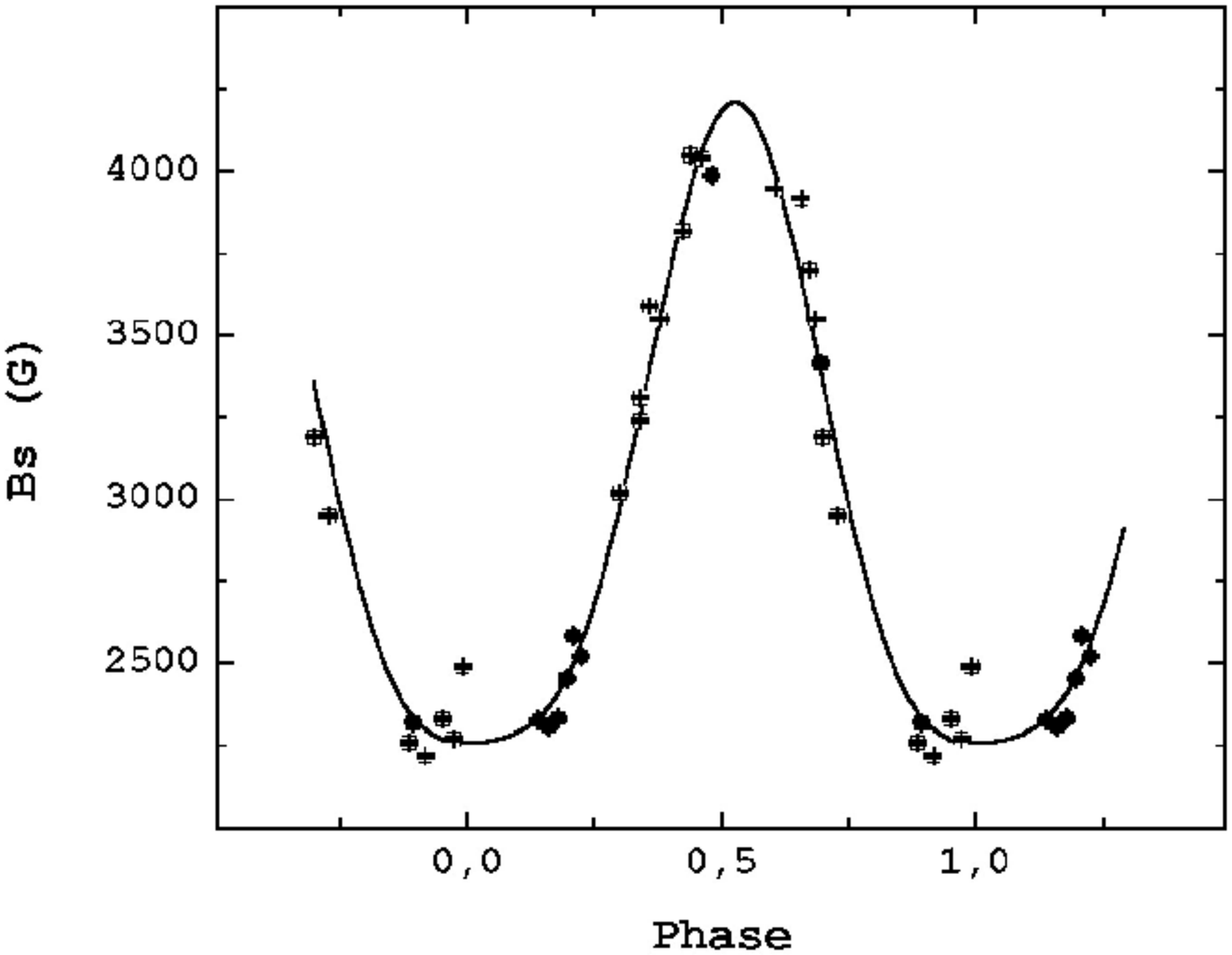}}
\vspace{-3.5mm}
\caption{ HD 59435 }
\label{fig:fig135}
\end{figure}

\begin{figure}
\resizebox{0.98\hsize}{!}{\includegraphics{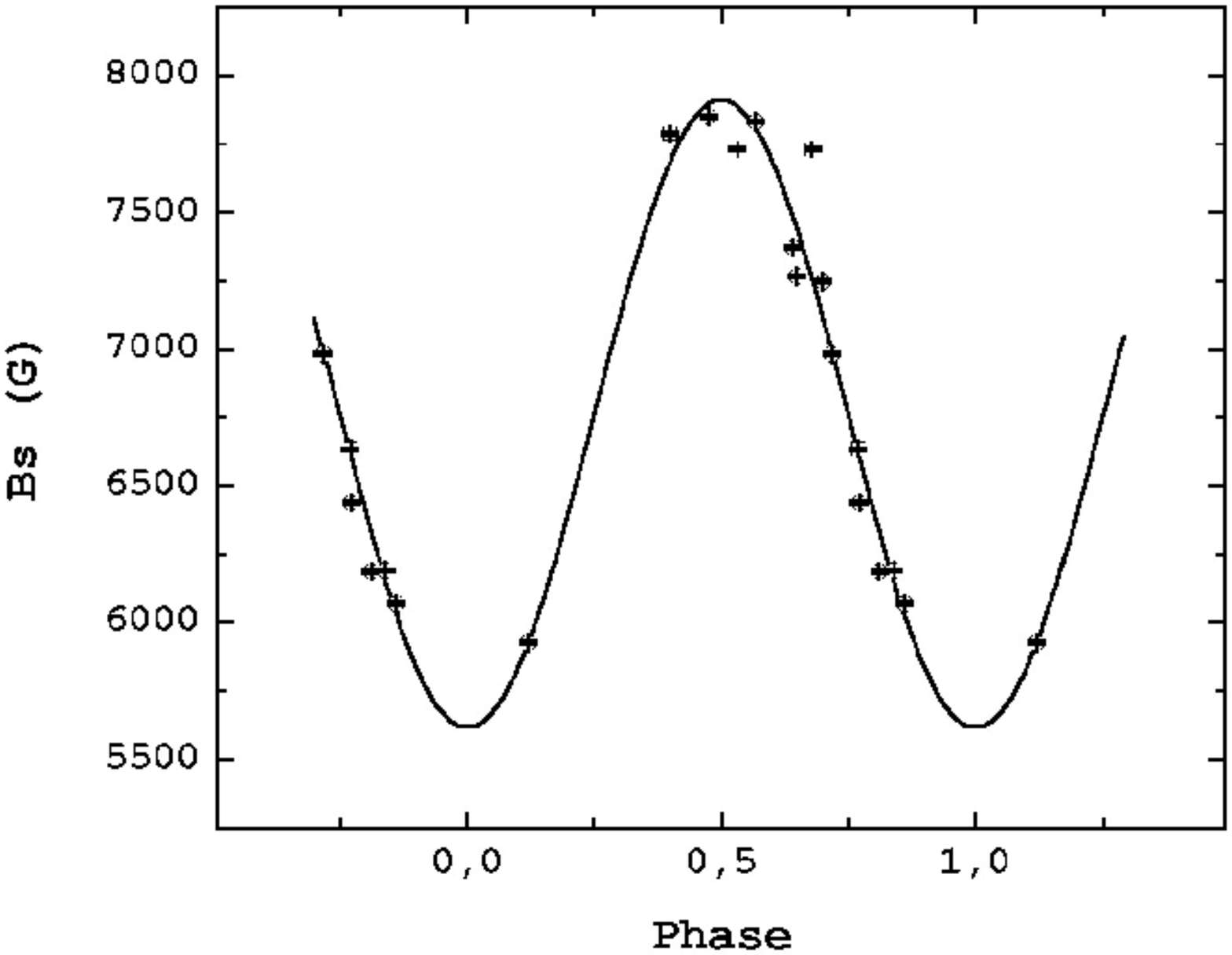}}
\vspace{-3.5mm}
\caption{ HD 61468 (1) }
\label{fig:fig136}
\end{figure}

\begin{figure}
\resizebox{0.98\hsize}{!}{\includegraphics{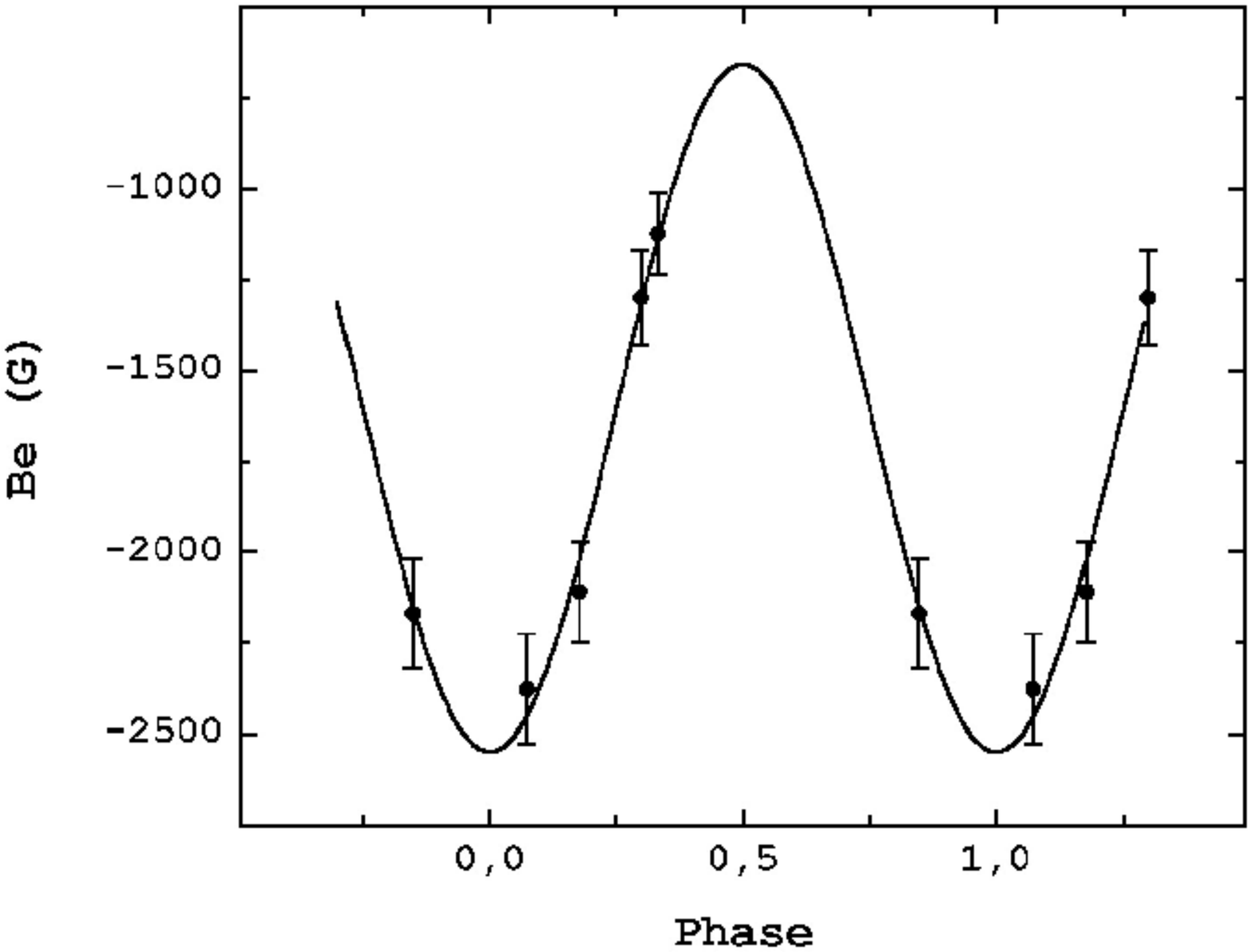}}
\vspace{-3.5mm}
\caption{ HD 61468 (2) }
\label{fig:fig137}
\end{figure}

\begin{figure}
\resizebox{0.98\hsize}{!}{\includegraphics{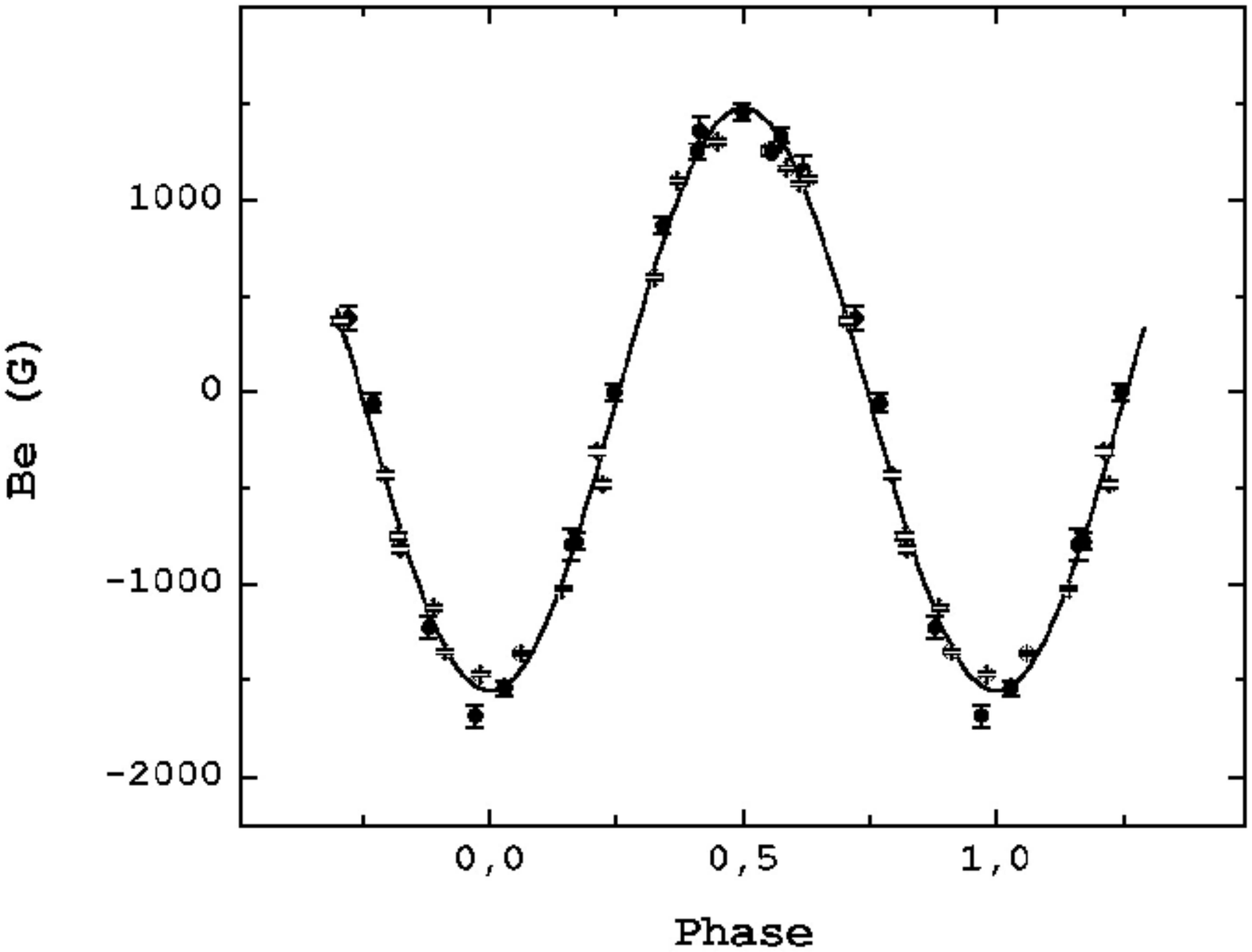}}
\vspace{-3.5mm}
\caption{ HD 62140 }
\label{fig:fig138}
\end{figure}

\begin{figure}
\resizebox{0.98\hsize}{!}{\includegraphics{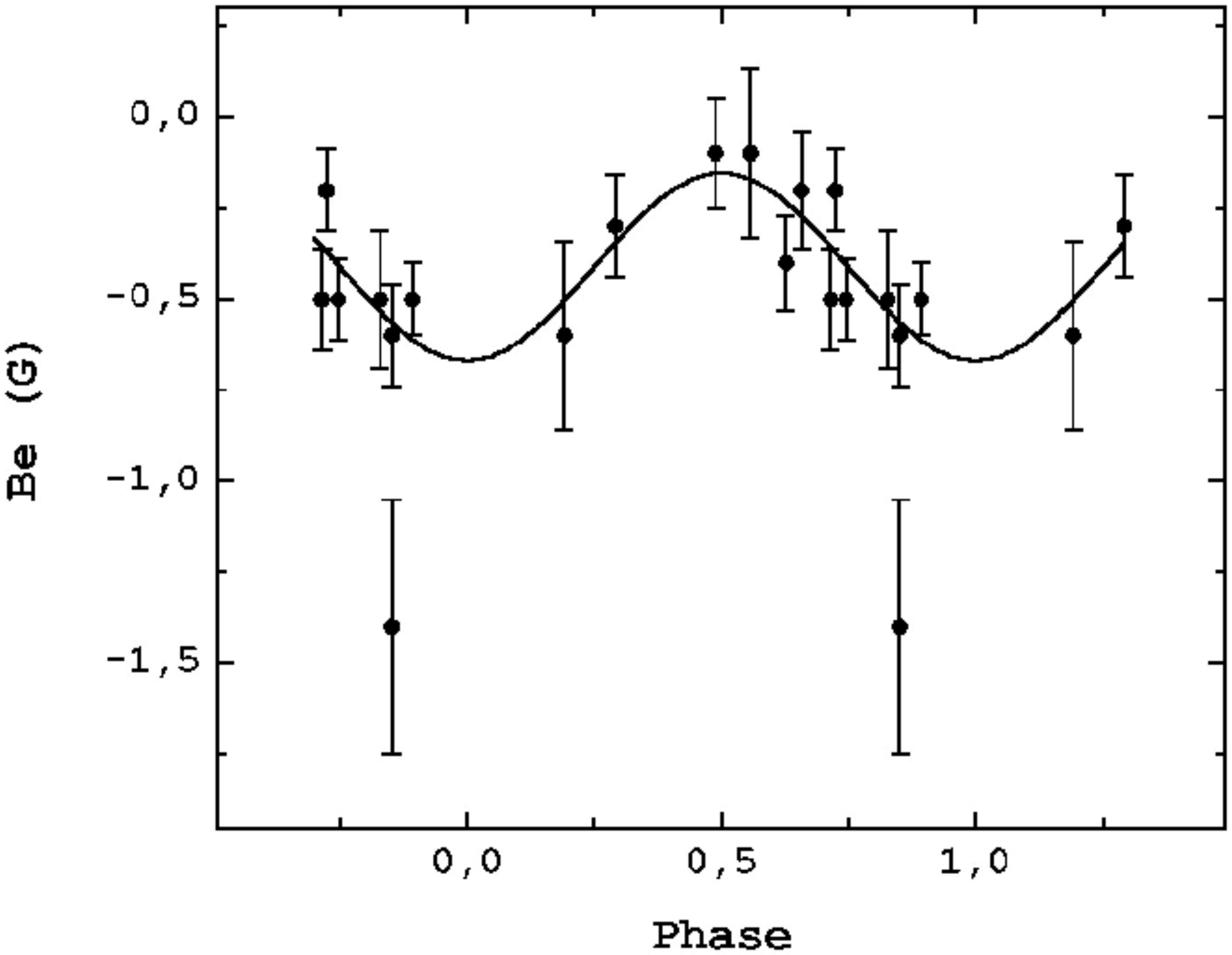}}
\vspace{-3.5mm}
\caption{ HD 62509 }
\label{fig:fig139}
\end{figure}

\begin{figure}
\resizebox{0.98\hsize}{!}{\includegraphics{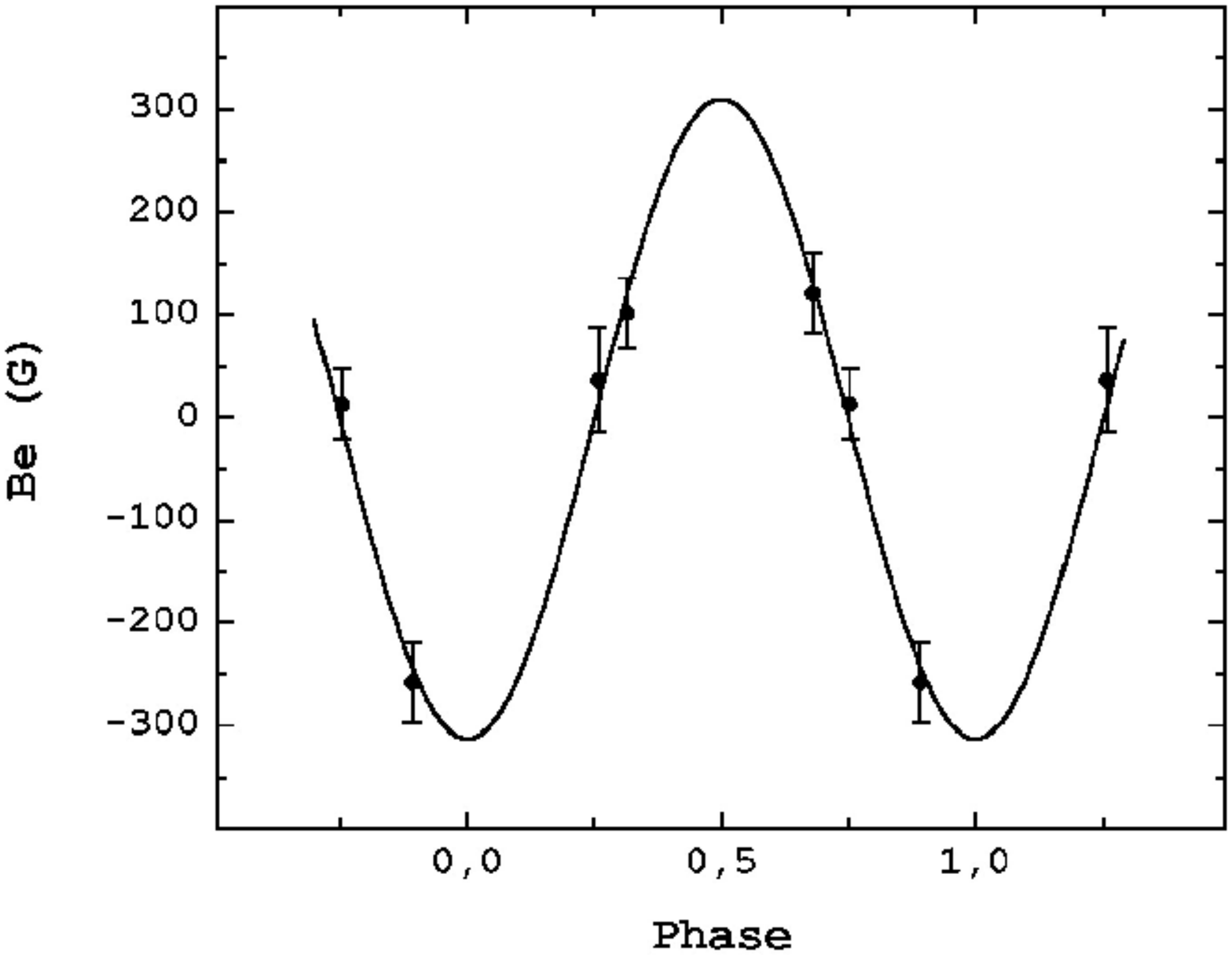}}
\vspace{-3.5mm}
\caption{ HD 62509 }
\label{fig:fig139}
\end{figure}

\clearpage
\newpage

\begin{figure}
\resizebox{0.98\hsize}{!}{\includegraphics{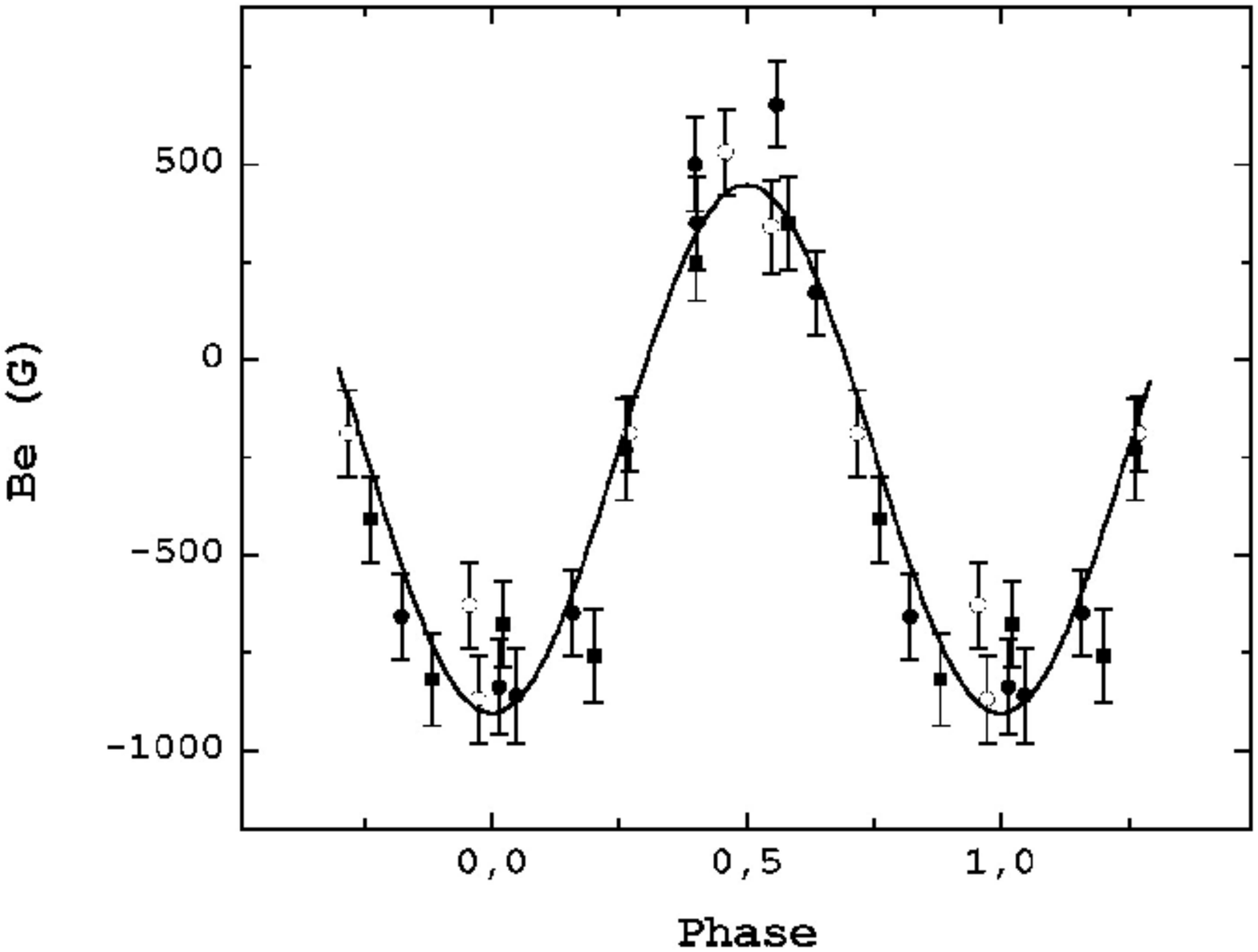}}
\vspace{-3.5mm}
\caption{ HD 64740 }
\label{fig:fig140}
\end{figure}

\begin{figure}
\resizebox{0.98\hsize}{!}{\includegraphics{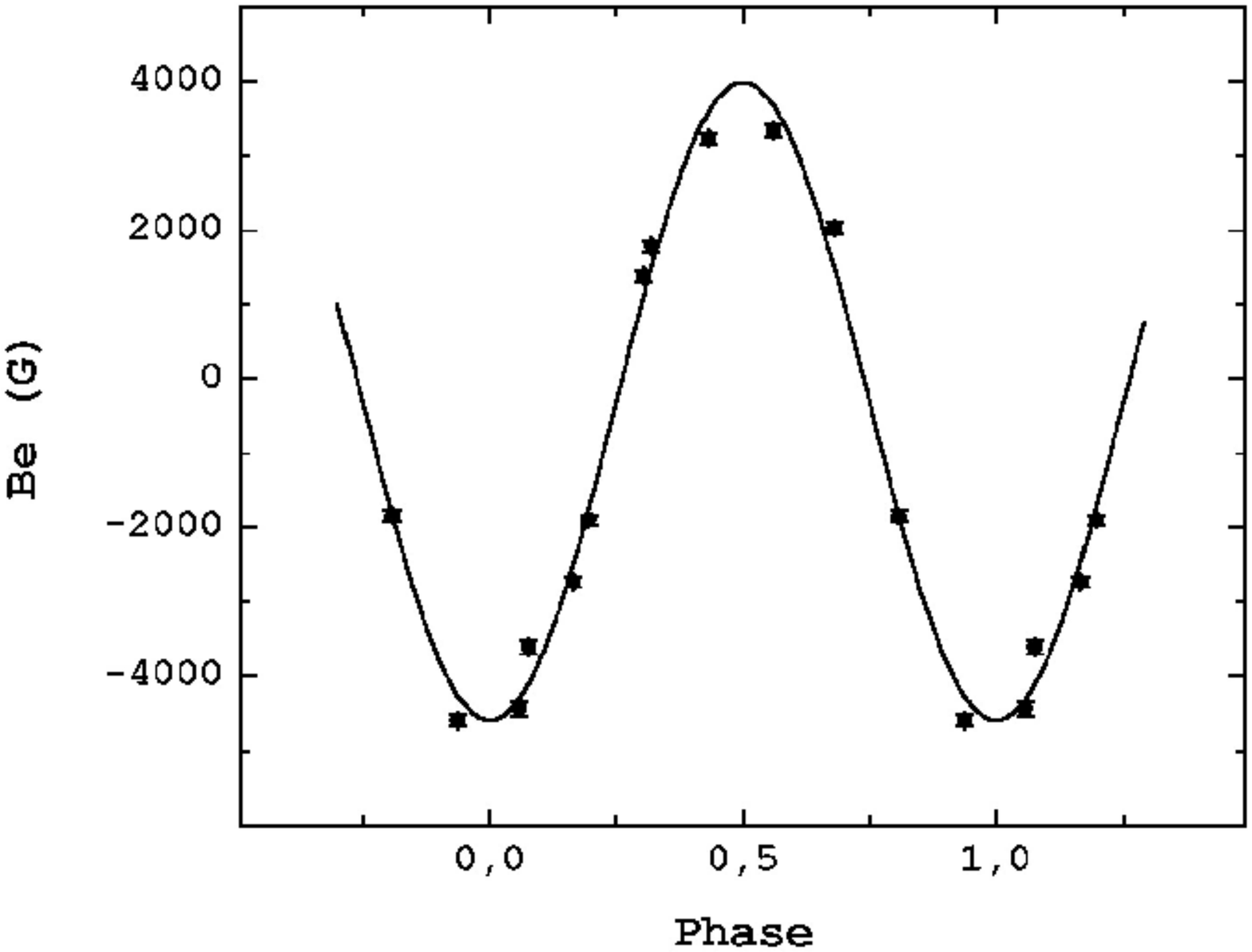}}
\vspace{-3.5mm}
\caption{ HD 65339 (1) }
\label{fig:fig141}
\end{figure}

\begin{figure}
\resizebox{0.98\hsize}{!}{\includegraphics{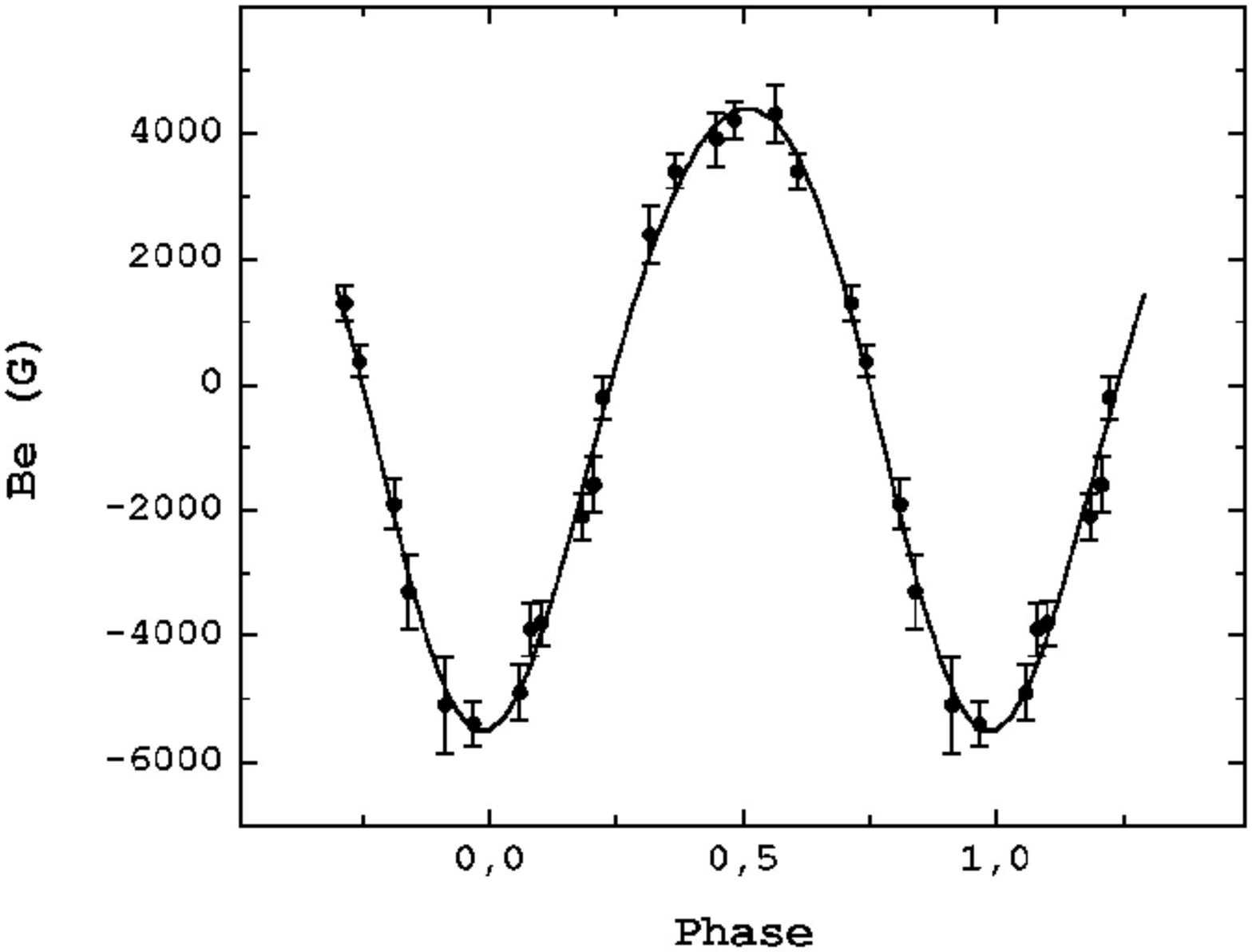}}
\vspace{-3.5mm}
\caption{ HD 65339 (2) }
\label{fig:fig142}
\end{figure}

\begin{figure}
\resizebox{0.98\hsize}{!}{\includegraphics{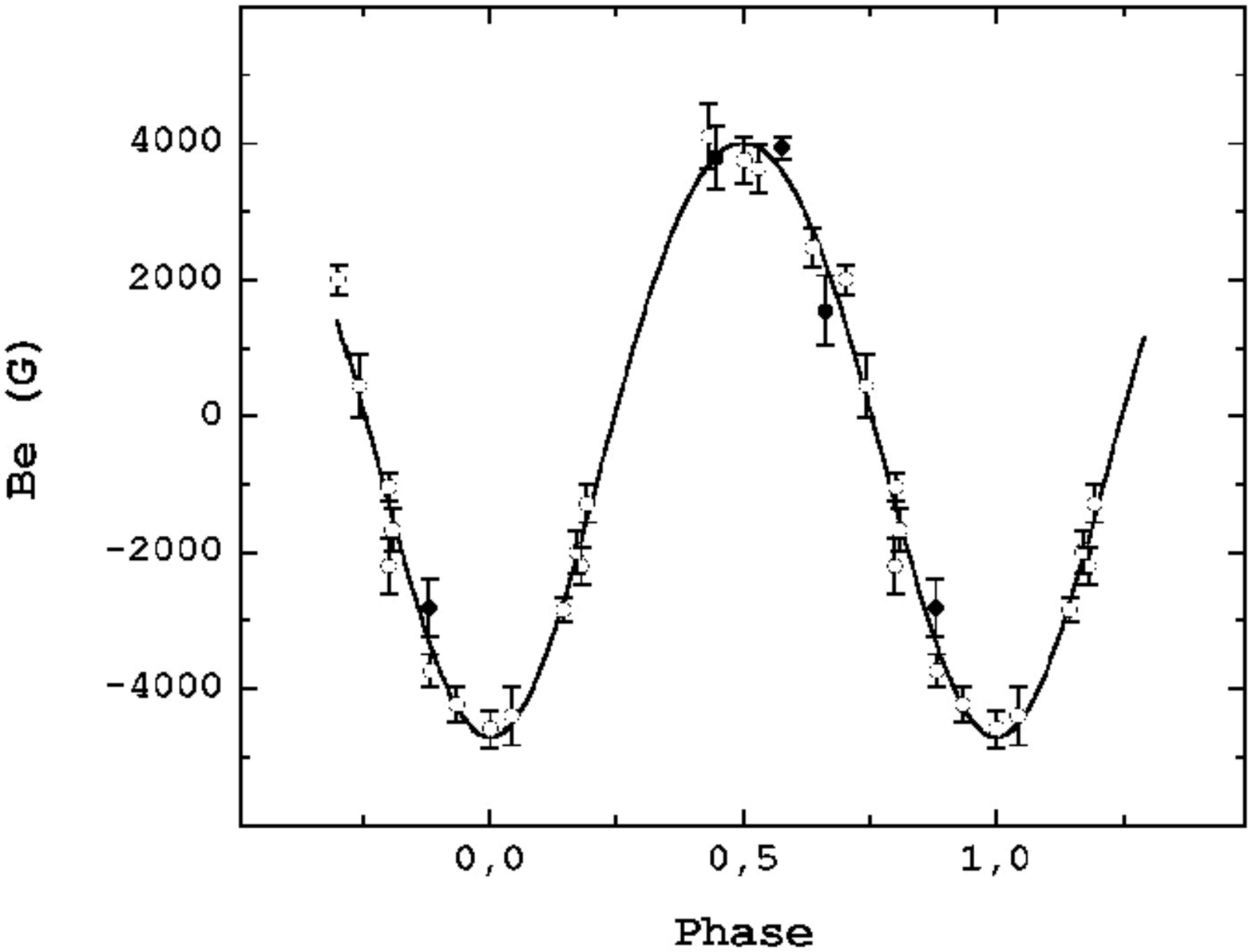}}
\vspace{-3.5mm}
\caption{ HD 65339 (3) }
\label{fig:fig143}
\end{figure}

\begin{figure}
\resizebox{0.98\hsize}{!}{\includegraphics{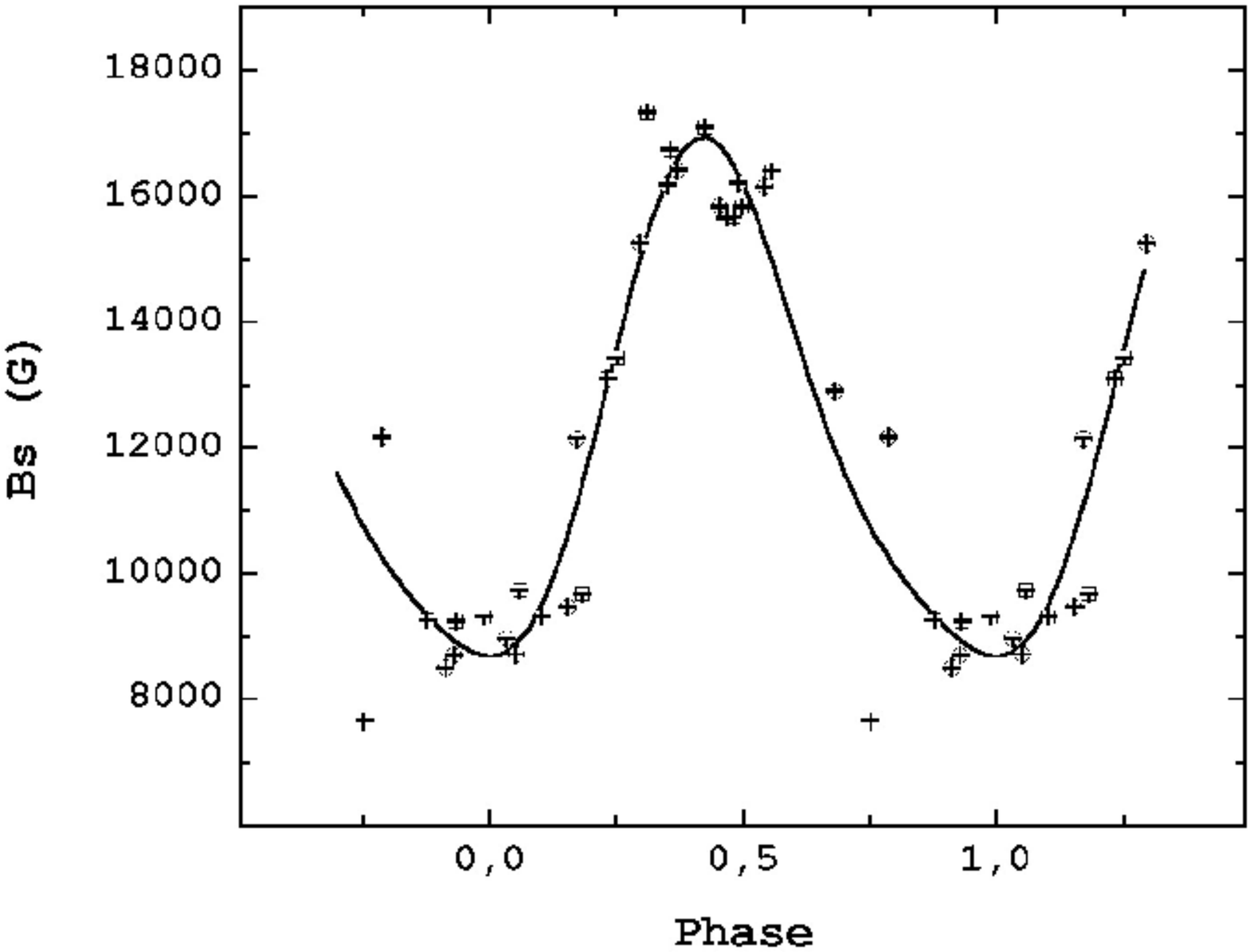}}
\vspace{-3.5mm}
\caption{ HD 65339 (4) }
\label{fig:fig144}
\end{figure}

\begin{figure}
\resizebox{0.98\hsize}{!}{\includegraphics{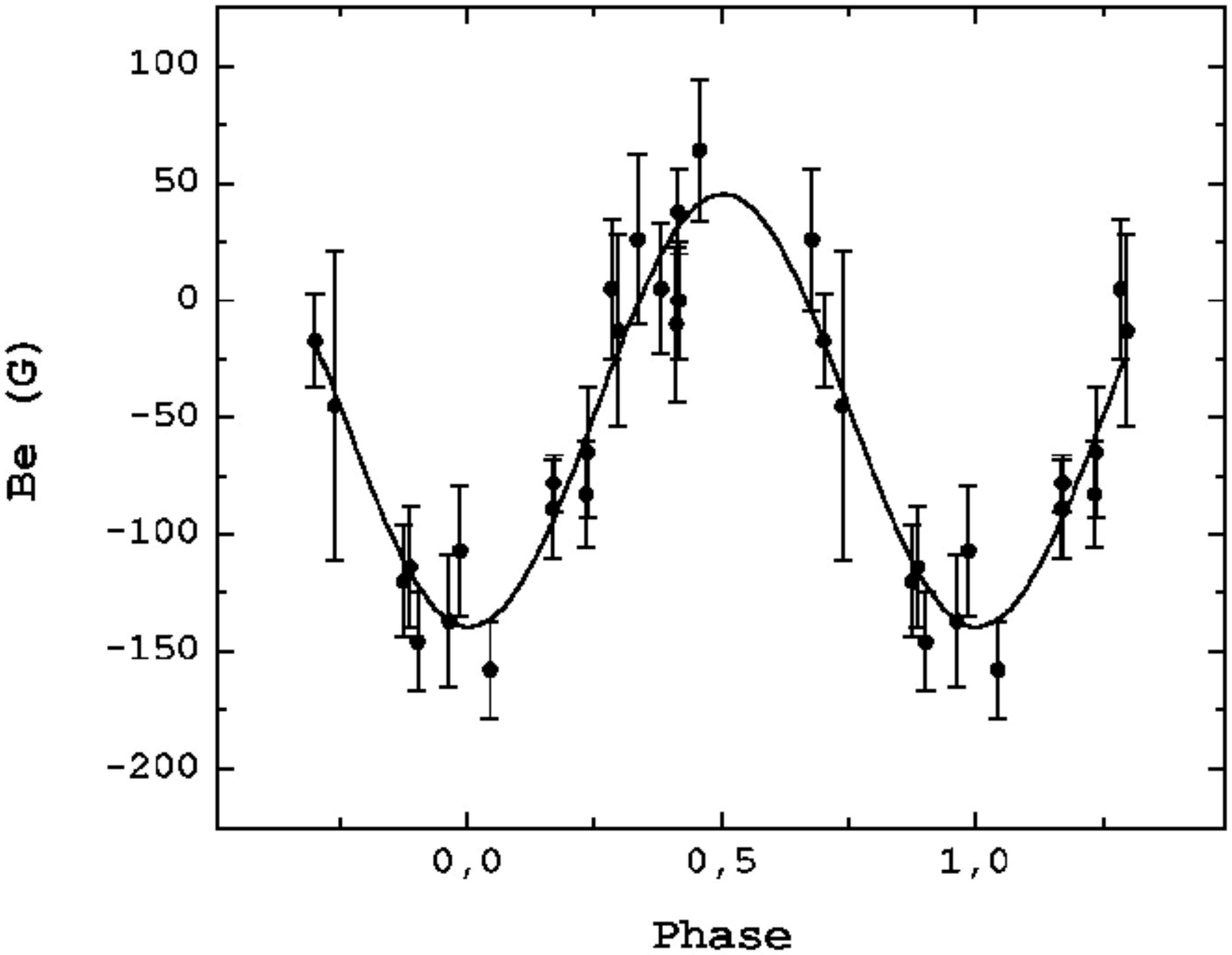}}
\vspace{-3.5mm}
\caption{ HD 66665 (1) }
\label{fig:fig145}
\end{figure}

\clearpage
\newpage

\begin{figure}
\resizebox{0.98\hsize}{!}{\includegraphics{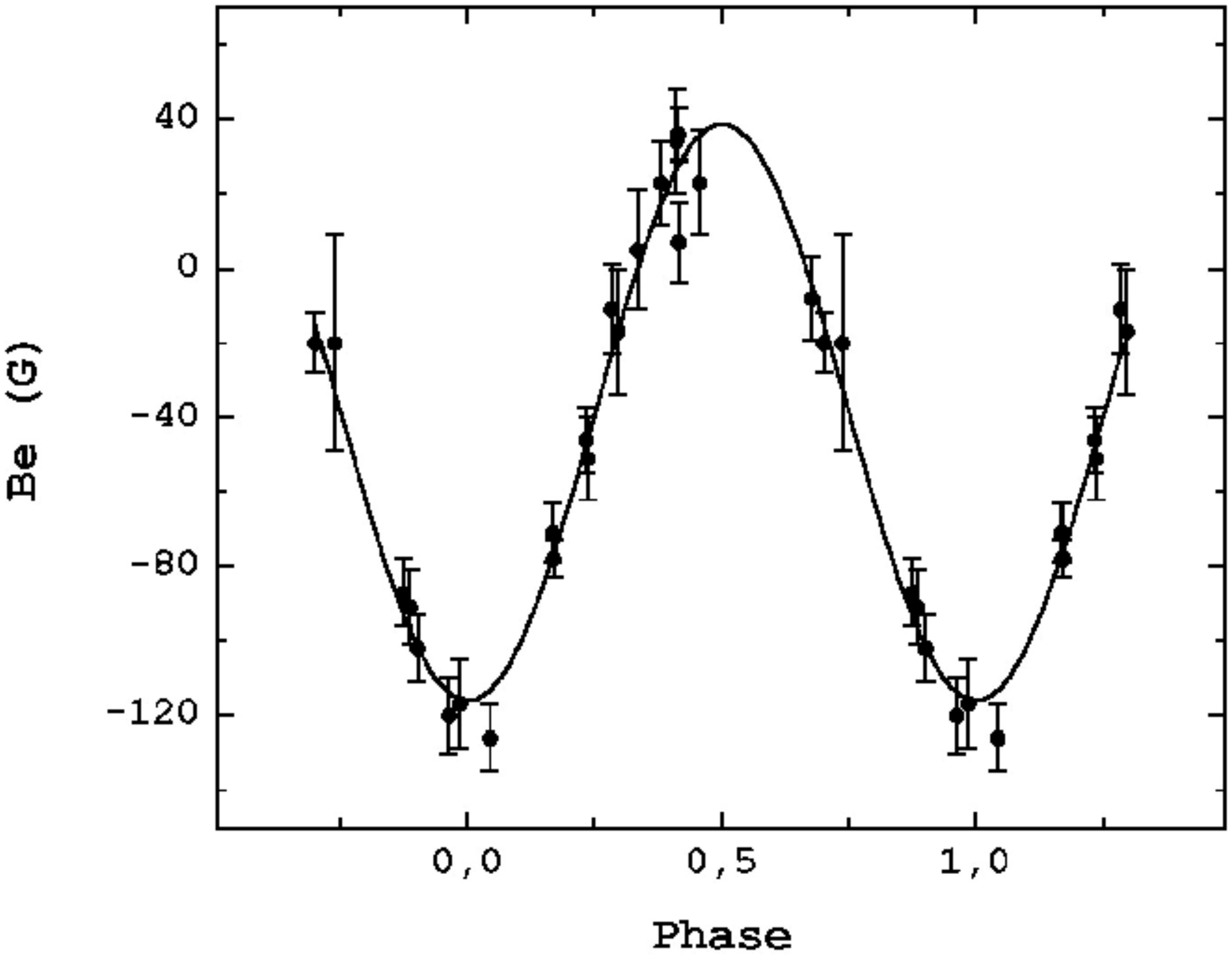}}
\vspace{-3.5mm}
\caption{ HD 66665 (2) }
\label{fig:fig145}
\end{figure}

\begin{figure}
\resizebox{0.98\hsize}{!}{\includegraphics{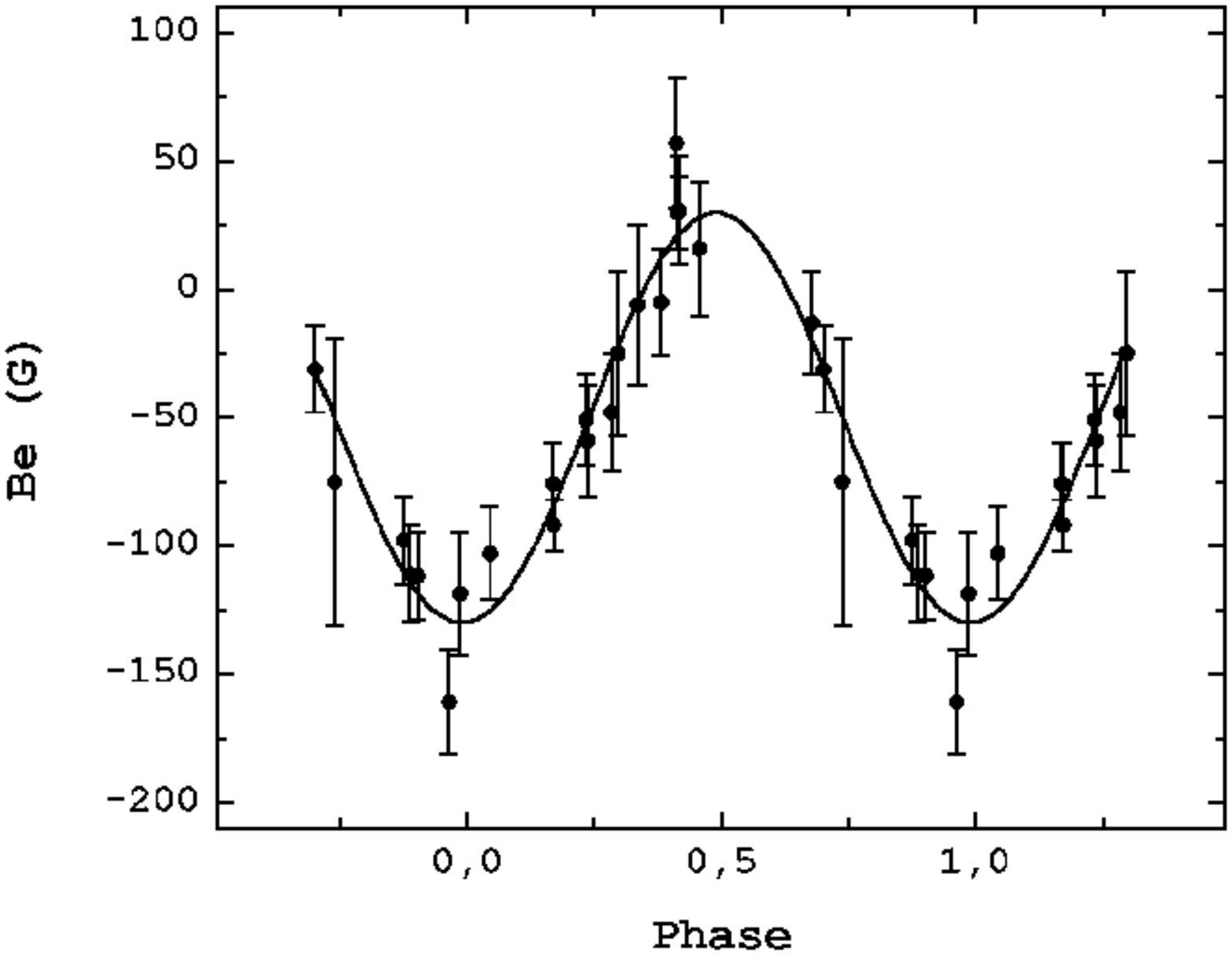}}
\vspace{-3.5mm}
\caption{ HD 66665 (3) }
\label{fig:fig145}
\end{figure}

\begin{figure}
\resizebox{0.98\hsize}{!}{\includegraphics{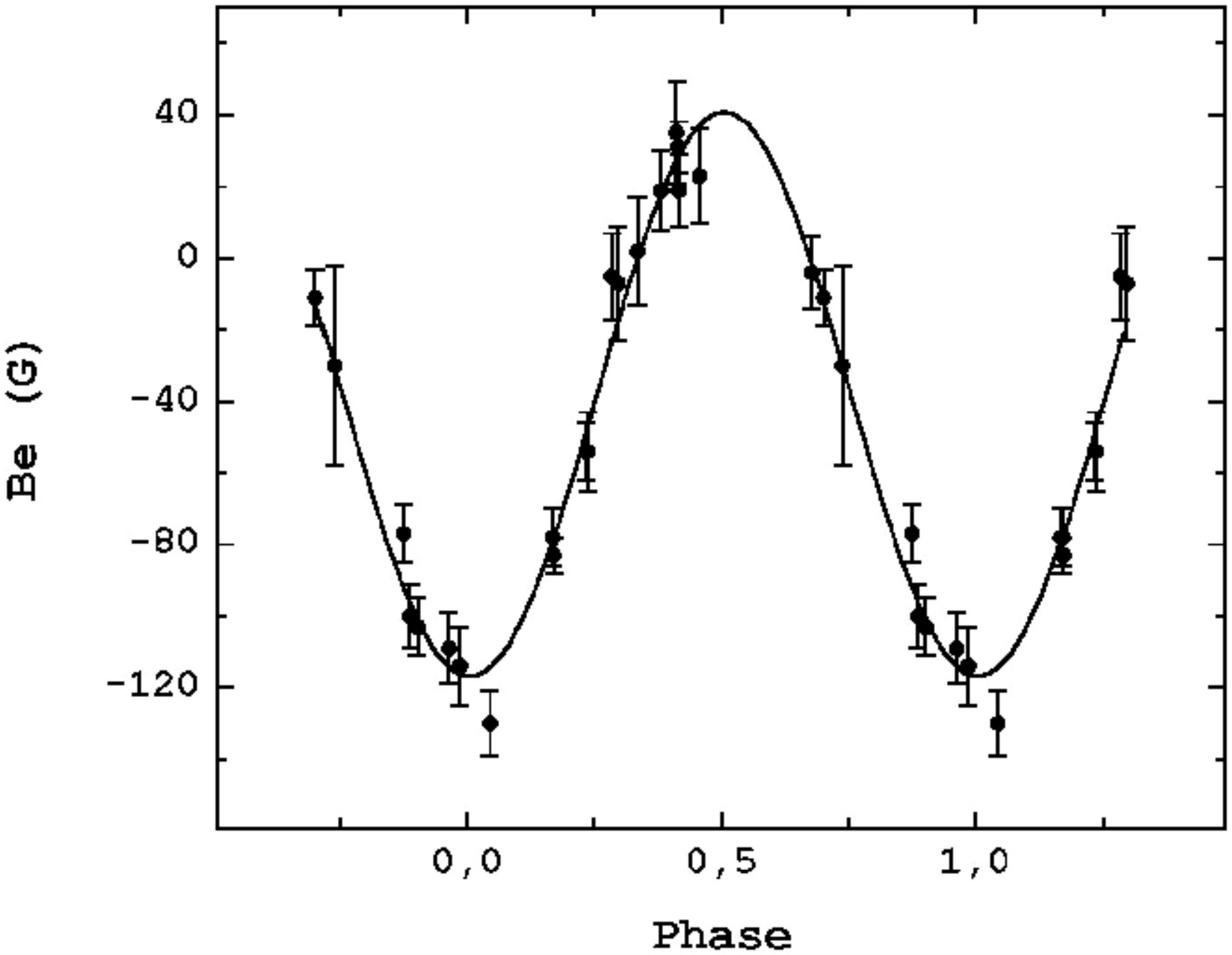}}
\vspace{-3.5mm}
\caption{ HD 66665 (4) }
\label{fig:fig145}
\end{figure}

\begin{figure}
\resizebox{0.98\hsize}{!}{\includegraphics{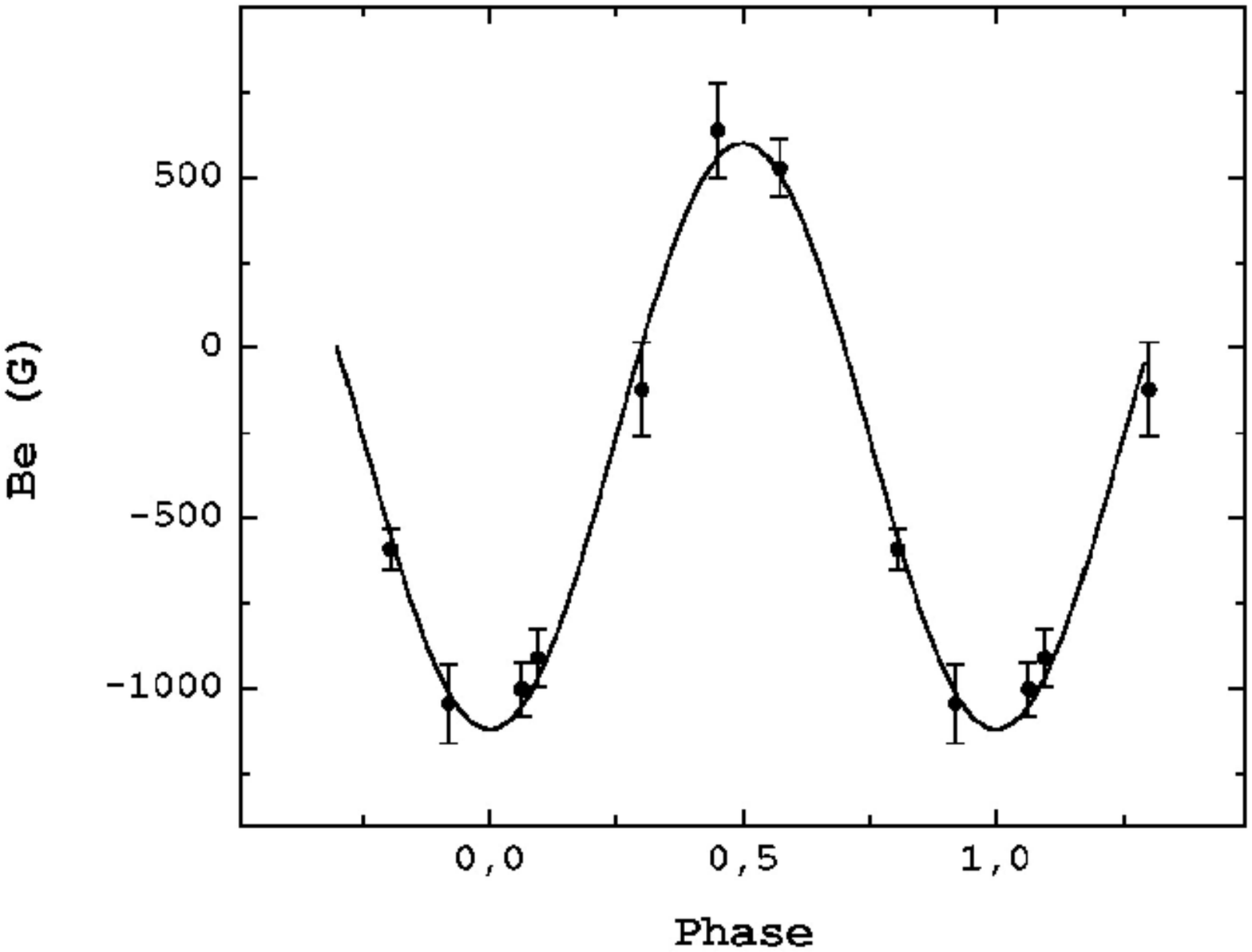}}
\vspace{-3.5mm}
\caption{ HD 66765 }
\label{fig:fig146}
\end{figure}

\begin{figure}
\resizebox{0.98\hsize}{!}{\includegraphics{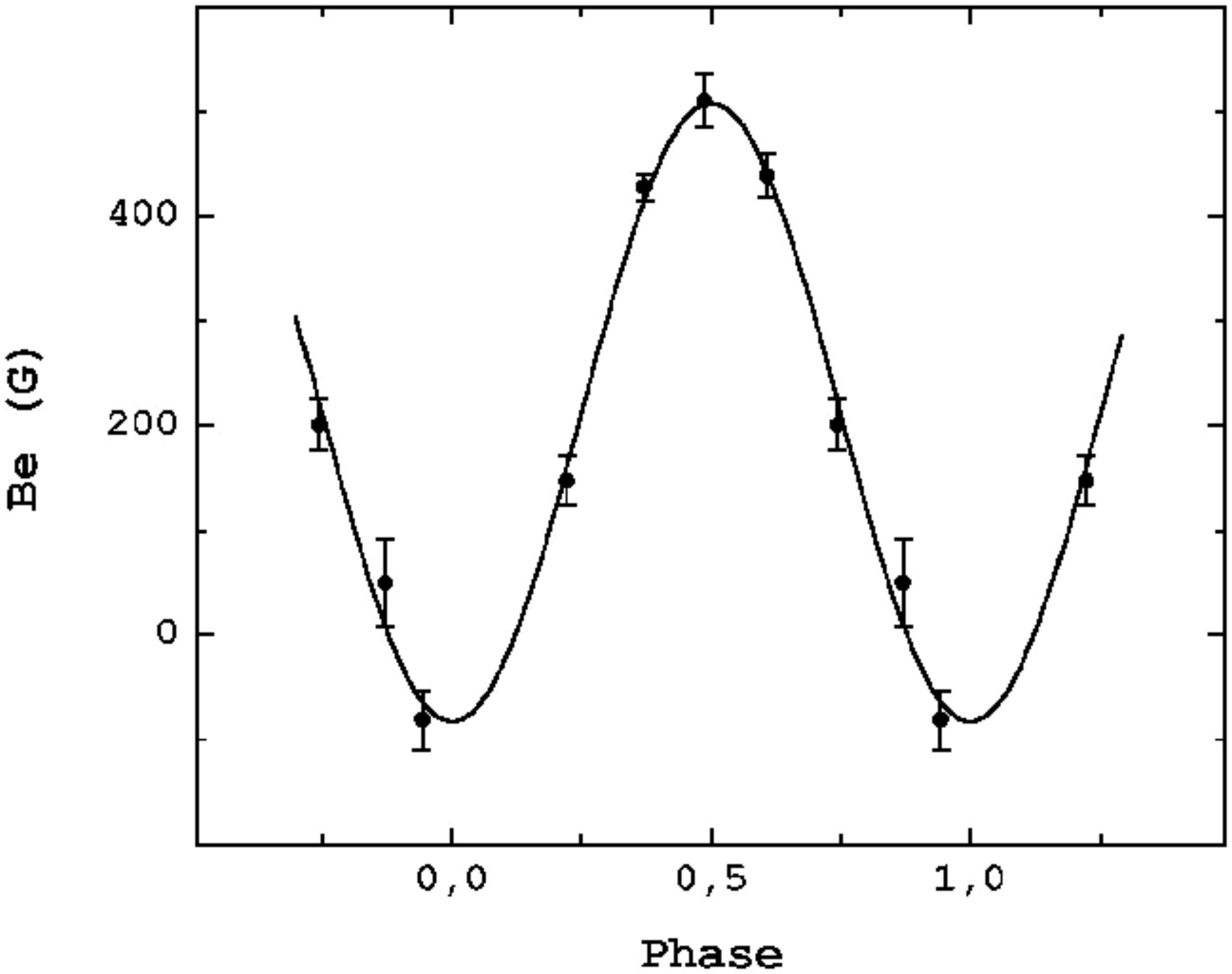}}
\vspace{-3.5mm}
\caption{ HD 67621 }
\label{fig:fig147}
\end{figure}

\begin{figure}
\resizebox{0.98\hsize}{!}{\includegraphics{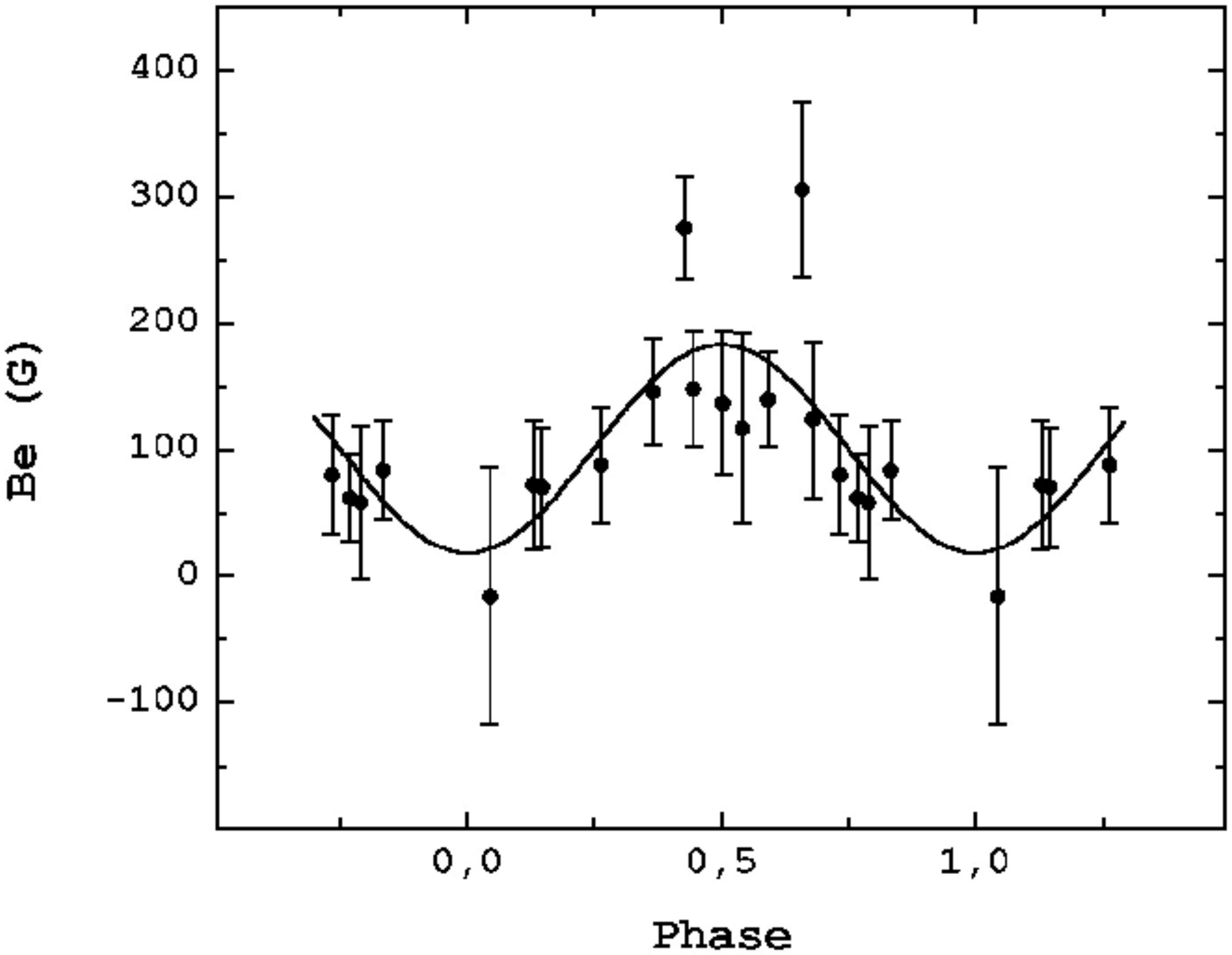}}
\vspace{-3.5mm}
\caption{ HD 68351 }
\label{fig:fig148}
\end{figure}

\clearpage
\newpage

\begin{figure}
\resizebox{0.98\hsize}{!}{\includegraphics{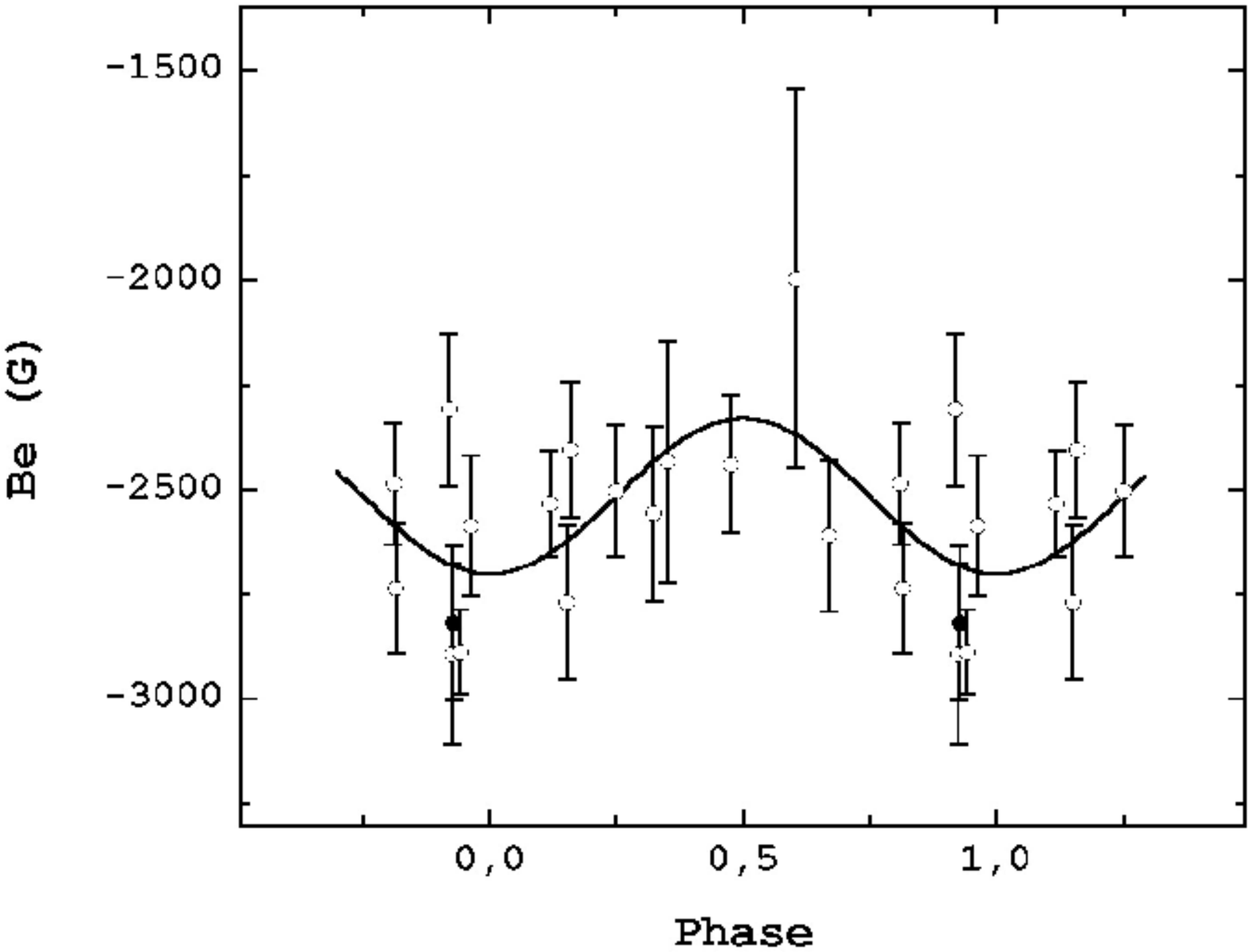}}
\vspace{-3.5mm}
\caption{ HD 70331 (1) }
\label{fig:fig149}
\end{figure}

\begin{figure}
\resizebox{0.98\hsize}{!}{\includegraphics{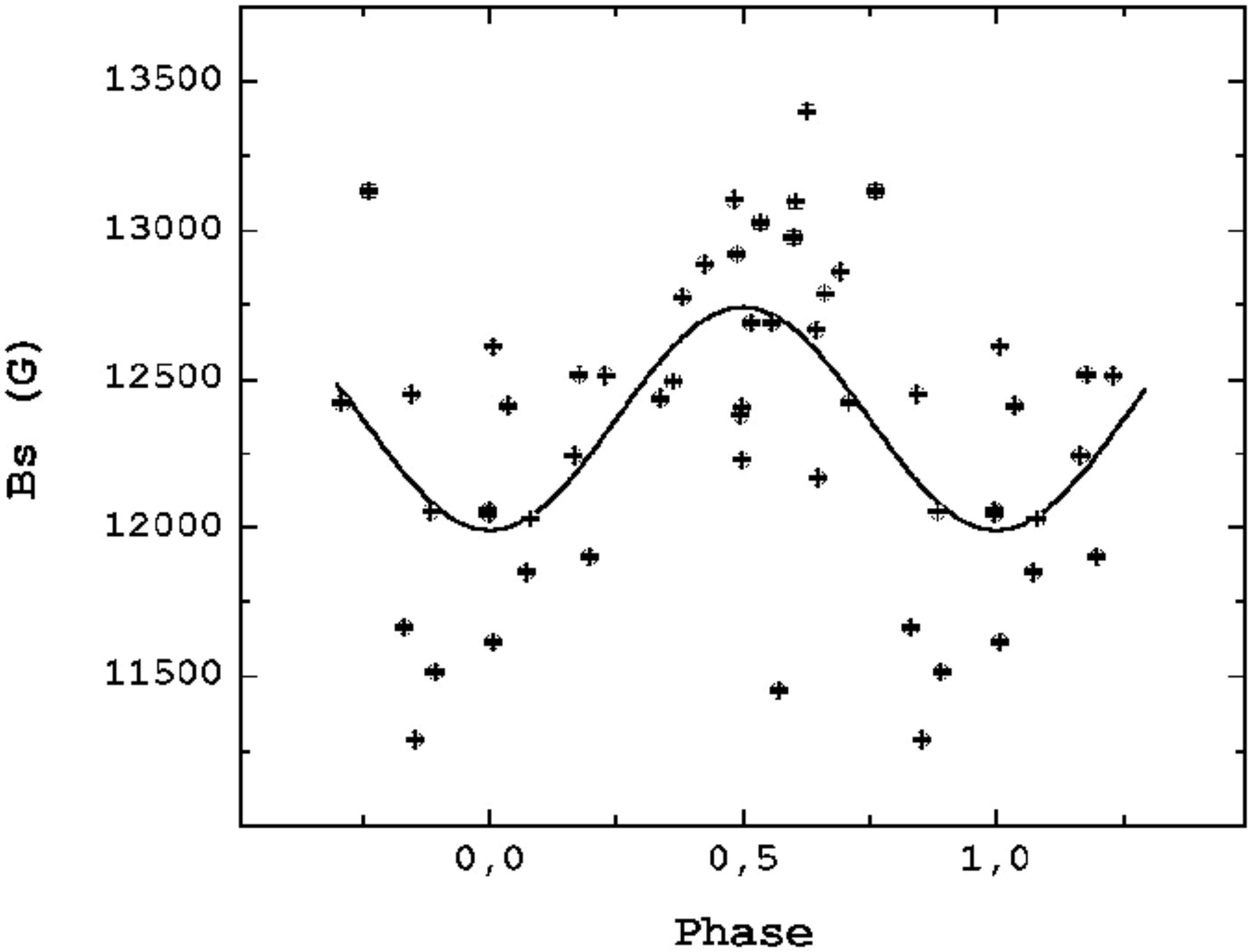}}
\vspace{-3.5mm}
\caption{ HD 70331 (2) }
\label{fig:fig150}
\end{figure}

\begin{figure}
\resizebox{0.98\hsize}{!}{\includegraphics{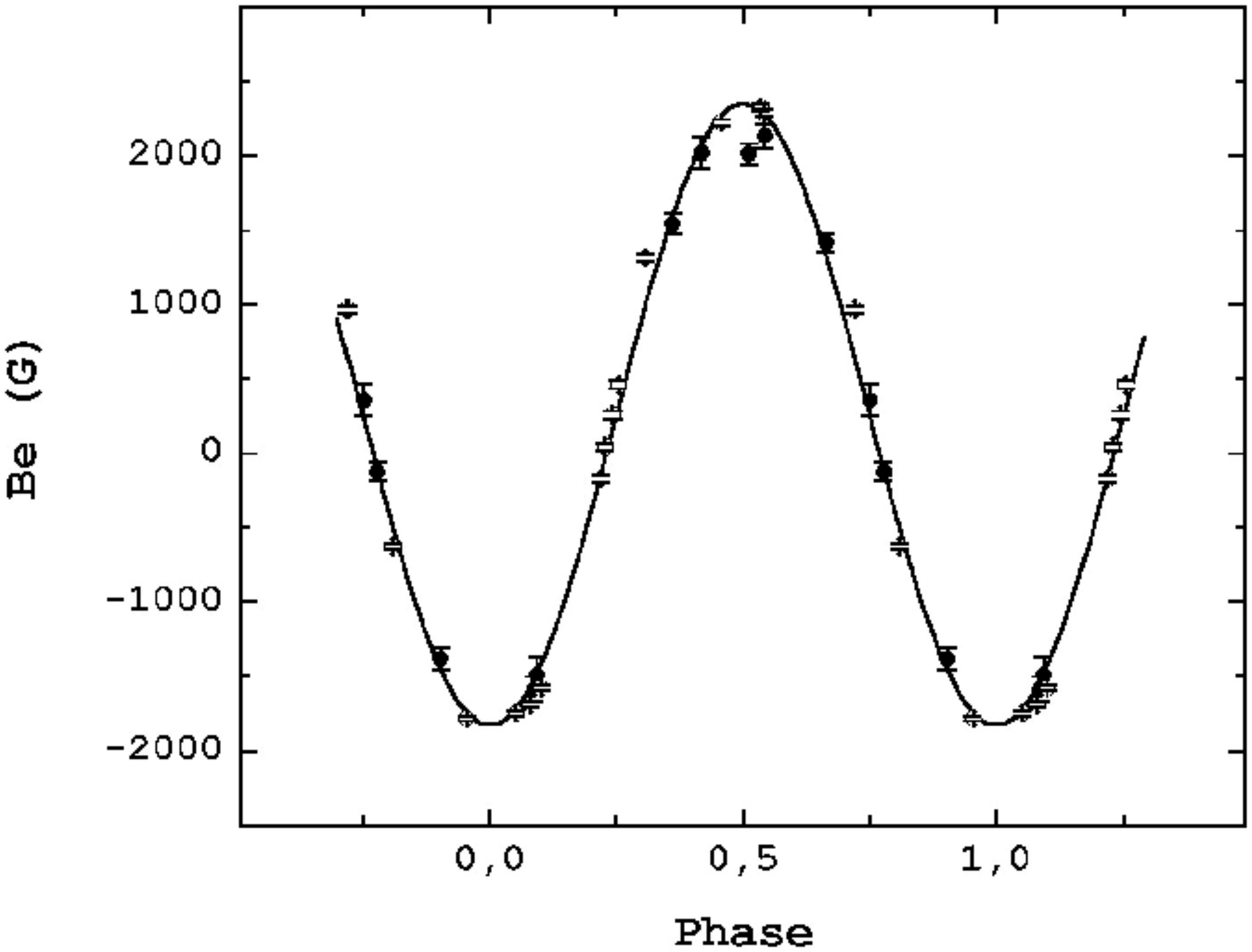}}
\vspace{-3.5mm}
\caption{ HD 71866 }
\label{fig:fig151}
\end{figure}

\begin{figure}
\resizebox{0.98\hsize}{!}{\includegraphics{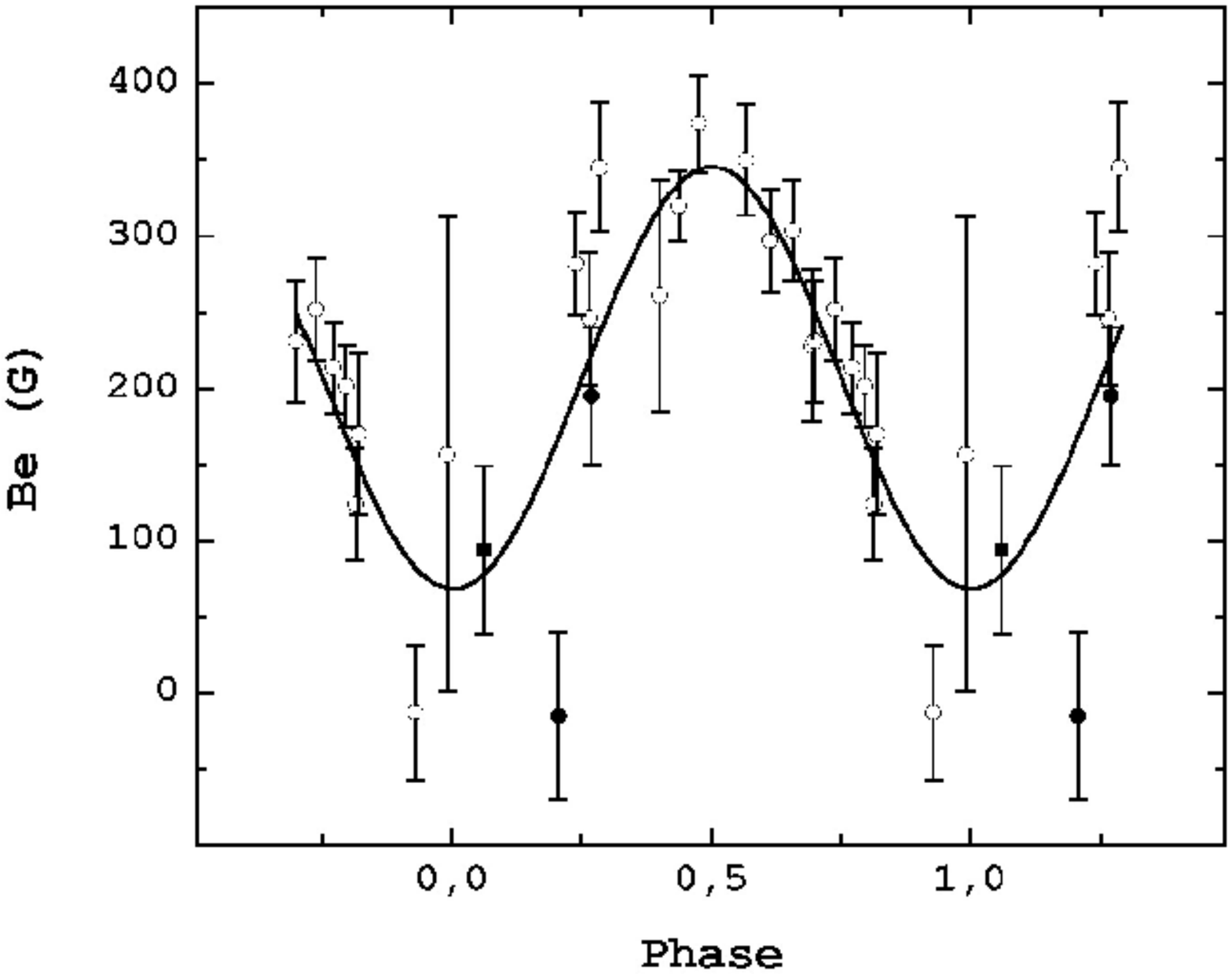}}
\vspace{-3.5mm}
\caption{ HD 72106 }
\label{fig:fig152}
\end{figure}

\begin{figure}
\resizebox{0.98\hsize}{!}{\includegraphics{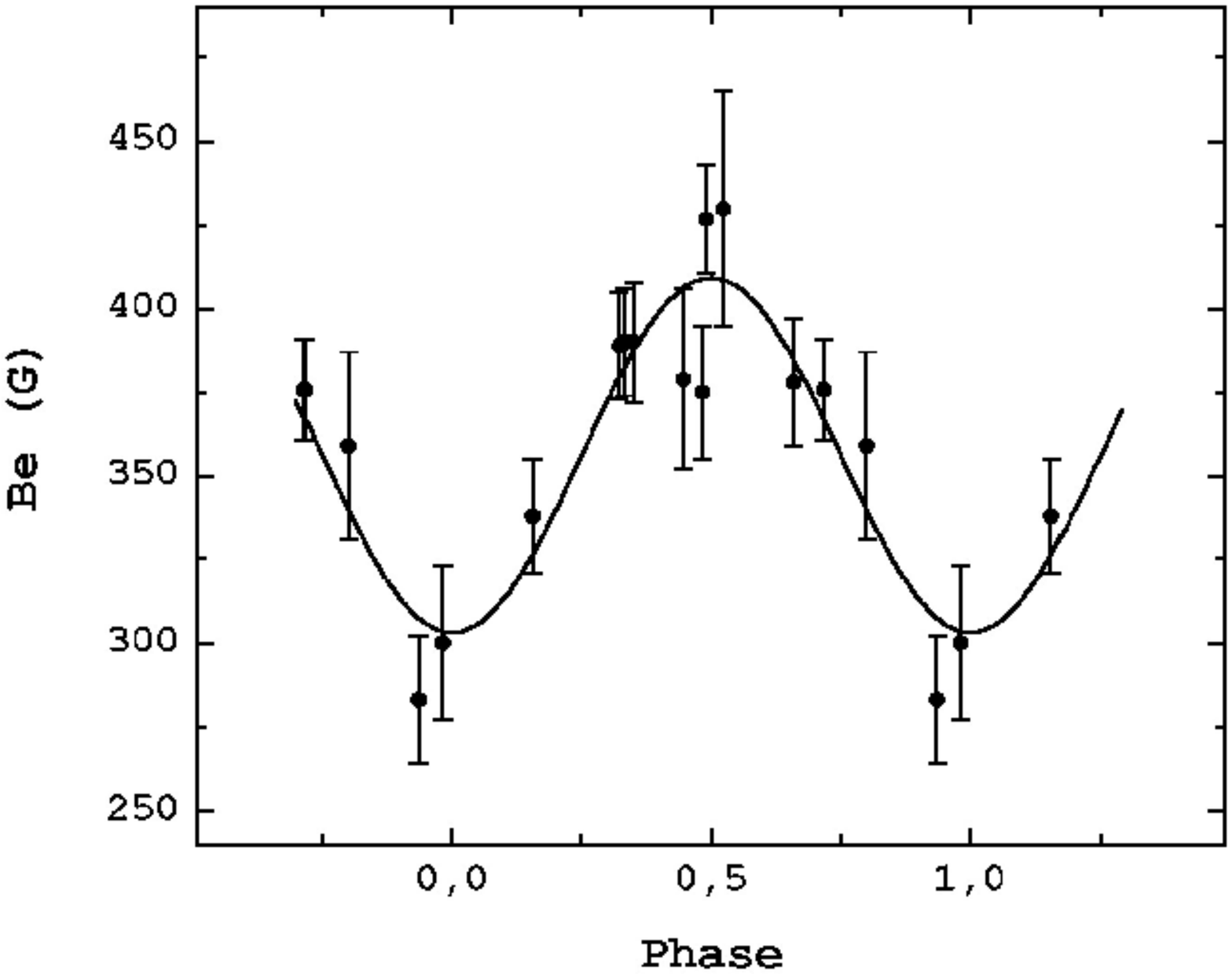}}
\vspace{-3.5mm}
\caption{ HD 72968 }
\label{fig:fig153}
\end{figure}

\begin{figure}
\resizebox{0.98\hsize}{!}{\includegraphics{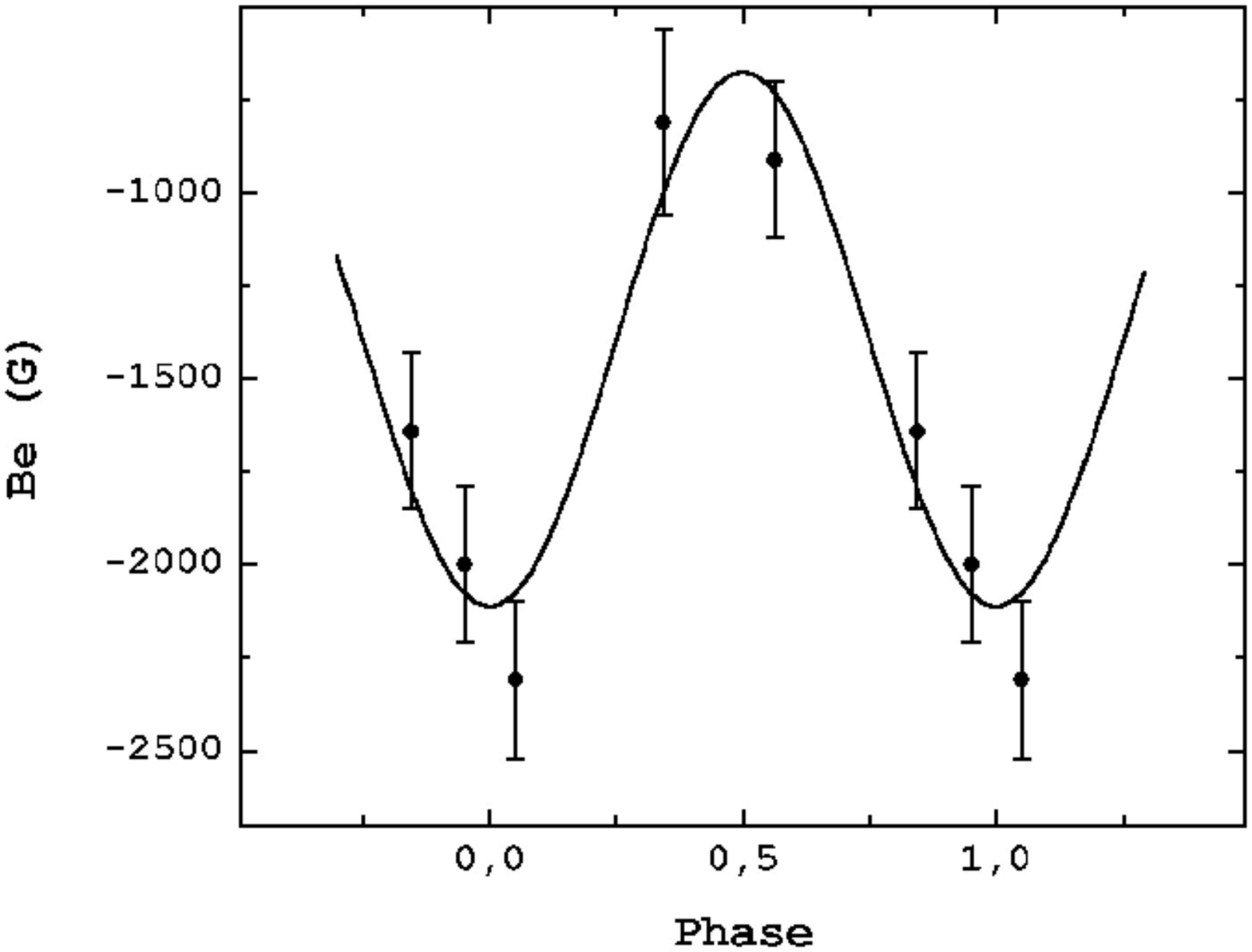}}
\vspace{-3.5mm}
\caption{ HD 73340 }
\label{fig:fig154}
\end{figure}

\clearpage
\newpage

\begin{figure}
\resizebox{0.98\hsize}{!}{\includegraphics{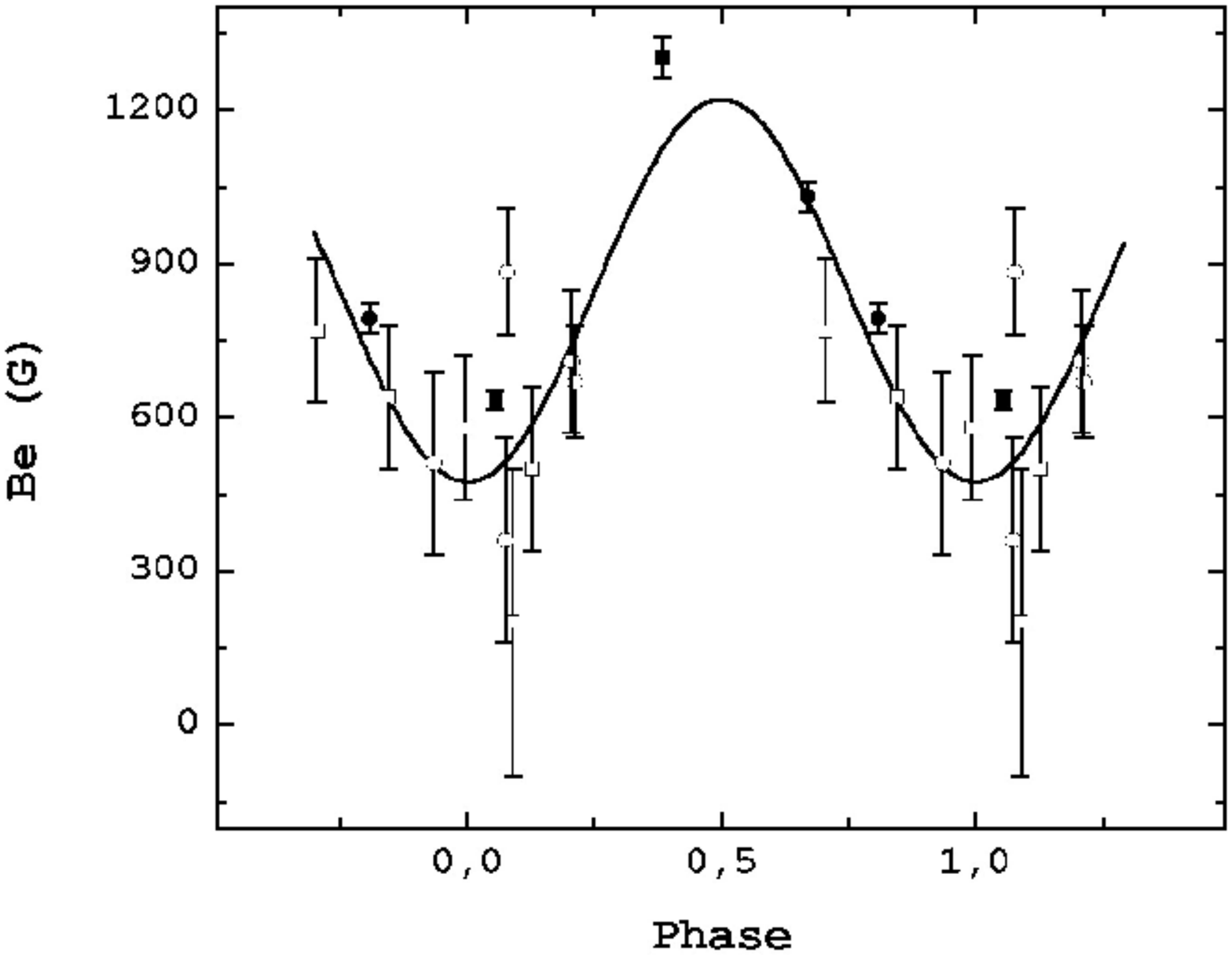}}
\vspace{-3.5mm}
\caption{ HD 74521 }
\label{fig:fig155}
\end{figure}

\begin{figure}
\resizebox{0.98\hsize}{!}{\includegraphics{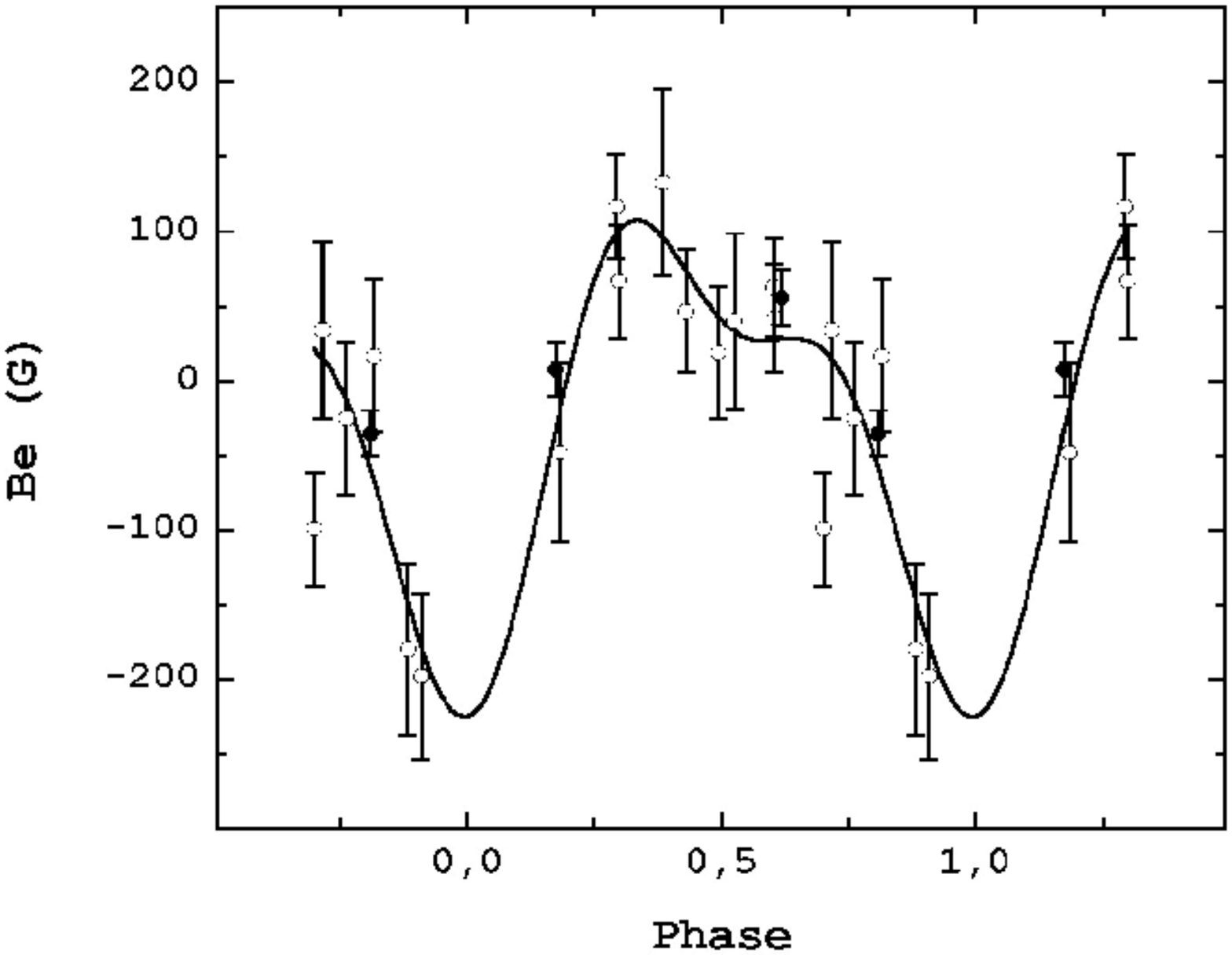}}
\vspace{-3.5mm}
\caption{ HD 74560 (1) }
\label{fig:fig156}
\end{figure}

\begin{figure}
\resizebox{0.98\hsize}{!}{\includegraphics{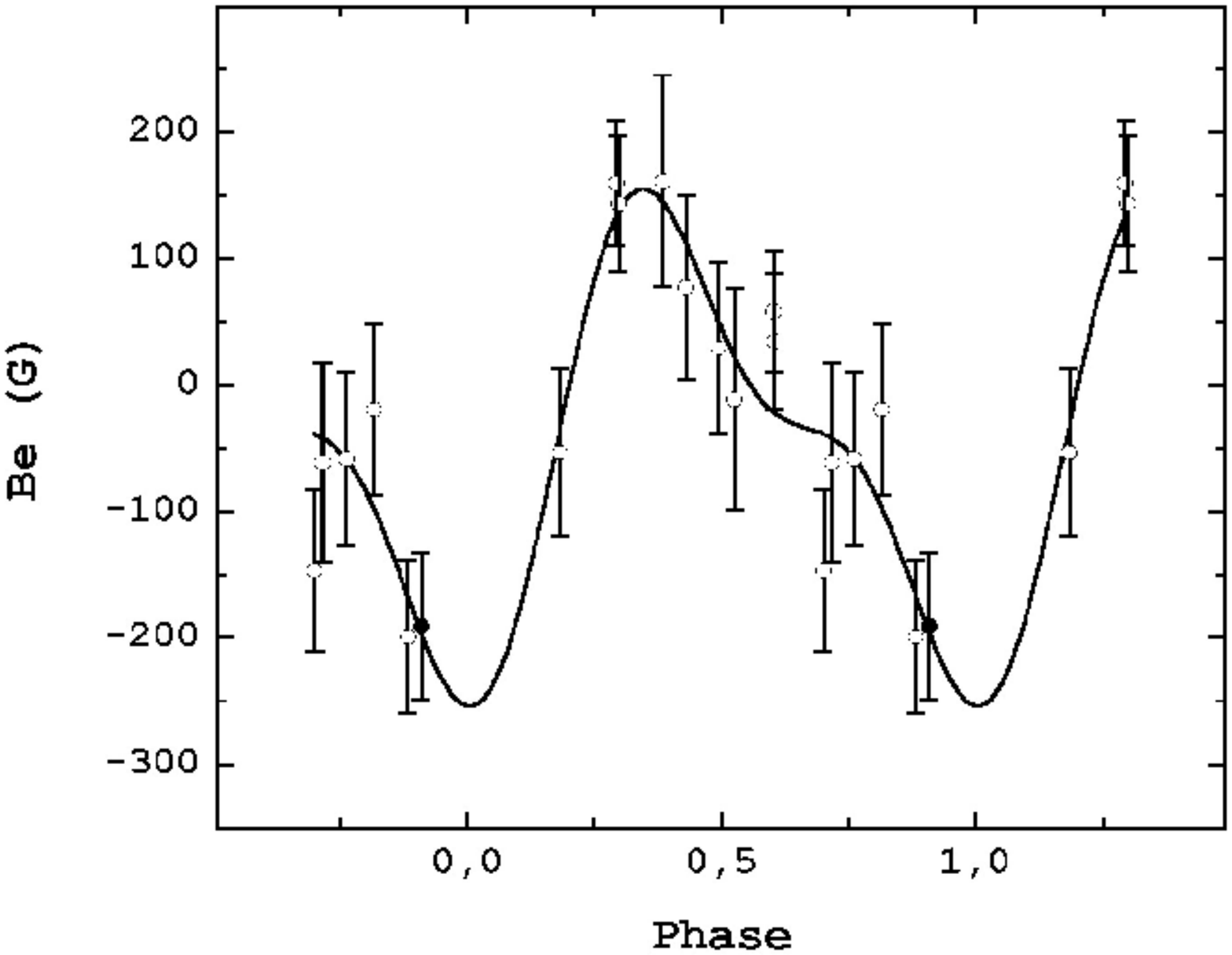}}
\vspace{-3.5mm}
\caption{ HD 74560 (2) }
\label{fig:fig157}
\end{figure}

\begin{figure}
\resizebox{0.98\hsize}{!}{\includegraphics{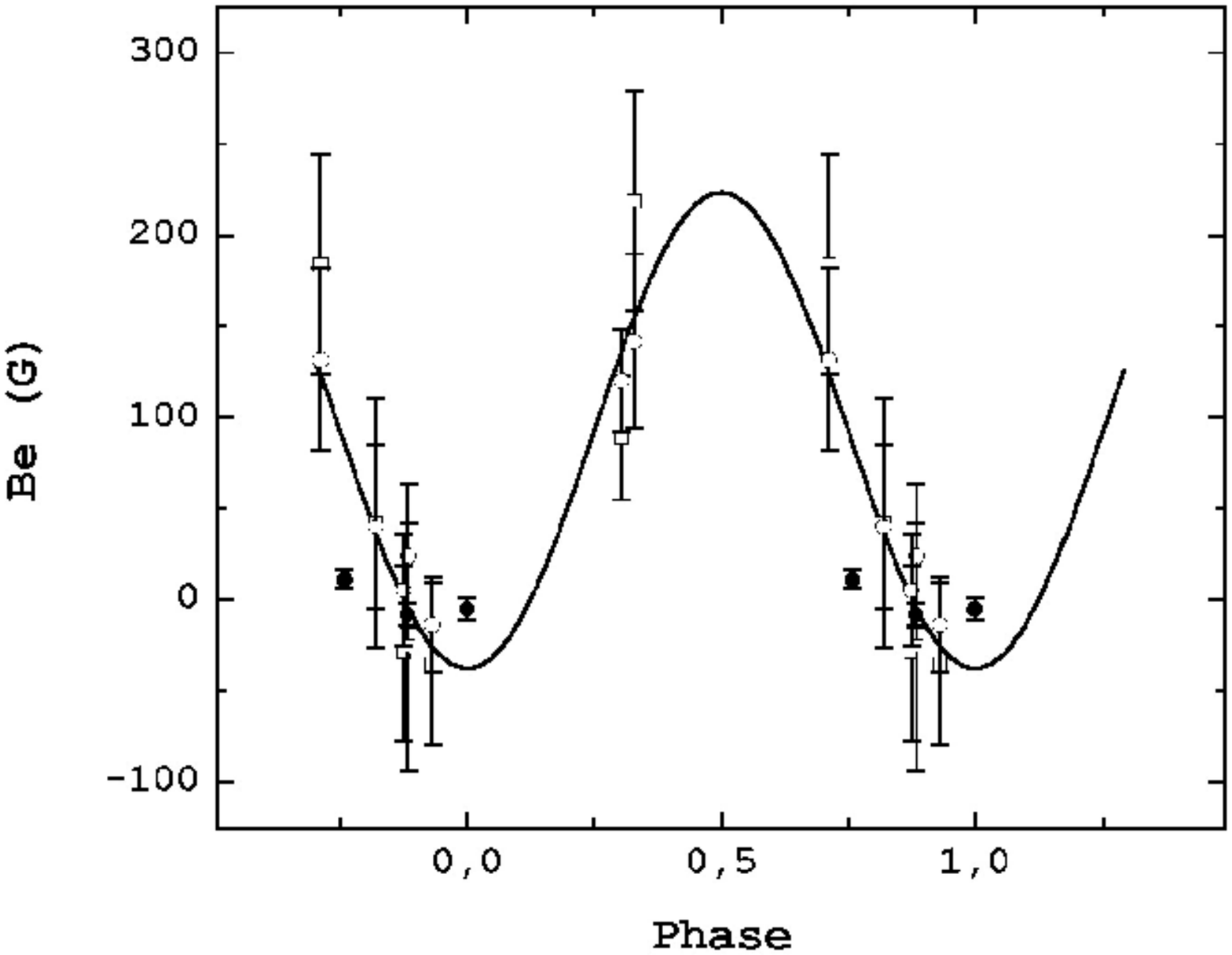}}
\vspace{-3.5mm}
\caption{ HD 74575 }
\label{fig:fig158}
\end{figure}

\begin{figure}
\resizebox{0.98\hsize}{!}{\includegraphics{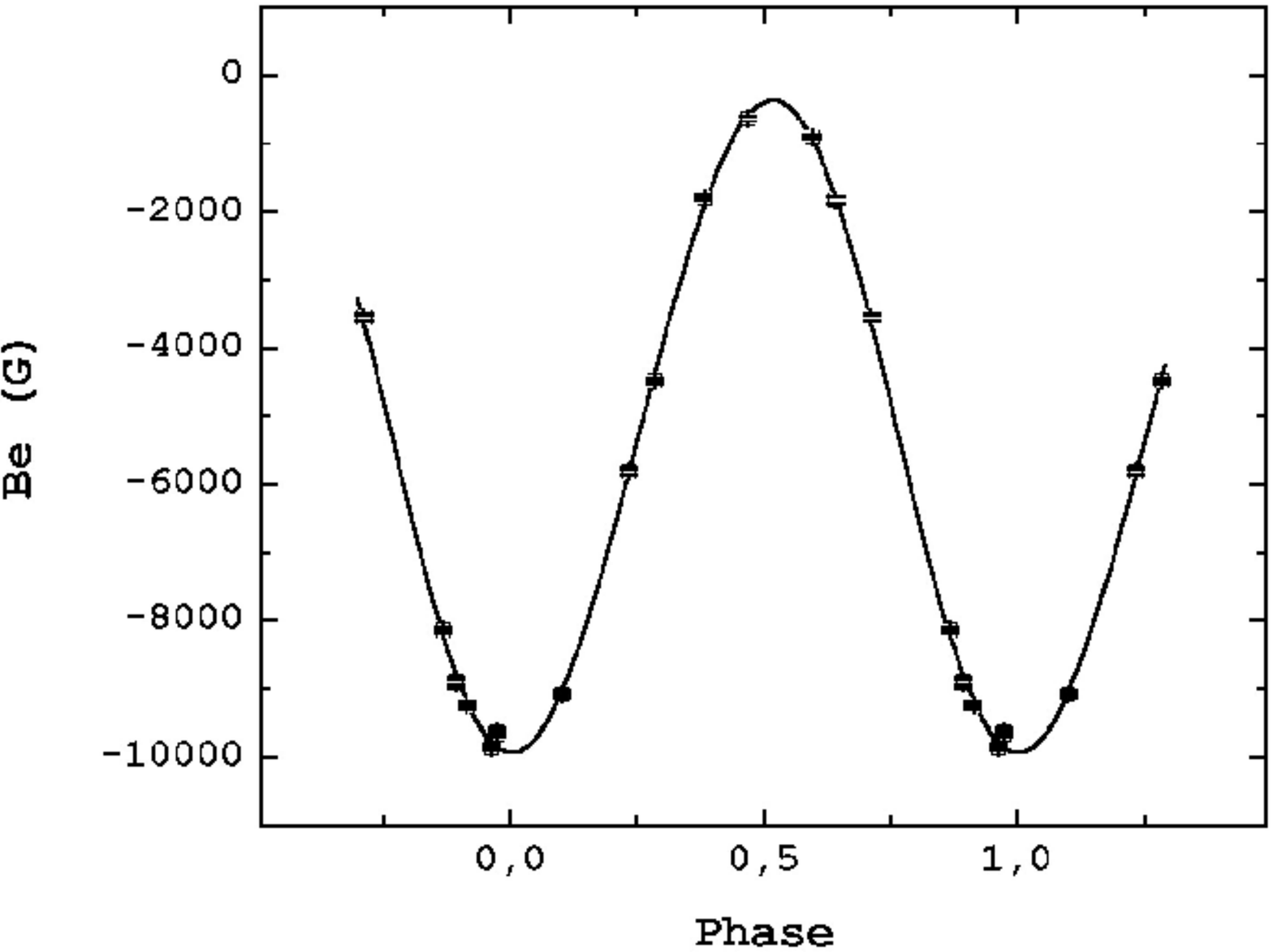}}
\vspace{-3.5mm}
\caption{ HD 75049 (1) }
\label{fig:fig159}
\end{figure}

\begin{figure}
\resizebox{0.98\hsize}{!}{\includegraphics{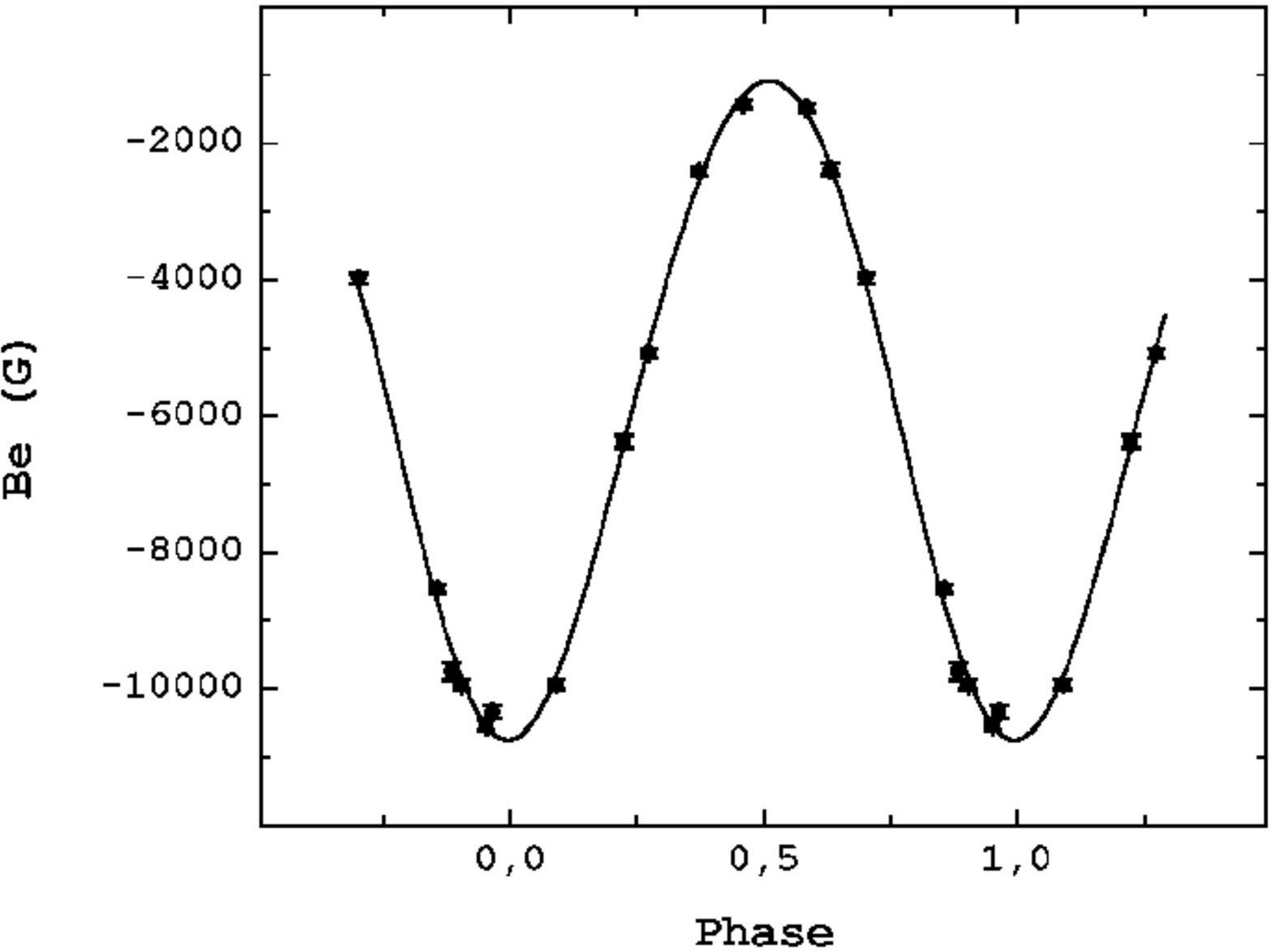}}
\vspace{-3.5mm}
\caption{ HD 75049 (2) }
\label{fig:fig160}
\end{figure}

\clearpage
\newpage

\begin{figure}
\resizebox{0.98\hsize}{!}{\includegraphics{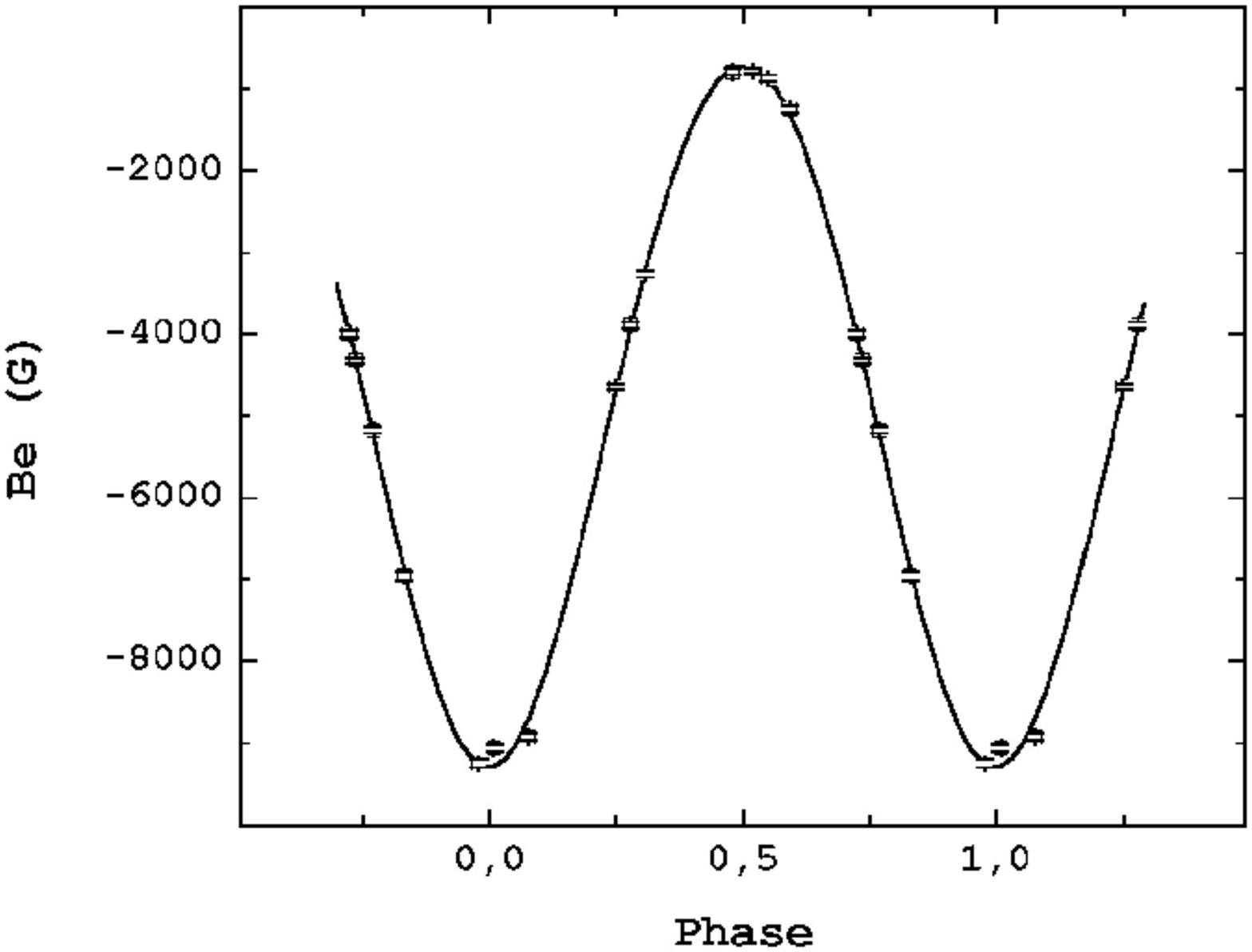}}
\vspace{-3.5mm}
\caption{ HD 75049 (3) }
\label{fig:fig161}
\end{figure}

\begin{figure}
\resizebox{0.98\hsize}{!}{\includegraphics{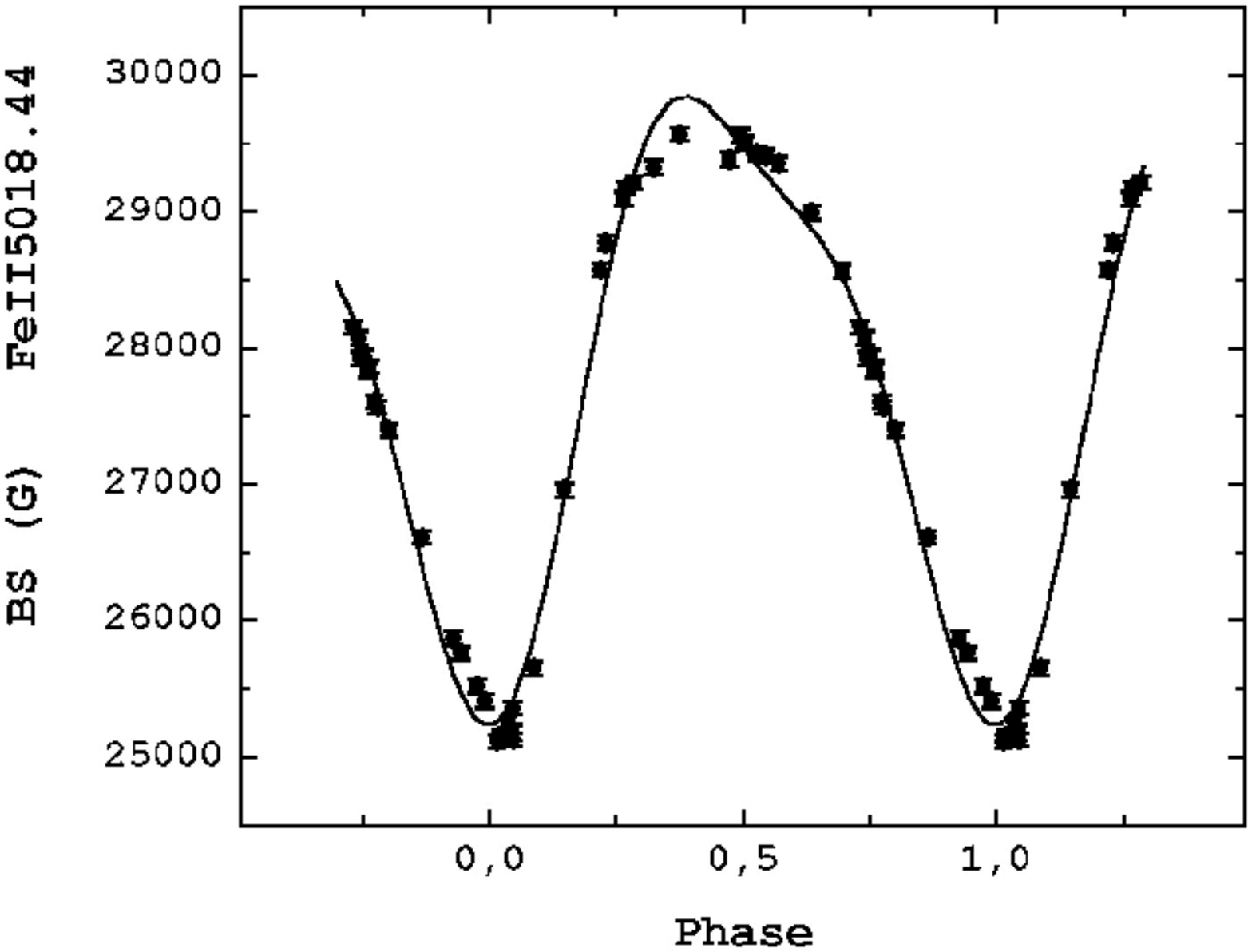}}
\vspace{-3.5mm}
\caption{ HD 75049 (4) }
\label{fig:fig162}
\end{figure}

\begin{figure}
\resizebox{0.98\hsize}{!}{\includegraphics{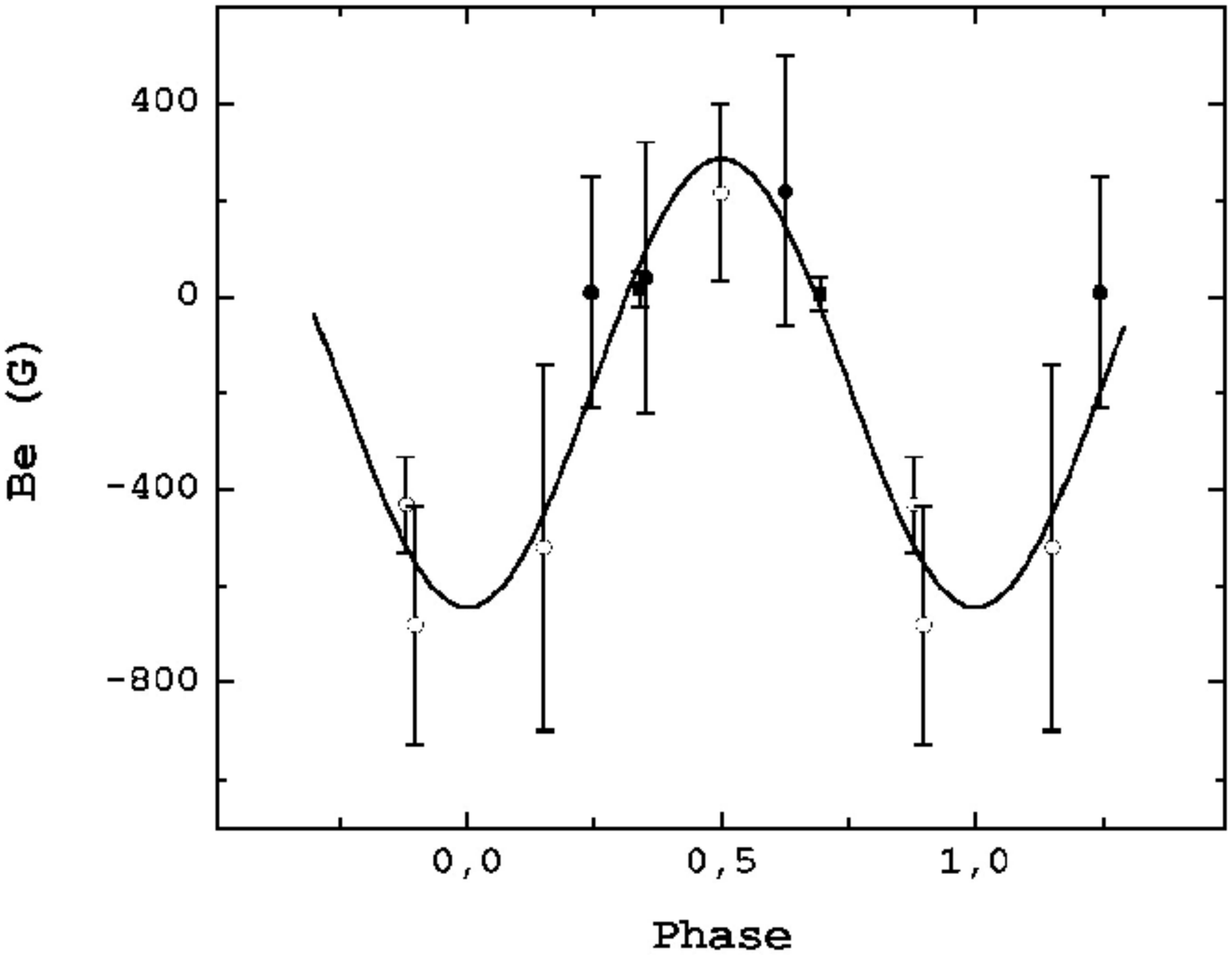}}
\vspace{-3.5mm}
\caption{ HD 77350 }
\label{fig:fig163}
\end{figure}

\begin{figure}
\resizebox{0.98\hsize}{!}{\includegraphics{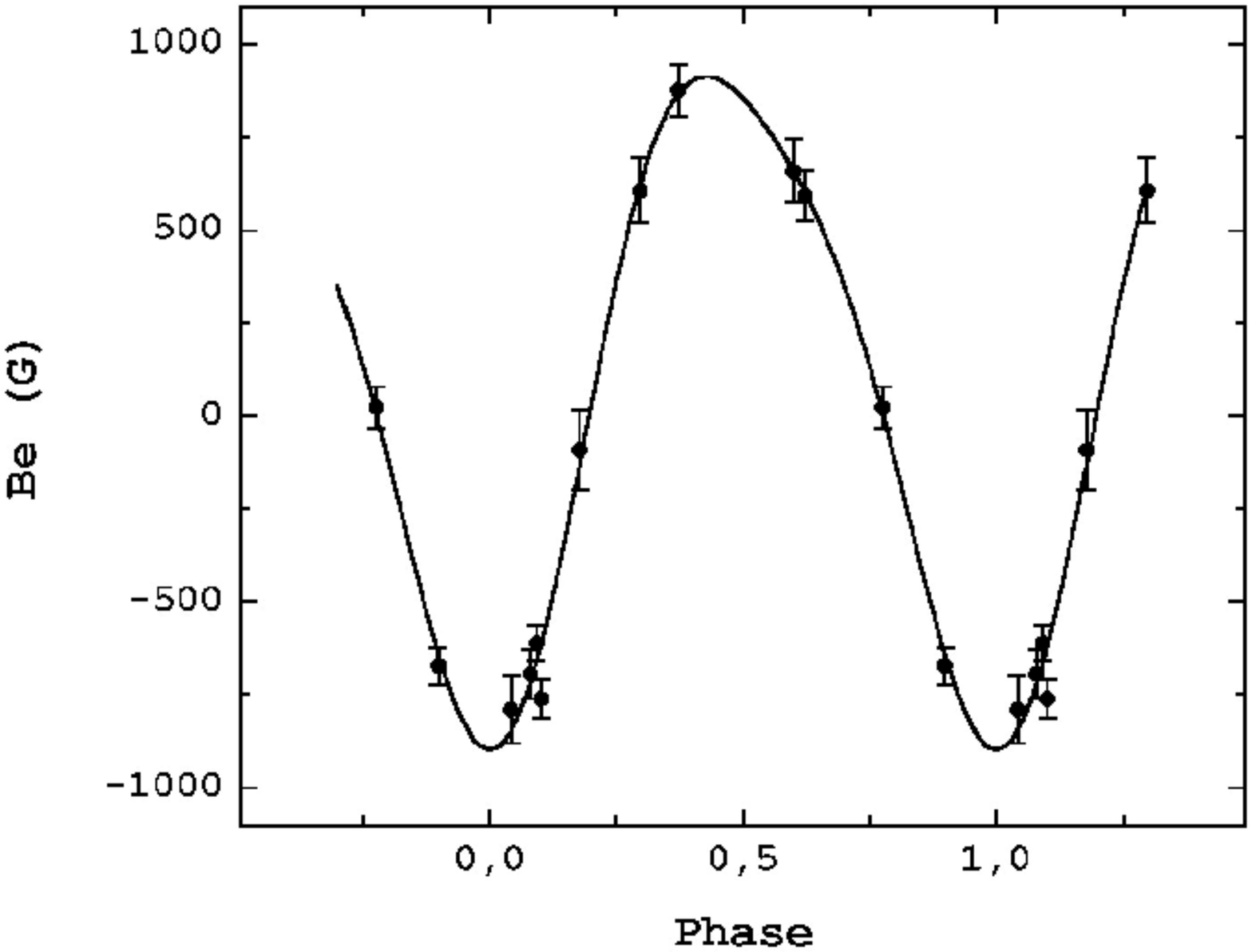}}
\vspace{-3.5mm}
\caption{ HD 79158 (1) }
\label{fig:fig164}
\end{figure}

\begin{figure}
\resizebox{0.98\hsize}{!}{\includegraphics{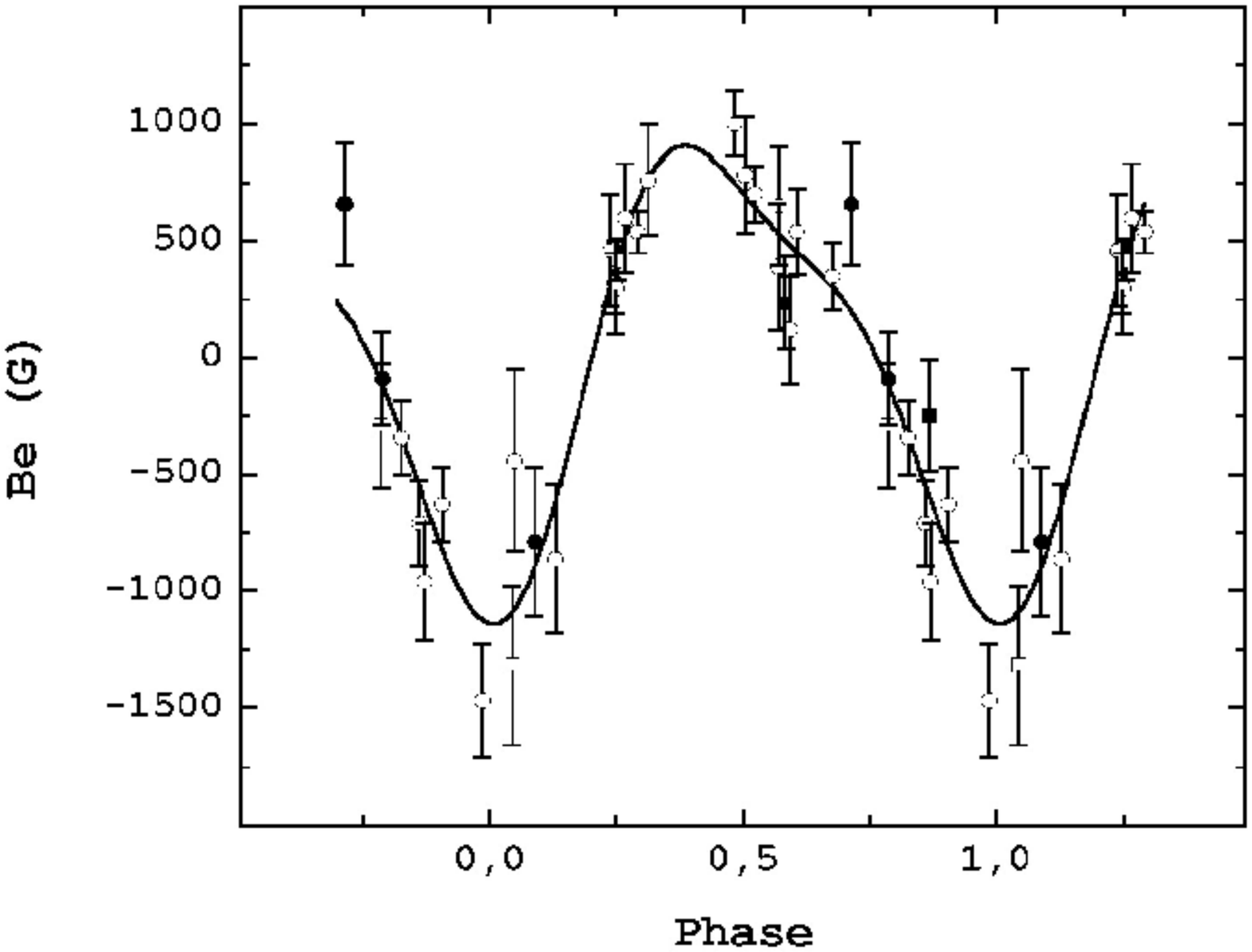}}
\vspace{-3.5mm}
\caption{ HD 79158 (2) }
\label{fig:fig165}
\end{figure}

\begin{figure}
\resizebox{0.98\hsize}{!}{\includegraphics{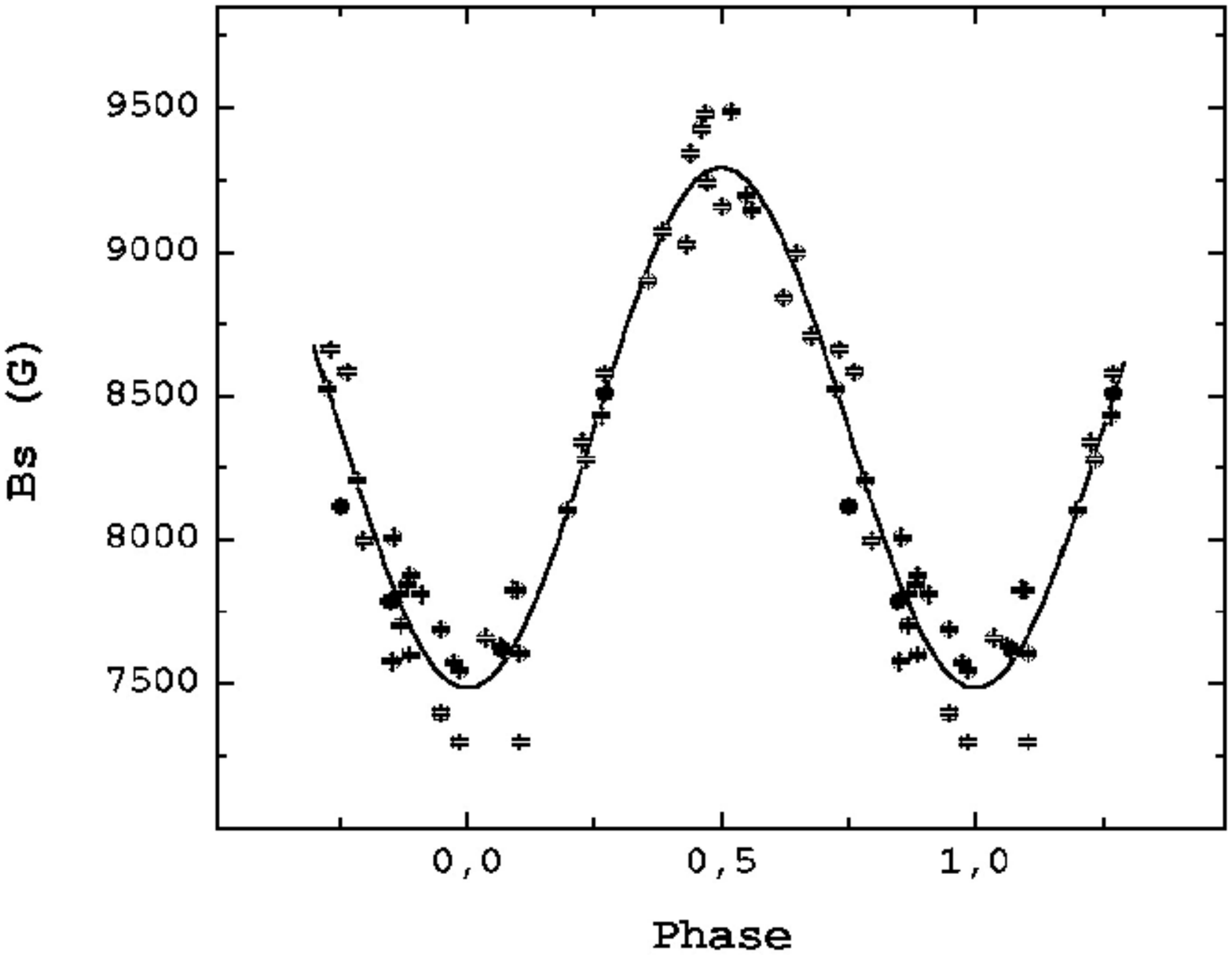}}
\vspace{-3.5mm}
\caption{ HD 81009 (1) }
\label{fig:fig166}
\end{figure}

\clearpage
\newpage

\begin{figure}
\resizebox{0.98\hsize}{!}{\includegraphics{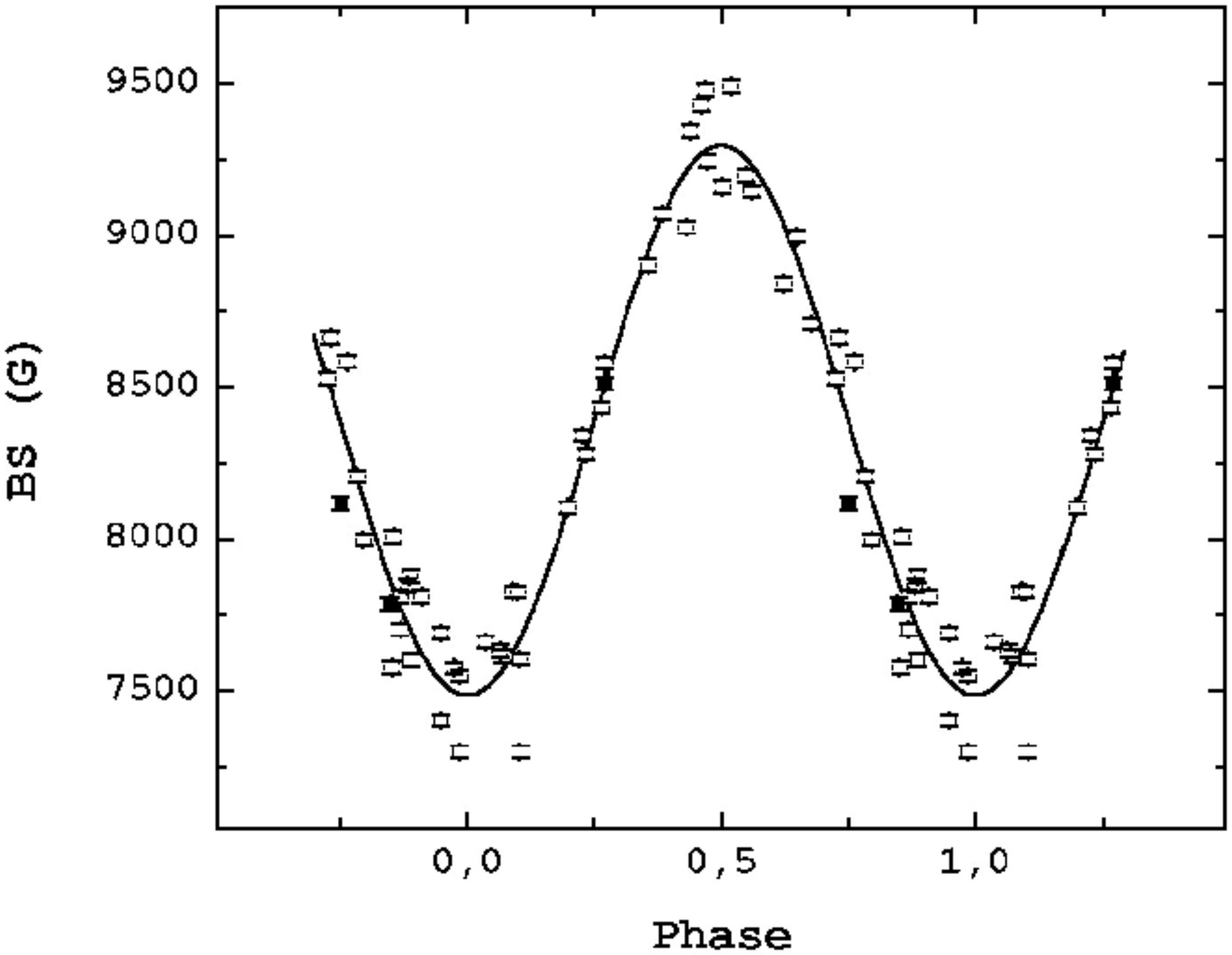}}
\vspace{-3.5mm}
\caption{ HD 81009 (2) }
\label{fig:fig167}
\end{figure}

\begin{figure}
\resizebox{0.98\hsize}{!}{\includegraphics{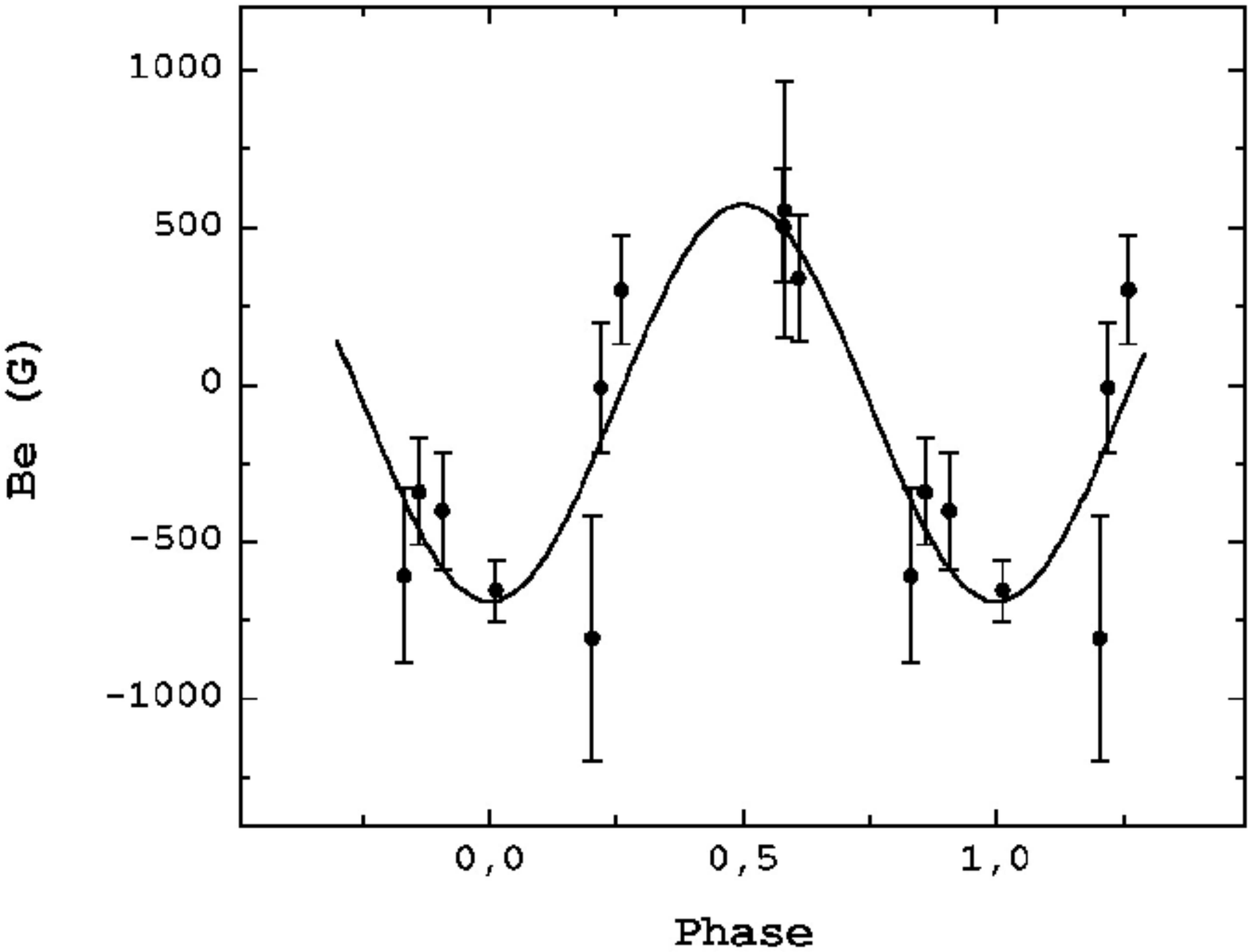}}
\vspace{-3.5mm}
\caption{ HD 83368 (1) }
\label{fig:fig168}
\end{figure}

\begin{figure}
\resizebox{0.98\hsize}{!}{\includegraphics{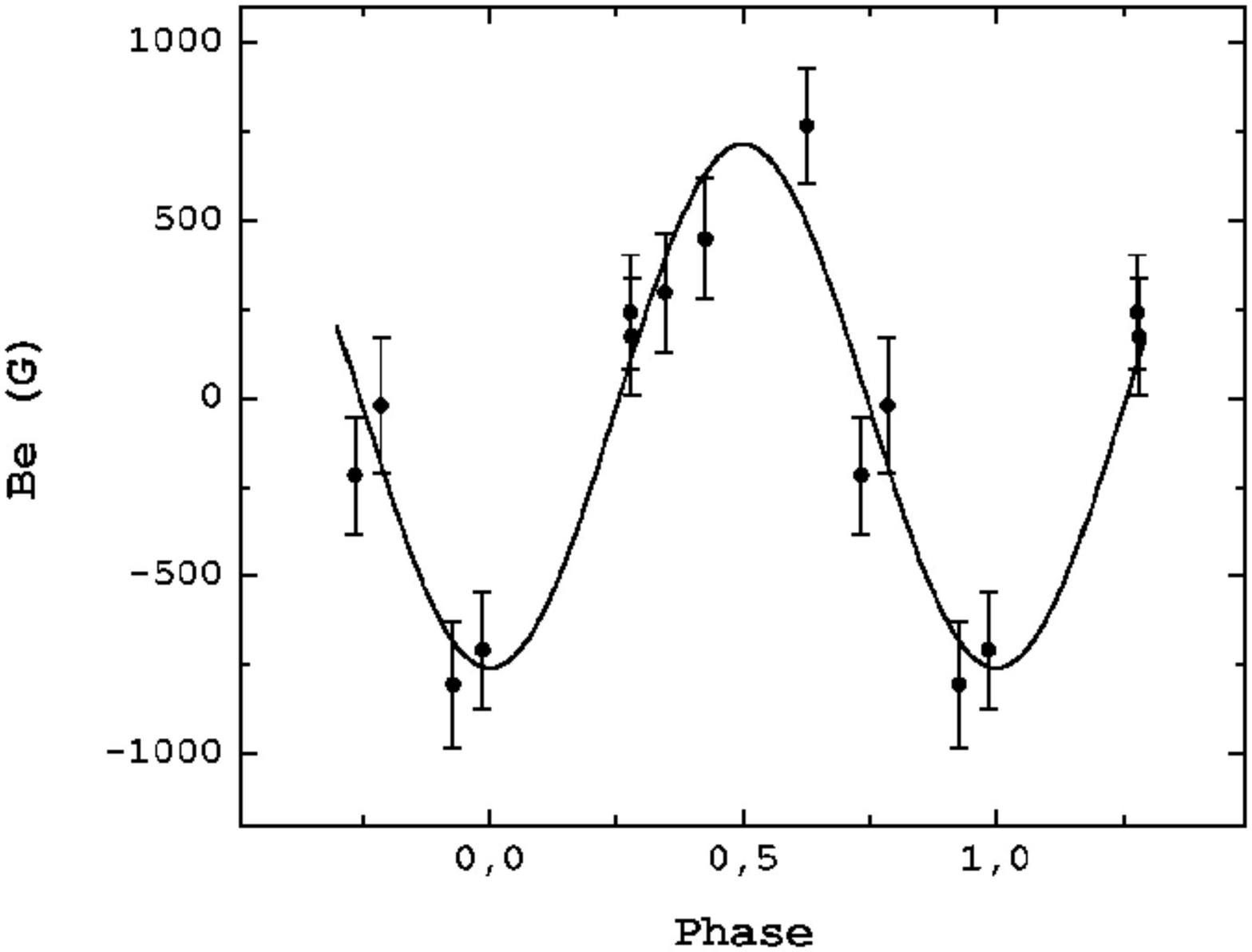}}
\vspace{-3.5mm}
\caption{ HD 83368 (2) }
\label{fig:fig169}
\end{figure}

\begin{figure}
\resizebox{0.98\hsize}{!}{\includegraphics{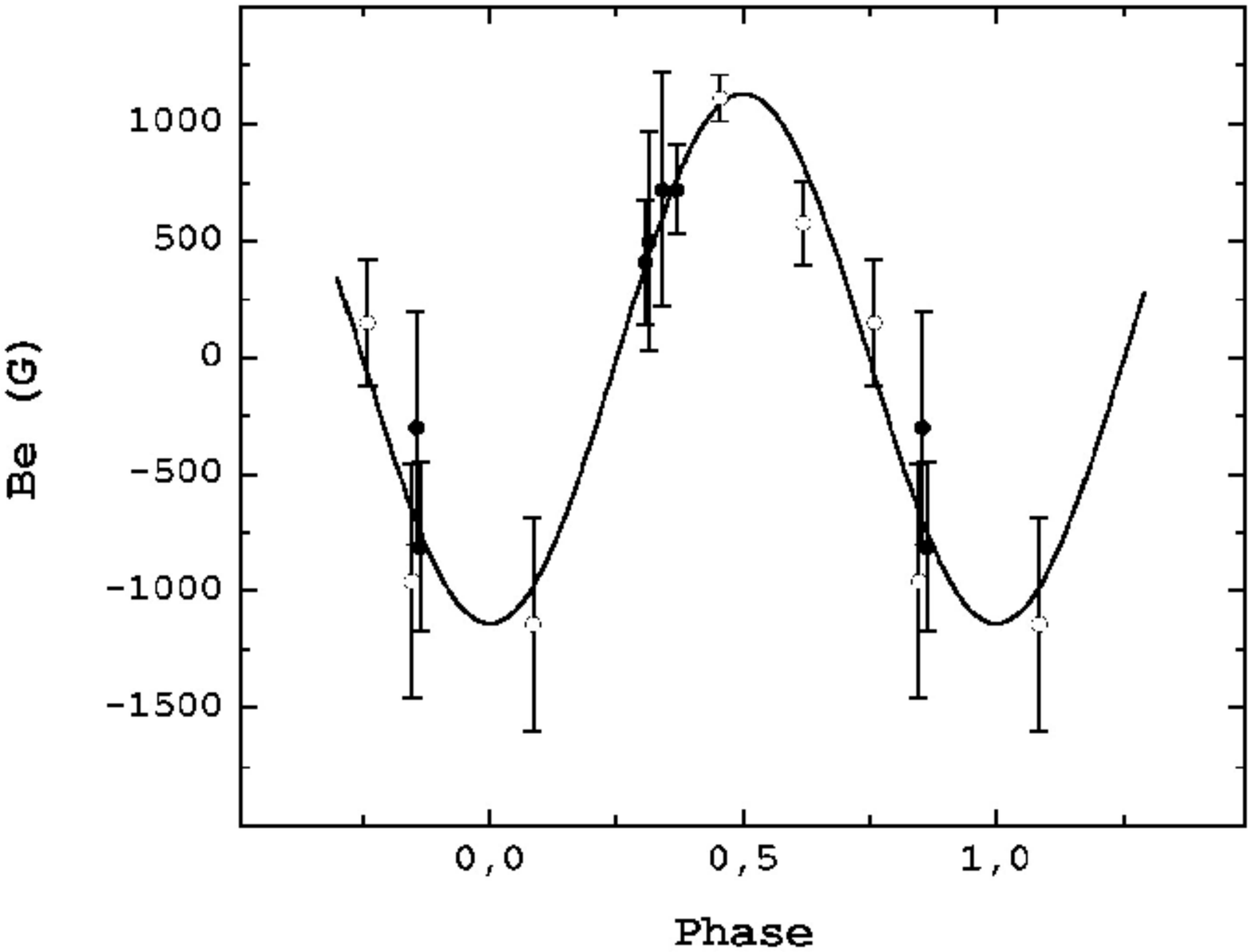}}
\vspace{-3.5mm}
\caption{ HD 90044 }
\label{fig:fig170}
\end{figure}

\begin{figure}
\resizebox{0.98\hsize}{!}{\includegraphics{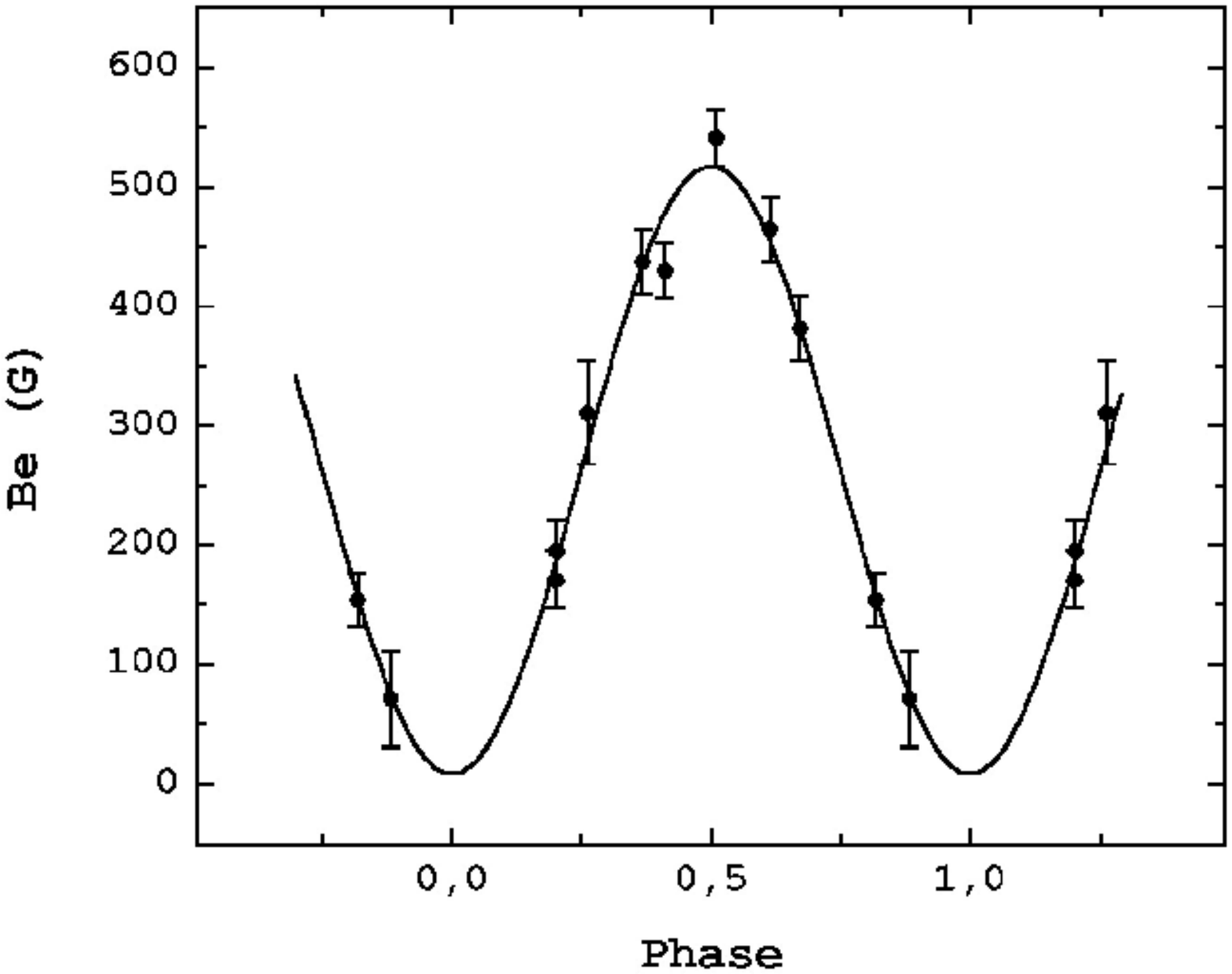}}
\vspace{-3.5mm}
\caption{ HD 90569 }
\label{fig:fig171}
\end{figure}

\begin{figure}
\resizebox{0.98\hsize}{!}{\includegraphics{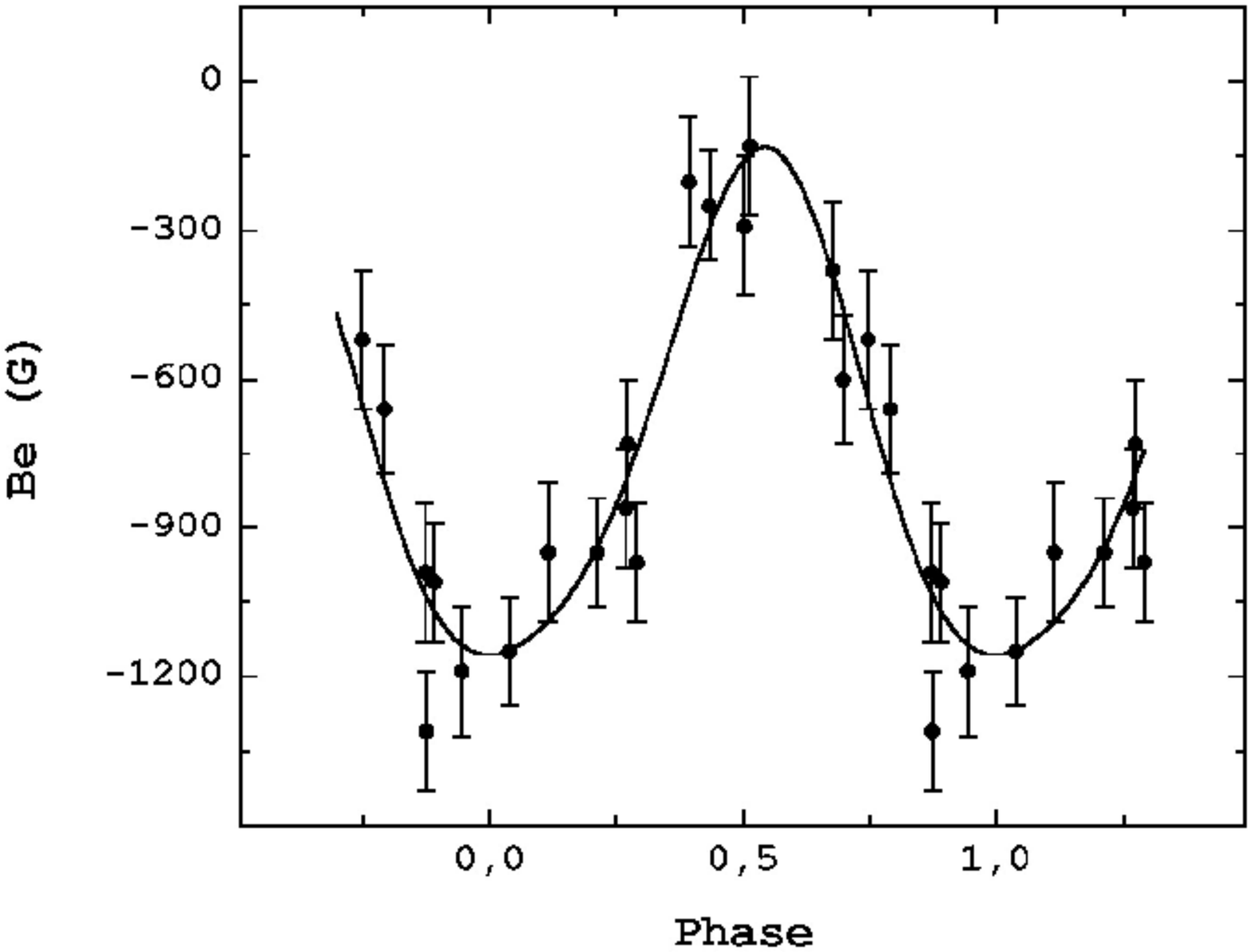}}
\vspace{-3.5mm}
\caption{ HD 92664 }
\label{fig:fig172}
\end{figure}

\clearpage
\newpage

\begin{figure}
\resizebox{0.98\hsize}{!}{\includegraphics{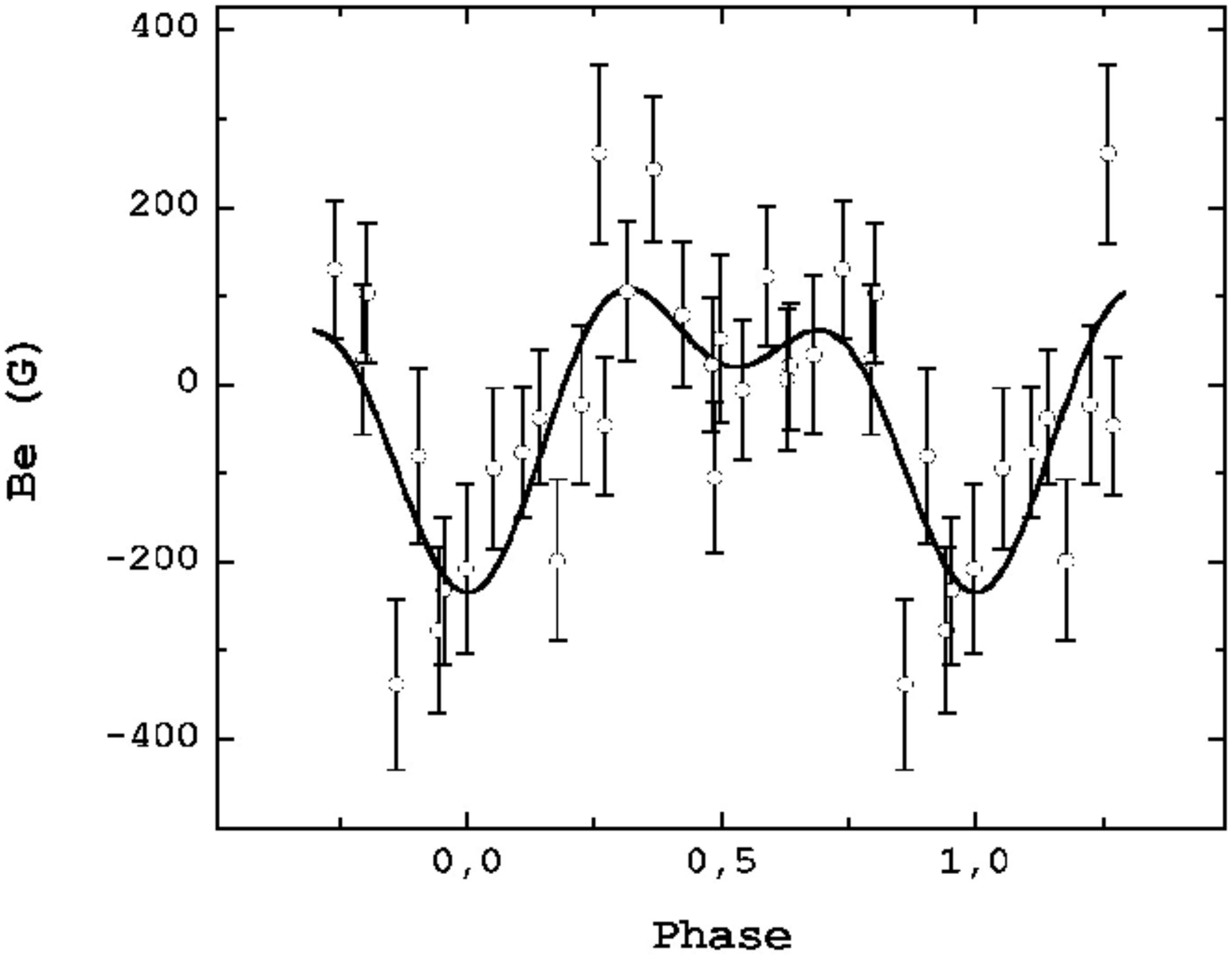}}
\vspace{-3.5mm}
\caption{ HD 93030 }
\label{fig:fig173}
\end{figure}

\begin{figure}
\resizebox{0.98\hsize}{!}{\includegraphics{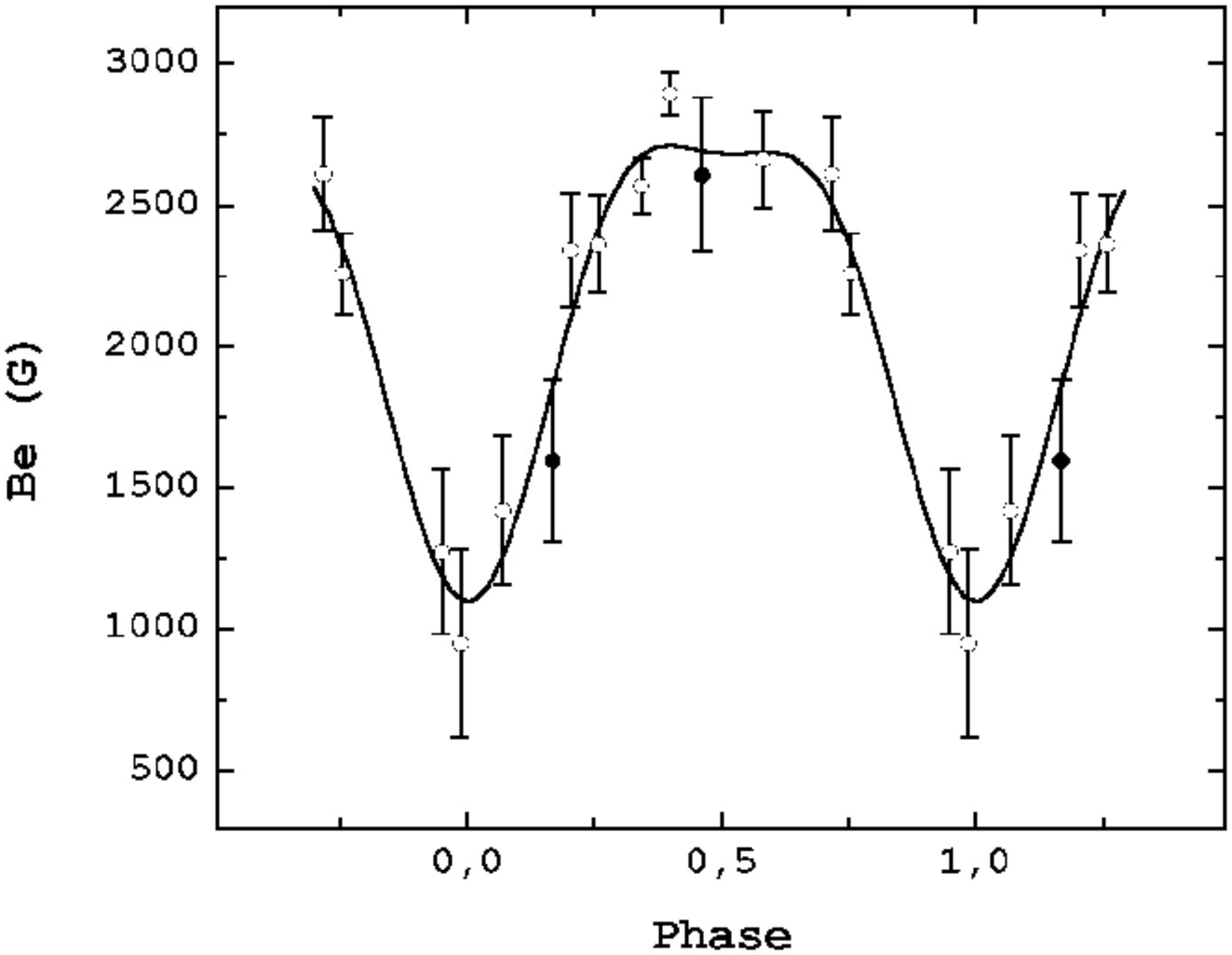}}
\vspace{-3.5mm}
\caption{ HD 93507 (1) }
\label{fig:fig174}
\end{figure}

\begin{figure}
\resizebox{0.98\hsize}{!}{\includegraphics{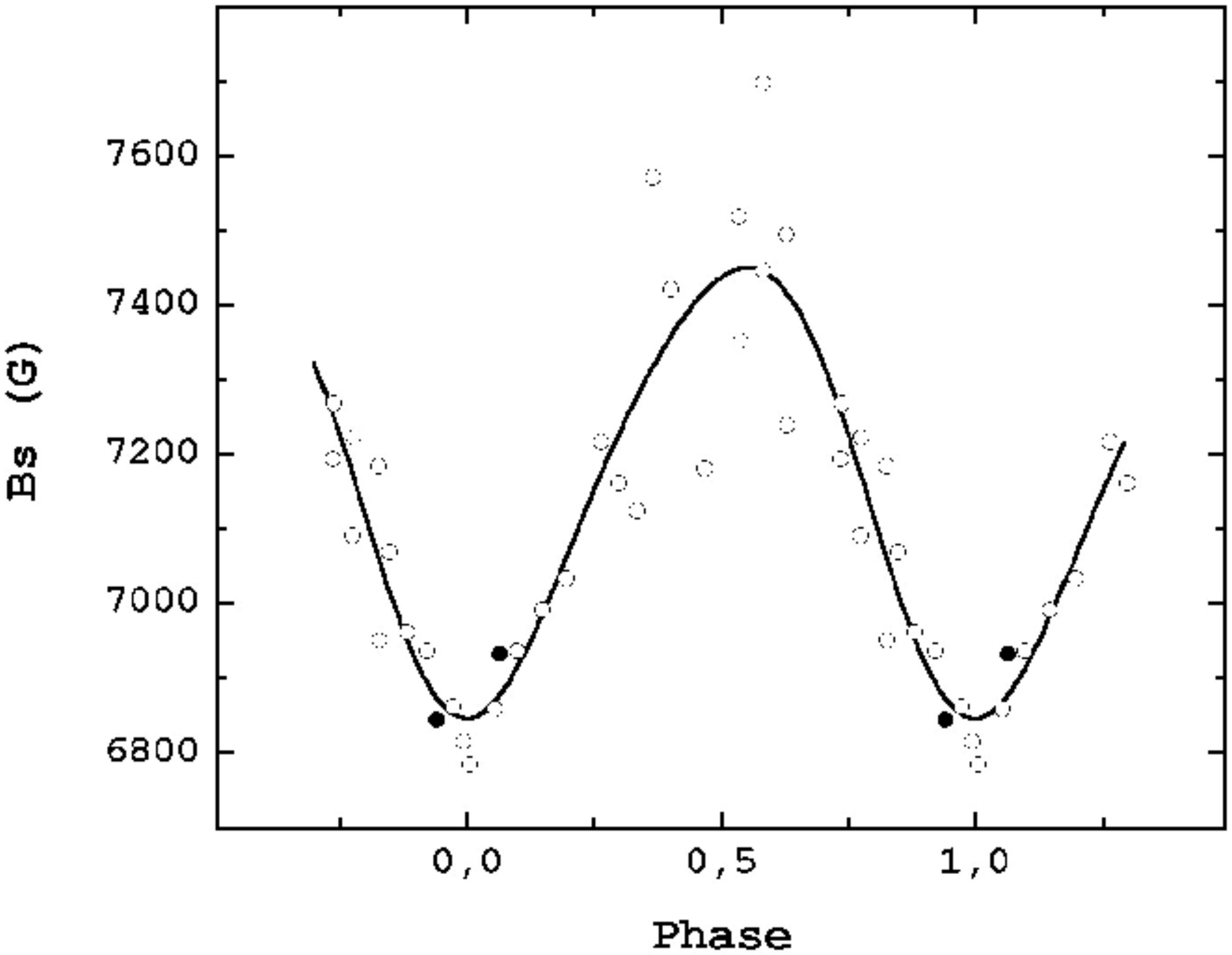}}
\vspace{-3.5mm}
\caption{ HD 93507 (2) }
\label{fig:fig175}
\end{figure}

\begin{figure}
\resizebox{0.98\hsize}{!}{\includegraphics{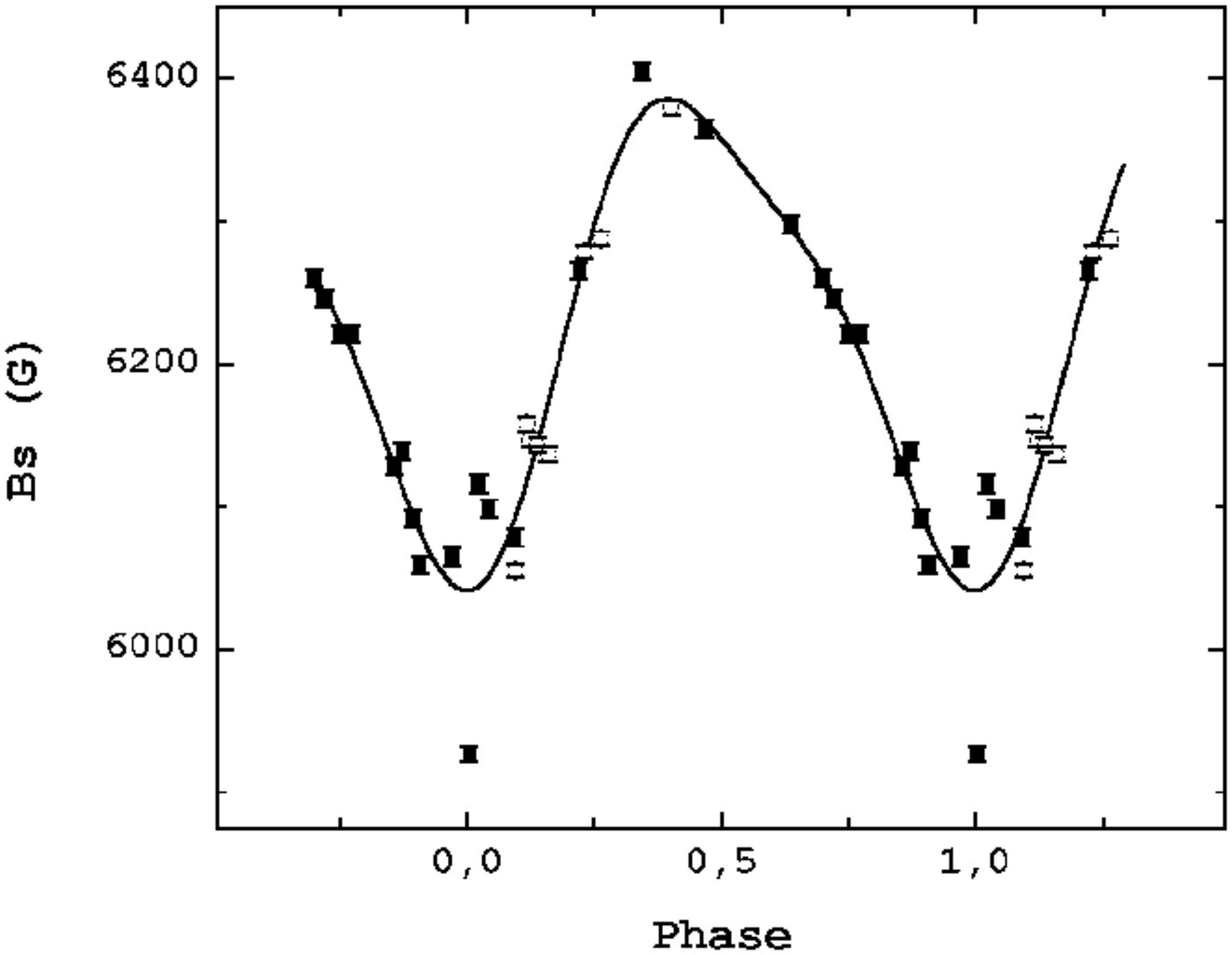}}
\vspace{-3.5mm}
\caption{ HD 94660 (1) }
\label{fig:fig176}
\end{figure}

\begin{figure}
\resizebox{0.98\hsize}{!}{\includegraphics{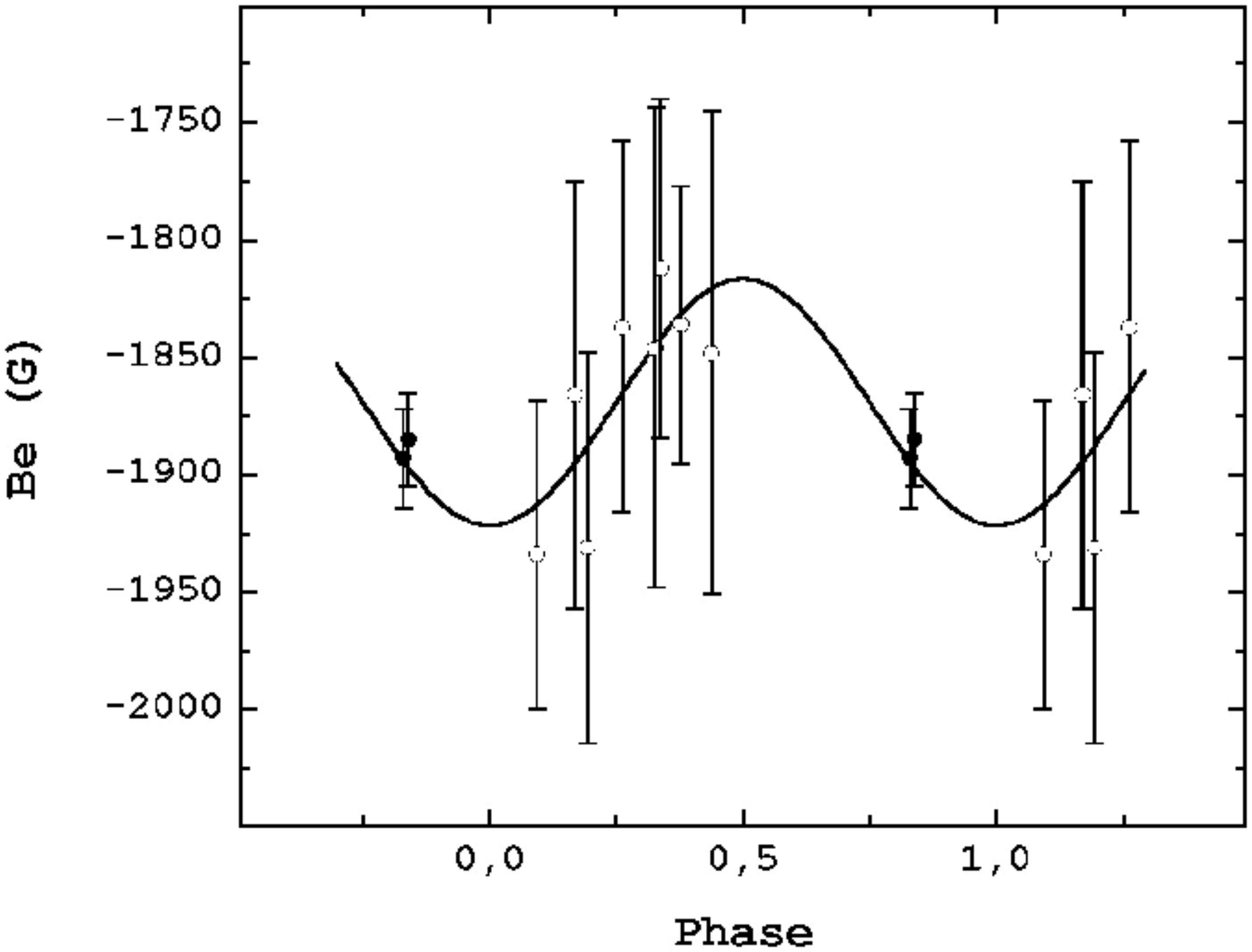}}
\vspace{-3.5mm}
\caption{ HD 94660 (2) }
\label{fig:fig177}
\end{figure}

\begin{figure}
\resizebox{0.98\hsize}{!}{\includegraphics{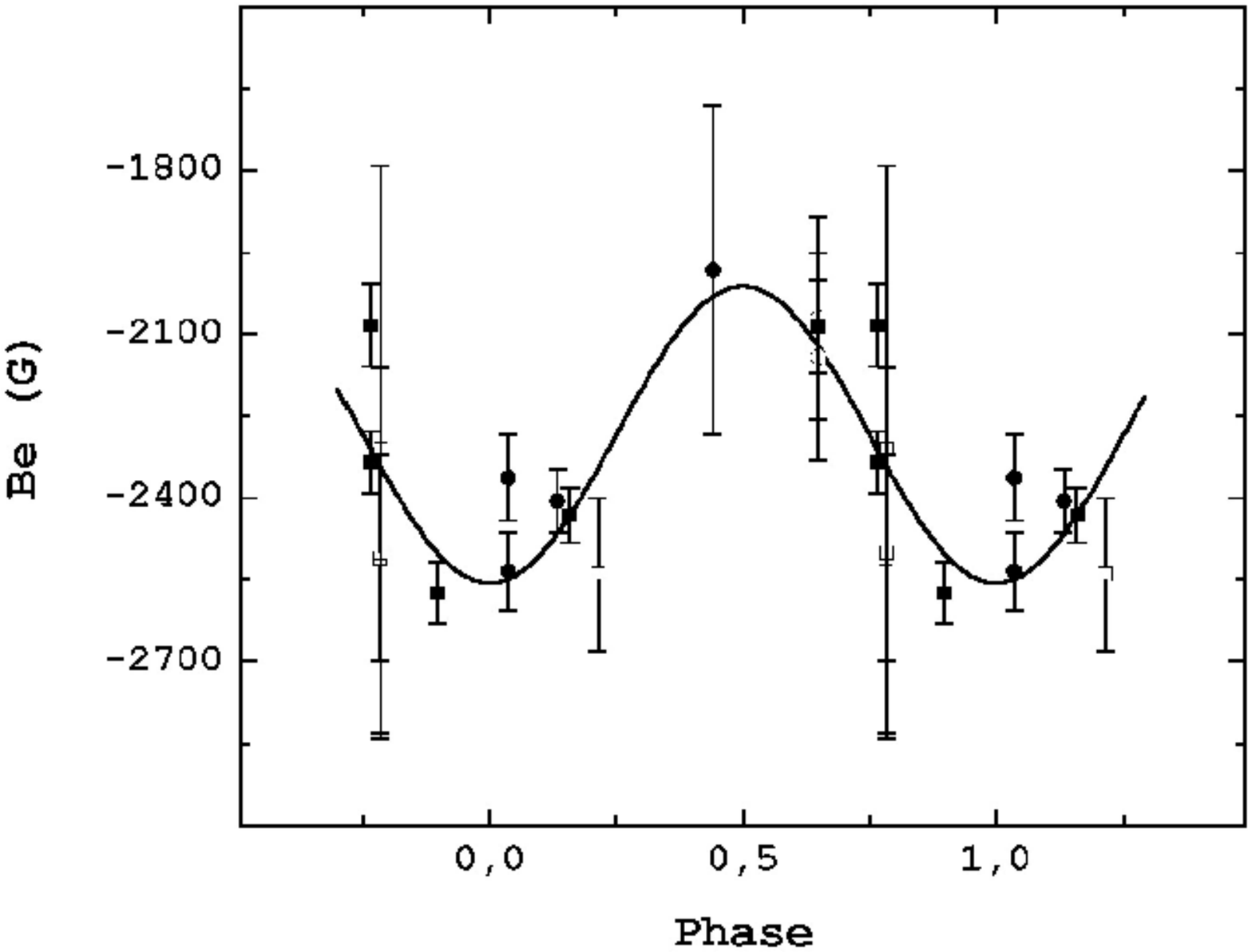}}
\vspace{-3.5mm}
\caption{ HD 94660 (3) }
\label{fig:fig178}
\end{figure}

\clearpage
\newpage

\begin{figure}
\resizebox{0.98\hsize}{!}{\includegraphics{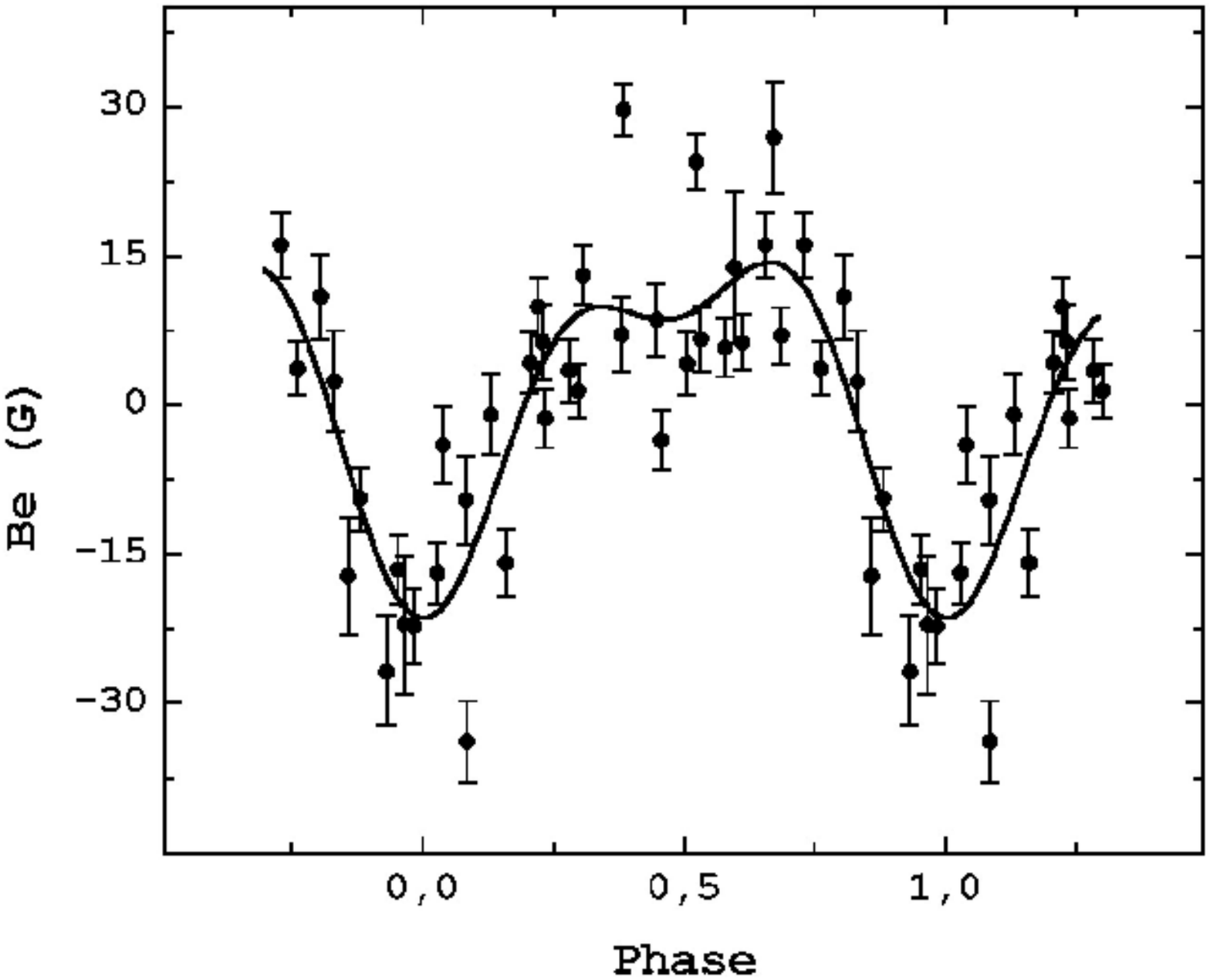}}
\vspace{-3.5mm}
\caption{ HD 95650 (1) }
\label{fig:fig179}
\end{figure}

\begin{figure}
\resizebox{0.98\hsize}{!}{\includegraphics{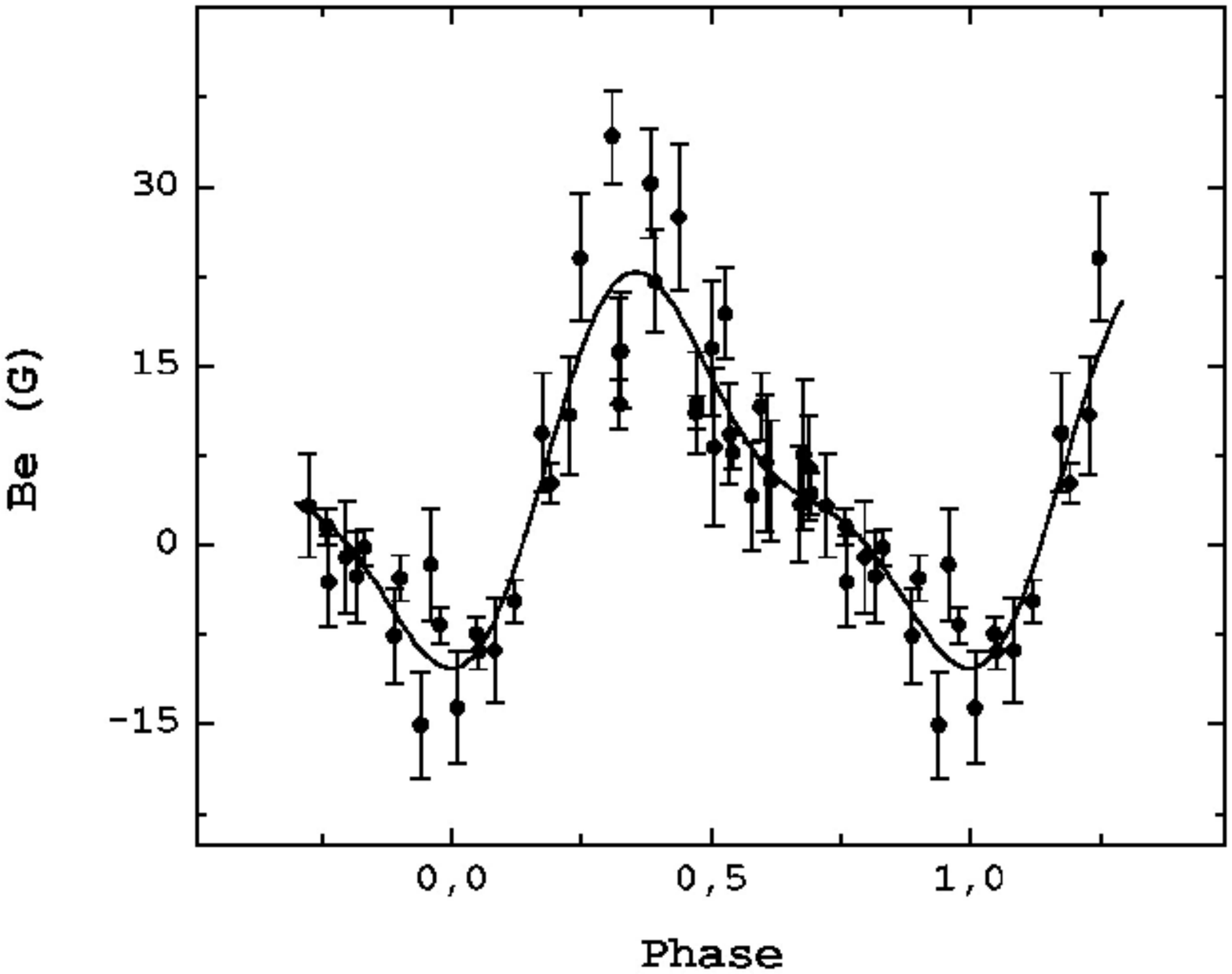}}
\vspace{-3.5mm}
\caption{ HD 95650 (2) }
\label{fig:fig180}
\end{figure}

\begin{figure}
\resizebox{0.98\hsize}{!}{\includegraphics{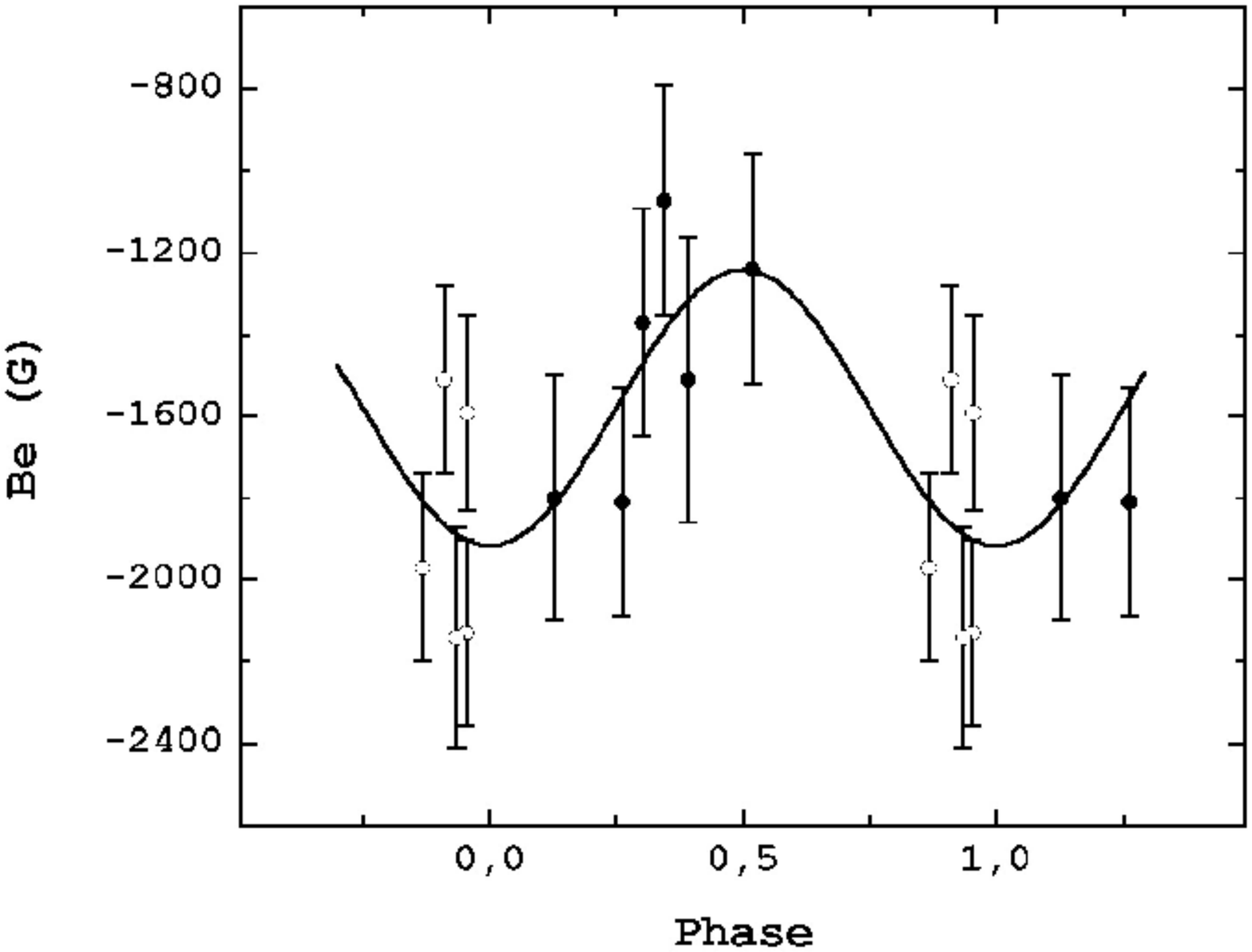}}
\vspace{-3.5mm}
\caption{ HD 96446 (1) }
\label{fig:fig181}
\end{figure}

\begin{figure}
\resizebox{0.98\hsize}{!}{\includegraphics{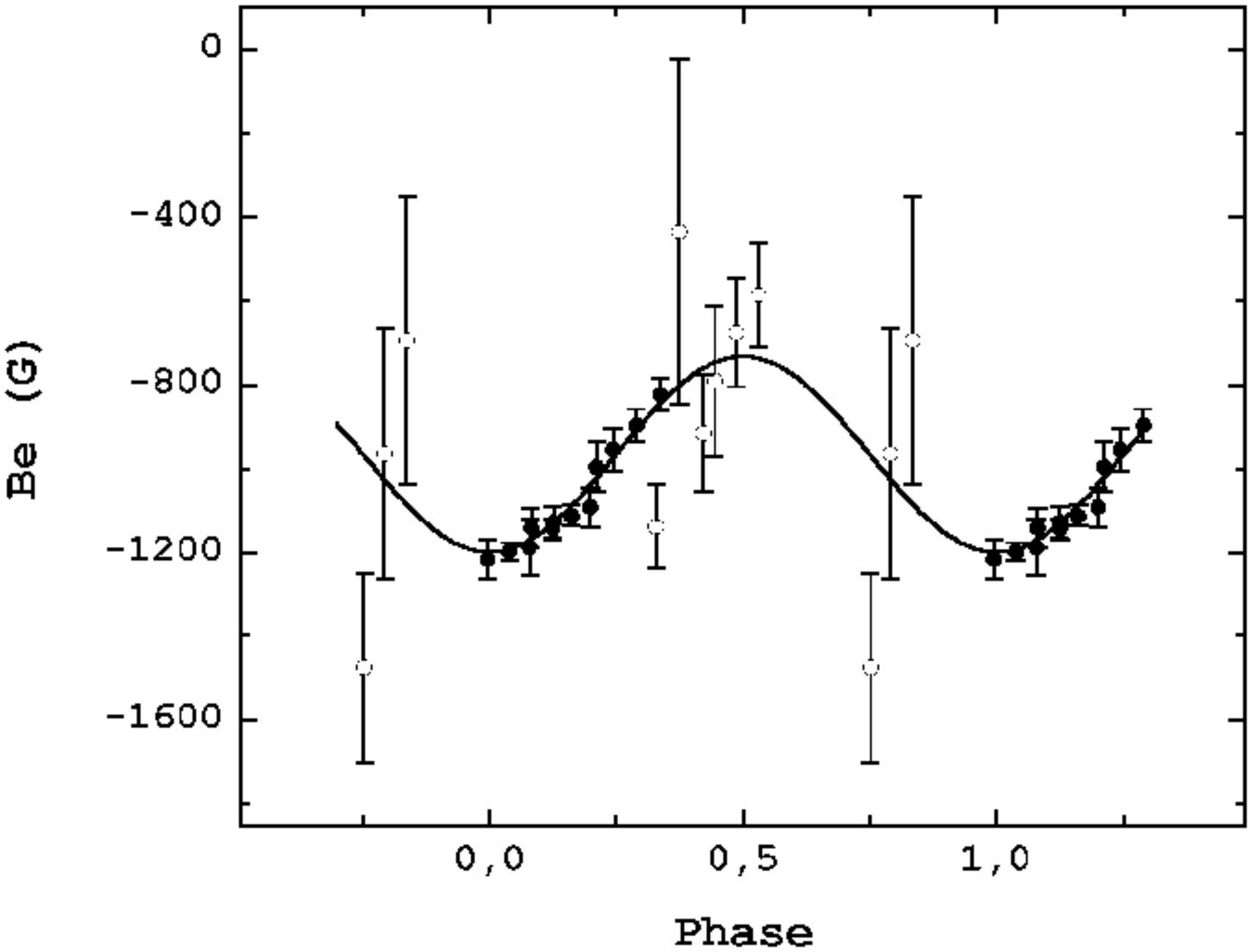}}
\vspace{-3.5mm}
\caption{ HD 96446 (2) }
\label{fig:fig182}
\end{figure}

\begin{figure}
\resizebox{0.98\hsize}{!}{\includegraphics{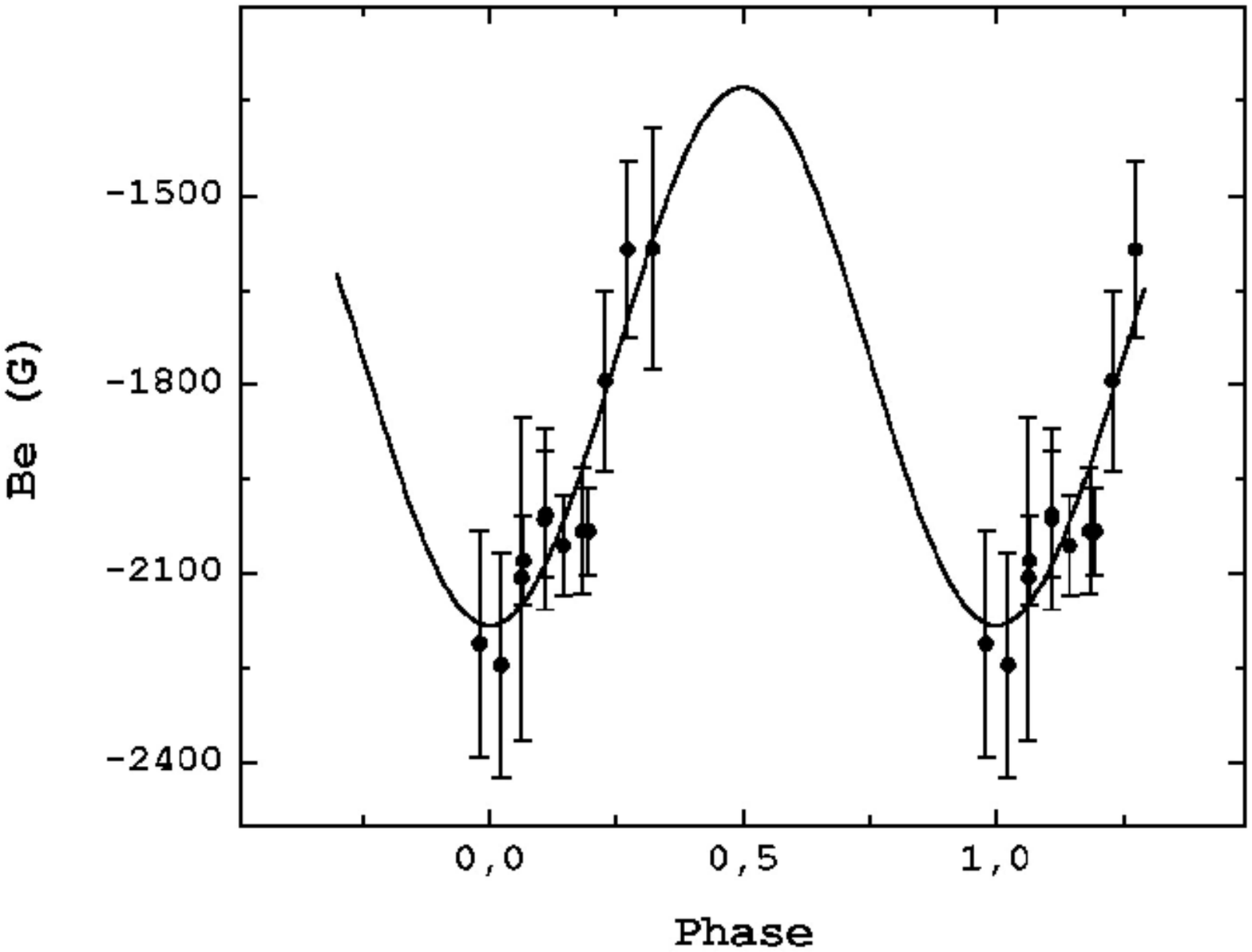}}
\vspace{-3.5mm}
\caption{ HD 96446 (3) }
\label{fig:fig183}
\end{figure}

\begin{figure}
\resizebox{0.98\hsize}{!}{\includegraphics{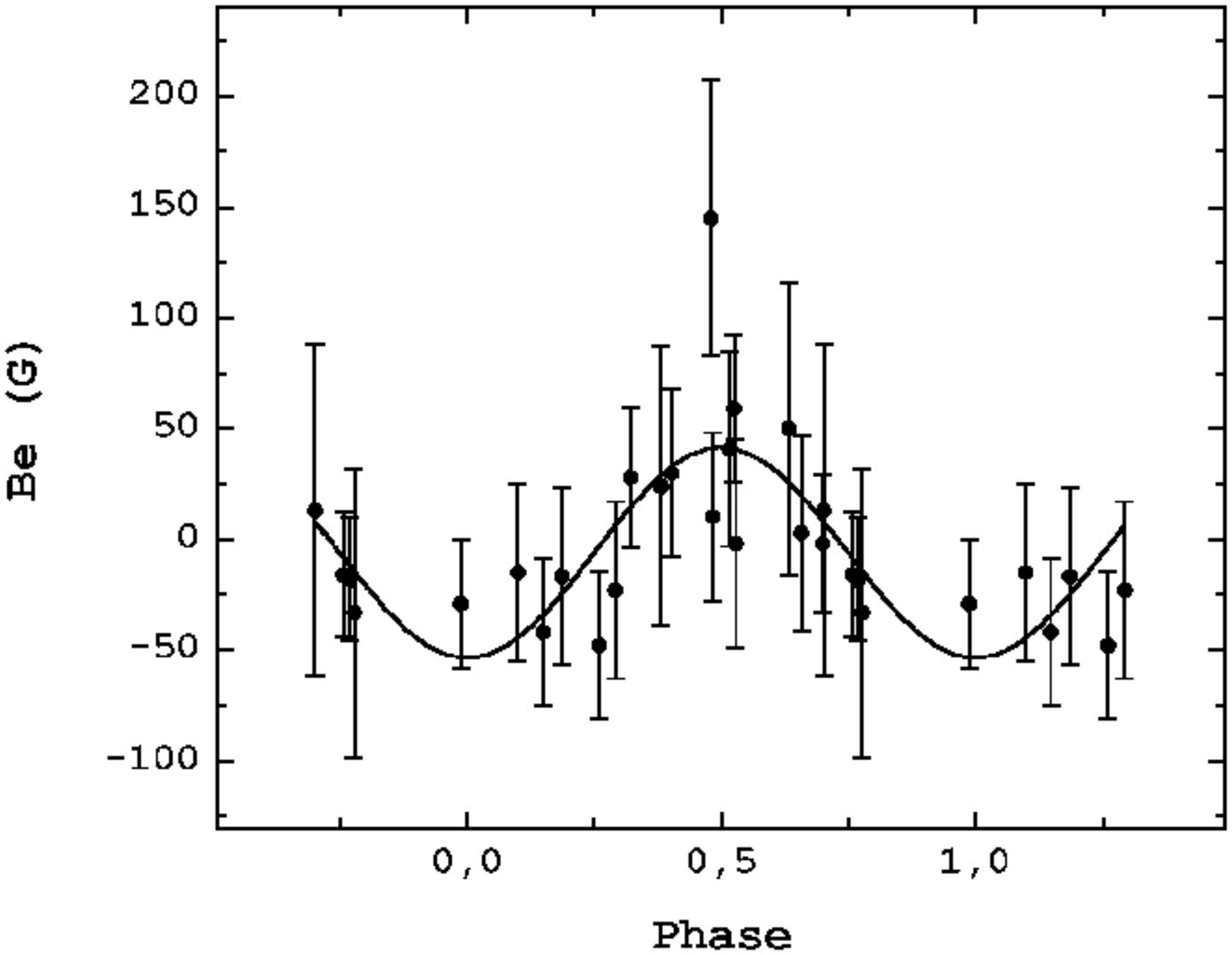}}
\vspace{-3.5mm}
\caption{ HD 96707 }
\label{fig:fig184}
\end{figure}

\clearpage
\newpage

\begin{figure}
\resizebox{0.98\hsize}{!}{\includegraphics{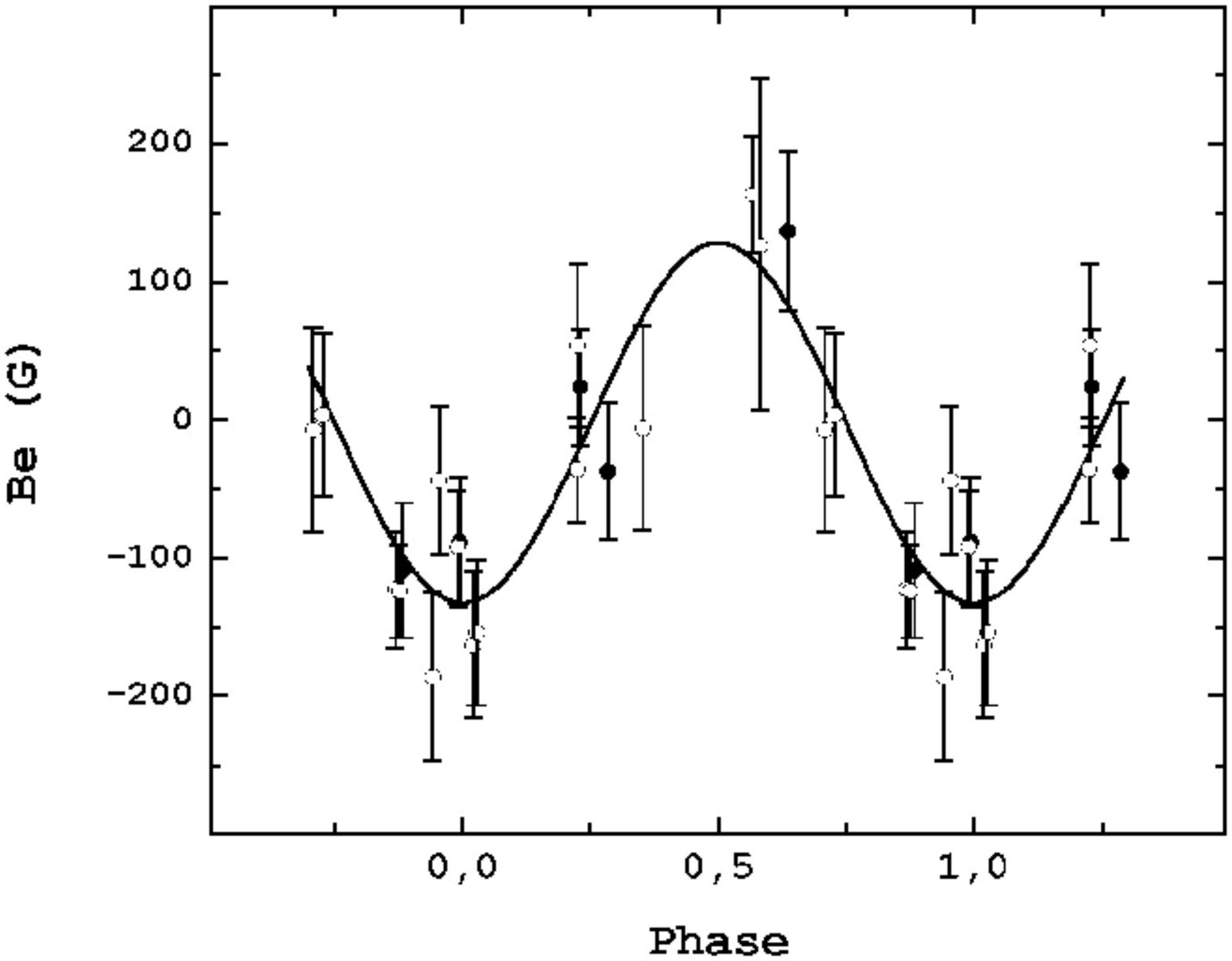}}
\vspace{-3.5mm}
\caption{ HD 97048 (1) }
\label{fig:fig185}
\end{figure}

\begin{figure}
\resizebox{0.98\hsize}{!}{\includegraphics{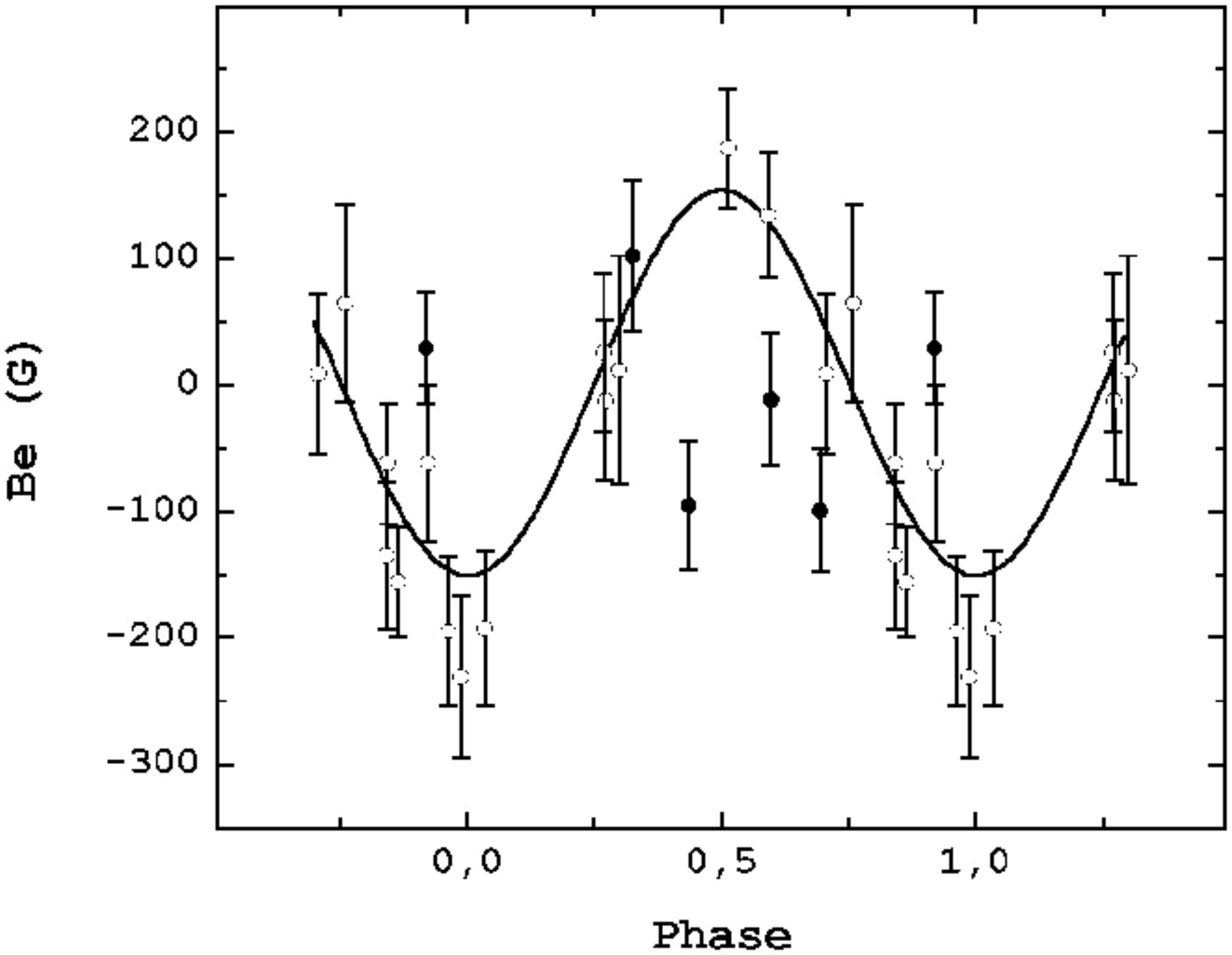}}
\vspace{-3.5mm}
\caption{ HD 97048 (2) }
\label{fig:fig186}
\end{figure}

\begin{figure}
\resizebox{0.98\hsize}{!}{\includegraphics{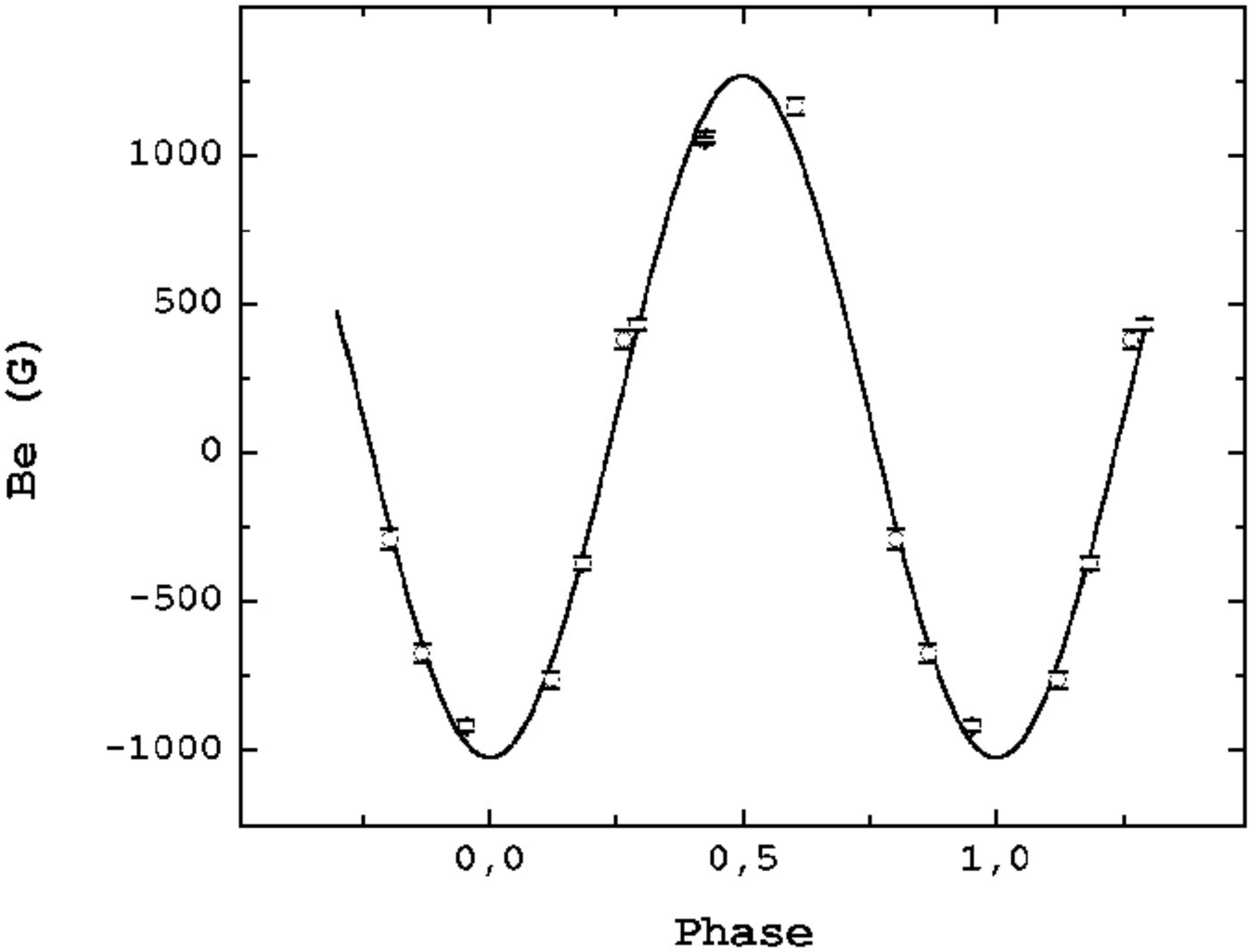}}
\vspace{-3.5mm}
\caption{ HD 98088 }
\label{fig:fig188}
\end{figure}

\begin{figure}
\resizebox{0.98\hsize}{!}{\includegraphics{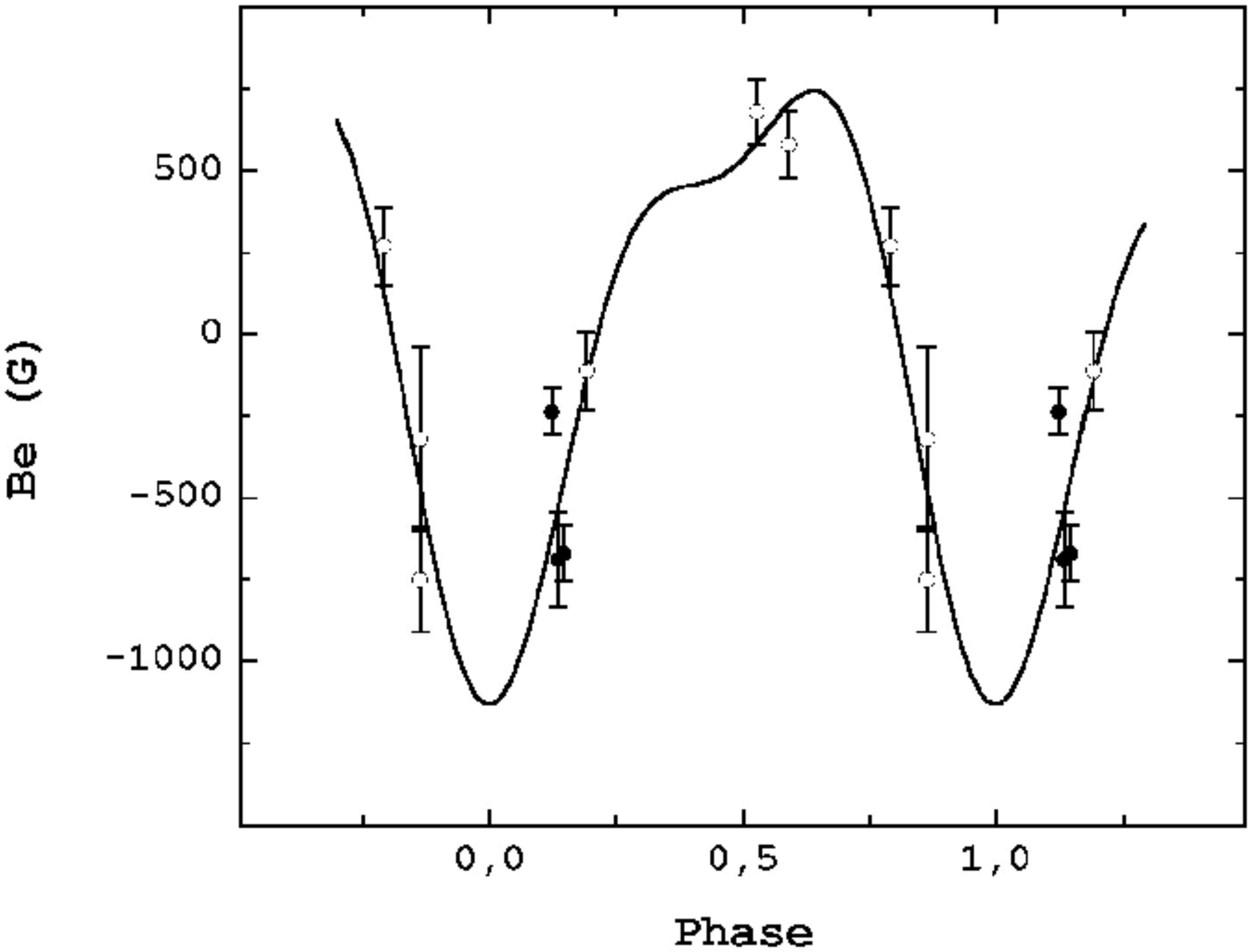}}
\vspace{-3.5mm}
\caption{ HD 99563 }
\label{fig:fig189}
\end{figure}

\begin{figure}
\resizebox{0.98\hsize}{!}{\includegraphics{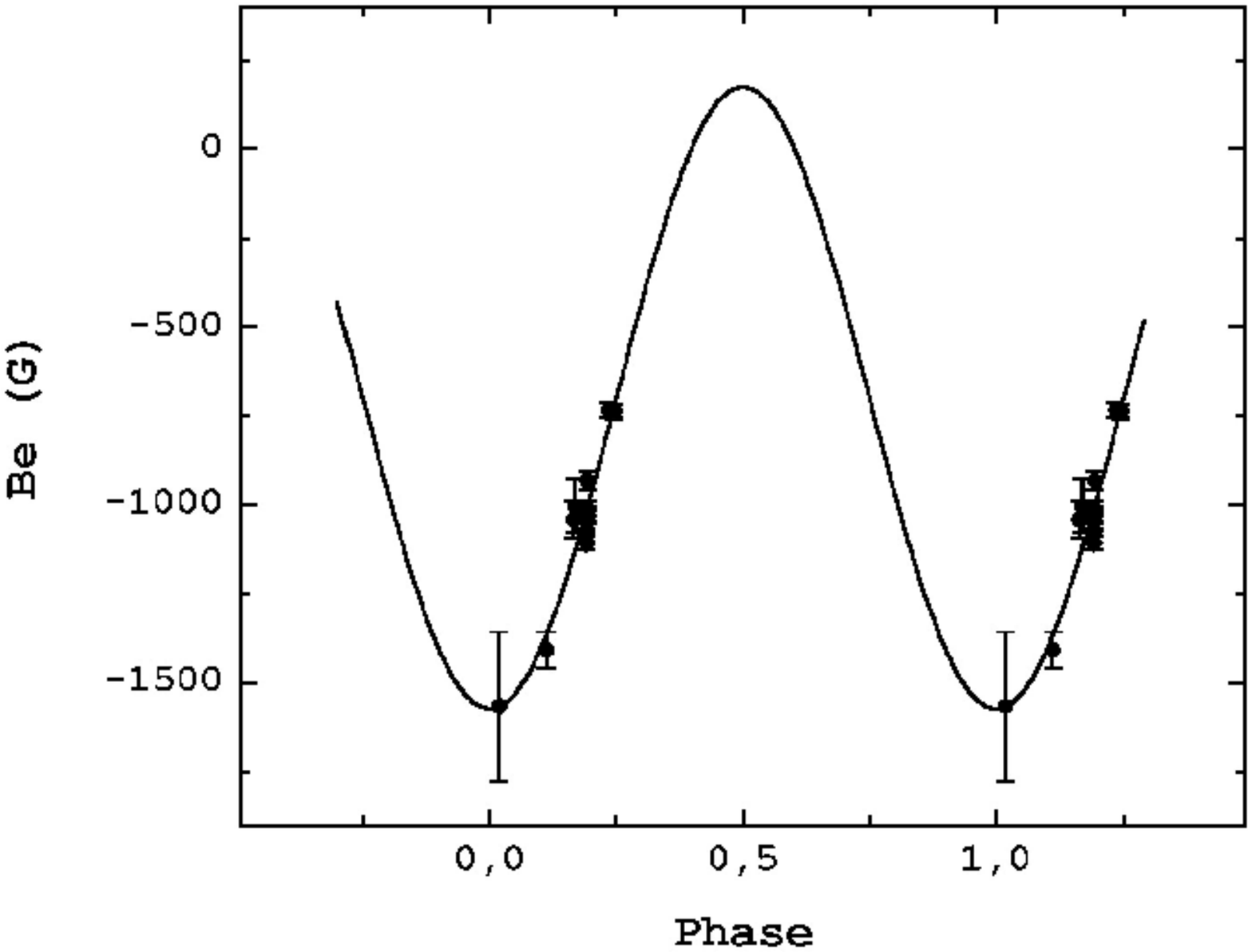}}
\vspace{-3.5mm}
\caption{ HD 101065 }
\label{fig:fig104}
\end{figure}

\begin{figure}
\resizebox{0.98\hsize}{!}{\includegraphics{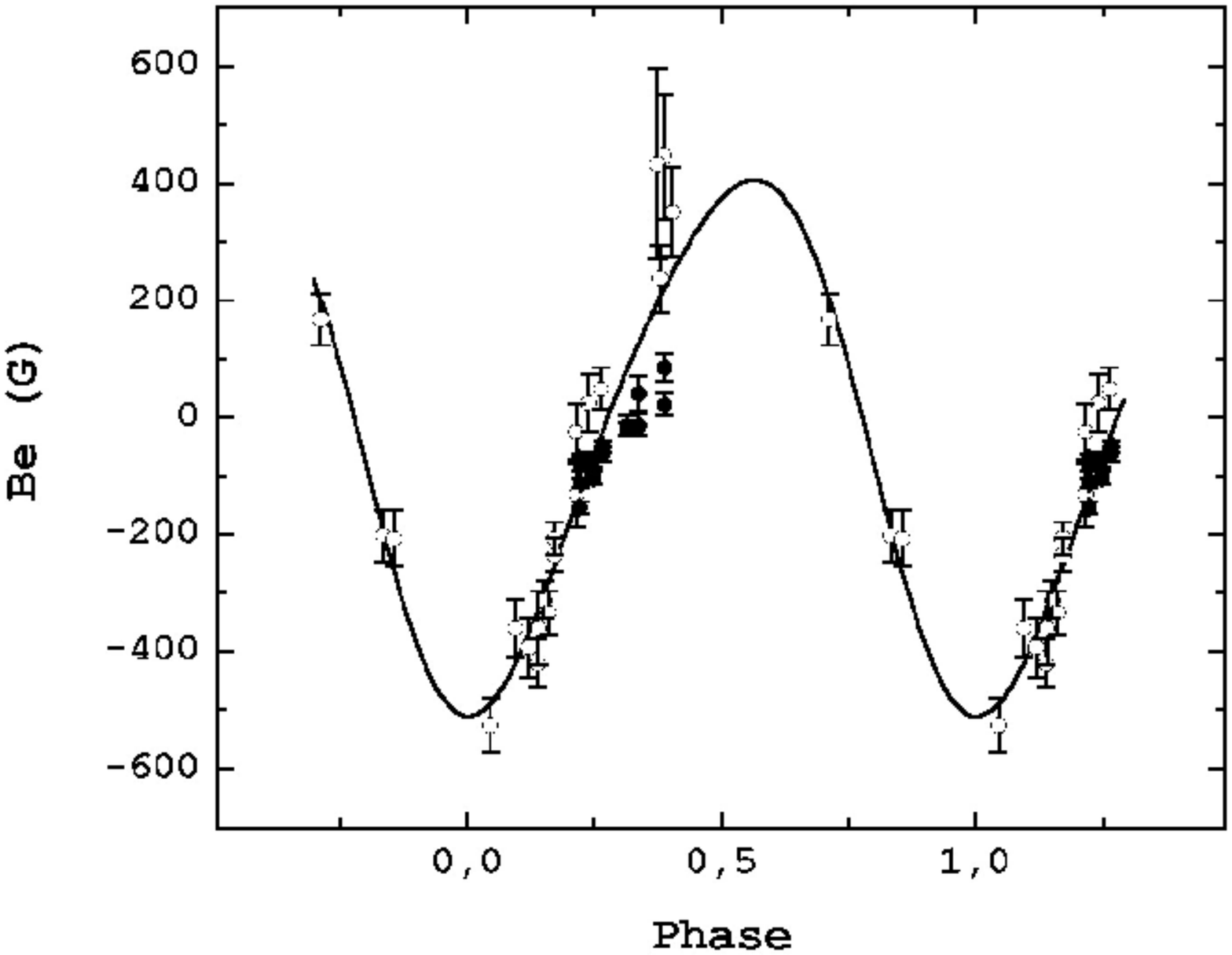}}
\vspace{-3.5mm}
\caption{ HD101412 (1) }
\label{fig:fig190}
\end{figure}

\clearpage
\newpage

\begin{figure}
\resizebox{0.98\hsize}{!}{\includegraphics{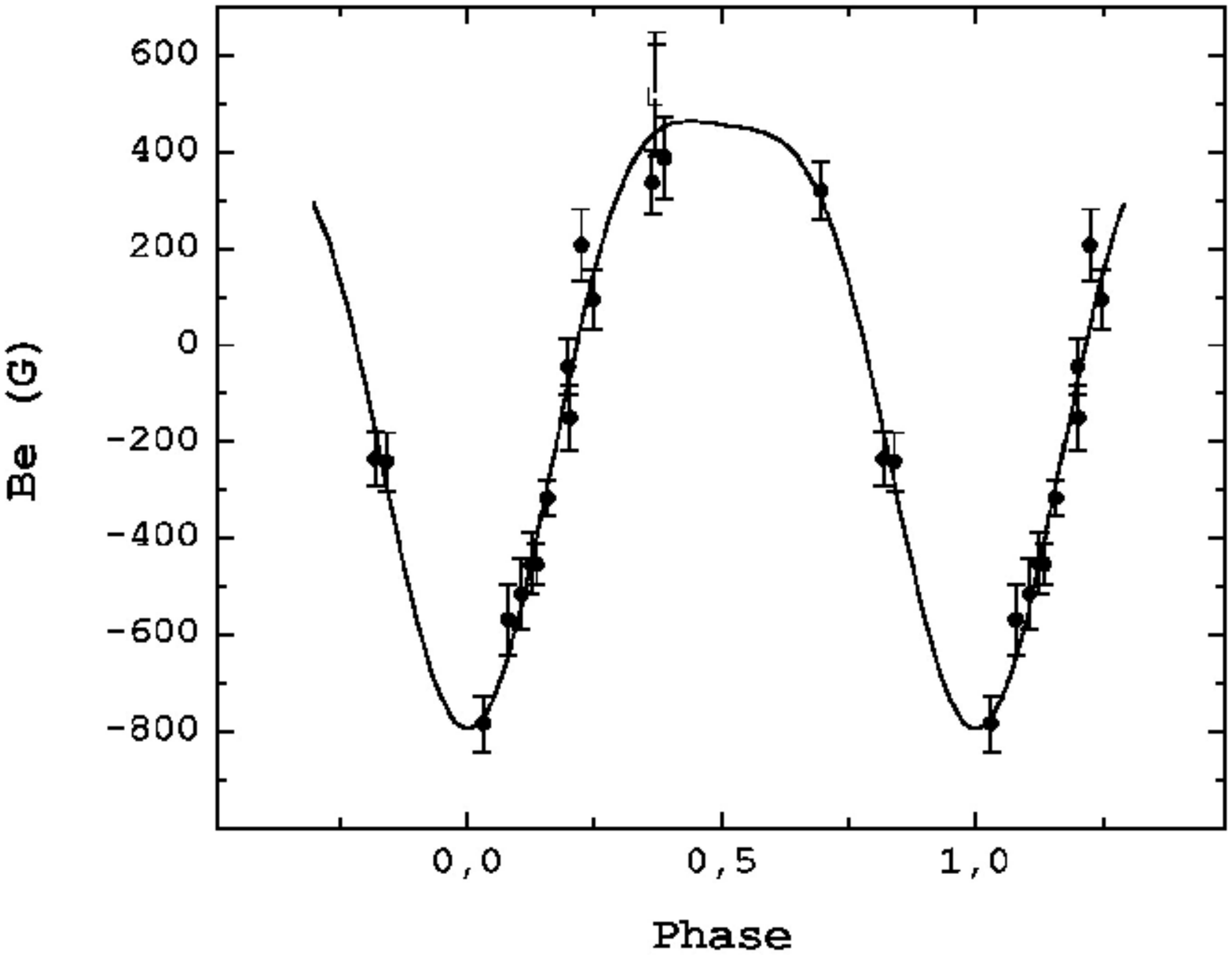}}
\vspace{-3.5mm}
\caption{ HD101412 (2) }
\label{fig:fig191}
\end{figure}

\begin{figure}
\resizebox{0.98\hsize}{!}{\includegraphics{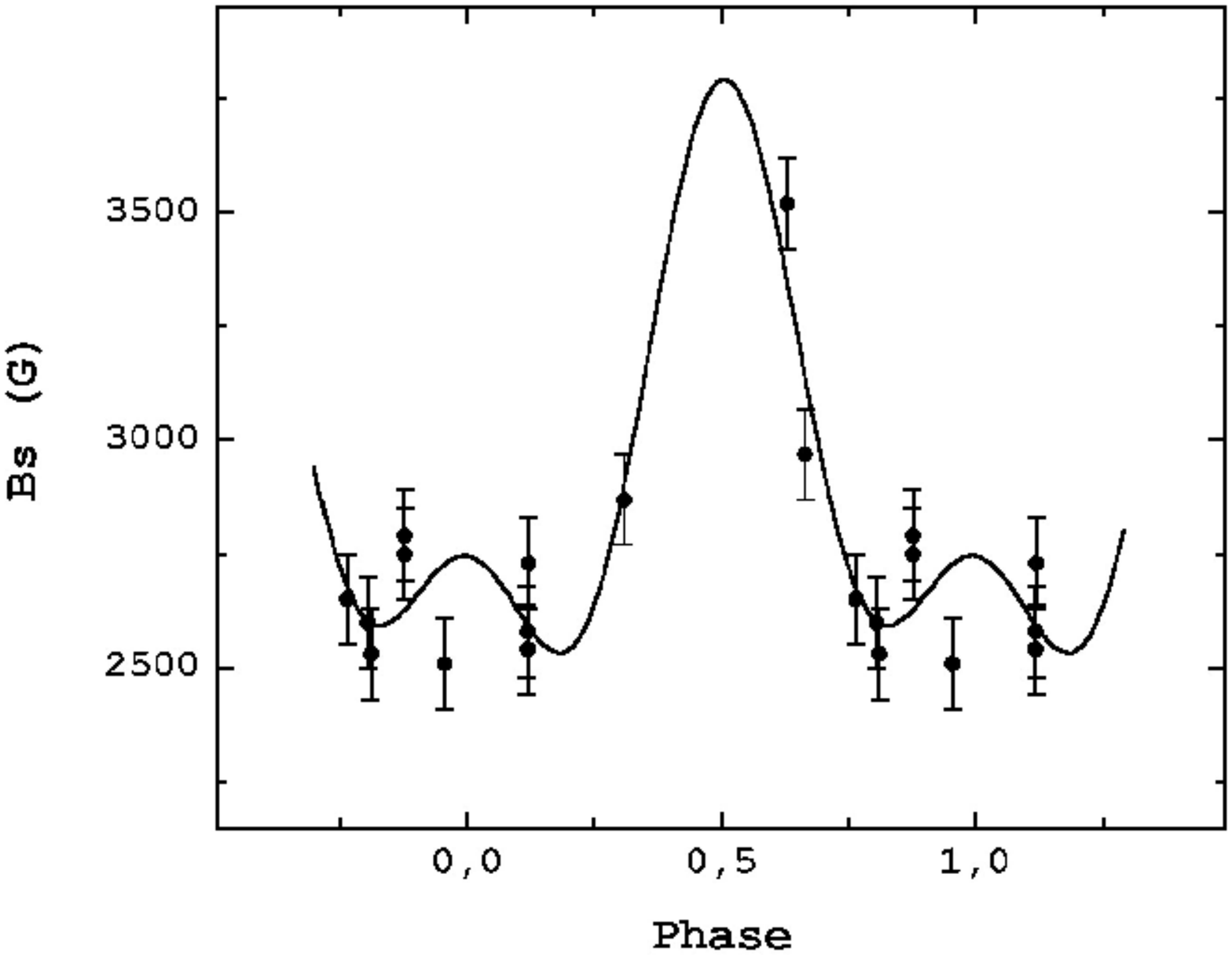}}
\vspace{-3.5mm}
\caption{ HD101412 (3) }
\label{fig:fig191}
\end{figure}

\begin{figure}
\resizebox{0.98\hsize}{!}{\includegraphics{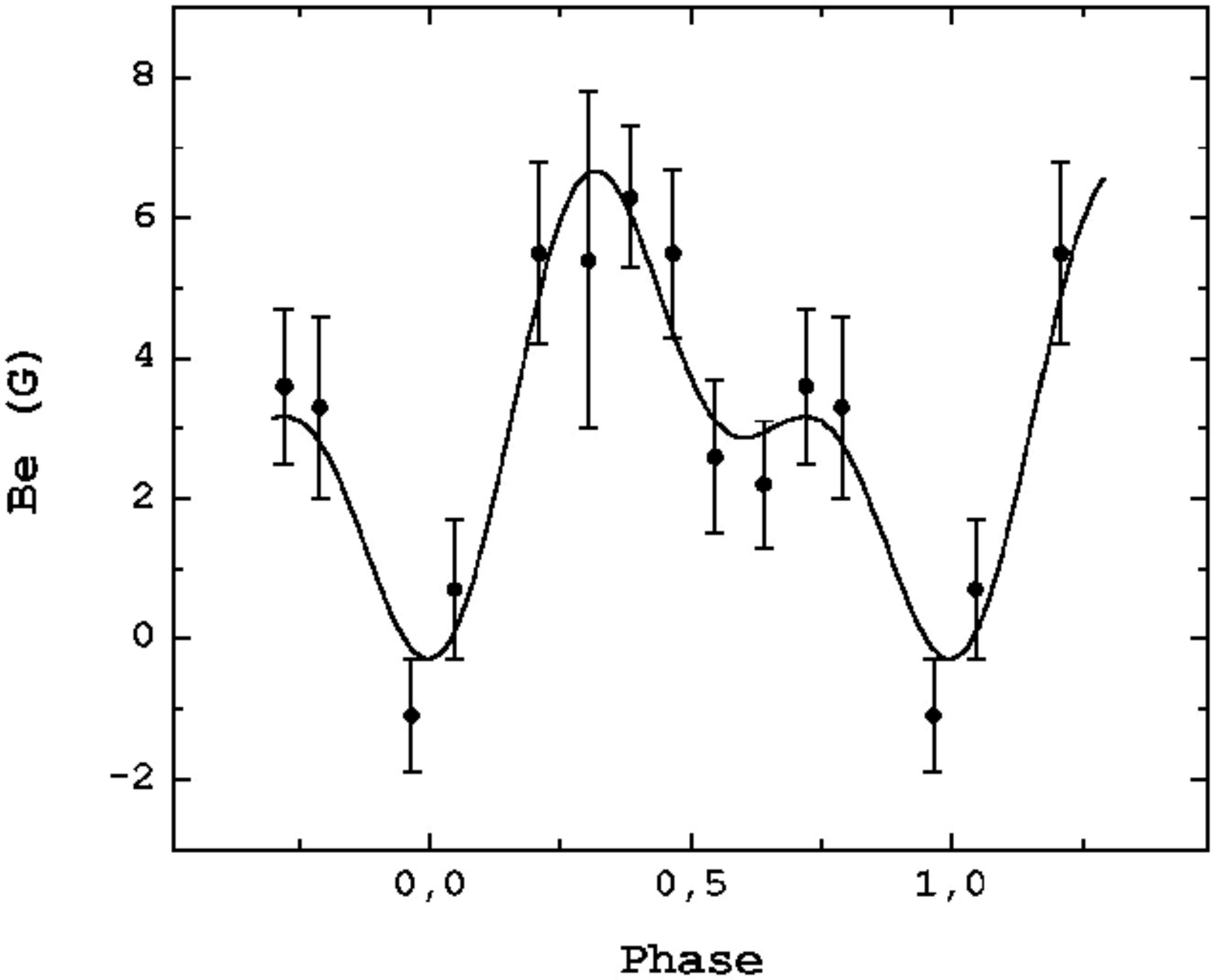}}
\vspace{-3.5mm}
\caption{ HD102195 }
\label{fig:fig192}
\end{figure}

\begin{figure}
\resizebox{0.98\hsize}{!}{\includegraphics{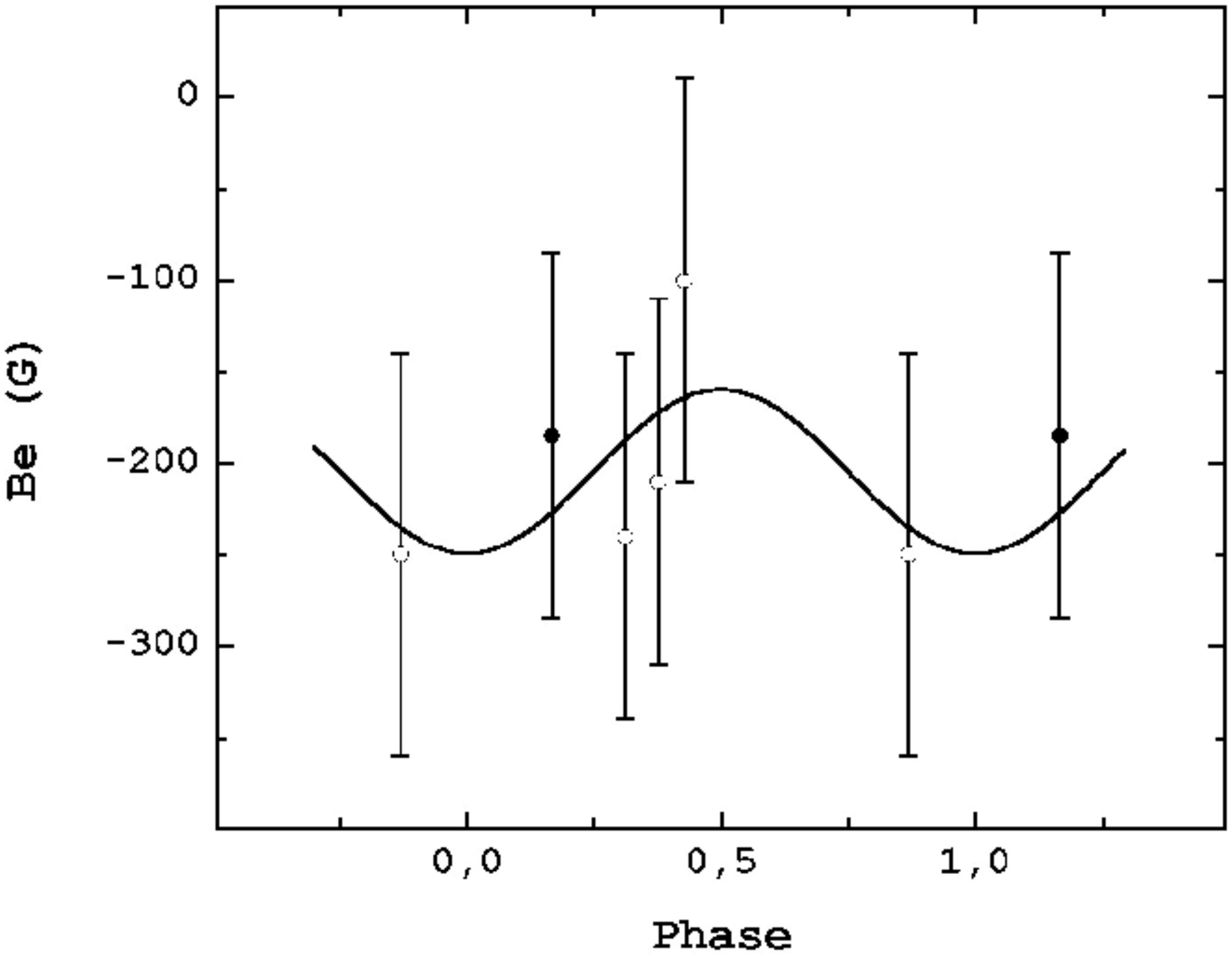}}
\vspace{-3.5mm}
\caption{ HD103192 }
\label{fig:fig193}
\end{figure}

\begin{figure}
\resizebox{0.98\hsize}{!}{\includegraphics{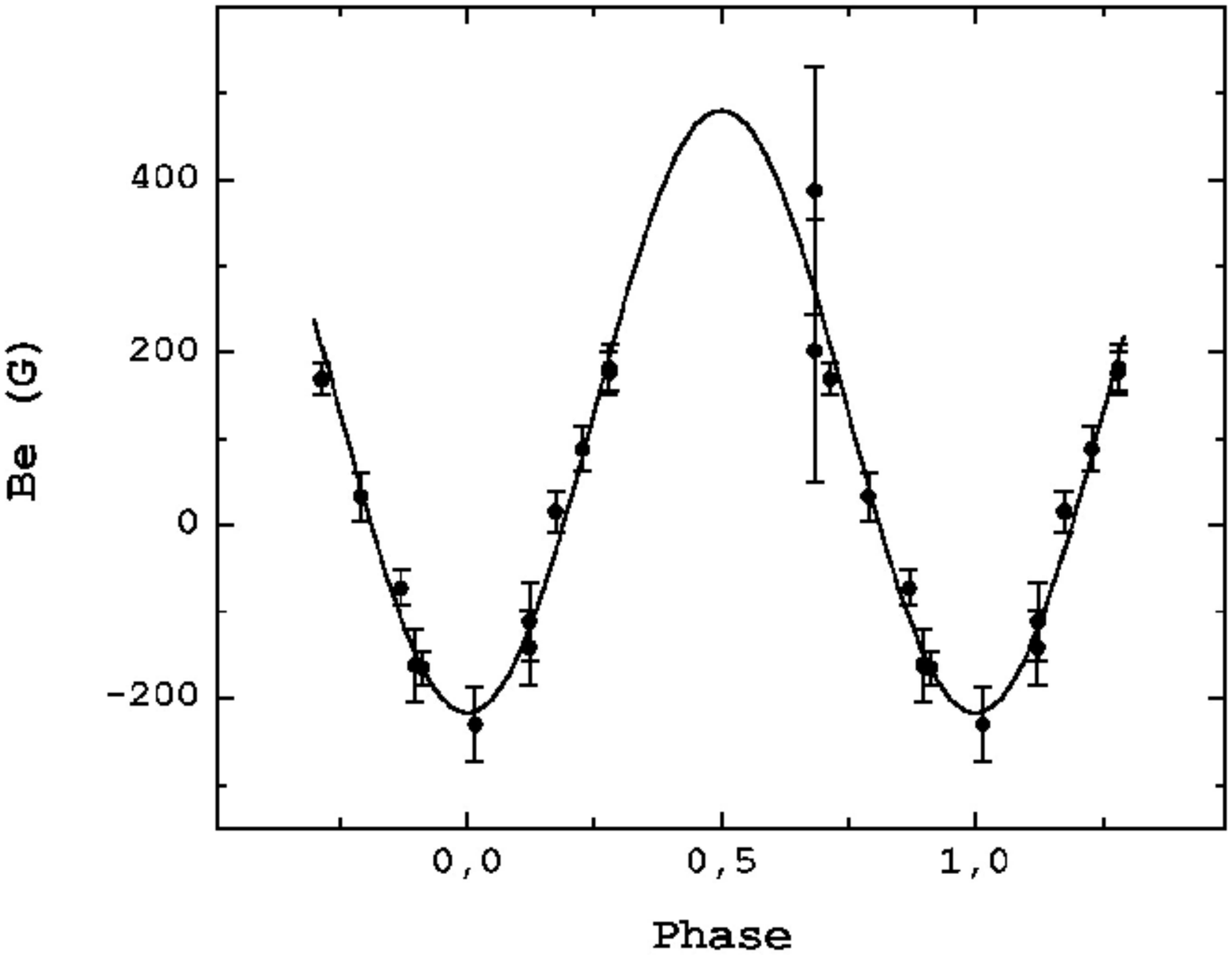}}
\vspace{-3.5mm}
\caption{ HD103498 }
\label{fig:fig194}
\end{figure}

\begin{figure}
\resizebox{0.98\hsize}{!}{\includegraphics{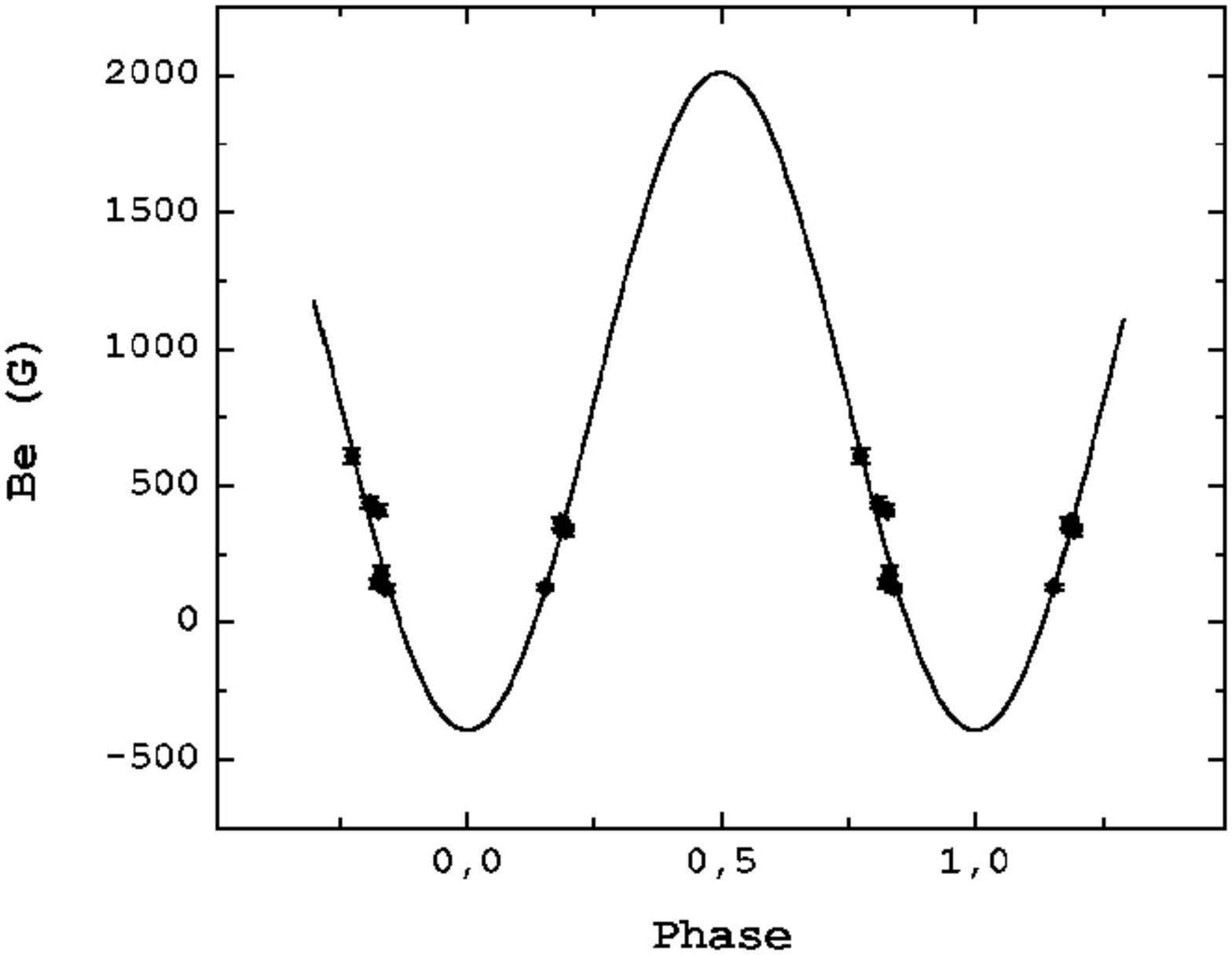}}
\vspace{-3.5mm}
\caption{ HD104237 }
\label{fig:fig191}
\end{figure}

\clearpage
\newpage

\begin{figure}
\resizebox{0.98\hsize}{!}{\includegraphics{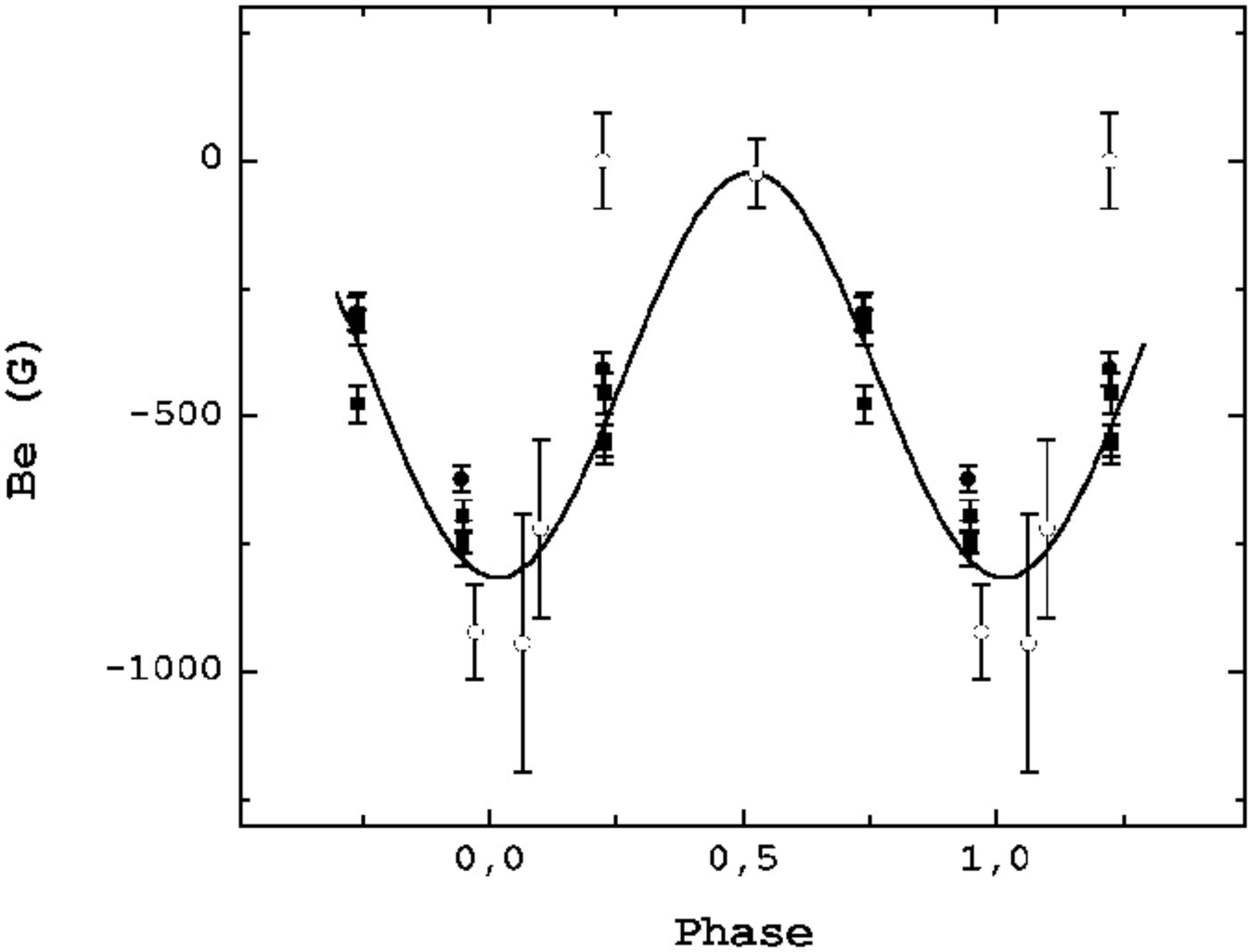}}
\vspace{-3.5mm}
\caption{ HD105382 }
\label{fig:fig195}
\end{figure}

\begin{figure}
\resizebox{0.98\hsize}{!}{\includegraphics{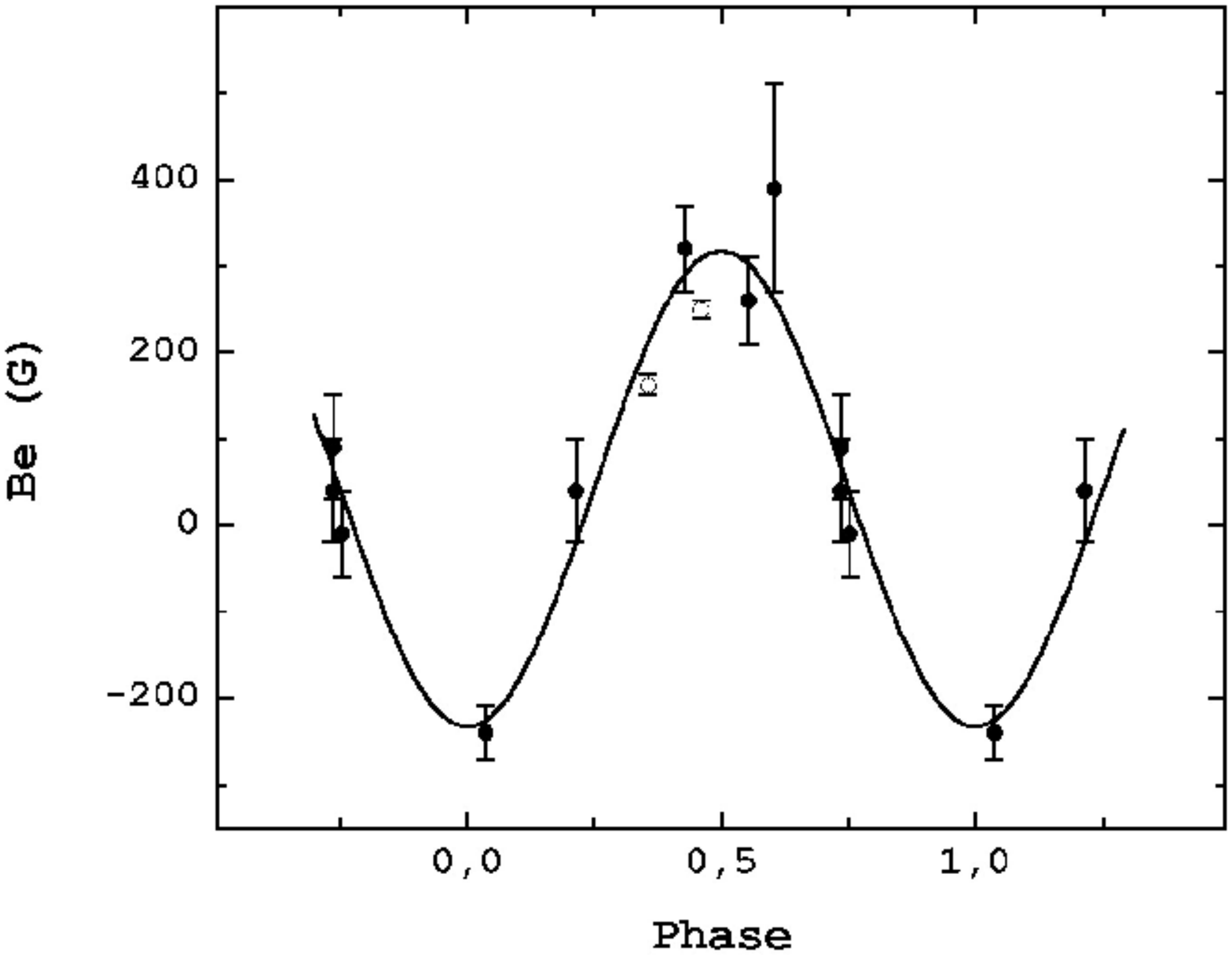}}
\vspace{-3.5mm}
\caption{ HD107000 }
\label{fig:fig191}
\end{figure}

\begin{figure}
\resizebox{0.98\hsize}{!}{\includegraphics{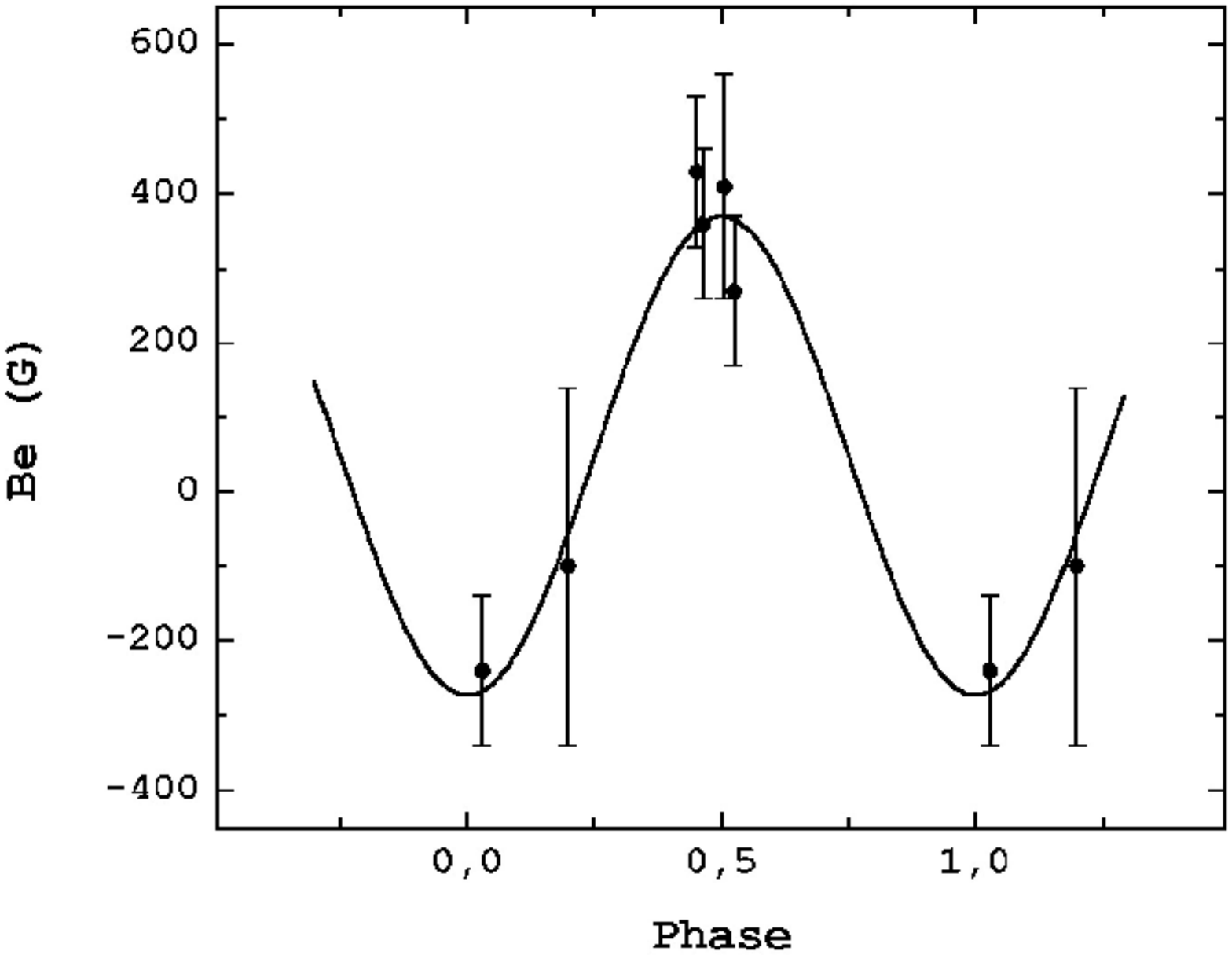}}
\vspace{-3.5mm}
\caption{ HD107612 }
\label{fig:fig196}
\end{figure}

\begin{figure}
\resizebox{0.98\hsize}{!}{\includegraphics{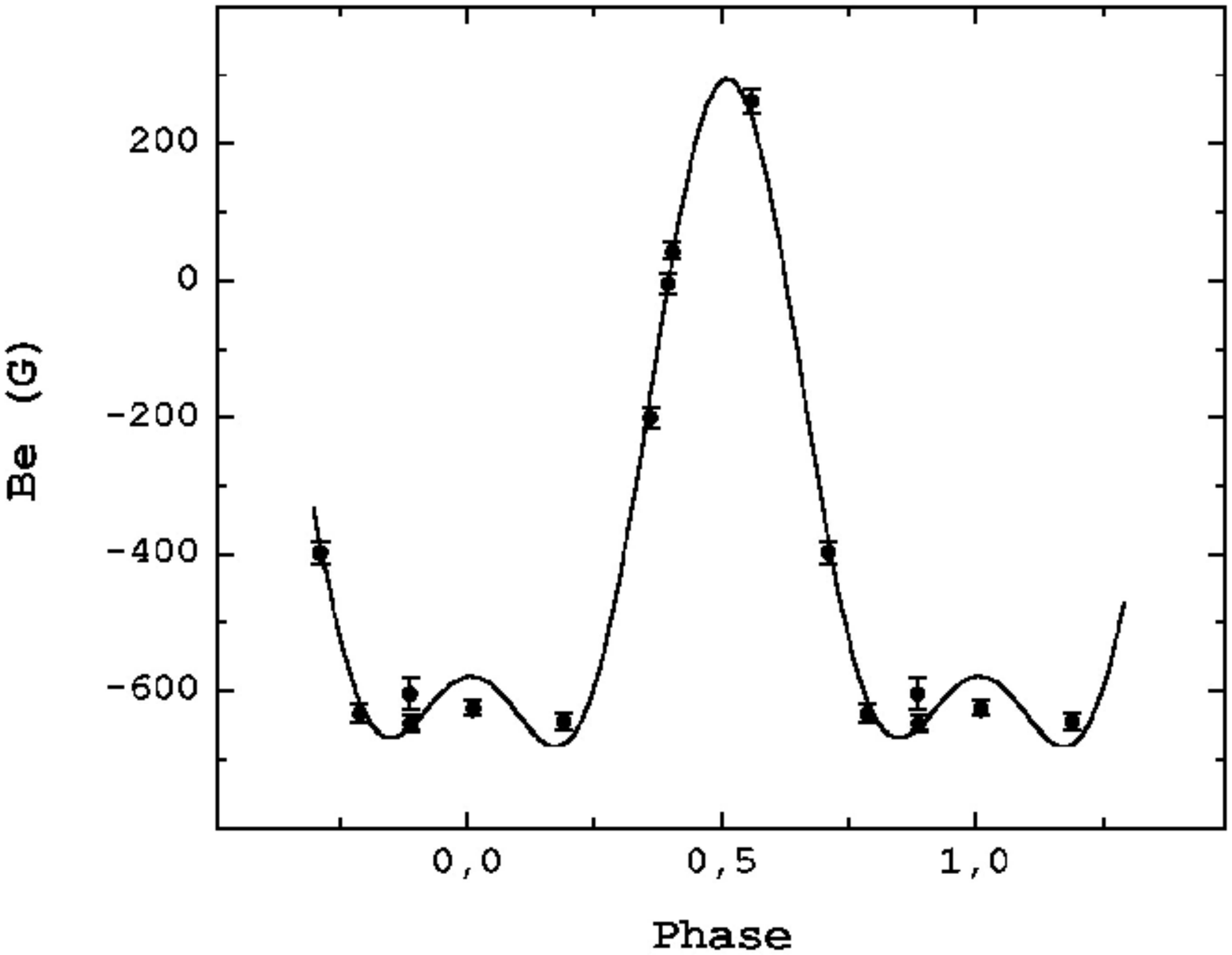}}
\vspace{-3.5mm}
\caption{ HD108662 }
\label{fig:fig197}
\end{figure}

\begin{figure}
\resizebox{0.98\hsize}{!}{\includegraphics{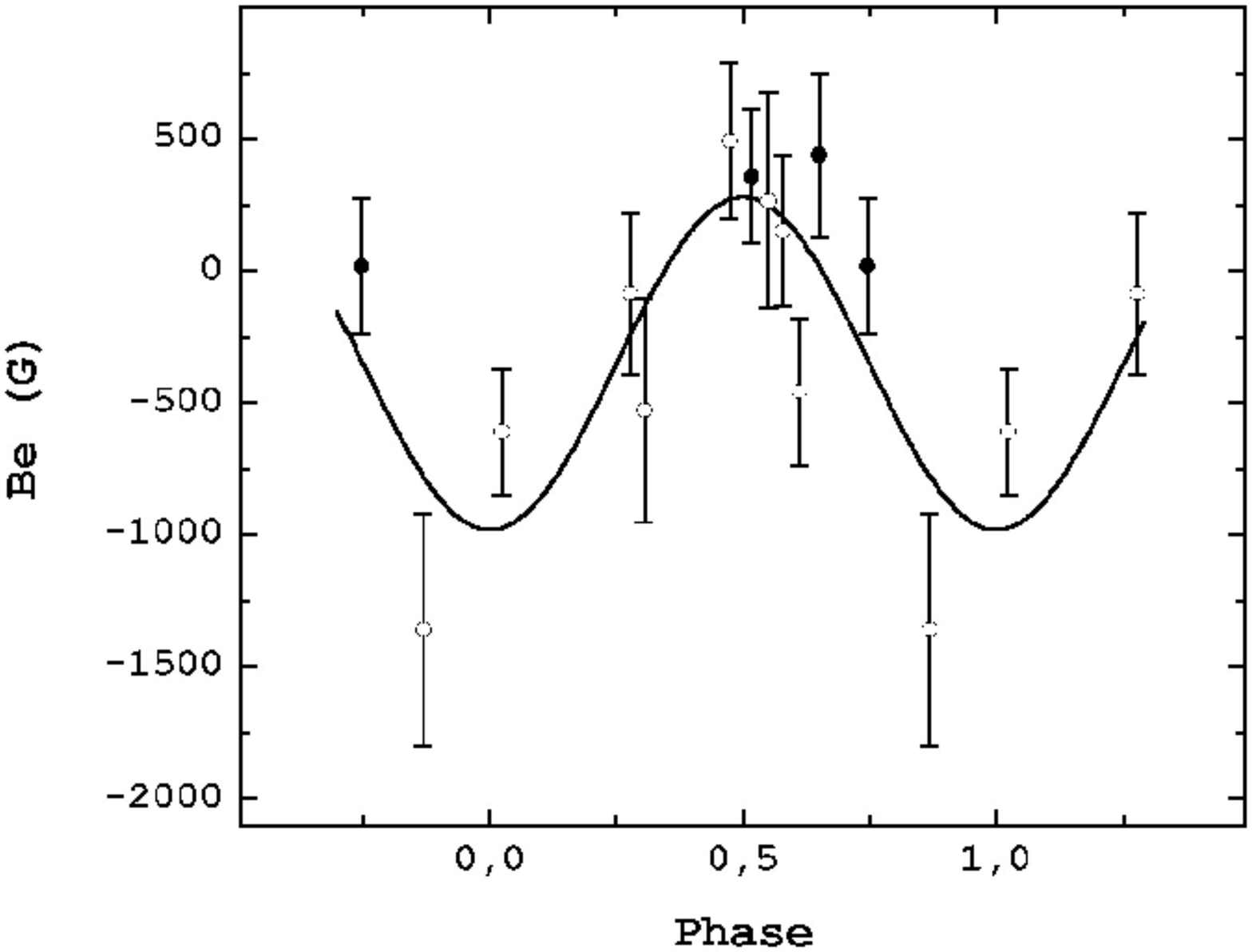}}
\vspace{-3.5mm}
\caption{ HD108945 (1) }
\label{fig:fig198}
\end{figure}

\begin{figure}
\resizebox{0.98\hsize}{!}{\includegraphics{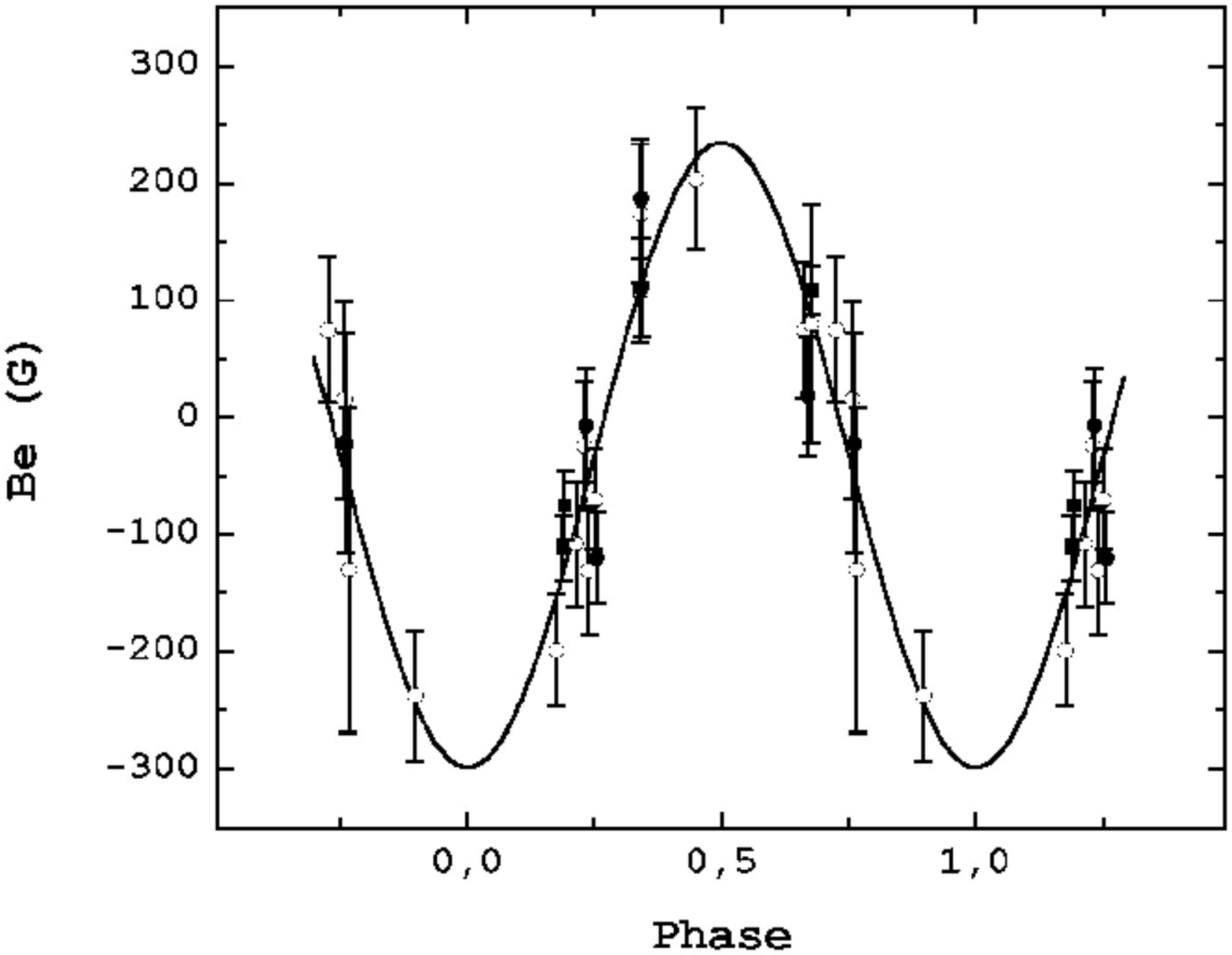}}
\vspace{-3.5mm}
\caption{ HD108945 (2) }
\label{fig:fig191}
\end{figure}

\clearpage
\newpage

\begin{figure}
\resizebox{0.98\hsize}{!}{\includegraphics{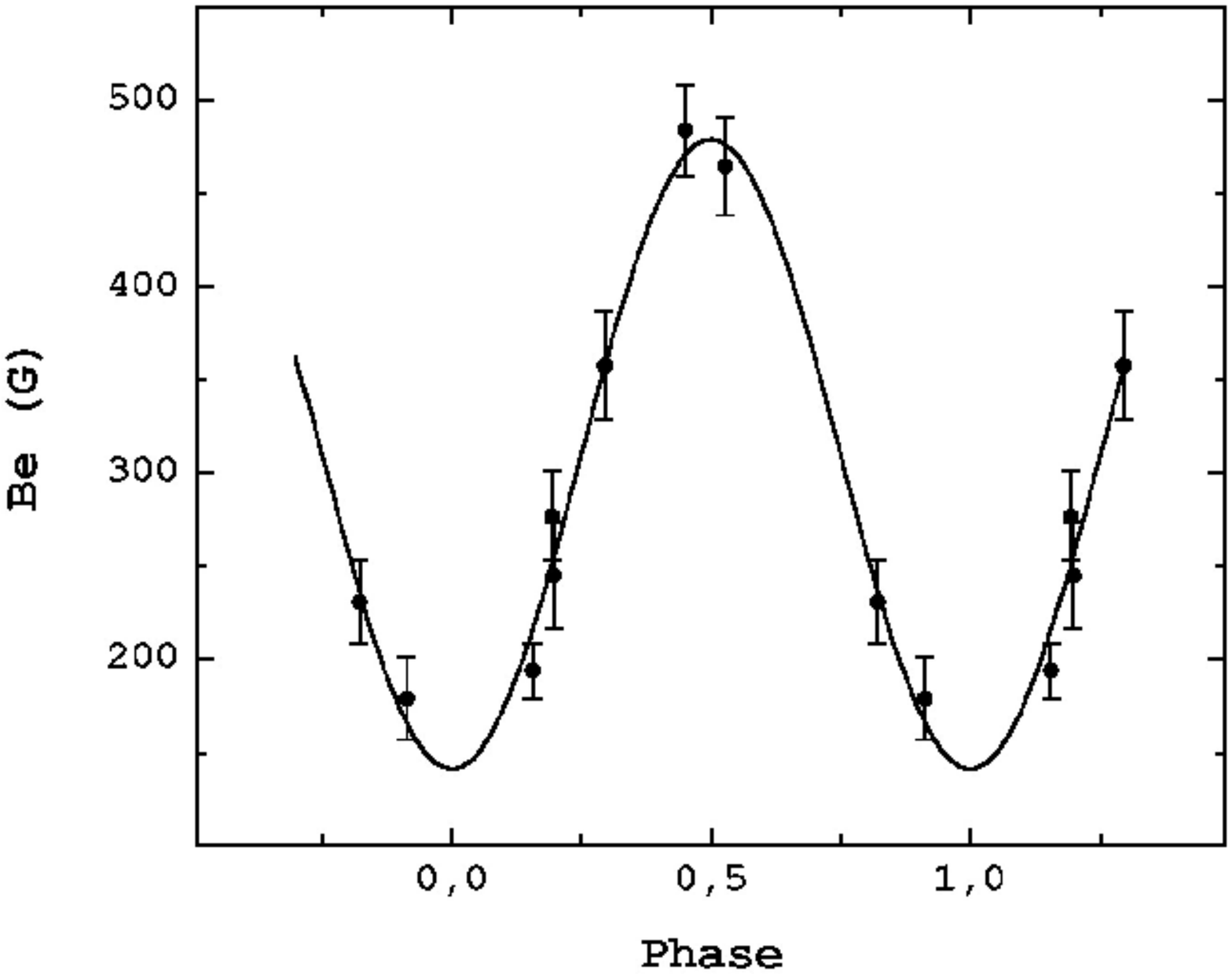}}
\vspace{-3.5mm}
\caption{ HD109026 }
\label{fig:fig199}
\end{figure}

\begin{figure}
\resizebox{0.98\hsize}{!}{\includegraphics{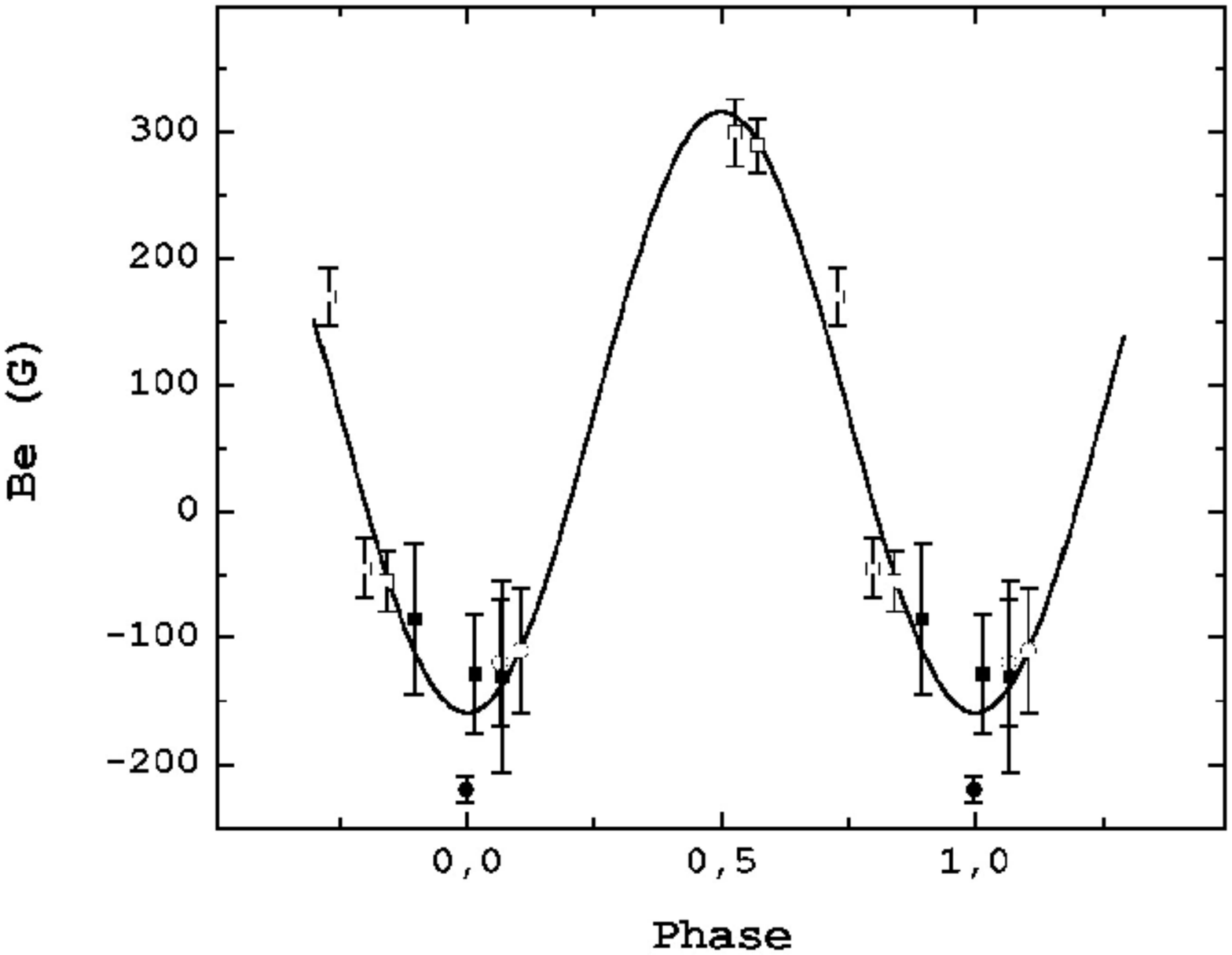}}
\vspace{-3.5mm}
\caption{ HD110066 }
\label{fig:fig200}
\end{figure}

\begin{figure}
\resizebox{0.98\hsize}{!}{\includegraphics{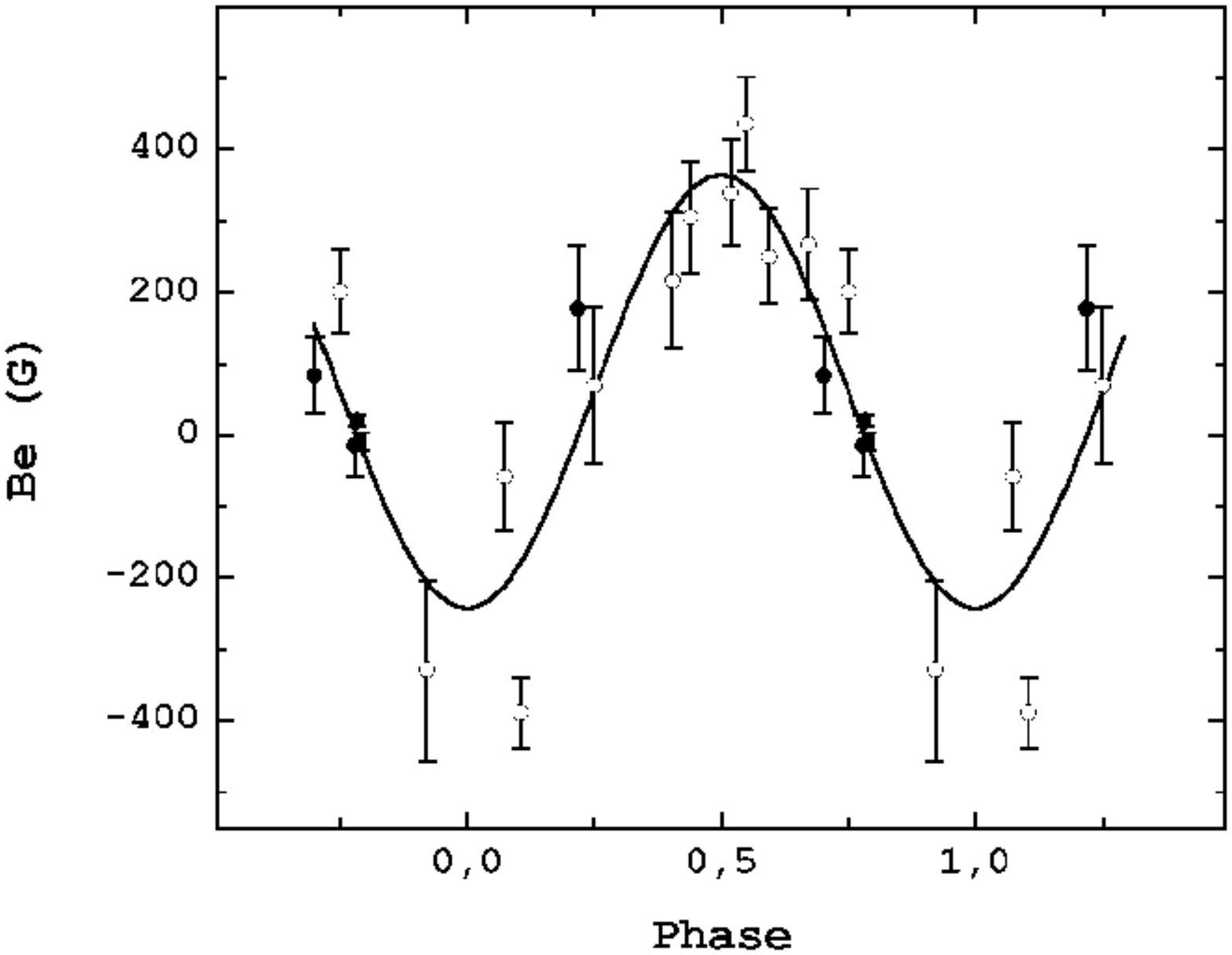}}
\vspace{-3.5mm}
\caption{ HD110379 }
\label{fig:fig201}
\end{figure}

\begin{figure}
\resizebox{0.98\hsize}{!}{\includegraphics{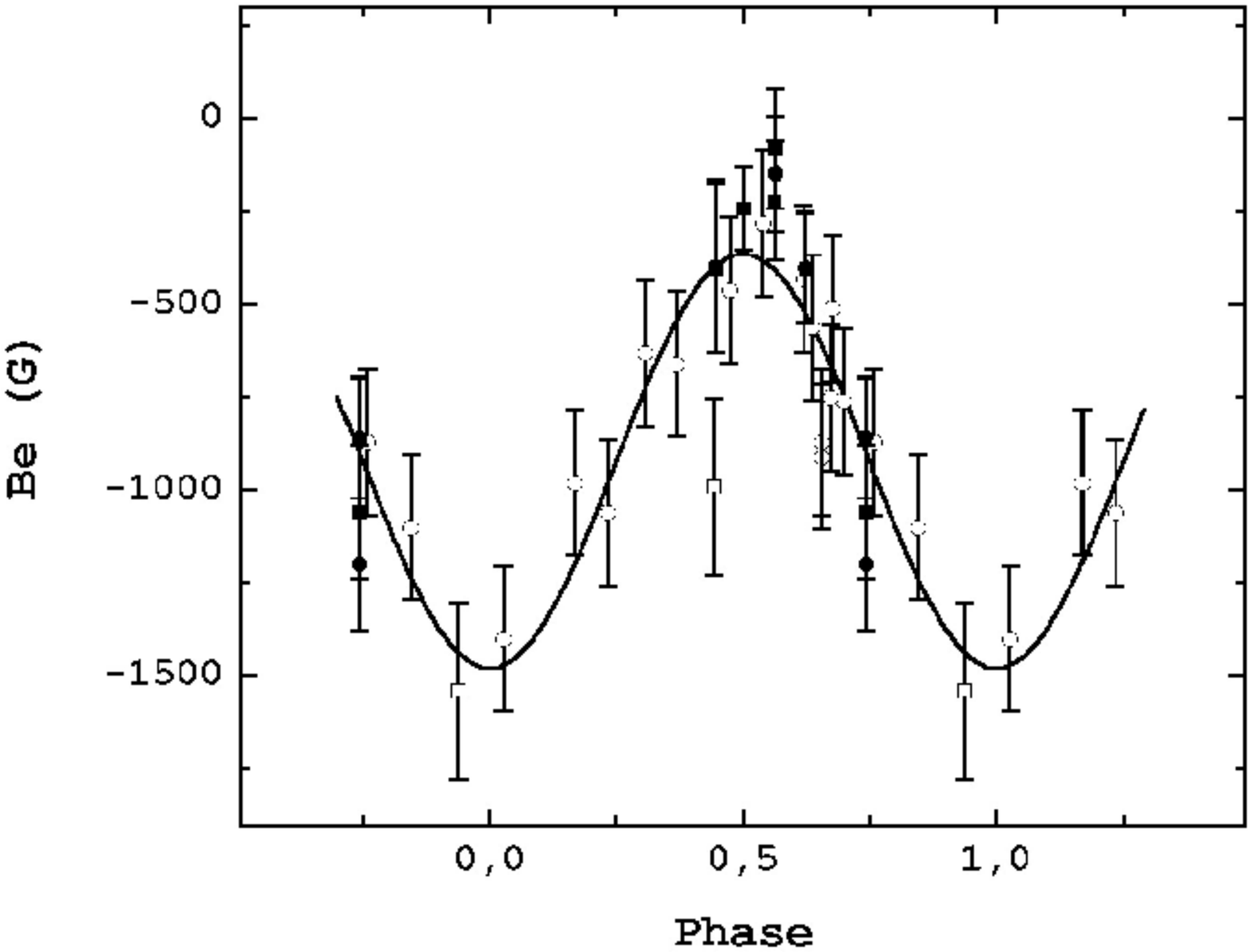}}
\vspace{-3.5mm}
\caption{ HD111133 }
\label{fig:fig203}
\end{figure}

\begin{figure}
\resizebox{0.98\hsize}{!}{\includegraphics{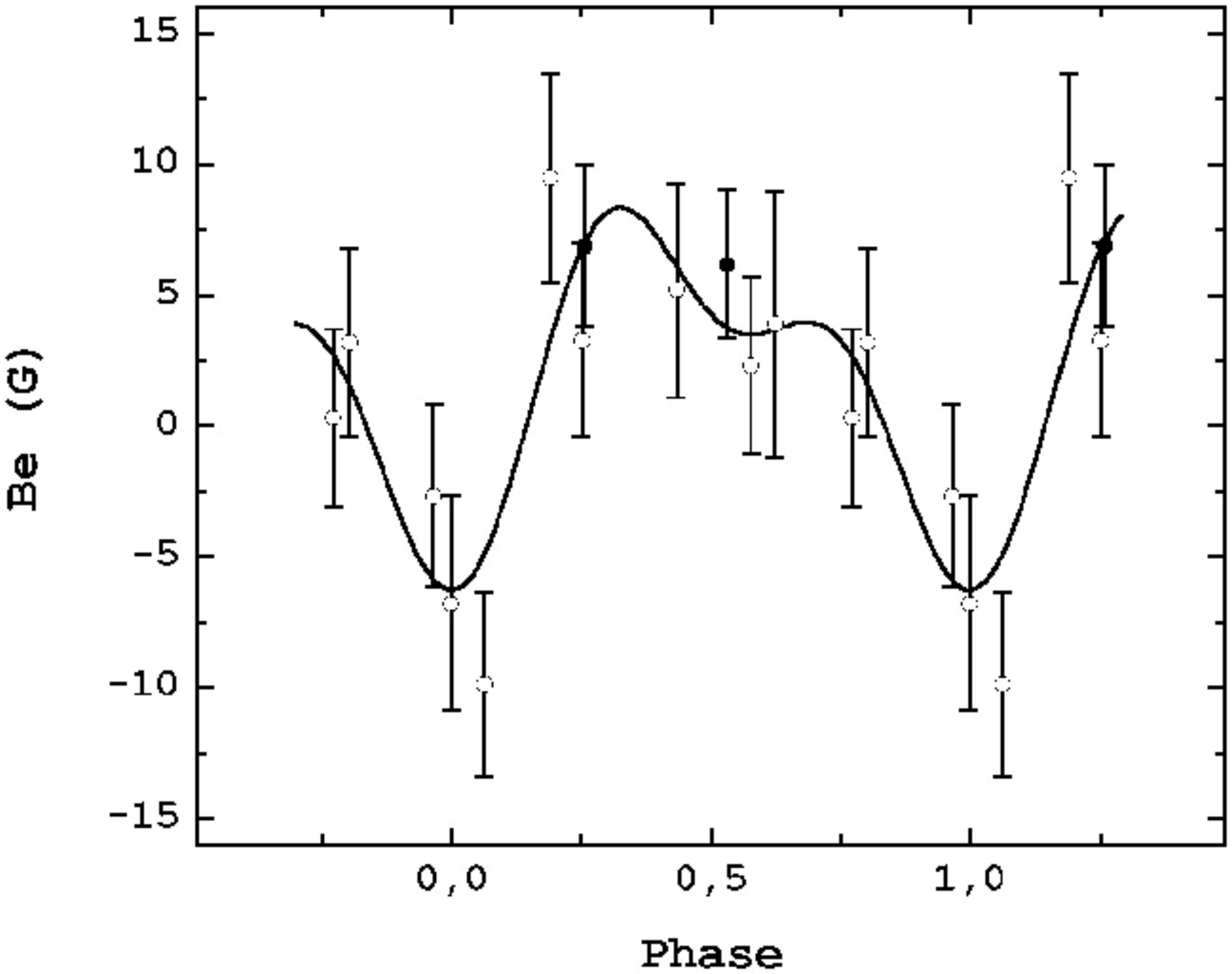}}
\vspace{-3.5mm}
\caption{ HD111812 }
\label{fig:fig202}
\end{figure}

\begin{figure}
\resizebox{0.98\hsize}{!}{\includegraphics{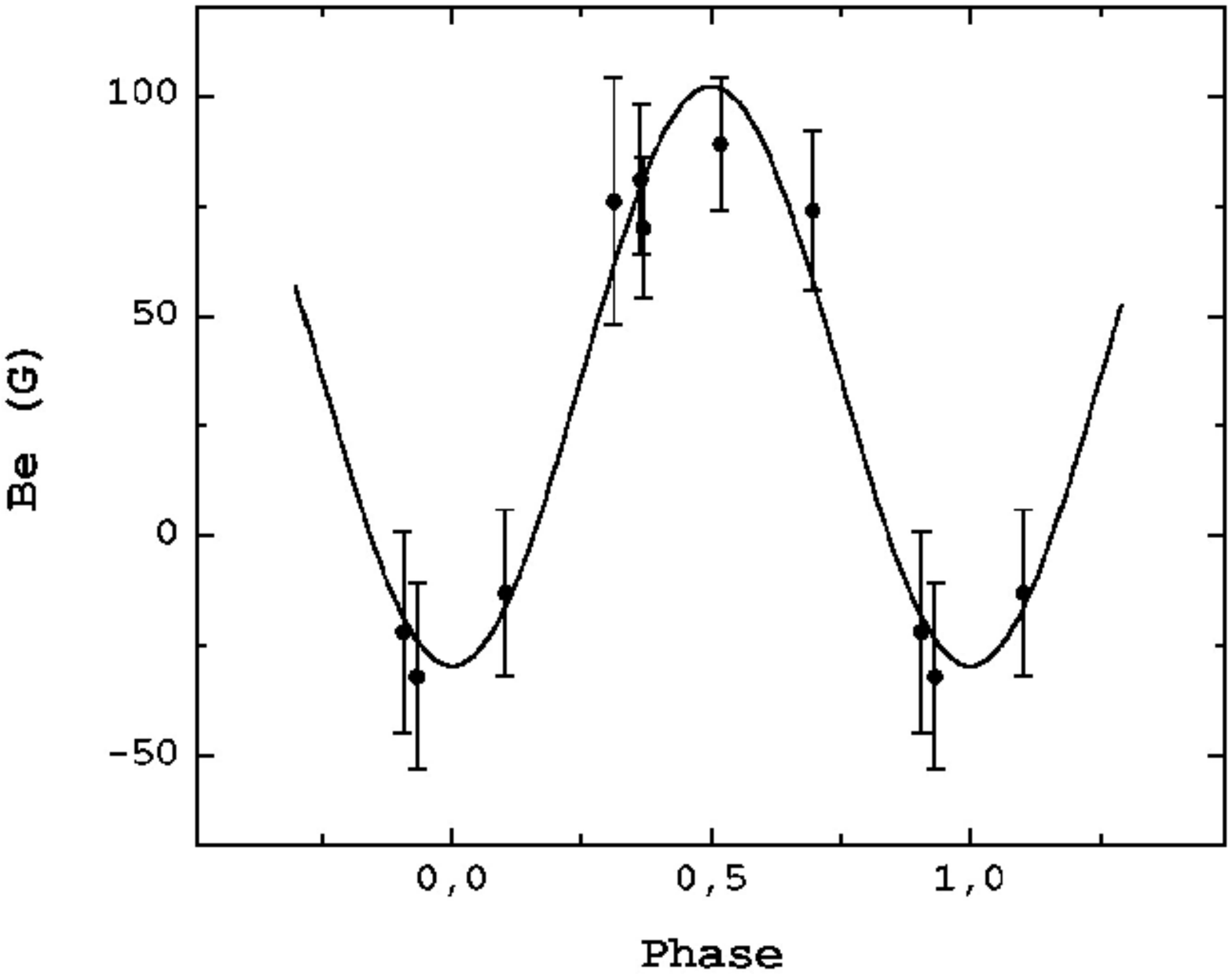}}
\vspace{-3.5mm}
\caption{ HD112185 (1) }
\label{fig:fig204}
\end{figure}

\clearpage
\newpage

\begin{figure}
\resizebox{0.98\hsize}{!}{\includegraphics{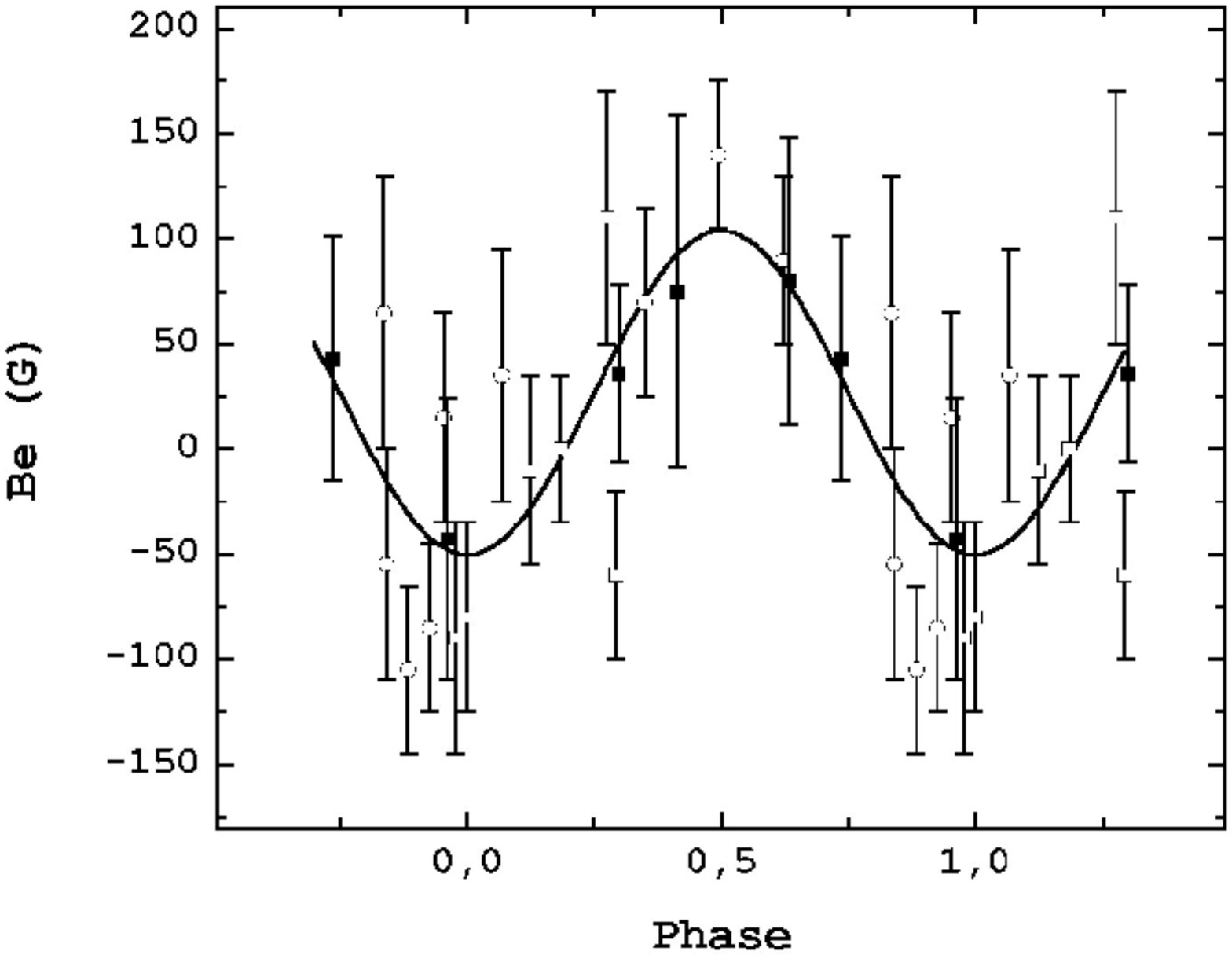}}
\vspace{-3.5mm}
\caption{ HD112185 (2) }
\label{fig:fig205}
\end{figure}

\begin{figure}
\resizebox{0.98\hsize}{!}{\includegraphics{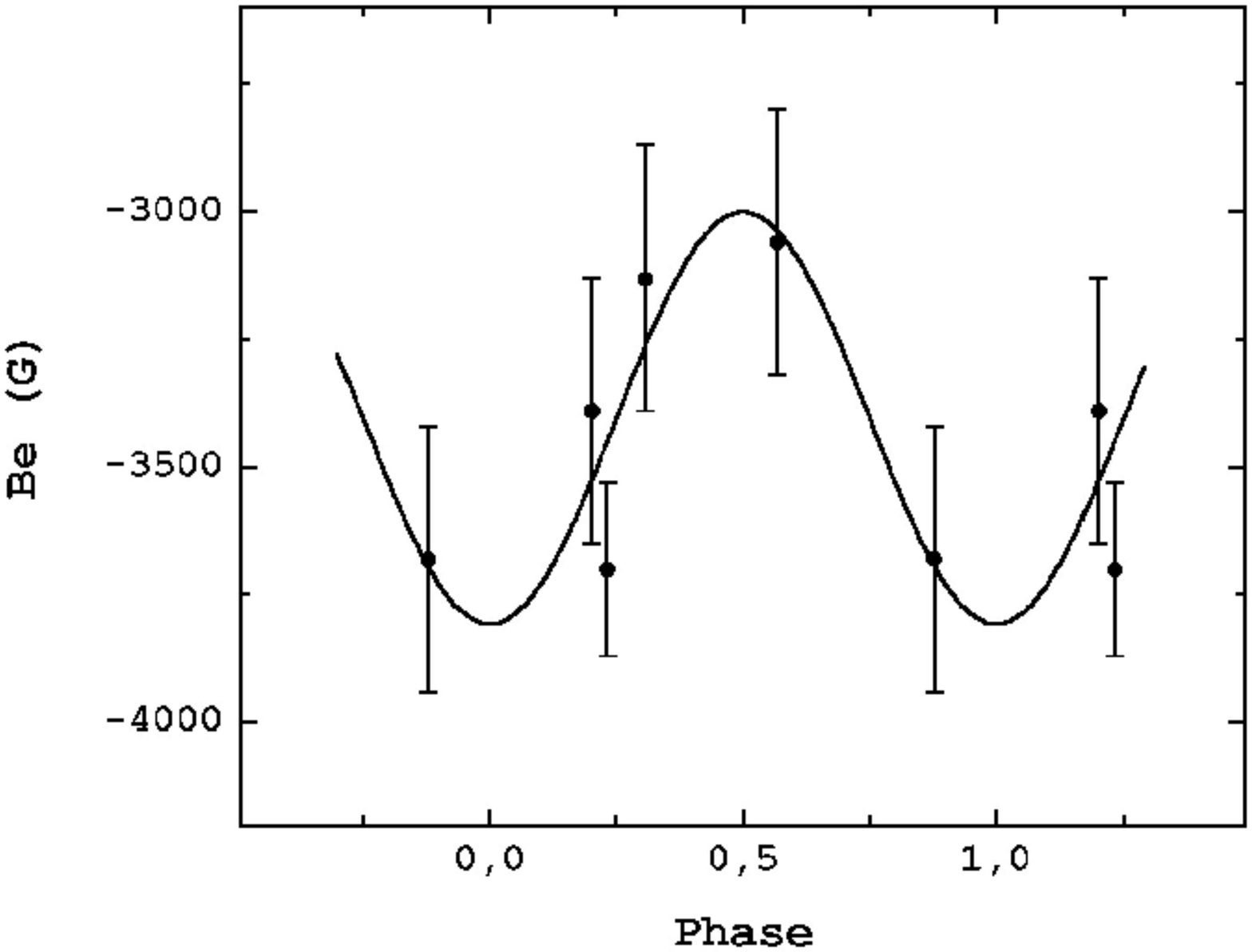}}
\vspace{-3.5mm}
\caption{ HD112381 }
\label{fig:fig206}
\end{figure}

\begin{figure}
\resizebox{0.98\hsize}{!}{\includegraphics{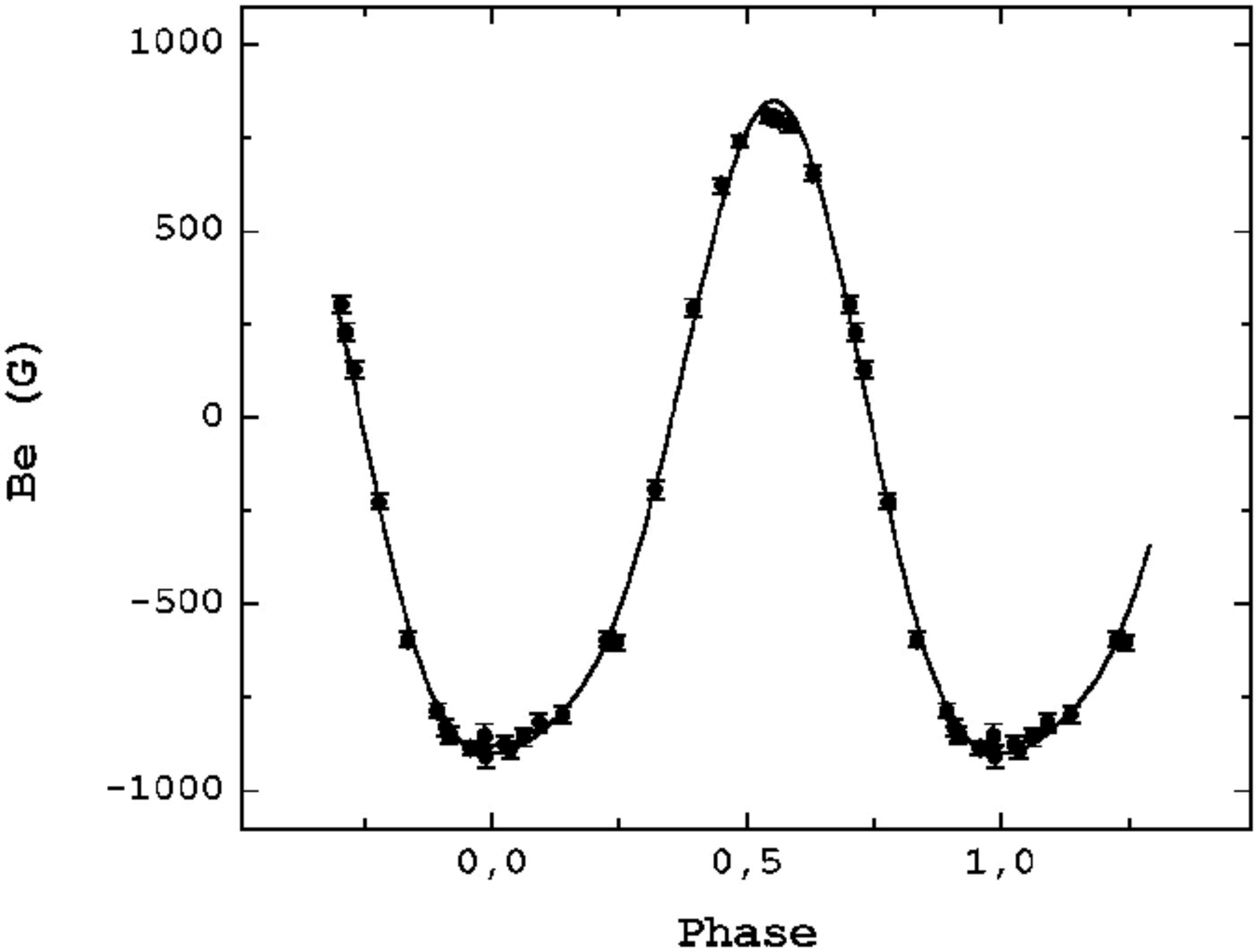}}
\vspace{-3.5mm}
\caption{ HD112413 (1) }
\label{fig:fig207}
\end{figure}

\begin{figure}
\resizebox{0.98\hsize}{!}{\includegraphics{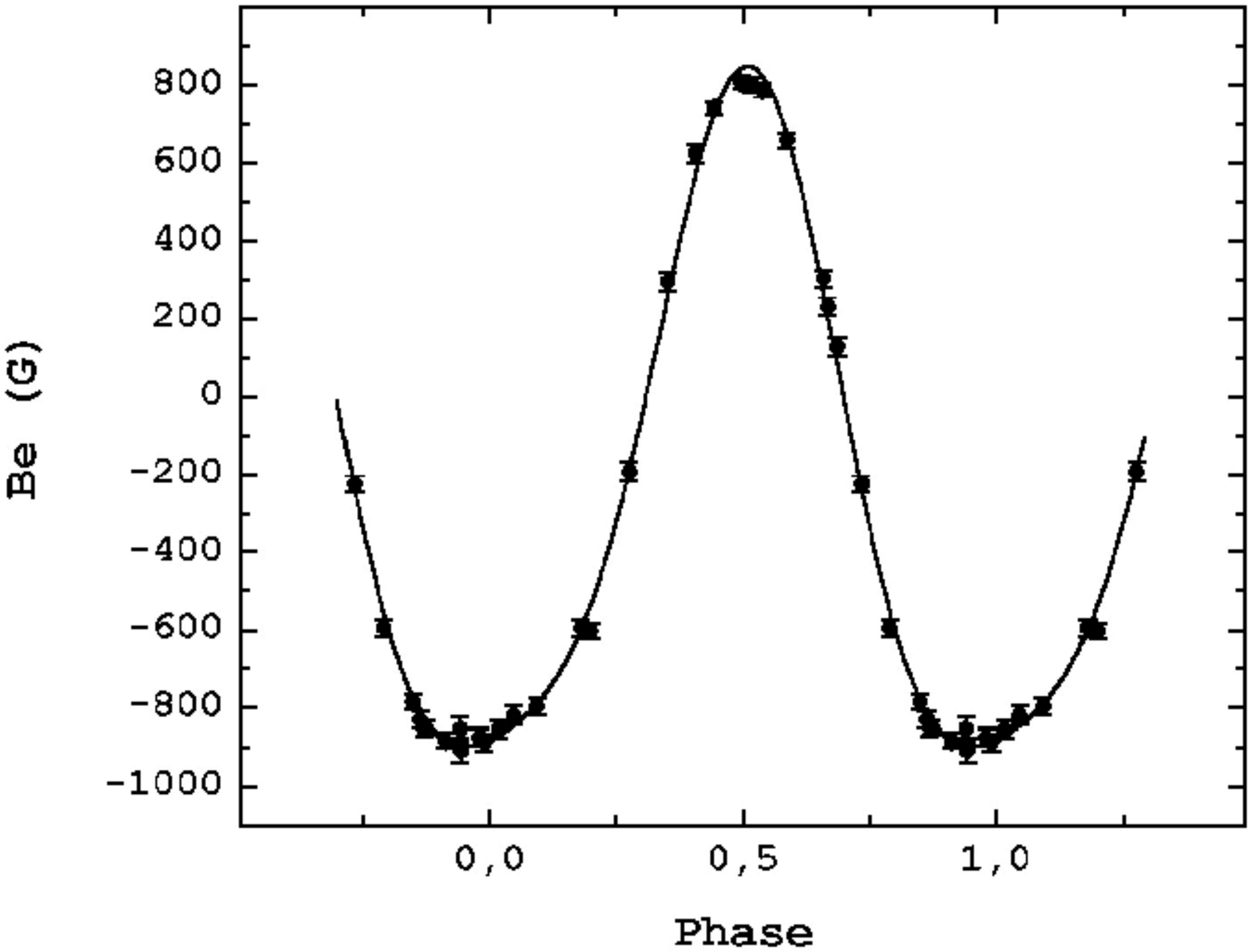}}
\vspace{-3.5mm}
\caption{ HD112413 (2) }
\label{fig:fig208}
\end{figure}

\begin{figure}
\resizebox{0.98\hsize}{!}{\includegraphics{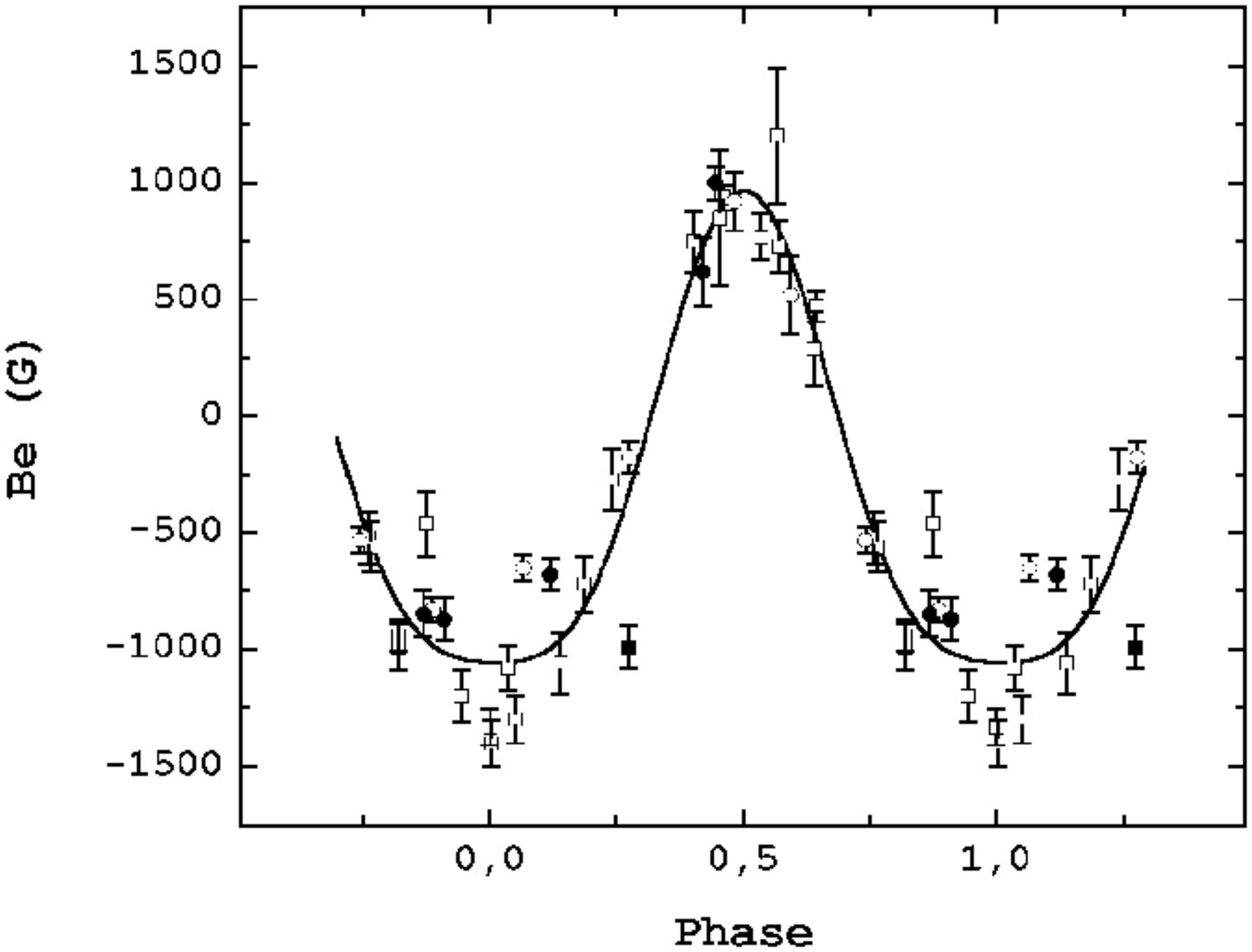}}
\vspace{-3.5mm}
\caption{ HD112413 (3) }
\label{fig:fig209}
\end{figure}

\begin{figure}
\resizebox{0.98\hsize}{!}{\includegraphics{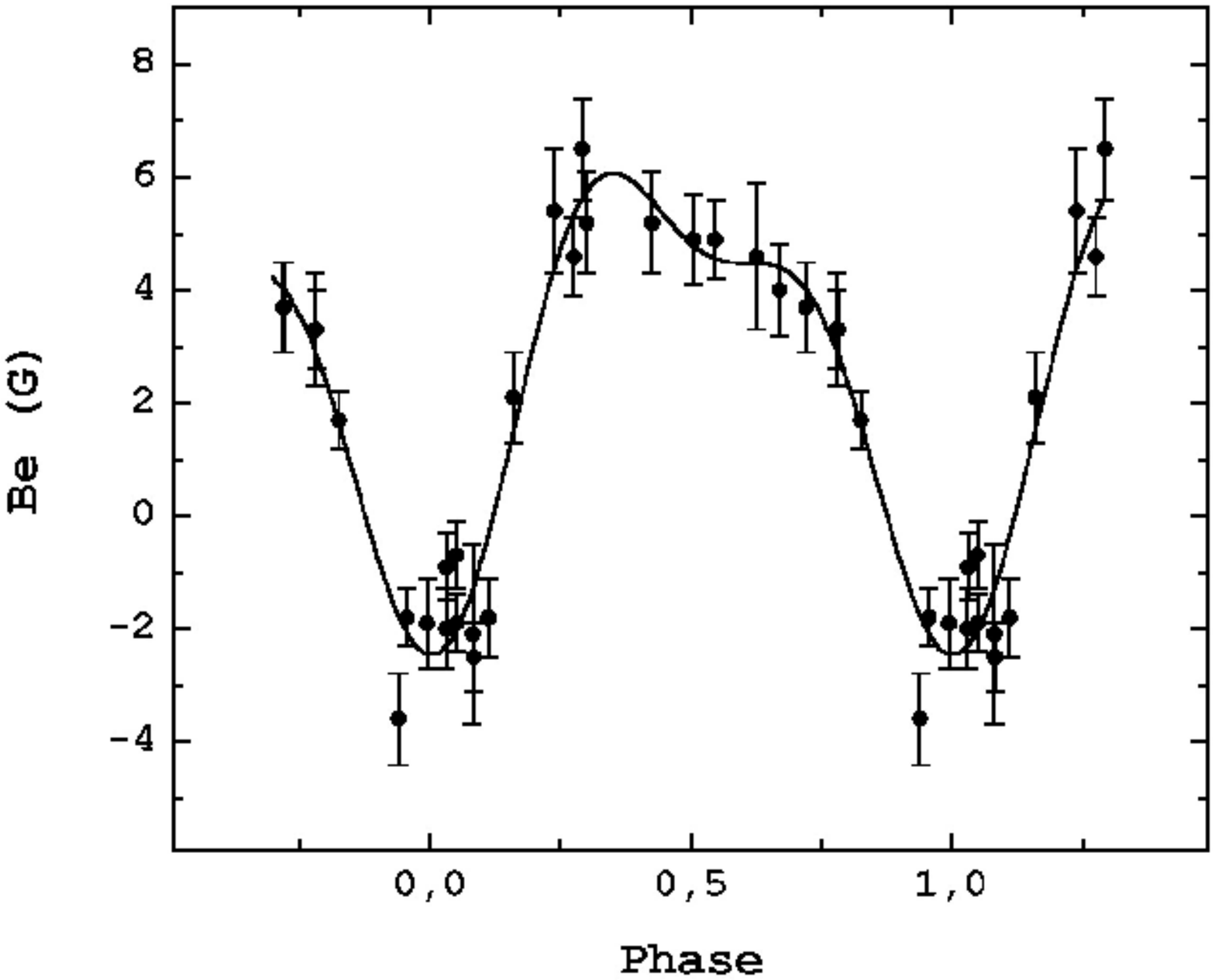}}
\vspace{-3.5mm}
\caption{ HD112989 }
\label{fig:fig210}
\end{figure}

\clearpage
\newpage

\begin{figure}
\resizebox{0.98\hsize}{!}{\includegraphics{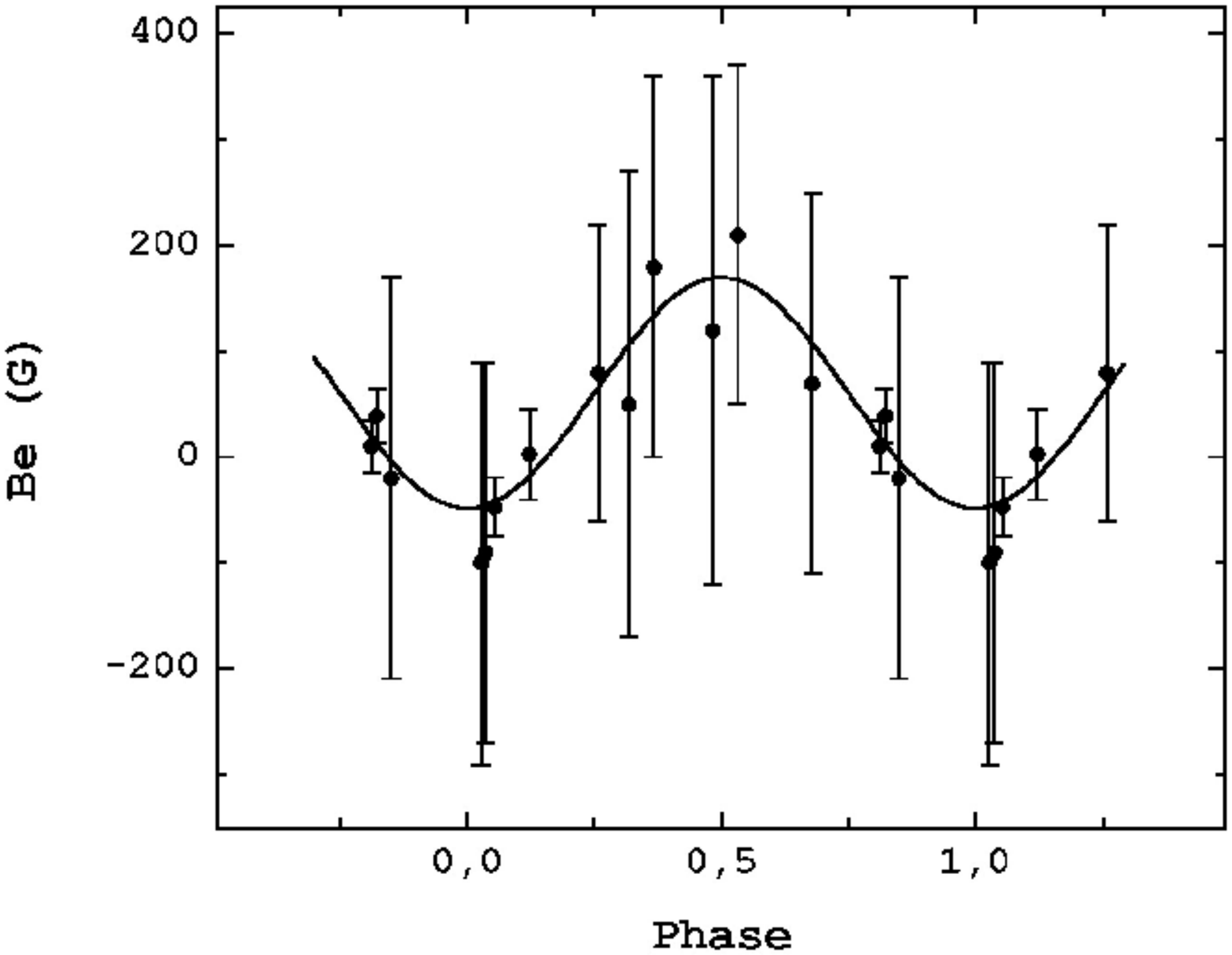}}
\vspace{-3.5mm}
\caption{ HD115735 }
\label{fig:fig203}
\end{figure}

\begin{figure}
\resizebox{0.98\hsize}{!}{\includegraphics{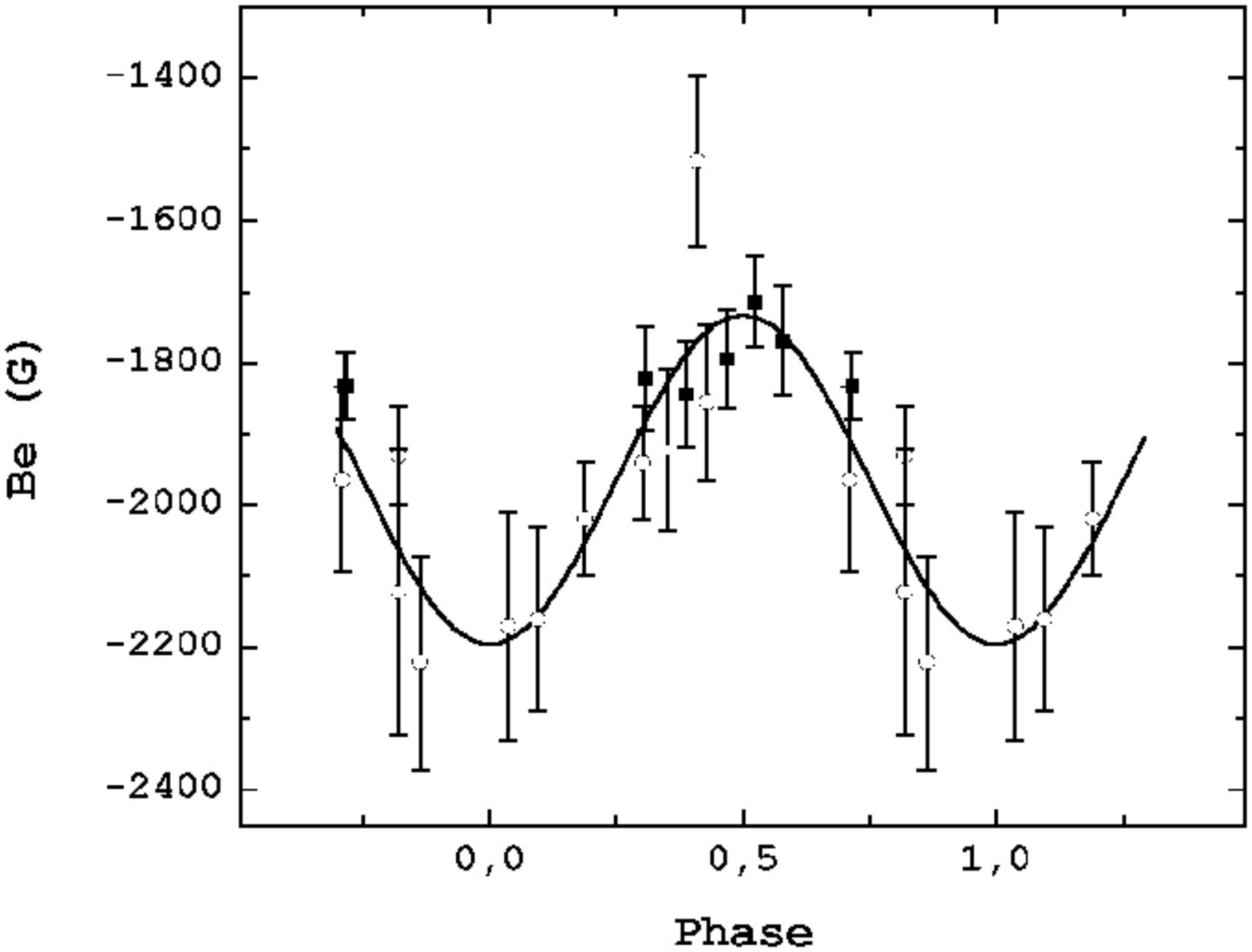}}
\vspace{-3.5mm}
\caption{ HD116114 }
\label{fig:fig211}
\end{figure}

\begin{figure}
\resizebox{0.98\hsize}{!}{\includegraphics{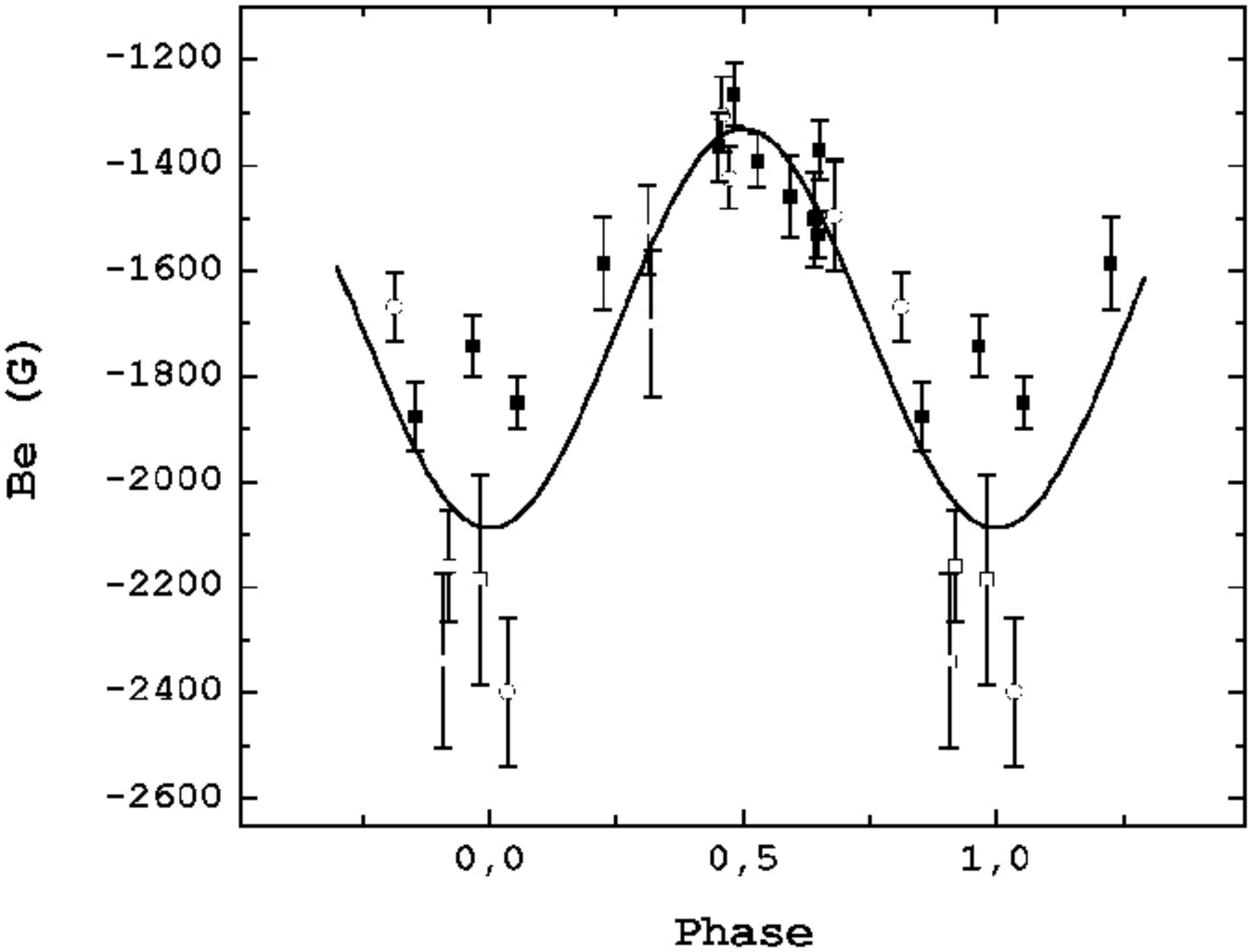}}
\vspace{-3.5mm}
\caption{ HD116458 }
\label{fig:fig212}
\end{figure}

\begin{figure}
\resizebox{0.98\hsize}{!}{\includegraphics{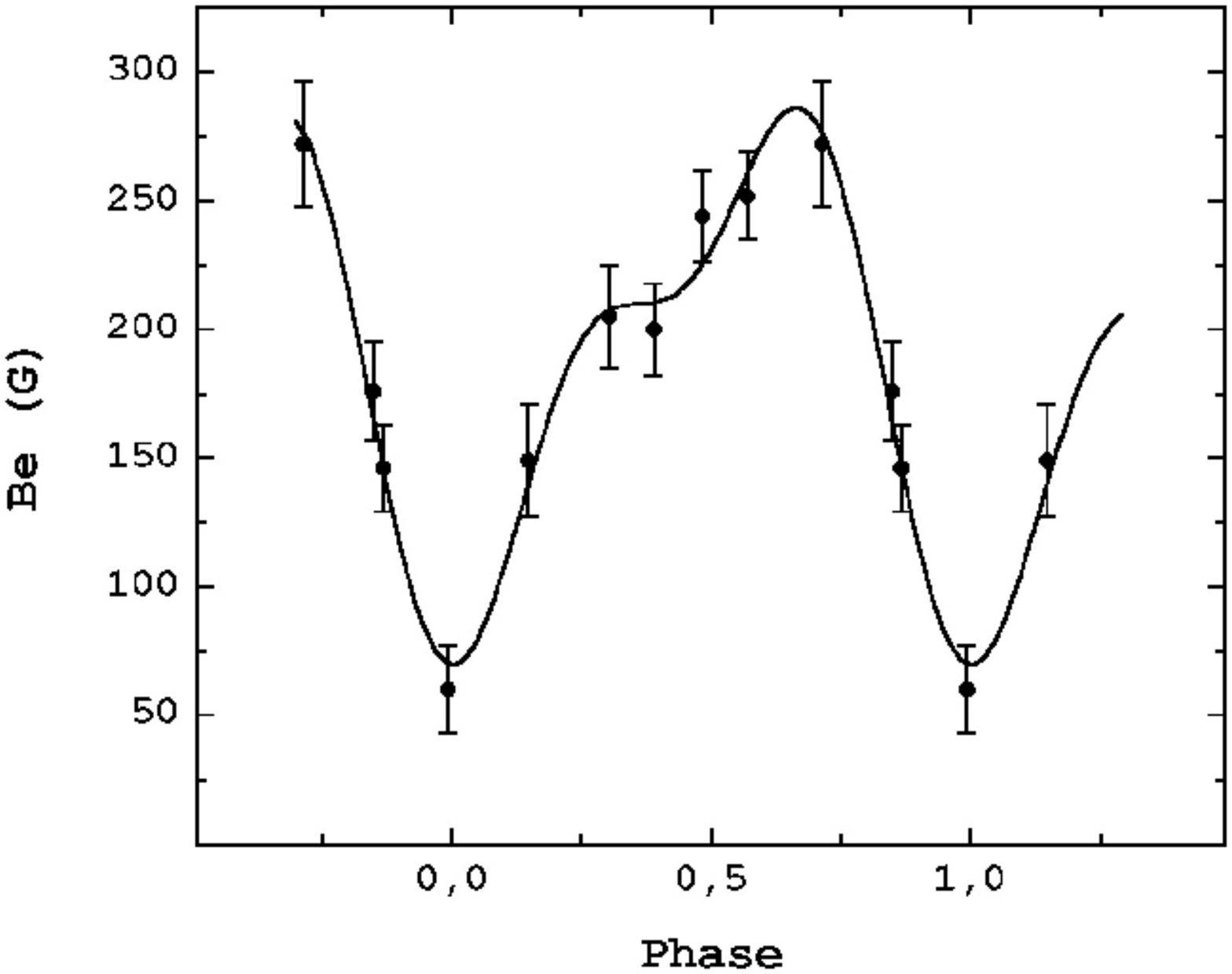}}
\vspace{-3.5mm}
\caption{ HD117555 (1) }
\label{fig:fig213}
\end{figure}

\begin{figure}
\resizebox{0.98\hsize}{!}{\includegraphics{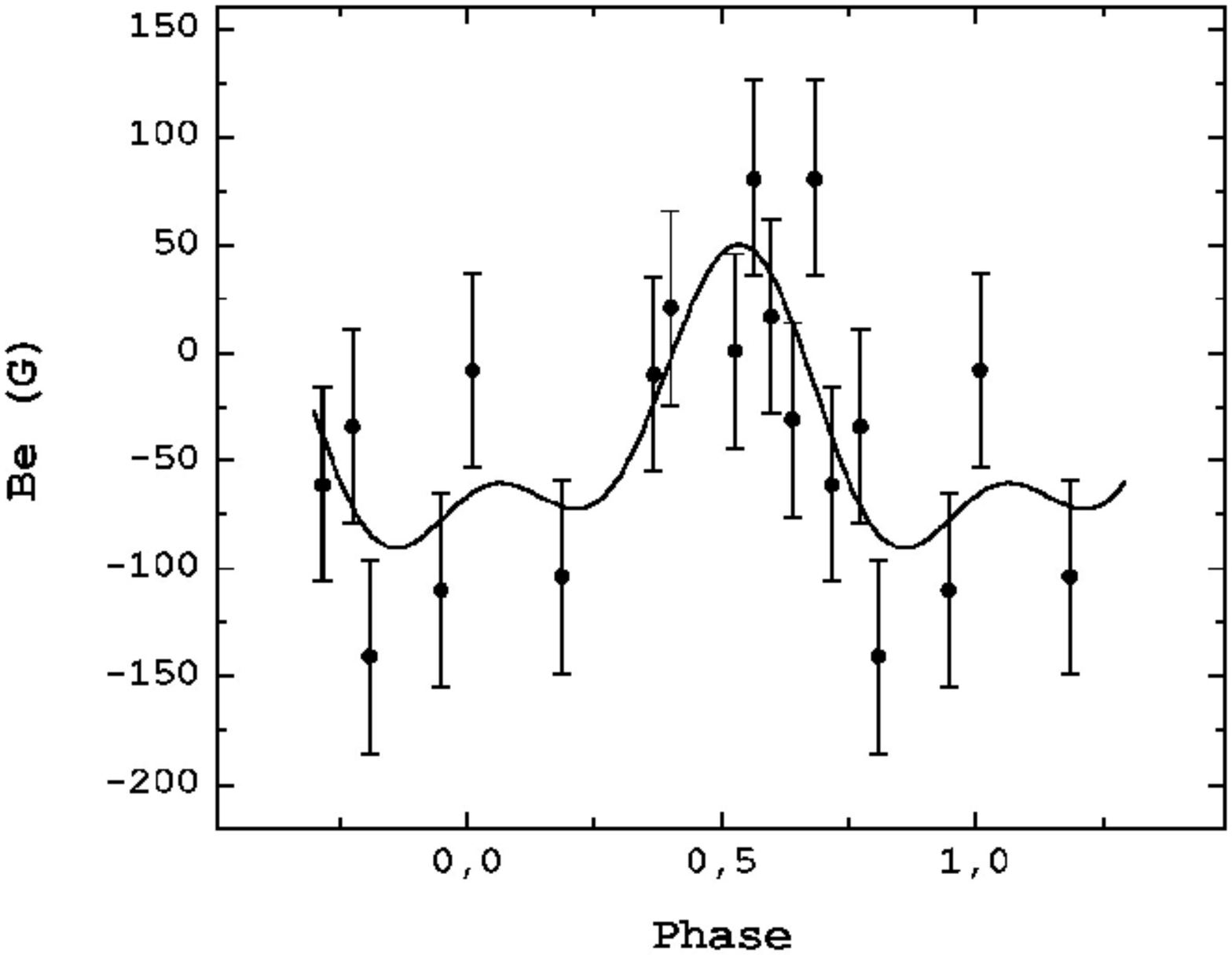}}
\vspace{-3.5mm}
\caption{ HD117555 (2) }
\label{fig:fig214}
\end{figure}

\begin{figure}
\resizebox{0.98\hsize}{!}{\includegraphics{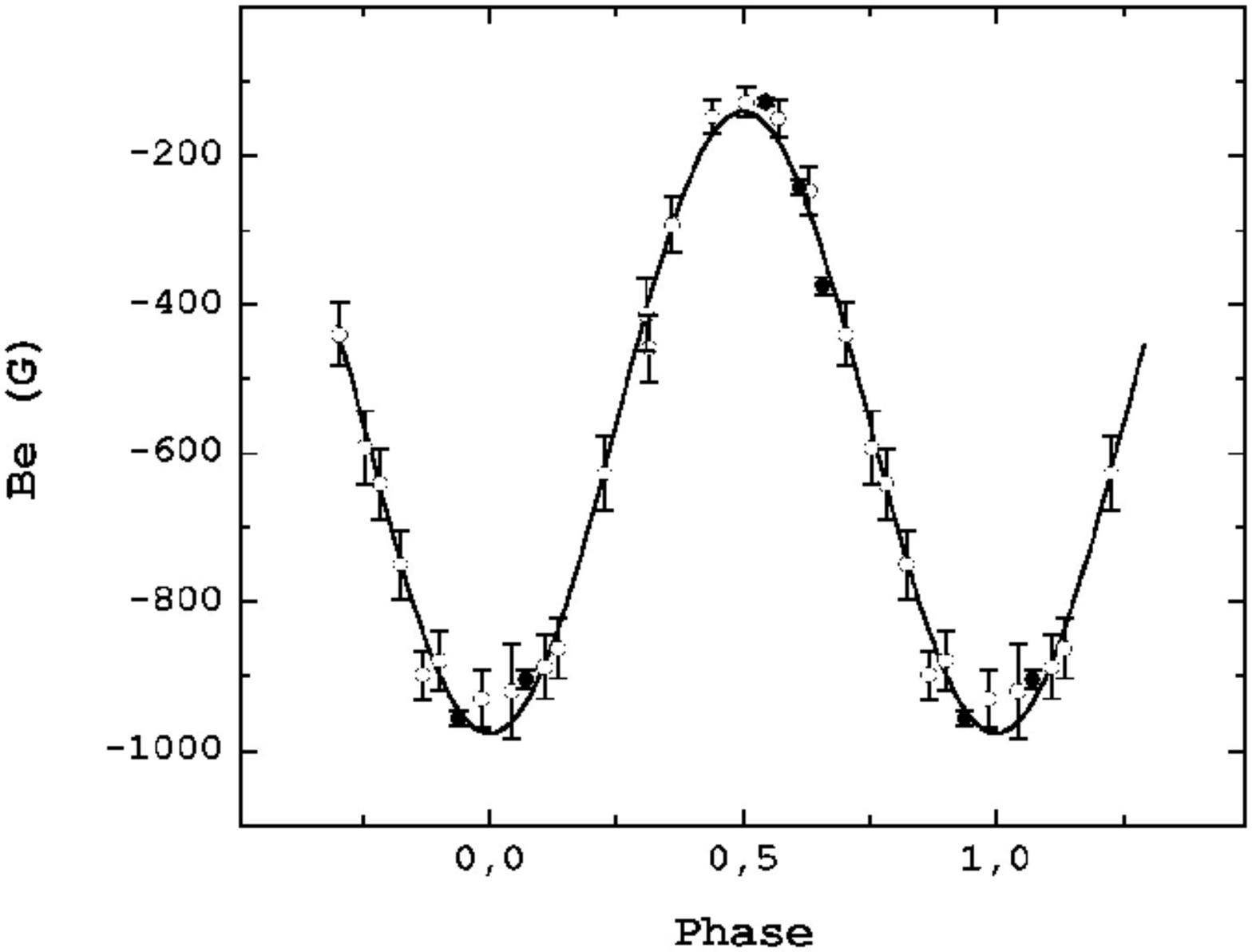}}
\vspace{-3.5mm}
\caption{ HD118022 (1) }
\label{fig:fig215}
\end{figure}

\clearpage
\newpage

\begin{figure}
\resizebox{0.98\hsize}{!}{\includegraphics{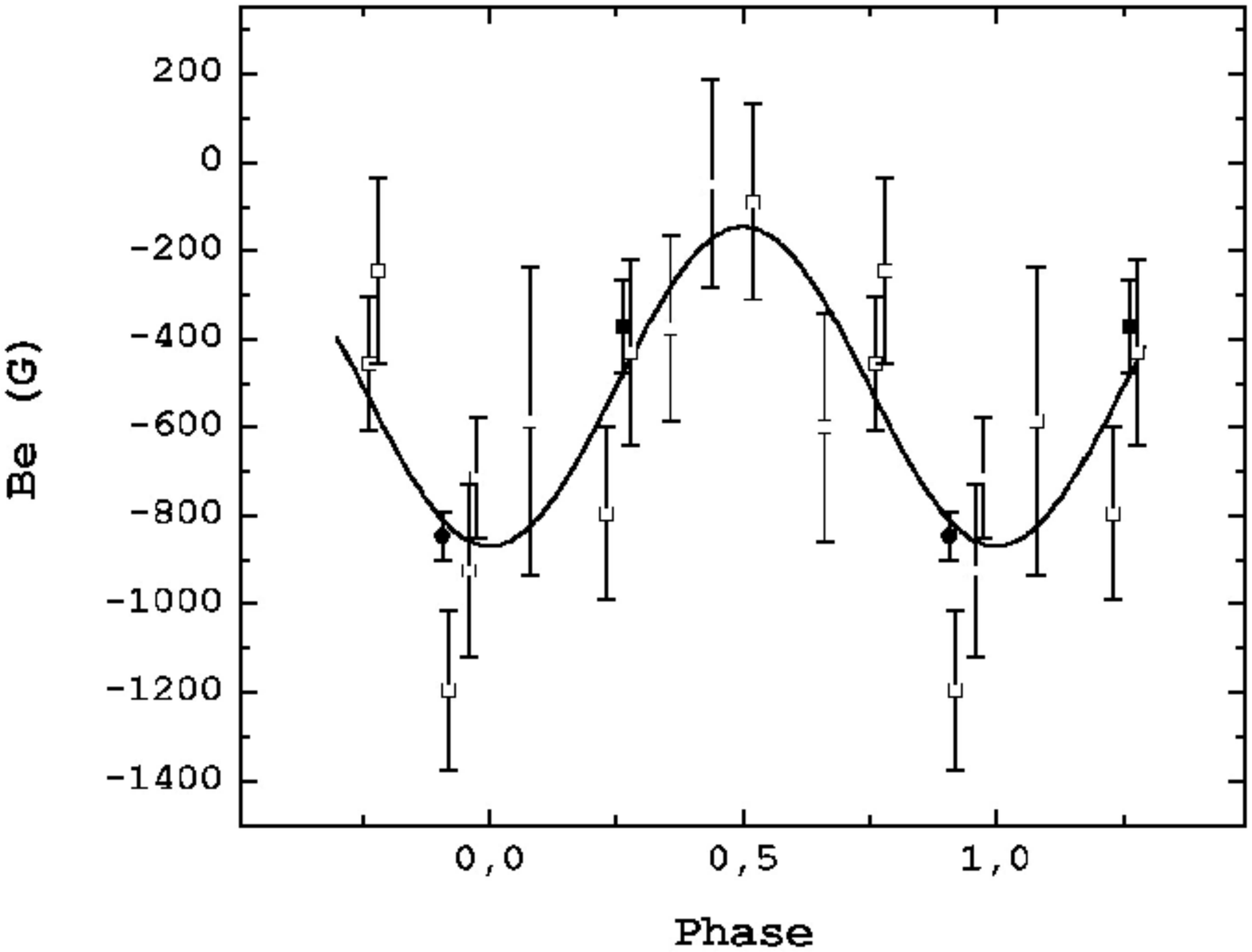}}
\vspace{-3.5mm}
\caption{ HD118022 (2) }
\label{fig:fig216}
\end{figure}

\begin{figure}
\resizebox{0.98\hsize}{!}{\includegraphics{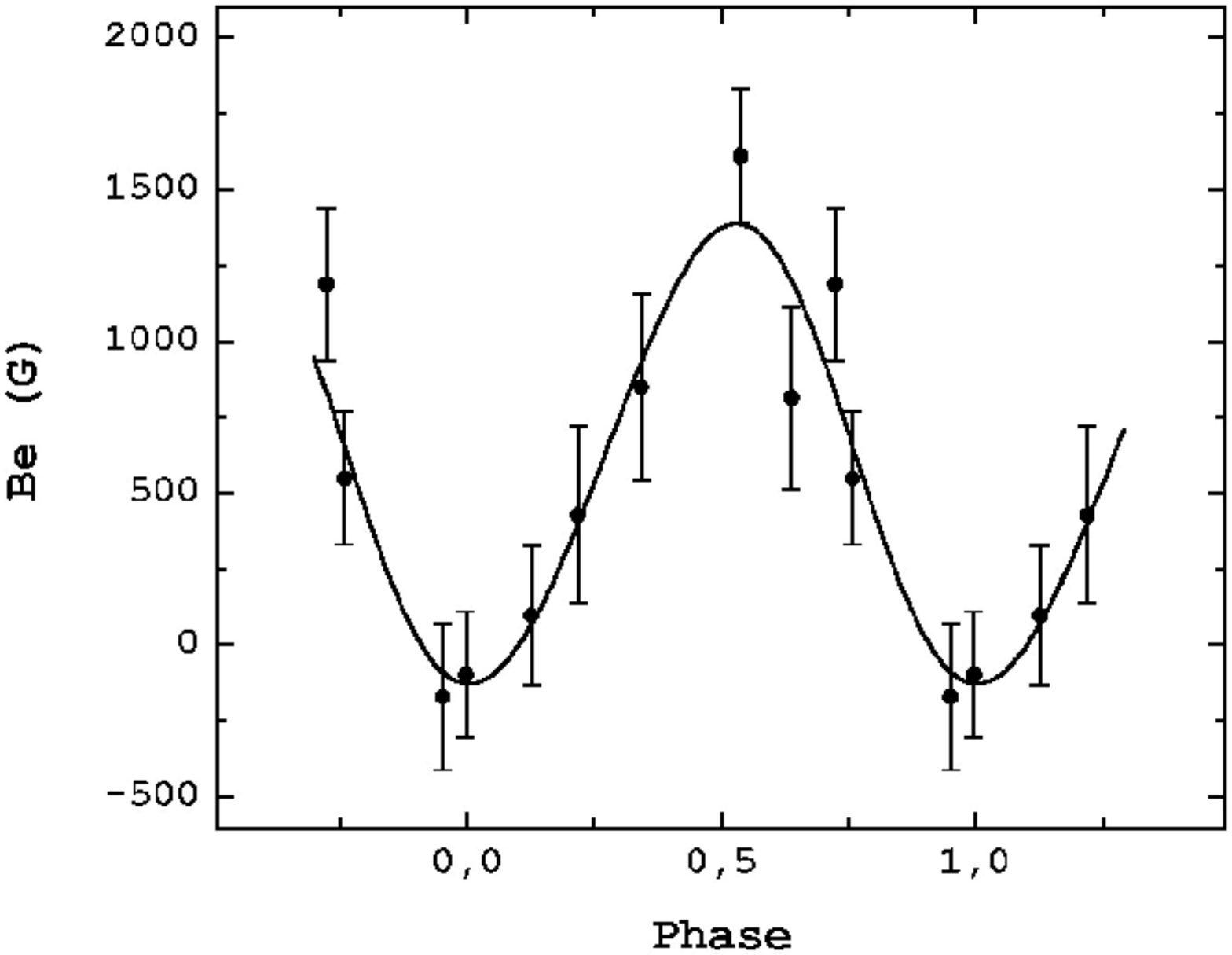}}
\vspace{-3.5mm}
\caption{ HD119213 (1) }
\label{fig:fig217}
\end{figure}

\begin{figure}
\resizebox{0.98\hsize}{!}{\includegraphics{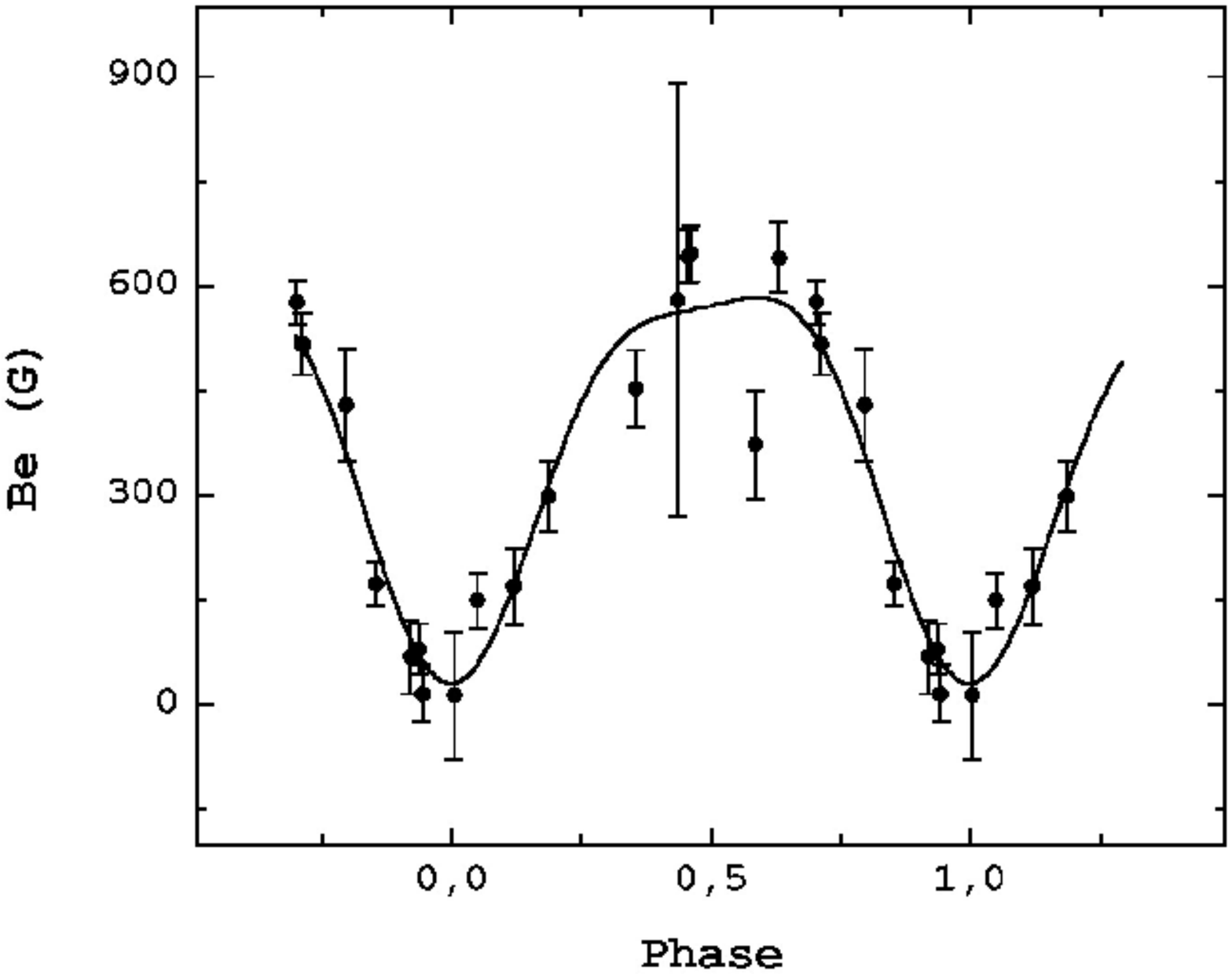}}
\vspace{-3.5mm}
\caption{ HD119213 (2) }
\label{fig:fig217}
\end{figure}

\begin{figure}
\resizebox{0.98\hsize}{!}{\includegraphics{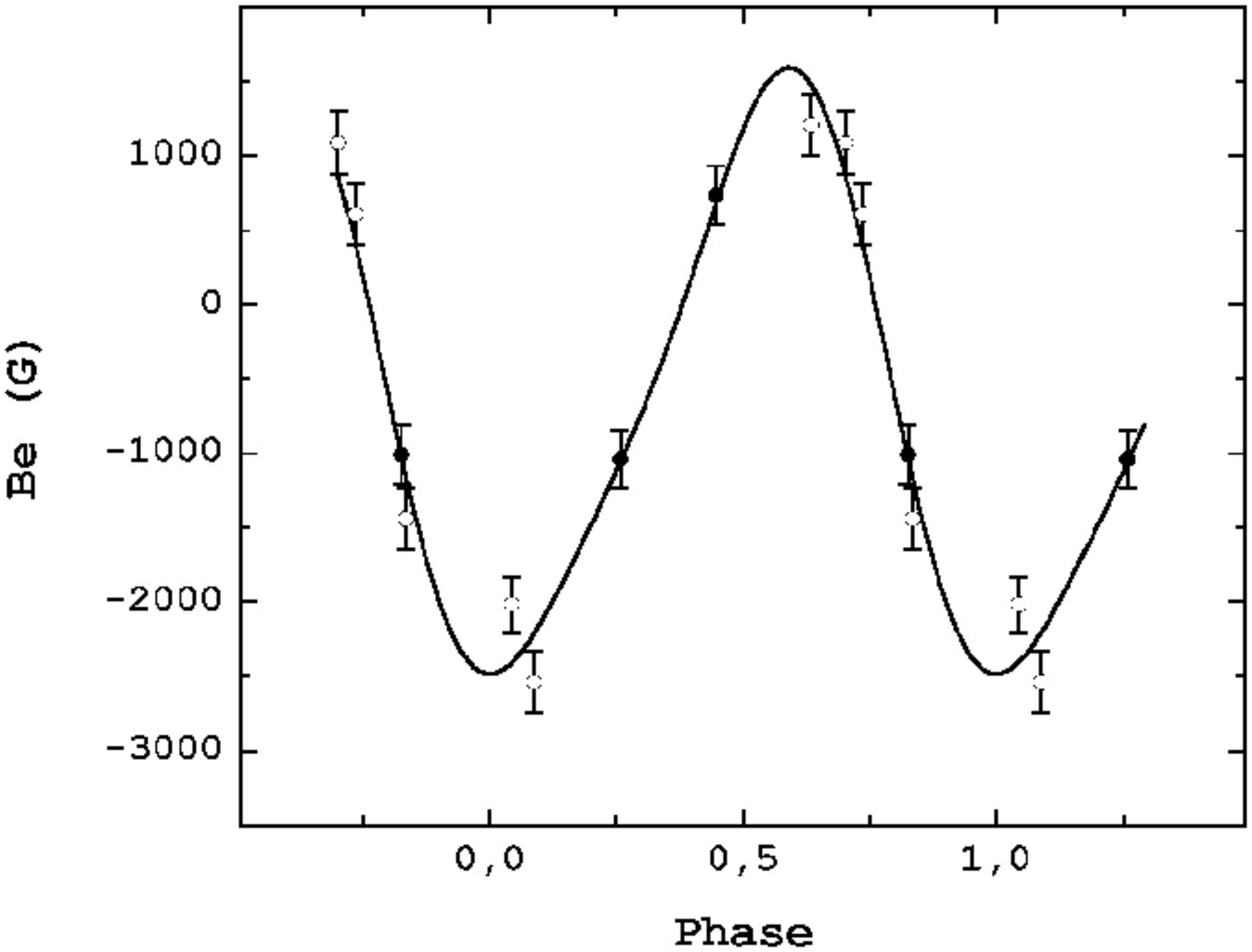}}
\vspace{-3.5mm}
\caption{ HD119419 (1) }
\label{fig:fig218}
\end{figure}

\begin{figure}
\resizebox{0.98\hsize}{!}{\includegraphics{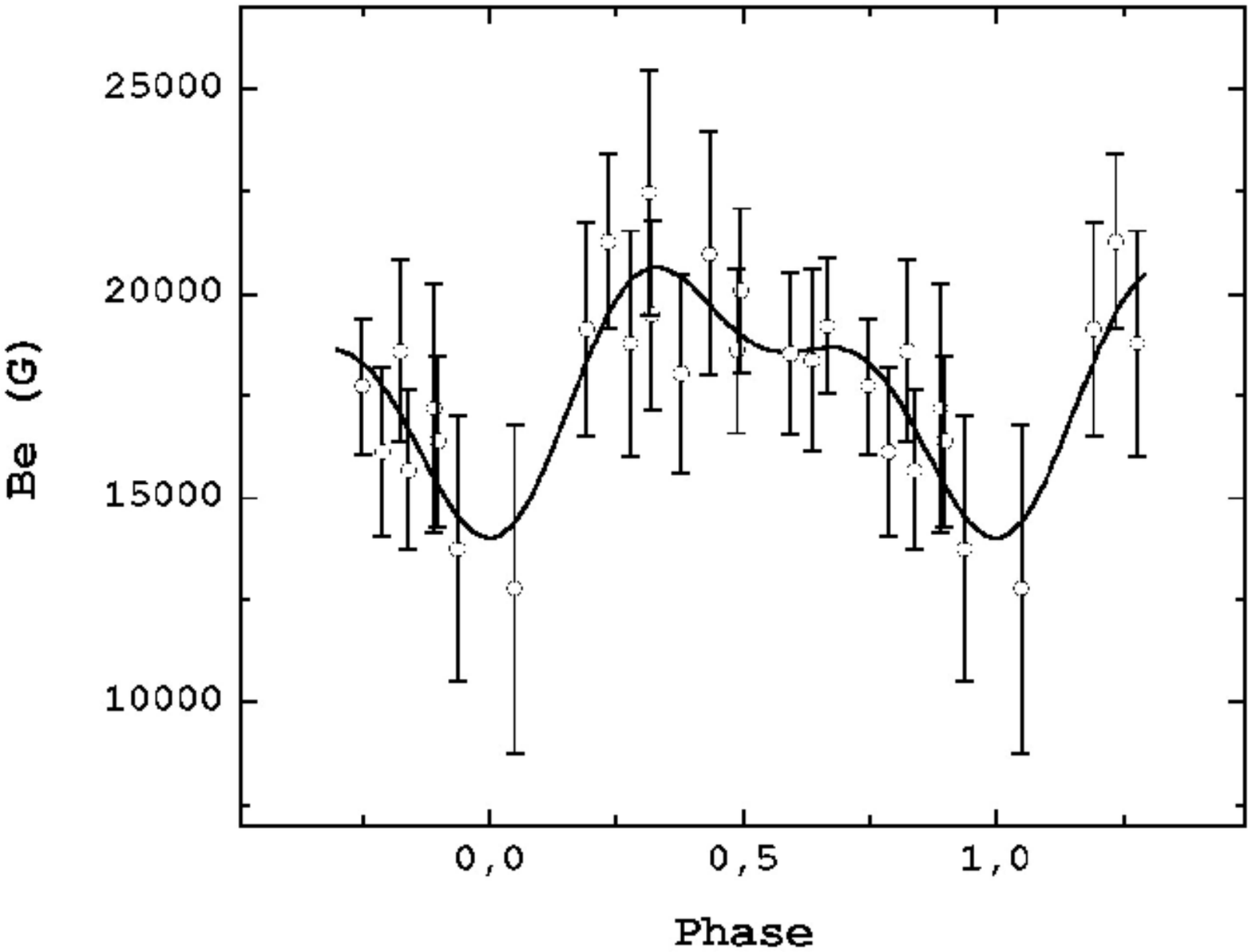}}
\vspace{-3.5mm}
\caption{ HD119419 (2) }
\label{fig:fig219}
\end{figure}

\begin{figure}
\resizebox{0.98\hsize}{!}{\includegraphics{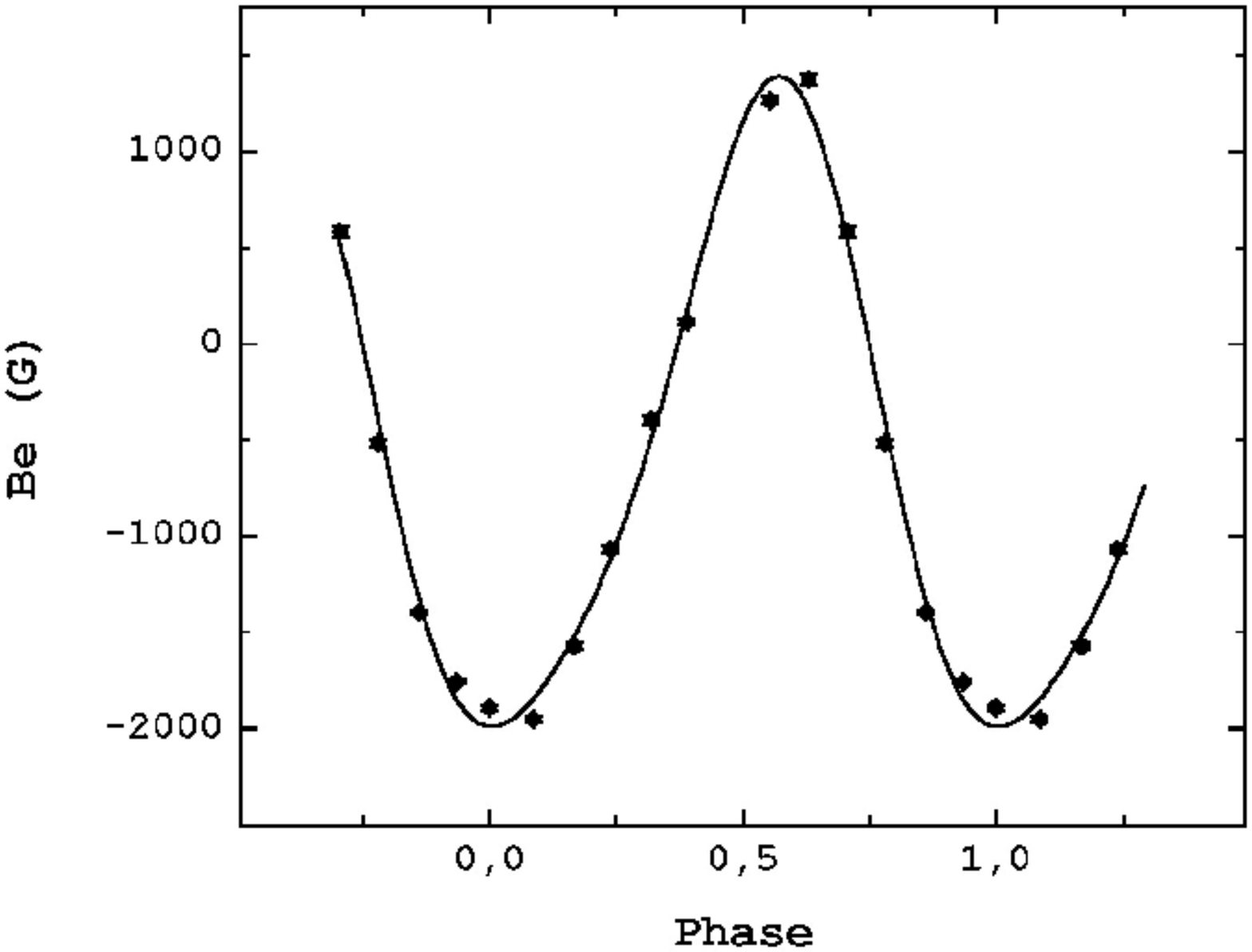}}
\vspace{-3.5mm}
\caption{ HD119419 (3) }
\label{fig:fig219}
\end{figure}

\clearpage
\newpage

\begin{figure}
\resizebox{0.98\hsize}{!}{\includegraphics{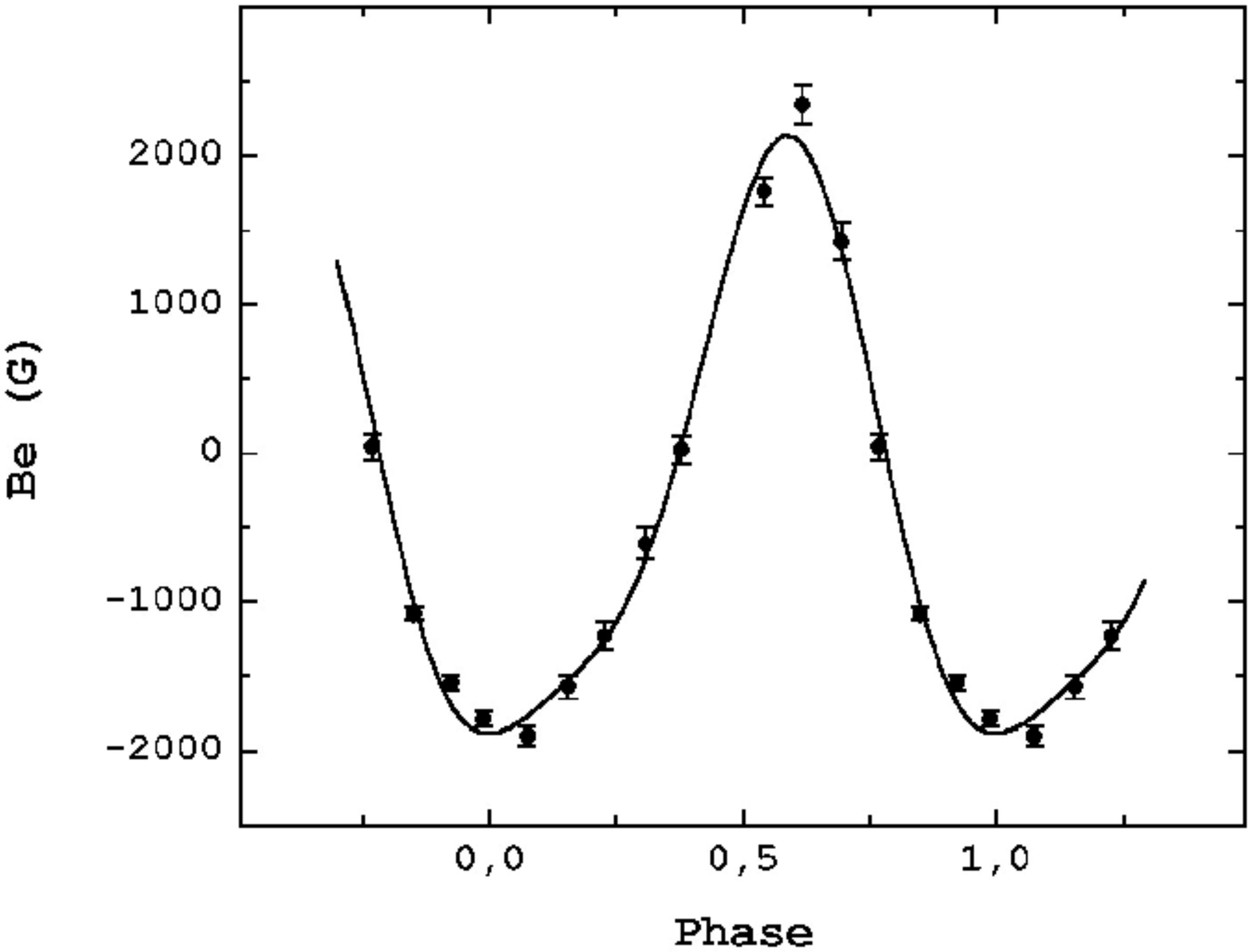}}
\vspace{-3.5mm}
\caption{ HD119419 (4) }
\label{fig:fig219}
\end{figure}

\begin{figure}
\resizebox{0.98\hsize}{!}{\includegraphics{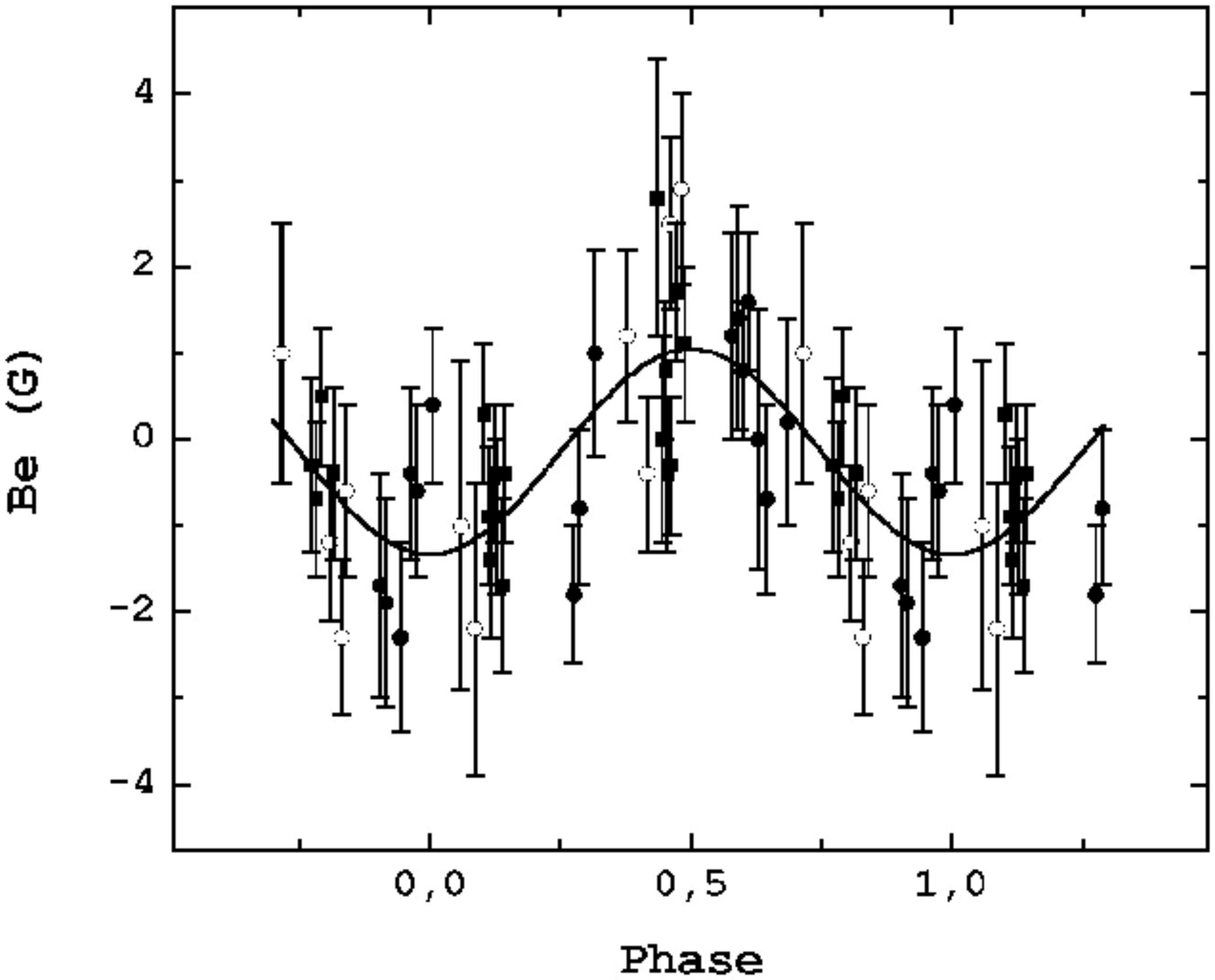}}
\vspace{-3.5mm}
\caption{ HD120136 }
\label{fig:fig220}
\end{figure}

\begin{figure}
\resizebox{0.98\hsize}{!}{\includegraphics{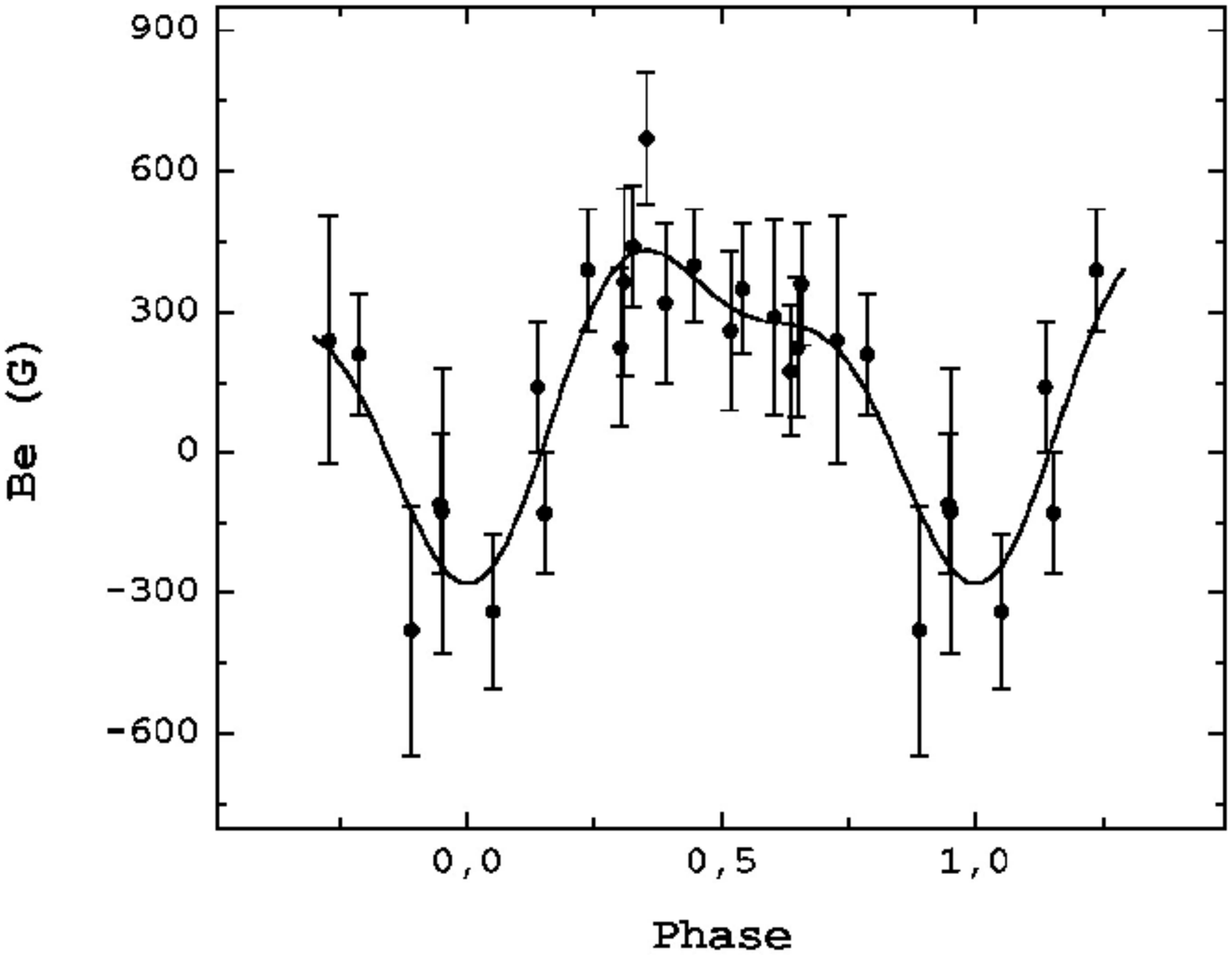}}
\vspace{-3.5mm}
\caption{ HD120198 }
\label{fig:fig221}
\end{figure}

\begin{figure}
\resizebox{0.98\hsize}{!}{\includegraphics{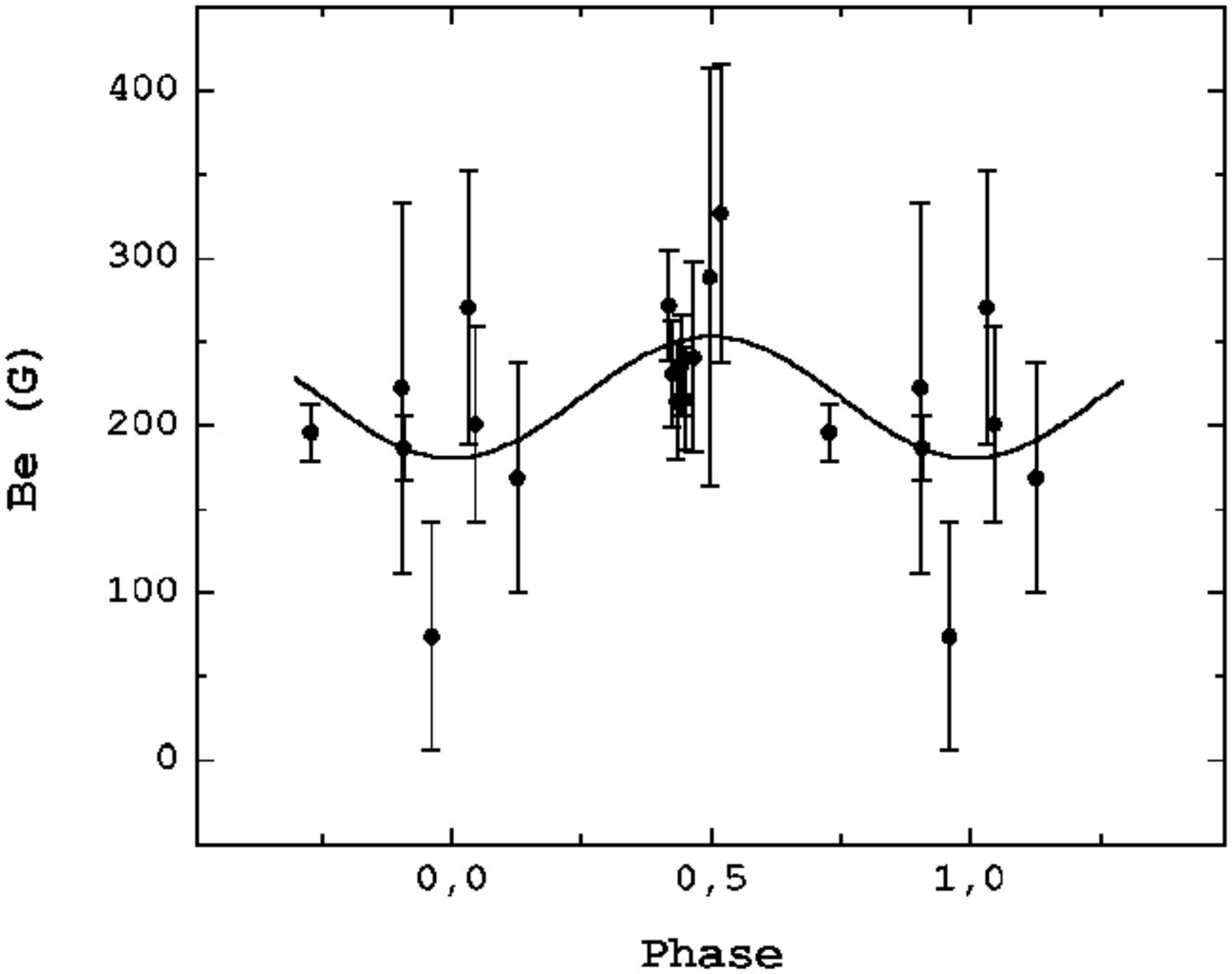}}
\vspace{-3.5mm}
\caption{ HD121743 (1) }
\label{fig:fig219}
\end{figure}

\begin{figure}
\resizebox{0.98\hsize}{!}{\includegraphics{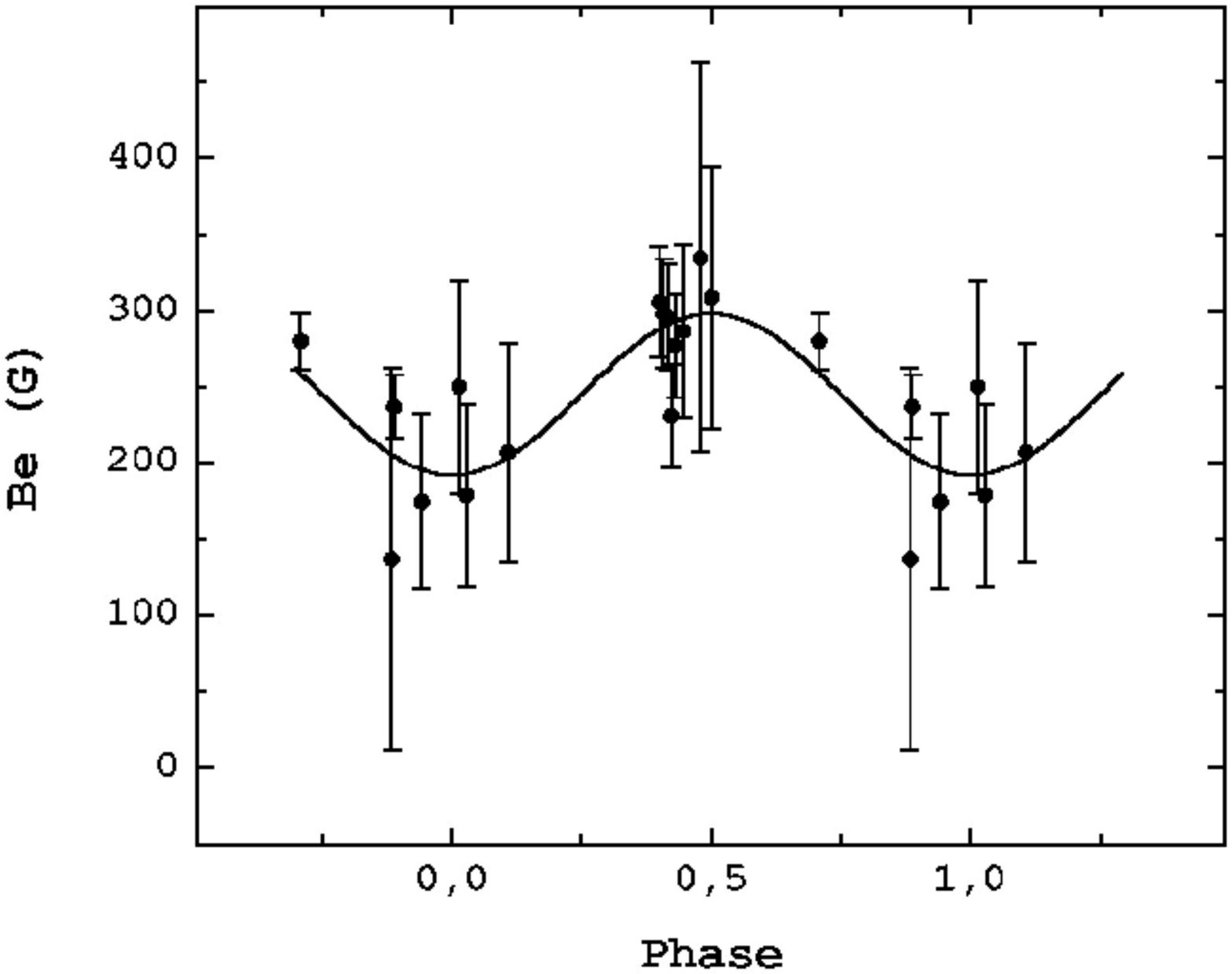}}
\vspace{-3.5mm}
\caption{ HD121743 (2) }
\label{fig:fig219}
\end{figure}

\begin{figure}
\resizebox{0.98\hsize}{!}{\includegraphics{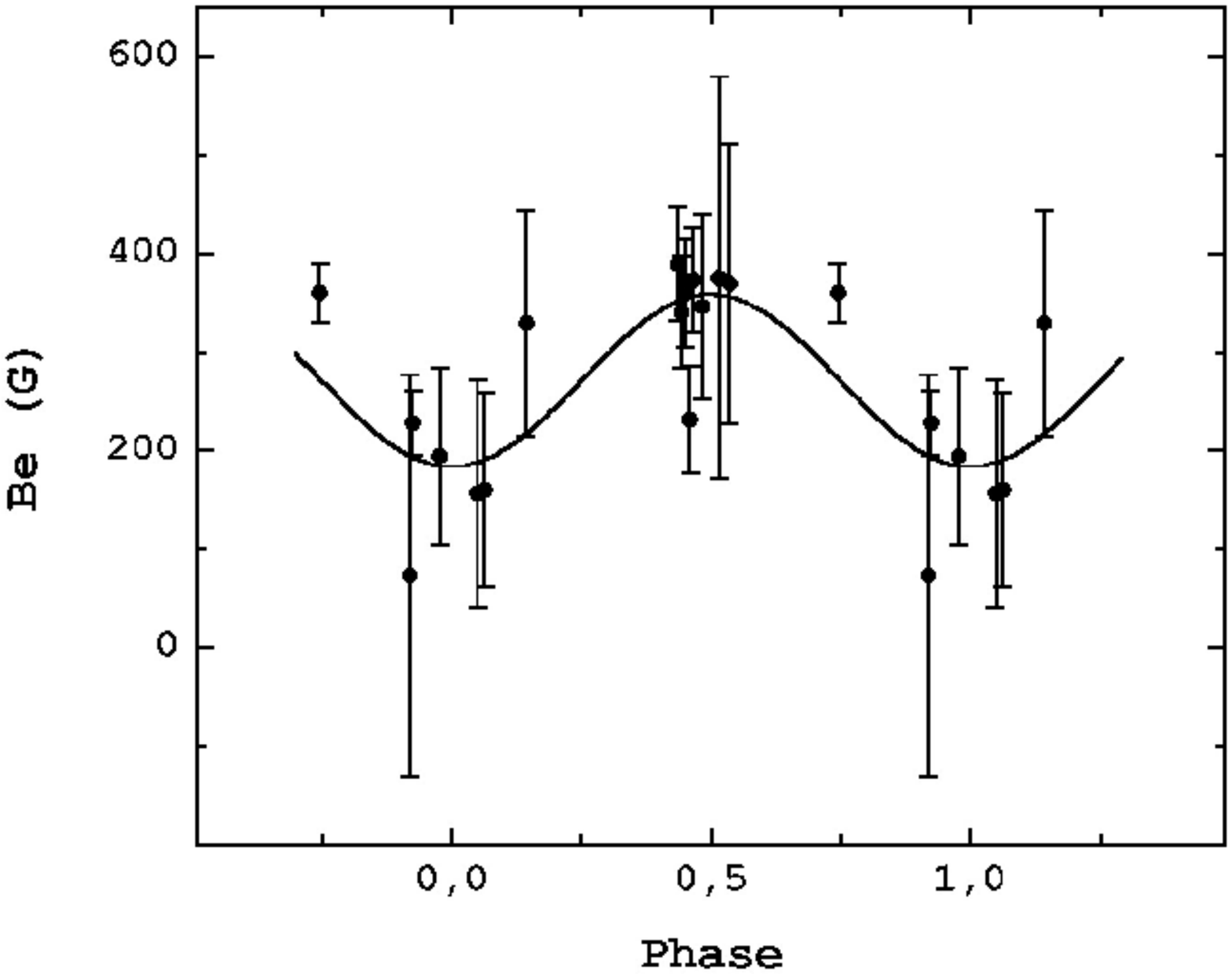}}
\vspace{-3.5mm}
\caption{ HD121743 (3) }
\label{fig:fig219}
\end{figure}

\clearpage
\newpage

\begin{figure}
\resizebox{0.98\hsize}{!}{\includegraphics{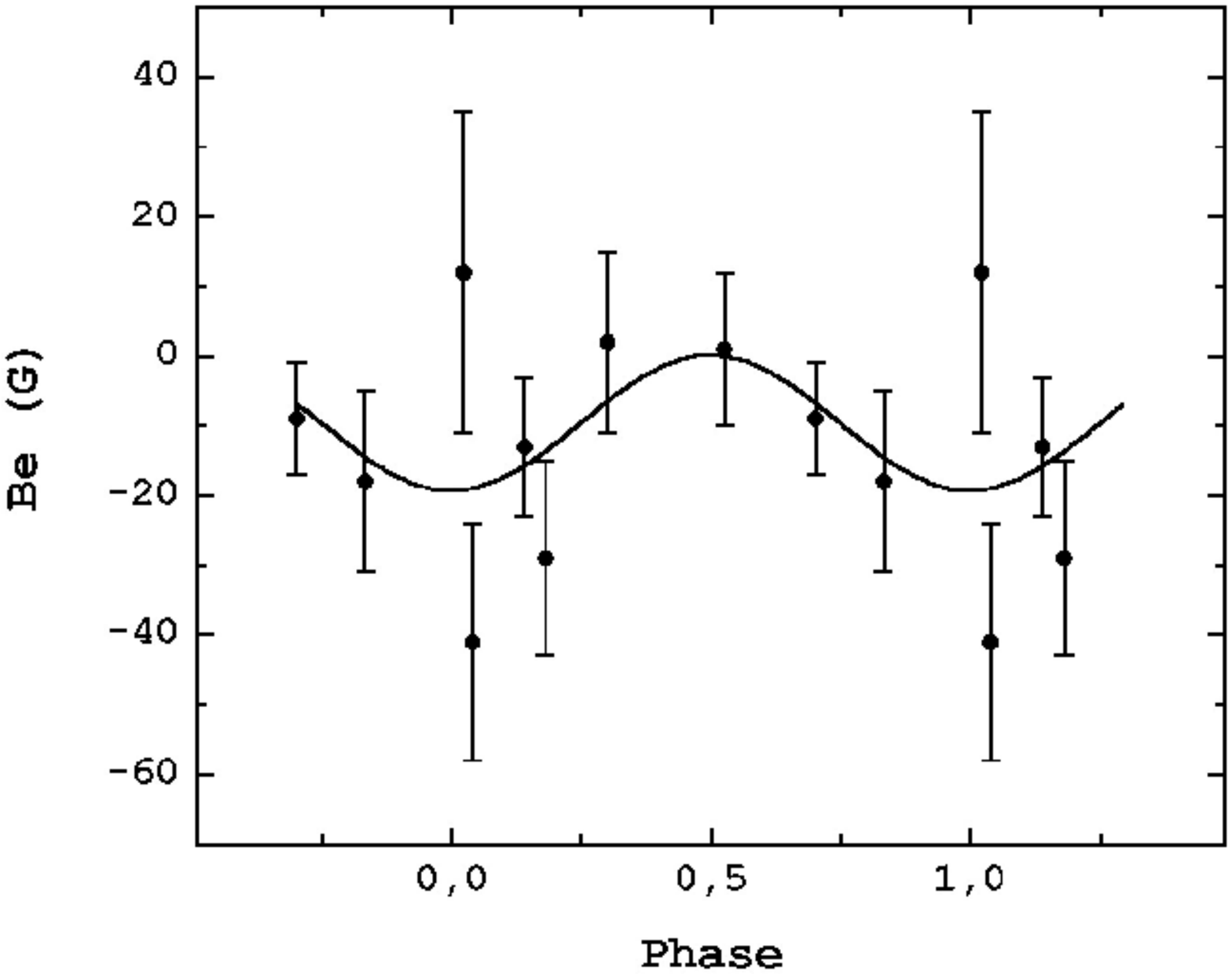}}
\vspace{-3.5mm}
\caption{ HD122451 }
\label{fig:fig219}
\end{figure}

\begin{figure}
\resizebox{0.98\hsize}{!}{\includegraphics{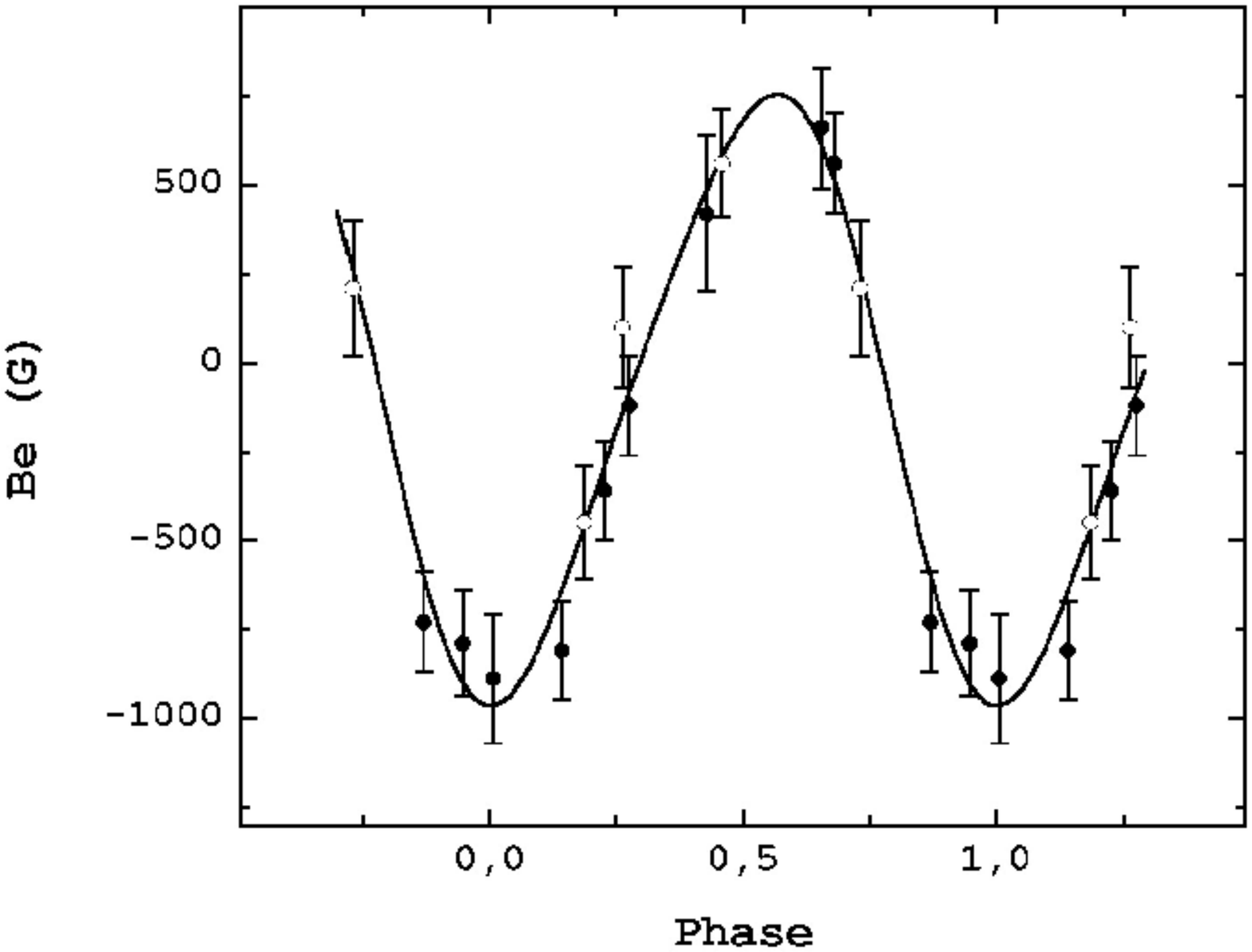}}
\vspace{-3.5mm}
\caption{ HD122532 (1) }
\label{fig:fig222}
\end{figure}

\begin{figure}
\resizebox{0.98\hsize}{!}{\includegraphics{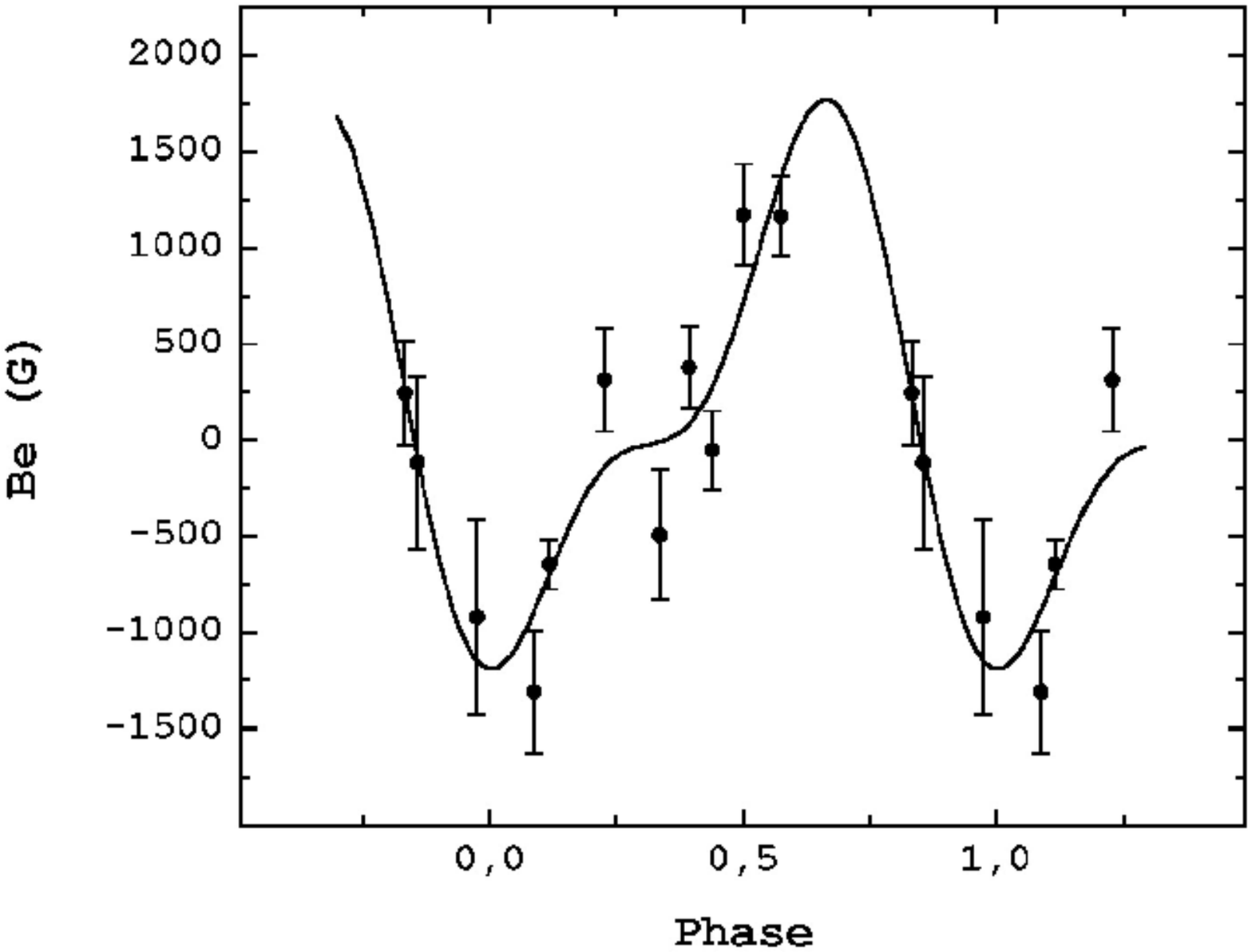}}
\vspace{-3.5mm}
\caption{ HD122532 (2) }
\label{fig:fig223}
\end{figure}

\begin{figure}
\resizebox{0.98\hsize}{!}{\includegraphics{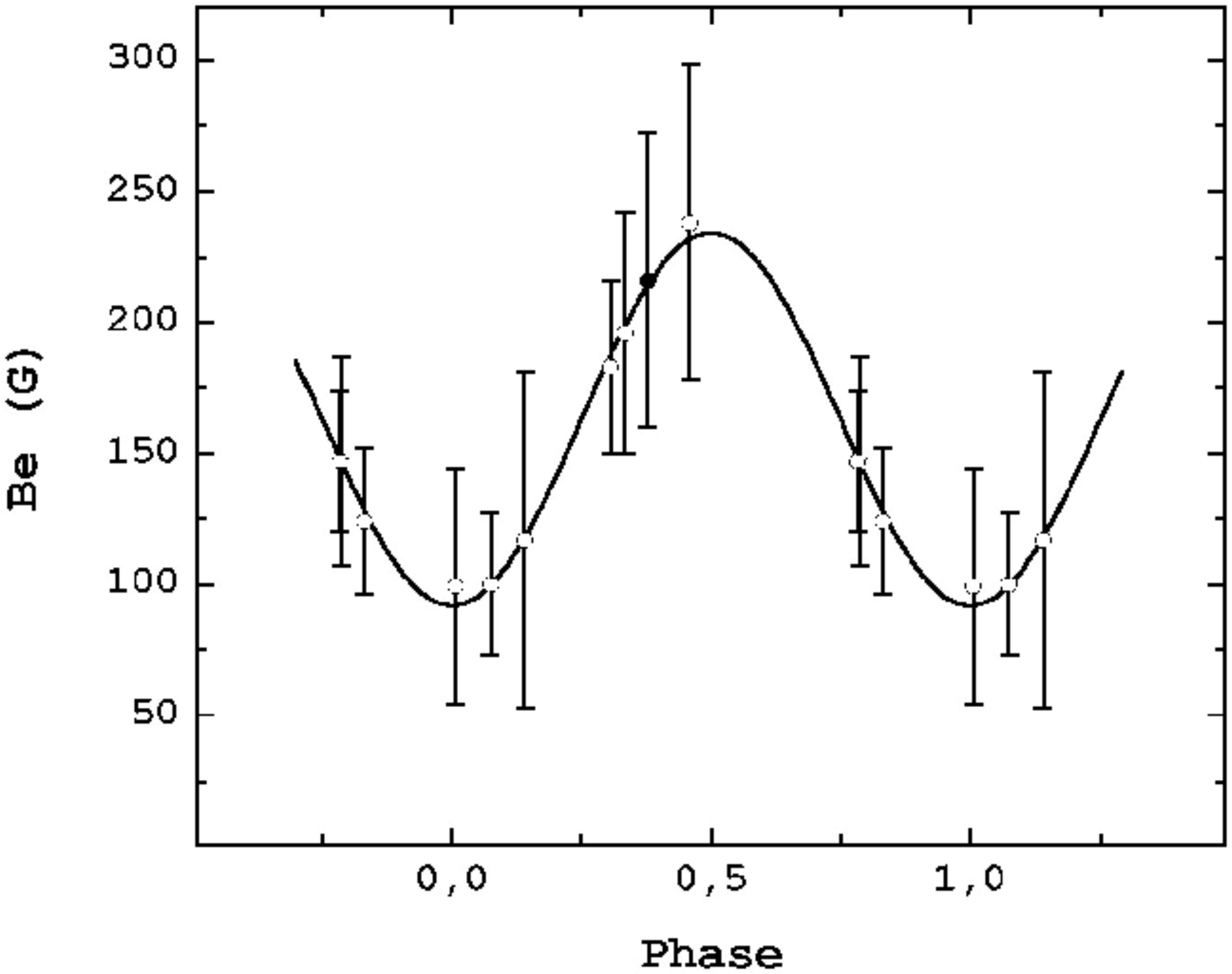}}
\vspace{-3.5mm}
\caption{ HD122970 }
\label{fig:fig224}
\end{figure}

\begin{figure}
\resizebox{0.98\hsize}{!}{\includegraphics{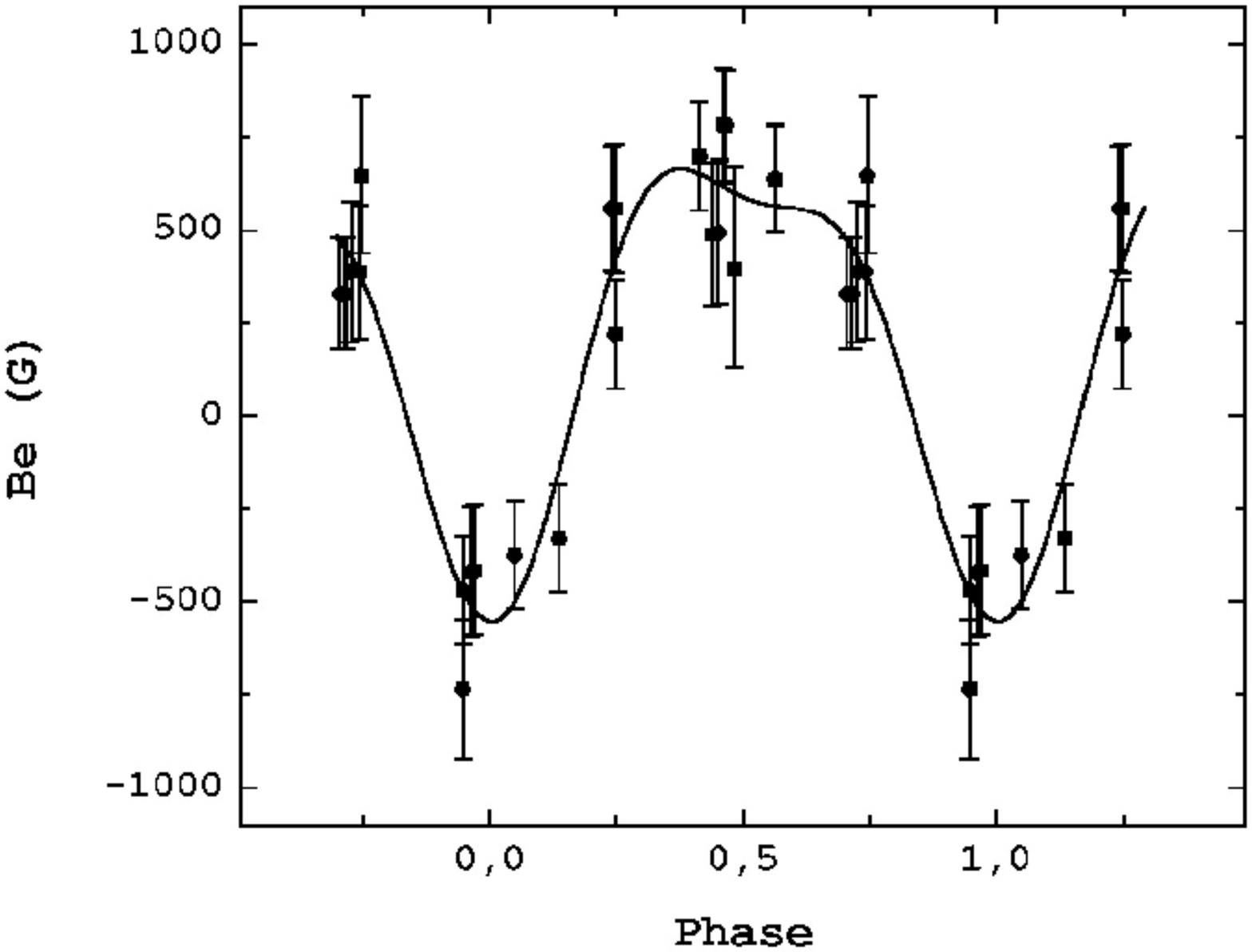}}
\vspace{-3.5mm}
\caption{ HD124224 (1) }
\label{fig:fig225}
\end{figure}

\begin{figure}
\resizebox{0.98\hsize}{!}{\includegraphics{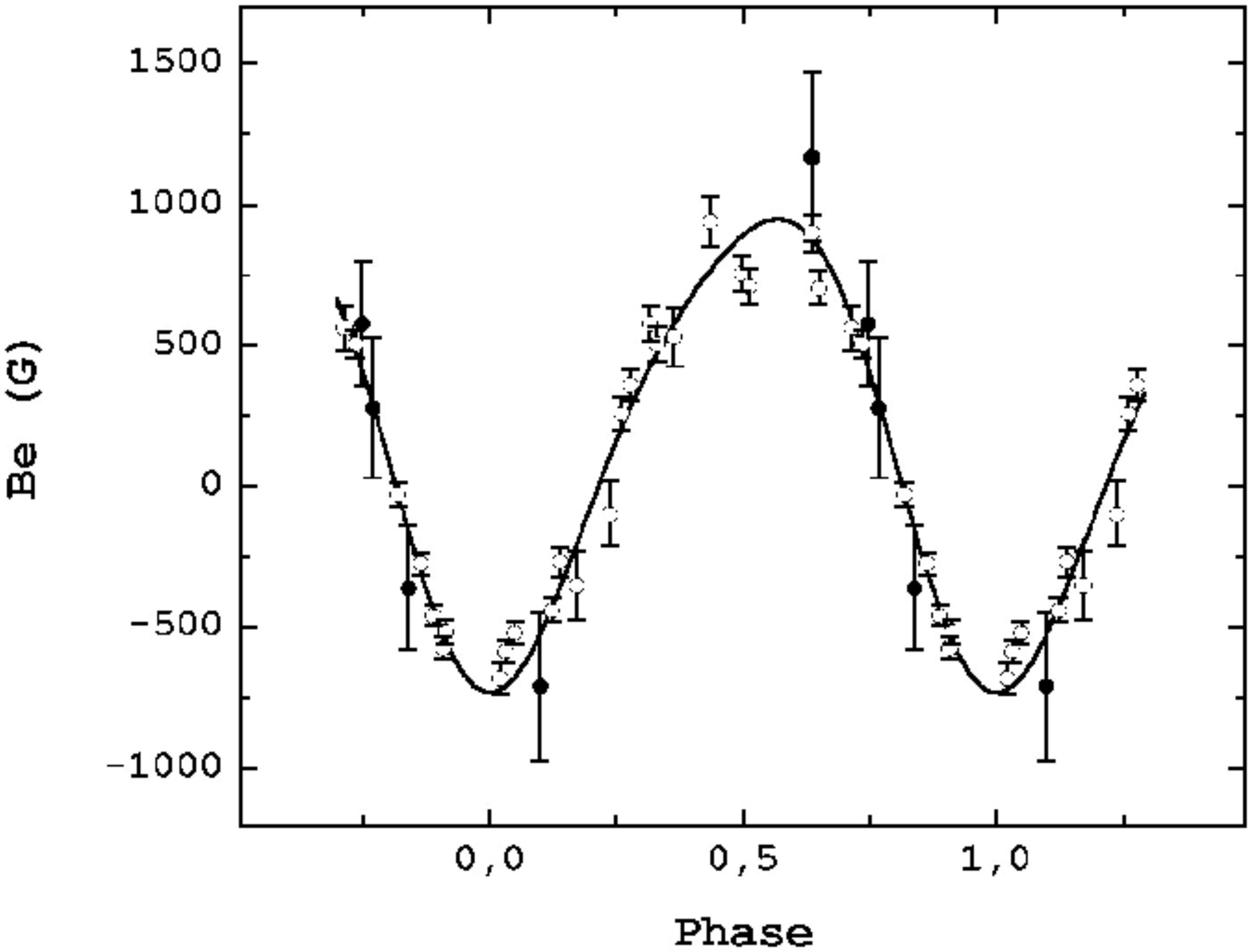}}
\vspace{-3.5mm}
\caption{ HD124224 (2) }
\label{fig:fig226}
\end{figure}

\clearpage
\newpage

\begin{figure}
\resizebox{0.98\hsize}{!}{\includegraphics{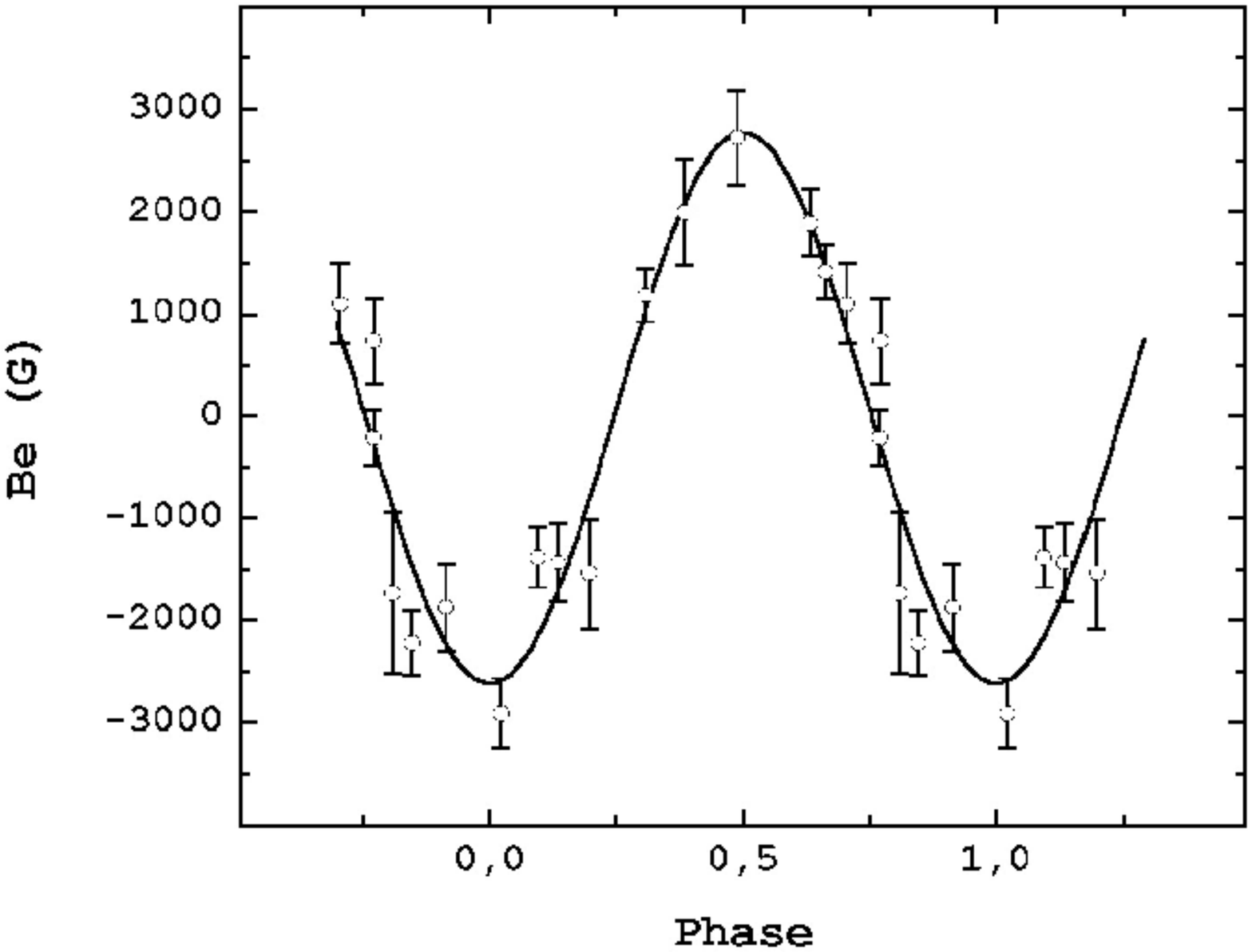}}
\vspace{-3.5mm}
\caption{ HD125248 (1) }
\label{fig:fig227}
\end{figure}

\begin{figure}
\resizebox{0.98\hsize}{!}{\includegraphics{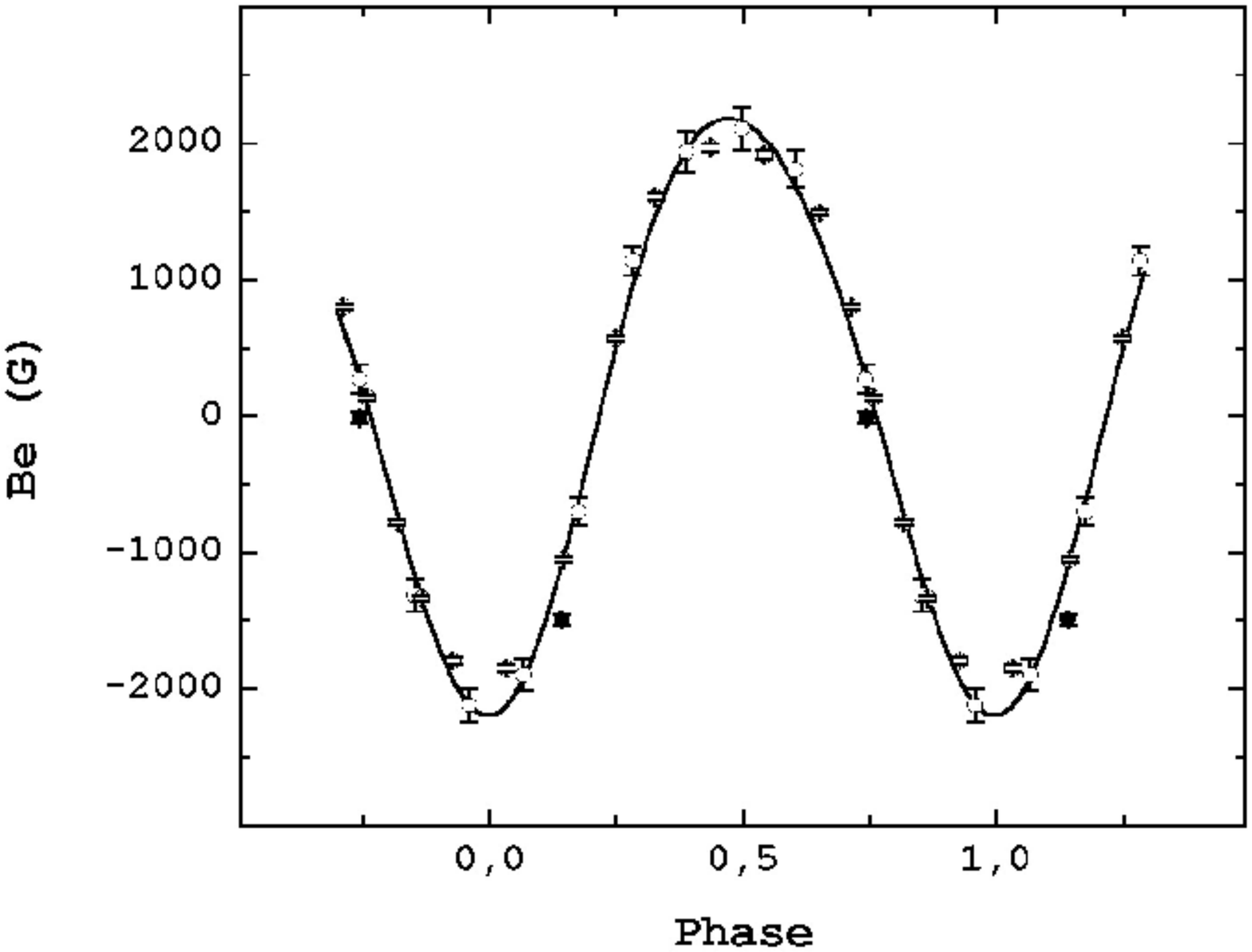}}
\vspace{-3.5mm}
\caption{ HD125248 (2) }
\label{fig:fig228}
\end{figure}

\begin{figure}
\resizebox{0.98\hsize}{!}{\includegraphics{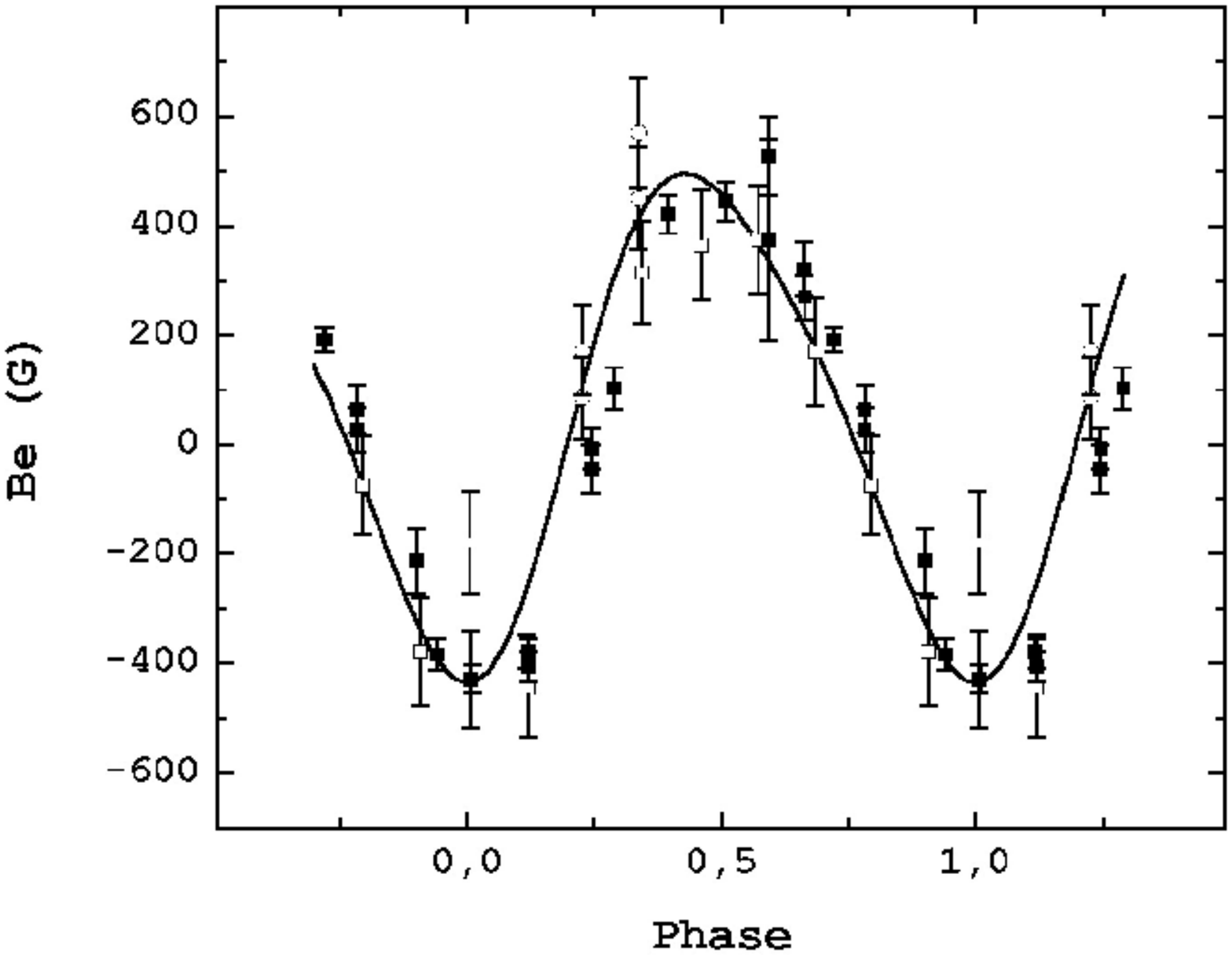}}
\vspace{-3.5mm}
\caption{ HD125823 (1) }
\label{fig:fig229}
\end{figure}

\begin{figure}
\resizebox{0.98\hsize}{!}{\includegraphics{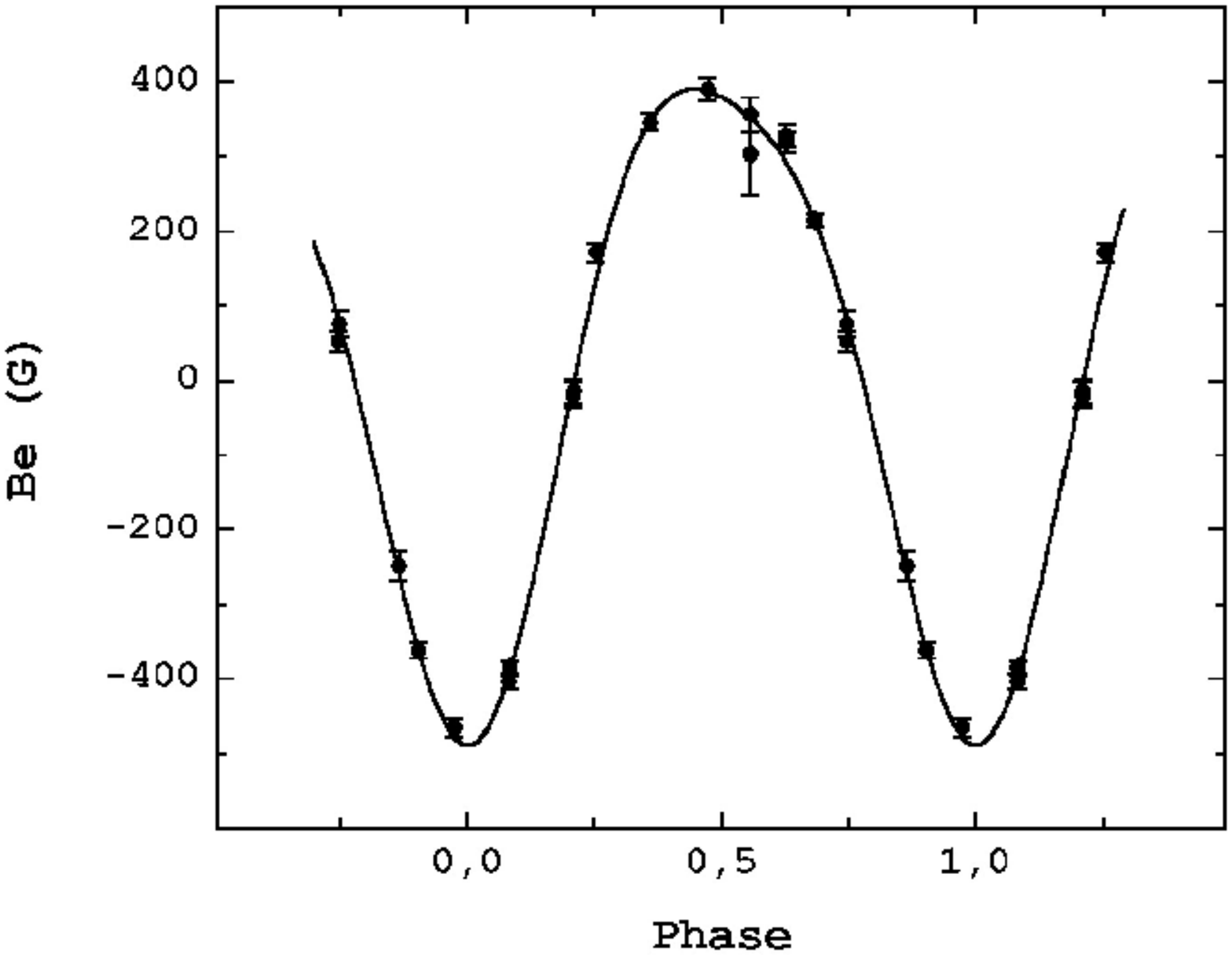}}
\vspace{-3.5mm}
\caption{ HD125823 (2) }
\label{fig:fig230}
\end{figure}

\begin{figure}
\resizebox{0.98\hsize}{!}{\includegraphics{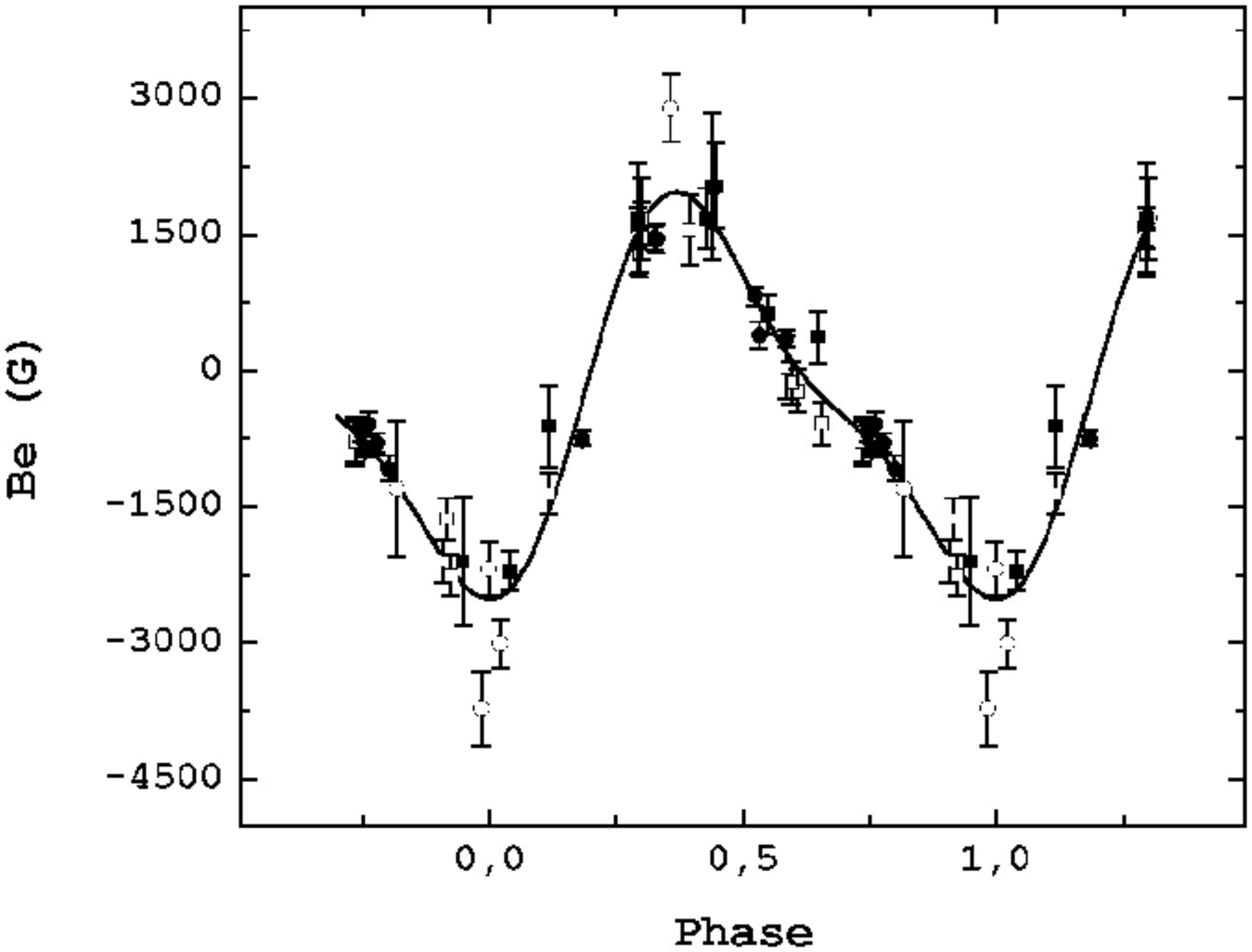}}
\vspace{-3.5mm}
\caption{ HD126515 }
\label{fig:fig231}
\end{figure}

\begin{figure}
\resizebox{0.98\hsize}{!}{\includegraphics{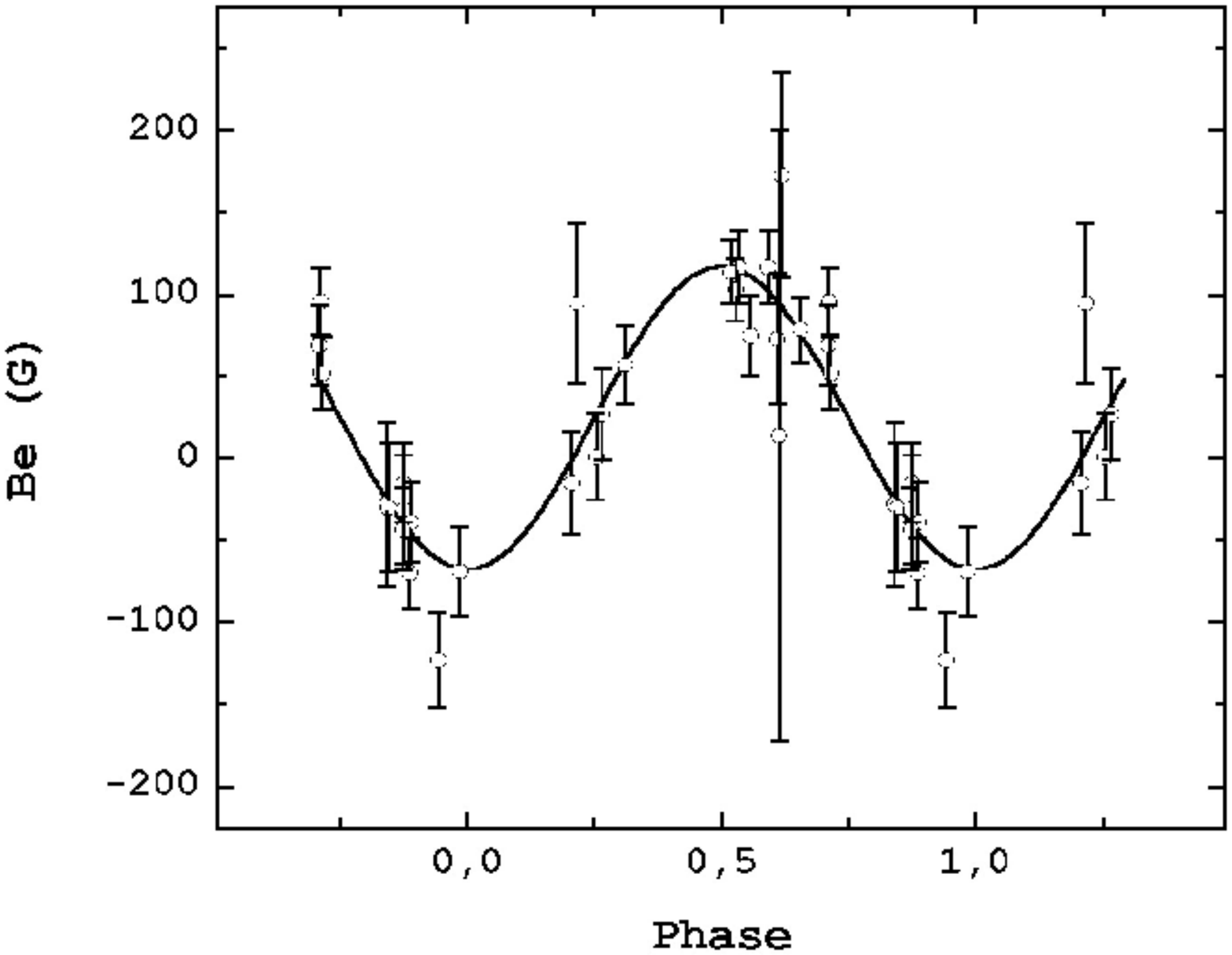}}
\vspace{-3.5mm}
\caption{ HD127381 }
\label{fig:fig232}
\end{figure}

\clearpage
\newpage

\begin{figure}
\resizebox{0.98\hsize}{!}{\includegraphics{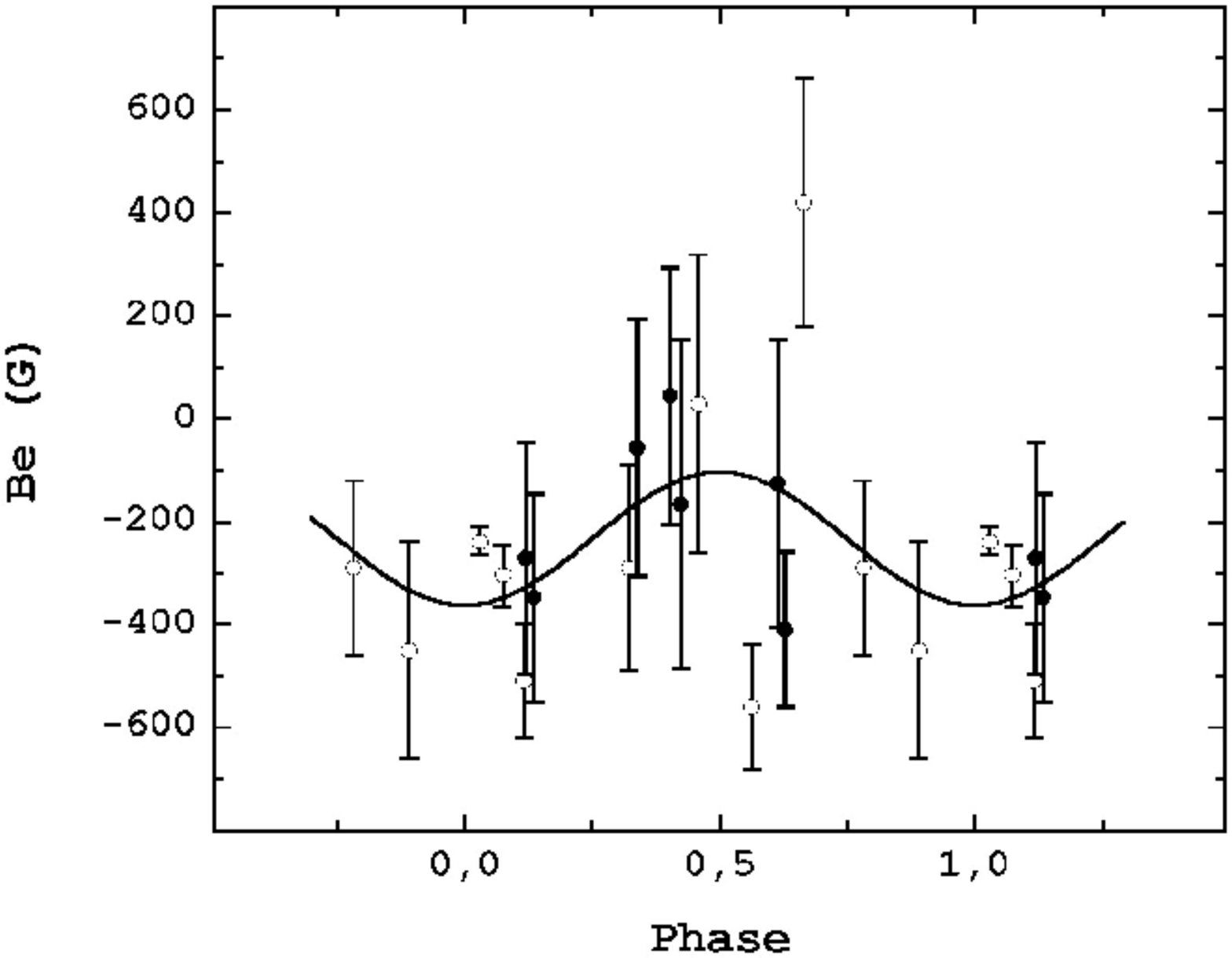}}
\vspace{-3.5mm}
\caption{ HD128898 }
\label{fig:fig233}
\end{figure}

\begin{figure}
\resizebox{0.98\hsize}{!}{\includegraphics{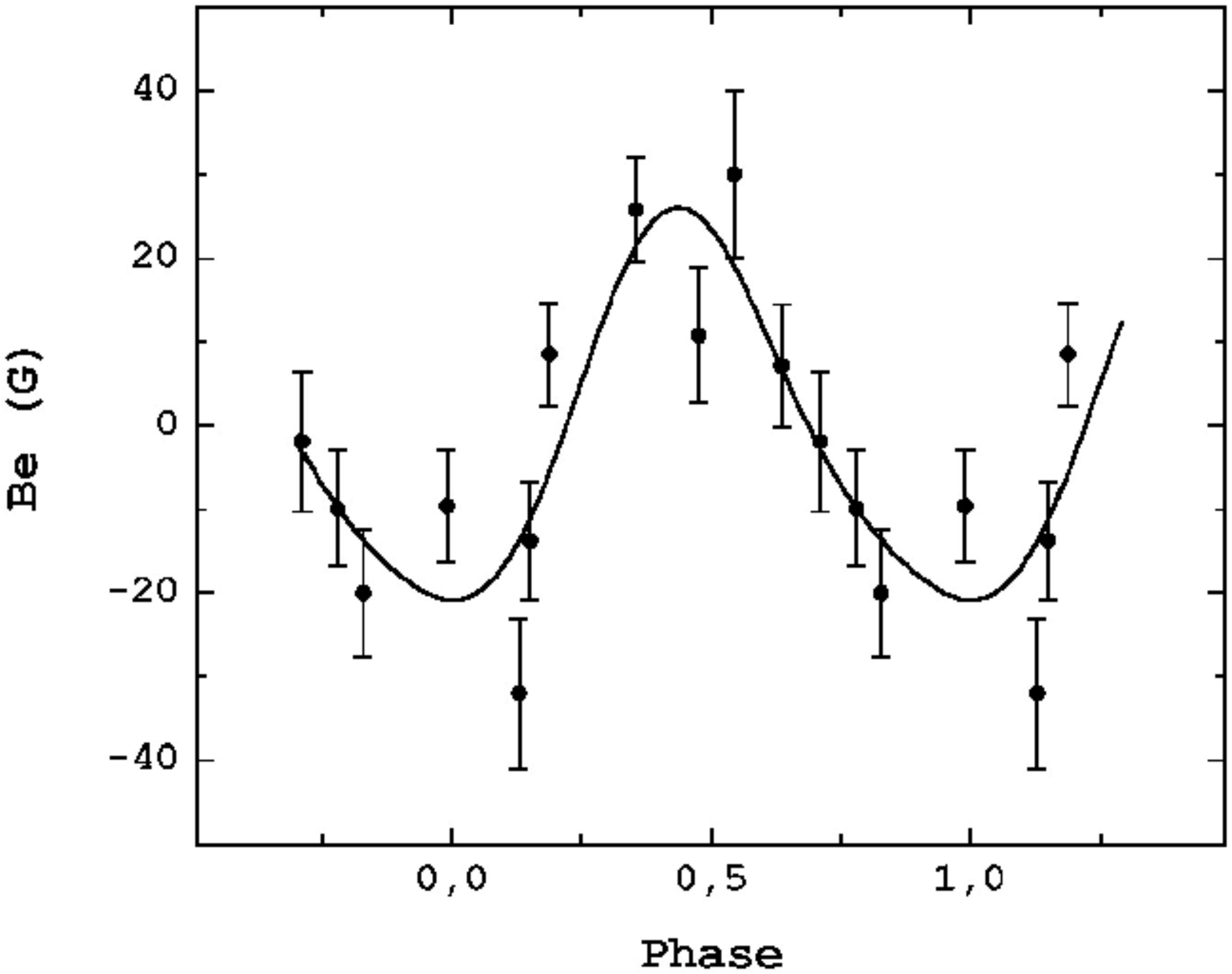}}
\vspace{-3.5mm}
\caption{ HD129333 (1) }
\label{fig:fig234}
\end{figure}

\begin{figure}
\resizebox{0.98\hsize}{!}{\includegraphics{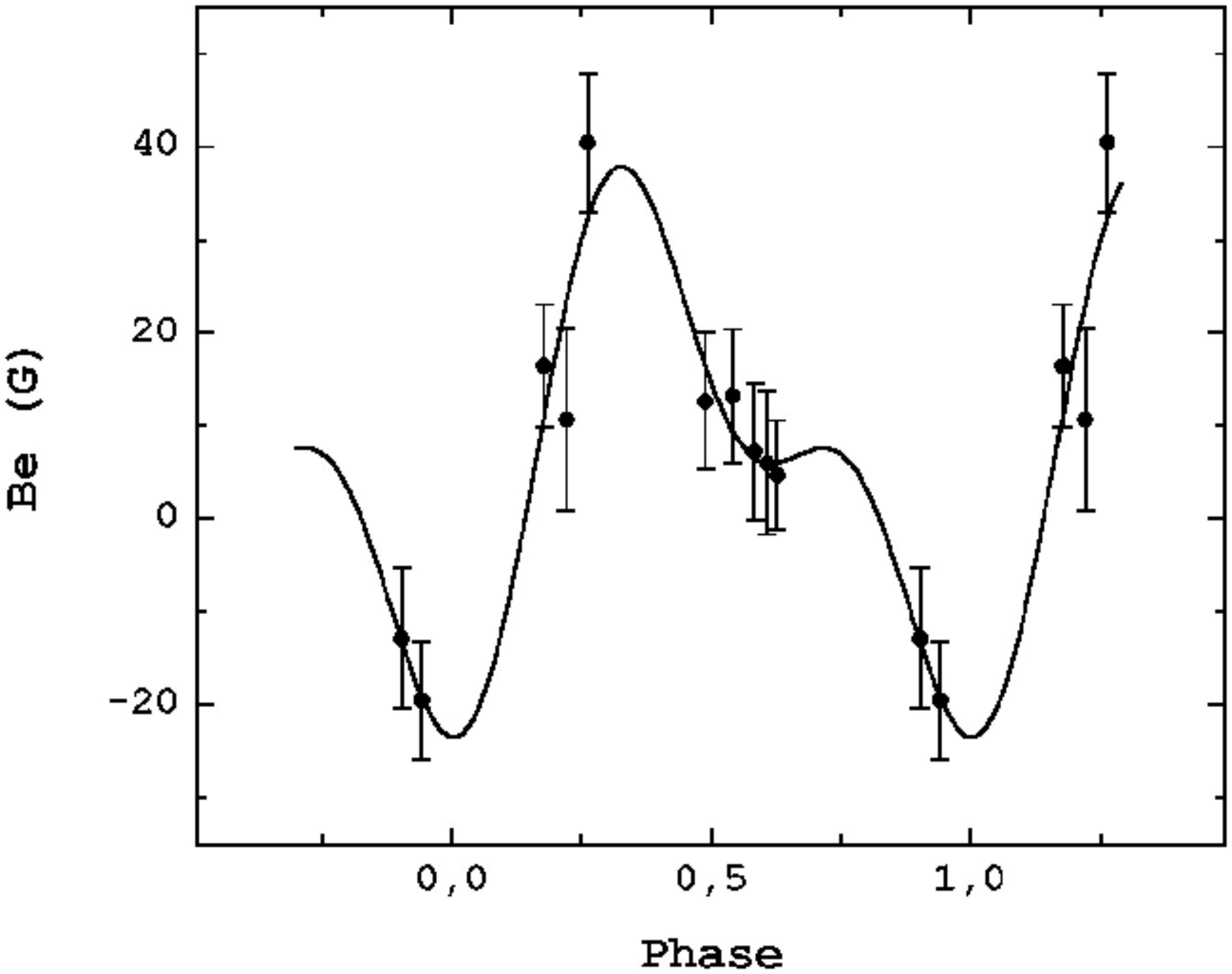}}
\vspace{-3.5mm}
\caption{ HD129333 (2) }
\label{fig:fig235}
\end{figure}

\begin{figure}
\resizebox{0.98\hsize}{!}{\includegraphics{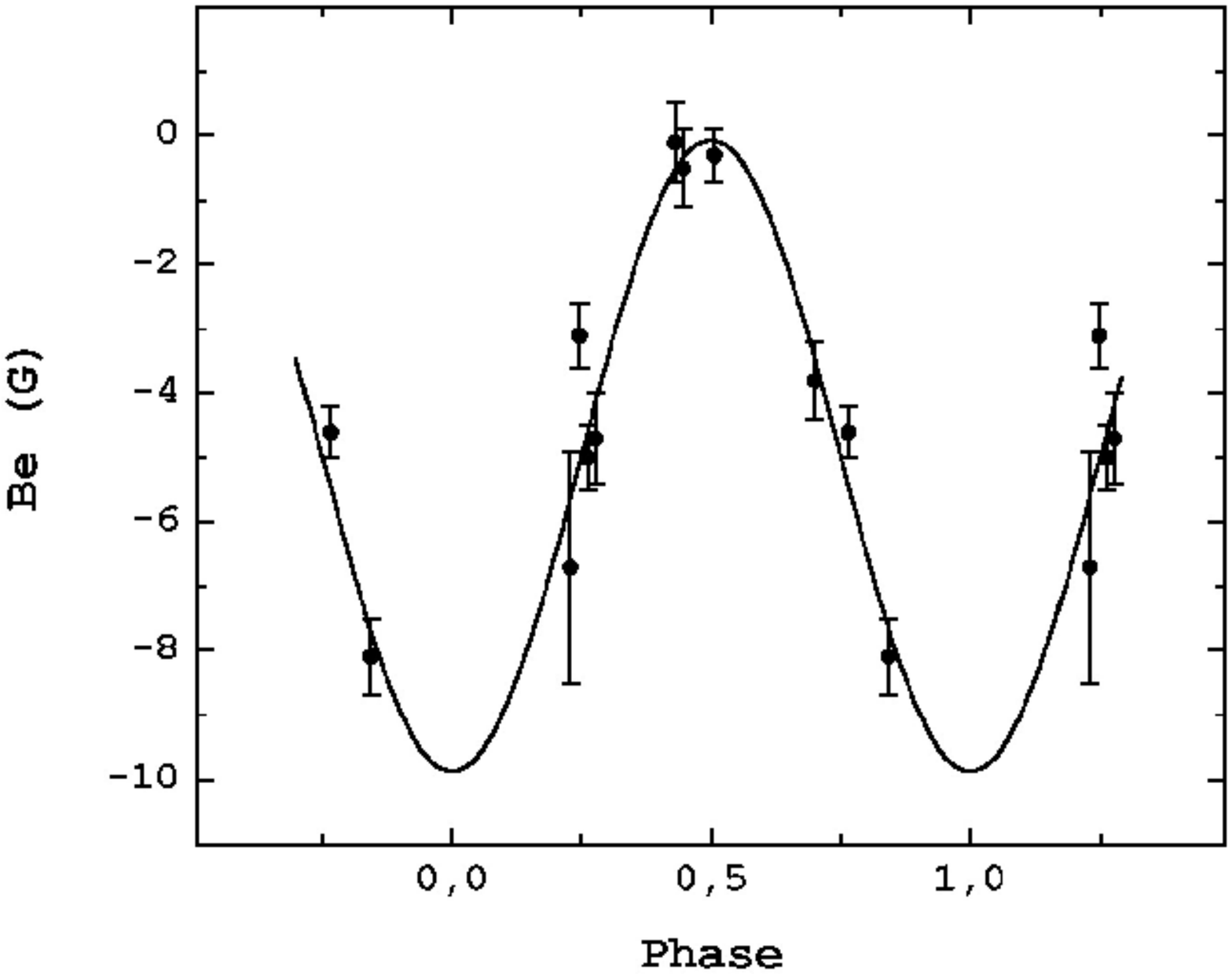}}
\vspace{-3.5mm}
\caption{ HD130144 }
\label{fig:fig236}
\end{figure}

\begin{figure}
\resizebox{0.98\hsize}{!}{\includegraphics{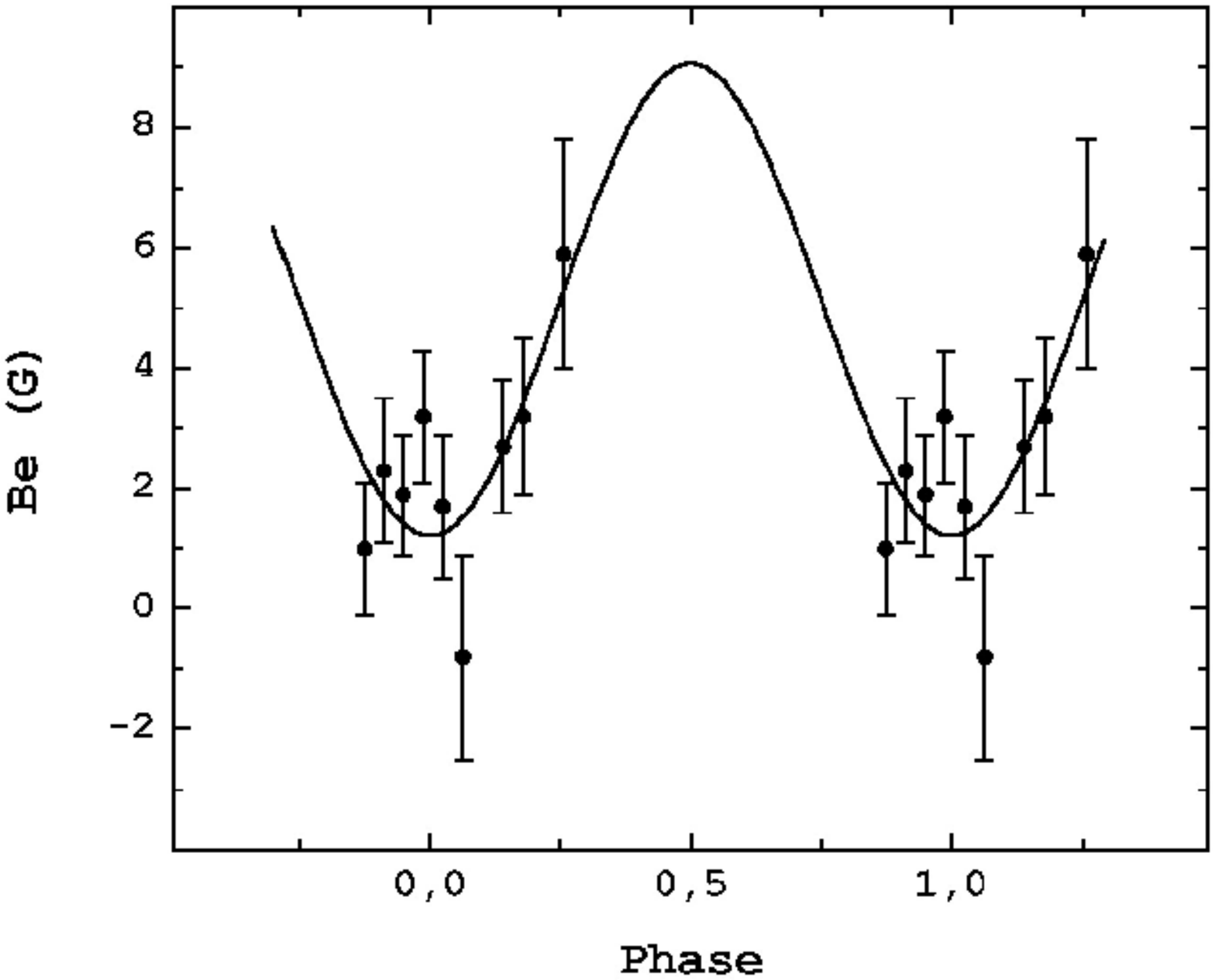}}
\vspace{-3.5mm}
\caption{ HD130322 }
\label{fig:fig237}
\end{figure}

\begin{figure}
\resizebox{0.98\hsize}{!}{\includegraphics{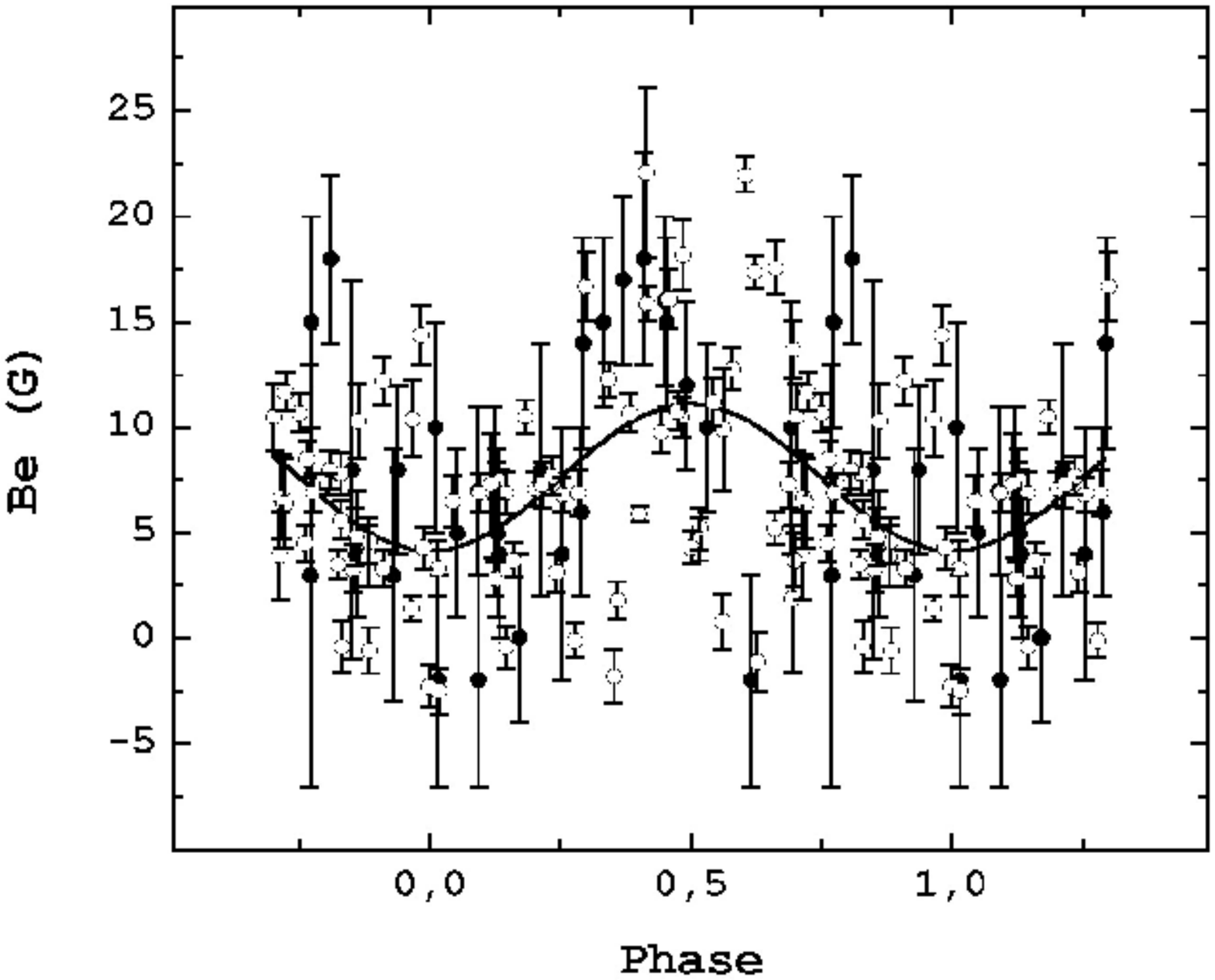}}
\vspace{-3.5mm}
\caption{ HD131156 }
\label{fig:fig238}
\end{figure}

\clearpage
\newpage

\begin{figure}
\resizebox{0.98\hsize}{!}{\includegraphics{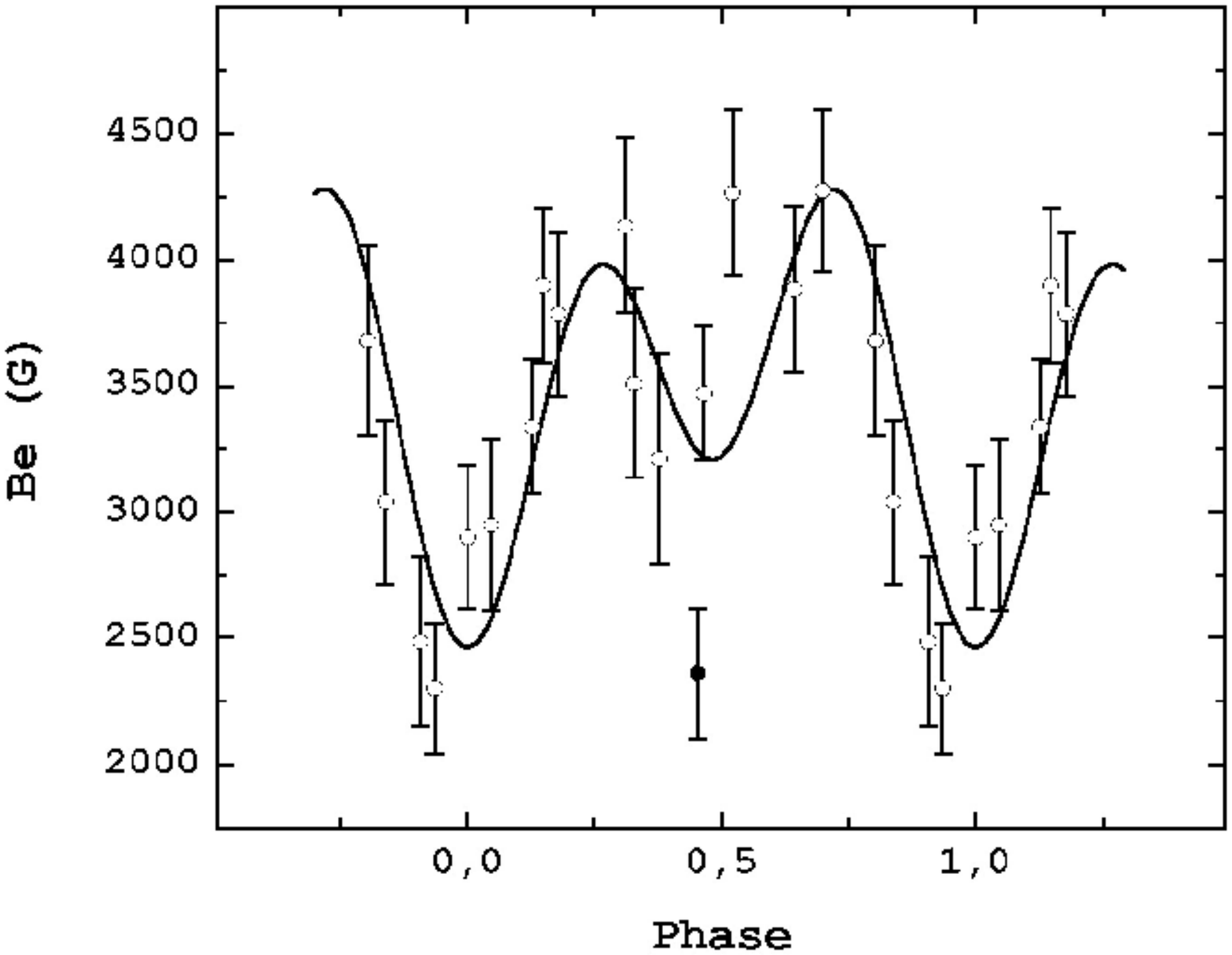}}
\vspace{-3.5mm}
\caption{ HD133029 (1) }
\label{fig:fig239}
\end{figure}

\begin{figure}
\resizebox{0.98\hsize}{!}{\includegraphics{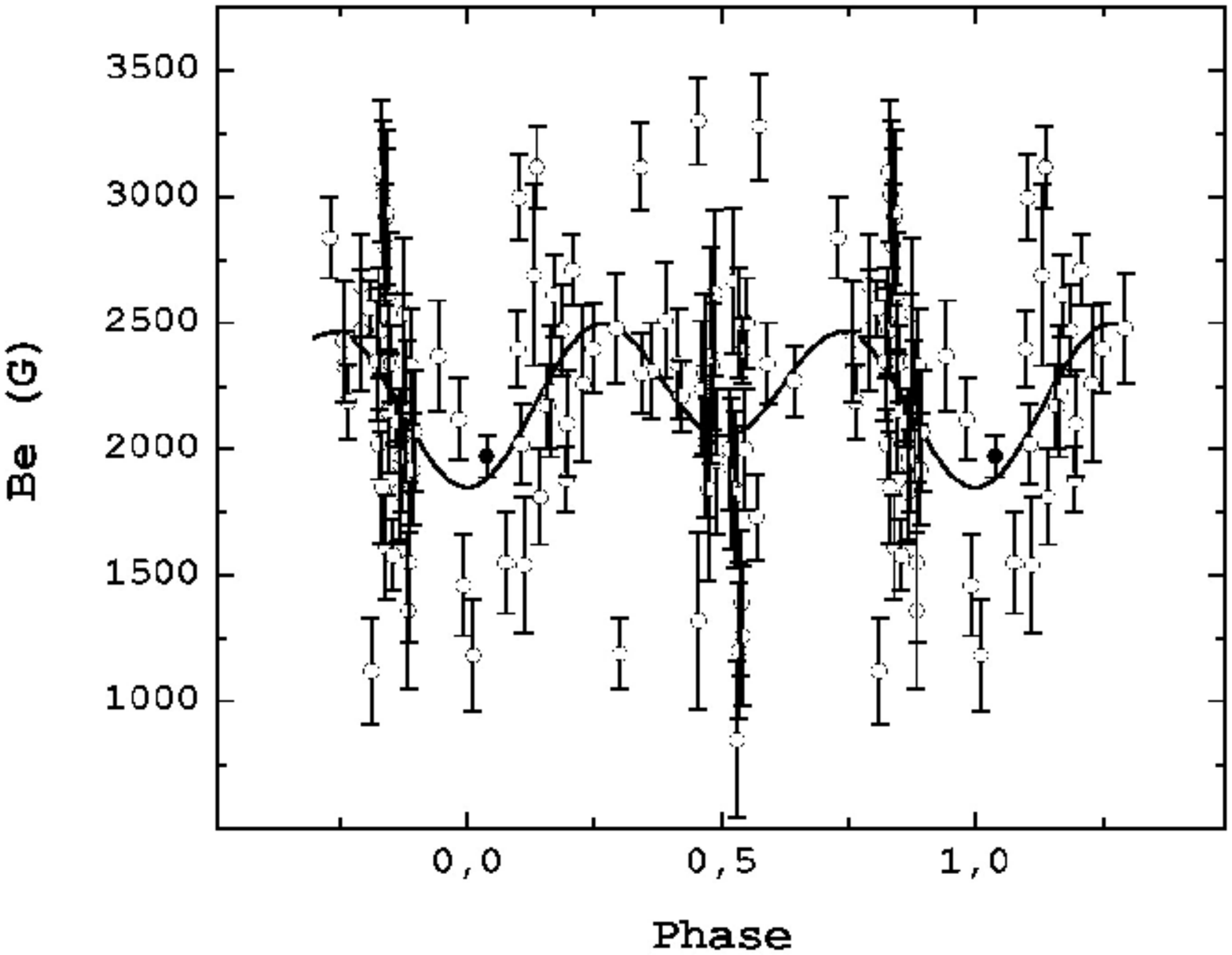}}
\vspace{-3.5mm}
\caption{ HD133029 (2) }
\label{fig:fig240}
\end{figure}

\begin{figure}
\resizebox{0.98\hsize}{!}{\includegraphics{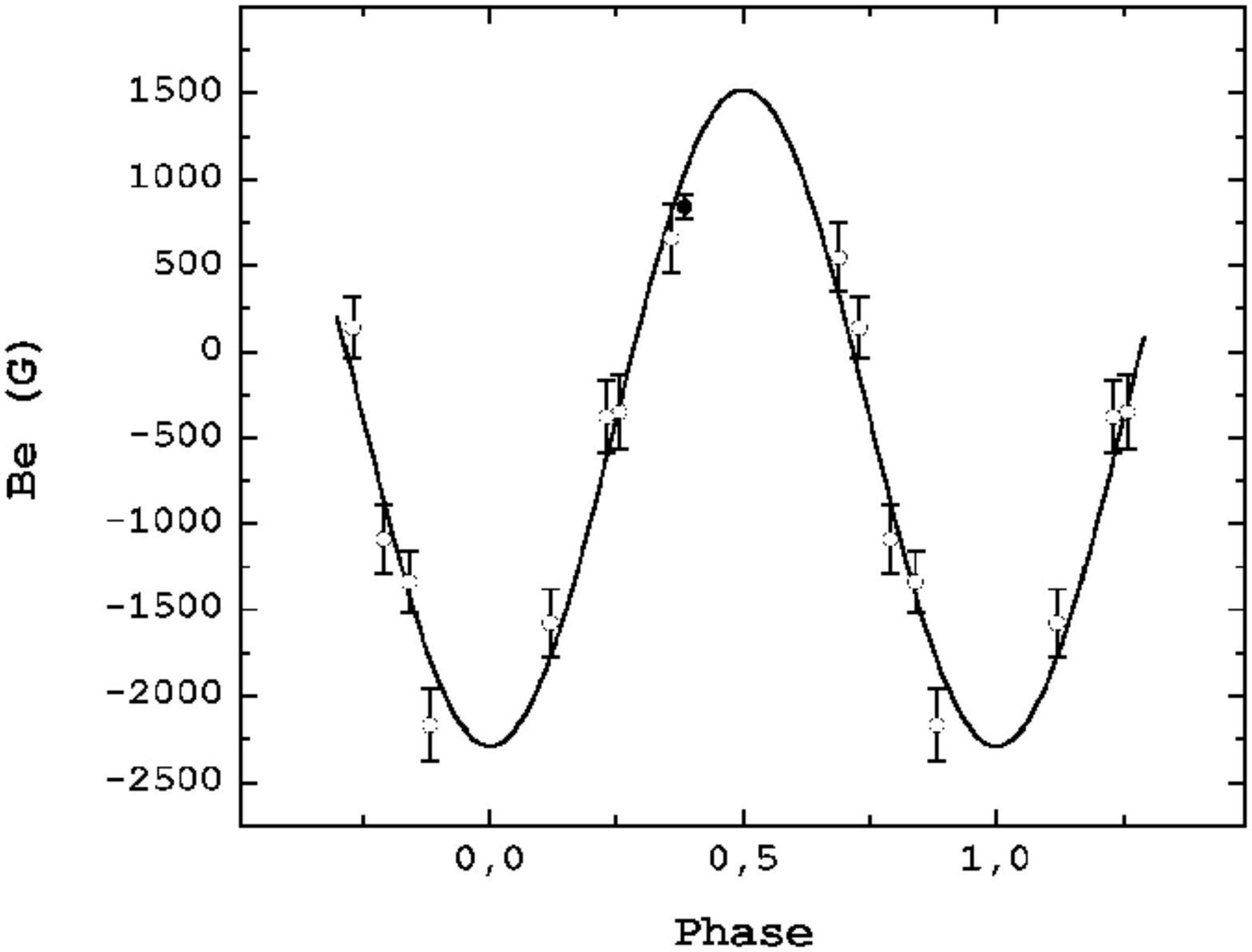}}
\vspace{-3.5mm}
\caption{ HD133652 }
\label{fig:fig241}
\end{figure}

\begin{figure}
\resizebox{0.98\hsize}{!}{\includegraphics{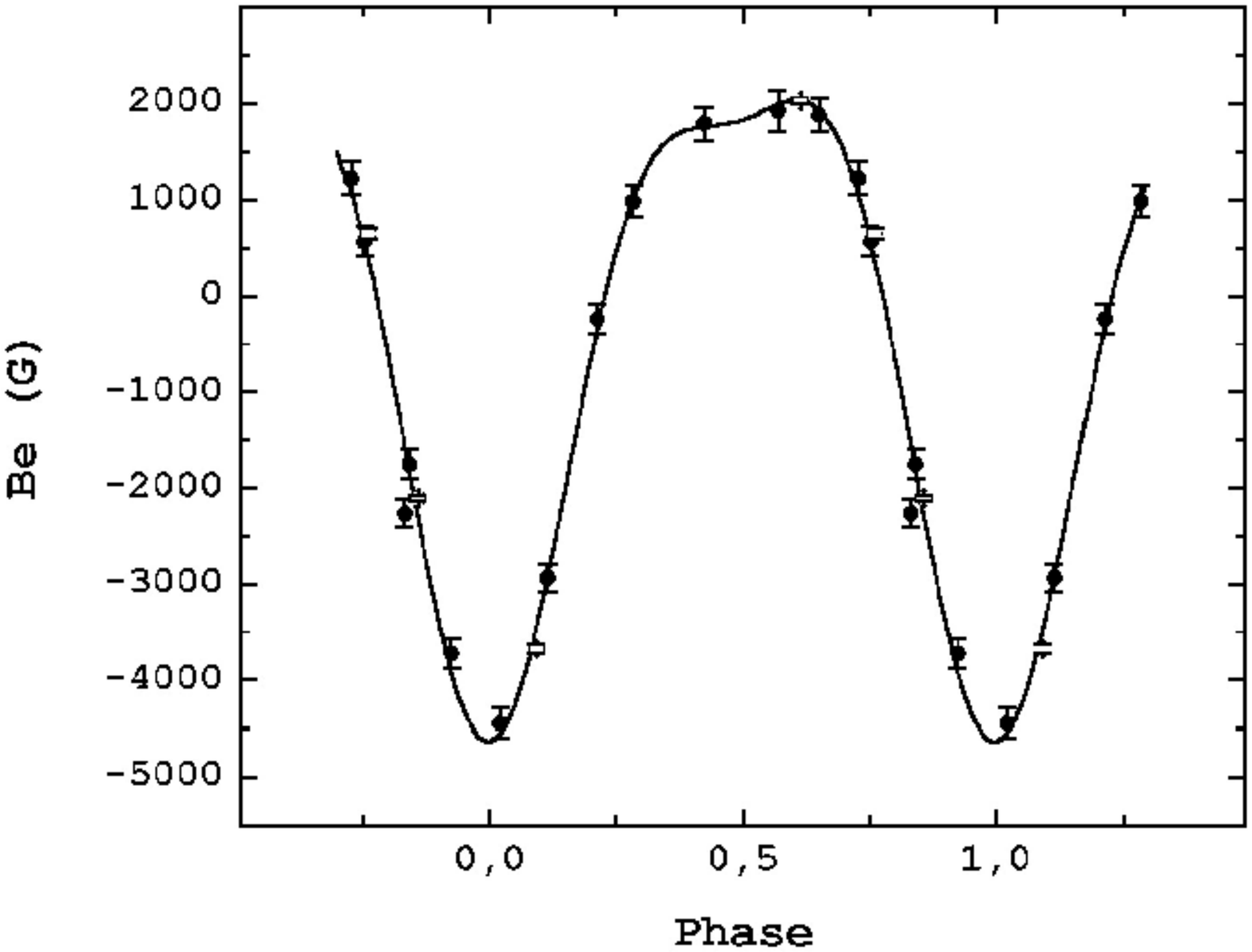}}
\vspace{-3.5mm}
\caption{ HD133880 }
\label{fig:fig242}
\end{figure}

\begin{figure}
\resizebox{0.98\hsize}{!}{\includegraphics{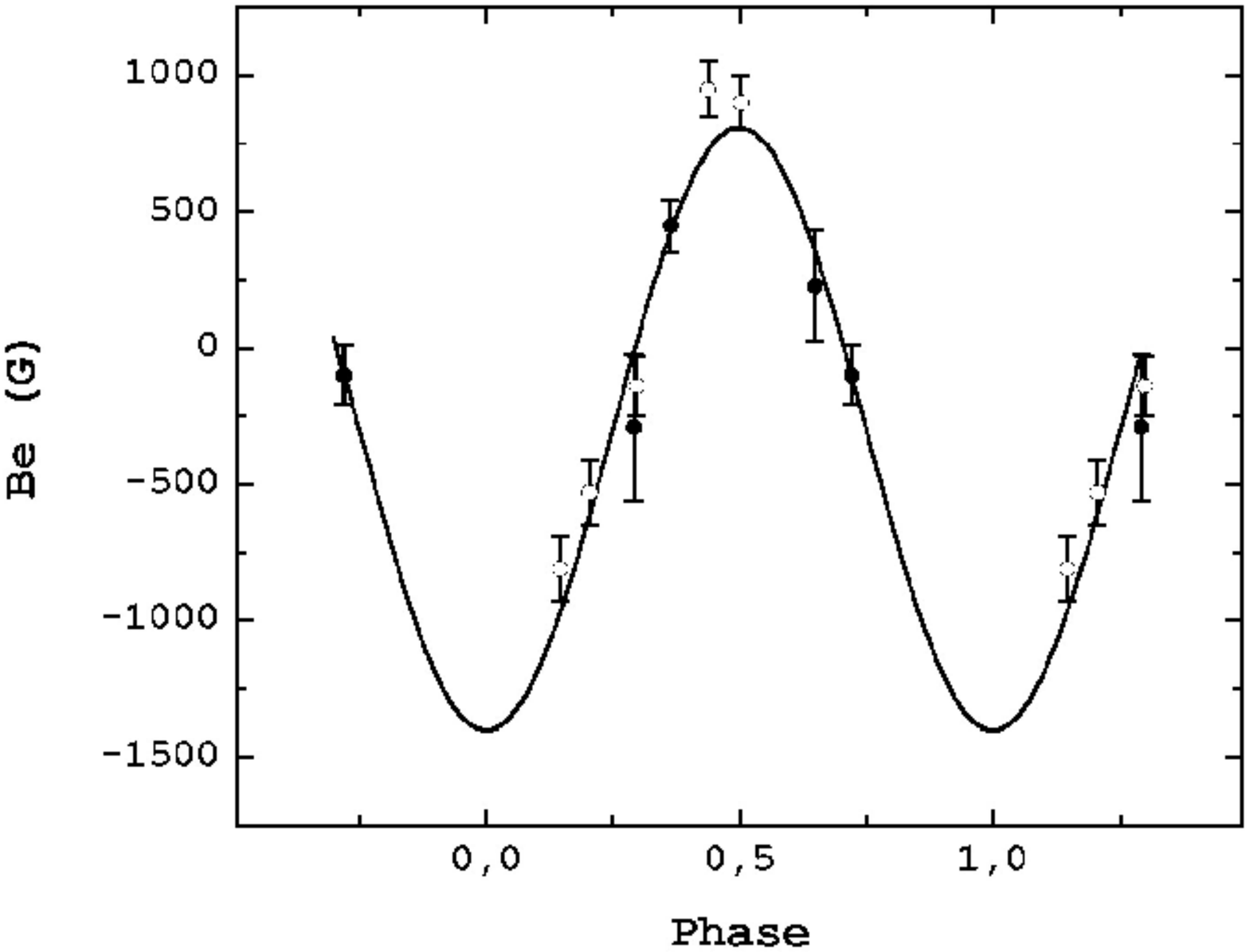}}
\vspace{-3.5mm}
\caption{ HD134793 }
\label{fig:fig243}
\end{figure}

\begin{figure}
\resizebox{0.98\hsize}{!}{\includegraphics{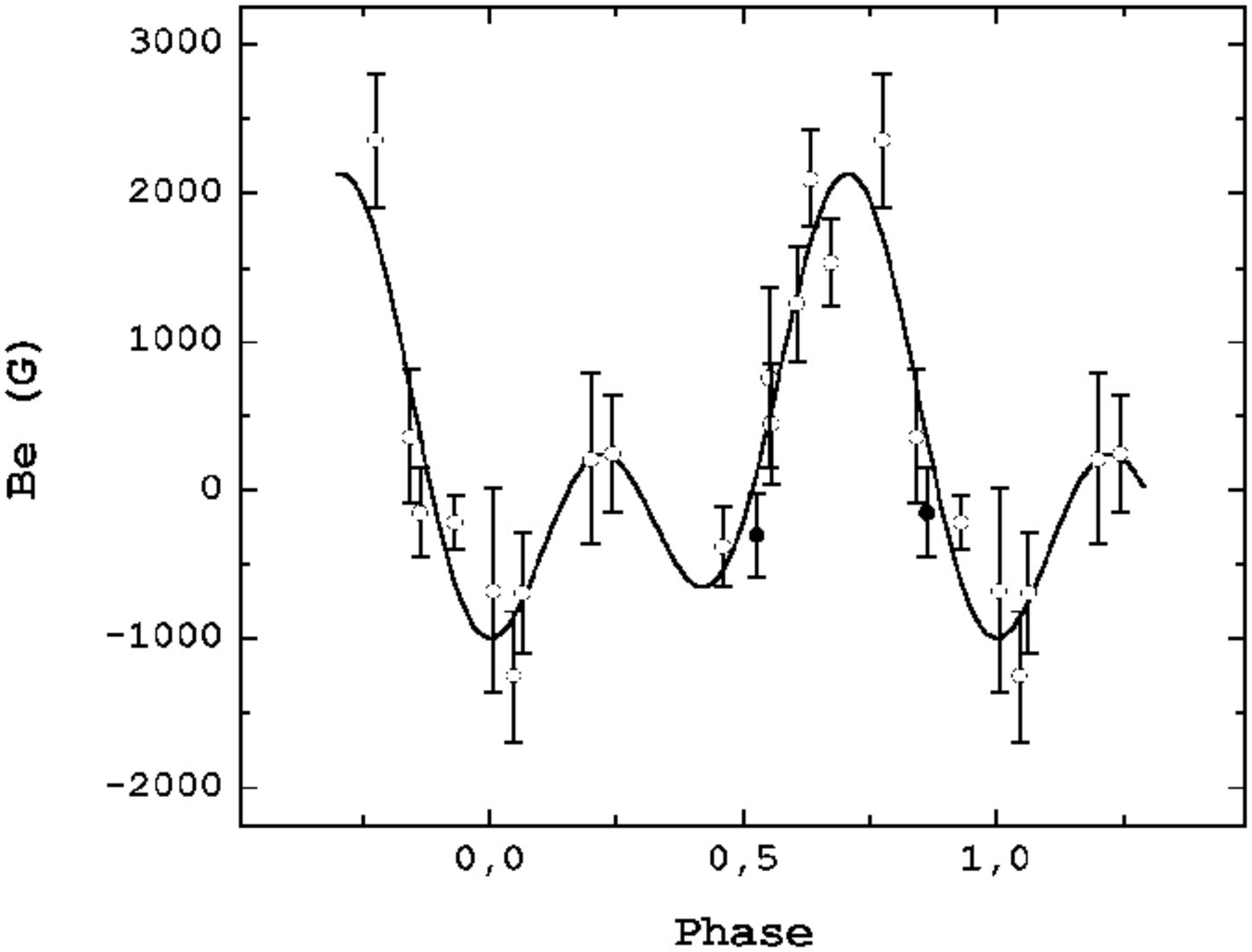}}
\vspace{-3.5mm}
\caption{ HD137509 }
\label{fig:fig244}
\end{figure}

\clearpage
\newpage

\begin{figure}
\resizebox{0.98\hsize}{!}{\includegraphics{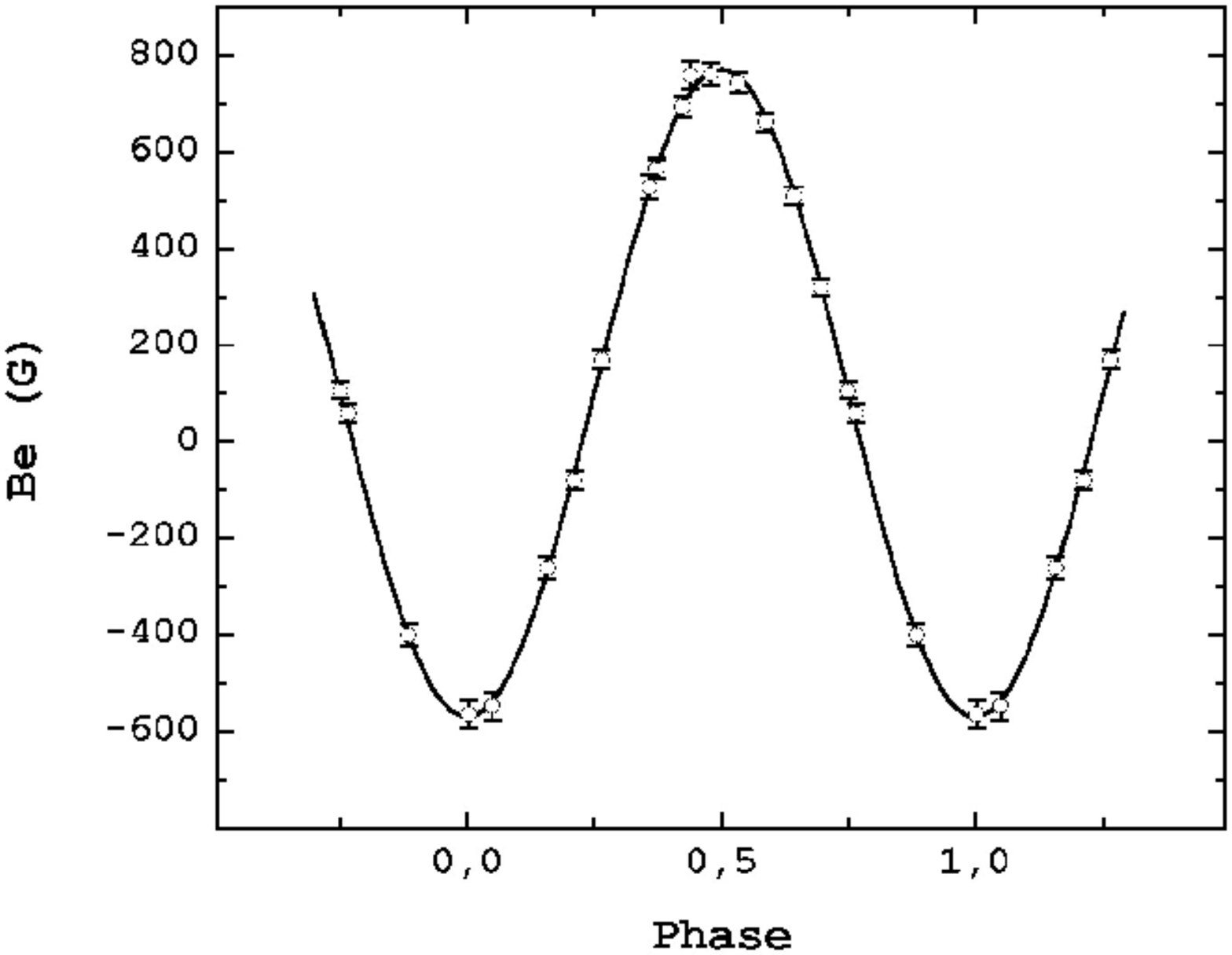}}
\vspace{-3.5mm}
\caption{ HD137909 (1) }
\label{fig:fig245}
\end{figure}

\begin{figure}
\resizebox{0.98\hsize}{!}{\includegraphics{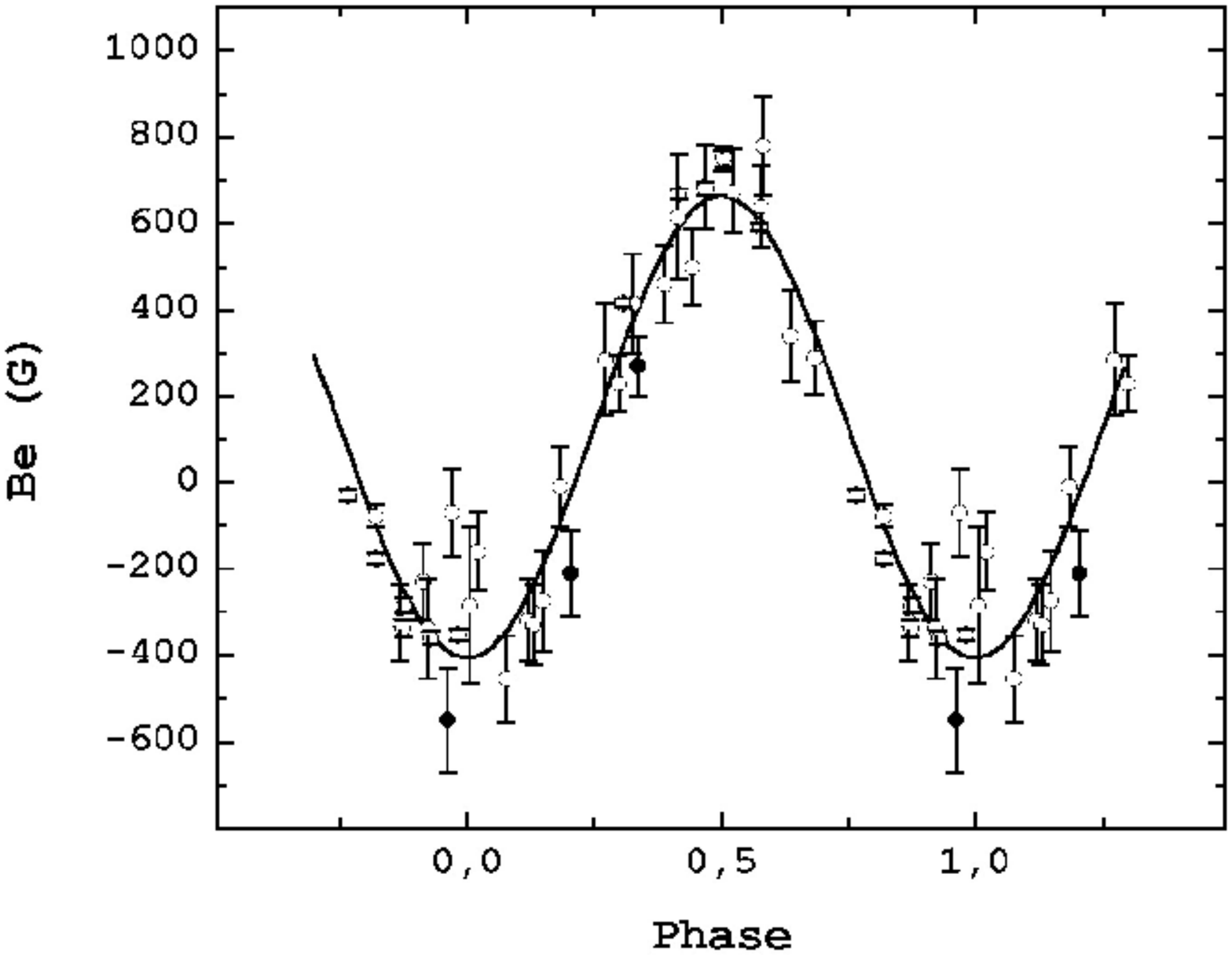}}
\vspace{-3.5mm}
\caption{ HD137909 (2) }
\label{fig:fig246}
\end{figure}

\begin{figure}
\resizebox{0.98\hsize}{!}{\includegraphics{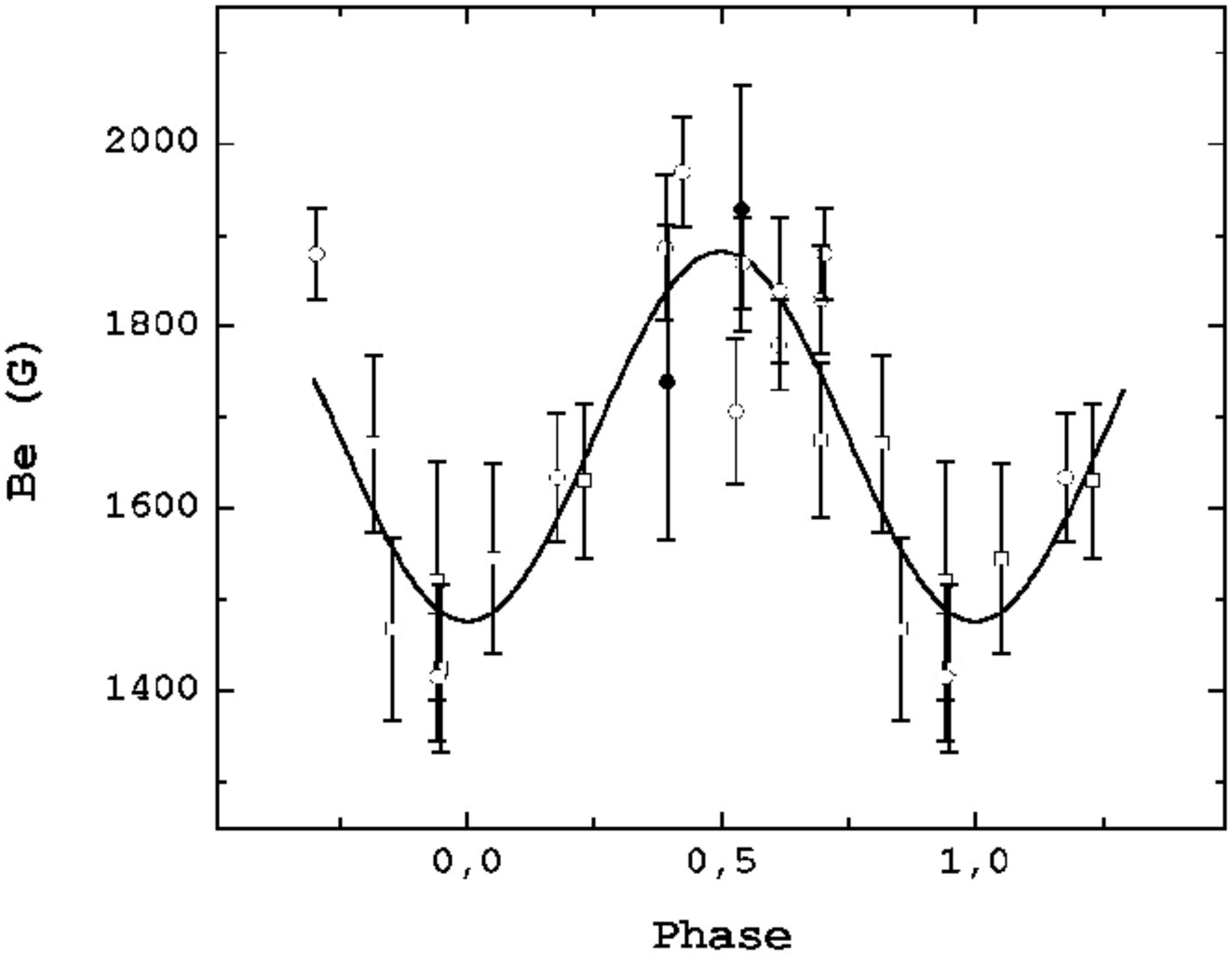}}
\vspace{-3.5mm}
\caption{ HD137949 }
\label{fig:fig247}
\end{figure}

\begin{figure}
\resizebox{0.98\hsize}{!}{\includegraphics{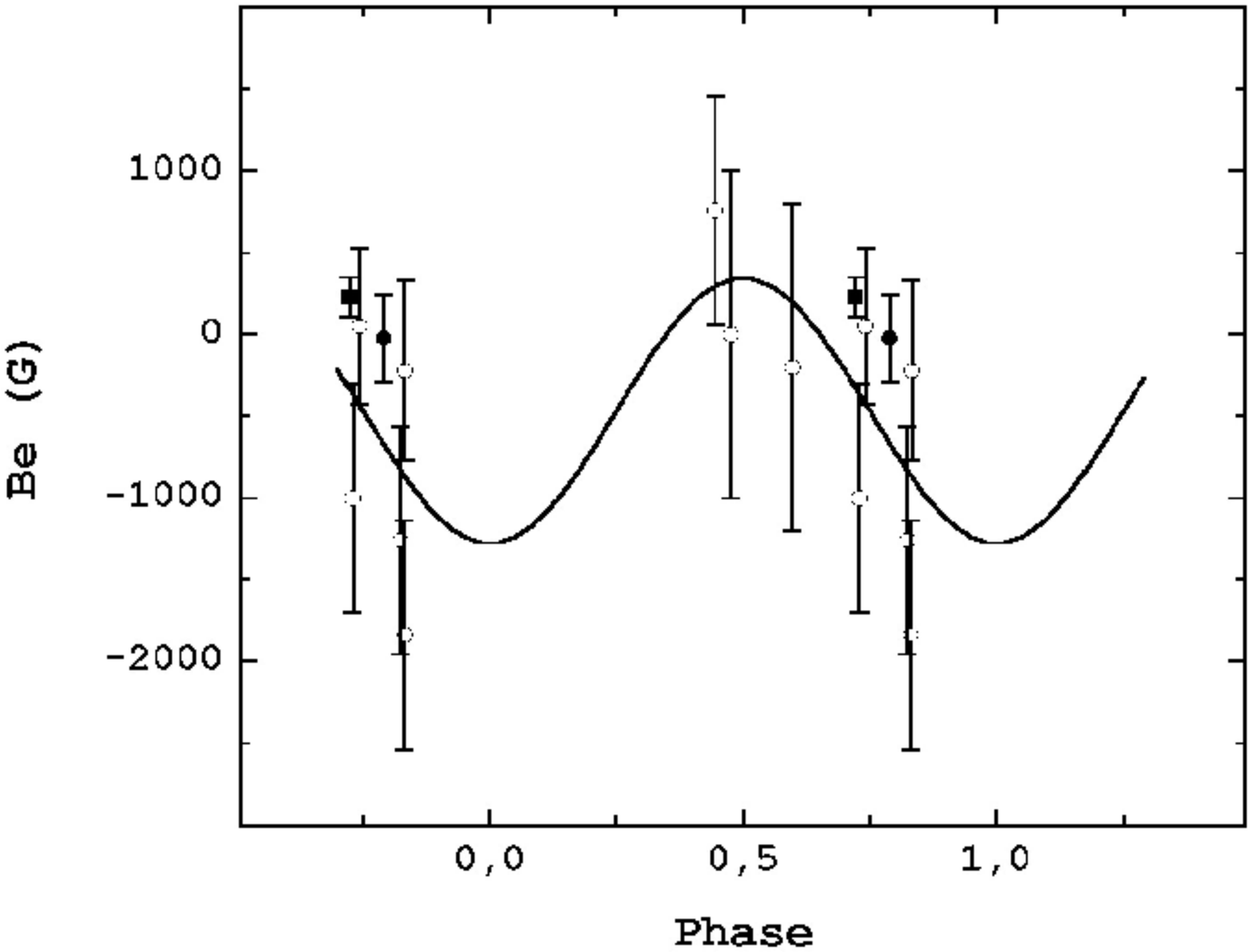}}
\vspace{-3.5mm}
\caption{ HD140160 }
\label{fig:fig248}
\end{figure}

\begin{figure}
\resizebox{0.98\hsize}{!}{\includegraphics{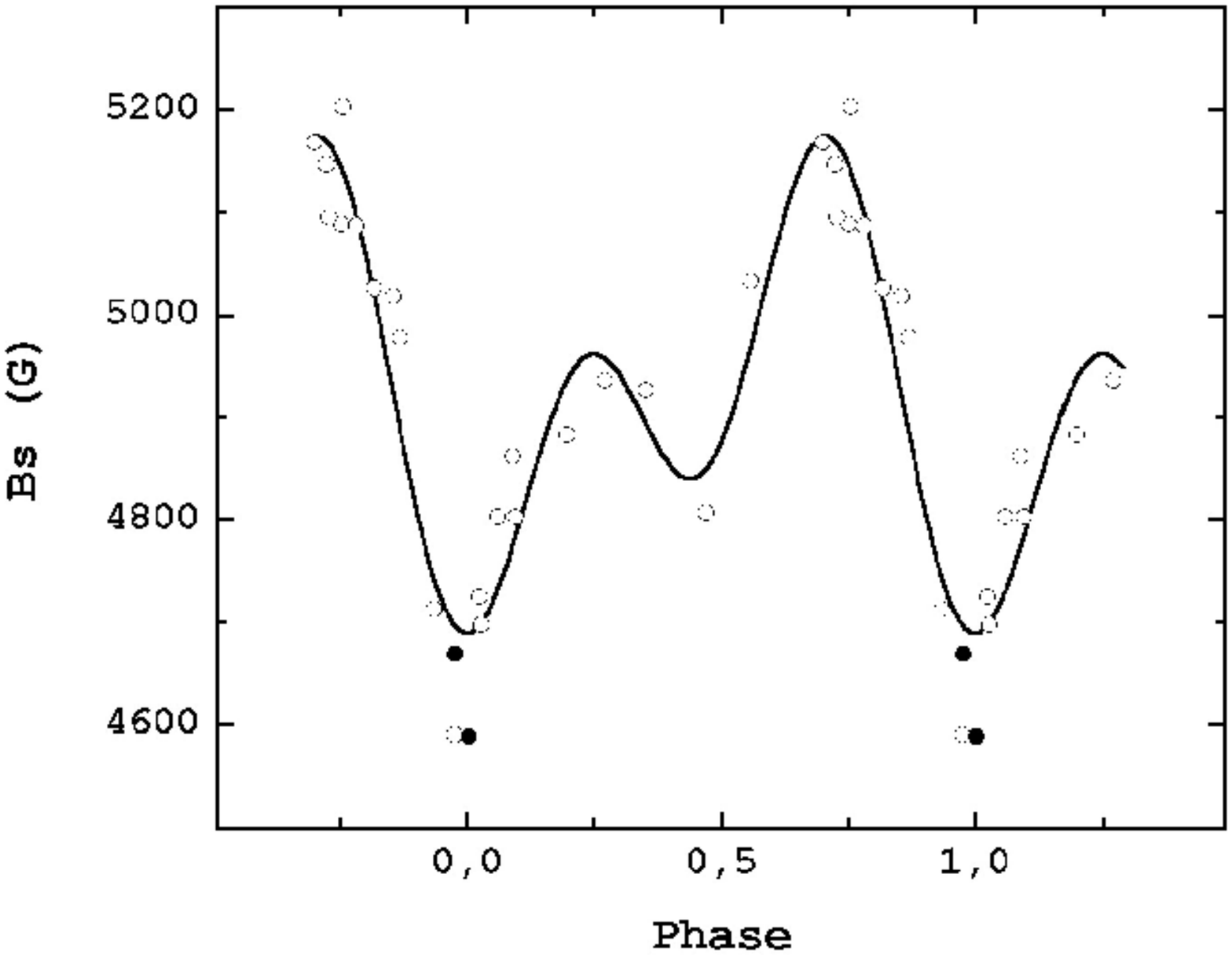}}
\vspace{-3.5mm}
\caption{ HD142070 (1) }
\label{fig:fig249}
\end{figure}

\begin{figure}
\resizebox{0.98\hsize}{!}{\includegraphics{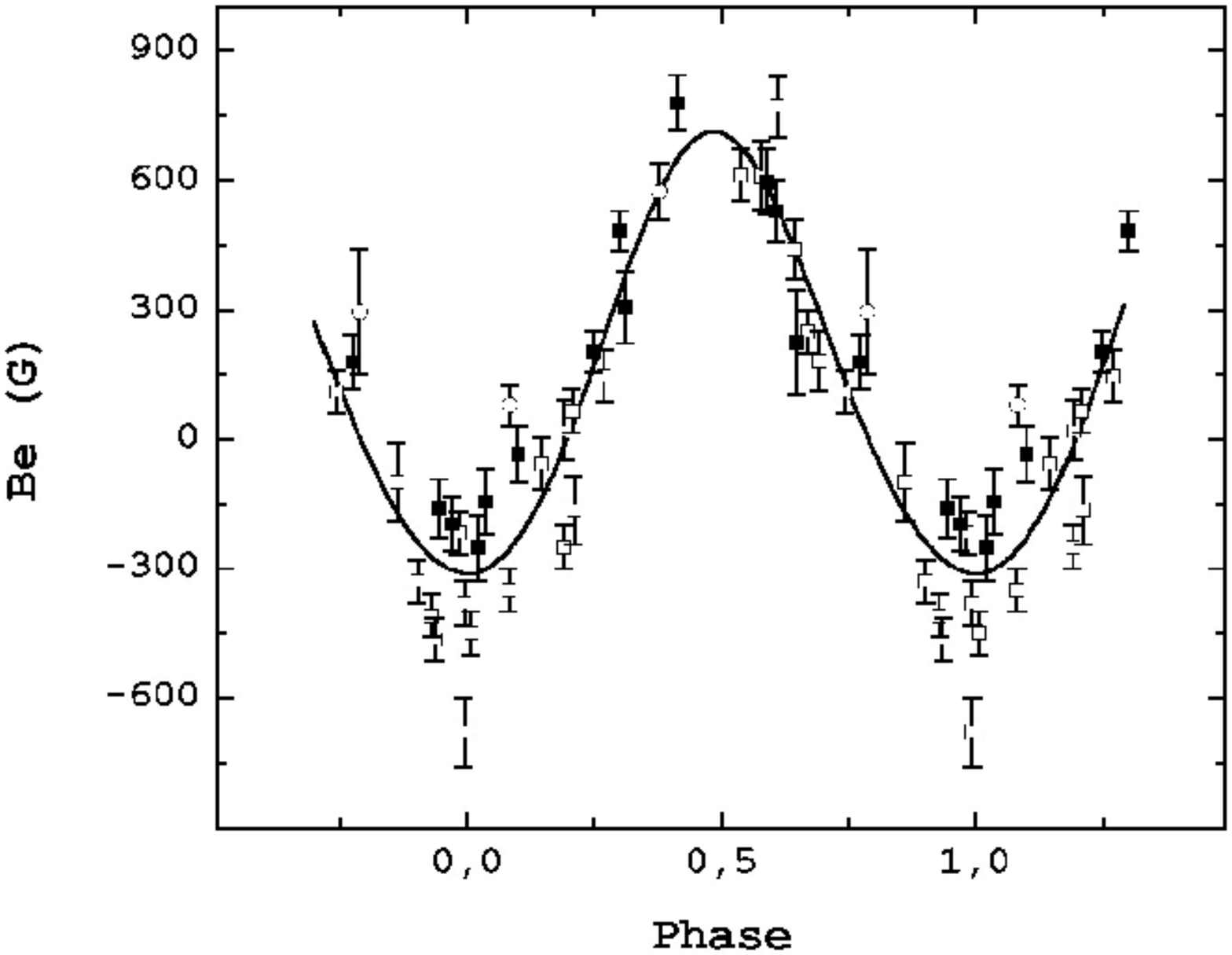}}
\vspace{-3.5mm}
\caption{ HD142070 (2) }
\label{fig:fig250}
\end{figure}

\clearpage
\newpage

\begin{figure}
\resizebox{0.98\hsize}{!}{\includegraphics{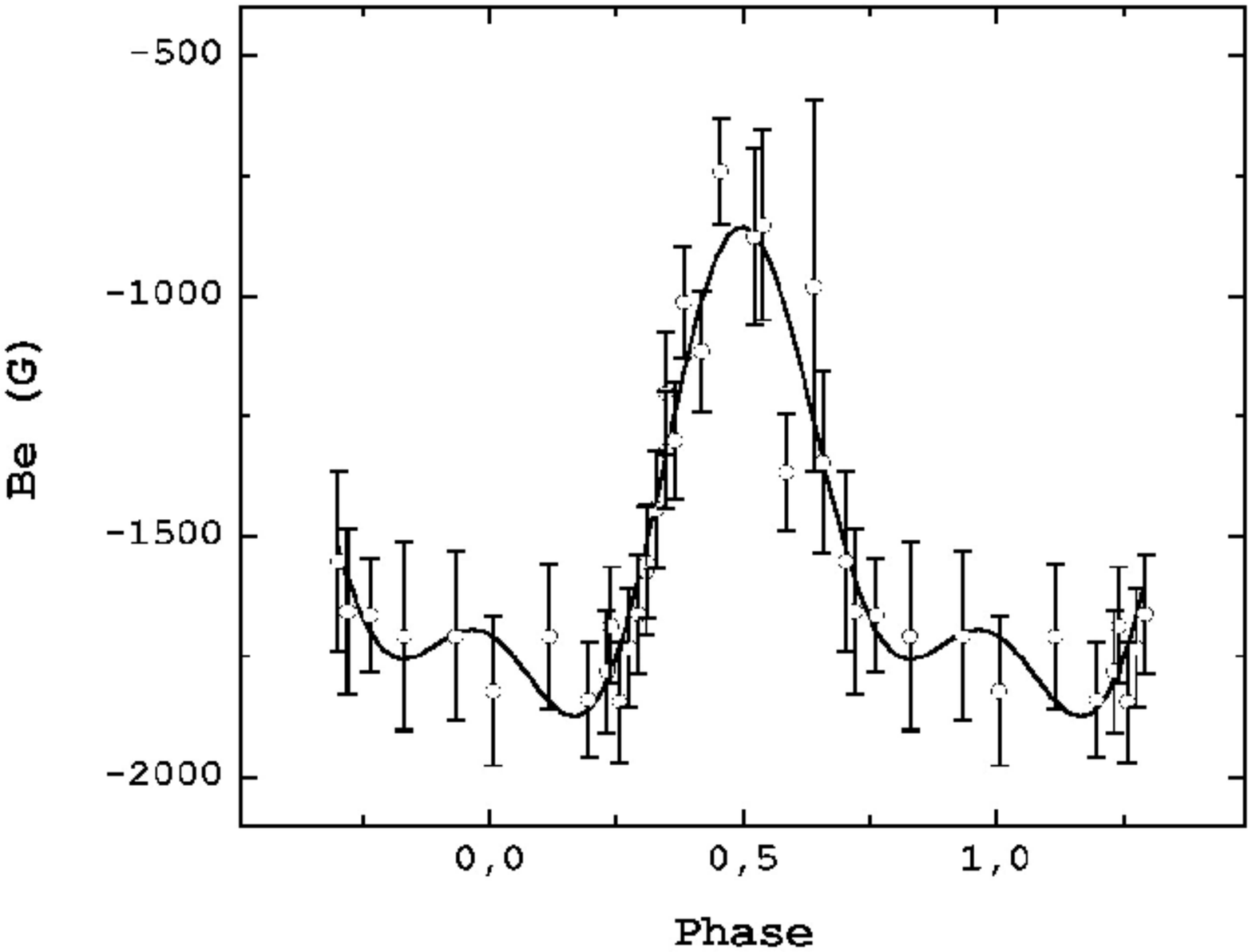}}
\vspace{-3.5mm}
\caption{ HD142184 }
\label{fig:fig251}
\end{figure}

\begin{figure}
\resizebox{0.98\hsize}{!}{\includegraphics{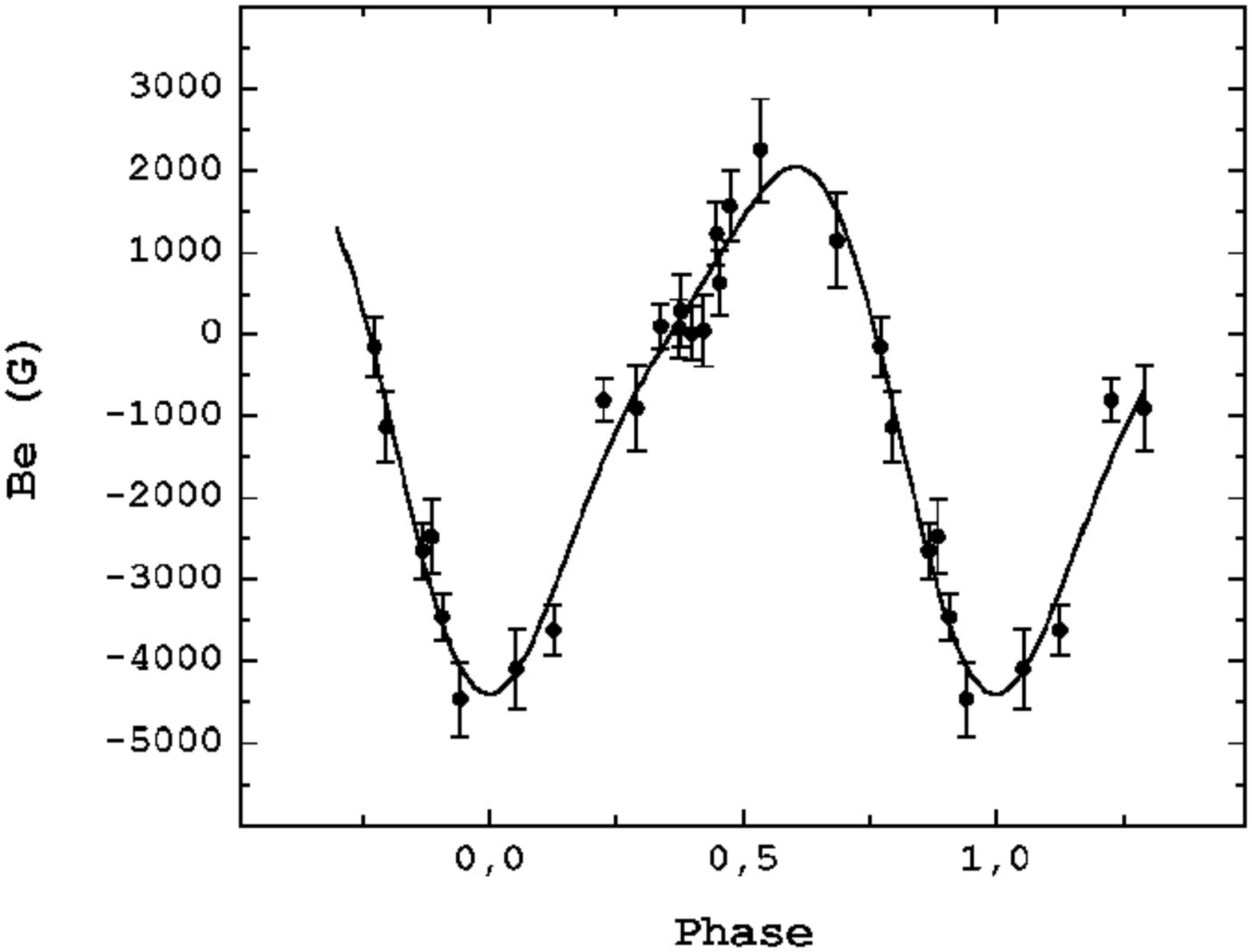}}
\vspace{-3.5mm}
\caption{ HD142301 }
\label{fig:fig252}
\end{figure}

\begin{figure}
\resizebox{0.98\hsize}{!}{\includegraphics{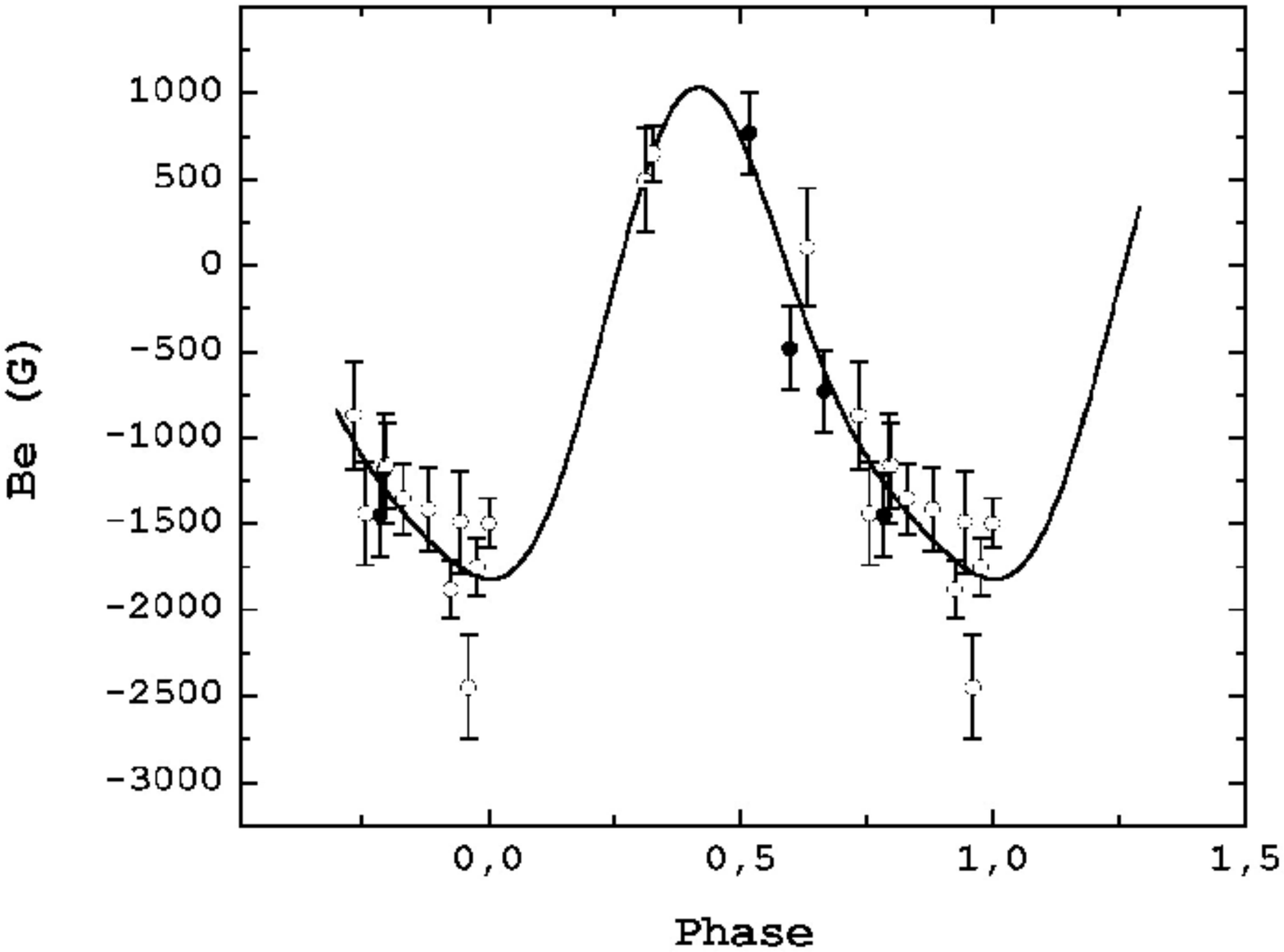}}
\vspace{-3.5mm}
\caption{ HD142990 }
\label{fig:fig253}
\end{figure}

\begin{figure}
\resizebox{0.98\hsize}{!}{\includegraphics{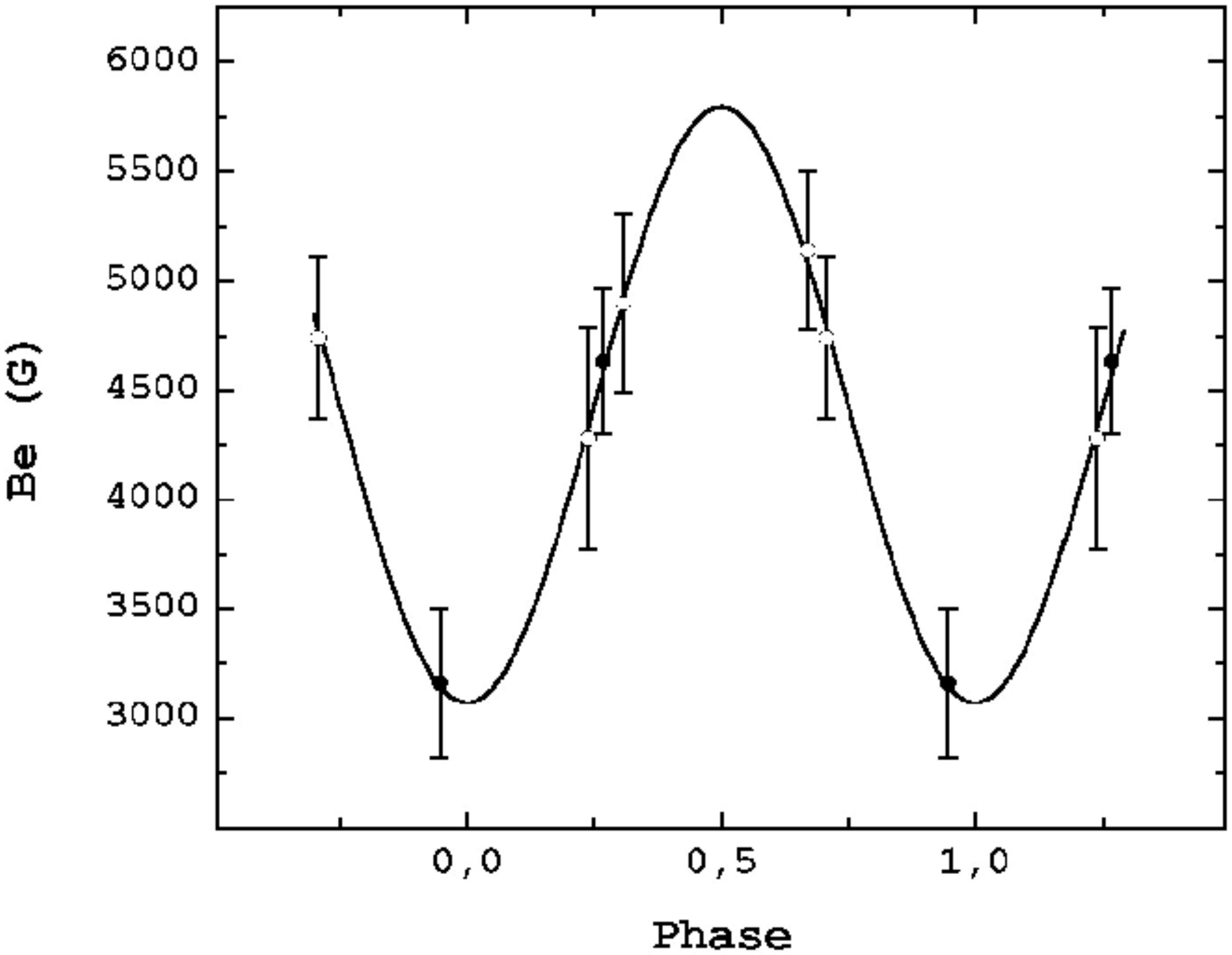}}
\vspace{-3.5mm}
\caption{ HD143473 }
\label{fig:fig254}
\end{figure}

\begin{figure}
\resizebox{0.98\hsize}{!}{\includegraphics{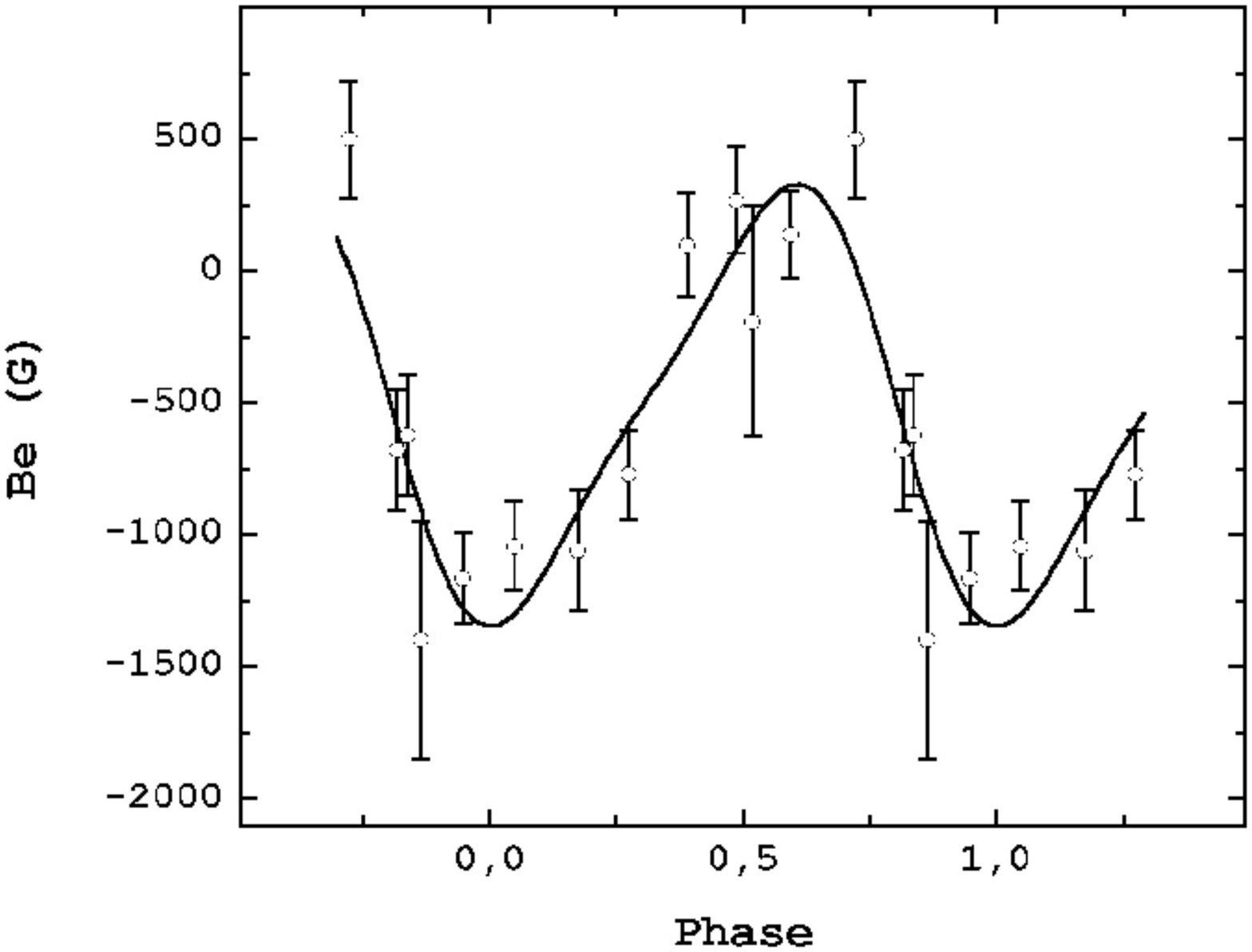}}
\vspace{-3.5mm}
\caption{ HD144334 }
\label{fig:fig255}
\end{figure}

\begin{figure}
\resizebox{0.98\hsize}{!}{\includegraphics{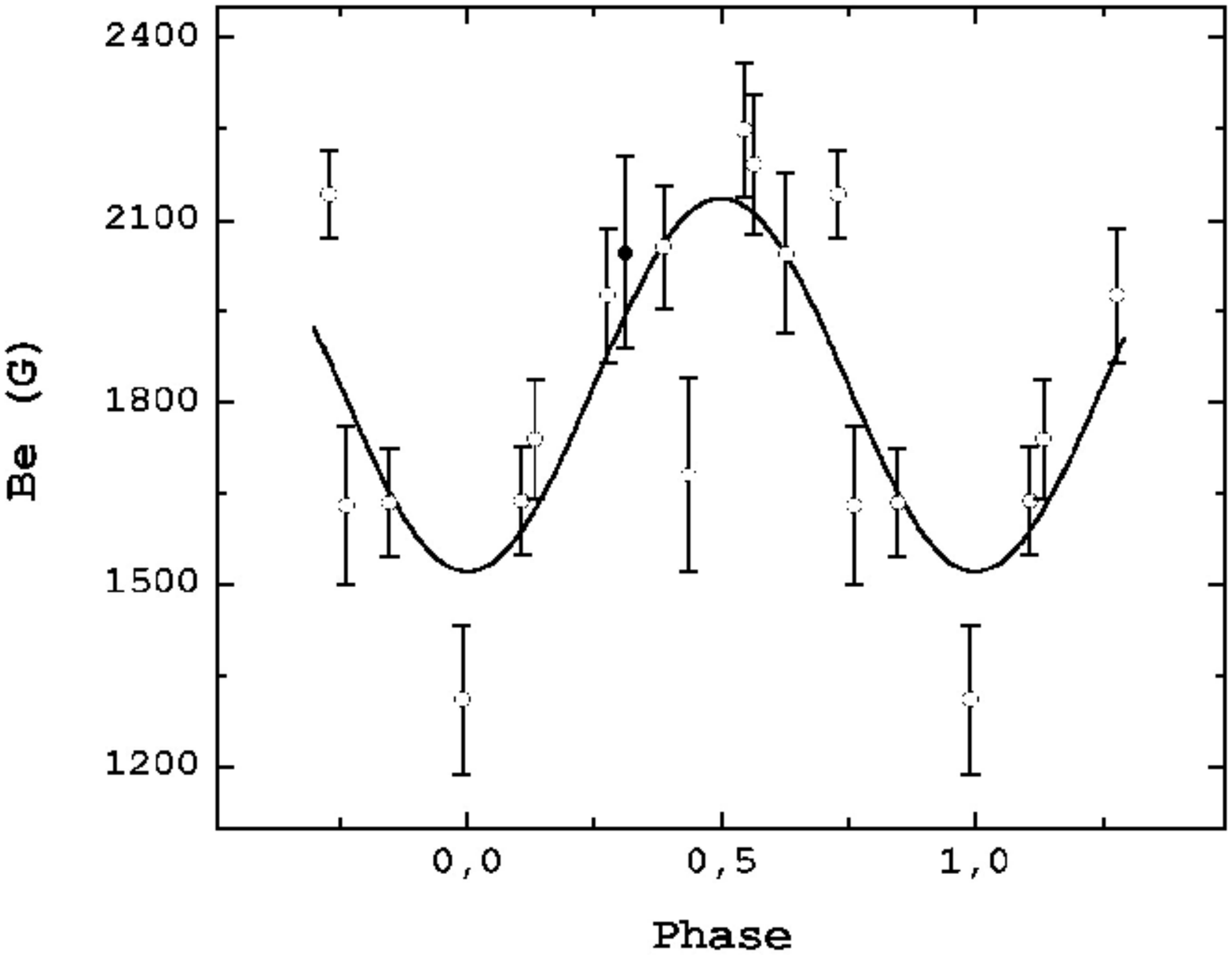}}
\vspace{-3.5mm}
\caption{ HD144897 (1) }
\label{fig:fig256}
\end{figure}

\clearpage
\newpage

\begin{figure}
\resizebox{0.98\hsize}{!}{\includegraphics{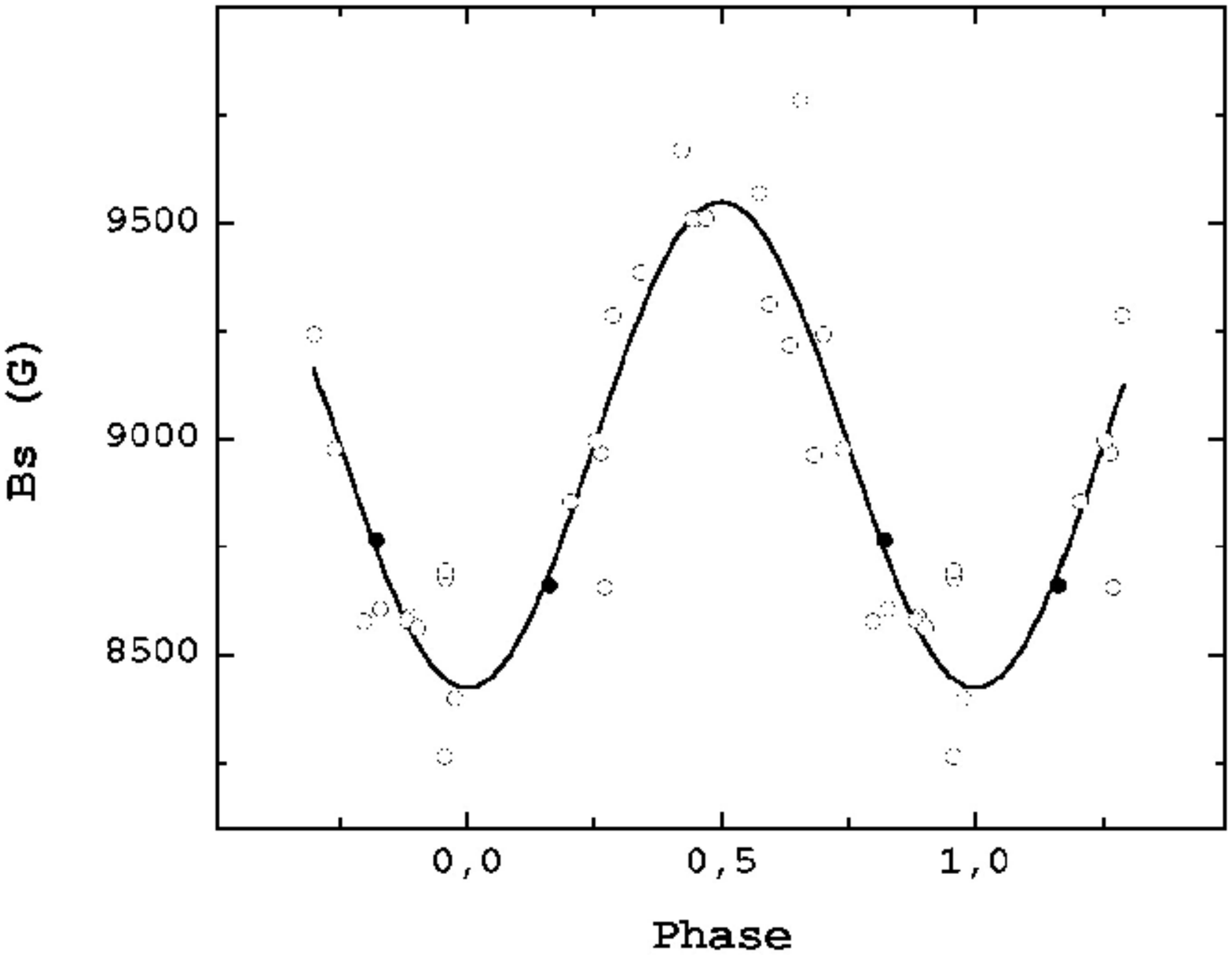}}
\vspace{-3.5mm}
\caption{ HD144897 (2) }
\label{fig:fig257}
\end{figure}

\begin{figure}
\resizebox{0.98\hsize}{!}{\includegraphics{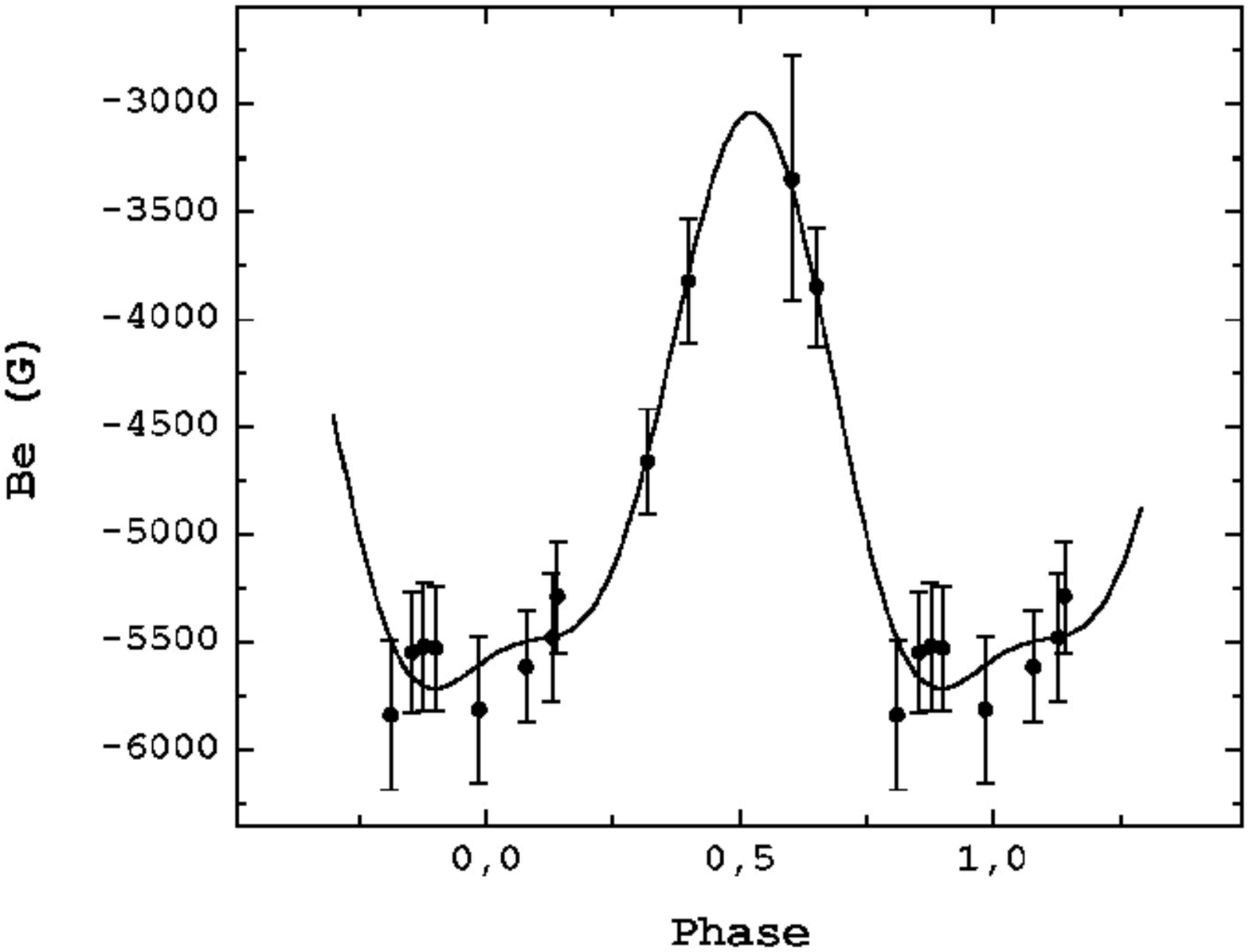}}
\vspace{-3.5mm}
\caption{ HD147010 (1) }
\label{fig:fig258}
\end{figure}

\begin{figure}
\resizebox{0.98\hsize}{!}{\includegraphics{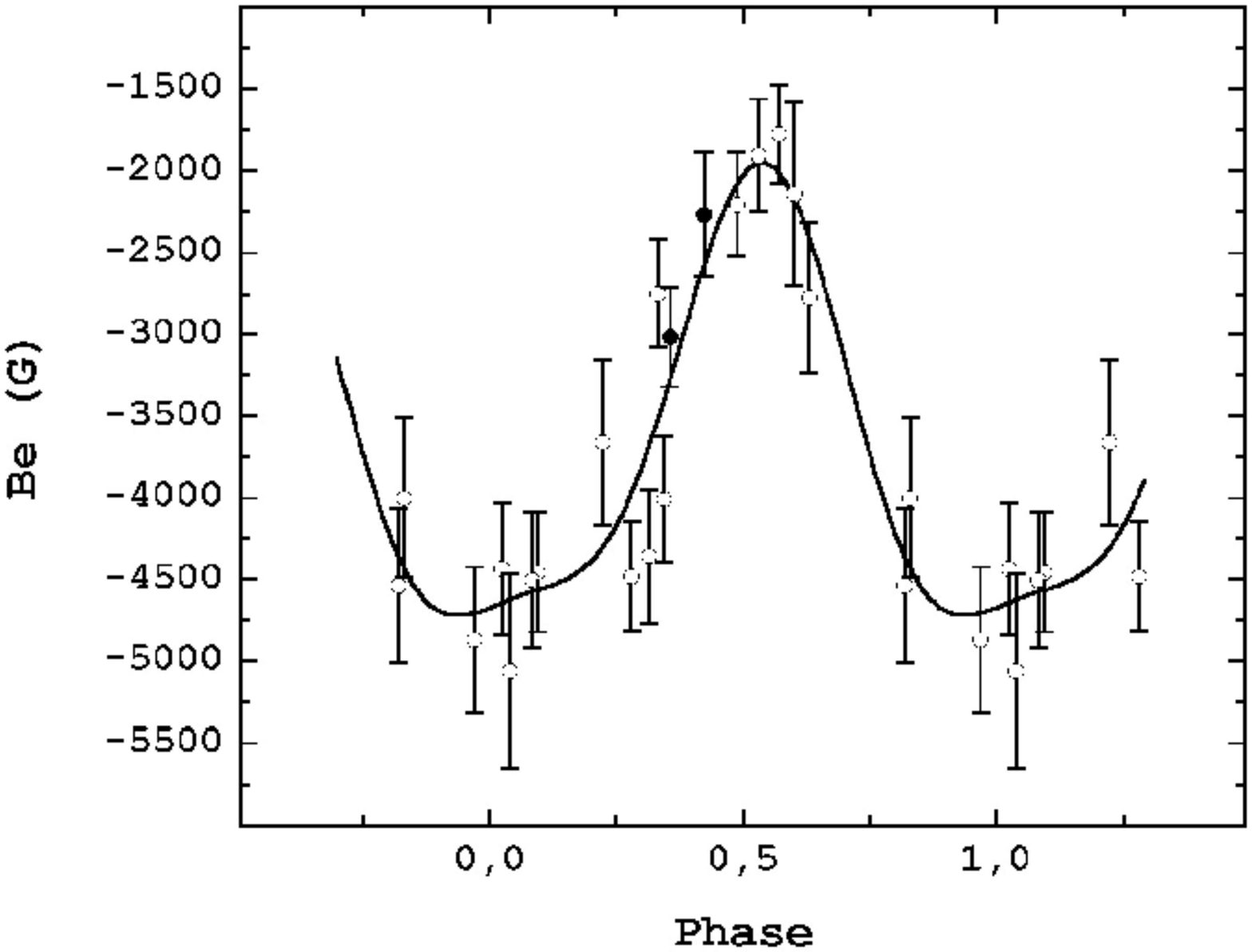}}
\vspace{-3.5mm}
\caption{ HD147010 (2) }
\label{fig:fig259}
\end{figure}

\begin{figure}
\resizebox{0.98\hsize}{!}{\includegraphics{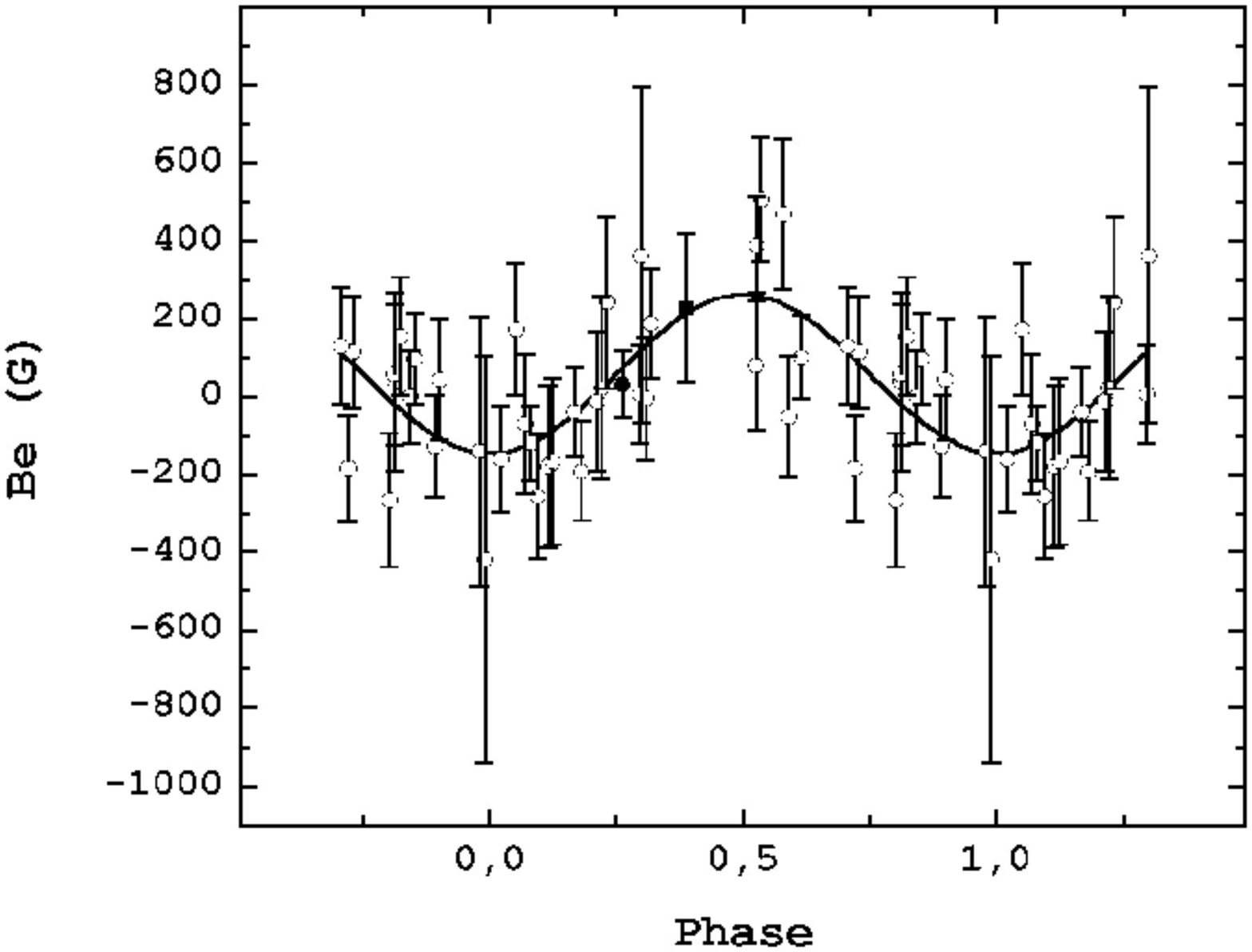}}
\vspace{-3.5mm}
\caption{ HD147394 }
\label{fig:fig260}
\end{figure}

\begin{figure}
\resizebox{0.98\hsize}{!}{\includegraphics{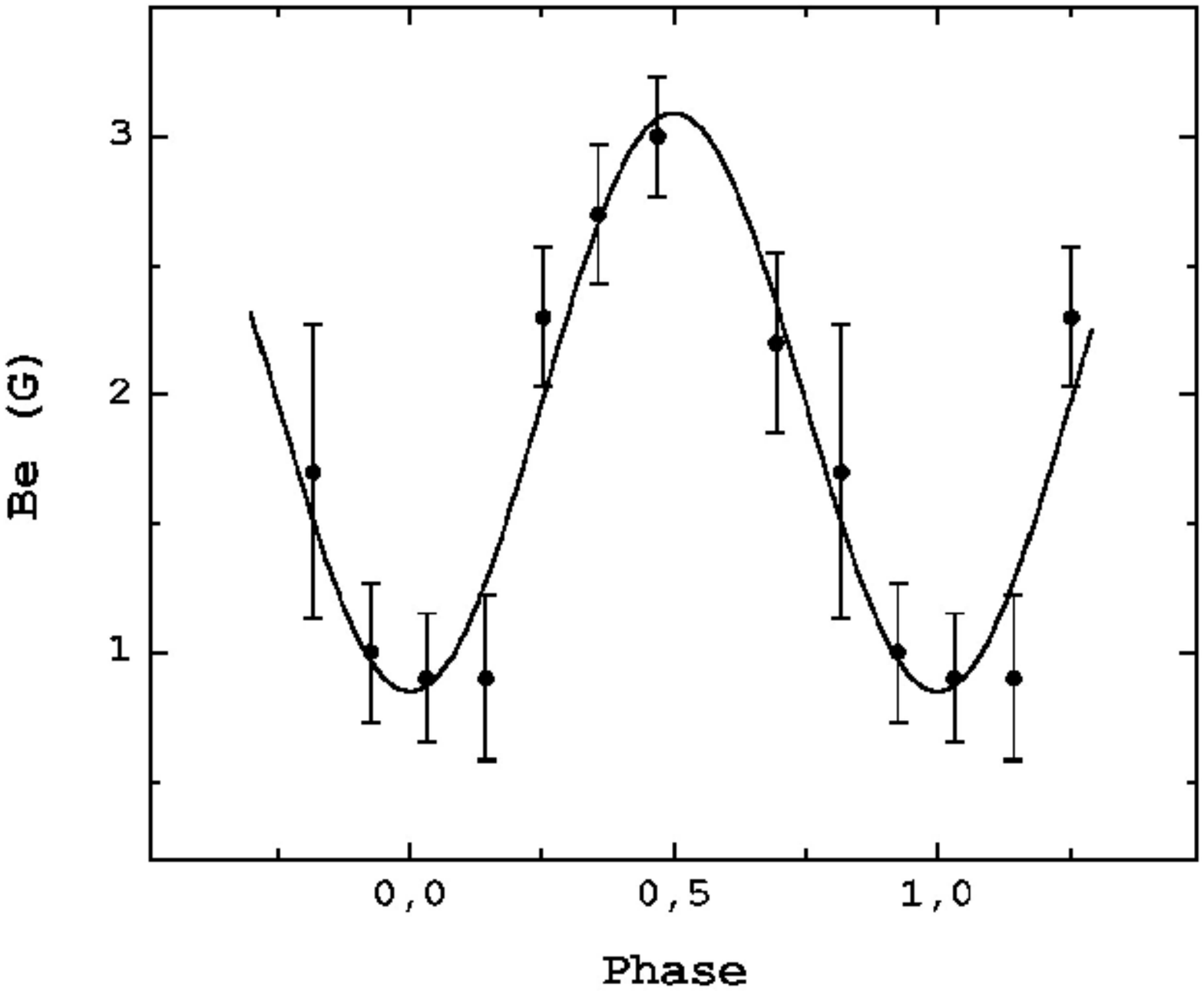}}
\vspace{-3.5mm}
\caption{ HD147513 }
\label{fig:fig261}
\end{figure}

\begin{figure}
\resizebox{0.98\hsize}{!}{\includegraphics{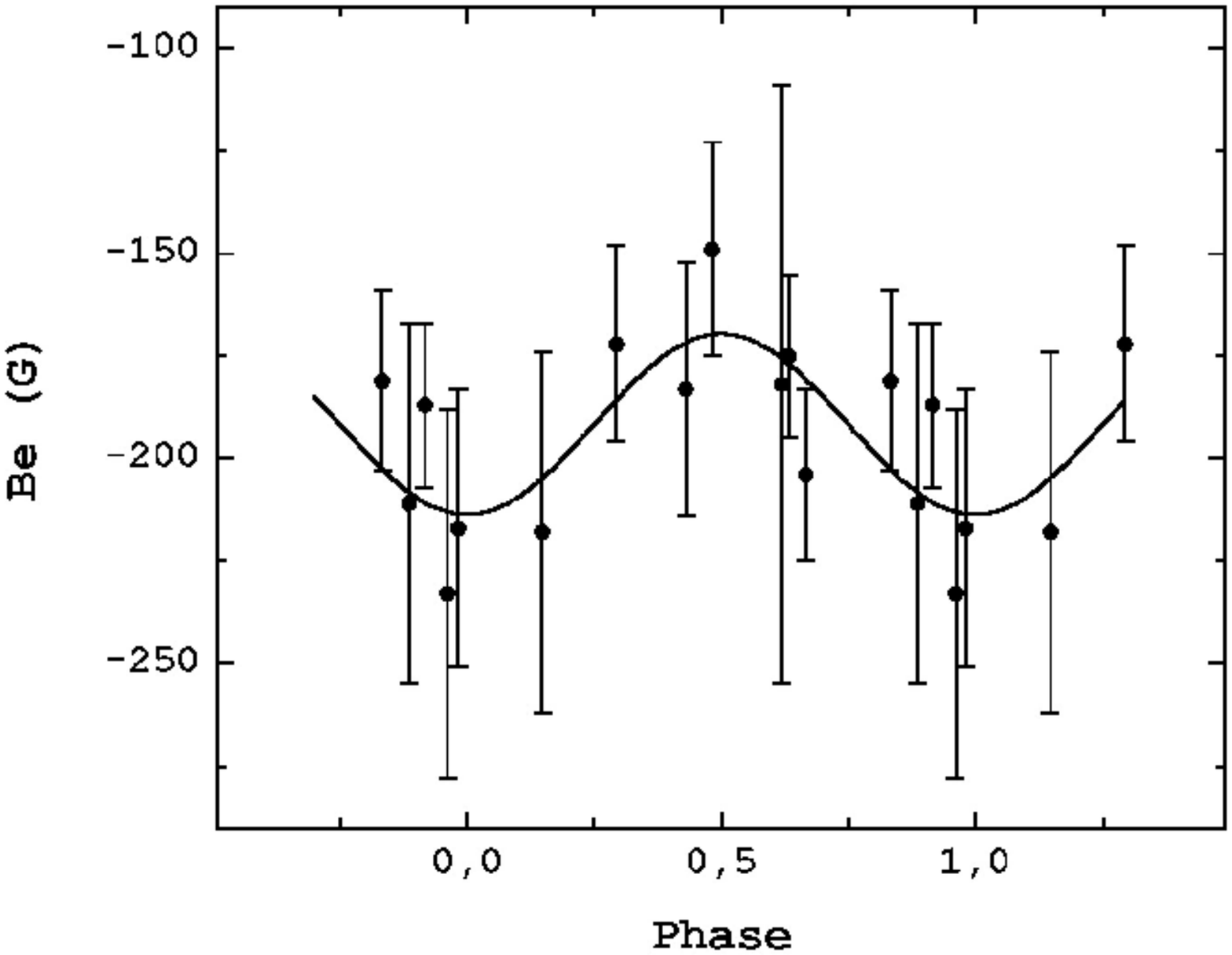}}
\vspace{-3.5mm}
\caption{ HD148112 }
\label{fig:fig262}
\end{figure}

\clearpage
\newpage

\begin{figure}
\resizebox{0.98\hsize}{!}{\includegraphics{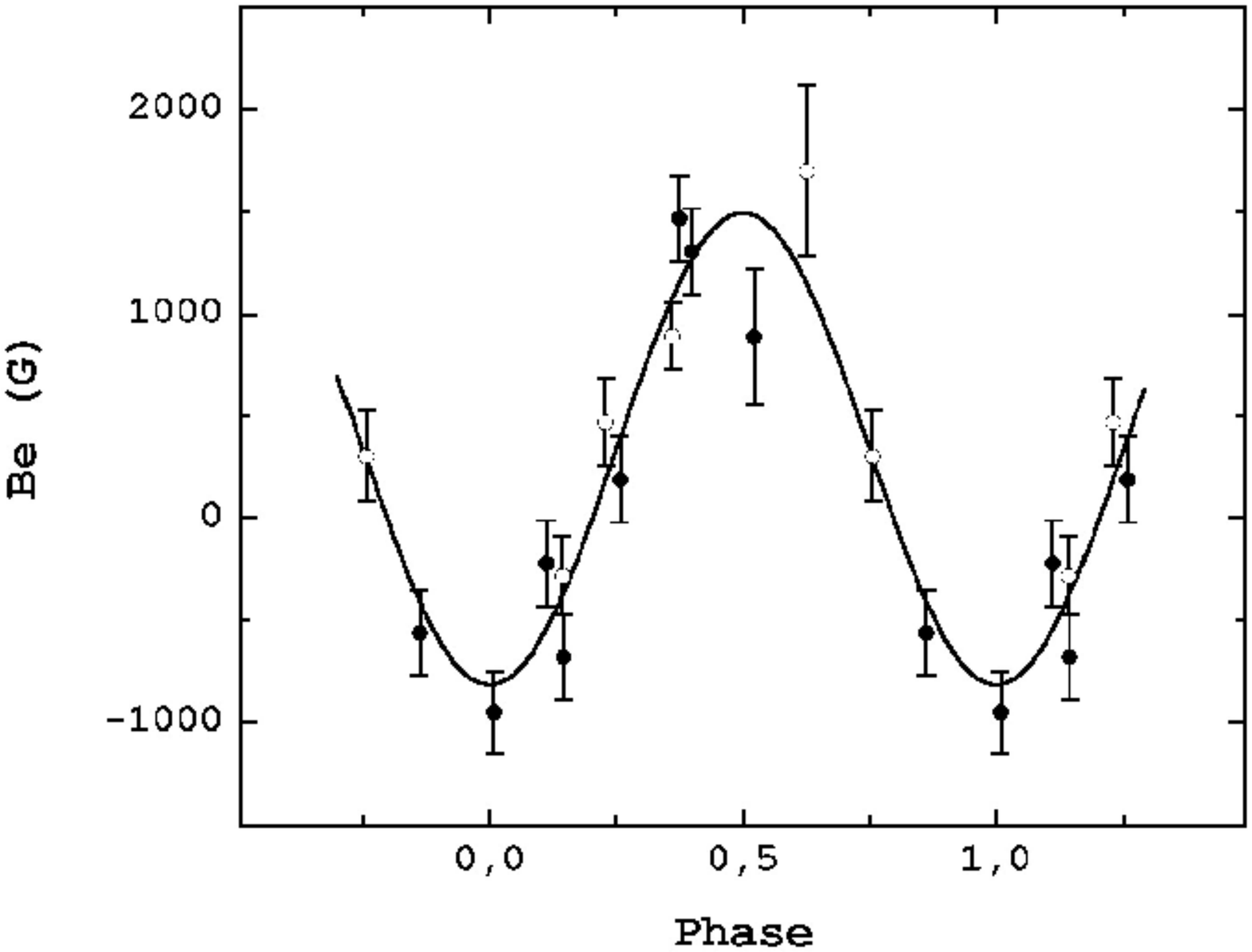}}
\vspace{-3.5mm}
\caption{ HD148199 }
\label{fig:fig263}
\end{figure}

\begin{figure}
\resizebox{0.98\hsize}{!}{\includegraphics{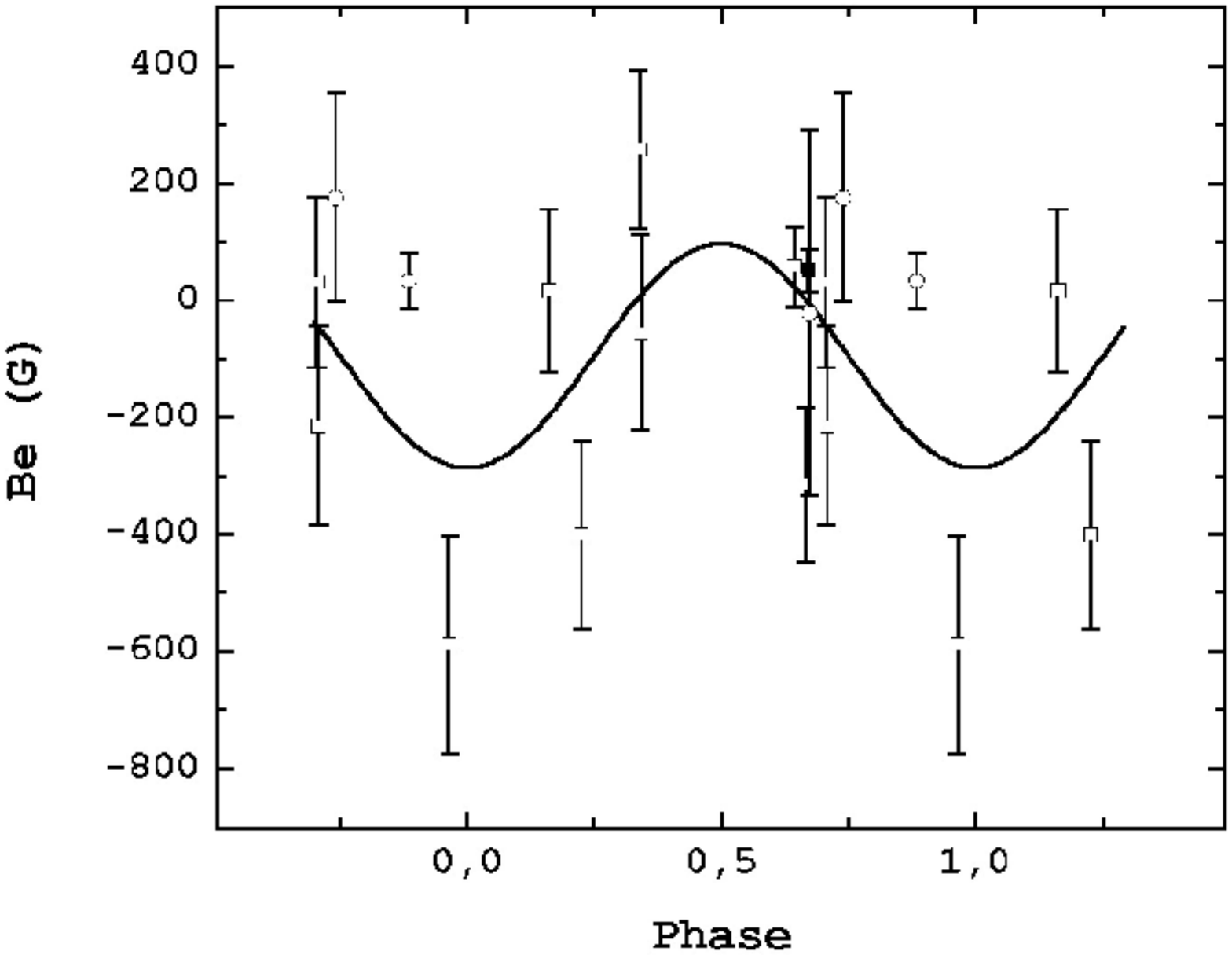}}
\vspace{-3.5mm}
\caption{ HD148330 }
\label{fig:fig264}
\end{figure}

\begin{figure}
\resizebox{0.98\hsize}{!}{\includegraphics{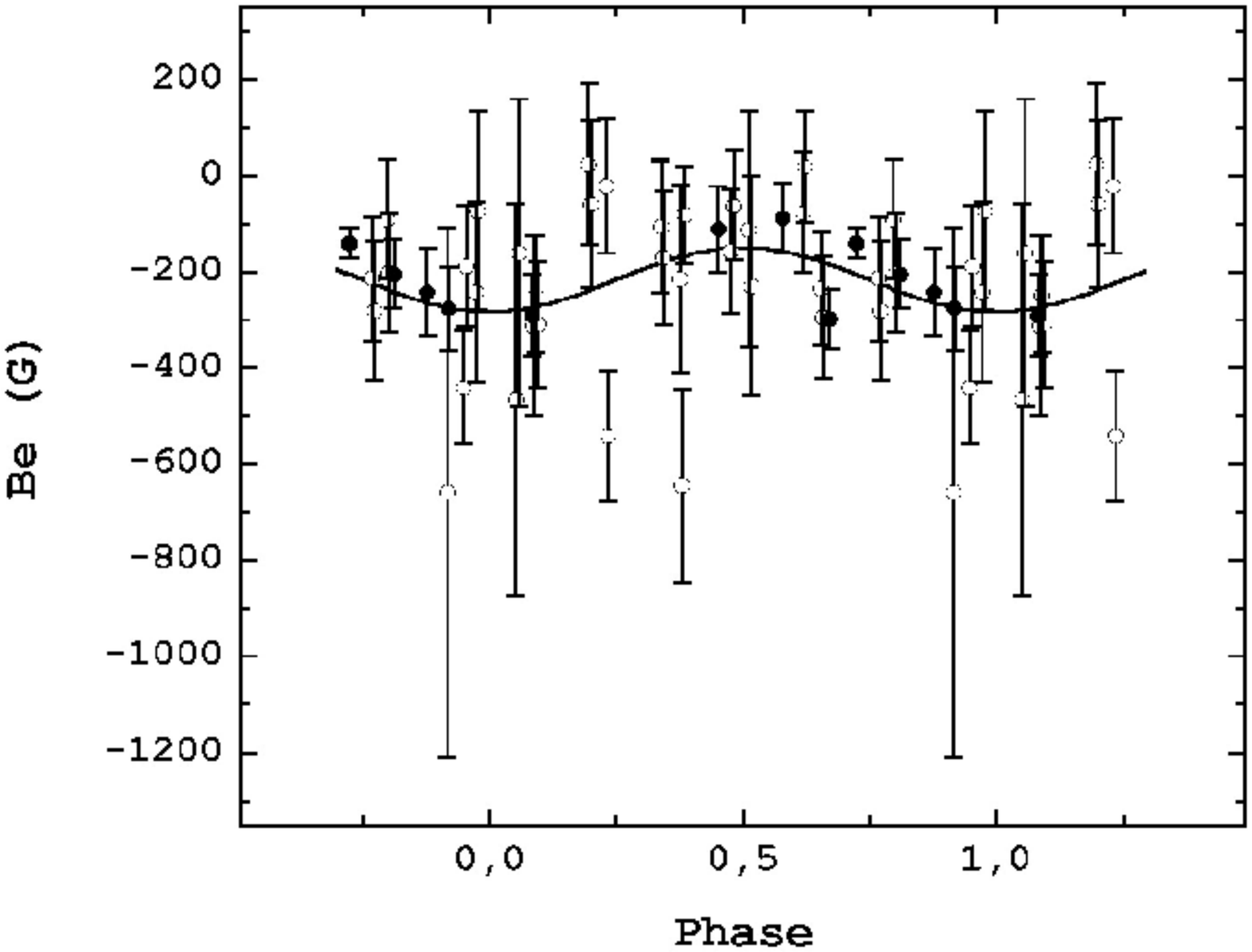}}
\vspace{-3.5mm}
\caption{ HD148937 }
\label{fig:fig265}
\end{figure}

\begin{figure}
\resizebox{0.98\hsize}{!}{\includegraphics{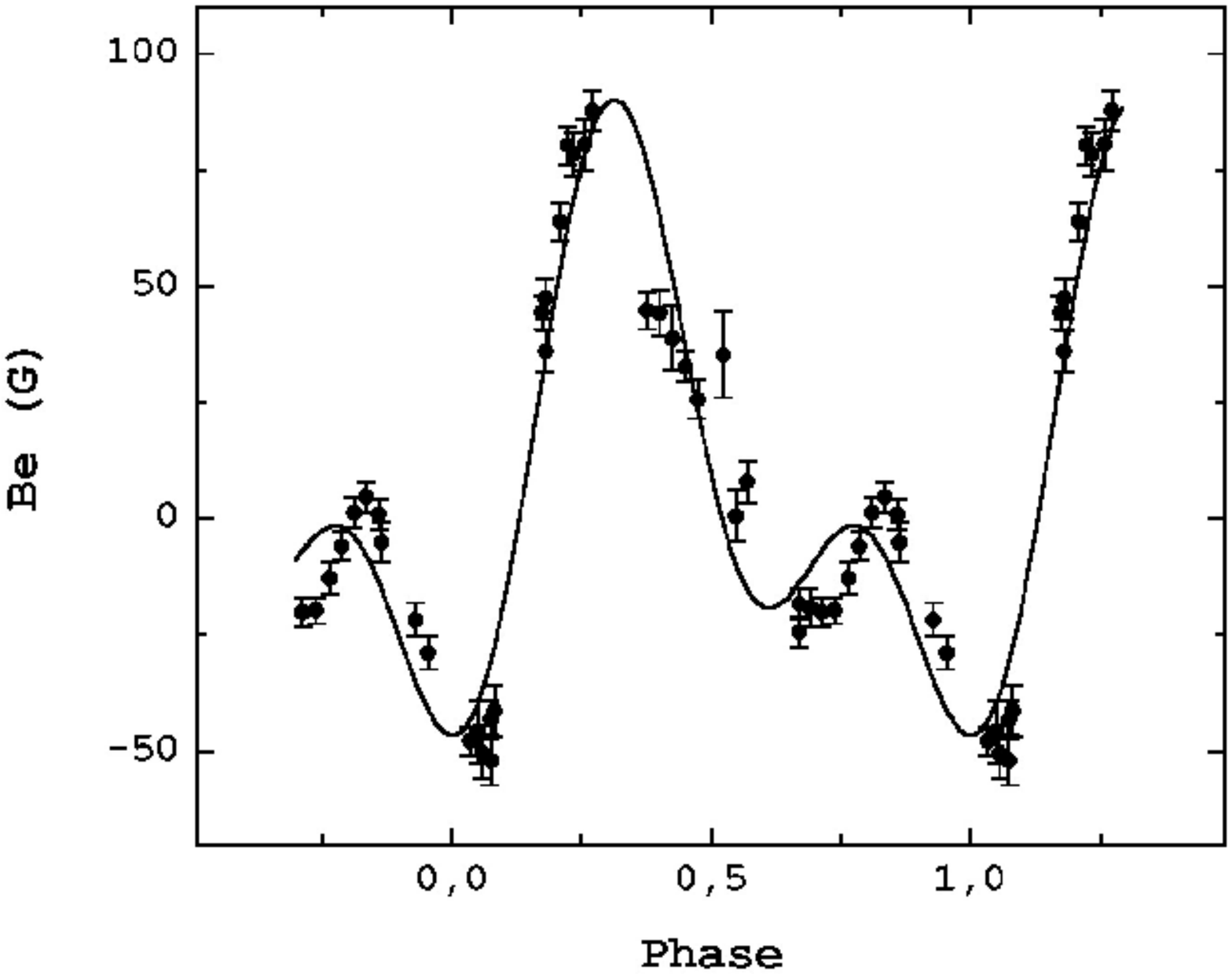}}
\vspace{-3.5mm}
\caption{ HD149438 }
\label{fig:fig266}
\end{figure}

\begin{figure}
\resizebox{0.98\hsize}{!}{\includegraphics{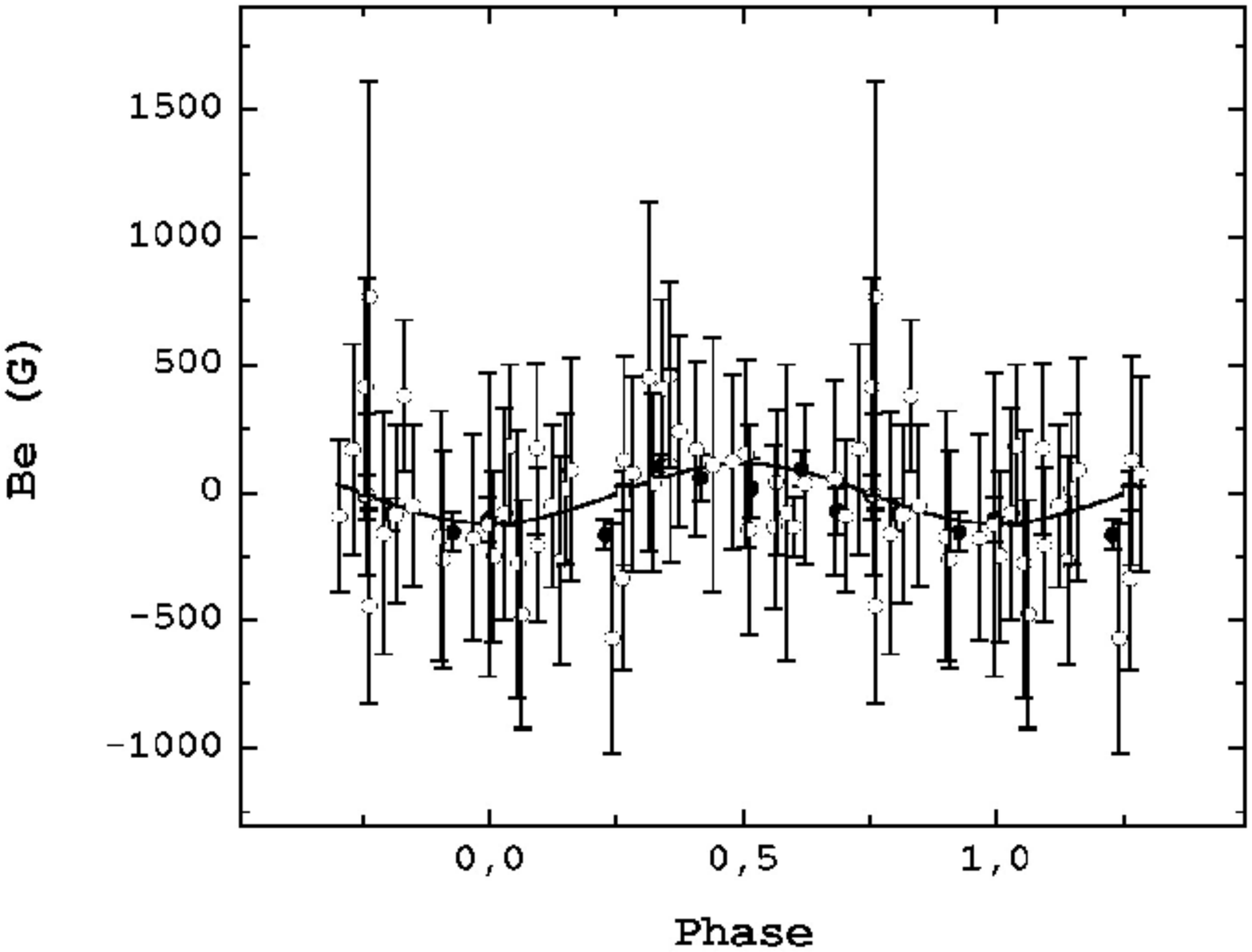}}
\vspace{-3.5mm}
\caption{ HD149757 }
\label{fig:fig267}
\end{figure}

\begin{figure}
\resizebox{0.98\hsize}{!}{\includegraphics{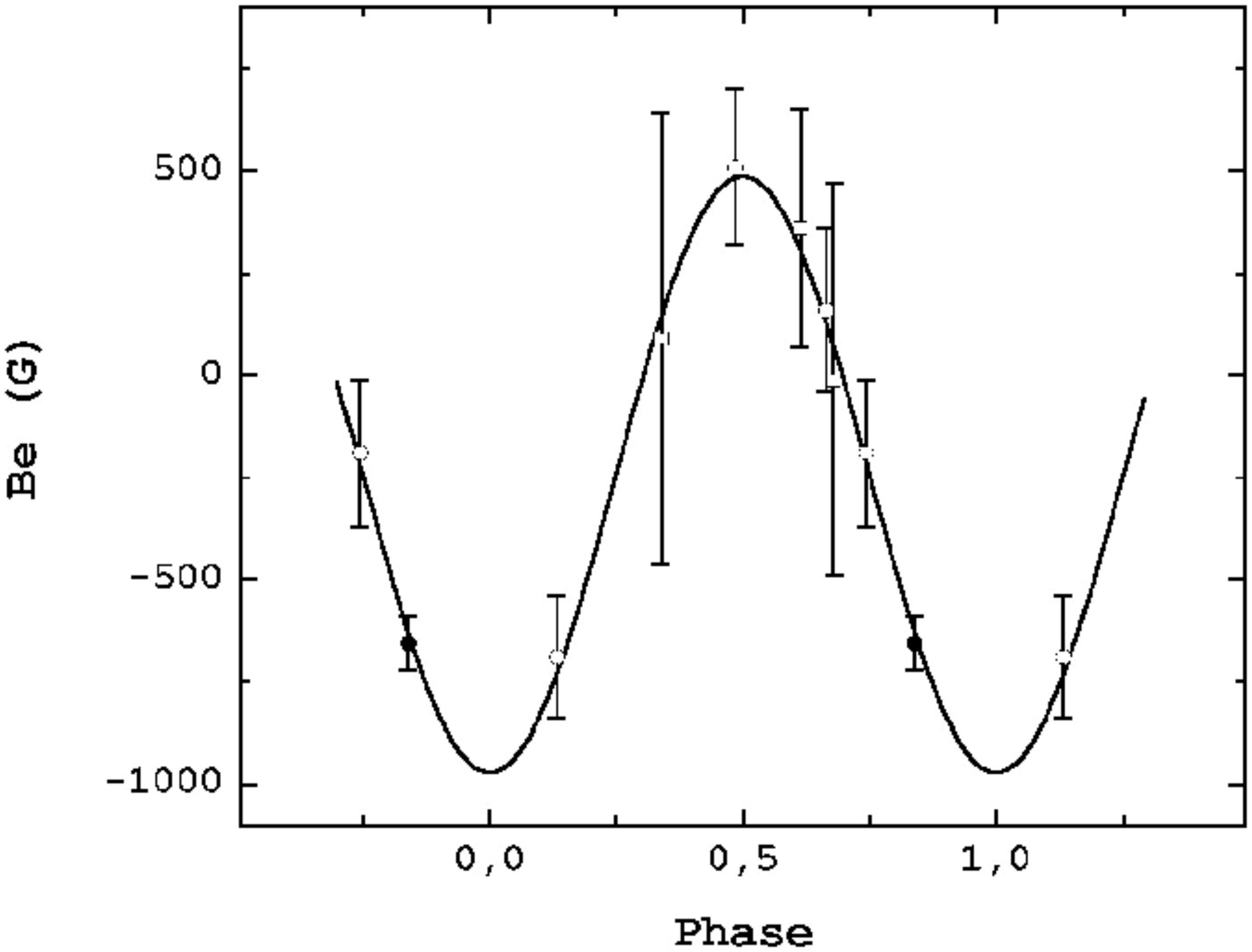}}
\vspace{-3.5mm}
\caption{ HD149822 }
\label{fig:fig268}
\end{figure}

\clearpage
\newpage

\begin{figure}
\resizebox{0.98\hsize}{!}{\includegraphics{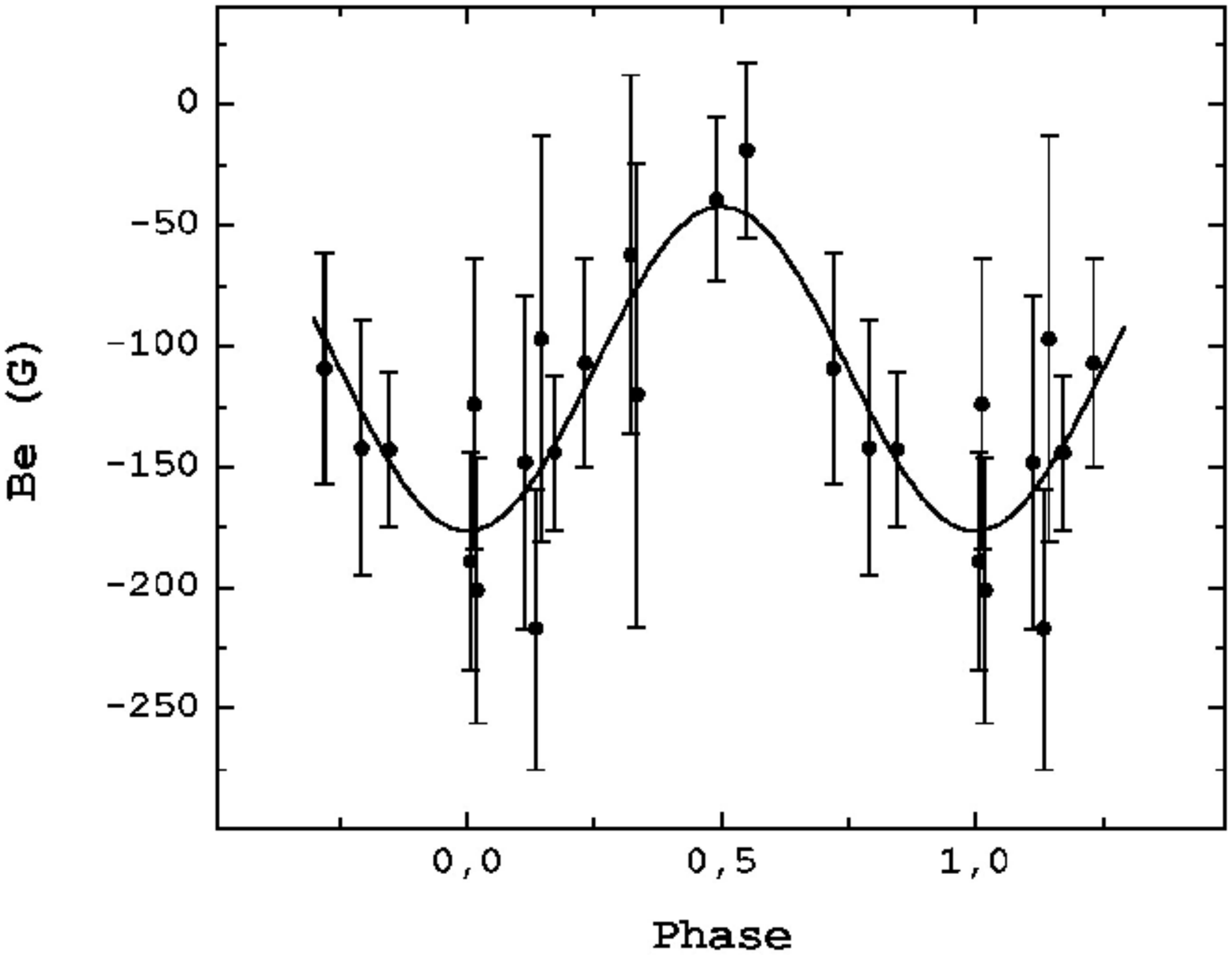}}
\vspace{-3.5mm}
\caption{ HD150193 (1) }
\label{fig:fig269}
\end{figure}

\begin{figure}
\resizebox{0.98\hsize}{!}{\includegraphics{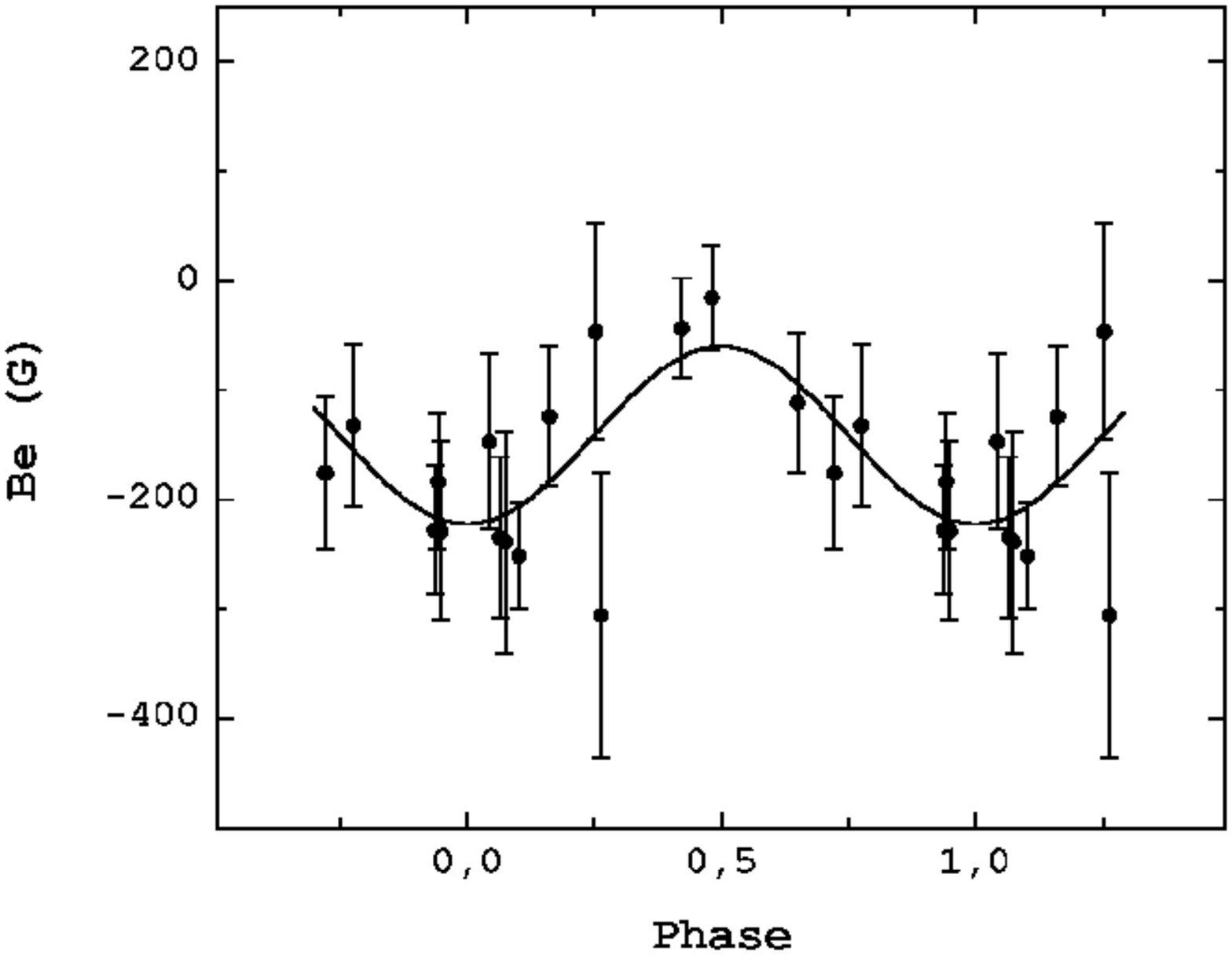}}
\vspace{-3.5mm}
\caption{ HD150193 (2) }
\label{fig:fig270}
\end{figure}

\begin{figure}
\resizebox{0.98\hsize}{!}{\includegraphics{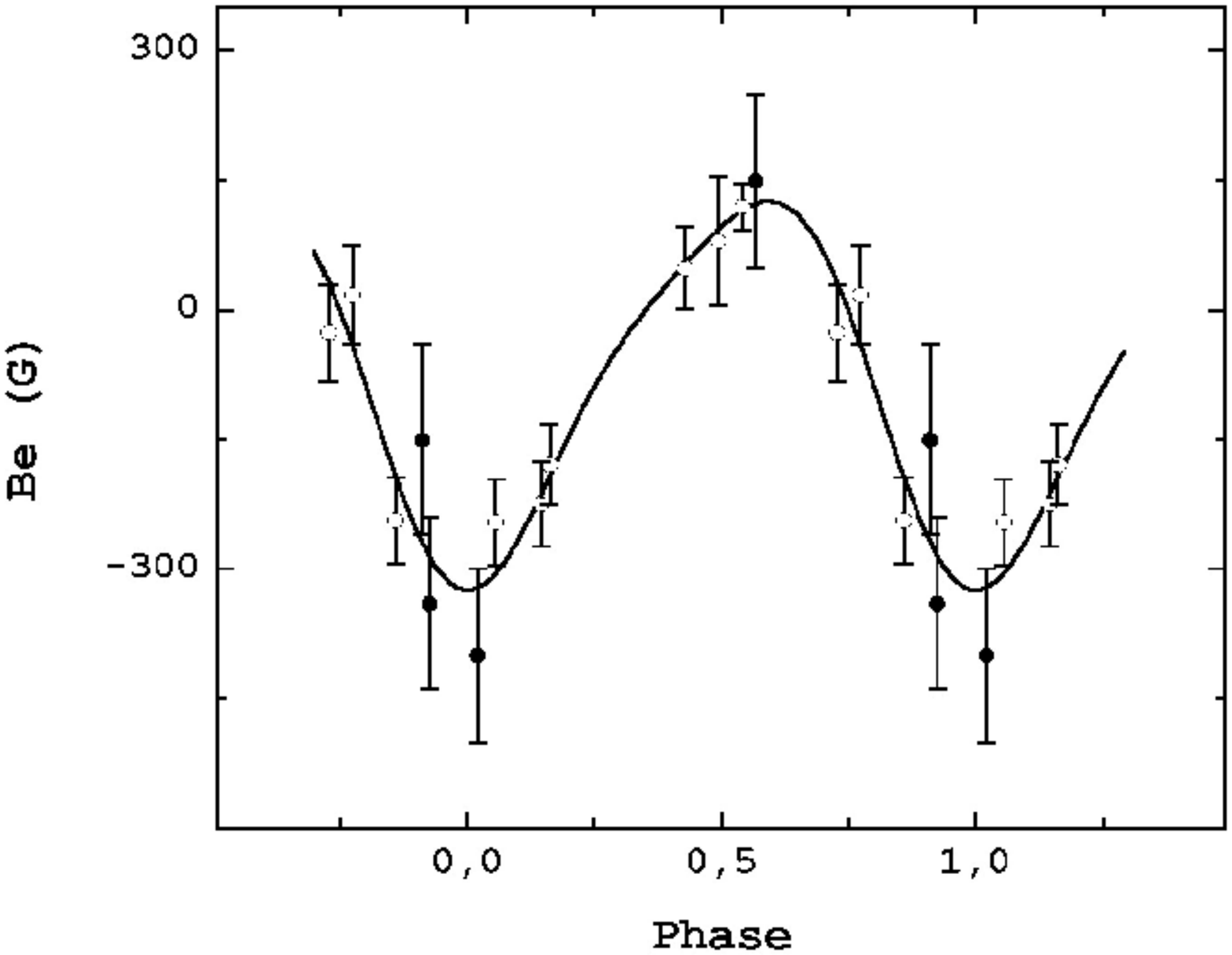}}
\vspace{-3.5mm}
\caption{ HD151199 }
\label{fig:fig219}
\end{figure}

\begin{figure}
\resizebox{0.98\hsize}{!}{\includegraphics{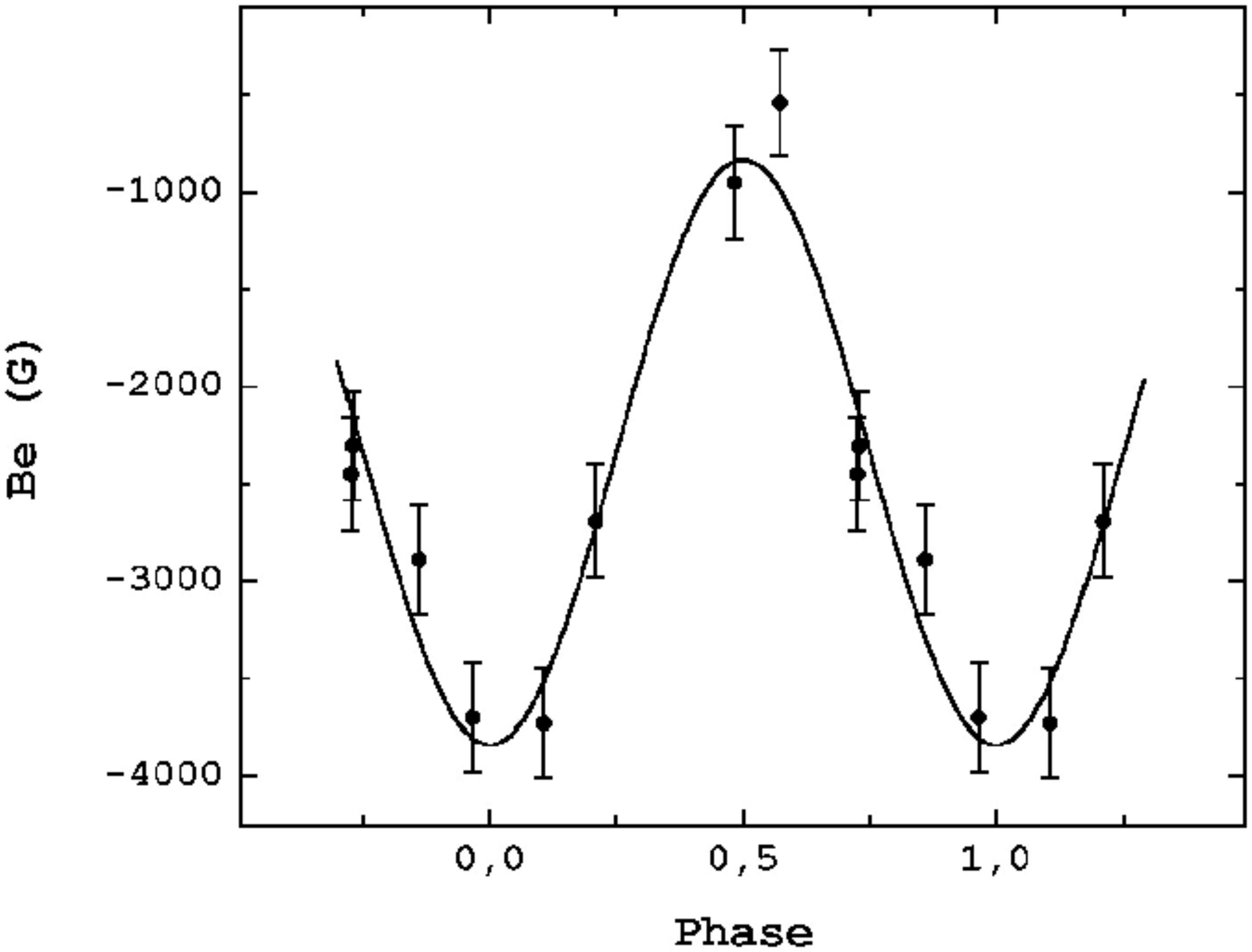}}
\vspace{-3.5mm}
\caption{ HD151965 }
\label{fig:fig271}
\end{figure}

\begin{figure}
\resizebox{0.98\hsize}{!}{\includegraphics{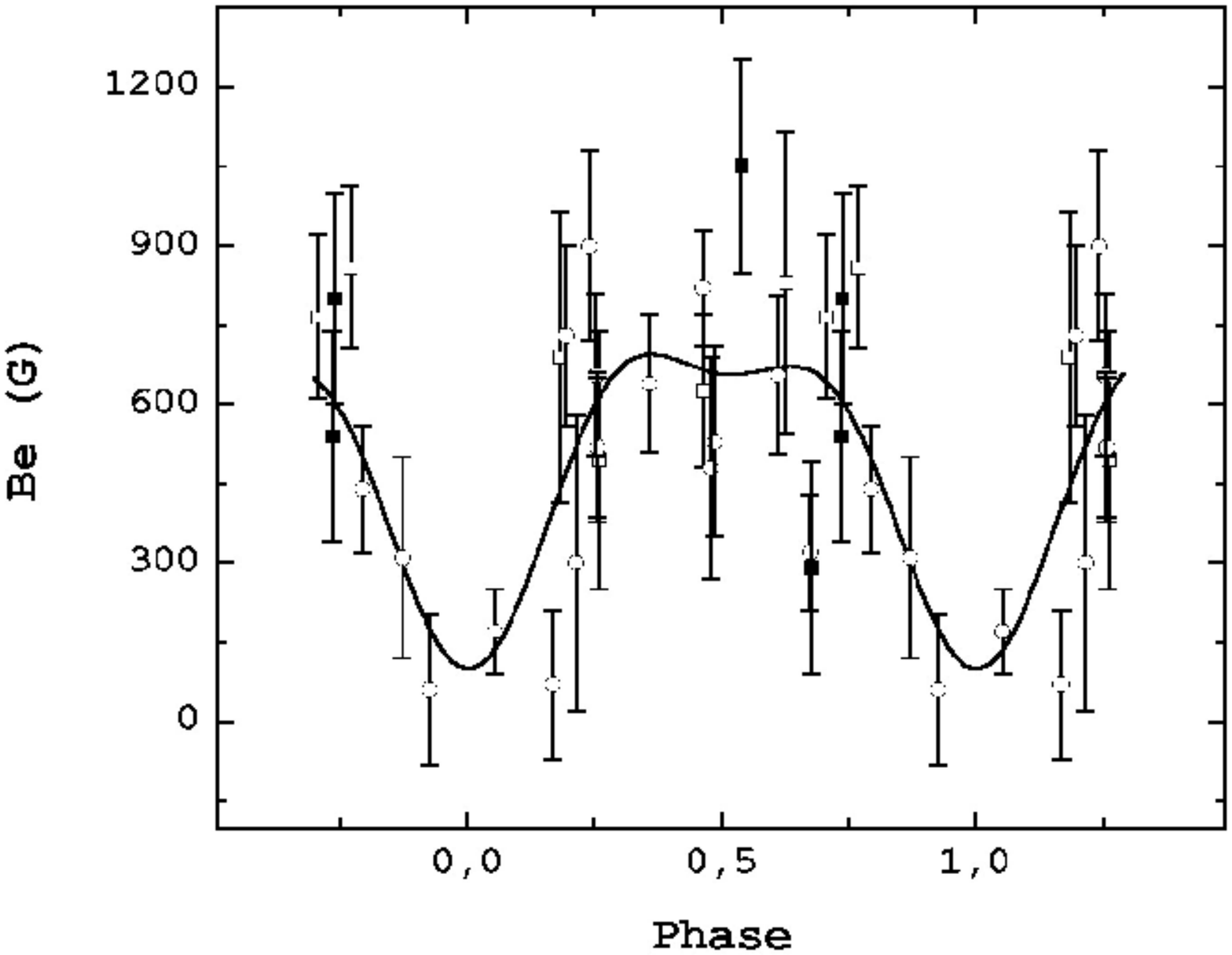}}
\vspace{-3.5mm}
\caption{ HD152107 (1) }
\label{fig:fig272}
\end{figure}

\begin{figure}
\resizebox{0.98\hsize}{!}{\includegraphics{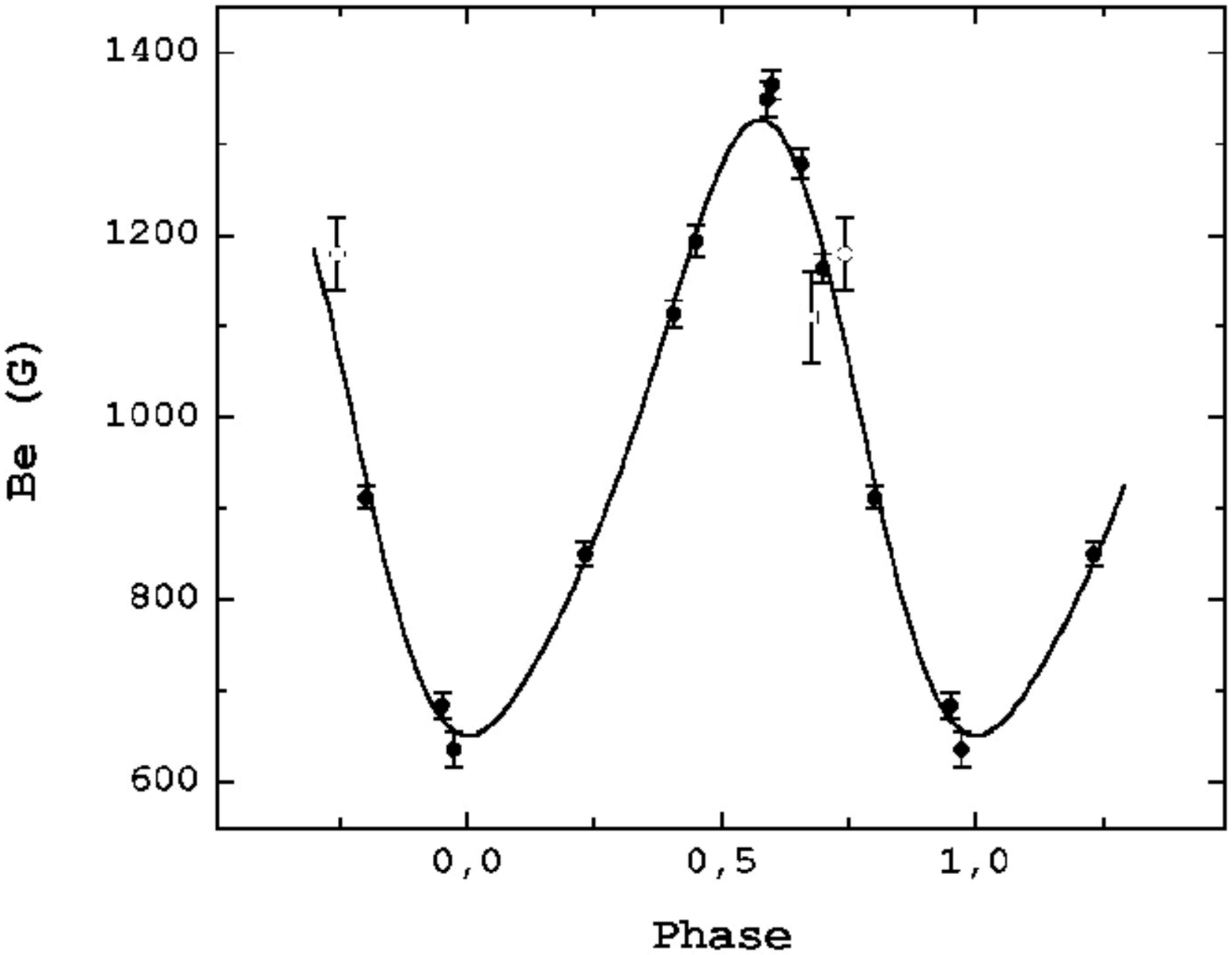}}
\vspace{-3.5mm}
\caption{ HD152107 (2) }
\label{fig:fig273}
\end{figure}

\clearpage
\newpage

\begin{figure}
\resizebox{0.98\hsize}{!}{\includegraphics{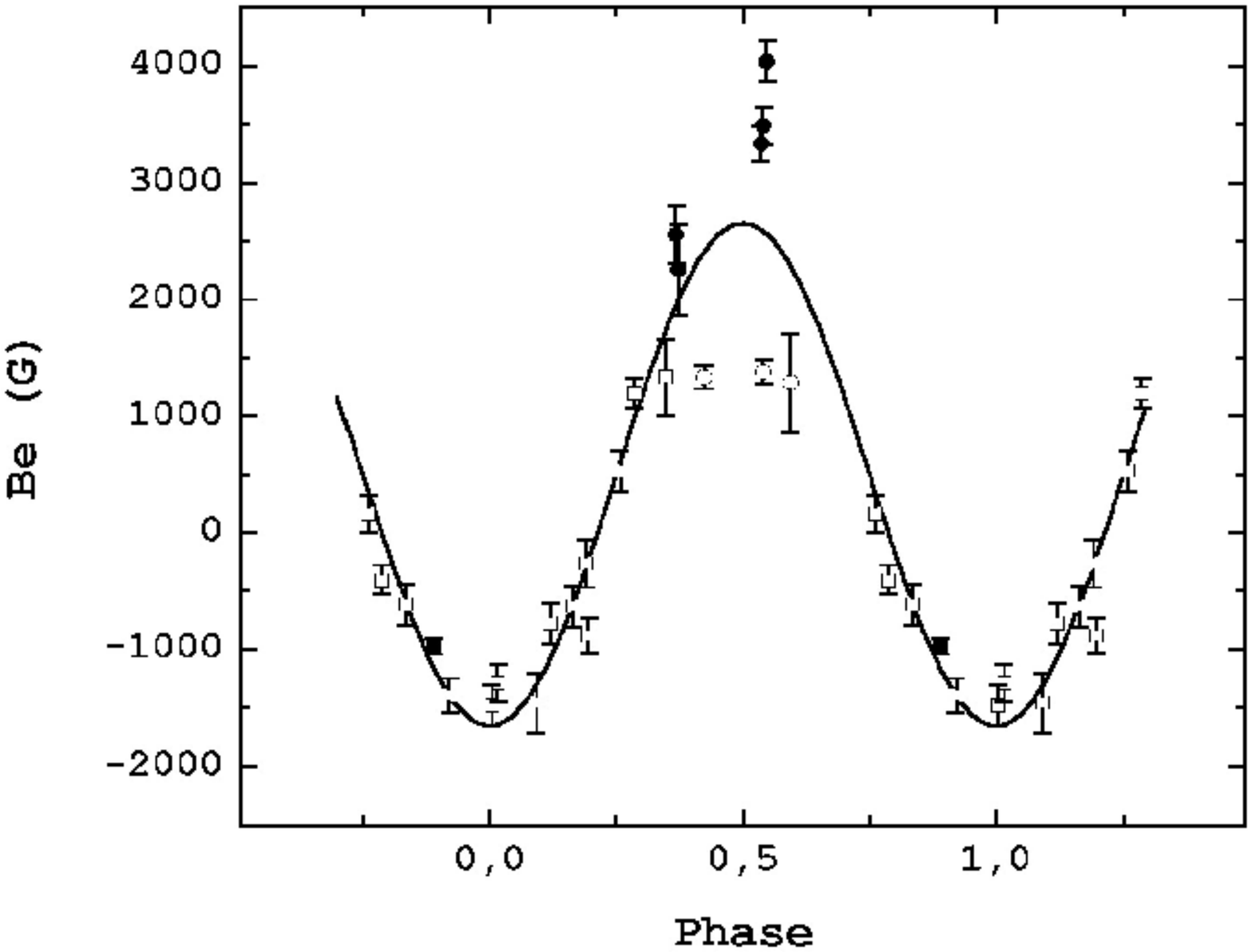}}
\vspace{-3.5mm}
\caption{ HD153882 }
\label{fig:fig274}
\end{figure}

\begin{figure}
\resizebox{0.98\hsize}{!}{\includegraphics{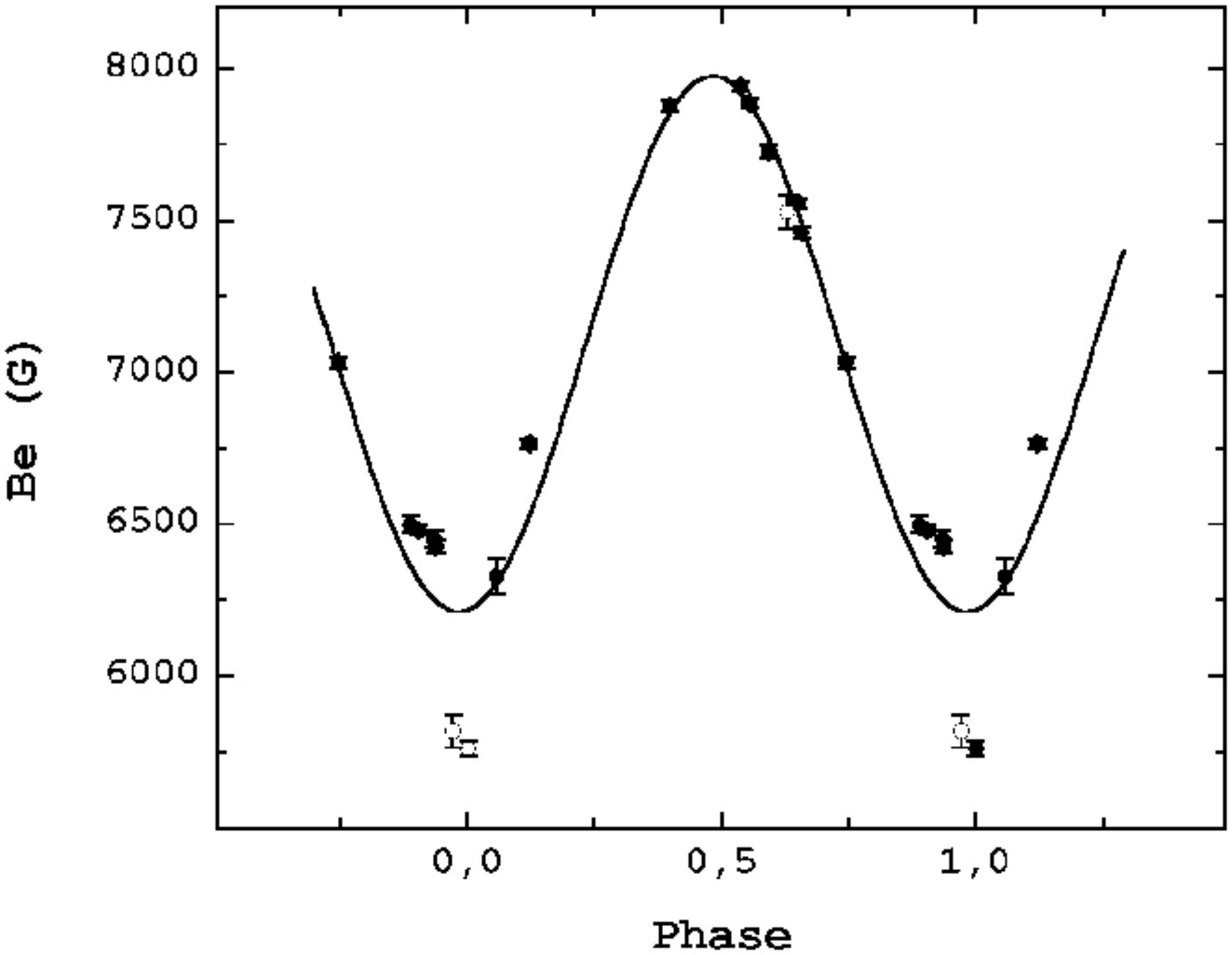}}
\vspace{-3.5mm}
\caption{ HD154708 (1) }
\label{fig:fig275}
\end{figure}

\begin{figure}
\resizebox{0.98\hsize}{!}{\includegraphics{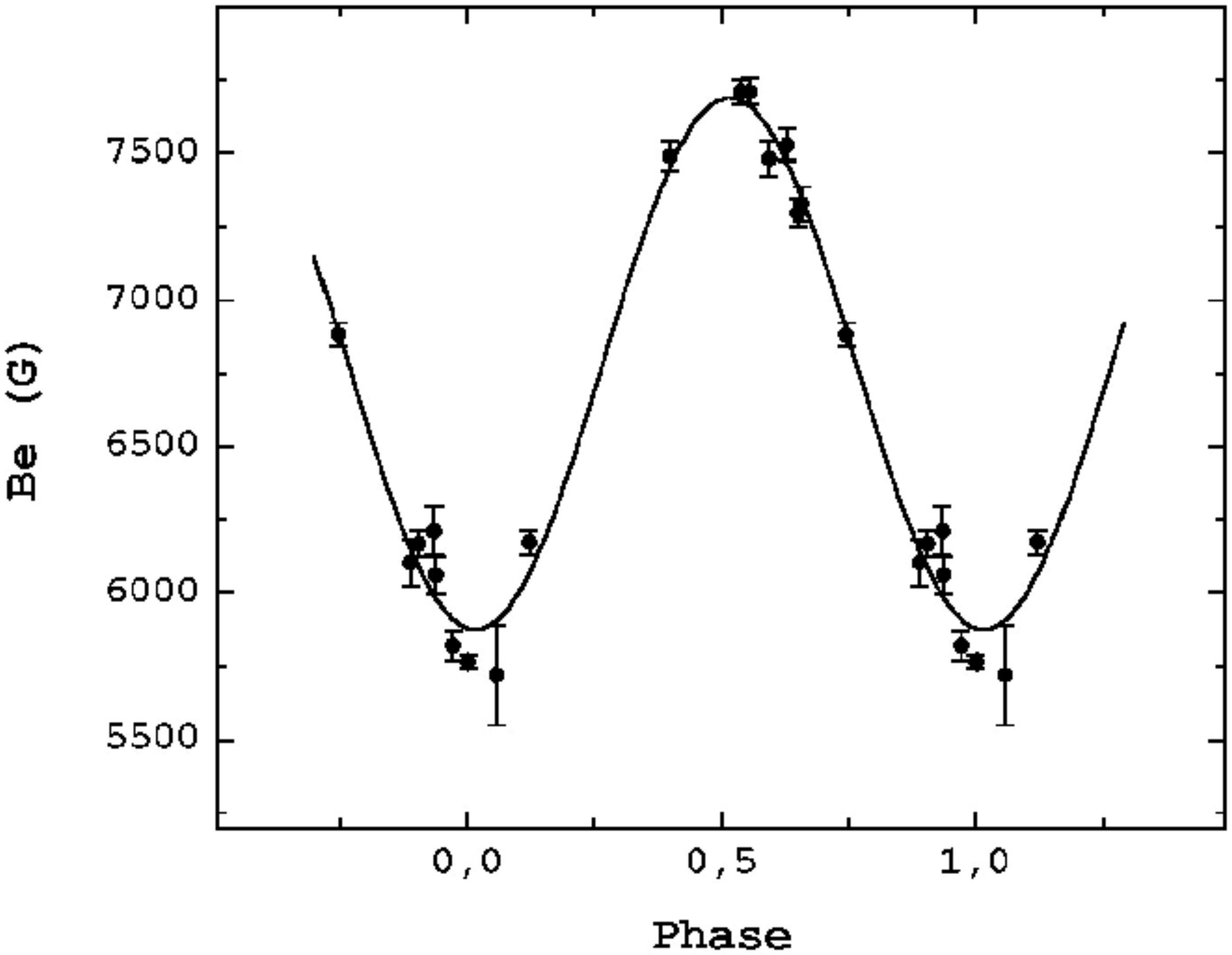}}
\vspace{-3.5mm}
\caption{ HD154708 (2) }
\label{fig:fig276}
\end{figure}

\begin{figure}
\resizebox{0.98\hsize}{!}{\includegraphics{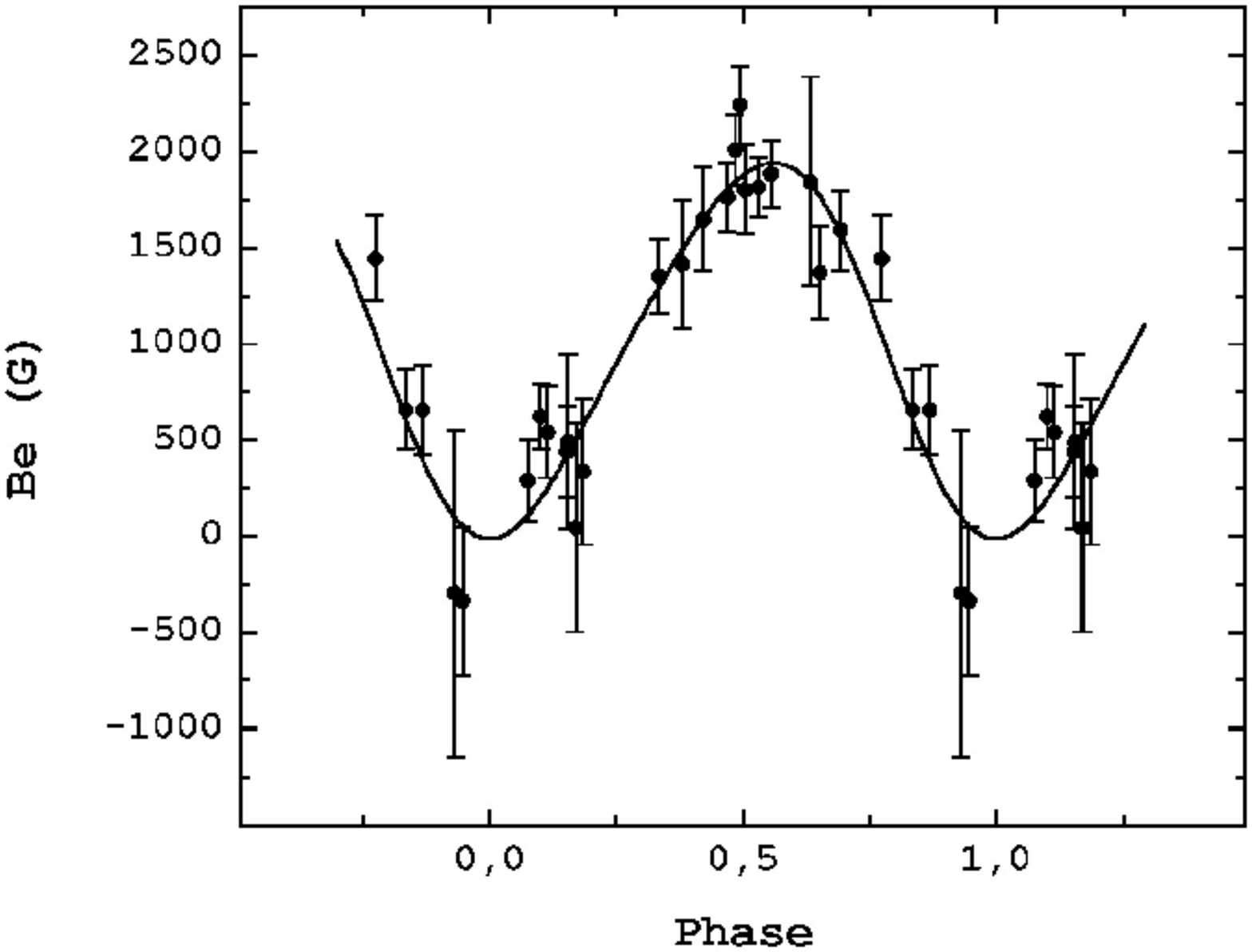}}
\vspace{-3.5mm}
\caption{ HD156324 (1) }
\label{fig:fig219}
\end{figure}

\begin{figure}
\resizebox{0.98\hsize}{!}{\includegraphics{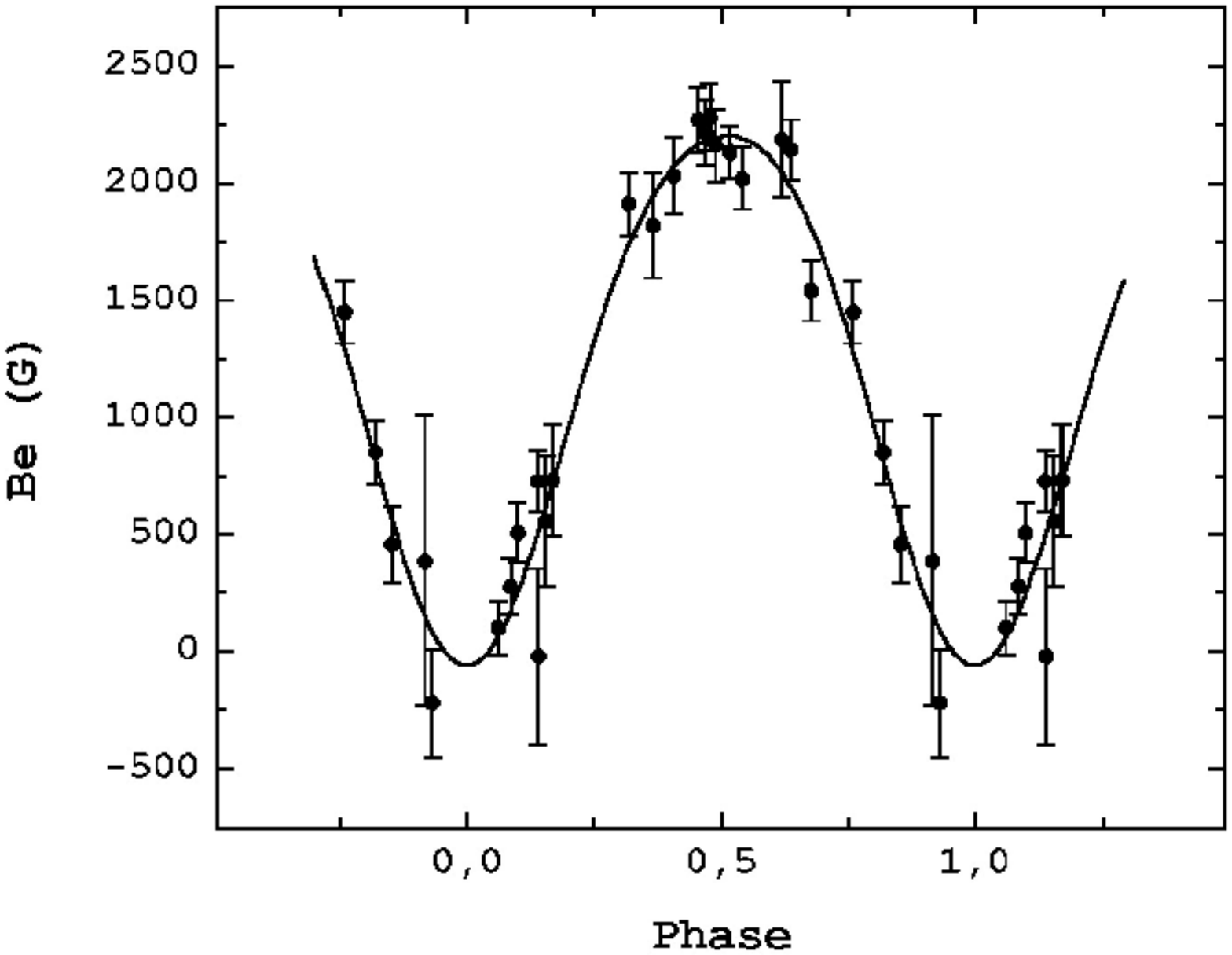}}
\vspace{-3.5mm}
\caption{ HD156324 (2) }
\label{fig:fig219}
\end{figure}

\begin{figure}
\resizebox{0.98\hsize}{!}{\includegraphics{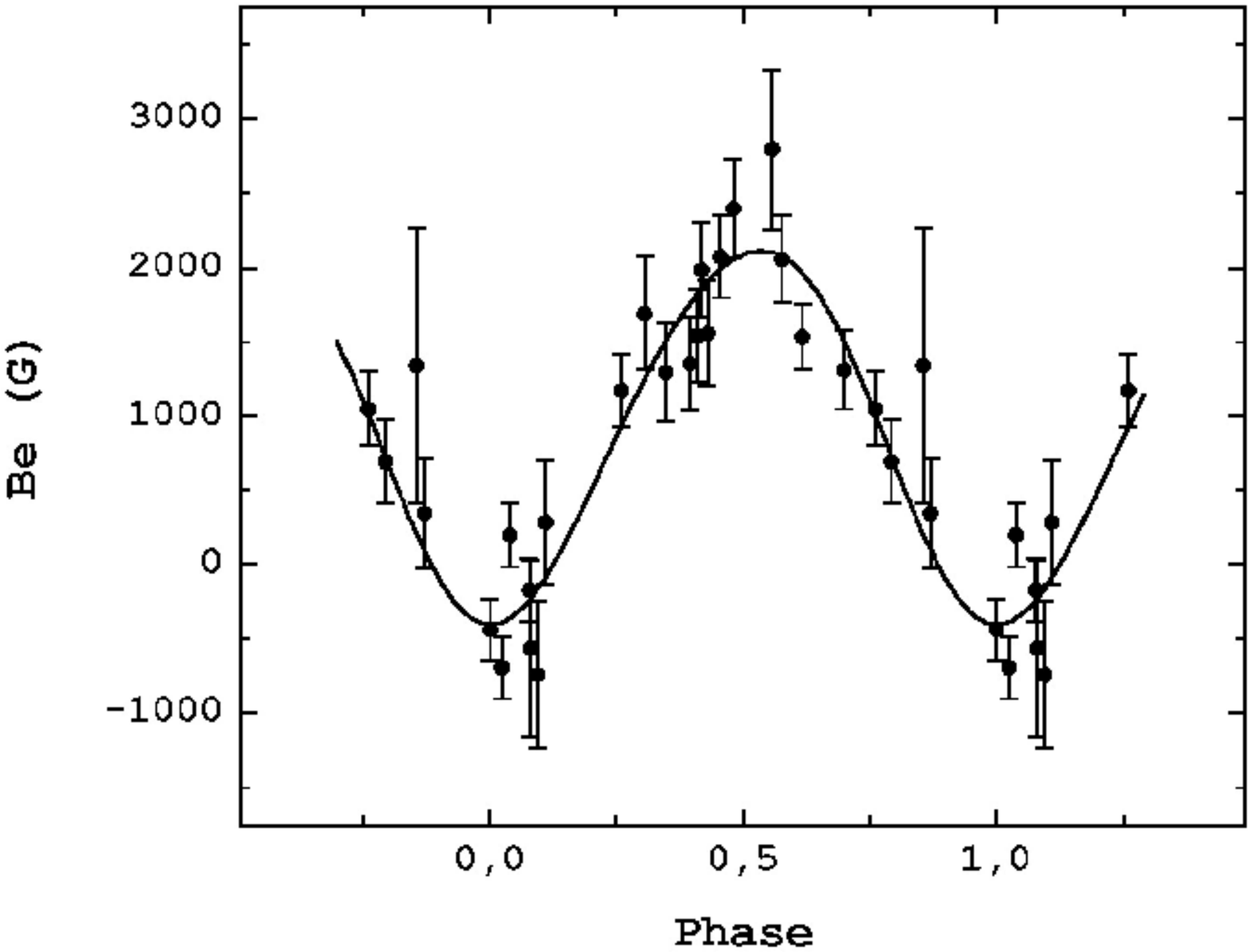}}
\vspace{-3.5mm}
\caption{ HD156324 (3) }
\label{fig:fig219}
\end{figure}

\clearpage
\newpage

\begin{figure}
\resizebox{0.98\hsize}{!}{\includegraphics{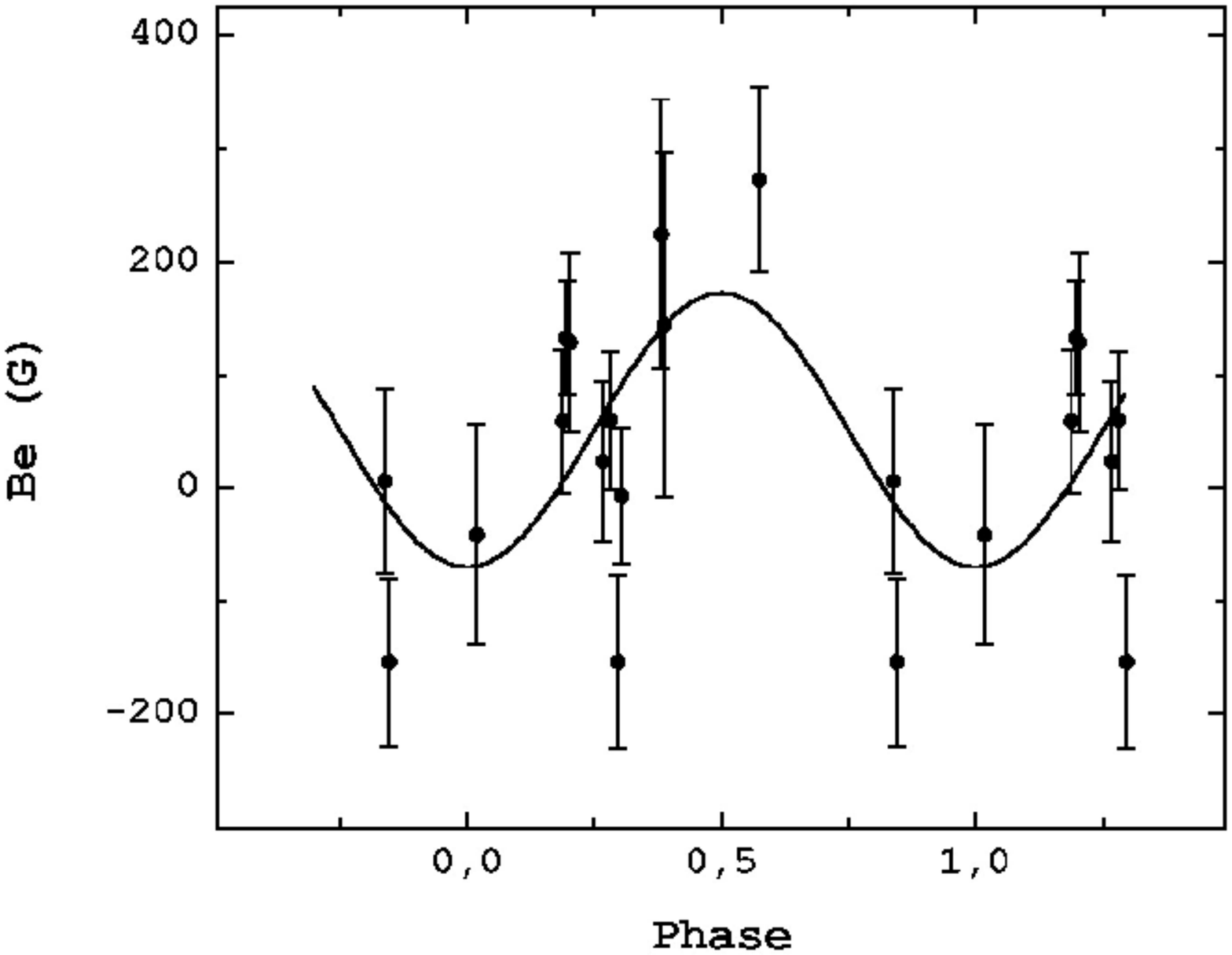}}
\vspace{-3.5mm}
\caption{ HD156424 (1) }
\label{fig:fig219}
\end{figure}

\begin{figure}
\resizebox{0.98\hsize}{!}{\includegraphics{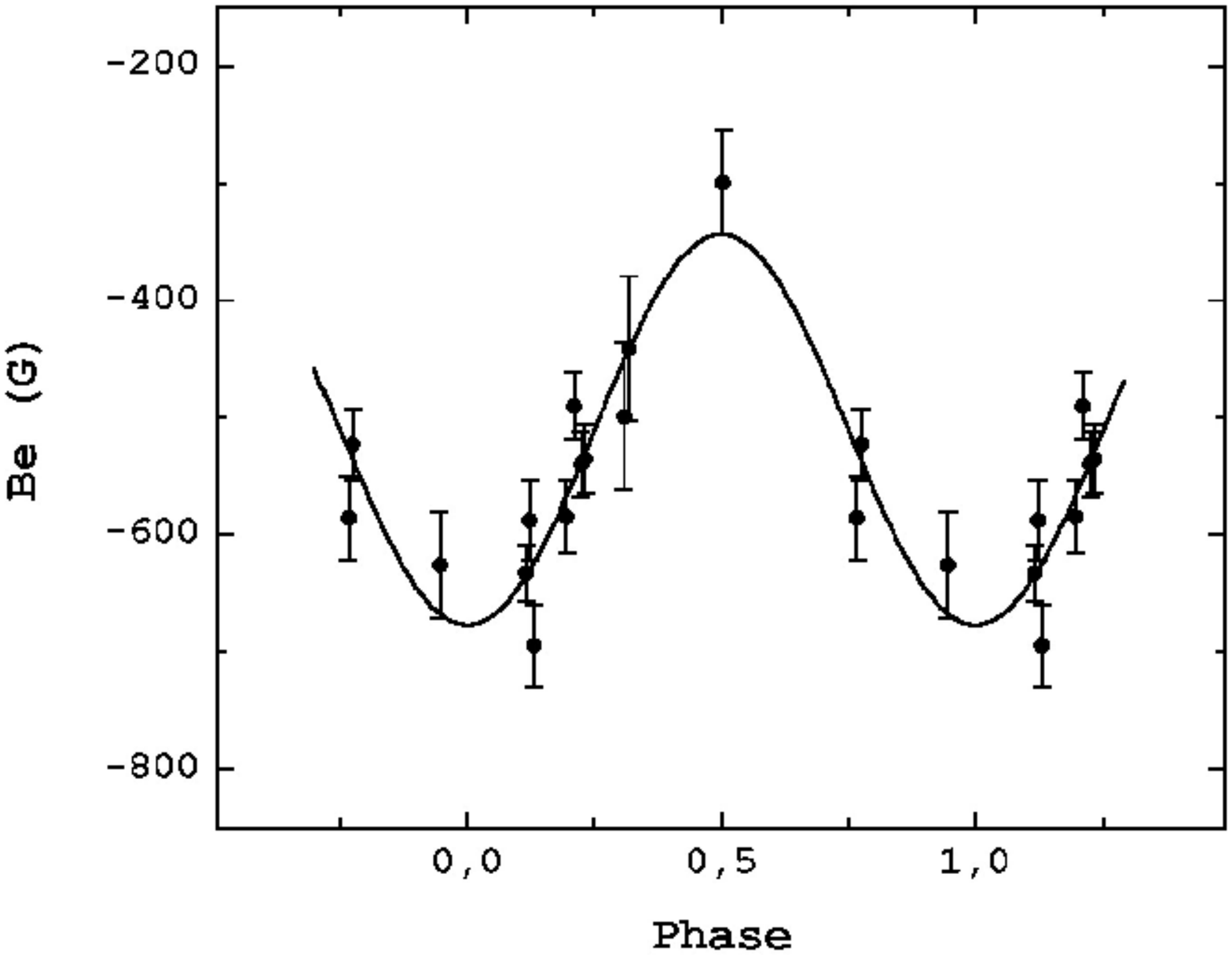}}
\vspace{-3.5mm}
\caption{ HD156424 (2) }
\label{fig:fig219}
\end{figure}

\begin{figure}
\resizebox{0.98\hsize}{!}{\includegraphics{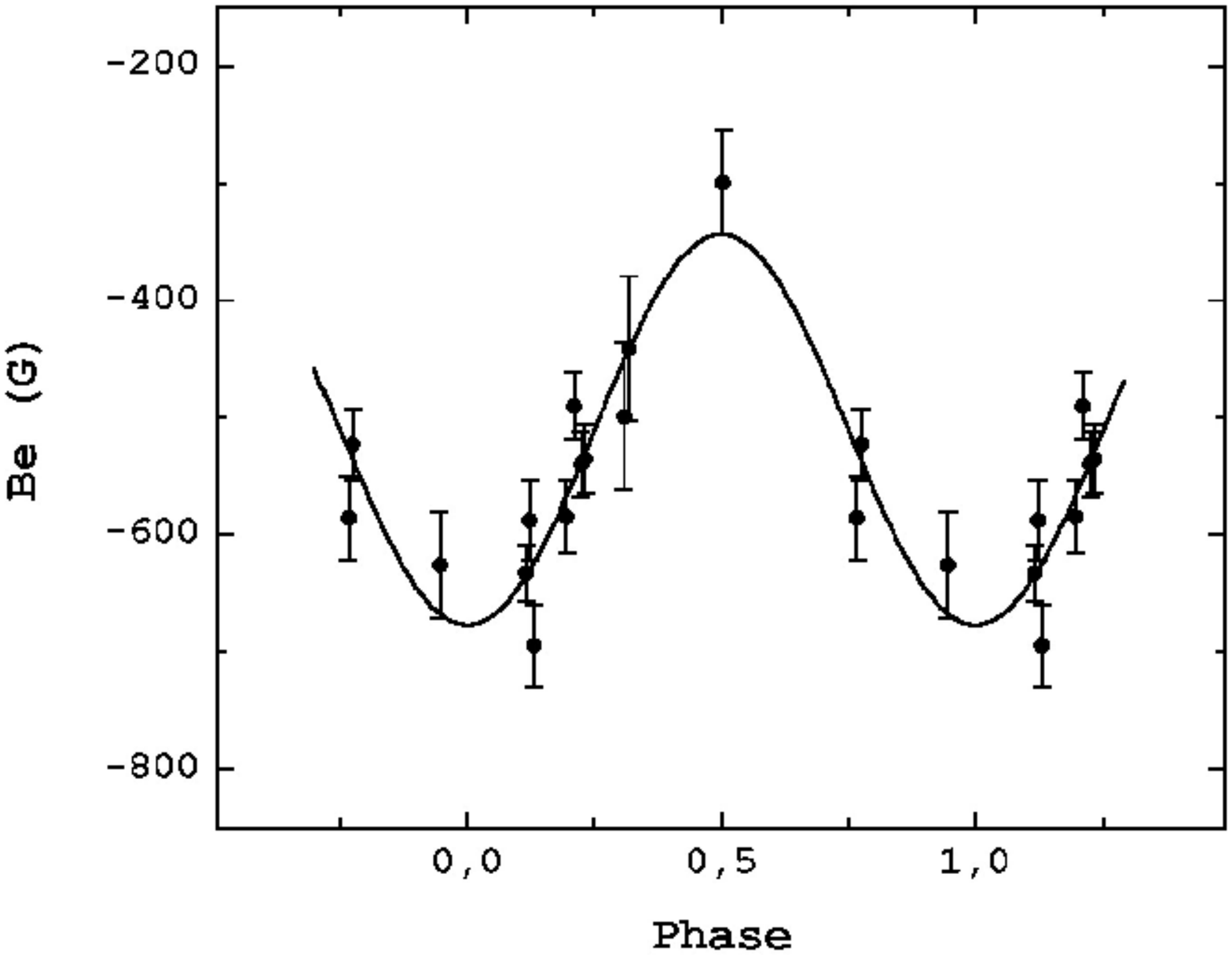}}
\vspace{-3.5mm}
\caption{ HD156424 (3) }
\label{fig:fig219}
\end{figure}

\begin{figure}
\resizebox{0.98\hsize}{!}{\includegraphics{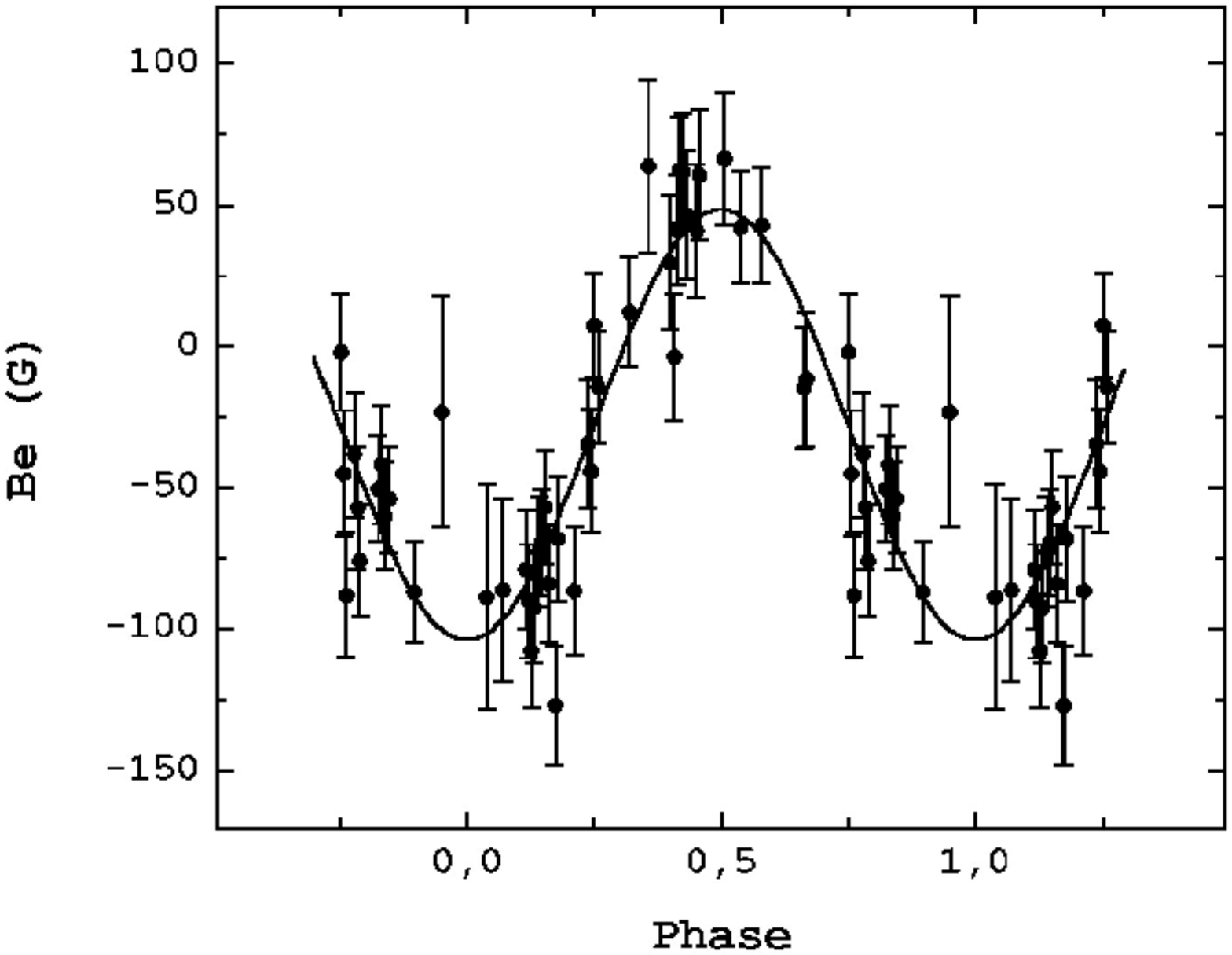}}
\vspace{-3.5mm}
\caption{ HD163472 (1) }
\label{fig:fig277}
\end{figure}

\begin{figure}
\resizebox{0.98\hsize}{!}{\includegraphics{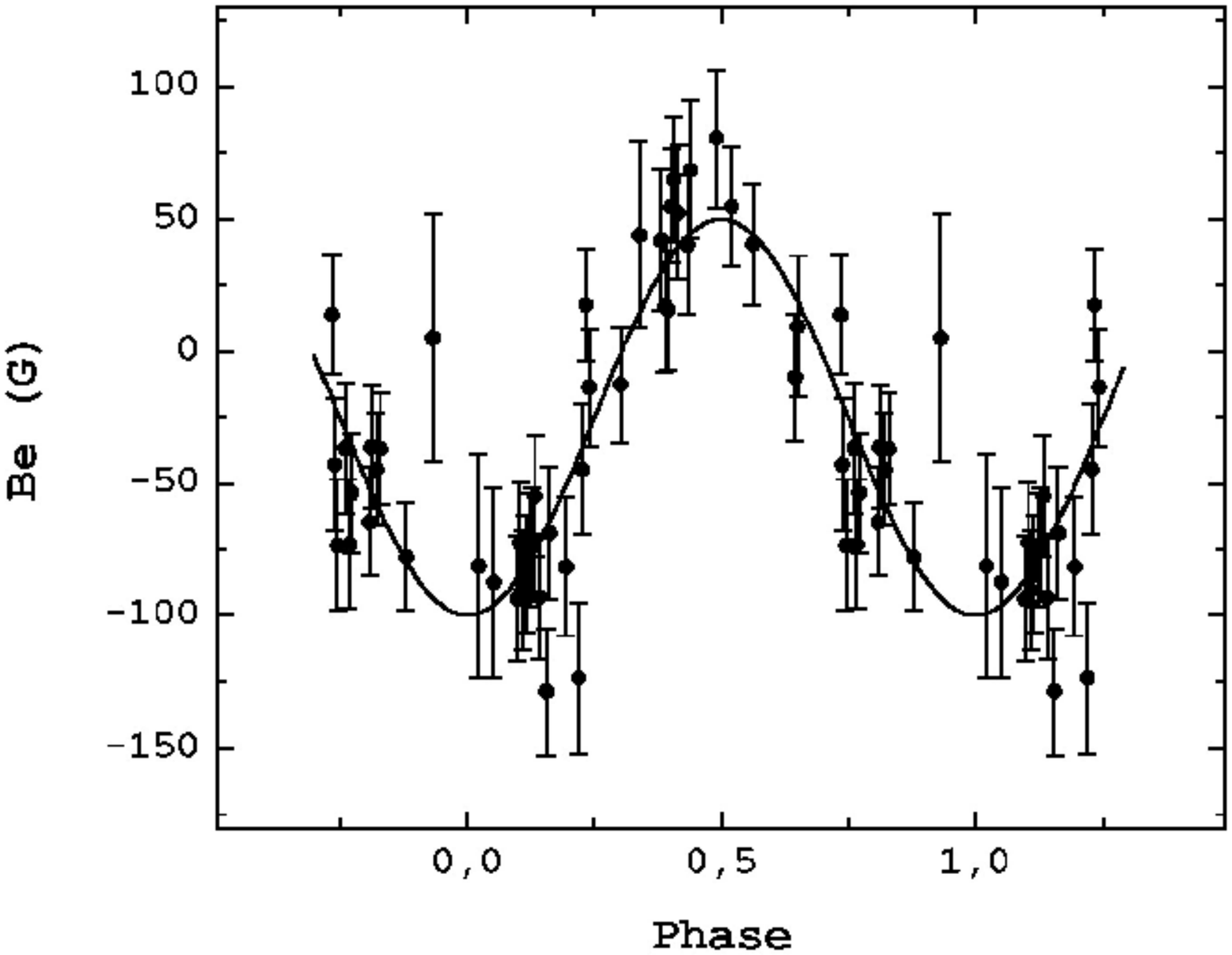}}
\vspace{-3.5mm}
\caption{ HD163472 (2) }
\label{fig:fig278}
\end{figure}

\begin{figure}
\resizebox{0.98\hsize}{!}{\includegraphics{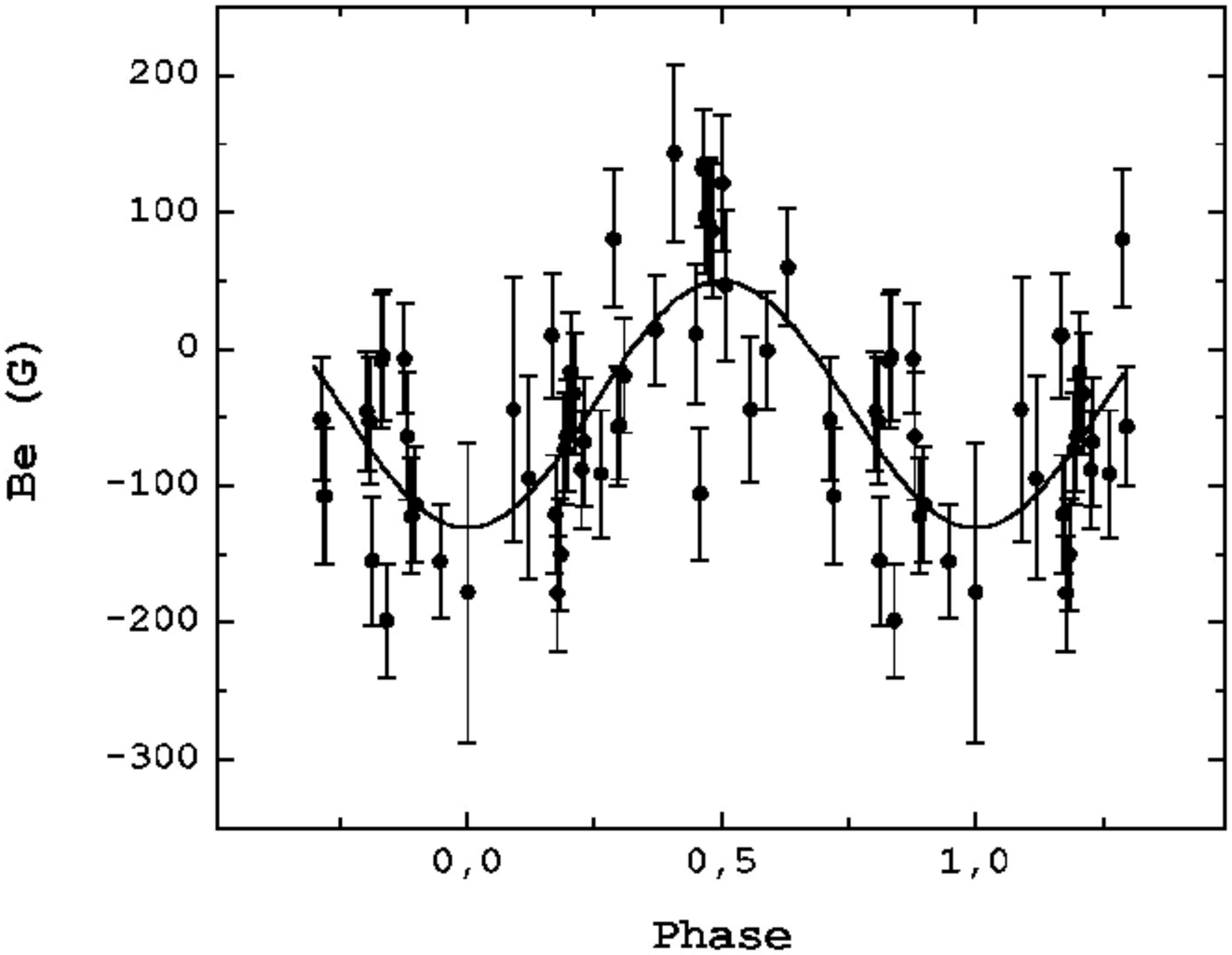}}
\vspace{-3.5mm}
\caption{ HD163472 (3) }
\label{fig:fig279}
\end{figure}

\clearpage
\newpage

\begin{figure}
\resizebox{0.98\hsize}{!}{\includegraphics{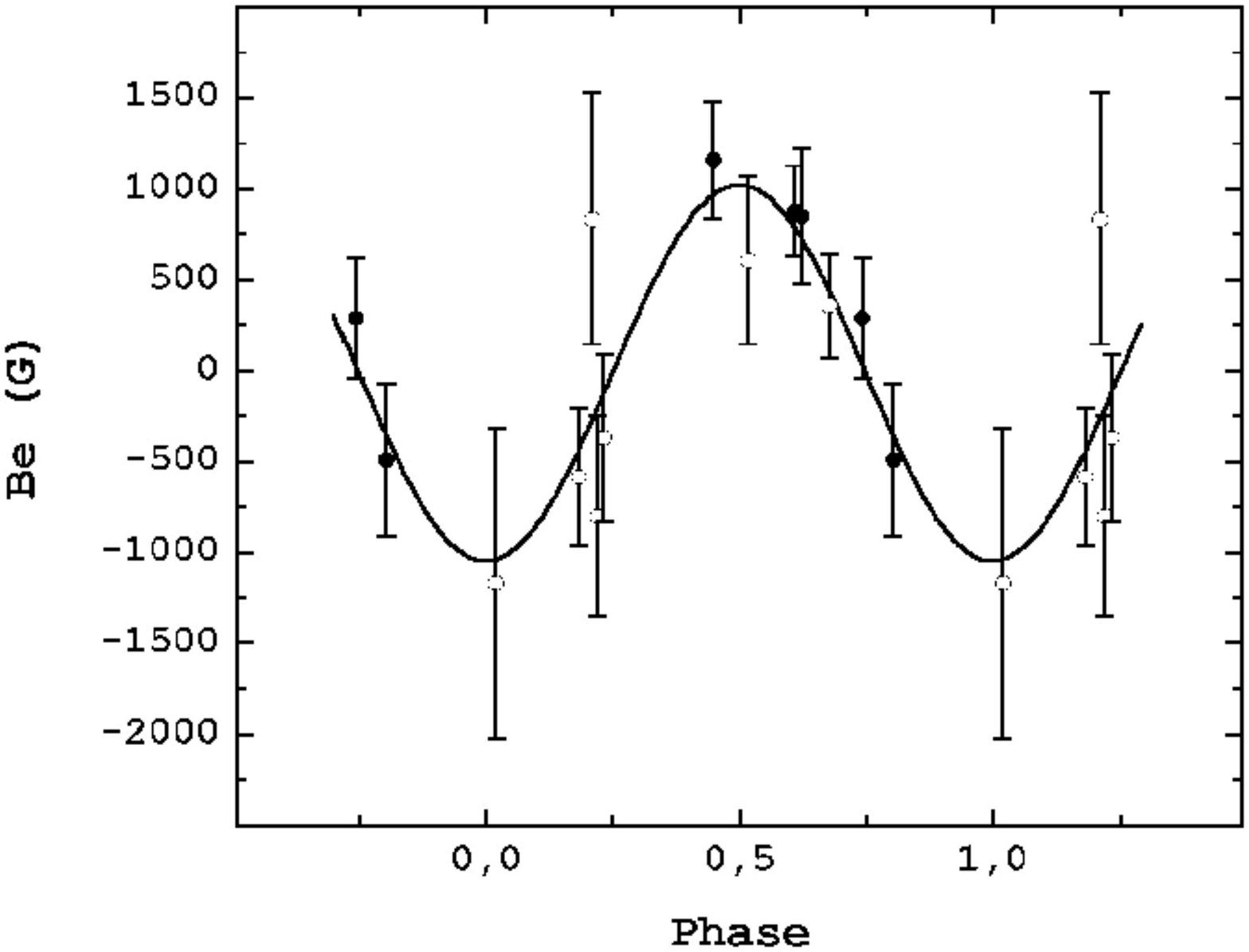}}
\vspace{-3.5mm}
\caption{ HD164258 }
\label{fig:fig280}
\end{figure}

\begin{figure}
\resizebox{0.98\hsize}{!}{\includegraphics{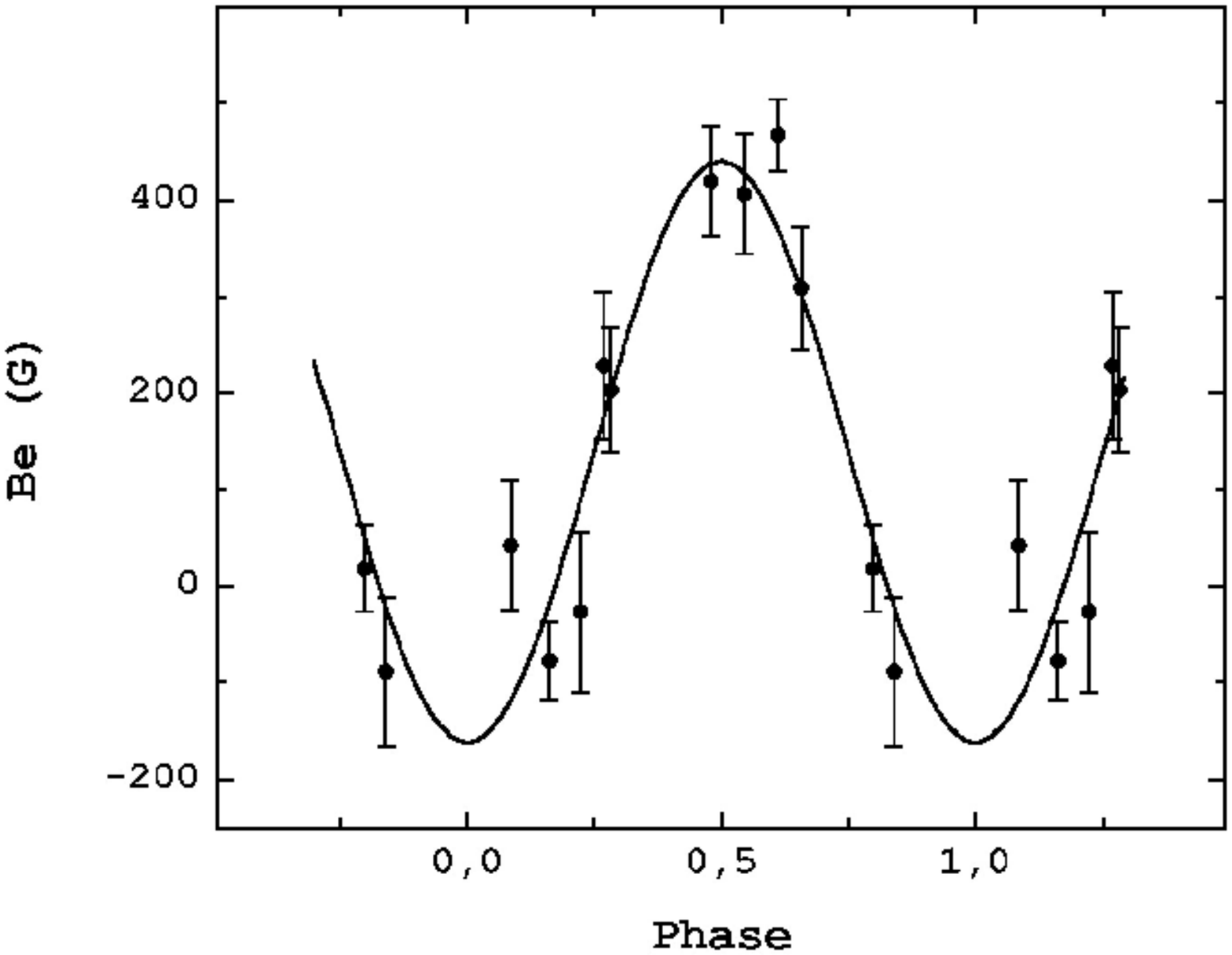}}
\vspace{-3.5mm}
\caption{ HD164492c (1) }
\label{fig:fig281}
\end{figure}

\begin{figure}
\resizebox{0.98\hsize}{!}{\includegraphics{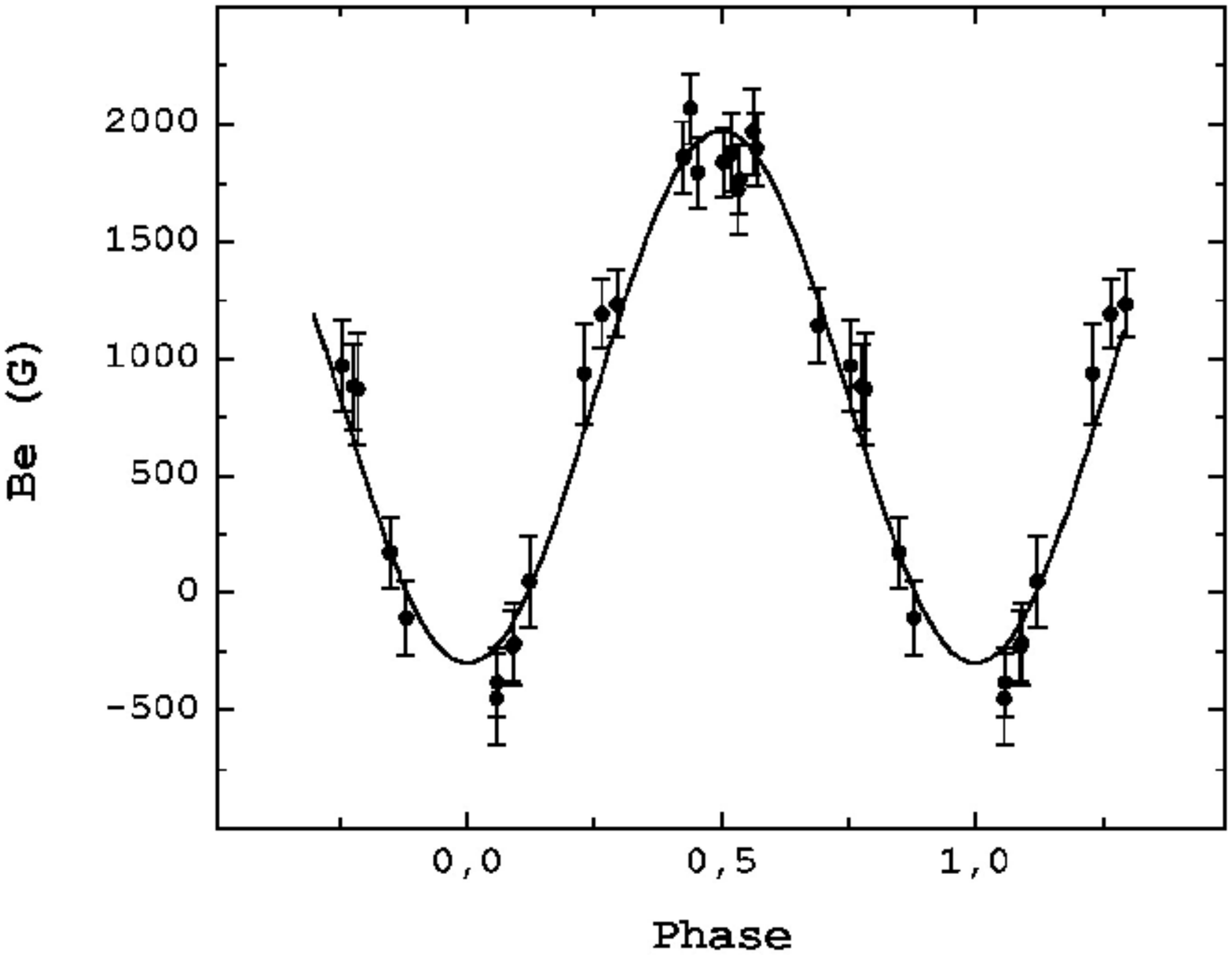}}
\vspace{-3.5mm}
\caption{ HD164492c (2) }
\label{fig:fig282}
\end{figure}

\begin{figure}
\resizebox{0.98\hsize}{!}{\includegraphics{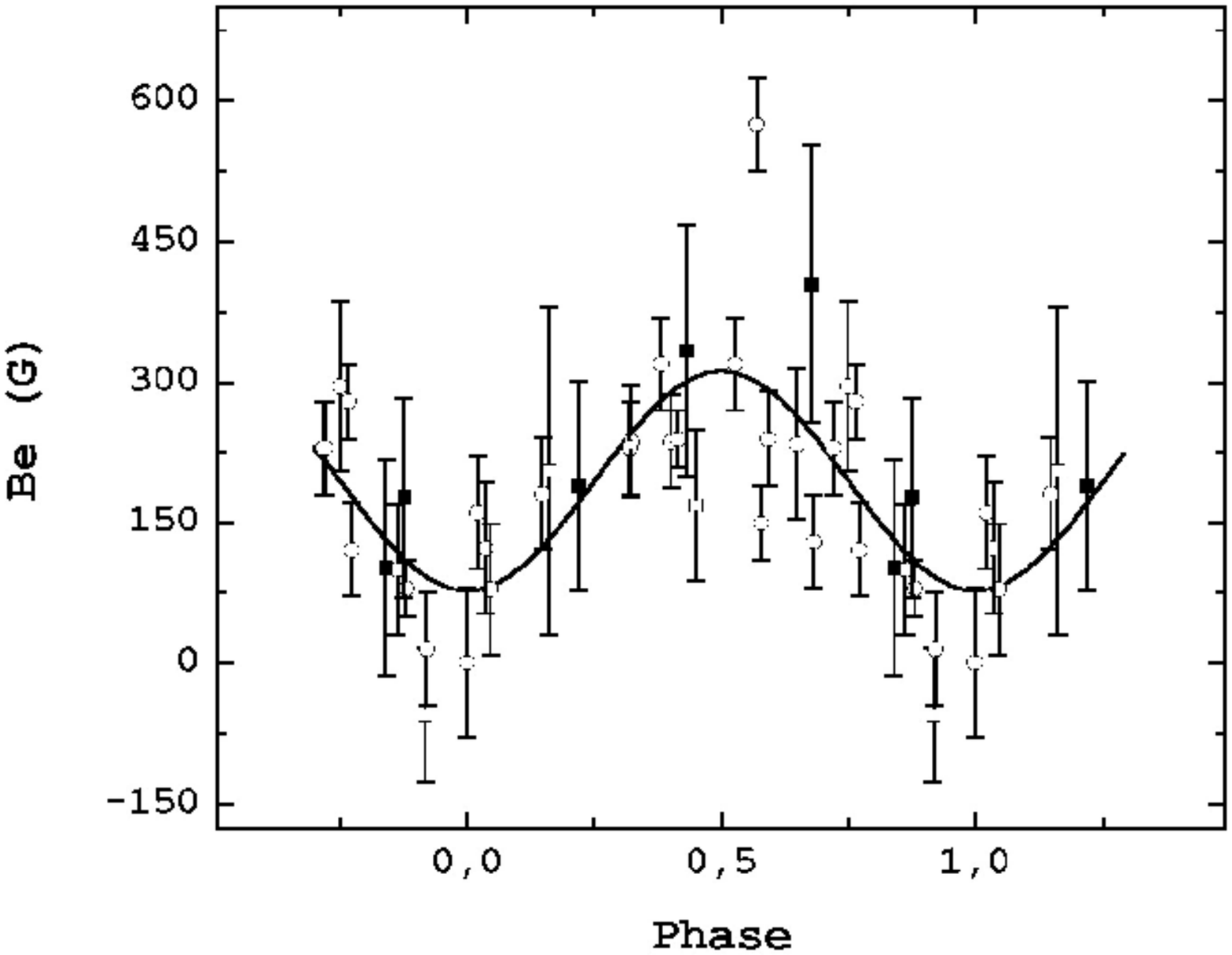}}
\vspace{-3.5mm}
\caption{ HD165474 }
\label{fig:fig283}
\end{figure}

\begin{figure}
\resizebox{0.98\hsize}{!}{\includegraphics{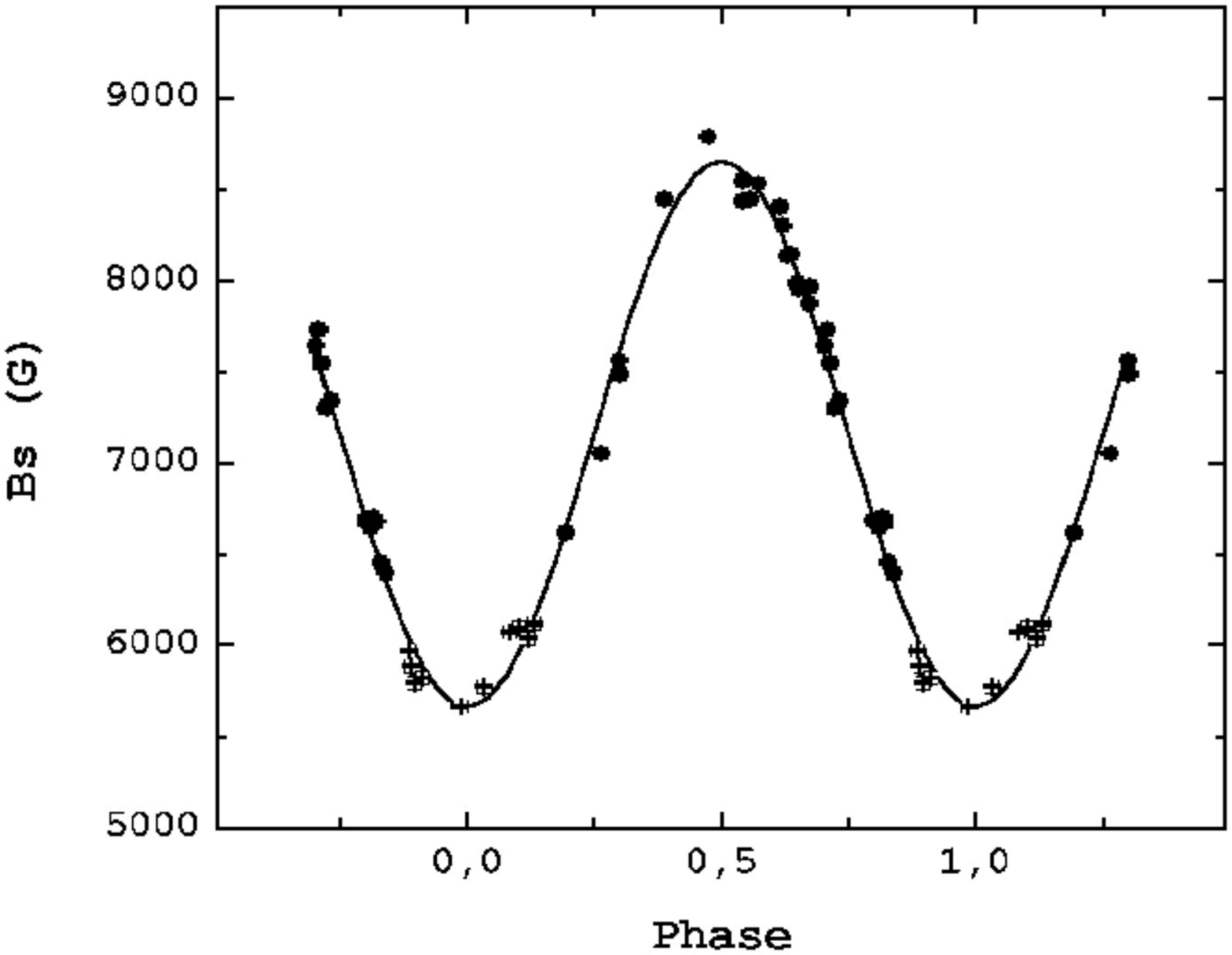}}
\vspace{-3.5mm}
\caption{ HD166473 (1) }
\label{fig:fig284}
\end{figure}

\begin{figure}
\resizebox{0.98\hsize}{!}{\includegraphics{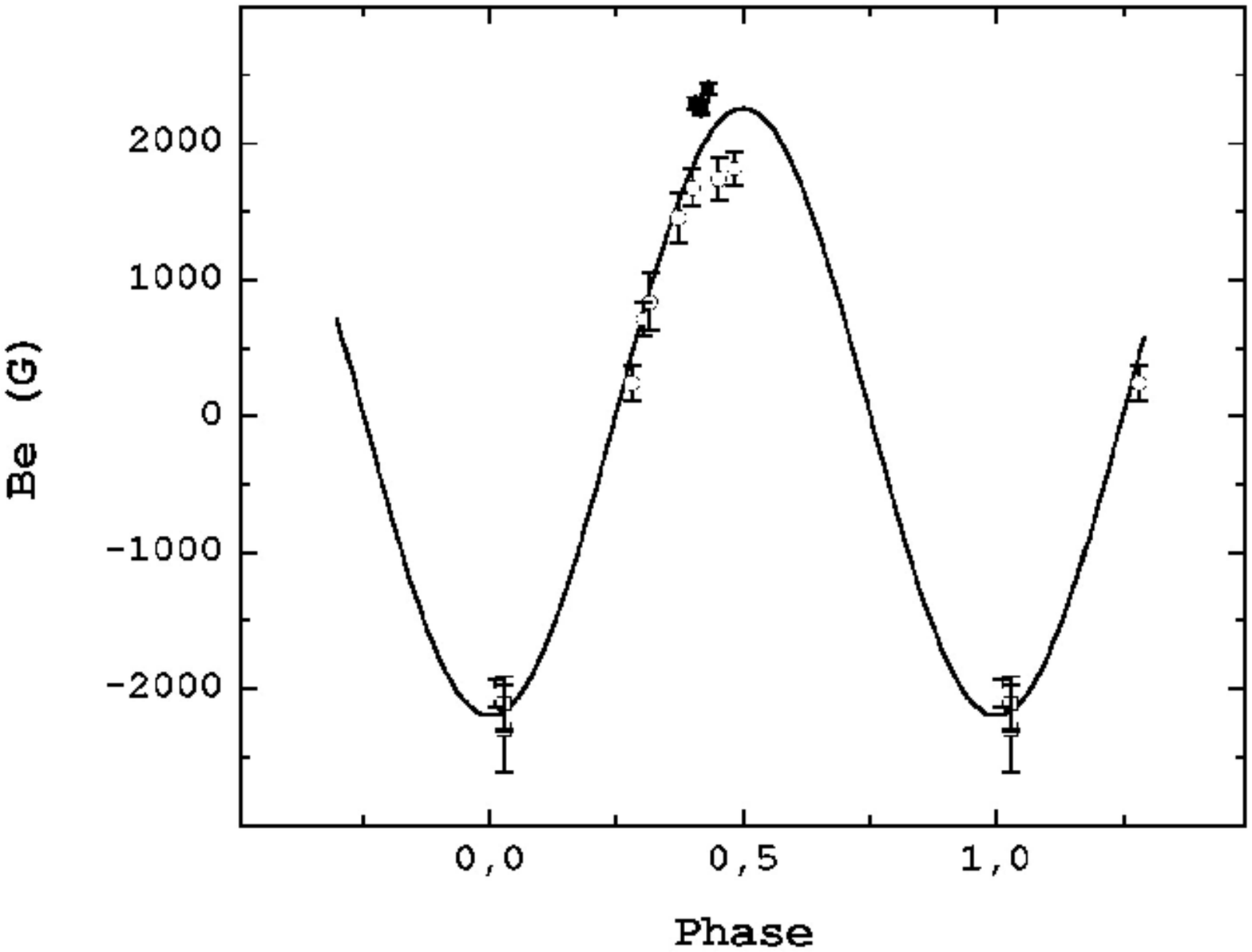}}
\vspace{-3.5mm}
\caption{ HD166473 (2) }
\label{fig:fig285}
\end{figure}

\clearpage
\newpage

\begin{figure}
\resizebox{0.98\hsize}{!}{\includegraphics{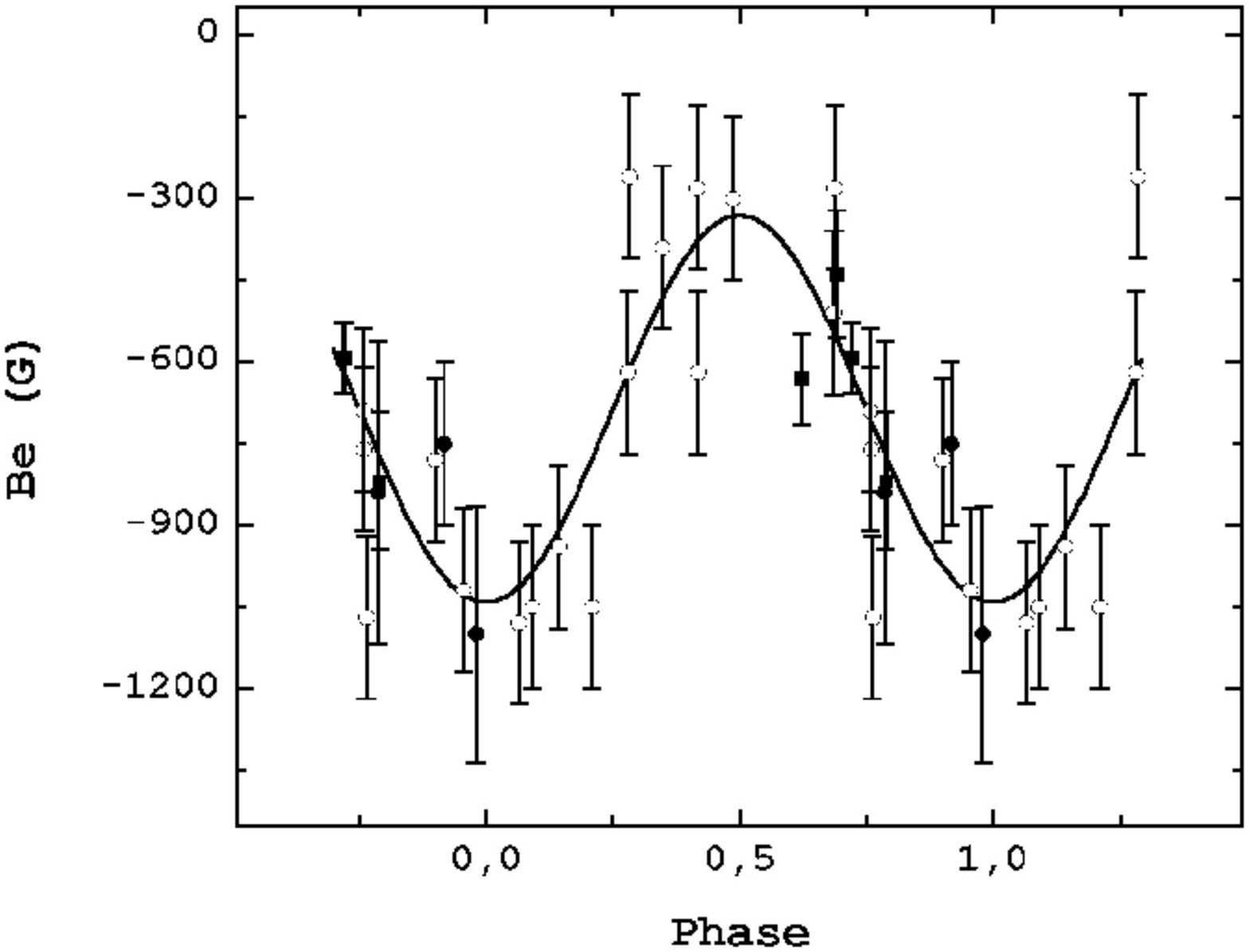}}
\vspace{-3.5mm}
\caption{ HD168733 }
\label{fig:fig286}
\end{figure}

\begin{figure}
\resizebox{0.98\hsize}{!}{\includegraphics{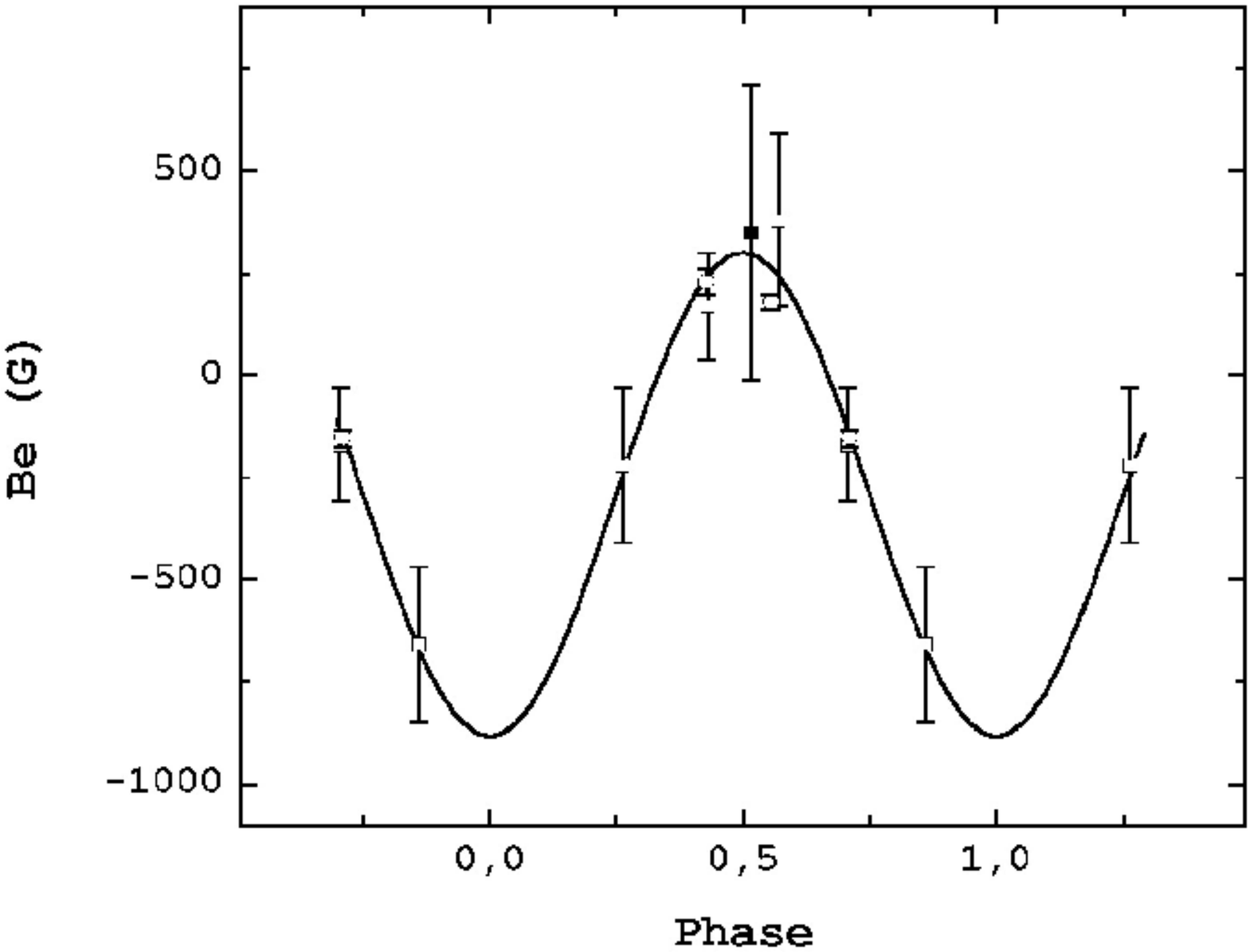}}
\vspace{-3.5mm}
\caption{ HD169842 }
\label{fig:fig287}
\end{figure}

\begin{figure}
\resizebox{0.98\hsize}{!}{\includegraphics{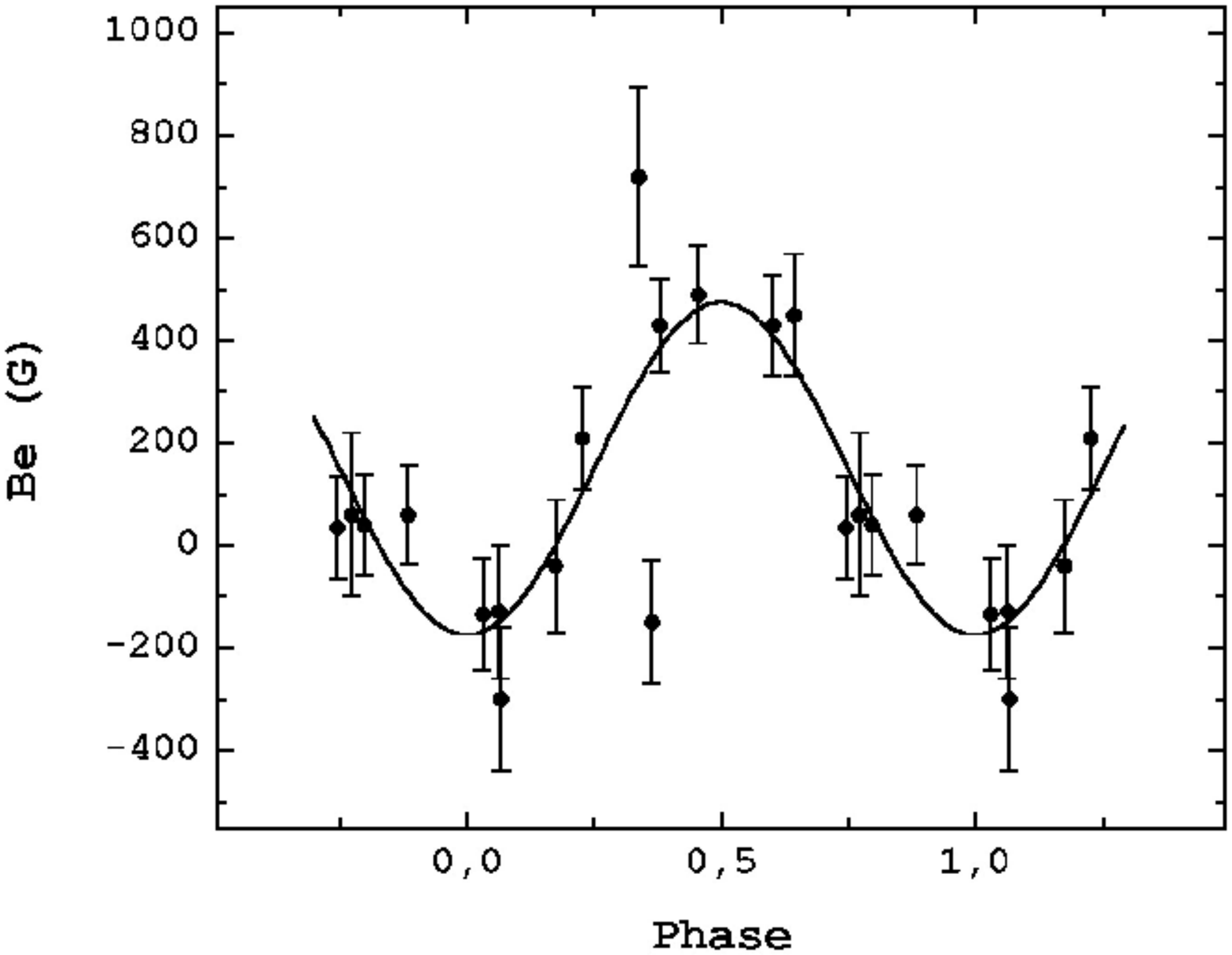}}
\vspace{-3.5mm}
\caption{ HD170000 (1) }
\label{fig:fig288}
\end{figure}

\begin{figure}
\resizebox{0.98\hsize}{!}{\includegraphics{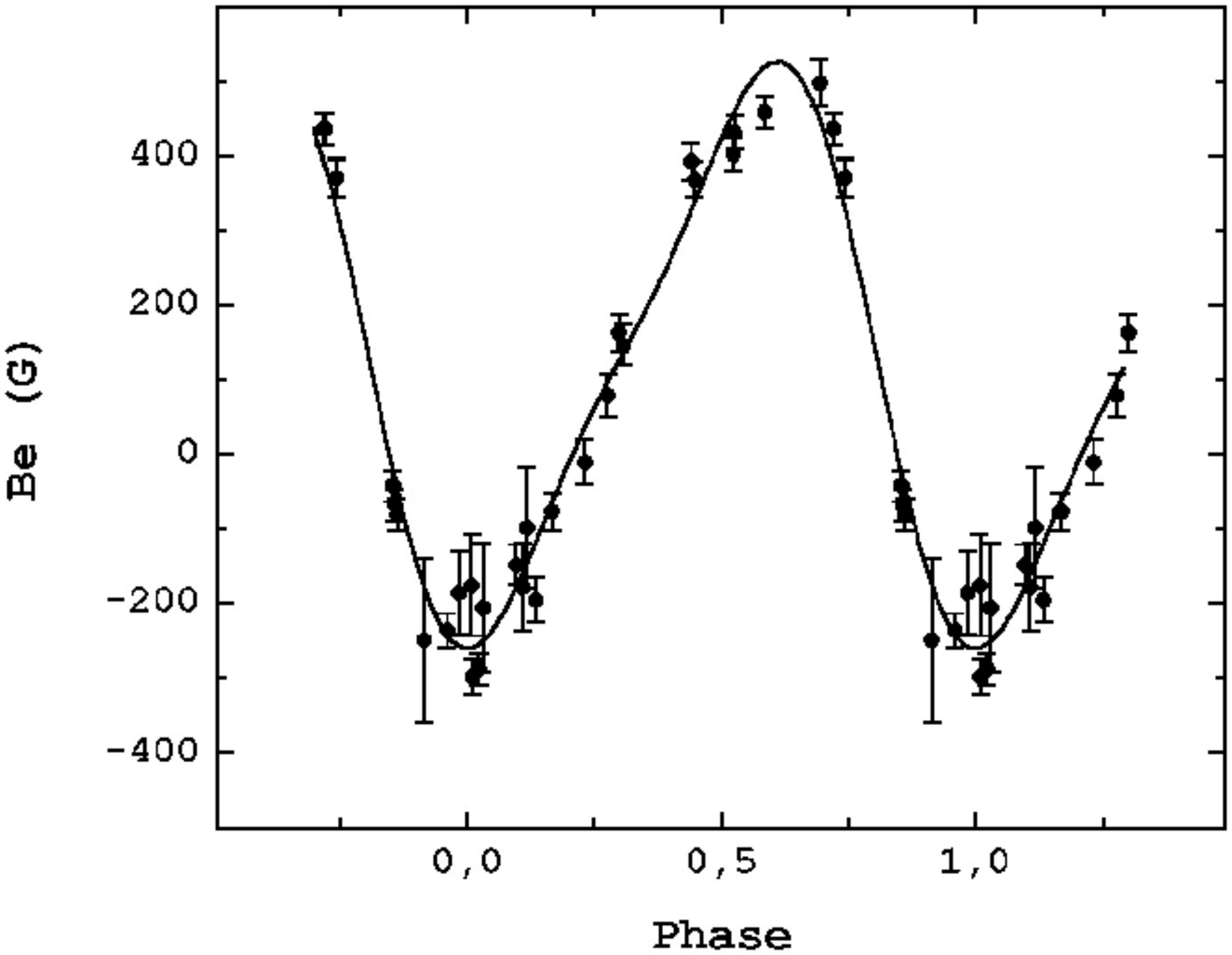}}
\vspace{-3.5mm}
\caption{ HD170000 (2) }
\label{fig:fig288}
\end{figure}

\begin{figure}
\resizebox{0.98\hsize}{!}{\includegraphics{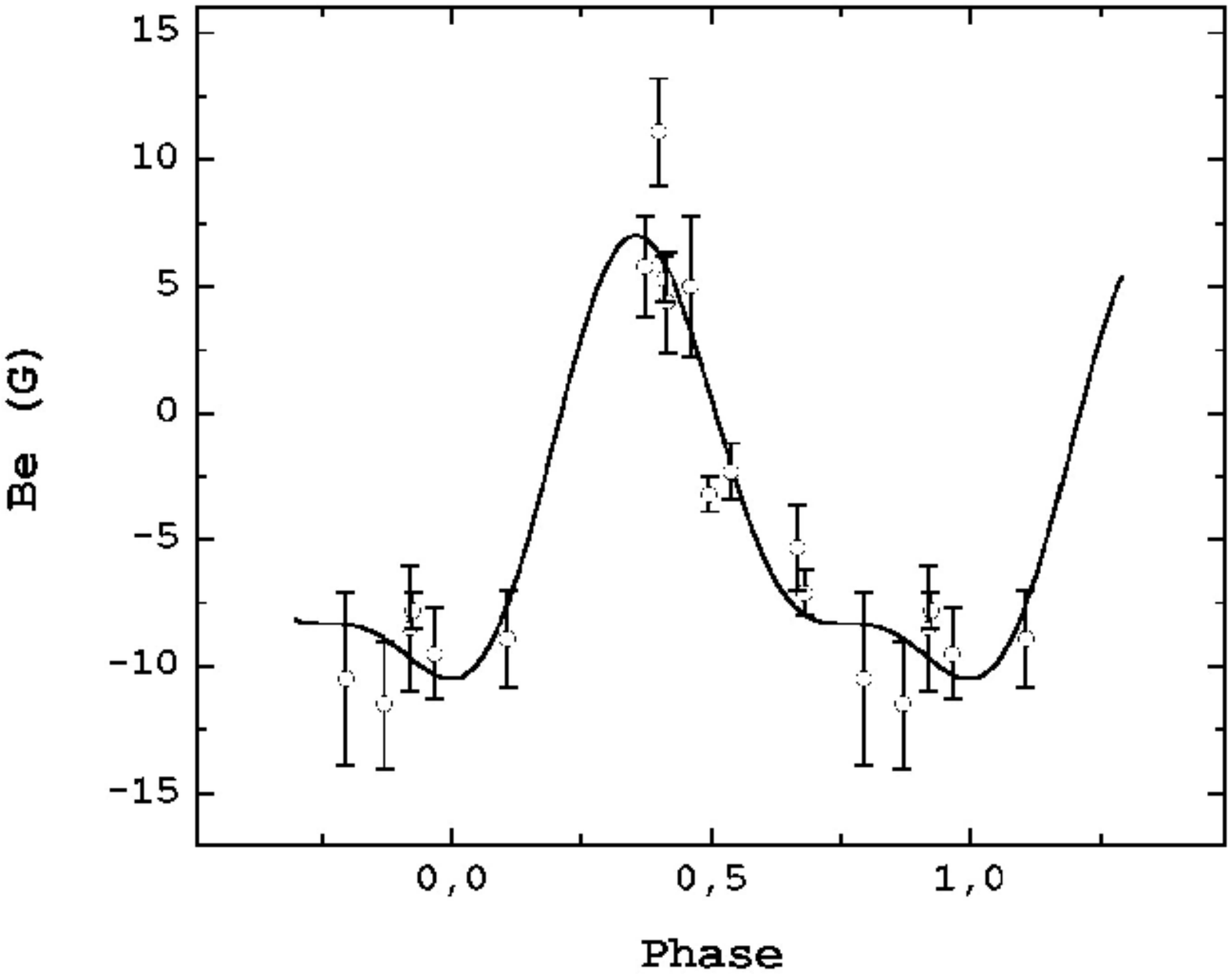}}
\vspace{-3.5mm}
\caption{ HD170153 }
\label{fig:fig289}
\end{figure}

\begin{figure}
\resizebox{0.98\hsize}{!}{\includegraphics{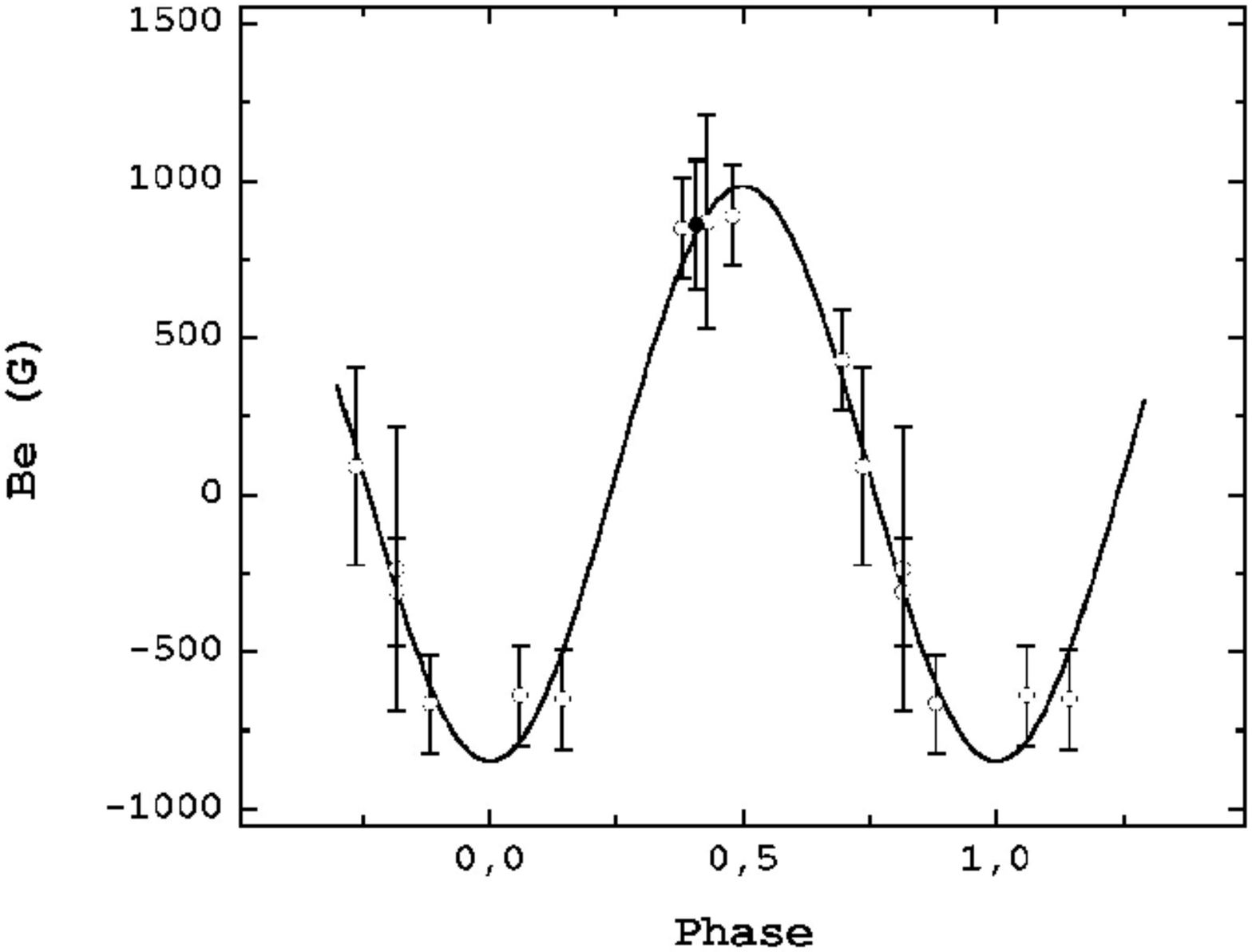}}
\vspace{-3.5mm}
\caption{ HD170397 }
\label{fig:fig291}
\end{figure}

\clearpage
\newpage

\begin{figure}
\resizebox{0.98\hsize}{!}{\includegraphics{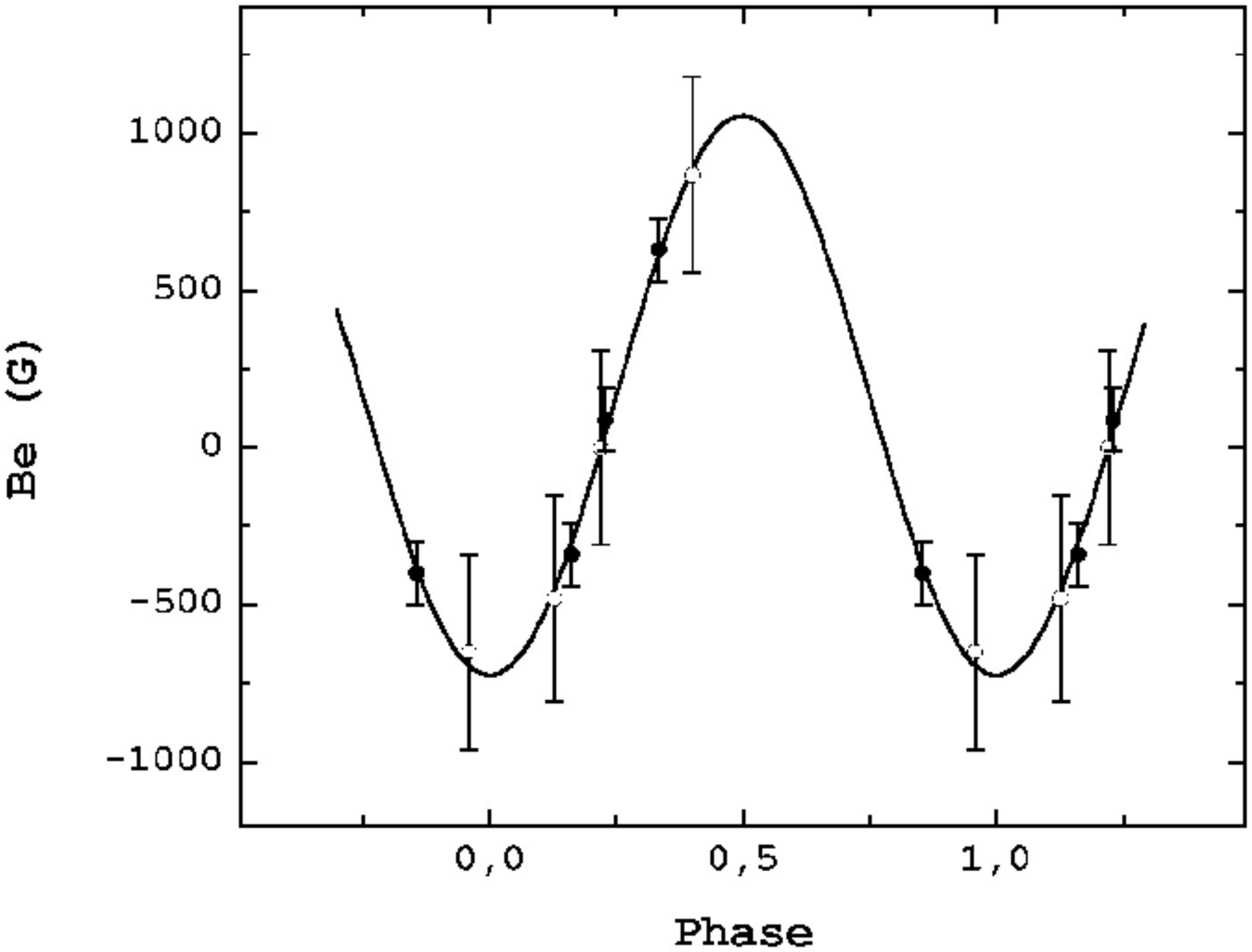}}
\vspace{-3.5mm}
\caption{ HD170973 }
\label{fig:fig292}
\end{figure}

\begin{figure}
\resizebox{0.98\hsize}{!}{\includegraphics{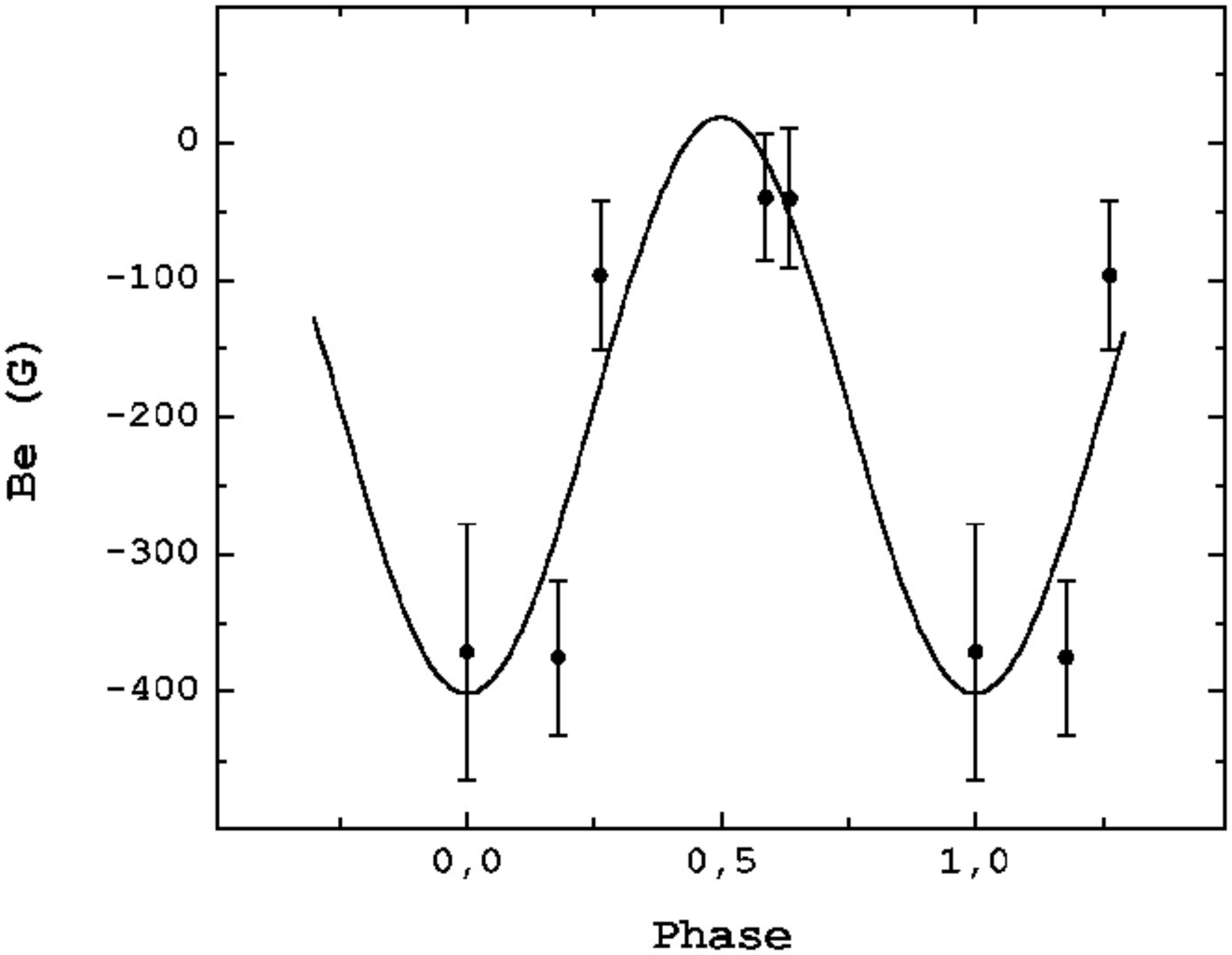}}
\vspace{-3.5mm}
\caption{ HD171586 }
\label{fig:fig293}
\end{figure}

\begin{figure}
\resizebox{0.98\hsize}{!}{\includegraphics{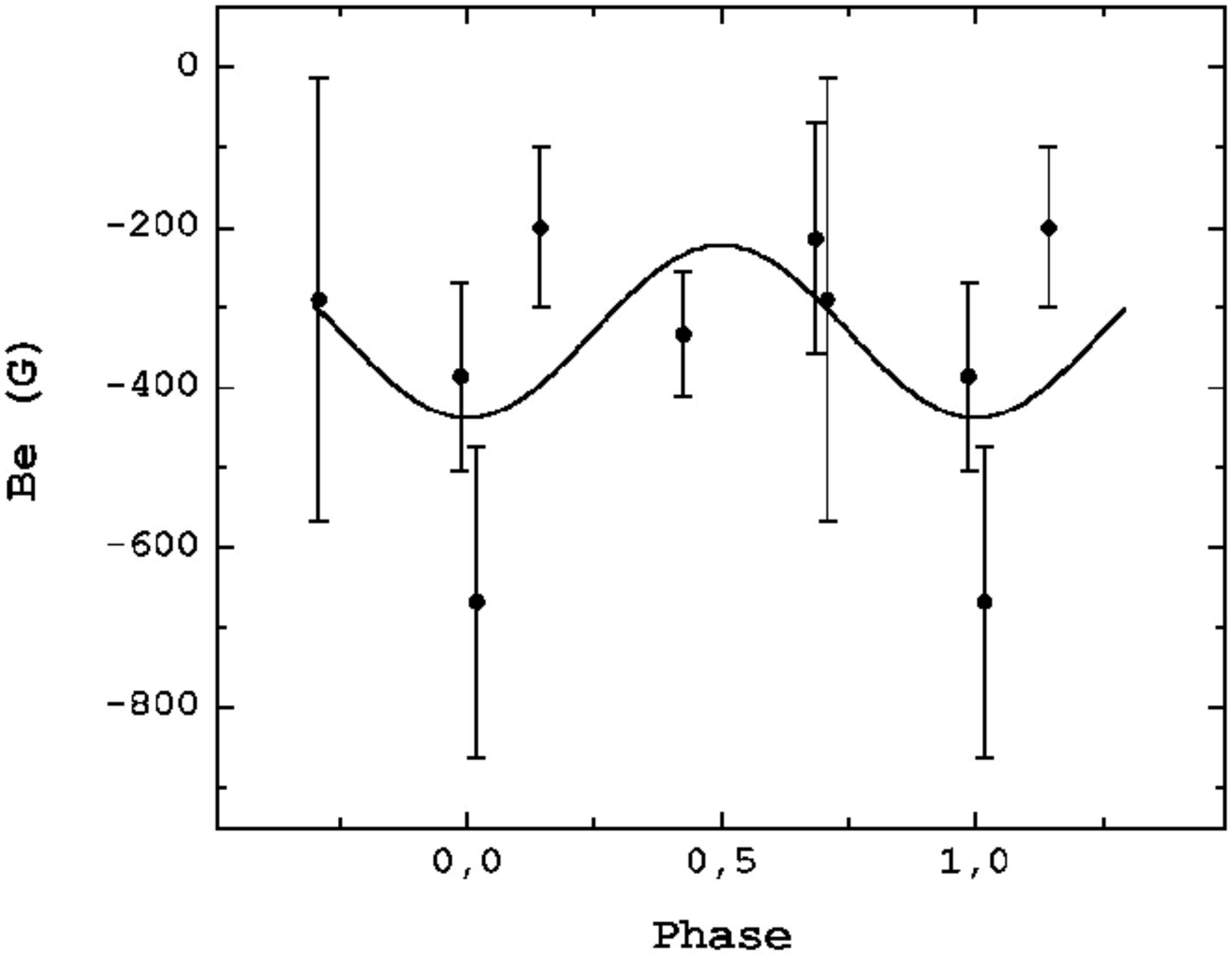}}
\vspace{-3.5mm}
\caption{ HD171782 }
\label{fig:fig294}
\end{figure}

\begin{figure}
\resizebox{0.98\hsize}{!}{\includegraphics{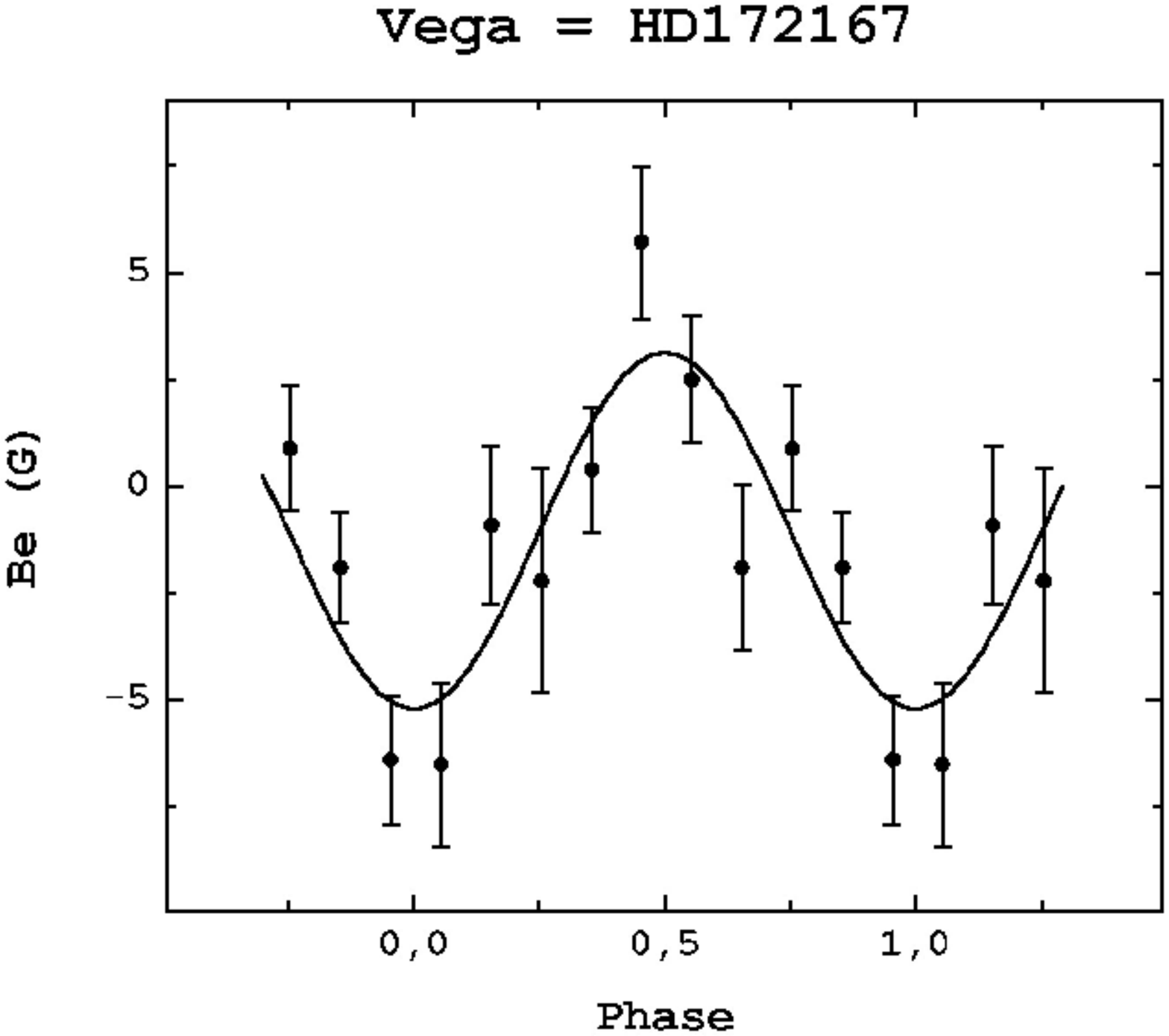}}
\vspace{-3.5mm}
\caption{ HD172167 }
\label{fig:fig295}
\end{figure}

\begin{figure}
\resizebox{0.98\hsize}{!}{\includegraphics{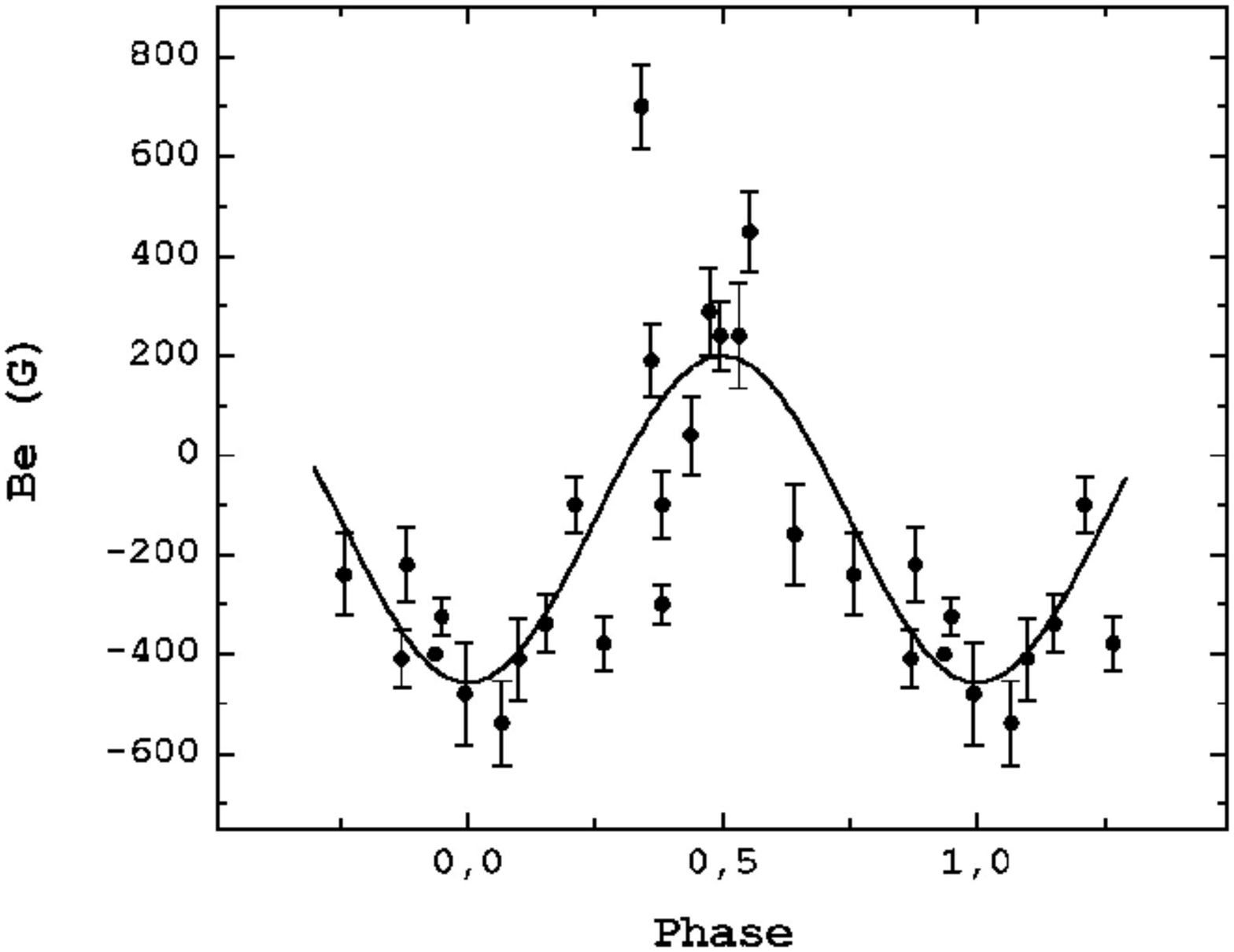}}
\vspace{-3.5mm}
\caption{ HD173650 }
\label{fig:fig296}
\end{figure}

\begin{figure}
\resizebox{0.98\hsize}{!}{\includegraphics{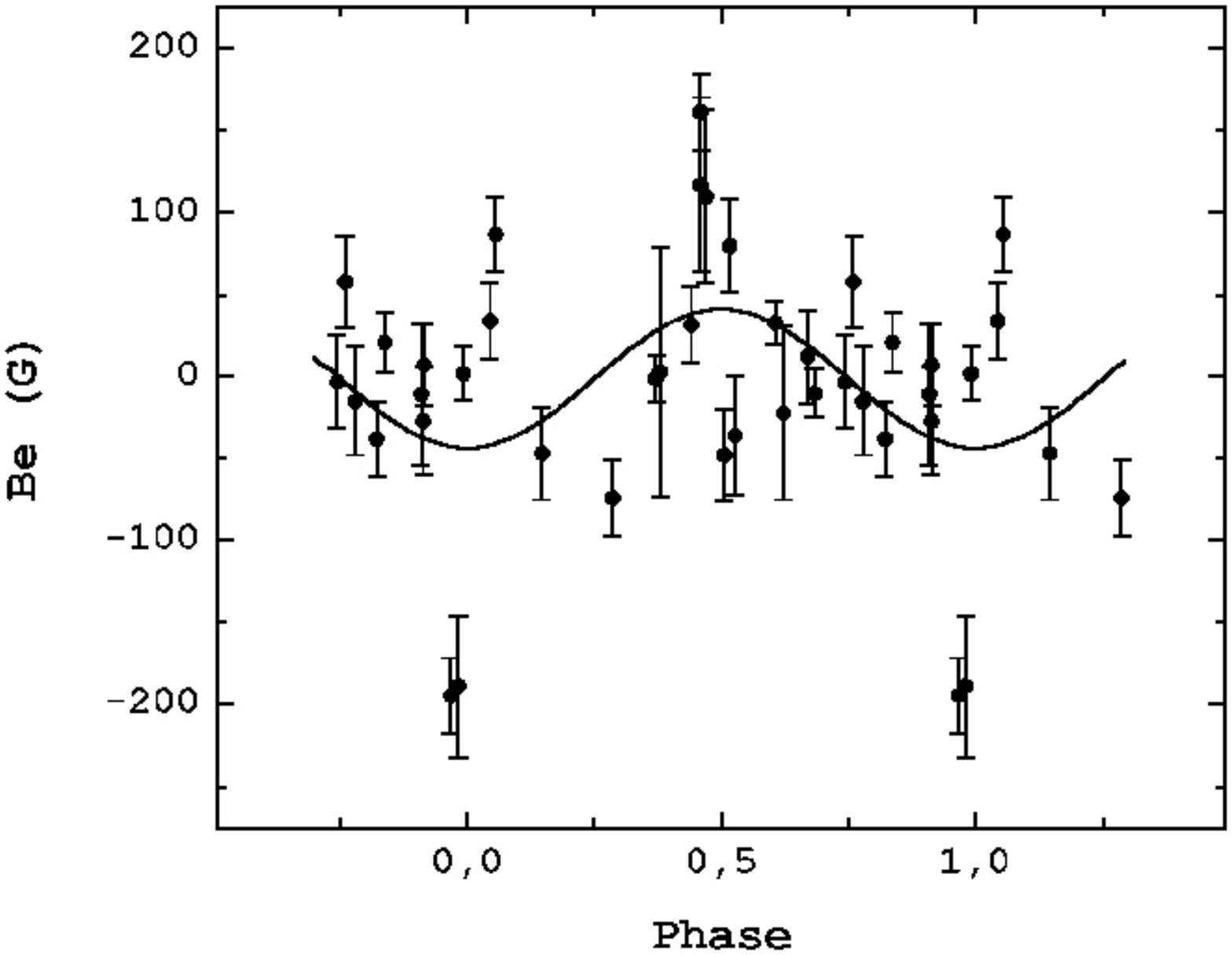}}
\vspace{-3.5mm}
\caption{ HD174638 }
\label{fig:fig297}
\end{figure}

\clearpage
\newpage

\begin{figure}
\resizebox{0.98\hsize}{!}{\includegraphics{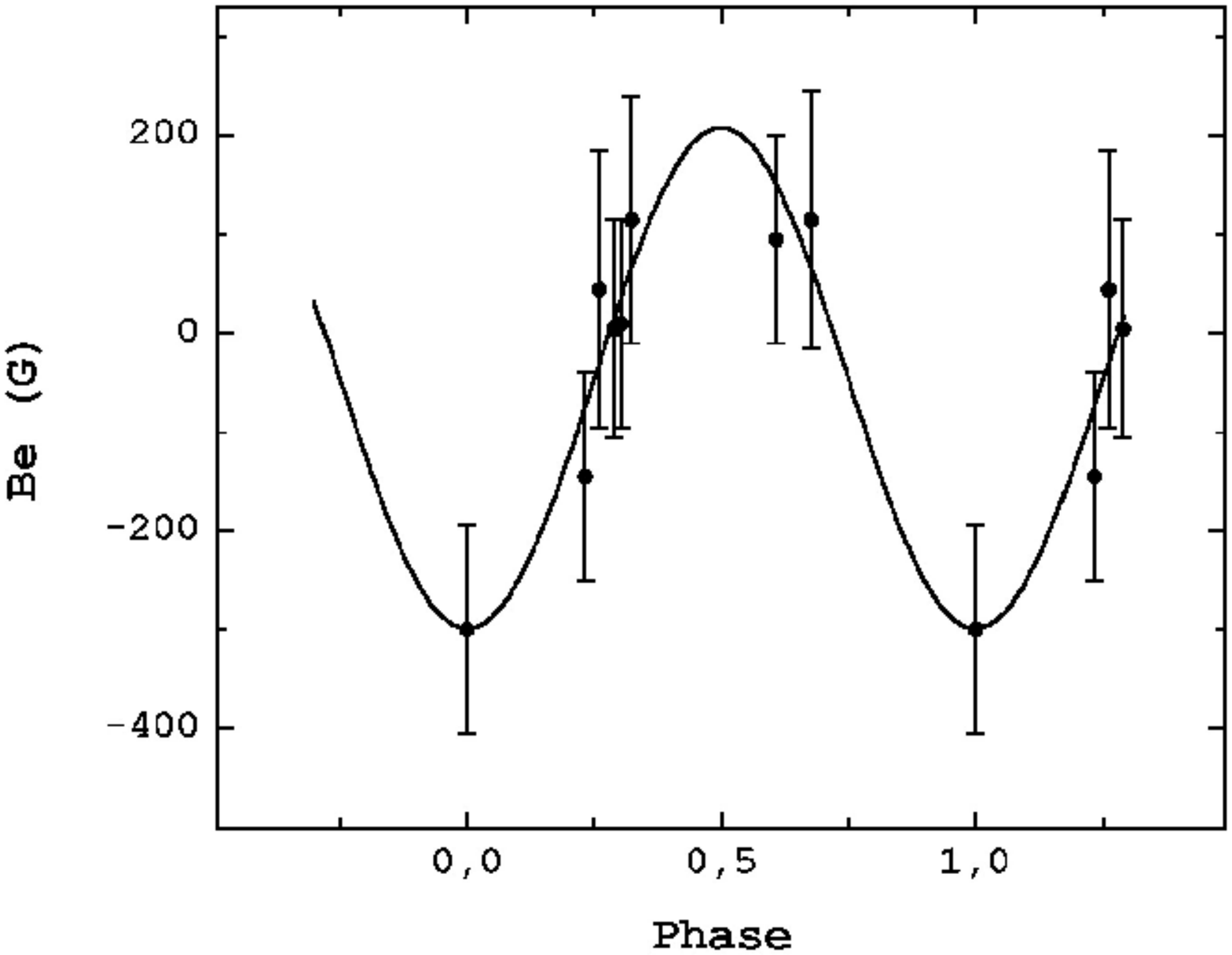}}
\vspace{-3.5mm}
\caption{ HD175156 }
\label{fig:fig298}
\end{figure}

\begin{figure}
\resizebox{0.98\hsize}{!}{\includegraphics{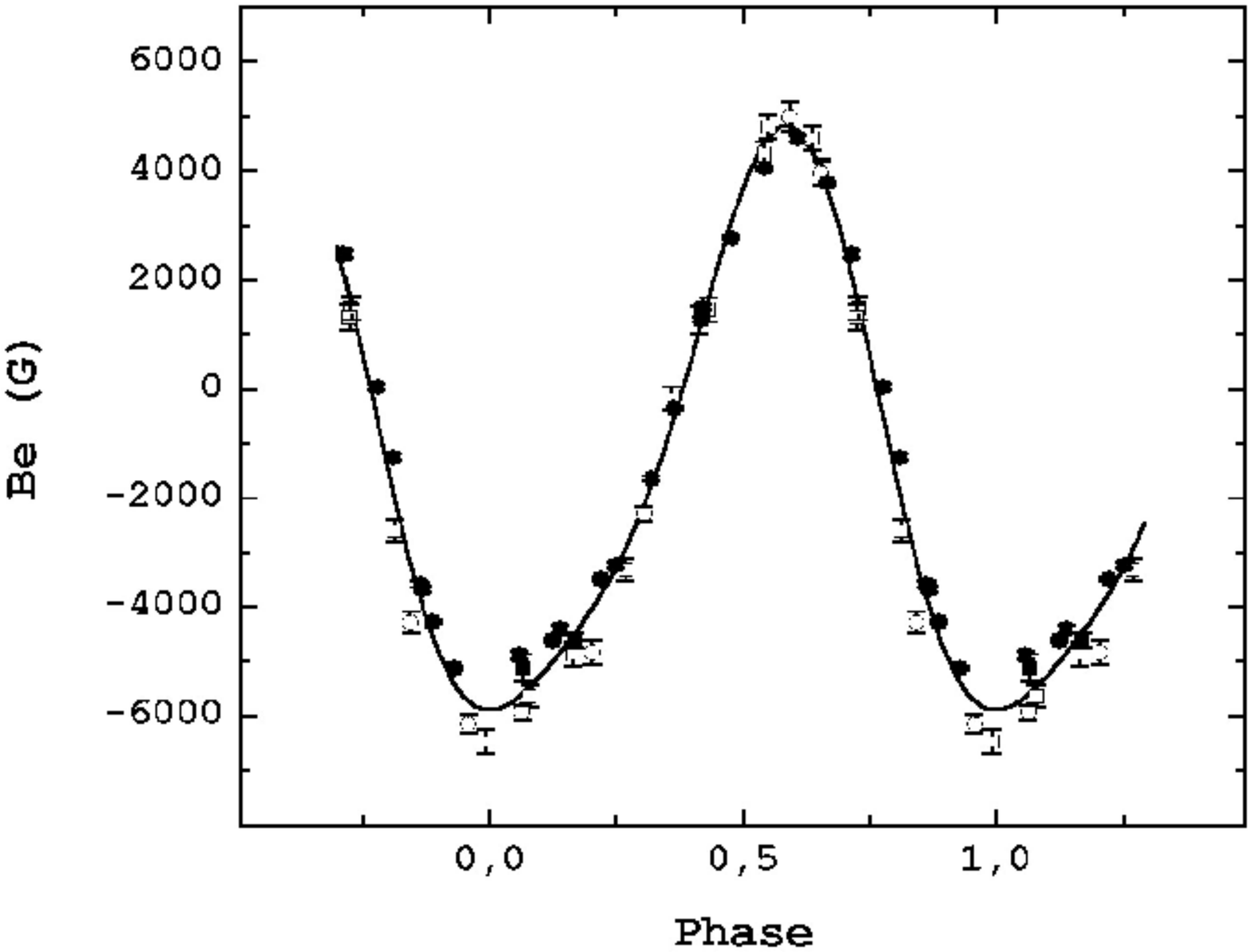}}
\vspace{-3.5mm}
\caption{ HD175362 (1) }
\label{fig:fig299}
\end{figure}

\begin{figure}
\resizebox{0.98\hsize}{!}{\includegraphics{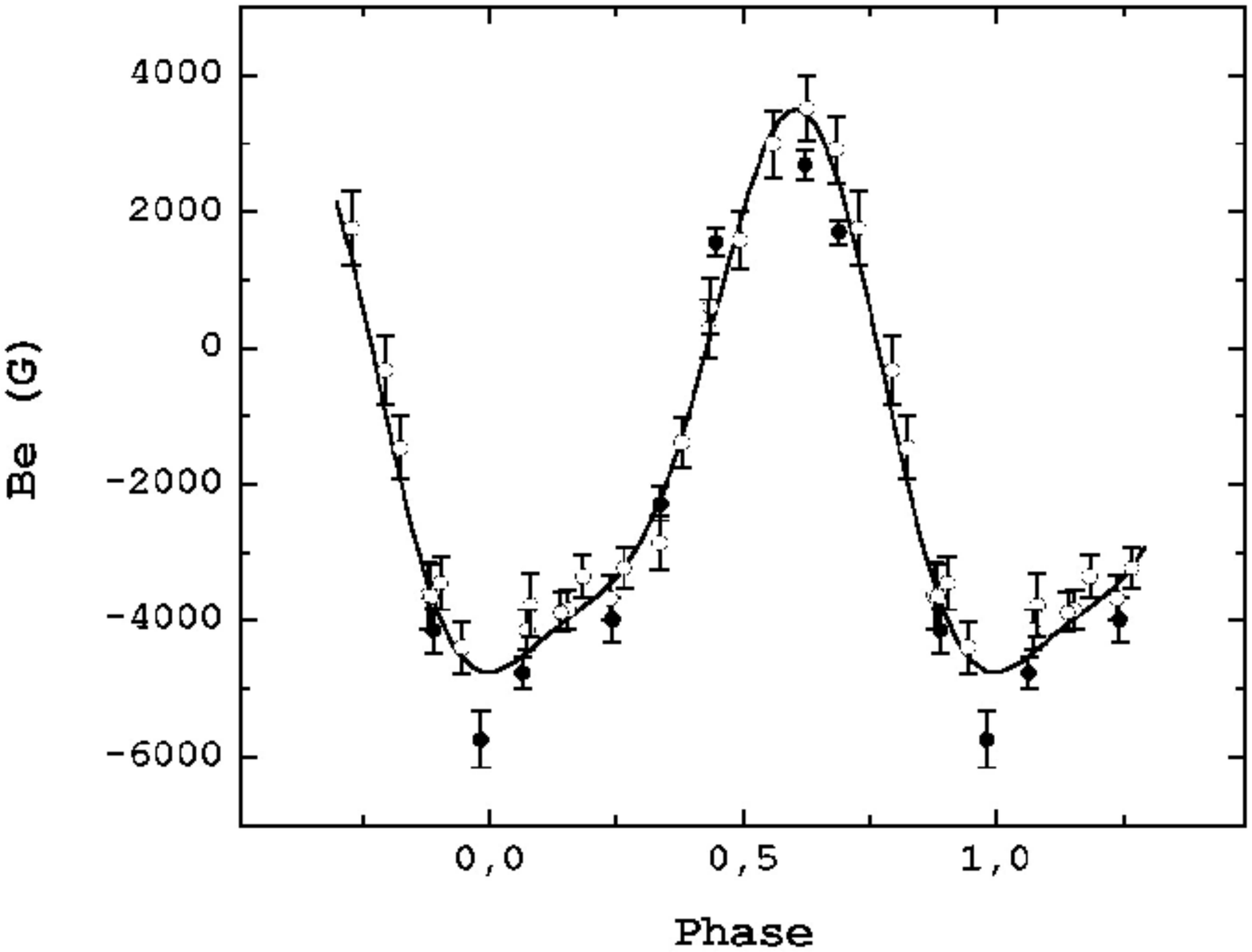}}
\vspace{-3.5mm}
\caption{ HD175362 (2) }
\label{fig:fig300}
\end{figure}

\begin{figure}
\resizebox{0.98\hsize}{!}{\includegraphics{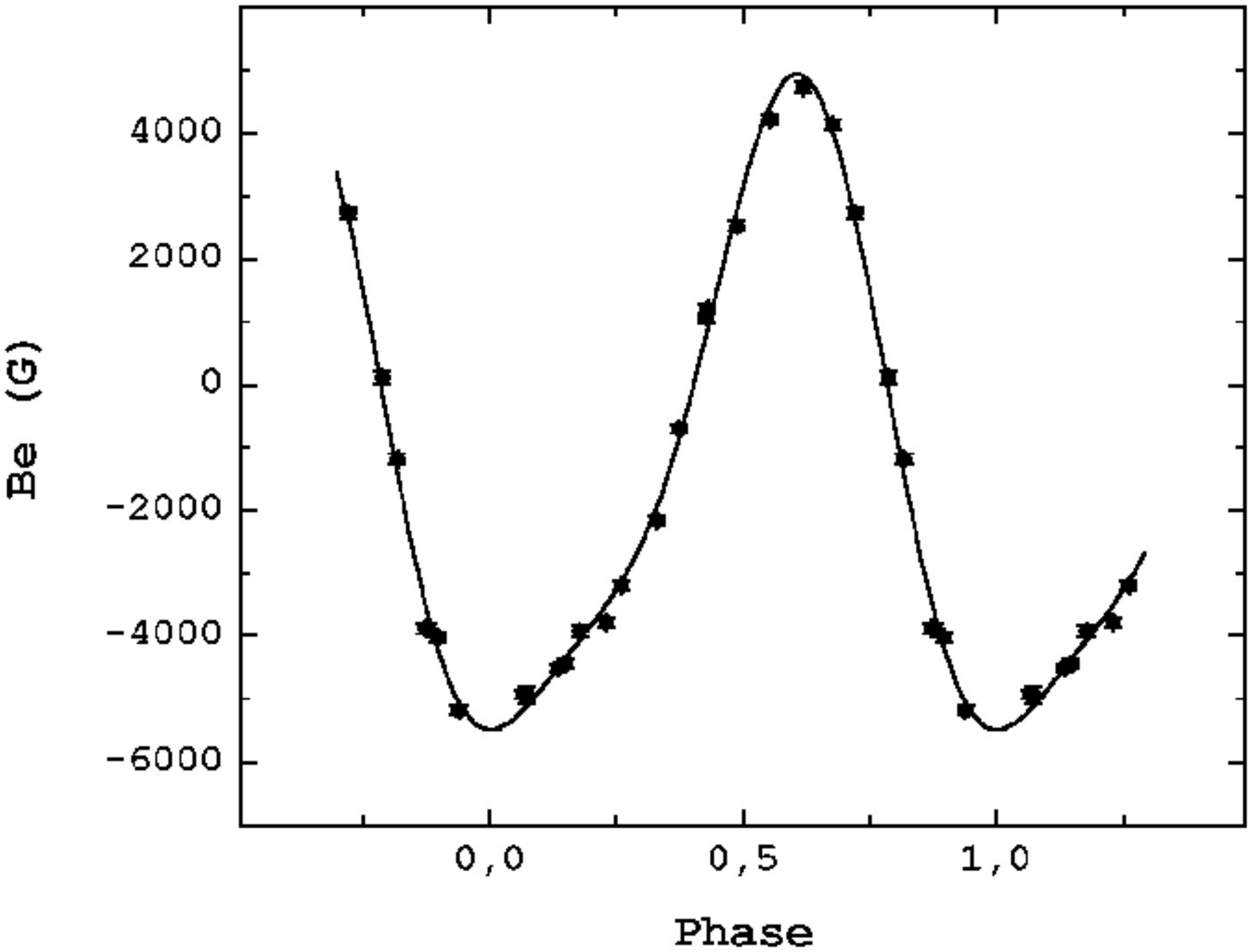}}
\vspace{-3.5mm}
\caption{ HD175362 (3) }
\label{fig:fig300}
\end{figure}

\begin{figure}
\resizebox{0.98\hsize}{!}{\includegraphics{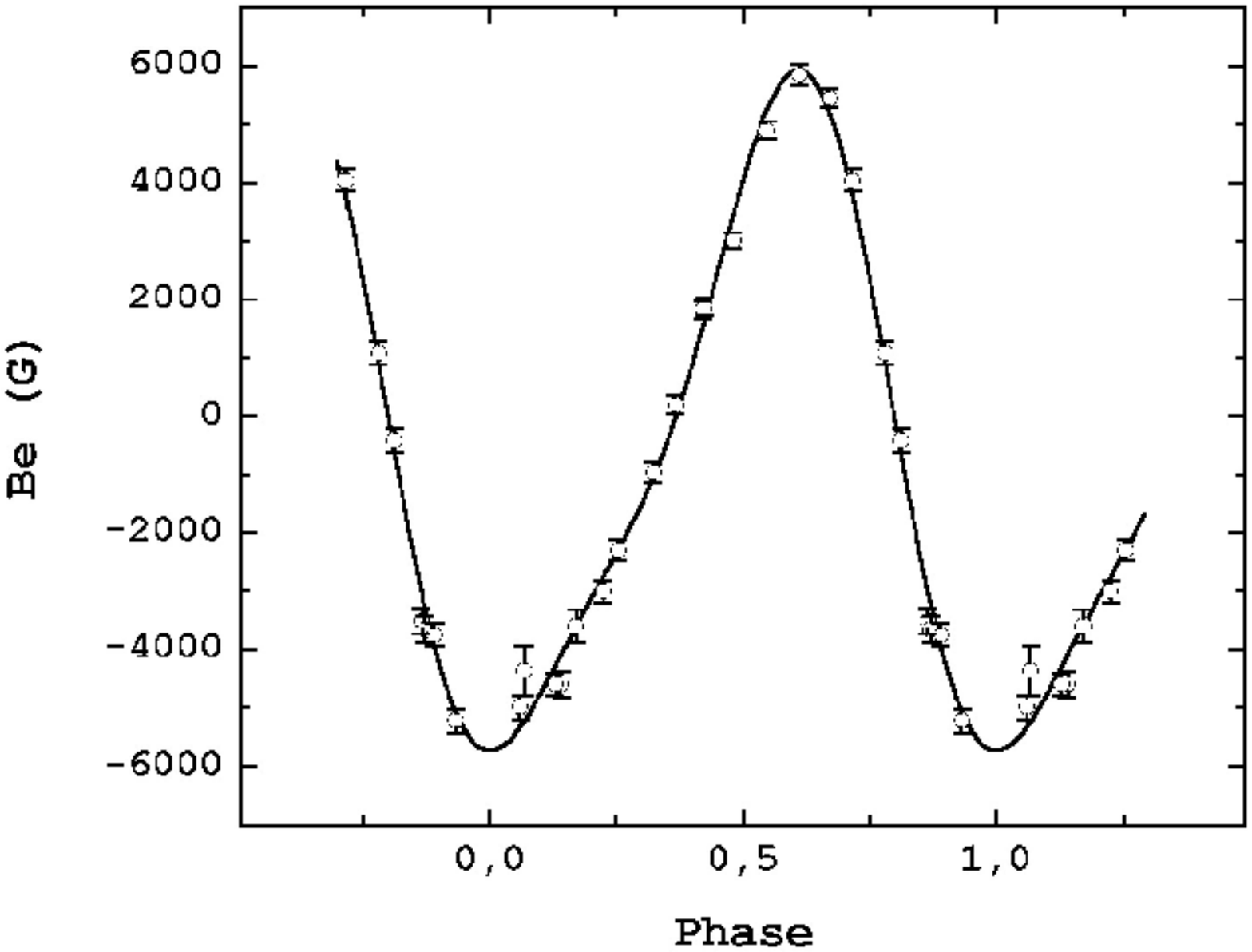}}
\vspace{-3.5mm}
\caption{ HD175362 (4) }
\label{fig:fig300}
\end{figure}

\begin{figure}
\resizebox{0.98\hsize}{!}{\includegraphics{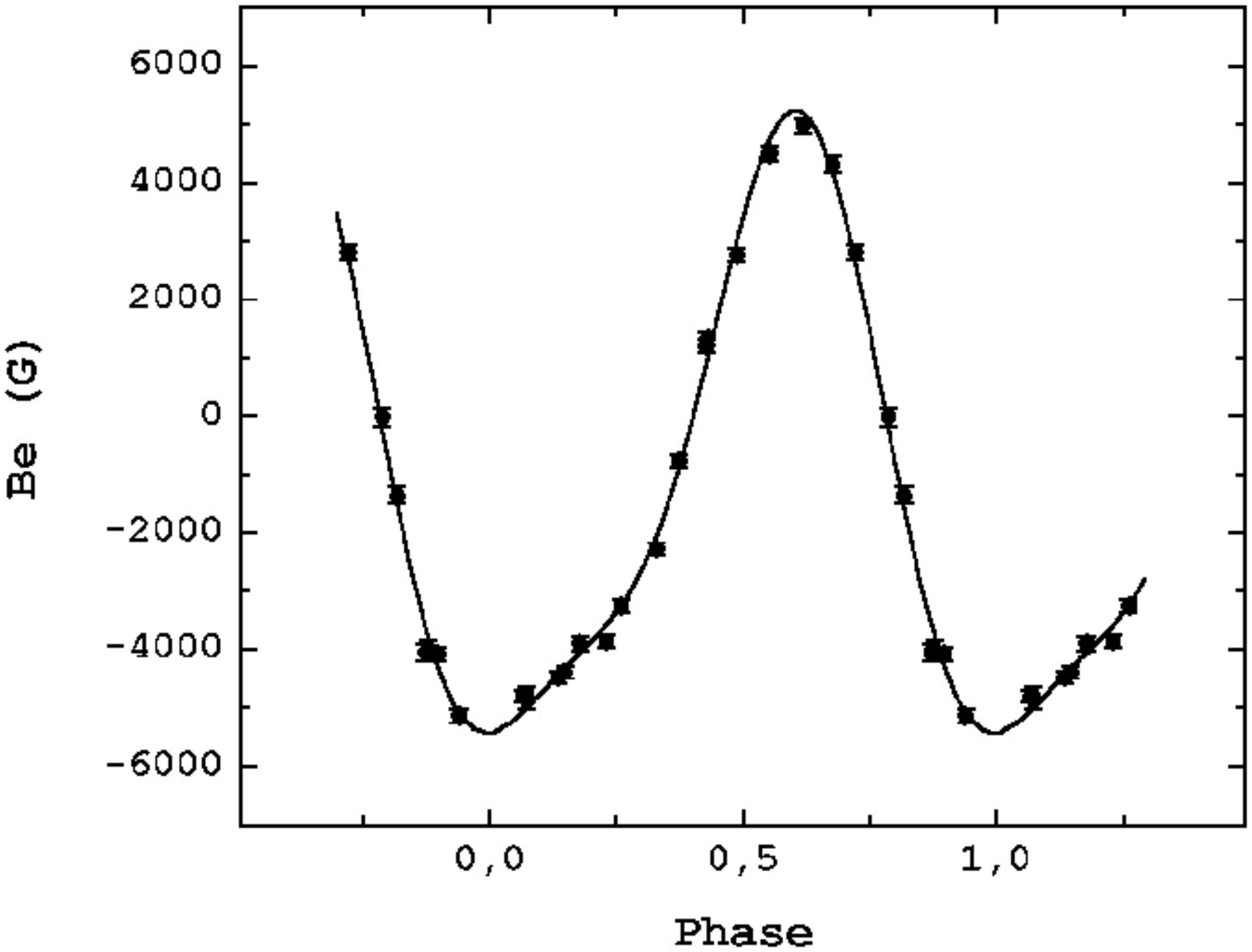}}
\vspace{-3.5mm}
\caption{ HD175362 (5) }
\label{fig:fig300}
\end{figure}

\clearpage
\newpage

\begin{figure}
\resizebox{0.98\hsize}{!}{\includegraphics{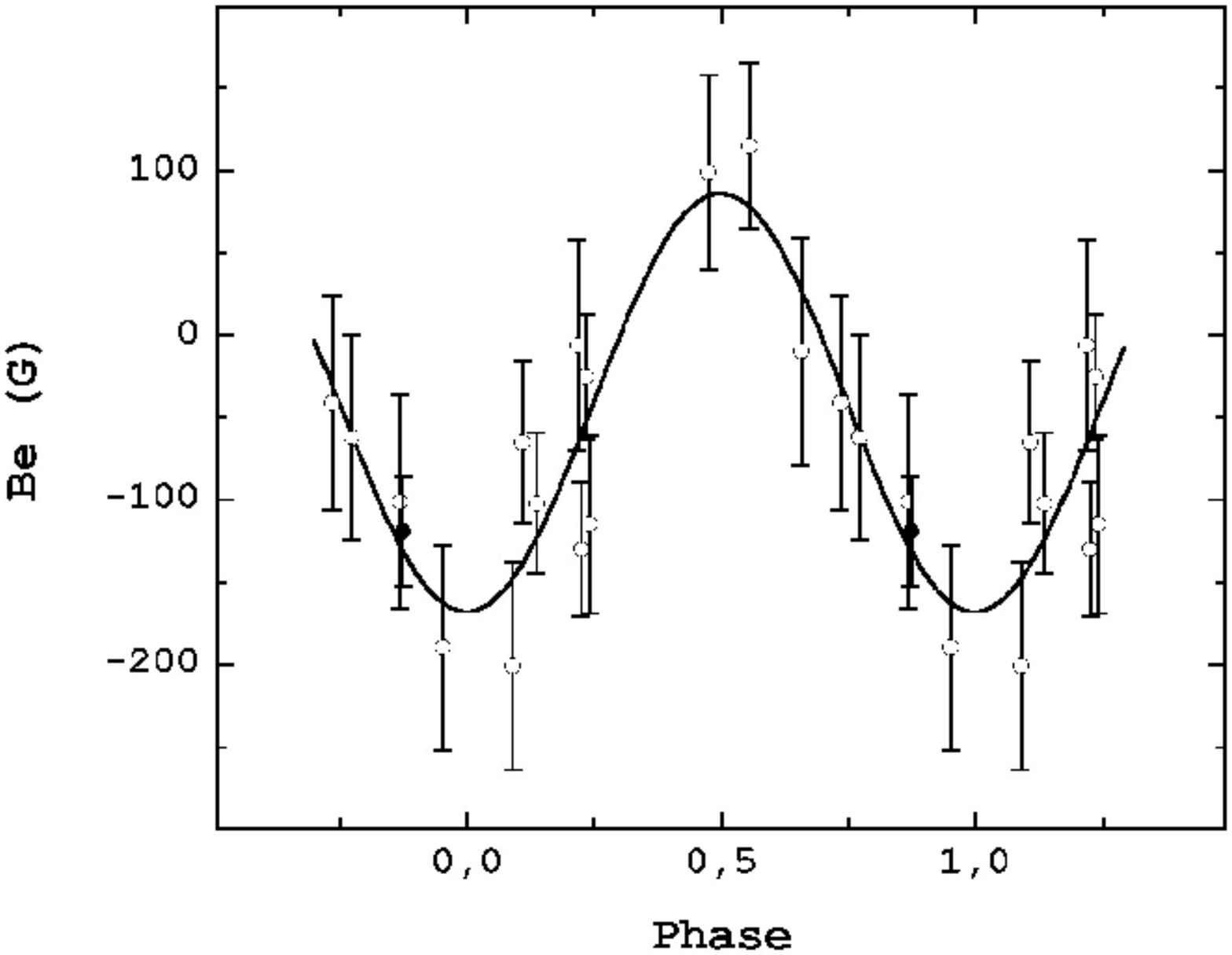}}
\vspace{-3.5mm}
\caption{ HD176386 (1) }
\label{fig:fig301}
\end{figure}

\begin{figure}
\resizebox{0.98\hsize}{!}{\includegraphics{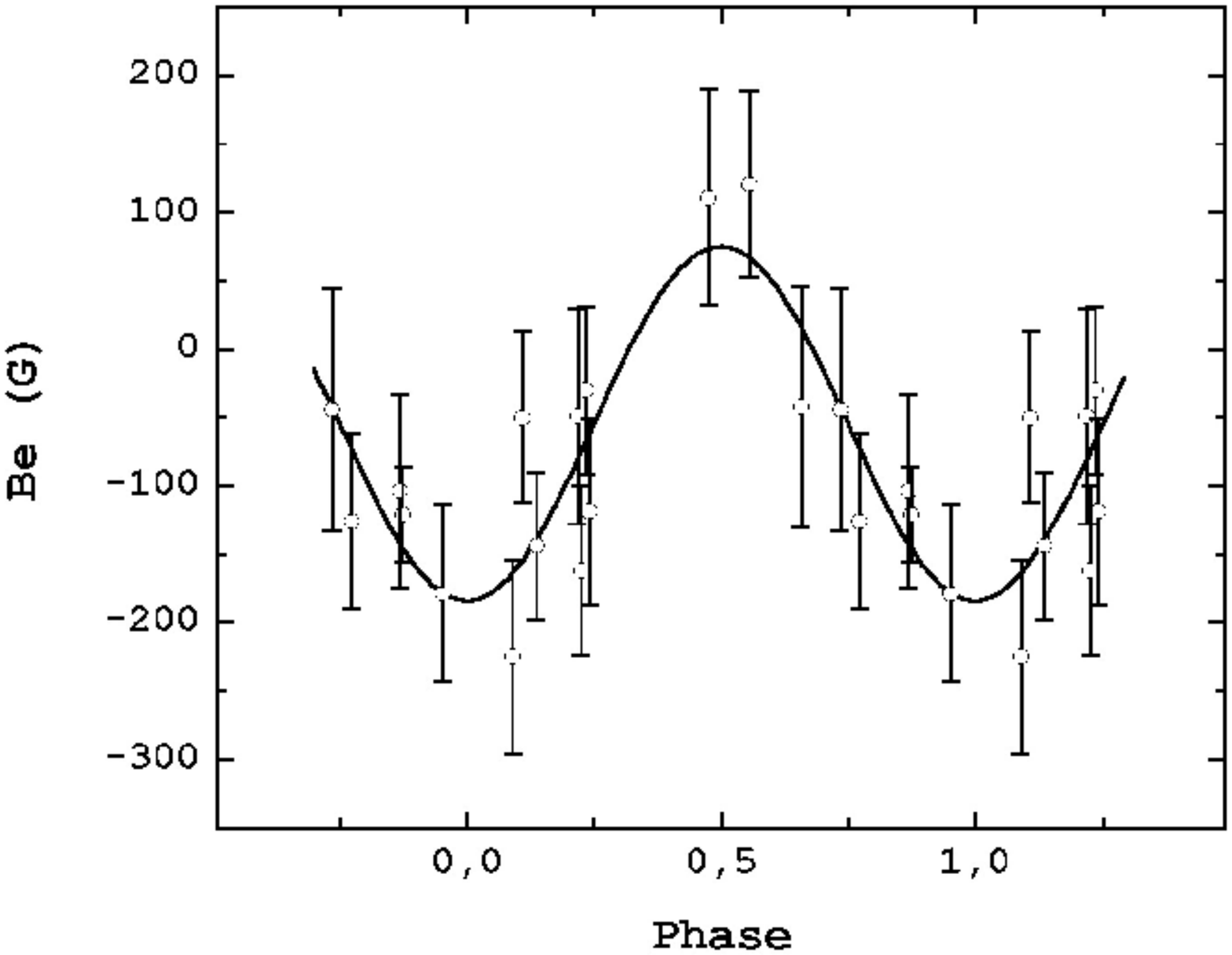}}
\vspace{-3.5mm}
\caption{ HD176386 (2) }
\label{fig:fig302}
\end{figure}

\begin{figure}
\resizebox{0.98\hsize}{!}{\includegraphics{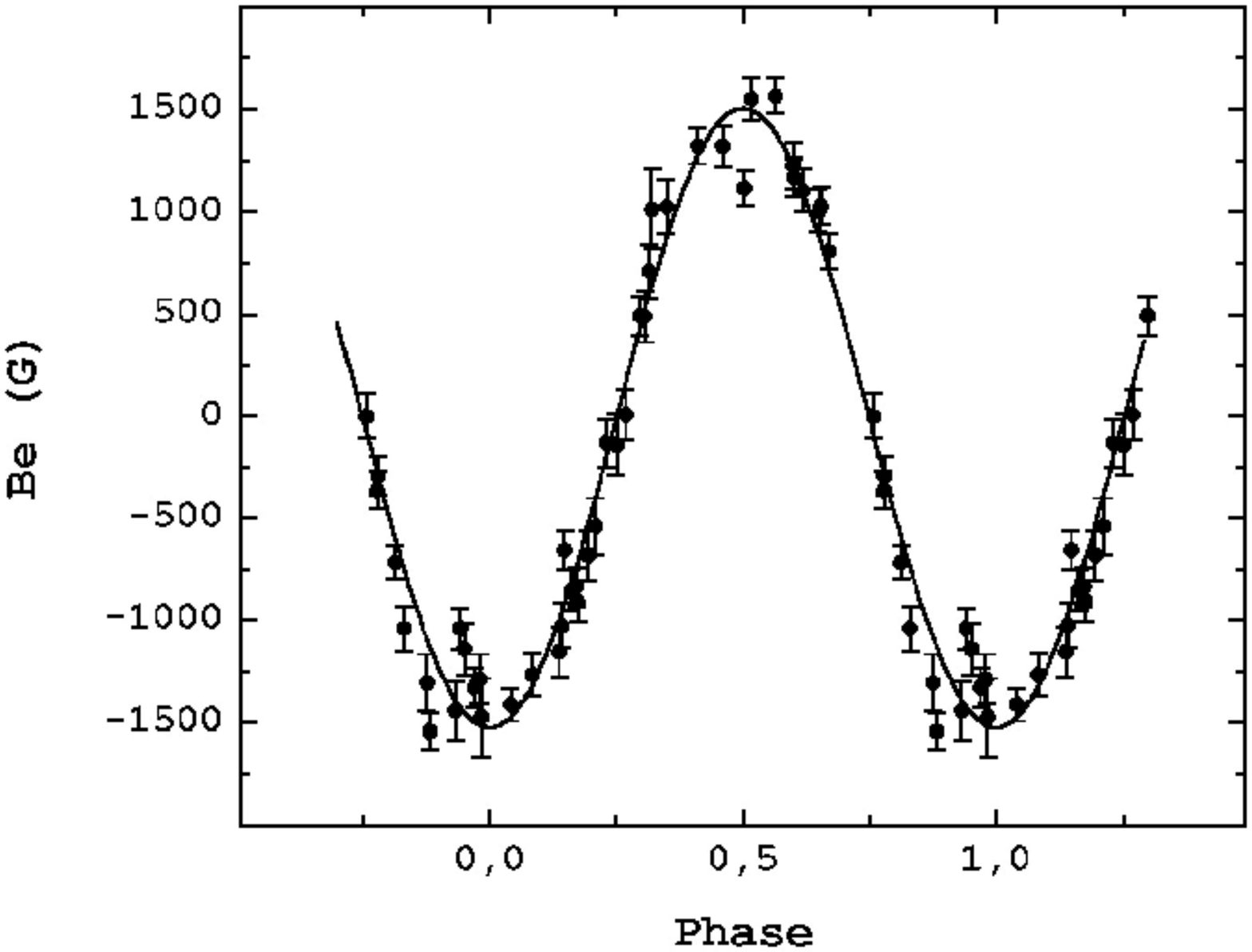}}
\vspace{-3.5mm}
\caption{ HD176582 (1) }
\label{fig:fig303}
\end{figure}

\begin{figure}
\resizebox{0.98\hsize}{!}{\includegraphics{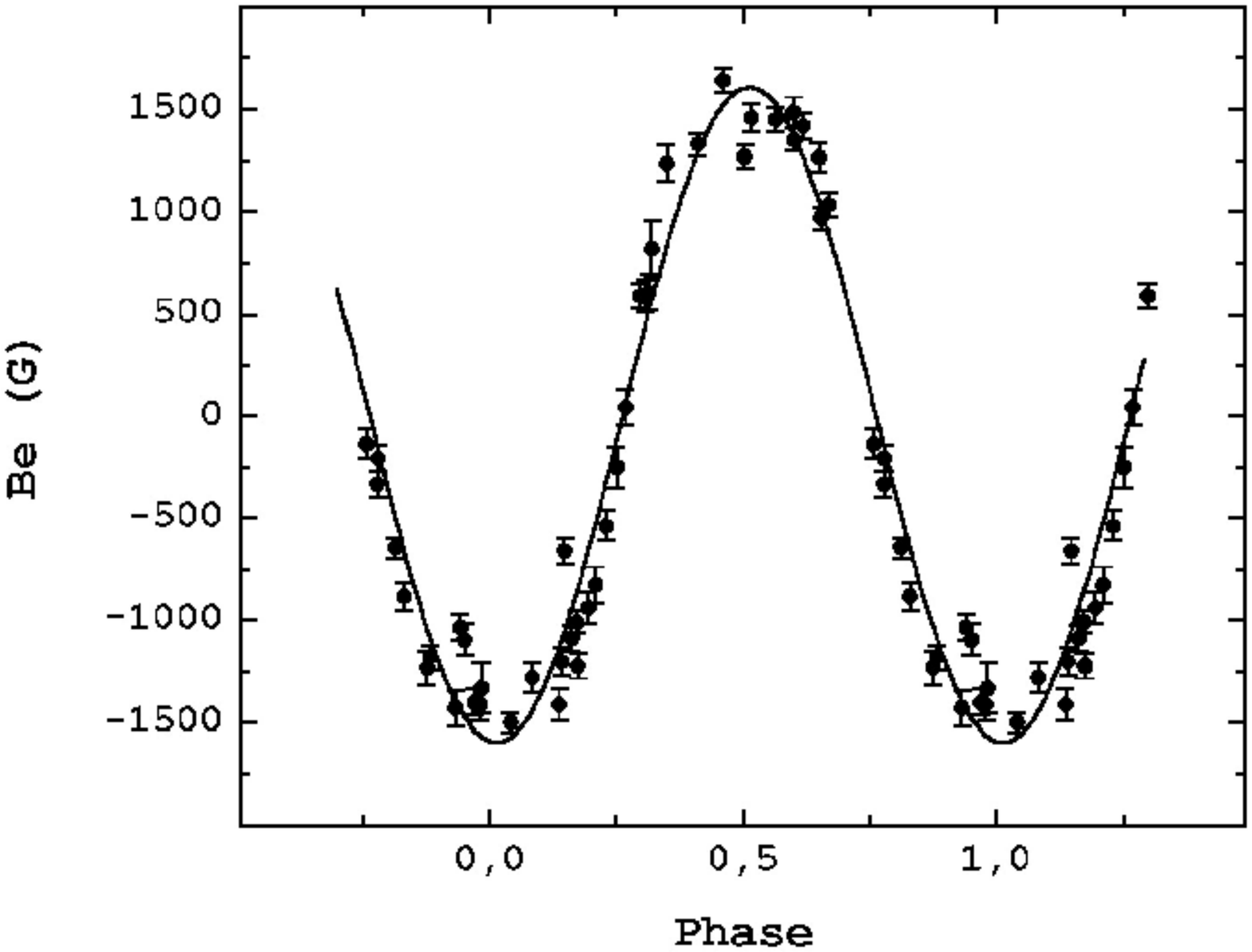}}
\vspace{-3.5mm}
\caption{ HD176582 (2) }
\label{fig:fig303}
\end{figure}

\begin{figure}
\resizebox{0.98\hsize}{!}{\includegraphics{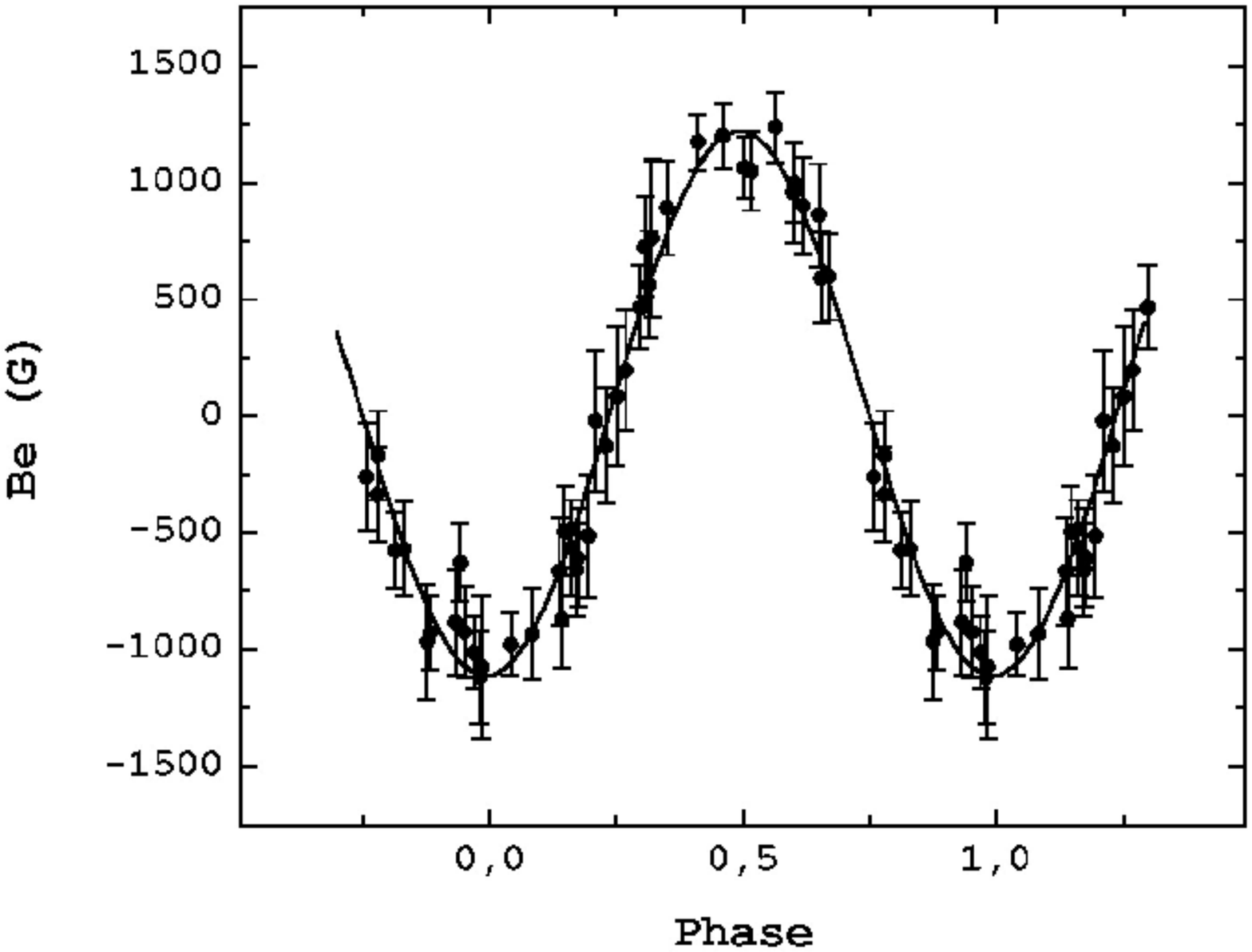}}
\vspace{-3.5mm}
\caption{ HD176582 (3) }
\label{fig:fig303}
\end{figure}

\begin{figure}
\resizebox{0.98\hsize}{!}{\includegraphics{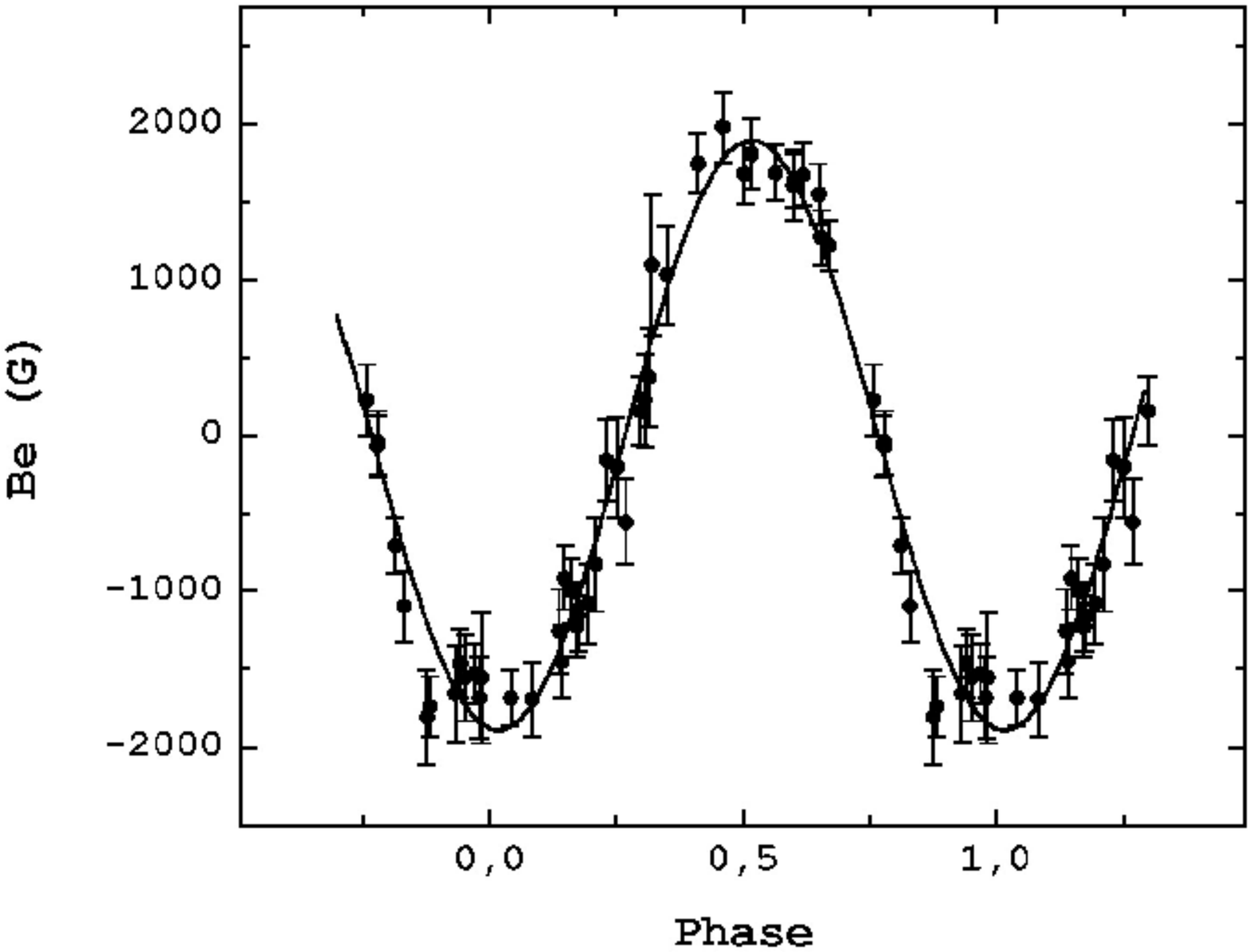}}
\vspace{-3.5mm}
\caption{ HD176582 (4) }
\label{fig:fig303}
\end{figure}

\clearpage
\newpage

\begin{figure}
\resizebox{0.98\hsize}{!}{\includegraphics{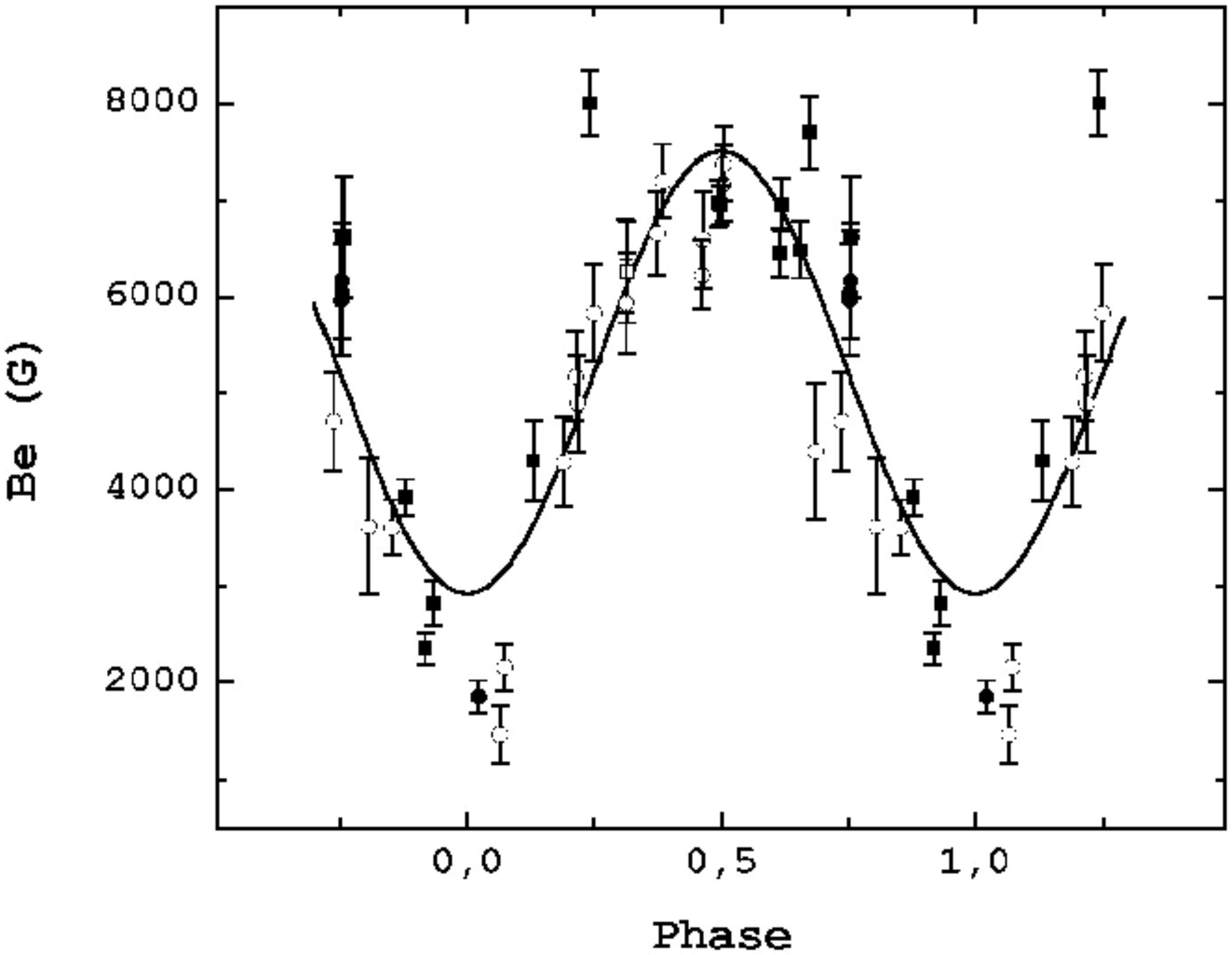}}
\vspace{-3.5mm}
\caption{ HD178892 (1) }
\label{fig:fig304}
\end{figure}

\begin{figure}
\resizebox{0.98\hsize}{!}{\includegraphics{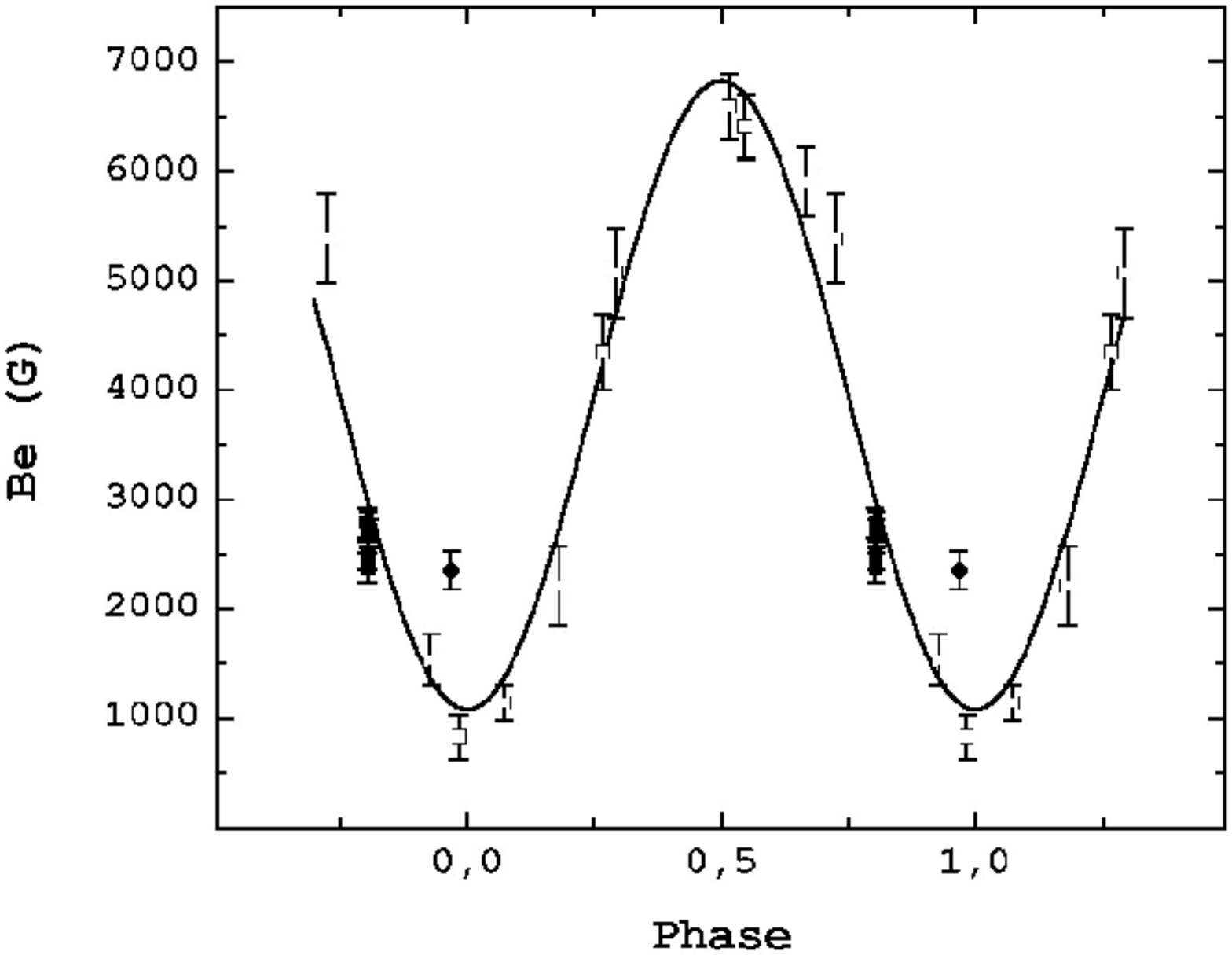}}
\vspace{-3.5mm}
\caption{ HD178892 (2) }
\label{fig:fig305}
\end{figure}

\begin{figure}
\resizebox{0.98\hsize}{!}{\includegraphics{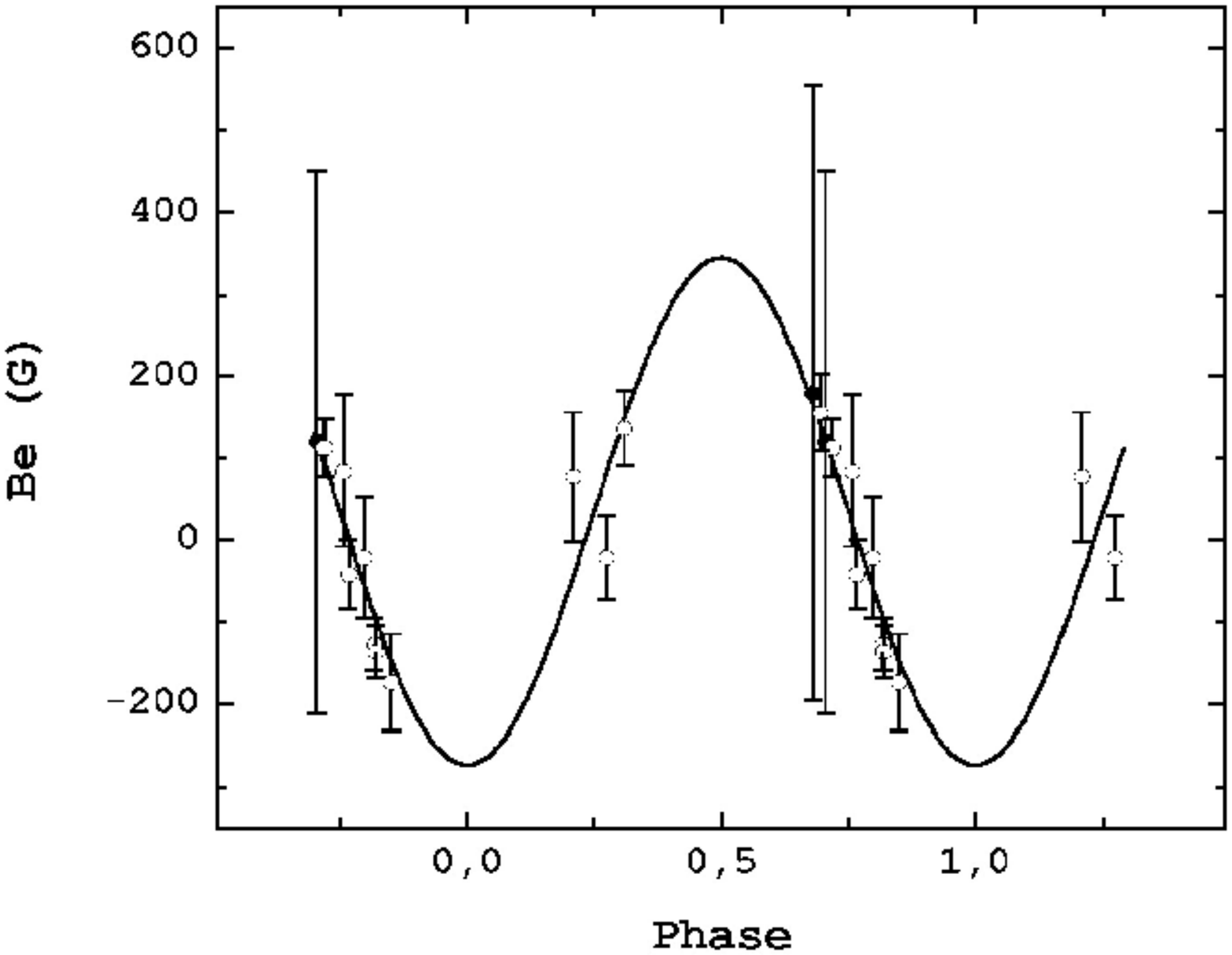}}
\vspace{-3.5mm}
\caption{ HD179527 }
\label{fig:fig306}
\end{figure}

\begin{figure}
\resizebox{0.98\hsize}{!}{\includegraphics{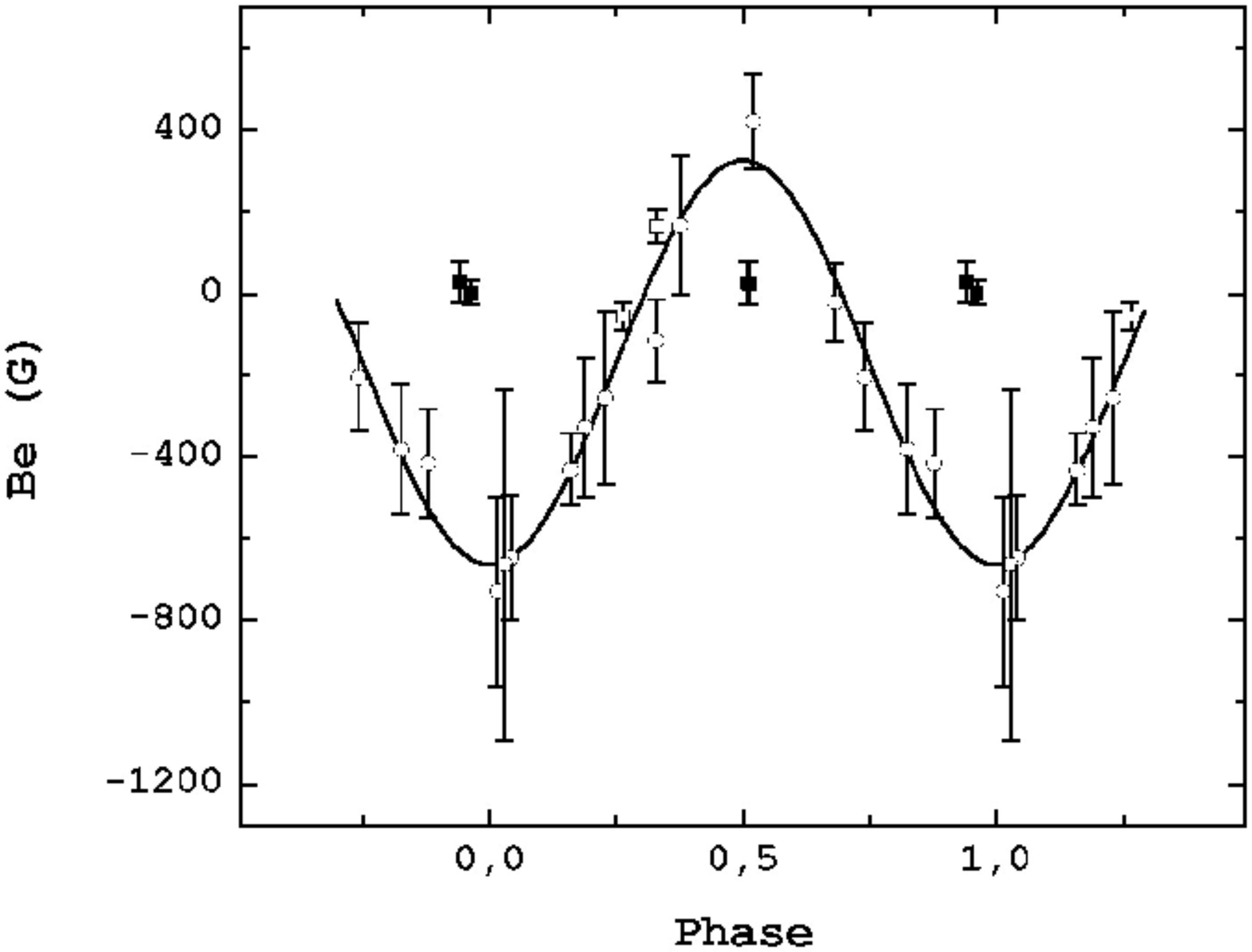}}
\vspace{-3.5mm}
\caption{ HD180642 }
\label{fig:fig307}
\end{figure}

\begin{figure}
\resizebox{0.98\hsize}{!}{\includegraphics{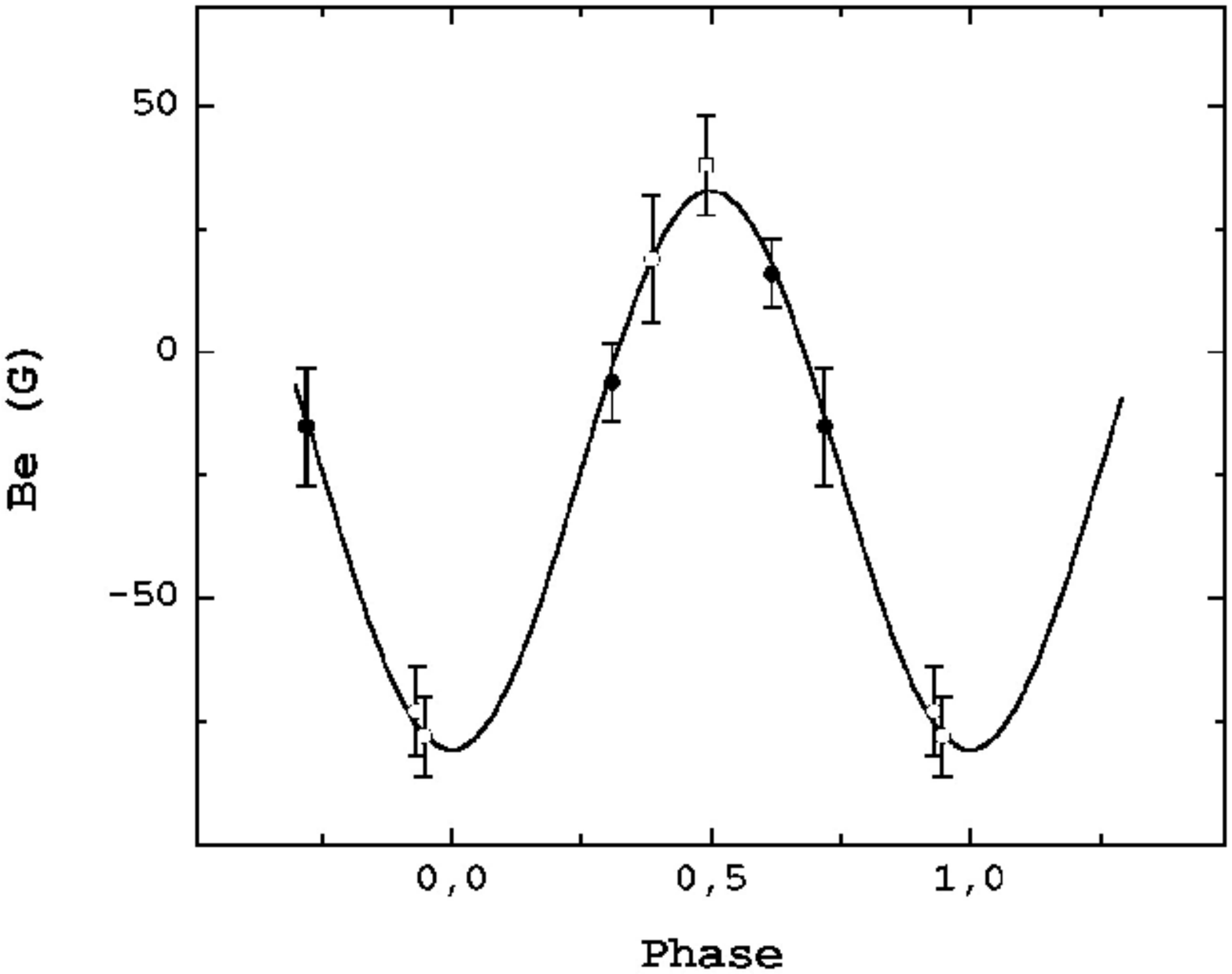}}
\vspace{-3.5mm}
\caption{ HD181615 }
\label{fig:fig308}
\end{figure}

\begin{figure}
\resizebox{0.98\hsize}{!}{\includegraphics{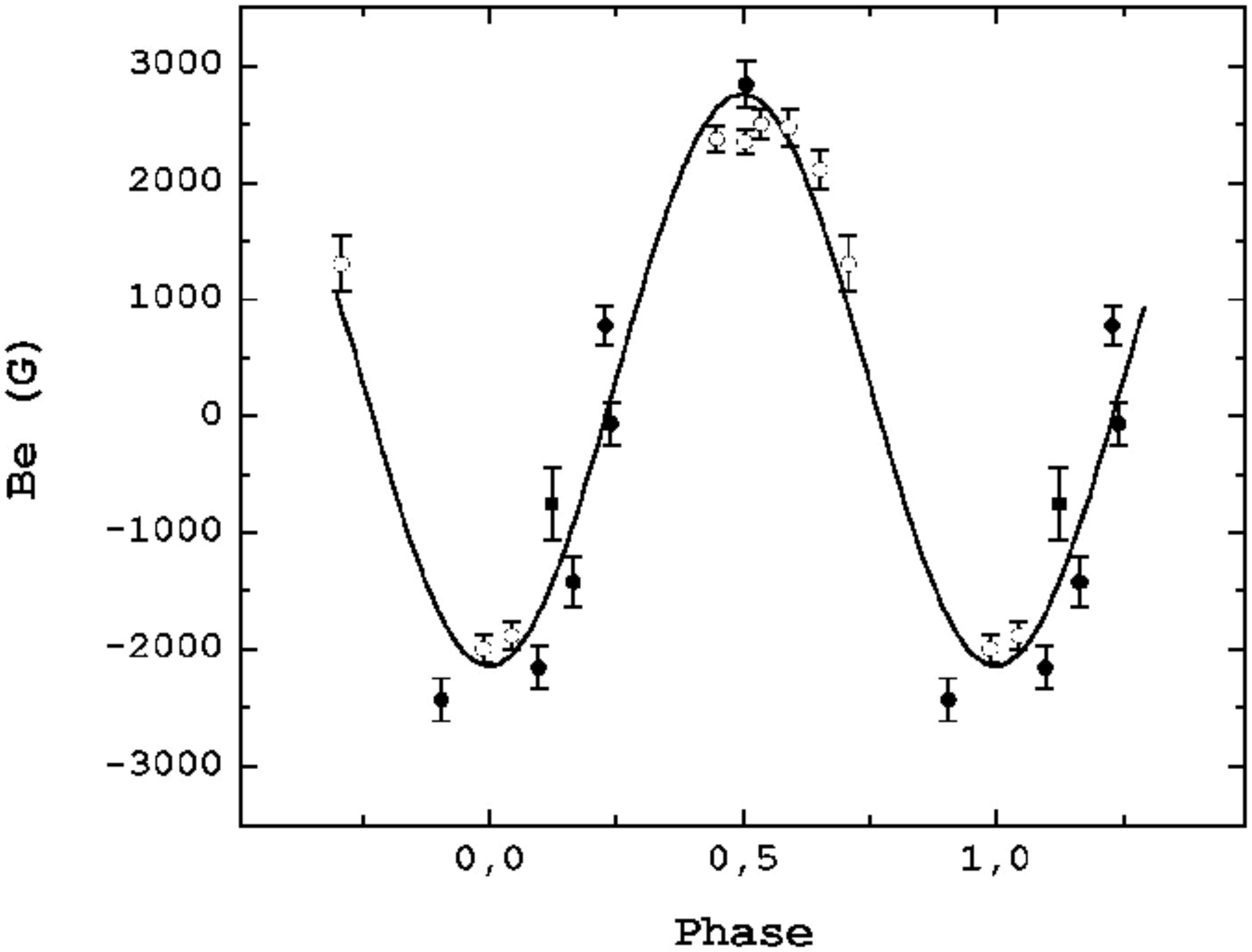}}
\vspace{-3.5mm}
\caption{ HD182180 }
\label{fig:fig309}
\end{figure}

\clearpage
\newpage

\begin{figure}
\resizebox{0.98\hsize}{!}{\includegraphics{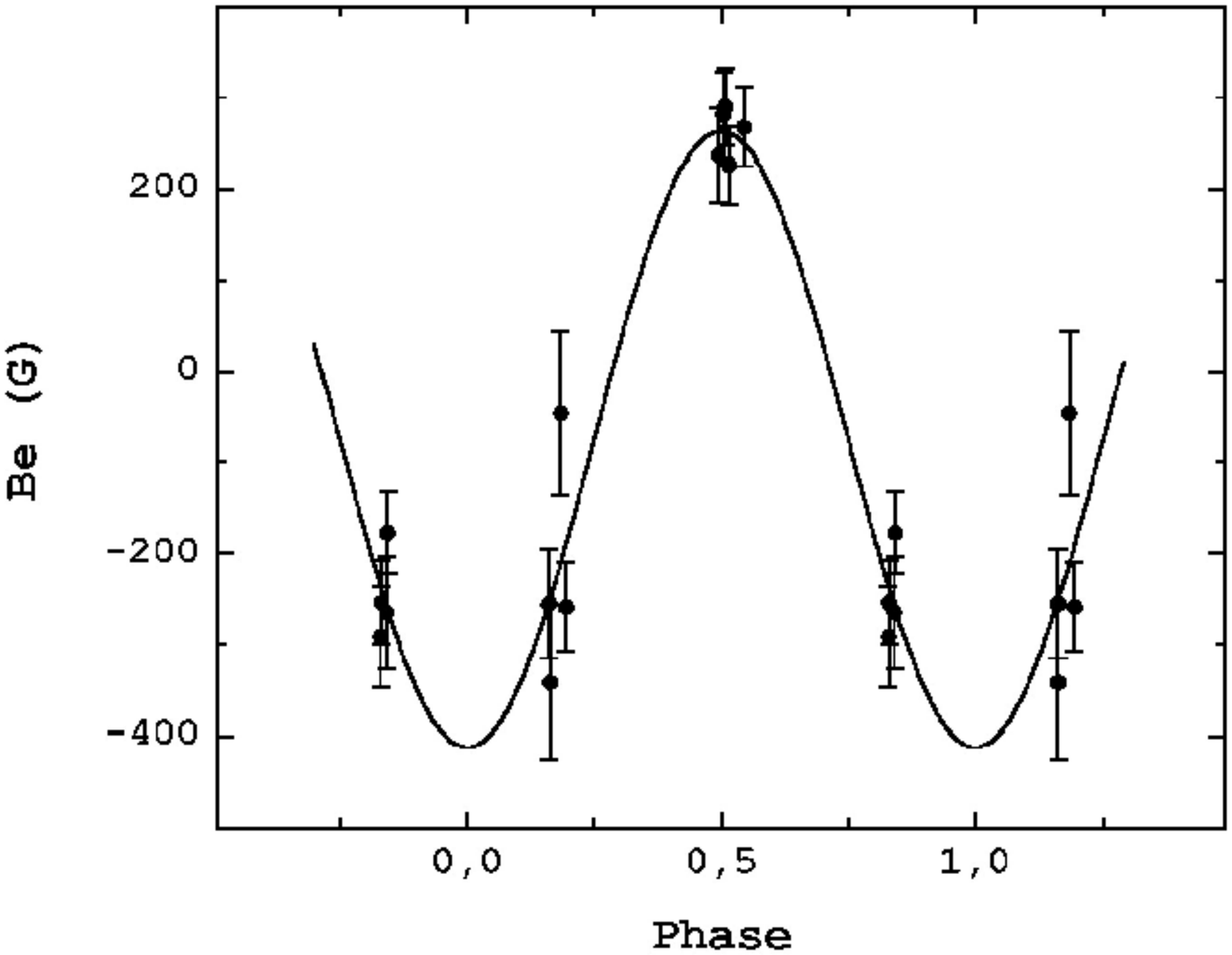}}
\vspace{-3.5mm}
\caption{ HD183056 }
\label{fig:fig310}
\end{figure}

\begin{figure}
\resizebox{0.98\hsize}{!}{\includegraphics{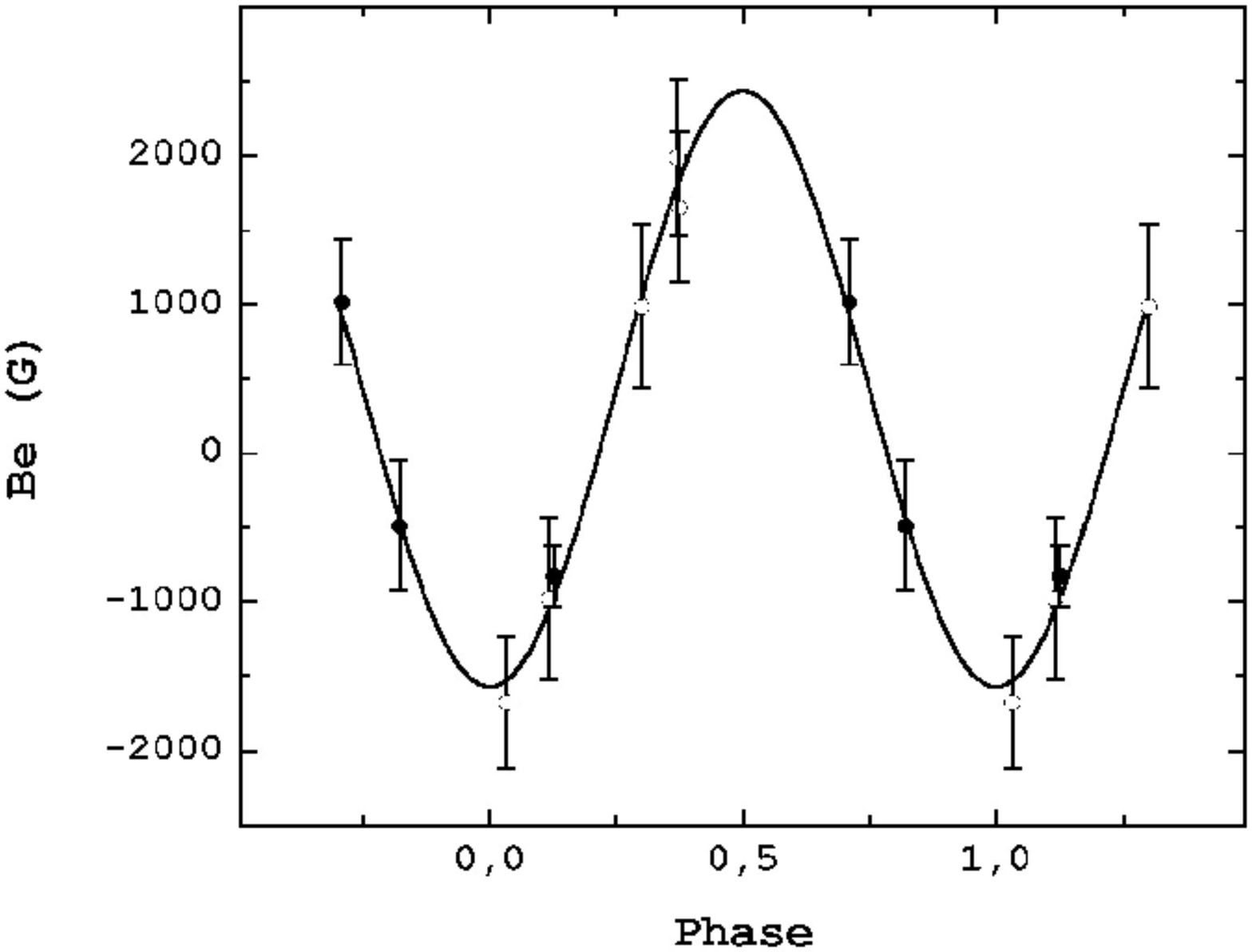}}
\vspace{-3.5mm}
\caption{ HD183339 }
\label{fig:fig311}
\end{figure}

\begin{figure}
\resizebox{0.98\hsize}{!}{\includegraphics{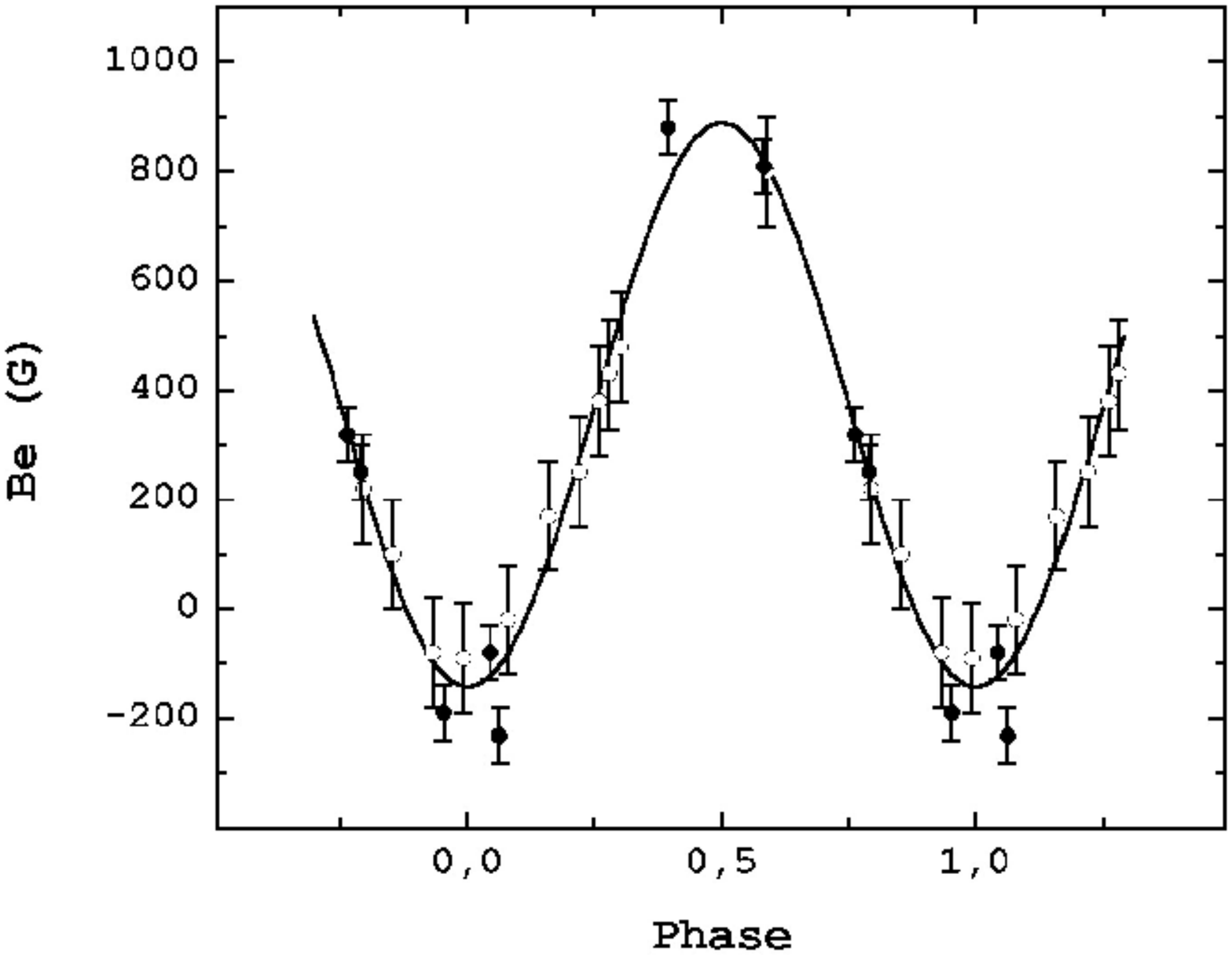}}
\vspace{-3.5mm}
\caption{ HD184471 (1) }
\label{fig:fig312}
\end{figure}

\begin{figure}
\resizebox{0.98\hsize}{!}{\includegraphics{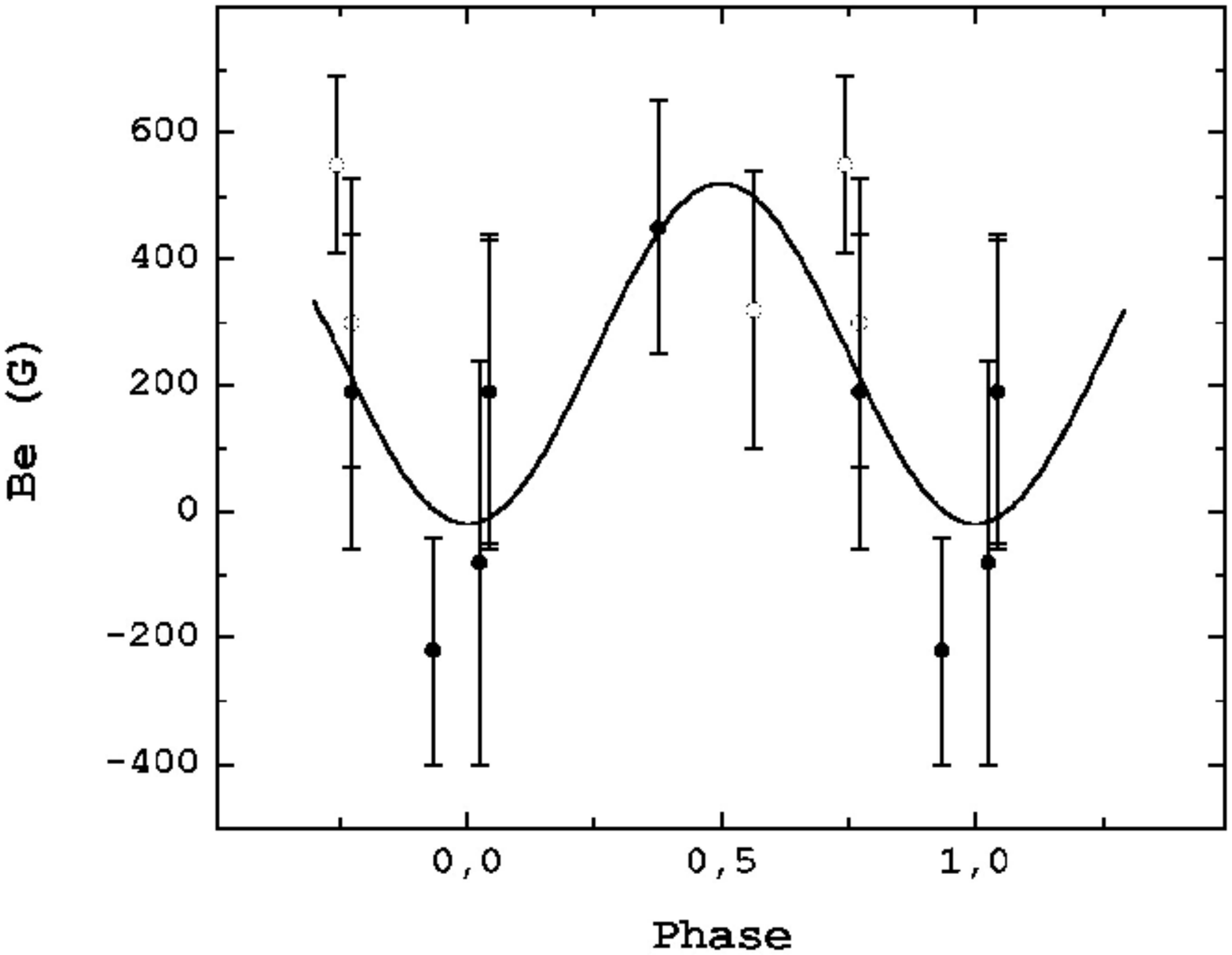}}
\vspace{-3.5mm}
\caption{ HD184471 (2) }
\label{fig:fig313}
\end{figure}

\begin{figure}
\resizebox{0.98\hsize}{!}{\includegraphics{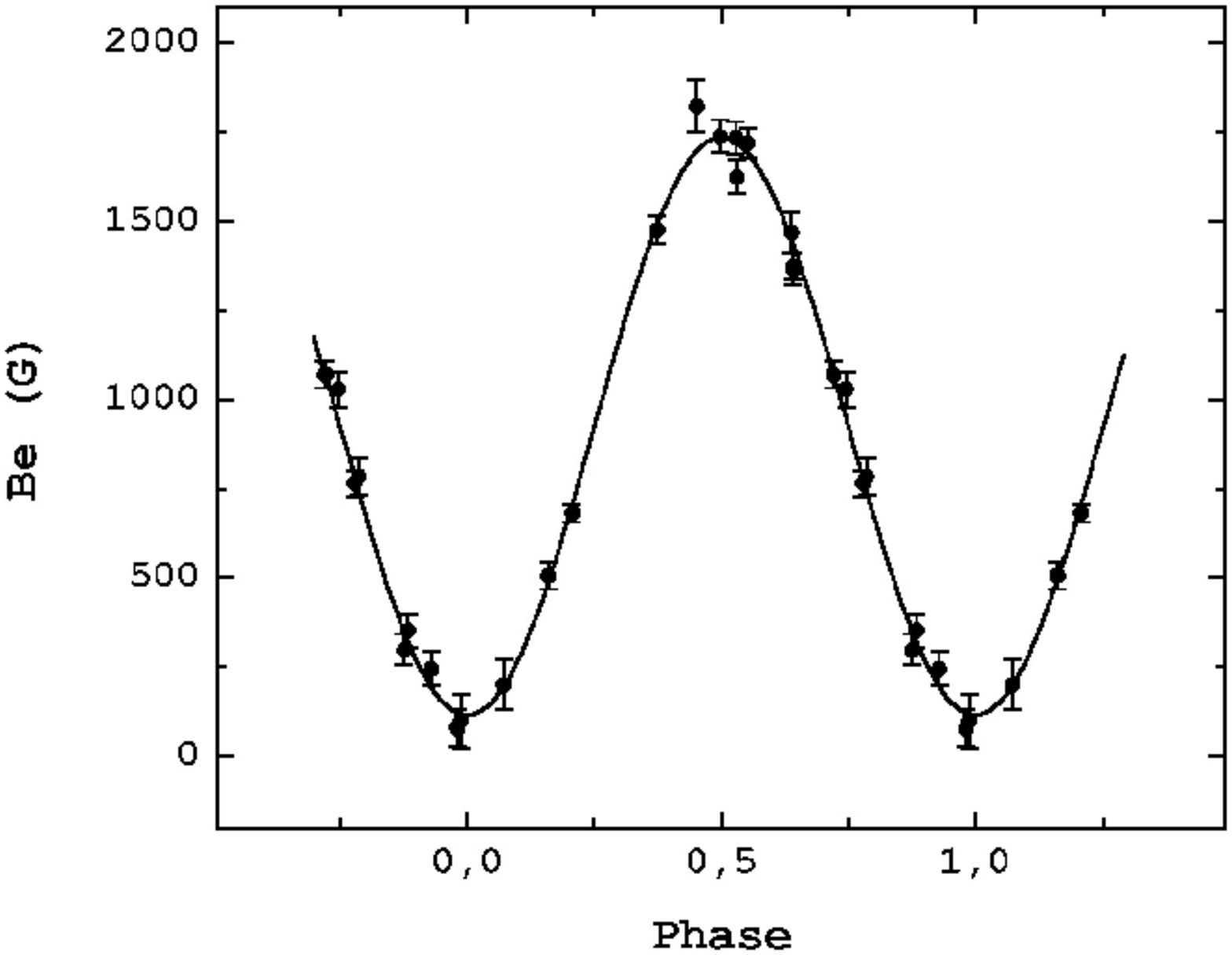}}
\vspace{-3.5mm}
\caption{ HD184927 (1) }
\label{fig:fig314}
\end{figure}

\begin{figure}
\resizebox{0.98\hsize}{!}{\includegraphics{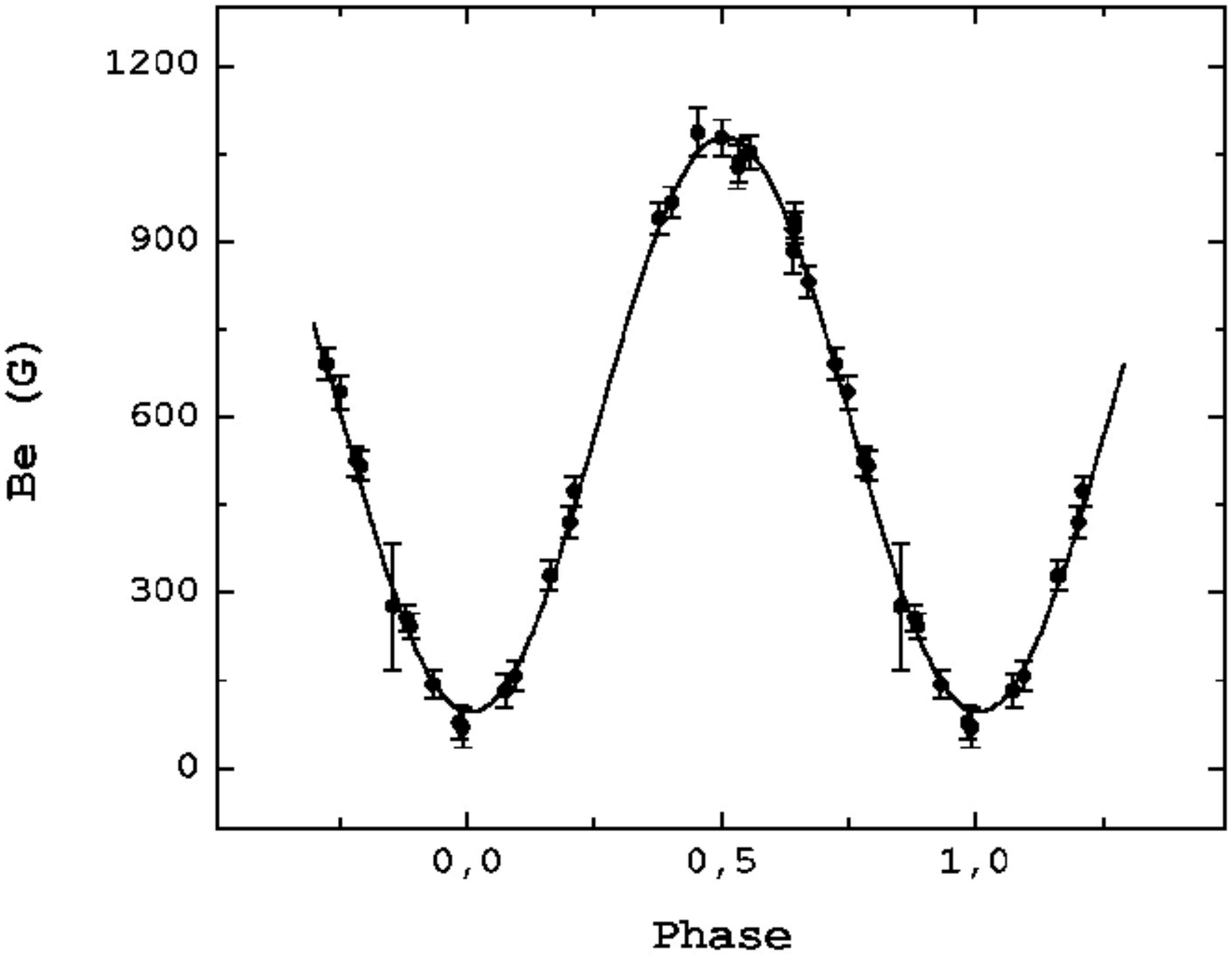}}
\vspace{-3.5mm}
\caption{ HD184927 (2) }
\label{fig:fig315}
\end{figure}

\clearpage
\newpage

\begin{figure}
\resizebox{0.98\hsize}{!}{\includegraphics{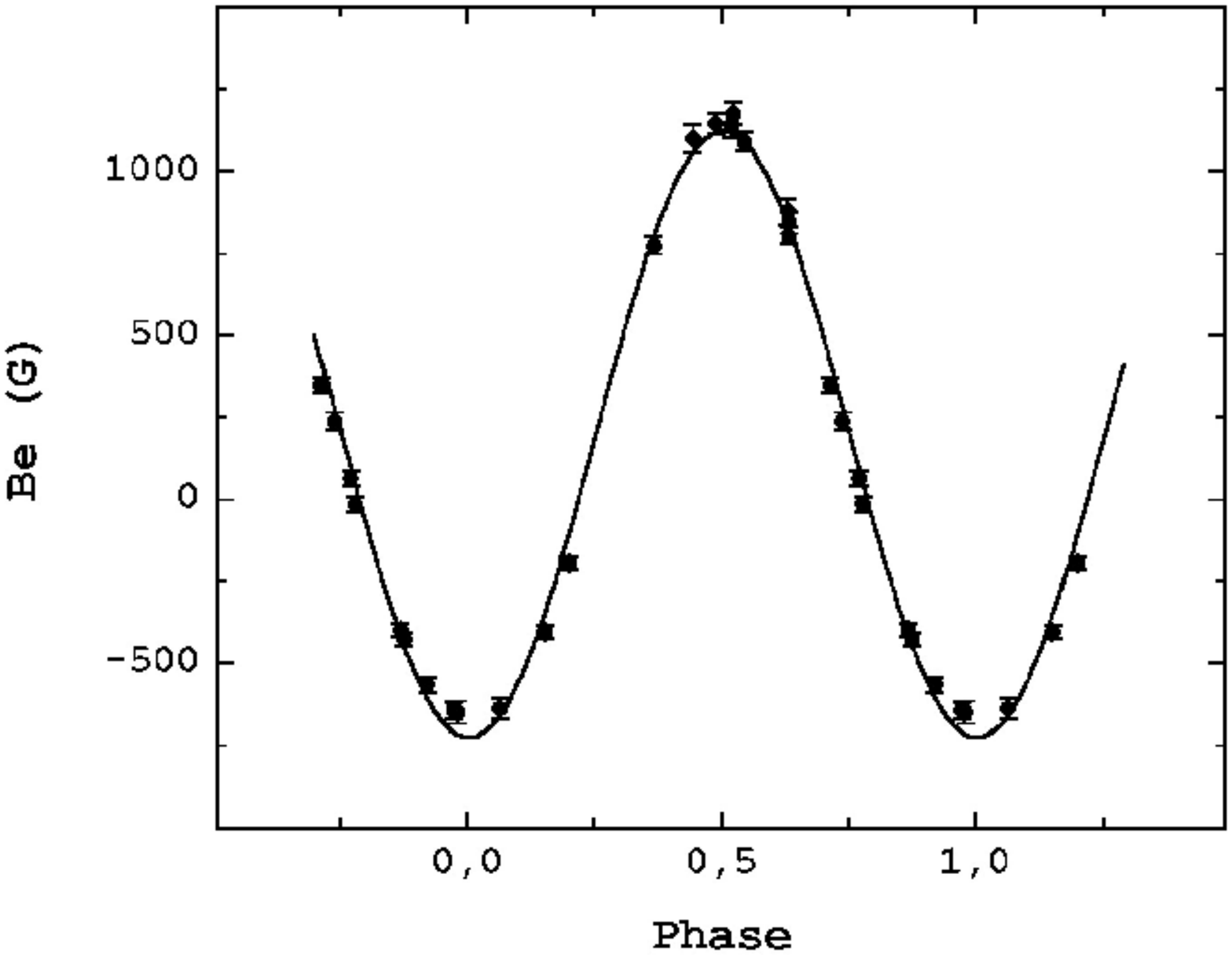}}
\vspace{-3.5mm}
\caption{ HD184927 (3) }
\label{fig:fig316}
\end{figure}

\begin{figure}
\resizebox{0.98\hsize}{!}{\includegraphics{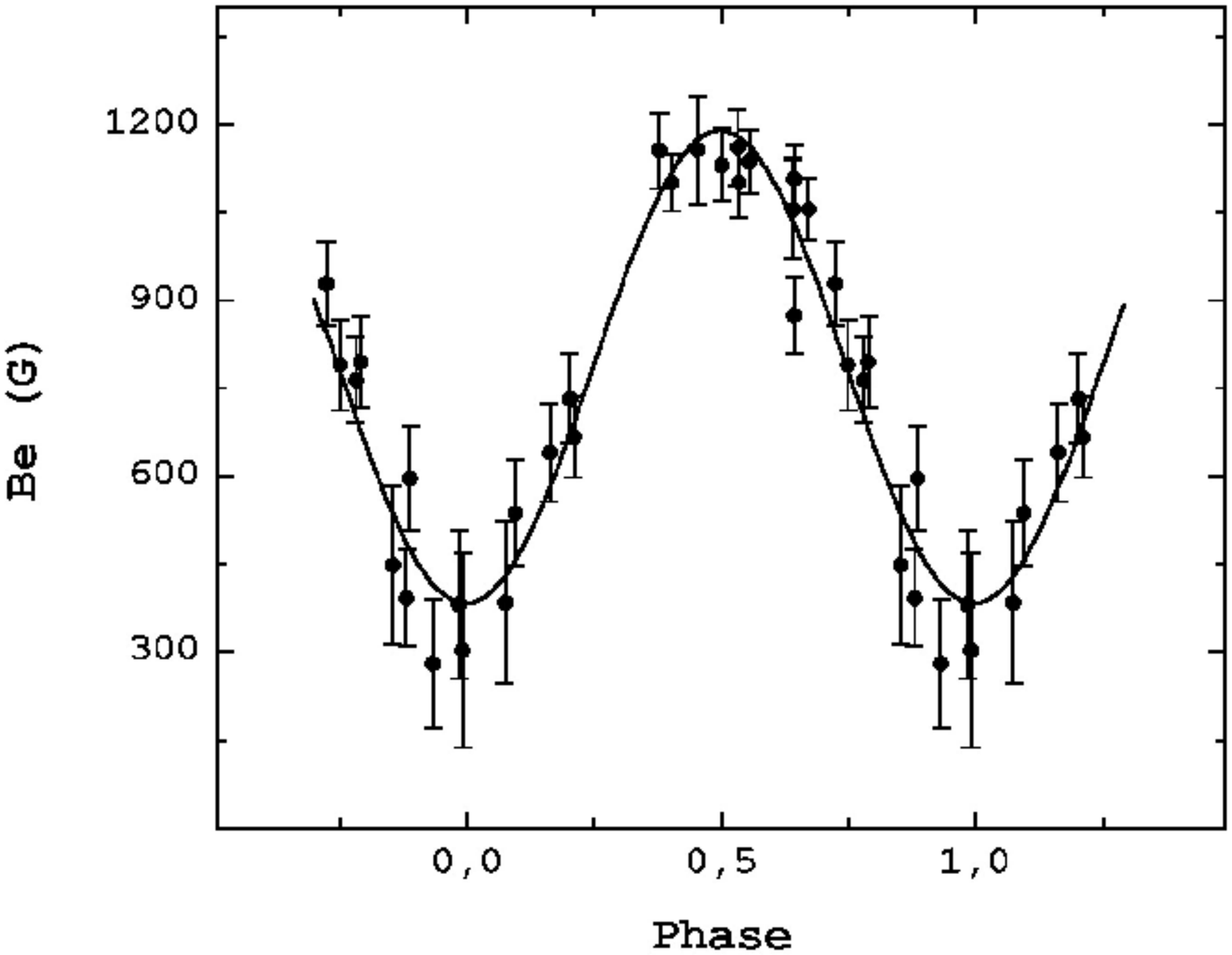}}
\vspace{-3.5mm}
\caption{ HD184927 (4) }
\label{fig:fig317}
\end{figure}

\begin{figure}
\resizebox{0.98\hsize}{!}{\includegraphics{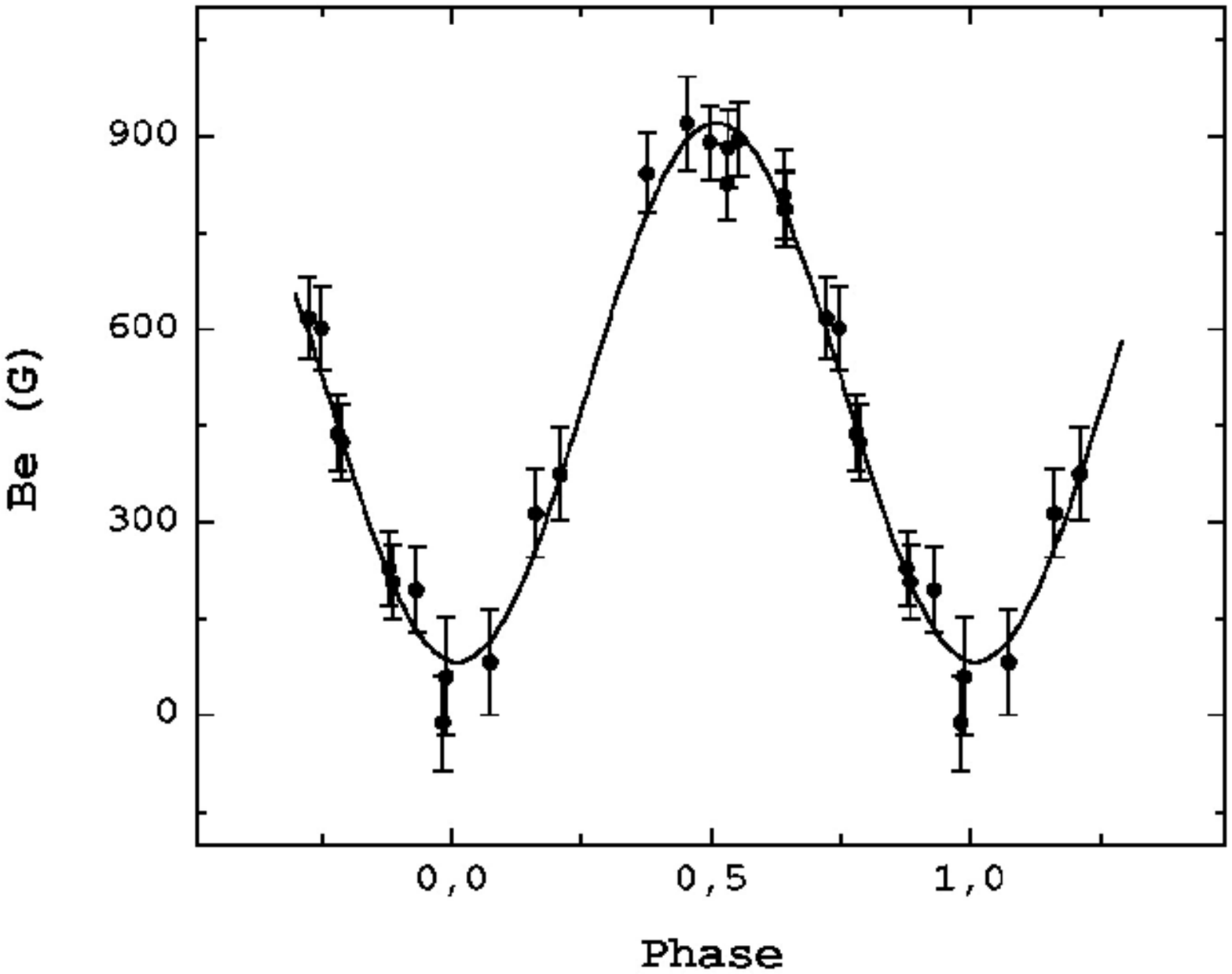}}
\vspace{-3.5mm}
\caption{ HD184927 (5) }
\label{fig:fig318}
\end{figure}

\begin{figure}
\resizebox{0.98\hsize}{!}{\includegraphics{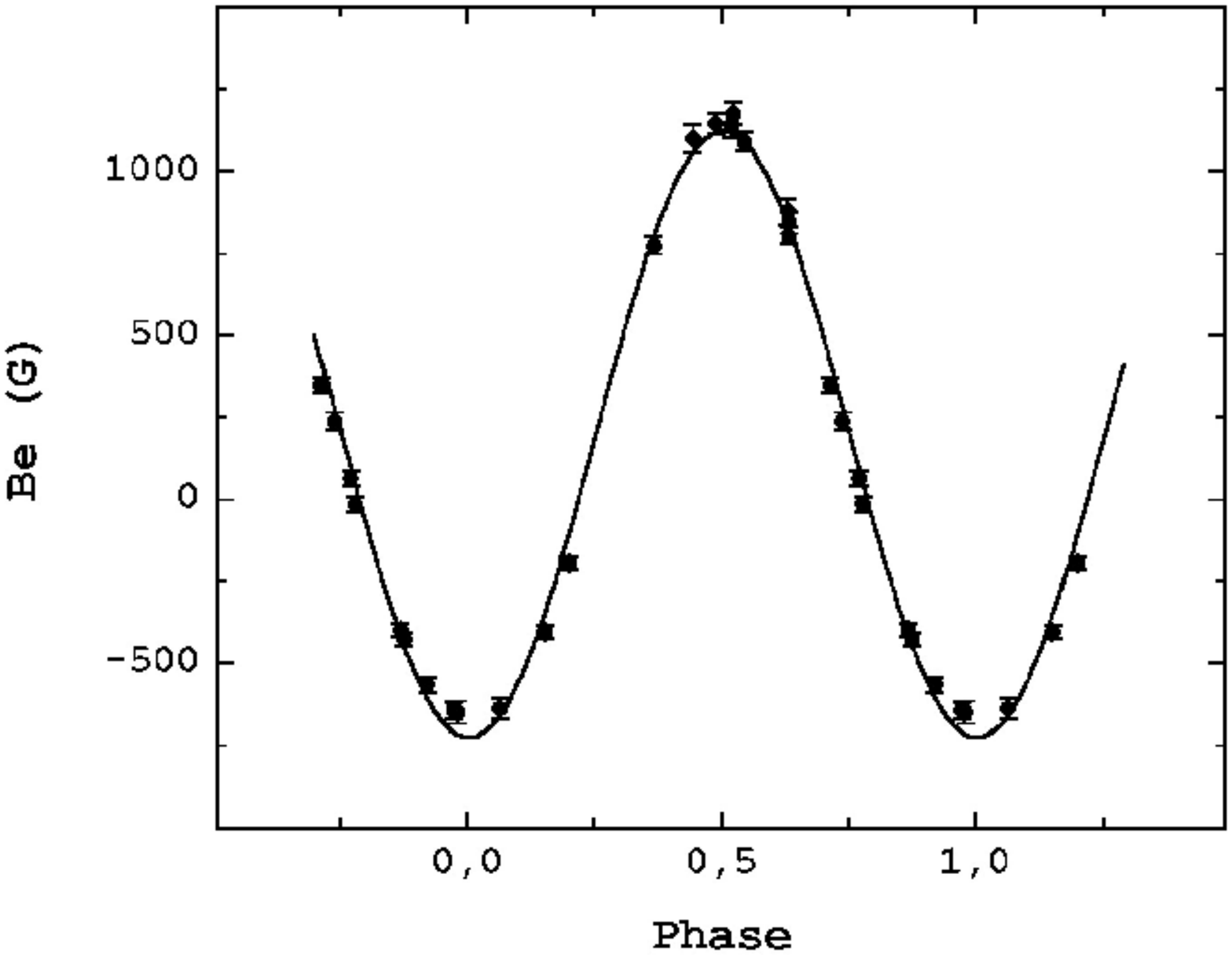}}
\vspace{-3.5mm}
\caption{ HD184927 (6) }
\label{fig:fig319}
\end{figure}

\begin{figure}
\resizebox{0.98\hsize}{!}{\includegraphics{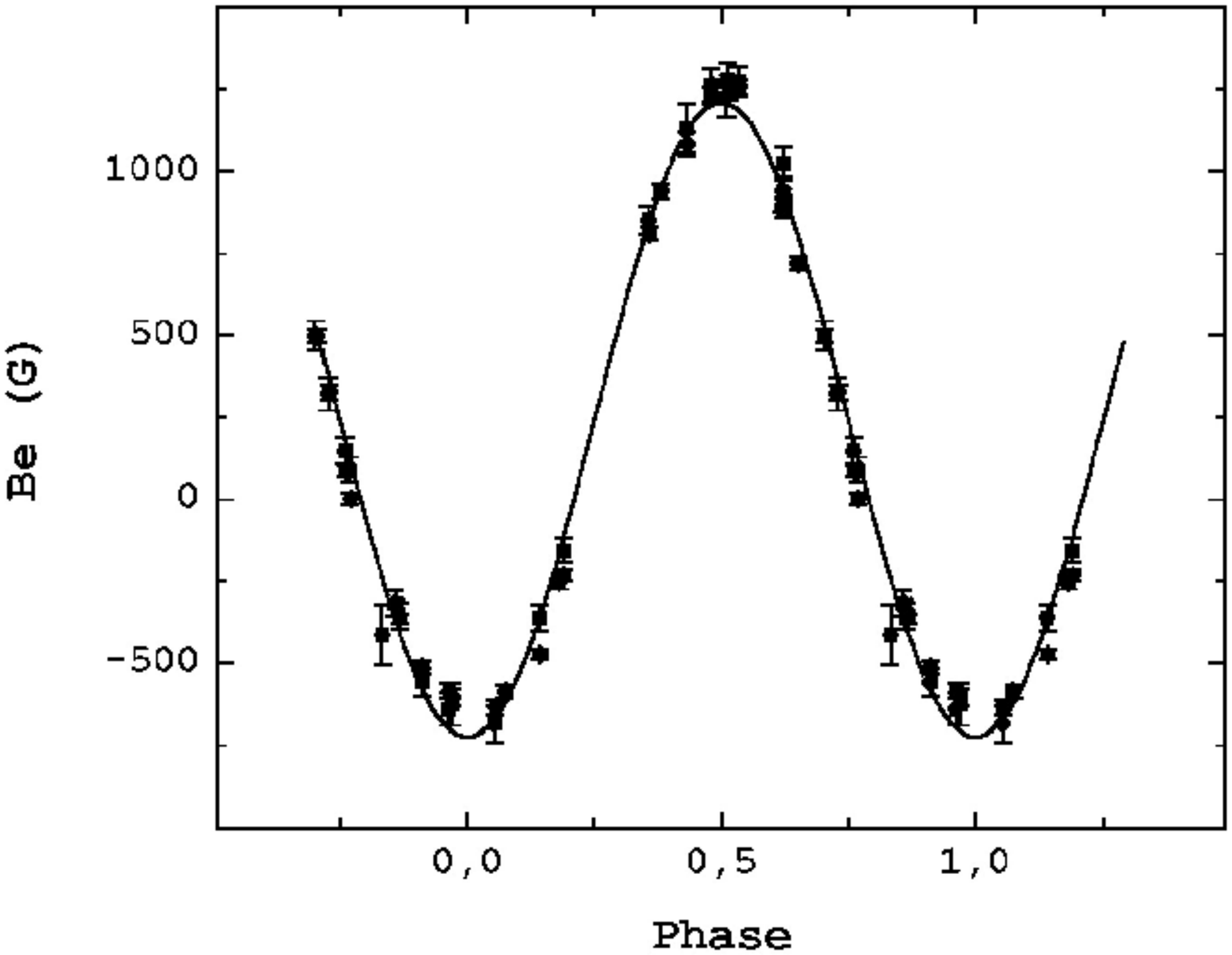}}
\vspace{-3.5mm}
\caption{ HD184927 (7) }
\label{fig:fig320}
\end{figure}

\begin{figure}
\resizebox{0.98\hsize}{!}{\includegraphics{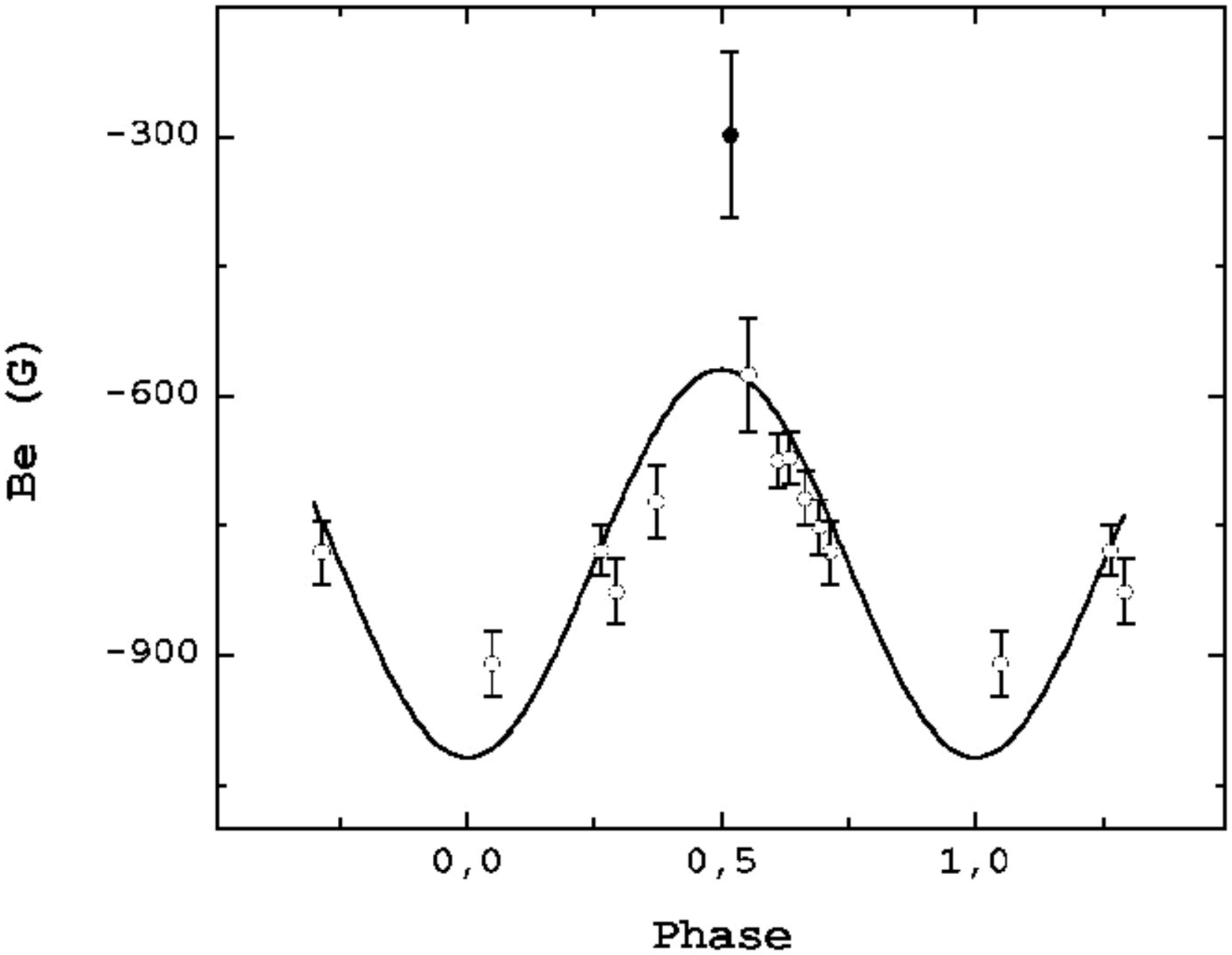}}
\vspace{-3.5mm}
\caption{ HD186205 (1) }
\label{fig:fig303}
\end{figure}

\clearpage
\newpage

\begin{figure}
\resizebox{0.98\hsize}{!}{\includegraphics{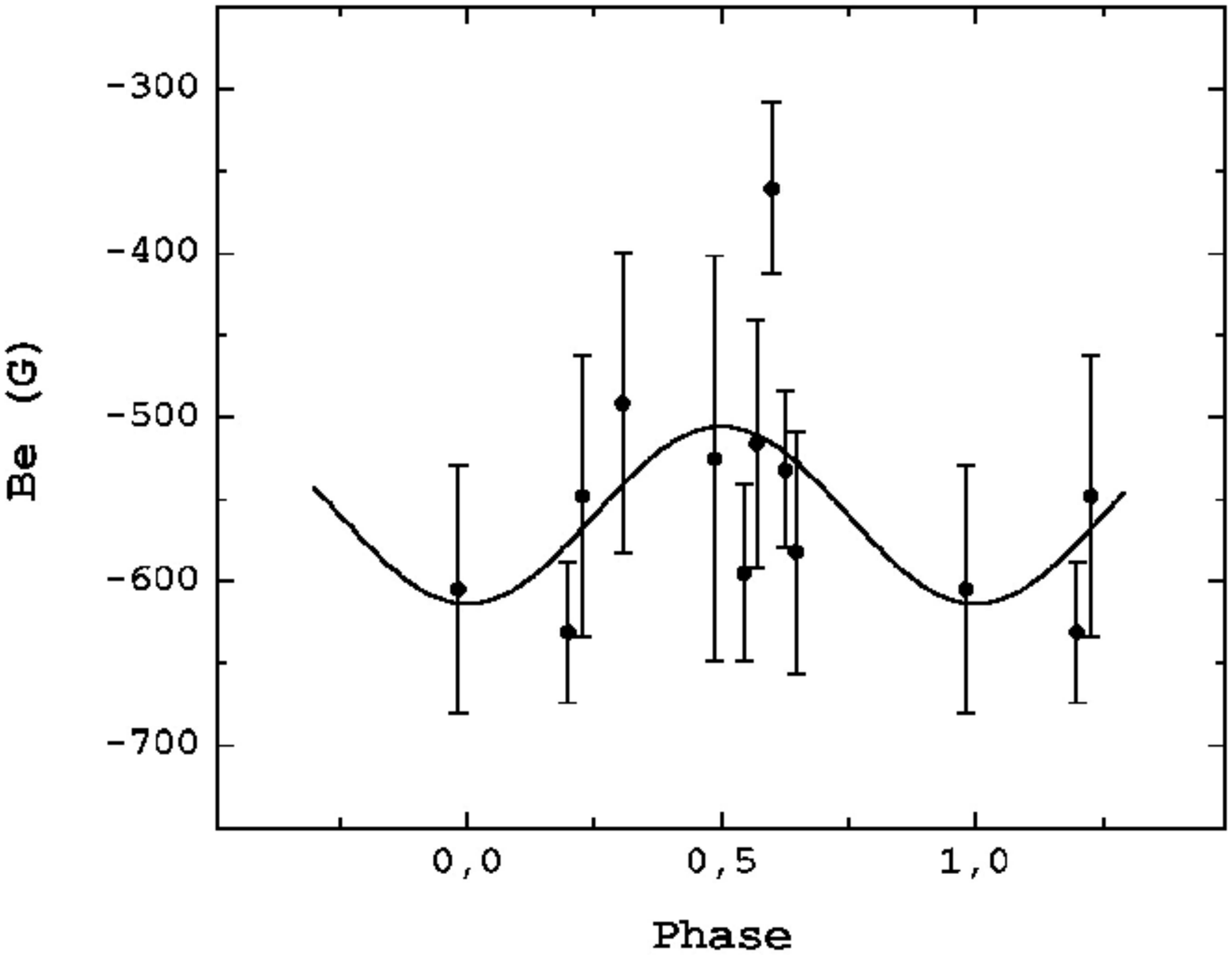}}
\vspace{-3.5mm}
\caption{ HD186205 (2) }
\label{fig:fig303}
\end{figure}

\begin{figure}
\resizebox{0.98\hsize}{!}{\includegraphics{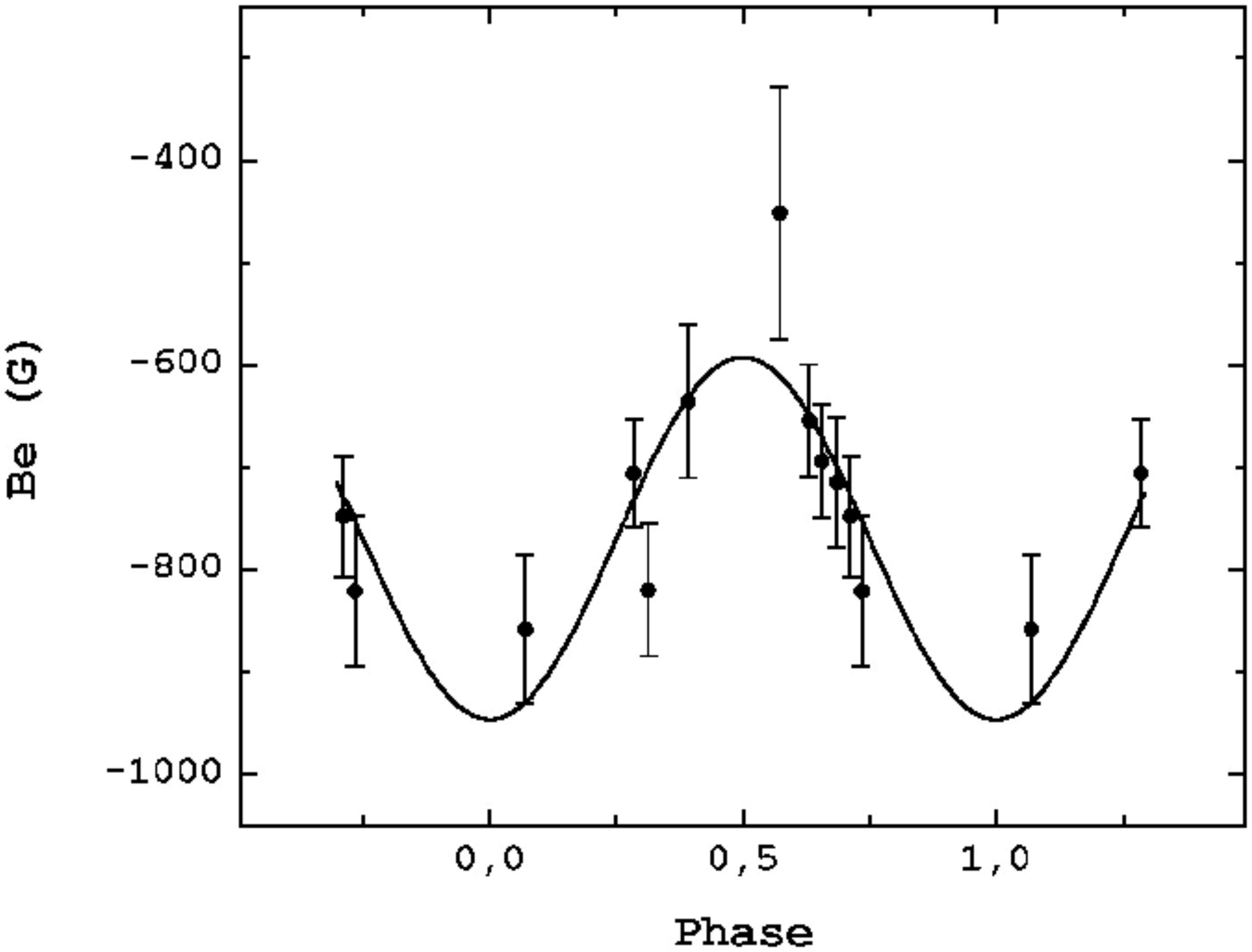}}
\vspace{-3.5mm}
\caption{ HD186205 (3) }
\label{fig:fig303}
\end{figure}

\begin{figure}
\resizebox{0.98\hsize}{!}{\includegraphics{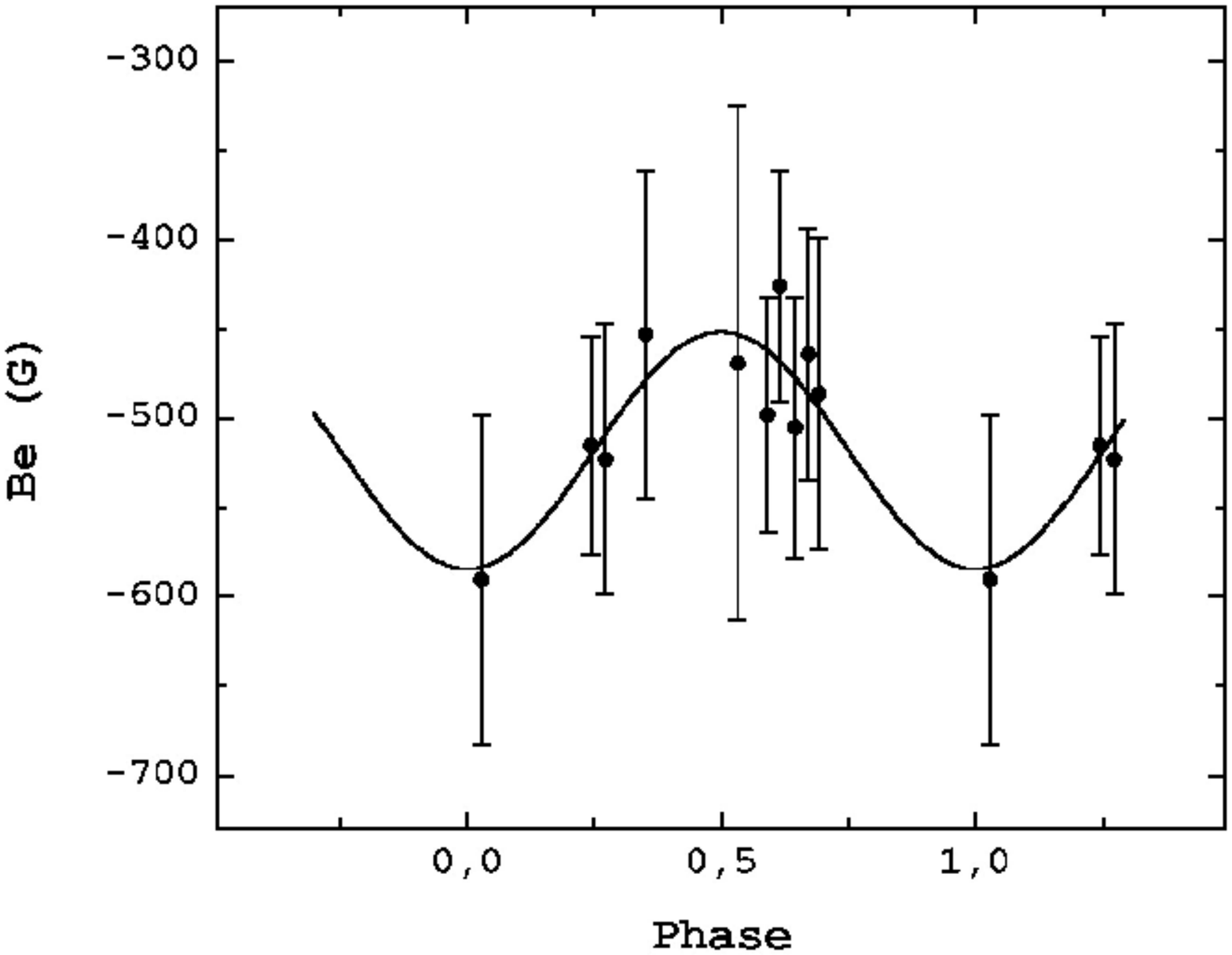}}
\vspace{-3.5mm}
\caption{ HD186205 (4) }
\label{fig:fig303}
\end{figure}

\begin{figure}
\resizebox{0.98\hsize}{!}{\includegraphics{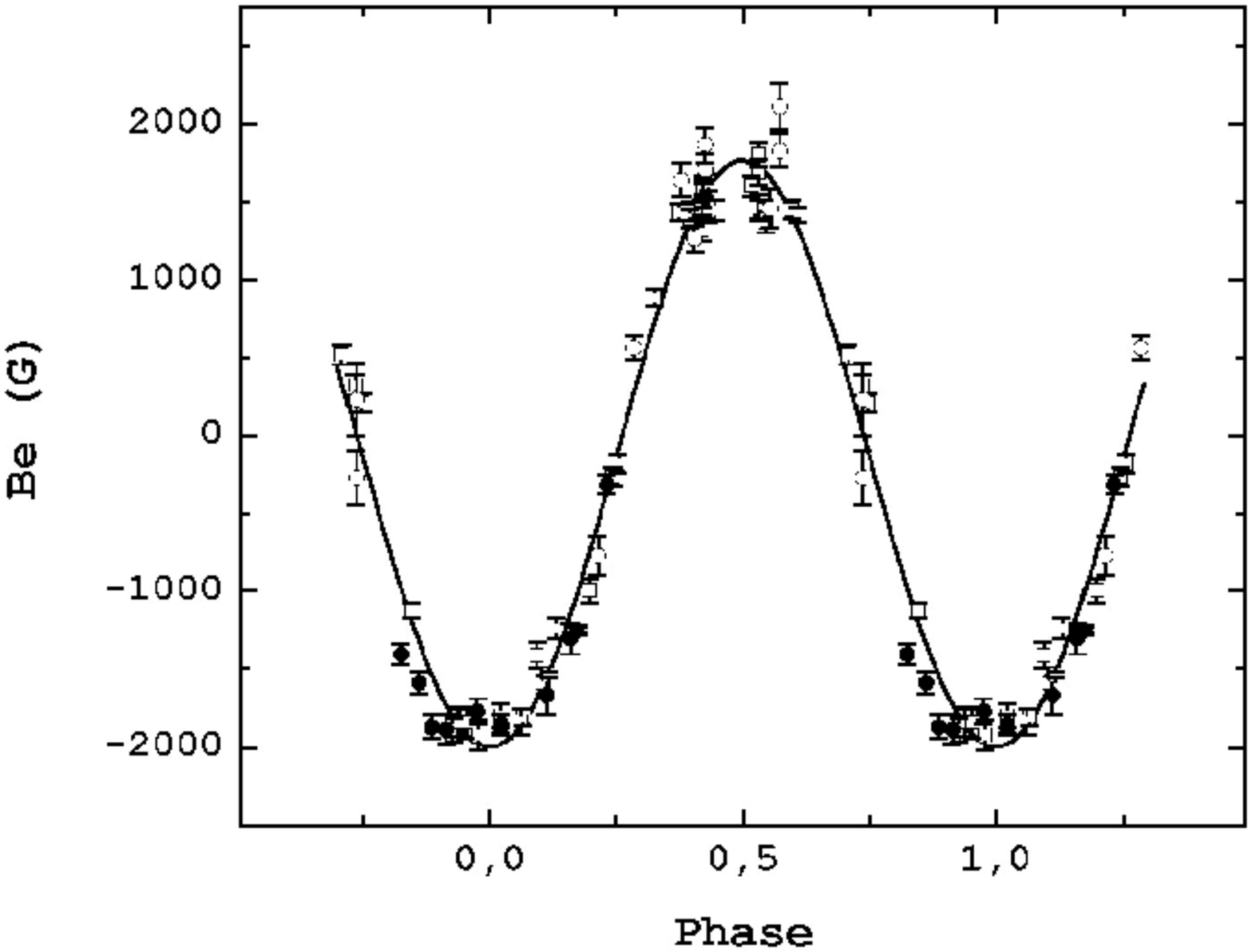}}
\vspace{-3.5mm}
\caption{ HD187474 (1)}
\label{fig:fig322}
\end{figure}

\begin{figure}
\resizebox{0.98\hsize}{!}{\includegraphics{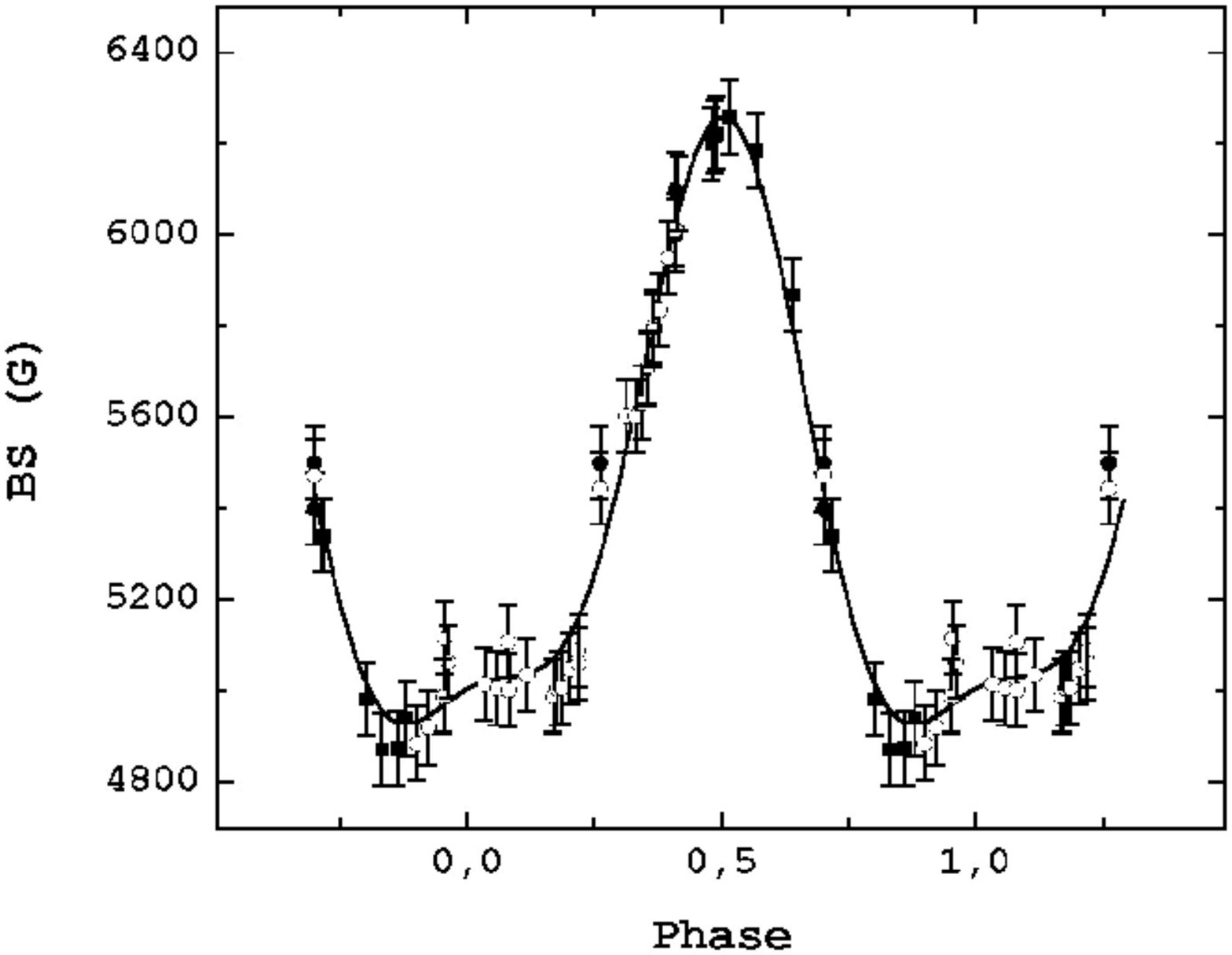}}
\vspace{-3.5mm}
\caption{ HD187474 (2) }
\label{fig:fig303}
\end{figure}

\begin{figure}
\resizebox{0.98\hsize}{!}{\includegraphics{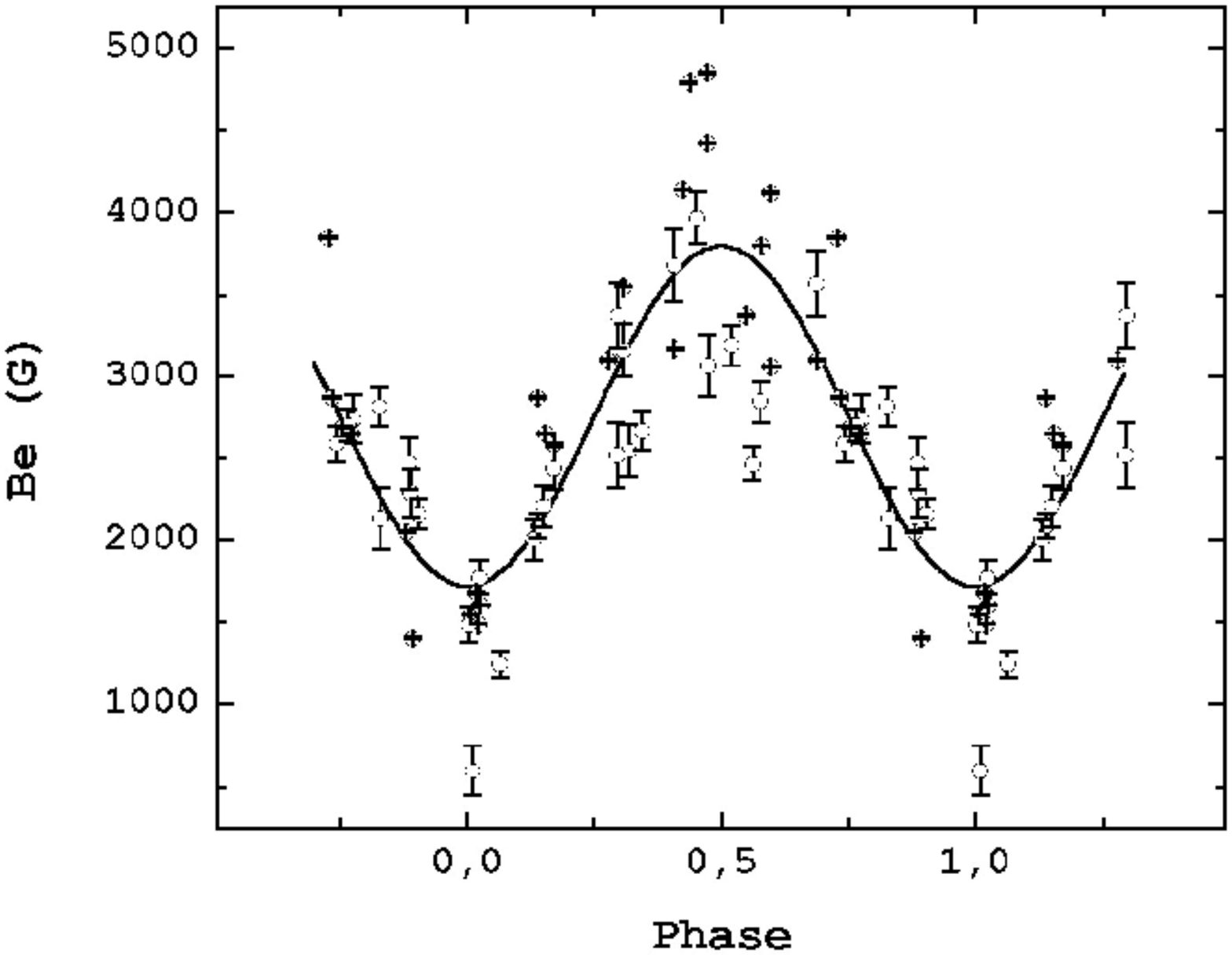}}
\vspace{-3.5mm}
\caption{ HD188041 }
\label{fig:fig323}
\end{figure}

\clearpage
\newpage

\begin{figure}
\resizebox{0.98\hsize}{!}{\includegraphics{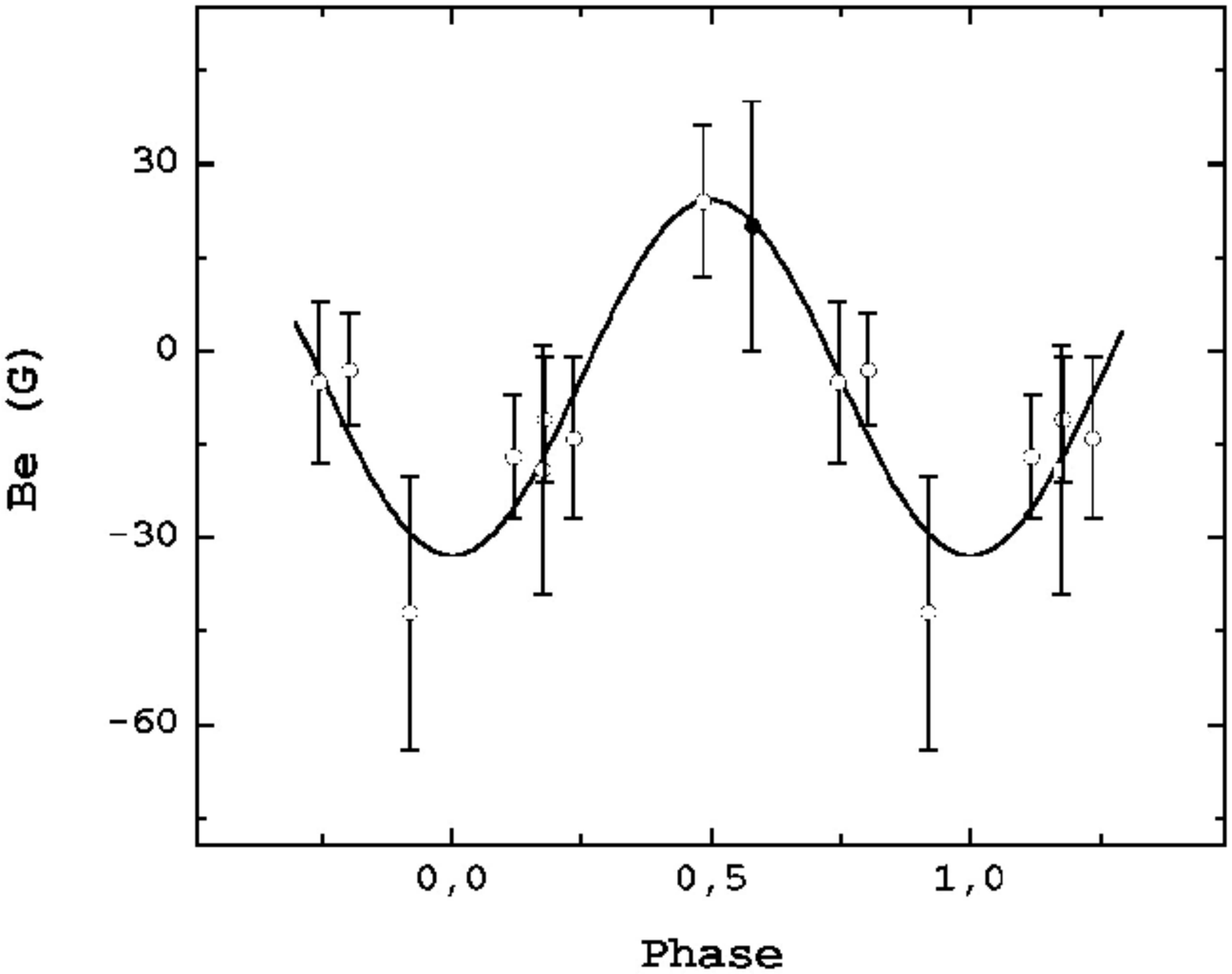}}
\vspace{-3.5mm}
\caption{ HD188209 }
\label{fig:fig303}
\end{figure}

\begin{figure}
\resizebox{0.98\hsize}{!}{\includegraphics{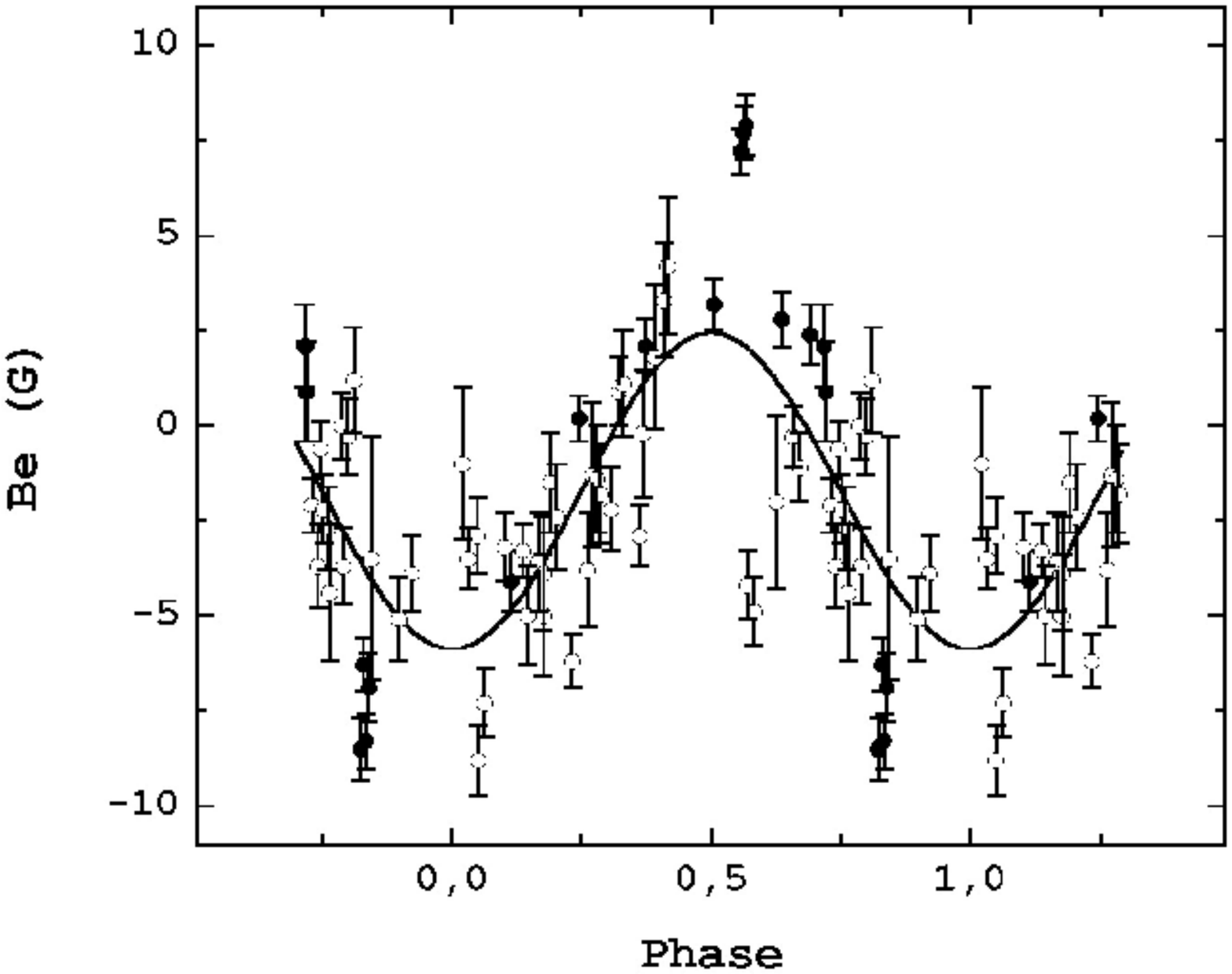}}
\vspace{-3.5mm}
\caption{ HD189733 }
\label{fig:fig324}
\end{figure}

\begin{figure}
\resizebox{0.98\hsize}{!}{\includegraphics{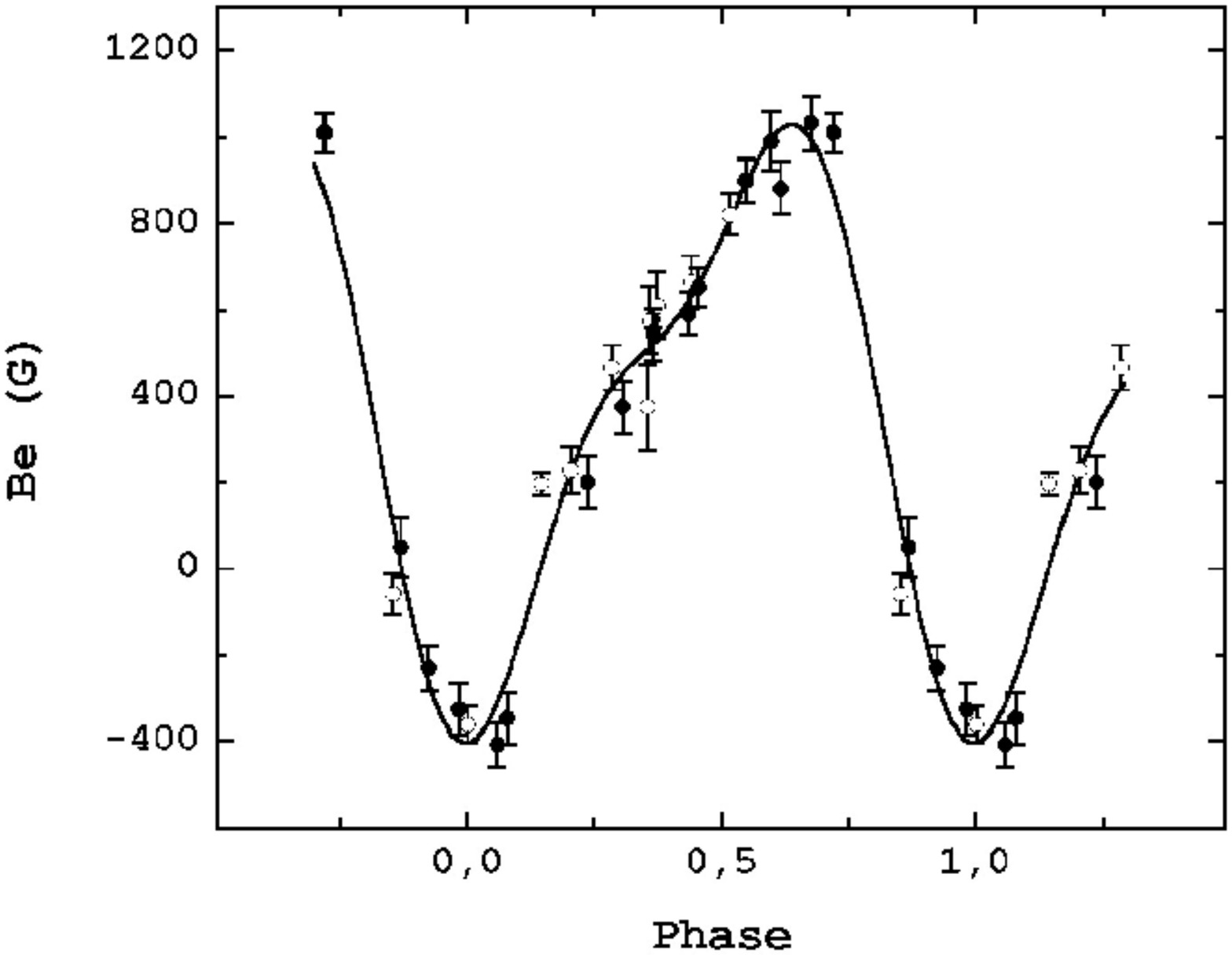}}
\vspace{-3.5mm}
\caption{ HD189775 (1) }
\label{fig:fig303}
\end{figure}

\begin{figure}
\resizebox{0.98\hsize}{!}{\includegraphics{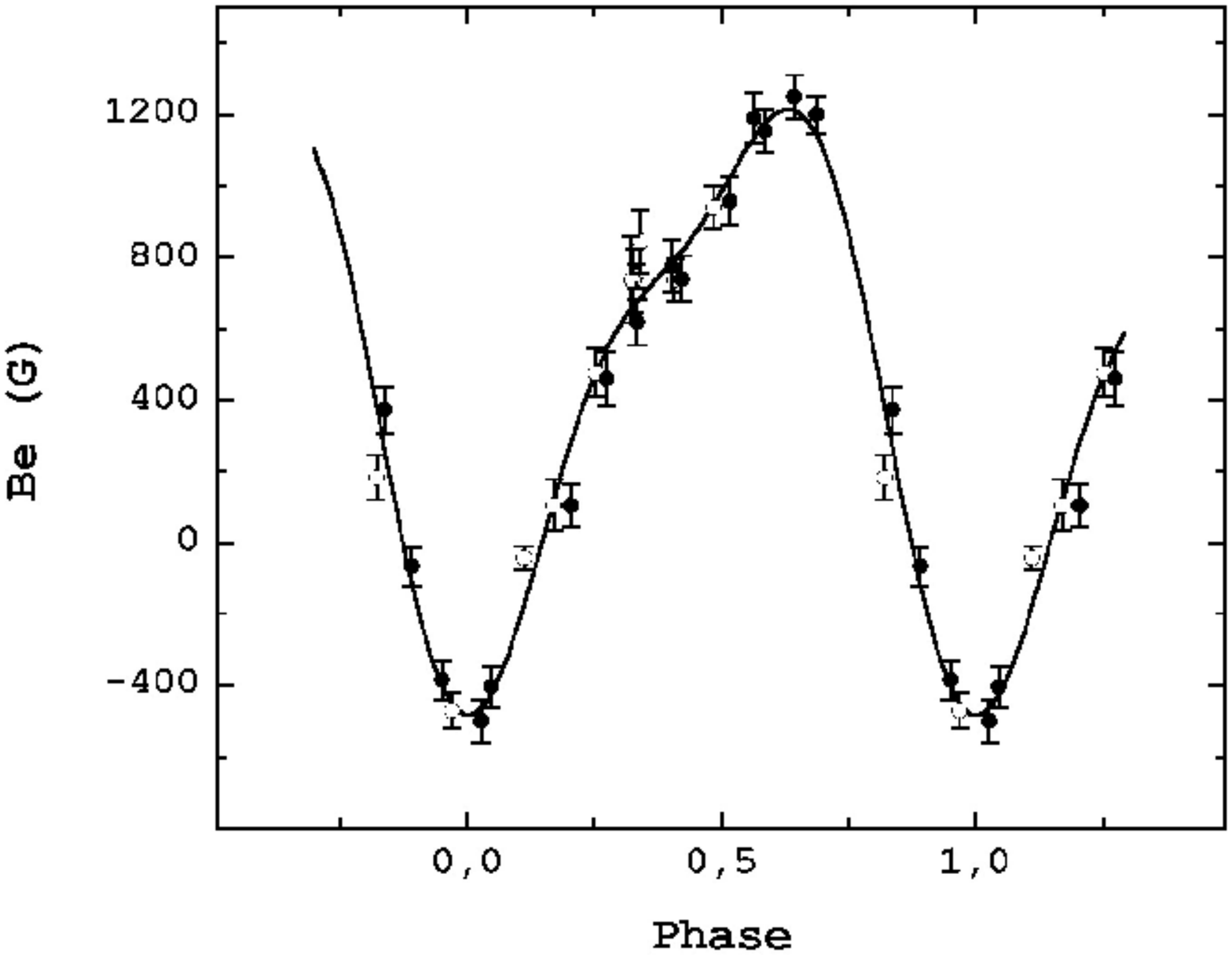}}
\vspace{-3.5mm}
\caption{ HD189775 (2) }
\label{fig:fig303}
\end{figure}

\begin{figure}
\resizebox{0.98\hsize}{!}{\includegraphics{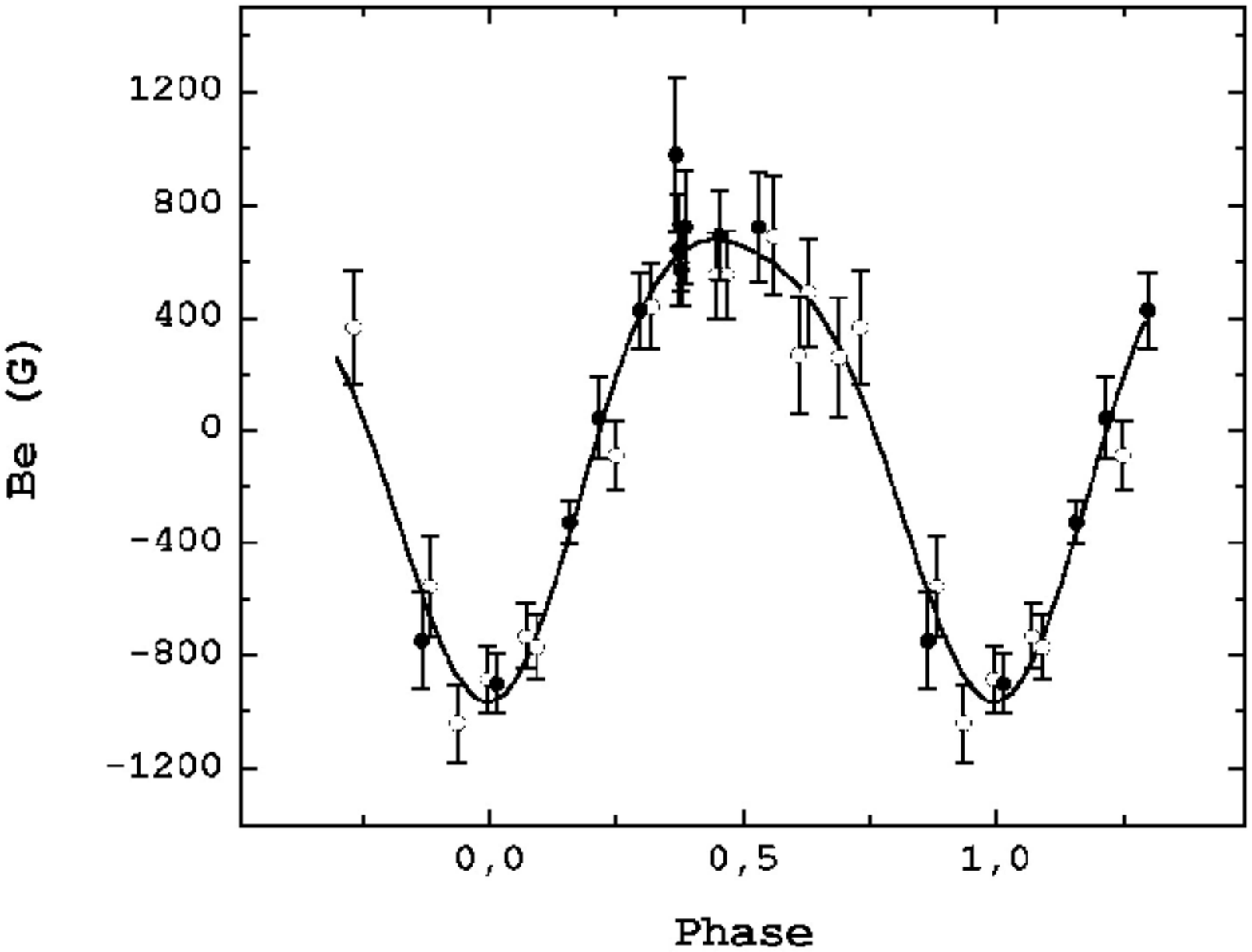}}
\vspace{-3.5mm}
\caption{ HD189775 (3) }
\label{fig:fig303}
\end{figure}

\begin{figure}
\resizebox{0.98\hsize}{!}{\includegraphics{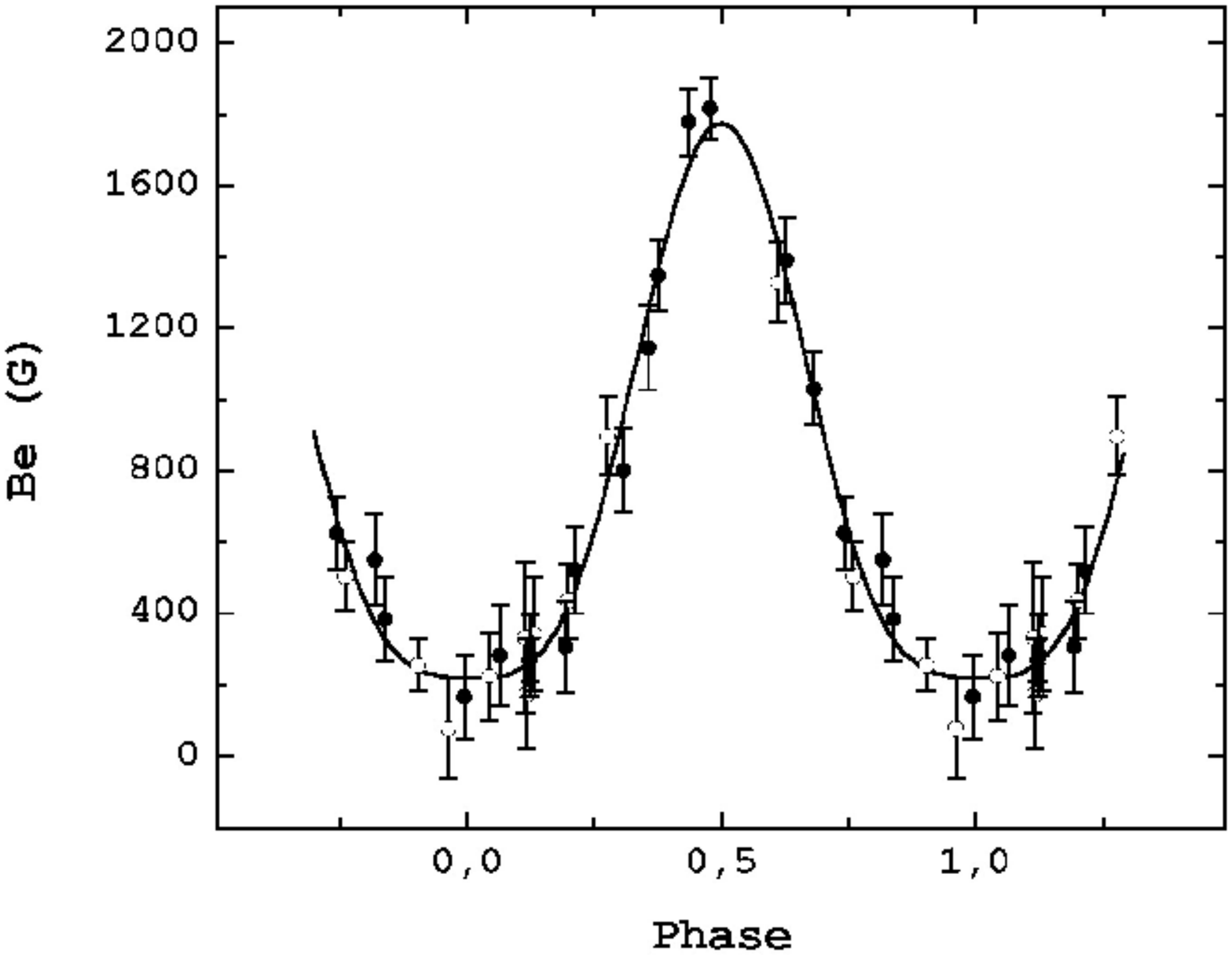}}
\vspace{-3.5mm}
\caption{ HD189775 (4) }
\label{fig:fig303}
\end{figure}

\clearpage
\newpage

\begin{figure}
\resizebox{0.98\hsize}{!}{\includegraphics{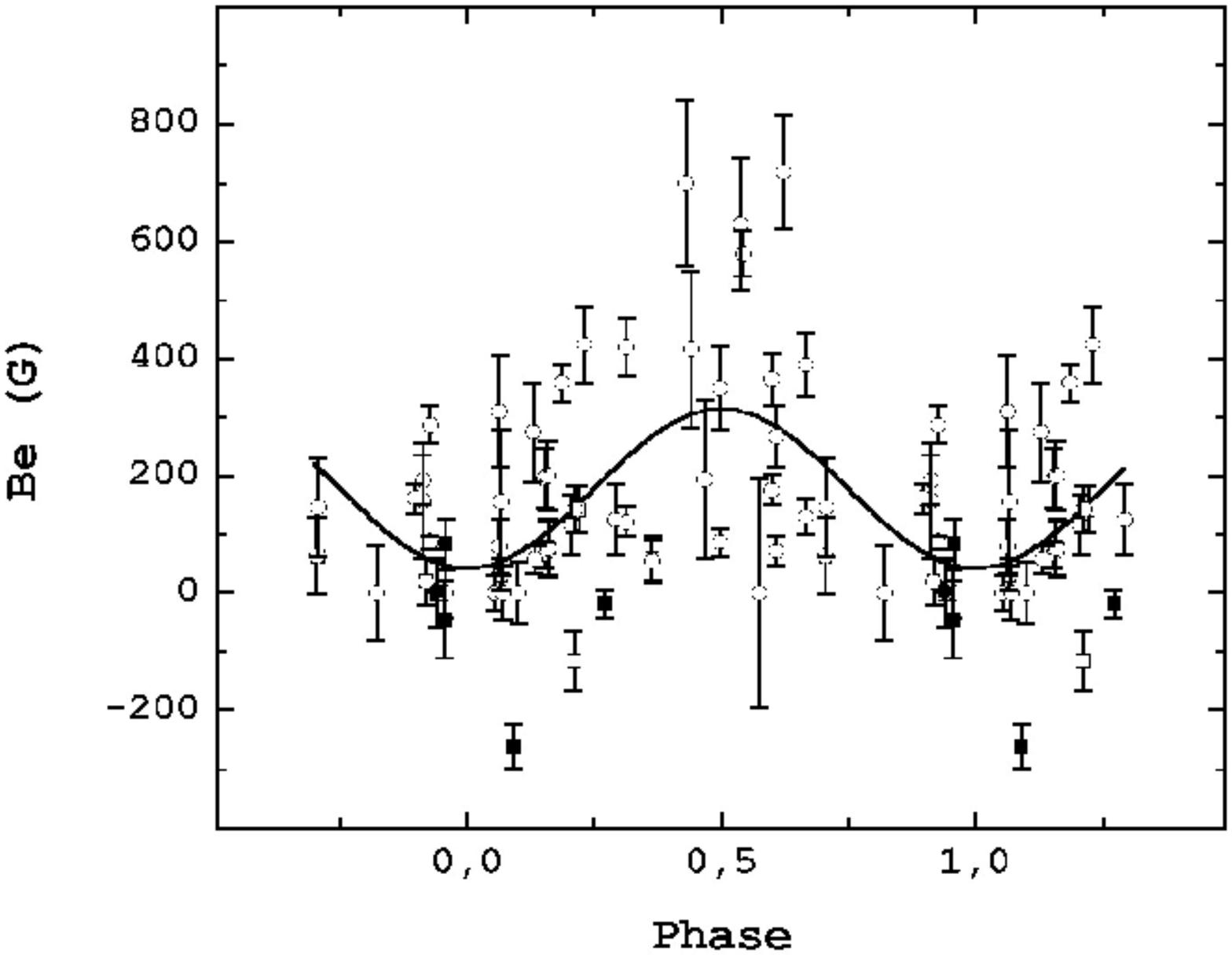}}
\vspace{-3.5mm}
\caption{ HD189849 }
\label{fig:fig325}
\end{figure}

\begin{figure}
\resizebox{0.98\hsize}{!}{\includegraphics{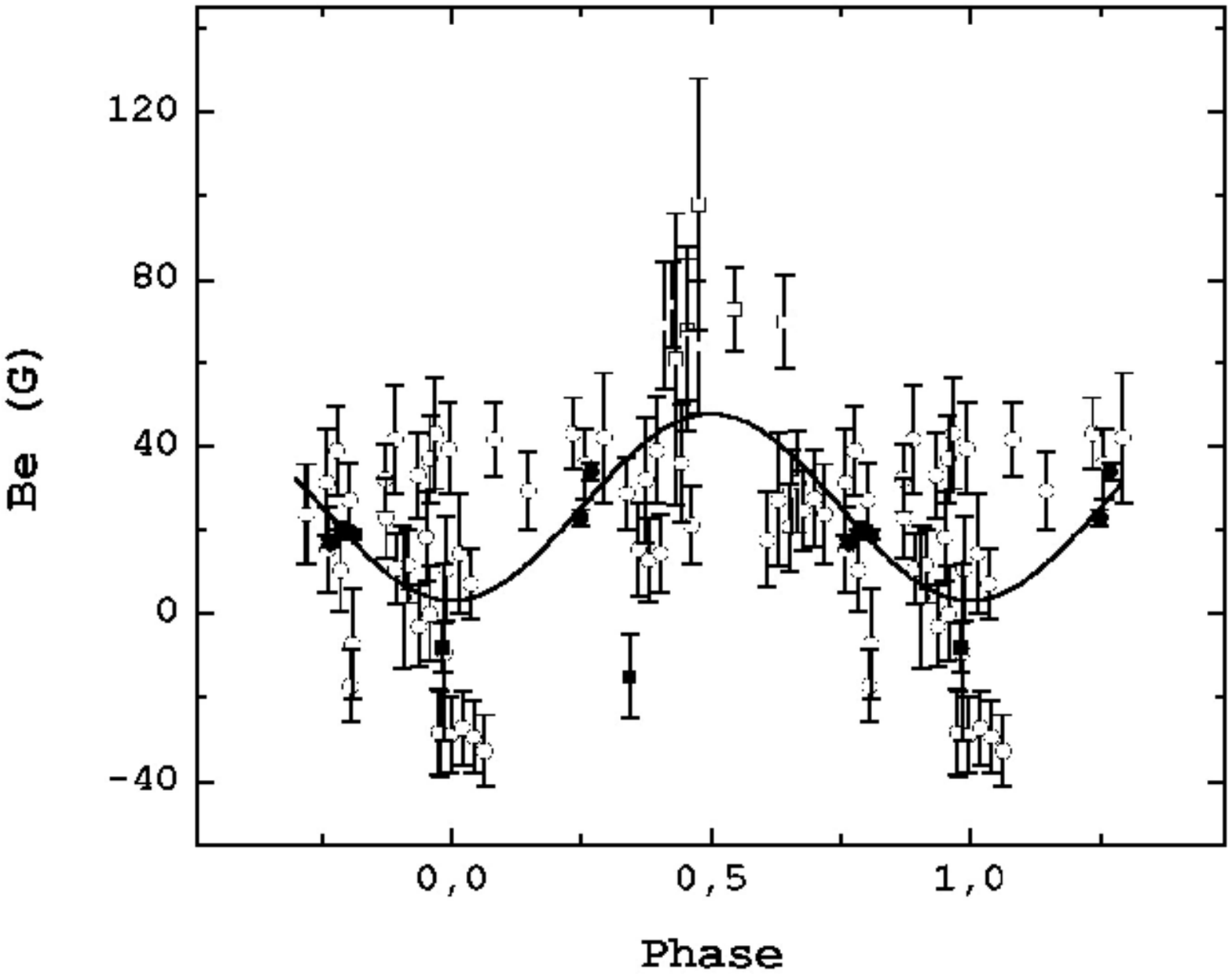}}
\vspace{-3.5mm}
\caption{ HD190073 }
\label{fig:fig326}
\end{figure}

\begin{figure}
\resizebox{0.98\hsize}{!}{\includegraphics{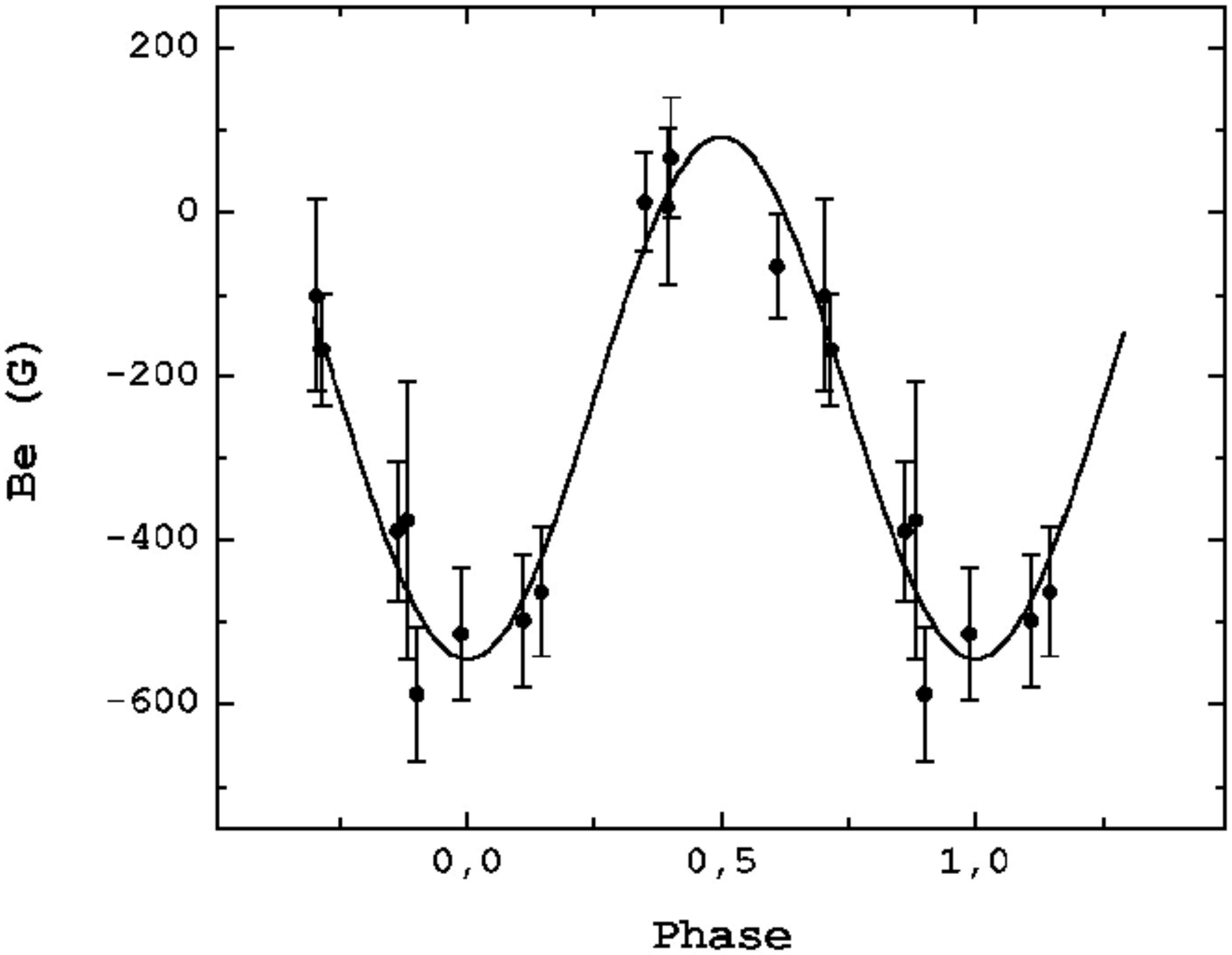}}
\vspace{-3.5mm}
\caption{ HD191612 }
\label{fig:fig327}
\end{figure}

\begin{figure}
\resizebox{0.98\hsize}{!}{\includegraphics{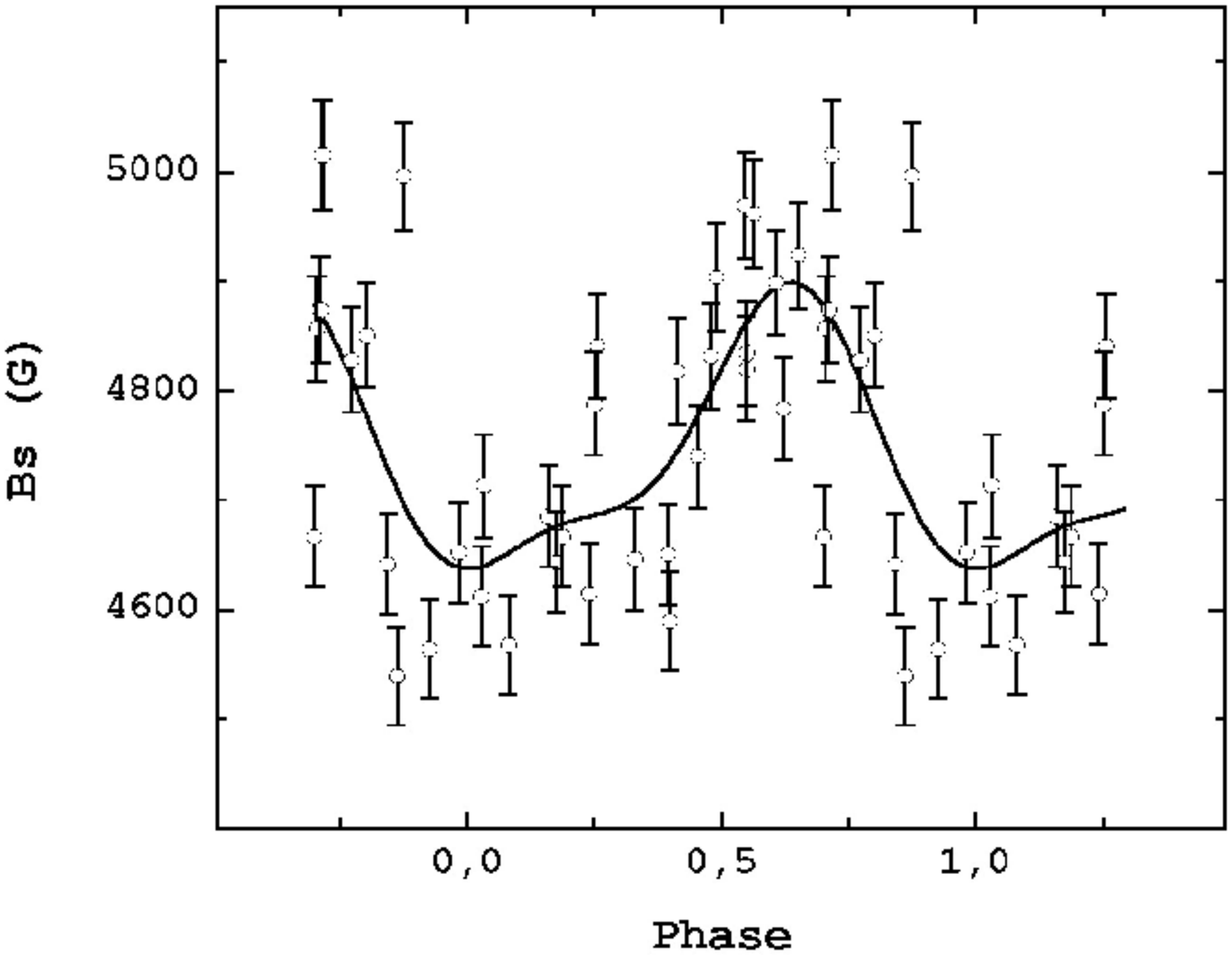}}
\vspace{-3.5mm}
\caption{ HD192678 (1) }
\label{fig:fig328}
\end{figure}

\begin{figure}
\resizebox{0.98\hsize}{!}{\includegraphics{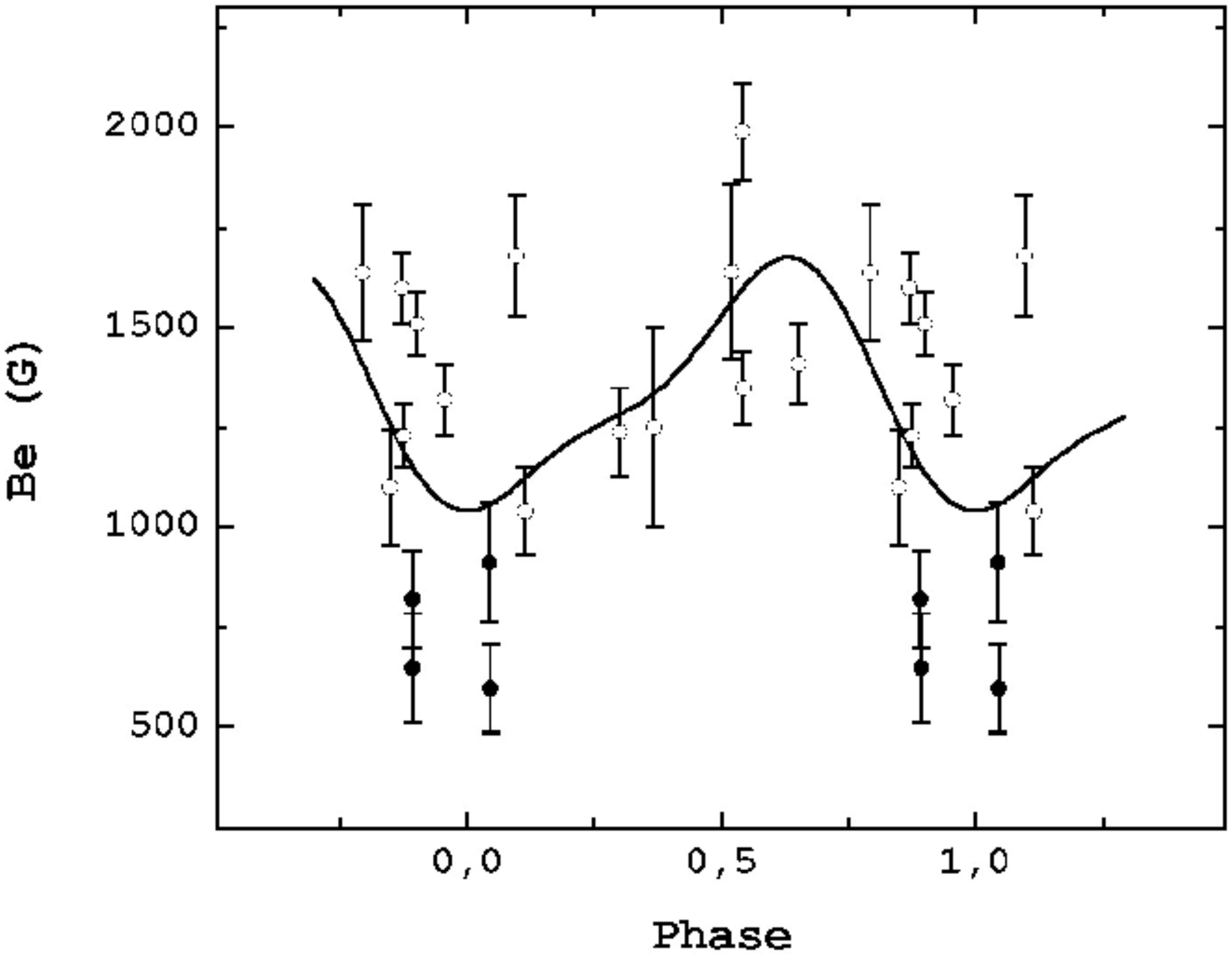}}
\vspace{-3.5mm}
\caption{ HD192678 (2) }
\label{fig:fig329}
\end{figure}

\begin{figure}
\resizebox{0.98\hsize}{!}{\includegraphics{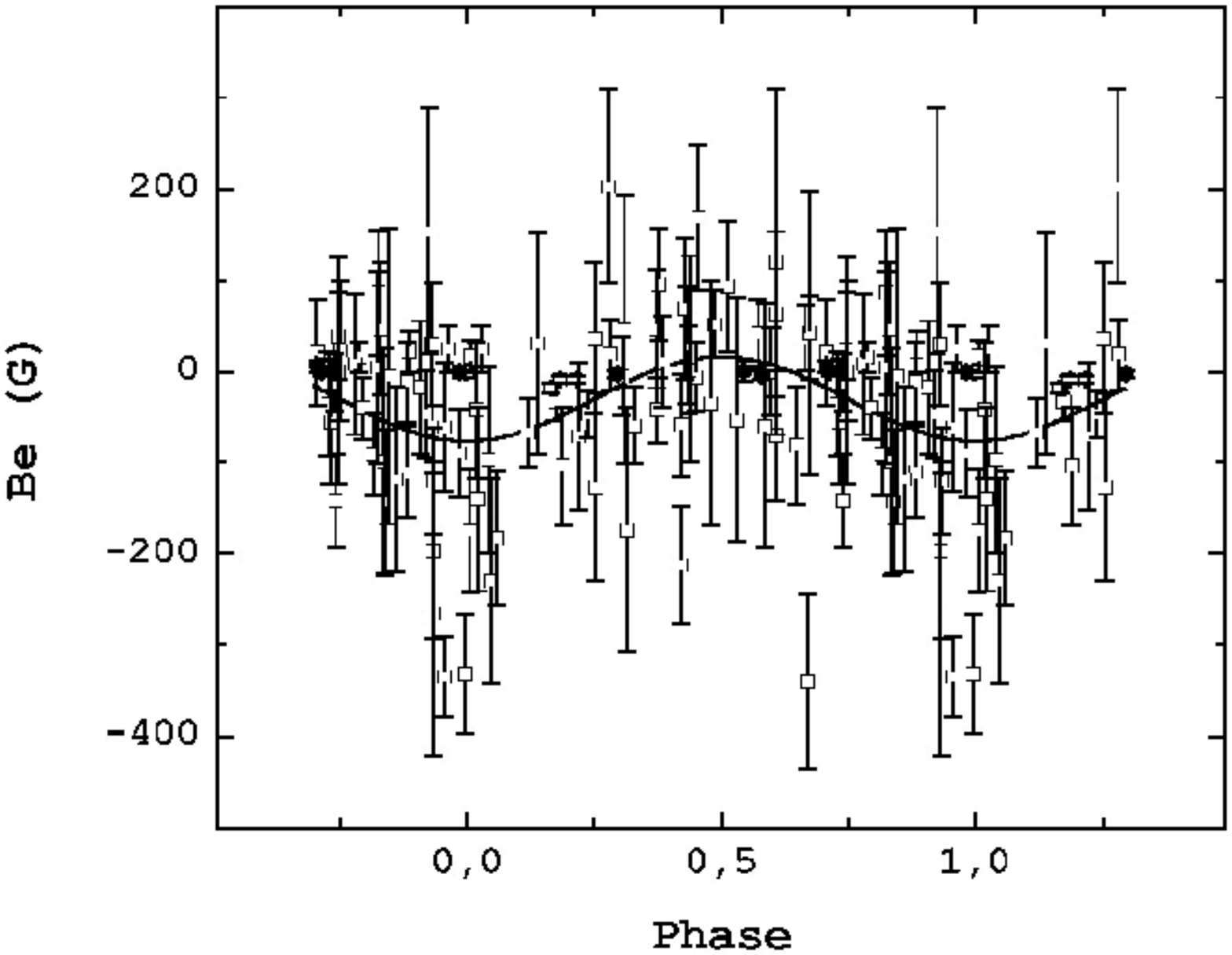}}
\vspace{-3.5mm}
\caption{ HD194093 }
\label{fig:fig330}
\end{figure}

\clearpage
\newpage

\begin{figure}
\resizebox{0.98\hsize}{!}{\includegraphics{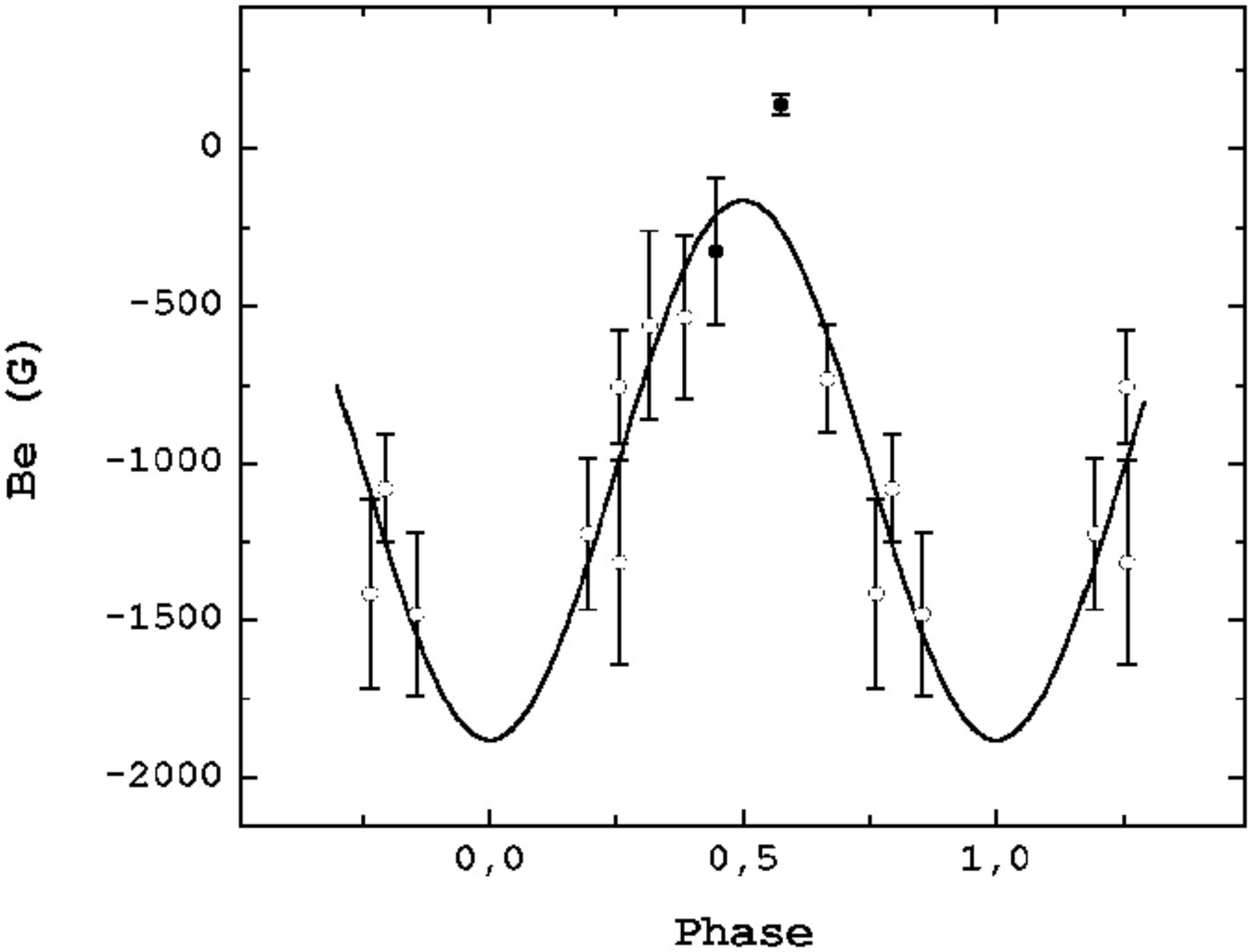}}
\vspace{-3.5mm}
\caption{ HD196178 }
\label{fig:fig331}
\end{figure}

\begin{figure}
\resizebox{0.98\hsize}{!}{\includegraphics{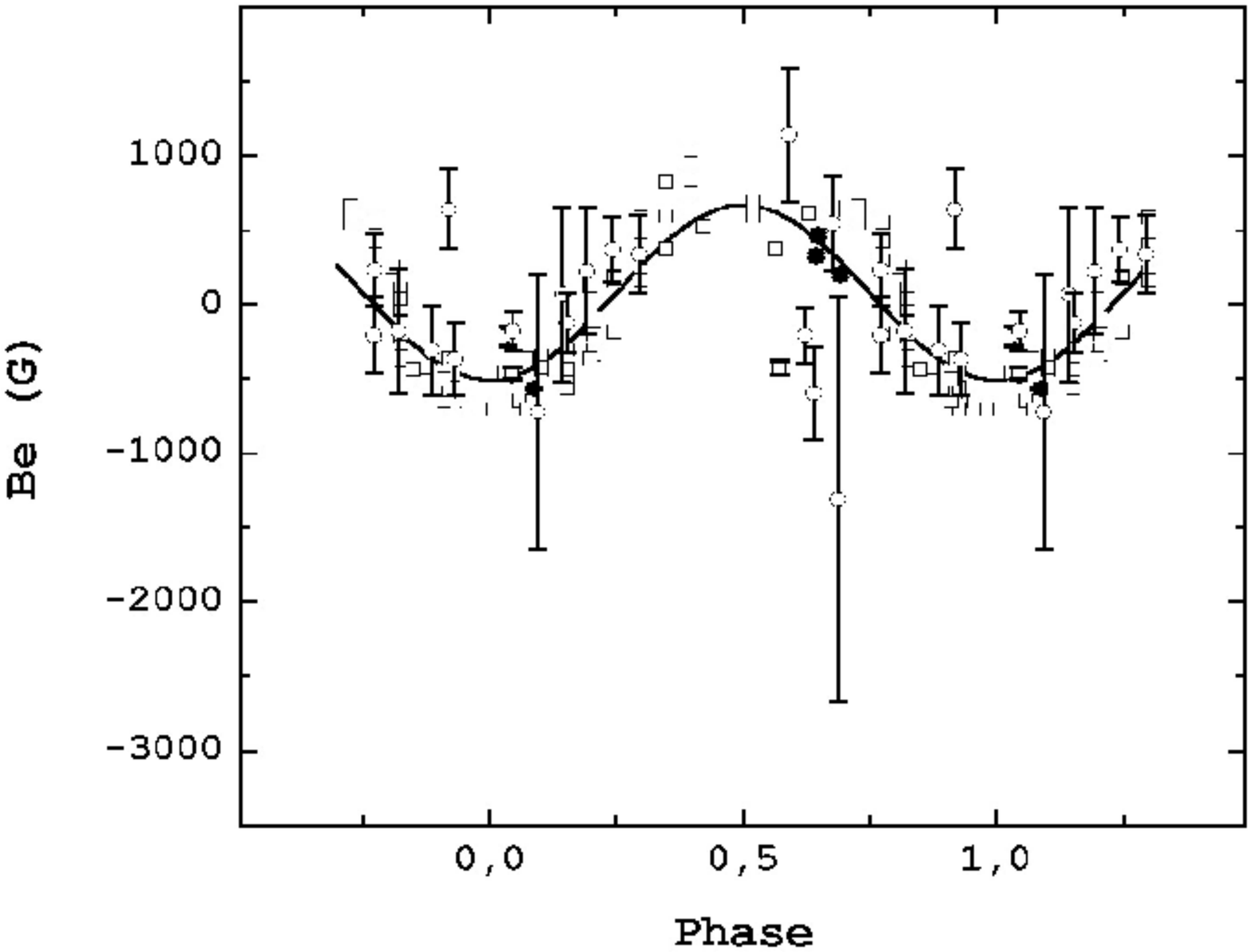}}
\vspace{-3.5mm}
\caption{ HD196502 }
\label{fig:fig332}
\end{figure}

\begin{figure}
\resizebox{0.98\hsize}{!}{\includegraphics{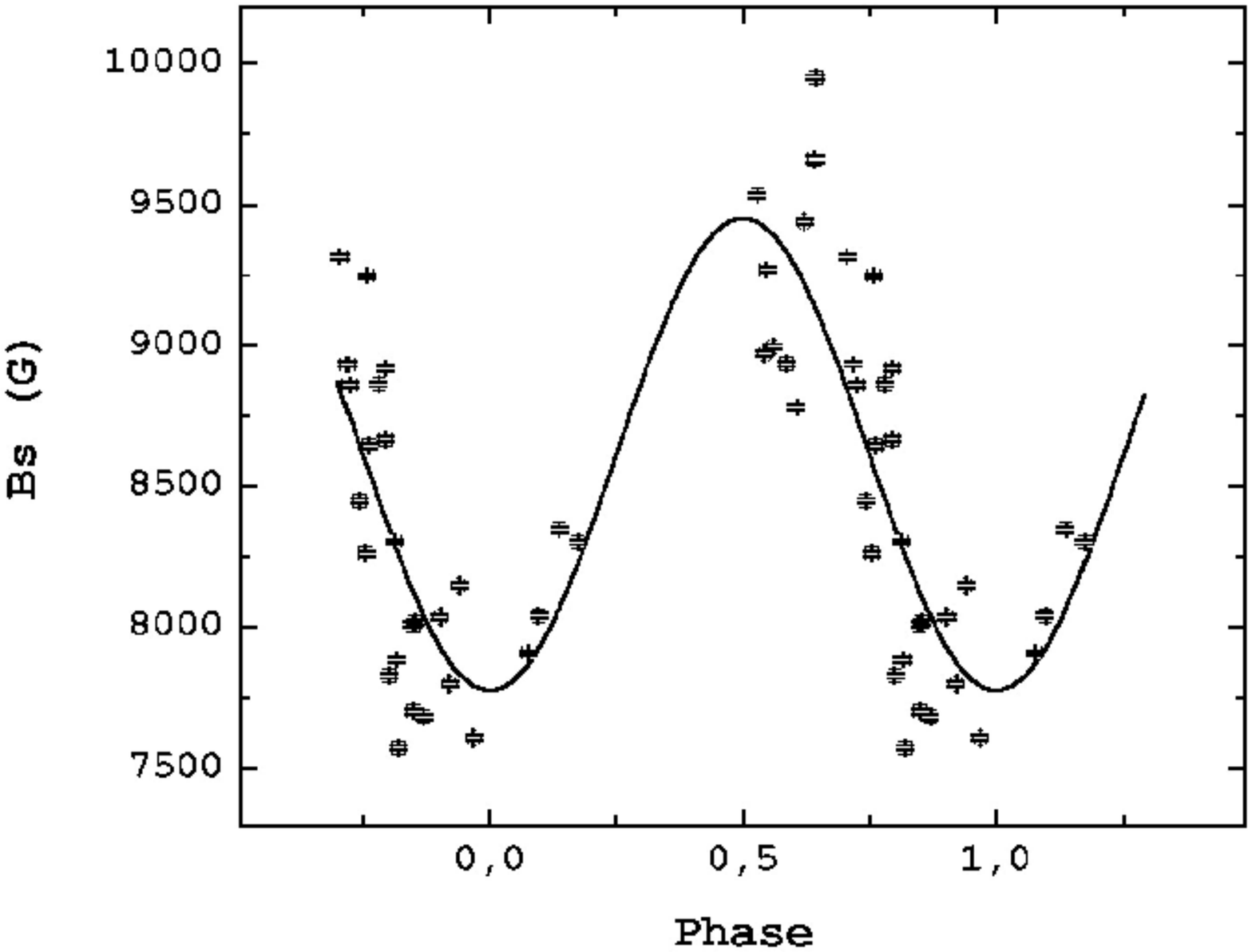}}
\vspace{-3.5mm}
\caption{ HD200311 (1) }
\label{fig:fig333}
\end{figure}

\begin{figure}
\resizebox{0.98\hsize}{!}{\includegraphics{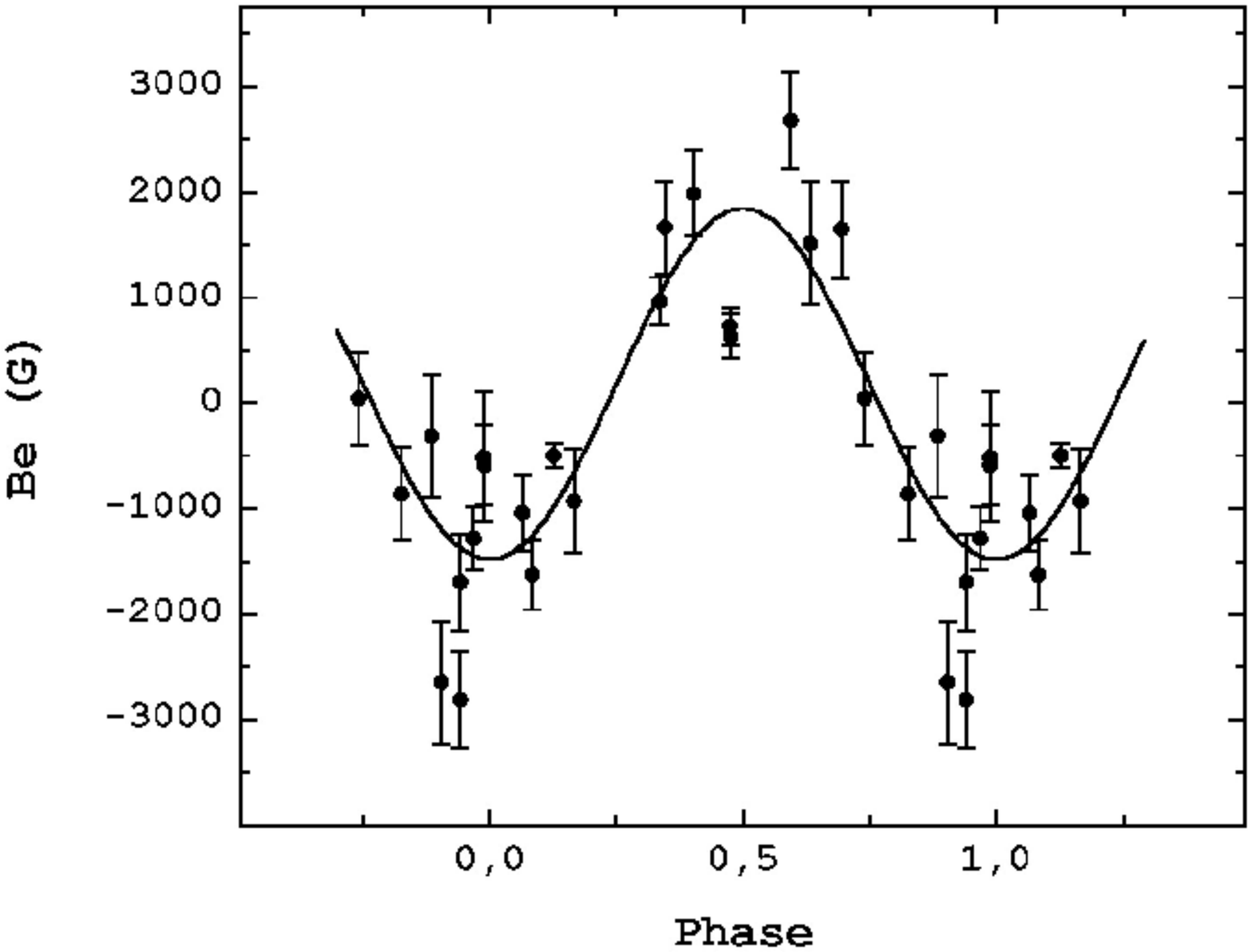}}
\vspace{-3.5mm}
\caption{ HD200311 (2) }
\label{fig:fig334}
\end{figure}

\begin{figure}
\resizebox{0.98\hsize}{!}{\includegraphics{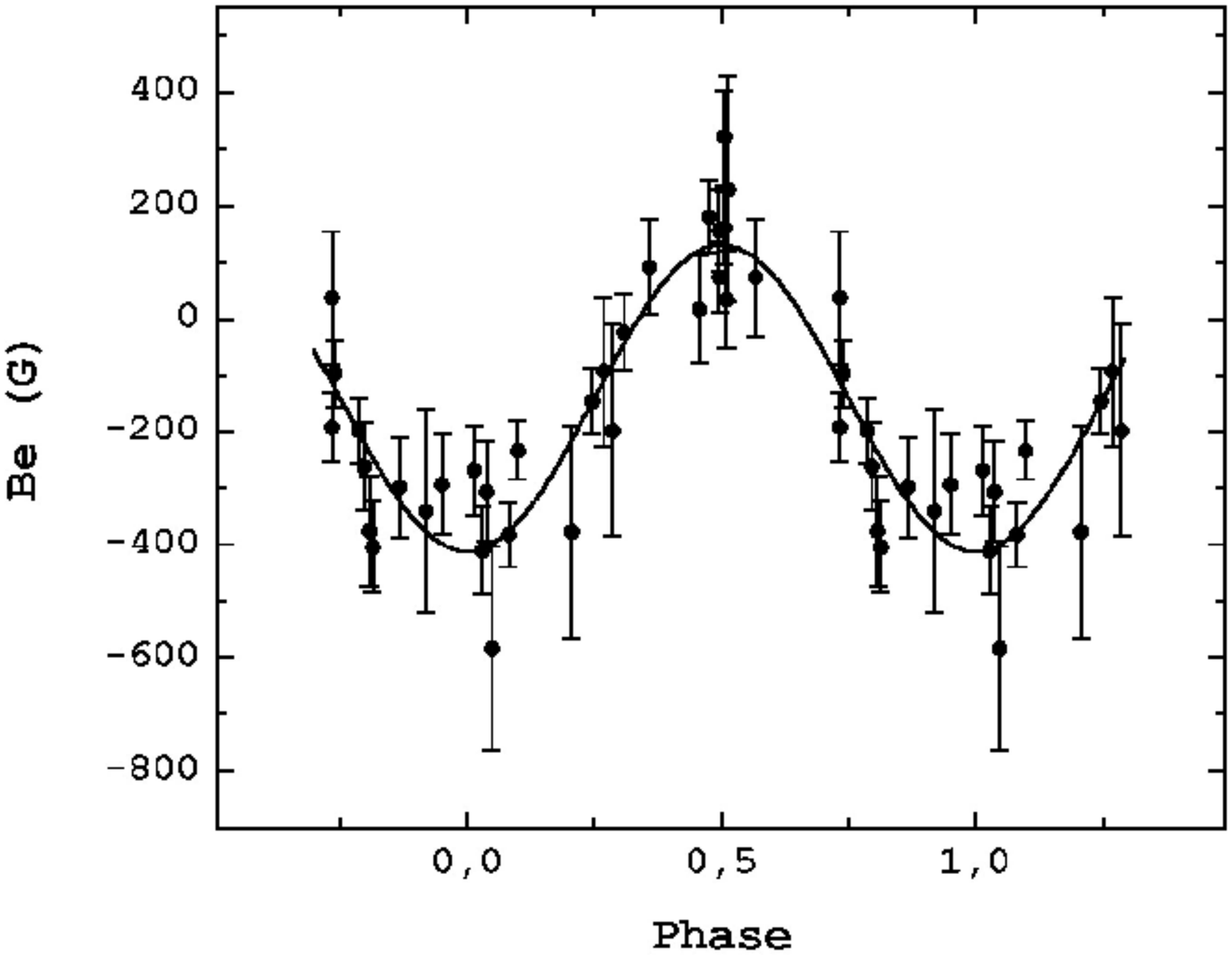}}
\vspace{-3.5mm}
\caption{ HD200775 }
\label{fig:fig335}
\end{figure}

\begin{figure}
\resizebox{0.98\hsize}{!}{\includegraphics{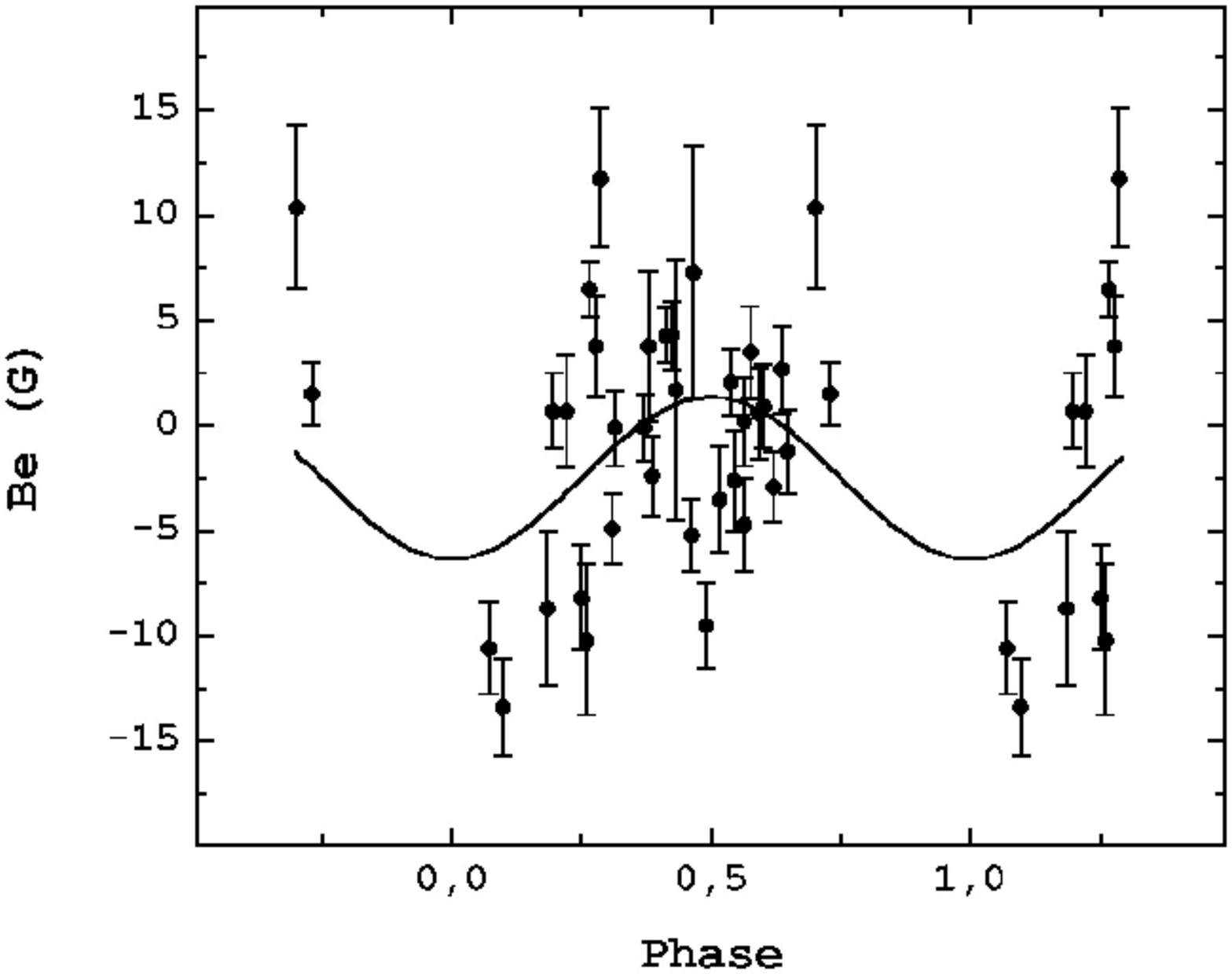}}
\vspace{-3.5mm}
\caption{ HD201091 (1) }
\label{fig:fig336}
\end{figure}

\clearpage
\newpage

\begin{figure}
\resizebox{0.98\hsize}{!}{\includegraphics{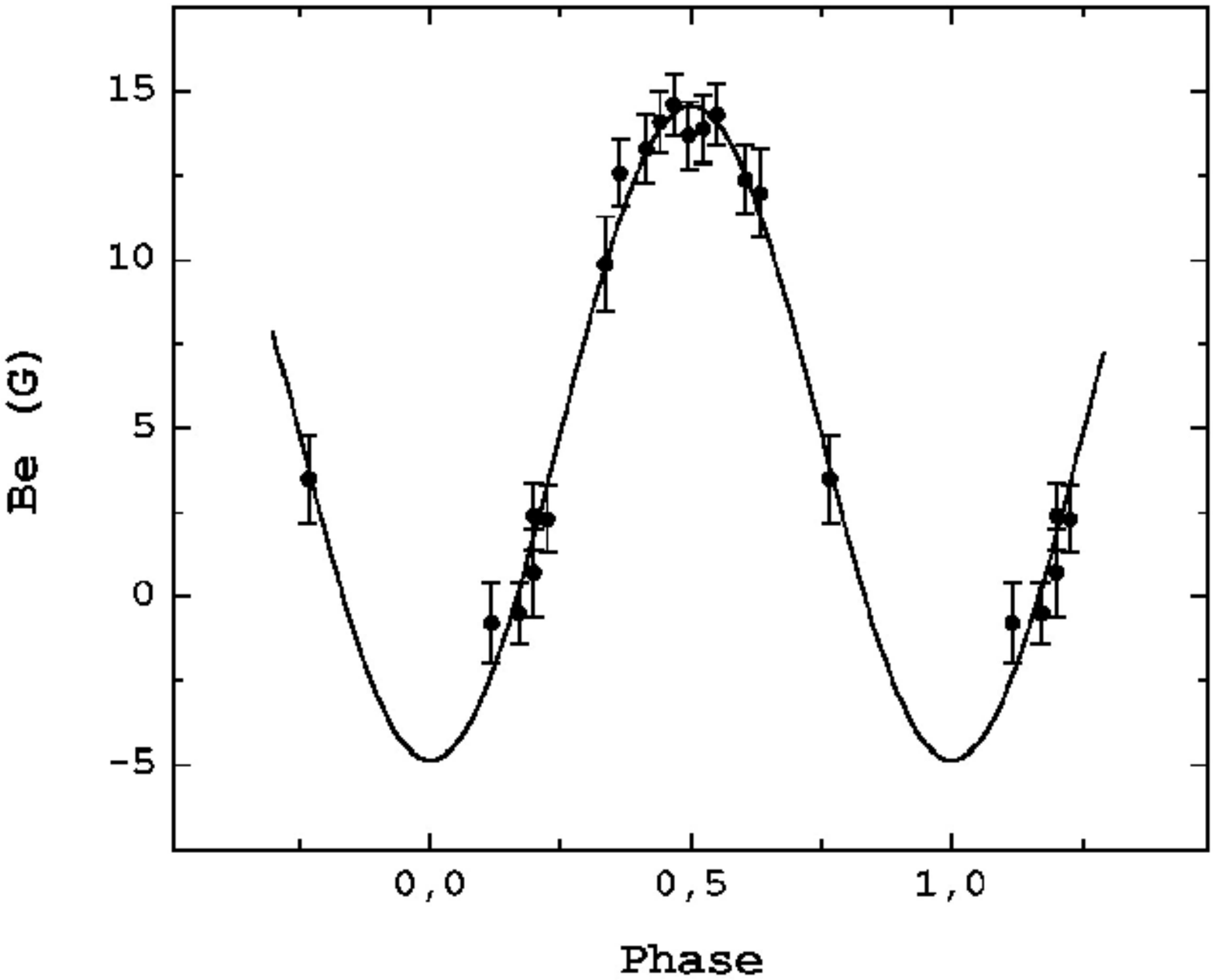}}
\vspace{-3.5mm}
\caption{ HD201091 (2) }
\label{fig:fig337}
\end{figure}

\begin{figure}
\resizebox{0.98\hsize}{!}{\includegraphics{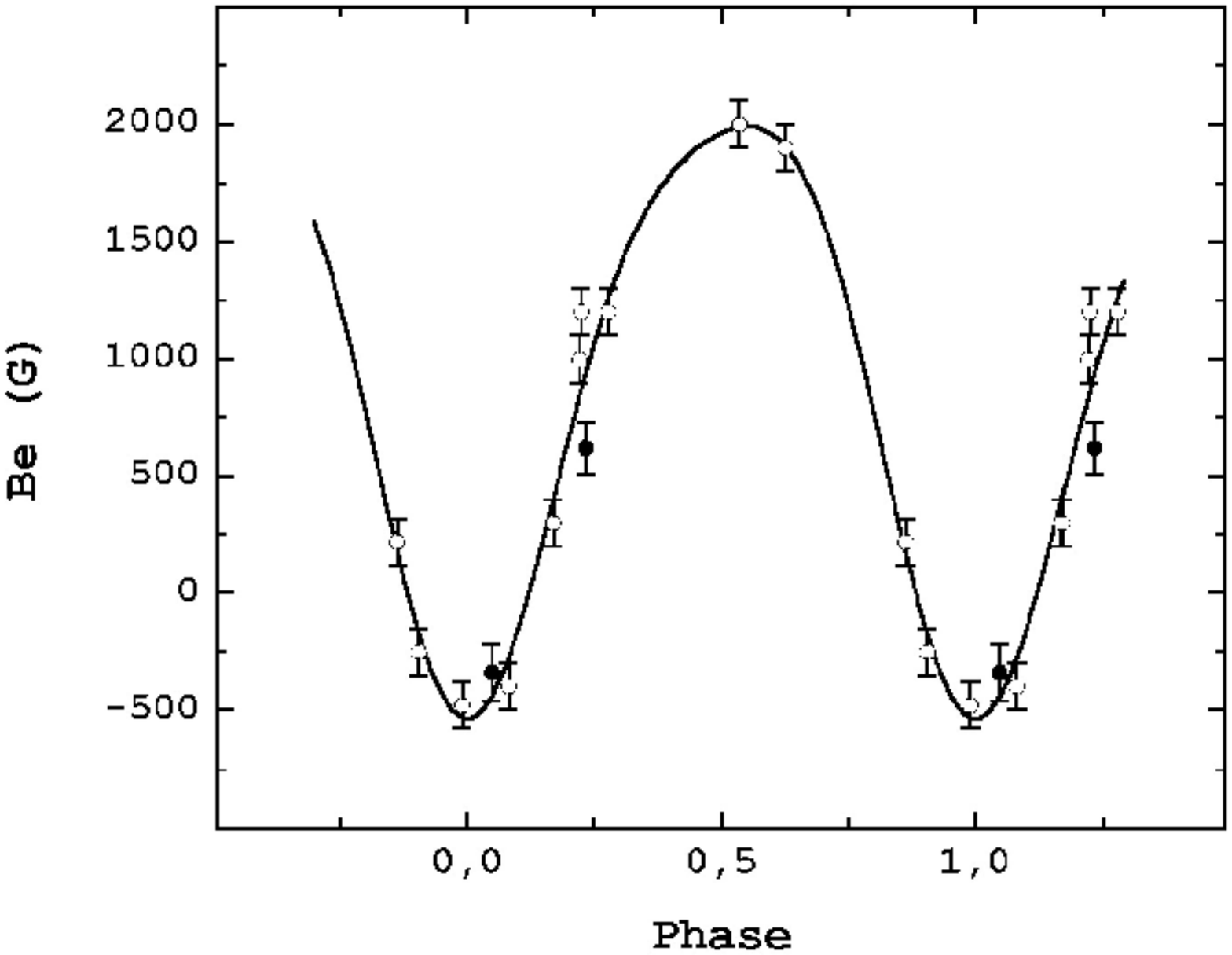}}
\vspace{-3.5mm}
\caption{ HD201174 (1) }
\label{fig:fig338}
\end{figure}

\begin{figure}
\resizebox{0.98\hsize}{!}{\includegraphics{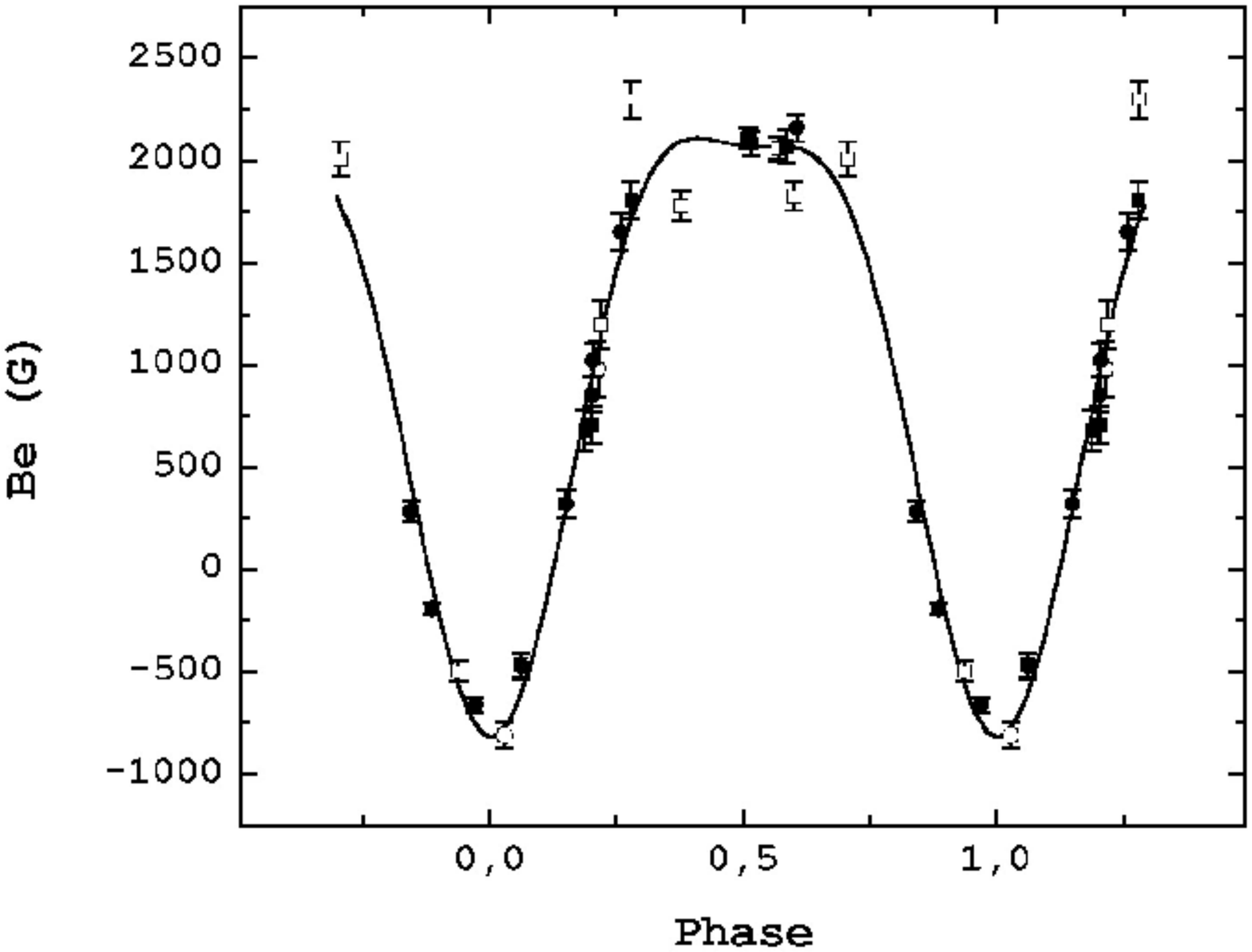}}
\vspace{-3.5mm}
\caption{ HD201174 (2) }
\label{fig:fig338}
\end{figure}

\begin{figure}
\resizebox{0.98\hsize}{!}{\includegraphics{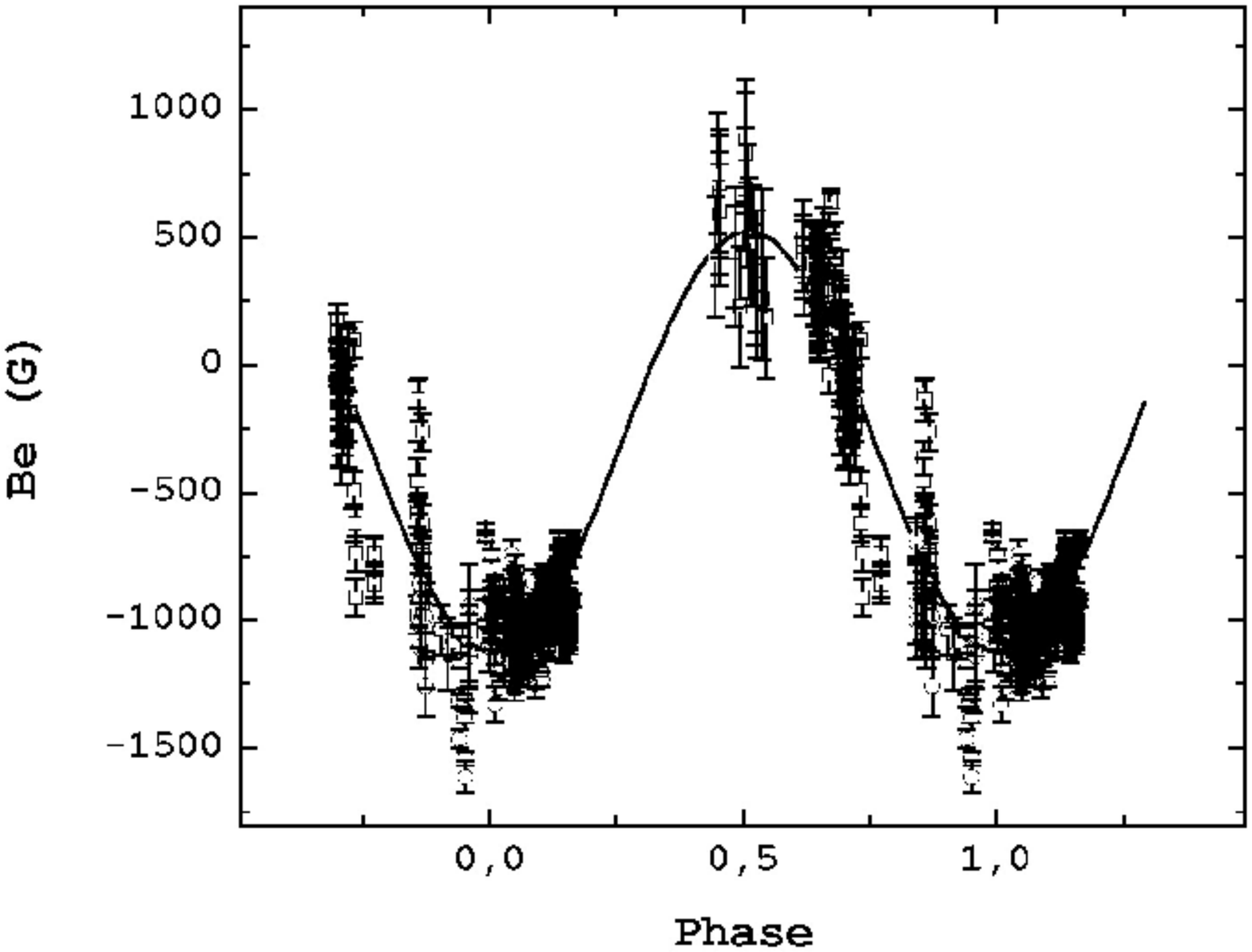}}
\vspace{-3.5mm}
\caption{ HD201601 }
\label{fig:fig339}
\end{figure}

\begin{figure}
\resizebox{0.98\hsize}{!}{\includegraphics{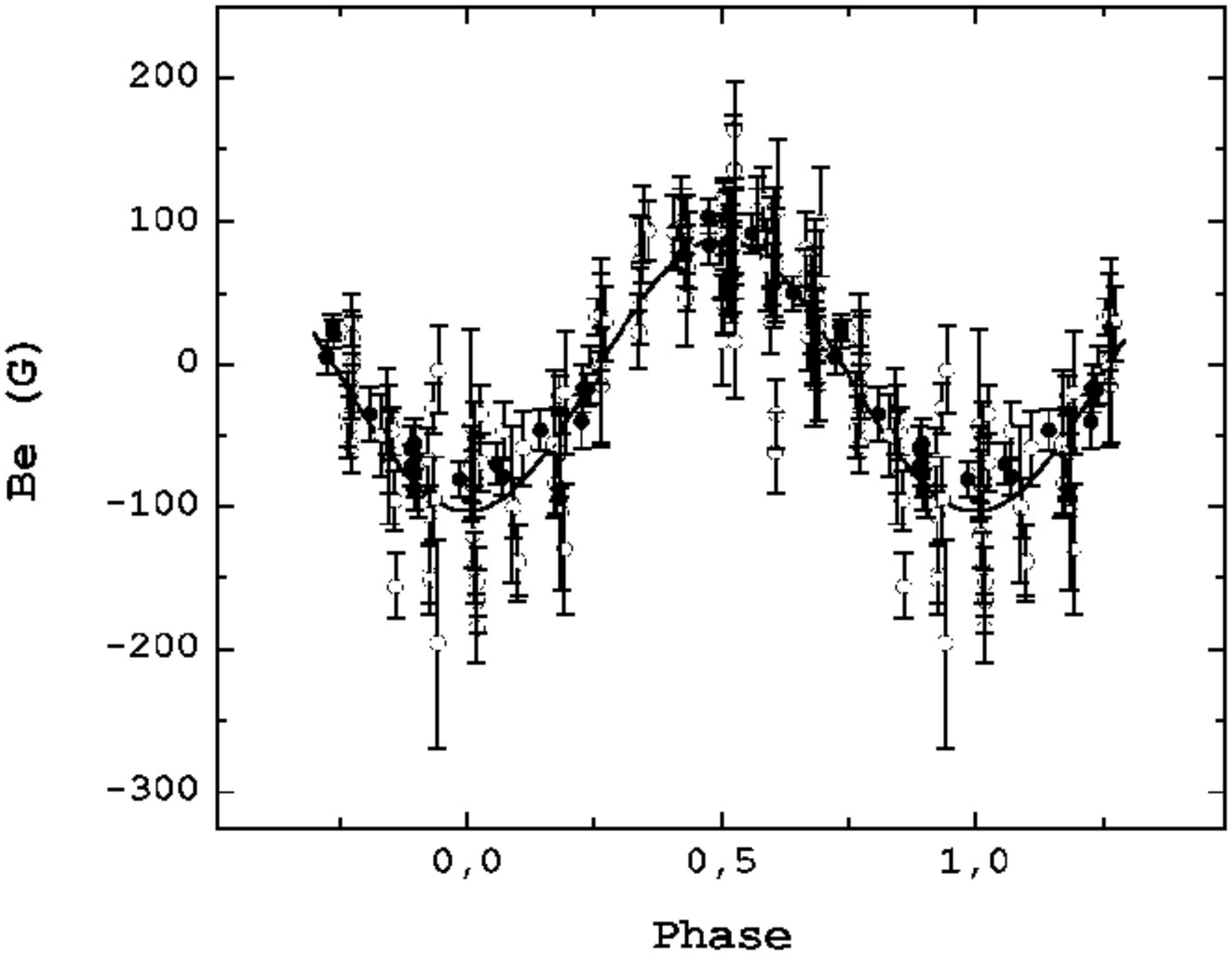}}
\vspace{-3.5mm}
\caption{ HD205021 (1) }
\label{fig:fig340}
\end{figure}

\begin{figure}
\resizebox{0.98\hsize}{!}{\includegraphics{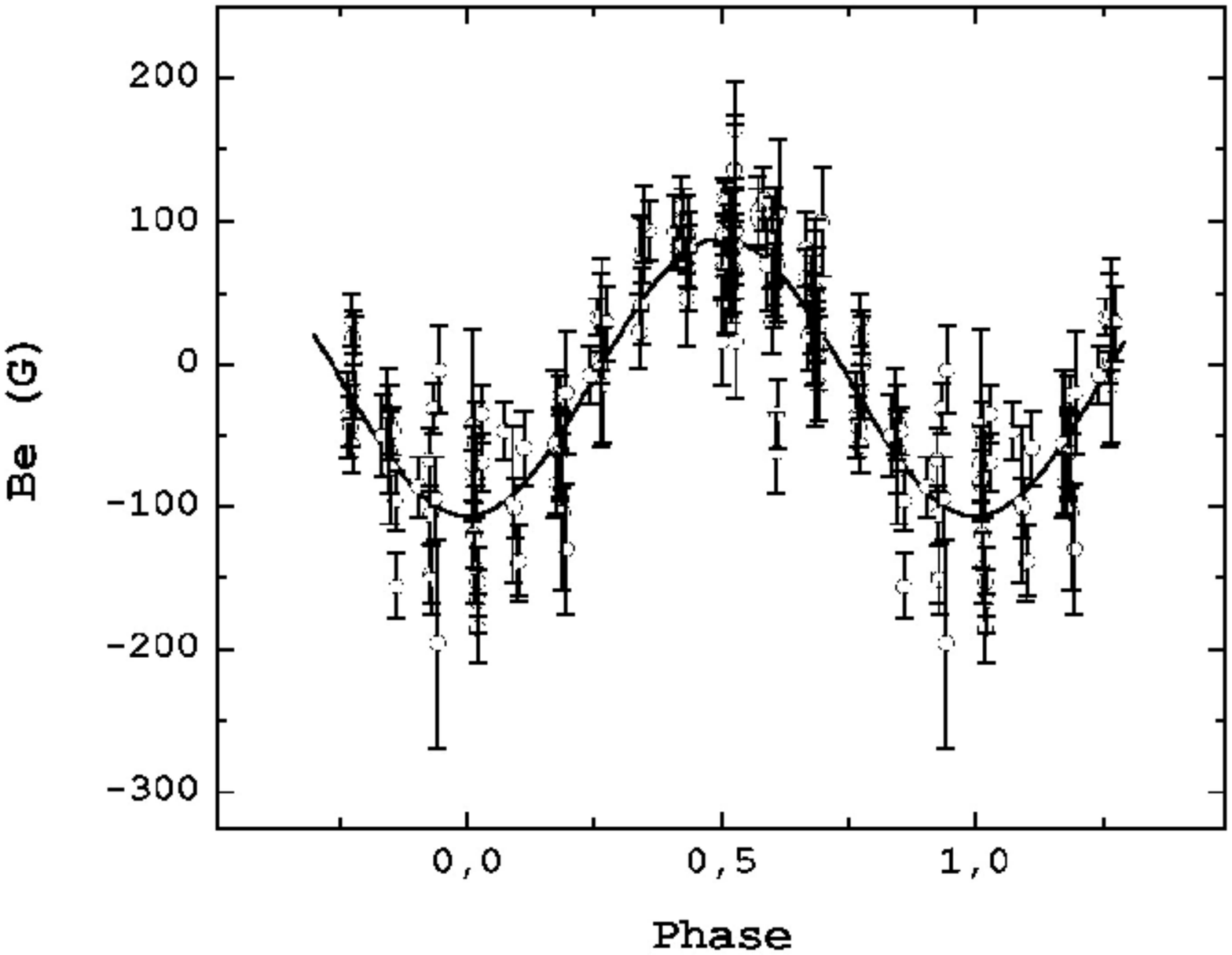}}
\vspace{-3.5mm}
\caption{ HD205021 (2) }
\label{fig:fig340}
\end{figure}

\clearpage
\newpage

\begin{figure}
\resizebox{0.98\hsize}{!}{\includegraphics{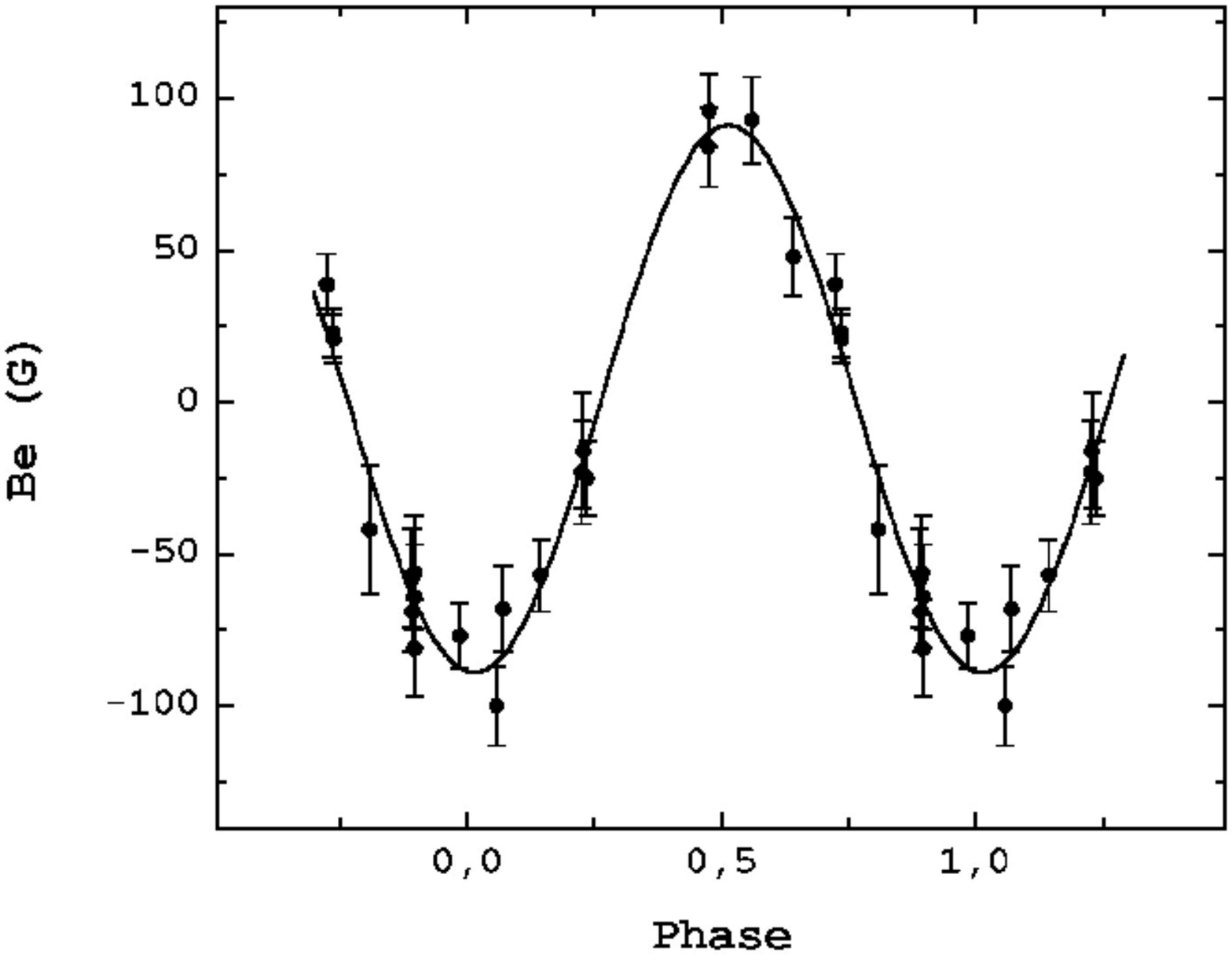}}
\vspace{-3.5mm}
\caption{ HD205021 (3) }
\label{fig:fig340}
\end{figure}

\begin{figure}
\resizebox{0.98\hsize}{!}{\includegraphics{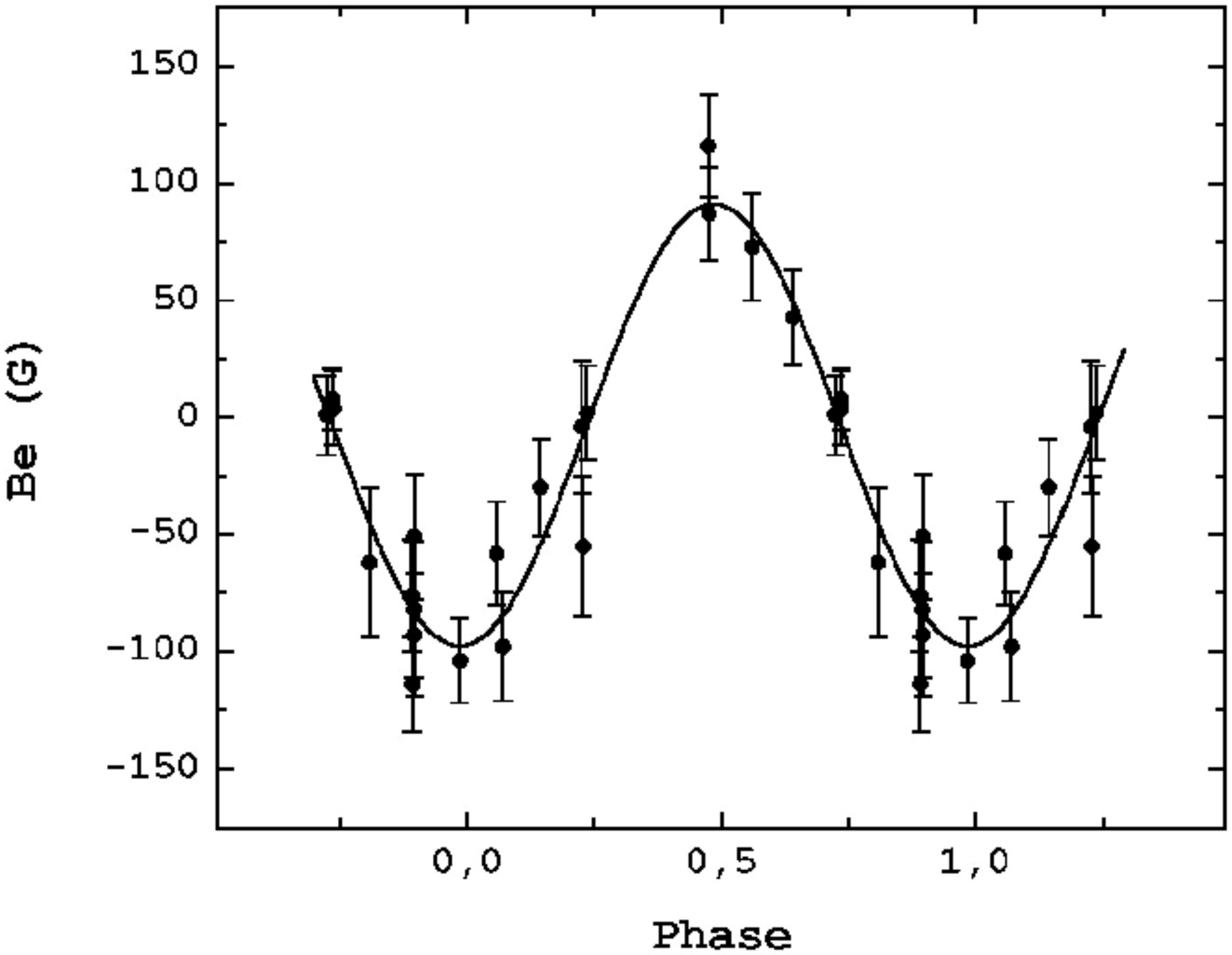}}
\vspace{-3.5mm}
\caption{ HD205021 (4) }
\label{fig:fig340}
\end{figure}

\begin{figure}
\resizebox{0.98\hsize}{!}{\includegraphics{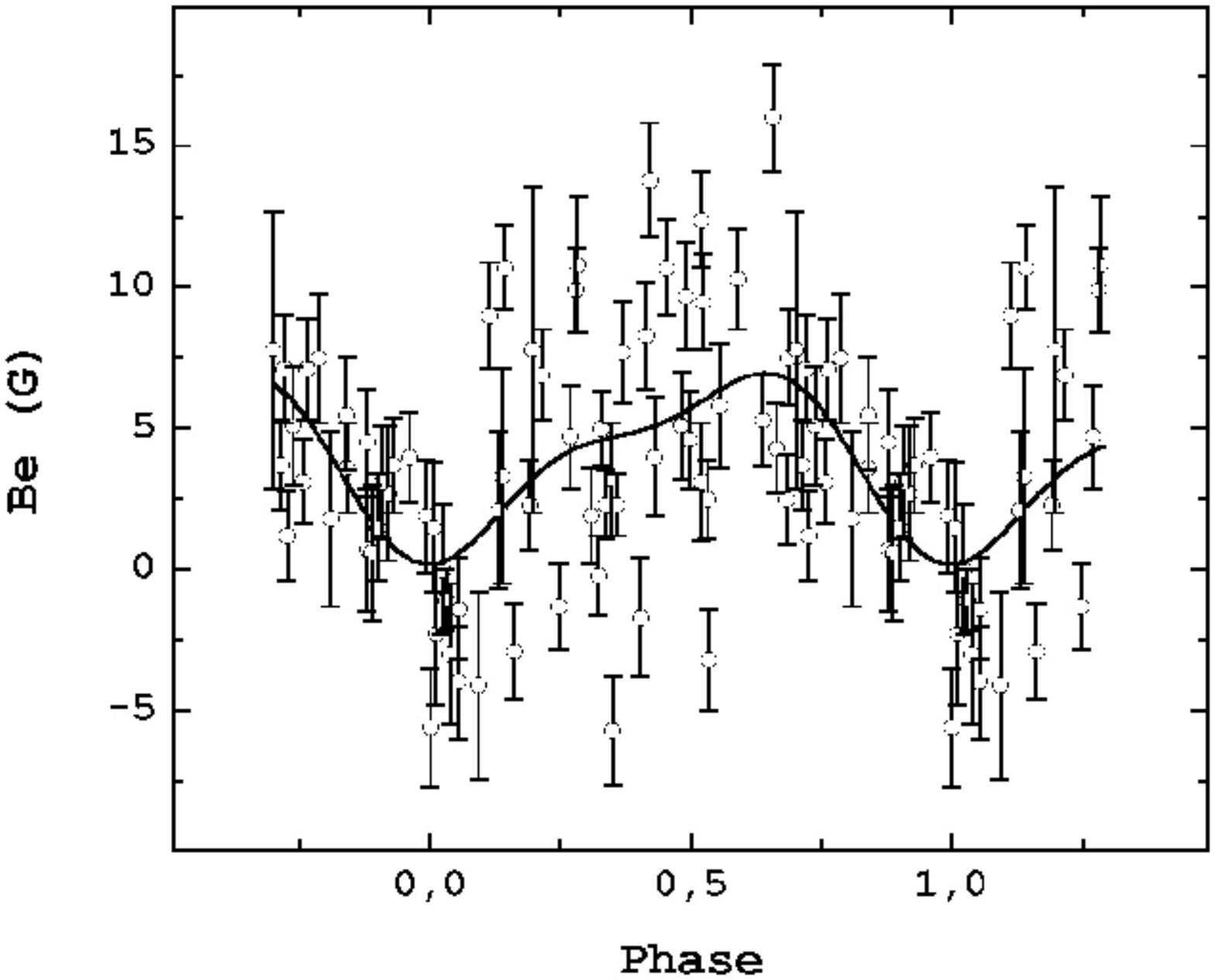}}
\vspace{-3.5mm}
\caption{ HD206860 }
\label{fig:fig341}
\end{figure}

\begin{figure}
\resizebox{0.98\hsize}{!}{\includegraphics{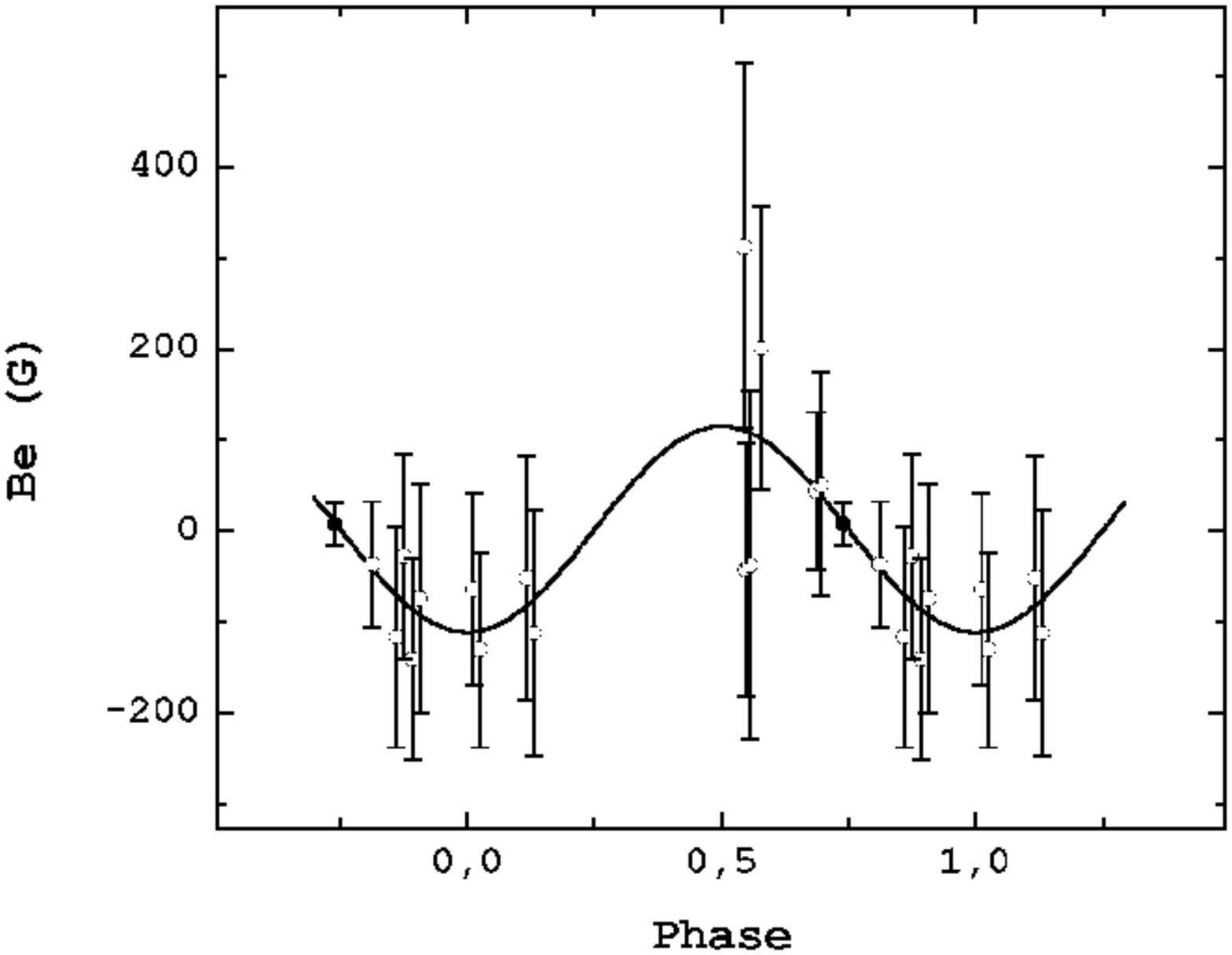}}
\vspace{-3.5mm}
\caption{ HD207330 }
\label{fig:fig342}
\end{figure}

\begin{figure}
\resizebox{0.98\hsize}{!}{\includegraphics{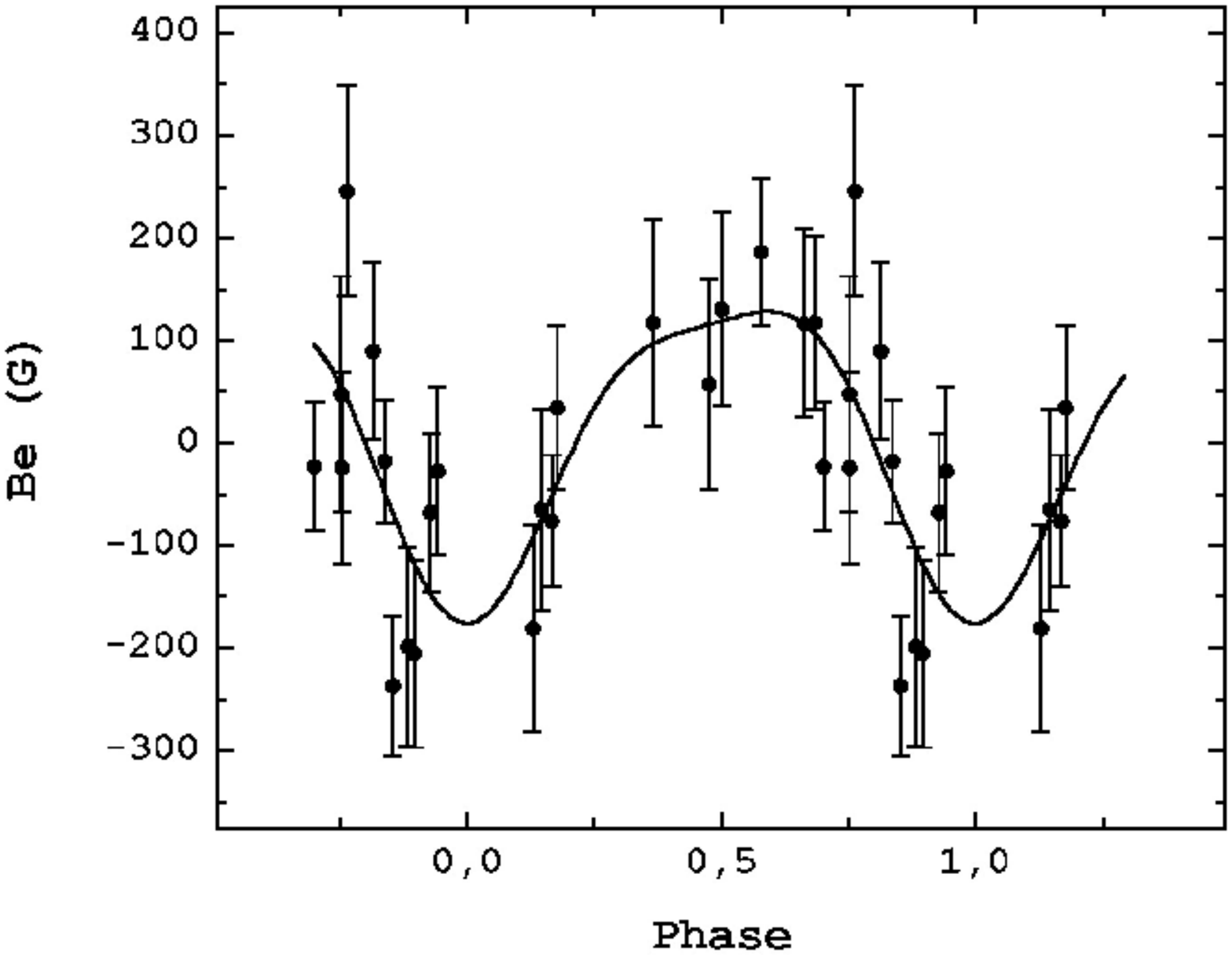}}
\vspace{-3.5mm}
\caption{ HD208057 (1) }
\label{fig:fig340}
\end{figure}

\begin{figure}
\resizebox{0.98\hsize}{!}{\includegraphics{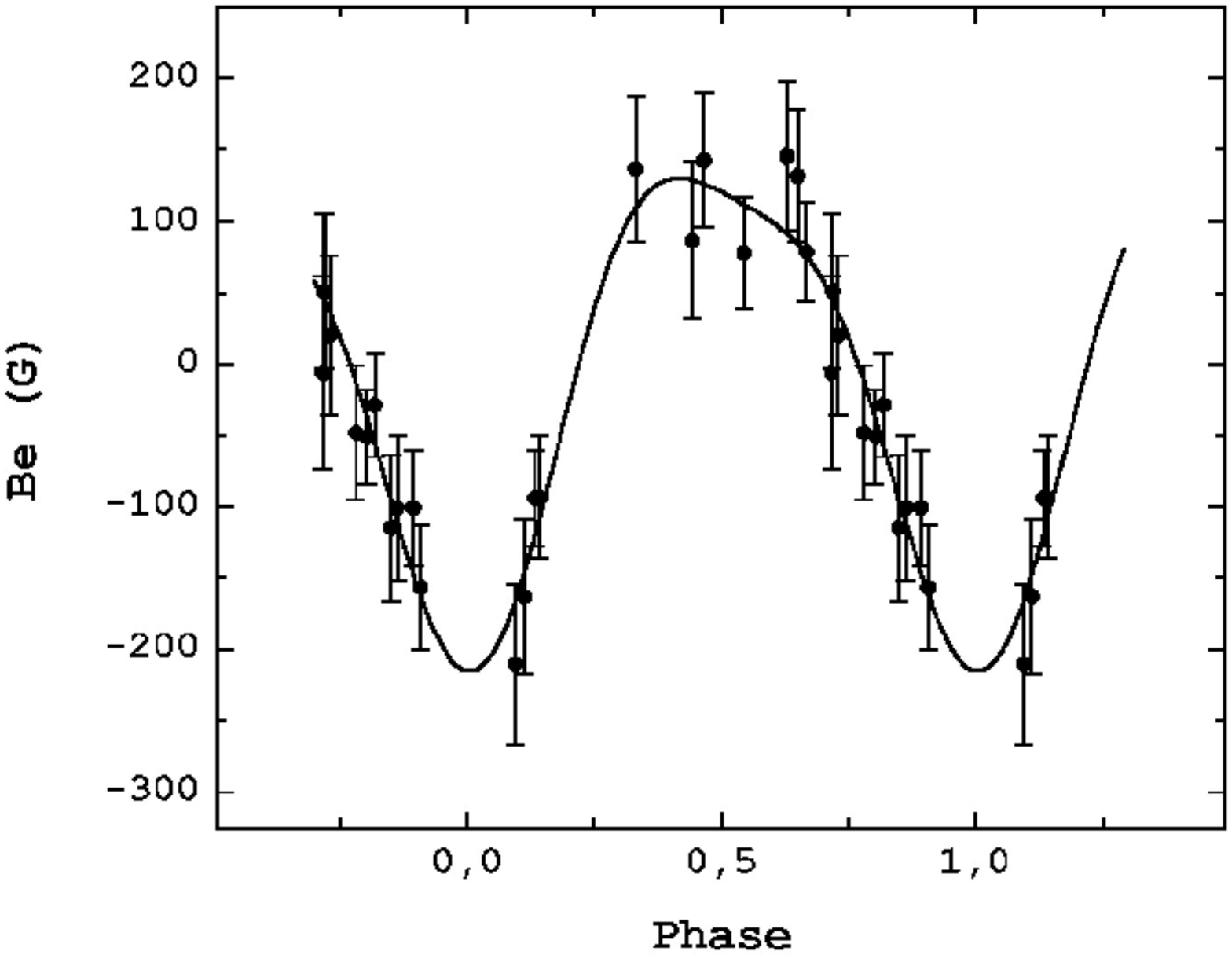}}
\vspace{-3.5mm}
\caption{ HD208057 (2) }
\label{fig:fig340}
\end{figure}

\clearpage
\newpage

\begin{figure}
\resizebox{0.98\hsize}{!}{\includegraphics{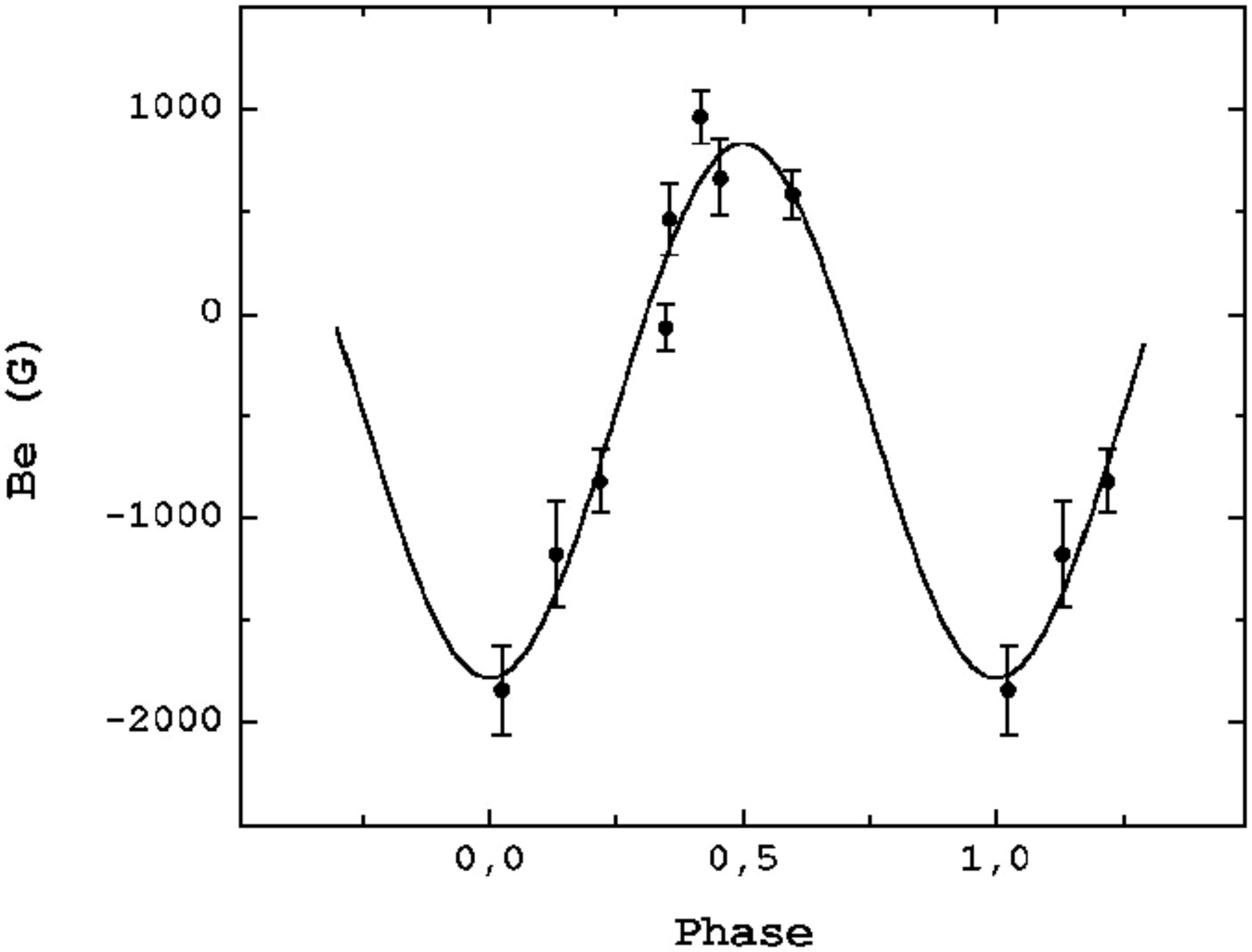}}
\vspace{-3.5mm}
\caption{ HD208217 }
\label{fig:fig343}
\end{figure}

\begin{figure}
\resizebox{0.98\hsize}{!}{\includegraphics{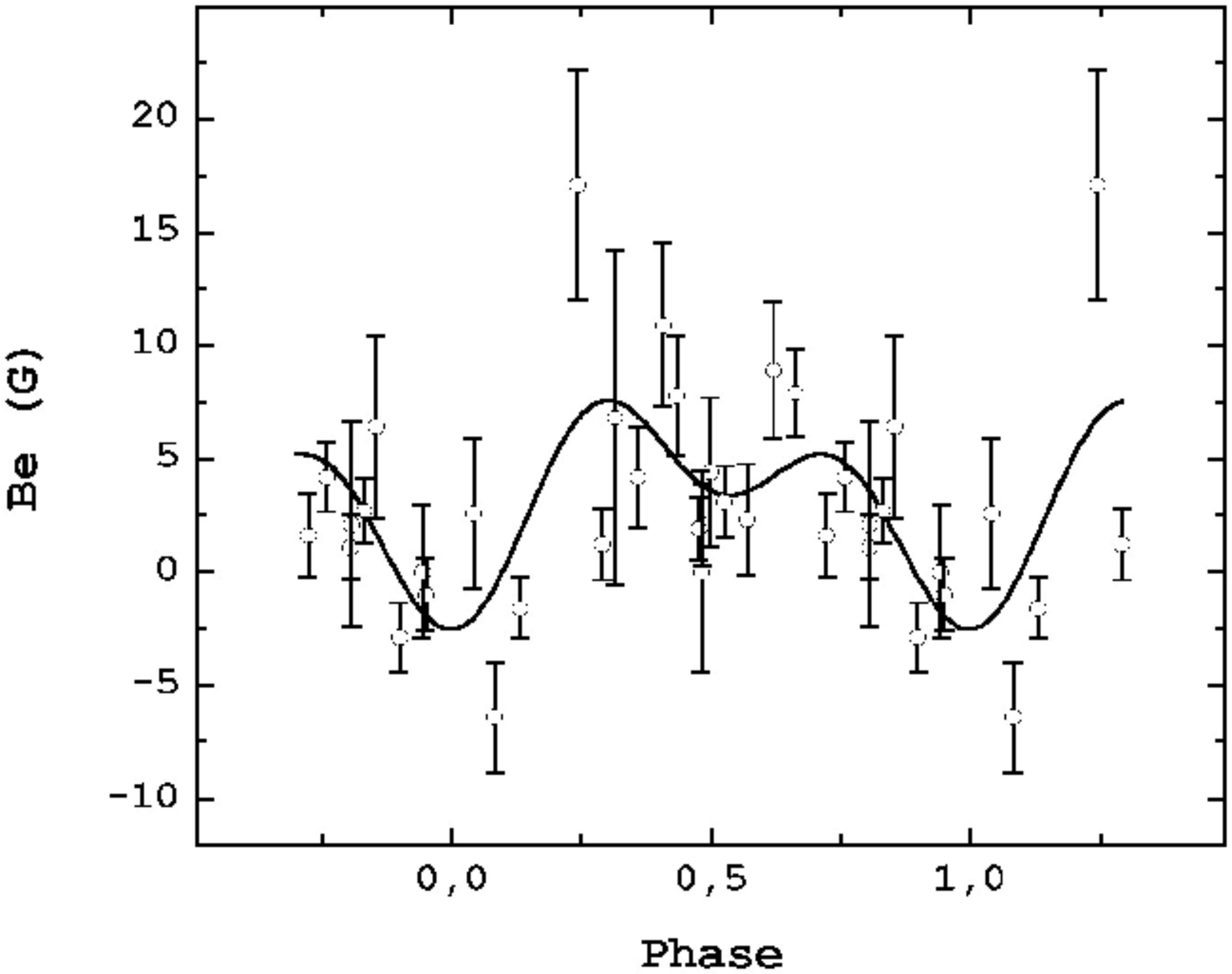}}
\vspace{-3.5mm}
\caption{ HD209290 }
\label{fig:fig344}
\end{figure}

\begin{figure}
\resizebox{0.98\hsize}{!}{\includegraphics{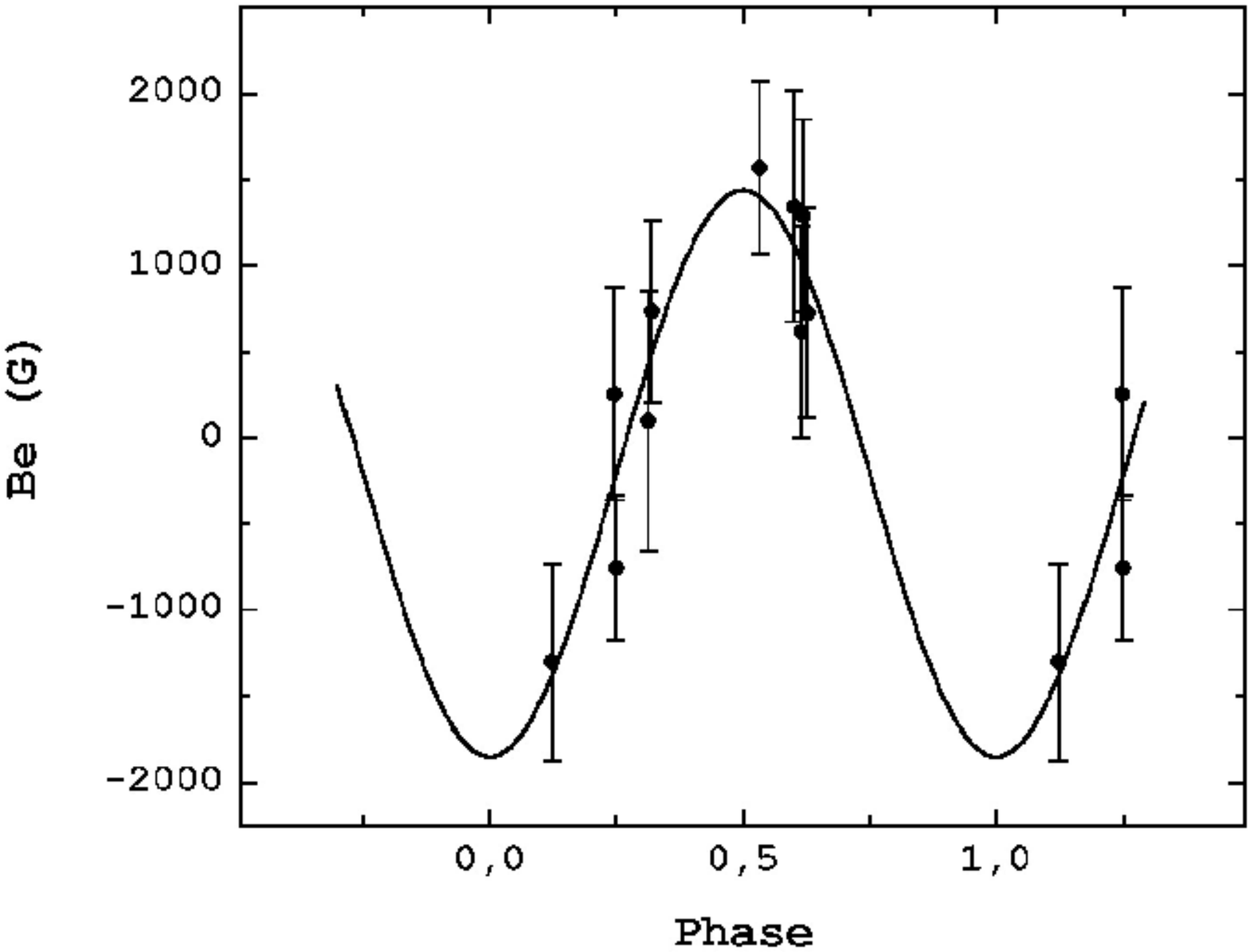}}
\vspace{-3.5mm}
\caption{ HD210873 }
\label{fig:fig345}
\end{figure}

\begin{figure}
\resizebox{0.98\hsize}{!}{\includegraphics{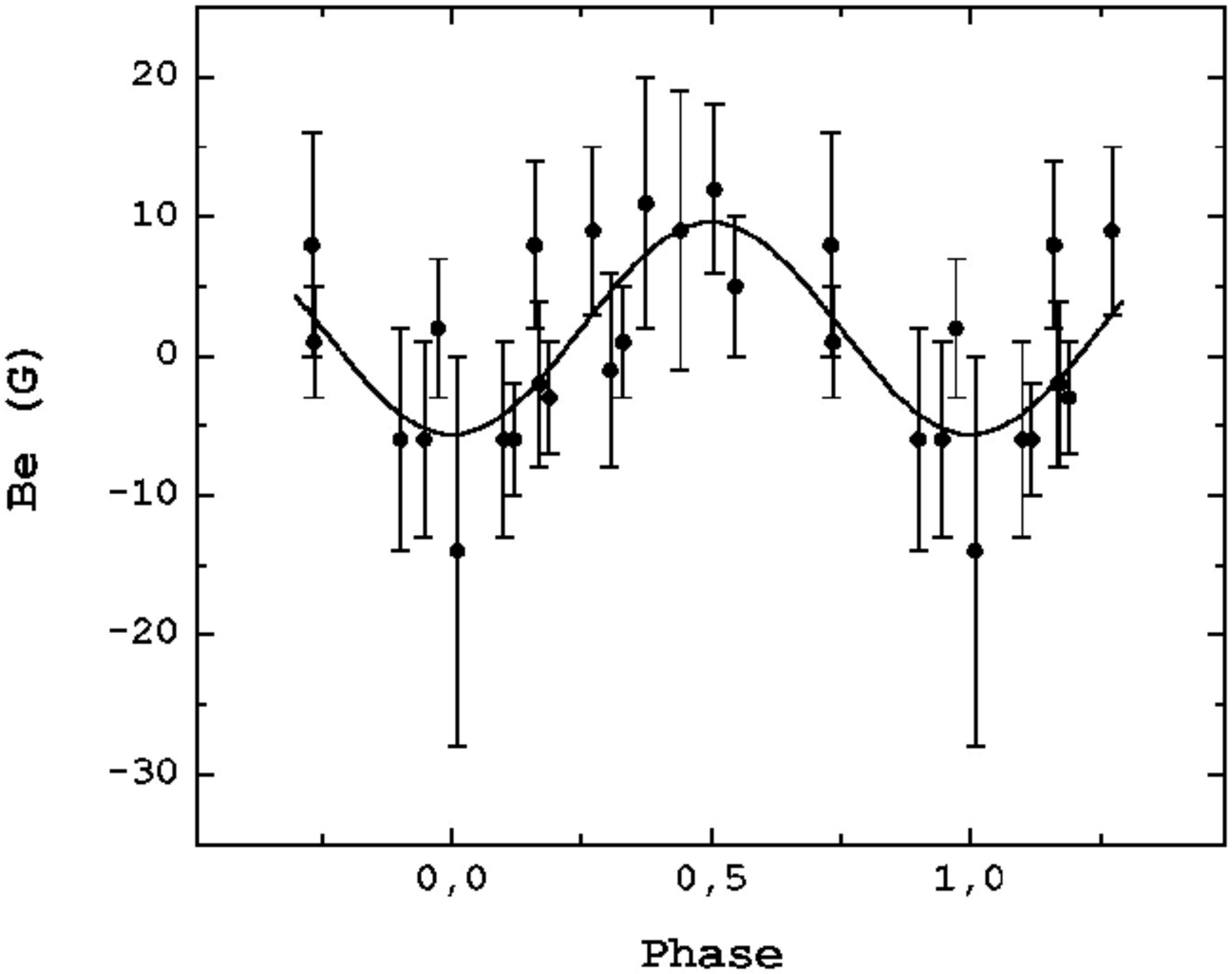}}
\vspace{-3.5mm}
\caption{ HD214680 }
\label{fig:fig346}
\end{figure}

\begin{figure}
\resizebox{0.98\hsize}{!}{\includegraphics{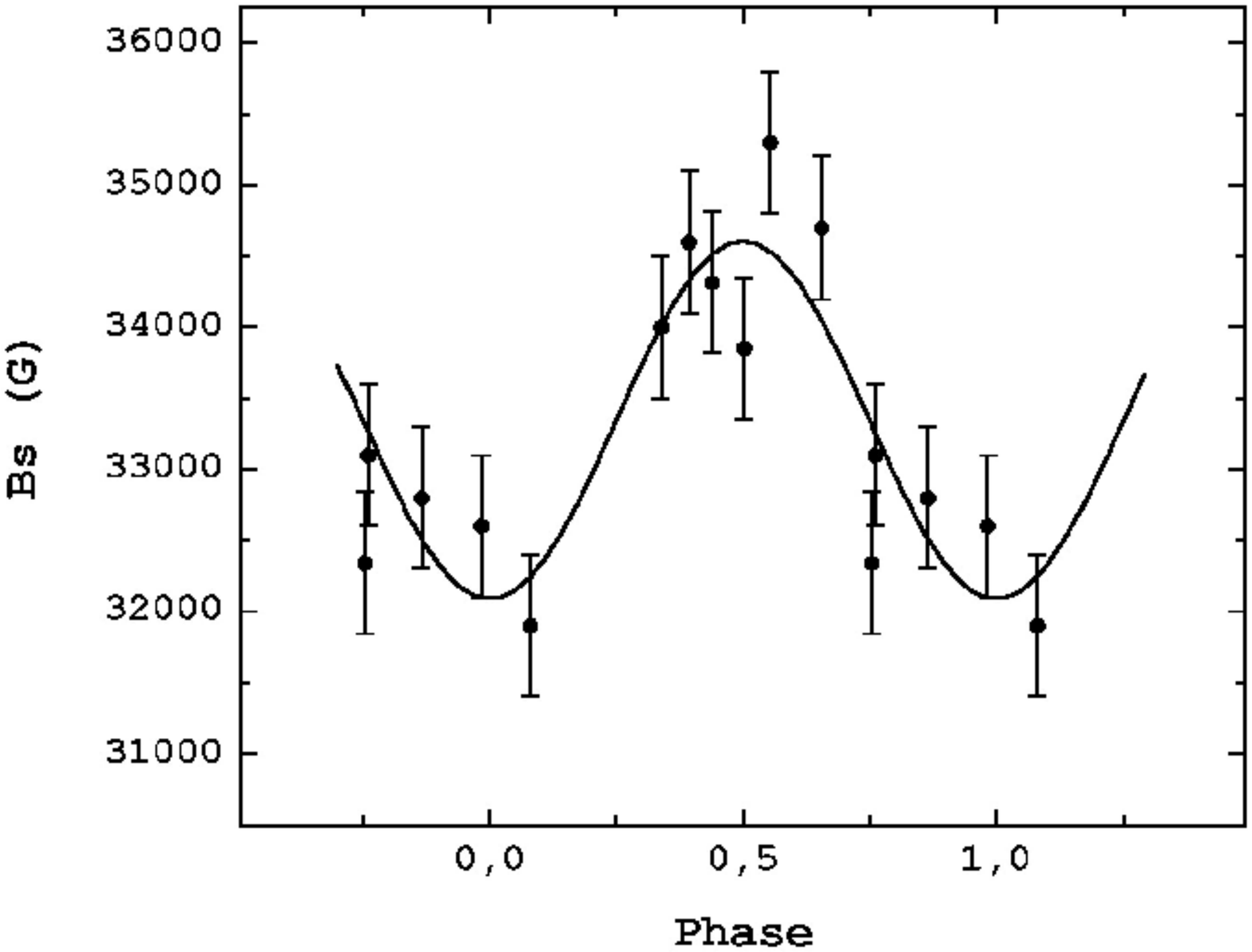}}
\vspace{-3.5mm}
\caption{ HD215441 (1) }
\label{fig:fig347}
\end{figure}

\begin{figure}
\resizebox{0.98\hsize}{!}{\includegraphics{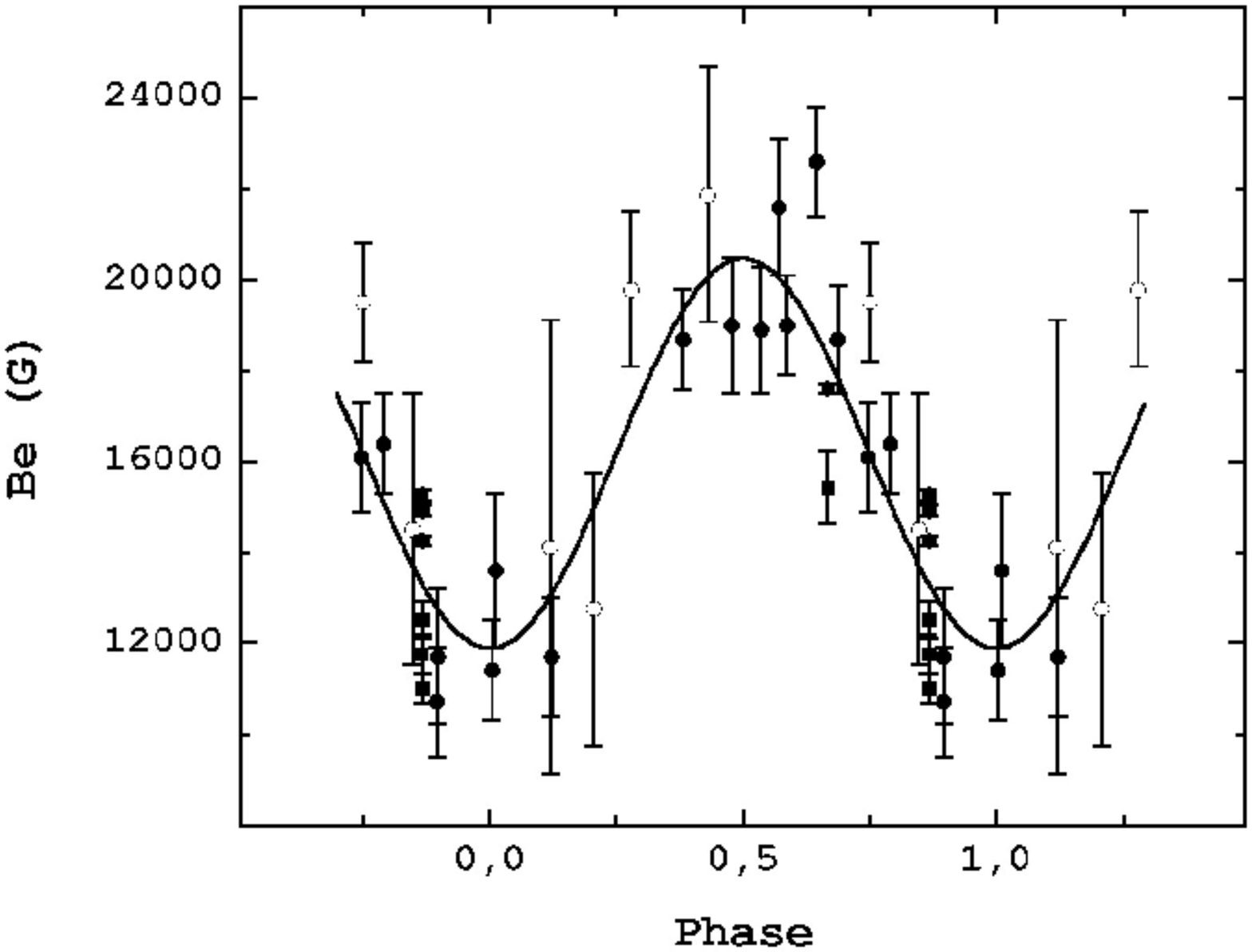}}
\vspace{-3.5mm}
\caption{ HD215441 (2) }
\label{fig:fig348}
\end{figure}

\clearpage
\newpage

\begin{figure}
\resizebox{0.98\hsize}{!}{\includegraphics{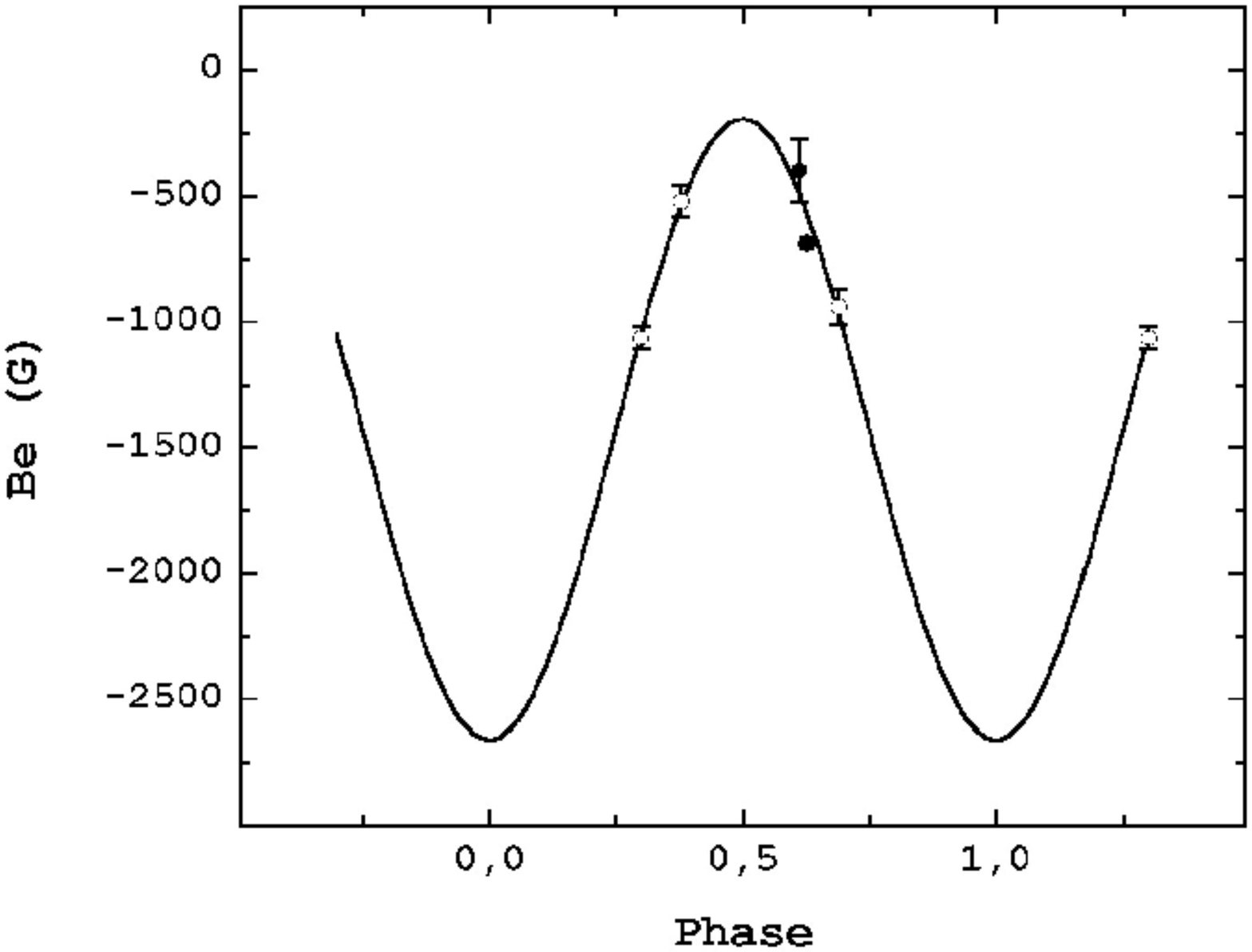}}
\vspace{-3.5mm}
\caption{ HD217522 }
\label{fig:fig340}
\end{figure}

\begin{figure}
\resizebox{0.98\hsize}{!}{\includegraphics{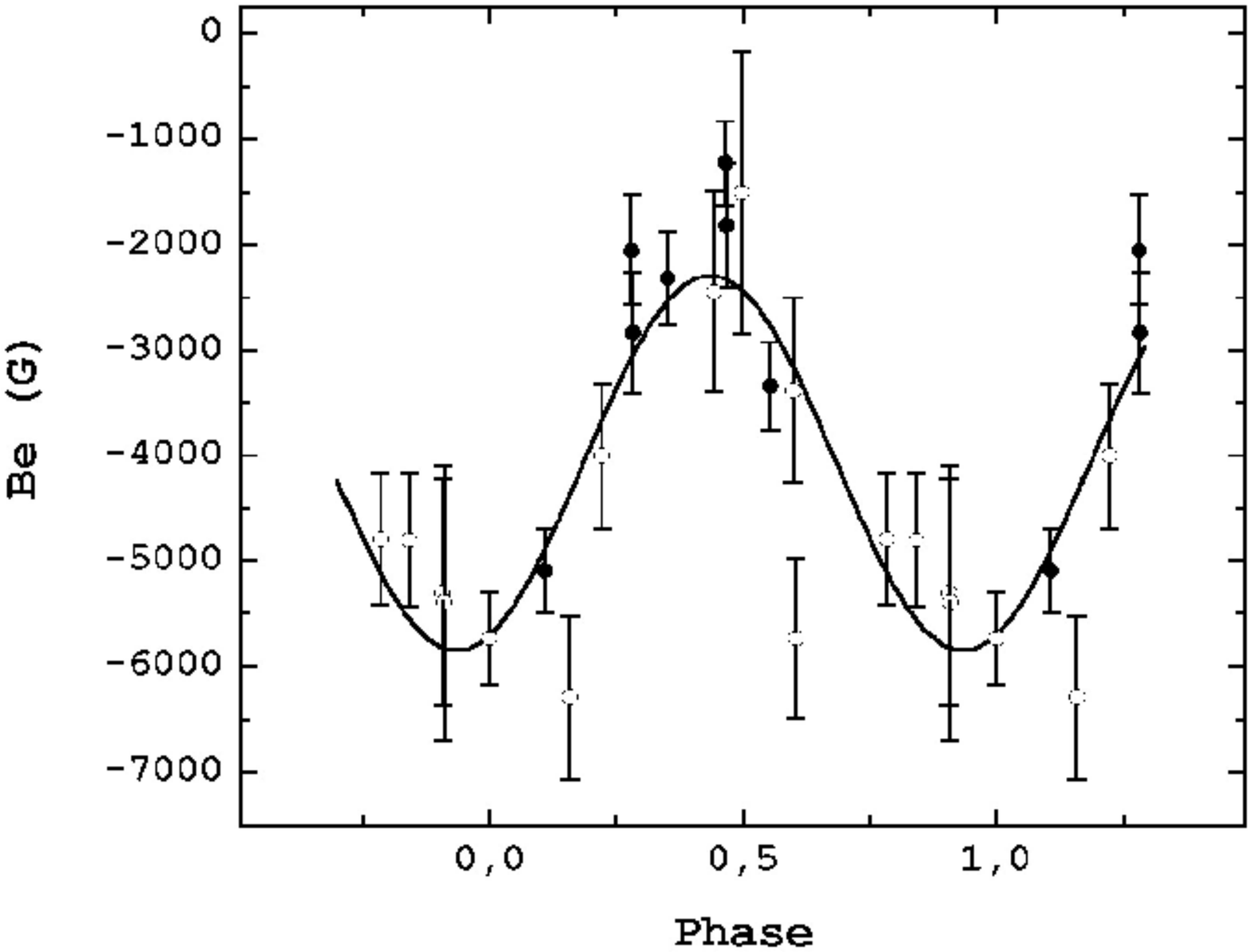}}
\vspace{-3.5mm}
\caption{ HD217833 (1) }
\label{fig:fig349}
\end{figure}

\begin{figure}
\resizebox{0.98\hsize}{!}{\includegraphics{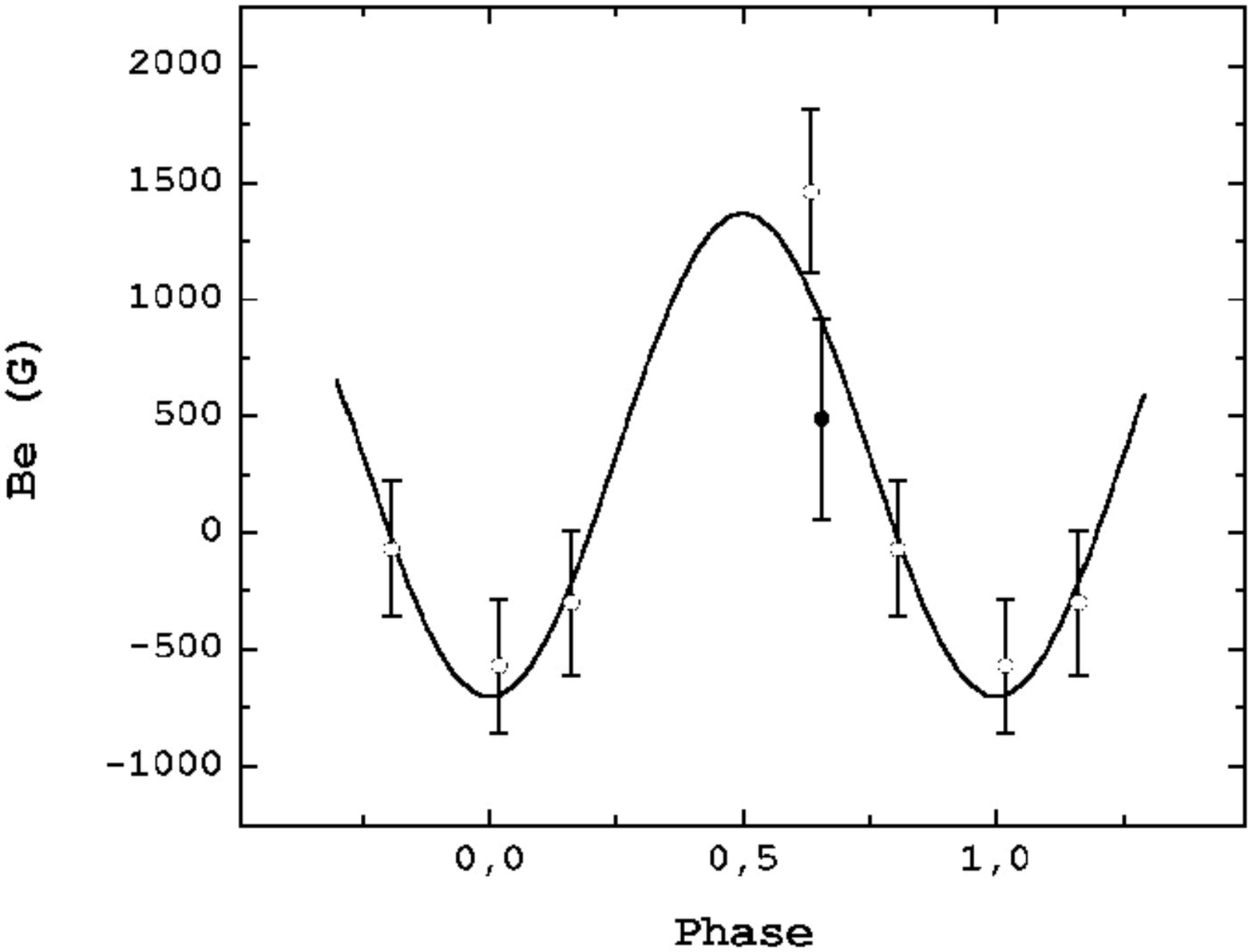}}
\vspace{-3.5mm}
\caption{ HD217833 (2) }
\label{fig:fig350}
\end{figure}

\begin{figure}
\resizebox{0.98\hsize}{!}{\includegraphics{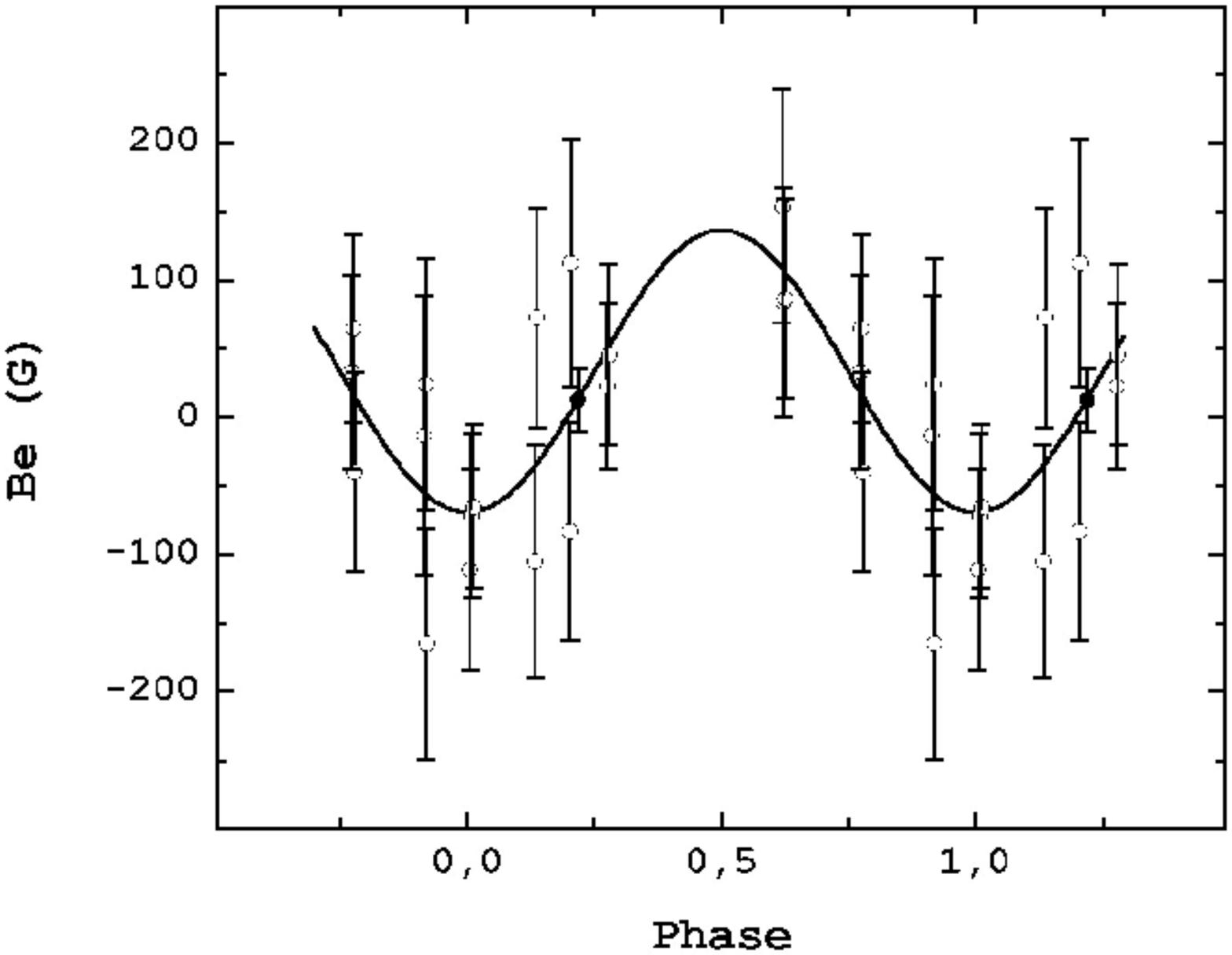}}
\vspace{-3.5mm}
\caption{ HD218376 }
\label{fig:fig351}
\end{figure}

\begin{figure}
\resizebox{0.98\hsize}{!}{\includegraphics{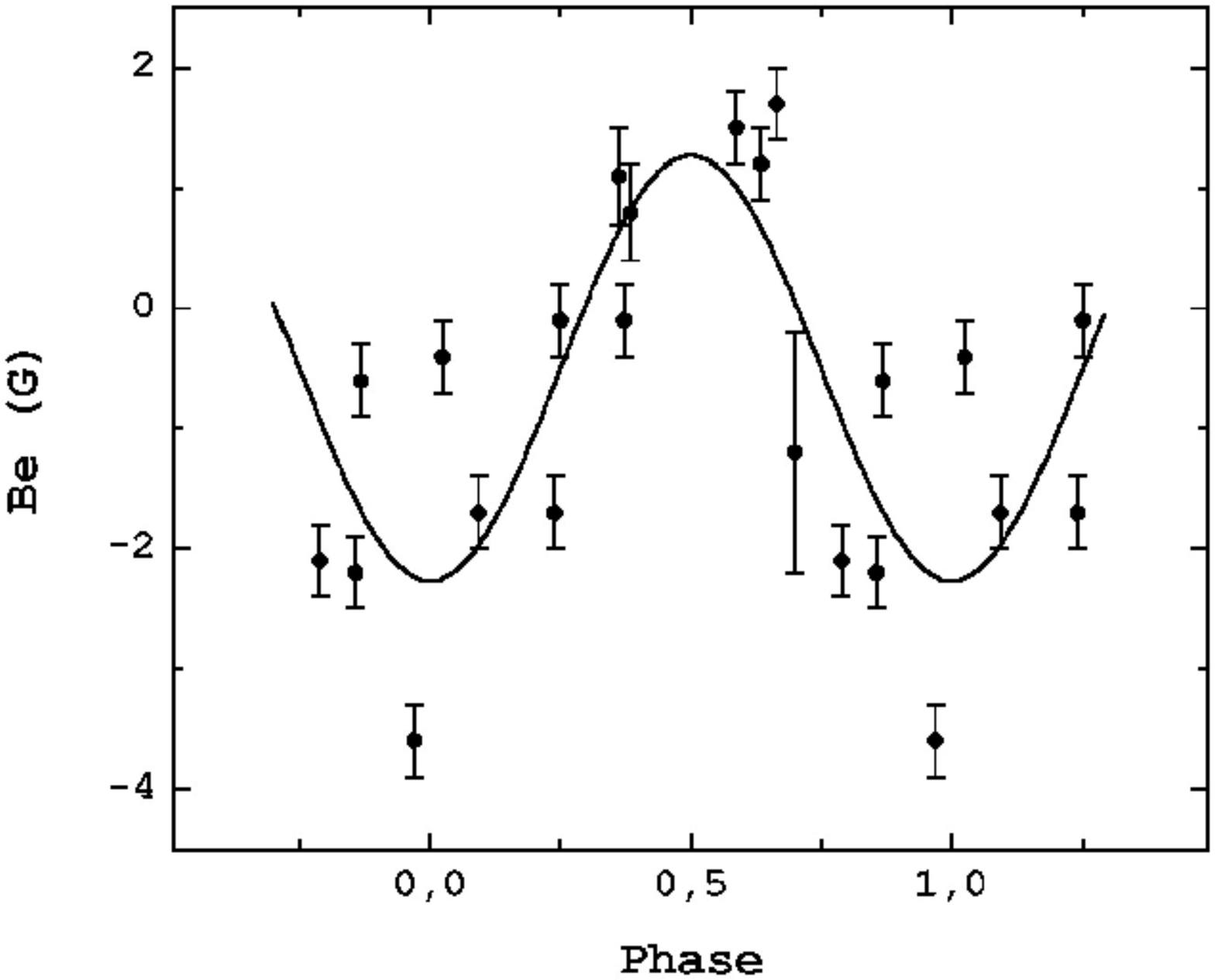}}
\vspace{-3.5mm}
\caption{ HD219134 }
\label{fig:fig340}
\end{figure}

\begin{figure}
\resizebox{0.98\hsize}{!}{\includegraphics{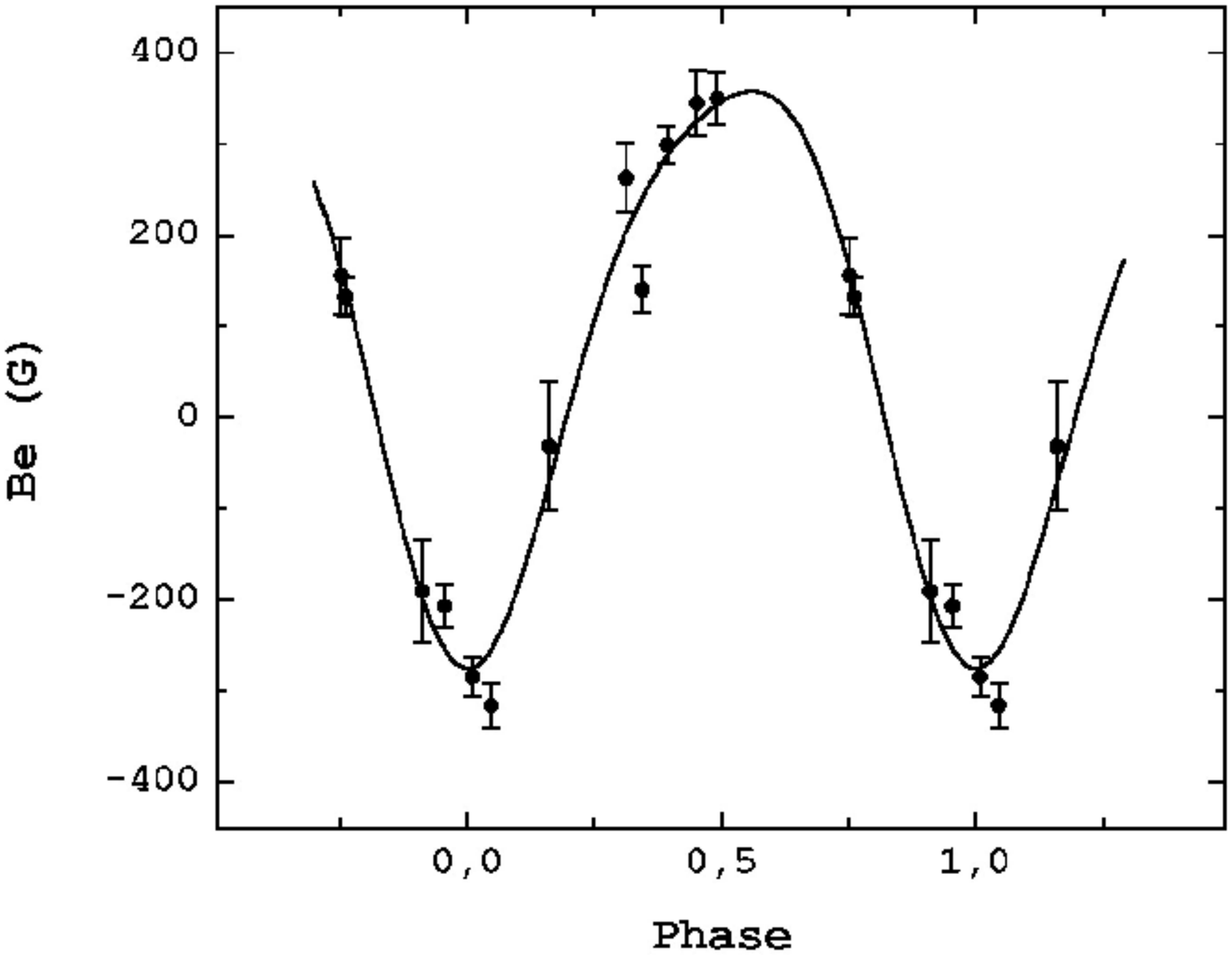}}
\vspace{-3.5mm}
\caption{ HD220825 }
\label{fig:fig352}
\end{figure}

\clearpage
\newpage

\begin{figure}
\resizebox{0.98\hsize}{!}{\includegraphics{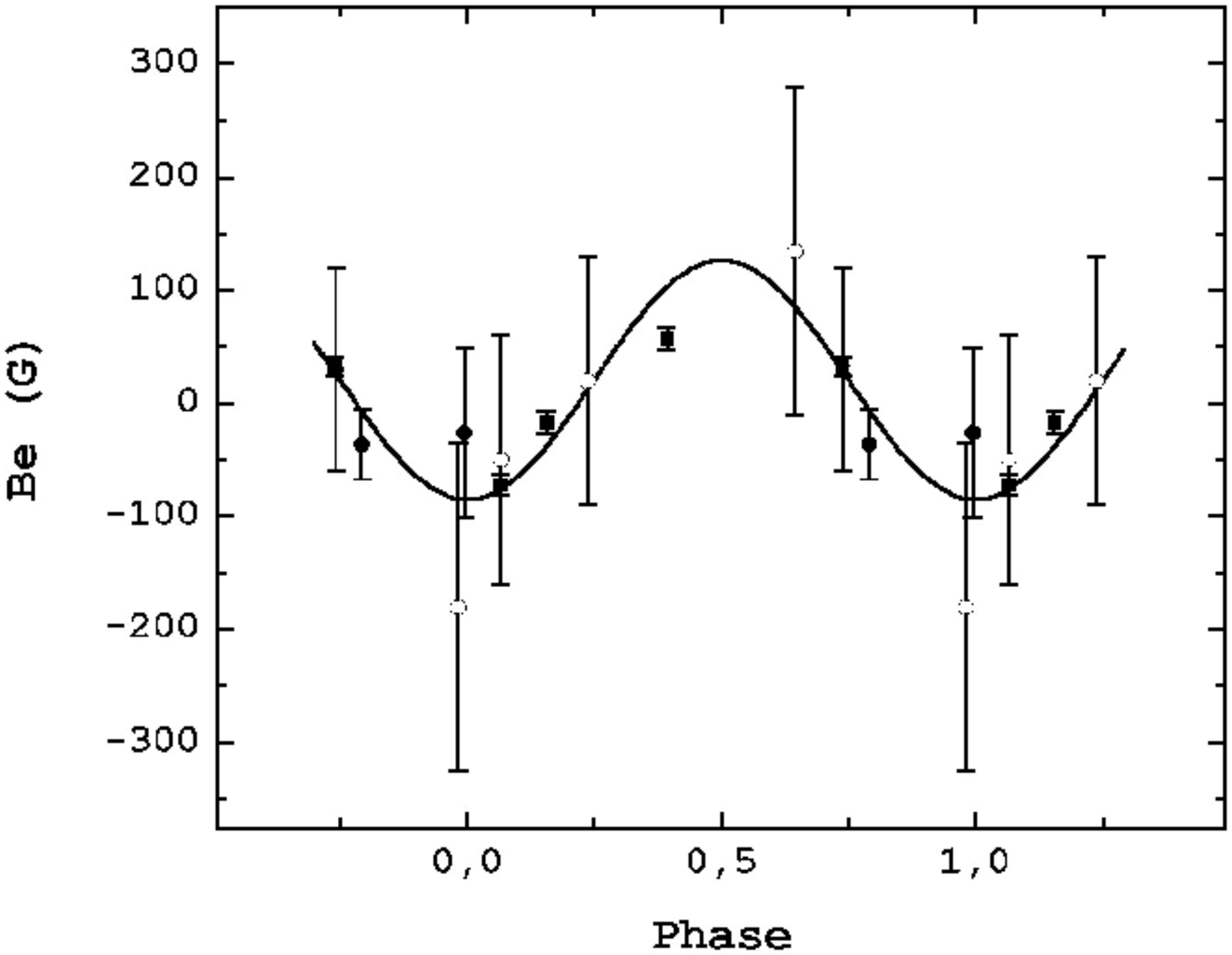}}
\vspace{-3.5mm}
\caption{ HD221760 }
\label{fig:fig340}
\end{figure}

\begin{figure}
\resizebox{0.98\hsize}{!}{\includegraphics{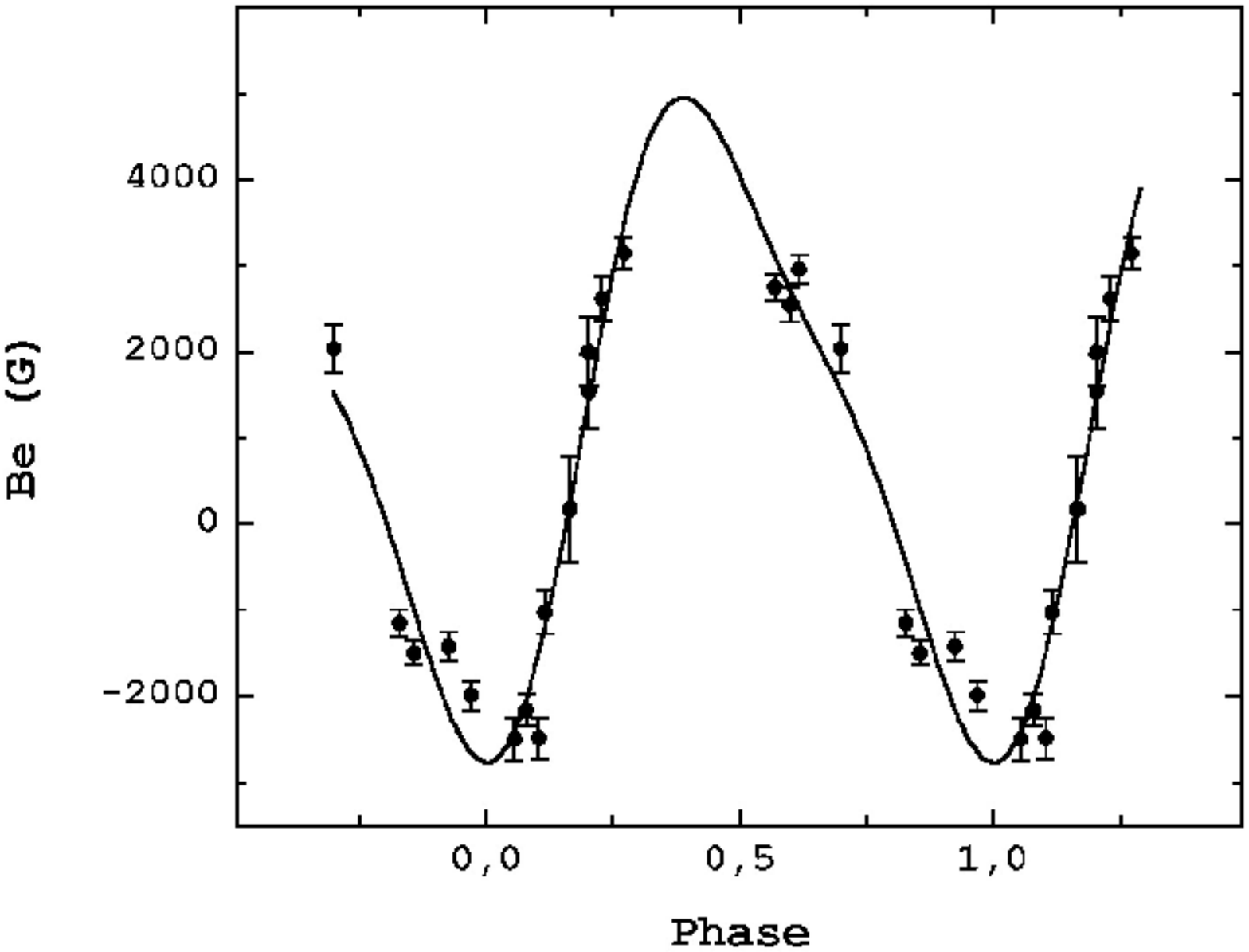}}
\vspace{-3.5mm}
\caption{ HD221936 }
\label{fig:fig353}
\end{figure}

\begin{figure}
\resizebox{0.98\hsize}{!}{\includegraphics{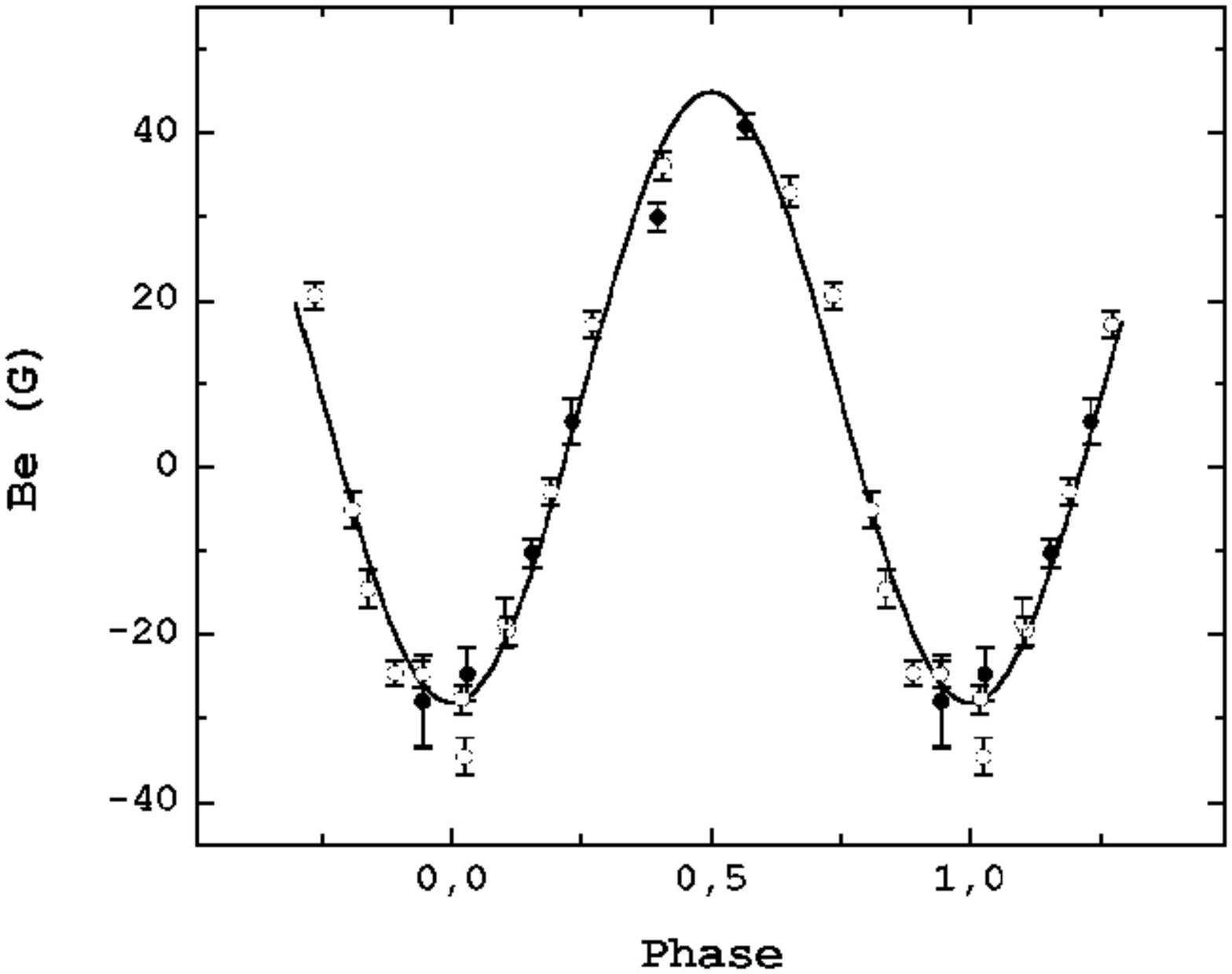}}
\vspace{-3.5mm}
\caption{ HD223460 }
\label{fig:fig354}
\end{figure}

\begin{figure}
\resizebox{0.98\hsize}{!}{\includegraphics{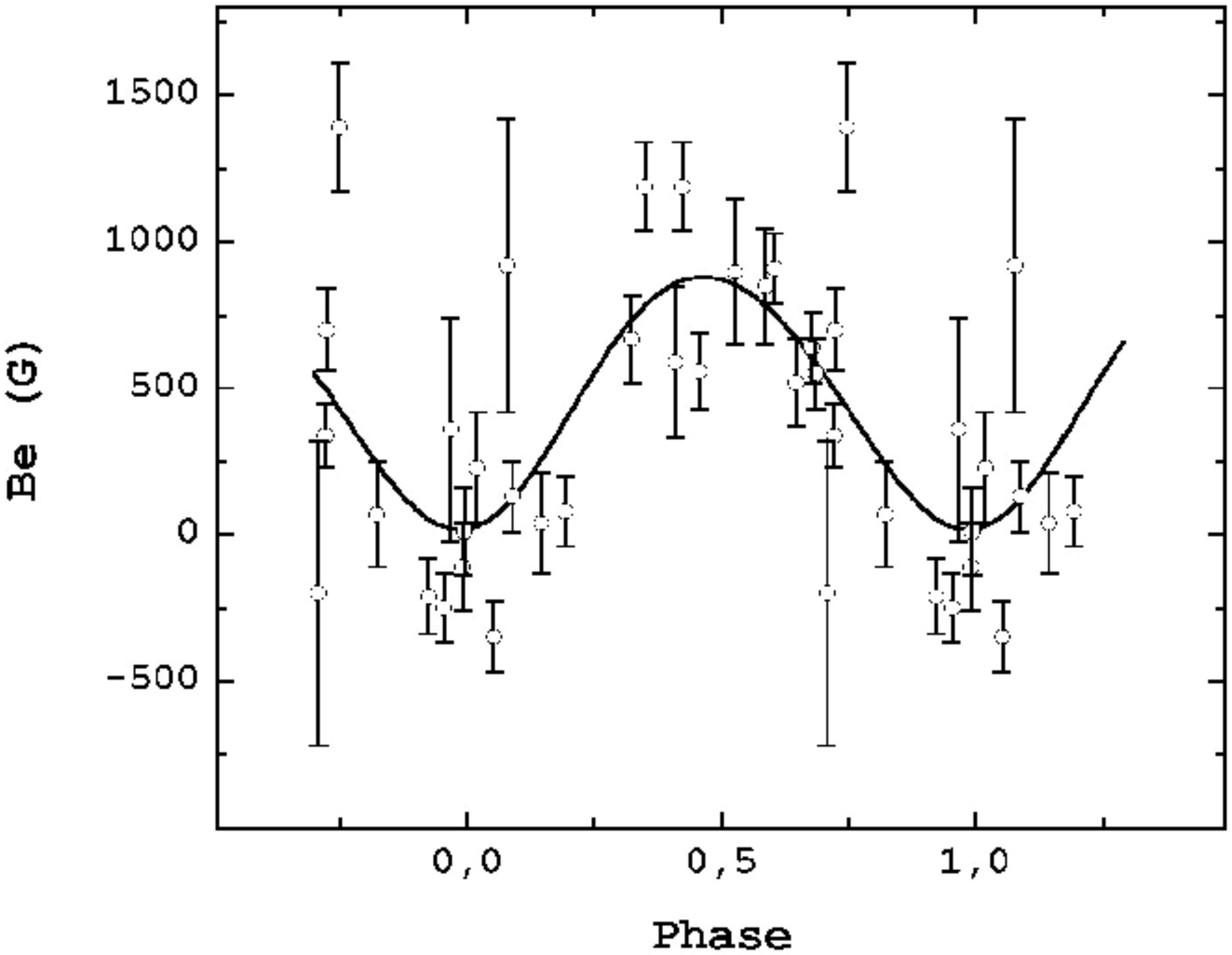}}
\vspace{-3.5mm}
\caption{ HD223640 (1) }
\label{fig:fig355}
\end{figure}

\begin{figure}
\resizebox{0.98\hsize}{!}{\includegraphics{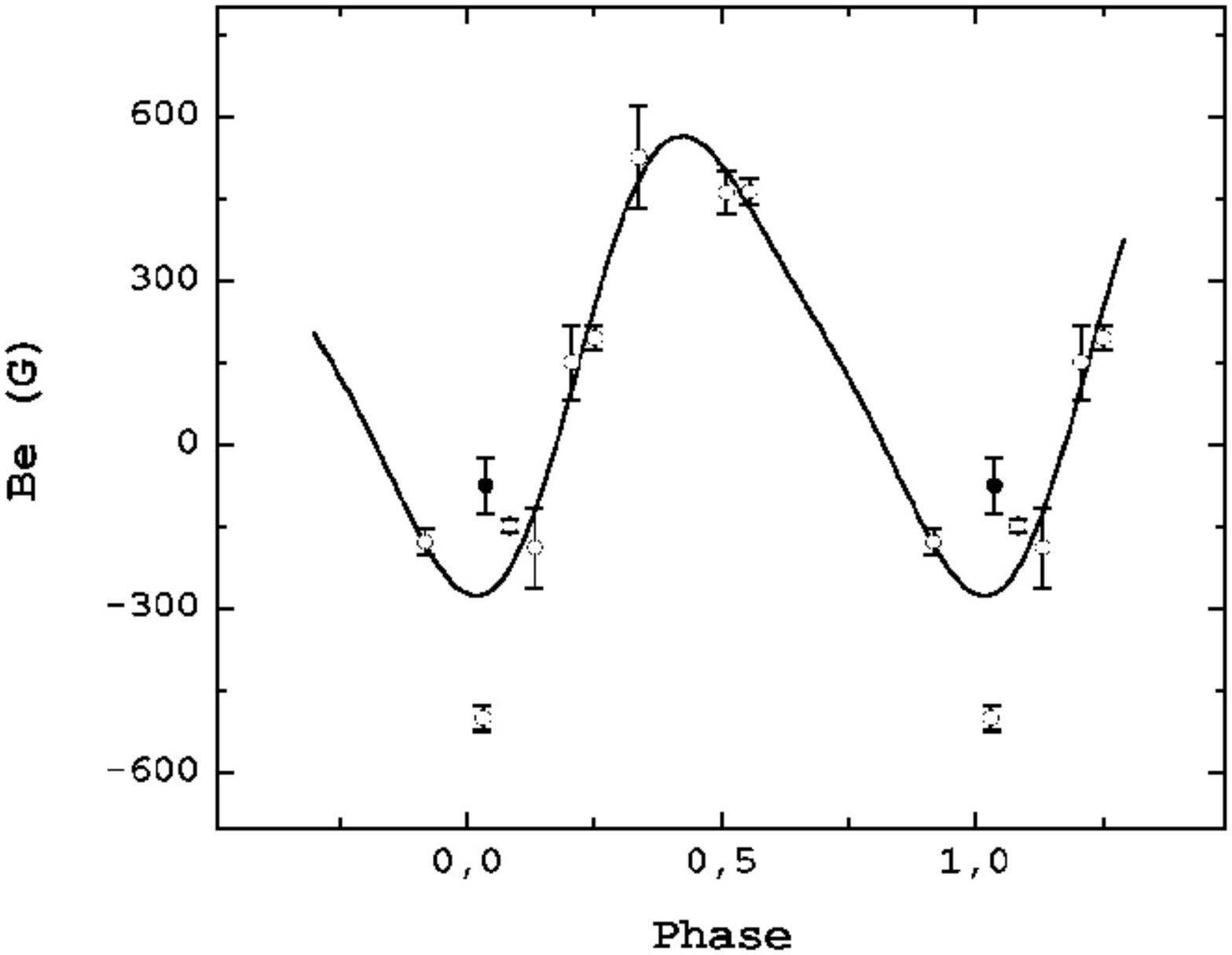}}
\vspace{-3.5mm}
\caption{ HD223640 (2) }
\label{fig:fig355}
\end{figure}

\begin{figure}
\resizebox{0.98\hsize}{!}{\includegraphics{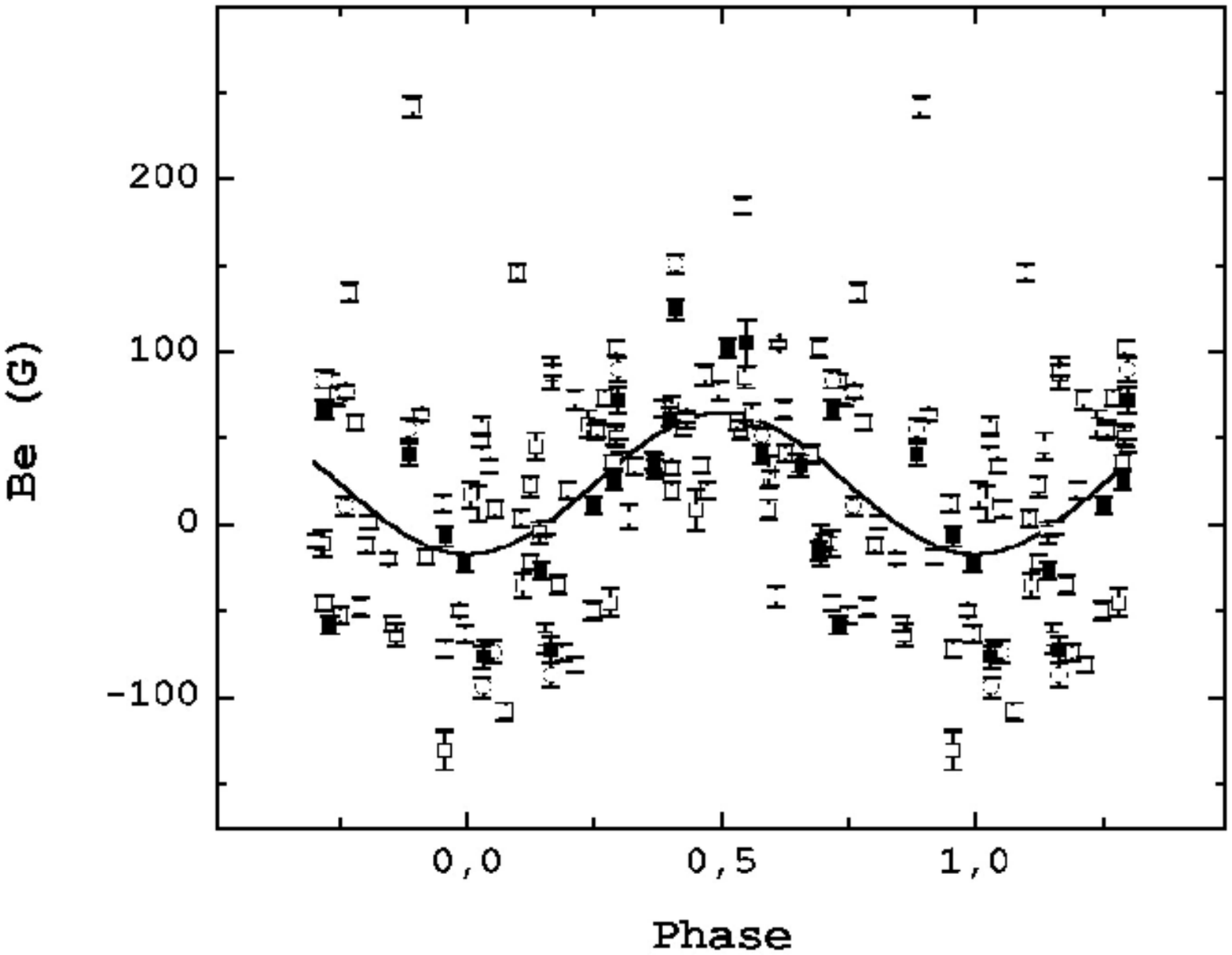}}
\vspace{-3.5mm}
\caption{ HD224085 }
\label{fig:fig356}
\end{figure}

\clearpage
\newpage

\begin{figure}
\resizebox{0.98\hsize}{!}{\includegraphics{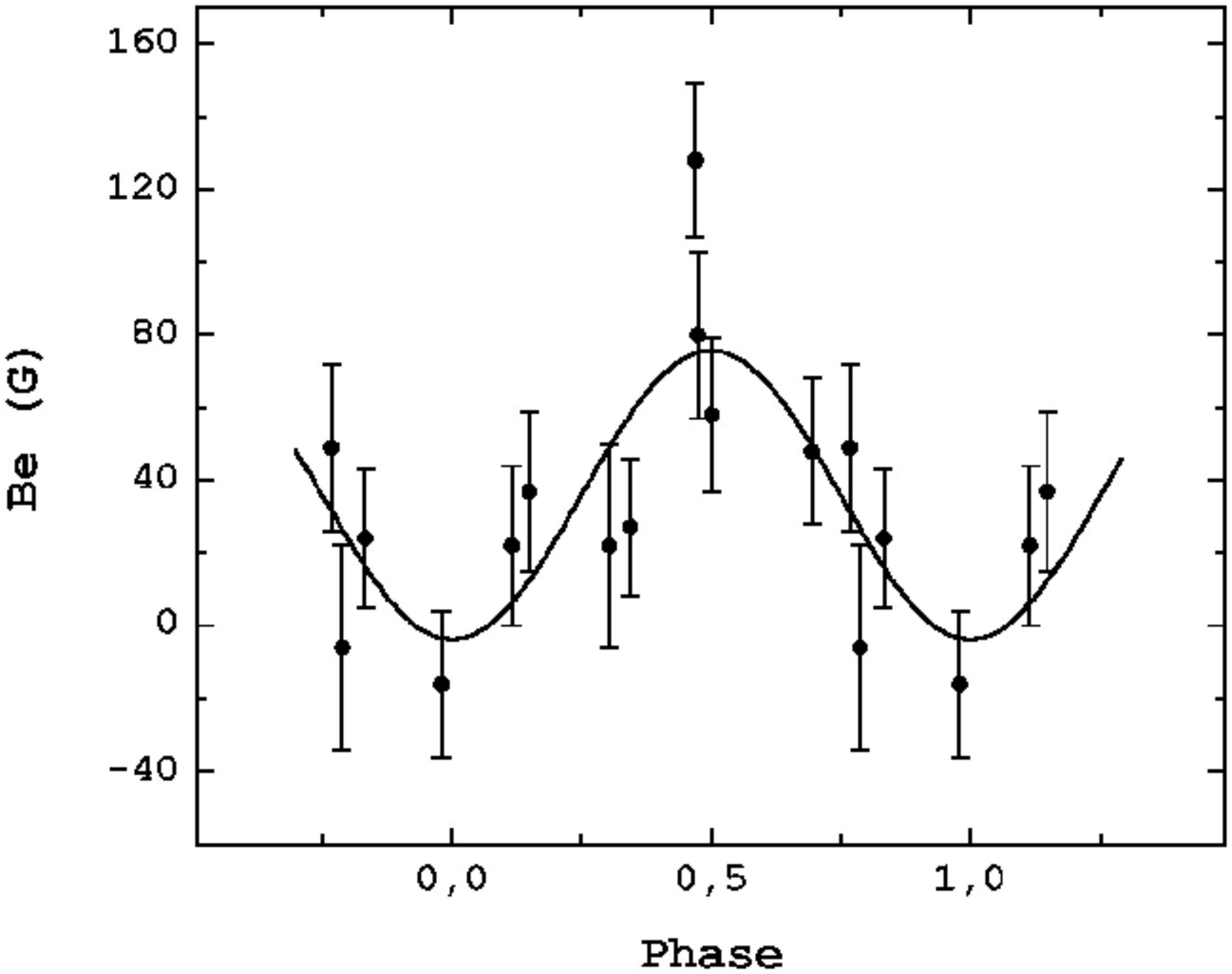}}
\vspace{-3.5mm}
\caption{ HD226868 }
\label{fig:fig357}
\end{figure}

\begin{figure}
\resizebox{0.98\hsize}{!}{\includegraphics{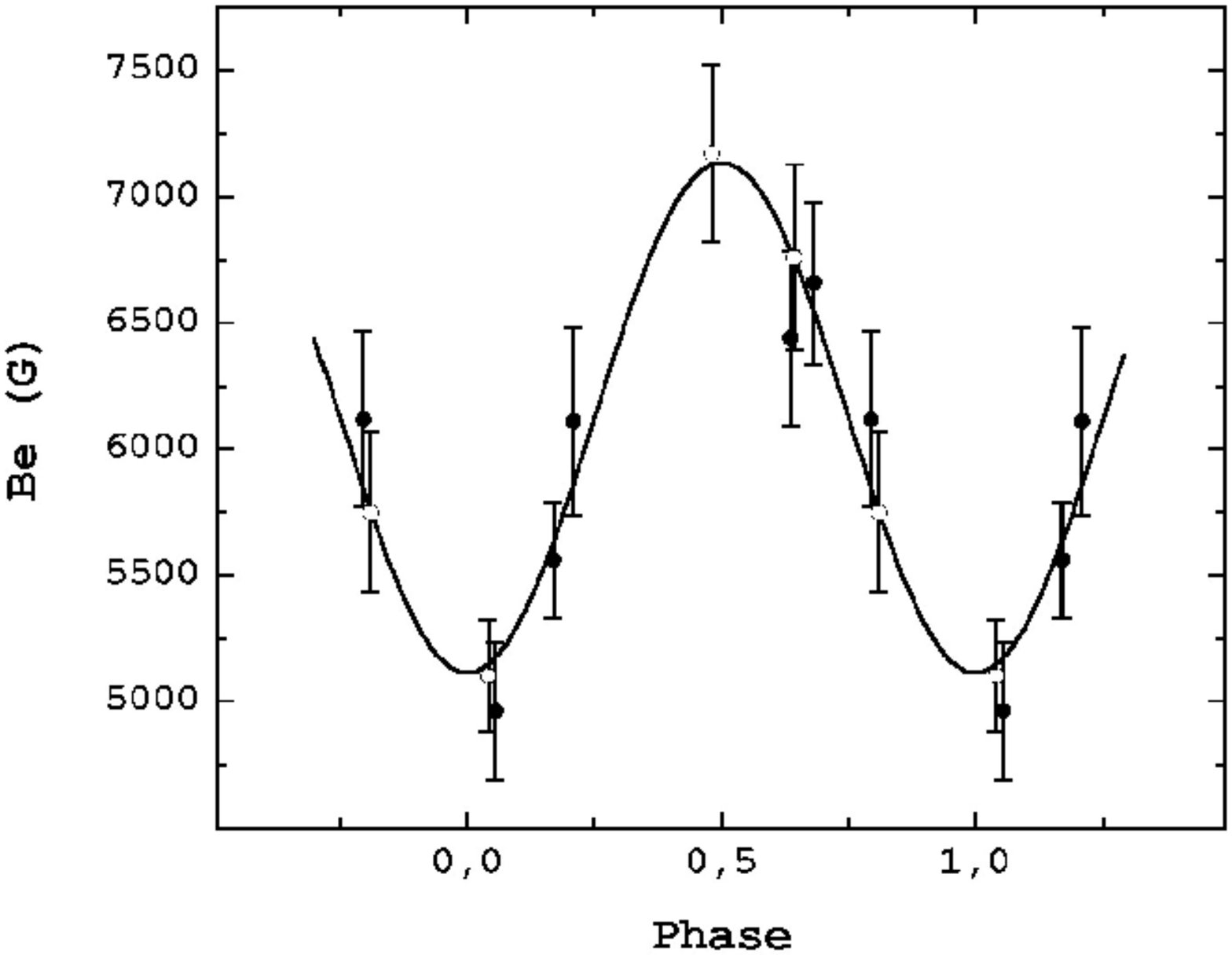}}
\vspace{-3.5mm}
\caption{ HD258686 }
\label{fig:fig358}
\end{figure}

\begin{figure}
\resizebox{0.98\hsize}{!}{\includegraphics{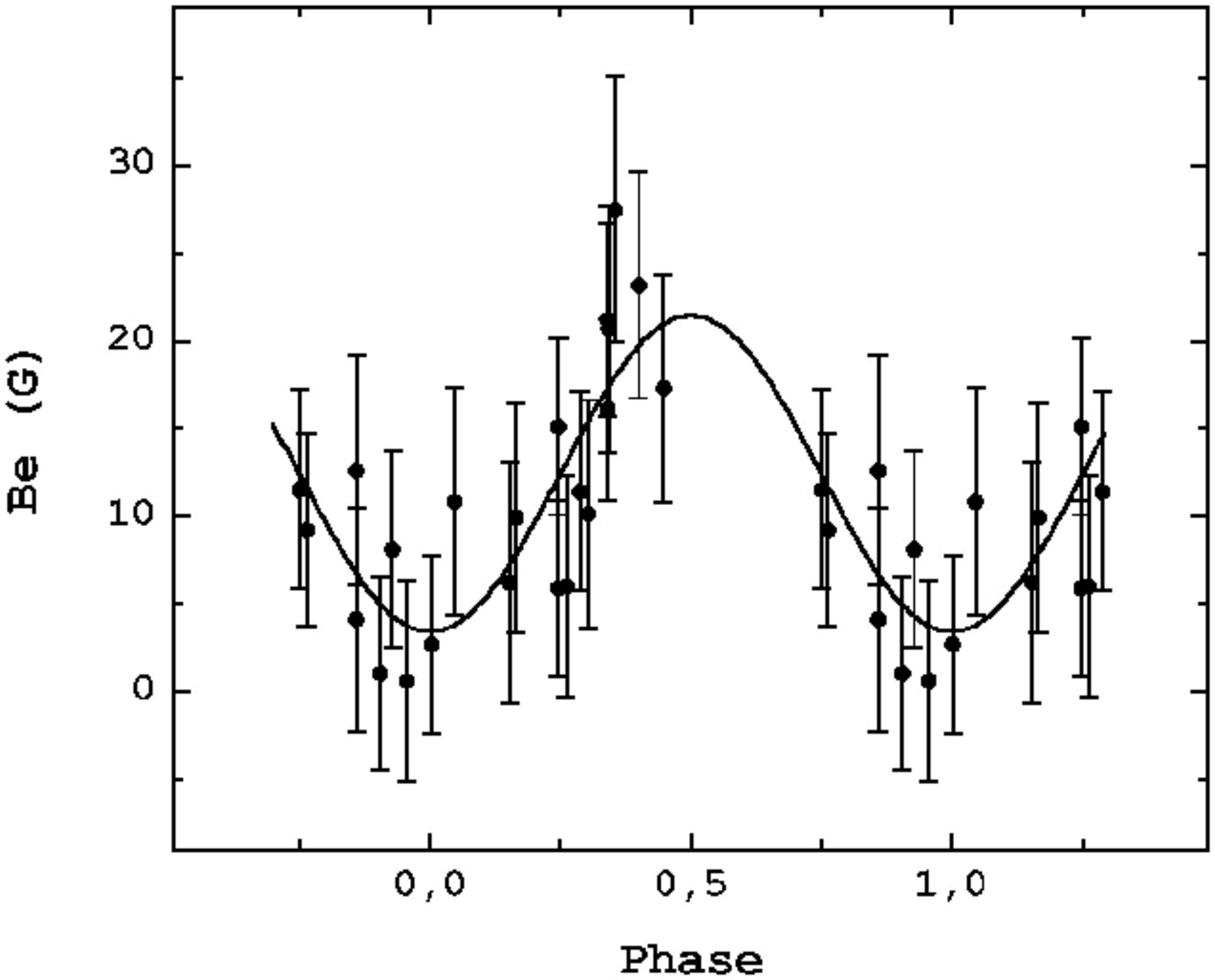}}
\vspace{-3.5mm}
\caption{ HD265866 }
\label{fig:fig355}
\end{figure}

\begin{figure}
\resizebox{0.98\hsize}{!}{\includegraphics{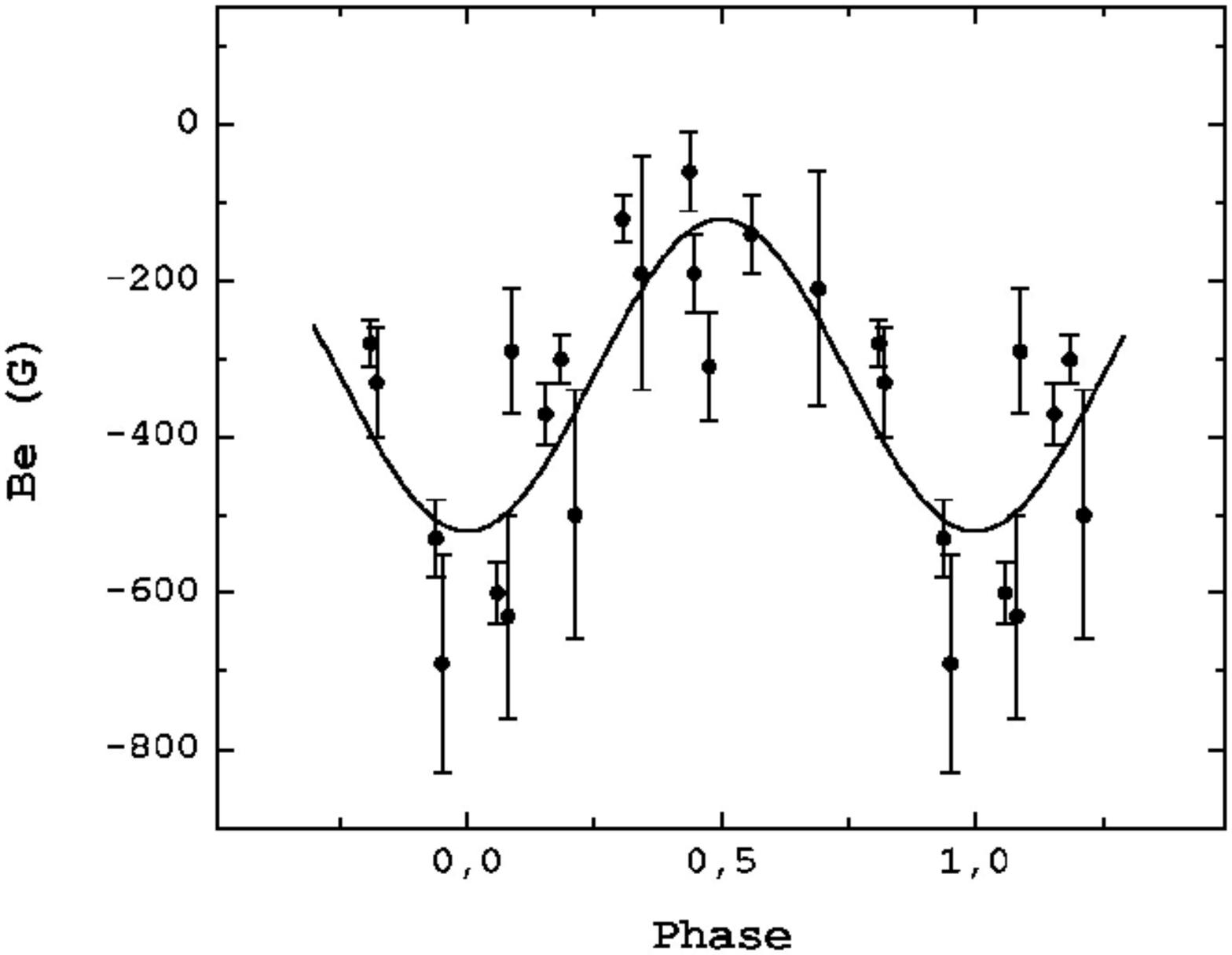}}
\vspace{-3.5mm}
\caption{ HD281934 }
\label{fig:fig359}
\end{figure}

\begin{figure}
\resizebox{0.98\hsize}{!}{\includegraphics{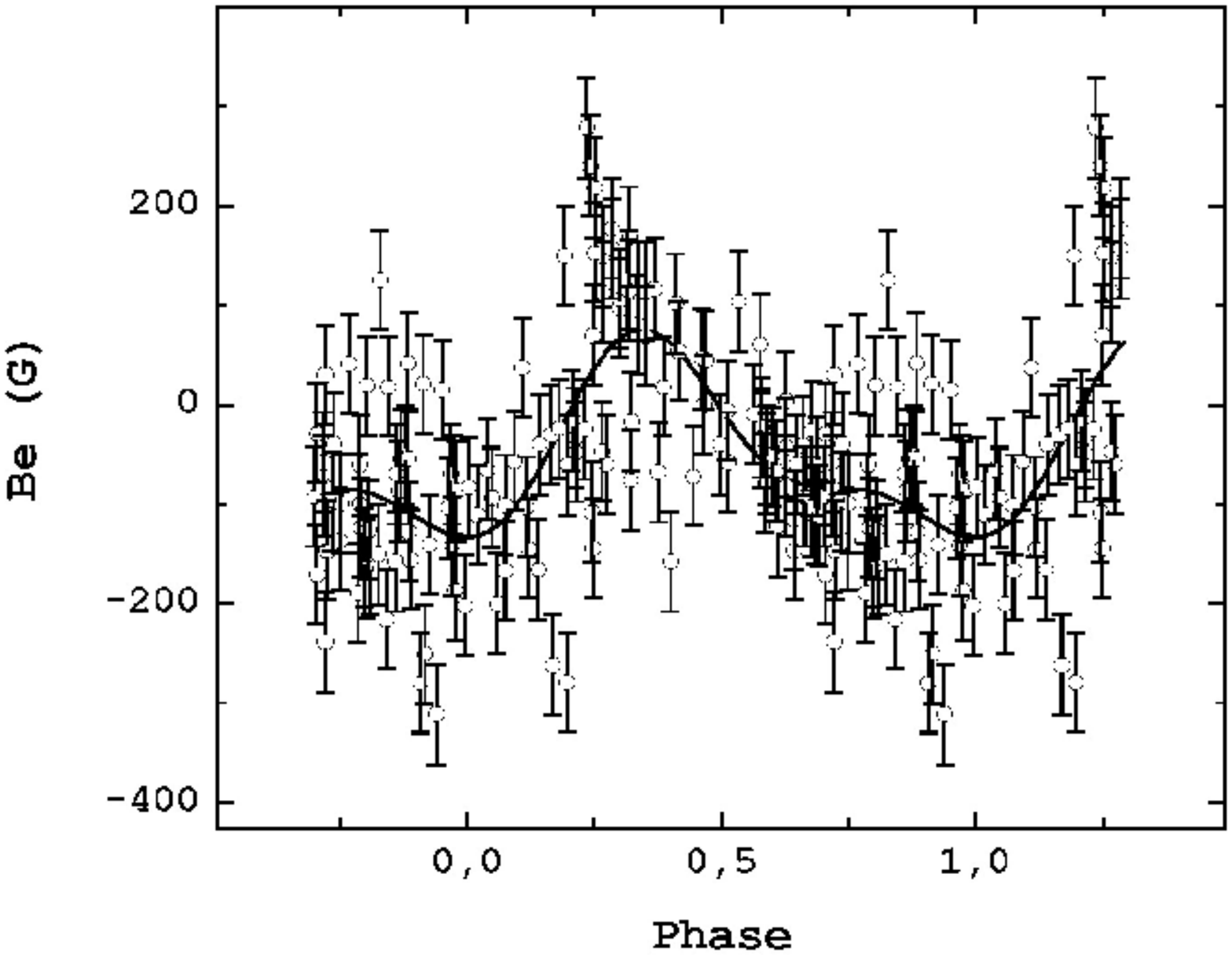}}
\vspace{-3.5mm}
\caption{ HD283518 }
\label{fig:fig355}
\end{figure}

\begin{figure}
\resizebox{0.98\hsize}{!}{\includegraphics{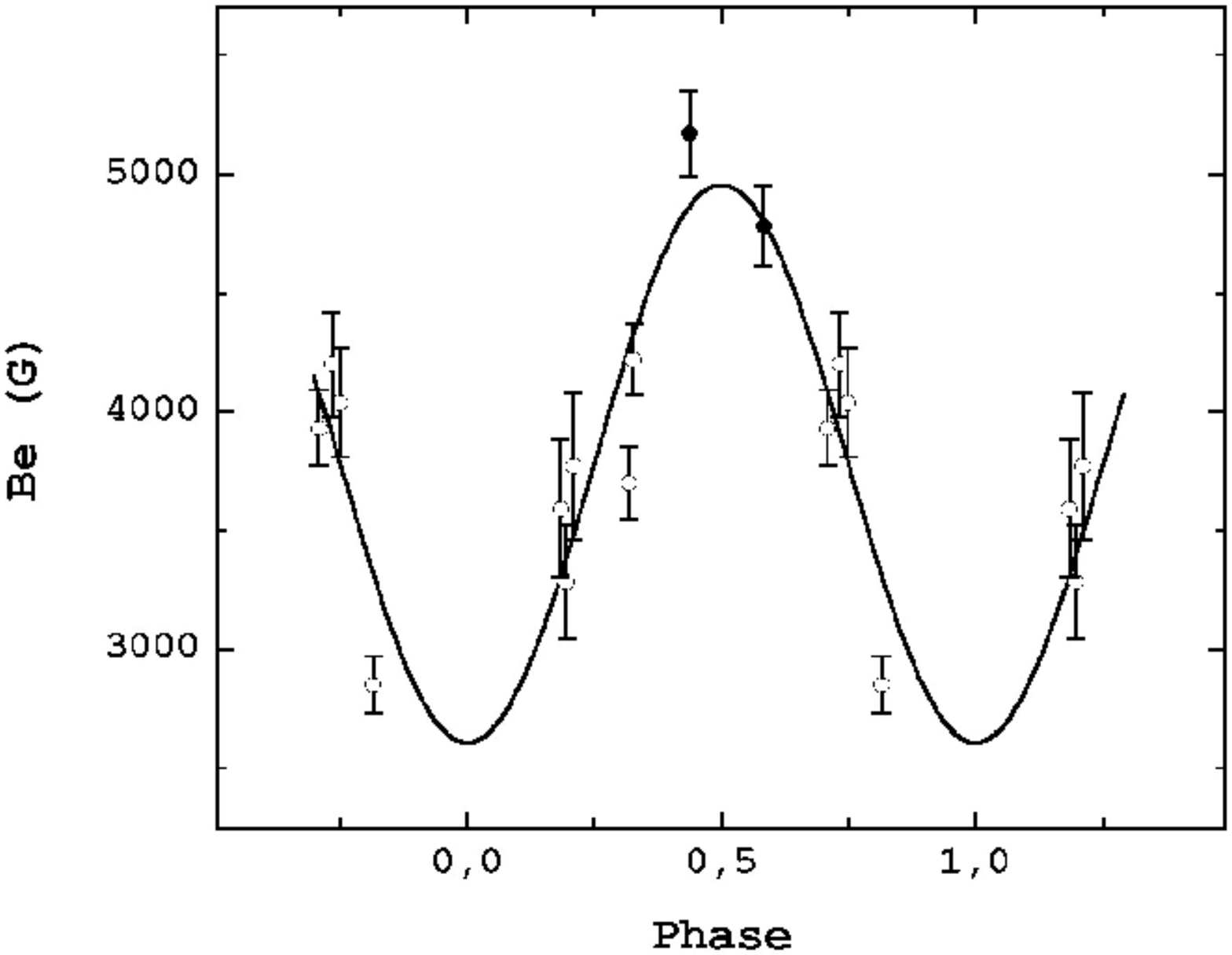}}
\vspace{-3.5mm}
\caption{ HD293764 }
\label{fig:fig360}
\end{figure}

\clearpage
\newpage

\begin{figure}
\resizebox{0.98\hsize}{!}{\includegraphics{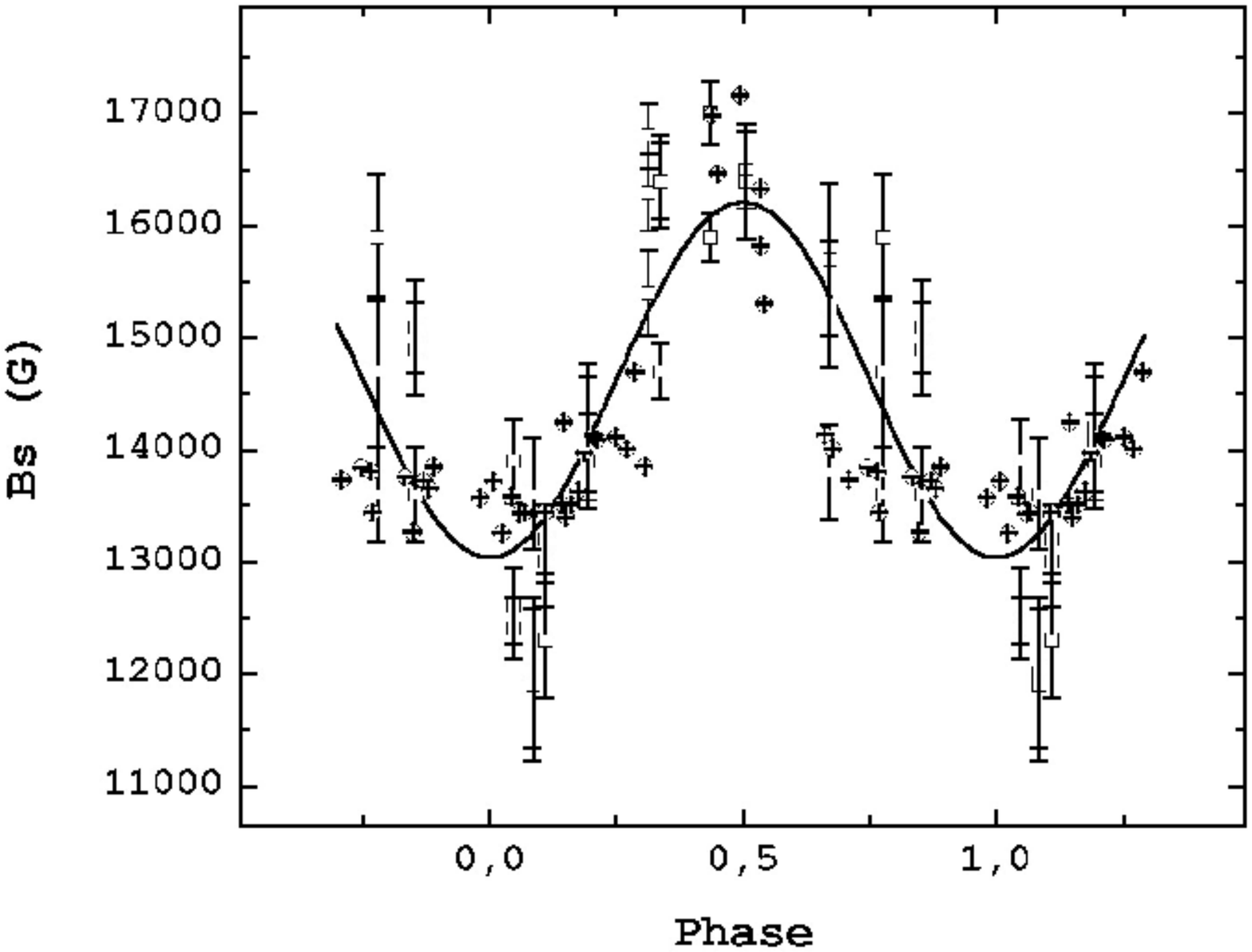}}
\vspace{-3.5mm}
\caption{ HD318107 (1) }
\label{fig:fig361}
\end{figure}

\begin{figure}
\resizebox{0.98\hsize}{!}{\includegraphics{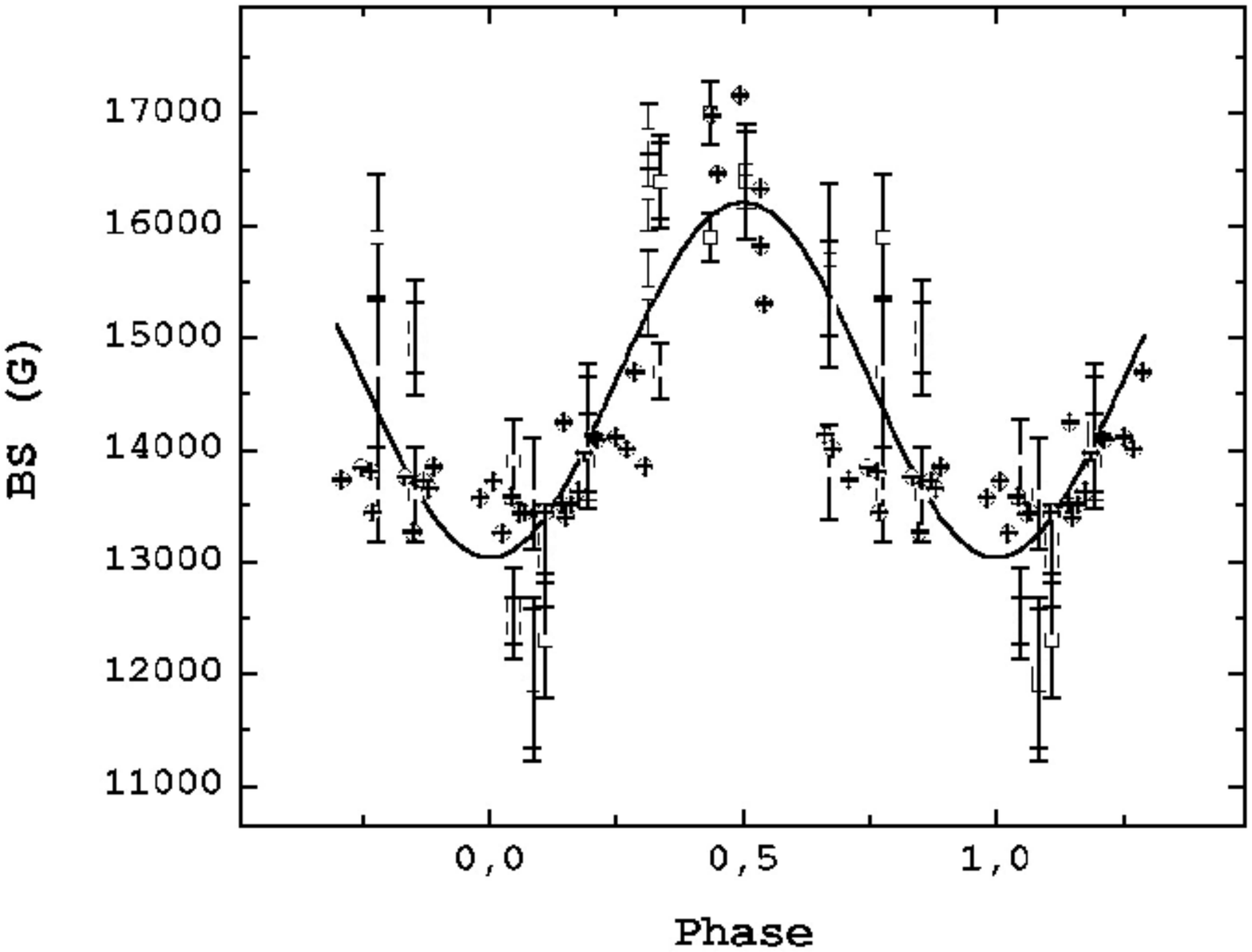}}
\vspace{-3.5mm}
\caption{ HD318107 (2) }
\label{fig:fig362}
\end{figure}

\begin{figure}
\resizebox{0.98\hsize}{!}{\includegraphics{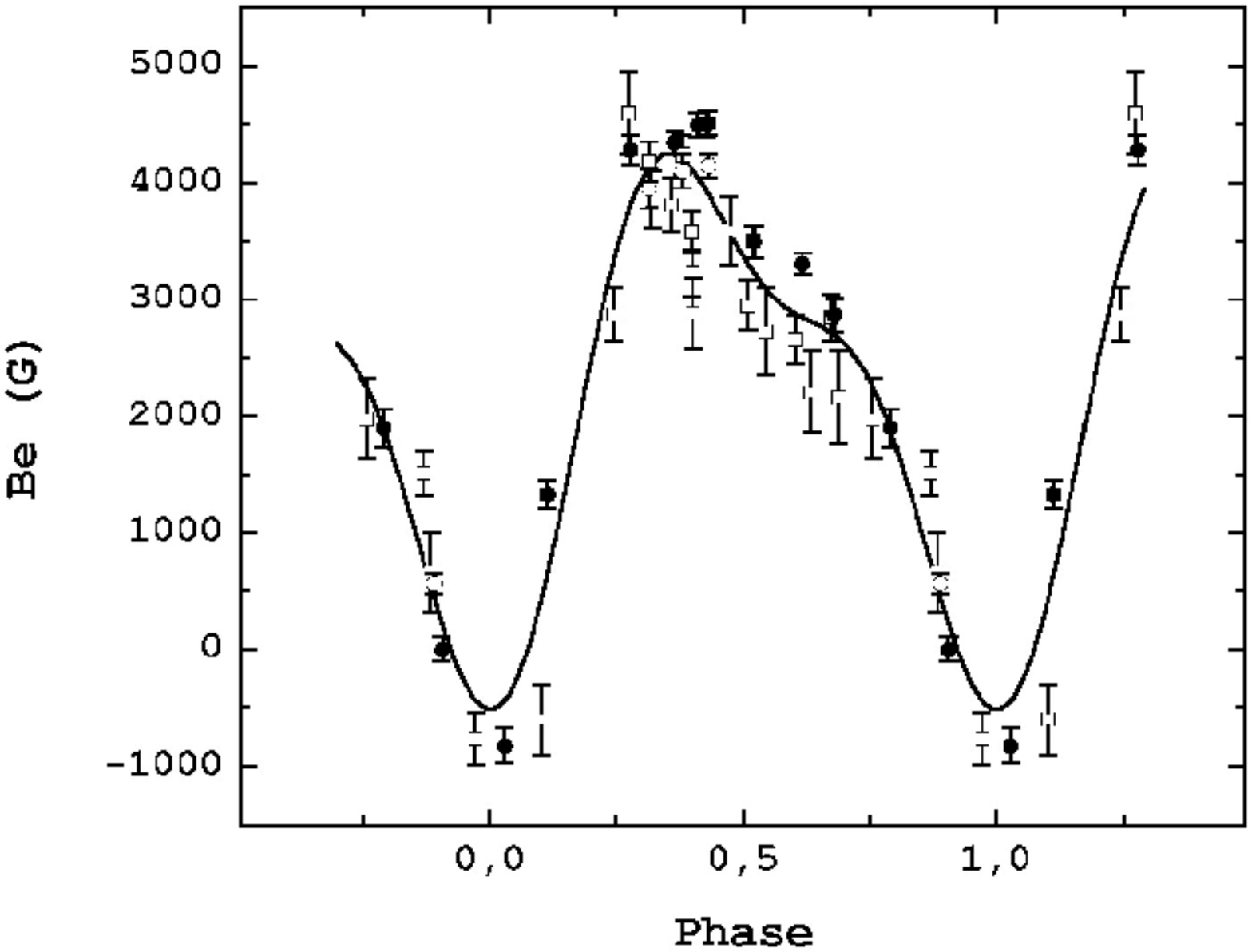}}
\vspace{-3.5mm}
\caption{ HD343872 (1) }
\label{fig:fig363}
\end{figure}

\begin{figure}
\resizebox{0.98\hsize}{!}{\includegraphics{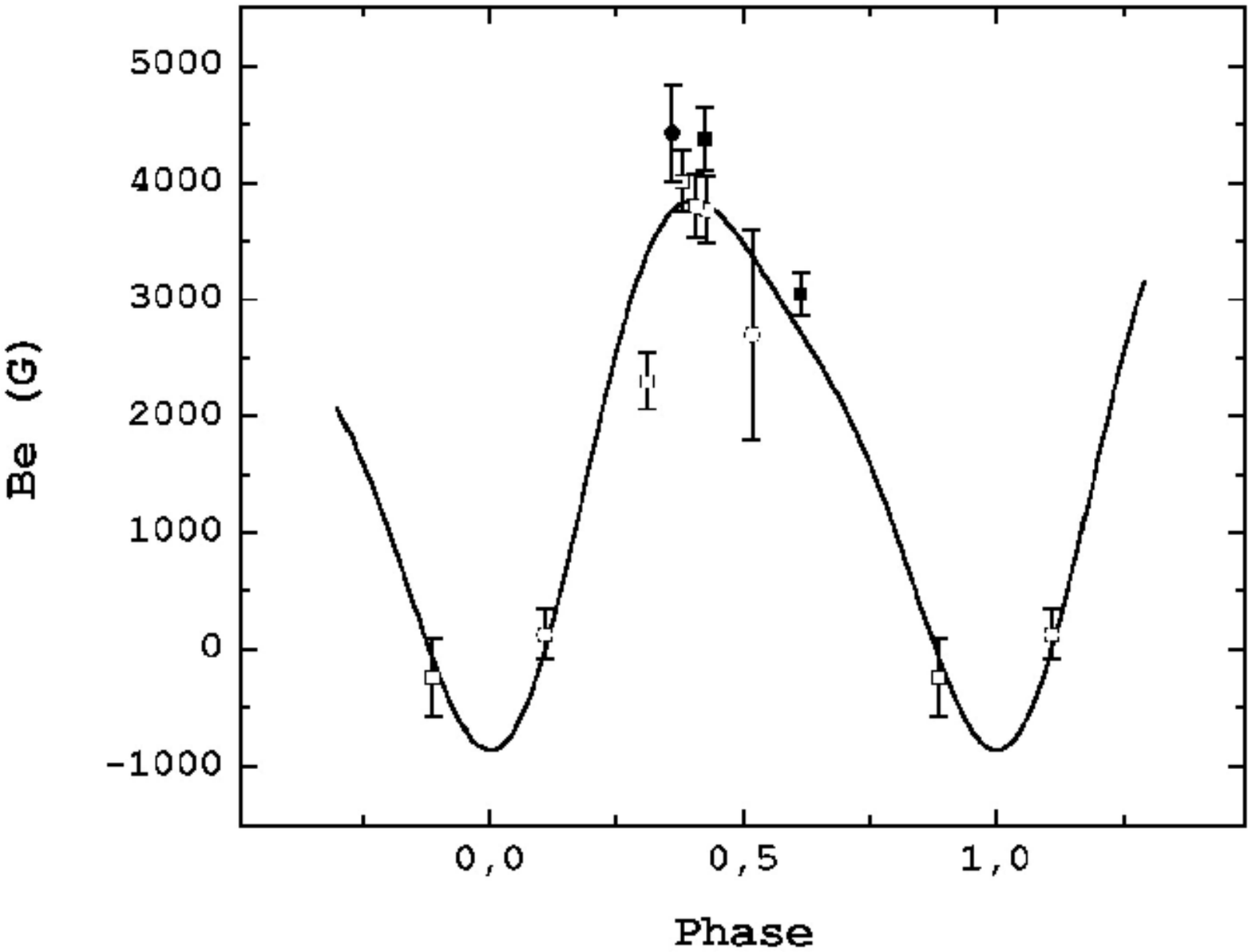}}
\vspace{-3.5mm}
\caption{ HD343872 (2) }
\label{fig:fig364}
\end{figure}

\begin{figure}
\resizebox{0.98\hsize}{!}{\includegraphics{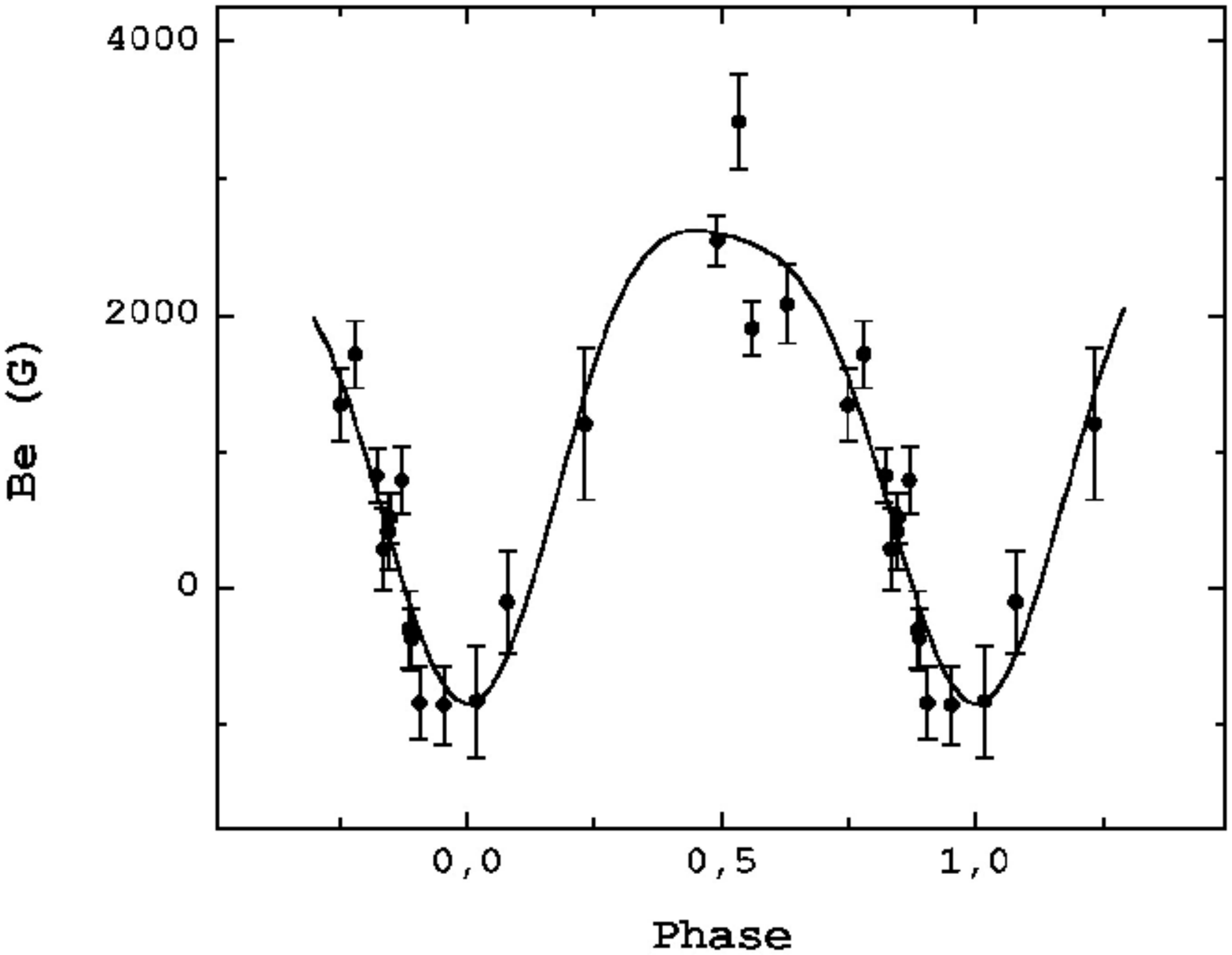}}
\vspace{-3.5mm}
\caption{ HD345439 (1) }
\label{fig:fig365}
\end{figure}

\begin{figure}
\resizebox{0.98\hsize}{!}{\includegraphics{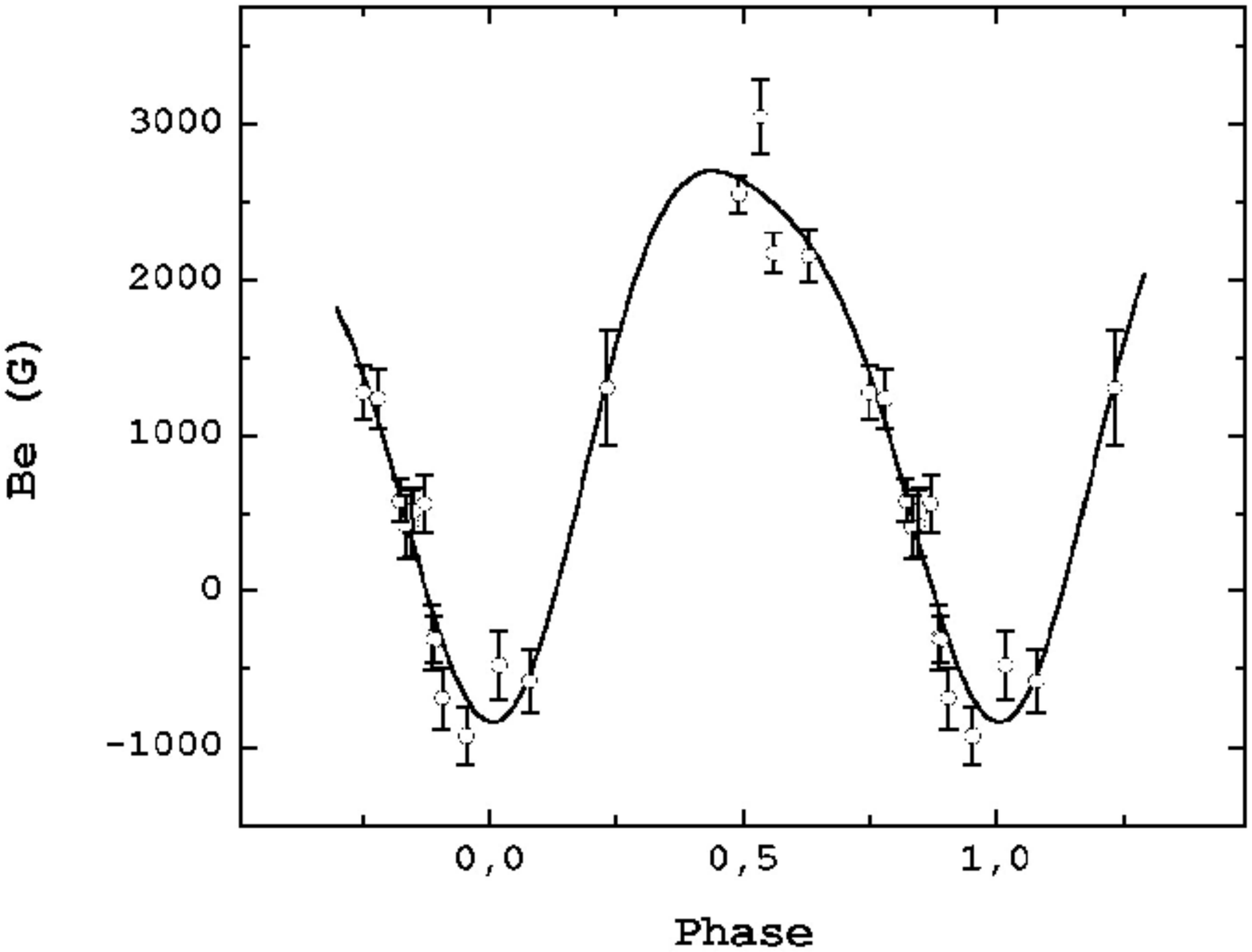}}
\vspace{-3.5mm}
\caption{ HD345439 (2) }
\label{fig:fig366}
\end{figure}

\clearpage
\newpage

\begin{figure}
\resizebox{0.98\hsize}{!}{\includegraphics{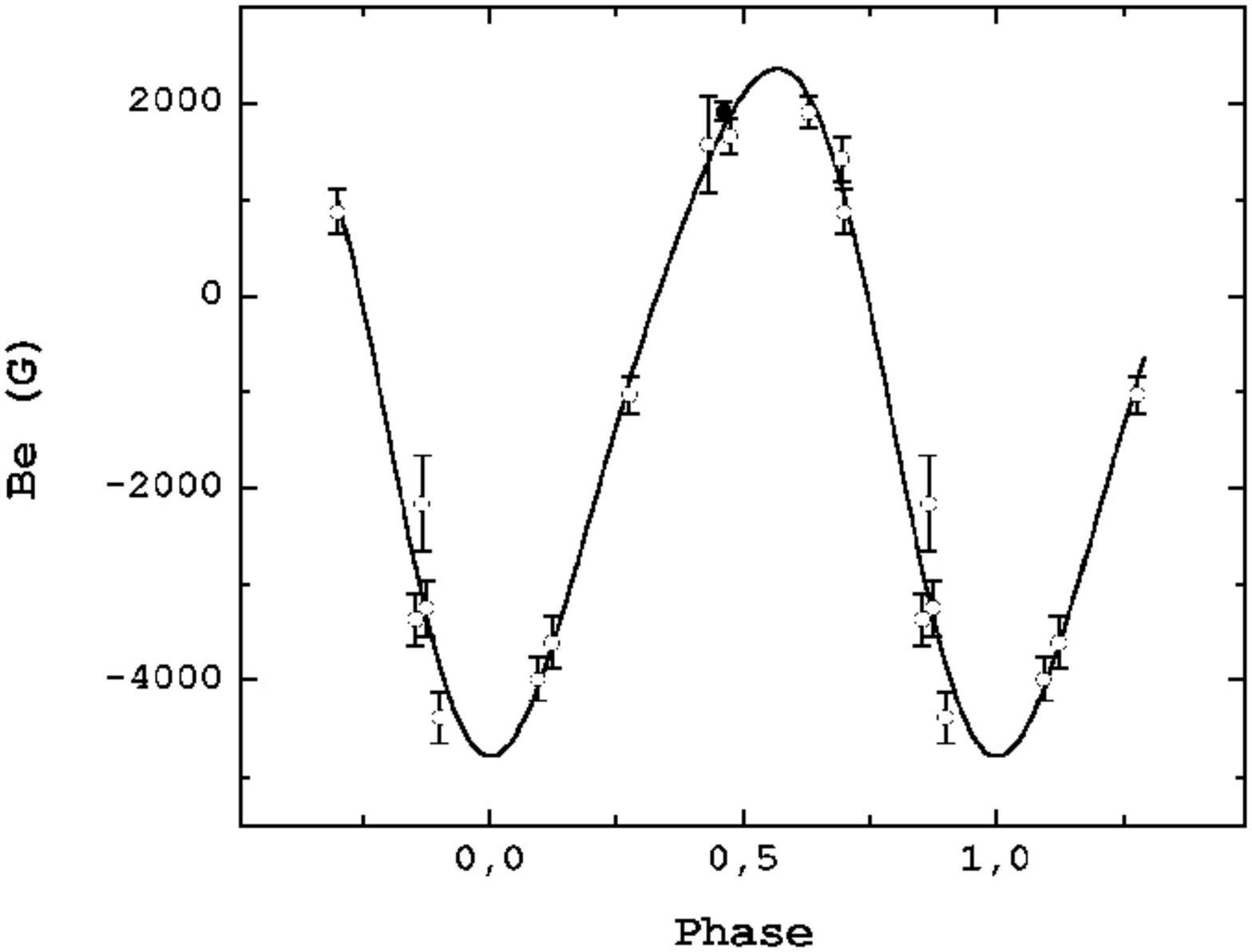}}
\vspace{-3.5mm}
\caption{ HD349321 }
\label{fig:fig367}
\end{figure}

\begin{figure}
\resizebox{0.98\hsize}{!}{\includegraphics{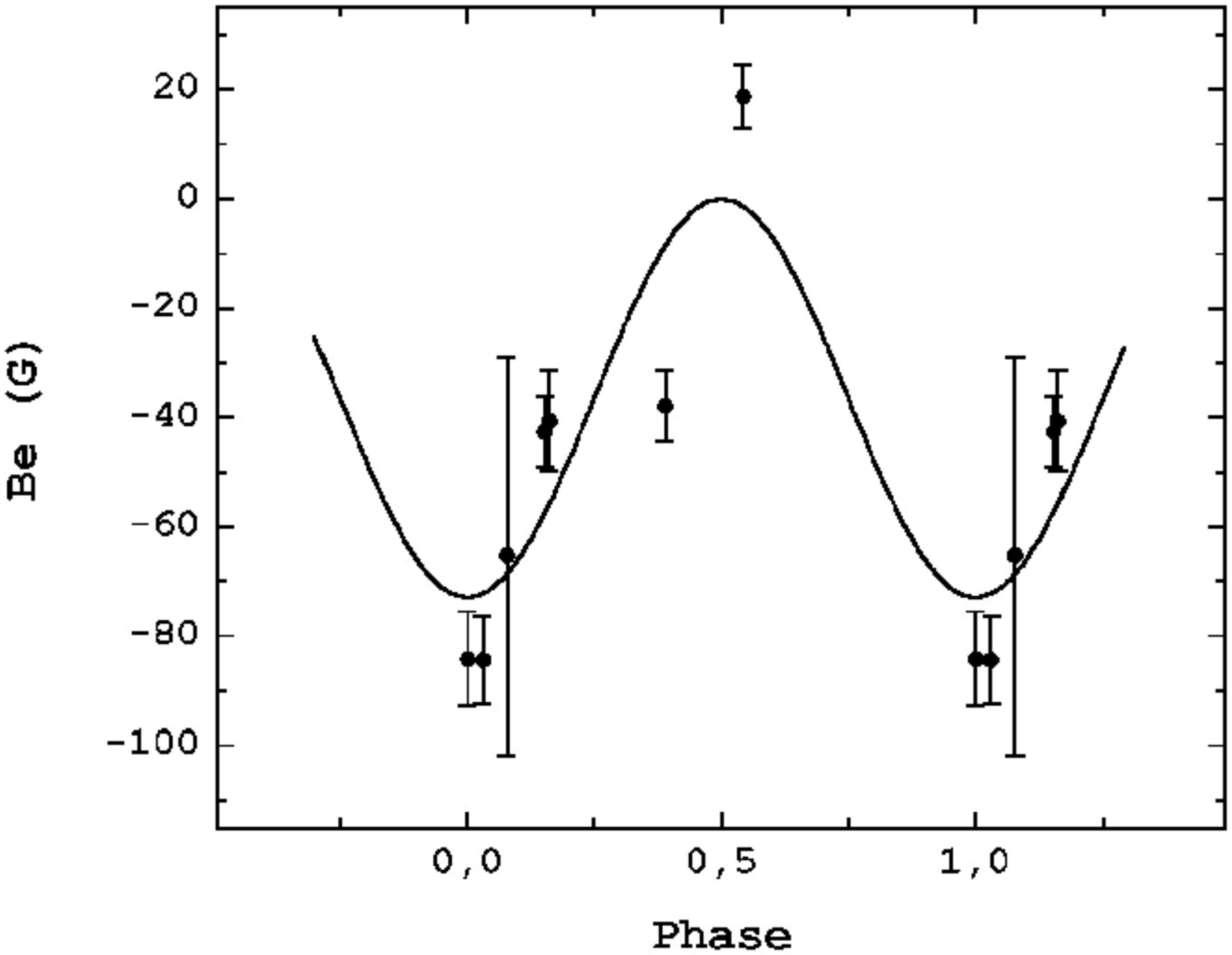}}
\vspace{-3.5mm}
\caption{ V1005 Ori }
\label{fig:fig368}
\end{figure}

\begin{figure}
\resizebox{0.98\hsize}{!}{\includegraphics{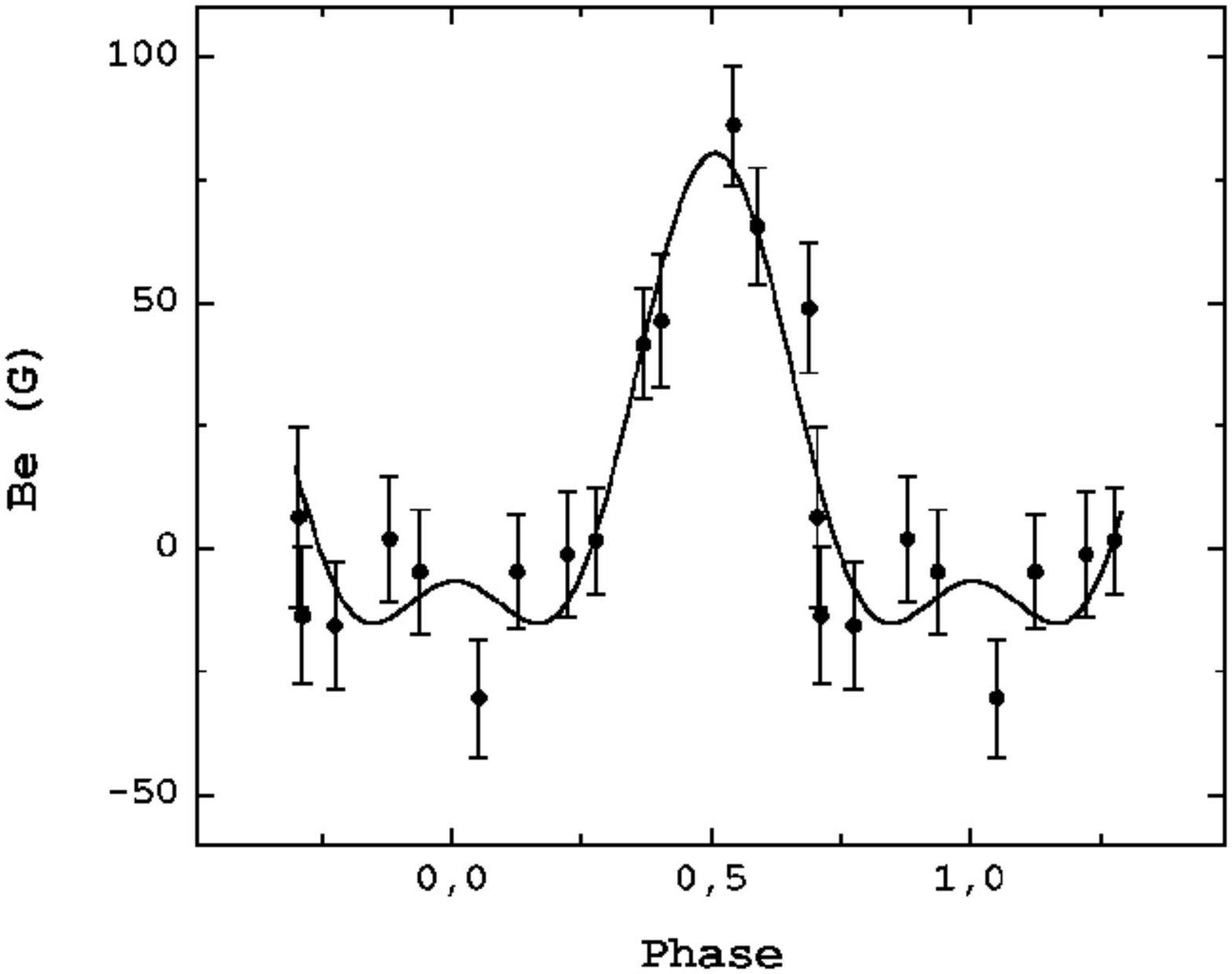}}
\vspace{-3.5mm}
\caption{ V1079 Tau }
\label{fig:fig368}
\end{figure}

\begin{figure}
\resizebox{0.98\hsize}{!}{\includegraphics{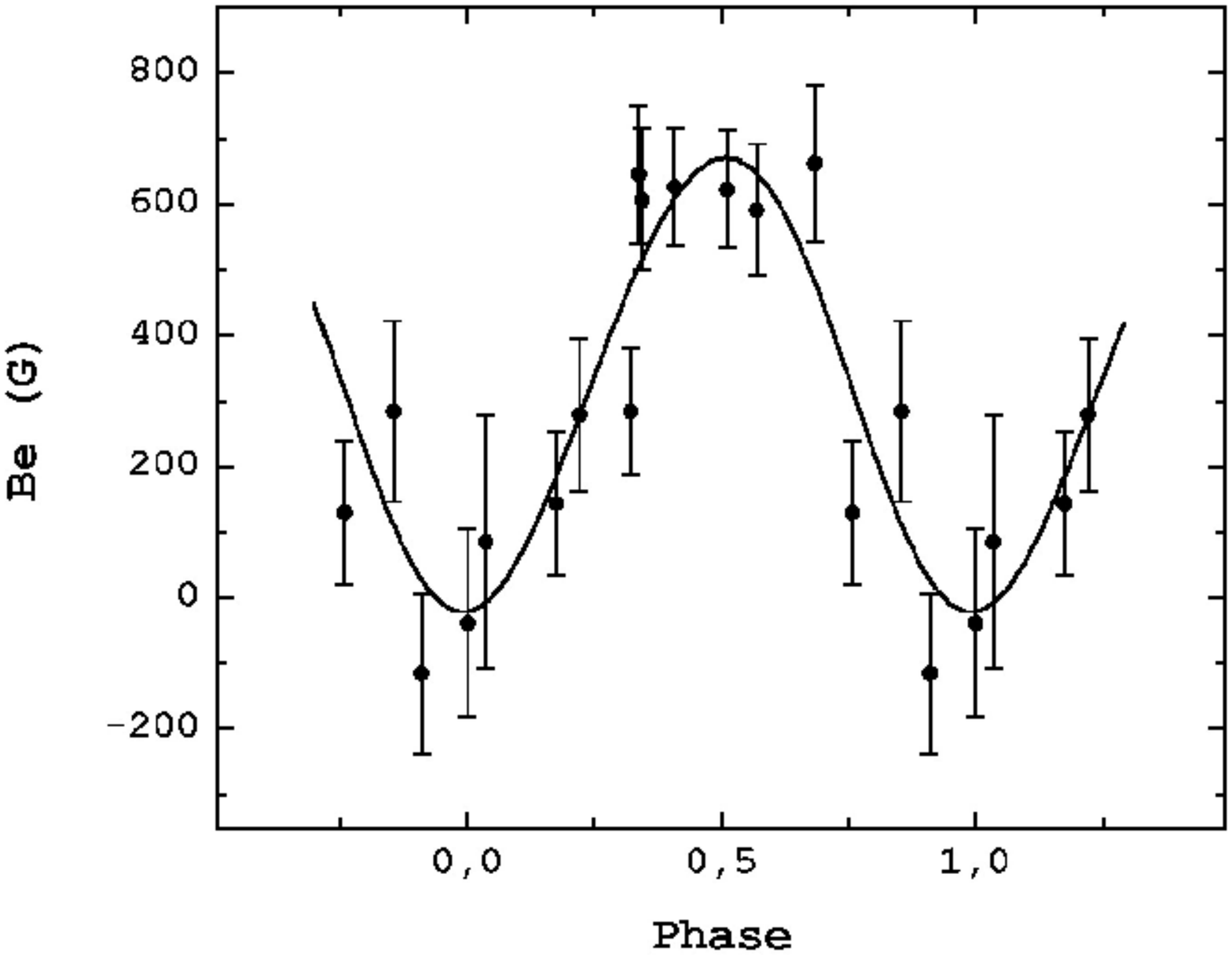}}
\vspace{-3.5mm}
\caption{ V1079 Tau }
\label{fig:fig368}
\end{figure}

\begin{figure}
\resizebox{0.98\hsize}{!}{\includegraphics{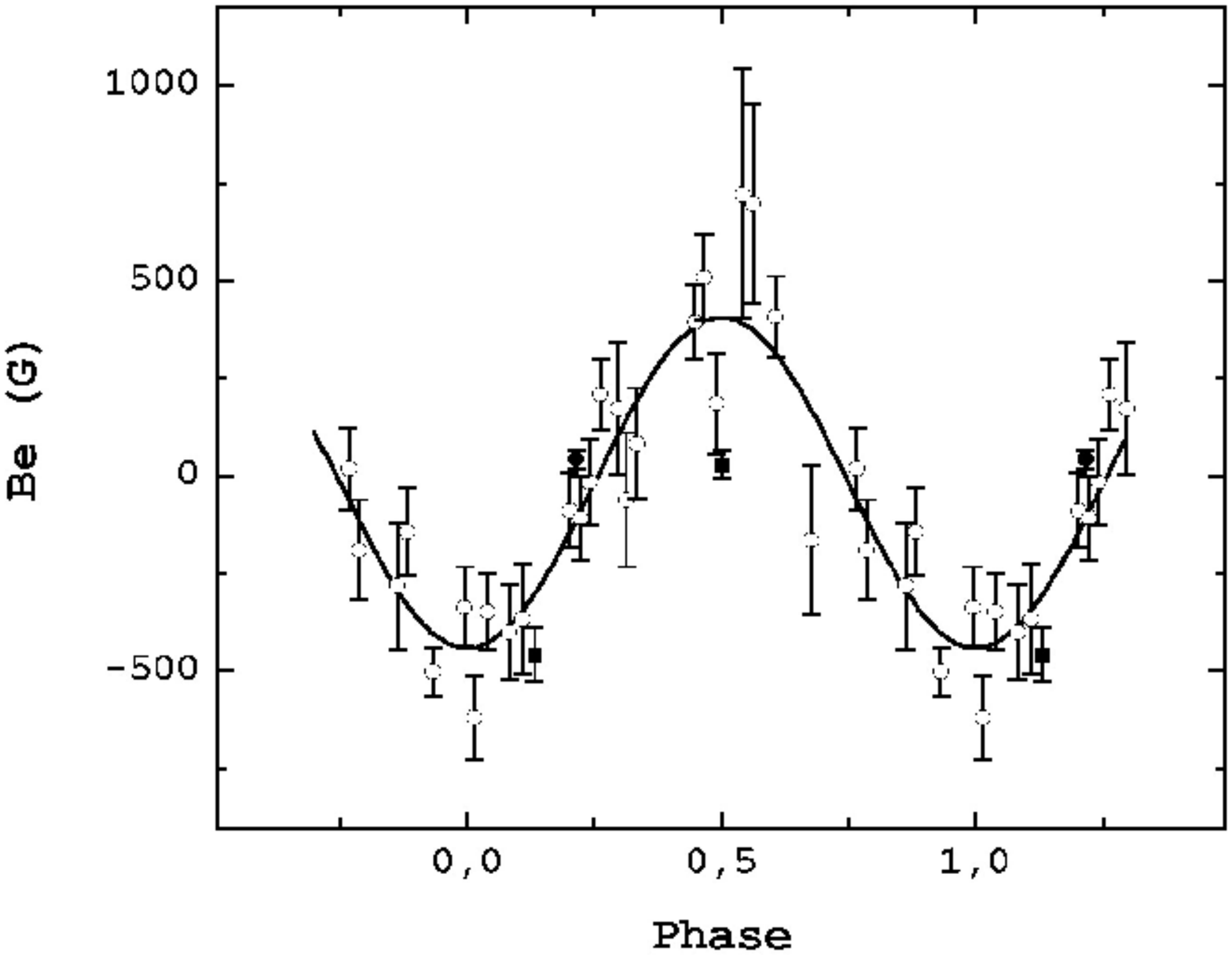}}
\vspace{-3.5mm}
\caption{ V380 Ori }
\label{fig:fig369}
\end{figure}

\begin{figure}
\resizebox{0.98\hsize}{!}{\includegraphics{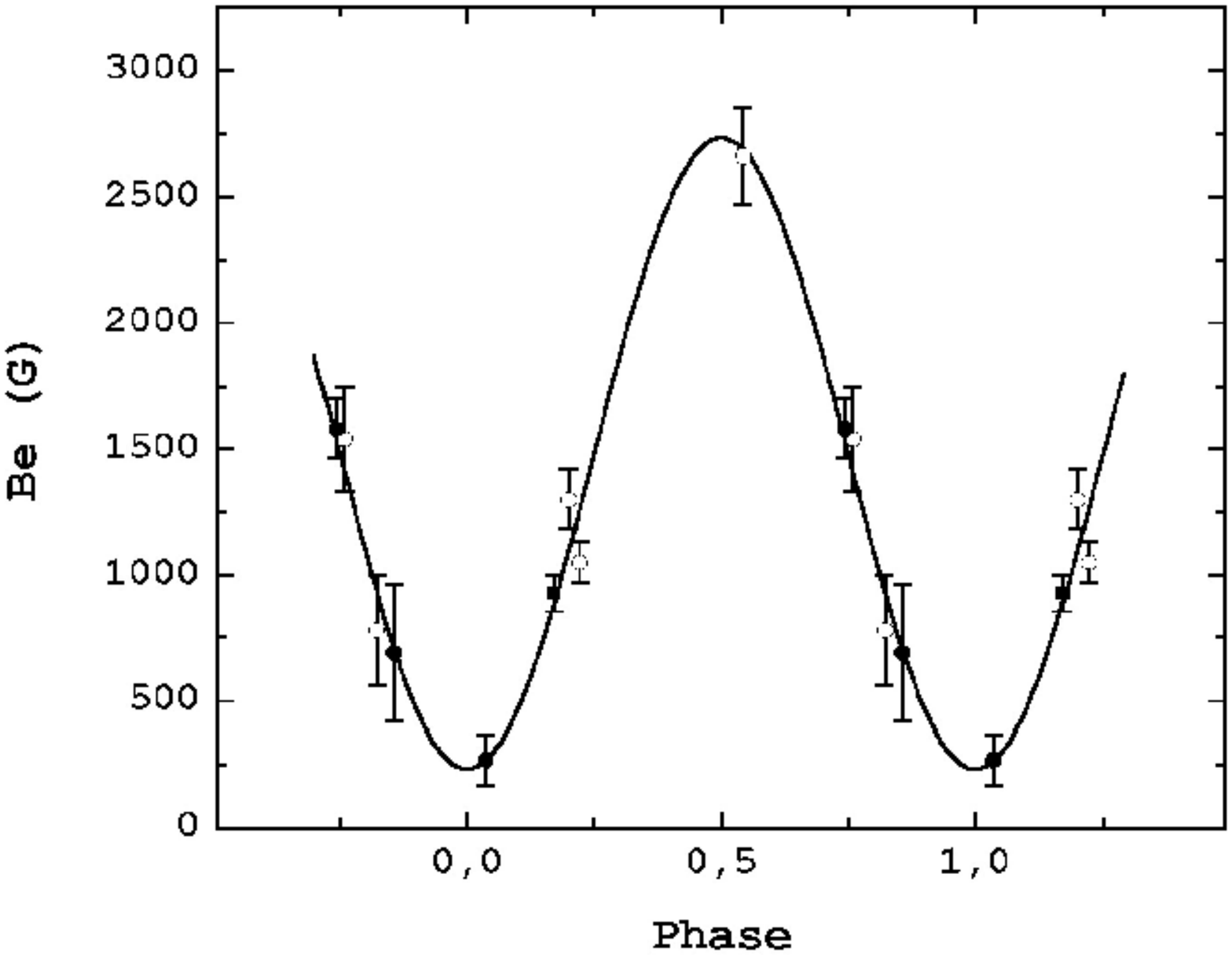}}
\vspace{-3.5mm}
\caption{ BD+40 175B }
\label{fig:fig370}
\end{figure}

\clearpage
\newpage

\begin{figure}
\resizebox{0.98\hsize}{!}{\includegraphics{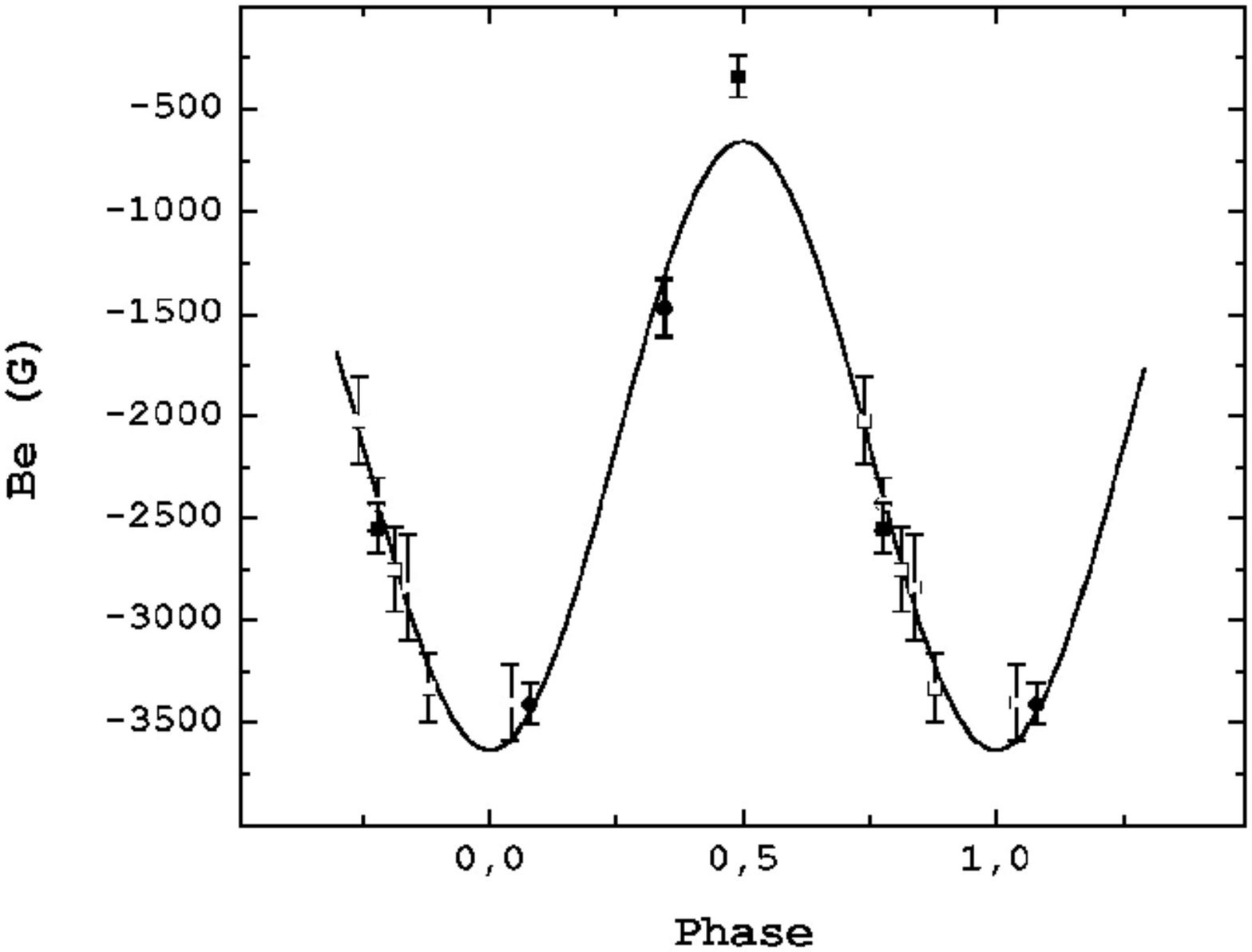}}
\vspace{-3.5mm}
\caption{ BD+40 175A }
\label{fig:fig368}
\end{figure}

\begin{figure}
\resizebox{0.98\hsize}{!}{\includegraphics{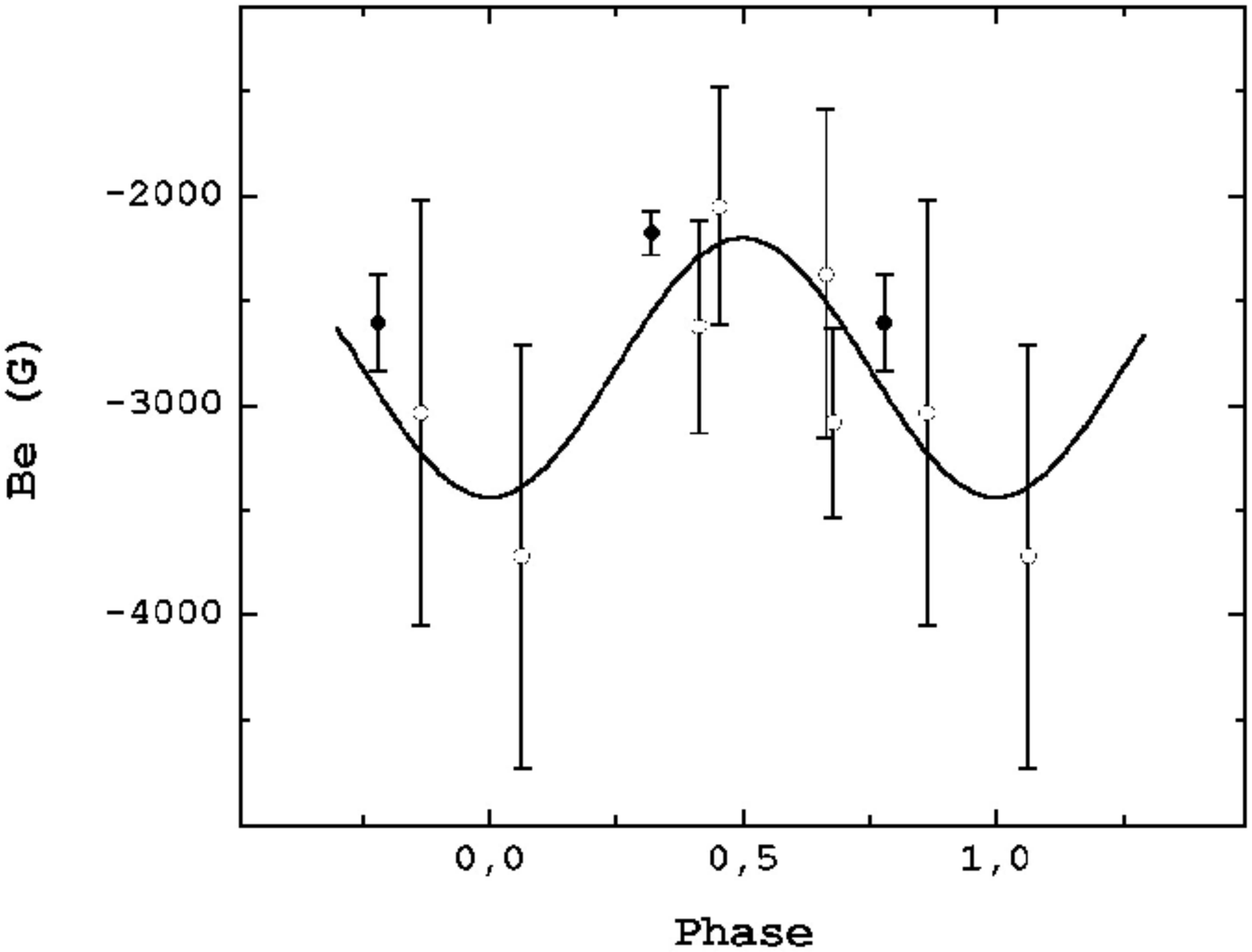}}
\vspace{-3.5mm}
\caption{ NGC 6193 17 (1) }
\label{fig:fig368}
\end{figure}

\begin{figure}
\resizebox{0.98\hsize}{!}{\includegraphics{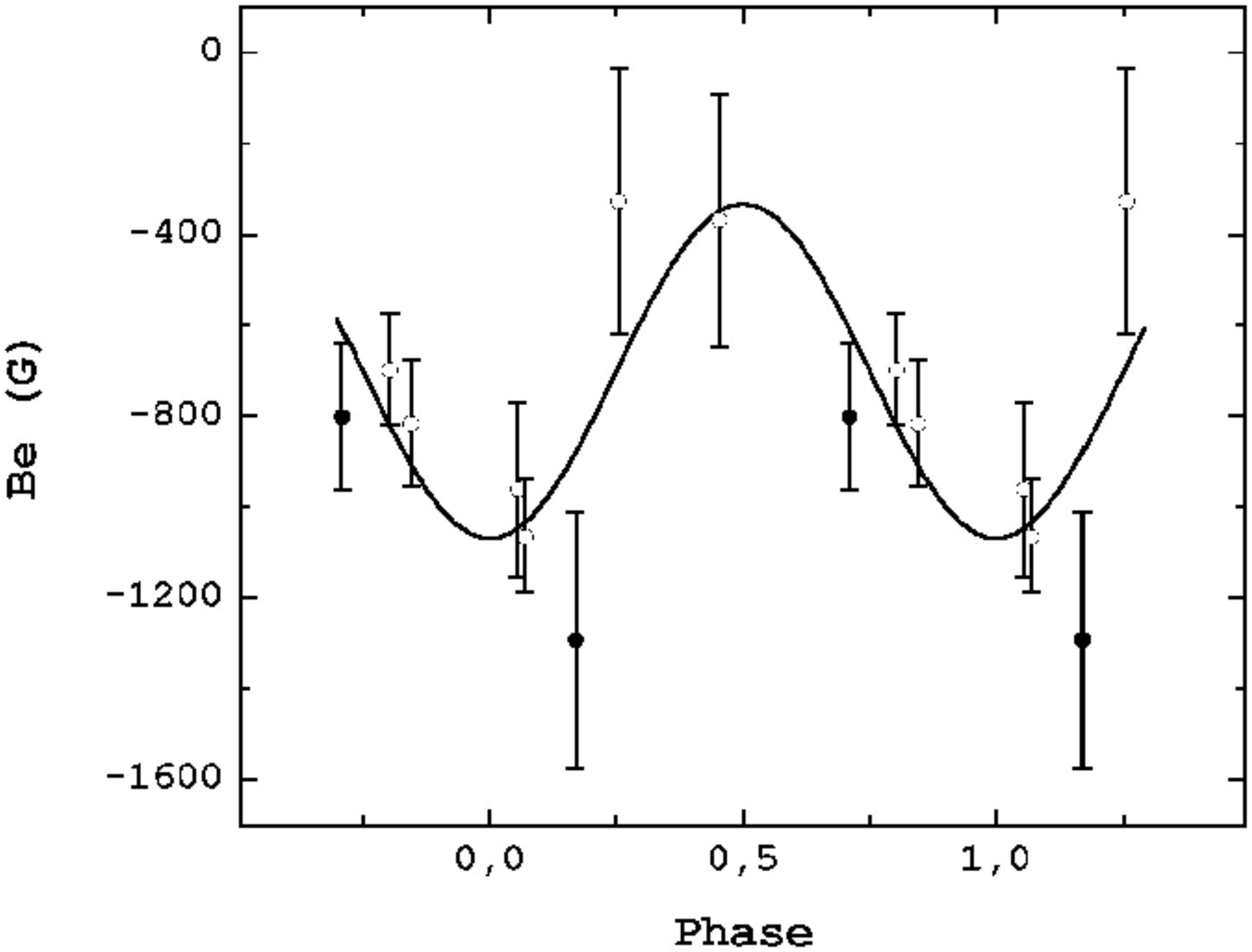}}
\vspace{-3.5mm}
\caption{ NGC 6193 17 (2) }
\label{fig:fig368}
\end{figure}

\begin{figure}
\resizebox{0.98\hsize}{!}{\includegraphics{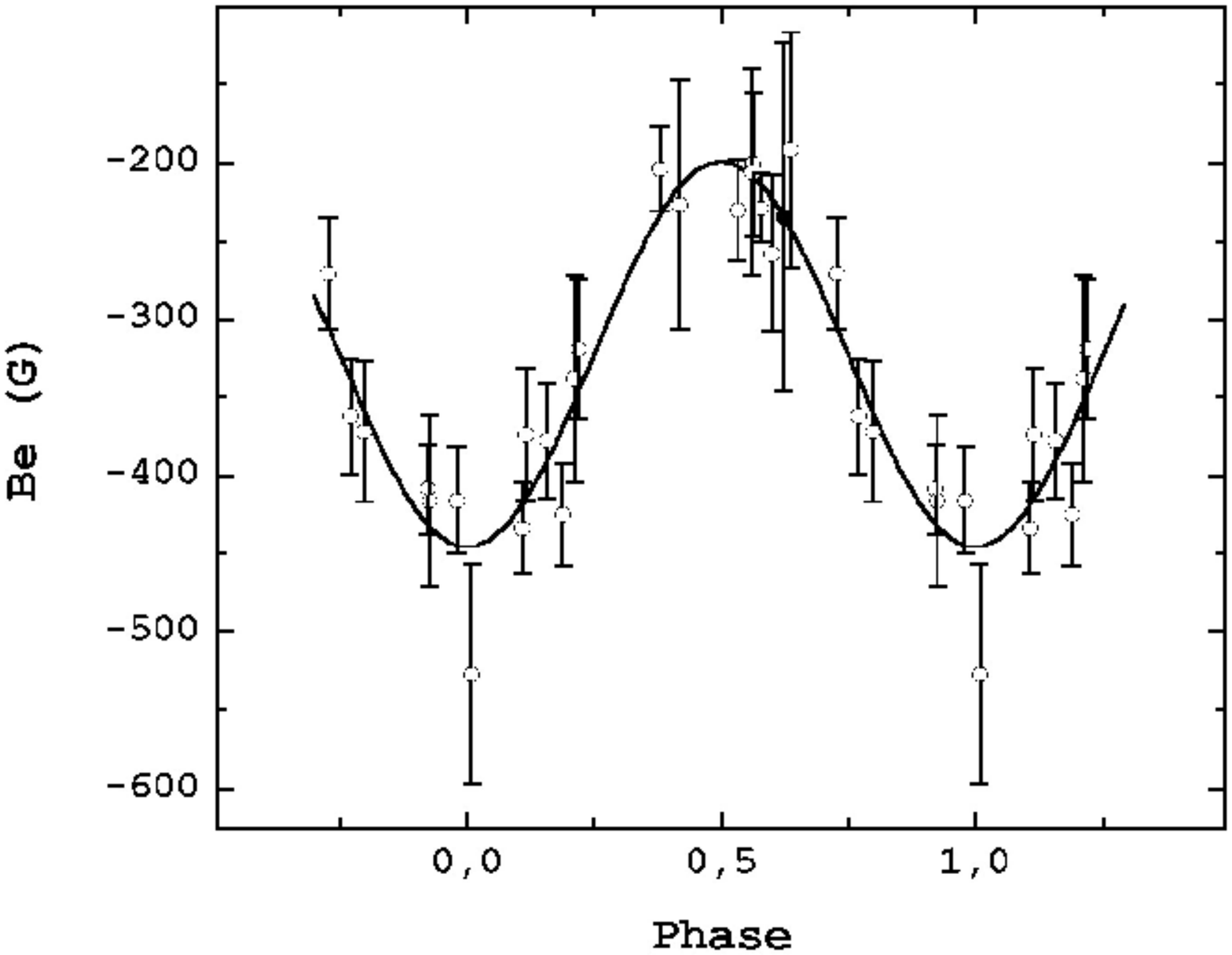}}
\vspace{-3.5mm}
\caption{ BD-19 5044L }
\label{fig:fig371}
\end{figure}

\begin{figure}
\resizebox{0.98\hsize}{!}{\includegraphics{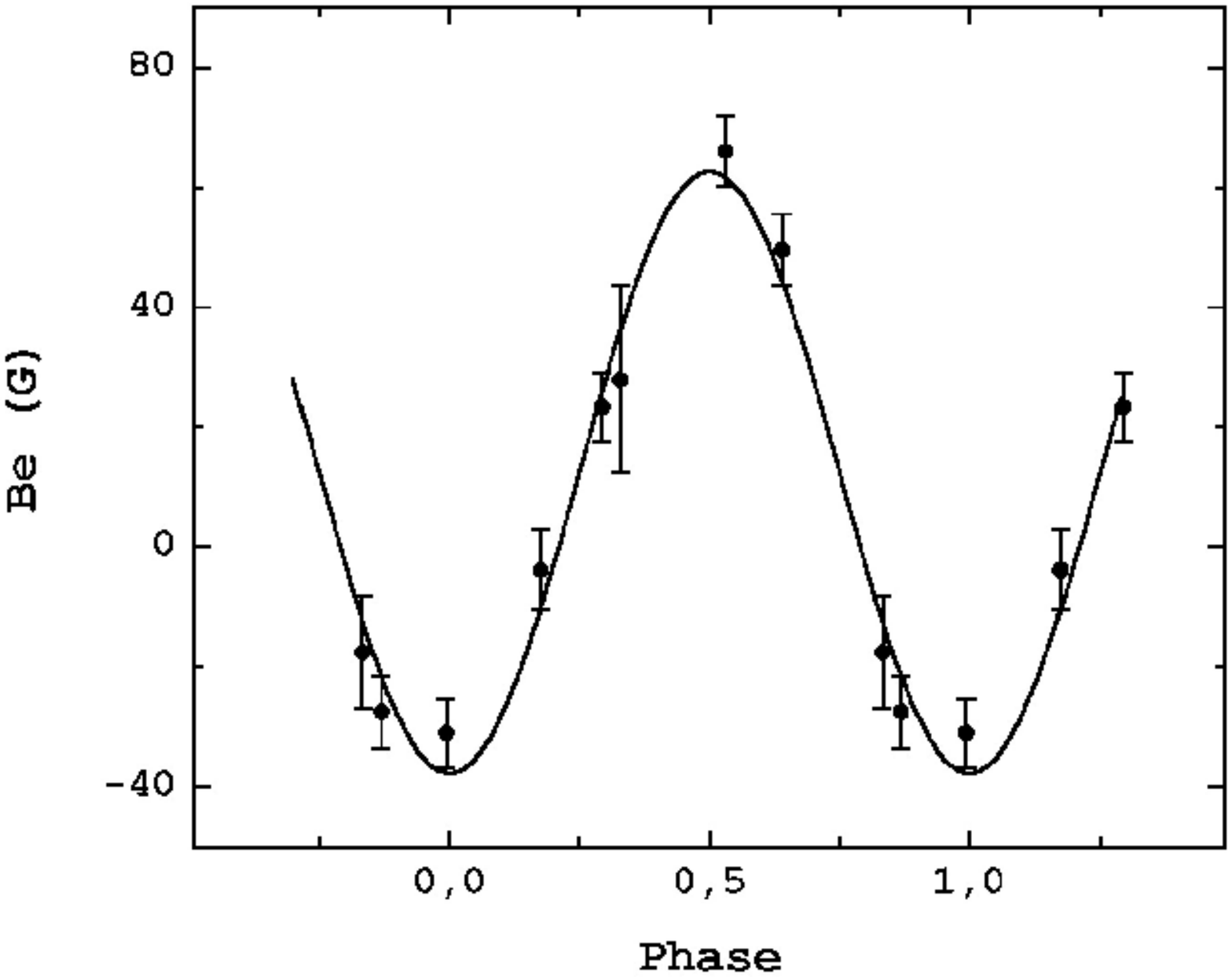}}
\vspace{-3.5mm}
\caption{ DT Vir (1) }
\label{fig:fig372}
\end{figure}

\begin{figure}
\resizebox{0.98\hsize}{!}{\includegraphics{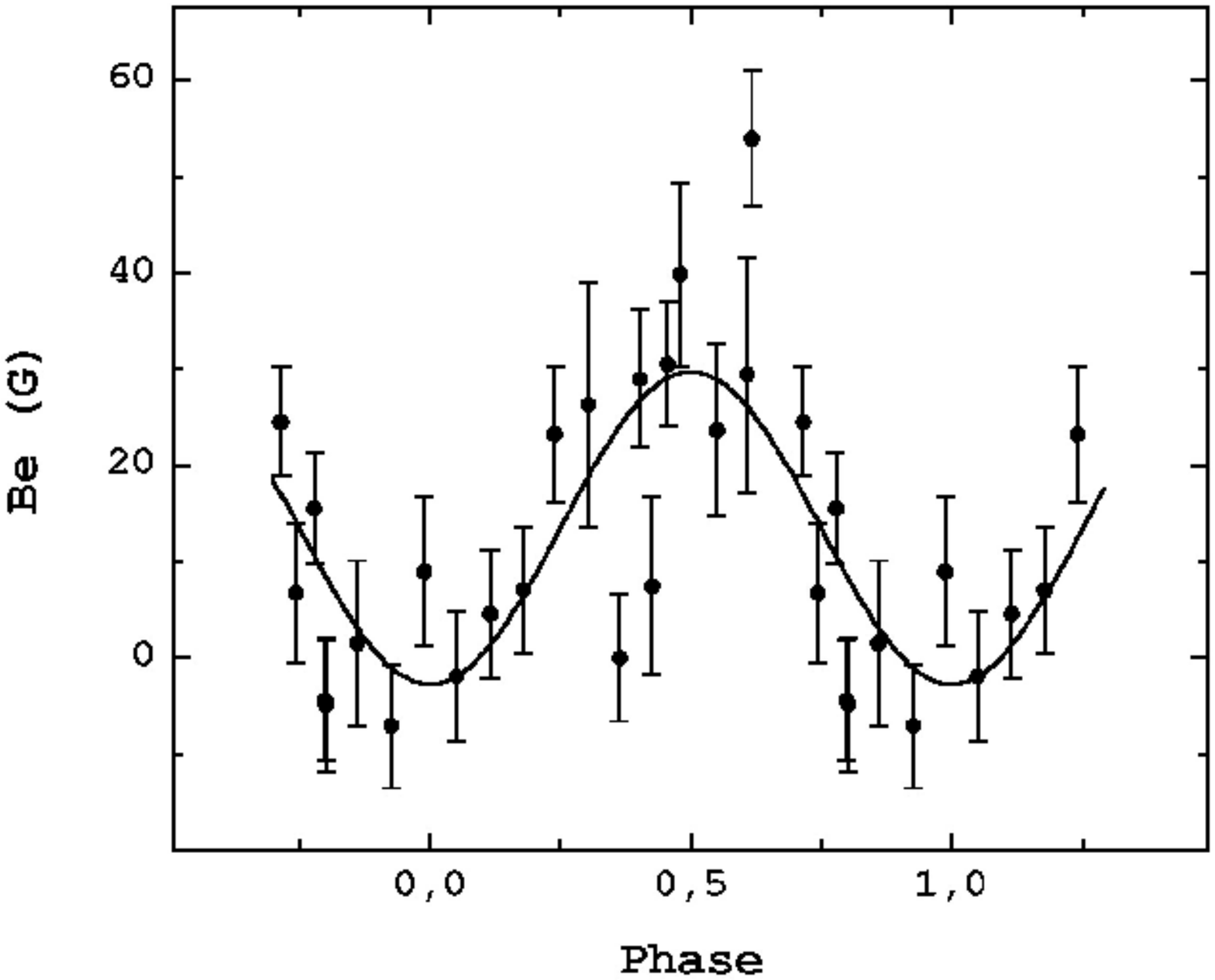}}
\vspace{-3.5mm}
\caption{ DT Vir (2) }
\label{fig:fig373}
\end{figure}

\clearpage
\newpage

\begin{figure}
\resizebox{0.98\hsize}{!}{\includegraphics{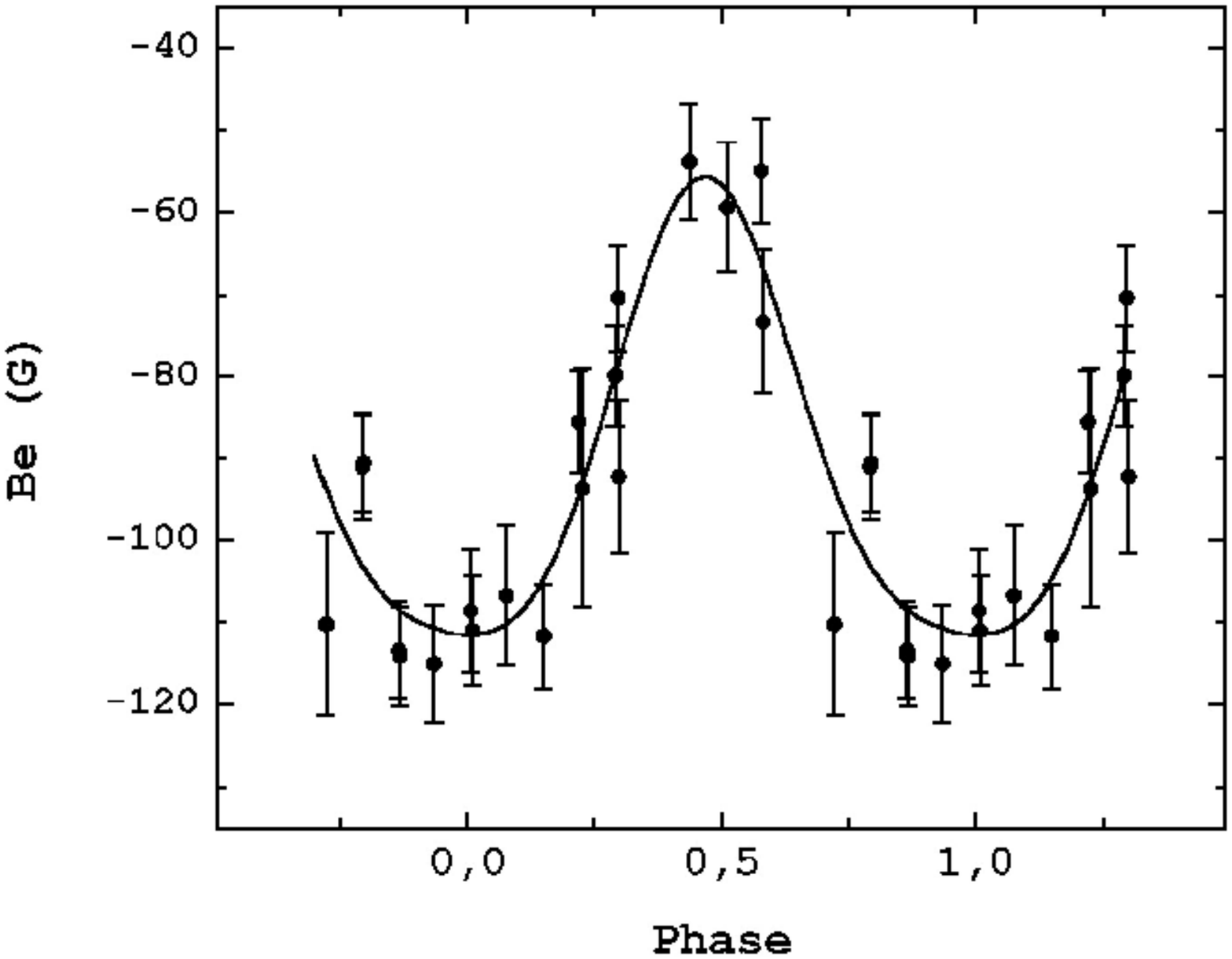}}
\vspace{-3.5mm}
\caption{ CE Boo }
\label{fig:fig374}
\end{figure}

\begin{figure}
\resizebox{0.98\hsize}{!}{\includegraphics{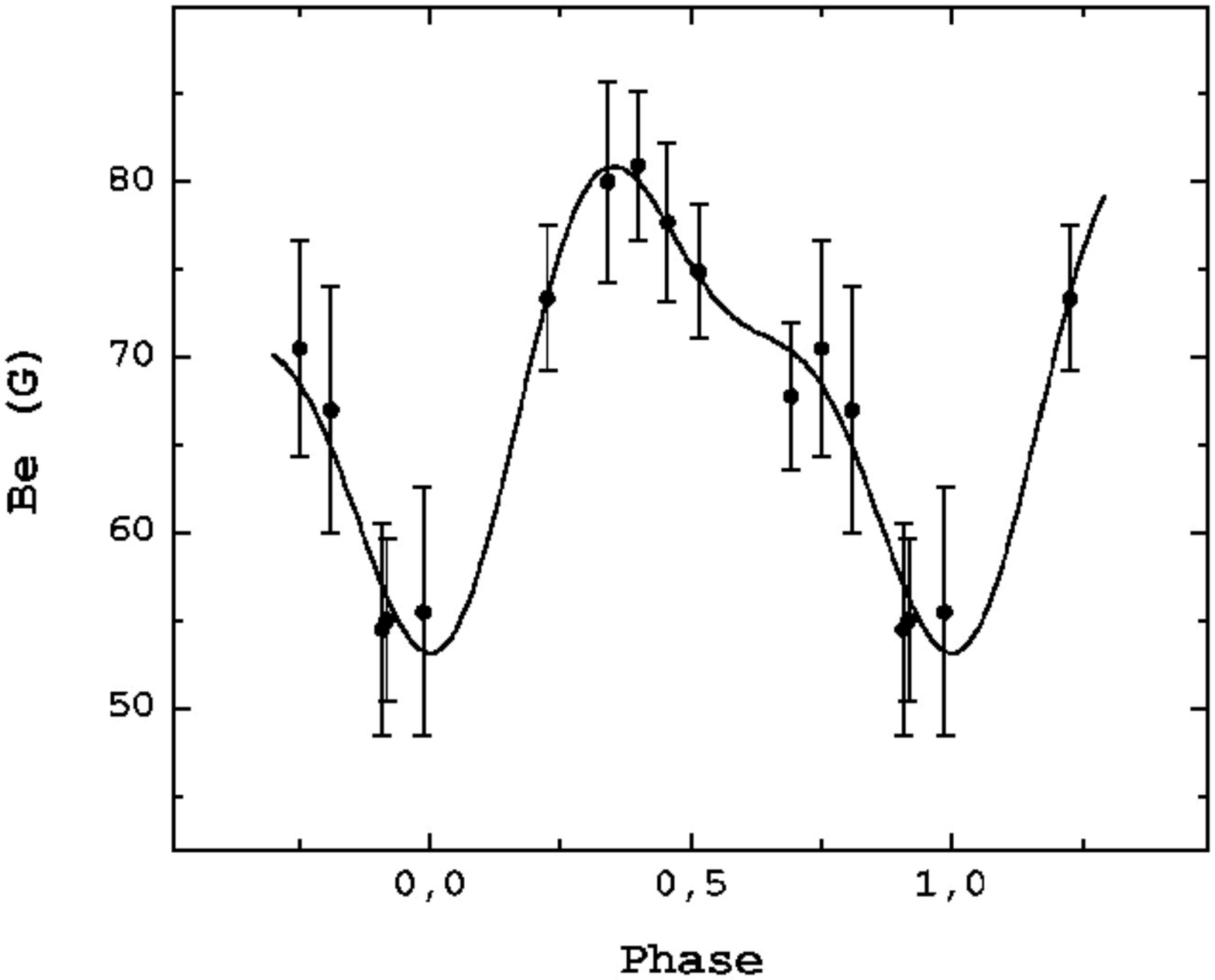}}
\vspace{-3.5mm}
\caption{ OT Ser (1) }
\label{fig:fig375}
\end{figure}

\begin{figure}
\resizebox{0.98\hsize}{!}{\includegraphics{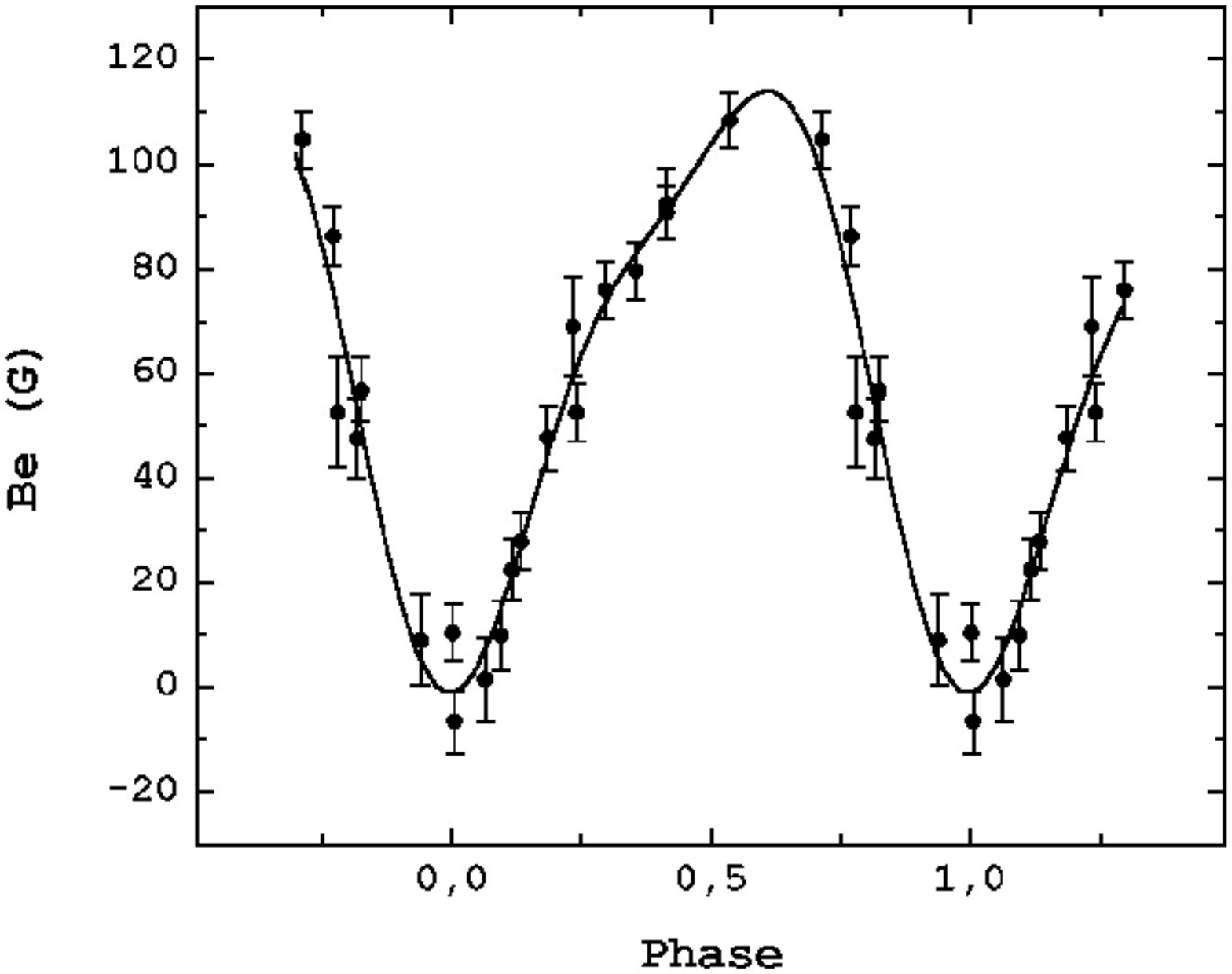}}
\vspace{-3.5mm}
\caption{ OT Ser (2) }
\label{fig:fig376}
\end{figure}

\begin{figure}
\resizebox{0.98\hsize}{!}{\includegraphics{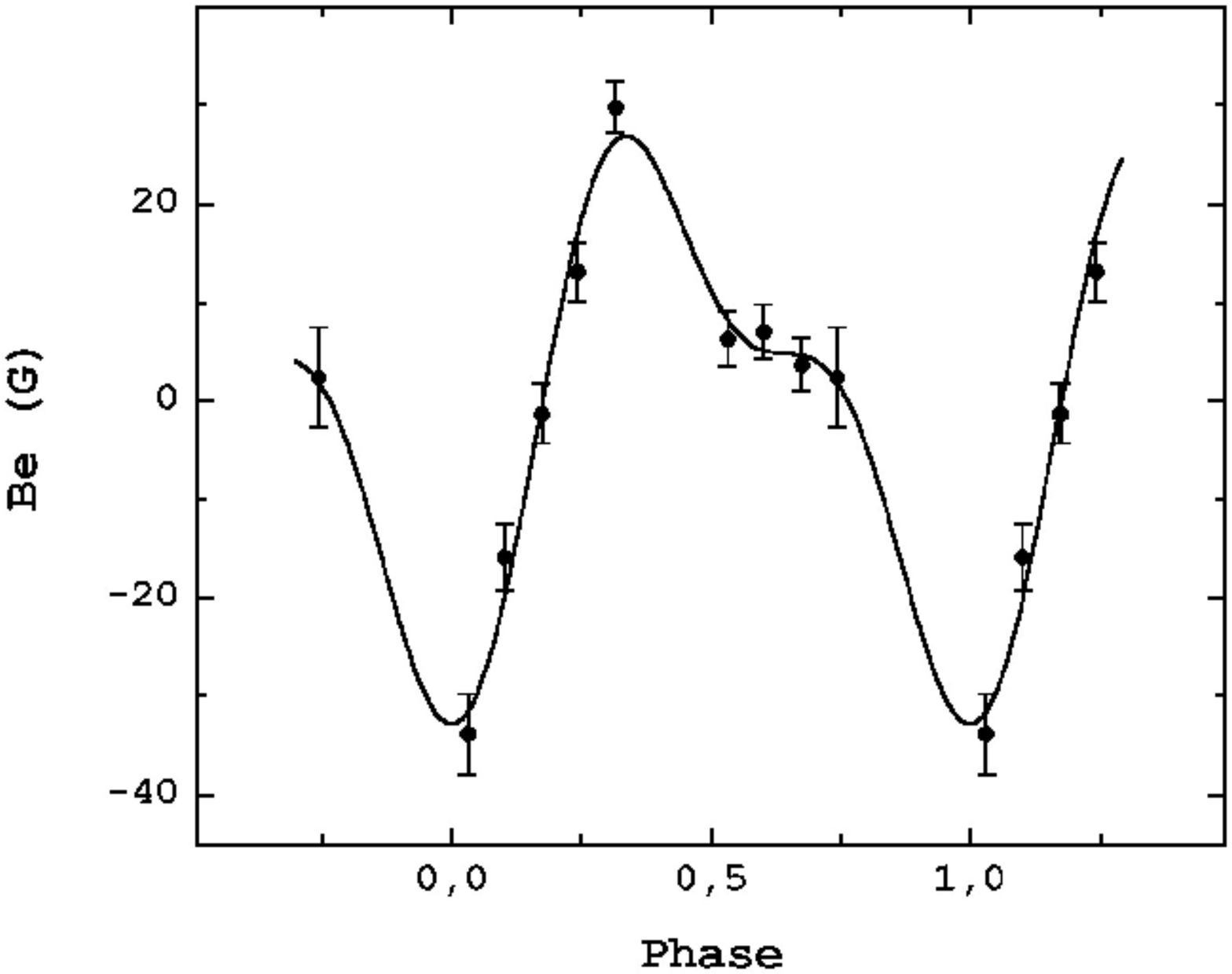}}
\vspace{-3.5mm}
\caption{ DS Leo (1) }
\label{fig:fig377}
\end{figure}

\begin{figure}
\resizebox{0.98\hsize}{!}{\includegraphics{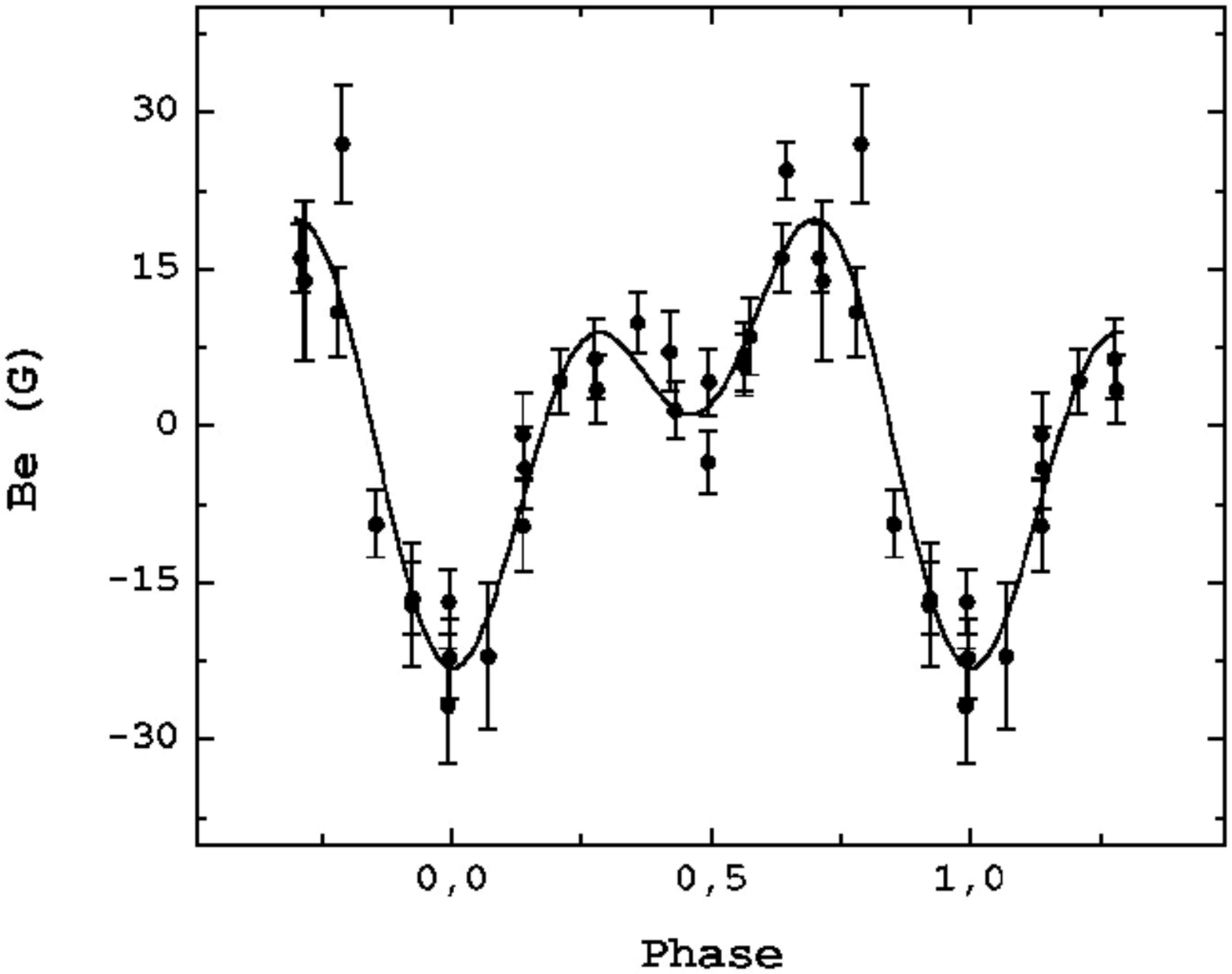}}
\vspace{-3.5mm}
\caption{ DS Leo (2) }
\label{fig:fig378}
\end{figure}

\begin{figure}
\resizebox{0.98\hsize}{!}{\includegraphics{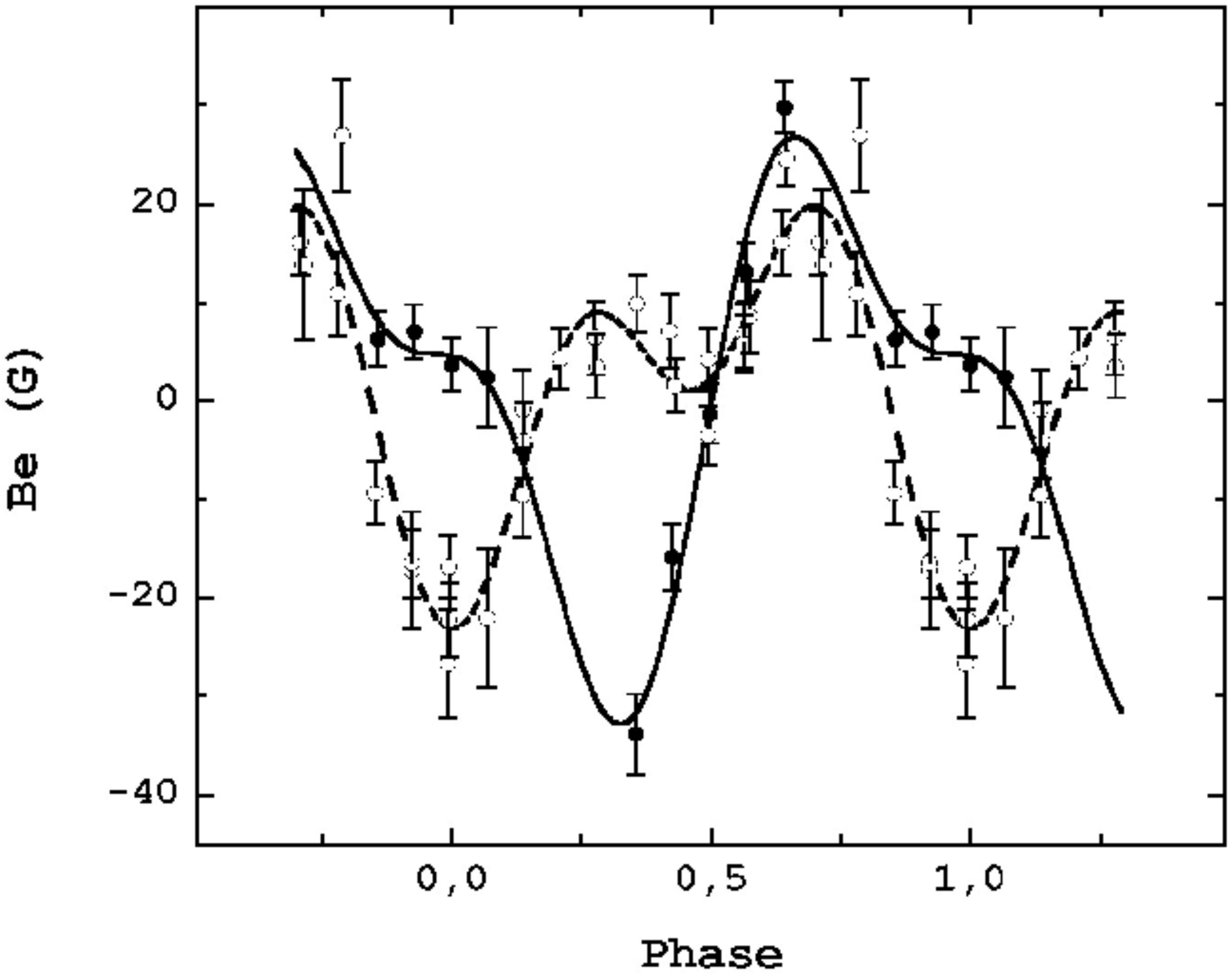}}
\vspace{-3.5mm}
\caption{ DS Leo (3) }
\label{fig:fig379}
\end{figure}

\clearpage
\newpage

\begin{figure}
\resizebox{0.98\hsize}{!}{\includegraphics{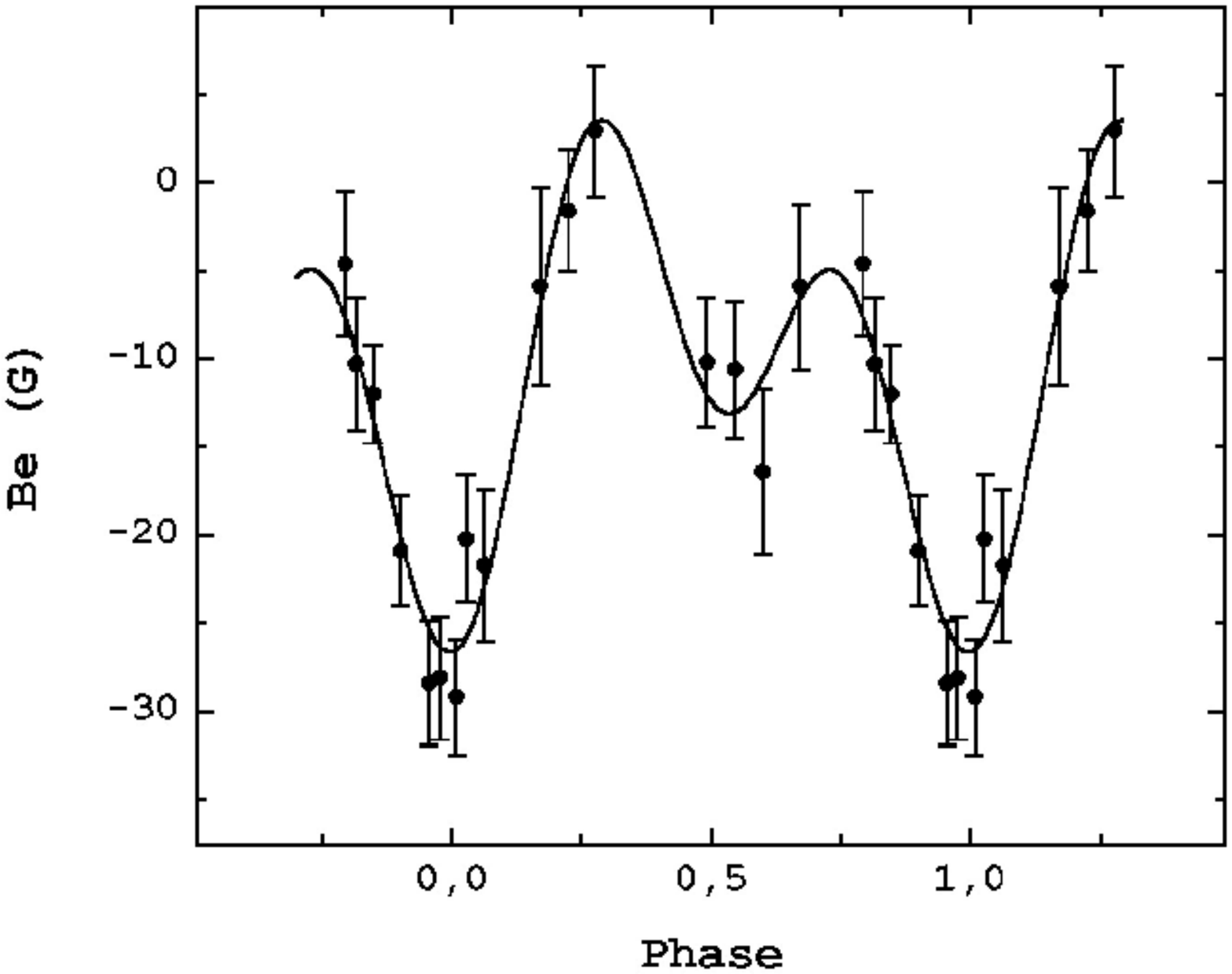}}
\vspace{-3.5mm}
\caption{ BD+61 195 }
\label{fig:fig380}
\end{figure}

\begin{figure}
\resizebox{0.98\hsize}{!}{\includegraphics{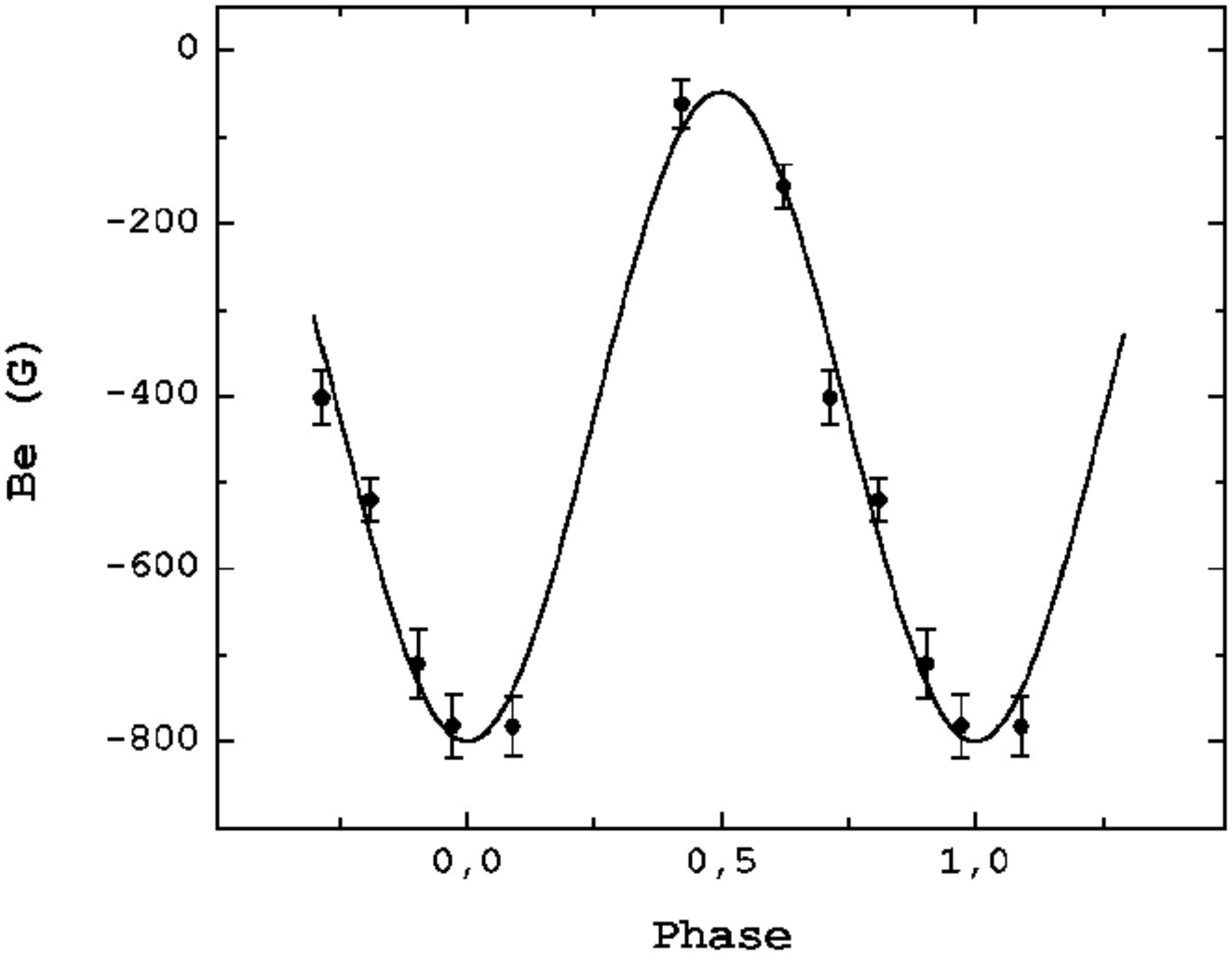}}
\vspace{-3.5mm}
\caption{ YZ CMi (1) }
\label{fig:fig381}
\end{figure}

\begin{figure}
\resizebox{0.98\hsize}{!}{\includegraphics{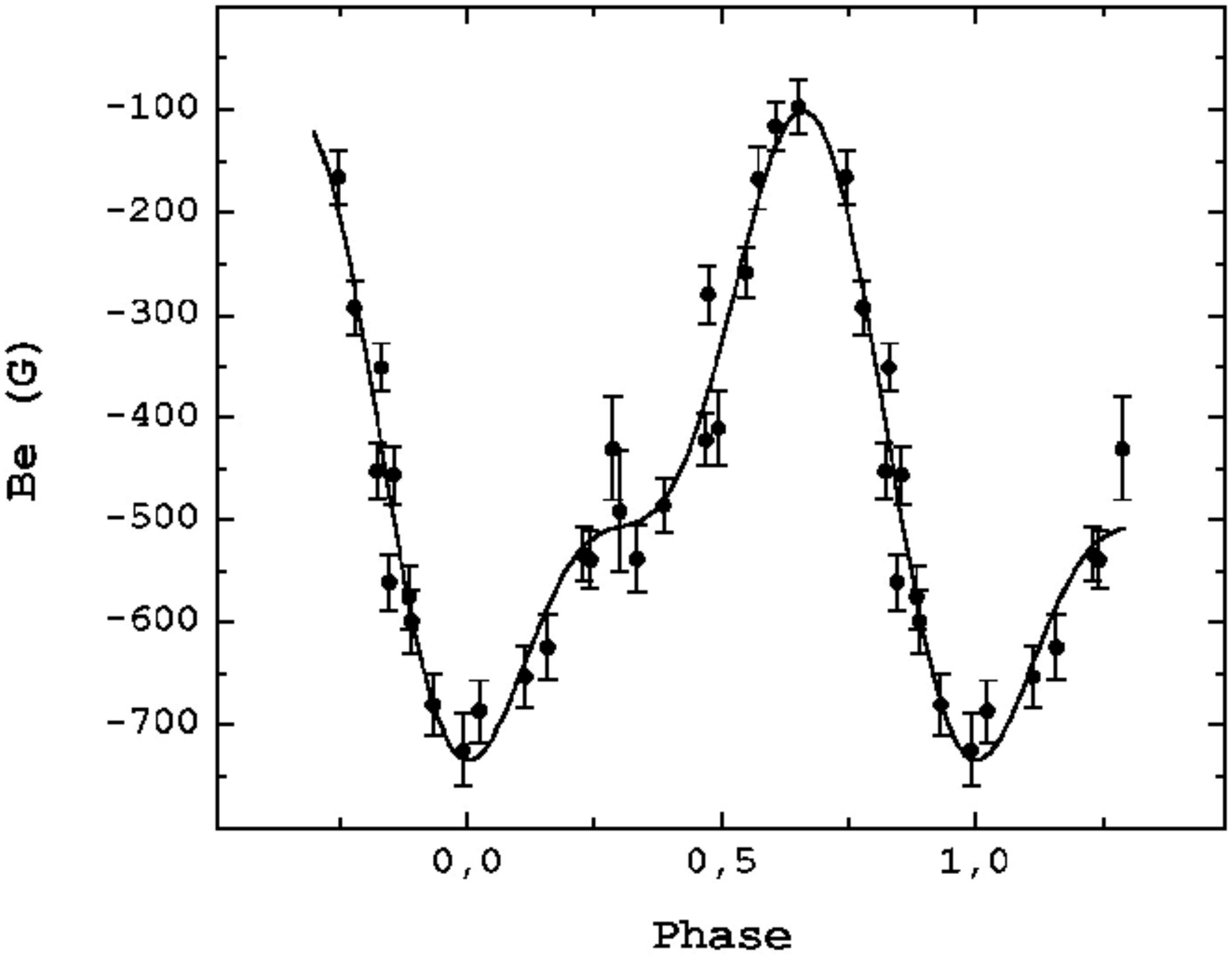}}
\vspace{-3.5mm}
\caption{ YZ CMi (2) }
\label{fig:fig382}
\end{figure}

\begin{figure}
\resizebox{0.98\hsize}{!}{\includegraphics{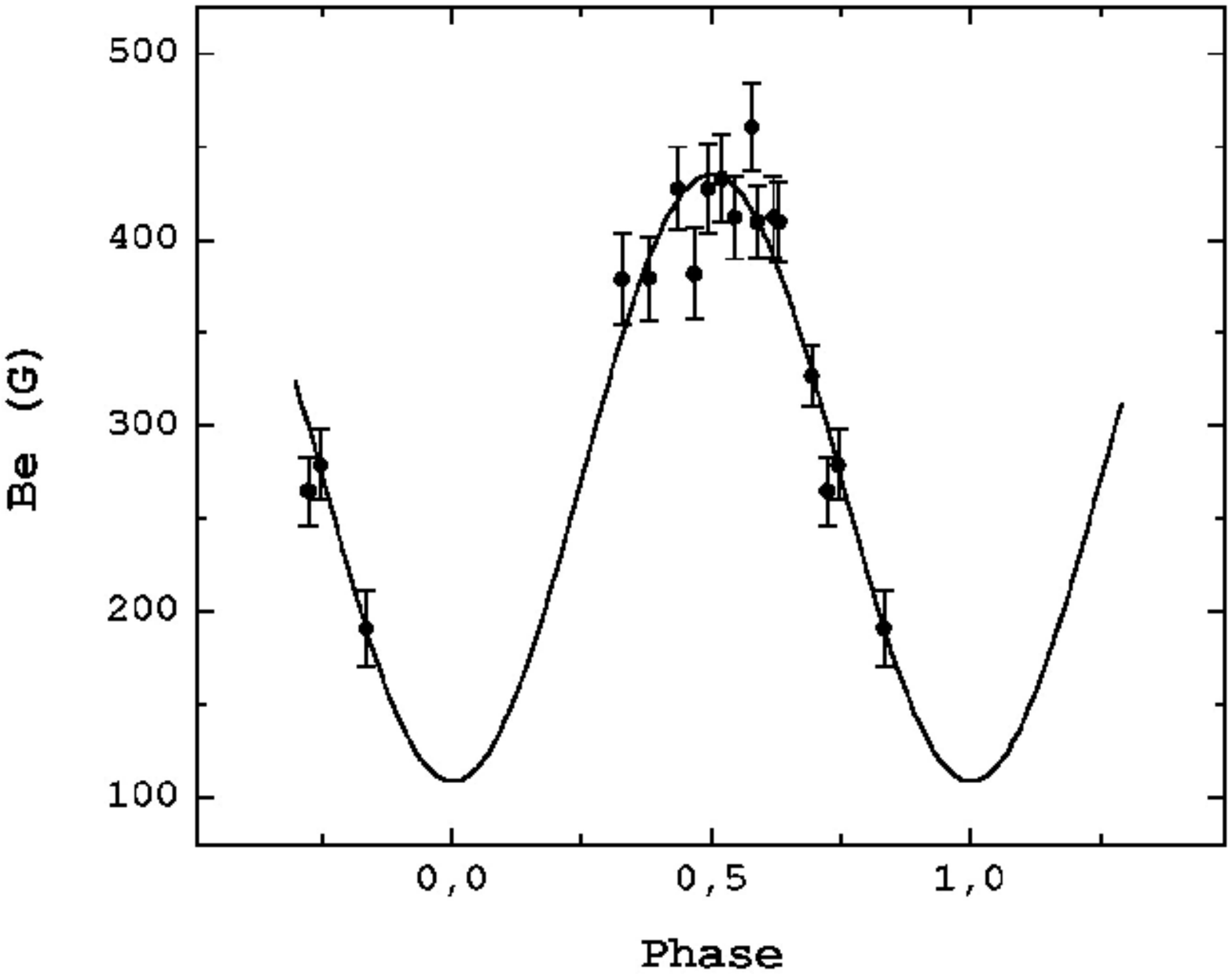}}
\vspace{-3.5mm}
\caption{ EQ Peg A }
\label{fig:fig383}
\end{figure}

\begin{figure}
\resizebox{0.98\hsize}{!}{\includegraphics{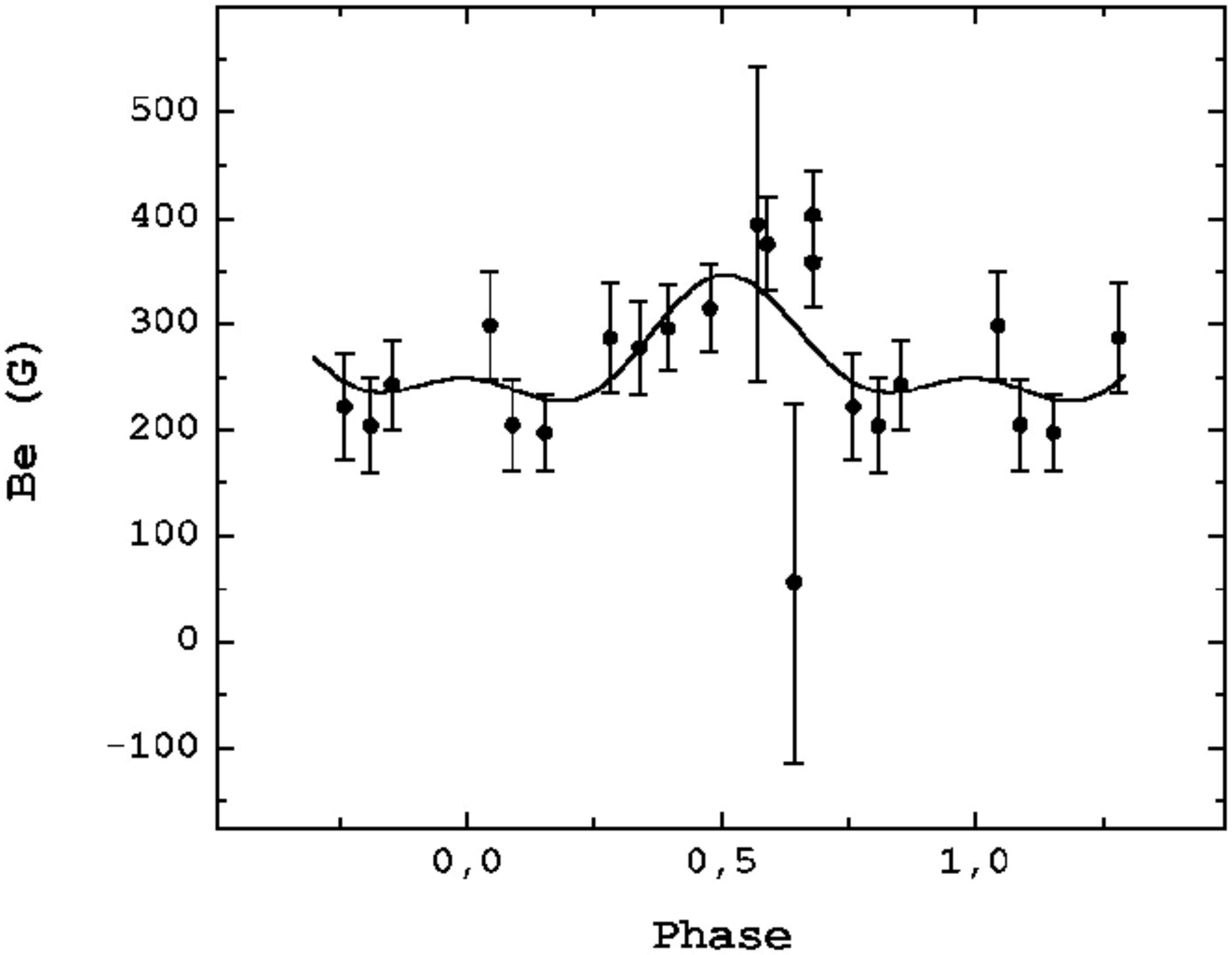}}
\vspace{-3.5mm}
\caption{ EQ Peg B }
\label{fig:fig384}
\end{figure}

\begin{figure}
\resizebox{0.98\hsize}{!}{\includegraphics{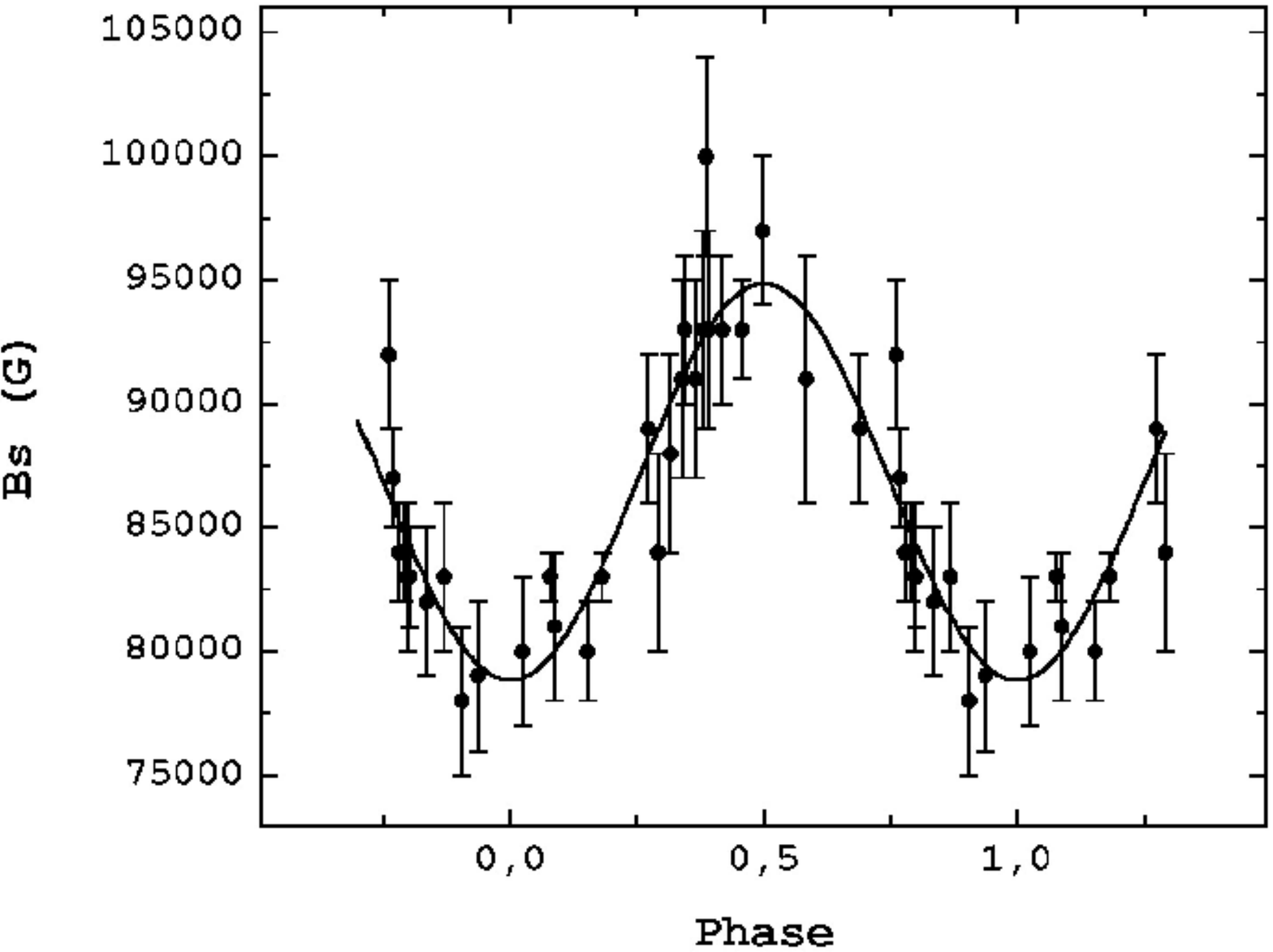}}
\vspace{-3.5mm}
\caption{ WD1953-011 }
\label{fig:fig385}
\end{figure}

\clearpage
\newpage

\begin{figure}
\resizebox{0.98\hsize}{!}{\includegraphics{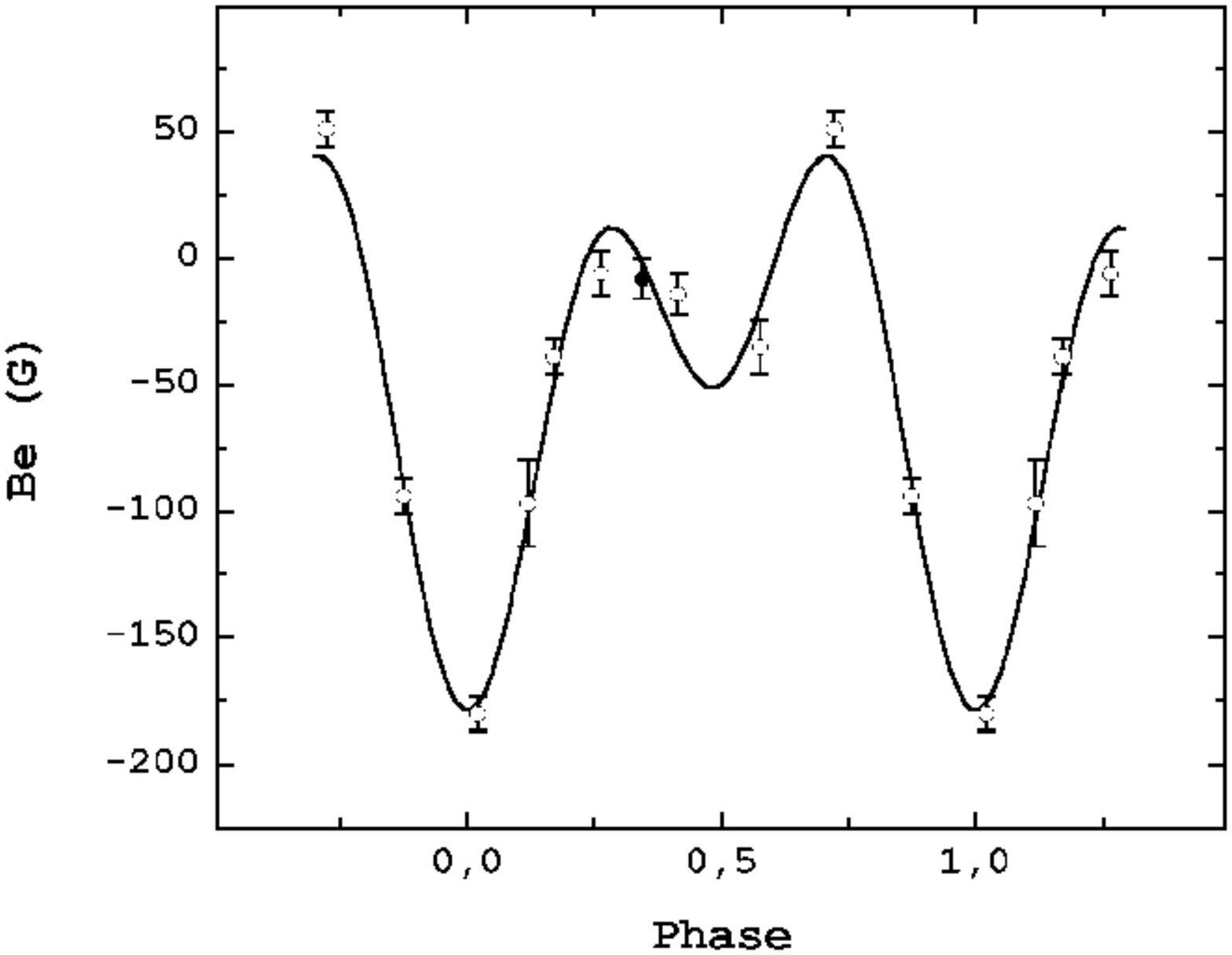}}
\vspace{-3.5mm}
\caption{ V2129 Oph (1) }
\label{fig:fig386}
\end{figure}

\begin{figure}
\resizebox{0.98\hsize}{!}{\includegraphics{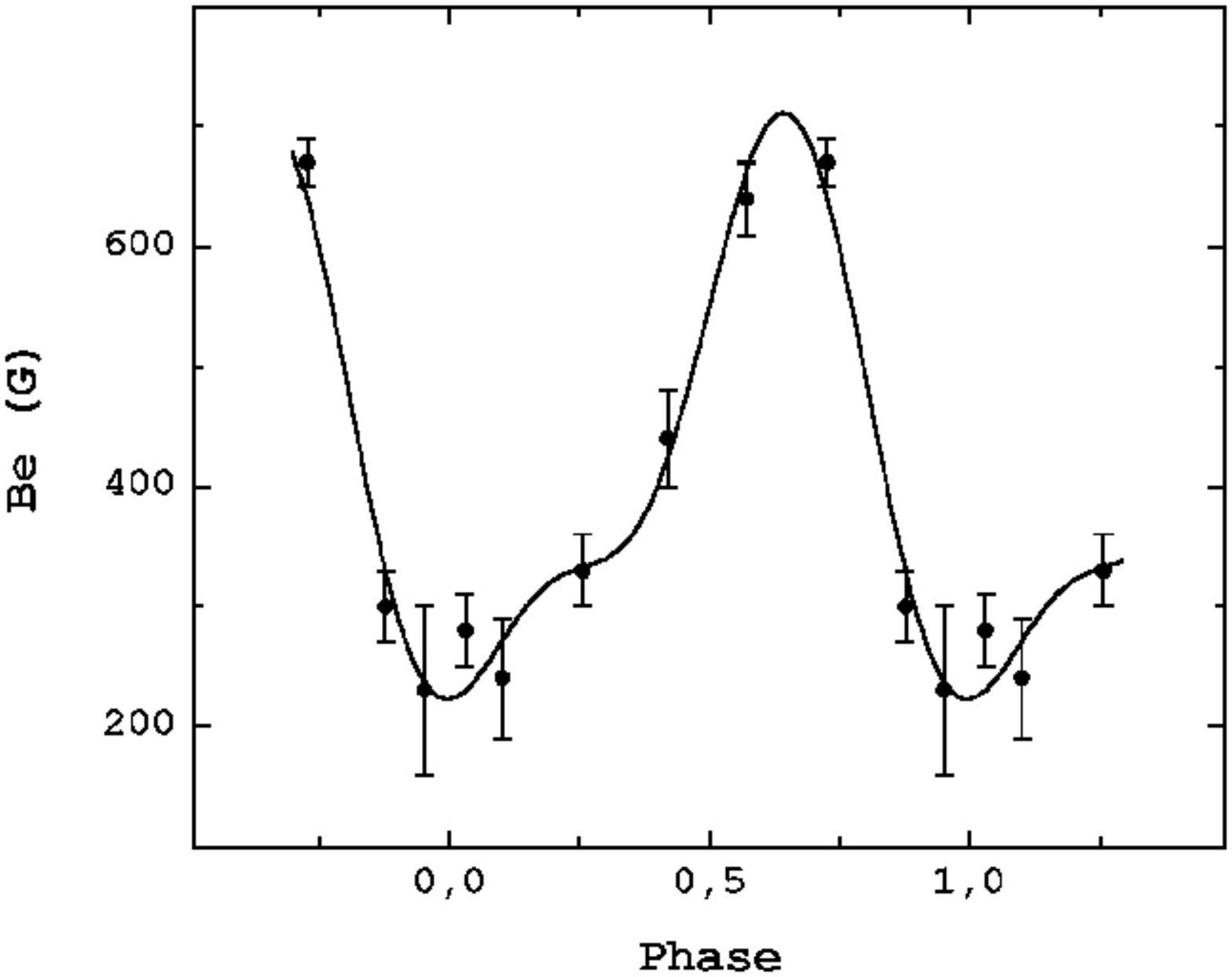}}
\vspace{-3.5mm}
\caption{ V2129 Oph (2) }
\label{fig:fig387}
\end{figure}

\begin{figure}
\resizebox{0.98\hsize}{!}{\includegraphics{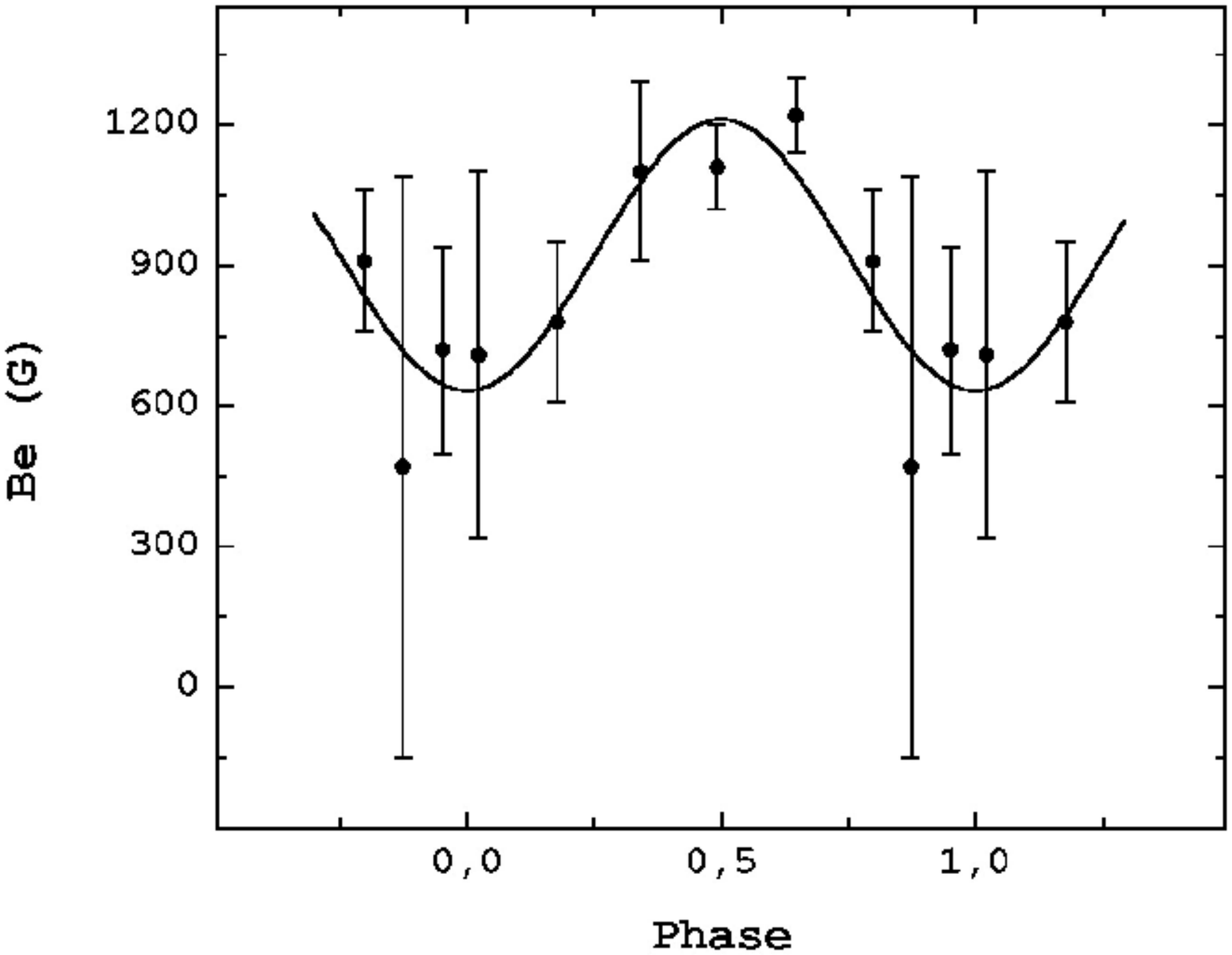}}
\vspace{-3.5mm}
\caption{ V2129 Oph (3) }
\label{fig:fig388}
\end{figure}

\begin{figure}
\resizebox{0.98\hsize}{!}{\includegraphics{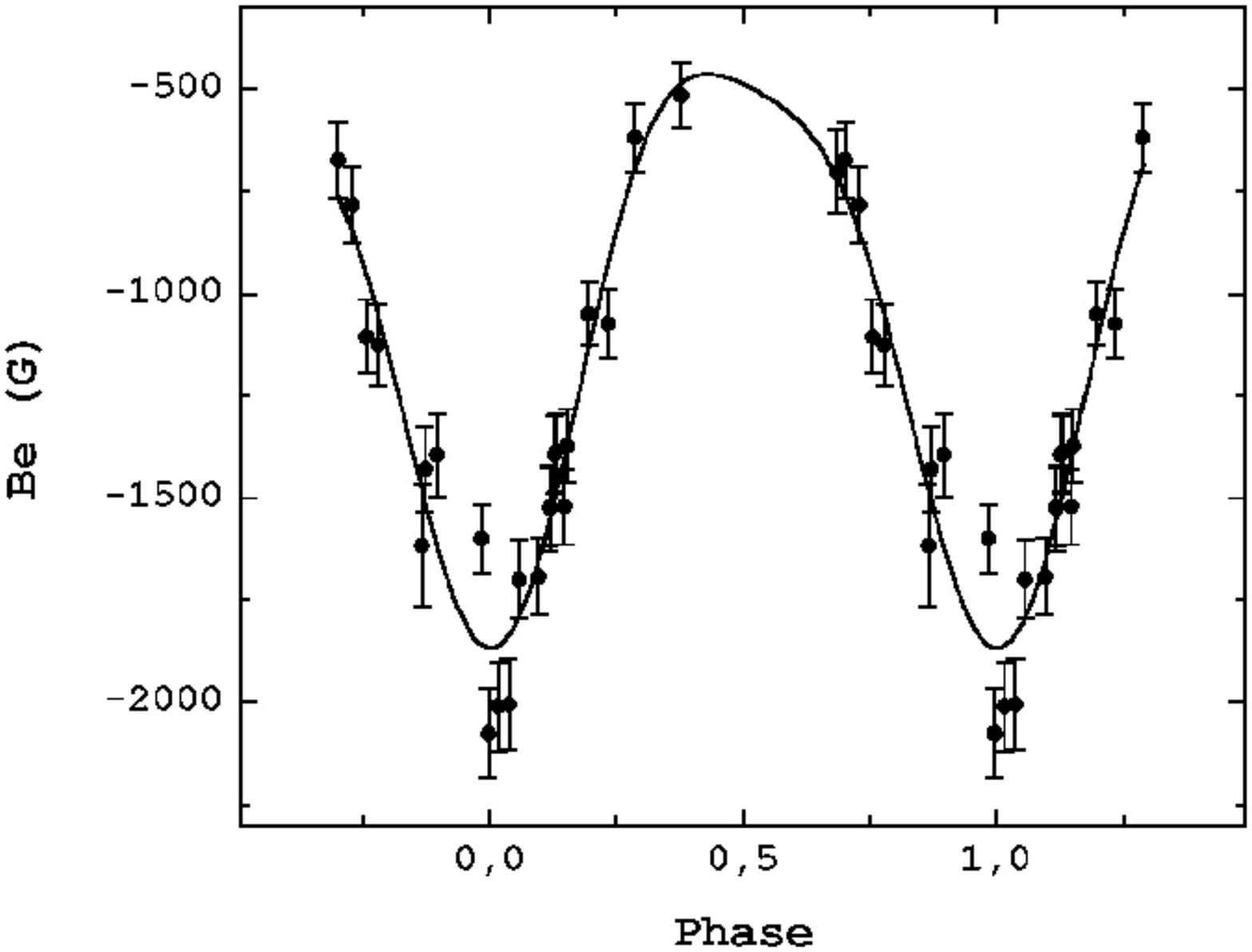}}
\vspace{-3.5mm}
\caption{ V388 Ori }
\label{fig:fig389}
\end{figure}

\begin{figure}
\resizebox{0.98\hsize}{!}{\includegraphics{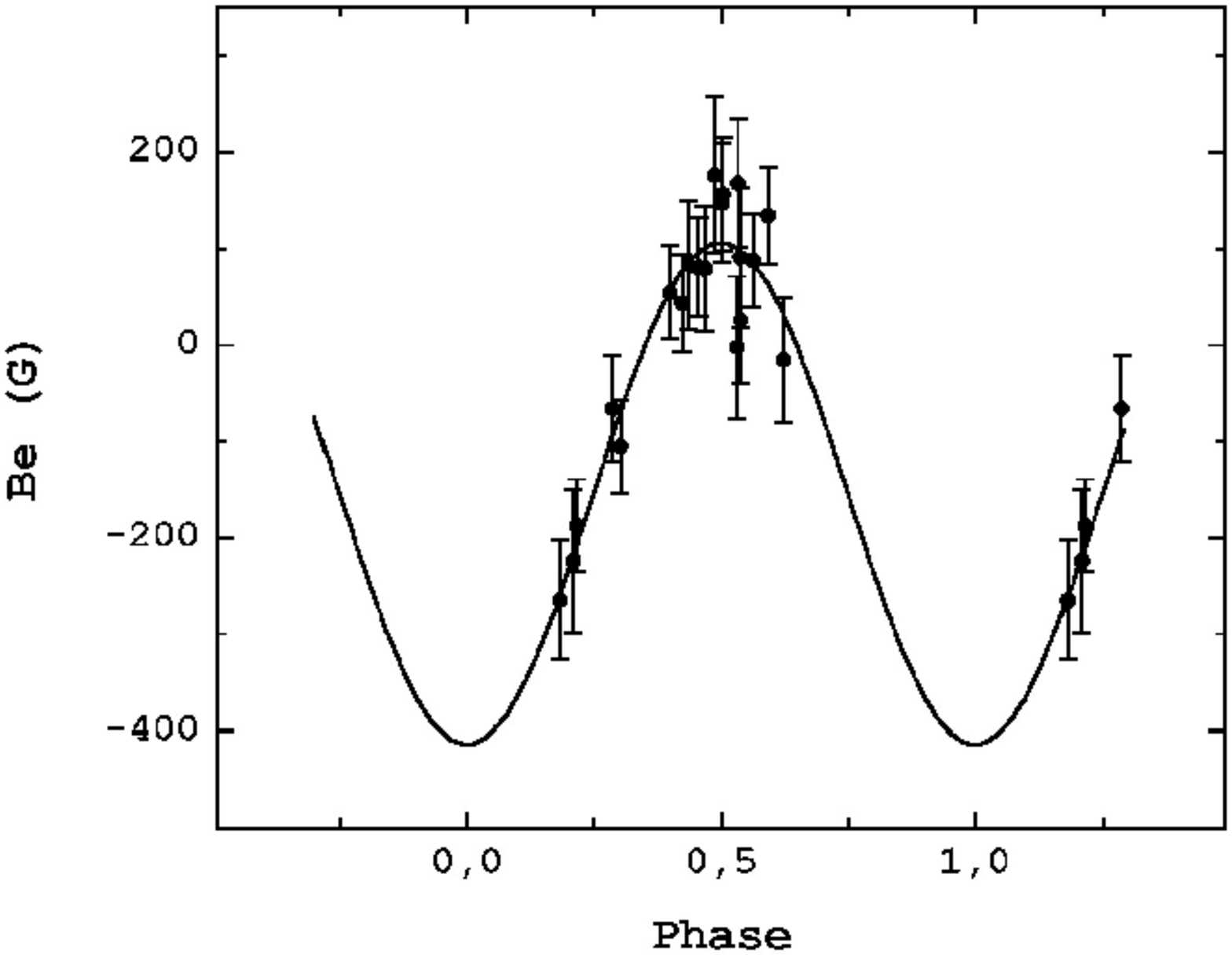}}
\vspace{-3.5mm}
\caption{ GL Vir }
\label{fig:fig390}
\end{figure}

\begin{figure}
\resizebox{0.98\hsize}{!}{\includegraphics{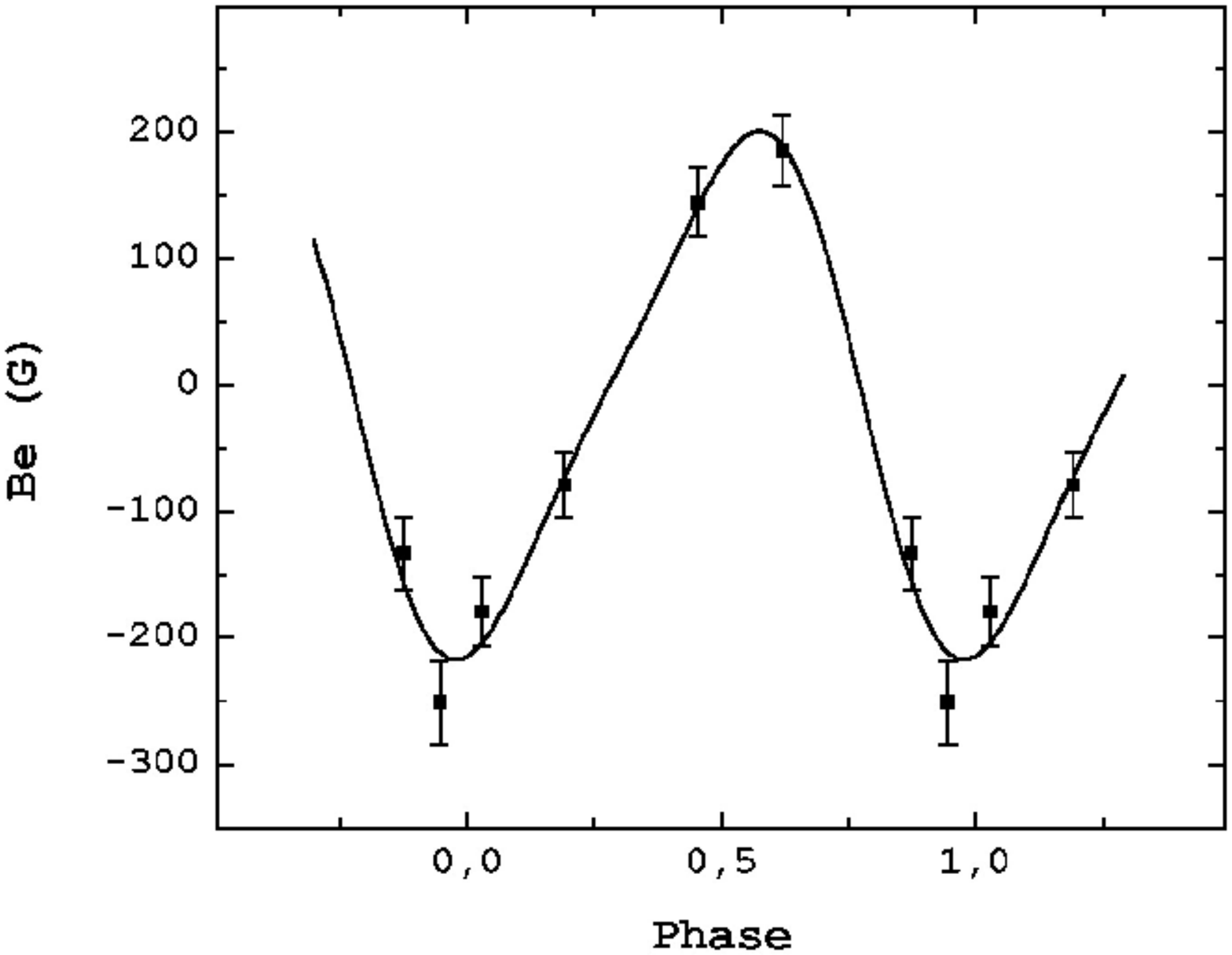}}
\vspace{-3.5mm}
\caption{ LHS 3495 (1) }
\label{fig:fig391}
\end{figure}

\clearpage
\newpage

\begin{figure}
\resizebox{0.98\hsize}{!}{\includegraphics{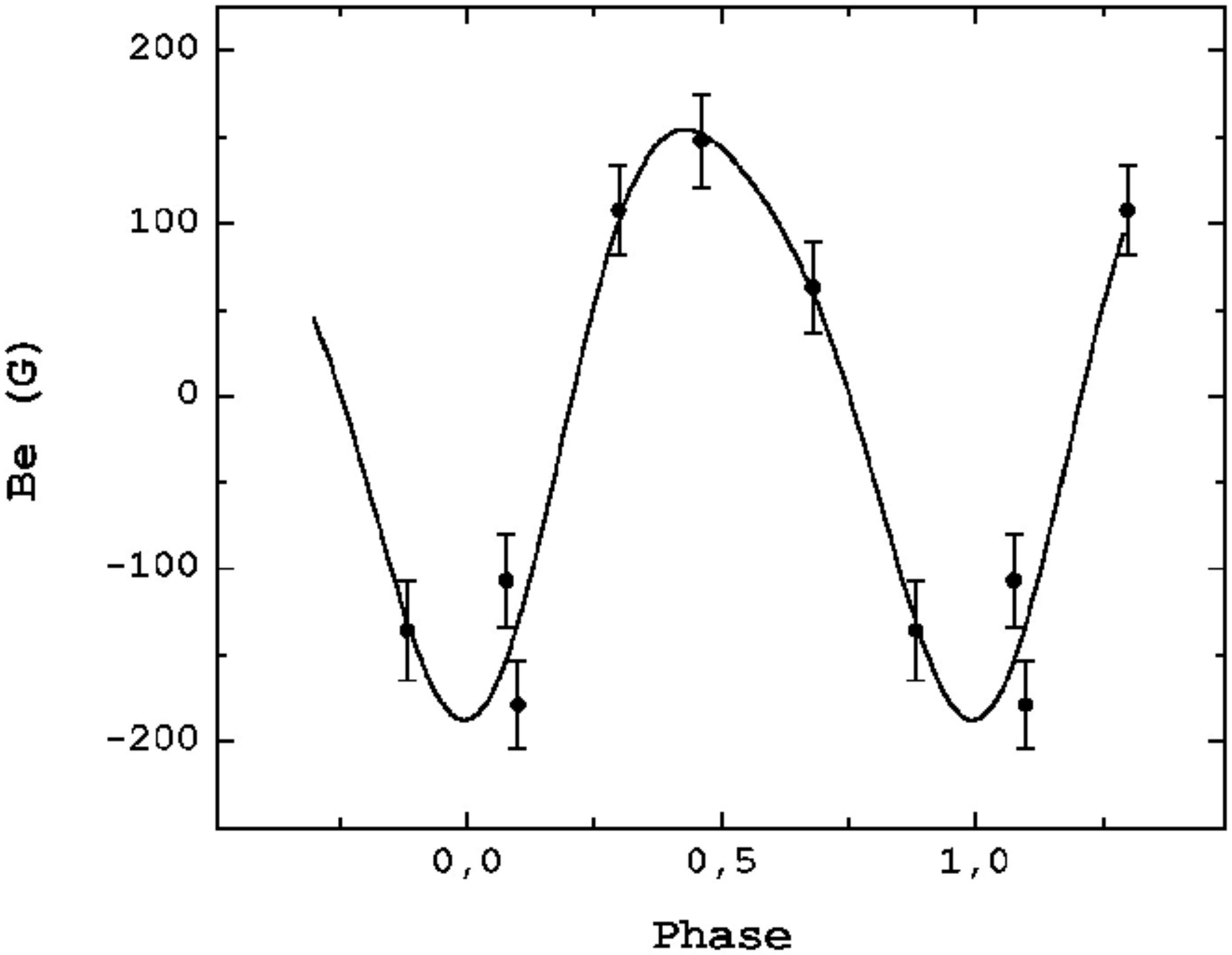}}
\vspace{-3.5mm}
\caption{ LHS 3495 (2) }
\label{fig:fig392}
\end{figure}

\begin{figure}
\resizebox{0.98\hsize}{!}{\includegraphics{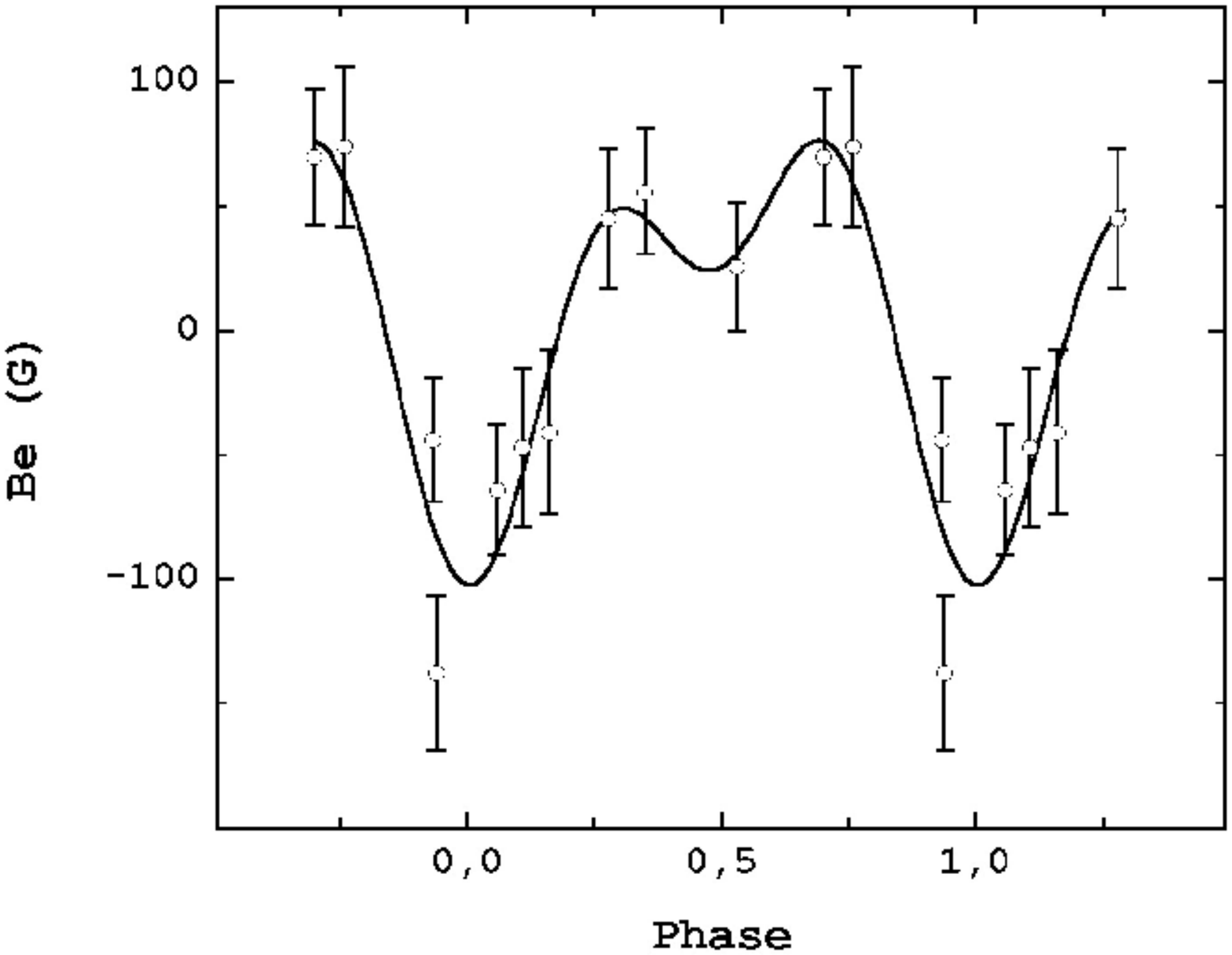}}
\vspace{-3.5mm}
\caption{ LHS 3495 (3) }
\label{fig:fig393}
\end{figure}

\begin{figure}
\resizebox{0.98\hsize}{!}{\includegraphics{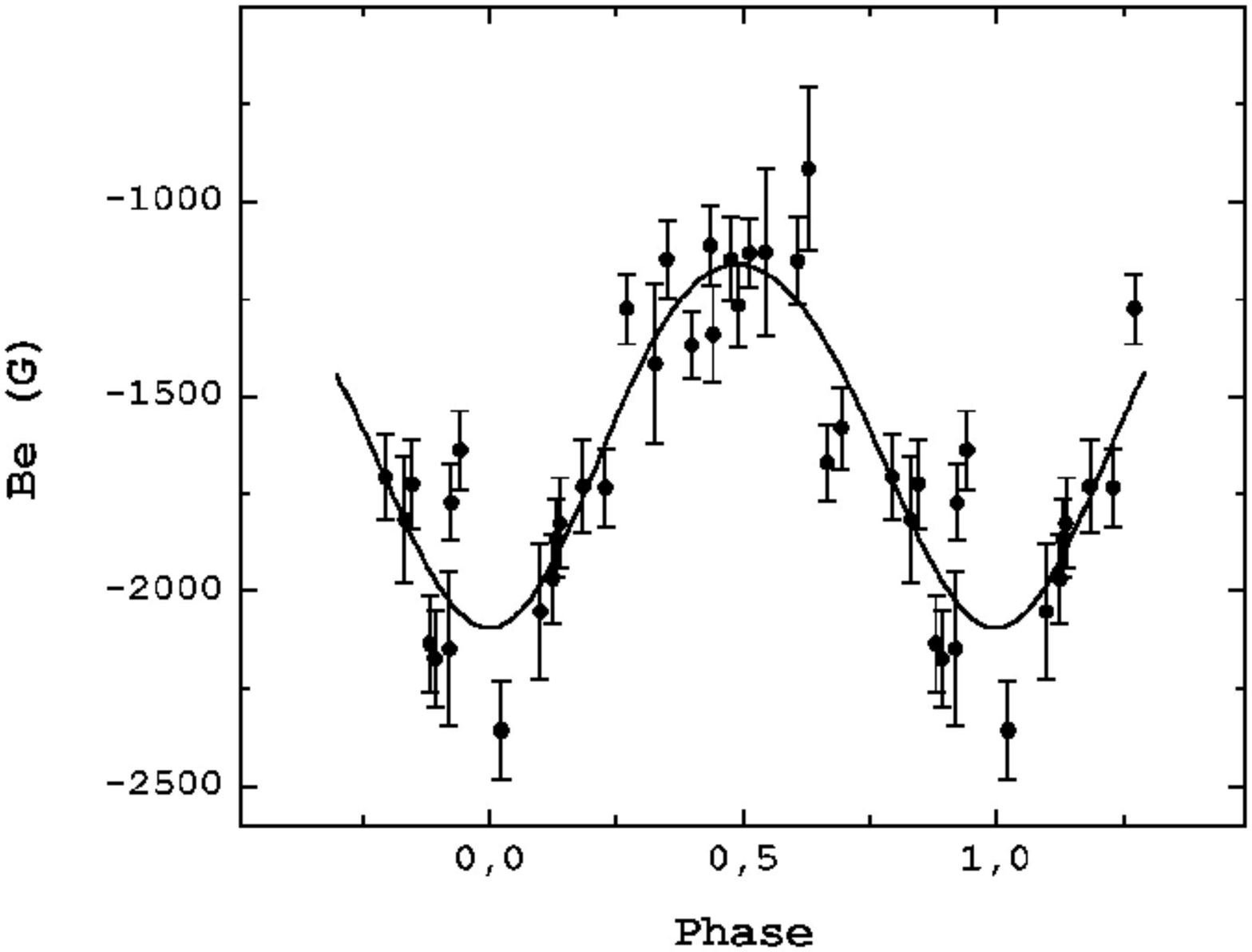}}
\vspace{-3.5mm}
\caption{ WX UMa }
\label{fig:fig394}
\end{figure}

\begin{figure}
\resizebox{0.98\hsize}{!}{\includegraphics{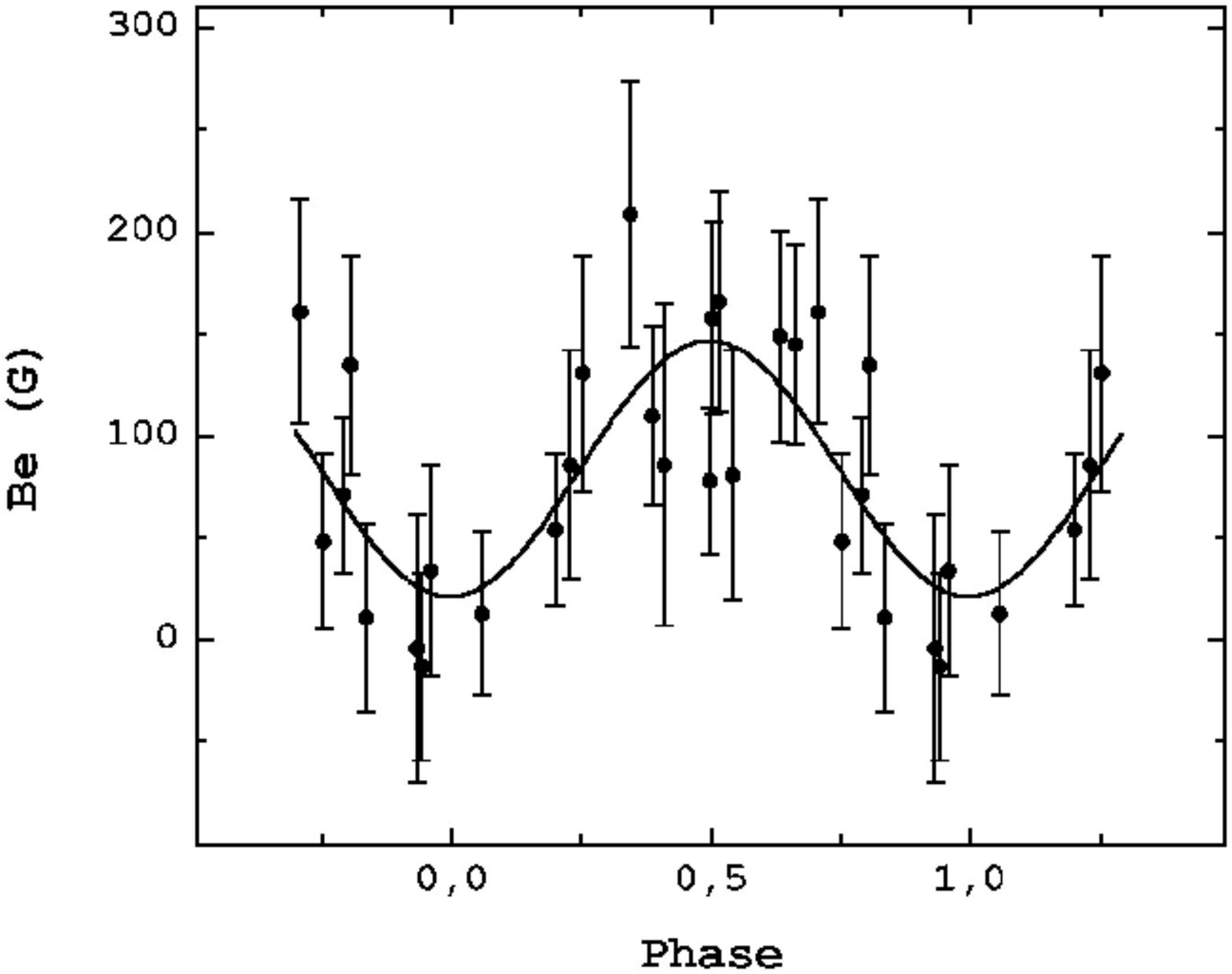}}
\vspace{-3.5mm}
\caption{ DX Cnc }
\label{fig:fig395}
\end{figure}

\begin{figure}
\resizebox{0.98\hsize}{!}{\includegraphics{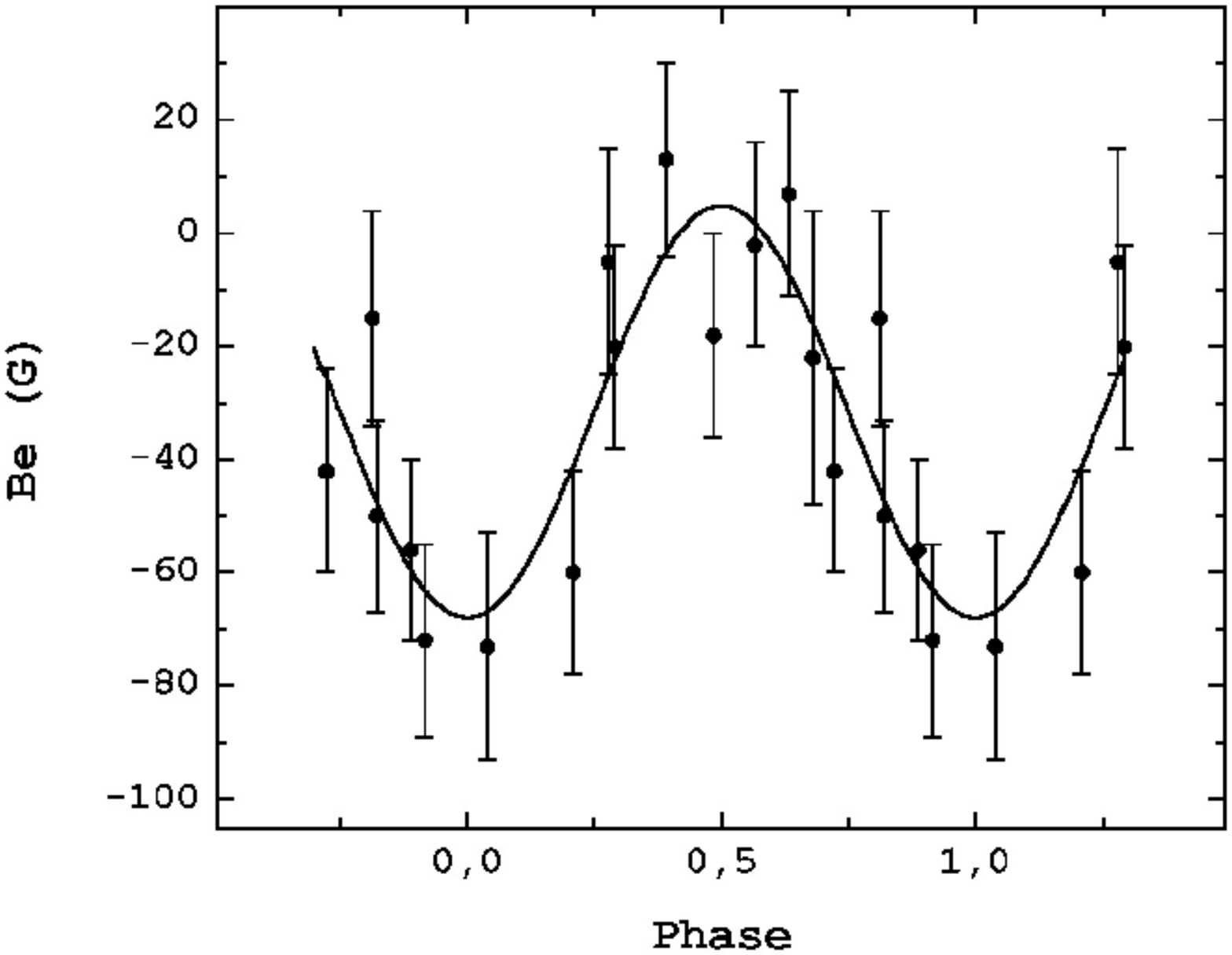}}
\vspace{-3.5mm}
\caption{ LHS 292 }
\label{fig:fig396}
\end{figure}

\begin{figure}
\resizebox{0.98\hsize}{!}{\includegraphics{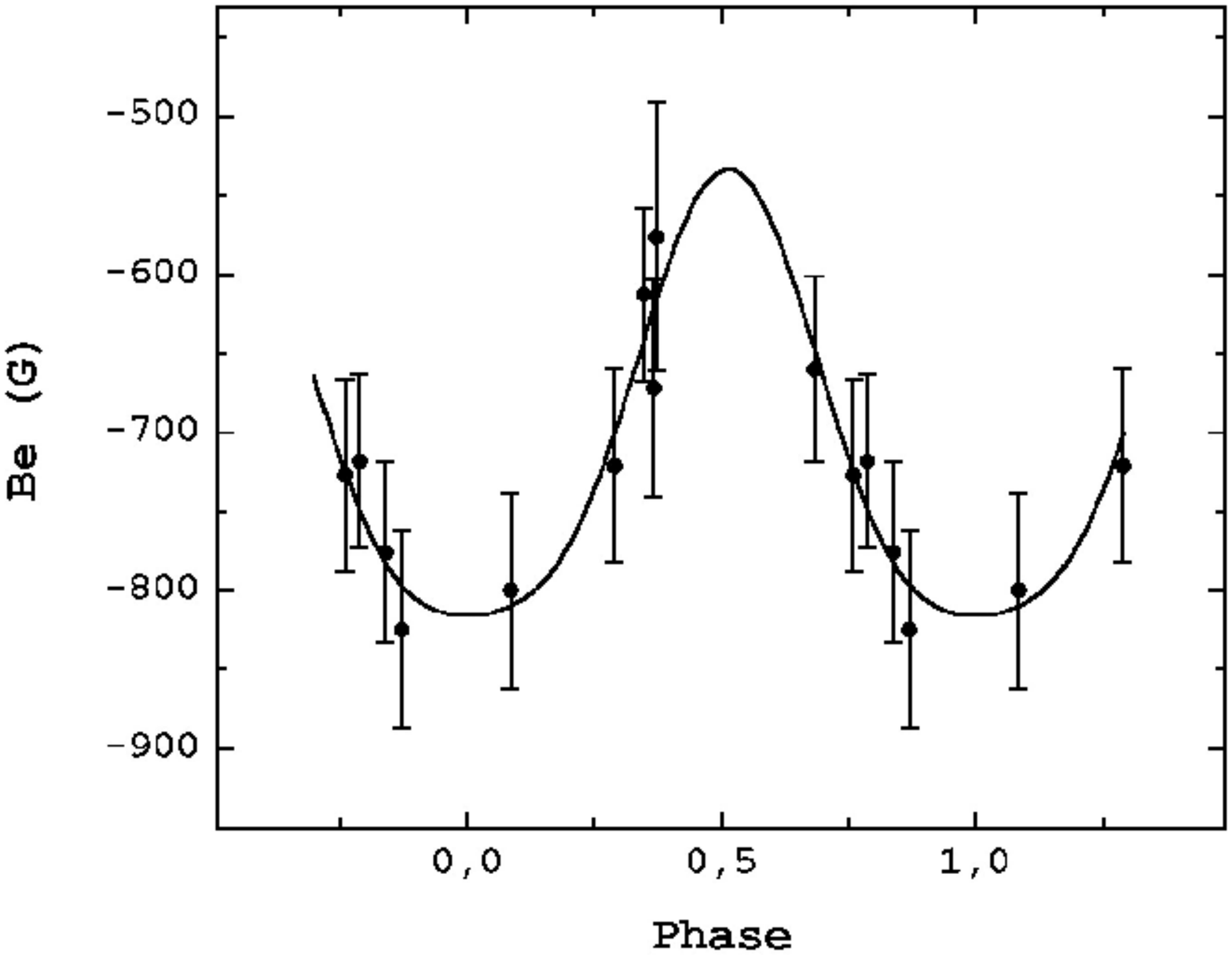}}
\vspace{-3.5mm}
\caption{ GJ 1154 A }
\label{fig:fig397}
\end{figure}

\clearpage
\newpage

\begin{figure}
\resizebox{0.98\hsize}{!}{\includegraphics{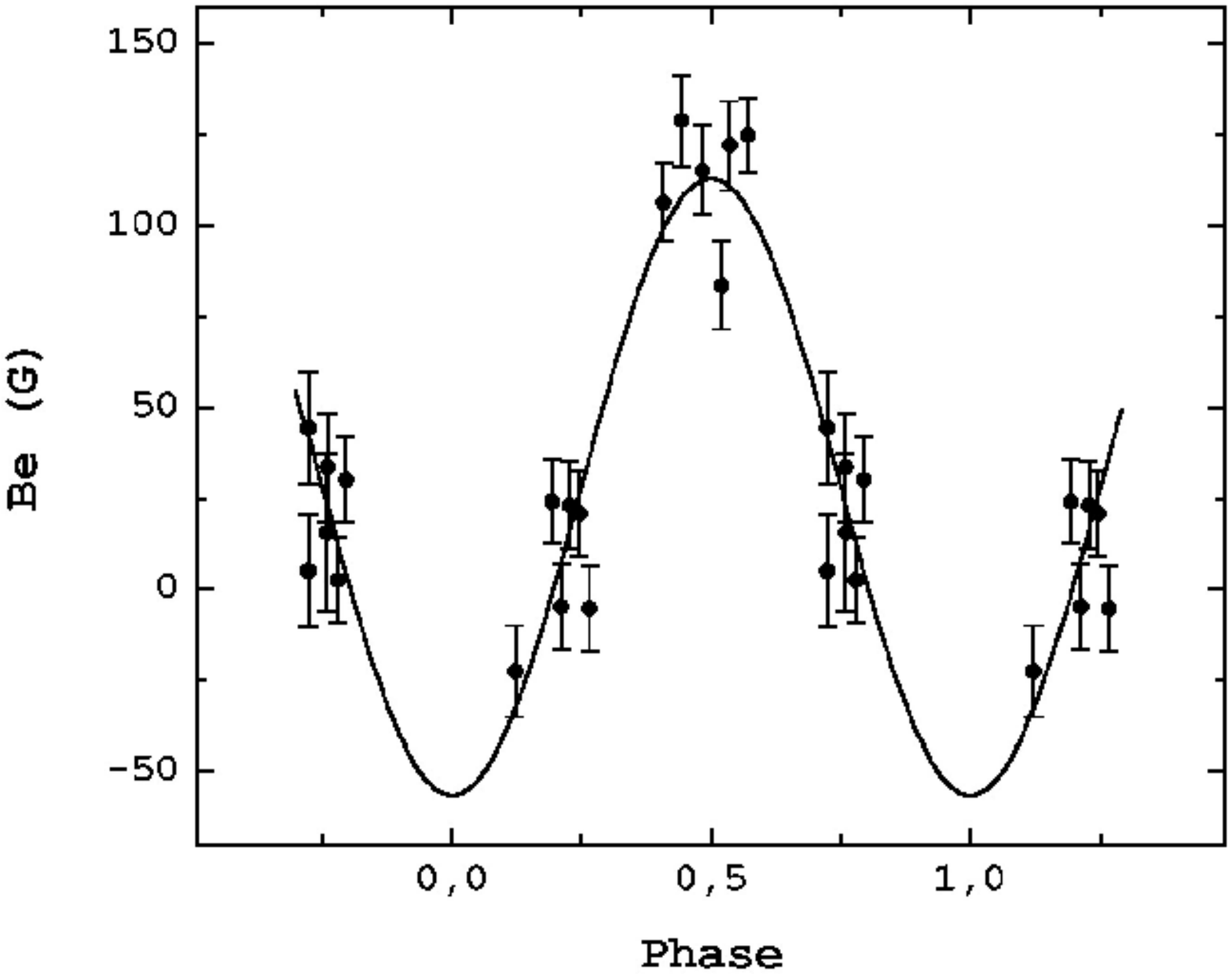}}
\vspace{-3.5mm}
\caption{ GJ 1224 }
\label{fig:fig398}
\end{figure}

\begin{figure}
\resizebox{0.98\hsize}{!}{\includegraphics{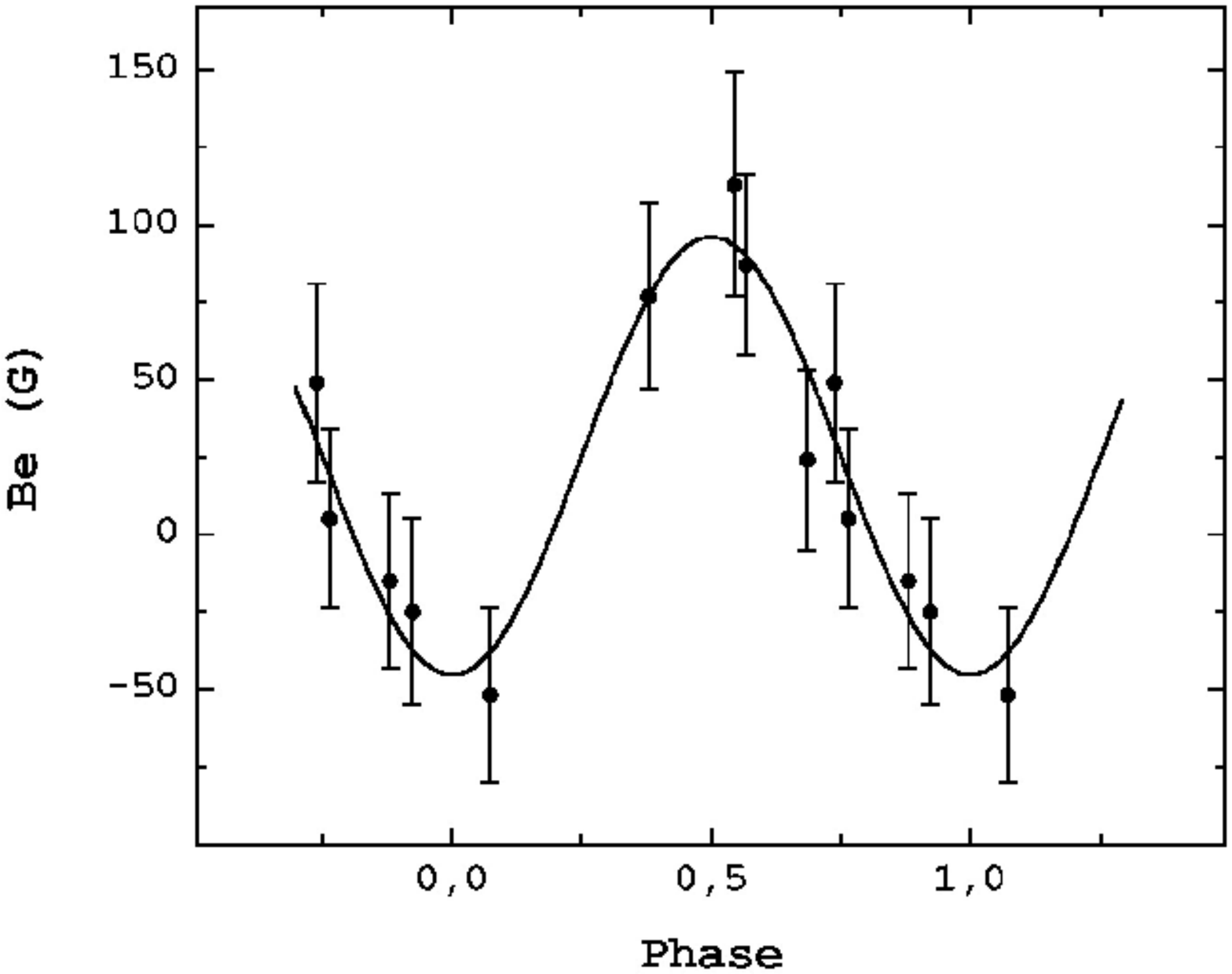}}
\vspace{-3.5mm}
\caption{ GJ 644 C }
\label{fig:fig399}
\end{figure}

\begin{figure}
\resizebox{0.98\hsize}{!}{\includegraphics{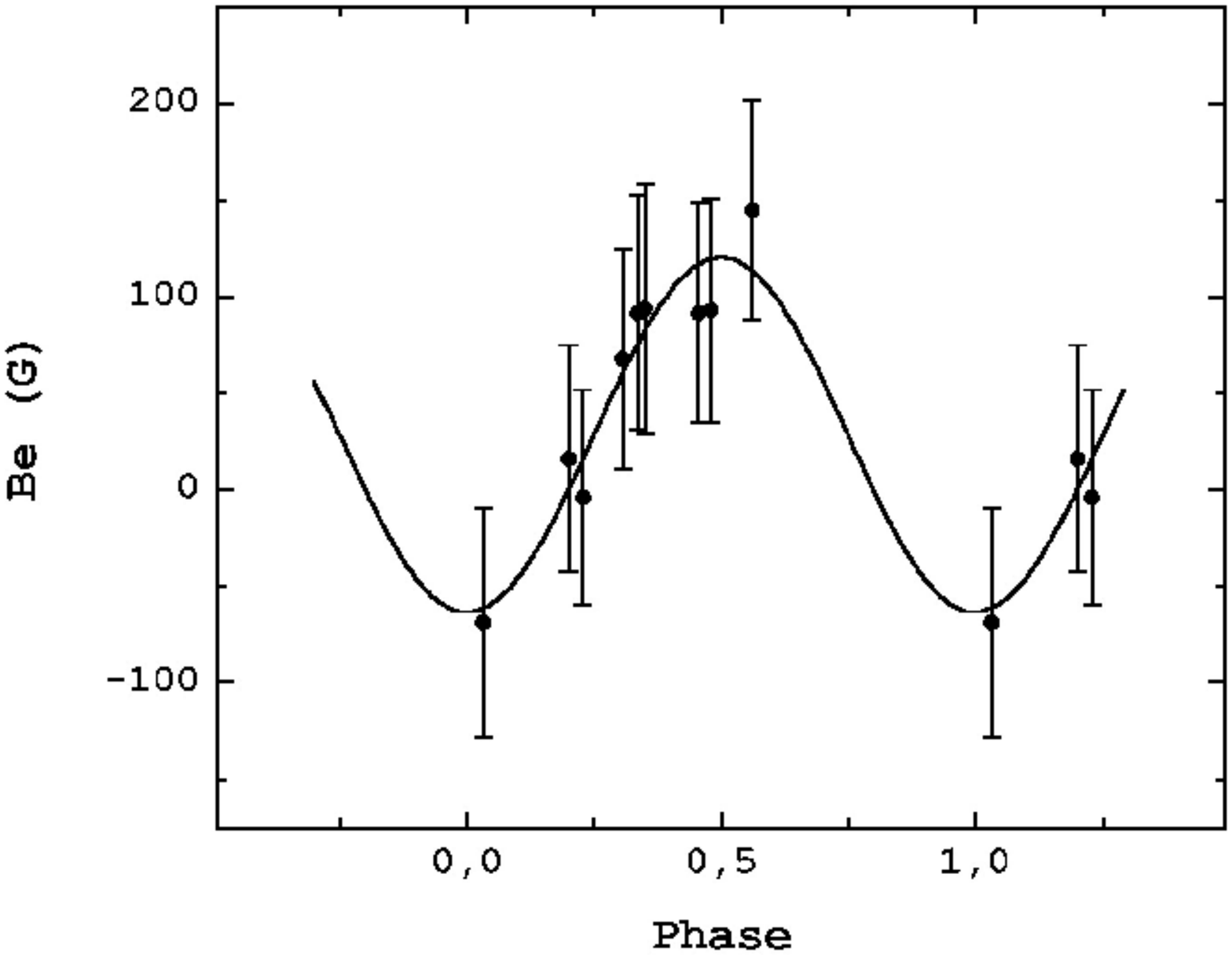}}
\vspace{-3.5mm}
\caption{ V1298 Aql }
\label{fig:fig400}
\end{figure}

\begin{figure}
\resizebox{0.98\hsize}{!}{\includegraphics{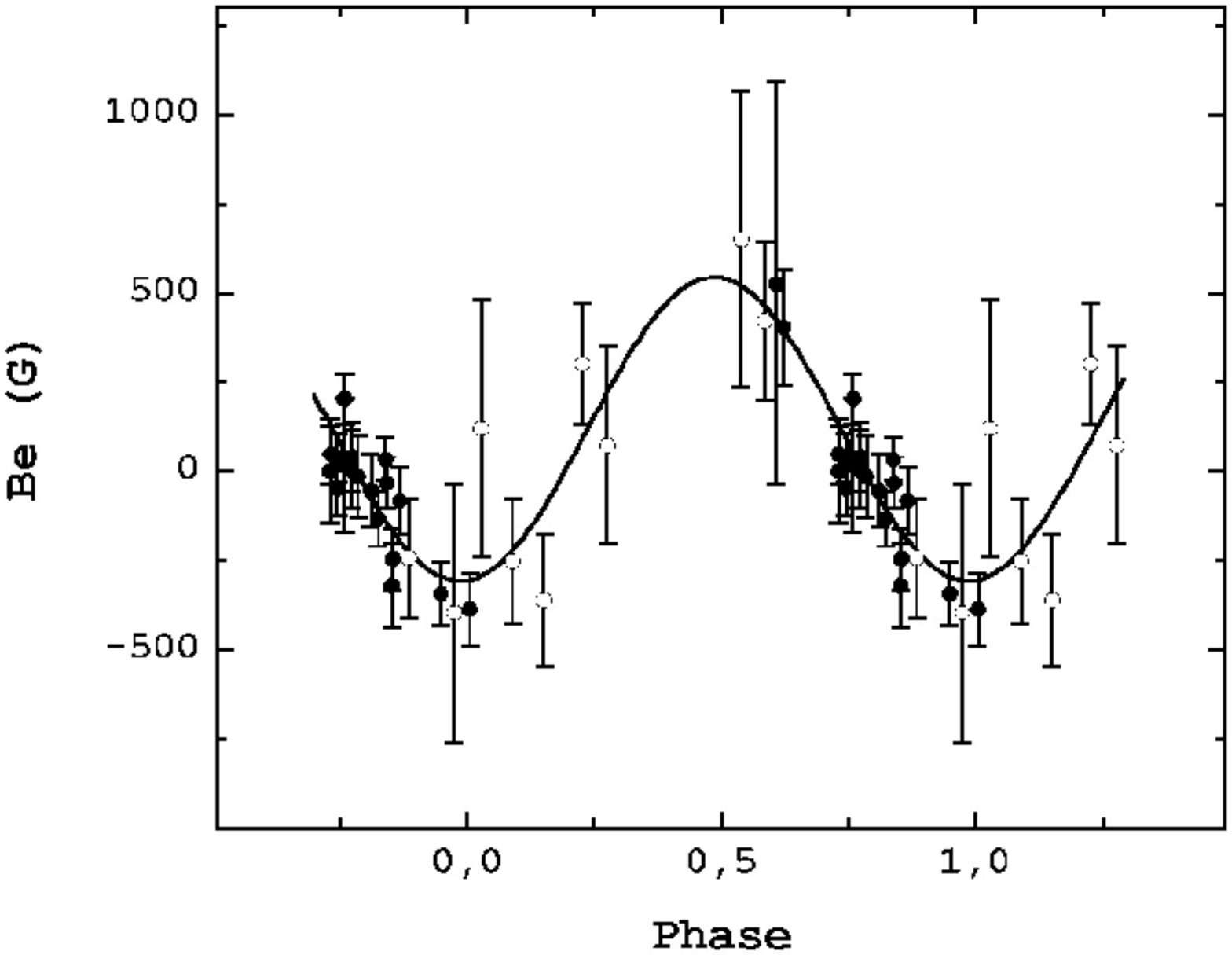}}
\vspace{-3.5mm}
\caption{ CPD-28 2561 (1) }
\label{fig:fig401}
\end{figure}

\begin{figure}
\resizebox{0.98\hsize}{!}{\includegraphics{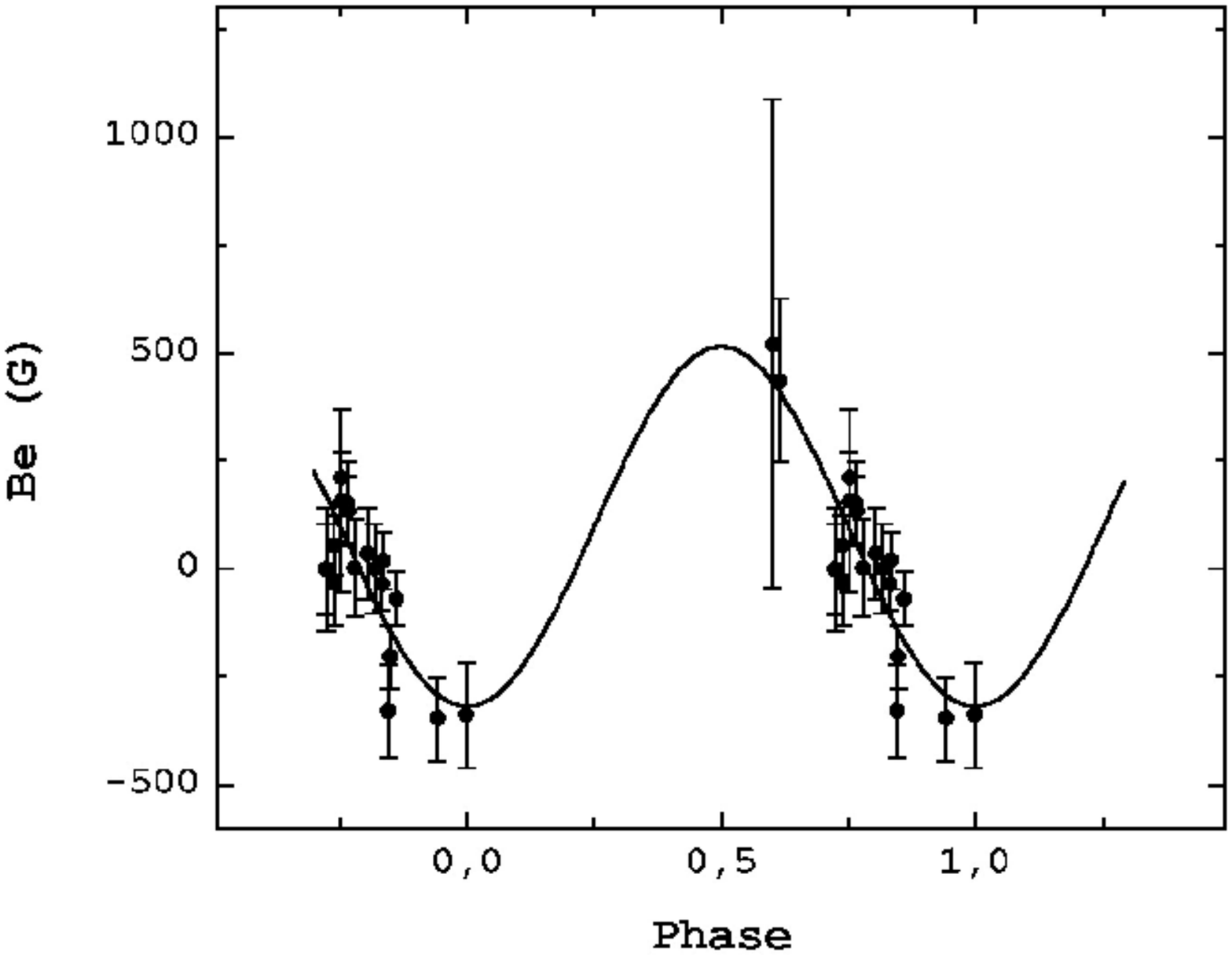}}
\vspace{-3.5mm}
\caption{ CPD-28 2561 (2) }
\label{fig:fig402}
\end{figure}

\begin{figure}
\resizebox{0.98\hsize}{!}{\includegraphics{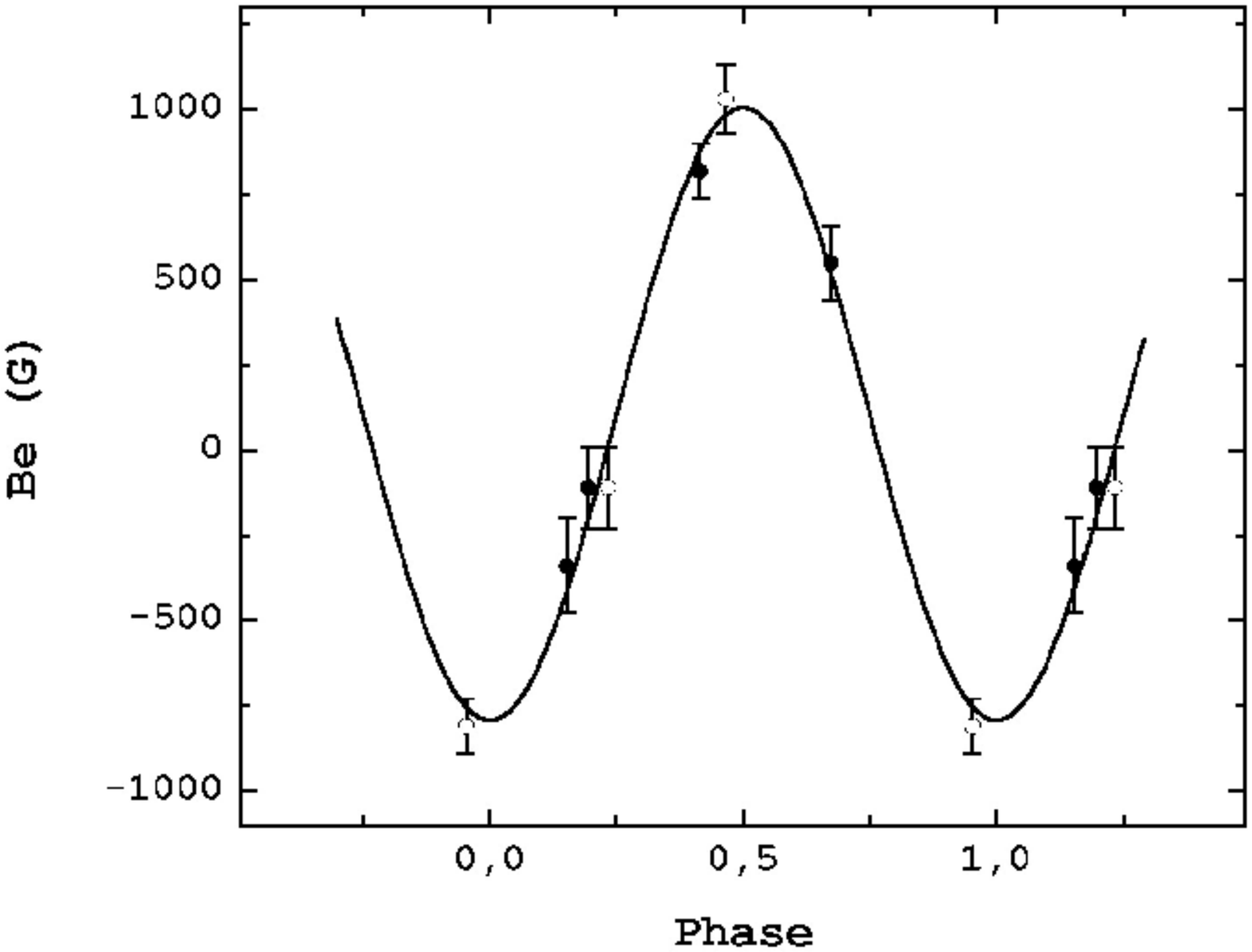}}
\vspace{-3.5mm}
\caption{ BD+53 1183 }
\label{fig:fig368}
\end{figure}

\clearpage
\newpage

\begin{figure}
\resizebox{0.98\hsize}{!}{\includegraphics{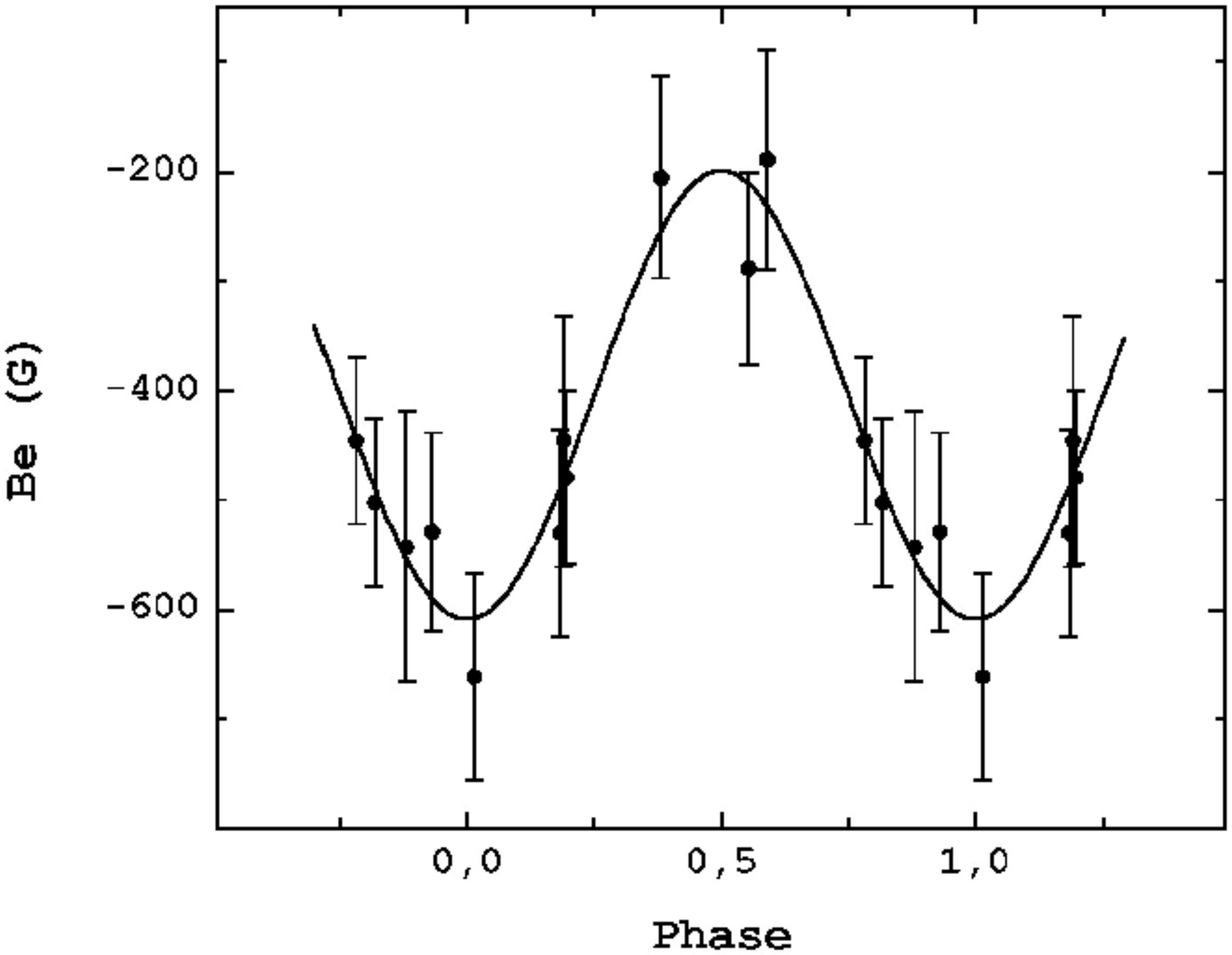}}
\vspace{-3.5mm}
\caption{ Tr16-22 }
\label{fig:fig403}
\end{figure}

\begin{figure}
\resizebox{0.98\hsize}{!}{\includegraphics{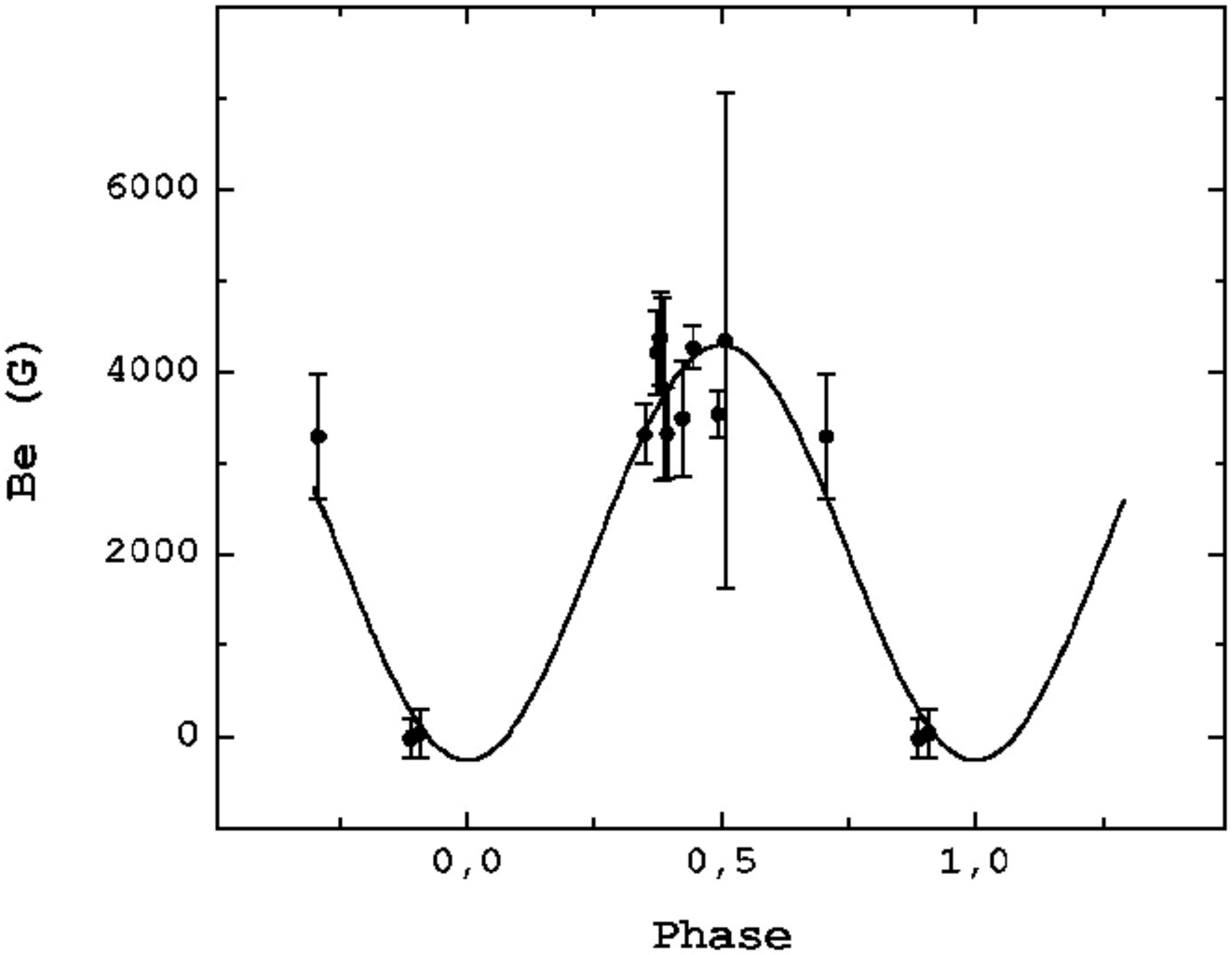}}
\vspace{-3.5mm}
\caption{ NGC1624-2 }
\label{fig:fig404}
\end{figure}

\begin{figure}
\resizebox{0.98\hsize}{!}{\includegraphics{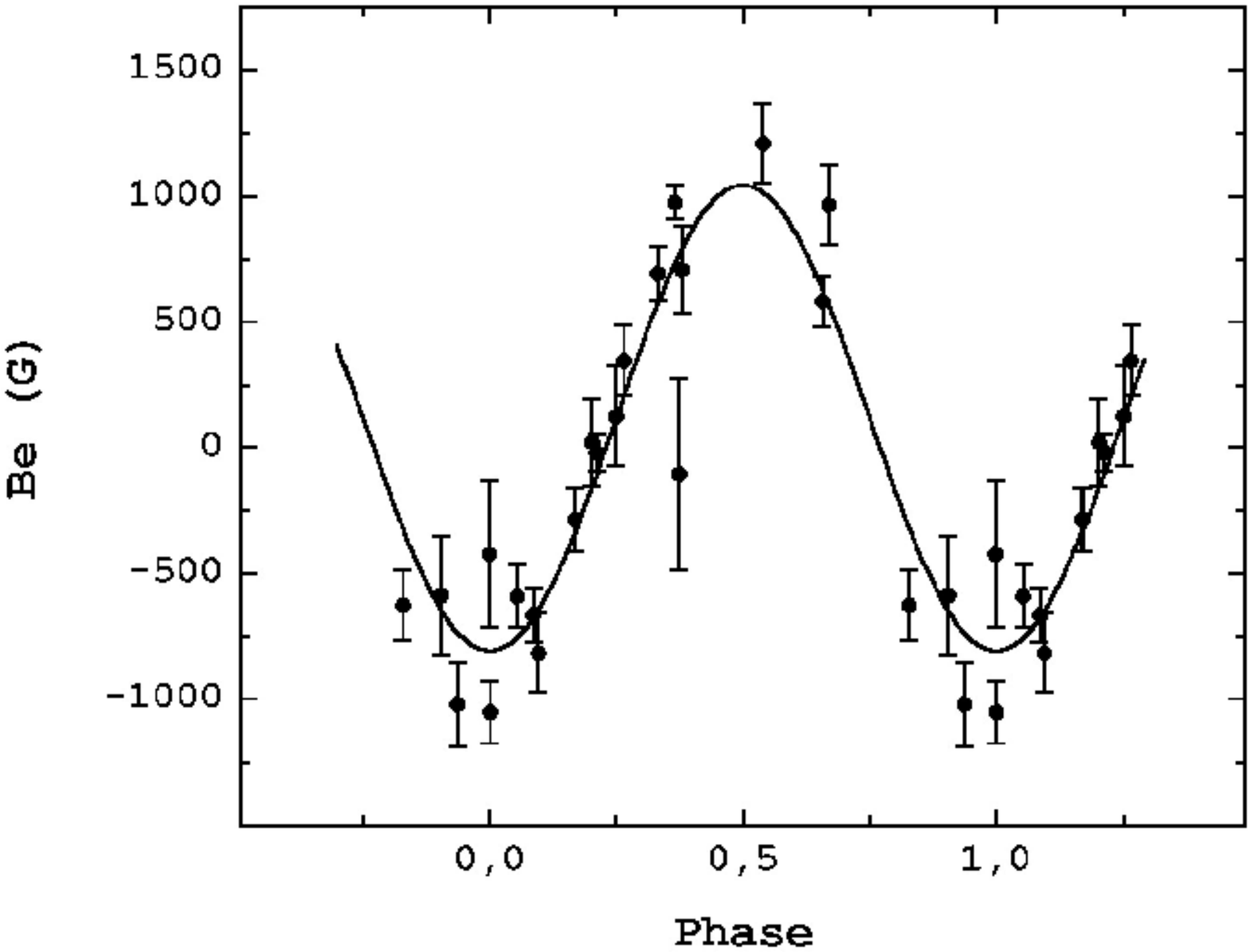}}
\vspace{-3.5mm}
\caption{ CPD-57 3509 (1) }
\label{fig:fig405}
\end{figure}

\begin{figure}
\resizebox{0.98\hsize}{!}{\includegraphics{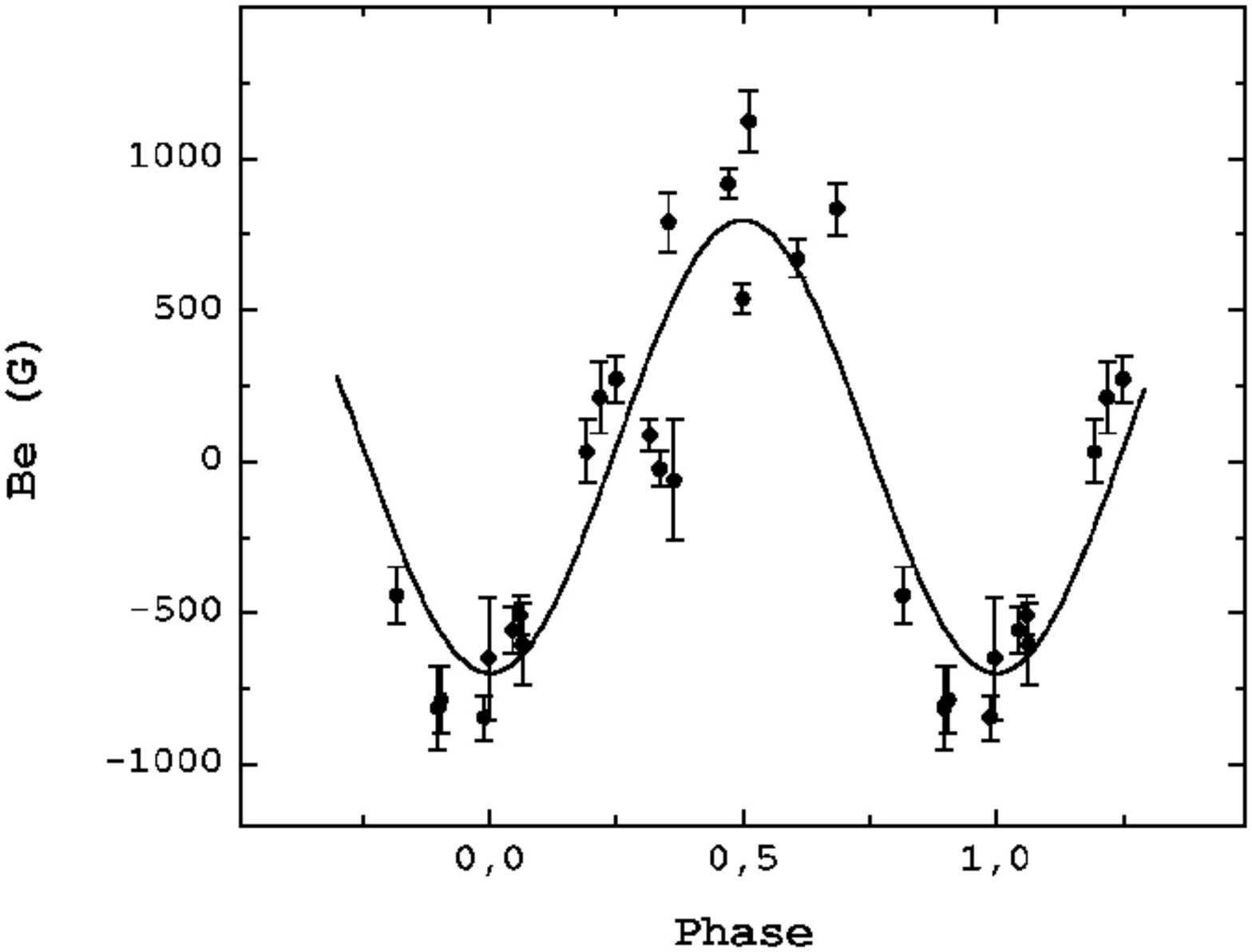}}
\vspace{-3.5mm}
\caption{ CPD-57 3509 (2) }
\label{fig:fig406}
\end{figure}

\begin{figure}
\resizebox{0.98\hsize}{!}{\includegraphics{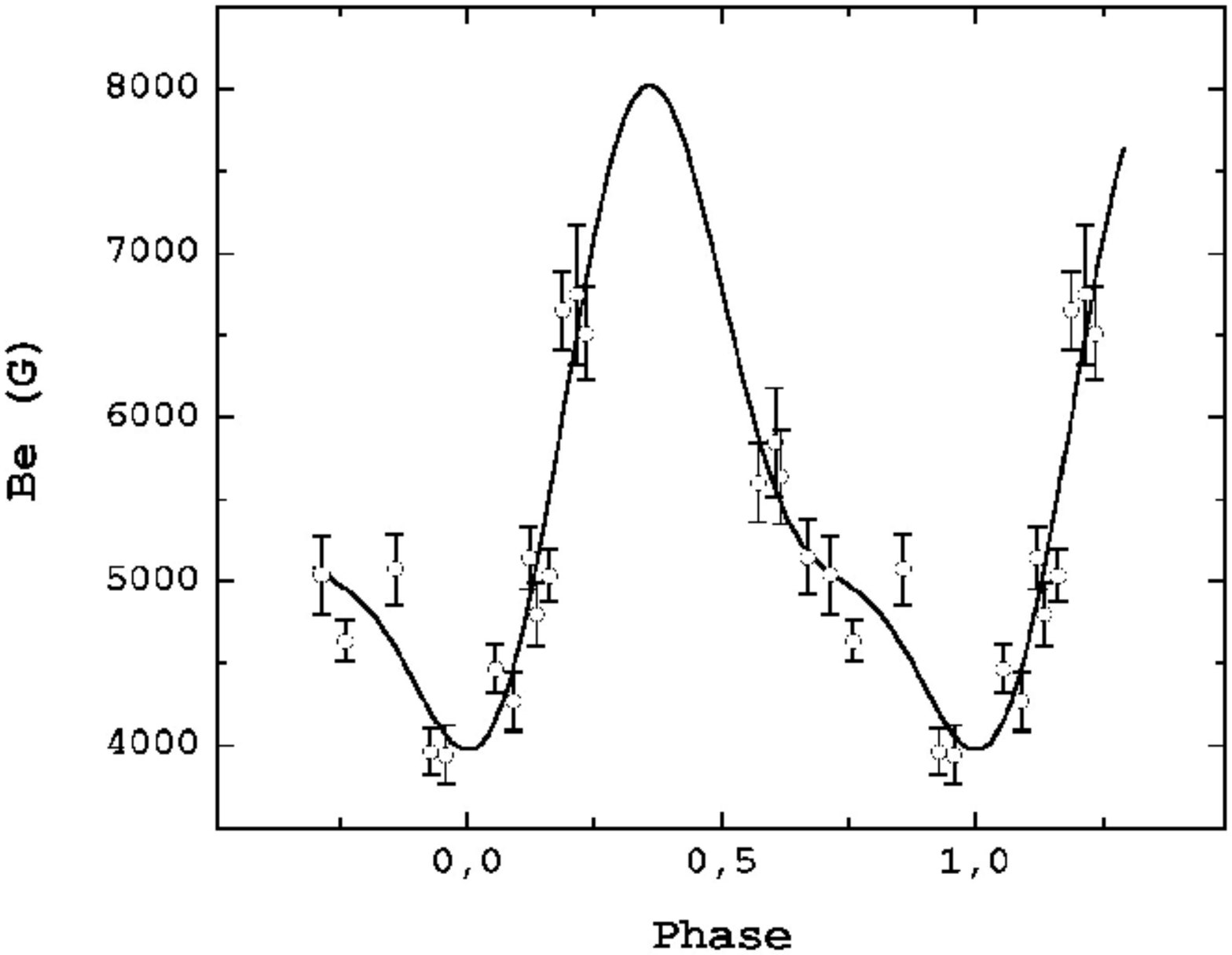}}
\vspace{-3.5mm}
\caption{ CPD-62 2124 (1) }
\label{fig:fig407}
\end{figure}

\begin{figure}
\resizebox{0.98\hsize}{!}{\includegraphics{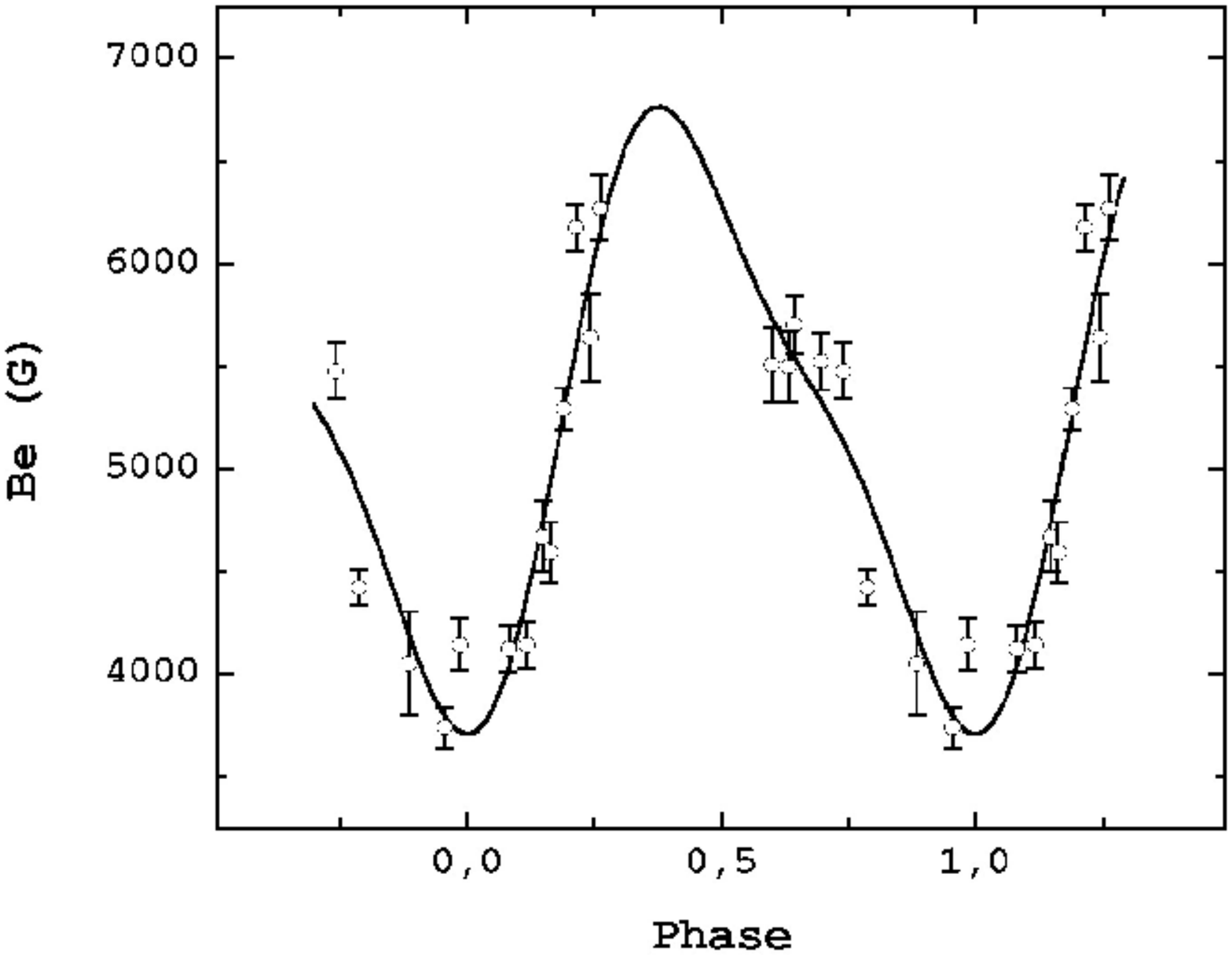}}
\vspace{-3.5mm}
\caption{ CPD-62 2124 (2) }
\label{fig:fig408}
\end{figure}

\clearpage
\newpage

\begin{figure}
\resizebox{0.98\hsize}{!}{\includegraphics{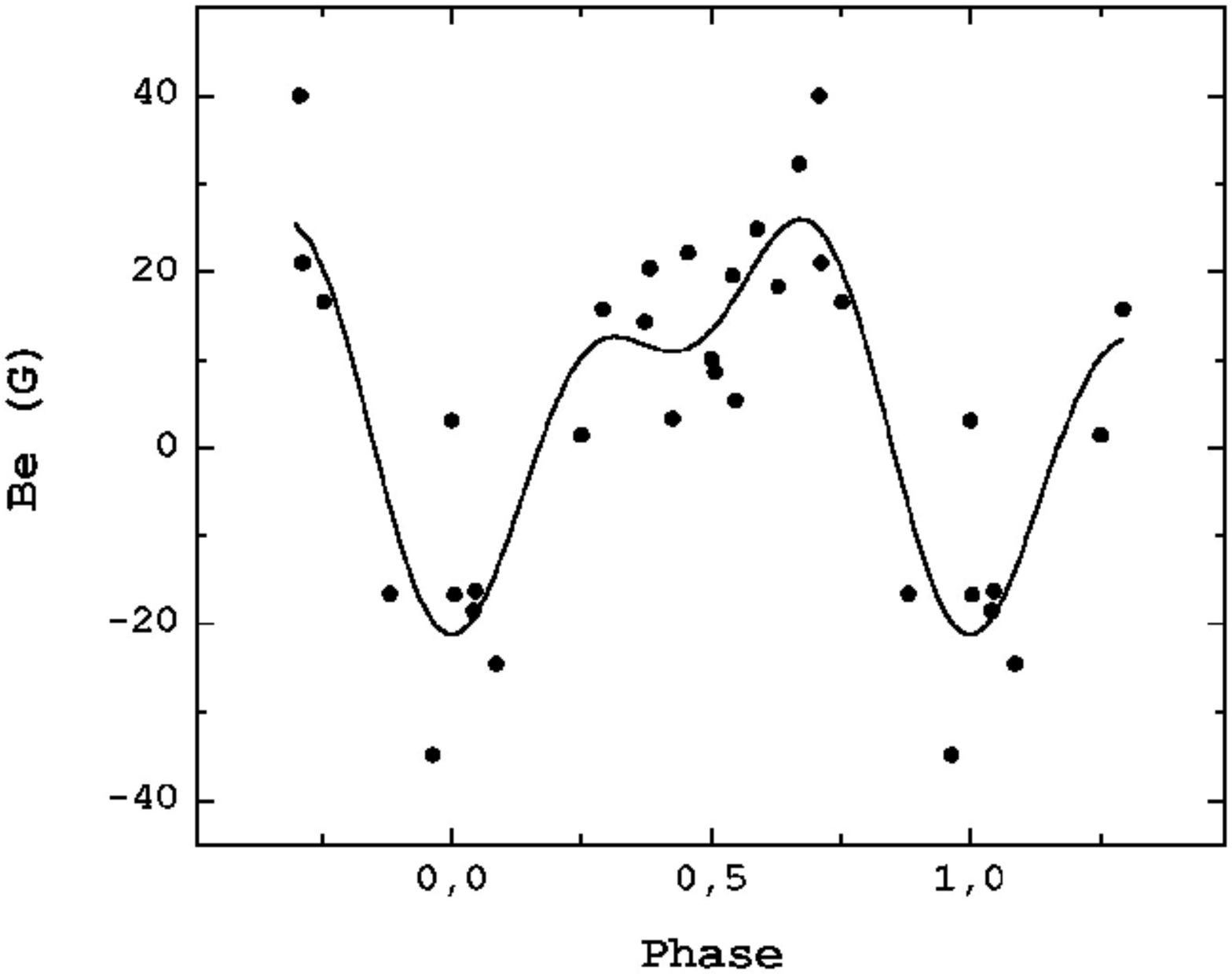}}
\vspace{-3.5mm}
\caption{ CD-51 6859 }
\label{fig:fig409}
\end{figure}

\begin{figure}
\resizebox{0.98\hsize}{!}{\includegraphics{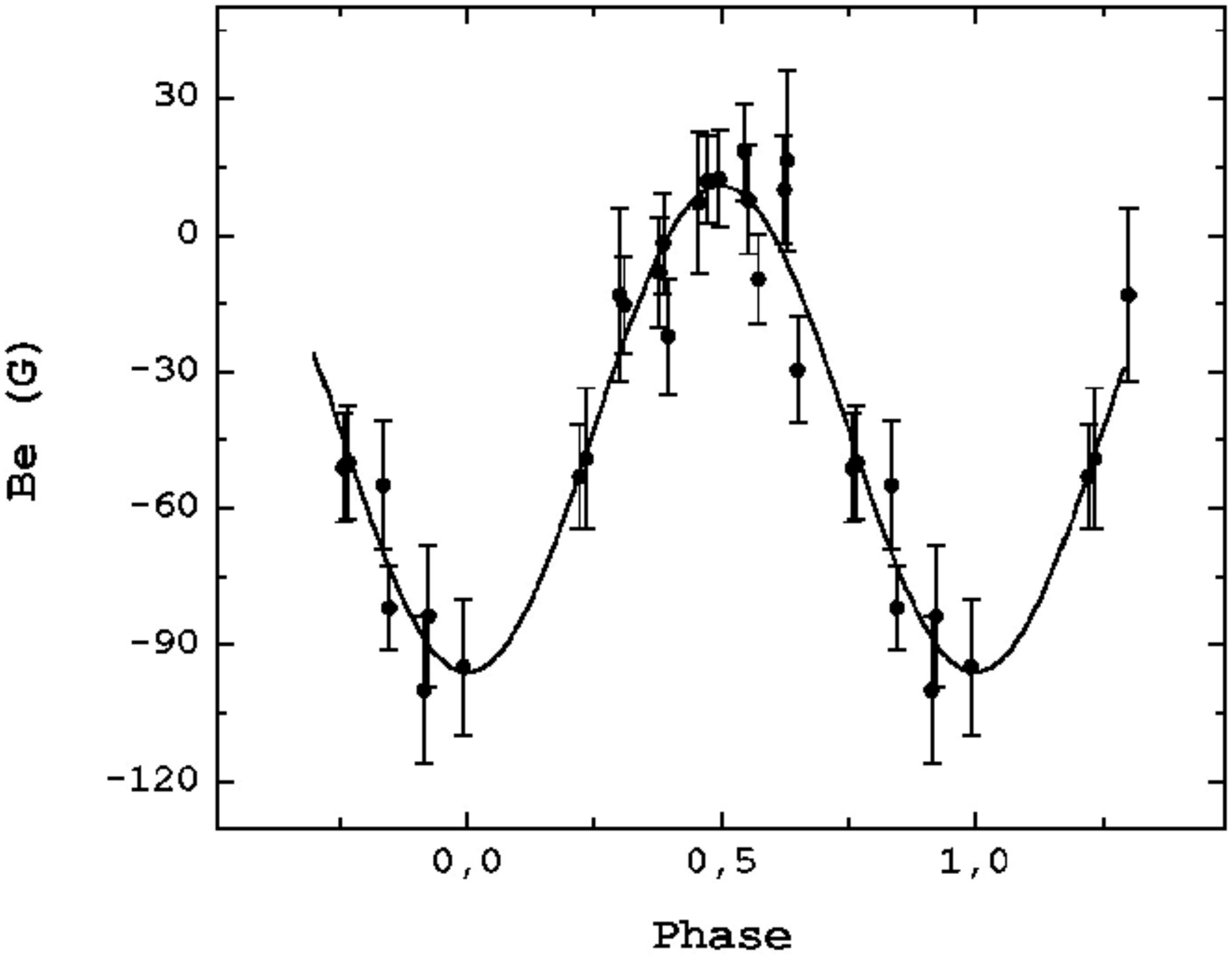}}
\vspace{-3.5mm}
\caption{ CD-40 5404 }
\label{fig:fig410}
\end{figure}

\begin{figure}
\resizebox{0.98\hsize}{!}{\includegraphics{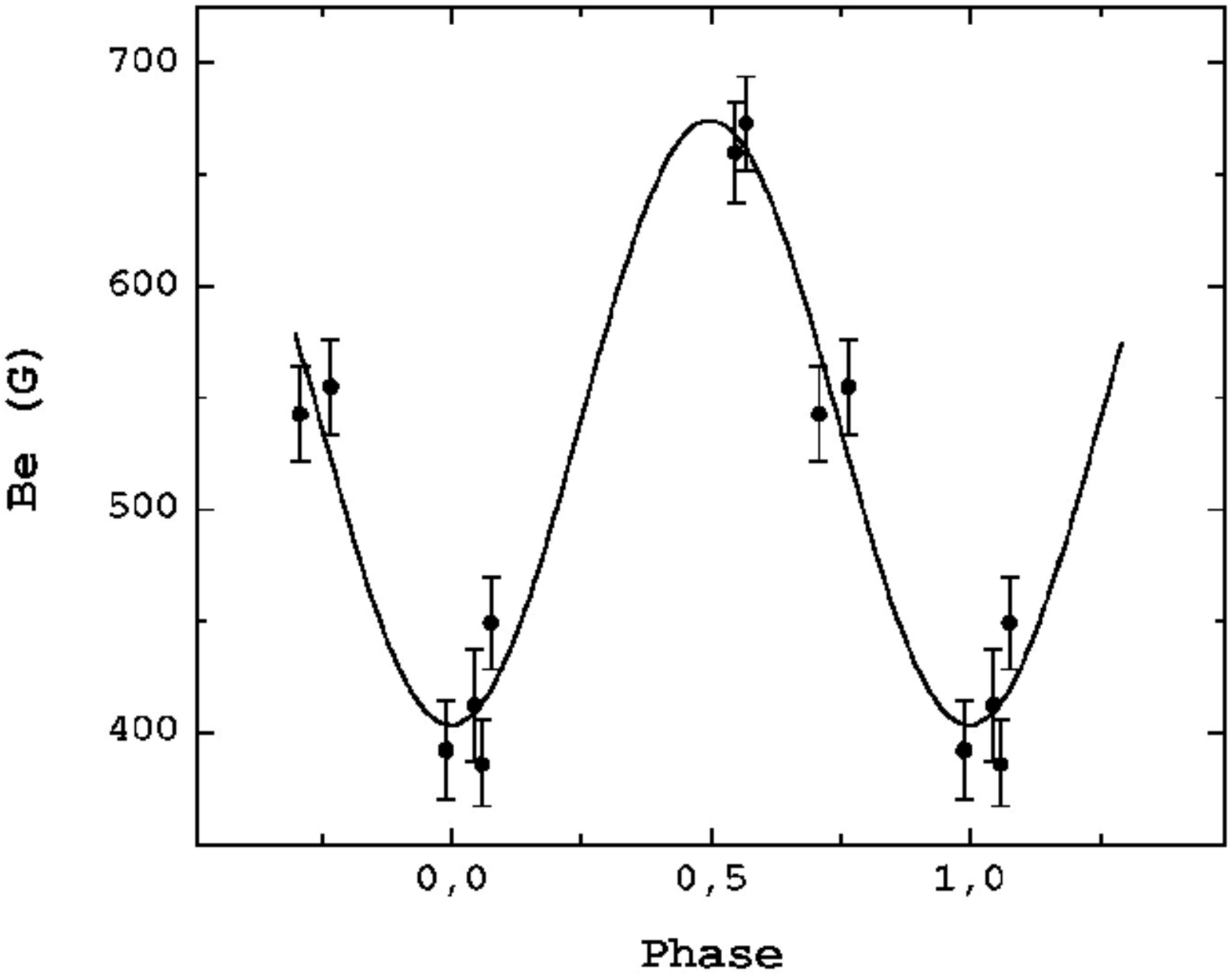}}
\vspace{-3.5mm}
\caption{ UV Cet }
\label{fig:fig411}
\end{figure}

\begin{figure}
\resizebox{0.98\hsize}{!}{\includegraphics{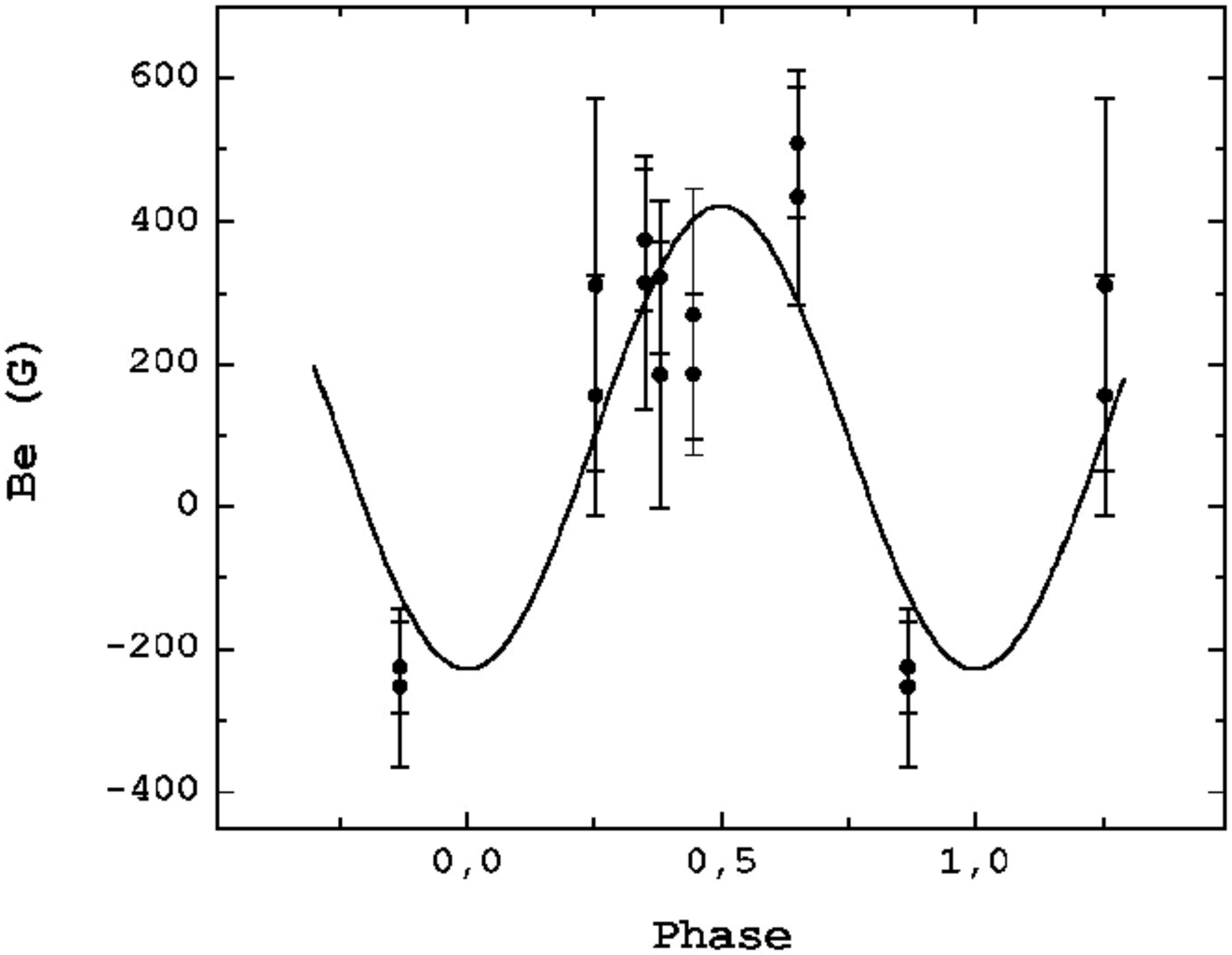}}
\vspace{-3.5mm}
\caption{ CPD-83 64B }
\label{fig:fig368}
\end{figure}

\begin{figure}
\resizebox{0.98\hsize}{!}{\includegraphics{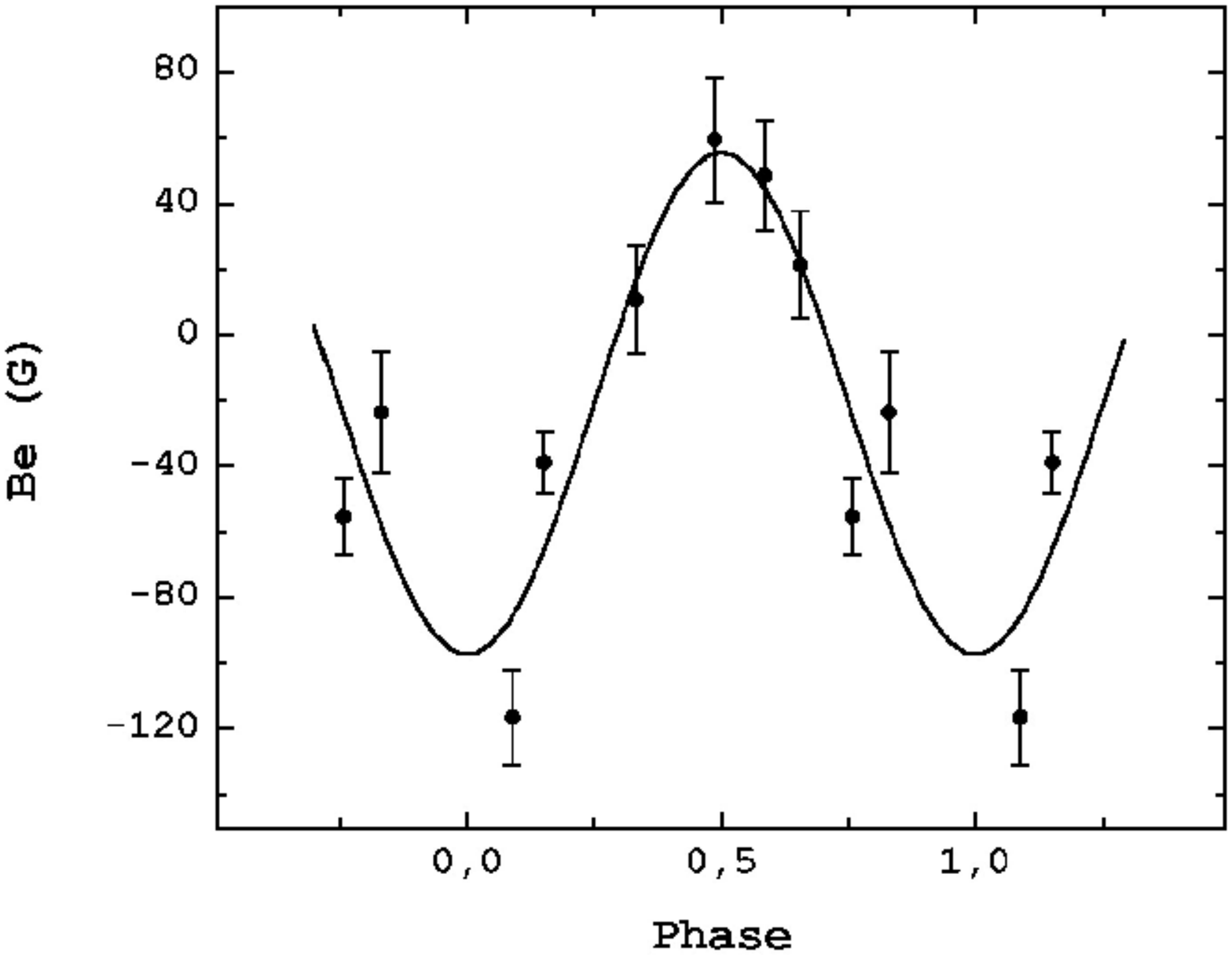}}
\vspace{-3.5mm}
\caption{ V1000 Sco }
\label{fig:fig368}
\end{figure}

\begin{figure}
\resizebox{0.98\hsize}{!}{\includegraphics{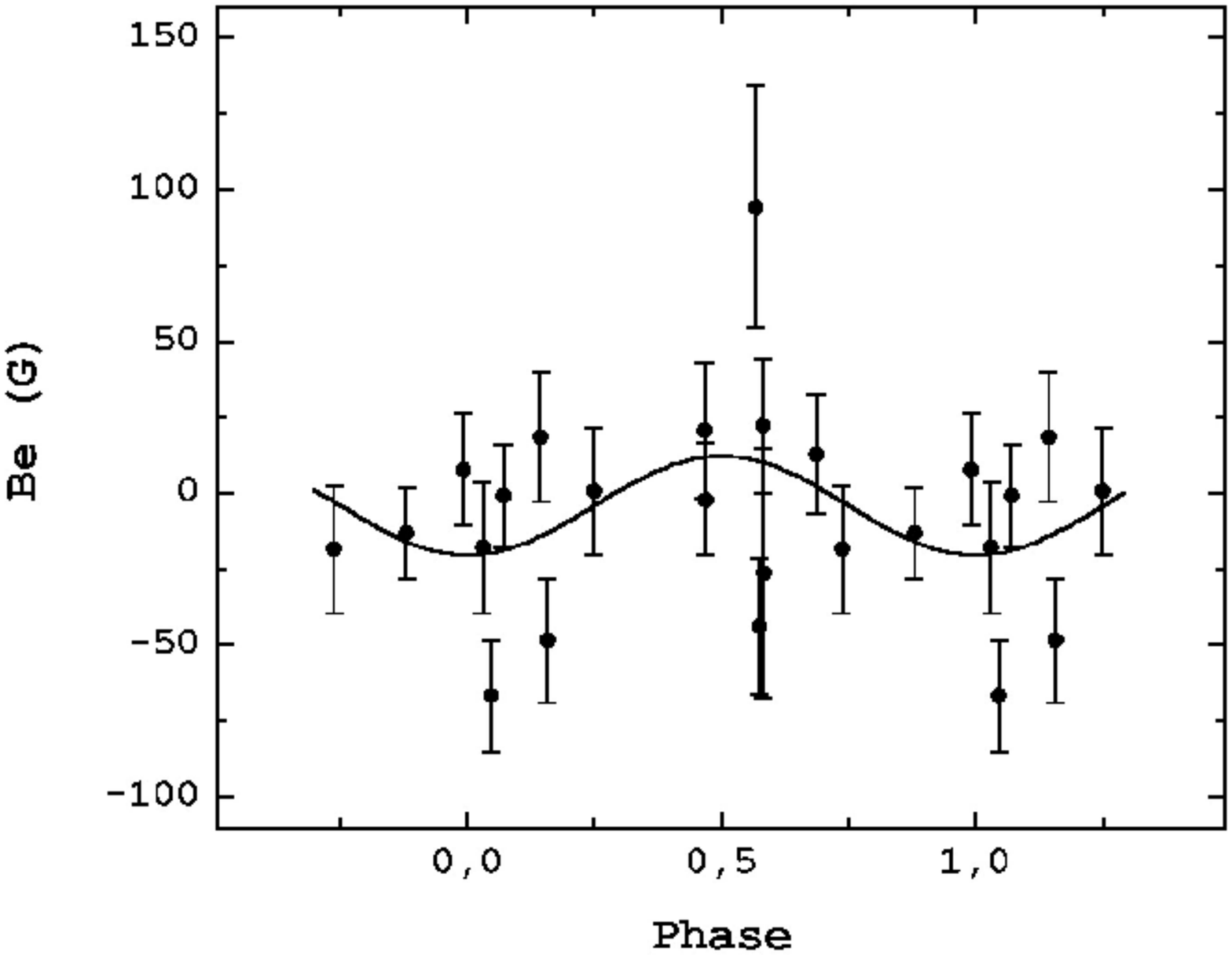}}
\vspace{-3.5mm}
\caption{ V1156 Sc0 }
\label{fig:fig368}
\end{figure}

\clearpage
\newpage

\begin{figure}
\resizebox{0.98\hsize}{!}{\includegraphics{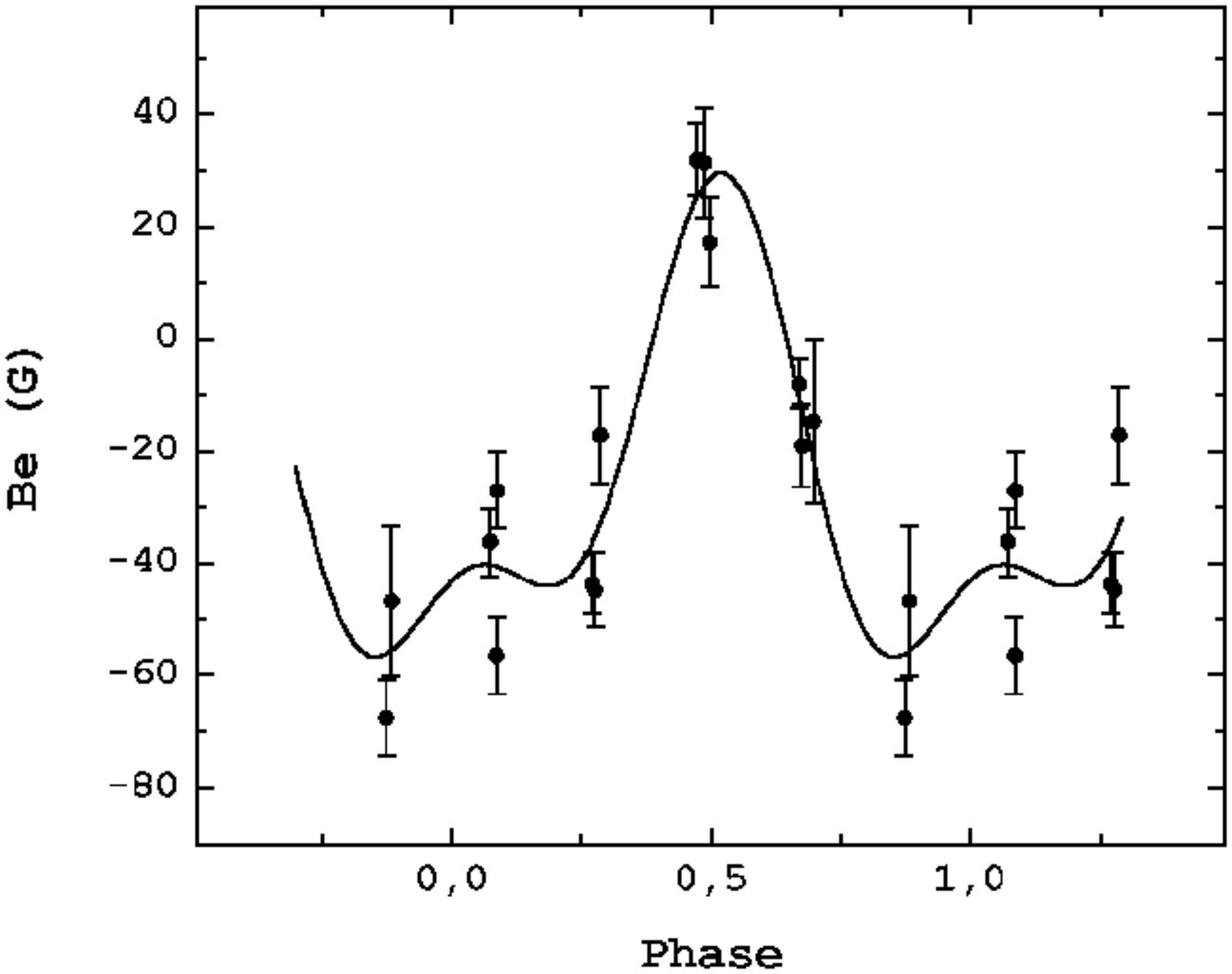}}
\vspace{-3.5mm}
\caption{ V1239 Cen }
\label{fig:fig368}
\end{figure}

\begin{figure}
\resizebox{0.98\hsize}{!}{\includegraphics{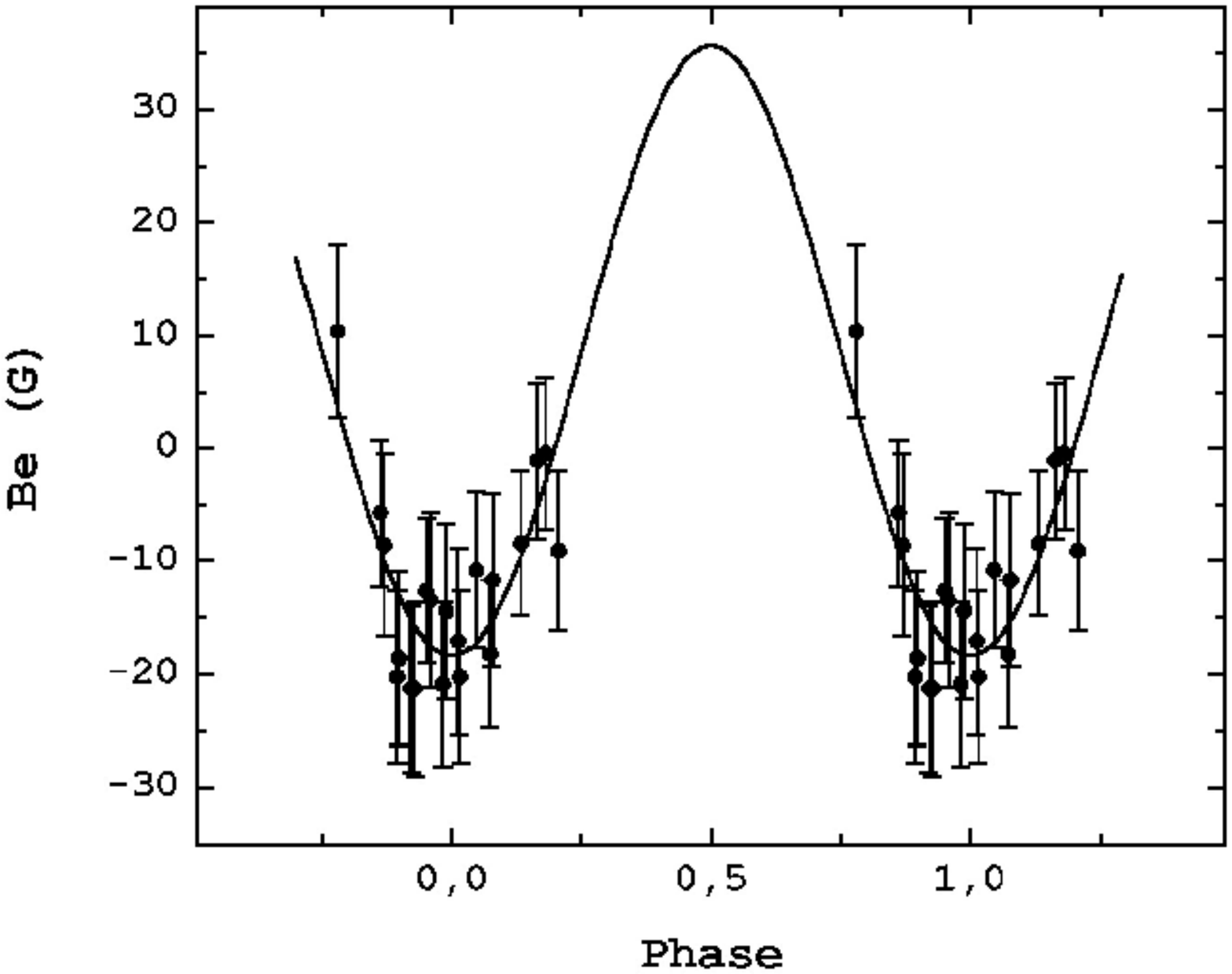}}
\vspace{-3.5mm}
\caption{ GJ 793 }
\label{fig:fig368}
\end{figure}

\clearpage
\newpage


\begin{thebibliography}{99}

\bibitem{} Alecian, E., Wade, G.A., Catala, C. et al. 2013, MNRAS, 429, 1001

\bibitem{} Borra, E.F., 1981, ApJ, 249, L39 

\bibitem{} Borra, E.F., Landstreet J.D., 1980, ApJS, 42, 421

\bibitem{} Bychkov, V.D., Bychkova, L.V., Madej, J. 2005, A\&A, 430, 1143

\bibitem{} Donati, J.F., Semel, M., Carter, B.D., Rees, D.E., Cameron, A.C.
              1997, MNRAS, 291, 658

\bibitem{} Donati, J.F., Howarth, I.D., Bouret, J.-C., Petit, P., 
              Catala, C., Landstreet, J. 2006, MNRAS, 365, L6
\bibitem{} Donati, J.F., Jardine, M.M., Gregory, S.G. et al. 2008, MNRAS, 386, 1234

\bibitem{} Landstreet J.D., Bagnulo S., Martin A., Valyavin G. 2016, A\&A, 591, A80
\bibitem{} Leone F. 2007, MNRAS, 382, 1690

\bibitem{} Preston G., 1971, ApJ, 164, 309

\bibitem{} Stibbs, D.W.N., 1950, MNRAS, 110, 395

\end{thebibliography}
\end{document}